# Ayurvedic Rasayana Restores Lipid Homeostasis in Liver: Insights from Systems Biology Studies

*Harshit B Joshi[1,4], Anupama Vallazhath[1], Priyotosh Sil[2,3], Rahul Tiwari[2,3], Krishna Raghava Hebbar[4], Swati Sharma[5], Satish Rao B S[6], Basavaraj Hadapad[4], Areejit Samal[2,3*], Manjunath B Joshi[1,*]*

[1] Department of Ageing Research, Manipal School of Life Sciences, Manipal Academy of Higher Education, Manipal, India
[2] The Institute of Mathematical Sciences (IMSc), Chennai, India
[3] Homi Bhabha National Institute (HBNI), Mumbai, India
[4] Division of Ayurveda, Centre for Integrative Medicine and Research, Manipal Academy of Higher Education, Manipal, India
[5] Department of Pathology, Kasturba Medical College, Manipal Academy of Higher Education, Manipal, India
[6] Department of Radiation Biology and Toxicology, Manipal School of Life Sciences, Manipal Academy of Higher Education, Manipal, India

*Corresponding author(s).
E-mail address: manjunath.joshi@manipal.edu (Manjunath B Joshi), asamal@imsc.res.in (Areejit Samal).

**Acknowledgments**
We thank Manipal Academy of Higher Education (MAHE), Manipal; DBT-BUILDER, Interdisciplinary Life Science Programme for Advance Research and Education (BT/1NF/22/SP43065/2021); and Fund for Improvement of S &T Infrastructure in Universities and Higher Educational Institutions (FIST- Level II-SR/FST/LS-II/2018/240) for providing infrastructure. HJ thank SATYAM, DST for providing fellowship. AS acknowledges funding from the Department of Atomic Energy (DAE), Government of India via Apex project Advanced Research and Education in Mathematical Sciences at IMSc [O.M. No. 2103/1/2025-R&D-II-DAE/IMSc/4885 dated 23.4.2026].

**Author contributions**

**Harshit B Joshi:** Investigation, Data Curation, Formal Analysis, Visualization, Writing – Original Draft. **Anupama Vallazhath:** Investigation, Data Curation. **Priyotosh Sil:** Software, Formal Analysis, Visualization, Writing – Review & Editing. **Rahul Tiwari:** Software, Formal Analysis, Visualization, Writing – Review & Editing. **Krishna Raghava Hebbar:** Resources, Investigation. **Swati Sharma:** Investigation, Histopathological Analysis, Validation. **Satish Rao B S:** Writing – Review & Editing. **Basavaraj Hadapad:** Resources, Supervision. **Areejit Samal:** Conceptualization, Methodology, Software, Formal Analysis, Supervision, Writing – Review & Editing. **Manjunath B Joshi:** Conceptualization, Resources, Supervision, Project Administration, Funding Acquisition, Writing – Review & Editing.

All authors read and approved the final manuscript.

**Conflicts of interest**
The authors declare no conflict of interest.

**Abstract**

*Background:* MASLD and obesity characterized by hepatic lipid accumulation are known risk factors to manifest into Type 2 diabetes (T2D). Increased hepatic lipids induce insulin resistance in multiple organs resulting in hyperglycaemia and such interorgan cross talk leads to dysregulation of various signalling and metabolic pathways including *de novo* lipid synthesis.

*Methods:* We evaluated two Ayurvedic formulations, Arogyavardhini Rasa (AVR) and Nishamalaki Rasayana (NA), known lipid reducing formulations to understand their mechanistic basis. Phytochemicals reported for the herbal components of the two formulations were curated from literature, and their human protein targets were compared with disease-associated genes for T2D, fatty liver disease and obesity using network medicine-based analyses. An integrated knowledge graph linking formulations, phytochemicals, protein targets, FDA-approved drugs and disease genes was constructed to enable systems-level exploration of these relationships. In parallel, male C57BL/6 mice were fed a high-fat diet to establish a prediabetic phenotype and subsequently treated with AVR or NA for 20 weeks. Multiple biochemical, histological and hepatic metabolomics analysis were performed.

*Results:* The analysis identified 128 phytochemicals in AVR and 182 in NA that met the predefined filtering criteria. The reported targets of both formulations showed significant network proximity to genes associated with all three metabolic conditions. Eight targets, namely PPARA, PPARG, AKT1, SIRT1, TNF, PPARD, SERPINE1, and ACE were common to both formulations and associated with all three disease conditions and biological pathways with disease-associated gene sets, including several targets of currently approved therapies. In the animal study, both formulations reduced high-fat-diet-associated increases in blood glucose, cholesterol, liver enzymes, macrovesicular and microvesicular hepatic steatosis. Untargeted liver metabolomics showed AVR and NA treatment restored sugar and nucleotide, pentose phosphate, bile and lipid metabolism pathways that were perturbed in response to high fat diet.

*Conclusions:* AVR and NA restored hepatic lipid homeostasis and inhibited manifestation of T2D.

## 1. Introduction

Metabolic dysfunction-associated steatotic liver disease (MASLD) is a primary contributor for the onset and progression of type 2 diabetes (T2D). Global prevalence of MASLD is 30-38% that increases risk for pathogenesis of T2D. A meta-analysis of 395 studies spanning 6,878,568 individuals across 40 countries found the prevalence of T2D among MASLD to be 28.3% (CI 25.2-31.6%), indicating significant hepatic dysfunction in T2D individuals[1]. Hyperlipidaemia in hepatic tissue leads to mitochondrial dysfunction, redox imbalance, inflammation and endoplasmic reticulum stress that contributes to hepatic insulin resistance[2]. Lipid accumulation in liver leads to insulin resistance upon impairing IRS-PI3K-AKT-FOXO1 signalling axis that maintains expression of G6PC and PCK1[3]. Diacylglycerols and ceramides impair insulin signalling pathway and activate PKC leading to reduced glucose uptake and enhanced hepatic glucose synthesis[4]. Free fatty acids (FFA) are also known to induce gluconeogenesis upon activation of PEPCK and G6Pase, rate limiting enzymes in gluconeogenesis[5]. Moreover, high fat diet feed in animal models leads to elevated PPARG levels that results in lipid accumulation in liver and subsequently NAFLD. Abrogating liver specific gene expression of PPARG reduces lipid accumulation, and consequently, decreases inflammatory response[6]. Excess fat in liver activates Kupfer cells, resident macrophages to release IL-6, IL-1β and TNF-α, and these pro-inflammatory mediators induce JNK mediated IRS1 phosphorylation and concomitant degradation, subsequently resulting in hepatic insulin resistance. Taken together, the existing mechanistic evidence indicate hyperlipidaemia significantly contributes to hepatic insulin resistance, uncontrolled gluconeogenesis, and concomitant development of T2D upon (in)activation of multiple signalling pathways.

Major limitations of current MASLD therapies include the existing drugs predominantly targeting individual components associated with a highly complex and multifactorial disease. MASLD manifests as consequence of interconnected metabolic, inflammatory, oxidative, and fibrotic pathways, and single-target pharmacological approaches often fail to provide comprehensive treatment and restore metabolic homeostasis. This limitation instigated us to explore network ethnopharmacology approach wherein multi-target therapeutic strategies can simultaneously modulate multiple pathological pathways involved in MASLD. Over the years, Ayurveda, the traditional Indian system of medicine, has demonstrated considerable potential in the effective management of metabolic disorders through its multi-targeted therapeutic approaches. However, underlying cellular and signalling mechanisms for majority of these formulations are sparsely understood. Hence, using a combination of *in vivo* validation and network medicine[7] based computational analysis, we explored two known hepatoprotective Ayurvedic formulations Arogyavardhini Rasa (AVR) and Nishamalaki Rasayana (NA) with ability to modulate metabolic and immune signalling, and often used for clinical management of MASLD. In the present study, for AVR and NA, we investigated whether the polyherbal formulations: (a) exhibit systems-level therapeutic relevance against the interconnected metabolic disorders of T2D, fatty liver disease, and obesity through their phytochemical-mediated multi-target action; (b) serve as a prophylactic drug to improve hepatic abnormalities in a prediabetic high-fat diet mouse model and inhibit progression to T2D; (c) restore lipid homeostasis in liver.

## 2. Materials and methods

### *2.1. Computational workflow integrating network medicine, network pharmacology and knowledge graph-based analyses*

Phytochemicals associated with the herbal constituents of two Ayurvedic formulations, Arogyavardhini Rasa (AVR) and Nishamalaki Rasayana (NA), were compiled from the IMPPAT database and the Ayurvedic Pharmacopoeia of India. Compounds that satisfied computational drug-likeness and bioavailability criteria were retained for downstream analyses.

Experimentally validated human protein targets of these compounds were retrieved from three different databases. Disease-associated genes for T2D, fatty liver-related disorders, and obesity were compiled by integrating information from multiple disease-gene databases with molecular targets of approved drugs for these conditions. Detailed data sources, identifiers, and filtering criteria are provided in the **SI text**.

Network proximity between formulation-derived targets and disease-related genes was assessed using the asymmetric closest-distance metric in an integrated human protein-protein interaction (PPI) network, with statistical significance evaluated using degree-aware randomization[7]. Furthermore, pathway enrichment analysis was performed using Enrichr[8]. A knowledge graph (KG) integrating formulations, medicinal plants, phytochemicals, protein targets, diseases, and approved drugs was constructed using Neo4j (https://neo4j.com/). An interactive HTML-based representation of the KG is provided as a Supplementary File (Interactive KG). The KG was used for several analyses, including quantifying target overlaps, identifying shared therapeutic targets, and exploring potential multi-component, multi-target associations for the formulations. Finally, STRING-based PPI networks were analyzed in Cytoscape to prioritize consensus hub proteins across five centrality measures. Additional methodological details, parameters, and software specifications are provided in the **SI text** (Supplementary Methods).

### *2.2. Animal study and ethical approval*

Ethical clearance was taken by the Institutional Animal Ethics Committee (IAEC), Kasturba Hospital, Manipal Academy of Higher Education (MAHE), Manipal, India (approval no. IAEC/KMC/29/2023), and all procedures followed the relevant guidelines. The experiments used male C57BL/6 mice (6–8 weeks old, 30–40 g).

### *2.3. Induction of pre-diabetic conditions in animal models and rasayana treatment*

Prediabetic condition was induced in animals upon feeding high fat diet (HFD) for two months. Chow and HFD were procured from Star Enterprises, Mangalore, India. Animals were randomly assigned to four groups; (1) Control group (n=4) fed with standard chow diet; (2) HFD group (n=4), 60% kcal fat diet; (3) HFD + AVR group (n=4); (4) HFD + NA group (n=4). Control mice were on chow diet during experimental period of six months. Diabetes group was fed with continuous HFD administration for 30 weeks. Animal groups were treated with either AVR or NA after 10 weeks of HFD upon confirming progressive obesity, hyperglycaemia, dyslipidaemia, and hepatic steatosis. Fasting blood glucose ≥ 150 mg/dL and ≥ 20% increase in body weight relative to controls confirmed the prediabetic phenotype before initiating treatment.

Following prediabetes confirmation, Rasayana treatment was initiated alongside continued HFD for 20 weeks. AVR was administered orally at 50 mg/kg/day by incorporation into HFD pellets. NA was prepared by combining powdered *Curcuma longa* rhizome with Amalaki Rasayana powder (Division of Ayurveda, CIMR, MAHE) and administered at 1000 mg/kg/day by incorporation into HFD pellets. Body weight was recorded weekly, and fasting blood glucose was measured at weeks 5, 10, 15, 20, 25, and 30 of diet exposure. Biochemical analysis to monitor liver function and histopathology of hepatics tissues were performed following standard protocols (Supplementary text 1.8 & 1.9).

### *2.4. Liver metabolomics analysis*

Once animals were sacrificed, liver was excised and subsequently perfused with saline. Frozen liver tissue (−80°C, ~50 mg) was homogenised in methanol:water (4:1, v/v) under chilled conditions, vortexed, and ultrasonicated at 60% power for 10 minutes. The homogenates were centrifuged at 14,000 rpm for 10 min at 4 °C and the supernatants retained. The pellets were re-extracted with pre-chilled solvent, and the pooled extracts were lyophilised and reconstituted in LC-MS-grade acetonitrile/water (1:1, v/v) with 0.1% formic acid. Liver extracts from the four groups were profiled by untargeted metabolomics on a Q Exactive hybrid quadrupole-Orbitrap high-resolution mass spectrometer with an ESI source (Thermo Fisher Scientific, USA). Data were acquired in full-scan mode over m/z 70–1050 with data-dependent MS/MS fragmentation. Raw files were processed in Compound Discoverer 3.1 (Thermo Fisher Scientific) for peak picking, retention-time alignment, background subtraction, and area normalisation. Annotation relied on spectral-library matching, accurate mass (<5 ppm), and isotope-pattern confirmation. Identified features were classified into HMDB chemical taxonomy classes.

Multivariate statistical analysis was performed in MetaboAnalyst 5.0. For PCA and PLS-DA analysis, metabolite intensities were $\log_{10}$-transformed and Pareto-scaled. Model quality was assessed by $R^2$ and $Q^2$ values via 10-fold cross-validation. Differential metabolites between groups were identified by volcano plot ($\log_2$ FC $\geq$ 1.5; $-\log_{10}$ p $\geq$ 1.3, t-test). Pathway enrichment analysis employed KEGG pathway maps and HMDB pathway databases with Fisher's exact test and pathway topology impact analysis.

## 3. Results

### *3.1. Compilation of phytochemicals, formulation-associated targets, and disease-associated genes*

The two formulations examined in this study consist of both herbal and non-herbal ingredients. Arogyavardhini (AVR) contains seven herbal constituents, while Nishamalaki (NA) is composed of two herbal components (**SI Table S1**). Among these, the fruit of *Phyllanthus emblica* is present in both formulations. The phytochemicals reported for each herbal ingredient were obtained from IMPPAT and API (**SI Text**; **SI Table S1**). The phytochemical profiles corresponding to each formulation were subsequently screened for computationally predicted drug-likeness and bioavailability criteria. After filtering, a total of 128 and 182 unique phytochemicals were retained for AVR and NA, respectively. The distribution of retained phytochemicals across the individual herbal constituents is provided in **SI Table S1**. Information for all retained phytochemicals, including chemical identifiers, names, and canonical SMILES strings, is listed in **SI Table S2**. We next retrieved information on human protein targets of the selected phytochemicals from published databases as described in **SI Text**. Protein target information was available for 58 phytochemicals in AVR and 66 phytochemicals in NA, corresponding to 657 and 618 unique human protein targets, respectively (**SI Table S1**). The phytochemicals with at least one reported target interaction, together with their associated targets and supporting source database(s), are listed in **SI Table S3**.

Disease-specific gene sets for T2D, fatty liver disease, and obesity were constructed by integrating FDA-approved drug targets with disease-related genes curated from several published databases (**SI Text**). For T2D, 23 approved drugs collectively mapped to 63 distinct therapeutic targets and combining it with database-derived genes resulted in 279 disease-related genes (**SI Table S4**). For fatty liver disease, eight approved drugs collectively mapped to four distinct therapeutic targets, yielding 149 disease-related genes after data integration and consensus-based selection (**SI Table S5**). Similarly, the eight approved drugs considered for obesity collectively mapped to ten distinct therapeutic targets, while integration of data from multiple sources identified 253 disease-related genes (**SI Table S6**).

### *3.2. Formulation-associated targets are proximal to disease-associated genes in the human interactome*

Network proximity analysis was performed to evaluate the systems-level association between formulation-associated targets and disease-associated gene sets within the human PPI network (**SI Text**). Both formulations exhibited statistically significant network proximity to the gene sets associated with T2D, fatty liver, and obesity (**SI Table S7**). For T2D-associated genes, the two formulations, AVR and NA, showed significant proximity with Z-scores of −5.58 and −5.78, respectively ($p < 0.001$ for both formulations). Similarly, significant proximity was observed for fatty liver-associated genes, with corresponding Z-scores of −4.05 ($p < 0.001$) and −3.72 ($p = 0.003$), respectively. Similarly for obesity, the formulations exhibited significant proximity, with Z-scores of −5.63 and −5.24, respectively ($p < 0.001$ for both formulations).

### *3.3. Functional overlap of enriched KEGG pathways between formulation-associated targets and disease genes*

To further examine the potential molecular processes associated with the therapeutic relevance of the two formulations in T2D, fatty liver disease, and obesity, we performed pathway enrichment analyses for the molecular targets derived from the formulations and the gene sets linked to the respective diseases (**SI Text**). The 20 most significantly enriched pathways from the KEGG database along with detailed enrichment statistics associated with each of the two formulations and the three disease conditions are provided in **SI Table S8**. Combining the top 20 significantly enriched pathways identified for each of the three disease conditions resulted in a nonredundant set of 33 pathways, among which nine pathways were common to all three conditions. Of the top 20 enriched pathways associated with AVR, 13 overlapped with disease-associated pathways, whereas 10 pathways associated with NA showed overlap with disease-associated pathways (**Figure 1**).

### *3.4. Knowledge graph-based integrative analysis of formulation-disease interactions in metabolic disorder*

While network proximity and pathway enrichment analyses identified systems-level associations between formulation-associated targets and disease genes, the integrated knowledge graph (KG; **SI Text**) enabled to explore connections linking the formulations and their phytochemical constituents to molecular targets, FDA-approved drugs, and disease-related genes.

#### *3.4.1. Overlap analysis of formulation-associated and disease-associated targets*

The integrated KG constructed in Neo4j was queried to investigate shared phytochemical- and target-level associations between the two formulations. The constructed KG contained 1510 nodes in total, comprising 2 formulations, 8 plants, 283 phytochemicals, 1175 distinct human proteins linked either to phytochemicals or to the disease conditions, 39 FDA-approved drugs related to T2D, fatty liver, and obesity, and 3 disease nodes representing T2D, fatty liver, and obesity. The network incorporated 4166 curated relationships, including 9 formulation-plant associations, 316 plant-phytochemical interactions, 3010 phytochemical-target associations, 681 target-disease connections, 111 drug-target connections, and 39 drug-disease links. A complete description of graph entities, their attributes, and the encoded associations is available in **SI Table S9**.

Among the phytochemicals associated with the two formulations that satisfied the drug-likeness and predicted bioavailability criteria, 27 phytochemicals were common to both formulations. Restricting the comparison to phytochemicals with available target information yielded an overlap of 19 compounds, representing less than one-third of the phytochemicals with known target associations in either formulation. In contrast, substantially greater convergence was observed at the target level, with 557 shared protein targets, corresponding to nearly 80-90% overlap between the formulation-associated target sets.

The KG was further used to examine overlap among disease-associated gene sets for T2D, fatty liver disease, and obesity **(see SI Table S10 for Cypher queries)**. The curated dataset comprised 279 genes associated with T2D, 149 associated with fatty liver disease, and 253 associated with obesity. Among these, 26 genes were common to all three conditions, while 41, 62, and 46 genes were shared between T2D-fatty liver disease, T2D-obesity, and fatty liver disease-obesity, respectively. All pairwise overlaps remained significant after Benjamini-Hochberg correction for multiple testing (FDR-adjusted p-values < 0.05[9]) **(SI Table S11**).

Significant overlap was also observed between targets of formulations and the gene sets associated with the three disease conditions (**SI Table S12).** Notably, eight targets, namely PPARA, PPARG, AKT1, SIRT1, TNF, PPARD, SERPINE1, and ACE, were common to both formulations and associated with all three disease conditions (**SI Table S13**). Among these, PPARG and PPARA are known therapeutic targets of FDA-approved drugs for T2D and fatty liver disease, respectively (**SI Tables S4-S5**). In addition, there were 13 targets shared between both formulations that were present in two of the three disease conditions (**SI Table S13**). Among these, HMGCR, INSR, and MC4R are known drug targets for fatty liver disease, T2D, and obesity respectively (**SI Tables S4-S6**).

### *3.4.2. Knowledge graph-driven investigation of shared therapeutic targets between Ayurvedic formulations and FDA-approved drugs in metabolic disorders*

The preceding analyses revealed that both AVR and NA map onto a substantial number of disease-associated proteins. Since the formulations consist of diverse phytochemical mixtures, their biological effects are likely mediated through simultaneous modulation of multiple targets and pathways. We therefore examined the extent to which formulation-associated targets coincide with the molecular targets of approved therapeutics across the three metabolic conditions. We used the KG, to identify proteins connected to both the FDA-approved drugs and formulation-derived phytochemicals (**Figure 2**). This analysis revealed that constituents of both formulations were associated with targets of several drugs spanning different pharmacological classes (see **SI Table S10** for Cypher queries). Both formulations displayed drug-combination like target coverage. All shared protein targets between the formulations and FDA-approved drugs through the intermediate constituent herbs and their associated phytochemicals are listed in **SI Table S14**. This analysis showed that both AVR and NA target multiple proteins that serve as therapeutic targets of approved drugs associated with the disease conditions. Importantly, these drugs act on distinct yet complementary aspects of disease pathophysiology.

### *3.5. PPI-guided network pharmacology analysis of shared formulation and disease targets*

Extending the KG-based overlap analyses that pinpointed disease-associated targets common to multiple formulation-disease pairs, we further characterized these relationships using network topology-based approaches (**SI Text**). Separate PPI networks based on the STRING database[10] were generated for each formulation-disease pair, enabling identification of recurrent topologically important proteins across different disease contexts. For each of the six formulation-disease combinations, we applied five centrality-based algorithms on the LCC of the PPI network (**SI Text**). For each network, a consensus hub set was established by retaining proteins ranked among the top ten by every one of the five centrality measures. Visual representations of the LCCs for each formulation-disease pair are shown in **Figure 3**, whereas the top-ranked proteins under each centrality metric, together with the derived consensus hub sets, are detailed in **SI Table S15**.

The union of all consensus hub sets across the six formulation-disease pairs resulted in a total of 12 unique proteins (**SI Table S16**). Among these, four proteins, namely PPARA, PPARG, AKT1, and TNF, were associated with all three disease conditions and were targeted by both formulations. Overall, seven proteins within the combined consensus hub sets were shared between the two formulations. SIRT1 and ESR1 were identified as common targets of both formulations in two disease contexts, specifically T2D and fatty liver disease for SIRT1, and fatty liver disease and obesity (only for NA) for ESR1.

### *3.6. Arogyavardhini Rasa and Nishamalaki Rasayana improved lipid levels and liver functions in HFD-induced prediabetic mouse models and delayed development of T2D*

Animals fed with high-fat diet (HFD) showed a progressive increase in body weight throughout the 30-week study, reaching 47.3 ± 5.5 g to the maximum compared with 37.8 ± 1.5 g in the control group. Rasayana treatment at week 10, with both AVR and NA did not allow the animals to gain weight in response to HFD. At week 30, body weight was significantly lower in the HFD+NA group (34.0 ± 7.1 g; $p = 0.026$ vs. HFD), while the HFD+AVR group reached 39.3 ± 6.5 g. Two-way ANOVA showed significant effects of treatment ($p = 0.0094$) and time ($p < 0.0001$), indicating distinct body weight trajectories among the groups (Figure 4A). HFD feeding also caused marked hyperglycemia, with fasting blood glucose increasing to 322.0 ± 22.6 mg/dl at week 30 compared with 104.0 ± 25.4 mg/dl in controls. In contrast, both Rasayana-treated groups-maintained glucose levels close to those of the control group (HFD+NA: 111.3 ± 14.2 mg/dl; HFD+AVR: 105.0 ± 14.5 mg/dl; both $p < 0.0001$ vs. HFD). A significant group-by-time interaction further confirmed the divergent glucose profiles between untreated and treated animals (Figure 4B). Animals fed with HFD significantly increased serum lipid levels. Total cholesterol increased to 158.0 ± 35.1 mg/dl compared with 75.5 ± 4.7 mg/dl in controls ($p = 0.0003$). Both treatments reduced cholesterol levels, with AVR restoring values close to control (82.5 ± 11.9 mg/dl), whereas NA also produced a significant reduction (109.5 ± 10.1 mg/dl; $p = 0.018$ vs. HFD) (Figure 4C). Triglyceride levels followed a similar trend, increasing to 235.0 ± 72.9 mg/dl in HFD mice compared with 80.7 ± 21.7 mg/dl in controls ($p = 0.0054$). AVR significantly reduced triglycerides (115.8 ± 60.3 mg/dl; $p = 0.029$), while NA showed a comparable reduction (130.6 ± 34.2 mg/dl), although this did not reach statistical significance ($p = 0.059$) (Figure 4D). HFD-induced liver injury was reflected by significant elevations in serum ALT and AST (Figures 4E and 4F). ALT increased from 21.9 ± 3.4 U/l in controls to 55.9 ± 7.3 U/l in HFD mice ($p < 0.0001$), whereas both NA (31.1 ± 3.1 U/l) and AVR (26.3 ± 6.6 U/l) significantly reduced enzyme levels. Similarly, AST increased to 110.4 ± 23.3 U/l in HFD animals compared with 43.0 ± 6.9 U/l in controls ($p = 0.0001$), and both treatments restored values close to the control range. Consistent with these findings, liver weight was significantly increased in the HFD group which were reduced upon administration of AVR and NA (Figure 4G). Although both treatments were effective, AVR consistently produced responses that were closer to control values across most metabolic and hepatic parameters. Overall, both formulations improved HFD-induced metabolic disturbances and hepatic injury.

### *3.7. Hepatic tissue metabolomics analysis revealed high fat diet significantly induced metabolic reprograming*

We next performed untargeted metabolomic profiling of liver tissue of multiple animal groups to understand influence of AVR and NA on HFD fed mouse and to mechanistically contextualize their observed prevention to develop T2D. Circular hierarchical clustering heatmap shown in Figure 5 indicated significant metabolic reprograming in response to HFD. LC-MS analysis revealed 2745 spectral features and 880 were annotated by compound discoverer. Further, endogenous metabolites were considered for analysis.

A total of 883 metabolite features were annotated across amino acid, lipid, nucleotide, carbohydrate, and cofactors. Partial least squares-discriminant analysis (PLS-DA) separated the four groups, with components 1 and 2 explaining 10.6% and 14.2% of the variance, respectively, and gave distinct clusters for Control, HFD, HFD+NA, and HFD+AVR (Figure S1A). The HFD group occupied a metabolic space clearly distinct from that of the Control group, confirming that chronic dietary fat exposure induced substantial perturbations in hepatic metabolism. Importantly, both Rasayana-treated groups aligned to distinct metabolic territories and demonstrated directional divergence from the untreated HFD cluster, with partial realignment toward the Control phenotype, indicating intervention-driven metabolic remodelling. Comparison of the Control and HFD groups revealed a separation, with Component 1 and Component 2 explaining 21.7% and 24.4% of the variance, respectively (Figure S1B), confirming a profound HFD-induced disruption of hepatic metabolism. At the metabolite level, 156 features were significantly upregulated and 144 were significantly downregulated ($|\log_2 FC| > 1$) in HFD liver relative to controls. Pairwise PLS-DA comparison of HFD and HFD+NA groups demonstrated clear metabolic discrimination (Component 1 = 19.5%, Component 2 = 25.5%; Figure S1C), with 183 metabolites upregulated and 175 downregulated relative to untreated HFD animals ($|\log_2 FC| > 1$). Similarly, a separation was observed between HFD and HFD+AVR groups (Component 1 = 22.1%, Component 2 = 25.5%; Figure S1D), with 209 metabolites upregulated and 162 downregulated.

To identify the liver specific biological pathways perturbed by chronic HFD feeding and further modulated by Rasayana interventions, metabolite set enrichment analysis (MSEA) was performed using MetaboAnalyst 6.0. Differentially regulated endogenous metabolites identified from volcano plots (Figure S2) based on a $\log_2$ fold change threshold of $|\log_2 FC| > 1$ were separately considered for pathway enrichment analysis. Significantly modulated metabolites in response to HFD and further treatment with NA and AVR along with biological implications have been provided in Table 1. Upregulated and downregulated metabolite lists were analyzed independently to distinguish activated and suppressed pathways (Figure 6). Among the 156 metabolites increased in HFD liver relative to controls, the most significantly enriched pathways included purine metabolism (enrichment ratio ~7, $p < 8\times10^{-5}$), folate biosynthesis (~13), one carbon pool by folate (~9), pantothenate and CoA biosynthesis and riboflavin metabolism. Metabolites associated with biosynthesis of unsaturated fatty acids, steroid hormone biosynthesis, butanoate metabolism, nicotinate and nicotinamide metabolism, citrate cycle (TCA cycle), propanoate metabolism, alanine/aspartate/glutamate metabolism, glutathione metabolism, pyrimidine metabolism, tryptophan metabolism, and primary bile acid biosynthesis were also elevated. Among the 144 decreased metabolites in HFD liver, the most significantly suppressed pathways included amino sugar and nucleotide sugar metabolism (enrichment ratio ~12, $p < 3\times10^{-3}$), pyrimidine metabolism (~11), pentose phosphate pathway (~10), purine metabolism (~9), and biosynthesis of various nucleotide sugars (~8).

### *3.8. AVR and NA treatment improved liver metabolism upon modulating sugar nucleotide metabolism and pentose phosphate pathways*

Total of 183 metabolites were upregulated by NA treatment relative to HFD. Strikingly, the top upregulated pathways in HFD+NA were amino sugar and nucleotide sugar metabolism (enrichment ratio ~25, $p < 1\times10^{-4}$) and pentose phosphate pathway (~20) which were severely suppressed pathways in HFD liver, strongly indicating that NA treatment restores HFD-impaired glycosylation, sugar nucleotide metabolism, and the antioxidant/ribose regeneration capacity of the pentose phosphate pathway. Metabolites aligning to pyrimidine metabolism, biosynthesis of various nucleotide sugars, galactose metabolism, fatty acid elongation, pantothenate and CoA biosynthesis, starch and sucrose metabolism, pentose and glucuronate interconversions, fructose and mannose metabolism, pyruvate metabolism, glycosylphosphatidylinositol (GPI) anchor biosynthesis, N-glycan biosynthesis, thiamine metabolism, sulfur metabolism, vitamin B6 metabolism, ascorbate and aldarate metabolism, primary bile acid biosynthesis, and fatty acid biosynthesis were also elevated in response to NA. Among the 175 metabolites downregulated by NA versus HFD, enriched pathways included purine metabolism (enrichment ratio ~25, $p < 1\times10^{-3}$), riboflavin metabolism (~25+, highest enrichment ratio across all comparisons), vitamin B5 metabolism, fatty acid degradation, pantothenate and CoA biosynthesis, phenylalanine/tyrosine/tryptophan metabolism, linoleic acid metabolism, lysine degradation, thiamine metabolism, phenylalanine metabolism, butanoate metabolism, glycerolipid metabolism, retinol metabolism and citrate cycle.

Among the 209 metabolites upregulated by AVR treatment relative to HFD, the most significantly enriched pathways were amino sugar and nucleotide sugar metabolism (enrichment ratio ~8.5, $p < 8 \times 10^{-6}$ the most statistically significant hit across all analyses), biosynthesis of various nucleotide sugars (~8), purine elongation, fructose and mannose metabolism, biosynthesis of unsaturated fatty acids, pantothenate and CoA biosynthesis, pyrimidine metabolism, fatty acid degradation, vitamin B6 metabolism, one carbon pool by folate, ascorbate and aldarate metabolism. Total of 162 metabolites downregulated by AVR versus HFD, the most enriched pathways included riboflavin metabolism (enrichment ratio ~40, $p < 8 \times 10^{-4}$ — the single highest enrichment ratio observed across all comparisons and groups), purine metabolism (~20), lysine degradation, phenylalanine/tyrosine/tryptophan metabolism (~15), linoleic acid metabolism, fatty acid degradation and pyruvate metabolism.

### *3.9. AVR and NA differentially reprogram lipid metabolism*

Metabolomics analysis showed multiple pathologically important lipid species were modulated by both AVR and NA (Table 2). Reduction of lipids in response to HFD such as UDP-N-acetylglucosamine, Phosphatidylcholine, Glycocholic acid, Palmitic acid, 2-docosahexaenoyl glycerol, platelet-activating factor, all-*cis*-4,7,10,13,16-docosapentaenoic acid, octanoyl-coenzyme A, cholic acid, 1,2-dioleoyl-sn-glycero-3-phospho-(1′-sn-glycerol-3′-phosphate) and 1-oleoyl-2-linoleoyl-sn-glycero-3-phosphocholine were restored in animals treated with AVR and NA.

Interestingly, AVR and NA reduced levels of pathologically important lipids which were increased in HFD animals. Phosphatidylinositol-1,2-dipalmitoyl, Docosahexaenoyl-phosphatidylserine, 1-[(9Z)-octadecenoyl]-2-tetradecanoyl-sn-glycero-3-phosphocholine, 3-(palmitoyloxy)-1,2-propanediyl bis-linoleate, Palmitoyl sphingomyelin, Prostaglandin K2, (S)-3-hydroxyhexanoyl-CoA, cyclohexanoyl-CoA, 2,3-dihydroxy-2,3-dihydrobenzoyl-CoA, 13,14-dihydro-19(R)-hydroxy prostaglandin E1, (6R,8Z)-6-hydroxy-3-oxo-8-tetradecenoic acid, O-propenoyl-D-carnitine, 3-hydroxytetradecanedioic acid, eicosatetraynoic acid, 11β-prostaglandin F2α, icosapentaenoic-acid-linked phospho-species 6β-hydroxytestosterone, 2-arachidonyl glycerol ether, crotonyl-CoA, 15-hydroperoxyeicosatetraenoic acid, linoleic acid, thromboxane B1 and allopregnanolone were decreased in AVR and NA treated animals.

Taurocholic acid, arachidonoyl amide, docosahexaenoic acid, stearic acid, docosahexaenoic acid methyl ester, ricinoleic acid, arachidonic acid and 16-hydroxyhexadecanoic acid, stearic acid, (20S)-20-hydroxypregn-4-en-3-one, threo-sphingosine and 1-(sn-glycero-3-phospho)-1D-myo-inositol and N-tridecanoylglycine were selectively elevated in response to AVR. Simultaneously, AVR uniquely downregulated 26 metabolites including 1-palmitoyl-2-[(9Z,12Z,15Z)-octadecatrienoyl]-sn-glycero-3-phosphocholine, 1-palmitoyl-2-oleoyl-sn-glycero-3-phosphocholine, 1-icosanoyl-sn-glycero-3-phosphoethanolamine, 2-methylbutyrylcarnitine, estradiol disulfate, 2-hydroxypropanoyl-CoA, ganglioside GM1 (18:1/20:0), palmitelaidic acid, (S)-3-hydroxyoctanoyl-CoA, 11-dehydro thromboxane B2 docosahexaenoyl-glycerol-phosphocholine species and myristoleic acid.

Accordingly, several lipids were selectively increased by NA including 2-Methoxyestrone Glycerophosphoglycerol, 1,3-dieicosapentaenoyl glycerol, glycerophospho-N-palmitoyl ethanolamine,1-eicosyl-2-docosanoyl-sn-glycero-3-phosphocholine, taurochenodeoxycholic acid sodium salt, 1-oleoyl lysophosphatidic acid and hydroxyhexanoylcarnitine. Lipids such as 11,12-epoxyeicosatrienoic acid, 8,9-dihydroxyeicosatrienoic acid, 1-[(8Z,11Z,14Z)-icosatrienoyl]-glycerophosphocholine, 11,12-dihydroxyeicosatrienoic acid (−1.52), 5,6-dihydroxyeicosatrienoic acid (−1.41), 10,11-epoxydocosahexaenoic acid were only reduced in response to NA but not in AVR treated animals.

## 4. Discussion

In the present study, we explored a systems biology approach, integrating network pharmacology and untargeted metabolomics, to investigate the mechanistic basis of Arogyavardhini Rasa (AVR) and Nishamalaki Rasayana (NA) mediated delay in T2D manifestation in a high-fat-diet model. We demonstrate: (a) AVR and NA contain phytochemicals with therapeutic potential that target signalling mediators regulating lipid signalling and subsequent manifestation of T2D, obesity and MASLD; (b) These formulations when treated in pre-diabetic conditions delayed progression of HFD induced diabetes; (c) AVR and NA restore metabolic homeostasis in liver upon altering sugar nucleotide metabolism and pentose phosphate pathways.

Network pharmacology has emerged as the standard framework for mechanistic reconstruction of drug action[21], and the framework has been applied extensively to explore traditional polyherbal formulations whose therapeutic effects can be understood via the combined, multi-pathway action of many constituents rather than one active ingredient[22,23]. The molecular targets and metabolic pathways mediating the known glycaemic- and lipid-lowering effects of AVR and NA had not been systematically mapped for either formulation. By integrating a network pharmacology approach (that predicts the bioactive constituents of each formulation and their putative signalling targets) with untargeted hepatic metabolomics in a chronic HFD mouse model, we sought to connect the phenotypic outcomes of these formulations to the specific metabolic pathways each engages at the tissue level.

In the network proximity analysis, the consistently negative Z-scores together with significant p-values indicate that the formulation-associated targets are positioned significantly closer to disease-associated genes in the interactome than expected by chance. Pathway enrichment analysis revealed substantial functional convergence among the metabolic disorders, highlighting shared molecular mechanisms underlying T2D, fatty liver disease, and obesity. Among the significantly enriched pathways were the non-alcoholic fatty liver disease (NAFLD) and Insulin resistance pathways, both of which are closely linked to the interconnected pathophysiology of these conditions. The NAFLD pathway reflects hepatic metabolic dysfunction while also exhibiting strong mechanistic associations with T2D and obesity[24]. In addition, the Insulin resistance pathway disrupts insulin signalling and contributes to disease progression through systemic metabolic dysregulation[16].

The significant intersection between formulation-associated and disease-related enriched pathways, indicates the plausible involvement of the formulations in biological processes relevant to the disease conditions. AVR showed overlap with alcoholic liver disease, non-alcoholic fatty liver disease (NAFLD), and insulin resistance pathways, suggesting associations with pathways involved in hepatic dysfunction and metabolic regulation. In contrast, NA exhibited enrichment in pathways such as AGE-RAGE signalling and FOXO signalling, which are associated with inflammation, oxidative stress, and metabolic dysregulation[25]. Together, these regulatory nodes are instrumental in mitigating the chronic inflammatory milieu and oxidative stress that structurally link the pathogenesis of T2D, fatty liver, and obesity.

The KG-based overlap and network pharmacology analyses identified several biologically relevant targets associated with the three metabolic disorders. Among these, Peroxisome proliferator-activated receptor gamma (PPARG) is a ligand-dependent nuclear receptor that regulates adipocyte differentiation, lipid metabolic processes, and glucose homeostasis[26]. Similarly, PPARA, which is highly expressed in the liver, is involved in the regulation of metabolism of fatty acid and hepatic lipid homeostasis[27]. Earlier studies have likewise demonstrated that PPARA activity is associated with reduced hepatic insulin resistance through modulation of diacylglycerol accumulation[28]. Furthermore, AKT1 emerged as another important regulatory node associated with metabolic disorders. AKT1 promotes cellular glucose uptake by facilitating the translocation of GLUT4 to the plasma membrane[29]. In addition, AKT1 regulates FOXO activity through phosphorylation-dependent cytoplasmic translocation, thereby suppressing gluconeogenesis and contributing to glucose homeostasis[30]. Another recurrent target identified in our analyses was SIRT1, a NAD+-dependent metabolic regulator whose activity is closely linked to cellular energy status[31]. Through sensing changes in nutrient availability, SIRT1 has been implicated in the regulation of lipid metabolism and inflammatory pathways associated with metabolic disorders[32].

Herbal formulations containing diverse phytochemicals demonstrated associations with multiple disease-related targets. Several of these targets are also modulated by FDA-approved drugs, indicating that the collective phytochemical composition may influence biologically and clinically relevant pathways. For instance, AVR formulation contains phytochemicals that are associated with targets such as GLP1R, DPP4, PPARG, MGAM, and INSR, all of which are established therapeutic drug targets. The observed multi-target coverage also shares similarities with clinically administered combination therapies, such as the Metformin-Sitagliptin combination[33], which collectively influences insulin resistance and insulin secretion through INSR- and DPP4-associated mechanisms. A comparable target association pattern was observed for AVR, where phytochemicals such as Ellagic acid and Gallic acid were linked to INSR and DPP4, respectively. Similarly, the clinically used Metformin-Pioglitazone combination[34] targeting PPARG-mediated metabolic regulation showed overlap with phytochemical-target associations involving compounds such as Ellagic acid and Quercetin. Collectively, these observations suggest that AVR may influence multiple disease-associated processes, including insulin resistance, insulin secretion, and adipose tissue regulation, through coordinated multi-target interactions. A similar pattern of overlap with clinically relevant targets was also observed for the Nishamalaki formulation.

Beyond these systems-level associations, both AVR and NA formulations used in this study have independently demonstrated experimental and clinical evidence of efficient glycaemic control and improved hepatic/lipid metabolism. NA has been evaluated in multiple clinical and preclinical trials for T2D and reported effects include reductions in fasting and postprandial blood glucose, improved HbA1c, favorable shifts in lipid profile, enhanced antioxidant enzyme activity, and inhibition of carbohydrate-digesting enzymes such as α-amylase and α-glucosidase[35,36], with comparative trials against standard agents like metformin and glyburide showed comparable efficacy and minimal adverse effects[37]. AVR a herbomineral formulation containing Katuki, Triphala, Guggulu, Shilajatu, and purified mineral bhasmas, have been used for the treatment of hepatic and lipid disorders. A prospective pilot cohort study in dyslipidemic patients reported significant reductions in total cholesterol, LDL, triglycerides, and CRP, with a corresponding rise in HDL, when AVR was given following Arjuna[38]. In Triton WR-1339-induced hyperlipidemic rats, AVR dose-dependently lowered cholesterol, triglycerides and LDL while raising HDL, with efficacy comparable to fenofibrate. AVR also reduced hepatic malondialdehyde while elevating glutathione indicating AVR as a potential antioxidant along with lowering lipids[35]. Recent studies have included AVR's evaluation to NAFLD in terms of Ayurveda, framing its actions in classical terms: correcting impaired Agni, eliminating Ama, stimulating bile secretion, improving lipid metabolism, and protecting hepatocytes from oxidative injury. These actions map onto measurable biochemical endpoints such as transaminases, bilirubin, and lipid panels in modern clinical evaluation[39].

The metabolomic signatures reframes the clinical observations mechanistically: AVR's established hepatoprotective and hypolipidemic effects are reflected in its stronger normalization of phosphatidylcholine[15,40], bile-acid[11], and steroid-hormone pathways. NA's established antihyperglycemic and antioxidant effects are reflected here in its restoration of amino-sugar/nucleotide-sugar metabolism, the pentose phosphate pathway, and suppression of eicosanoid[20] and endocannabinoid[41] signalling. This convergence between clinical/preclinical phenotypic data and the present pathway-resolved metabolomic evidence corroborates the network pharmacology analyses described above.

The present study possesses limitations. The experimental design in the present study captured only a single treatment endpoint and the sustained effects in metabolic improvements in response to AVR and NA withdrawal were not assessed. A dedicated relapse study with a treatment-cessation arm may be required to determine this and further studies are warranted. While network pharmacology and metabolomics provide a systems-level, hypothesis-generating view of the pathways affected by AVR and NA, these observations are inherently correlative rather than causal. Further, it would be interesting to examine influence of AVR and NA on restorative effects in other metabolically relevant organs, pancreatic islets, skeletal muscle, adipose tissue, and kidney that are equally implicated in the pathophysiology T2D.

**5. Conclusions:** Upon combining network medicine and in vivo validation along with hepatic metabolomics analysis, this study provides insights into the likely mechanisms regulating the metabolic benefits of Arogyavardhini Rasa (AVR) and Nishamalaki Rasayana (NA). Both formulations lay close to T2D, fatty liver, and obesity genes in the interactome, pointing to shared targets in metabolic regulation. In HFD-induced prediabetic mice, both improved glycemic control, serum lipids, liver enzymes, and liver histology. Metabolomics analysis showed remodeling of sugar and nucleotide metabolism, the pentose phosphate pathway, bile acid metabolism, and lipid homeostasis. Together, the data suggest that AVR and NA may counter HFD-induced metabolic dysfunction through coordinated multi-target effects on hepatic metabolism.

**Figures:**

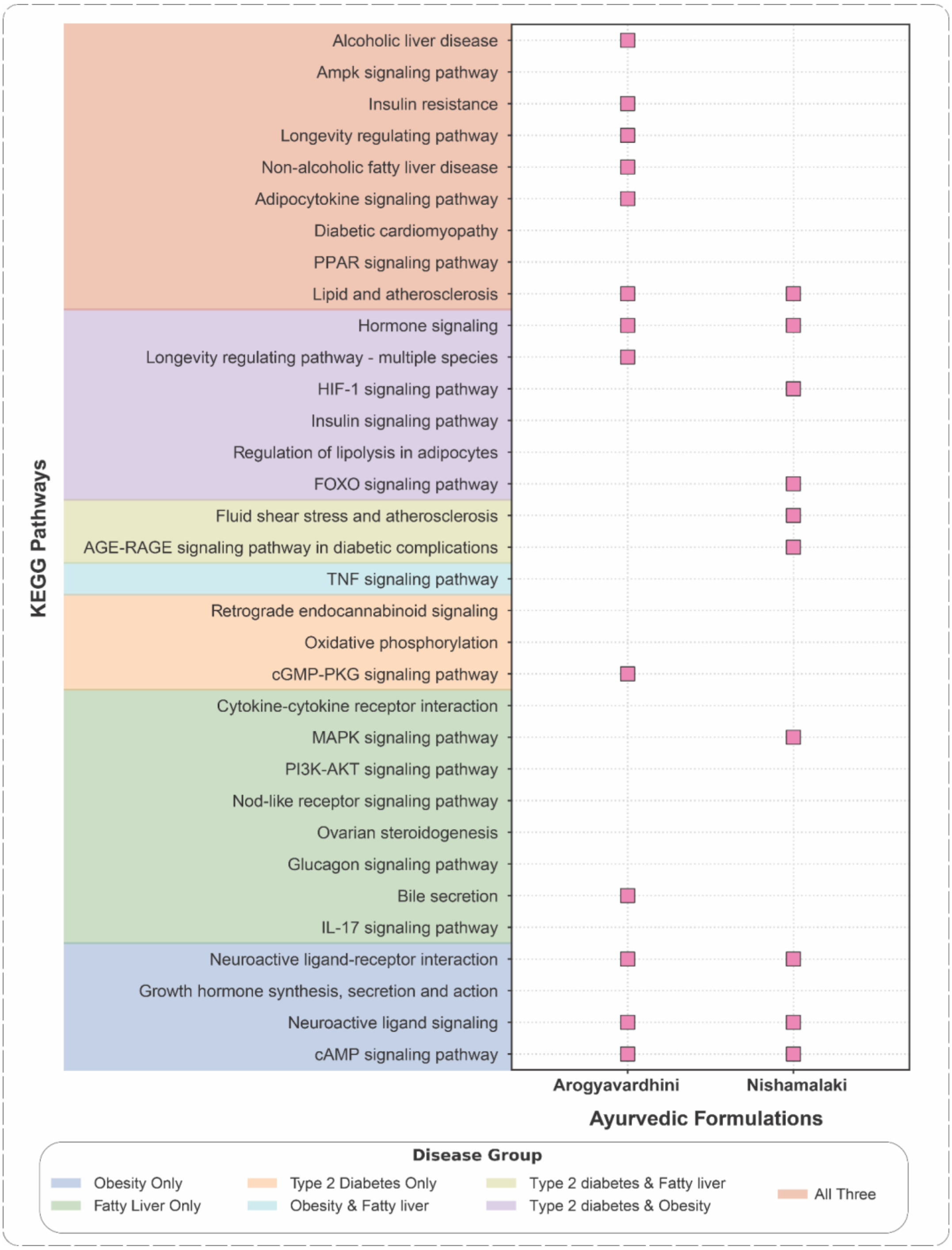

**Figure 1. Overlap of enriched KEGG pathways between formulation-derived targets and disease-related gene sets.** The vertical axis lists the 33 distinct pathways obtained by consolidating the 20 highest-ranked enriched pathways for each of three disease conditions (type 2 diabetes, fatty liver disease, and obesity) using the curated disease gene sets. Pathways common to all three diseases, shared between disease pairs, or specific to individual conditions are distinguished using different color annotations as indicated in the legend. The horizontal axis represents the two formulations, where filled squares indicate pathways that are included in the top 20 enriched pathways for the corresponding formulation. This figure shows a considerable functional convergence between the pathways associated with the formulations and the disease conditions.

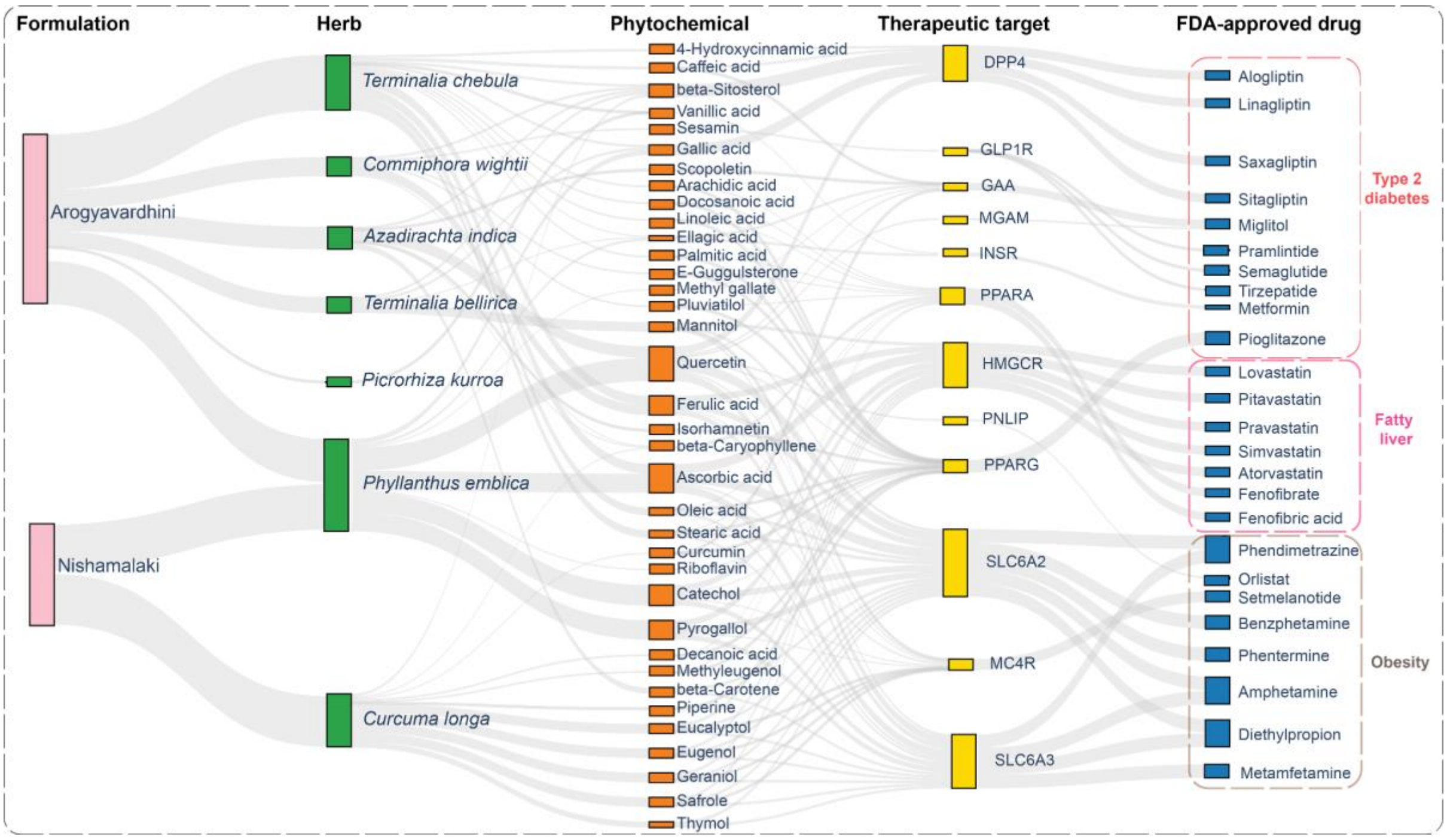

**Figure 2. Sankey diagram showing the connections among the two formulations, their constituent plants, associated phytochemical constituents, protein targets, and approved drugs.** Links between the formulations and approved drugs are established through overlapping protein targets, illustrating coverage of multiple, potentially complementary targets, overlapping target profiles, and possible mechanistic parallels between the formulations and multidrug regimens involving clinically administered drugs.

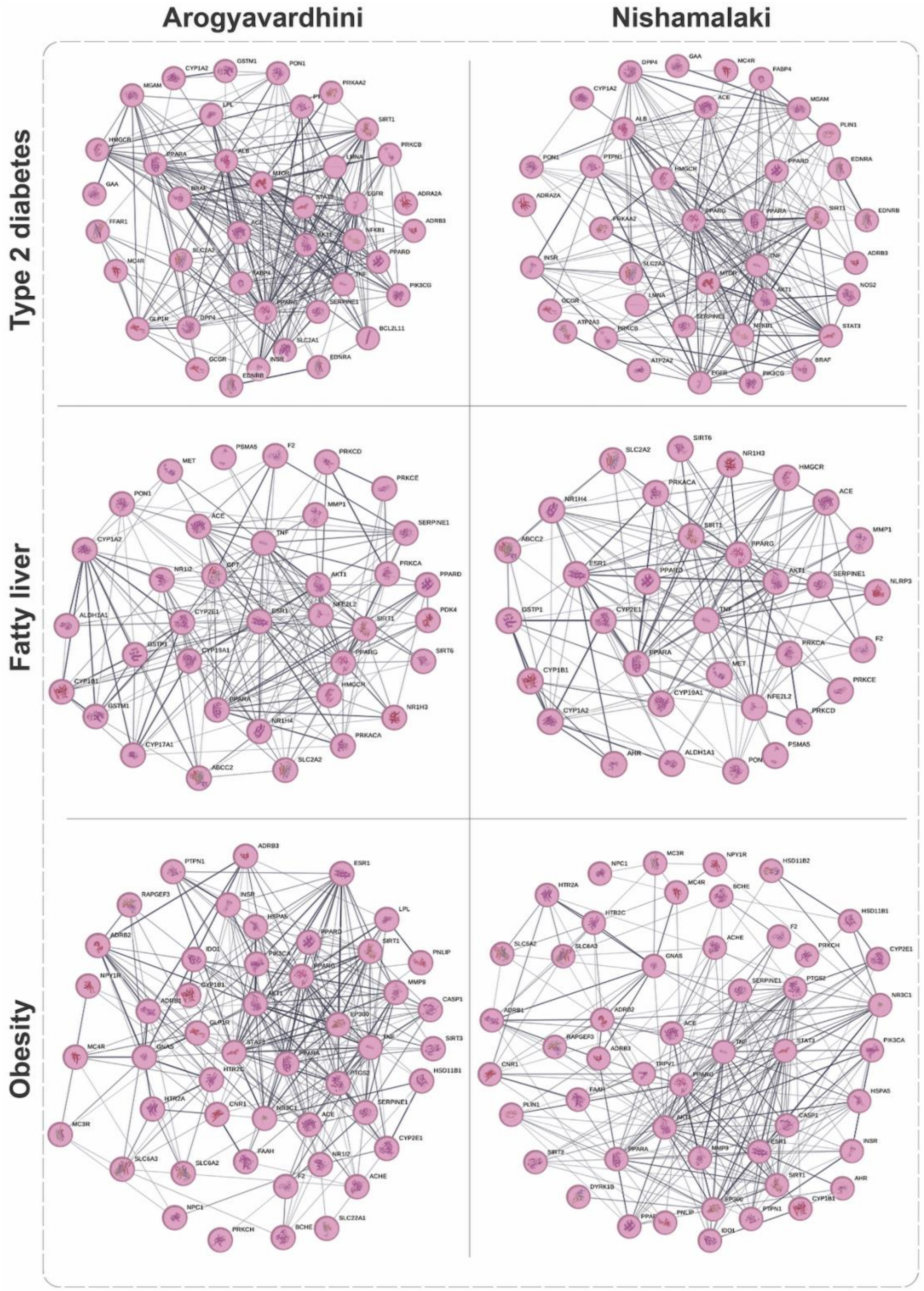

**Figure 3. Network representation of the largest connected components (LCCs) extracted from protein-protein interaction (PPI) networks constructed using overlapping formulation-associated and disease-associated targets.** The displayed networks correspond to proteins common between the gene sets associated with the diseases and targets modulated by the two formulations (Arogyavardhini and Nishamalaki). The top, middle, and bottom panels represent the LCCs corresponding to type 2 diabetes, fatty liver, and obesity, respectively.

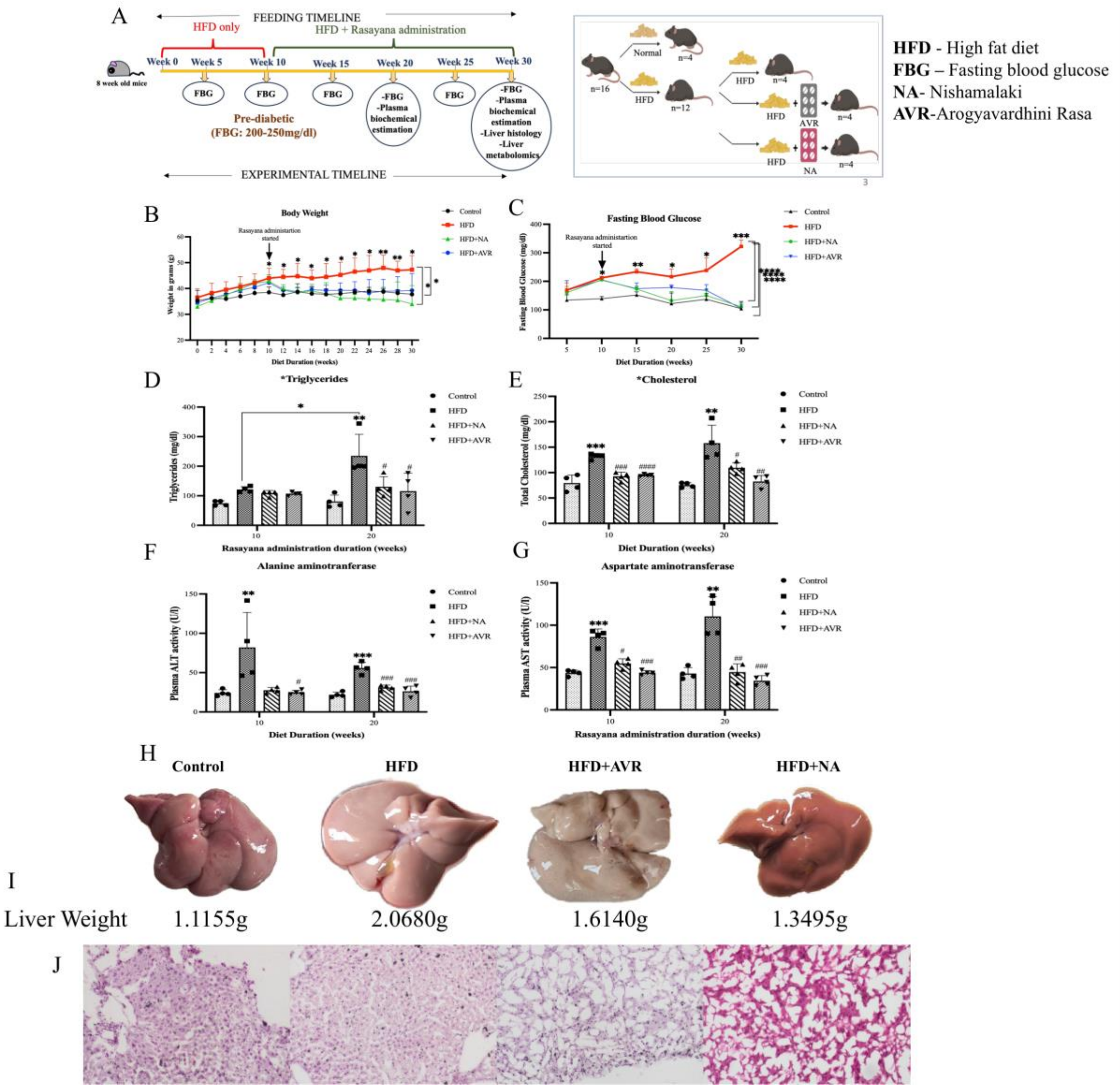

**Figure 4. Effects of Nishamalaki Rasayana (NA) and Arogyavardhini Rasa (AVR) on metabolic parameters, liver function, and hepatic histopathology in high-fat diet (HFD)-induced prediabetic mice. (A)** Study design and feeding timeline. Mice were fed HFD alone for the first 10 weeks to establish a pre-diabetic state (fasting blood glucose 200–250 mg/dL), after which Rasayana administration (AVR or NA) began and continued alongside HFD feeding through week 30. Fasting blood glucose (FBG) was monitored every 5 weeks; plasma biochemistry was assessed at weeks 20 and 30; liver histology and metabolomics were performed at week 30. Group allocation: Control (n = 4); HFD (n = 12), subsequently split into HFD-only (n = 4), HFD+AVR (n = 4), and HFD+NA (n = 4). **(B)** Longitudinal body weight across the 30-week diet duration for Control, HFD, HFD+NA, and HFD+AVR groups. Arrow indicates initiation of Rasayana administration at week 10. **(C)** Fasting blood glucose measured at 5-weekly intervals across the 30-week diet duration. Arrow indicates initiation of Rasayana administration. **(D)** Plasma triglyceride concentrations at weeks 10 and 20 of Rasayana administration. **(E)** Total plasma cholesterol concentrations at weeks 10 and 20. **(F)** Plasma alanine aminotransferase (ALT) activity at weeks 10 and 20, as an index of hepatocellular injury. **(G)** Plasma aspartate aminotransferase (AST) activity at weeks 10 and 20. **(H)** Representative gross liver morphology at sacrifice (Control, HFD, HFD+AVR, HFD+NA). **(I)** Terminal liver weights (g) corresponding to the livers shown in (H). **(J)** Representative haematoxylin and eosin

(H&E)-stained liver sections, low power, illustrating overall hepatic architecture across groups. **(K)** Microvesicular steatosis grade (0–3, corresponding to <5%, 5–33%, 34–66%, and >66% of hepatocytes affected) across groups. **(L)** Macrovesicular steatosis grade, scored on the same 0–3 band system as (K). Data in (B)–(G) are presented as mean ± SEM; significance was determined by one-way ANOVA with Tukey's multiple comparisons test (*$p < 0.05$, **$p < 0.01$, ***$p < 0.001$, ****$p < 0.0001$ vs. Control; #$p < 0.05$, ##$p < 0.01$, ###$p < 0.001$, ####$p < 0.0001$ vs. HFD). Data in (K) and (L) are presented as mean ± SEM; significance was determined by Kruskal-Wallis test with Dunn's post-hoc test (Bonferroni-corrected) for pairwise comparisons, shown as direct brackets between the two compared groups (*$p < 0.05$) this is a different statistical test from (B)–(G) because (K)/(L) are ordinal histology grades with small group sizes ($n = 4$), not continuous biochemical measurements.

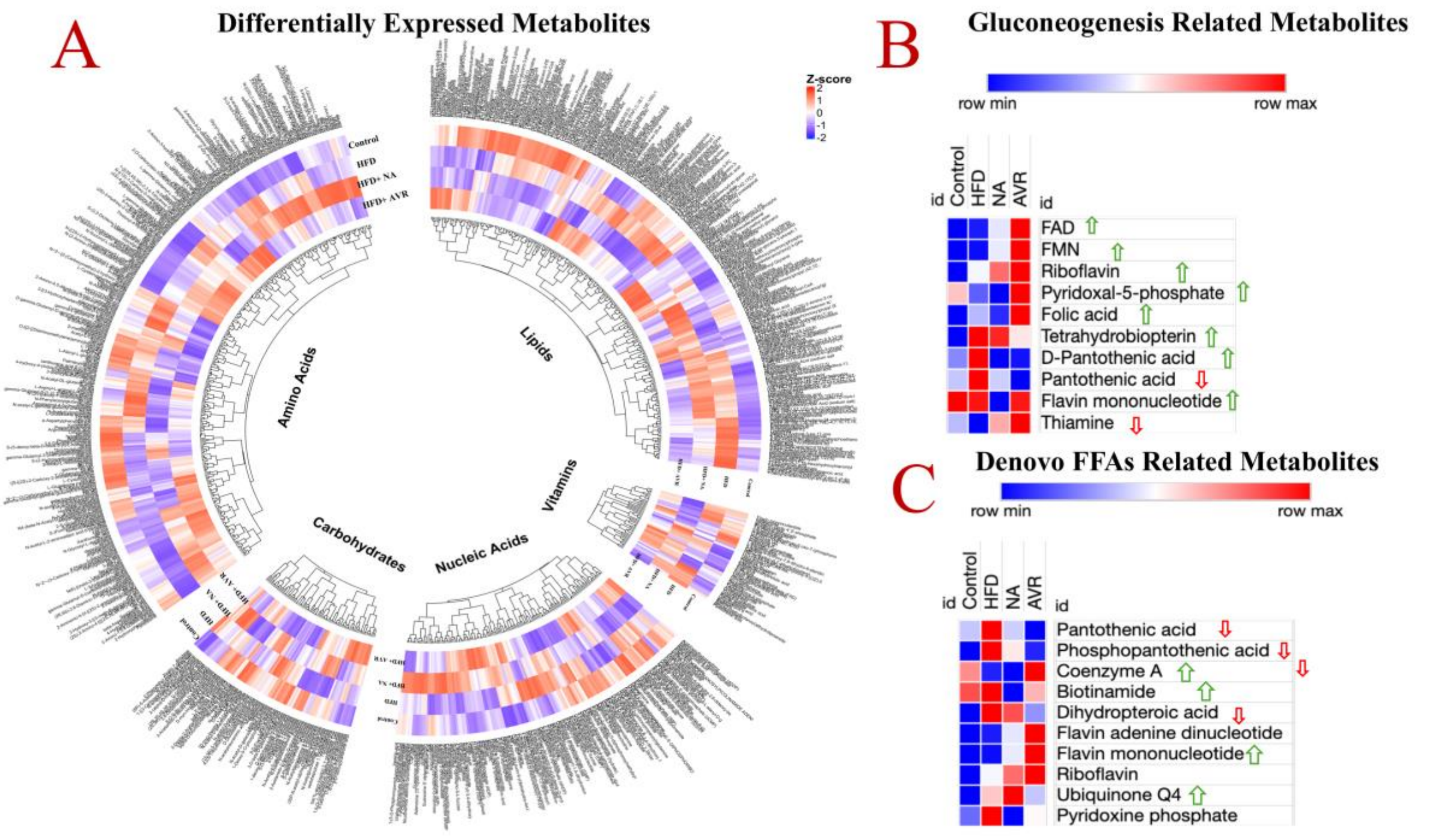

**Figure 5. Untargeted metabolomic profiling of Control, HFD, HFD+NA, and HFD+AVR groups: (A)** Circular hierarchical clustering heatmap showing relative abundance patterns of significantly altered metabolites across Control, HFD, HFD+Nishamalaki Rasayana (HFD+NA), and HFD+Arogyavardhini Rasa (HFD+AVR) groups. Metabolites are organized according to major biochemical classes, including amino acids, lipids, carbohydrates, nucleic acids, and vitamins. Color intensity represents normalized metabolite abundance expressed as Z-scores, with red indicating higher abundance and blue indicating lower abundance.**(B)** Gluconeogenesis related metabolites classes were significantly altered following **Arogyavardhini Rasa (AVR)** treatment relative to the HFD group. Green arrows indicate upregulated metabolites, whereas red arrows indicate downregulated metabolites. **(C)** DenovoFFAs-related metabolite classes significantly altered following **Nishamalaki Rasayana (NA)** treatment relative to the HFD group. Green arrows indicate upregulated metabolites, whereas red arrows indicate downregulated metabolites. Data are represented as normalized metabolite abundances obtained from untargeted metabolomic analysis. Hierarchical clustering was performed using Euclidean distance and complete linkage methods. Z-score normalization was applied before visualization.

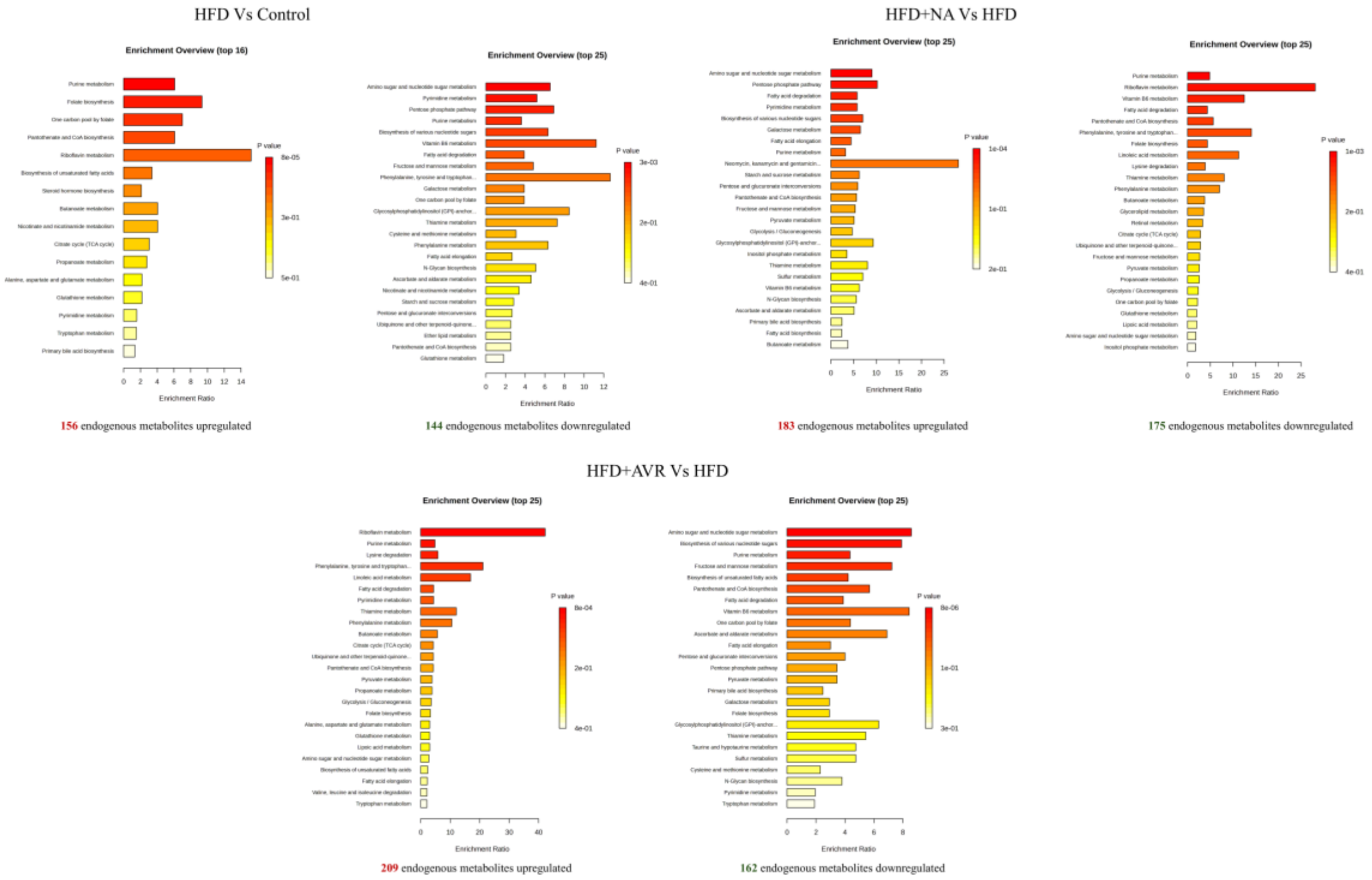

**Figure 6. Metabolite set enrichment analysis (MSEA) of differentially regulated endogenous metabolites across experimental comparisons.** Pathway enrichment analysis was performed using MetaboAnalyst 6.0 on upregulated and downregulated endogenous metabolite lists derived from volcano plot-based differential analysis ($|\log_2 FC| > 1$) for each pairwise comparison. Bar plots display enrichment ratio on the x-axis, with pathways ranked by significance and colour-coded by p-value (red = most significant, yellow = least significant), as indicated by the colour scale bar in each panel. **(A)** Top 16 enriched pathways among 156 endogenous metabolites upregulated in HFD liver relative to Control, highlighting activation of purine metabolism, folate biosynthesis, one-carbon pool by folate, and riboflavin metabolism. **(B)** Top 25 enriched pathways among 144 endogenous metabolites downregulated in HFD liver relative to Control, demonstrating suppression of amino sugar and nucleotide sugar metabolism, pyrimidine metabolism, pentose phosphate pathway, purine metabolism, and biosynthesis of various nucleotide sugars. **(C)** Top 25 enriched pathways among 183 endogenous metabolites upregulated in HFD+Nishamalaki Rasayana (HFD+NA) liver relative to HFD, revealing restoration of amino sugar and nucleotide sugar metabolism, pentose phosphate pathway, glycolysis/gluconeogenesis, and galactose metabolism. **(D)** Top 25 enriched pathways among 175 endogenous metabolites downregulated in HFD+NA liver relative to HFD, indicating suppression of purine metabolism and riboflavin metabolism — pathways aberrantly elevated by HFD feeding. **(E)** Top 25 enriched pathways among 209 endogenous metabolites upregulated in HFD+Arogyavardhini Rasa (HFD+AVR) liver relative to HFD, demonstrating broad restoration of amino sugar and nucleotide sugar metabolism, biosynthesis of various nucleotide sugars, fructose and mannose metabolism, pantothenate and CoA biosynthesis, and pentose phosphate pathway. **(F)** Top 25 enriched pathways among 162 endogenous metabolites downregulated in HFD+AVR liver relative to HFD, showing the strongest suppression of riboflavin metabolism (enrichment ratio ~40) and purine metabolism observed across all comparisons, indicative of potent normalization of HFD-driven cofactor and nucleotide dysregulation.

**Tables:**

**Table 1. Top 10 Differentially abundant metabolites in response to HFD and subsequent Rasayana treatment**

**UPREGULATED METABOLITES**

| Control vs HFD Metabolite | Biological implication | HFD vs AVR Metabolite | Biological implication | HFD vs NA Metabolite | Biological implication |
|---|---|---|---|---|---|
| cys-gln | GSH-pathway dipeptide; protein turnover and oxidative-stress-linked cysteine/glutamine handling. | N6-Me-Adenosine | m6A nucleoside; RNA methylation/turnover product. | 2-Methoxyestrone | Catechol-estrogen metabolite reflecting hepatic estrogen catabolism. |
| 5-Methyltetrahydrofolic acid | Active folate; one-carbon metabolism, methylation and homocysteine remethylation. | Uridine diphosphate glucose | UDP-glucose; donor for glycogen synthesis and hepatic glucuronidation. | 3-Hydroxy-3-[(3-methylbutanoyl)oxy]-4-(trimethylammonio)butanoate | Isovaleryl-carnitine-like acylcarnitine; BCAA catabolism and mitochondrial acyl-CoA handling. |
| 3-O-Methylglycyrol | Plant-derived furanocoumarin xenobiotic; dietary/xenobiotic biotransformation. | Long steroidal-glycoside name* | Complex steroidal glycoside; likely dietary/plant-derived xenobiotic. | N6-Me-Adenosine | m6A nucleoside; RNA methylation/turnover product. |
| 1-palmitoyl-2-oleoyl-sn-glycero-3-phosphocholine | Major membrane phosphatidylcholine; membrane remodeling/ER stress. | Leucylasparagine | Circulating dipeptide; protein digestion/turnover. | Lys-phe | Lysylphenylalanine; protein digestion/turnover. |
| ser-arg | Serylarginine; proteolytic turnover. | Taurocholic acid | Taurine-conjugated bile acid; FXR/TGR5 signalling and glucose/lipid regulation. | tyr-asn | Tyrosylasparagine; protein digestion/turnover. |
| Glutaminylcysteine | GSH precursor supporting hepatic antioxidant synthesis. | Uridine 5'-diphosphogalactose | UDP-galactose; glycoprotein/glycolipid galactosylation. | GDP-beta-L-colitose | Rare bacterial-type nucleotide deoxysugar; likely gut-microbial origin. |
| S-hydroxymethylglutathione | GSH-formaldehyde conjugate; aldehyde detoxification/redox handling. | Lys-phe | Lysylphenylalanine; protein digestion/turnover. | (2R,6S)-2,6-Diamino-3-hydroxyheptanedioic acid | Diaminopimelate-pathway-related metabolite; bacterial lysine biosynthesis/gut-microbial signature. |
| L-gamma-Glutamyl-L-leucine | Gamma-glutamyl dipeptide/GGT activity; oxidative stress and GSH cycle. | tyr-asn | Tyrosylasparagine; protein digestion/turnover. | epsilon-(gamma-Glutamyl)-lysine | Transglutaminase-generated isopeptide; protein cross-linking/turnover. |
| 1-Linoleoyl glycerol | Monoacylglycerol; lipolysis/endocannabinoid intermediate. | UDP-N-acetylglucosamine | UDP-GlcNAc; HBP end-product/O-GlcNAc substrate; nutrient sensing. | D-(+)-Tryptophan | D-tryptophan; atypical in mammals, often microbial origin. |
| val-arg | Valylarginine; protein digestion/turnover. | Orotidine 5'-monophosphate | OMP; penultimate de novo pyrimidine intermediate. | gamma-Glutamyl-3-(sulfosulfanyl)alanylglycine | Modified glutathione conjugate; sulfur/GSH conjugation chemistry. |

## DOWNREGULATED METABOLITES

| Control vs HFD Metabolite | Biological implication | HFD vs AVR Metabolite | Biological implication | HFD vs NA Metabolite | Biological implication |
|---|---|---|---|---|---|
| tyr-asn | Tyrosylasparagine; protein turnover. | Kyotorphin | Tyr-Arg neuropeptide; neuromodulatory/hepatic stress signalling. | NADPH | Central reducing cofactor for antioxidant defence and reductive biosynthesis. |
| 2-Amino-1,3,4-octadecanetriol | Sphingolipid/ceramide precursor; insulin/apoptotic signalling. | Acetyl-L-carnitine | Mitochondrial acetyl-CoA buffer; fatty-acid oxidation. | Kyotorphin | Tyr-Arg neuropeptide; neuromodulatory/hepatic stress signalling. |
| hypaphorine | Tryptophan-betaine plant/gut-microbial alkaloid; xenobiotic exposure. | 1-palmitoyl-2-[(9Z,12Z,15Z)-octadecatrienoyl]-sn-glycero-3-phosphocholine | PUFA phosphatidylcholine; hepatic membrane remodeling. | AICA ribonucleotide | ZMP; purine-biosynthesis intermediate and AMPK-activating nucleotide. |
| Uridine 5'-diphosphogalactose | UDP-galactose; glycosylation donor. | dihydrobiopterin | Oxidized BH4 precursor; NOS/aromatic amino-acid hydroxylase cofactor cycle. | 7,8-Dihydrobiopterin | Oxidized BH4-cycle form; NOS and hydroxylase cofactor pathway. |
| N2-(1-Carboxyethyl)-N5-(diaminomethylene)ornithine | Modified arginine/ornithine derivative; urea-cycle nitrogen handling. | Phosphatidylinositol-1,2-dipalmitoyl | PI species; precursor of PI3K/Akt insulin-signalling messengers. | Acetyl-L-carnitine | Mitochondrial acetyl-CoA buffer; fatty-acid oxidation. |
| S-(5-deoxy-beta-D-ribos-5-yl)-L-homocysteine | Methionine-salvage/polyamine intermediate; gut-microbial one-carbon metabolism. | Docosahexaenoyl-phosphatidylserine with oleoyl | DHA-enriched phosphatidylserine; apoptotic/inflammatory signalling. | Adenosine diphosphate ribose | ADP-ribose from NAD+-consuming enzymes; NAD+ turnover/redox/DNA-damage signalling. |
| Uridine diphosphate glucose | UDP-glucose; glycogen synthesis/glucuronidation donor. | val-arg | Valylarginine; protein turnover. | Maltodecaose (glucose decamer) | Glucose oligomer; glycogen/carbohydrate handling. |
| uridine 5′-diphosphate | UDP; glycosyl-transfer/glycogen turnover. | Glycyl-L-leucine | Circulating dipeptide; protein turnover. | 3-(Palmitoyloxy)-1,2-propanediyl (9E,12E,9'E,12'E)bis(-9,12-octadecadienoate) | Linoleate-rich glycerolipid; hepatic triglyceride/fatty-acid storage pool. |
| UDP-N-acetylglucosamine | UDP-GlcNAc; HBP/O-GlcNAc nutrient-sensing metabolite. | 1-[(9Z)-octadecenoyl]-2-tetradecanoyl-sn-glycero-3-phosphocholine | Mixed-acyl phosphatidylcholine; membrane remodeling. | riboflavin cyclic 4',5'-phosphate | Vitamin B2-pathway derivative; FAD/FMN and mitochondrial flavoprotein metabolism. |
| Threonylglutamine | Circulating dipeptide; amino-acid/protein turnover. | Indoxyl sulfate | Gut-microbial tryptophan indole; gut-liver-axis toxin/oxidative stress. | N-acetyl-L-gamma-glutamyl phosphate | Acetylated-glutamate/ornithine pathway intermediate; arginine biosynthesis/nitrogen handling. |

**Table 1. Top 10 Differentially abundant metabolites in response to HFD and subsequent Rasayana treatment**

**Table 2. Differentially expressed lipid-associated metabolites following HFD feeding and Rasayana treatment.**

| Metabolite | Family | log2FC HFD vs Ctrl | log2FC NA vs HFD | log2FC AVR vs HFD | Biological implication | Ref. |
|---|---|---|---|---|---|---|
| Taurocholic acid | Bile acid | +0.93 | +0.70 | +6.65 | Bile-acid species signalling via FXR/TGR5; regulates hepatic lipid and glucose metabolism. | [1] |
| Acetyl-L-carnitine | Acylcarnitine | +0.32 | -5.32 | -5.91 | Carnitine-conjugated mitochondrial fatty-acid β-oxidation intermediate. | [2] |
| 2-Amino-1,3,4-octadecanetriol | Ceramide/Sphingolipid | -6.26 | +5.20 | +4.50 | Sphingolipid precursor regulating ceramide synthesis, apoptosis and insulin signalling. | [3] |
| 2-Arachidonoyl glycerol | Monoacylglycerol (MG) | -0.25 | +0.31 | +5.08 | Lipolysis/endocannabinoid-pathway intermediate. | [4] |
| Phosphatidylcholine | Phosphatidylcholine (PC) | -2.48 | +2.24 | +4.97 | Major membrane phospholipid; precursor for hepatic VLDL/bile secretion. | [5] |
| 1-palmitoyl-2-[(9Z,12Z,15Z)-octadecatrienoyl]-sn-glycero-3-phosphocholine | Phosphatidylcholine (PC) | +5.29 | +0.01 | -4.95 | Major membrane phospholipid; precursor for hepatic VLDL/bile secretion. | [5] |
| Phosphatidylinositol-1,2-dipalmitoyl | Phosphatidylinositol (PI) | +2.94 | -1.05 | -4.22 | Precursor of PI3K/Akt second messengers in hepatic insulin signalling. | [6] |
| Docosahexaenoyl-phosphatidylserine with oleoyl | Phosphatidylserine (PS) | +4.28 | -2.81 | -4.13 | Membrane phospholipid involved in apoptotic/inflammatory signalling. | [7] |
| Glycocholic acid | Bile acid | -1.53 | +1.03 | +3.72 | Bile-acid species signalling via FXR/TGR5; regulates hepatic lipid and glucose metabolism. | [1] |
| Palmitic acid | Free fatty acid | -1.96 | +3.69 | +3.08 | Non-esterified fatty acid reflecting hepatic lipolysis, uptake and β-oxidation. | [8] |
| Palmitoyl sphingomyelin | Sphingomyelin (SM) | +2.73 | -1.60 | -3.01 | Membrane sphingolipid whose hydrolysis generates ceramide and hepatic lipotoxicity. | [3] |
| 1-Linoleoyl glycerol | Monoacylglycerol (MG) | +6.13 | -2.89 | +1.45 | Lipolysis/endocannabinoid-pathway intermediate lipid. | [4] |
| Docosahexaenoic acid | Free fatty acid | +1.80 | +0.48 | +2.81 | Non-esterified fatty acid reflecting hepatic lipolysis, uptake and β-oxidation. | [8] |
| Glutarylcarnitine | Acylcarnitine | +2.06 | +2.73 | +1.10 | Carnitine-conjugated mitochondrial fatty-acid β-oxidation intermediate. | [2] |
| 1-heptadecanoyl-sn-glycero-3-phosphocholine | Lyso-PC | +2.25 | -2.72 | +0.24 | PLA2-derived lipid signal linked to hepatic inflammation and insulin resistance. | [5,9] |
| Thromboxane B1 | Eicosanoid/Oxylipin | -1.62 | -2.58 | -1.17 | Arachidonic-acid-derived lipid mediator associated with hepatic inflammation. | [10] |
| 5,15-Dihydroxyeicosatetraenoic acid | Eicosanoid/Oxylipin | -0.58 | -2.56 | -0.86 | Arachidonic-acid-derived lipid mediator associated with hepatic inflammation. | [10] |
| Cervonoyl ethanolamide | Endocannabinoid | +1.88 | -2.48 | -0.93 | Endocannabinoid lipid acting via CB1/CB2; regulates hepatic lipogenesis and energy balance. | [4] |
| 1-icosanoyl-sn-glycero-3-phosphoethanolamine | Lyso-PE | +1.70 | -0.21 | -2.47 | Lysophospholipid signalling species generated during membrane turnover. | [5,9] |
| THREO-SPHINGOSINE, (-)- | Ceramide/Sphingolipid | -2.31 | -0.36 | +2.38 | Sphingolipid precursor regulating ceramide synthesis, apoptosis and insulin signalling. | [3] |

**Table 2. Differentially expressed lipid-associated metabolites following HFD feeding and Rasayana treatment.** Selected metabolites representing bile acids, acylcarnitines, sphingolipids, phospholipids, fatty acids, lysophospholipids, eicosanoids, and endocannabinoids are presented as $\log_2$ fold changes for HFD vs Control, NA vs HFD, and AVR vs HFD. Positive values indicate increased abundance and negative values indicate decreased abundance. The selected metabolites highlight treatment-associated alterations in bile-acid, fatty-acid, membrane-lipid, sphingolipid, and lipid-signalling pathways. References indicate supporting literature for the biological implications.

## Supplementary Text

**Systems biology insights into protective effects of Arogyavardhini Rasa and Nishamalaki Rasayana in restoring hepatic lipid homeostasis in diet induced Type 2 Diabetes**

Harshit B Joshi[1,4], Anupama Vallazhath[1], Priyotosh Sil[2,3], Rahul Tiwari[2,3], Krishna Raghava Hebbar[4], Swati Sharma[5], Satish Rao B S[6], Basavaraj Hadapad[4], Areejit Samal[2,3*], Manjunath B Joshi[1,*]

[1] Department of Ageing Research, Manipal School of Life Sciences, Manipal Academy of Higher Education, Manipal, India

[2] The Institute of Mathematical Sciences (IMSc), Chennai, India

[3] Homi Bhabha National Institute (HBNI), Mumbai, India

[4] Division of Ayurveda, Centre for Integrative Medicine and Research, Manipal Academy of Higher Education, Manipal, India

[5] Department of Pathology, Kasturba Medical College, Manipal Academy of Higher Education, Manipal, India

[6] Department of Radiation Biology and Toxicology, Manipal School of Life Sciences, Manipal Academy of Higher Education, Manipal, India

**Corresponding Authors**

1. Dr. Areejit Samal
   Professor, Computational Biology GroupThe Institute of Mathematical Sciences (IMSc), Chennai, 600113, Tamil Nadu, India; asamal@imsc.res.in
2. Dr. Manjunath B Joshi
   Professor, Department of Ageing Research, Manipal School of Life Sciences, Manipal Academy of Higher Education, Manipal, 576104, Karnataka, India; manjunath.joshi@manipal.edu

# 1. Supplementary Methods

## 1.1 Compilation of phytochemicals associated with the two Ayurvedic formulations Arogyavardhini Rasa and Nishamalaki Rasayana

We carried out a computational investigation of two Ayurvedic formulations, Arogyavardhini Rasa (AVR) and Nishamalaki Rasayana (NA), to explore their potential therapeutic relevance in the context of multiple coexisting metabolic disorders, including T2D, fatty liver disease and obesity. AVR was procured from Doothpapeshwar, Pune, India and constituent ingredients were identified from the accompanying product information leaflet. NA was freshly prepared by mixing equal parts of dried rhizome of *Curcuma longa* and fruit pericarp of *Emblica officinalis*. For the present analysis, we restricted our focus exclusively to the herbal ingredients, specifically considering the corresponding plant parts used in each formulation, while excluding non-herbal constituents.

For each herbal ingredient in the formulations, information on associated phytochemicals was retrieved from the in-house IMPPAT database (Mohanraj et al., 2018; Vivek-Ananth et al., 2023), with the search limited to the specific plant part(s) reported to be used in preparation of the formulations. In addition, phytochemicals documented through thin-layer chromatography (TLC) data in the Ayurvedic Pharmacopoeia of India (API) were also incorporated whenever valid chemical identifiers, such as PubChem CIDs, were available. The pharmacokinetic suitability of the identified phytochemicals was subsequently assessed using the SwissADME web server (Daina et al., 2017). Phytochemicals were retained for downstream analyses if they satisfied standard drug-likeness criteria, defined here as having no more than one violation of Lipinski's Rule of Five (Lipinski et al., 2001) together with a predicted bioavailability score of at least 0.50.

## 1.2 Identification and curation of phytochemical-regulated and disease-associated human protein targets

Phytochemicals that satisfied the computational criteria for drug-likeness and oral bioavailability were subsequently considered for the identification of their known experimentally validated human protein targets. Target information was compiled from NPASS (Zhao et al., 2023), BindingDB (Gilson et al., 2016), and ChEMBL (Zdrazil et al., 2024), considering only human proteins. For ChEMBL-derived interactions, entries annotated in the comment field as 'inactive', 'not active', 'not determined', 'inconclusive', 'undetermined', 'not evaluated', or 'no data' were excluded. Records containing non-empty entries in the data validity comment field were also discarded. No additional filtering criteria were applied to interactions obtained from NPASS or BindingDB. Disease-associated target sets for T2D, fatty liver and obesity-related disorders were assembled using an integrative multi-database strategy. Disease-relevant drugs were first identified from DrugBank (Knox et al., 2024) by screening the indication field using keywords such as 'diabetes', 'diabetic', 'hyperglycemia', 'insulin', 'glycaemic', 'obesity', and 'weight loss'. The retrieved drugs were manually curated to retain only non-withdrawn US Food and Drug Administration (FDA)-approved small molecules or peptides with established relevance to the diseases under investigation. Therapeutic targets corresponding to these drugs were primarily obtained from the Therapeutic Target Database (TTD) (Zhou et al., 2024) and further supplemented using the curated therapeutic target compilation reported by Santos et al. (2017), which integrates information from DrugCentral (Ursu et al., 2017), ChEMBL (Bento et al., 2014), and canSAR (Tym et al., 2016). For drugs not covered by these resources, target annotations were manually retrieved from DrugCentral, retaining only those explicitly marked with a 'Green Tick' in the Mechanism of Action field.

Disease-associated genes were collected from multiple complementary resources. High-confidence associations were first retrieved from DisGeNET (Piñero et al., 2017). For T2D (CUI:C0011860) and obesity (CUI:C0028754), single Concept Unique Identifiers (CUIs) corresponding to each disease were used, whereas for fatty liver-related disorders, multiple CUIs corresponding to fatty liver disease (CUI:C0015695), non-alcoholic fatty liver disease (CUI:C0400966), metabolic dysfunction-associated fatty liver disease (CUI:C5577648), and nonalcoholic steatohepatitis (CUI:C3241937) were considered to comprehensively capture the disease spectrum. Only protein-coding genes with a Gene-Disease Association (GDA) score ≥ 0.40 and an Evidence Index (EI) > 0.5 in DisGeNET were retained. Additional disease-associated genes were obtained from the DISEASES database (Grissa et al., 2022) using the corresponding Disease Ontology identifiers. While single identifiers were used for T2D (DOID:9352) and obesity (DOID:9970), multiple ontology terms representing different fatty liver-associated conditions were included for fatty liver-related disorders (DOID:0080208, DOID:9452, DOID:0080547). Genes with relevance scores ≥ 3 in DISEASES database were retained. To ensure consistency across datasets, Ensembl protein identifiers were converted to Entrez Gene identifiers using the mygene Python package (version 3.2.2), and unresolved mappings were manually curated using NCBI Gene resources (Sayers et al., 2020). The disease-associated gene sets were further expanded using information in the Comparative Toxicogenomics Database (CTD) (Davis et al., 2021) with disease-specific Medical Subject Headings (MeSH) terms corresponding to T2D (MESH:D003924), fatty liver-associated disorders (MESH:D065626, MESH:D005234), and obesity (MESH:D009765). Only human-associated entries were retained following SynGO-based filtering (Koopmans et al., 2019). Additional genes were also retrieved from GeneCards (Stelzer et al., 2016) using the queries 'Type 2 diabetes', 'Fatty liver disease', and 'Obesity' retaining only protein-coding genes with relevance scores ≥ 15. Finally, therapeutic targets of FDA-approved drugs were integrated with the compiled disease-associated genes to construct high-confidence disease gene sets. Drug targets were retained directly, whereas additional candidate genes were included only if supported by at least two independent resources among DisGeNET, DISEASES, CTD, and GeneCards. Overall, this integrative framework was used to generate robust and biologically relevant disease-associated gene sets for subsequent analyses.

### 1.3 Network proximity analysis

Network proximity analysis was performed to investigate the systems-level association between formulation-associated targets and disease-associated genes within the human protein-protein interaction (PPI) network.

#### *1.3.1 Construction of protein-protein interaction network*

A large-scale background PPI network was assembled for network proximity analysis by integrating multiple curated human interactome resources to achieve broad interaction coverage. First, the human PPI network reported by Gan *et al.* (18,505 nodes and 327,924 edges) (Gan et al., 2023) was included. Next, the HI-union interactome was obtained from the Human Interactome Atlas (https://interactome-atlas.org/) (Luck et al., 2020). Since the interaction data in this resource were represented using Ensembl gene identifiers, they were converted to Entrez Gene identifiers using the Syngo web server (Koopmans et al., 2019). Following removal of duplicate interactions and self-loop edges, the processed HI-union interactome contained 9,053 proteins connected through 63,669 unique interactions. In addition, the curated human PPI network compiled by Ruiz *et al.* (17,660 nodes and 387,626 edges) (Ruiz et al., 2021), which integrates interaction evidence from multiple databases, was incorporated into the analysis. The three interactomes were subsequently combined by taking the union of all proteins and interaction pairs, followed by elimination of redundant edges. The resulting integrated human PPI comprised 19,572 unique proteins and 541,092 high-confidence interactions, and was used as the background PPI network for the network proximity analysis.

### *1.3.2 Computation of network proximity*

Network proximity analysis was performed to assess the topological relationship between formulation-associated targets and the disease-associated gene set using NetMedPy (version 0.1.171) in Python 3.11 (Aldana et al., 2025). All analyses were restricted to the largest connected component (LCC; 19,545 nodes and 541,065 edges) of the integrated PPI network to enable valid shortest-path distance calculations.

For a formulation, all nonredundant protein targets associated with its constituent phytochemicals were combined to generate a formulation-specific target set $F_i$. The disease-associated gene set $G$ was defined as described previously. Network proximity between $F_i$ and $G$ was quantified using the asymmetric closest-distance metric based on average shortest-path length, following the established network proximity framework (Guney et al., 2016). The closest distance between the formulation target set and the disease gene set was calculated as:

$$d_c(F_i, G) = \frac{1}{|G|} \sum_{y \epsilon G} min_{x \epsilon F_i} d(x, y) \quad (1)$$

where $d(x, y)$ represents the shortest-path distance between nodes $x$ and $y$ in the PPI network. Smaller values of $d_c$ indicate closer network proximity between formulation-associated targets and disease-associated genes.

### *1.3.3 Statistical significance assessment*

The significance of the observed proximity values was evaluated using a degree-aware null model implemented in NetMedPy (Aldana et al., 2025). Random target sets were generated using a logarithmic degree-binning strategy to approximately preserve the degree distribution and size of the original formulation-specific target sets. For each formulation, 1000 randomized target sets were generated, and the corresponding proximity to the disease-associated gene set was computed using the same distance metric. These randomized measurements were used to construct an empirical null distribution. The statistical significance of the observed proximity was quantified using a Z-score defined as:

$$Z = \frac{d_c(F_i, G) - \mu_{rand}}{\sigma_{rand}} \quad (2)$$

where $\mu_{rand}$ and $\sigma_{rand}$ denote the mean and standard deviation of the randomized proximity values, respectively. A one-tailed empirical p-value was calculated as the fraction of randomized proximity scores less than or equal to the observed value. Negative Z-scores together with $p < 0.05$ were considered indicative of statistically significant network proximity between formulation-associated targets and disease-associated genes.

## 1.4 Pathway enrichment analysis

Biological pathway enrichment analysis was performed to characterize the functional processes associated with formulation-associated targets and disease-associated gene sets. Enrichment analysis was conducted using the KEGG 2026 Human pathway database available through the Enrichr platform (Xie et al., 2021), where the default background gene set provided by Enrichr was used. For each gene set, significantly enriched KEGG pathways were identified. To enhance robustness and minimize bias arising from small pathways, only KEGG pathways containing more than 50 genes were considered. Pathways with a Benjamini-Hochberg false discovery rate (FDR)-adjusted p-value below 0.01 were regarded as significantly enriched (Benjamini & Hochberg, 1995). Significant pathways were subsequently mapped to their corresponding hierarchical categories using annotations from the KEGG BRITE database (https://www.genome.jp/kegg/brite.html). Pathways associated with broad disease categories not directly relevant to the pathophysiology of T2D, fatty liver, or obesity including cancer, neurodegenerative, excretory system, environmental adaptation, immune disease pathways, and infectious disease pathways, were excluded from downstream analyses. For comparative interpretation, the top 20 enriched KEGG pathways were selected for each gene set.

### 1.5 Knowledge graph construction

A comprehensive knowledge graph (KG) representing the relationships among formulations, phytochemicals, protein targets, and metabolic disorders was developed using Neo4j (version 5.26.18) (https://neo4j.com/). Information compiled from the earlier curation and integration stages was converted into a structured triplet format consisting of a source identifier, source node type, relationship, relationship properties, target identifier, and target node type.

The KG incorporated several types of nodes, including the two Ayurvedic formulations, their constituent medicinal plants represented by botanical names, phytochemicals that passed the predefined drug-likeness and bioavailability filters, human gene/protein targets associated either with phytochemical interactions or disease conditions, the three disease conditions considered in this study (T2D, fatty liver disease, and obesity), and FDA-approved drugs for these conditions. Connections between nodes were established according to known biological or pharmacological associations. The formulations were linked to their constituent plants through 'CONTAINS_HERB' relationship, while plants were connected to phytochemicals through a 'CONTAINS_CHEMICAL' relationship that additionally stored metadata such as the plant part from which the compound was reported. Phytochemicals were associated with protein targets through 'INTERACTS_WITH' relationships, disease-associated targets were linked to disease nodes through 'CORRESPONDS_TO', approved drugs were connected to diseases using 'TREATS', and drug-target interactions were represented using 'AFFECTS' relationships. This graph-based organization enabled efficient querying, integration, and exploration of interconnected entities within the dataset. To support interactive visualization and manual inspection of the network structure, an HTML-based representation of the KG was generated using PyVis (version 0.3.2) and is provided as Supplementary File (Interactive KG).

### 1.6 Analysis based on the constructed knowledge graph

The constructed knowledge graph (KG) was subsequently used for downstream integrative analyses and relational exploration of the curated entities. The KG framework enabled systematic examination of associations among formulations, plants, phytochemicals, protein targets, disease conditions, and FDA-approved drugs. Formulation-associated target sets encoded in the KG were used to evaluate the extent of target overlap between the two formulations. In addition, overlaps among the gene sets associated with T2D, fatty liver, and obesity were quantified. Overlap analyses were also performed between formulation-associated targets and disease-associated genes for each condition. Statistical significance of the observed overlaps was assessed using a right-tailed hypergeometric test using the 19,653 human protein-coding genes annotated in Ensembl (Cunningham et al., 2022) as the background universe.

The KG was further utilized to compare formulation-associated targets with therapeutic targets of FDA-approved drugs used for T2D, fatty liver, and obesity. Shared targets were identified by examining overlaps between the protein targets associated with formulation-derived phytochemicals and those linked to approved drugs. In addition, the network structure of the KG was analyzed to identify instances in which multiple phytochemical constituents within the same formulation are collectively mapped to groups of proteins that are simultaneously targeted by clinically used drug combinations. Such patterns could suggest a potential multi-component and multi-target mechanism underlying the action of these formulations, resembling the conceptual framework of combination pharmacotherapy.

### 1.7 Protein-protein interaction (PPI)-based network pharmacology analysis

Network pharmacology analysis was performed to identify and prioritize key proteins within the overlapping target space between formulation-associated targets and disease-associated genes. For this analysis, PPI networks were constructed using the STRING database (version 12.0) (Szklarczyk et al., 2023), considering only human protein interactions and applying a minimum interaction confidence score of 0.4. For each formulation-disease combination, the corresponding PPI network was generated using the common proteins shared between the formulation-specific target set and the disease-associated gene set.

The LCCs of the resulting networks were analyzed in Cytoscape (Shannon et al., 2003) using the CytoHubba plugin (version 0.1) (Chin et al., 2014) to identify topologically important proteins. Five centrality-based metrics were independently applied: Maximal Clique Centrality (MCC), Maximum Neighborhood Component (MNC), degree centrality, closeness centrality, and betweenness centrality. For each metric, the top ten ranked proteins were extracted. A consensus hub set for each formulation-disease network was subsequently defined by selecting proteins that appeared among the top-ranked nodes across all five centrality measures. This consensus-based approach was used to prioritize proteins exhibiting consistent topological importance across multiple network measures. Finally, hub proteins shared between the two formulations and associated with multiple disease conditions were identified.

### 1.8 Physiological and Plasma Biochemical Assessment

Peripheral blood was collected by retro-orbital sinus puncture under fasting conditions at weeks 10 and 20 of treatment. Plasma was obtained by centrifugation at 2000 rpm for 15 minutes. Fasting blood glucose was measured by Accu-Chek Instant glucometer (Roche Diagnostics). Plasma triglycerides and total cholesterol were quantified using enzymatic colorimetric kits (Aspen Laboratories Pvt. Ltd., New Delhi) on a STAR-20 semi-autoanalyzer. Plasma ALT and AST activities were measured using commercially available kinetic assay kits (Aspen Laboratories Pvt. Ltd.) on a STAR-20 semi-autoanalyzer, following manufacturer protocols. ALT and AST are established markers of hepatocellular injury and are elevated in NAFLD, hepatitis, and drug-induced liver injury. Liver weight was recorded immediately upon sacrifice.

### 1.9 Histopathological Evaluation

Liver tissues were fixed in 10% neutral buffered formalin, processed through graded alcohol and xylene, embedded in paraffin, sectioned at 5 µm, and stained with haematoxylin and eosin (H&E). Sections were examined by light microscopy (Olympus BX53) at 10× and 40× magnification. Hepatic steatosis was assessed based on the proportion of hepatocytes containing lipid vacuoles and graded according to the steatosis component of the Nonalcoholic Steatohepatitis Clinical Research Network (NASH CRN) histological scoring system: grade 0 (S0), less than 5% of hepatocytes affected; grade 1 (S1), 5–33%; grade 2 (S2), 34–66%; and grade 3 (S3), greater than 66%. Multiple representative microscopic fields were examined from each liver section to assess the distribution and severity of steatotic changes (**Table S17**).

**Table 2 References:**

**Supplementary Figures:**

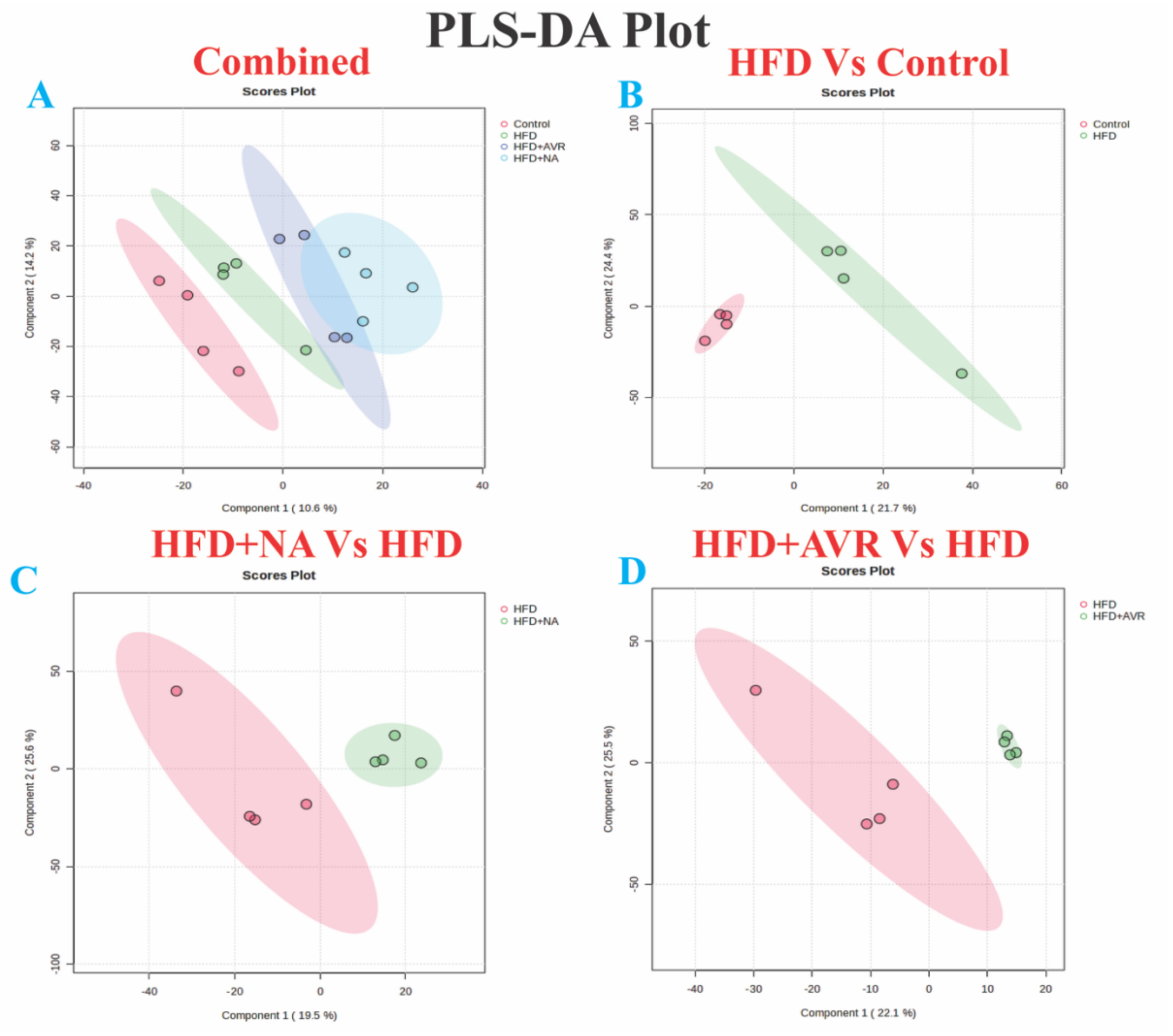

**Supplementary Figure 1.** Partial least squares discriminant analysis (PLS-DA) score plots showing separation of metabolic profiles among experimental groups. (A) Multigroup PLS-DA score plot showing the overall distribution of Control, high-fat diet (HFD), HFD supplemented with Nishamalaki Rasayana (HFD+NA), and HFD supplemented with Arogyavardhini Rasa (HFD+AVR) groups. (B) PLS-DA score plot comparing Control and HFD groups. (C) PLS-DA score plot comparing HFD and HFD+NA groups. (D) PLS-DA score plot comparing HFD and HFD+AVR groups. Each point represents an individual biological sample, and shaded ellipses represent the corresponding group distribution. The distinct clustering and separation between groups indicate differences in their overall metabolic profiles following HFD feeding and Rasayana treatment.

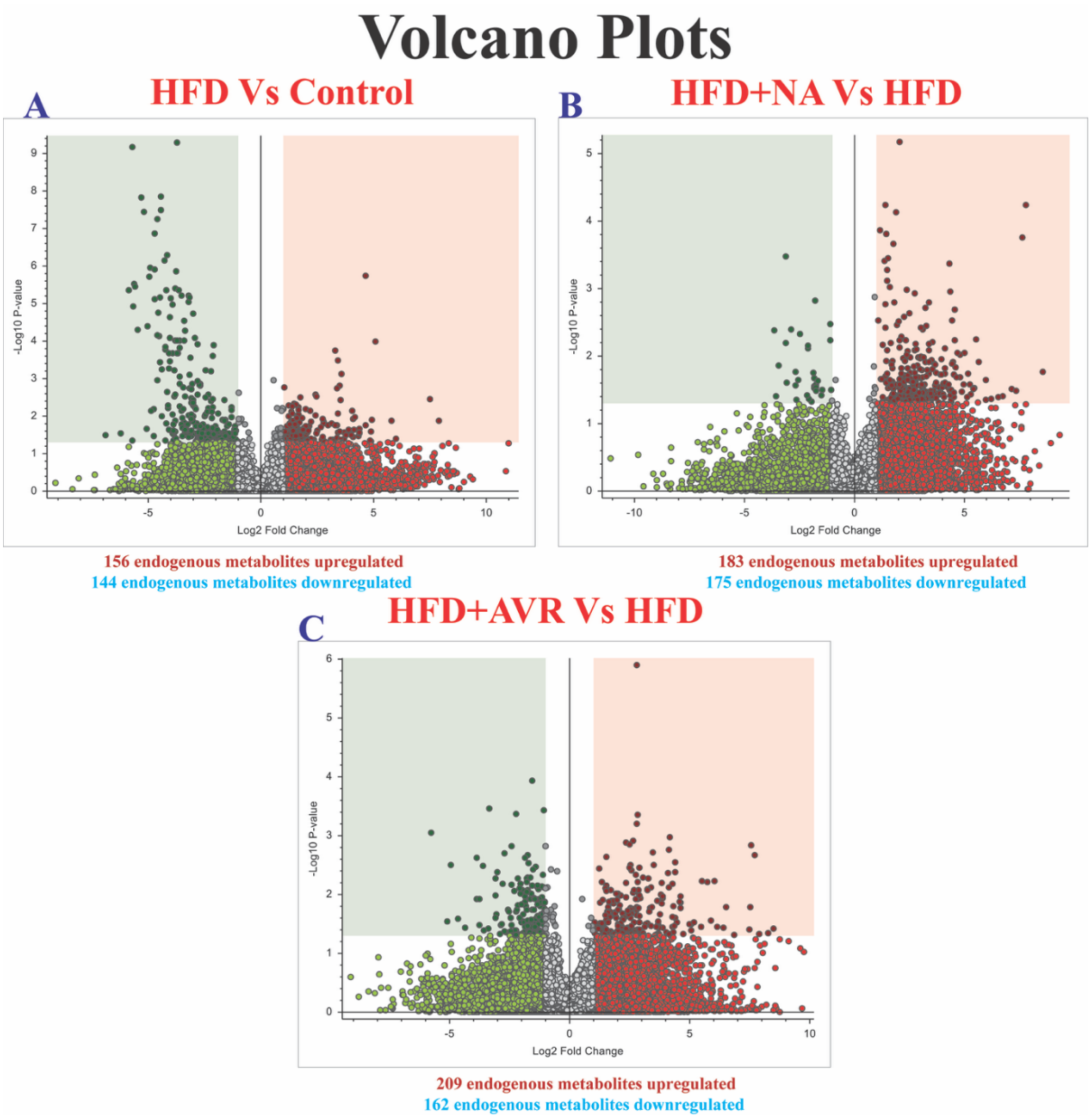

**Supplementary Figure 2: Volcano plots showing differential metabolite abundance across experimental groups.** Volcano plots illustrate the differential abundance of endogenous metabolites in (A) high-fat diet (HFD) versus control, (B) HFD supplemented with Nishamalaki Rasayana (HFD+NA) versus HFD, and (C) HFD supplemented with Arogyavardhini Rasa (HFD+AVR) versus HFD. The x-axis represents the $\log_2$ fold change, while the y-axis represents $-\log_{10}(p\text{-value})$.

**Table S1: Phytochemical and target profiling of the two Ayurvedic formulations Arogyavardhini and Nishamalaki. This table summarizes the sequential filtering of phytochemicals derived from the constituent plant parts of the two formulations ArogyaVardhini and Nishamalaki. It includes the number of phytochemicals before and after drug-likeness and bioavailability filtering, the subset of filtered phytochemicals with known human targets, and the corresponding target counts at both the individual plant and overall formulation levels.**

| Formulation | Herb name | Plant part | Phytochemicals before filtering | Phytohemicals after drug-likeness and bioavailability filtering | Filtered phytochemicals with known human targets | Total filtered phytohemicals for per formulation | Total filtered phytochemicals with known target(s) per formulations | Number of human targets | Total number of human targets per formulation |
|---|---|---|---|---|---|---|---|---|---|
| Arogyavardhini | Terminalia chebula | Fruit | 34 | 23 | 21 | 128 | 58 | 408 | 657 |
| | Terminalia bellirica | Fruit | 26 | 19 | 9 | | | 282 | |
| | Phyllanthus emblica | Fruit | 27 | 14 | 12 | | | 516 | |
| | Commiphora wightii | Plant exudate | 21 | 21 | 11 | | | 195 | |
| | Plumbago zeylanica | Root | 12 | 12 | 3 | | | 64 | |
| | Picrorhiza kurroa | Rhizome | 21 | 11 | 6 | | | 49 | |
| | Azadirachta indica | Leaf | 55 | 48 | 14 | | | 402 | |
| Nishamalaki | Curcuma longa | Rhizome | 168 | 168 | 54 | 182 | 66 | 351 | 618 |
| | Phyllanthus emblica | Fruit | 27 | 14 | 12 | | | 516 | |

**Table S2: Phytochemical profiles of constituent herbs in the two formulations. The table provides a comprehensive mapping of formulations, constituent herbs, and corresponding plant parts to their associated phytochemical constituents. For each phytochemical, the table includes chemical identifiers, compound names, IMPPAT/API references, and canonical SMILES representations.**

| Formulation | Herb name | Part | Chemical identifier | Chemical name | Source | Canonical SMILES |
|---|---|---|---|---|---|---|
| Arogyavardhini | Azadirachta indica | Leaf | CAS_105377-74-0 | Isomargosinolide | IMPPAT | COC(=O)C[C@H]1[C@]2(C)C3=C(C)[C@@H](C[C@H]3O[C@@H]2[C@@H]2[C@@H]3[C@]1(C)C=CC(=O)[C@@]3(C)CO2)C1=CC(=O)OC1O |
| Arogyavardhini | Azadirachta indica | Leaf | CHEMSPIDER_10298763 | Margosinolide | IMPPAT | COC(=O)C[C@H]1[C@@]2(C)[C@H](O[C@H]3C2=C(C)[C@@H](C3)C2=CC(OC2=O)O)[C@H]2[C@@H]3[C@]1(C)C=CC(=O)[C@@]3(C)CO2 |
| Arogyavardhini | Azadirachta indica | Leaf | CID_100017 | Methyl 2-[6-(furan-3-yl)-7,9,11,15-tetramethyl-12,16-dioxo-3,17-dioxapentacyclo[9.6.1.02,9.04,8.015,18]octadeca-7,13-dien-10-yl]acetate | IMPPAT | COC(=O)CC1C2(C)C(OC3C2=C(C)C(C3)c2cocc2)C2C3C1(C)C(=O)C=CC3(C)C(=O)O2 |
| Arogyavardhini | Azadirachta indica | Leaf | CID_101289833 | Meldenin | IMPPAT | CC(=O)O[C@H]1[C@@H](O)[C@]2(C)[C@@H]([C@@]3([C@@H]1C(C)(C)C(=O)CC3)C)CC[C@@]1(C2=CC[C@H]1c1ccoc1)C |
| Arogyavardhini | Azadirachta indica | Leaf | CID_101425842 | Isoazadirolide | IMPPAT | COC(=O)CC1C2(C)C(OC3C2=C(C)C(C3)C2=CC(=O)OC2O)C2C3C1(C)C(OC(=O)C=C(C)C)CC(C3(C)CO2)O |
| Arogyavardhini | Azadirachta indica | Leaf | CID_101650343 | (1S)-1-[(2R,4S,5R)-5-hydroxy-4-[(3S,5R,9R,10R,13S,14S,17S)-3-hydroxy-4,4,10,13,14-pentamethyl-2,3,5,6,9,11,12,15,16,17-decahydro-1H-cyclopenta[a]phenanthren-17-yl]oxolan-2-yl]-2-methylpropane-1,2-diol | IMPPAT | O[C@@H]1O[C@H](C[C@H]1[C@@H]1CC[C@]2([C@@]1(C)CC[C@H]1C2=CC[C@@H]2[C@]1(C)CC[C@@H](C2(C)C)O)C)[C@@H](C(O)(C)C)O |
| Arogyavardhini | Azadirachta indica | Leaf | CID_102090424 | Vilasinin | IMPPAT | O[C@H]1C[C@@H](O)[C@@]2([C@H]3[C@@]1(C)[C@H]1CC[C@@]4(C(=CC[C@H]4c4ccoc4)[C@]1(C)[C@@H]([C@@H]3OC2)O)C)C |
| Arogyavardhini | Azadirachta indica | Leaf | CID_10505484 | 6-Deacetylnimbin | IMPPAT | COC(=O)C[C@H]1[C@@]2(C)[C@H](O[C@H]3C2=C(C)[C@@H](C3)c2cocc2)[C@@H]([C@@H]2[C@]1(C)C(=O)C=C[C@@]2(C)C(=O)OC)O |
| Arogyavardhini | Azadirachta indica | Leaf | CID_1107 | Sterol | IMPPAT | OC1CCC2C(C1)CCC1C2CCC2C1CCC2 |
| Arogyavardhini | Azadirachta indica | Leaf | CID_12309055 | beta-Sitosterol-beta-D-glucoside | IMPPAT | CC[C@@H](C(C)C)CC[C@H]([C@H]1CC[C@@H]2[C@]1(C)CC[C@H]1[C@@H]2CC=C2[C@@]1(C)CC[C@@H](C2)O[C@H]1O[C@H](CO)[C@H]([C@@H]([C@H]1O)O)O)C |
| Arogyavardhini | Azadirachta indica | Leaf | CID_12309449 | delta-Elemene | IMPPAT | C=C[C@@]1(C)CCC(=C[C@@H]1C(=C)C)C(C)C |
| Arogyavardhini | Azadirachta indica | Leaf | CID_12313020 | gamma-Muurolene | IMPPAT | CC1=C[C@@H]2[C@H](CC1)C(=C)CC[C@H]2C(C)C |
| Arogyavardhini | Azadirachta indica | Leaf | CID_12409 | Nonacosane | IMPPAT | CCCCCCCCCCCCCCCCCCCCCCCCCCCCC |
| Arogyavardhini | Azadirachta indica | Leaf | CID_13875774 | Nimbocinolide | IMPPAT | CC(=O)O[C@@H]1C[C@@H]2[C@]([C@@H]3[C@]1(C)C1=CC[C@H]([C@@]1([C@H]([C@H]3OC(=O)C(O)(C)C)O)C)C1=CC(OC1=O)O)(C)C=CC(=O)C2(C)C |
| Arogyavardhini | Azadirachta indica | Leaf | CID_14458886 | 3-Deacetylsalannin | IMPPAT | COC(=O)C[C@@H]1[C@@]2(C)[C@@H](OC(=O)/C(=C/C)/C)C[C@H]([C@@]3([C@@H]2[C@H]([C@@H]2[C@@]1(C)C1=C(C)[C@@H](C[C@H]1O2)c1ccoc1)OC3)C)O |
| Arogyavardhini | Azadirachta indica | Leaf | CID_14467538 | 28-Deoxonimbolide | IMPPAT | COC(=O)C[C@H]1[C@@]2(C)[C@H](O[C@H]3C2=C(C)[C@@H](C3)c2cocc2)[C@H]2[C@@H]3[C@]1(C)C(=O)C=C[C@@]3(C)CO2 |
| Arogyavardhini | Azadirachta indica | Leaf | CID_14492795 | 5,7-Dihydroxy-4'-methoxy-8,3'-di-C-prenylflavanone | IMPPAT | COc1ccc(cc1CC=C(C)C)C1CC(=O)c2c(O1)c(CC=C(C)C)c(cc2O)O |
| Arogyavardhini | Azadirachta indica | Leaf | CID_157277 | 2H-Cyclopenta(b)naphtho(2,3-d)furan-10-carboxylic acid,2-(3-furanyl)-3,3a,4a,5,5a,6,9,9a,10,10a-decahydro-5,6-dihydroxy-1,6,9a,10a-tetramethyl-9-oxo-, methyl ester, (2R,3aS,4aS,5R,5aS,6R,9aR,10S,10aR)- | IMPPAT | COC(=O)C[C@H]1[C@@]2(C)[C@H](O[C@@H]3C2=C(C)[C@@H](C3)c2cocc2)[C@@H]([C@@H]2[C@@]1(C)C(=O)C=C[C@@]2(C)O)O |
| Arogyavardhini | Azadirachta indica | Leaf | CID_178770 | Nimocinol | IMPPAT | CC(=O)O[C@@H]1[C@H](O)[C@@H]2[C@]([C@@H]3C1(C)C1=CC[C@H]([C@@]1(CC3)C)c1ccoc1)(C)C=CC(=O)C2(C)C |
| Arogyavardhini | Azadirachta indica | Leaf | CID_184310 | [(5R,6R,7S,9R,10R,13S,17R)-6-hydroxy-17-(2-hydroxy-5-oxo-2H-furan-3-yl)-4,4,8,10,13-pentamethyl-3-oxo-5,6,7,9,11,12,16,17-octahydrocyclopenta[a]phenanthren-7-yl] acetate | IMPPAT | CC(=O)O[C@@H]1[C@H](O)[C@@H]2[C@]([C@@H]3C1(C)C1=CC[C@H]([C@]1(C)CC3)C1=CC(=O)OC1O)(C)C=CC(=O)C2(C)C |
| Arogyavardhini | Azadirachta indica | Leaf | CID_184503 | (9R,10R,13S,14S,17R)-17-[(2R)-2-[(1R,2S)-1,3-dihydroxy-2-methylpropyl]-2,3-dihydrofuran-4-yl]-4,4,10,13,14-pentamethyl-1,2,5,6,9,11,12,15,16,17-decahydrocyclopenta[a]phenanthren-3-one | IMPPAT | OC[C@@H]([C@H]([C@@H]1OC=C(C1)[C@@H]1CC[C@]2([C@@]1(C)CC[C@H]1C2=CCC2[C@]1(C)CCC(=O)C2(C)C)C)O)C |
| Arogyavardhini | Azadirachta indica | Leaf | CID_222284 | beta-Sitosterol | IMPPAT | CC[C@@H](C(C)C)CC[C@H]([C@H]1CC[C@@H]2[C@]1(C)CC[C@H]1[C@H]2CC=C2[C@]1(C)CC[C@@H](C2)O)C |
| Arogyavardhini | Azadirachta indica | Leaf | CID_243696 | 1-Nonacosanol | IMPPAT | CCCCCCCCCCCCCCCCCCCCCCCCCCCCCO |
| Arogyavardhini | Azadirachta indica | Leaf | CID_31291 | Tetradecanal | IMPPAT | CCCCCCCCCCCCCC=O |
| Arogyavardhini | Azadirachta indica | Leaf | CID_42608158 | Allo-Aromadendrene | IMPPAT | C[C@@H]1CC[C@H]2[C@@H]1C1C(C1(C)C)CCC2=C |
| Arogyavardhini | Azadirachta indica | Leaf | CID_44715635 | Nimbinene | IMPPAT | COC(=O)C[C@H]1[C@@]2(C)[C@H](OC3C2=C(C)[C@@H](C3)c2cocc2)[C@@H](C2[C@]1(C)C(=O)CC=C2C)OC(=O)C |
| Arogyavardhini | Azadirachta indica | Leaf | CID_521710 | alpha-Patchoulene | IMPPAT | CC1CCC23C1CC(C3(C)C)CC=C2C |
| Arogyavardhini | Azadirachta indica | Leaf | CID_5280343 | Quercetin | IMPPAT | Oc1cc(O)c2c(c1)oc(c(c2=O)O)c1ccc(c(c1)O)O |
| Arogyavardhini | Azadirachta indica | Leaf | CID_5280460 | Scopoletin | IMPPAT | COc1cc2ccc(=O)oc2cc1O |
| Arogyavardhini | Azadirachta indica | Leaf | CID_5280489 | beta-Carotene | IMPPAT | C/C(=C\C=C\C=C(\C=C\C=C(\C=C\C1=C(C)CCCC1(C)C)/C)/C)/C=C/C=C(/C=C/C1=C(C)CCCC1(C)C)\C |
| Arogyavardhini | Azadirachta indica | Leaf | CID_5280794 | Stigmasterol | IMPPAT | CC[C@@H](C(C)C)/C=C/[C@H]([C@H]1CC[C@@H]2[C@]1(C)CC[C@H]1[C@H]2CC=C2[C@]1(C)CC[C@@H](C2)O)C |
| Arogyavardhini | Azadirachta indica | Leaf | CID_5281243 | Lutein | IMPPAT | C/C(=C\C=C\C=C(\C=C\C=C(\C=C\C1=C(C)C[C@H](CC1(C)C)O)/C)/C)/C=C/C=C(/C=C/[C@H]1C(=C[C@@H](CC1(C)C)O)C)\C |
| Arogyavardhini | Azadirachta indica | Leaf | CID_5281515 | beta-Caryophyllene | IMPPAT | C/C/1=C\CCC(=C)[C@@H]2[C@@H](CC1)C(C2)(C)C |
| Arogyavardhini | Azadirachta indica | Leaf | CID_5281519 | Germacrene B | IMPPAT | C/C/1=C\CC/C(=C/CC(=C(C)C)CC1)/C |
| Arogyavardhini | Azadirachta indica | Leaf | CID_5281520 | Humulene | IMPPAT | C/C/1=C\CC(C)(C)/C=C/C/C(=C/CC1)/C |
| Arogyavardhini | Azadirachta indica | Leaf | CID_5281654 | Isorhamnetin | IMPPAT | COc1cc(ccc1O)c1oc2cc(O)cc(c2c(=O)c1O)O |
| Arogyavardhini | Azadirachta indica | Leaf | CID_5742590 | Daucosterol | IMPPAT | CC[C@@H](C(C)C)CC[C@H]([C@H]1CC[C@@H]2[C@]1(C)CC[C@H]1[C@H]2CC=C2[C@]1(C)CC[C@@H](C2)O[C@@H]1O[C@H](CO)[C@H]([C@@H]([C@H]1O)O)O)C |
| Arogyavardhini | Azadirachta indica | Leaf | CID_6432312 | gamma-Elemene | IMPPAT | C=C[C@]1(C)CCC(=C(C)C)C[C@H]1C(=C)C |
| Arogyavardhini | Azadirachta indica | Leaf | CID_6442906 | [(4S,5R,6R,11R,12R,15S,16R,17Z)-17-(2,2-dihydroxyethylidene)-5-hydroxy-3,7,7,11,15-pentamethyl-8,18-dioxo-19-oxapentacyclo[14.3.1.02,15.03,12.06,11]icosa-1,9-dien-4-yl] acetate | IMPPAT | CC(=O)O[C@@H]1[C@H](O)[C@@H]2[C@]([C@@H]3C1(C)C1=C4OC(=O)/C(=C\C(O)O)/[C@@H]([C@@]1(CC3)C)C4)(C)C=CC(=O)C2(C)C |
| Arogyavardhini | Azadirachta indica | Leaf | CID_68171 | 1-Hexacosanol | IMPPAT | CCCCCCCCCCCCCCCCCCCCCCCCCO |
| Arogyavardhini | Azadirachta indica | Leaf | CID_6918391 | beta-Elemene | IMPPAT | C=C[C@]1(C)CC[C@H](C[C@H]1C(=C)C)C(=C)C |
| Arogyavardhini | Azadirachta indica | Leaf | CID_6918774 | Corosolic acid | IMPPAT | C[C@@H]1CC[C@]2([C@@H]([C@H]1C)C1=CC[C@H]3[C@@]([C@@]1(CC2)C)(C)CC[C@@H]1[C@]3(C)C[C@@H](O)[C@@H](C1(C)C)O)C(=O)O |
| Arogyavardhini | Azadirachta indica | Leaf | CID_70678558 | alpha-Copaene | IMPPAT | CC([C@@H]1CC[C@]2([C@@H]3[C@H]1C2C(=CC3)C)C)C |
| Arogyavardhini | Azadirachta indica | Leaf | CID_71597391 | Triterpenoids | IMPPAT | O[C@H]1C[C@@]2(C)C(=CC[C@@]3([C@@H]2CC=C2[C@@]3(C)CC[C@@]3([C@H]2CC(C)(C)CC3)C(=O)O)C)[C@]([C@H]1O)(C)O |
| Arogyavardhini | Azadirachta indica | Leaf | CID_76316558 | Isomeldenin | IMPPAT | CC(=O)OC1[C@H](O)[C@@H]2[C@]([C@@H]3[C@]1(C)C1=CC[C@H]([C@@]1(CC3)C)c1ccoc1)(C)CCC(=O)C2(C)C |
| Arogyavardhini | Azadirachta indica | Leaf | CID_91723653 | (S,1Z,6Z)-8-Isopropyl-1-methyl-5-methylenecyclodeca-1,6-diene | IMPPAT | C/C/1=C/CCC(=C)/C=C\[C@@H](CC1)C(C)C |
| Arogyavardhini | Azadirachta indica | Leaf | CID_91886694 | 2',3'-Dehydrosalannol | IMPPAT | COC(=O)C[C@@H]1[C@@]2(C)[C@@H](OC(=O)C=C(C)C)C[C@H]([C@@]3([C@@H]2[C@H]([C@@H]2[C@@]1(C)C1=C(C)[C@@H](C[C@H]1O2)c1ccoc1)OC3)C)O |
| Arogyavardhini | Azadirachta indica | Leaf | CID_984 | Hexadecanal | IMPPAT | CCCCCCCCCCCCCCCC=O |
| Arogyavardhini | Commiphora wightii | Plant exudate | CID_101297673 | Guggulsterol I | IMPPAT | CC(CC[C@H](C([C@H]1[C@@H](O)C[C@@H]2[C@]1(C)CC[C@H]1[C@H]2CCC2=CC(=O)CC[C@]12C)(O)C)O)C |
| Arogyavardhini | Commiphora wightii | Plant exudate | CID_101297674 | Guggulsterol II | IMPPAT | CC(CCCC([C@H]1[C@@H](O)C[C@@H]2[C@]1(C)CC[C@H]1[C@H]2CC=C2[C@]1(C)CC[C@@H](C2)O)(O)C)C |

**Table S2: Phytochemical profiles of constituent herbs in the two formulations. The table provides a comprehensive mapping of formulations, constituent herbs, and corresponding plant parts to their associated phytochemical constituents. For each phytochemical, the table includes chemical identifiers, compound names, IMPPAT/API references, and canonical SMILES representations.**

| Formulation | Herb name | Part | Chemical identifier | Chemical name | Source | Canonical SMILES |
|---|---|---|---|---|---|---|
| Arogyavardhini | Commiphora wightii | Plant exudate | CID_101297675 | Guggulsterol III | IMPPAT | CC(CCCC([C@H]1CC[C@@H]2[C@]1(C)CC[C@H]1[C@H]2C[C@H](C2=CC(=O)CC[C@]12C)O)(O)C)C |
| Arogyavardhini | Commiphora wightii | Plant exudate | CID_101750 | alpha-Camphorene | IMPPAT | CC(=CCCC1=CCC(CC1)C(=C)CCC=C(C)C)C |
| Arogyavardhini | Commiphora wightii | Plant exudate | CID_11747713 | Thunbergene | IMPPAT | C/C/1=C/CC/C(=C/C/C=C(\C=C\[C@@H](CC1)C(C)C)/C)/C |
| Arogyavardhini | Commiphora wightii | Plant exudate | CID_122173119 | (8S,9R,10R,13R,14S,17Z)-17-ethylidene-10,13-dimethyl-1,2,6,7,8,9,11,12,14,15-decahydrocyclopenta[a]phenanthrene-3,16-dione | IMPPAT | C/C=C/1\C(=O)C[C@@H]2[C@@]1(C)CC[C@@H]1[C@@H]2CCC2=CC(=O)CC[C@]12C |
| Arogyavardhini | Commiphora wightii | Plant exudate | CID_13255923 | Mukulol | IMPPAT | C/C/1=C\CC/C(=C/[C@H]([C@@H](CC/C(=C/CC1)/C)C(C)C)O)/C |
| Arogyavardhini | Commiphora wightii | Plant exudate | CID_161109 | 20-Hydroxypregn-4-en-3-one | IMPPAT | O=C1CC[C@]2(C(=C1)CC[C@@H]1[C@@H]2CC[C@]2([C@H]1CC[C@@H]2C(O)C)C)C |
| Arogyavardhini | Commiphora wightii | Plant exudate | CID_222284 | beta-Sitosterol | IMPPAT | CC[C@@H](C(C)C)CC[C@H]([C@H]1CC[C@@H]2[C@]1(C)CC[C@H]1[C@H]2CC=C2[C@]1(C)CC[C@@H](C2)O)C |
| Arogyavardhini | Commiphora wightii | Plant exudate | CID_31253 | Myrcene | IMPPAT | C=CC(=C)CCC=C(C)C |
| Arogyavardhini | Commiphora wightii | Plant exudate | CID_445858 | Ferulic acid | IMPPAT | COc1cc(/C=C/C(=O)O)ccc1O |
| Arogyavardhini | Commiphora wightii | Plant exudate | CID_5204 | 5-[3-(1,3-Benzodioxol-5-yl)-1,3,3a,4,6,6a-hexahydrofuro[3,4-c]furan-6-yl]-1,3-benzodioxole | IMPPAT | C1Oc2c(O1)cc(cc2)C1OCC2C1COC2c1ccc2c(c1)OCO2 |
| Arogyavardhini | Commiphora wightii | Plant exudate | CID_5280794 | Stigmasterol | IMPPAT | CC[C@@H](C(C)C)/C=C/[C@H]([C@H]1CC[C@@H]2[C@]1(C)CC[C@H]1[C@H]2CC=C2[C@]1(C)CC[C@@H](C2)O)C |
| Arogyavardhini | Commiphora wightii | Plant exudate | CID_5281384 | Neocembrene | IMPPAT | C/C/1=C\C[C@@H](CC/C(=C/CC/C(=C/CC1)/C)/C)C(=C)C |
| Arogyavardhini | Commiphora wightii | Plant exudate | CID_5281515 | beta-Caryophyllene | IMPPAT | C/C/1=C\CCC(=C)[C@@H]2[C@@H](CC1)C(C2)(C)C |
| Arogyavardhini | Commiphora wightii | Plant exudate | CID_5368823 | (2E,6Z,10E)-3,7,11-trimethyl-14-propan-2-ylcyclotetradeca-2,6,10-trien-1-ol | IMPPAT | C/C/1=C\CC/C(=C\CC/C(=C/C(C(CC1)C(C)C)O)/C)/C |
| Arogyavardhini | Commiphora wightii | Plant exudate | CID_5997 | Cholesterol | IMPPAT | CC(CCC[C@H]([C@H]1CC[C@@H]2[C@]1(C)CC[C@H]1[C@H]2CC=C2[C@]1(C)CC[C@@H](C2)O)C)C |
| Arogyavardhini | Commiphora wightii | Plant exudate | CID_6439929 | E-Guggulsterone | IMPPAT | C/C=C\1/C(=O)C[C@@H]2[C@]1(C)CC[C@H]1[C@H]2CCC2=CC(=O)CC[C@]12C |
| Arogyavardhini | Commiphora wightii | Plant exudate | CID_68972 | 1-Triacontanol | IMPPAT | CCCCCCCCCCCCCCCCCCCCCCCCCCCCCCO |
| Arogyavardhini | Commiphora wightii | Plant exudate | CID_70695727 | Pluviatilol | IMPPAT | COc1cc(ccc1O)[C@H]1OC[C@@H]2[C@H]1CO[C@H]2c1ccc2c(c1)OCO2 |
| Arogyavardhini | Commiphora wightii | Plant exudate | CID_92747 | (20R)-20-Hydroxypregn-4-en-3-one | IMPPAT | O=C1CC[C@]2(C(=C1)CC[C@@H]1[C@@H]2CC[C@]2([C@H]1CC[C@@H]2[C@H](O)C)C)C |
| Arogyavardhini | Phyllanthus emblica | Fruit | CID_1057 | Pyrogallol | IMPPAT | Oc1c(O)cccc1O |
| Arogyavardhini | Phyllanthus emblica | Fruit | CID_12314211 | Phyllantidine | API & IMPPAT | O=C1C=C2[C@@]3(O1)C[C@@H](C=C2)ON1[C@H]3CCCC1 |
| Arogyavardhini | Phyllanthus emblica | Fruit | CID_124021 | beta-Glucogallin | IMPPAT | OC[C@H]1O[C@@H](OC(=O)c2cc(O)c(c(c2)O)O)[C@@H]([C@H]([C@@H]1O)O)O |
| Arogyavardhini | Phyllanthus emblica | Fruit | CID_13250 | Ethyl gallate | IMPPAT | CCOC(=O)c1cc(O)c(c(c1)O)O |
| Arogyavardhini | Phyllanthus emblica | Fruit | CID_193552 | Phyllanthine | API | CO[C@@H]1CCN2[C@H](C1)C13OC(=O)C=C3C=C[C@@H]2C1 |
| Arogyavardhini | Phyllanthus emblica | Fruit | CID_289 | Catechol | IMPPAT | Oc1ccccc1O |
| Arogyavardhini | Phyllanthus emblica | Fruit | CID_359 | Phloroglucinol | IMPPAT | Oc1cc(O)cc(c1)O |
| Arogyavardhini | Phyllanthus emblica | Fruit | CID_370 | Gallic acid | API & IMPPAT | OC(=O)c1cc(O)c(c(c1)O)O |
| Arogyavardhini | Phyllanthus emblica | Fruit | CID_449093 | trans-Zeatin | IMPPAT | OC/C(=C/CNc1ncnc2c1[nH]cn2)/C |
| Arogyavardhini | Phyllanthus emblica | Fruit | CID_493570 | Riboflavin | IMPPAT | OC[C@H]([C@H]([C@H](Cn1c2-c(nc3c1cc(C)c(c3)C)c(=O)[nH]c(=O)n2)O)O)O |
| Arogyavardhini | Phyllanthus emblica | Fruit | CID_5280343 | Quercetin | IMPPAT | Oc1cc(O)c2c(c1)oc(c(c2=O)O)c1ccc(c(c1)O)O |
| Arogyavardhini | Phyllanthus emblica | Fruit | CID_5281855 | Ellagic acid | API & IMPPAT | Oc1cc2c(=O)oc3c4c2c(c1O)oc(=O)c4cc(c3O)O |
| Arogyavardhini | Phyllanthus emblica | Fruit | CID_54670067 | Ascorbic acid | API & IMPPAT | OC[C@@H]([C@H]1OC(=O)C(=C1O)O)O |
| Arogyavardhini | Phyllanthus emblica | Fruit | CID_7428 | Methyl gallate | IMPPAT | COC(=O)c1cc(O)c(c(c1)O)O |
| Arogyavardhini | Picrorhiza kurroa | Rhizome | CID_14081907 | Bartsioside | IMPPAT | OCC1=CC[C@@H]2[C@H]1[C@@H](OC=C2)O[C@@H]1O[C@H](CO)[C@H]([C@@H]([C@H]1O)O)O |
| Arogyavardhini | Picrorhiza kurroa | Rhizome | CID_155942 | 8-Epiiridotrial glucoside | IMPPAT | O=CC1=CO[C@H]([C@H]2[C@@H]1CC[C@H]2C)O[C@@H]1O[C@H](CO)[C@H]([C@@H]([C@H]1O)O)O |
| Arogyavardhini | Picrorhiza kurroa | Rhizome | CID_164648 | Androsin | API | OC[C@H]1O[C@@H](Oc2ccc(cc2OC)C(=O)C)[C@@H]([C@H]([C@@H]1O)O)O |
| Arogyavardhini | Picrorhiza kurroa | Rhizome | CID_2214 | Acetovanillone | API & IMPPAT | COc1cc(ccc1O)C(=O)C |
| Arogyavardhini | Picrorhiza kurroa | Rhizome | CID_443353 | 10-Deoxygeniposidic acid | IMPPAT | OC[C@H]1O[C@@H](O[C@@H]2OC=C([C@@H]3[C@H]2C(=CC3)C)C(=O)O)[C@@H]([C@H]([C@@H]1O)O)O |
| Arogyavardhini | Picrorhiza kurroa | Rhizome | CID_6440892 | Picroside I | API | OC[C@@]12O[C@H]2[C@H]([C@H]2[C@@H]1[C@@H](OC=C2)O[C@@H]1O[C@H](COC(=O)/C=C/c2ccccc2)[C@H]([C@@H]([C@H]1O)O)O)O |
| Arogyavardhini | Picrorhiza kurroa | Rhizome | CID_8468 | Vanillic acid | API | COc1cc(ccc1O)C(=O)O |
| Arogyavardhini | Picrorhiza kurroa | Rhizome | CID_91458 | Aucubin | IMPPAT | OCC1=C[C@H]([C@H]2[C@@H]1[C@@H](OC=C2)O[C@@H]1O[C@H](CO)[C@H]([C@@H]([C@H]1O)O)O)O |
| Arogyavardhini | Picrorhiza kurroa | Rhizome | CID_91520 | Catalpol | IMPPAT | OC[C@H]1O[C@@H](O[C@@H]2OC=C[C@@H]3[C@H]2[C@@]2(CO)O[C@H]2[C@@H]3O)[C@@H]([C@H]([C@@H]1O)O)O |
| Arogyavardhini | Picrorhiza kurroa | Rhizome | CID_92123 | Picein | API | OC[C@H]1O[C@@H](Oc2ccc(cc2)C(=O)C)[C@@H]([C@H]([C@@H]1O)O)O |
| Arogyavardhini | Picrorhiza kurroa | Rhizome | CID_ 119287 | cucurbitacins | API | C[C@H]1[C@H]2[C@@H](C[C@@]3([C@@]2(CC(=O)[C@@]4([C@H]3CC=C5[C@H]4C=C(C(=O)C5(C)C)O)C)C)C)O[C@@H](CC1=O)C(C)(C)O |
| Arogyavardhini | Plumbago zeylanica | Root | CID_100947536 | Isozeylanone | IMPPAT | CC1=C(CC2=CC(=O)c3c(C2=O)cccc3O)C(=O)c2c(C1=O)cccc2O |
| Arogyavardhini | Plumbago zeylanica | Root | CID_100947539 | Plumbazeylanone | IMPPAT | O=C1c2cccc(c2C(=O)C2([C@H]1C[C@@]1([C@@]2(C)C(=O)c2c(C1=O)c(O)ccc2)C)C1=C(C)C(=O)c2c(C1=O)c(O)ccc2)O |
| Arogyavardhini | Plumbago zeylanica | Root | CID_10205 | Plumbagin | API & IMPPAT | CC1=CC(=O)c2c(C1=O)cccc2O |
| Arogyavardhini | Plumbago zeylanica | Root | CID_146680 | (2,2'-Binaphthalene)-5,5',8,8'-tetrone, 1,1'-dihydroxy-6,6'-dimethyl- | IMPPAT | O=C1C=C(C)C(=O)c2c1c(O)c(cc2)c1ccc2c(c1O)C(=O)C=C(C2=O)C |
| Arogyavardhini | Plumbago zeylanica | Root | CID_271296 | Methylnaphthazarin | IMPPAT | CC1=CC(=O)c2c(C1=O)c(O)ccc2O |
| Arogyavardhini | Plumbago zeylanica | Root | CID_2723872 | D-Fructose | IMPPAT | OCC1(O)OC[C@H]([C@H]([C@@H]1O)O)O |
| Arogyavardhini | Plumbago zeylanica | Root | CID_338719 | 3-Chloro-5-hydroxy-2-methylnaphthalene-1,4-dione | IMPPAT | CC1=C(Cl)C(=O)c2c(C1=O)cccc2O |
| Arogyavardhini | Plumbago zeylanica | Root | CID_442739 | Droserone | IMPPAT | CC1=C(O)c2cccc(c2C(=O)C1=O)O |
| Arogyavardhini | Plumbago zeylanica | Root | CID_5276618 | Zeylanone | IMPPAT | Oc1cccc2c1C(=O)C1=C(C2=O)CC2C1(C)C(=O)c1c(C2=O)c(O)ccc1 |
| Arogyavardhini | Plumbago zeylanica | Root | CID_5793 | D-Glucose | IMPPAT | OC[C@H]1OC(O)[C@@H]([C@H]([C@@H]1O)O)O |
| Arogyavardhini | Plumbago zeylanica | Root | CID_633024 | 5-Hydroxy-8-(4-hydroxy-7-methyl-5,8-dioxonaphthalen-1-yl)-2-methylnaphthalene-1,4-dione | IMPPAT | CC1=CC(=O)c2c(C1=O)c(ccc2O)c1ccc(c2c1C(=O)C(=CC2=O)C)O |
| Arogyavardhini | Plumbago zeylanica | Root | CID_633072 | Chitranone | IMPPAT | O=C1C(=CC(=O)c2c1ccc(c2O)C1=C(C)C(=O)c2c(C1=O)c(O)ccc2)C |
| Arogyavardhini | Terminalia bellirica | Fruit | CID_124021 | beta-Glucogallin | IMPPAT | OC[C@H]1O[C@@H](OC(=O)c2cc(O)c(c(c2)O)O)[C@@H]([C@H]([C@@H]1O)O)O |
| Arogyavardhini | Terminalia bellirica | Fruit | CID_12411 | Tritriacontane | IMPPAT | CCCCCCCCCCCCCCCCCCCCCCCCCCCCCCCCC |
| Arogyavardhini | Terminalia bellirica | Fruit | CID_12444386 | Arjungenin | IMPPAT | OC[C@]1(C)[C@@H](O)[C@H](O)C[C@]2([C@H]1CC[C@@]1([C@@H]2CC=C2[C@@]1(C)CC[C@@]1([C@H]2[C@H](O)C(CC1)(C)C)C(=O)O)C)C |
| Arogyavardhini | Terminalia bellirica | Fruit | CID_129688132 | Tritriacontan-9-one | IMPPAT | CCCCCCCCCCCCCCCCCCCCCCCC(=O)CCCCCCCC |
| Arogyavardhini | Terminalia bellirica | Fruit | CID_13250 | Ethyl gallate | API & IMPPAT | CCOC(=O)c1cc(O)c(c(c1)O)O |

**Table S2: Phytochemical profiles of constituent herbs in the two formulations. The table provides a comprehensive mapping of formulations, constituent herbs, and corresponding plant parts to their associated phytochemical constituents. For each phytochemical, the table includes chemical identifiers, compound names, IMPPAT/API references, and canonical SMILES representations.**

| Formulation | Herb name | Part | Chemical identifier | Chemical name | Source | Canonical SMILES |
|---|---|---|---|---|---|---|
| Arogyavardhini | Terminalia bellirica | Fruit | CID_14105933 | (4aS,6aR,6aS,6bR,8aR,10R,11R,12aR,14bS)-10,11-dihydroxy-9,9-bis(hydroxymethyl)-2,2,6a,6b,12a-pentamethyl-1,3,4,5,6,6a,7,8,8a,10,11,12,13,14b-tetradecahydropicene-4a-carboxylic acid | IMPPAT | OCC1(CO)[C@@H](O)[C@H](O)C[C@]2([C@H]1CC[C@@]1([C@@H]2CC=C2[C@@]1(C)CC[C@@]1([C@H]2CC(CC1)(C)C)C(=O)O)C)C |
| Arogyavardhini | Terminalia bellirica | Fruit | CID_21726781 | (4aS,6aR,6aR,6bR,7R,8aR,9R,10R,11R,12aR,14bS)-7,10,11-trihydroxy-9-(hydroxymethyl)-2,2,6a,6b,9,12a-hexamethyl-1,3,4,5,6,6a,7,8,8a,10,11,12,13,14b-tetradecahydropicene-4a-carboxylic acid | API & IMPPAT | OC[C@]1(C)[C@@H](O)[C@H](O)C[C@]2([C@H]1C[C@@H](O)[C@@]1([C@@H]2CC=C2[C@@]1(C)CC[C@@]1([C@H]2CC(CC1)(C)C)C(=O)O)C)C |
| Arogyavardhini | Terminalia bellirica | Fruit | CID_222284 | beta-Sitosterol | IMPPAT | CC[C@@H](C(C)C)CC[C@H]([C@H]1CC[C@@H]2[C@]1(C)CC[C@H]1[C@H]2CC=C2[C@]1(C)CC[C@@H](C2)O)C |
| Arogyavardhini | Terminalia bellirica | Fruit | CID_25310 | L-Rhamnose | IMPPAT | O[C@H]1[C@H](C)OC([C@@H]([C@@H]1O)O)O |
| Arogyavardhini | Terminalia bellirica | Fruit | CID_26519 | Tetratriacontane | IMPPAT | CCCCCCCCCCCCCCCCCCCCCCCCCCCCCCCCCC |
| Arogyavardhini | Terminalia bellirica | Fruit | CID_2723872 | D-Fructose | IMPPAT | OCC1(O)OC[C@H]([C@H]([C@@H]1O)O)O |
| Arogyavardhini | Terminalia bellirica | Fruit | CID_370 | Gallic acid | API & IMPPAT | OC(=O)c1cc(O)c(c(c1)O)O |
| Arogyavardhini | Terminalia bellirica | Fruit | CID_466076 | Termilignan | API & IMPPAT | COc1ccc(c(c1)O)CC(=C)C(=C)Cc1ccc(cc1)O |
| Arogyavardhini | Terminalia bellirica | Fruit | CID_466078 | 7-Hydroxy-3',4'-methylenedioxyflavan | IMPPAT | Oc1ccc2c(c1)O[C@@H](CC2)c1ccc2c(c1)OCO2 |
| Arogyavardhini | Terminalia bellirica | Fruit | CID_5281855 | Ellagic acid | API & IMPPAT | Oc1cc2c(=O)oc3c4c2c(c1O)oc(=O)c4cc(c3O)O |
| Arogyavardhini | Terminalia bellirica | Fruit | CID_5793 | D-Glucose | IMPPAT | OC[C@H]1OC(O)[C@@H]([C@H]([C@@H]1O)O)O |
| Arogyavardhini | Terminalia bellirica | Fruit | CID_6036 | D-Galactose | IMPPAT | OC[C@H]1OC(O)[C@@H]([C@H]([C@H]1O)O)O |
| Arogyavardhini | Terminalia bellirica | Fruit | CID_6251 | Mannitol | IMPPAT | OC[C@H]([C@H]([C@@H]([C@@H](CO)O)O)O)O |
| Arogyavardhini | Terminalia bellirica | Fruit | CID_72388 | Anolignan B | IMPPAT | C=C(C(=C)Cc1ccc(cc1)O)Cc1ccc(cc1)O |
| Arogyavardhini | Terminalia chebula | Fruit | CID_10467 | Arachidic acid | IMPPAT | CCCCCCCCCCCCCCCCCCCC(=O)O |
| Arogyavardhini | Terminalia chebula | Fruit | CID_10471 | Triacontanoic acid | IMPPAT | CCCCCCCCCCCCCCCCCCCCCCCCCCCCCC(=O)O |
| Arogyavardhini | Terminalia chebula | Fruit | CID_10742 | Syringic acid | API & IMPPAT | COc1cc(cc(c1O)OC)C(=O)O |
| Arogyavardhini | Terminalia chebula | Fruit | CID_12444386 | Arjungenin | IMPPAT | OC[C@]1(C)[C@@H](O)[C@H](O)C[C@]2([C@H]1CC[C@@]1([C@@H]2CC=C2[C@@]1(C)CC[C@@]1([C@H]2[C@H](O)C(CC1)(C)C)C(=O)O)C)C |
| Arogyavardhini | Terminalia chebula | Fruit | CID_13250 | Ethyl gallate | IMPPAT | CCOC(=O)c1cc(O)c(c(c1)O)O |
| Arogyavardhini | Terminalia chebula | Fruit | CID_222284 | beta-Sitosterol | IMPPAT | CC[C@@H](C(C)C)CC[C@H]([C@H]1CC[C@@H]2[C@]1(C)CC[C@H]1[C@H]2CC=C2[C@]1(C)CC[C@@H](C2)O)C |
| Arogyavardhini | Terminalia chebula | Fruit | CID_370 | Gallic acid | API & IMPPAT | OC(=O)c1cc(O)c(c(c1)O)O |
| Arogyavardhini | Terminalia chebula | Fruit | CID_445639 | Oleic acid | IMPPAT | CCCCCCCC/C=C\CCCCCCCC(=O)O |
| Arogyavardhini | Terminalia chebula | Fruit | CID_445858 | Ferulic acid | IMPPAT | COc1cc(/C=C/C(=O)O)ccc1O |
| Arogyavardhini | Terminalia chebula | Fruit | CID_5280450 | Linoleic acid | IMPPAT | CCCCC/C=C\C/C=C\CCCCCCCC(=O)O |
| Arogyavardhini | Terminalia chebula | Fruit | CID_5281 | Stearic acid | IMPPAT | CCCCCCCCCCCCCCCCCC(=O)O |
| Arogyavardhini | Terminalia chebula | Fruit | CID_5281855 | Ellagic acid | API & IMPPAT | Oc1cc2c(=O)oc3c4c2c(c1O)oc(=O)c4cc(c3O)O |
| Arogyavardhini | Terminalia chebula | Fruit | CID_54670067 | Ascorbic acid | IMPPAT | OC[C@@H]([C@H]1OC(=O)C(=C1O)O)O |
| Arogyavardhini | Terminalia chebula | Fruit | CID_5742590 | Daucosterol | IMPPAT | CC[C@@H](C(C)C)CC[C@H]([C@H]1CC[C@@H]2[C@]1(C)CC[C@H]1[C@H]2CC=C2[C@]1(C)CC[C@@H](C2)O[C@@H]1O[C@H](CO)[C@H]([C@@H]([C@H]1O)O)O)C |
| Arogyavardhini | Terminalia chebula | Fruit | CID_637542 | 4-Hydroxycinnamic acid | IMPPAT | OC(=O)/C=C/c1ccc(cc1)O |
| Arogyavardhini | Terminalia chebula | Fruit | CID_6780 | Anthraquinone | API & IMPPAT | O=C1c2ccccc2C(=O)c2c1cccc2 |
| Arogyavardhini | Terminalia chebula | Fruit | CID_689043 | Caffeic acid | IMPPAT | OC(=O)/C=C/c1ccc(c(c1)O)O |
| Arogyavardhini | Terminalia chebula | Fruit | CID_73641 | Arjunolic acid | IMPPAT | OC[C@]1(C)[C@@H](O)[C@H](O)C[C@]2([C@H]1CC[C@@]1([C@@H]2CC=C2[C@@]1(C)CC[C@@]1([C@H]2CC(CC1)(C)C)C(=O)O)C)C |
| Arogyavardhini | Terminalia chebula | Fruit | CID_73659 | Maslinic acid | IMPPAT | O[C@@H]1C[C@@]2(C)[C@H](C([C@H]1O)(C)C)CC[C@@]1([C@@H]2CC=C2[C@@]1(C)CC[C@@]1([C@H]2CC(C)(C)CC1)C(=O)O)C |
| Arogyavardhini | Terminalia chebula | Fruit | CID_8215 | Docosanoic acid | IMPPAT | CCCCCCCCCCCCCCCCCCCCCC(=O)O |
| Arogyavardhini | Terminalia chebula | Fruit | CID_8468 | Vanillic acid | IMPPAT | COc1cc(ccc1O)C(=O)O |
| Arogyavardhini | Terminalia chebula | Fruit | CID_8742 | Shikimic acid | IMPPAT | O[C@@H]1CC(=C[C@H]([C@H]1O)O)C(=O)O |
| Arogyavardhini | Terminalia chebula | Fruit | CID_985 | Palmitic acid | IMPPAT | CCCCCCCCCCCCCCCC(=O)O |
| Nishamalaki | Curcuma longa | Rhizome | CHEMSPIDER_16737047 | Ar-dihydro-turmerone | IMPPAT | CC(CC(=O)C[C@@H](c1ccc(cc1)C)C)C |
| Nishamalaki | Curcuma longa | Rhizome | CID_10104370 | beta-Bisabolene | IMPPAT | CC(=CCCC(=C)[C@H]1CCC(=CC1)C)C |
| Nishamalaki | Curcuma longa | Rhizome | CID_101297706 | Humuladienone | IMPPAT | C/C/1=C/CC(C)(C)/C=C\CC(C(=O)CC1)C |
| Nishamalaki | Curcuma longa | Rhizome | CID_101680 | Camphene hydrate | IMPPAT | CC1(C)C2CCC(C1(C)O)C2 |
| Nishamalaki | Curcuma longa | Rhizome | CID_101926857 | Bisabola-3,10-diene-2-one | IMPPAT | CC(=CCCC(C1CC=C(C(=O)C1)C)C)C |
| Nishamalaki | Curcuma longa | Rhizome | CID_102024 | 2-Cyclohexen-1-ol, 2-methyl-5-(1-methylethenyl)-, acetate, cis- | IMPPAT | CC(=O)O[C@@H]1C[C@@H](CC=C1C)C(=C)C |
| Nishamalaki | Curcuma longa | Rhizome | CID_10248 | Elemicin | IMPPAT | C=CCc1cc(OC)c(c(c1)OC)OC |
| Nishamalaki | Curcuma longa | Rhizome | CID_10263440 | Epiprocurcumenol | IMPPAT | CC1=CC(=O)C(=C(C)C)C[C@H]2[C@@H]1CC[C@]2(C)O |
| Nishamalaki | Curcuma longa | Rhizome | CID_10364 | Carvacrol | IMPPAT | CC(c1ccc(c(c1)O)C)C |
| Nishamalaki | Curcuma longa | Rhizome | CID_10421034 | 1-Bisabolone | IMPPAT | CC(=CCC[C@@H]([C@H]1CCC(=CC1=O)C)C)C |
| Nishamalaki | Curcuma longa | Rhizome | CID_10429233 | (1E)-1,7-bis(4-hydroxy-3-methoxyphenyl)hept-1-ene-3,5-dione | IMPPAT | COc1cc(CCC(=O)CC(=O)/C=C/c2ccc(c(c2)OC)O)ccc1O |
| Nishamalaki | Curcuma longa | Rhizome | CID_10469828 | (1E,4E)-1-(4-Hydroxy-3-methoxyphenyl)-5-(4-hydroxyphenyl)-1,4-pentadiene-3-one | IMPPAT | COc1cc(/C=C/C(=O)/C=C/c2ccc(cc2)O)ccc1O |
| Nishamalaki | Curcuma longa | Rhizome | CID_10582 | Myrtenol | IMPPAT | OCC1=CCC2CC1C2(C)C |
| Nishamalaki | Curcuma longa | Rhizome | CID_10703 | O-Cymene | IMPPAT | CC(c1ccccc1C)C |
| Nishamalaki | Curcuma longa | Rhizome | CID_10704181 | Humulene epoxide II | IMPPAT | C/C/1=C\CC(C)(C)/C=C/C[C@@]2([C@@H](CC1)O2)C |
| Nishamalaki | Curcuma longa | Rhizome | CID_10856614 | alpha-Selinene | IMPPAT | CC1=CCC[C@]2([C@H]1C[C@@H](CC2)C(=C)C)C |
| Nishamalaki | Curcuma longa | Rhizome | CID_10955433 | 2-Hepten-4-one, 6-(2-hydroxy-4-methylphenyl)-2-methyl- | IMPPAT | CC(=CC(=O)CC(c1ccc(cc1O)C)C)C |
| Nishamalaki | Curcuma longa | Rhizome | CID_10976 | 2-Heptanol | IMPPAT | CCCCCC(O)C |
| Nishamalaki | Curcuma longa | Rhizome | CID_11116492 | 2-Furanmethanol, 5-ethenyltetrahydro-alpha,alpha,5-trimethyl-, cis- | IMPPAT | C=C[C@@]1(C)CC[C@H](O1)C(O)(C)C |
| Nishamalaki | Curcuma longa | Rhizome | CID_11230 | 4-Carvomenthenol | IMPPAT | CC1=CCC(CC1)(O)C(C)C |
| Nishamalaki | Curcuma longa | Rhizome | CID_1140 | Toluene | IMPPAT | Cc1ccccc1 |
| Nishamalaki | Curcuma longa | Rhizome | CID_11463 | Terpinolene | IMPPAT | CC1=CCC(=C(C)C)CC1 |
| Nishamalaki | Curcuma longa | Rhizome | CID_11467 | gamma-Terpineol | IMPPAT | CC(=C1CCC(CC1)(C)O)C |
| Nishamalaki | Curcuma longa | Rhizome | CID_11586487 | beta-Himachalene | IMPPAT | CC1=C[C@H]2C(=C(C)CCCC2(C)C)CC1 |
| Nishamalaki | Curcuma longa | Rhizome | CID_11716 | 3-Buten-2-OL | IMPPAT | CC(C=C)O |
| Nishamalaki | Curcuma longa | Rhizome | CID_11996452 | Viridiflorol | IMPPAT | C[C@@H]1CC[C@H]2[C@@H]1[C@H]1[C@H](C1(C)C)CC[C@]2(C)O |
| Nishamalaki | Curcuma longa | Rhizome | CID_122130015 | Ethene;2-methoxyphenol;phenol;propanedial | IMPPAT | Oc1ccccc1.COc1ccccc1O.O=CCC=O.C=C.C=C |
| Nishamalaki | Curcuma longa | Rhizome | CID_12233170 | cis-Carvotanacetol | IMPPAT | CC([C@H]1CC=C([C@H](C1)O)C)C |

**Table S2: Phytochemical profiles of constituent herbs in the two formulations. The table provides a comprehensive mapping of formulations, constituent herbs, and corresponding plant parts to their associated phytochemical constituents. For each phytochemical, the table includes chemical identifiers, compound names, IMPPAT/API references, and canonical SMILES representations.**

| Formulation | Herb name | Part | Chemical identifier | Chemical name | Source | Canonical SMILES |
|---|---|---|---|---|---|---|
| Nishamalaki | Curcuma longa | Rhizome | CID_122484 | 2-Cyclohexen-1-ol, 1-methyl-4-(1-methylethyl)-, trans- | IMPPAT | CC([C@@H]1CC[C@@](C=C1)(C)O)C |
| Nishamalaki | Curcuma longa | Rhizome | CID_12299867 | alpha-Atlantone | IMPPAT | CC(=CC(=O)/C=C(\[C@@H]1CCC(=CC1)C)/C)C |
| Nishamalaki | Curcuma longa | Rhizome | CID_12300148 | 6-Epi-beta-bisabolol | IMPPAT | CC(=CCC[C@@H]([C@@]1(O)CCC(=CC1)C)C)C |
| Nishamalaki | Curcuma longa | Rhizome | CID_12302222 | tau-Cadinol | IMPPAT | CC1=C[C@H]2[C@H](CC1)[C@](C)(O)CC[C@H]2C(C)C |
| Nishamalaki | Curcuma longa | Rhizome | CID_12306047 | alpha-Muurolene | IMPPAT | CC1=C[C@@H]2[C@H](CC1)C(=CC[C@H]2C(C)C)C |
| Nishamalaki | Curcuma longa | Rhizome | CID_12306048 | (+)-alpha-Cadinene | IMPPAT | CC1=C[C@@H]2[C@@H](CC1)C(=CC[C@H]2C(C)C)C |
| Nishamalaki | Curcuma longa | Rhizome | CID_12309449 | delta-Elemene | IMPPAT | C=C[C@@]1(C)CCC(=C[C@@H]1C(=C)C)C(C)C |
| Nishamalaki | Curcuma longa | Rhizome | CID_12310886 | (1S,2S,5S,8S)-2-methyl-6-methylidene-9-propan-2-ylidene-11-oxatricyclo[6.2.1.01,5]undecan-8-ol | IMPPAT | CC(=C1C[C@]23O[C@@]1(O)CC(=C)[C@@H]3CC[C@@H]2C)C |
| Nishamalaki | Curcuma longa | Rhizome | CID_12367 | 2-Nonanol | IMPPAT | CCCCCCCC(O)C |
| Nishamalaki | Curcuma longa | Rhizome | CID_12388 | Tridecane | IMPPAT | CCCCCCCCCCCCC |
| Nishamalaki | Curcuma longa | Rhizome | CID_12389 | Tetradecane | IMPPAT | CCCCCCCCCCCCCC |
| Nishamalaki | Curcuma longa | Rhizome | CID_14014430 | (-)-beta-Curcumene | IMPPAT | CC(=CCC[C@H](C1=CCC(=CC1)C)C)C |
| Nishamalaki | Curcuma longa | Rhizome | CID_14191392 | (6Z)-6,10-dimethyl-3-propan-2-ylidenecyclodec-6-ene-1,4-dione | IMPPAT | C/C/1=C/CCC(C)C(=O)CC(=C(C)C)C(=O)C1 |
| Nishamalaki | Curcuma longa | Rhizome | CID_14240392 | Curcumol | API & IMPPAT | CC([C@@H]1C[C@]23O[C@]1(O)CC(=C)[C@@H]3CC[C@@H]2C)C |
| Nishamalaki | Curcuma longa | Rhizome | CID_14254 | 2-Decanol | IMPPAT | CCCCCCCCC(O)C |
| Nishamalaki | Curcuma longa | Rhizome | CID_14287397 | Bisacurone | IMPPAT | CC(=CC(=O)C[C@@H]([C@@H]1C=C[C@]([C@H](C1)O)(C)O)C)C |
| Nishamalaki | Curcuma longa | Rhizome | CID_14529 | 2-(4-Methylphenyl)propan-2-ol | IMPPAT | Cc1ccc(cc1)C(O)(C)C |
| Nishamalaki | Curcuma longa | Rhizome | CID_14543198 | Isoprocurcumenol | IMPPAT | C=C1CC(=O)C(=C(C)C)C[C@H]2[C@H]1CC[C@]2(C)O |
| Nishamalaki | Curcuma longa | Rhizome | CID_14632996 | alpha-Turmerone | API & IMPPAT | CC(=CC(=O)CC(C1CC=C(C=C1)C)C)C |
| Nishamalaki | Curcuma longa | Rhizome | CID_14633011 | Procurcumadiol | IMPPAT | CC1=CC(=O)C(=C(C)C)CC2(C1CCC2(C)O)O |
| Nishamalaki | Curcuma longa | Rhizome | CID_14896 | beta-Pinene | IMPPAT | C=C1CCC2CC1C2(C)C |
| Nishamalaki | Curcuma longa | Rhizome | CID_153845 | Bicyclo(4.1.0)heptan-3-one, 1-methyl-4-(1-methylethylidene)-7-(3-oxobutyl)-, (1S,6R,7R)- | IMPPAT | CC(=O)CC[C@@H]1[C@@H]2[C@@]1(C)CC(=O)C(=C(C)C)C2 |
| Nishamalaki | Curcuma longa | Rhizome | CID_15448 | 2-Undecanol | IMPPAT | CCCCCCCCCC(O)C |
| Nishamalaki | Curcuma longa | Rhizome | CID_1549026 | Geranyl acetate | IMPPAT | C/C(=C\COC(=O)C)/CCC=C(C)C |
| Nishamalaki | Curcuma longa | Rhizome | CID_1550890 | Cinnamyl cinnamate | IMPPAT | O=C(/C=C/c1ccccc1)OC/C=C/c1ccccc1 |
| Nishamalaki | Curcuma longa | Rhizome | CID_15858385 | Turmeronol A | IMPPAT | CC(=CC(=O)CC(c1ccc(c(c1)O)C)C)C |
| Nishamalaki | Curcuma longa | Rhizome | CID_160512 | ar-Turmerone | API & IMPPAT | CC(=CC(=O)C[C@@H](c1ccc(cc1)C)C)C |
| Nishamalaki | Curcuma longa | Rhizome | CID_167812 | (+)-Curcumenol | IMPPAT | CC(=C1C[C@]23O[C@]1(O)C=C([C@@H]3CC[C@@H]2C)C)C |
| Nishamalaki | Curcuma longa | Rhizome | CID_17100 | alpha-Terpineol | IMPPAT | CC1=CCC(CC1)C(O)(C)C |
| Nishamalaki | Curcuma longa | Rhizome | CID_173183 | Campesterol | IMPPAT | O[C@H]1CC[C@]2(C(=CC[C@@H]3[C@@H]2CC[C@]2([C@H]3CC[C@@H]2[C@@H](CC[C@H](C(C)C)C)C)C)C1)C |
| Nishamalaki | Curcuma longa | Rhizome | CID_1742210 | Caryophyllene oxide | IMPPAT | C=C1CC[C@H]2O[C@@]2(CC[C@@H]2[C@@H]1CC2(C)C)C |
| Nishamalaki | Curcuma longa | Rhizome | CID_176983 | p-Mentha-1,3,8-triene | IMPPAT | CC1=CC=C(CC1)C(=C)C |
| Nishamalaki | Curcuma longa | Rhizome | CID_18818 | Sabinene | IMPPAT | C=C1CCC2(C1C2)C(C)C |
| Nishamalaki | Curcuma longa | Rhizome | CID_189061 | Procurcumenol | IMPPAT | CC1=CC(=O)C(=C(C)C)C[C@H]2[C@H]1CC[C@]2(C)O |
| Nishamalaki | Curcuma longa | Rhizome | CID_196216 | Curlone | API & IMPPAT | CC(C1CCC(=C)C=C1)CC(=O)C=C(C)C |
| Nishamalaki | Curcuma longa | Rhizome | CID_20055538 | Turmerol | IMPPAT | CC(=CC(CC(C1=CCC(=CC1)C)C)O)C |
| Nishamalaki | Curcuma longa | Rhizome | CID_20083 | 2-Octanol | IMPPAT | CCCCCCC(O)C |
| Nishamalaki | Curcuma longa | Rhizome | CID_20928 | 1-Hexen-3-OL | IMPPAT | CCCC(C=C)O |
| Nishamalaki | Curcuma longa | Rhizome | CID_222284 | beta-Sitosterol | IMPPAT | CC[C@@H](C(C)C)CC[C@H]([C@H]1CC[C@@H]2[C@]1(C)CC[C@H]1[C@H]2CC=C2[C@]1(C)CC[C@@H](C2)O)C |
| Nishamalaki | Curcuma longa | Rhizome | CID_22311 | Limonene | IMPPAT | CC1=CCC(CC1)C(=C)C |
| Nishamalaki | Curcuma longa | Rhizome | CID_247573 | Isobornyl acetate | IMPPAT | CC(=O)O[C@H]1C[C@@H]2C([C@]1(C)CC2)(C)C |
| Nishamalaki | Curcuma longa | Rhizome | CID_24834047 | 4,10-Epizedoarondiol | IMPPAT | CC(=C1C[C@H]2[C@H]([C@@](CC1=O)(C)O)CC[C@@]2(C)O)C |
| Nishamalaki | Curcuma longa | Rhizome | CID_25202482 | Sesquisabinene | IMPPAT | CC(=CCCC(C12CCC(=C)C2C1)C)C |
| Nishamalaki | Curcuma longa | Rhizome | CID_2537 | Camphor | IMPPAT | O=C1CC2C(C1(C)CC2)(C)C |
| Nishamalaki | Curcuma longa | Rhizome | CID_26049 | 3-Carene | IMPPAT | CC1=CCC2C(C1)C2(C)C |
| Nishamalaki | Curcuma longa | Rhizome | CID_2758 | Eucalyptol | IMPPAT | CC12CCC(CC1)C(O2)(C)C |
| Nishamalaki | Curcuma longa | Rhizome | CID_28930 | alpha-Fenchene | IMPPAT | C=C1CC2C(C1CC2)(C)C |
| Nishamalaki | Curcuma longa | Rhizome | CID_2969 | Decanoic acid | IMPPAT | CCCCCCCCCC(=O)O |
| Nishamalaki | Curcuma longa | Rhizome | CID_3081930 | Curzerenone | IMPPAT | C=C[C@@]1(C)Cc2occ(c2C(=O)[C@@H]1C(=C)C)C |
| Nishamalaki | Curcuma longa | Rhizome | CID_31253 | Myrcene | IMPPAT | C=CC(=C)CCC=C(C)C |
| Nishamalaki | Curcuma longa | Rhizome | CID_31291 | Tetradecanal | IMPPAT | CCCCCCCCCCCCCC=O |
| Nishamalaki | Curcuma longa | Rhizome | CID_325 | 4-Isopropylbenzyl alcohol | IMPPAT | OCc1ccc(cc1)C(C)C |
| Nishamalaki | Curcuma longa | Rhizome | CID_330573 | (-)-cis-Carveol | IMPPAT | CC(=C)[C@@H]1CC=C([C@@H](C1)O)C |
| Nishamalaki | Curcuma longa | Rhizome | CID_3314 | Eugenol | IMPPAT | C=CCc1ccc(c(c1)OC)O |
| Nishamalaki | Curcuma longa | Rhizome | CID_360253 | Curcuphenol | IMPPAT | CC(=CCC[C@H](c1ccc(cc1O)C)C)C |
| Nishamalaki | Curcuma longa | Rhizome | CID_42608158 | Allo-Aromadendrene | IMPPAT | C[C@@H]1CC[C@H]2[C@@H]1C1C(C1(C)C)CCC2=C |
| Nishamalaki | Curcuma longa | Rhizome | CID_439711 | alpha-Fenchol | IMPPAT | O[C@H]1[C@@]2(C)CC[C@@H](C1(C)C)C2 |
| Nishamalaki | Curcuma longa | Rhizome | CID_441005 | (+)-delta-Cadinene | IMPPAT | CC1=C[C@@H]2C(=C(C)CC[C@H]2C(C)C)CC1 |
| Nishamalaki | Curcuma longa | Rhizome | CID_442343 | Levomenol | IMPPAT | CC(=CCC[C@@]([C@H]1CCC(=CC1)C)(O)C)C |
| Nishamalaki | Curcuma longa | Rhizome | CID_442484 | (+)-beta-Phellandrene | IMPPAT | CC([C@@H]1CCC(=C)C=C1)C |
| Nishamalaki | Curcuma longa | Rhizome | CID_442495 | Pulegone | IMPPAT | C[C@@H]1CCC(=C(C)C)C(=O)C1 |
| Nishamalaki | Curcuma longa | Rhizome | CID_445070 | Farnesol | IMPPAT | OC/C=C(/CC/C=C(/CCC=C(C)C)\C)\C |
| Nishamalaki | Curcuma longa | Rhizome | CID_5144 | Safrole | IMPPAT | C=CCc1ccc2c(c1)OCO2 |
| Nishamalaki | Curcuma longa | Rhizome | CID_519764 | 3-(1,5-Dimethyl-4-hexenyl)-6-methylene-1-cyclohexene | IMPPAT | CC(C1CCC(=C)C=C1)CCC=C(C)C |
| Nishamalaki | Curcuma longa | Rhizome | CID_520453 | Bisabolane | IMPPAT | CC(CCCC(C1CCC(CC1)C)C)C |
| Nishamalaki | Curcuma longa | Rhizome | CID_522296 | 3,7(11)-Eudesmadiene | IMPPAT | CC1=CCCC2(C1CC(=C(C)C)CC2)C |
| Nishamalaki | Curcuma longa | Rhizome | CID_5280794 | Stigmasterol | IMPPAT | CC[C@@H](C(C)C)/C=C/[C@H]([C@H]1CC[C@@H]2[C@]1(C)CC[C@H]1[C@H]2CC=C2[C@]1(C)CC[C@@H](C2)O)C |
| Nishamalaki | Curcuma longa | Rhizome | CID_5281167 | cis-3-Hexen-1-ol | IMPPAT | OCC/C=C\CC |
| Nishamalaki | Curcuma longa | Rhizome | CID_5281515 | beta-Caryophyllene | IMPPAT | C/C/1=C\CCC(=C)[C@@H]2[C@@H](CC1)C(C2)(C)C |

**Table S2: Phytochemical profiles of constituent herbs in the two formulations. The table provides a comprehensive mapping of formulations, constituent herbs, and corresponding plant parts to their associated phytochemical constituents. For each phytochemical, the table includes chemical identifiers, compound names, IMPPAT/API references, and canonical SMILES representations.**

| Formulation | Herb name | Part | Chemical identifier | Chemical name | Source | Canonical SMILES |
|---|---|---|---|---|---|---|
| Nishamalaki | Curcuma longa | Rhizome | CID_5281517 | beta-Farnesene | IMPPAT | C=CC(=C)CC/C=C(/CCC=C(C)C)\C |
| Nishamalaki | Curcuma longa | Rhizome | CID_5281519 | Germacrene B | IMPPAT | C/C/1=C\CC/C(=C/CC(=C(C)C)CC1)/C |
| Nishamalaki | Curcuma longa | Rhizome | CID_5281520 | Humulene | IMPPAT | C/C/1=C\CC(C)(C)/C=C/C/C(=C/CC1)/C |
| Nishamalaki | Curcuma longa | Rhizome | CID_5281553 | (E)-beta-ocimene | IMPPAT | C=C/C(=C/CC=C(C)C)/C |
| Nishamalaki | Curcuma longa | Rhizome | CID_5282109 | Geranyl formate | IMPPAT | O=COC/C=C(/CCC=C(C)C)\C |
| Nishamalaki | Curcuma longa | Rhizome | CID_5284507 | Nerolidol | IMPPAT | C=CC(CC/C=C(/CCC=C(C)C)\C)(O)C |
| Nishamalaki | Curcuma longa | Rhizome | CID_530426 | Isoledene | IMPPAT | CC1CCC2C(C3=C1CCC3C)C2(C)C |
| Nishamalaki | Curcuma longa | Rhizome | CID_5315469 | Bisacumol | IMPPAT | CC(=CC(CC(c1ccc(cc1)C)C)O)C |
| Nishamalaki | Curcuma longa | Rhizome | CID_5315472 | Bisdemethoxycurcumin | IMPPAT | O=C(CC(=O)/C=C/c1ccc(cc1)O)/C=C/c1ccc(cc1)O |
| Nishamalaki | Curcuma longa | Rhizome | CID_5317319 | cis-beta-Farnesene | IMPPAT | C=CC(=C)CC/C=C(\CCC=C(C)C)/C |
| Nishamalaki | Curcuma longa | Rhizome | CID_5317844 | alpha-Guaiene | IMPPAT | CC(=C)[C@@H]1CC[C@@H](C2=C(C1)[C@@H](C)CC2)C |
| Nishamalaki | Curcuma longa | Rhizome | CID_5320250 | (Z)-beta-Ocimene | IMPPAT | C=C/C(=C\CC=C(C)C)/C |
| Nishamalaki | Curcuma longa | Rhizome | CID_5355856 | Geranyl butyrate | IMPPAT | CCCC(=O)OC/C=C(/CCC=C(C)C)\C |
| Nishamalaki | Curcuma longa | Rhizome | CID_5365992 | Geranyl hexanoate | IMPPAT | CCCCCC(=O)OC/C=C(/CCC=C(C)C)\C |
| Nishamalaki | Curcuma longa | Rhizome | CID_5469424 | Demethoxycurcumin | IMPPAT | COc1cc(/C=C/C(=O)CC(=O)/C=C/c2ccc(cc2)O)ccc1O |
| Nishamalaki | Curcuma longa | Rhizome | CID_572766 | Curzerene | IMPPAT | C=CC1(C)Cc2occ(c2CC1C(=C)C)C |
| Nishamalaki | Curcuma longa | Rhizome | CID_57386731 | delta-Curcumene | IMPPAT | CC(=CCC[C@@H](C1CC=C(C=C1)C)C)C |
| Nishamalaki | Curcuma longa | Rhizome | CID_5997 | Cholesterol | IMPPAT | CC(CCC[C@H]([C@H]1CC[C@@H]2[C@]1(C)CC[C@H]1[C@H]2CC=C2[C@]1(C)CC[C@@H](C2)O)C)C |
| Nishamalaki | Curcuma longa | Rhizome | CID_61060 | d-Borneol | IMPPAT | O[C@@H]1C[C@H]2C([C@@]1(C)CC2)(C)C |
| Nishamalaki | Curcuma longa | Rhizome | CID_61098 | Linalyl propionate | IMPPAT | CCC(=O)OC(CCC=C(C)C)(C=C)C |
| Nishamalaki | Curcuma longa | Rhizome | CID_61130 | Myrtenal | IMPPAT | O=CC1=CCC2CC1C2(C)C |
| Nishamalaki | Curcuma longa | Rhizome | CID_62385 | 1-Methyl-4-(prop-1-en-2-yl)benzene | IMPPAT | Cc1ccc(cc1)C(=C)C |
| Nishamalaki | Curcuma longa | Rhizome | CID_637563 | Anethole | IMPPAT | C/C=C/c1ccc(cc1)OC |
| Nishamalaki | Curcuma longa | Rhizome | CID_637566 | Geraniol | IMPPAT | OC/C=C(/CCC=C(C)C)\C |
| Nishamalaki | Curcuma longa | Rhizome | CID_638011 | Citral | IMPPAT | O=C/C=C(/CCC=C(C)C)\C |
| Nishamalaki | Curcuma longa | Rhizome | CID_638024 | Piperine | IMPPAT | O=C(N1CCCCC1)/C=C/C=C/c1ccc2c(c1)OCO2 |
| Nishamalaki | Curcuma longa | Rhizome | CID_6428435 | 7-epi-cis-Sesquisabinene hydrate | IMPPAT | CC(=CCCC([C@]12CC[C@@](C2C1)(C)O)C)C |
| Nishamalaki | Curcuma longa | Rhizome | CID_6428444 | trans-Sesquisabinene hydrate | IMPPAT | CC(=CCCC([C@]12CC[C@](C2C1)(C)O)C)C |
| Nishamalaki | Curcuma longa | Rhizome | CID_6429302 | trans-alpha-Bergamotene | IMPPAT | CC(=CCC[C@]1(C)[C@H]2CC=C([C@@H]1C2)C)C |
| Nishamalaki | Curcuma longa | Rhizome | CID_6430754 | 10-epi-gamma-Eudesmol | IMPPAT | CC1=C2C[C@@H](CC[C@@]2(CCC1)C)C(O)(C)C |
| Nishamalaki | Curcuma longa | Rhizome | CID_6432254 | trans-Linalool oxide | IMPPAT | C=C[C@]1(C)CC[C@H](O1)C(O)(C)C |
| Nishamalaki | Curcuma longa | Rhizome | CID_6432312 | gamma-Elemene | IMPPAT | C=C[C@]1(C)CCC(=C(C)C)C[C@H]1C(=C)C |
| Nishamalaki | Curcuma longa | Rhizome | CID_6436348 | Germacrone | IMPPAT | C/C/1=C\CC(=C(C)C)C(=O)C/C(=C/CC1)/C |
| Nishamalaki | Curcuma longa | Rhizome | CID_643779 | Neral | IMPPAT | O=C/C=C(\CCC=C(C)C)/C |
| Nishamalaki | Curcuma longa | Rhizome | CID_6441391 | Germacr-1(10)-ene-5,8-dione | IMPPAT | C/C/1=C\CC[C@H](C)C(=O)C[C@H](C(=O)C1)C(C)C |
| Nishamalaki | Curcuma longa | Rhizome | CID_6442374 | Furanodienon | IMPPAT | C/C/1=C/C(=O)c2c(C/C(=C\CC1)/C)occ2C |
| Nishamalaki | Curcuma longa | Rhizome | CID_6451618 | (1R)-2-methyl-5-propan-2-ylbicyclo[3.1.0]hex-2-ene | IMPPAT | CC1=CCC2([C@@H]1C2)C(C)C |
| Nishamalaki | Curcuma longa | Rhizome | CID_6532 | Linalyl isobutyrate | IMPPAT | C=CC(OC(=O)C(C)C)(CCC=C(C)C)C |
| Nishamalaki | Curcuma longa | Rhizome | CID_6549 | Linalool | IMPPAT | C=CC(CCC=C(C)C)(O)C |
| Nishamalaki | Curcuma longa | Rhizome | CID_6616 | Camphene | IMPPAT | C=C1C2CCC(C1(C)C)C2 |
| Nishamalaki | Curcuma longa | Rhizome | CID_6654 | alpha-Pinene | IMPPAT | CC1=CCC2CC1C2(C)C |
| Nishamalaki | Curcuma longa | Rhizome | CID_68252 | Thymol acetate | IMPPAT | CC(=O)Oc1cc(C)ccc1C(C)C |
| Nishamalaki | Curcuma longa | Rhizome | CID_68381 | Perilla ketone | IMPPAT | CC(CCC(=O)c1cocc1)C |
| Nishamalaki | Curcuma longa | Rhizome | CID_6918391 | beta-Elemene | IMPPAT | C=C[C@]1(C)CC[C@H](C[C@H]1C(=C)C)C(=C)C |
| Nishamalaki | Curcuma longa | Rhizome | CID_69879809 | Cyclocurcumin | IMPPAT | COc1cc(/C=C/C2=CC(=O)CC(O2)c2ccc(c(c2)OC)O)ccc1O |
| Nishamalaki | Curcuma longa | Rhizome | CID_6989 | Thymol | IMPPAT | Cc1ccc(c(c1)O)C(C)C |
| Nishamalaki | Curcuma longa | Rhizome | CID_7127 | Methyleugenol | IMPPAT | C=CCc1ccc(c(c1)OC)OC |
| Nishamalaki | Curcuma longa | Rhizome | CID_71346280 | 1,7-Bis(4-hydroxyphenyl)-1,4,6-heptatrien-3-one | IMPPAT | O=C(C=Cc1ccc(cc1)O)C=CC=Cc1ccc(cc1)O |
| Nishamalaki | Curcuma longa | Rhizome | CID_7439 | Carvone | IMPPAT | CC(=C)C1CC=C(C(=O)C1)C |
| Nishamalaki | Curcuma longa | Rhizome | CID_7460 | alpha-Phellandrene | IMPPAT | CC1=CCC(C=C1)C(C)C |
| Nishamalaki | Curcuma longa | Rhizome | CID_7461 | gamma-Terpinene | IMPPAT | CC1=CCC(=CC1)C(C)C |
| Nishamalaki | Curcuma longa | Rhizome | CID_7462 | alpha-Terpinene | IMPPAT | CC1=CC=C(CC1)C(C)C |
| Nishamalaki | Curcuma longa | Rhizome | CID_7463 | p-Cymene | IMPPAT | Cc1ccc(cc1)C(C)C |
| Nishamalaki | Curcuma longa | Rhizome | CID_79035 | Tricyclene | IMPPAT | CC12C3C1CC(C2(C)C)C3 |
| Nishamalaki | Curcuma longa | Rhizome | CID_80420 | Heptyl salicylate | IMPPAT | CCCCCCCOC(=O)c1ccccc1O |
| Nishamalaki | Curcuma longa | Rhizome | CID_8051 | 2-Heptanone | IMPPAT | CCCCCC(=O)C |
| Nishamalaki | Curcuma longa | Rhizome | CID_8163 | 2-Undecanone | IMPPAT | CCCCCCCCCC(=O)C |
| Nishamalaki | Curcuma longa | Rhizome | CID_8184 | 1-Undecanol | IMPPAT | CCCCCCCCCCCO |
| Nishamalaki | Curcuma longa | Rhizome | CID_8294 | Linalyl acetate | IMPPAT | C=CC(OC(=O)C)(CCC=C(C)C)C |
| Nishamalaki | Curcuma longa | Rhizome | CID_8500 | 4'-Methylacetophenone | IMPPAT | Cc1ccc(cc1)C(=O)C |
| Nishamalaki | Curcuma longa | Rhizome | CID_87014 | 2-(Hydroxymethyl)anthraquinone | IMPPAT | OCc1ccc2c(c1)C(=O)c1c(C2=O)cccc1 |
| Nishamalaki | Curcuma longa | Rhizome | CID_91457 | beta-Eudesmol | IMPPAT | C=C1CCC[C@]2([C@H]1C[C@@H](CC2)C(O)(C)C)C |
| Nishamalaki | Curcuma longa | Rhizome | CID_91723653 | (S,1Z,6Z)-8-Isopropyl-1-methyl-5-methylenecyclodeca-1,6-diene | IMPPAT | C/C/1=C/CCC(=C)/C=C\[C@@H](CC1)C(C)C |
| Nishamalaki | Curcuma longa | Rhizome | CID_91753231 | (4S,5S)-Germacrone-4,5-epoxide | IMPPAT | CC1=CCC[C@@]2(C)O[C@H]2CC(=C(C)C)C(=O)C1 |
| Nishamalaki | Curcuma longa | Rhizome | CID_91753502 | cis-alpha-Bergamotene | IMPPAT | CC(=CCCC1(C)[C@@H]2CC=C([C@H]1C2)C)C |
| Nishamalaki | Curcuma longa | Rhizome | CID_92139 | alpha-Curcumene | IMPPAT | CC(=CCCC(c1ccc(cc1)C)C)C |
| Nishamalaki | Curcuma longa | Rhizome | CID_92776 | Zingiberene | IMPPAT | CC(=CCC[C@@H]([C@H]1CC=C(C=C1)C)C)C |
| Nishamalaki | Curcuma longa | Rhizome | CID_93135 | Xanthorrhizol | IMPPAT | CC(=CCC[C@H](c1ccc(c(c1)O)C)C)C |
| Nishamalaki | Curcuma longa | Rhizome | CID_94147 | (+)-cis-Sabinol | IMPPAT | CC([C@]12C[C@@H]2C(=C)[C@@H](C1)O)C |
| Nishamalaki | Curcuma longa | Rhizome | CID_94266 | Sabinyl acetate | IMPPAT | CC(=O)OC1CC2(C(C1=C)C2)C(C)C |

**Table S2: Phytochemical profiles of constituent herbs in the two formulations. The table provides a comprehensive mapping of formulations, constituent herbs, and corresponding plant parts to their associated phytochemical constituents. For each phytochemical, the table includes chemical identifiers, compound names, IMPPAT/API references, and canonical SMILES representations.**

| Formulation | Herb name | Part | Chemical identifier | Chemical name | Source | Canonical SMILES |
|---|---|---|---|---|---|---|
| Nishamalaki | Curcuma longa | Rhizome | CID_969516 | Curcumin | API & IMPPAT | COc1cc(/C=C/C(=O)CC(=O)/C=C/c2ccc(c(c2)OC)O)ccc1O |
| Nishamalaki | Phyllanthus emblica | Fruit | CID_1057 | Pyrogallol | IMPPAT | Oc1c(O)cccc1O |
| Nishamalaki | Phyllanthus emblica | Fruit | CID_12314211 | Phyllantidine | API & IMPPAT | O=C1C=C2[C@@]3(O1)C[C@@H](C=C2)ON1[C@H]3CCCC1 |
| Nishamalaki | Phyllanthus emblica | Fruit | CID_124021 | beta-Glucogallin | IMPPAT | OC[C@H]1O[C@@H](OC(=O)c2cc(O)c(c(c2)O)O)[C@@H]([C@H]([C@@H]1O)O)O |
| Nishamalaki | Phyllanthus emblica | Fruit | CID_13250 | Ethyl gallate | IMPPAT | CCOC(=O)c1cc(O)c(c(c1)O)O |
| Nishamalaki | Phyllanthus emblica | Fruit | CID_193552 | Phyllanthine | API | CO[C@@H]1CCN2[C@H](C1)C13OC(=O)C=C3C=C[C@@H]2C1 |
| Nishamalaki | Phyllanthus emblica | Fruit | CID_289 | Catechol | IMPPAT | Oc1ccccc1O |
| Nishamalaki | Phyllanthus emblica | Fruit | CID_359 | Phloroglucinol | IMPPAT | Oc1cc(O)cc(c1)O |
| Nishamalaki | Phyllanthus emblica | Fruit | CID_370 | Gallic acid | API & IMPPAT | OC(=O)c1cc(O)c(c(c1)O)O |
| Nishamalaki | Phyllanthus emblica | Fruit | CID_449093 | trans-Zeatin | IMPPAT | OC/C(=C/CNc1ncnc2c1[nH]cn2)/C |
| Nishamalaki | Phyllanthus emblica | Fruit | CID_493570 | Riboflavin | IMPPAT | OC[C@H]([C@H]([C@H](Cn1c2-c(nc3c1cc(C)c(c3)C)c(=O)[nH]c(=O)n2)O)O)O |
| Nishamalaki | Phyllanthus emblica | Fruit | CID_5280343 | Quercetin | IMPPAT | Oc1cc(O)c2c(c1)oc(c(c2=O)O)c1ccc(c(c1)O)O |
| Nishamalaki | Phyllanthus emblica | Fruit | CID_5281855 | Ellagic acid | API & IMPPAT | Oc1cc2c(=O)oc3c4c2c(c1O)oc(=O)c4cc(c3O)O |
| Nishamalaki | Phyllanthus emblica | Fruit | CID_54670067 | Ascorbic acid | API & IMPPAT | OC[C@@H]([C@H]1OC(=O)C(=C1O)O)O |
| Nishamalaki | Phyllanthus emblica | Fruit | CID_7428 | Methyl gallate | IMPPAT | COC(=O)c1cc(O)c(c(c1)O)O |

**Table S3: Associations between formulations, constituent plants, phytochemicals, and human molecular targets. The table lists the formulations, plant name and plant part, phytochemical identifiers, associated human target genes (Entrez Gene ID and gene symbol), and the database sources supporting each phytochemical-target interaction.**

| Formulation | Plant name | Part | PubChem CID | Entrez Gene ID | Gene Symbol | Source database(s) |
|---|---|---|---|---|---|---|
| Arogyavardhini | Azadirachta indica | Leaf | CID_14458886 | 3320 | HSP90AA1 | BindingDB \| ChEMBL \| NPASS |
| Arogyavardhini | Azadirachta indica | Leaf | CID_222284 | 10013 | HDAC6 | ChEMBL |
| Arogyavardhini | Azadirachta indica | Leaf | CID_222284 | 10599 | SLCO1B1 | ChEMBL \| NPASS |
| Arogyavardhini | Azadirachta indica | Leaf | CID_222284 | 1565 | CYP2D6 | ChEMBL \| NPASS |
| Arogyavardhini | Azadirachta indica | Leaf | CID_222284 | 1576 | CYP3A4 | BindingDB \| ChEMBL \| NPASS |
| Arogyavardhini | Azadirachta indica | Leaf | CID_222284 | 1803 | DPP4 | ChEMBL \| NPASS |
| Arogyavardhini | Azadirachta indica | Leaf | CID_222284 | 2147 | F2 | ChEMBL \| NPASS |
| Arogyavardhini | Azadirachta indica | Leaf | CID_222284 | 28234 | SLCO1B3 | ChEMBL \| NPASS |
| Arogyavardhini | Azadirachta indica | Leaf | CID_222284 | 2932 | GSK3B | ChEMBL |
| Arogyavardhini | Azadirachta indica | Leaf | CID_222284 | 3417 | IDH1 | ChEMBL \| NPASS |
| Arogyavardhini | Azadirachta indica | Leaf | CID_222284 | 51422 | PRKAG2 | ChEMBL |
| Arogyavardhini | Azadirachta indica | Leaf | CID_222284 | 5243 | ABCB1 | ChEMBL |
| Arogyavardhini | Azadirachta indica | Leaf | CID_222284 | 53632 | PRKAG3 | ChEMBL |
| Arogyavardhini | Azadirachta indica | Leaf | CID_222284 | 5423 | POLB | BindingDB \| ChEMBL \| NPASS |
| Arogyavardhini | Azadirachta indica | Leaf | CID_222284 | 5562 | PRKAA1 | ChEMBL |
| Arogyavardhini | Azadirachta indica | Leaf | CID_222284 | 5563 | PRKAA2 | ChEMBL |
| Arogyavardhini | Azadirachta indica | Leaf | CID_222284 | 5564 | PRKAB1 | ChEMBL |
| Arogyavardhini | Azadirachta indica | Leaf | CID_222284 | 5565 | PRKAB2 | ChEMBL |
| Arogyavardhini | Azadirachta indica | Leaf | CID_222284 | 5571 | PRKAG1 | ChEMBL |
| Arogyavardhini | Azadirachta indica | Leaf | CID_222284 | 7299 | TYR | ChEMBL \| NPASS |
| Arogyavardhini | Azadirachta indica | Leaf | CID_5280343 | 10013 | HDAC6 | ChEMBL |
| Arogyavardhini | Azadirachta indica | Leaf | CID_5280343 | 10072 | DPP3 | BindingDB |
| Arogyavardhini | Azadirachta indica | Leaf | CID_5280343 | 1017 | CDK2 | BindingDB \| ChEMBL \| NPASS |
| Arogyavardhini | Azadirachta indica | Leaf | CID_5280343 | 1019 | CDK4 | BindingDB \| ChEMBL |
| Arogyavardhini | Azadirachta indica | Leaf | CID_5280343 | 1020 | CDK5 | ChEMBL |
| Arogyavardhini | Azadirachta indica | Leaf | CID_5280343 | 1021 | CDK6 | BindingDB \| ChEMBL \| NPASS |
| Arogyavardhini | Azadirachta indica | Leaf | CID_5280343 | 10257 | ABCC4 | BindingDB \| ChEMBL \| NPASS |
| Arogyavardhini | Azadirachta indica | Leaf | CID_5280343 | 10298 | PAK4 | ChEMBL \| NPASS |
| Arogyavardhini | Azadirachta indica | Leaf | CID_5280343 | 10494 | STK25 | ChEMBL \| NPASS |
| Arogyavardhini | Azadirachta indica | Leaf | CID_5280343 | 10599 | SLCO1B1 | ChEMBL \| NPASS |
| Arogyavardhini | Azadirachta indica | Leaf | CID_5280343 | 10645 | CAMKK2 | ChEMBL \| NPASS |
| Arogyavardhini | Azadirachta indica | Leaf | CID_5280343 | 106821730 | BUB1B-PAK6 | ChEMBL |
| Arogyavardhini | Azadirachta indica | Leaf | CID_5280343 | 10733 | PLK4 | ChEMBL \| NPASS |
| Arogyavardhini | Azadirachta indica | Leaf | CID_5280343 | 10783 | NEK6 | BindingDB \| ChEMBL \| NPASS |
| Arogyavardhini | Azadirachta indica | Leaf | CID_5280343 | 1080 | CFTR | ChEMBL |
| Arogyavardhini | Azadirachta indica | Leaf | CID_5280343 | 10919 | EHMT2 | ChEMBL \| NPASS |
| Arogyavardhini | Azadirachta indica | Leaf | CID_5280343 | 10951 | CBX1 | NPASS |
| Arogyavardhini | Azadirachta indica | Leaf | CID_5280343 | 11040 | PIM2 | ChEMBL \| NPASS |
| Arogyavardhini | Azadirachta indica | Leaf | CID_5280343 | 1111 | CHEK1 | ChEMBL \| NPASS |
| Arogyavardhini | Azadirachta indica | Leaf | CID_5280343 | 11200 | CHEK2 | ChEMBL \| NPASS |
| Arogyavardhini | Azadirachta indica | Leaf | CID_5280343 | 11201 | POLI | ChEMBL \| NPASS |
| Arogyavardhini | Azadirachta indica | Leaf | CID_5280343 | 11238 | CA5B | BindingDB \| ChEMBL \| NPASS |
| Arogyavardhini | Azadirachta indica | Leaf | CID_5280343 | 11255 | HRH3 | ChEMBL |
| Arogyavardhini | Azadirachta indica | Leaf | CID_5280343 | 1128 | CHRM1 | ChEMBL |
| Arogyavardhini | Azadirachta indica | Leaf | CID_5280343 | 1129 | CHRM2 | ChEMBL |
| Arogyavardhini | Azadirachta indica | Leaf | CID_5280343 | 11309 | SLCO2B1 | ChEMBL \| NPASS |
| Arogyavardhini | Azadirachta indica | Leaf | CID_5280343 | 1131 | CHRM3 | BindingDB \| ChEMBL |
| Arogyavardhini | Azadirachta indica | Leaf | CID_5280343 | 11329 | STK38 | ChEMBL \| NPASS |
| Arogyavardhini | Azadirachta indica | Leaf | CID_5280343 | 1133 | CHRM5 | BindingDB |
| Arogyavardhini | Azadirachta indica | Leaf | CID_5280343 | 11330 | CTRC | BindingDB \| ChEMBL \| NPASS |
| Arogyavardhini | Azadirachta indica | Leaf | CID_5280343 | 1137 | CHRNA4 | ChEMBL |
| Arogyavardhini | Azadirachta indica | Leaf | CID_5280343 | 118429 | ANTXR2 | NPASS |
| Arogyavardhini | Azadirachta indica | Leaf | CID_5280343 | 1195 | CLK1 | ChEMBL \| NPASS |
| Arogyavardhini | Azadirachta indica | Leaf | CID_5280343 | 1196 | CLK2 | ChEMBL \| NPASS |
| Arogyavardhini | Azadirachta indica | Leaf | CID_5280343 | 1198 | CLK3 | ChEMBL \| NPASS |
| Arogyavardhini | Azadirachta indica | Leaf | CID_5280343 | 1233 | CCR4 | ChEMBL \| NPASS |
| Arogyavardhini | Azadirachta indica | Leaf | CID_5280343 | 1244 | ABCC2 | ChEMBL \| NPASS |
| Arogyavardhini | Azadirachta indica | Leaf | CID_5280343 | 1268 | CNR1 | ChEMBL |
| Arogyavardhini | Azadirachta indica | Leaf | CID_5280343 | 134 | ADORA1 | ChEMBL |
| Arogyavardhini | Azadirachta indica | Leaf | CID_5280343 | 135 | ADORA2A | ChEMBL |
| Arogyavardhini | Azadirachta indica | Leaf | CID_5280343 | 1394 | CRHR1 | ChEMBL |
| Arogyavardhini | Azadirachta indica | Leaf | CID_5280343 | 1395 | CRHR2 | ChEMBL |
| Arogyavardhini | Azadirachta indica | Leaf | CID_5280343 | 140 | ADORA3 | BindingDB \| ChEMBL \| NPASS |
| Arogyavardhini | Azadirachta indica | Leaf | CID_5280343 | 1432 | MAPK14 | BindingDB \| ChEMBL \| NPASS |
| Arogyavardhini | Azadirachta indica | Leaf | CID_5280343 | 1455 | CSNK1G2 | ChEMBL \| NPASS |
| Arogyavardhini | Azadirachta indica | Leaf | CID_5280343 | 1456 | CSNK1G3 | ChEMBL \| NPASS |
| Arogyavardhini | Azadirachta indica | Leaf | CID_5280343 | 1457 | CSNK2A1 | BindingDB \| ChEMBL \| NPASS |
| Arogyavardhini | Azadirachta indica | Leaf | CID_5280343 | 1459 | CSNK2A2 | ChEMBL \| NPASS |
| Arogyavardhini | Azadirachta indica | Leaf | CID_5280343 | 1460 | CSNK2B | ChEMBL \| NPASS |
| Arogyavardhini | Azadirachta indica | Leaf | CID_5280343 | 148 | ADRA1A | BindingDB \| ChEMBL |

**Table S3: Associations between formulations, constituent plants, phytochemicals, and human molecular targets. The table lists the formulations, plant name and plant part, phytochemical identifiers, associated human target genes (Entrez Gene ID and gene symbol), and the database sources supporting each phytochemical-target interaction.**

| Formulation | Plant name | Part | PubChem CID | Entrez Gene ID | Gene Symbol | Source database(s) |
|---|---|---|---|---|---|---|
| Arogyavardhini | Azadirachta indica | Leaf | CID_5280343 | 1499 | CTNNB1 | BindingDB \| ChEMBL |
| Arogyavardhini | Azadirachta indica | Leaf | CID_5280343 | 150 | ADRA2A | ChEMBL |
| Arogyavardhini | Azadirachta indica | Leaf | CID_5280343 | 1504 | CTRB1 | BindingDB \| ChEMBL \| NPASS |
| Arogyavardhini | Azadirachta indica | Leaf | CID_5280343 | 151 | ADRA2B | ChEMBL |
| Arogyavardhini | Azadirachta indica | Leaf | CID_5280343 | 1514 | CTSL | BindingDB |
| Arogyavardhini | Azadirachta indica | Leaf | CID_5280343 | 152 | ADRA2C | ChEMBL |
| Arogyavardhini | Azadirachta indica | Leaf | CID_5280343 | 153 | ADRB1 | ChEMBL |
| Arogyavardhini | Azadirachta indica | Leaf | CID_5280343 | 154 | ADRB2 | ChEMBL |
| Arogyavardhini | Azadirachta indica | Leaf | CID_5280343 | 1543 | CYP1A1 | BindingDB \| ChEMBL \| NPASS |
| Arogyavardhini | Azadirachta indica | Leaf | CID_5280343 | 1544 | CYP1A2 | BindingDB \| ChEMBL \| NPASS |
| Arogyavardhini | Azadirachta indica | Leaf | CID_5280343 | 1545 | CYP1B1 | BindingDB \| ChEMBL \| NPASS |
| Arogyavardhini | Azadirachta indica | Leaf | CID_5280343 | 155 | ADRB3 | ChEMBL |
| Arogyavardhini | Azadirachta indica | Leaf | CID_5280343 | 1557 | CYP2C19 | BindingDB \| ChEMBL \| NPASS |
| Arogyavardhini | Azadirachta indica | Leaf | CID_5280343 | 1558 | CYP2C8 | ChEMBL \| NPASS |
| Arogyavardhini | Azadirachta indica | Leaf | CID_5280343 | 1559 | CYP2C9 | BindingDB \| ChEMBL \| NPASS |
| Arogyavardhini | Azadirachta indica | Leaf | CID_5280343 | 1565 | CYP2D6 | ChEMBL \| NPASS |
| Arogyavardhini | Azadirachta indica | Leaf | CID_5280343 | 1573 | CYP2J2 | ChEMBL \| NPASS |
| Arogyavardhini | Azadirachta indica | Leaf | CID_5280343 | 1576 | CYP3A4 | BindingDB \| ChEMBL \| NPASS |
| Arogyavardhini | Azadirachta indica | Leaf | CID_5280343 | 1577 | CYP3A5 | ChEMBL |
| Arogyavardhini | Azadirachta indica | Leaf | CID_5280343 | 1588 | CYP19A1 | BindingDB \| ChEMBL \| NPASS |
| Arogyavardhini | Azadirachta indica | Leaf | CID_5280343 | 1612 | DAPK1 | BindingDB \| ChEMBL \| NPASS |
| Arogyavardhini | Azadirachta indica | Leaf | CID_5280343 | 1613 | DAPK3 | ChEMBL \| NPASS |
| Arogyavardhini | Azadirachta indica | Leaf | CID_5280343 | 1760 | DMPK | ChEMBL \| NPASS |
| Arogyavardhini | Azadirachta indica | Leaf | CID_5280343 | 1803 | DPP4 | BindingDB \| ChEMBL \| NPASS |
| Arogyavardhini | Azadirachta indica | Leaf | CID_5280343 | 1812 | DRD1 | ChEMBL \| NPASS |
| Arogyavardhini | Azadirachta indica | Leaf | CID_5280343 | 1813 | DRD2 | ChEMBL |
| Arogyavardhini | Azadirachta indica | Leaf | CID_5280343 | 1814 | DRD3 | ChEMBL |
| Arogyavardhini | Azadirachta indica | Leaf | CID_5280343 | 1815 | DRD4 | BindingDB \| ChEMBL \| NPASS |
| Arogyavardhini | Azadirachta indica | Leaf | CID_5280343 | 185 | AGTR1 | ChEMBL |
| Arogyavardhini | Azadirachta indica | Leaf | CID_5280343 | 1909 | EDNRA | ChEMBL |
| Arogyavardhini | Azadirachta indica | Leaf | CID_5280343 | 1910 | EDNRB | ChEMBL |
| Arogyavardhini | Azadirachta indica | Leaf | CID_5280343 | 1956 | EGFR | BindingDB \| ChEMBL \| NPASS |
| Arogyavardhini | Azadirachta indica | Leaf | CID_5280343 | 1991 | ELANE | BindingDB \| ChEMBL \| NPASS |
| Arogyavardhini | Azadirachta indica | Leaf | CID_5280343 | 1994 | ELAVL1 | BindingDB \| ChEMBL \| NPASS |
| Arogyavardhini | Azadirachta indica | Leaf | CID_5280343 | 1995 | ELAVL3 | ChEMBL \| NPASS |
| Arogyavardhini | Azadirachta indica | Leaf | CID_5280343 | 2030 | SLC29A1 | ChEMBL |
| Arogyavardhini | Azadirachta indica | Leaf | CID_5280343 | 2053 | EPHX2 | ChEMBL \| NPASS |
| Arogyavardhini | Azadirachta indica | Leaf | CID_5280343 | 207 | AKT1 | ChEMBL \| NPASS |
| Arogyavardhini | Azadirachta indica | Leaf | CID_5280343 | 2099 | ESR1 | ChEMBL \| NPASS |
| Arogyavardhini | Azadirachta indica | Leaf | CID_5280343 | 2100 | ESR2 | ChEMBL |
| Arogyavardhini | Azadirachta indica | Leaf | CID_5280343 | 2147 | F2 | ChEMBL \| NPASS |
| Arogyavardhini | Azadirachta indica | Leaf | CID_5280343 | 216 | ALDH1A1 | ChEMBL \| NPASS |
| Arogyavardhini | Azadirachta indica | Leaf | CID_5280343 | 2167 | FABP4 | ChEMBL |
| Arogyavardhini | Azadirachta indica | Leaf | CID_5280343 | 2203 | FBP1 | ChEMBL \| NPASS |
| Arogyavardhini | Azadirachta indica | Leaf | CID_5280343 | 2237 | FEN1 | ChEMBL \| NPASS |
| Arogyavardhini | Azadirachta indica | Leaf | CID_5280343 | 22933 | SIRT2 | ChEMBL |
| Arogyavardhini | Azadirachta indica | Leaf | CID_5280343 | 23028 | KDM1A | BindingDB \| ChEMBL \| NPASS |
| Arogyavardhini | Azadirachta indica | Leaf | CID_5280343 | 23043 | TNIK | ChEMBL \| NPASS |
| Arogyavardhini | Azadirachta indica | Leaf | CID_5280343 | 231 | AKR1B1 | ChEMBL \| NPASS |
| Arogyavardhini | Azadirachta indica | Leaf | CID_5280343 | 2322 | FLT3 | ChEMBL \| NPASS |
| Arogyavardhini | Azadirachta indica | Leaf | CID_5280343 | 23408 | SIRT5 | ChEMBL |
| Arogyavardhini | Azadirachta indica | Leaf | CID_5280343 | 23410 | SIRT3 | ChEMBL |
| Arogyavardhini | Azadirachta indica | Leaf | CID_5280343 | 23411 | SIRT1 | ChEMBL |
| Arogyavardhini | Azadirachta indica | Leaf | CID_5280343 | 23621 | BACE1 | BindingDB \| ChEMBL \| NPASS |
| Arogyavardhini | Azadirachta indica | Leaf | CID_5280343 | 23632 | CA14 | BindingDB \| ChEMBL \| NPASS |
| Arogyavardhini | Azadirachta indica | Leaf | CID_5280343 | 238 | ALK | BindingDB \| ChEMBL \| NPASS |
| Arogyavardhini | Azadirachta indica | Leaf | CID_5280343 | 239 | ALOX12 | BindingDB \| ChEMBL \| NPASS |
| Arogyavardhini | Azadirachta indica | Leaf | CID_5280343 | 240 | ALOX5 | BindingDB \| ChEMBL \| NPASS |
| Arogyavardhini | Azadirachta indica | Leaf | CID_5280343 | 246 | ALOX15 | BindingDB \| ChEMBL \| NPASS |
| Arogyavardhini | Azadirachta indica | Leaf | CID_5280343 | 247 | ALOX15B | BindingDB \| ChEMBL \| NPASS |
| Arogyavardhini | Azadirachta indica | Leaf | CID_5280343 | 253430 | IPMK | ChEMBL \| NPASS |
| Arogyavardhini | Azadirachta indica | Leaf | CID_5280343 | 2548 | GAA | ChEMBL \| NPASS |
| Arogyavardhini | Azadirachta indica | Leaf | CID_5280343 | 2584 | GALK1 | ChEMBL \| NPASS |
| Arogyavardhini | Azadirachta indica | Leaf | CID_5280343 | 2629 | GBA1 | NPASS |
| Arogyavardhini | Azadirachta indica | Leaf | CID_5280343 | 2642 | GCGR | ChEMBL |
| Arogyavardhini | Azadirachta indica | Leaf | CID_5280343 | 2693 | GHSR | ChEMBL |
| Arogyavardhini | Azadirachta indica | Leaf | CID_5280343 | 2696 | GIPR | ChEMBL |
| Arogyavardhini | Azadirachta indica | Leaf | CID_5280343 | 2739 | GLO1 | ChEMBL \| NPASS |
| Arogyavardhini | Azadirachta indica | Leaf | CID_5280343 | 276 | AMY1A | ChEMBL |
| Arogyavardhini | Azadirachta indica | Leaf | CID_5280343 | 277 | AMY1B | BindingDB |

| Table S3: Associations between formulations, constituent plants, phytochemicals, and human molecular targets. The table lists the formulations, plant name and plant part, phytochemical identifiers, associated human target genes (Entrez Gene ID and gene symbol), and the database sources supporting each phytochemical-target interaction. | | | | | | |
|---|---|---|---|---|---|---|
| Formulation | Plant name | Part | PubChem CID | Entrez Gene ID | Gene Symbol | Source database(s) |
| Arogyavardhini | Azadirachta indica | Leaf | CID_5280343 | 28234 | SLCO1B3 | ChEMBL \| NPASS |
| Arogyavardhini | Azadirachta indica | Leaf | CID_5280343 | 283106 | CSNK2A3 | ChEMBL \| NPASS |
| Arogyavardhini | Azadirachta indica | Leaf | CID_5280343 | 2859 | GPR35 | BindingDB \| ChEMBL \| NPASS |
| Arogyavardhini | Azadirachta indica | Leaf | CID_5280343 | 2862 | MLNR | ChEMBL |
| Arogyavardhini | Azadirachta indica | Leaf | CID_5280343 | 2908 | NR3C1 | ChEMBL |
| Arogyavardhini | Azadirachta indica | Leaf | CID_5280343 | 2931 | GSK3A | ChEMBL |
| Arogyavardhini | Azadirachta indica | Leaf | CID_5280343 | 2932 | GSK3B | BindingDB \| ChEMBL \| NPASS |
| Arogyavardhini | Azadirachta indica | Leaf | CID_5280343 | 2936 | GSR | NPASS |
| Arogyavardhini | Azadirachta indica | Leaf | CID_5280343 | 3028 | HSD17B10 | ChEMBL \| NPASS |
| Arogyavardhini | Azadirachta indica | Leaf | CID_5280343 | 3091 | HIF1A | ChEMBL \| NPASS |
| Arogyavardhini | Azadirachta indica | Leaf | CID_5280343 | 3248 | HPGD | ChEMBL \| NPASS |
| Arogyavardhini | Azadirachta indica | Leaf | CID_5280343 | 3269 | HRH1 | ChEMBL |
| Arogyavardhini | Azadirachta indica | Leaf | CID_5280343 | 3274 | HRH2 | ChEMBL |
| Arogyavardhini | Azadirachta indica | Leaf | CID_5280343 | 328 | APEX1 | ChEMBL \| NPASS |
| Arogyavardhini | Azadirachta indica | Leaf | CID_5280343 | 3290 | HSD11B1 | ChEMBL \| NPASS |
| Arogyavardhini | Azadirachta indica | Leaf | CID_5280343 | 3292 | HSD17B1 | ChEMBL \| NPASS |
| Arogyavardhini | Azadirachta indica | Leaf | CID_5280343 | 3294 | HSD17B2 | BindingDB \| ChEMBL \| NPASS |
| Arogyavardhini | Azadirachta indica | Leaf | CID_5280343 | 3297 | HSF1 | BindingDB \| ChEMBL \| NPASS |
| Arogyavardhini | Azadirachta indica | Leaf | CID_5280343 | 3309 | HSPA5 | NPASS |
| Arogyavardhini | Azadirachta indica | Leaf | CID_5280343 | 3315 | HSPB1 | ChEMBL |
| Arogyavardhini | Azadirachta indica | Leaf | CID_5280343 | 3350 | HTR1A | ChEMBL \| NPASS |
| Arogyavardhini | Azadirachta indica | Leaf | CID_5280343 | 3352 | HTR1D | BindingDB |
| Arogyavardhini | Azadirachta indica | Leaf | CID_5280343 | 3356 | HTR2A | ChEMBL |
| Arogyavardhini | Azadirachta indica | Leaf | CID_5280343 | 3357 | HTR2B | ChEMBL |
| Arogyavardhini | Azadirachta indica | Leaf | CID_5280343 | 3358 | HTR2C | ChEMBL |
| Arogyavardhini | Azadirachta indica | Leaf | CID_5280343 | 3359 | HTR3A | BindingDB \| ChEMBL |
| Arogyavardhini | Azadirachta indica | Leaf | CID_5280343 | 3362 | HTR6 | BindingDB |
| Arogyavardhini | Azadirachta indica | Leaf | CID_5280343 | 3363 | HTR7 | BindingDB |
| Arogyavardhini | Azadirachta indica | Leaf | CID_5280343 | 3480 | IGF1R | BindingDB \| ChEMBL \| NPASS |
| Arogyavardhini | Azadirachta indica | Leaf | CID_5280343 | 351 | APP | BindingDB \| ChEMBL \| NPASS |
| Arogyavardhini | Azadirachta indica | Leaf | CID_5280343 | 3577 | CXCR1 | ChEMBL \| NPASS |
| Arogyavardhini | Azadirachta indica | Leaf | CID_5280343 | 367 | AR | NPASS |
| Arogyavardhini | Azadirachta indica | Leaf | CID_5280343 | 3716 | JAK1 | ChEMBL \| NPASS |
| Arogyavardhini | Azadirachta indica | Leaf | CID_5280343 | 3725 | JUN | NPASS |
| Arogyavardhini | Azadirachta indica | Leaf | CID_5280343 | 3757 | KCNH2 | ChEMBL |
| Arogyavardhini | Azadirachta indica | Leaf | CID_5280343 | 3791 | KDR | BindingDB \| ChEMBL \| NPASS |
| Arogyavardhini | Azadirachta indica | Leaf | CID_5280343 | 387129 | NPSR1 | NPASS |
| Arogyavardhini | Azadirachta indica | Leaf | CID_5280343 | 390245 | KDM4E | ChEMBL \| NPASS |
| Arogyavardhini | Azadirachta indica | Leaf | CID_5280343 | 3932 | LCK | ChEMBL \| NPASS |
| Arogyavardhini | Azadirachta indica | Leaf | CID_5280343 | 4000 | LMNA | NPASS |
| Arogyavardhini | Azadirachta indica | Leaf | CID_5280343 | 4088 | SMAD3 | NPASS |
| Arogyavardhini | Azadirachta indica | Leaf | CID_5280343 | 410 | ARSA | NPASS |
| Arogyavardhini | Azadirachta indica | Leaf | CID_5280343 | 4128 | MAOA | ChEMBL \| NPASS |
| Arogyavardhini | Azadirachta indica | Leaf | CID_5280343 | 4129 | MAOB | ChEMBL \| NPASS |
| Arogyavardhini | Azadirachta indica | Leaf | CID_5280343 | 4137 | MAPT | ChEMBL \| NPASS |
| Arogyavardhini | Azadirachta indica | Leaf | CID_5280343 | 415116 | PIM3 | ChEMBL \| NPASS |
| Arogyavardhini | Azadirachta indica | Leaf | CID_5280343 | 4154 | MBNL1 | NPASS |
| Arogyavardhini | Azadirachta indica | Leaf | CID_5280343 | 4159 | MC3R | ChEMBL |
| Arogyavardhini | Azadirachta indica | Leaf | CID_5280343 | 4160 | MC4R | ChEMBL |
| Arogyavardhini | Azadirachta indica | Leaf | CID_5280343 | 4217 | MAP3K5 | ChEMBL \| NPASS |
| Arogyavardhini | Azadirachta indica | Leaf | CID_5280343 | 4233 | MET | BindingDB \| ChEMBL \| NPASS |
| Arogyavardhini | Azadirachta indica | Leaf | CID_5280343 | 4297 | KMT2A | NPASS |
| Arogyavardhini | Azadirachta indica | Leaf | CID_5280343 | 43 | ACHE | BindingDB \| ChEMBL \| NPASS |
| Arogyavardhini | Azadirachta indica | Leaf | CID_5280343 | 4312 | MMP1 | ChEMBL |
| Arogyavardhini | Azadirachta indica | Leaf | CID_5280343 | 4313 | MMP2 | ChEMBL \| NPASS |
| Arogyavardhini | Azadirachta indica | Leaf | CID_5280343 | 4314 | MMP3 | ChEMBL \| NPASS |
| Arogyavardhini | Azadirachta indica | Leaf | CID_5280343 | 4318 | MMP9 | BindingDB \| ChEMBL \| NPASS |
| Arogyavardhini | Azadirachta indica | Leaf | CID_5280343 | 4321 | MMP12 | BindingDB \| ChEMBL \| NPASS |
| Arogyavardhini | Azadirachta indica | Leaf | CID_5280343 | 4322 | MMP13 | BindingDB \| ChEMBL \| NPASS |
| Arogyavardhini | Azadirachta indica | Leaf | CID_5280343 | 4350 | MPG | ChEMBL \| NPASS |
| Arogyavardhini | Azadirachta indica | Leaf | CID_5280343 | 4353 | MPO | BindingDB \| ChEMBL \| NPASS |
| Arogyavardhini | Azadirachta indica | Leaf | CID_5280343 | 4363 | ABCC1 | ChEMBL \| NPASS |
| Arogyavardhini | Azadirachta indica | Leaf | CID_5280343 | 4751 | NEK2 | ChEMBL \| NPASS |
| Arogyavardhini | Azadirachta indica | Leaf | CID_5280343 | 4759 | NEU2 | BindingDB \| NPASS |
| Arogyavardhini | Azadirachta indica | Leaf | CID_5280343 | 4790 | NFKB1 | ChEMBL \| NPASS |
| Arogyavardhini | Azadirachta indica | Leaf | CID_5280343 | 4791 | NFKB2 | ChEMBL |
| Arogyavardhini | Azadirachta indica | Leaf | CID_5280343 | 4886 | NPY1R | ChEMBL |
| Arogyavardhini | Azadirachta indica | Leaf | CID_5280343 | 4887 | NPY2R | ChEMBL |
| Arogyavardhini | Azadirachta indica | Leaf | CID_5280343 | 4907 | NT5E | BindingDB \| ChEMBL |
| Arogyavardhini | Azadirachta indica | Leaf | CID_5280343 | 4923 | NTSR1 | ChEMBL |

**Table S3: Associations between formulations, constituent plants, phytochemicals, and human molecular targets. The table lists the formulations, plant name and plant part, phytochemical identifiers, associated human target genes (Entrez Gene ID and gene symbol), and the database sources supporting each phytochemical-target interaction.**

| Formulation | Plant name | Part | PubChem CID | Entrez Gene ID | Gene Symbol | Source database(s) |
|---|---|---|---|---|---|---|
| Arogyavardhini | Azadirachta indica | Leaf | CID_5280343 | 4985 | OPRD1 | ChEMBL |
| Arogyavardhini | Azadirachta indica | Leaf | CID_5280343 | 4986 | OPRK1 | BindingDB \| ChEMBL |
| Arogyavardhini | Azadirachta indica | Leaf | CID_5280343 | 4988 | OPRM1 | BindingDB \| ChEMBL |
| Arogyavardhini | Azadirachta indica | Leaf | CID_5280343 | 50507 | NOX4 | BindingDB \| ChEMBL \| NPASS |
| Arogyavardhini | Azadirachta indica | Leaf | CID_5280343 | 51231 | VRK3 | ChEMBL \| NPASS |
| Arogyavardhini | Azadirachta indica | Leaf | CID_5280343 | 5127 | CDK16 | ChEMBL \| NPASS |
| Arogyavardhini | Azadirachta indica | Leaf | CID_5280343 | 5139 | PDE3A | ChEMBL |
| Arogyavardhini | Azadirachta indica | Leaf | CID_5280343 | 51426 | POLK | ChEMBL \| NPASS |
| Arogyavardhini | Azadirachta indica | Leaf | CID_5280343 | 5144 | PDE4D | ChEMBL |
| Arogyavardhini | Azadirachta indica | Leaf | CID_5280343 | 51447 | IP6K2 | ChEMBL \| NPASS |
| Arogyavardhini | Azadirachta indica | Leaf | CID_5280343 | 51548 | SIRT6 | ChEMBL |
| Arogyavardhini | Azadirachta indica | Leaf | CID_5280343 | 5163 | PDK1 | ChEMBL \| NPASS |
| Arogyavardhini | Azadirachta indica | Leaf | CID_5280343 | 5170 | PDPK1 | ChEMBL \| NPASS |
| Arogyavardhini | Azadirachta indica | Leaf | CID_5280343 | 51765 | STK26 | ChEMBL \| NPASS |
| Arogyavardhini | Azadirachta indica | Leaf | CID_5280343 | 5241 | PGR | ChEMBL |
| Arogyavardhini | Azadirachta indica | Leaf | CID_5280343 | 5243 | ABCB1 | ChEMBL \| NPASS |
| Arogyavardhini | Azadirachta indica | Leaf | CID_5280343 | 5290 | PIK3CA | ChEMBL |
| Arogyavardhini | Azadirachta indica | Leaf | CID_5280343 | 5292 | PIM1 | ChEMBL \| NPASS |
| Arogyavardhini | Azadirachta indica | Leaf | CID_5280343 | 5294 | PIK3CG | BindingDB \| ChEMBL \| NPASS |
| Arogyavardhini | Azadirachta indica | Leaf | CID_5280343 | 5295 | PIK3R1 | ChEMBL \| NPASS |
| Arogyavardhini | Azadirachta indica | Leaf | CID_5280343 | 5300 | PIN1 | NPASS |
| Arogyavardhini | Azadirachta indica | Leaf | CID_5280343 | 5319 | PLA2G1B | BindingDB \| ChEMBL \| NPASS |
| Arogyavardhini | Azadirachta indica | Leaf | CID_5280343 | 5328 | PLAU | BindingDB \| ChEMBL \| NPASS |
| Arogyavardhini | Azadirachta indica | Leaf | CID_5280343 | 5340 | PLG | ChEMBL \| NPASS |
| Arogyavardhini | Azadirachta indica | Leaf | CID_5280343 | 5347 | PLK1 | BindingDB \| ChEMBL \| NPASS |
| Arogyavardhini | Azadirachta indica | Leaf | CID_5280343 | 5376 | PMP22 | NPASS |
| Arogyavardhini | Azadirachta indica | Leaf | CID_5280343 | 53944 | CSNK1G1 | ChEMBL \| NPASS |
| Arogyavardhini | Azadirachta indica | Leaf | CID_5280343 | 5406 | PNLIP | BindingDB \| ChEMBL |
| Arogyavardhini | Azadirachta indica | Leaf | CID_5280343 | 5423 | POLB | ChEMBL \| NPASS |
| Arogyavardhini | Azadirachta indica | Leaf | CID_5280343 | 5429 | POLH | BindingDB \| ChEMBL \| NPASS |
| Arogyavardhini | Azadirachta indica | Leaf | CID_5280343 | 5444 | PON1 | ChEMBL \| NPASS |
| Arogyavardhini | Azadirachta indica | Leaf | CID_5280343 | 54575 | UGT1A10 | ChEMBL \| NPASS |
| Arogyavardhini | Azadirachta indica | Leaf | CID_5280343 | 54576 | UGT1A8 | ChEMBL \| NPASS |
| Arogyavardhini | Azadirachta indica | Leaf | CID_5280343 | 54578 | UGT1A6 | ChEMBL \| NPASS |
| Arogyavardhini | Azadirachta indica | Leaf | CID_5280343 | 54600 | UGT1A9 | ChEMBL \| NPASS |
| Arogyavardhini | Azadirachta indica | Leaf | CID_5280343 | 54657 | UGT1A4 | ChEMBL \| NPASS |
| Arogyavardhini | Azadirachta indica | Leaf | CID_5280343 | 54658 | UGT1A1 | ChEMBL \| NPASS |
| Arogyavardhini | Azadirachta indica | Leaf | CID_5280343 | 54659 | UGT1A3 | ChEMBL \| NPASS |
| Arogyavardhini | Azadirachta indica | Leaf | CID_5280343 | 5467 | PPARD | NPASS |
| Arogyavardhini | Azadirachta indica | Leaf | CID_5280343 | 5468 | PPARG | BindingDB \| ChEMBL \| NPASS |
| Arogyavardhini | Azadirachta indica | Leaf | CID_5280343 | 54737 | MPHOSPH8 | ChEMBL \| NPASS |
| Arogyavardhini | Azadirachta indica | Leaf | CID_5280343 | 552 | AVPR1A | ChEMBL |
| Arogyavardhini | Azadirachta indica | Leaf | CID_5280343 | 554 | AVPR2 | ChEMBL \| NPASS |
| Arogyavardhini | Azadirachta indica | Leaf | CID_5280343 | 5550 | PREP | ChEMBL \| NPASS |
| Arogyavardhini | Azadirachta indica | Leaf | CID_5280343 | 5563 | PRKAA2 | ChEMBL \| NPASS |
| Arogyavardhini | Azadirachta indica | Leaf | CID_5280343 | 5566 | PRKACA | ChEMBL \| NPASS |
| Arogyavardhini | Azadirachta indica | Leaf | CID_5280343 | 55662 | HIF1AN | ChEMBL \| NPASS |
| Arogyavardhini | Azadirachta indica | Leaf | CID_5280343 | 55775 | TDP1 | ChEMBL \| NPASS |
| Arogyavardhini | Azadirachta indica | Leaf | CID_5280343 | 5578 | PRKCA | ChEMBL \| NPASS |
| Arogyavardhini | Azadirachta indica | Leaf | CID_5280343 | 55781 | RIOK2 | ChEMBL \| NPASS |
| Arogyavardhini | Azadirachta indica | Leaf | CID_5280343 | 558 | AXL | ChEMBL \| NPASS |
| Arogyavardhini | Azadirachta indica | Leaf | CID_5280343 | 5585 | PKN1 | BindingDB \| ChEMBL \| NPASS |
| Arogyavardhini | Azadirachta indica | Leaf | CID_5280343 | 55872 | PBK | ChEMBL \| NPASS |
| Arogyavardhini | Azadirachta indica | Leaf | CID_5280343 | 5594 | MAPK1 | ChEMBL \| NPASS |
| Arogyavardhini | Azadirachta indica | Leaf | CID_5280343 | 5595 | MAPK3 | ChEMBL \| NPASS |
| Arogyavardhini | Azadirachta indica | Leaf | CID_5280343 | 5597 | MAPK6 | ChEMBL \| NPASS |
| Arogyavardhini | Azadirachta indica | Leaf | CID_5280343 | 5599 | MAPK8 | ChEMBL \| NPASS |
| Arogyavardhini | Azadirachta indica | Leaf | CID_5280343 | 5600 | MAPK11 | ChEMBL \| NPASS |
| Arogyavardhini | Azadirachta indica | Leaf | CID_5280343 | 5601 | MAPK9 | ChEMBL |
| Arogyavardhini | Azadirachta indica | Leaf | CID_5280343 | 5602 | MAPK10 | BindingDB |
| Arogyavardhini | Azadirachta indica | Leaf | CID_5280343 | 5603 | MAPK13 | ChEMBL \| NPASS |
| Arogyavardhini | Azadirachta indica | Leaf | CID_5280343 | 5604 | MAP2K1 | ChEMBL \| NPASS |
| Arogyavardhini | Azadirachta indica | Leaf | CID_5280343 | 5605 | MAP2K2 | ChEMBL \| NPASS |
| Arogyavardhini | Azadirachta indica | Leaf | CID_5280343 | 5608 | MAP2K6 | ChEMBL \| NPASS |
| Arogyavardhini | Azadirachta indica | Leaf | CID_5280343 | 56339 | METTL3 | ChEMBL |
| Arogyavardhini | Azadirachta indica | Leaf | CID_5280343 | 5644 | PRSS1 | BindingDB \| ChEMBL |
| Arogyavardhini | Azadirachta indica | Leaf | CID_5280343 | 5645 | PRSS2 | ChEMBL |
| Arogyavardhini | Azadirachta indica | Leaf | CID_5280343 | 5646 | PRSS3 | ChEMBL \| NPASS |
| Arogyavardhini | Azadirachta indica | Leaf | CID_5280343 | 56924 | PAK6 | NPASS |
| Arogyavardhini | Azadirachta indica | Leaf | CID_5280343 | 5693 | PSMB5 | BindingDB \| ChEMBL \| NPASS |

Table S3: Associations between formulations, constituent plants, phytochemicals, and human molecular targets. The table lists the formulations, plant name and plant part, phytochemical identifiers, associated human target genes (Entrez Gene ID and gene symbol), and the database sources supporting each phytochemical-target interaction.

| Formulation | Plant name | Part | PubChem CID | Entrez Gene ID | Gene Symbol | Source database(s) |
|---|---|---|---|---|---|---|
| Arogyavardhini | Azadirachta indica | Leaf | CID_5280343 | 57016 | AKR1B10 | ChEMBL \| NPASS |
| Arogyavardhini | Azadirachta indica | Leaf | CID_5280343 | 57118 | CAMK1D | ChEMBL \| NPASS |
| Arogyavardhini | Azadirachta indica | Leaf | CID_5280343 | 57144 | PAK5 | ChEMBL \| NPASS |
| Arogyavardhini | Azadirachta indica | Leaf | CID_5280343 | 57172 | CAMK1G | ChEMBL \| NPASS |
| Arogyavardhini | Azadirachta indica | Leaf | CID_5280343 | 5742 | PTGS1 | ChEMBL \| NPASS |
| Arogyavardhini | Azadirachta indica | Leaf | CID_5280343 | 5743 | PTGS2 | BindingDB \| ChEMBL \| NPASS |
| Arogyavardhini | Azadirachta indica | Leaf | CID_5280343 | 5747 | PTK2 | BindingDB \| ChEMBL \| NPASS |
| Arogyavardhini | Azadirachta indica | Leaf | CID_5280343 | 5770 | PTPN1 | BindingDB \| ChEMBL \| NPASS |
| Arogyavardhini | Azadirachta indica | Leaf | CID_5280343 | 57721 | METTL14 | ChEMBL |
| Arogyavardhini | Azadirachta indica | Leaf | CID_5280343 | 5836 | PYGL | BindingDB \| ChEMBL \| NPASS |
| Arogyavardhini | Azadirachta indica | Leaf | CID_5280343 | 590 | BCHE | BindingDB \| ChEMBL \| NPASS |
| Arogyavardhini | Azadirachta indica | Leaf | CID_5280343 | 595 | CCND1 | ChEMBL \| NPASS |
| Arogyavardhini | Azadirachta indica | Leaf | CID_5280343 | 5965 | RECQL | ChEMBL \| NPASS |
| Arogyavardhini | Azadirachta indica | Leaf | CID_5280343 | 5970 | RELA | ChEMBL |
| Arogyavardhini | Azadirachta indica | Leaf | CID_5280343 | 5979 | RET | BindingDB \| ChEMBL \| NPASS |
| Arogyavardhini | Azadirachta indica | Leaf | CID_5280343 | 5999 | RGS4 | NPASS |
| Arogyavardhini | Azadirachta indica | Leaf | CID_5280343 | 607 | BCL9 | BindingDB \| ChEMBL |
| Arogyavardhini | Azadirachta indica | Leaf | CID_5280343 | 6197 | RPS6KA3 | ChEMBL \| NPASS |
| Arogyavardhini | Azadirachta indica | Leaf | CID_5280343 | 6198 | RPS6KB1 | ChEMBL \| NPASS |
| Arogyavardhini | Azadirachta indica | Leaf | CID_5280343 | 623 | BDKRB1 | ChEMBL |
| Arogyavardhini | Azadirachta indica | Leaf | CID_5280343 | 624 | BDKRB2 | ChEMBL |
| Arogyavardhini | Azadirachta indica | Leaf | CID_5280343 | 6300 | MAPK12 | ChEMBL \| NPASS |
| Arogyavardhini | Azadirachta indica | Leaf | CID_5280343 | 641 | BLM | ChEMBL \| NPASS |
| Arogyavardhini | Azadirachta indica | Leaf | CID_5280343 | 6446 | SGK1 | ChEMBL \| NPASS |
| Arogyavardhini | Azadirachta indica | Leaf | CID_5280343 | 6476 | SI | ChEMBL \| NPASS |
| Arogyavardhini | Azadirachta indica | Leaf | CID_5280343 | 6514 | SLC2A2 | ChEMBL \| NPASS |
| Arogyavardhini | Azadirachta indica | Leaf | CID_5280343 | 6530 | SLC6A2 | BindingDB \| ChEMBL |
| Arogyavardhini | Azadirachta indica | Leaf | CID_5280343 | 6531 | SLC6A3 | ChEMBL |
| Arogyavardhini | Azadirachta indica | Leaf | CID_5280343 | 6532 | SLC6A4 | BindingDB \| ChEMBL |
| Arogyavardhini | Azadirachta indica | Leaf | CID_5280343 | 65975 | STK33 | BindingDB |
| Arogyavardhini | Azadirachta indica | Leaf | CID_5280343 | 6606 | SMN1 | NPASS |
| Arogyavardhini | Azadirachta indica | Leaf | CID_5280343 | 6622 | SNCA | ChEMBL \| NPASS |
| Arogyavardhini | Azadirachta indica | Leaf | CID_5280343 | 6652 | SORD | BindingDB \| ChEMBL \| NPASS |
| Arogyavardhini | Azadirachta indica | Leaf | CID_5280343 | 6714 | SRC | ChEMBL \| NPASS |
| Arogyavardhini | Azadirachta indica | Leaf | CID_5280343 | 673 | BRAF | ChEMBL \| NPASS |
| Arogyavardhini | Azadirachta indica | Leaf | CID_5280343 | 6778 | STAT6 | NPASS |
| Arogyavardhini | Azadirachta indica | Leaf | CID_5280343 | 6789 | STK4 | ChEMBL \| NPASS |
| Arogyavardhini | Azadirachta indica | Leaf | CID_5280343 | 6793 | STK10 | ChEMBL \| NPASS |
| Arogyavardhini | Azadirachta indica | Leaf | CID_5280343 | 6850 | SYK | BindingDB \| ChEMBL \| NPASS |
| Arogyavardhini | Azadirachta indica | Leaf | CID_5280343 | 6915 | TBXA2R | ChEMBL |
| Arogyavardhini | Azadirachta indica | Leaf | CID_5280343 | 7068 | THRB | NPASS |
| Arogyavardhini | Azadirachta indica | Leaf | CID_5280343 | 7150 | TOP1 | ChEMBL |
| Arogyavardhini | Azadirachta indica | Leaf | CID_5280343 | 7153 | TOP2A | ChEMBL \| NPASS |
| Arogyavardhini | Azadirachta indica | Leaf | CID_5280343 | 7155 | TOP2B | ChEMBL \| NPASS |
| Arogyavardhini | Azadirachta indica | Leaf | CID_5280343 | 7157 | TP53 | NPASS |
| Arogyavardhini | Azadirachta indica | Leaf | CID_5280343 | 7224 | TRPC5 | ChEMBL \| NPASS |
| Arogyavardhini | Azadirachta indica | Leaf | CID_5280343 | 7253 | TSHR | NPASS |
| Arogyavardhini | Azadirachta indica | Leaf | CID_5280343 | 7257 | TSNAX | ChEMBL \| NPASS |
| Arogyavardhini | Azadirachta indica | Leaf | CID_5280343 | 7299 | TYR | BindingDB \| ChEMBL \| NPASS |
| Arogyavardhini | Azadirachta indica | Leaf | CID_5280343 | 7366 | UGT2B15 | ChEMBL \| NPASS |
| Arogyavardhini | Azadirachta indica | Leaf | CID_5280343 | 7398 | USP1 | NPASS |
| Arogyavardhini | Azadirachta indica | Leaf | CID_5280343 | 7443 | VRK1 | ChEMBL \| NPASS |
| Arogyavardhini | Azadirachta indica | Leaf | CID_5280343 | 7444 | VRK2 | ChEMBL \| NPASS |
| Arogyavardhini | Azadirachta indica | Leaf | CID_5280343 | 7498 | XDH | BindingDB \| ChEMBL \| NPASS |
| Arogyavardhini | Azadirachta indica | Leaf | CID_5280343 | 7525 | YES1 | BindingDB \| ChEMBL \| NPASS |
| Arogyavardhini | Azadirachta indica | Leaf | CID_5280343 | 759 | CA1 | BindingDB \| ChEMBL \| NPASS |
| Arogyavardhini | Azadirachta indica | Leaf | CID_5280343 | 760 | CA2 | BindingDB \| ChEMBL \| NPASS |
| Arogyavardhini | Azadirachta indica | Leaf | CID_5280343 | 761 | CA3 | BindingDB \| ChEMBL \| NPASS |
| Arogyavardhini | Azadirachta indica | Leaf | CID_5280343 | 762 | CA4 | BindingDB \| ChEMBL \| NPASS |
| Arogyavardhini | Azadirachta indica | Leaf | CID_5280343 | 763 | CA5A | BindingDB \| ChEMBL \| NPASS |
| Arogyavardhini | Azadirachta indica | Leaf | CID_5280343 | 765 | CA6 | ChEMBL \| NPASS |
| Arogyavardhini | Azadirachta indica | Leaf | CID_5280343 | 766 | CA7 | BindingDB \| ChEMBL \| NPASS |
| Arogyavardhini | Azadirachta indica | Leaf | CID_5280343 | 768 | CA9 | BindingDB \| ChEMBL \| NPASS |
| Arogyavardhini | Azadirachta indica | Leaf | CID_5280343 | 771 | CA12 | BindingDB \| ChEMBL \| NPASS |
| Arogyavardhini | Azadirachta indica | Leaf | CID_5280343 | 79915 | ATAD5 | ChEMBL \| NPASS |
| Arogyavardhini | Azadirachta indica | Leaf | CID_5280343 | 805 | CALM2 | ChEMBL |
| Arogyavardhini | Azadirachta indica | Leaf | CID_5280343 | 814 | CAMK4 | ChEMBL \| NPASS |
| Arogyavardhini | Azadirachta indica | Leaf | CID_5280343 | 815 | CAMK2A | ChEMBL \| NPASS |
| Arogyavardhini | Azadirachta indica | Leaf | CID_5280343 | 816 | CAMK2B | BindingDB \| ChEMBL \| NPASS |
| Arogyavardhini | Azadirachta indica | Leaf | CID_5280343 | 817 | CAMK2D | ChEMBL \| NPASS |

**Table S3: Associations between formulations, constituent plants, phytochemicals, and human molecular targets. The table lists the formulations, plant name and plant part, phytochemical identifiers, associated human target genes (Entrez Gene ID and gene symbol), and the database sources supporting each phytochemical-target interaction.**

| Formulation | Plant name | Part | PubChem CID | Entrez Gene ID | Gene Symbol | Source database(s) |
|---|---|---|---|---|---|---|
| Arogyavardhini | Azadirachta indica | Leaf | CID_5280343 | 818 | CAMK2G | ChEMBL \| NPASS |
| Arogyavardhini | Azadirachta indica | Leaf | CID_5280343 | 85417 | CCNB3 | ChEMBL |
| Arogyavardhini | Azadirachta indica | Leaf | CID_5280343 | 8550 | MAPKAPK5 | ChEMBL \| NPASS |
| Arogyavardhini | Azadirachta indica | Leaf | CID_5280343 | 8576 | STK16 | ChEMBL \| NPASS |
| Arogyavardhini | Azadirachta indica | Leaf | CID_5280343 | 8647 | ABCB11 | NPASS |
| Arogyavardhini | Azadirachta indica | Leaf | CID_5280343 | 8654 | PDE5A | ChEMBL \| NPASS |
| Arogyavardhini | Azadirachta indica | Leaf | CID_5280343 | 8714 | ABCC3 | BindingDB \| NPASS |
| Arogyavardhini | Azadirachta indica | Leaf | CID_5280343 | 873 | CBR1 | ChEMBL \| NPASS |
| Arogyavardhini | Azadirachta indica | Leaf | CID_5280343 | 874 | CBR3 | ChEMBL \| NPASS |
| Arogyavardhini | Azadirachta indica | Leaf | CID_5280343 | 8814 | CDKL1 | ChEMBL \| NPASS |
| Arogyavardhini | Azadirachta indica | Leaf | CID_5280343 | 8851 | CDK5R1 | ChEMBL \| NPASS |
| Arogyavardhini | Azadirachta indica | Leaf | CID_5280343 | 886 | CCKAR | ChEMBL |
| Arogyavardhini | Azadirachta indica | Leaf | CID_5280343 | 887 | CCKBR | ChEMBL |
| Arogyavardhini | Azadirachta indica | Leaf | CID_5280343 | 8877 | SPHK1 | ChEMBL |
| Arogyavardhini | Azadirachta indica | Leaf | CID_5280343 | 891 | CCNB1 | ChEMBL \| NPASS |
| Arogyavardhini | Azadirachta indica | Leaf | CID_5280343 | 8972 | MGAM | ChEMBL \| NPASS |
| Arogyavardhini | Azadirachta indica | Leaf | CID_5280343 | 9099 | USP2 | NPASS |
| Arogyavardhini | Azadirachta indica | Leaf | CID_5280343 | 9133 | CCNB2 | ChEMBL \| NPASS |
| Arogyavardhini | Azadirachta indica | Leaf | CID_5280343 | 9212 | AURKB | BindingDB \| ChEMBL \| NPASS |
| Arogyavardhini | Azadirachta indica | Leaf | CID_5280343 | 9252 | RPS6KA5 | ChEMBL \| NPASS |
| Arogyavardhini | Azadirachta indica | Leaf | CID_5280343 | 9261 | MAPKAPK2 | ChEMBL \| NPASS |
| Arogyavardhini | Azadirachta indica | Leaf | CID_5280343 | 9263 | STK17A | ChEMBL \| NPASS |
| Arogyavardhini | Azadirachta indica | Leaf | CID_5280343 | 93432 | MGAM2 | ChEMBL \| NPASS |
| Arogyavardhini | Azadirachta indica | Leaf | CID_5280343 | 9429 | ABCG2 | BindingDB \| ChEMBL \| NPASS |
| Arogyavardhini | Azadirachta indica | Leaf | CID_5280343 | 952 | CD38 | BindingDB \| ChEMBL \| NPASS |
| Arogyavardhini | Azadirachta indica | Leaf | CID_5280343 | 9748 | SLK | ChEMBL \| NPASS |
| Arogyavardhini | Azadirachta indica | Leaf | CID_5280343 | 983 | CDK1 | BindingDB \| ChEMBL \| NPASS |
| Arogyavardhini | Azadirachta indica | Leaf | CID_5280343 | 9891 | NUAK1 | BindingDB \| ChEMBL \| NPASS |
| Arogyavardhini | Azadirachta indica | Leaf | CID_5280343 | 9943 | OXSR1 | ChEMBL \| NPASS |
| Arogyavardhini | Azadirachta indica | Leaf | CID_5280460 | 10073 | SNUPN | NPASS |
| Arogyavardhini | Azadirachta indica | Leaf | CID_5280460 | 10599 | SLCO1B1 | ChEMBL \| NPASS |
| Arogyavardhini | Azadirachta indica | Leaf | CID_5280460 | 10919 | EHMT2 | NPASS |
| Arogyavardhini | Azadirachta indica | Leaf | CID_5280460 | 10941 | UGT2A1 | ChEMBL |
| Arogyavardhini | Azadirachta indica | Leaf | CID_5280460 | 1544 | CYP1A2 | NPASS |
| Arogyavardhini | Azadirachta indica | Leaf | CID_5280460 | 216 | ALDH1A1 | NPASS |
| Arogyavardhini | Azadirachta indica | Leaf | CID_5280460 | 231 | AKR1B1 | ChEMBL \| NPASS |
| Arogyavardhini | Azadirachta indica | Leaf | CID_5280460 | 240 | ALOX5 | BindingDB \| ChEMBL \| NPASS |
| Arogyavardhini | Azadirachta indica | Leaf | CID_5280460 | 2548 | GAA | NPASS |
| Arogyavardhini | Azadirachta indica | Leaf | CID_5280460 | 2597 | GAPDH | BindingDB \| NPASS |
| Arogyavardhini | Azadirachta indica | Leaf | CID_5280460 | 2629 | GBA1 | NPASS |
| Arogyavardhini | Azadirachta indica | Leaf | CID_5280460 | 2648 | KAT2A | ChEMBL \| NPASS |
| Arogyavardhini | Azadirachta indica | Leaf | CID_5280460 | 2671 | GFER | ChEMBL \| NPASS |
| Arogyavardhini | Azadirachta indica | Leaf | CID_5280460 | 2717 | GLA | NPASS |
| Arogyavardhini | Azadirachta indica | Leaf | CID_5280460 | 2778 | GNAS | NPASS |
| Arogyavardhini | Azadirachta indica | Leaf | CID_5280460 | 28234 | SLCO1B3 | ChEMBL \| NPASS |
| Arogyavardhini | Azadirachta indica | Leaf | CID_5280460 | 3028 | HSD17B10 | ChEMBL \| NPASS |
| Arogyavardhini | Azadirachta indica | Leaf | CID_5280460 | 3091 | HIF1A | NPASS |
| Arogyavardhini | Azadirachta indica | Leaf | CID_5280460 | 3248 | HPGD | NPASS |
| Arogyavardhini | Azadirachta indica | Leaf | CID_5280460 | 328 | APEX1 | NPASS |
| Arogyavardhini | Azadirachta indica | Leaf | CID_5280460 | 377677 | CA13 | ChEMBL \| NPASS |
| Arogyavardhini | Azadirachta indica | Leaf | CID_5280460 | 3837 | KPNB1 | NPASS |
| Arogyavardhini | Azadirachta indica | Leaf | CID_5280460 | 390245 | KDM4E | NPASS |
| Arogyavardhini | Azadirachta indica | Leaf | CID_5280460 | 4128 | MAOA | ChEMBL \| NPASS |
| Arogyavardhini | Azadirachta indica | Leaf | CID_5280460 | 4129 | MAOB | ChEMBL \| NPASS |
| Arogyavardhini | Azadirachta indica | Leaf | CID_5280460 | 4170 | MCL1 | BindingDB \| ChEMBL \| NPASS |
| Arogyavardhini | Azadirachta indica | Leaf | CID_5280460 | 4297 | KMT2A | NPASS |
| Arogyavardhini | Azadirachta indica | Leaf | CID_5280460 | 43 | ACHE | BindingDB \| NPASS |
| Arogyavardhini | Azadirachta indica | Leaf | CID_5280460 | 51053 | GMNN | ChEMBL \| NPASS |
| Arogyavardhini | Azadirachta indica | Leaf | CID_5280460 | 51752 | ERAP1 | ChEMBL \| NPASS |
| Arogyavardhini | Azadirachta indica | Leaf | CID_5280460 | 5300 | PIN1 | NPASS |
| Arogyavardhini | Azadirachta indica | Leaf | CID_5280460 | 54575 | UGT1A10 | ChEMBL \| NPASS |
| Arogyavardhini | Azadirachta indica | Leaf | CID_5280460 | 54576 | UGT1A8 | ChEMBL \| NPASS |
| Arogyavardhini | Azadirachta indica | Leaf | CID_5280460 | 54600 | UGT1A9 | ChEMBL |
| Arogyavardhini | Azadirachta indica | Leaf | CID_5280460 | 54657 | UGT1A4 | ChEMBL \| NPASS |
| Arogyavardhini | Azadirachta indica | Leaf | CID_5280460 | 54658 | UGT1A1 | ChEMBL \| NPASS |
| Arogyavardhini | Azadirachta indica | Leaf | CID_5280460 | 54659 | UGT1A3 | ChEMBL \| NPASS |
| Arogyavardhini | Azadirachta indica | Leaf | CID_5280460 | 55775 | TDP1 | NPASS |
| Arogyavardhini | Azadirachta indica | Leaf | CID_5280460 | 5743 | PTGS2 | BindingDB \| ChEMBL \| NPASS |
| Arogyavardhini | Azadirachta indica | Leaf | CID_5280460 | 5901 | RAN | NPASS |
| Arogyavardhini | Azadirachta indica | Leaf | CID_5280460 | 5965 | RECQL | NPASS |

**Table S3: Associations between formulations, constituent plants, phytochemicals, and human molecular targets. The table lists the formulations, plant name and plant part, phytochemical identifiers, associated human target genes (Entrez Gene ID and gene symbol), and the database sources supporting each phytochemical-target interaction.**

| Formulation | Plant name | Part | PubChem CID | Entrez Gene ID | Gene Symbol | Source database(s) |
|---|---|---|---|---|---|---|
| Arogyavardhini | Azadirachta indica | Leaf | CID_5280460 | 5970 | RELA | ChEMBL \| NPASS |
| Arogyavardhini | Azadirachta indica | Leaf | CID_5280460 | 6311 | ATXN2 | NPASS |
| Arogyavardhini | Azadirachta indica | Leaf | CID_5280460 | 64167 | ERAP2 | ChEMBL \| NPASS |
| Arogyavardhini | Azadirachta indica | Leaf | CID_5280460 | 6652 | SORD | BindingDB \| NPASS |
| Arogyavardhini | Azadirachta indica | Leaf | CID_5280460 | 7366 | UGT2B15 | ChEMBL \| NPASS |
| Arogyavardhini | Azadirachta indica | Leaf | CID_5280460 | 7367 | UGT2B17 | ChEMBL |
| Arogyavardhini | Azadirachta indica | Leaf | CID_5280460 | 759 | CA1 | BindingDB \| ChEMBL \| NPASS |
| Arogyavardhini | Azadirachta indica | Leaf | CID_5280460 | 760 | CA2 | BindingDB \| ChEMBL \| NPASS |
| Arogyavardhini | Azadirachta indica | Leaf | CID_5280460 | 766 | CA7 | BindingDB \| ChEMBL \| NPASS |
| Arogyavardhini | Azadirachta indica | Leaf | CID_5280460 | 768 | CA9 | BindingDB \| ChEMBL \| NPASS |
| Arogyavardhini | Azadirachta indica | Leaf | CID_5280460 | 771 | CA12 | BindingDB \| ChEMBL \| NPASS |
| Arogyavardhini | Azadirachta indica | Leaf | CID_5280460 | 834 | CASP1 | NPASS |
| Arogyavardhini | Azadirachta indica | Leaf | CID_5280460 | 840 | CASP7 | NPASS |
| Arogyavardhini | Azadirachta indica | Leaf | CID_5280489 | 10013 | HDAC6 | ChEMBL |
| Arogyavardhini | Azadirachta indica | Leaf | CID_5280489 | 10599 | SLCO1B1 | ChEMBL \| NPASS |
| Arogyavardhini | Azadirachta indica | Leaf | CID_5280489 | 11255 | HRH3 | ChEMBL |
| Arogyavardhini | Azadirachta indica | Leaf | CID_5280489 | 1128 | CHRM1 | ChEMBL |
| Arogyavardhini | Azadirachta indica | Leaf | CID_5280489 | 1129 | CHRM2 | ChEMBL |
| Arogyavardhini | Azadirachta indica | Leaf | CID_5280489 | 140 | ADORA3 | ChEMBL |
| Arogyavardhini | Azadirachta indica | Leaf | CID_5280489 | 148 | ADRA1A | ChEMBL |
| Arogyavardhini | Azadirachta indica | Leaf | CID_5280489 | 150 | ADRA2A | ChEMBL |
| Arogyavardhini | Azadirachta indica | Leaf | CID_5280489 | 153 | ADRB1 | ChEMBL |
| Arogyavardhini | Azadirachta indica | Leaf | CID_5280489 | 154 | ADRB2 | ChEMBL |
| Arogyavardhini | Azadirachta indica | Leaf | CID_5280489 | 155 | ADRB3 | ChEMBL \| NPASS |
| Arogyavardhini | Azadirachta indica | Leaf | CID_5280489 | 1812 | DRD1 | ChEMBL |
| Arogyavardhini | Azadirachta indica | Leaf | CID_5280489 | 1814 | DRD3 | ChEMBL |
| Arogyavardhini | Azadirachta indica | Leaf | CID_5280489 | 2099 | ESR1 | ChEMBL |
| Arogyavardhini | Azadirachta indica | Leaf | CID_5280489 | 2147 | F2 | ChEMBL |
| Arogyavardhini | Azadirachta indica | Leaf | CID_5280489 | 216 | ALDH1A1 | NPASS |
| Arogyavardhini | Azadirachta indica | Leaf | CID_5280489 | 2648 | KAT2A | NPASS |
| Arogyavardhini | Azadirachta indica | Leaf | CID_5280489 | 28234 | SLCO1B3 | ChEMBL \| NPASS |
| Arogyavardhini | Azadirachta indica | Leaf | CID_5280489 | 3350 | HTR1A | ChEMBL |
| Arogyavardhini | Azadirachta indica | Leaf | CID_5280489 | 3357 | HTR2B | ChEMBL |
| Arogyavardhini | Azadirachta indica | Leaf | CID_5280489 | 367 | AR | ChEMBL |
| Arogyavardhini | Azadirachta indica | Leaf | CID_5280489 | 3757 | KCNH2 | ChEMBL |
| Arogyavardhini | Azadirachta indica | Leaf | CID_5280489 | 3791 | KDR | ChEMBL |
| Arogyavardhini | Azadirachta indica | Leaf | CID_5280489 | 4000 | LMNA | NPASS |
| Arogyavardhini | Azadirachta indica | Leaf | CID_5280489 | 4128 | MAOA | ChEMBL |
| Arogyavardhini | Azadirachta indica | Leaf | CID_5280489 | 43 | ACHE | ChEMBL |
| Arogyavardhini | Azadirachta indica | Leaf | CID_5280489 | 4988 | OPRM1 | ChEMBL |
| Arogyavardhini | Azadirachta indica | Leaf | CID_5280489 | 51053 | GMNN | NPASS |
| Arogyavardhini | Azadirachta indica | Leaf | CID_5280489 | 5139 | PDE3A | ChEMBL |
| Arogyavardhini | Azadirachta indica | Leaf | CID_5280489 | 5141 | PDE4A | ChEMBL |
| Arogyavardhini | Azadirachta indica | Leaf | CID_5280489 | 5241 | PGR | ChEMBL |
| Arogyavardhini | Azadirachta indica | Leaf | CID_5280489 | 55775 | TDP1 | NPASS |
| Arogyavardhini | Azadirachta indica | Leaf | CID_5280489 | 5742 | PTGS1 | ChEMBL |
| Arogyavardhini | Azadirachta indica | Leaf | CID_5280489 | 6530 | SLC6A2 | ChEMBL |
| Arogyavardhini | Azadirachta indica | Leaf | CID_5280489 | 6531 | SLC6A3 | ChEMBL |
| Arogyavardhini | Azadirachta indica | Leaf | CID_5280489 | 6532 | SLC6A4 | ChEMBL |
| Arogyavardhini | Azadirachta indica | Leaf | CID_5280489 | 6915 | TBXA2R | ChEMBL |
| Arogyavardhini | Azadirachta indica | Leaf | CID_5280489 | 8647 | ABCB11 | ChEMBL \| NPASS |
| Arogyavardhini | Azadirachta indica | Leaf | CID_5280794 | 10013 | HDAC6 | ChEMBL |
| Arogyavardhini | Azadirachta indica | Leaf | CID_5280794 | 10599 | SLCO1B1 | ChEMBL \| NPASS |
| Arogyavardhini | Azadirachta indica | Leaf | CID_5280794 | 28234 | SLCO1B3 | ChEMBL \| NPASS |
| Arogyavardhini | Azadirachta indica | Leaf | CID_5280794 | 3363 | HTR7 | ChEMBL \| NPASS |
| Arogyavardhini | Azadirachta indica | Leaf | CID_5280794 | 3417 | IDH1 | ChEMBL \| NPASS |
| Arogyavardhini | Azadirachta indica | Leaf | CID_5280794 | 5423 | POLB | BindingDB \| ChEMBL \| NPASS |
| Arogyavardhini | Azadirachta indica | Leaf | CID_5281243 | 10013 | HDAC6 | ChEMBL |
| Arogyavardhini | Azadirachta indica | Leaf | CID_5281243 | 367 | AR | NPASS |
| Arogyavardhini | Azadirachta indica | Leaf | CID_5281243 | 55775 | TDP1 | NPASS |
| Arogyavardhini | Azadirachta indica | Leaf | CID_5281515 | 10013 | HDAC6 | ChEMBL |
| Arogyavardhini | Azadirachta indica | Leaf | CID_5281515 | 10599 | SLCO1B1 | ChEMBL \| NPASS |
| Arogyavardhini | Azadirachta indica | Leaf | CID_5281515 | 1269 | CNR2 | BindingDB \| ChEMBL \| NPASS |
| Arogyavardhini | Azadirachta indica | Leaf | CID_5281515 | 28234 | SLCO1B3 | ChEMBL \| NPASS |
| Arogyavardhini | Azadirachta indica | Leaf | CID_5281515 | 4297 | KMT2A | NPASS |
| Arogyavardhini | Azadirachta indica | Leaf | CID_5281515 | 5465 | PPARA | ChEMBL |
| Arogyavardhini | Azadirachta indica | Leaf | CID_5281520 | 10599 | SLCO1B1 | ChEMBL \| NPASS |
| Arogyavardhini | Azadirachta indica | Leaf | CID_5281520 | 28234 | SLCO1B3 | ChEMBL \| NPASS |
| Arogyavardhini | Azadirachta indica | Leaf | CID_5281654 | 1432 | MAPK14 | BindingDB \| ChEMBL |
| Arogyavardhini | Azadirachta indica | Leaf | CID_5281654 | 1543 | CYP1A1 | BindingDB \| ChEMBL \| NPASS |

**Table S3: Associations between formulations, constituent plants, phytochemicals, and human molecular targets. The table lists the formulations, plant name and plant part, phytochemical identifiers, associated human target genes (Entrez Gene ID and gene symbol), and the database sources supporting each phytochemical-target interaction.**

| Formulation | Plant name | Part | PubChem CID | Entrez Gene ID | Gene Symbol | Source database(s) |
|---|---|---|---|---|---|---|
| Arogyavardhini | Azadirachta indica | Leaf | CID_5281654 | 1544 | CYP1A2 | BindingDB \| ChEMBL \| NPASS |
| Arogyavardhini | Azadirachta indica | Leaf | CID_5281654 | 1545 | CYP1B1 | BindingDB \| ChEMBL \| NPASS |
| Arogyavardhini | Azadirachta indica | Leaf | CID_5281654 | 2237 | FEN1 | NPASS |
| Arogyavardhini | Azadirachta indica | Leaf | CID_5281654 | 253430 | IPMK | ChEMBL \| NPASS |
| Arogyavardhini | Azadirachta indica | Leaf | CID_5281654 | 276 | AMY1A | ChEMBL |
| Arogyavardhini | Azadirachta indica | Leaf | CID_5281654 | 277 | AMY1B | BindingDB |
| Arogyavardhini | Azadirachta indica | Leaf | CID_5281654 | 328 | APEX1 | NPASS |
| Arogyavardhini | Azadirachta indica | Leaf | CID_5281654 | 4297 | KMT2A | NPASS |
| Arogyavardhini | Azadirachta indica | Leaf | CID_5281654 | 4790 | NFKB1 | ChEMBL |
| Arogyavardhini | Azadirachta indica | Leaf | CID_5281654 | 4791 | NFKB2 | ChEMBL |
| Arogyavardhini | Azadirachta indica | Leaf | CID_5281654 | 51426 | POLK | NPASS |
| Arogyavardhini | Azadirachta indica | Leaf | CID_5281654 | 51447 | IP6K2 | ChEMBL \| NPASS |
| Arogyavardhini | Azadirachta indica | Leaf | CID_5281654 | 5468 | PPARG | BindingDB \| ChEMBL |
| Arogyavardhini | Azadirachta indica | Leaf | CID_5281654 | 5594 | MAPK1 | ChEMBL |
| Arogyavardhini | Azadirachta indica | Leaf | CID_5281654 | 5595 | MAPK3 | ChEMBL \| NPASS |
| Arogyavardhini | Azadirachta indica | Leaf | CID_5281654 | 5600 | MAPK11 | ChEMBL |
| Arogyavardhini | Azadirachta indica | Leaf | CID_5281654 | 5602 | MAPK10 | BindingDB |
| Arogyavardhini | Azadirachta indica | Leaf | CID_5281654 | 5603 | MAPK13 | ChEMBL |
| Arogyavardhini | Azadirachta indica | Leaf | CID_5281654 | 5970 | RELA | ChEMBL |
| Arogyavardhini | Azadirachta indica | Leaf | CID_5281654 | 6300 | MAPK12 | ChEMBL \| NPASS |
| Arogyavardhini | Azadirachta indica | Leaf | CID_5281654 | 7498 | XDH | BindingDB \| ChEMBL \| NPASS |
| Arogyavardhini | Azadirachta indica | Leaf | CID_5281654 | 759 | CA1 | BindingDB \| ChEMBL \| NPASS |
| Arogyavardhini | Azadirachta indica | Leaf | CID_5281654 | 760 | CA2 | BindingDB \| ChEMBL \| NPASS |
| Arogyavardhini | Azadirachta indica | Leaf | CID_5281654 | 762 | CA4 | BindingDB \| ChEMBL \| NPASS |
| Arogyavardhini | Azadirachta indica | Leaf | CID_5281654 | 766 | CA7 | BindingDB \| ChEMBL \| NPASS |
| Arogyavardhini | Azadirachta indica | Leaf | CID_5281654 | 771 | CA12 | BindingDB \| ChEMBL \| NPASS |
| Arogyavardhini | Azadirachta indica | Leaf | CID_5281654 | 9212 | AURKB | BindingDB |
| Arogyavardhini | Azadirachta indica | Leaf | CID_5742590 | 10013 | HDAC6 | ChEMBL |
| Arogyavardhini | Azadirachta indica | Leaf | CID_5742590 | 2932 | GSK3B | ChEMBL |
| Arogyavardhini | Azadirachta indica | Leaf | CID_5742590 | 3417 | IDH1 | ChEMBL \| NPASS |
| Arogyavardhini | Azadirachta indica | Leaf | CID_5742590 | 51422 | PRKAG2 | ChEMBL |
| Arogyavardhini | Azadirachta indica | Leaf | CID_5742590 | 53632 | PRKAG3 | ChEMBL |
| Arogyavardhini | Azadirachta indica | Leaf | CID_5742590 | 5423 | POLB | BindingDB \| ChEMBL \| NPASS |
| Arogyavardhini | Azadirachta indica | Leaf | CID_5742590 | 5562 | PRKAA1 | ChEMBL |
| Arogyavardhini | Azadirachta indica | Leaf | CID_5742590 | 5563 | PRKAA2 | ChEMBL |
| Arogyavardhini | Azadirachta indica | Leaf | CID_5742590 | 5564 | PRKAB1 | ChEMBL |
| Arogyavardhini | Azadirachta indica | Leaf | CID_5742590 | 5565 | PRKAB2 | ChEMBL |
| Arogyavardhini | Azadirachta indica | Leaf | CID_5742590 | 5571 | PRKAG1 | ChEMBL |
| Arogyavardhini | Azadirachta indica | Leaf | CID_6918391 | 10013 | HDAC6 | ChEMBL |
| Arogyavardhini | Azadirachta indica | Leaf | CID_6918391 | 5315 | PKM | ChEMBL |
| Arogyavardhini | Azadirachta indica | Leaf | CID_6918774 | 10013 | HDAC6 | ChEMBL |
| Arogyavardhini | Azadirachta indica | Leaf | CID_6918774 | 10951 | CBX1 | NPASS |
| Arogyavardhini | Azadirachta indica | Leaf | CID_6918774 | 11069 | RAPGEF4 | NPASS |
| Arogyavardhini | Azadirachta indica | Leaf | CID_6918774 | 240 | ALOX5 | BindingDB \| ChEMBL \| NPASS |
| Arogyavardhini | Azadirachta indica | Leaf | CID_6918774 | 247 | ALOX15B | ChEMBL \| NPASS |
| Arogyavardhini | Azadirachta indica | Leaf | CID_6918774 | 29843 | SENP1 | ChEMBL |
| Arogyavardhini | Azadirachta indica | Leaf | CID_6918774 | 3064 | HTT | NPASS |
| Arogyavardhini | Azadirachta indica | Leaf | CID_6918774 | 3290 | HSD11B1 | BindingDB \| ChEMBL \| NPASS |
| Arogyavardhini | Azadirachta indica | Leaf | CID_6918774 | 4088 | SMAD3 | NPASS |
| Arogyavardhini | Azadirachta indica | Leaf | CID_6918774 | 4780 | NFE2L2 | NPASS |
| Arogyavardhini | Azadirachta indica | Leaf | CID_6918774 | 51053 | GMNN | NPASS |
| Arogyavardhini | Azadirachta indica | Leaf | CID_6918774 | 51422 | PRKAG2 | ChEMBL |
| Arogyavardhini | Azadirachta indica | Leaf | CID_6918774 | 53632 | PRKAG3 | ChEMBL |
| Arogyavardhini | Azadirachta indica | Leaf | CID_6918774 | 5423 | POLB | BindingDB \| ChEMBL \| NPASS |
| Arogyavardhini | Azadirachta indica | Leaf | CID_6918774 | 5562 | PRKAA1 | ChEMBL |
| Arogyavardhini | Azadirachta indica | Leaf | CID_6918774 | 5563 | PRKAA2 | ChEMBL |
| Arogyavardhini | Azadirachta indica | Leaf | CID_6918774 | 5564 | PRKAB1 | ChEMBL |
| Arogyavardhini | Azadirachta indica | Leaf | CID_6918774 | 5565 | PRKAB2 | ChEMBL |
| Arogyavardhini | Azadirachta indica | Leaf | CID_6918774 | 5571 | PRKAG1 | ChEMBL |
| Arogyavardhini | Azadirachta indica | Leaf | CID_6918774 | 55775 | TDP1 | ChEMBL \| NPASS |
| Arogyavardhini | Azadirachta indica | Leaf | CID_6918774 | 5743 | PTGS2 | BindingDB \| ChEMBL \| NPASS |
| Arogyavardhini | Azadirachta indica | Leaf | CID_6918774 | 5770 | PTPN1 | BindingDB \| ChEMBL \| NPASS |
| Arogyavardhini | Azadirachta indica | Leaf | CID_6918774 | 5771 | PTPN2 | BindingDB \| ChEMBL \| NPASS |
| Arogyavardhini | Azadirachta indica | Leaf | CID_6918774 | 5777 | PTPN6 | BindingDB \| ChEMBL \| NPASS |
| Arogyavardhini | Azadirachta indica | Leaf | CID_6918774 | 5786 | PTPRA | ChEMBL \| NPASS |
| Arogyavardhini | Azadirachta indica | Leaf | CID_6918774 | 5791 | PTPRE | ChEMBL \| NPASS |
| Arogyavardhini | Azadirachta indica | Leaf | CID_6918774 | 5792 | PTPRF | BindingDB \| ChEMBL \| NPASS |
| Arogyavardhini | Azadirachta indica | Leaf | CID_6918774 | 5837 | PYGM | BindingDB \| ChEMBL \| NPASS |
| Arogyavardhini | Azadirachta indica | Leaf | CID_6918774 | 6311 | ATXN2 | NPASS |
| Arogyavardhini | Azadirachta indica | Leaf | CID_6918774 | 7874 | USP7 | BindingDB \| ChEMBL |

**Table S3: Associations between formulations, constituent plants, phytochemicals, and human molecular targets. The table lists the formulations, plant name and plant part, phytochemical identifiers, associated human target genes (Entrez Gene ID and gene symbol), and the database sources supporting each phytochemical-target interaction.**

| Formulation | Plant name | Part | PubChem CID | Entrez Gene ID | Gene Symbol | Source database(s) |
|---|---|---|---|---|---|---|
| Arogyavardhini | Azadirachta indica | Leaf | CID_6918774 | 79915 | ATAD5 | NPASS |
| Arogyavardhini | Azadirachta indica | Leaf | CID_984 | 10013 | HDAC6 | ChEMBL |
| Arogyavardhini | Azadirachta indica | Leaf | CID_984 | 4023 | LPL | ChEMBL \| NPASS |
| Arogyavardhini | Azadirachta indica | Leaf | CID_984 | 9388 | LIPG | ChEMBL \| NPASS |
| Arogyavardhini | Commiphora wightii | Plant exudate | CID_222284 | 10013 | HDAC6 | ChEMBL |
| Arogyavardhini | Commiphora wightii | Plant exudate | CID_222284 | 10599 | SLCO1B1 | ChEMBL \| NPASS |
| Arogyavardhini | Commiphora wightii | Plant exudate | CID_222284 | 1565 | CYP2D6 | ChEMBL \| NPASS |
| Arogyavardhini | Commiphora wightii | Plant exudate | CID_222284 | 1576 | CYP3A4 | BindingDB \| ChEMBL \| NPASS |
| Arogyavardhini | Commiphora wightii | Plant exudate | CID_222284 | 1803 | DPP4 | ChEMBL \| NPASS |
| Arogyavardhini | Commiphora wightii | Plant exudate | CID_222284 | 2147 | F2 | ChEMBL \| NPASS |
| Arogyavardhini | Commiphora wightii | Plant exudate | CID_222284 | 28234 | SLCO1B3 | ChEMBL \| NPASS |
| Arogyavardhini | Commiphora wightii | Plant exudate | CID_222284 | 2932 | GSK3B | ChEMBL |
| Arogyavardhini | Commiphora wightii | Plant exudate | CID_222284 | 3417 | IDH1 | ChEMBL \| NPASS |
| Arogyavardhini | Commiphora wightii | Plant exudate | CID_222284 | 51422 | PRKAG2 | ChEMBL |
| Arogyavardhini | Commiphora wightii | Plant exudate | CID_222284 | 5243 | ABCB1 | ChEMBL |
| Arogyavardhini | Commiphora wightii | Plant exudate | CID_222284 | 53632 | PRKAG3 | ChEMBL |
| Arogyavardhini | Commiphora wightii | Plant exudate | CID_222284 | 5423 | POLB | BindingDB \| ChEMBL \| NPASS |
| Arogyavardhini | Commiphora wightii | Plant exudate | CID_222284 | 5562 | PRKAA1 | ChEMBL |
| Arogyavardhini | Commiphora wightii | Plant exudate | CID_222284 | 5563 | PRKAA2 | ChEMBL |
| Arogyavardhini | Commiphora wightii | Plant exudate | CID_222284 | 5564 | PRKAB1 | ChEMBL |
| Arogyavardhini | Commiphora wightii | Plant exudate | CID_222284 | 5565 | PRKAB2 | ChEMBL |
| Arogyavardhini | Commiphora wightii | Plant exudate | CID_222284 | 5571 | PRKAG1 | ChEMBL |
| Arogyavardhini | Commiphora wightii | Plant exudate | CID_222284 | 7299 | TYR | ChEMBL \| NPASS |
| Arogyavardhini | Commiphora wightii | Plant exudate | CID_31253 | 2908 | NR3C1 | NPASS |
| Arogyavardhini | Commiphora wightii | Plant exudate | CID_31253 | 3315 | HSPB1 | NPASS |
| Arogyavardhini | Commiphora wightii | Plant exudate | CID_31253 | 3725 | JUN | NPASS |
| Arogyavardhini | Commiphora wightii | Plant exudate | CID_31253 | 4790 | NFKB1 | NPASS |
| Arogyavardhini | Commiphora wightii | Plant exudate | CID_31253 | 5467 | PPARD | NPASS |
| Arogyavardhini | Commiphora wightii | Plant exudate | CID_31253 | 7253 | TSHR | NPASS |
| Arogyavardhini | Commiphora wightii | Plant exudate | CID_31253 | 9971 | NR1H4 | NPASS |
| Arogyavardhini | Commiphora wightii | Plant exudate | CID_445858 | 10013 | HDAC6 | ChEMBL |
| Arogyavardhini | Commiphora wightii | Plant exudate | CID_445858 | 10014 | HDAC5 | ChEMBL |
| Arogyavardhini | Commiphora wightii | Plant exudate | CID_445858 | 10280 | SIGMAR1 | ChEMBL |
| Arogyavardhini | Commiphora wightii | Plant exudate | CID_445858 | 10599 | SLCO1B1 | ChEMBL \| NPASS |
| Arogyavardhini | Commiphora wightii | Plant exudate | CID_445858 | 10800 | CYSLTR1 | ChEMBL |
| Arogyavardhini | Commiphora wightii | Plant exudate | CID_445858 | 11238 | CA5B | BindingDB \| ChEMBL \| NPASS |
| Arogyavardhini | Commiphora wightii | Plant exudate | CID_445858 | 1128 | CHRM1 | ChEMBL |
| Arogyavardhini | Commiphora wightii | Plant exudate | CID_445858 | 1129 | CHRM2 | ChEMBL |
| Arogyavardhini | Commiphora wightii | Plant exudate | CID_445858 | 1131 | CHRM3 | ChEMBL |
| Arogyavardhini | Commiphora wightii | Plant exudate | CID_445858 | 1132 | CHRM4 | ChEMBL |
| Arogyavardhini | Commiphora wightii | Plant exudate | CID_445858 | 1133 | CHRM5 | ChEMBL |
| Arogyavardhini | Commiphora wightii | Plant exudate | CID_445858 | 1233 | CCR4 | ChEMBL |
| Arogyavardhini | Commiphora wightii | Plant exudate | CID_445858 | 1234 | CCR5 | ChEMBL |
| Arogyavardhini | Commiphora wightii | Plant exudate | CID_445858 | 1268 | CNR1 | ChEMBL |
| Arogyavardhini | Commiphora wightii | Plant exudate | CID_445858 | 134 | ADORA1 | ChEMBL |
| Arogyavardhini | Commiphora wightii | Plant exudate | CID_445858 | 135 | ADORA2A | ChEMBL |
| Arogyavardhini | Commiphora wightii | Plant exudate | CID_445858 | 140 | ADORA3 | ChEMBL |
| Arogyavardhini | Commiphora wightii | Plant exudate | CID_445858 | 1432 | MAPK14 | ChEMBL |
| Arogyavardhini | Commiphora wightii | Plant exudate | CID_445858 | 1457 | CSNK2A1 | ChEMBL |
| Arogyavardhini | Commiphora wightii | Plant exudate | CID_445858 | 1459 | CSNK2A2 | ChEMBL |
| Arogyavardhini | Commiphora wightii | Plant exudate | CID_445858 | 146 | ADRA1D | ChEMBL |
| Arogyavardhini | Commiphora wightii | Plant exudate | CID_445858 | 1460 | CSNK2B | ChEMBL |
| Arogyavardhini | Commiphora wightii | Plant exudate | CID_445858 | 150 | ADRA2A | ChEMBL |
| Arogyavardhini | Commiphora wightii | Plant exudate | CID_445858 | 151 | ADRA2B | ChEMBL |
| Arogyavardhini | Commiphora wightii | Plant exudate | CID_445858 | 1511 | CTSG | ChEMBL |
| Arogyavardhini | Commiphora wightii | Plant exudate | CID_445858 | 152 | ADRA2C | ChEMBL |
| Arogyavardhini | Commiphora wightii | Plant exudate | CID_445858 | 153 | ADRB1 | ChEMBL |
| Arogyavardhini | Commiphora wightii | Plant exudate | CID_445858 | 154 | ADRB2 | ChEMBL |
| Arogyavardhini | Commiphora wightii | Plant exudate | CID_445858 | 1544 | CYP1A2 | ChEMBL |
| Arogyavardhini | Commiphora wightii | Plant exudate | CID_445858 | 1548 | CYP2A6 | ChEMBL |
| Arogyavardhini | Commiphora wightii | Plant exudate | CID_445858 | 155 | ADRB3 | ChEMBL |
| Arogyavardhini | Commiphora wightii | Plant exudate | CID_445858 | 1557 | CYP2C19 | ChEMBL |
| Arogyavardhini | Commiphora wightii | Plant exudate | CID_445858 | 1559 | CYP2C9 | ChEMBL |
| Arogyavardhini | Commiphora wightii | Plant exudate | CID_445858 | 1565 | CYP2D6 | ChEMBL |
| Arogyavardhini | Commiphora wightii | Plant exudate | CID_445858 | 1571 | CYP2E1 | ChEMBL |
| Arogyavardhini | Commiphora wightii | Plant exudate | CID_445858 | 1576 | CYP3A4 | ChEMBL |
| Arogyavardhini | Commiphora wightii | Plant exudate | CID_445858 | 1812 | DRD1 | ChEMBL |
| Arogyavardhini | Commiphora wightii | Plant exudate | CID_445858 | 1813 | DRD2 | ChEMBL |
| Arogyavardhini | Commiphora wightii | Plant exudate | CID_445858 | 1814 | DRD3 | ChEMBL |
| Arogyavardhini | Commiphora wightii | Plant exudate | CID_445858 | 1815 | DRD4 | ChEMBL |

**Table S3: Associations between formulations, constituent plants, phytochemicals, and human molecular targets. The table lists the formulations, plant name and plant part, phytochemical identifiers, associated human target genes (Entrez Gene ID and gene symbol), and the database sources supporting each phytochemical-target interaction.**

| Formulation | Plant name | Part | PubChem CID | Entrez Gene ID | Gene Symbol | Source database(s) |
|---|---|---|---|---|---|---|
| Arogyavardhini | Commiphora wightii | Plant exudate | CID_445858 | 186 | AGTR2 | ChEMBL |
| Arogyavardhini | Commiphora wightii | Plant exudate | CID_445858 | 1909 | EDNRA | ChEMBL |
| Arogyavardhini | Commiphora wightii | Plant exudate | CID_445858 | 1956 | EGFR | ChEMBL |
| Arogyavardhini | Commiphora wightii | Plant exudate | CID_445858 | 1991 | ELANE | ChEMBL |
| Arogyavardhini | Commiphora wightii | Plant exudate | CID_445858 | 2064 | ERBB2 | ChEMBL |
| Arogyavardhini | Commiphora wightii | Plant exudate | CID_445858 | 2099 | ESR1 | ChEMBL \| NPASS |
| Arogyavardhini | Commiphora wightii | Plant exudate | CID_445858 | 2100 | ESR2 | ChEMBL |
| Arogyavardhini | Commiphora wightii | Plant exudate | CID_445858 | 2260 | FGFR1 | ChEMBL \| NPASS |
| Arogyavardhini | Commiphora wightii | Plant exudate | CID_445858 | 2263 | FGFR2 | ChEMBL \| NPASS |
| Arogyavardhini | Commiphora wightii | Plant exudate | CID_445858 | 2321 | FLT1 | ChEMBL |
| Arogyavardhini | Commiphora wightii | Plant exudate | CID_445858 | 23632 | CA14 | BindingDB \| ChEMBL \| NPASS |
| Arogyavardhini | Commiphora wightii | Plant exudate | CID_445858 | 2534 | FYN | ChEMBL |
| Arogyavardhini | Commiphora wightii | Plant exudate | CID_445858 | 259285 | TAS2R39 | ChEMBL \| NPASS |
| Arogyavardhini | Commiphora wightii | Plant exudate | CID_445858 | 28234 | SLCO1B3 | ChEMBL \| NPASS |
| Arogyavardhini | Commiphora wightii | Plant exudate | CID_445858 | 283106 | CSNK2A3 | BindingDB \| ChEMBL |
| Arogyavardhini | Commiphora wightii | Plant exudate | CID_445858 | 2908 | NR3C1 | ChEMBL |
| Arogyavardhini | Commiphora wightii | Plant exudate | CID_445858 | 3065 | HDAC1 | ChEMBL |
| Arogyavardhini | Commiphora wightii | Plant exudate | CID_445858 | 3066 | HDAC2 | ChEMBL |
| Arogyavardhini | Commiphora wightii | Plant exudate | CID_445858 | 3156 | HMGCR | ChEMBL |
| Arogyavardhini | Commiphora wightii | Plant exudate | CID_445858 | 3269 | HRH1 | ChEMBL |
| Arogyavardhini | Commiphora wightii | Plant exudate | CID_445858 | 3274 | HRH2 | ChEMBL |
| Arogyavardhini | Commiphora wightii | Plant exudate | CID_445858 | 3356 | HTR2A | ChEMBL |
| Arogyavardhini | Commiphora wightii | Plant exudate | CID_445858 | 3357 | HTR2B | ChEMBL |
| Arogyavardhini | Commiphora wightii | Plant exudate | CID_445858 | 3358 | HTR2C | ChEMBL |
| Arogyavardhini | Commiphora wightii | Plant exudate | CID_445858 | 3362 | HTR6 | ChEMBL |
| Arogyavardhini | Commiphora wightii | Plant exudate | CID_445858 | 3363 | HTR7 | ChEMBL \| NPASS |
| Arogyavardhini | Commiphora wightii | Plant exudate | CID_445858 | 3373 | HYAL1 | BindingDB \| NPASS |
| Arogyavardhini | Commiphora wightii | Plant exudate | CID_445858 | 338442 | HCAR2 | ChEMBL \| NPASS |
| Arogyavardhini | Commiphora wightii | Plant exudate | CID_445858 | 351 | APP | BindingDB \| ChEMBL \| NPASS |
| Arogyavardhini | Commiphora wightii | Plant exudate | CID_445858 | 3577 | CXCR1 | ChEMBL |
| Arogyavardhini | Commiphora wightii | Plant exudate | CID_445858 | 3579 | CXCR2 | ChEMBL |
| Arogyavardhini | Commiphora wightii | Plant exudate | CID_445858 | 3757 | KCNH2 | ChEMBL |
| Arogyavardhini | Commiphora wightii | Plant exudate | CID_445858 | 390245 | KDM4E | NPASS |
| Arogyavardhini | Commiphora wightii | Plant exudate | CID_445858 | 3932 | LCK | ChEMBL |
| Arogyavardhini | Commiphora wightii | Plant exudate | CID_445858 | 4128 | MAOA | ChEMBL |
| Arogyavardhini | Commiphora wightii | Plant exudate | CID_445858 | 4159 | MC3R | ChEMBL |
| Arogyavardhini | Commiphora wightii | Plant exudate | CID_445858 | 4160 | MC4R | ChEMBL |
| Arogyavardhini | Commiphora wightii | Plant exudate | CID_445858 | 4161 | MC5R | ChEMBL |
| Arogyavardhini | Commiphora wightii | Plant exudate | CID_445858 | 43 | ACHE | BindingDB \| ChEMBL \| NPASS |
| Arogyavardhini | Commiphora wightii | Plant exudate | CID_445858 | 4312 | MMP1 | ChEMBL |
| Arogyavardhini | Commiphora wightii | Plant exudate | CID_445858 | 4318 | MMP9 | ChEMBL |
| Arogyavardhini | Commiphora wightii | Plant exudate | CID_445858 | 4886 | NPY1R | ChEMBL |
| Arogyavardhini | Commiphora wightii | Plant exudate | CID_445858 | 4887 | NPY2R | ChEMBL |
| Arogyavardhini | Commiphora wightii | Plant exudate | CID_445858 | 4985 | OPRD1 | ChEMBL |
| Arogyavardhini | Commiphora wightii | Plant exudate | CID_445858 | 4986 | OPRK1 | ChEMBL |
| Arogyavardhini | Commiphora wightii | Plant exudate | CID_445858 | 4988 | OPRM1 | ChEMBL |
| Arogyavardhini | Commiphora wightii | Plant exudate | CID_445858 | 51053 | GMNN | NPASS |
| Arogyavardhini | Commiphora wightii | Plant exudate | CID_445858 | 51564 | HDAC7 | ChEMBL |
| Arogyavardhini | Commiphora wightii | Plant exudate | CID_445858 | 5315 | PKM | ChEMBL |
| Arogyavardhini | Commiphora wightii | Plant exudate | CID_445858 | 54657 | UGT1A4 | ChEMBL \| NPASS |
| Arogyavardhini | Commiphora wightii | Plant exudate | CID_445858 | 54658 | UGT1A1 | ChEMBL \| NPASS |
| Arogyavardhini | Commiphora wightii | Plant exudate | CID_445858 | 552 | AVPR1A | ChEMBL |
| Arogyavardhini | Commiphora wightii | Plant exudate | CID_445858 | 5530 | PPP3CA | ChEMBL |
| Arogyavardhini | Commiphora wightii | Plant exudate | CID_445858 | 55775 | TDP1 | NPASS |
| Arogyavardhini | Commiphora wightii | Plant exudate | CID_445858 | 5578 | PRKCA | ChEMBL |
| Arogyavardhini | Commiphora wightii | Plant exudate | CID_445858 | 55869 | HDAC8 | ChEMBL |
| Arogyavardhini | Commiphora wightii | Plant exudate | CID_445858 | 5594 | MAPK1 | ChEMBL |
| Arogyavardhini | Commiphora wightii | Plant exudate | CID_445858 | 5595 | MAPK3 | ChEMBL |
| Arogyavardhini | Commiphora wightii | Plant exudate | CID_445858 | 5724 | PTAFR | ChEMBL |
| Arogyavardhini | Commiphora wightii | Plant exudate | CID_445858 | 5742 | PTGS1 | ChEMBL |
| Arogyavardhini | Commiphora wightii | Plant exudate | CID_445858 | 5743 | PTGS2 | ChEMBL |
| Arogyavardhini | Commiphora wightii | Plant exudate | CID_445858 | 5745 | PTH1R | NPASS |
| Arogyavardhini | Commiphora wightii | Plant exudate | CID_445858 | 5788 | PTPRC | ChEMBL |
| Arogyavardhini | Commiphora wightii | Plant exudate | CID_445858 | 590 | BCHE | BindingDB \| ChEMBL \| NPASS |
| Arogyavardhini | Commiphora wightii | Plant exudate | CID_445858 | 624 | BDKRB2 | ChEMBL |
| Arogyavardhini | Commiphora wightii | Plant exudate | CID_445858 | 6530 | SLC6A2 | ChEMBL |
| Arogyavardhini | Commiphora wightii | Plant exudate | CID_445858 | 6531 | SLC6A3 | ChEMBL |
| Arogyavardhini | Commiphora wightii | Plant exudate | CID_445858 | 6532 | SLC6A4 | ChEMBL |
| Arogyavardhini | Commiphora wightii | Plant exudate | CID_445858 | 6622 | SNCA | ChEMBL \| NPASS |
| Arogyavardhini | Commiphora wightii | Plant exudate | CID_445858 | 6865 | TACR2 | ChEMBL |

**Table S3: Associations between formulations, constituent plants, phytochemicals, and human molecular targets. The table lists the formulations, plant name and plant part, phytochemical identifiers, associated human target genes (Entrez Gene ID and gene symbol), and the database sources supporting each phytochemical-target interaction.**

| Formulation | Plant name | Part | PubChem CID | Entrez Gene ID | Gene Symbol | Source database(s) |
|---|---|---|---|---|---|---|
| Arogyavardhini | Commiphora wightii | Plant exudate | CID_445858 | 6869 | TACR1 | ChEMBL |
| Arogyavardhini | Commiphora wightii | Plant exudate | CID_445858 | 6916 | TBXAS1 | ChEMBL |
| Arogyavardhini | Commiphora wightii | Plant exudate | CID_445858 | 729230 | CCR2 | ChEMBL |
| Arogyavardhini | Commiphora wightii | Plant exudate | CID_445858 | 7299 | TYR | ChEMBL \| NPASS |
| Arogyavardhini | Commiphora wightii | Plant exudate | CID_445858 | 7366 | UGT2B15 | ChEMBL \| NPASS |
| Arogyavardhini | Commiphora wightii | Plant exudate | CID_445858 | 7433 | VIPR1 | ChEMBL |
| Arogyavardhini | Commiphora wightii | Plant exudate | CID_445858 | 759 | CA1 | BindingDB \| ChEMBL \| NPASS |
| Arogyavardhini | Commiphora wightii | Plant exudate | CID_445858 | 760 | CA2 | BindingDB \| ChEMBL \| NPASS |
| Arogyavardhini | Commiphora wightii | Plant exudate | CID_445858 | 761 | CA3 | BindingDB \| ChEMBL \| NPASS |
| Arogyavardhini | Commiphora wightii | Plant exudate | CID_445858 | 762 | CA4 | BindingDB \| ChEMBL \| NPASS |
| Arogyavardhini | Commiphora wightii | Plant exudate | CID_445858 | 763 | CA5A | BindingDB \| ChEMBL \| NPASS |
| Arogyavardhini | Commiphora wightii | Plant exudate | CID_445858 | 765 | CA6 | ChEMBL \| NPASS |
| Arogyavardhini | Commiphora wightii | Plant exudate | CID_445858 | 766 | CA7 | BindingDB \| ChEMBL \| NPASS |
| Arogyavardhini | Commiphora wightii | Plant exudate | CID_445858 | 768 | CA9 | BindingDB \| ChEMBL \| NPASS |
| Arogyavardhini | Commiphora wightii | Plant exudate | CID_445858 | 771 | CA12 | BindingDB \| ChEMBL \| NPASS |
| Arogyavardhini | Commiphora wightii | Plant exudate | CID_445858 | 79885 | HDAC11 | ChEMBL |
| Arogyavardhini | Commiphora wightii | Plant exudate | CID_445858 | 799 | CALCR | ChEMBL |
| Arogyavardhini | Commiphora wightii | Plant exudate | CID_445858 | 834 | CASP1 | ChEMBL |
| Arogyavardhini | Commiphora wightii | Plant exudate | CID_445858 | 83933 | HDAC10 | ChEMBL \| NPASS |
| Arogyavardhini | Commiphora wightii | Plant exudate | CID_445858 | 8654 | PDE5A | ChEMBL |
| Arogyavardhini | Commiphora wightii | Plant exudate | CID_445858 | 8841 | HDAC3 | ChEMBL |
| Arogyavardhini | Commiphora wightii | Plant exudate | CID_445858 | 886 | CCKAR | ChEMBL |
| Arogyavardhini | Commiphora wightii | Plant exudate | CID_445858 | 9734 | HDAC9 | ChEMBL |
| Arogyavardhini | Commiphora wightii | Plant exudate | CID_445858 | 9759 | HDAC4 | ChEMBL |
| Arogyavardhini | Commiphora wightii | Plant exudate | CID_5204 | 10951 | CBX1 | NPASS |
| Arogyavardhini | Commiphora wightii | Plant exudate | CID_5204 | 11201 | POLI | NPASS |
| Arogyavardhini | Commiphora wightii | Plant exudate | CID_5204 | 23435 | TARDBP | NPASS |
| Arogyavardhini | Commiphora wightii | Plant exudate | CID_5204 | 2740 | GLP1R | NPASS |
| Arogyavardhini | Commiphora wightii | Plant exudate | CID_5204 | 51053 | GMNN | NPASS |
| Arogyavardhini | Commiphora wightii | Plant exudate | CID_5204 | 55775 | TDP1 | NPASS |
| Arogyavardhini | Commiphora wightii | Plant exudate | CID_5204 | 5745 | PTH1R | NPASS |
| Arogyavardhini | Commiphora wightii | Plant exudate | CID_5280794 | 10013 | HDAC6 | ChEMBL |
| Arogyavardhini | Commiphora wightii | Plant exudate | CID_5280794 | 10599 | SLCO1B1 | ChEMBL \| NPASS |
| Arogyavardhini | Commiphora wightii | Plant exudate | CID_5280794 | 28234 | SLCO1B3 | ChEMBL \| NPASS |
| Arogyavardhini | Commiphora wightii | Plant exudate | CID_5280794 | 3363 | HTR7 | ChEMBL \| NPASS |
| Arogyavardhini | Commiphora wightii | Plant exudate | CID_5280794 | 3417 | IDH1 | ChEMBL \| NPASS |
| Arogyavardhini | Commiphora wightii | Plant exudate | CID_5280794 | 5423 | POLB | BindingDB \| ChEMBL \| NPASS |
| Arogyavardhini | Commiphora wightii | Plant exudate | CID_5281384 | 5319 | PLA2G1B | ChEMBL \| NPASS |
| Arogyavardhini | Commiphora wightii | Plant exudate | CID_5281515 | 10013 | HDAC6 | ChEMBL |
| Arogyavardhini | Commiphora wightii | Plant exudate | CID_5281515 | 10599 | SLCO1B1 | ChEMBL \| NPASS |
| Arogyavardhini | Commiphora wightii | Plant exudate | CID_5281515 | 1269 | CNR2 | BindingDB \| ChEMBL \| NPASS |
| Arogyavardhini | Commiphora wightii | Plant exudate | CID_5281515 | 28234 | SLCO1B3 | ChEMBL \| NPASS |
| Arogyavardhini | Commiphora wightii | Plant exudate | CID_5281515 | 4297 | KMT2A | NPASS |
| Arogyavardhini | Commiphora wightii | Plant exudate | CID_5281515 | 5465 | PPARA | ChEMBL |
| Arogyavardhini | Commiphora wightii | Plant exudate | CID_5997 | 10013 | HDAC6 | ChEMBL |
| Arogyavardhini | Commiphora wightii | Plant exudate | CID_5997 | 10062 | NR1H3 | ChEMBL \| NPASS |
| Arogyavardhini | Commiphora wightii | Plant exudate | CID_5997 | 10599 | SLCO1B1 | ChEMBL \| NPASS |
| Arogyavardhini | Commiphora wightii | Plant exudate | CID_5997 | 10858 | CYP46A1 | ChEMBL \| NPASS |
| Arogyavardhini | Commiphora wightii | Plant exudate | CID_5997 | 1595 | CYP51A1 | NPASS |
| Arogyavardhini | Commiphora wightii | Plant exudate | CID_5997 | 23762 | OSBP2 | BindingDB \| ChEMBL |
| Arogyavardhini | Commiphora wightii | Plant exudate | CID_5997 | 28234 | SLCO1B3 | ChEMBL \| NPASS |
| Arogyavardhini | Commiphora wightii | Plant exudate | CID_5997 | 2908 | NR3C1 | NPASS |
| Arogyavardhini | Commiphora wightii | Plant exudate | CID_5997 | 3417 | IDH1 | ChEMBL \| NPASS |
| Arogyavardhini | Commiphora wightii | Plant exudate | CID_5997 | 367 | AR | NPASS |
| Arogyavardhini | Commiphora wightii | Plant exudate | CID_5997 | 4000 | LMNA | NPASS |
| Arogyavardhini | Commiphora wightii | Plant exudate | CID_5997 | 4780 | NFE2L2 | ChEMBL \| NPASS |
| Arogyavardhini | Commiphora wightii | Plant exudate | CID_5997 | 5007 | OSBP | ChEMBL |
| Arogyavardhini | Commiphora wightii | Plant exudate | CID_5997 | 5243 | ABCB1 | ChEMBL \| NPASS |
| Arogyavardhini | Commiphora wightii | Plant exudate | CID_5997 | 5422 | POLA1 | BindingDB \| NPASS |
| Arogyavardhini | Commiphora wightii | Plant exudate | CID_5997 | 5467 | PPARD | NPASS |
| Arogyavardhini | Commiphora wightii | Plant exudate | CID_5997 | 6097 | RORC | ChEMBL \| NPASS |
| Arogyavardhini | Commiphora wightii | Plant exudate | CID_5997 | 6323 | SCN1A | ChEMBL |
| Arogyavardhini | Commiphora wightii | Plant exudate | CID_5997 | 6326 | SCN2A | ChEMBL |
| Arogyavardhini | Commiphora wightii | Plant exudate | CID_5997 | 6328 | SCN3A | ChEMBL \| NPASS |
| Arogyavardhini | Commiphora wightii | Plant exudate | CID_5997 | 6622 | SNCA | ChEMBL \| NPASS |
| Arogyavardhini | Commiphora wightii | Plant exudate | CID_5997 | 6820 | SULT2B1 | ChEMBL |
| Arogyavardhini | Commiphora wightii | Plant exudate | CID_5997 | 7155 | TOP2B | NPASS |
| Arogyavardhini | Commiphora wightii | Plant exudate | CID_5997 | 9429 | ABCG2 | ChEMBL |
| Arogyavardhini | Commiphora wightii | Plant exudate | CID_6439929 | 10062 | NR1H3 | ChEMBL \| NPASS |
| Arogyavardhini | Commiphora wightii | Plant exudate | CID_6439929 | 1576 | CYP3A4 | ChEMBL \| NPASS |

**Table S3: Associations between formulations, constituent plants, phytochemicals, and human molecular targets. The table lists the formulations, plant name and plant part, phytochemical identifiers, associated human target genes (Entrez Gene ID and gene symbol), and the database sources supporting each phytochemical-target interaction.**

| Formulation | Plant name | Part | PubChem CID | Entrez Gene ID | Gene Symbol | Source database(s) |
|---|---|---|---|---|---|---|
| Arogyavardhini | Commiphora wightii | Plant exudate | CID_6439929 | 2099 | ESR1 | ChEMBL \| NPASS |
| Arogyavardhini | Commiphora wightii | Plant exudate | CID_6439929 | 2100 | ESR2 | ChEMBL |
| Arogyavardhini | Commiphora wightii | Plant exudate | CID_6439929 | 216 | ALDH1A1 | NPASS |
| Arogyavardhini | Commiphora wightii | Plant exudate | CID_6439929 | 2908 | NR3C1 | ChEMBL \| NPASS |
| Arogyavardhini | Commiphora wightii | Plant exudate | CID_6439929 | 3315 | HSPB1 | NPASS |
| Arogyavardhini | Commiphora wightii | Plant exudate | CID_6439929 | 3576 | CXCL8 | NPASS |
| Arogyavardhini | Commiphora wightii | Plant exudate | CID_6439929 | 367 | AR | BindingDB \| ChEMBL \| NPASS |
| Arogyavardhini | Commiphora wightii | Plant exudate | CID_6439929 | 3757 | KCNH2 | NPASS |
| Arogyavardhini | Commiphora wightii | Plant exudate | CID_6439929 | 4306 | NR3C2 | BindingDB \| ChEMBL \| NPASS |
| Arogyavardhini | Commiphora wightii | Plant exudate | CID_6439929 | 4780 | NFE2L2 | ChEMBL \| NPASS |
| Arogyavardhini | Commiphora wightii | Plant exudate | CID_6439929 | 4790 | NFKB1 | NPASS |
| Arogyavardhini | Commiphora wightii | Plant exudate | CID_6439929 | 5241 | PGR | ChEMBL \| NPASS |
| Arogyavardhini | Commiphora wightii | Plant exudate | CID_6439929 | 5465 | PPARA | ChEMBL \| NPASS |
| Arogyavardhini | Commiphora wightii | Plant exudate | CID_6439929 | 5467 | PPARD | ChEMBL \| NPASS |
| Arogyavardhini | Commiphora wightii | Plant exudate | CID_6439929 | 5468 | PPARG | ChEMBL \| NPASS |
| Arogyavardhini | Commiphora wightii | Plant exudate | CID_6439929 | 5594 | MAPK1 | NPASS |
| Arogyavardhini | Commiphora wightii | Plant exudate | CID_6439929 | 5914 | RARA | ChEMBL \| NPASS |
| Arogyavardhini | Commiphora wightii | Plant exudate | CID_6439929 | 5915 | RARB | ChEMBL \| NPASS |
| Arogyavardhini | Commiphora wightii | Plant exudate | CID_6439929 | 5916 | RARG | ChEMBL \| NPASS |
| Arogyavardhini | Commiphora wightii | Plant exudate | CID_6439929 | 6097 | RORC | ChEMBL \| NPASS |
| Arogyavardhini | Commiphora wightii | Plant exudate | CID_6439929 | 6256 | RXRA | ChEMBL \| NPASS |
| Arogyavardhini | Commiphora wightii | Plant exudate | CID_6439929 | 6606 | SMN1 | NPASS |
| Arogyavardhini | Commiphora wightii | Plant exudate | CID_6439929 | 7067 | THRA | ChEMBL |
| Arogyavardhini | Commiphora wightii | Plant exudate | CID_6439929 | 7068 | THRB | ChEMBL \| NPASS |
| Arogyavardhini | Commiphora wightii | Plant exudate | CID_6439929 | 7157 | TP53 | NPASS |
| Arogyavardhini | Commiphora wightii | Plant exudate | CID_6439929 | 7253 | TSHR | NPASS |
| Arogyavardhini | Commiphora wightii | Plant exudate | CID_6439929 | 7376 | NR1H2 | ChEMBL \| NPASS |
| Arogyavardhini | Commiphora wightii | Plant exudate | CID_6439929 | 7421 | VDR | ChEMBL \| NPASS |
| Arogyavardhini | Commiphora wightii | Plant exudate | CID_6439929 | 865 | CBFB | NPASS |
| Arogyavardhini | Commiphora wightii | Plant exudate | CID_6439929 | 8856 | NR1I2 | ChEMBL \| NPASS |
| Arogyavardhini | Commiphora wightii | Plant exudate | CID_6439929 | 9971 | NR1H4 | BindingDB \| ChEMBL \| NPASS |
| Arogyavardhini | Commiphora wightii | Plant exudate | CID_70695727 | 5465 | PPARA | ChEMBL \| NPASS |
| Arogyavardhini | Commiphora wightii | Plant exudate | CID_70695727 | 5467 | PPARD | ChEMBL \| NPASS |
| Arogyavardhini | Commiphora wightii | Plant exudate | CID_70695727 | 5468 | PPARG | BindingDB \| ChEMBL \| NPASS |
| Arogyavardhini | Commiphora wightii | Plant exudate | CID_92747 | 1586 | CYP17A1 | BindingDB \| ChEMBL \| NPASS |
| Arogyavardhini | Commiphora wightii | Plant exudate | CID_92747 | 5241 | PGR | ChEMBL \| NPASS |
| Arogyavardhini | Phyllanthus emblica | Fruit | CID_1057 | 1017 | CDK2 | BindingDB \| ChEMBL |
| Arogyavardhini | Phyllanthus emblica | Fruit | CID_1057 | 10280 | SIGMAR1 | ChEMBL |
| Arogyavardhini | Phyllanthus emblica | Fruit | CID_1057 | 10800 | CYSLTR1 | ChEMBL |
| Arogyavardhini | Phyllanthus emblica | Fruit | CID_1057 | 10919 | EHMT2 | ChEMBL \| NPASS |
| Arogyavardhini | Phyllanthus emblica | Fruit | CID_1057 | 10988 | METAP2 | ChEMBL \| NPASS |
| Arogyavardhini | Phyllanthus emblica | Fruit | CID_1057 | 11096 | ADAMTS5 | BindingDB \| ChEMBL |
| Arogyavardhini | Phyllanthus emblica | Fruit | CID_1057 | 1128 | CHRM1 | ChEMBL |
| Arogyavardhini | Phyllanthus emblica | Fruit | CID_1057 | 1129 | CHRM2 | ChEMBL |
| Arogyavardhini | Phyllanthus emblica | Fruit | CID_1057 | 1131 | CHRM3 | ChEMBL |
| Arogyavardhini | Phyllanthus emblica | Fruit | CID_1057 | 1132 | CHRM4 | ChEMBL |
| Arogyavardhini | Phyllanthus emblica | Fruit | CID_1057 | 1133 | CHRM5 | ChEMBL |
| Arogyavardhini | Phyllanthus emblica | Fruit | CID_1057 | 1233 | CCR4 | ChEMBL |
| Arogyavardhini | Phyllanthus emblica | Fruit | CID_1057 | 1234 | CCR5 | ChEMBL |
| Arogyavardhini | Phyllanthus emblica | Fruit | CID_1057 | 1268 | CNR1 | ChEMBL |
| Arogyavardhini | Phyllanthus emblica | Fruit | CID_1057 | 134 | ADORA1 | ChEMBL |
| Arogyavardhini | Phyllanthus emblica | Fruit | CID_1057 | 135 | ADORA2A | ChEMBL |
| Arogyavardhini | Phyllanthus emblica | Fruit | CID_1057 | 140 | ADORA3 | ChEMBL |
| Arogyavardhini | Phyllanthus emblica | Fruit | CID_1057 | 1432 | MAPK14 | ChEMBL |
| Arogyavardhini | Phyllanthus emblica | Fruit | CID_1057 | 146 | ADRA1D | ChEMBL |
| Arogyavardhini | Phyllanthus emblica | Fruit | CID_1057 | 150 | ADRA2A | ChEMBL |
| Arogyavardhini | Phyllanthus emblica | Fruit | CID_1057 | 151 | ADRA2B | ChEMBL |
| Arogyavardhini | Phyllanthus emblica | Fruit | CID_1057 | 1511 | CTSG | ChEMBL |
| Arogyavardhini | Phyllanthus emblica | Fruit | CID_1057 | 152 | ADRA2C | ChEMBL |
| Arogyavardhini | Phyllanthus emblica | Fruit | CID_1057 | 153 | ADRB1 | ChEMBL |
| Arogyavardhini | Phyllanthus emblica | Fruit | CID_1057 | 154 | ADRB2 | ChEMBL |
| Arogyavardhini | Phyllanthus emblica | Fruit | CID_1057 | 1544 | CYP1A2 | ChEMBL |
| Arogyavardhini | Phyllanthus emblica | Fruit | CID_1057 | 1548 | CYP2A6 | ChEMBL |
| Arogyavardhini | Phyllanthus emblica | Fruit | CID_1057 | 155 | ADRB3 | ChEMBL |
| Arogyavardhini | Phyllanthus emblica | Fruit | CID_1057 | 1557 | CYP2C19 | ChEMBL |
| Arogyavardhini | Phyllanthus emblica | Fruit | CID_1057 | 1559 | CYP2C9 | ChEMBL |
| Arogyavardhini | Phyllanthus emblica | Fruit | CID_1057 | 1565 | CYP2D6 | ChEMBL |
| Arogyavardhini | Phyllanthus emblica | Fruit | CID_1057 | 1571 | CYP2E1 | ChEMBL |
| Arogyavardhini | Phyllanthus emblica | Fruit | CID_1057 | 1576 | CYP3A4 | ChEMBL |
| Arogyavardhini | Phyllanthus emblica | Fruit | CID_1057 | 1812 | DRD1 | ChEMBL |

**Table S3: Associations between formulations, constituent plants, phytochemicals, and human molecular targets. The table lists the formulations, plant name and plant part, phytochemical identifiers, associated human target genes (Entrez Gene ID and gene symbol), and the database sources supporting each phytochemical-target interaction.**

| Formulation | Plant name | Part | PubChem CID | Entrez Gene ID | Gene Symbol | Source database(s) |
|---|---|---|---|---|---|---|
| Arogyavardhini | Phyllanthus emblica | Fruit | CID_1057 | 1813 | DRD2 | ChEMBL |
| Arogyavardhini | Phyllanthus emblica | Fruit | CID_1057 | 1814 | DRD3 | ChEMBL |
| Arogyavardhini | Phyllanthus emblica | Fruit | CID_1057 | 1815 | DRD4 | ChEMBL |
| Arogyavardhini | Phyllanthus emblica | Fruit | CID_1057 | 186 | AGTR2 | ChEMBL |
| Arogyavardhini | Phyllanthus emblica | Fruit | CID_1057 | 1909 | EDNRA | ChEMBL |
| Arogyavardhini | Phyllanthus emblica | Fruit | CID_1057 | 1956 | EGFR | ChEMBL \| NPASS |
| Arogyavardhini | Phyllanthus emblica | Fruit | CID_1057 | 1991 | ELANE | ChEMBL |
| Arogyavardhini | Phyllanthus emblica | Fruit | CID_1057 | 2064 | ERBB2 | ChEMBL |
| Arogyavardhini | Phyllanthus emblica | Fruit | CID_1057 | 2099 | ESR1 | ChEMBL \| NPASS |
| Arogyavardhini | Phyllanthus emblica | Fruit | CID_1057 | 2100 | ESR2 | ChEMBL |
| Arogyavardhini | Phyllanthus emblica | Fruit | CID_1057 | 213 | ALB | ChEMBL |
| Arogyavardhini | Phyllanthus emblica | Fruit | CID_1057 | 216 | ALDH1A1 | ChEMBL \| NPASS |
| Arogyavardhini | Phyllanthus emblica | Fruit | CID_1057 | 2321 | FLT1 | ChEMBL |
| Arogyavardhini | Phyllanthus emblica | Fruit | CID_1057 | 246 | ALOX15 | ChEMBL \| NPASS |
| Arogyavardhini | Phyllanthus emblica | Fruit | CID_1057 | 2534 | FYN | ChEMBL \| NPASS |
| Arogyavardhini | Phyllanthus emblica | Fruit | CID_1057 | 2744 | GLS | NPASS |
| Arogyavardhini | Phyllanthus emblica | Fruit | CID_1057 | 2908 | NR3C1 | ChEMBL |
| Arogyavardhini | Phyllanthus emblica | Fruit | CID_1057 | 3028 | HSD17B10 | ChEMBL \| NPASS |
| Arogyavardhini | Phyllanthus emblica | Fruit | CID_1057 | 3156 | HMGCR | ChEMBL |
| Arogyavardhini | Phyllanthus emblica | Fruit | CID_1057 | 3248 | HPGD | ChEMBL \| NPASS |
| Arogyavardhini | Phyllanthus emblica | Fruit | CID_1057 | 3269 | HRH1 | ChEMBL |
| Arogyavardhini | Phyllanthus emblica | Fruit | CID_1057 | 3274 | HRH2 | ChEMBL |
| Arogyavardhini | Phyllanthus emblica | Fruit | CID_1057 | 328 | APEX1 | ChEMBL \| NPASS |
| Arogyavardhini | Phyllanthus emblica | Fruit | CID_1057 | 3356 | HTR2A | ChEMBL |
| Arogyavardhini | Phyllanthus emblica | Fruit | CID_1057 | 3357 | HTR2B | ChEMBL |
| Arogyavardhini | Phyllanthus emblica | Fruit | CID_1057 | 3358 | HTR2C | ChEMBL |
| Arogyavardhini | Phyllanthus emblica | Fruit | CID_1057 | 3362 | HTR6 | ChEMBL |
| Arogyavardhini | Phyllanthus emblica | Fruit | CID_1057 | 3576 | CXCL8 | ChEMBL \| NPASS |
| Arogyavardhini | Phyllanthus emblica | Fruit | CID_1057 | 3577 | CXCR1 | ChEMBL |
| Arogyavardhini | Phyllanthus emblica | Fruit | CID_1057 | 3579 | CXCR2 | ChEMBL |
| Arogyavardhini | Phyllanthus emblica | Fruit | CID_1057 | 367 | AR | NPASS |
| Arogyavardhini | Phyllanthus emblica | Fruit | CID_1057 | 3757 | KCNH2 | ChEMBL \| NPASS |
| Arogyavardhini | Phyllanthus emblica | Fruit | CID_1057 | 383 | ARG1 | ChEMBL \| NPASS |
| Arogyavardhini | Phyllanthus emblica | Fruit | CID_1057 | 3932 | LCK | ChEMBL |
| Arogyavardhini | Phyllanthus emblica | Fruit | CID_1057 | 4000 | LMNA | NPASS |
| Arogyavardhini | Phyllanthus emblica | Fruit | CID_1057 | 4128 | MAOA | ChEMBL |
| Arogyavardhini | Phyllanthus emblica | Fruit | CID_1057 | 4159 | MC3R | ChEMBL |
| Arogyavardhini | Phyllanthus emblica | Fruit | CID_1057 | 4160 | MC4R | ChEMBL |
| Arogyavardhini | Phyllanthus emblica | Fruit | CID_1057 | 4161 | MC5R | ChEMBL |
| Arogyavardhini | Phyllanthus emblica | Fruit | CID_1057 | 43 | ACHE | ChEMBL |
| Arogyavardhini | Phyllanthus emblica | Fruit | CID_1057 | 4312 | MMP1 | ChEMBL |
| Arogyavardhini | Phyllanthus emblica | Fruit | CID_1057 | 4313 | MMP2 | ChEMBL |
| Arogyavardhini | Phyllanthus emblica | Fruit | CID_1057 | 4317 | MMP8 | BindingDB \| ChEMBL |
| Arogyavardhini | Phyllanthus emblica | Fruit | CID_1057 | 4318 | MMP9 | BindingDB \| ChEMBL |
| Arogyavardhini | Phyllanthus emblica | Fruit | CID_1057 | 4321 | MMP12 | BindingDB \| ChEMBL |
| Arogyavardhini | Phyllanthus emblica | Fruit | CID_1057 | 4780 | NFE2L2 | ChEMBL \| NPASS |
| Arogyavardhini | Phyllanthus emblica | Fruit | CID_1057 | 4886 | NPY1R | ChEMBL |
| Arogyavardhini | Phyllanthus emblica | Fruit | CID_1057 | 4887 | NPY2R | ChEMBL |
| Arogyavardhini | Phyllanthus emblica | Fruit | CID_1057 | 4985 | OPRD1 | ChEMBL |
| Arogyavardhini | Phyllanthus emblica | Fruit | CID_1057 | 4986 | OPRK1 | ChEMBL |
| Arogyavardhini | Phyllanthus emblica | Fruit | CID_1057 | 4988 | OPRM1 | ChEMBL |
| Arogyavardhini | Phyllanthus emblica | Fruit | CID_1057 | 51426 | POLK | ChEMBL \| NPASS |
| Arogyavardhini | Phyllanthus emblica | Fruit | CID_1057 | 54658 | UGT1A1 | ChEMBL \| NPASS |
| Arogyavardhini | Phyllanthus emblica | Fruit | CID_1057 | 552 | AVPR1A | ChEMBL |
| Arogyavardhini | Phyllanthus emblica | Fruit | CID_1057 | 5530 | PPP3CA | ChEMBL |
| Arogyavardhini | Phyllanthus emblica | Fruit | CID_1057 | 55775 | TDP1 | ChEMBL \| NPASS |
| Arogyavardhini | Phyllanthus emblica | Fruit | CID_1057 | 5578 | PRKCA | ChEMBL |
| Arogyavardhini | Phyllanthus emblica | Fruit | CID_1057 | 5594 | MAPK1 | ChEMBL |
| Arogyavardhini | Phyllanthus emblica | Fruit | CID_1057 | 5595 | MAPK3 | ChEMBL |
| Arogyavardhini | Phyllanthus emblica | Fruit | CID_1057 | 5724 | PTAFR | ChEMBL |
| Arogyavardhini | Phyllanthus emblica | Fruit | CID_1057 | 5742 | PTGS1 | ChEMBL |
| Arogyavardhini | Phyllanthus emblica | Fruit | CID_1057 | 5743 | PTGS2 | BindingDB \| ChEMBL \| NPASS |
| Arogyavardhini | Phyllanthus emblica | Fruit | CID_1057 | 5788 | PTPRC | ChEMBL |
| Arogyavardhini | Phyllanthus emblica | Fruit | CID_1057 | 5965 | RECQL | ChEMBL \| NPASS |
| Arogyavardhini | Phyllanthus emblica | Fruit | CID_1057 | 624 | BDKRB2 | ChEMBL |
| Arogyavardhini | Phyllanthus emblica | Fruit | CID_1057 | 6401 | SELE | ChEMBL \| NPASS |
| Arogyavardhini | Phyllanthus emblica | Fruit | CID_1057 | 6402 | SELL | ChEMBL \| NPASS |
| Arogyavardhini | Phyllanthus emblica | Fruit | CID_1057 | 6403 | SELP | ChEMBL \| NPASS |
| Arogyavardhini | Phyllanthus emblica | Fruit | CID_1057 | 6530 | SLC6A2 | ChEMBL |
| Arogyavardhini | Phyllanthus emblica | Fruit | CID_1057 | 6531 | SLC6A3 | ChEMBL |

**Table S3: Associations between formulations, constituent plants, phytochemicals, and human molecular targets. The table lists the formulations, plant name and plant part, phytochemical identifiers, associated human target genes (Entrez Gene ID and gene symbol), and the database sources supporting each phytochemical-target interaction.**

| Formulation | Plant name | Part | PubChem CID | Entrez Gene ID | Gene Symbol | Source database(s) |
|---|---|---|---|---|---|---|
| Arogyavardhini | Phyllanthus emblica | Fruit | CID_1057 | 6532 | SLC6A4 | ChEMBL |
| Arogyavardhini | Phyllanthus emblica | Fruit | CID_1057 | 6865 | TACR2 | ChEMBL |
| Arogyavardhini | Phyllanthus emblica | Fruit | CID_1057 | 6869 | TACR1 | ChEMBL |
| Arogyavardhini | Phyllanthus emblica | Fruit | CID_1057 | 6916 | TBXAS1 | ChEMBL |
| Arogyavardhini | Phyllanthus emblica | Fruit | CID_1057 | 729230 | CCR2 | ChEMBL |
| Arogyavardhini | Phyllanthus emblica | Fruit | CID_1057 | 7299 | TYR | ChEMBL \| NPASS |
| Arogyavardhini | Phyllanthus emblica | Fruit | CID_1057 | 7366 | UGT2B15 | ChEMBL \| NPASS |
| Arogyavardhini | Phyllanthus emblica | Fruit | CID_1057 | 7421 | VDR | ChEMBL \| NPASS |
| Arogyavardhini | Phyllanthus emblica | Fruit | CID_1057 | 7433 | VIPR1 | ChEMBL |
| Arogyavardhini | Phyllanthus emblica | Fruit | CID_1057 | 759 | CA1 | BindingDB \| ChEMBL \| NPASS |
| Arogyavardhini | Phyllanthus emblica | Fruit | CID_1057 | 760 | CA2 | BindingDB \| ChEMBL \| NPASS |
| Arogyavardhini | Phyllanthus emblica | Fruit | CID_1057 | 762 | CA4 | BindingDB |
| Arogyavardhini | Phyllanthus emblica | Fruit | CID_1057 | 765 | CA6 | ChEMBL \| NPASS |
| Arogyavardhini | Phyllanthus emblica | Fruit | CID_1057 | 799 | CALCR | ChEMBL |
| Arogyavardhini | Phyllanthus emblica | Fruit | CID_1057 | 834 | CASP1 | ChEMBL |
| Arogyavardhini | Phyllanthus emblica | Fruit | CID_1057 | 861 | RUNX1 | ChEMBL |
| Arogyavardhini | Phyllanthus emblica | Fruit | CID_1057 | 865 | CBFB | ChEMBL \| NPASS |
| Arogyavardhini | Phyllanthus emblica | Fruit | CID_1057 | 8654 | PDE5A | ChEMBL |
| Arogyavardhini | Phyllanthus emblica | Fruit | CID_1057 | 886 | CCKAR | ChEMBL |
| Arogyavardhini | Phyllanthus emblica | Fruit | CID_1057 | 9507 | ADAMTS4 | BindingDB \| ChEMBL |
| Arogyavardhini | Phyllanthus emblica | Fruit | CID_124021 | 231 | AKR1B1 | ChEMBL \| NPASS |
| Arogyavardhini | Phyllanthus emblica | Fruit | CID_124021 | 23683 | PRKD3 | ChEMBL |
| Arogyavardhini | Phyllanthus emblica | Fruit | CID_124021 | 5578 | PRKCA | BindingDB \| ChEMBL |
| Arogyavardhini | Phyllanthus emblica | Fruit | CID_124021 | 5579 | PRKCB | ChEMBL |
| Arogyavardhini | Phyllanthus emblica | Fruit | CID_124021 | 5580 | PRKCD | ChEMBL |
| Arogyavardhini | Phyllanthus emblica | Fruit | CID_124021 | 5581 | PRKCE | ChEMBL |
| Arogyavardhini | Phyllanthus emblica | Fruit | CID_124021 | 5582 | PRKCG | ChEMBL |
| Arogyavardhini | Phyllanthus emblica | Fruit | CID_124021 | 5583 | PRKCH | ChEMBL |
| Arogyavardhini | Phyllanthus emblica | Fruit | CID_124021 | 5584 | PRKCI | ChEMBL |
| Arogyavardhini | Phyllanthus emblica | Fruit | CID_124021 | 5587 | PRKD1 | ChEMBL |
| Arogyavardhini | Phyllanthus emblica | Fruit | CID_124021 | 5588 | PRKCQ | ChEMBL \| NPASS |
| Arogyavardhini | Phyllanthus emblica | Fruit | CID_124021 | 5590 | PRKCZ | ChEMBL |
| Arogyavardhini | Phyllanthus emblica | Fruit | CID_13250 | 2099 | ESR1 | ChEMBL |
| Arogyavardhini | Phyllanthus emblica | Fruit | CID_13250 | 22943 | DKK1 | ChEMBL |
| Arogyavardhini | Phyllanthus emblica | Fruit | CID_13250 | 4040 | LRP6 | ChEMBL |
| Arogyavardhini | Phyllanthus emblica | Fruit | CID_289 | 10280 | SIGMAR1 | ChEMBL |
| Arogyavardhini | Phyllanthus emblica | Fruit | CID_289 | 10800 | CYSLTR1 | ChEMBL |
| Arogyavardhini | Phyllanthus emblica | Fruit | CID_289 | 10919 | EHMT2 | ChEMBL \| NPASS |
| Arogyavardhini | Phyllanthus emblica | Fruit | CID_289 | 10951 | CBX1 | NPASS |
| Arogyavardhini | Phyllanthus emblica | Fruit | CID_289 | 11201 | POLI | ChEMBL \| NPASS |
| Arogyavardhini | Phyllanthus emblica | Fruit | CID_289 | 11238 | CA5B | BindingDB \| ChEMBL \| NPASS |
| Arogyavardhini | Phyllanthus emblica | Fruit | CID_289 | 1128 | CHRM1 | ChEMBL |
| Arogyavardhini | Phyllanthus emblica | Fruit | CID_289 | 1129 | CHRM2 | ChEMBL |
| Arogyavardhini | Phyllanthus emblica | Fruit | CID_289 | 1131 | CHRM3 | ChEMBL |
| Arogyavardhini | Phyllanthus emblica | Fruit | CID_289 | 1132 | CHRM4 | ChEMBL |
| Arogyavardhini | Phyllanthus emblica | Fruit | CID_289 | 1133 | CHRM5 | ChEMBL |
| Arogyavardhini | Phyllanthus emblica | Fruit | CID_289 | 1233 | CCR4 | ChEMBL |
| Arogyavardhini | Phyllanthus emblica | Fruit | CID_289 | 1234 | CCR5 | ChEMBL |
| Arogyavardhini | Phyllanthus emblica | Fruit | CID_289 | 1268 | CNR1 | ChEMBL |
| Arogyavardhini | Phyllanthus emblica | Fruit | CID_289 | 134 | ADORA1 | ChEMBL |
| Arogyavardhini | Phyllanthus emblica | Fruit | CID_289 | 135 | ADORA2A | ChEMBL |
| Arogyavardhini | Phyllanthus emblica | Fruit | CID_289 | 140 | ADORA3 | ChEMBL |
| Arogyavardhini | Phyllanthus emblica | Fruit | CID_289 | 1432 | MAPK14 | ChEMBL |
| Arogyavardhini | Phyllanthus emblica | Fruit | CID_289 | 146 | ADRA1D | ChEMBL |
| Arogyavardhini | Phyllanthus emblica | Fruit | CID_289 | 150 | ADRA2A | ChEMBL |
| Arogyavardhini | Phyllanthus emblica | Fruit | CID_289 | 151 | ADRA2B | ChEMBL |
| Arogyavardhini | Phyllanthus emblica | Fruit | CID_289 | 1511 | CTSG | ChEMBL |
| Arogyavardhini | Phyllanthus emblica | Fruit | CID_289 | 152 | ADRA2C | ChEMBL |
| Arogyavardhini | Phyllanthus emblica | Fruit | CID_289 | 153 | ADRB1 | ChEMBL |
| Arogyavardhini | Phyllanthus emblica | Fruit | CID_289 | 154 | ADRB2 | ChEMBL |
| Arogyavardhini | Phyllanthus emblica | Fruit | CID_289 | 1544 | CYP1A2 | ChEMBL |
| Arogyavardhini | Phyllanthus emblica | Fruit | CID_289 | 1548 | CYP2A6 | ChEMBL |
| Arogyavardhini | Phyllanthus emblica | Fruit | CID_289 | 155 | ADRB3 | ChEMBL |
| Arogyavardhini | Phyllanthus emblica | Fruit | CID_289 | 1557 | CYP2C19 | ChEMBL |
| Arogyavardhini | Phyllanthus emblica | Fruit | CID_289 | 1559 | CYP2C9 | ChEMBL |
| Arogyavardhini | Phyllanthus emblica | Fruit | CID_289 | 1565 | CYP2D6 | ChEMBL |
| Arogyavardhini | Phyllanthus emblica | Fruit | CID_289 | 1571 | CYP2E1 | ChEMBL |
| Arogyavardhini | Phyllanthus emblica | Fruit | CID_289 | 1576 | CYP3A4 | ChEMBL \| NPASS |
| Arogyavardhini | Phyllanthus emblica | Fruit | CID_289 | 1812 | DRD1 | ChEMBL |
| Arogyavardhini | Phyllanthus emblica | Fruit | CID_289 | 1813 | DRD2 | ChEMBL |

Table S3: Associations between formulations, constituent plants, phytochemicals, and human molecular targets. The table lists the formulations, plant name and plant part, phytochemical identifiers, associated human target genes (Entrez Gene ID and gene symbol), and the database sources supporting each phytochemical-target interaction.

| Formulation | Plant name | Part | PubChem CID | Entrez Gene ID | Gene Symbol | Source database(s) |
|---|---|---|---|---|---|---|
| Arogyavardhini | Phyllanthus emblica | Fruit | CID_289 | 1814 | DRD3 | ChEMBL |
| Arogyavardhini | Phyllanthus emblica | Fruit | CID_289 | 1815 | DRD4 | ChEMBL |
| Arogyavardhini | Phyllanthus emblica | Fruit | CID_289 | 186 | AGTR2 | ChEMBL |
| Arogyavardhini | Phyllanthus emblica | Fruit | CID_289 | 1909 | EDNRA | ChEMBL |
| Arogyavardhini | Phyllanthus emblica | Fruit | CID_289 | 1956 | EGFR | ChEMBL \| NPASS |
| Arogyavardhini | Phyllanthus emblica | Fruit | CID_289 | 1991 | ELANE | ChEMBL |
| Arogyavardhini | Phyllanthus emblica | Fruit | CID_289 | 2064 | ERBB2 | ChEMBL |
| Arogyavardhini | Phyllanthus emblica | Fruit | CID_289 | 2099 | ESR1 | ChEMBL \| NPASS |
| Arogyavardhini | Phyllanthus emblica | Fruit | CID_289 | 2100 | ESR2 | ChEMBL |
| Arogyavardhini | Phyllanthus emblica | Fruit | CID_289 | 216 | ALDH1A1 | ChEMBL \| NPASS |
| Arogyavardhini | Phyllanthus emblica | Fruit | CID_289 | 2237 | FEN1 | ChEMBL \| NPASS |
| Arogyavardhini | Phyllanthus emblica | Fruit | CID_289 | 2321 | FLT1 | ChEMBL |
| Arogyavardhini | Phyllanthus emblica | Fruit | CID_289 | 23632 | CA14 | BindingDB \| ChEMBL \| NPASS |
| Arogyavardhini | Phyllanthus emblica | Fruit | CID_289 | 240 | ALOX5 | ChEMBL \| NPASS |
| Arogyavardhini | Phyllanthus emblica | Fruit | CID_289 | 246 | ALOX15 | ChEMBL \| NPASS |
| Arogyavardhini | Phyllanthus emblica | Fruit | CID_289 | 247 | ALOX15B | NPASS |
| Arogyavardhini | Phyllanthus emblica | Fruit | CID_289 | 2475 | MTOR | ChEMBL \| NPASS |
| Arogyavardhini | Phyllanthus emblica | Fruit | CID_289 | 2534 | FYN | ChEMBL \| NPASS |
| Arogyavardhini | Phyllanthus emblica | Fruit | CID_289 | 25797 | QPCT | BindingDB \| ChEMBL |
| Arogyavardhini | Phyllanthus emblica | Fruit | CID_289 | 26013 | L3MBTL1 | ChEMBL \| NPASS |
| Arogyavardhini | Phyllanthus emblica | Fruit | CID_289 | 2744 | GLS | NPASS |
| Arogyavardhini | Phyllanthus emblica | Fruit | CID_289 | 2908 | NR3C1 | ChEMBL \| NPASS |
| Arogyavardhini | Phyllanthus emblica | Fruit | CID_289 | 3028 | HSD17B10 | ChEMBL \| NPASS |
| Arogyavardhini | Phyllanthus emblica | Fruit | CID_289 | 3091 | HIF1A | ChEMBL \| NPASS |
| Arogyavardhini | Phyllanthus emblica | Fruit | CID_289 | 3156 | HMGCR | ChEMBL |
| Arogyavardhini | Phyllanthus emblica | Fruit | CID_289 | 3248 | HPGD | ChEMBL \| NPASS |
| Arogyavardhini | Phyllanthus emblica | Fruit | CID_289 | 3269 | HRH1 | ChEMBL |
| Arogyavardhini | Phyllanthus emblica | Fruit | CID_289 | 3274 | HRH2 | ChEMBL |
| Arogyavardhini | Phyllanthus emblica | Fruit | CID_289 | 328 | APEX1 | ChEMBL \| NPASS |
| Arogyavardhini | Phyllanthus emblica | Fruit | CID_289 | 3309 | HSPA5 | NPASS |
| Arogyavardhini | Phyllanthus emblica | Fruit | CID_289 | 3315 | HSPB1 | NPASS |
| Arogyavardhini | Phyllanthus emblica | Fruit | CID_289 | 3356 | HTR2A | ChEMBL |
| Arogyavardhini | Phyllanthus emblica | Fruit | CID_289 | 3357 | HTR2B | ChEMBL |
| Arogyavardhini | Phyllanthus emblica | Fruit | CID_289 | 3358 | HTR2C | ChEMBL |
| Arogyavardhini | Phyllanthus emblica | Fruit | CID_289 | 3362 | HTR6 | ChEMBL |
| Arogyavardhini | Phyllanthus emblica | Fruit | CID_289 | 3417 | IDH1 | NPASS |
| Arogyavardhini | Phyllanthus emblica | Fruit | CID_289 | 3576 | CXCL8 | NPASS |
| Arogyavardhini | Phyllanthus emblica | Fruit | CID_289 | 3577 | CXCR1 | ChEMBL |
| Arogyavardhini | Phyllanthus emblica | Fruit | CID_289 | 3579 | CXCR2 | ChEMBL |
| Arogyavardhini | Phyllanthus emblica | Fruit | CID_289 | 3620 | IDO1 | ChEMBL \| NPASS |
| Arogyavardhini | Phyllanthus emblica | Fruit | CID_289 | 367 | AR | NPASS |
| Arogyavardhini | Phyllanthus emblica | Fruit | CID_289 | 3725 | JUN | NPASS |
| Arogyavardhini | Phyllanthus emblica | Fruit | CID_289 | 3757 | KCNH2 | ChEMBL |
| Arogyavardhini | Phyllanthus emblica | Fruit | CID_289 | 390245 | KDM4E | ChEMBL \| NPASS |
| Arogyavardhini | Phyllanthus emblica | Fruit | CID_289 | 3932 | LCK | ChEMBL |
| Arogyavardhini | Phyllanthus emblica | Fruit | CID_289 | 4000 | LMNA | NPASS |
| Arogyavardhini | Phyllanthus emblica | Fruit | CID_289 | 4128 | MAOA | ChEMBL |
| Arogyavardhini | Phyllanthus emblica | Fruit | CID_289 | 4137 | MAPT | NPASS |
| Arogyavardhini | Phyllanthus emblica | Fruit | CID_289 | 4159 | MC3R | ChEMBL |
| Arogyavardhini | Phyllanthus emblica | Fruit | CID_289 | 4160 | MC4R | ChEMBL |
| Arogyavardhini | Phyllanthus emblica | Fruit | CID_289 | 4161 | MC5R | ChEMBL |
| Arogyavardhini | Phyllanthus emblica | Fruit | CID_289 | 4193 | MDM2 | BindingDB \| ChEMBL \| NPASS |
| Arogyavardhini | Phyllanthus emblica | Fruit | CID_289 | 43 | ACHE | ChEMBL |
| Arogyavardhini | Phyllanthus emblica | Fruit | CID_289 | 4312 | MMP1 | ChEMBL \| NPASS |
| Arogyavardhini | Phyllanthus emblica | Fruit | CID_289 | 4313 | MMP2 | ChEMBL \| NPASS |
| Arogyavardhini | Phyllanthus emblica | Fruit | CID_289 | 4314 | MMP3 | ChEMBL \| NPASS |
| Arogyavardhini | Phyllanthus emblica | Fruit | CID_289 | 4317 | MMP8 | ChEMBL \| NPASS |
| Arogyavardhini | Phyllanthus emblica | Fruit | CID_289 | 4318 | MMP9 | ChEMBL \| NPASS |
| Arogyavardhini | Phyllanthus emblica | Fruit | CID_289 | 4780 | NFE2L2 | ChEMBL \| NPASS |
| Arogyavardhini | Phyllanthus emblica | Fruit | CID_289 | 4790 | NFKB1 | NPASS |
| Arogyavardhini | Phyllanthus emblica | Fruit | CID_289 | 4864 | NPC1 | NPASS |
| Arogyavardhini | Phyllanthus emblica | Fruit | CID_289 | 4886 | NPY1R | ChEMBL |
| Arogyavardhini | Phyllanthus emblica | Fruit | CID_289 | 4887 | NPY2R | ChEMBL |
| Arogyavardhini | Phyllanthus emblica | Fruit | CID_289 | 4985 | OPRD1 | ChEMBL |
| Arogyavardhini | Phyllanthus emblica | Fruit | CID_289 | 4986 | OPRK1 | ChEMBL |
| Arogyavardhini | Phyllanthus emblica | Fruit | CID_289 | 4988 | OPRM1 | ChEMBL |
| Arogyavardhini | Phyllanthus emblica | Fruit | CID_289 | 51053 | GMNN | NPASS |
| Arogyavardhini | Phyllanthus emblica | Fruit | CID_289 | 51426 | POLK | ChEMBL \| NPASS |
| Arogyavardhini | Phyllanthus emblica | Fruit | CID_289 | 5290 | PIK3CA | BindingDB \| ChEMBL \| NPASS |
| Arogyavardhini | Phyllanthus emblica | Fruit | CID_289 | 5300 | PIN1 | NPASS |

**Table S3: Associations between formulations, constituent plants, phytochemicals, and human molecular targets. The table lists the formulations, plant name and plant part, phytochemical identifiers, associated human target genes (Entrez Gene ID and gene symbol), and the database sources supporting each phytochemical-target interaction.**

| Formulation | Plant name | Part | PubChem CID | Entrez Gene ID | Gene Symbol | Source database(s) |
|---|---|---|---|---|---|---|
| Arogyavardhini | Phyllanthus emblica | Fruit | CID_289 | 5376 | PMP22 | NPASS |
| Arogyavardhini | Phyllanthus emblica | Fruit | CID_289 | 5423 | POLB | ChEMBL \| NPASS |
| Arogyavardhini | Phyllanthus emblica | Fruit | CID_289 | 5429 | POLH | NPASS |
| Arogyavardhini | Phyllanthus emblica | Fruit | CID_289 | 5467 | PPARD | NPASS |
| Arogyavardhini | Phyllanthus emblica | Fruit | CID_289 | 5468 | PPARG | NPASS |
| Arogyavardhini | Phyllanthus emblica | Fruit | CID_289 | 54737 | MPHOSPH8 | ChEMBL \| NPASS |
| Arogyavardhini | Phyllanthus emblica | Fruit | CID_289 | 552 | AVPR1A | ChEMBL |
| Arogyavardhini | Phyllanthus emblica | Fruit | CID_289 | 5530 | PPP3CA | ChEMBL |
| Arogyavardhini | Phyllanthus emblica | Fruit | CID_289 | 55775 | TDP1 | NPASS |
| Arogyavardhini | Phyllanthus emblica | Fruit | CID_289 | 5578 | PRKCA | ChEMBL |
| Arogyavardhini | Phyllanthus emblica | Fruit | CID_289 | 5594 | MAPK1 | ChEMBL |
| Arogyavardhini | Phyllanthus emblica | Fruit | CID_289 | 5595 | MAPK3 | ChEMBL |
| Arogyavardhini | Phyllanthus emblica | Fruit | CID_289 | 5724 | PTAFR | ChEMBL |
| Arogyavardhini | Phyllanthus emblica | Fruit | CID_289 | 5742 | PTGS1 | ChEMBL |
| Arogyavardhini | Phyllanthus emblica | Fruit | CID_289 | 5743 | PTGS2 | ChEMBL |
| Arogyavardhini | Phyllanthus emblica | Fruit | CID_289 | 5745 | PTH1R | NPASS |
| Arogyavardhini | Phyllanthus emblica | Fruit | CID_289 | 5788 | PTPRC | ChEMBL |
| Arogyavardhini | Phyllanthus emblica | Fruit | CID_289 | 5965 | RECQL | ChEMBL \| NPASS |
| Arogyavardhini | Phyllanthus emblica | Fruit | CID_289 | 5999 | RGS4 | NPASS |
| Arogyavardhini | Phyllanthus emblica | Fruit | CID_289 | 60489 | APOBEC3G | NPASS |
| Arogyavardhini | Phyllanthus emblica | Fruit | CID_289 | 6097 | RORC | ChEMBL \| NPASS |
| Arogyavardhini | Phyllanthus emblica | Fruit | CID_289 | 624 | BDKRB2 | ChEMBL |
| Arogyavardhini | Phyllanthus emblica | Fruit | CID_289 | 6311 | ATXN2 | NPASS |
| Arogyavardhini | Phyllanthus emblica | Fruit | CID_289 | 641 | BLM | NPASS |
| Arogyavardhini | Phyllanthus emblica | Fruit | CID_289 | 6530 | SLC6A2 | ChEMBL |
| Arogyavardhini | Phyllanthus emblica | Fruit | CID_289 | 6531 | SLC6A3 | ChEMBL |
| Arogyavardhini | Phyllanthus emblica | Fruit | CID_289 | 6532 | SLC6A4 | ChEMBL |
| Arogyavardhini | Phyllanthus emblica | Fruit | CID_289 | 6606 | SMN1 | NPASS |
| Arogyavardhini | Phyllanthus emblica | Fruit | CID_289 | 6865 | TACR2 | ChEMBL |
| Arogyavardhini | Phyllanthus emblica | Fruit | CID_289 | 6869 | TACR1 | ChEMBL |
| Arogyavardhini | Phyllanthus emblica | Fruit | CID_289 | 6916 | TBXAS1 | ChEMBL |
| Arogyavardhini | Phyllanthus emblica | Fruit | CID_289 | 7066 | THPO | NPASS |
| Arogyavardhini | Phyllanthus emblica | Fruit | CID_289 | 729230 | CCR2 | ChEMBL |
| Arogyavardhini | Phyllanthus emblica | Fruit | CID_289 | 7433 | VIPR1 | ChEMBL |
| Arogyavardhini | Phyllanthus emblica | Fruit | CID_289 | 7498 | XDH | ChEMBL \| NPASS |
| Arogyavardhini | Phyllanthus emblica | Fruit | CID_289 | 759 | CA1 | BindingDB \| ChEMBL \| NPASS |
| Arogyavardhini | Phyllanthus emblica | Fruit | CID_289 | 760 | CA2 | BindingDB \| ChEMBL \| NPASS |
| Arogyavardhini | Phyllanthus emblica | Fruit | CID_289 | 761 | CA3 | BindingDB \| ChEMBL \| NPASS |
| Arogyavardhini | Phyllanthus emblica | Fruit | CID_289 | 762 | CA4 | BindingDB \| ChEMBL \| NPASS |
| Arogyavardhini | Phyllanthus emblica | Fruit | CID_289 | 763 | CA5A | BindingDB \| ChEMBL \| NPASS |
| Arogyavardhini | Phyllanthus emblica | Fruit | CID_289 | 765 | CA6 | NPASS |
| Arogyavardhini | Phyllanthus emblica | Fruit | CID_289 | 766 | CA7 | BindingDB \| NPASS |
| Arogyavardhini | Phyllanthus emblica | Fruit | CID_289 | 768 | CA9 | BindingDB \| NPASS |
| Arogyavardhini | Phyllanthus emblica | Fruit | CID_289 | 771 | CA12 | BindingDB \| ChEMBL \| NPASS |
| Arogyavardhini | Phyllanthus emblica | Fruit | CID_289 | 799 | CALCR | ChEMBL |
| Arogyavardhini | Phyllanthus emblica | Fruit | CID_289 | 834 | CASP1 | ChEMBL |
| Arogyavardhini | Phyllanthus emblica | Fruit | CID_289 | 8654 | PDE5A | ChEMBL |
| Arogyavardhini | Phyllanthus emblica | Fruit | CID_289 | 886 | CCKAR | ChEMBL |
| Arogyavardhini | Phyllanthus emblica | Fruit | CID_289 | 9367 | RAB9A | NPASS |
| Arogyavardhini | Phyllanthus emblica | Fruit | CID_289 | 9682 | KDM4A | ChEMBL \| NPASS |
| Arogyavardhini | Phyllanthus emblica | Fruit | CID_289 | 9971 | NR1H4 | NPASS |
| Arogyavardhini | Phyllanthus emblica | Fruit | CID_359 | 10013 | HDAC6 | ChEMBL |
| Arogyavardhini | Phyllanthus emblica | Fruit | CID_359 | 23621 | BACE1 | BindingDB \| ChEMBL \| NPASS |
| Arogyavardhini | Phyllanthus emblica | Fruit | CID_359 | 51426 | POLK | NPASS |
| Arogyavardhini | Phyllanthus emblica | Fruit | CID_359 | 7299 | TYR | ChEMBL \| NPASS |
| Arogyavardhini | Phyllanthus emblica | Fruit | CID_370 | 10013 | HDAC6 | ChEMBL |
| Arogyavardhini | Phyllanthus emblica | Fruit | CID_370 | 10599 | SLCO1B1 | ChEMBL \| NPASS |
| Arogyavardhini | Phyllanthus emblica | Fruit | CID_370 | 10919 | EHMT2 | ChEMBL \| NPASS |
| Arogyavardhini | Phyllanthus emblica | Fruit | CID_370 | 11238 | CA5B | BindingDB \| ChEMBL \| NPASS |
| Arogyavardhini | Phyllanthus emblica | Fruit | CID_370 | 118429 | ANTXR2 | NPASS |
| Arogyavardhini | Phyllanthus emblica | Fruit | CID_370 | 1588 | CYP19A1 | ChEMBL \| NPASS |
| Arogyavardhini | Phyllanthus emblica | Fruit | CID_370 | 1636 | ACE | BindingDB \| NPASS |
| Arogyavardhini | Phyllanthus emblica | Fruit | CID_370 | 1803 | DPP4 | BindingDB \| ChEMBL \| NPASS |
| Arogyavardhini | Phyllanthus emblica | Fruit | CID_370 | 2099 | ESR1 | ChEMBL \| NPASS |
| Arogyavardhini | Phyllanthus emblica | Fruit | CID_370 | 216 | ALDH1A1 | ChEMBL \| NPASS |
| Arogyavardhini | Phyllanthus emblica | Fruit | CID_370 | 231 | AKR1B1 | NPASS |
| Arogyavardhini | Phyllanthus emblica | Fruit | CID_370 | 23632 | CA14 | BindingDB \| ChEMBL \| NPASS |
| Arogyavardhini | Phyllanthus emblica | Fruit | CID_370 | 2526 | FUT4 | ChEMBL \| NPASS |
| Arogyavardhini | Phyllanthus emblica | Fruit | CID_370 | 2529 | FUT7 | ChEMBL \| NPASS |
| Arogyavardhini | Phyllanthus emblica | Fruit | CID_370 | 2590 | GALNT2 | ChEMBL \| NPASS |

**Table S3: Associations between formulations, constituent plants, phytochemicals, and human molecular targets. The table lists the formulations, plant name and plant part, phytochemical identifiers, associated human target genes (Entrez Gene ID and gene symbol), and the database sources supporting each phytochemical-target interaction.**

| Formulation | Plant name | Part | PubChem CID | Entrez Gene ID | Gene Symbol | Source database(s) |
|---|---|---|---|---|---|---|
| Arogyavardhini | Phyllanthus emblica | Fruit | CID_370 | 259285 | TAS2R39 | ChEMBL \| NPASS |
| Arogyavardhini | Phyllanthus emblica | Fruit | CID_370 | 2597 | GAPDH | ChEMBL \| NPASS |
| Arogyavardhini | Phyllanthus emblica | Fruit | CID_370 | 2648 | KAT2A | NPASS |
| Arogyavardhini | Phyllanthus emblica | Fruit | CID_370 | 2744 | GLS | ChEMBL \| NPASS |
| Arogyavardhini | Phyllanthus emblica | Fruit | CID_370 | 28234 | SLCO1B3 | ChEMBL \| NPASS |
| Arogyavardhini | Phyllanthus emblica | Fruit | CID_370 | 2908 | NR3C1 | NPASS |
| Arogyavardhini | Phyllanthus emblica | Fruit | CID_370 | 3028 | HSD17B10 | ChEMBL \| NPASS |
| Arogyavardhini | Phyllanthus emblica | Fruit | CID_370 | 3248 | HPGD | ChEMBL \| NPASS |
| Arogyavardhini | Phyllanthus emblica | Fruit | CID_370 | 328 | APEX1 | NPASS |
| Arogyavardhini | Phyllanthus emblica | Fruit | CID_370 | 3315 | HSPB1 | NPASS |
| Arogyavardhini | Phyllanthus emblica | Fruit | CID_370 | 3417 | IDH1 | ChEMBL \| NPASS |
| Arogyavardhini | Phyllanthus emblica | Fruit | CID_370 | 351 | APP | BindingDB \| ChEMBL |
| Arogyavardhini | Phyllanthus emblica | Fruit | CID_370 | 3576 | CXCL8 | NPASS |
| Arogyavardhini | Phyllanthus emblica | Fruit | CID_370 | 367 | AR | NPASS |
| Arogyavardhini | Phyllanthus emblica | Fruit | CID_370 | 3725 | JUN | NPASS |
| Arogyavardhini | Phyllanthus emblica | Fruit | CID_370 | 390245 | KDM4E | ChEMBL \| NPASS |
| Arogyavardhini | Phyllanthus emblica | Fruit | CID_370 | 4000 | LMNA | NPASS |
| Arogyavardhini | Phyllanthus emblica | Fruit | CID_370 | 4137 | MAPT | ChEMBL \| NPASS |
| Arogyavardhini | Phyllanthus emblica | Fruit | CID_370 | 4316 | MMP7 | BindingDB |
| Arogyavardhini | Phyllanthus emblica | Fruit | CID_370 | 4780 | NFE2L2 | ChEMBL \| NPASS |
| Arogyavardhini | Phyllanthus emblica | Fruit | CID_370 | 5054 | SERPINE1 | BindingDB |
| Arogyavardhini | Phyllanthus emblica | Fruit | CID_370 | 51426 | POLK | ChEMBL \| NPASS |
| Arogyavardhini | Phyllanthus emblica | Fruit | CID_370 | 54658 | UGT1A1 | ChEMBL \| NPASS |
| Arogyavardhini | Phyllanthus emblica | Fruit | CID_370 | 55775 | TDP1 | ChEMBL \| NPASS |
| Arogyavardhini | Phyllanthus emblica | Fruit | CID_370 | 5594 | MAPK1 | NPASS |
| Arogyavardhini | Phyllanthus emblica | Fruit | CID_370 | 5693 | PSMB5 | BindingDB \| NPASS |
| Arogyavardhini | Phyllanthus emblica | Fruit | CID_370 | 5743 | PTGS2 | ChEMBL \| NPASS |
| Arogyavardhini | Phyllanthus emblica | Fruit | CID_370 | 6401 | SELE | BindingDB \| NPASS |
| Arogyavardhini | Phyllanthus emblica | Fruit | CID_370 | 6402 | SELL | BindingDB \| ChEMBL \| NPASS |
| Arogyavardhini | Phyllanthus emblica | Fruit | CID_370 | 6403 | SELP | BindingDB \| ChEMBL \| NPASS |
| Arogyavardhini | Phyllanthus emblica | Fruit | CID_370 | 6606 | SMN1 | NPASS |
| Arogyavardhini | Phyllanthus emblica | Fruit | CID_370 | 6622 | SNCA | ChEMBL \| NPASS |
| Arogyavardhini | Phyllanthus emblica | Fruit | CID_370 | 6868 | ADAM17 | ChEMBL \| NPASS |
| Arogyavardhini | Phyllanthus emblica | Fruit | CID_370 | 7157 | TP53 | ChEMBL \| NPASS |
| Arogyavardhini | Phyllanthus emblica | Fruit | CID_370 | 7421 | VDR | ChEMBL \| NPASS |
| Arogyavardhini | Phyllanthus emblica | Fruit | CID_370 | 759 | CA1 | BindingDB \| ChEMBL \| NPASS |
| Arogyavardhini | Phyllanthus emblica | Fruit | CID_370 | 760 | CA2 | BindingDB \| ChEMBL \| NPASS |
| Arogyavardhini | Phyllanthus emblica | Fruit | CID_370 | 761 | CA3 | BindingDB \| ChEMBL \| NPASS |
| Arogyavardhini | Phyllanthus emblica | Fruit | CID_370 | 762 | CA4 | BindingDB \| ChEMBL \| NPASS |
| Arogyavardhini | Phyllanthus emblica | Fruit | CID_370 | 763 | CA5A | BindingDB \| ChEMBL \| NPASS |
| Arogyavardhini | Phyllanthus emblica | Fruit | CID_370 | 765 | CA6 | ChEMBL \| NPASS |
| Arogyavardhini | Phyllanthus emblica | Fruit | CID_370 | 766 | CA7 | BindingDB \| ChEMBL \| NPASS |
| Arogyavardhini | Phyllanthus emblica | Fruit | CID_370 | 768 | CA9 | BindingDB \| ChEMBL \| NPASS |
| Arogyavardhini | Phyllanthus emblica | Fruit | CID_370 | 771 | CA12 | BindingDB \| ChEMBL \| NPASS |
| Arogyavardhini | Phyllanthus emblica | Fruit | CID_449093 | 10013 | HDAC6 | ChEMBL |
| Arogyavardhini | Phyllanthus emblica | Fruit | CID_449093 | 328 | APEX1 | NPASS |
| Arogyavardhini | Phyllanthus emblica | Fruit | CID_449093 | 5745 | PTH1R | NPASS |
| Arogyavardhini | Phyllanthus emblica | Fruit | CID_449093 | 5965 | RECQL | NPASS |
| Arogyavardhini | Phyllanthus emblica | Fruit | CID_449093 | 7253 | TSHR | NPASS |
| Arogyavardhini | Phyllanthus emblica | Fruit | CID_493570 | 10013 | HDAC6 | ChEMBL |
| Arogyavardhini | Phyllanthus emblica | Fruit | CID_493570 | 10599 | SLCO1B1 | ChEMBL \| NPASS |
| Arogyavardhini | Phyllanthus emblica | Fruit | CID_493570 | 10919 | EHMT2 | ChEMBL |
| Arogyavardhini | Phyllanthus emblica | Fruit | CID_493570 | 11330 | CTRC | ChEMBL \| NPASS |
| Arogyavardhini | Phyllanthus emblica | Fruit | CID_493570 | 122706 | PSMB11 | ChEMBL |
| Arogyavardhini | Phyllanthus emblica | Fruit | CID_493570 | 143471 | PSMA8 | ChEMBL |
| Arogyavardhini | Phyllanthus emblica | Fruit | CID_493570 | 1544 | CYP1A2 | ChEMBL |
| Arogyavardhini | Phyllanthus emblica | Fruit | CID_493570 | 1576 | CYP3A4 | NPASS |
| Arogyavardhini | Phyllanthus emblica | Fruit | CID_493570 | 216 | ALDH1A1 | NPASS |
| Arogyavardhini | Phyllanthus emblica | Fruit | CID_493570 | 246 | ALOX15 | NPASS |
| Arogyavardhini | Phyllanthus emblica | Fruit | CID_493570 | 28234 | SLCO1B3 | ChEMBL \| NPASS |
| Arogyavardhini | Phyllanthus emblica | Fruit | CID_493570 | 3028 | HSD17B10 | ChEMBL \| NPASS |
| Arogyavardhini | Phyllanthus emblica | Fruit | CID_493570 | 3248 | HPGD | NPASS |
| Arogyavardhini | Phyllanthus emblica | Fruit | CID_493570 | 3315 | HSPB1 | NPASS |
| Arogyavardhini | Phyllanthus emblica | Fruit | CID_493570 | 3329 | HSPD1 | ChEMBL \| NPASS |
| Arogyavardhini | Phyllanthus emblica | Fruit | CID_493570 | 3336 | HSPE1 | ChEMBL \| NPASS |
| Arogyavardhini | Phyllanthus emblica | Fruit | CID_493570 | 367 | AR | NPASS |
| Arogyavardhini | Phyllanthus emblica | Fruit | CID_493570 | 3725 | JUN | NPASS |
| Arogyavardhini | Phyllanthus emblica | Fruit | CID_493570 | 4780 | NFE2L2 | NPASS |
| Arogyavardhini | Phyllanthus emblica | Fruit | CID_493570 | 4790 | NFKB1 | NPASS |
| Arogyavardhini | Phyllanthus emblica | Fruit | CID_493570 | 5467 | PPARD | NPASS |

**Table S3: Associations between formulations, constituent plants, phytochemicals, and human molecular targets. The table lists the formulations, plant name and plant part, phytochemical identifiers, associated human target genes (Entrez Gene ID and gene symbol), and the database sources supporting each phytochemical-target interaction.**

| Formulation | Plant name | Part | PubChem CID | Entrez Gene ID | Gene Symbol | Source database(s) |
|---|---|---|---|---|---|---|
| Arogyavardhini | Phyllanthus emblica | Fruit | CID_493570 | 5468 | PPARG | NPASS |
| Arogyavardhini | Phyllanthus emblica | Fruit | CID_493570 | 55775 | TDP1 | NPASS |
| Arogyavardhini | Phyllanthus emblica | Fruit | CID_493570 | 5594 | MAPK1 | NPASS |
| Arogyavardhini | Phyllanthus emblica | Fruit | CID_493570 | 5682 | PSMA1 | ChEMBL |
| Arogyavardhini | Phyllanthus emblica | Fruit | CID_493570 | 5683 | PSMA2 | ChEMBL |
| Arogyavardhini | Phyllanthus emblica | Fruit | CID_493570 | 5684 | PSMA3 | ChEMBL |
| Arogyavardhini | Phyllanthus emblica | Fruit | CID_493570 | 5685 | PSMA4 | ChEMBL |
| Arogyavardhini | Phyllanthus emblica | Fruit | CID_493570 | 5686 | PSMA5 | ChEMBL |
| Arogyavardhini | Phyllanthus emblica | Fruit | CID_493570 | 5687 | PSMA6 | ChEMBL |
| Arogyavardhini | Phyllanthus emblica | Fruit | CID_493570 | 5688 | PSMA7 | ChEMBL |
| Arogyavardhini | Phyllanthus emblica | Fruit | CID_493570 | 5689 | PSMB1 | ChEMBL |
| Arogyavardhini | Phyllanthus emblica | Fruit | CID_493570 | 5690 | PSMB2 | ChEMBL |
| Arogyavardhini | Phyllanthus emblica | Fruit | CID_493570 | 5691 | PSMB3 | ChEMBL |
| Arogyavardhini | Phyllanthus emblica | Fruit | CID_493570 | 5692 | PSMB4 | ChEMBL |
| Arogyavardhini | Phyllanthus emblica | Fruit | CID_493570 | 5693 | PSMB5 | ChEMBL |
| Arogyavardhini | Phyllanthus emblica | Fruit | CID_493570 | 5694 | PSMB6 | ChEMBL |
| Arogyavardhini | Phyllanthus emblica | Fruit | CID_493570 | 5695 | PSMB7 | ChEMBL |
| Arogyavardhini | Phyllanthus emblica | Fruit | CID_493570 | 5696 | PSMB8 | ChEMBL |
| Arogyavardhini | Phyllanthus emblica | Fruit | CID_493570 | 5698 | PSMB9 | ChEMBL |
| Arogyavardhini | Phyllanthus emblica | Fruit | CID_493570 | 5699 | PSMB10 | ChEMBL |
| Arogyavardhini | Phyllanthus emblica | Fruit | CID_493570 | 5931 | RBBP7 | ChEMBL |
| Arogyavardhini | Phyllanthus emblica | Fruit | CID_493570 | 5965 | RECQL | NPASS |
| Arogyavardhini | Phyllanthus emblica | Fruit | CID_493570 | 6256 | RXRA | NPASS |
| Arogyavardhini | Phyllanthus emblica | Fruit | CID_493570 | 672 | BRCA1 | ChEMBL \| NPASS |
| Arogyavardhini | Phyllanthus emblica | Fruit | CID_493570 | 7068 | THRB | NPASS |
| Arogyavardhini | Phyllanthus emblica | Fruit | CID_493570 | 7124 | TNF | NPASS |
| Arogyavardhini | Phyllanthus emblica | Fruit | CID_493570 | 7263 | TST | ChEMBL \| NPASS |
| Arogyavardhini | Phyllanthus emblica | Fruit | CID_493570 | 7398 | USP1 | ChEMBL |
| Arogyavardhini | Phyllanthus emblica | Fruit | CID_493570 | 7421 | VDR | NPASS |
| Arogyavardhini | Phyllanthus emblica | Fruit | CID_493570 | 834 | CASP1 | NPASS |
| Arogyavardhini | Phyllanthus emblica | Fruit | CID_493570 | 840 | CASP7 | NPASS |
| Arogyavardhini | Phyllanthus emblica | Fruit | CID_493570 | 8520 | HAT1 | ChEMBL |
| Arogyavardhini | Phyllanthus emblica | Fruit | CID_493570 | 9971 | NR1H4 | NPASS |
| Arogyavardhini | Phyllanthus emblica | Fruit | CID_5280343 | 10013 | HDAC6 | ChEMBL |
| Arogyavardhini | Phyllanthus emblica | Fruit | CID_5280343 | 10072 | DPP3 | BindingDB |
| Arogyavardhini | Phyllanthus emblica | Fruit | CID_5280343 | 1017 | CDK2 | BindingDB \| ChEMBL \| NPASS |
| Arogyavardhini | Phyllanthus emblica | Fruit | CID_5280343 | 1019 | CDK4 | BindingDB \| ChEMBL |
| Arogyavardhini | Phyllanthus emblica | Fruit | CID_5280343 | 1020 | CDK5 | ChEMBL |
| Arogyavardhini | Phyllanthus emblica | Fruit | CID_5280343 | 1021 | CDK6 | BindingDB \| ChEMBL \| NPASS |
| Arogyavardhini | Phyllanthus emblica | Fruit | CID_5280343 | 10257 | ABCC4 | BindingDB \| ChEMBL \| NPASS |
| Arogyavardhini | Phyllanthus emblica | Fruit | CID_5280343 | 10298 | PAK4 | ChEMBL \| NPASS |
| Arogyavardhini | Phyllanthus emblica | Fruit | CID_5280343 | 10494 | STK25 | ChEMBL \| NPASS |
| Arogyavardhini | Phyllanthus emblica | Fruit | CID_5280343 | 10599 | SLCO1B1 | ChEMBL \| NPASS |
| Arogyavardhini | Phyllanthus emblica | Fruit | CID_5280343 | 10645 | CAMKK2 | ChEMBL \| NPASS |
| Arogyavardhini | Phyllanthus emblica | Fruit | CID_5280343 | 106821730 | BUB1B-PAK6 | ChEMBL |
| Arogyavardhini | Phyllanthus emblica | Fruit | CID_5280343 | 10733 | PLK4 | ChEMBL \| NPASS |
| Arogyavardhini | Phyllanthus emblica | Fruit | CID_5280343 | 10783 | NEK6 | BindingDB \| ChEMBL \| NPASS |
| Arogyavardhini | Phyllanthus emblica | Fruit | CID_5280343 | 1080 | CFTR | ChEMBL |
| Arogyavardhini | Phyllanthus emblica | Fruit | CID_5280343 | 10919 | EHMT2 | ChEMBL \| NPASS |
| Arogyavardhini | Phyllanthus emblica | Fruit | CID_5280343 | 10951 | CBX1 | NPASS |
| Arogyavardhini | Phyllanthus emblica | Fruit | CID_5280343 | 11040 | PIM2 | ChEMBL \| NPASS |
| Arogyavardhini | Phyllanthus emblica | Fruit | CID_5280343 | 1111 | CHEK1 | ChEMBL \| NPASS |
| Arogyavardhini | Phyllanthus emblica | Fruit | CID_5280343 | 11200 | CHEK2 | ChEMBL \| NPASS |
| Arogyavardhini | Phyllanthus emblica | Fruit | CID_5280343 | 11201 | POLI | ChEMBL \| NPASS |
| Arogyavardhini | Phyllanthus emblica | Fruit | CID_5280343 | 11238 | CA5B | BindingDB \| ChEMBL \| NPASS |
| Arogyavardhini | Phyllanthus emblica | Fruit | CID_5280343 | 11255 | HRH3 | ChEMBL |
| Arogyavardhini | Phyllanthus emblica | Fruit | CID_5280343 | 1128 | CHRM1 | ChEMBL |
| Arogyavardhini | Phyllanthus emblica | Fruit | CID_5280343 | 1129 | CHRM2 | ChEMBL |
| Arogyavardhini | Phyllanthus emblica | Fruit | CID_5280343 | 11309 | SLCO2B1 | ChEMBL \| NPASS |
| Arogyavardhini | Phyllanthus emblica | Fruit | CID_5280343 | 1131 | CHRM3 | BindingDB \| ChEMBL |
| Arogyavardhini | Phyllanthus emblica | Fruit | CID_5280343 | 11329 | STK38 | ChEMBL \| NPASS |
| Arogyavardhini | Phyllanthus emblica | Fruit | CID_5280343 | 1133 | CHRM5 | BindingDB |
| Arogyavardhini | Phyllanthus emblica | Fruit | CID_5280343 | 11330 | CTRC | BindingDB \| ChEMBL \| NPASS |
| Arogyavardhini | Phyllanthus emblica | Fruit | CID_5280343 | 1137 | CHRNA4 | ChEMBL |
| Arogyavardhini | Phyllanthus emblica | Fruit | CID_5280343 | 118429 | ANTXR2 | NPASS |
| Arogyavardhini | Phyllanthus emblica | Fruit | CID_5280343 | 1195 | CLK1 | ChEMBL \| NPASS |
| Arogyavardhini | Phyllanthus emblica | Fruit | CID_5280343 | 1196 | CLK2 | ChEMBL \| NPASS |
| Arogyavardhini | Phyllanthus emblica | Fruit | CID_5280343 | 1198 | CLK3 | ChEMBL \| NPASS |
| Arogyavardhini | Phyllanthus emblica | Fruit | CID_5280343 | 1233 | CCR4 | ChEMBL \| NPASS |
| Arogyavardhini | Phyllanthus emblica | Fruit | CID_5280343 | 1244 | ABCC2 | ChEMBL \| NPASS |

**Table S3: Associations between formulations, constituent plants, phytochemicals, and human molecular targets. The table lists the formulations, plant name and plant part, phytochemical identifiers, associated human target genes (Entrez Gene ID and gene symbol), and the database sources supporting each phytochemical-target interaction.**

| Formulation | Plant name | Part | PubChem CID | Entrez Gene ID | Gene Symbol | Source database(s) |
|---|---|---|---|---|---|---|
| Arogyavardhini | Phyllanthus emblica | Fruit | CID_5280343 | 1268 | CNR1 | ChEMBL |
| Arogyavardhini | Phyllanthus emblica | Fruit | CID_5280343 | 134 | ADORA1 | ChEMBL |
| Arogyavardhini | Phyllanthus emblica | Fruit | CID_5280343 | 135 | ADORA2A | ChEMBL |
| Arogyavardhini | Phyllanthus emblica | Fruit | CID_5280343 | 1394 | CRHR1 | ChEMBL |
| Arogyavardhini | Phyllanthus emblica | Fruit | CID_5280343 | 1395 | CRHR2 | ChEMBL |
| Arogyavardhini | Phyllanthus emblica | Fruit | CID_5280343 | 140 | ADORA3 | BindingDB \| ChEMBL \| NPASS |
| Arogyavardhini | Phyllanthus emblica | Fruit | CID_5280343 | 1432 | MAPK14 | BindingDB \| ChEMBL \| NPASS |
| Arogyavardhini | Phyllanthus emblica | Fruit | CID_5280343 | 1455 | CSNK1G2 | ChEMBL \| NPASS |
| Arogyavardhini | Phyllanthus emblica | Fruit | CID_5280343 | 1456 | CSNK1G3 | ChEMBL \| NPASS |
| Arogyavardhini | Phyllanthus emblica | Fruit | CID_5280343 | 1457 | CSNK2A1 | BindingDB \| ChEMBL \| NPASS |
| Arogyavardhini | Phyllanthus emblica | Fruit | CID_5280343 | 1459 | CSNK2A2 | ChEMBL \| NPASS |
| Arogyavardhini | Phyllanthus emblica | Fruit | CID_5280343 | 1460 | CSNK2B | ChEMBL \| NPASS |
| Arogyavardhini | Phyllanthus emblica | Fruit | CID_5280343 | 148 | ADRA1A | BindingDB \| ChEMBL |
| Arogyavardhini | Phyllanthus emblica | Fruit | CID_5280343 | 1499 | CTNNB1 | BindingDB \| ChEMBL |
| Arogyavardhini | Phyllanthus emblica | Fruit | CID_5280343 | 150 | ADRA2A | ChEMBL |
| Arogyavardhini | Phyllanthus emblica | Fruit | CID_5280343 | 1504 | CTRB1 | BindingDB \| ChEMBL \| NPASS |
| Arogyavardhini | Phyllanthus emblica | Fruit | CID_5280343 | 151 | ADRA2B | ChEMBL |
| Arogyavardhini | Phyllanthus emblica | Fruit | CID_5280343 | 1514 | CTSL | BindingDB |
| Arogyavardhini | Phyllanthus emblica | Fruit | CID_5280343 | 152 | ADRA2C | ChEMBL |
| Arogyavardhini | Phyllanthus emblica | Fruit | CID_5280343 | 153 | ADRB1 | ChEMBL |
| Arogyavardhini | Phyllanthus emblica | Fruit | CID_5280343 | 154 | ADRB2 | ChEMBL |
| Arogyavardhini | Phyllanthus emblica | Fruit | CID_5280343 | 1543 | CYP1A1 | BindingDB \| ChEMBL \| NPASS |
| Arogyavardhini | Phyllanthus emblica | Fruit | CID_5280343 | 1544 | CYP1A2 | BindingDB \| ChEMBL \| NPASS |
| Arogyavardhini | Phyllanthus emblica | Fruit | CID_5280343 | 1545 | CYP1B1 | BindingDB \| ChEMBL \| NPASS |
| Arogyavardhini | Phyllanthus emblica | Fruit | CID_5280343 | 155 | ADRB3 | ChEMBL |
| Arogyavardhini | Phyllanthus emblica | Fruit | CID_5280343 | 1557 | CYP2C19 | BindingDB \| ChEMBL \| NPASS |
| Arogyavardhini | Phyllanthus emblica | Fruit | CID_5280343 | 1558 | CYP2C8 | ChEMBL \| NPASS |
| Arogyavardhini | Phyllanthus emblica | Fruit | CID_5280343 | 1559 | CYP2C9 | BindingDB \| ChEMBL \| NPASS |
| Arogyavardhini | Phyllanthus emblica | Fruit | CID_5280343 | 1565 | CYP2D6 | ChEMBL \| NPASS |
| Arogyavardhini | Phyllanthus emblica | Fruit | CID_5280343 | 1573 | CYP2J2 | ChEMBL \| NPASS |
| Arogyavardhini | Phyllanthus emblica | Fruit | CID_5280343 | 1576 | CYP3A4 | BindingDB \| ChEMBL \| NPASS |
| Arogyavardhini | Phyllanthus emblica | Fruit | CID_5280343 | 1577 | CYP3A5 | ChEMBL |
| Arogyavardhini | Phyllanthus emblica | Fruit | CID_5280343 | 1588 | CYP19A1 | BindingDB \| ChEMBL \| NPASS |
| Arogyavardhini | Phyllanthus emblica | Fruit | CID_5280343 | 1612 | DAPK1 | BindingDB \| ChEMBL \| NPASS |
| Arogyavardhini | Phyllanthus emblica | Fruit | CID_5280343 | 1613 | DAPK3 | ChEMBL \| NPASS |
| Arogyavardhini | Phyllanthus emblica | Fruit | CID_5280343 | 1760 | DMPK | ChEMBL \| NPASS |
| Arogyavardhini | Phyllanthus emblica | Fruit | CID_5280343 | 1803 | DPP4 | BindingDB \| ChEMBL \| NPASS |
| Arogyavardhini | Phyllanthus emblica | Fruit | CID_5280343 | 1812 | DRD1 | ChEMBL \| NPASS |
| Arogyavardhini | Phyllanthus emblica | Fruit | CID_5280343 | 1813 | DRD2 | ChEMBL |
| Arogyavardhini | Phyllanthus emblica | Fruit | CID_5280343 | 1814 | DRD3 | ChEMBL |
| Arogyavardhini | Phyllanthus emblica | Fruit | CID_5280343 | 1815 | DRD4 | BindingDB \| ChEMBL \| NPASS |
| Arogyavardhini | Phyllanthus emblica | Fruit | CID_5280343 | 185 | AGTR1 | ChEMBL |
| Arogyavardhini | Phyllanthus emblica | Fruit | CID_5280343 | 1909 | EDNRA | ChEMBL |
| Arogyavardhini | Phyllanthus emblica | Fruit | CID_5280343 | 1910 | EDNRB | ChEMBL |
| Arogyavardhini | Phyllanthus emblica | Fruit | CID_5280343 | 1956 | EGFR | BindingDB \| ChEMBL \| NPASS |
| Arogyavardhini | Phyllanthus emblica | Fruit | CID_5280343 | 1991 | ELANE | BindingDB \| ChEMBL \| NPASS |
| Arogyavardhini | Phyllanthus emblica | Fruit | CID_5280343 | 1994 | ELAVL1 | BindingDB \| ChEMBL \| NPASS |
| Arogyavardhini | Phyllanthus emblica | Fruit | CID_5280343 | 1995 | ELAVL3 | ChEMBL \| NPASS |
| Arogyavardhini | Phyllanthus emblica | Fruit | CID_5280343 | 2030 | SLC29A1 | ChEMBL |
| Arogyavardhini | Phyllanthus emblica | Fruit | CID_5280343 | 2053 | EPHX2 | ChEMBL \| NPASS |
| Arogyavardhini | Phyllanthus emblica | Fruit | CID_5280343 | 207 | AKT1 | ChEMBL \| NPASS |
| Arogyavardhini | Phyllanthus emblica | Fruit | CID_5280343 | 2099 | ESR1 | ChEMBL \| NPASS |
| Arogyavardhini | Phyllanthus emblica | Fruit | CID_5280343 | 2100 | ESR2 | ChEMBL |
| Arogyavardhini | Phyllanthus emblica | Fruit | CID_5280343 | 2147 | F2 | ChEMBL \| NPASS |
| Arogyavardhini | Phyllanthus emblica | Fruit | CID_5280343 | 216 | ALDH1A1 | ChEMBL \| NPASS |
| Arogyavardhini | Phyllanthus emblica | Fruit | CID_5280343 | 2167 | FABP4 | ChEMBL |
| Arogyavardhini | Phyllanthus emblica | Fruit | CID_5280343 | 2203 | FBP1 | ChEMBL \| NPASS |
| Arogyavardhini | Phyllanthus emblica | Fruit | CID_5280343 | 2237 | FEN1 | ChEMBL \| NPASS |
| Arogyavardhini | Phyllanthus emblica | Fruit | CID_5280343 | 22933 | SIRT2 | ChEMBL |
| Arogyavardhini | Phyllanthus emblica | Fruit | CID_5280343 | 23028 | KDM1A | BindingDB \| ChEMBL \| NPASS |
| Arogyavardhini | Phyllanthus emblica | Fruit | CID_5280343 | 23043 | TNIK | ChEMBL \| NPASS |
| Arogyavardhini | Phyllanthus emblica | Fruit | CID_5280343 | 231 | AKR1B1 | ChEMBL \| NPASS |
| Arogyavardhini | Phyllanthus emblica | Fruit | CID_5280343 | 2322 | FLT3 | ChEMBL \| NPASS |
| Arogyavardhini | Phyllanthus emblica | Fruit | CID_5280343 | 23408 | SIRT5 | ChEMBL |
| Arogyavardhini | Phyllanthus emblica | Fruit | CID_5280343 | 23410 | SIRT3 | ChEMBL |
| Arogyavardhini | Phyllanthus emblica | Fruit | CID_5280343 | 23411 | SIRT1 | ChEMBL |
| Arogyavardhini | Phyllanthus emblica | Fruit | CID_5280343 | 23621 | BACE1 | BindingDB \| ChEMBL \| NPASS |
| Arogyavardhini | Phyllanthus emblica | Fruit | CID_5280343 | 23632 | CA14 | BindingDB \| ChEMBL \| NPASS |
| Arogyavardhini | Phyllanthus emblica | Fruit | CID_5280343 | 238 | ALK | BindingDB \| ChEMBL \| NPASS |
| Arogyavardhini | Phyllanthus emblica | Fruit | CID_5280343 | 239 | ALOX12 | BindingDB \| ChEMBL \| NPASS |

Table S3: Associations between formulations, constituent plants, phytochemicals, and human molecular targets. The table lists the formulations, plant name and plant part, phytochemical identifiers, associated human target genes (Entrez Gene ID and gene symbol), and the database sources supporting each phytochemical-target interaction.

| Formulation | Plant name | Part | PubChem CID | Entrez Gene ID | Gene Symbol | Source database(s) |
|---|---|---|---|---|---|---|
| Arogyavardhini | Phyllanthus emblica | Fruit | CID_5280343 | 240 | ALOX5 | BindingDB \| ChEMBL \| NPASS |
| Arogyavardhini | Phyllanthus emblica | Fruit | CID_5280343 | 246 | ALOX15 | BindingDB \| ChEMBL \| NPASS |
| Arogyavardhini | Phyllanthus emblica | Fruit | CID_5280343 | 247 | ALOX15B | BindingDB \| ChEMBL \| NPASS |
| Arogyavardhini | Phyllanthus emblica | Fruit | CID_5280343 | 253430 | IPMK | ChEMBL \| NPASS |
| Arogyavardhini | Phyllanthus emblica | Fruit | CID_5280343 | 2548 | GAA | ChEMBL \| NPASS |
| Arogyavardhini | Phyllanthus emblica | Fruit | CID_5280343 | 2584 | GALK1 | ChEMBL \| NPASS |
| Arogyavardhini | Phyllanthus emblica | Fruit | CID_5280343 | 2629 | GBA1 | NPASS |
| Arogyavardhini | Phyllanthus emblica | Fruit | CID_5280343 | 2642 | GCGR | ChEMBL |
| Arogyavardhini | Phyllanthus emblica | Fruit | CID_5280343 | 2693 | GHSR | ChEMBL |
| Arogyavardhini | Phyllanthus emblica | Fruit | CID_5280343 | 2696 | GIPR | ChEMBL |
| Arogyavardhini | Phyllanthus emblica | Fruit | CID_5280343 | 2739 | GLO1 | ChEMBL \| NPASS |
| Arogyavardhini | Phyllanthus emblica | Fruit | CID_5280343 | 276 | AMY1A | ChEMBL |
| Arogyavardhini | Phyllanthus emblica | Fruit | CID_5280343 | 277 | AMY1B | BindingDB |
| Arogyavardhini | Phyllanthus emblica | Fruit | CID_5280343 | 28234 | SLCO1B3 | ChEMBL \| NPASS |
| Arogyavardhini | Phyllanthus emblica | Fruit | CID_5280343 | 283106 | CSNK2A3 | ChEMBL \| NPASS |
| Arogyavardhini | Phyllanthus emblica | Fruit | CID_5280343 | 2859 | GPR35 | BindingDB \| ChEMBL \| NPASS |
| Arogyavardhini | Phyllanthus emblica | Fruit | CID_5280343 | 2862 | MLNR | ChEMBL |
| Arogyavardhini | Phyllanthus emblica | Fruit | CID_5280343 | 2908 | NR3C1 | ChEMBL |
| Arogyavardhini | Phyllanthus emblica | Fruit | CID_5280343 | 2931 | GSK3A | ChEMBL |
| Arogyavardhini | Phyllanthus emblica | Fruit | CID_5280343 | 2932 | GSK3B | BindingDB \| ChEMBL \| NPASS |
| Arogyavardhini | Phyllanthus emblica | Fruit | CID_5280343 | 2936 | GSR | NPASS |
| Arogyavardhini | Phyllanthus emblica | Fruit | CID_5280343 | 3028 | HSD17B10 | ChEMBL \| NPASS |
| Arogyavardhini | Phyllanthus emblica | Fruit | CID_5280343 | 3091 | HIF1A | ChEMBL \| NPASS |
| Arogyavardhini | Phyllanthus emblica | Fruit | CID_5280343 | 3248 | HPGD | ChEMBL \| NPASS |
| Arogyavardhini | Phyllanthus emblica | Fruit | CID_5280343 | 3269 | HRH1 | ChEMBL |
| Arogyavardhini | Phyllanthus emblica | Fruit | CID_5280343 | 3274 | HRH2 | ChEMBL |
| Arogyavardhini | Phyllanthus emblica | Fruit | CID_5280343 | 328 | APEX1 | ChEMBL \| NPASS |
| Arogyavardhini | Phyllanthus emblica | Fruit | CID_5280343 | 3290 | HSD11B1 | ChEMBL \| NPASS |
| Arogyavardhini | Phyllanthus emblica | Fruit | CID_5280343 | 3292 | HSD17B1 | ChEMBL \| NPASS |
| Arogyavardhini | Phyllanthus emblica | Fruit | CID_5280343 | 3294 | HSD17B2 | BindingDB \| ChEMBL \| NPASS |
| Arogyavardhini | Phyllanthus emblica | Fruit | CID_5280343 | 3297 | HSF1 | BindingDB \| ChEMBL \| NPASS |
| Arogyavardhini | Phyllanthus emblica | Fruit | CID_5280343 | 3309 | HSPA5 | NPASS |
| Arogyavardhini | Phyllanthus emblica | Fruit | CID_5280343 | 3315 | HSPB1 | ChEMBL |
| Arogyavardhini | Phyllanthus emblica | Fruit | CID_5280343 | 3350 | HTR1A | ChEMBL \| NPASS |
| Arogyavardhini | Phyllanthus emblica | Fruit | CID_5280343 | 3352 | HTR1D | BindingDB |
| Arogyavardhini | Phyllanthus emblica | Fruit | CID_5280343 | 3356 | HTR2A | ChEMBL |
| Arogyavardhini | Phyllanthus emblica | Fruit | CID_5280343 | 3357 | HTR2B | ChEMBL |
| Arogyavardhini | Phyllanthus emblica | Fruit | CID_5280343 | 3358 | HTR2C | ChEMBL |
| Arogyavardhini | Phyllanthus emblica | Fruit | CID_5280343 | 3359 | HTR3A | BindingDB \| ChEMBL |
| Arogyavardhini | Phyllanthus emblica | Fruit | CID_5280343 | 3362 | HTR6 | BindingDB |
| Arogyavardhini | Phyllanthus emblica | Fruit | CID_5280343 | 3363 | HTR7 | BindingDB |
| Arogyavardhini | Phyllanthus emblica | Fruit | CID_5280343 | 3480 | IGF1R | BindingDB \| ChEMBL \| NPASS |
| Arogyavardhini | Phyllanthus emblica | Fruit | CID_5280343 | 351 | APP | BindingDB \| ChEMBL \| NPASS |
| Arogyavardhini | Phyllanthus emblica | Fruit | CID_5280343 | 3577 | CXCR1 | ChEMBL \| NPASS |
| Arogyavardhini | Phyllanthus emblica | Fruit | CID_5280343 | 367 | AR | NPASS |
| Arogyavardhini | Phyllanthus emblica | Fruit | CID_5280343 | 3716 | JAK1 | ChEMBL \| NPASS |
| Arogyavardhini | Phyllanthus emblica | Fruit | CID_5280343 | 3725 | JUN | NPASS |
| Arogyavardhini | Phyllanthus emblica | Fruit | CID_5280343 | 3757 | KCNH2 | ChEMBL |
| Arogyavardhini | Phyllanthus emblica | Fruit | CID_5280343 | 3791 | KDR | BindingDB \| ChEMBL \| NPASS |
| Arogyavardhini | Phyllanthus emblica | Fruit | CID_5280343 | 387129 | NPSR1 | NPASS |
| Arogyavardhini | Phyllanthus emblica | Fruit | CID_5280343 | 390245 | KDM4E | ChEMBL \| NPASS |
| Arogyavardhini | Phyllanthus emblica | Fruit | CID_5280343 | 3932 | LCK | ChEMBL \| NPASS |
| Arogyavardhini | Phyllanthus emblica | Fruit | CID_5280343 | 4000 | LMNA | NPASS |
| Arogyavardhini | Phyllanthus emblica | Fruit | CID_5280343 | 4088 | SMAD3 | NPASS |
| Arogyavardhini | Phyllanthus emblica | Fruit | CID_5280343 | 410 | ARSA | NPASS |
| Arogyavardhini | Phyllanthus emblica | Fruit | CID_5280343 | 4128 | MAOA | ChEMBL \| NPASS |
| Arogyavardhini | Phyllanthus emblica | Fruit | CID_5280343 | 4129 | MAOB | ChEMBL \| NPASS |
| Arogyavardhini | Phyllanthus emblica | Fruit | CID_5280343 | 4137 | MAPT | ChEMBL \| NPASS |
| Arogyavardhini | Phyllanthus emblica | Fruit | CID_5280343 | 415116 | PIM3 | ChEMBL \| NPASS |
| Arogyavardhini | Phyllanthus emblica | Fruit | CID_5280343 | 4154 | MBNL1 | NPASS |
| Arogyavardhini | Phyllanthus emblica | Fruit | CID_5280343 | 4159 | MC3R | ChEMBL |
| Arogyavardhini | Phyllanthus emblica | Fruit | CID_5280343 | 4160 | MC4R | ChEMBL |
| Arogyavardhini | Phyllanthus emblica | Fruit | CID_5280343 | 4217 | MAP3K5 | ChEMBL \| NPASS |
| Arogyavardhini | Phyllanthus emblica | Fruit | CID_5280343 | 4233 | MET | BindingDB \| ChEMBL \| NPASS |
| Arogyavardhini | Phyllanthus emblica | Fruit | CID_5280343 | 4297 | KMT2A | NPASS |
| Arogyavardhini | Phyllanthus emblica | Fruit | CID_5280343 | 43 | ACHE | BindingDB \| ChEMBL \| NPASS |
| Arogyavardhini | Phyllanthus emblica | Fruit | CID_5280343 | 4312 | MMP1 | ChEMBL |
| Arogyavardhini | Phyllanthus emblica | Fruit | CID_5280343 | 4313 | MMP2 | ChEMBL \| NPASS |
| Arogyavardhini | Phyllanthus emblica | Fruit | CID_5280343 | 4314 | MMP3 | ChEMBL \| NPASS |
| Arogyavardhini | Phyllanthus emblica | Fruit | CID_5280343 | 4318 | MMP9 | BindingDB \| ChEMBL \| NPASS |

| Table S3: Associations between formulations, constituent plants, phytochemicals, and human molecular targets. The table lists the formulations, plant name and plant part, phytochemical identifiers, associated human target genes (Entrez Gene ID and gene symbol), and the database sources supporting each phytochemical-target interaction. | | | | | | |
|---|---|---|---|---|---|---|
| **Formulation** | **Plant name** | **Part** | **PubChem CID** | **Entrez Gene ID** | **Gene Symbol** | **Source database(s)** |
| Arogyavardhini | Phyllanthus emblica | Fruit | CID_5280343 | 4321 | MMP12 | BindingDB \| ChEMBL \| NPASS |
| Arogyavardhini | Phyllanthus emblica | Fruit | CID_5280343 | 4322 | MMP13 | BindingDB \| ChEMBL \| NPASS |
| Arogyavardhini | Phyllanthus emblica | Fruit | CID_5280343 | 4350 | MPG | ChEMBL \| NPASS |
| Arogyavardhini | Phyllanthus emblica | Fruit | CID_5280343 | 4353 | MPO | BindingDB \| ChEMBL \| NPASS |
| Arogyavardhini | Phyllanthus emblica | Fruit | CID_5280343 | 4363 | ABCC1 | ChEMBL \| NPASS |
| Arogyavardhini | Phyllanthus emblica | Fruit | CID_5280343 | 4751 | NEK2 | ChEMBL \| NPASS |
| Arogyavardhini | Phyllanthus emblica | Fruit | CID_5280343 | 4759 | NEU2 | BindingDB \| NPASS |
| Arogyavardhini | Phyllanthus emblica | Fruit | CID_5280343 | 4790 | NFKB1 | ChEMBL \| NPASS |
| Arogyavardhini | Phyllanthus emblica | Fruit | CID_5280343 | 4791 | NFKB2 | ChEMBL |
| Arogyavardhini | Phyllanthus emblica | Fruit | CID_5280343 | 4886 | NPY1R | ChEMBL |
| Arogyavardhini | Phyllanthus emblica | Fruit | CID_5280343 | 4887 | NPY2R | ChEMBL |
| Arogyavardhini | Phyllanthus emblica | Fruit | CID_5280343 | 4907 | NT5E | BindingDB \| ChEMBL |
| Arogyavardhini | Phyllanthus emblica | Fruit | CID_5280343 | 4923 | NTSR1 | ChEMBL |
| Arogyavardhini | Phyllanthus emblica | Fruit | CID_5280343 | 4985 | OPRD1 | ChEMBL |
| Arogyavardhini | Phyllanthus emblica | Fruit | CID_5280343 | 4986 | OPRK1 | BindingDB \| ChEMBL |
| Arogyavardhini | Phyllanthus emblica | Fruit | CID_5280343 | 4988 | OPRM1 | BindingDB \| ChEMBL |
| Arogyavardhini | Phyllanthus emblica | Fruit | CID_5280343 | 50507 | NOX4 | BindingDB \| ChEMBL \| NPASS |
| Arogyavardhini | Phyllanthus emblica | Fruit | CID_5280343 | 51231 | VRK3 | ChEMBL \| NPASS |
| Arogyavardhini | Phyllanthus emblica | Fruit | CID_5280343 | 5127 | CDK16 | ChEMBL \| NPASS |
| Arogyavardhini | Phyllanthus emblica | Fruit | CID_5280343 | 5139 | PDE3A | ChEMBL |
| Arogyavardhini | Phyllanthus emblica | Fruit | CID_5280343 | 51426 | POLK | ChEMBL \| NPASS |
| Arogyavardhini | Phyllanthus emblica | Fruit | CID_5280343 | 5144 | PDE4D | ChEMBL |
| Arogyavardhini | Phyllanthus emblica | Fruit | CID_5280343 | 51447 | IP6K2 | ChEMBL \| NPASS |
| Arogyavardhini | Phyllanthus emblica | Fruit | CID_5280343 | 51548 | SIRT6 | ChEMBL |
| Arogyavardhini | Phyllanthus emblica | Fruit | CID_5280343 | 5163 | PDK1 | ChEMBL \| NPASS |
| Arogyavardhini | Phyllanthus emblica | Fruit | CID_5280343 | 5170 | PDPK1 | ChEMBL \| NPASS |
| Arogyavardhini | Phyllanthus emblica | Fruit | CID_5280343 | 51765 | STK26 | ChEMBL \| NPASS |
| Arogyavardhini | Phyllanthus emblica | Fruit | CID_5280343 | 5241 | PGR | ChEMBL |
| Arogyavardhini | Phyllanthus emblica | Fruit | CID_5280343 | 5243 | ABCB1 | ChEMBL \| NPASS |
| Arogyavardhini | Phyllanthus emblica | Fruit | CID_5280343 | 5290 | PIK3CA | ChEMBL |
| Arogyavardhini | Phyllanthus emblica | Fruit | CID_5280343 | 5292 | PIM1 | ChEMBL \| NPASS |
| Arogyavardhini | Phyllanthus emblica | Fruit | CID_5280343 | 5294 | PIK3CG | BindingDB \| ChEMBL \| NPASS |
| Arogyavardhini | Phyllanthus emblica | Fruit | CID_5280343 | 5295 | PIK3R1 | ChEMBL \| NPASS |
| Arogyavardhini | Phyllanthus emblica | Fruit | CID_5280343 | 5300 | PIN1 | NPASS |
| Arogyavardhini | Phyllanthus emblica | Fruit | CID_5280343 | 5319 | PLA2G1B | BindingDB \| ChEMBL \| NPASS |
| Arogyavardhini | Phyllanthus emblica | Fruit | CID_5280343 | 5328 | PLAU | BindingDB \| ChEMBL \| NPASS |
| Arogyavardhini | Phyllanthus emblica | Fruit | CID_5280343 | 5340 | PLG | ChEMBL \| NPASS |
| Arogyavardhini | Phyllanthus emblica | Fruit | CID_5280343 | 5347 | PLK1 | BindingDB \| ChEMBL \| NPASS |
| Arogyavardhini | Phyllanthus emblica | Fruit | CID_5280343 | 5376 | PMP22 | NPASS |
| Arogyavardhini | Phyllanthus emblica | Fruit | CID_5280343 | 53944 | CSNK1G1 | ChEMBL \| NPASS |
| Arogyavardhini | Phyllanthus emblica | Fruit | CID_5280343 | 5406 | PNLIP | BindingDB \| ChEMBL |
| Arogyavardhini | Phyllanthus emblica | Fruit | CID_5280343 | 5423 | POLB | ChEMBL \| NPASS |
| Arogyavardhini | Phyllanthus emblica | Fruit | CID_5280343 | 5429 | POLH | BindingDB \| ChEMBL \| NPASS |
| Arogyavardhini | Phyllanthus emblica | Fruit | CID_5280343 | 5444 | PON1 | ChEMBL \| NPASS |
| Arogyavardhini | Phyllanthus emblica | Fruit | CID_5280343 | 54575 | UGT1A10 | ChEMBL \| NPASS |
| Arogyavardhini | Phyllanthus emblica | Fruit | CID_5280343 | 54576 | UGT1A8 | ChEMBL \| NPASS |
| Arogyavardhini | Phyllanthus emblica | Fruit | CID_5280343 | 54578 | UGT1A6 | ChEMBL \| NPASS |
| Arogyavardhini | Phyllanthus emblica | Fruit | CID_5280343 | 54600 | UGT1A9 | ChEMBL \| NPASS |
| Arogyavardhini | Phyllanthus emblica | Fruit | CID_5280343 | 54657 | UGT1A4 | ChEMBL \| NPASS |
| Arogyavardhini | Phyllanthus emblica | Fruit | CID_5280343 | 54658 | UGT1A1 | ChEMBL \| NPASS |
| Arogyavardhini | Phyllanthus emblica | Fruit | CID_5280343 | 54659 | UGT1A3 | ChEMBL \| NPASS |
| Arogyavardhini | Phyllanthus emblica | Fruit | CID_5280343 | 5467 | PPARD | NPASS |
| Arogyavardhini | Phyllanthus emblica | Fruit | CID_5280343 | 5468 | PPARG | BindingDB \| ChEMBL \| NPASS |
| Arogyavardhini | Phyllanthus emblica | Fruit | CID_5280343 | 54737 | MPHOSPH8 | ChEMBL \| NPASS |
| Arogyavardhini | Phyllanthus emblica | Fruit | CID_5280343 | 552 | AVPR1A | ChEMBL |
| Arogyavardhini | Phyllanthus emblica | Fruit | CID_5280343 | 554 | AVPR2 | ChEMBL \| NPASS |
| Arogyavardhini | Phyllanthus emblica | Fruit | CID_5280343 | 5550 | PREP | ChEMBL \| NPASS |
| Arogyavardhini | Phyllanthus emblica | Fruit | CID_5280343 | 5563 | PRKAA2 | ChEMBL \| NPASS |
| Arogyavardhini | Phyllanthus emblica | Fruit | CID_5280343 | 5566 | PRKACA | ChEMBL \| NPASS |
| Arogyavardhini | Phyllanthus emblica | Fruit | CID_5280343 | 55662 | HIF1AN | ChEMBL \| NPASS |
| Arogyavardhini | Phyllanthus emblica | Fruit | CID_5280343 | 55775 | TDP1 | ChEMBL \| NPASS |
| Arogyavardhini | Phyllanthus emblica | Fruit | CID_5280343 | 5578 | PRKCA | ChEMBL \| NPASS |
| Arogyavardhini | Phyllanthus emblica | Fruit | CID_5280343 | 55781 | RIOK2 | ChEMBL \| NPASS |
| Arogyavardhini | Phyllanthus emblica | Fruit | CID_5280343 | 558 | AXL | ChEMBL \| NPASS |
| Arogyavardhini | Phyllanthus emblica | Fruit | CID_5280343 | 5585 | PKN1 | BindingDB \| ChEMBL \| NPASS |
| Arogyavardhini | Phyllanthus emblica | Fruit | CID_5280343 | 55872 | PBK | ChEMBL \| NPASS |
| Arogyavardhini | Phyllanthus emblica | Fruit | CID_5280343 | 5594 | MAPK1 | ChEMBL \| NPASS |
| Arogyavardhini | Phyllanthus emblica | Fruit | CID_5280343 | 5595 | MAPK3 | ChEMBL \| NPASS |
| Arogyavardhini | Phyllanthus emblica | Fruit | CID_5280343 | 5597 | MAPK6 | ChEMBL \| NPASS |
| Arogyavardhini | Phyllanthus emblica | Fruit | CID_5280343 | 5599 | MAPK8 | ChEMBL \| NPASS |

**Table S3: Associations between formulations, constituent plants, phytochemicals, and human molecular targets. The table lists the formulations, plant name and plant part, phytochemical identifiers, associated human target genes (Entrez Gene ID and gene symbol), and the database sources supporting each phytochemical-target interaction.**

| Formulation | Plant name | Part | PubChem CID | Entrez Gene ID | Gene Symbol | Source database(s) |
|---|---|---|---|---|---|---|
| Arogyavardhini | Phyllanthus emblica | Fruit | CID_5280343 | 5600 | MAPK11 | ChEMBL \| NPASS |
| Arogyavardhini | Phyllanthus emblica | Fruit | CID_5280343 | 5601 | MAPK9 | ChEMBL |
| Arogyavardhini | Phyllanthus emblica | Fruit | CID_5280343 | 5602 | MAPK10 | BindingDB |
| Arogyavardhini | Phyllanthus emblica | Fruit | CID_5280343 | 5603 | MAPK13 | ChEMBL \| NPASS |
| Arogyavardhini | Phyllanthus emblica | Fruit | CID_5280343 | 5604 | MAP2K1 | ChEMBL \| NPASS |
| Arogyavardhini | Phyllanthus emblica | Fruit | CID_5280343 | 5605 | MAP2K2 | ChEMBL \| NPASS |
| Arogyavardhini | Phyllanthus emblica | Fruit | CID_5280343 | 5608 | MAP2K6 | ChEMBL \| NPASS |
| Arogyavardhini | Phyllanthus emblica | Fruit | CID_5280343 | 56339 | METTL3 | ChEMBL |
| Arogyavardhini | Phyllanthus emblica | Fruit | CID_5280343 | 5644 | PRSS1 | BindingDB \| ChEMBL |
| Arogyavardhini | Phyllanthus emblica | Fruit | CID_5280343 | 5645 | PRSS2 | ChEMBL |
| Arogyavardhini | Phyllanthus emblica | Fruit | CID_5280343 | 5646 | PRSS3 | ChEMBL \| NPASS |
| Arogyavardhini | Phyllanthus emblica | Fruit | CID_5280343 | 56924 | PAK6 | NPASS |
| Arogyavardhini | Phyllanthus emblica | Fruit | CID_5280343 | 5693 | PSMB5 | BindingDB \| ChEMBL \| NPASS |
| Arogyavardhini | Phyllanthus emblica | Fruit | CID_5280343 | 57016 | AKR1B10 | ChEMBL \| NPASS |
| Arogyavardhini | Phyllanthus emblica | Fruit | CID_5280343 | 57118 | CAMK1D | ChEMBL \| NPASS |
| Arogyavardhini | Phyllanthus emblica | Fruit | CID_5280343 | 57144 | PAK5 | ChEMBL \| NPASS |
| Arogyavardhini | Phyllanthus emblica | Fruit | CID_5280343 | 57172 | CAMK1G | ChEMBL \| NPASS |
| Arogyavardhini | Phyllanthus emblica | Fruit | CID_5280343 | 5742 | PTGS1 | ChEMBL \| NPASS |
| Arogyavardhini | Phyllanthus emblica | Fruit | CID_5280343 | 5743 | PTGS2 | BindingDB \| ChEMBL \| NPASS |
| Arogyavardhini | Phyllanthus emblica | Fruit | CID_5280343 | 5747 | PTK2 | BindingDB \| ChEMBL \| NPASS |
| Arogyavardhini | Phyllanthus emblica | Fruit | CID_5280343 | 5770 | PTPN1 | BindingDB \| ChEMBL \| NPASS |
| Arogyavardhini | Phyllanthus emblica | Fruit | CID_5280343 | 57721 | METTL14 | ChEMBL |
| Arogyavardhini | Phyllanthus emblica | Fruit | CID_5280343 | 5836 | PYGL | BindingDB \| ChEMBL \| NPASS |
| Arogyavardhini | Phyllanthus emblica | Fruit | CID_5280343 | 590 | BCHE | BindingDB \| ChEMBL \| NPASS |
| Arogyavardhini | Phyllanthus emblica | Fruit | CID_5280343 | 595 | CCND1 | ChEMBL \| NPASS |
| Arogyavardhini | Phyllanthus emblica | Fruit | CID_5280343 | 5965 | RECQL | ChEMBL \| NPASS |
| Arogyavardhini | Phyllanthus emblica | Fruit | CID_5280343 | 5970 | RELA | ChEMBL |
| Arogyavardhini | Phyllanthus emblica | Fruit | CID_5280343 | 5979 | RET | BindingDB \| ChEMBL \| NPASS |
| Arogyavardhini | Phyllanthus emblica | Fruit | CID_5280343 | 5999 | RGS4 | NPASS |
| Arogyavardhini | Phyllanthus emblica | Fruit | CID_5280343 | 607 | BCL9 | BindingDB \| ChEMBL |
| Arogyavardhini | Phyllanthus emblica | Fruit | CID_5280343 | 6197 | RPS6KA3 | ChEMBL \| NPASS |
| Arogyavardhini | Phyllanthus emblica | Fruit | CID_5280343 | 6198 | RPS6KB1 | ChEMBL \| NPASS |
| Arogyavardhini | Phyllanthus emblica | Fruit | CID_5280343 | 623 | BDKRB1 | ChEMBL |
| Arogyavardhini | Phyllanthus emblica | Fruit | CID_5280343 | 624 | BDKRB2 | ChEMBL |
| Arogyavardhini | Phyllanthus emblica | Fruit | CID_5280343 | 6300 | MAPK12 | ChEMBL \| NPASS |
| Arogyavardhini | Phyllanthus emblica | Fruit | CID_5280343 | 641 | BLM | ChEMBL \| NPASS |
| Arogyavardhini | Phyllanthus emblica | Fruit | CID_5280343 | 6446 | SGK1 | ChEMBL \| NPASS |
| Arogyavardhini | Phyllanthus emblica | Fruit | CID_5280343 | 6476 | SI | ChEMBL \| NPASS |
| Arogyavardhini | Phyllanthus emblica | Fruit | CID_5280343 | 6514 | SLC2A2 | ChEMBL \| NPASS |
| Arogyavardhini | Phyllanthus emblica | Fruit | CID_5280343 | 6530 | SLC6A2 | BindingDB \| ChEMBL |
| Arogyavardhini | Phyllanthus emblica | Fruit | CID_5280343 | 6531 | SLC6A3 | ChEMBL |
| Arogyavardhini | Phyllanthus emblica | Fruit | CID_5280343 | 6532 | SLC6A4 | BindingDB \| ChEMBL |
| Arogyavardhini | Phyllanthus emblica | Fruit | CID_5280343 | 65975 | STK33 | BindingDB |
| Arogyavardhini | Phyllanthus emblica | Fruit | CID_5280343 | 6606 | SMN1 | NPASS |
| Arogyavardhini | Phyllanthus emblica | Fruit | CID_5280343 | 6622 | SNCA | ChEMBL \| NPASS |
| Arogyavardhini | Phyllanthus emblica | Fruit | CID_5280343 | 6652 | SORD | BindingDB \| ChEMBL \| NPASS |
| Arogyavardhini | Phyllanthus emblica | Fruit | CID_5280343 | 6714 | SRC | ChEMBL \| NPASS |
| Arogyavardhini | Phyllanthus emblica | Fruit | CID_5280343 | 673 | BRAF | ChEMBL \| NPASS |
| Arogyavardhini | Phyllanthus emblica | Fruit | CID_5280343 | 6778 | STAT6 | NPASS |
| Arogyavardhini | Phyllanthus emblica | Fruit | CID_5280343 | 6789 | STK4 | ChEMBL \| NPASS |
| Arogyavardhini | Phyllanthus emblica | Fruit | CID_5280343 | 6793 | STK10 | ChEMBL \| NPASS |
| Arogyavardhini | Phyllanthus emblica | Fruit | CID_5280343 | 6850 | SYK | BindingDB \| ChEMBL \| NPASS |
| Arogyavardhini | Phyllanthus emblica | Fruit | CID_5280343 | 6915 | TBXA2R | ChEMBL |
| Arogyavardhini | Phyllanthus emblica | Fruit | CID_5280343 | 7068 | THRB | NPASS |
| Arogyavardhini | Phyllanthus emblica | Fruit | CID_5280343 | 7150 | TOP1 | ChEMBL |
| Arogyavardhini | Phyllanthus emblica | Fruit | CID_5280343 | 7153 | TOP2A | ChEMBL \| NPASS |
| Arogyavardhini | Phyllanthus emblica | Fruit | CID_5280343 | 7155 | TOP2B | ChEMBL \| NPASS |
| Arogyavardhini | Phyllanthus emblica | Fruit | CID_5280343 | 7157 | TP53 | NPASS |
| Arogyavardhini | Phyllanthus emblica | Fruit | CID_5280343 | 7224 | TRPC5 | ChEMBL \| NPASS |
| Arogyavardhini | Phyllanthus emblica | Fruit | CID_5280343 | 7253 | TSHR | NPASS |
| Arogyavardhini | Phyllanthus emblica | Fruit | CID_5280343 | 7257 | TSNAX | ChEMBL \| NPASS |
| Arogyavardhini | Phyllanthus emblica | Fruit | CID_5280343 | 7299 | TYR | BindingDB \| ChEMBL \| NPASS |
| Arogyavardhini | Phyllanthus emblica | Fruit | CID_5280343 | 7366 | UGT2B15 | ChEMBL \| NPASS |
| Arogyavardhini | Phyllanthus emblica | Fruit | CID_5280343 | 7398 | USP1 | NPASS |
| Arogyavardhini | Phyllanthus emblica | Fruit | CID_5280343 | 7443 | VRK1 | ChEMBL \| NPASS |
| Arogyavardhini | Phyllanthus emblica | Fruit | CID_5280343 | 7444 | VRK2 | ChEMBL \| NPASS |
| Arogyavardhini | Phyllanthus emblica | Fruit | CID_5280343 | 7498 | XDH | BindingDB \| ChEMBL \| NPASS |
| Arogyavardhini | Phyllanthus emblica | Fruit | CID_5280343 | 7525 | YES1 | BindingDB \| ChEMBL \| NPASS |
| Arogyavardhini | Phyllanthus emblica | Fruit | CID_5280343 | 759 | CA1 | BindingDB \| ChEMBL \| NPASS |
| Arogyavardhini | Phyllanthus emblica | Fruit | CID_5280343 | 760 | CA2 | BindingDB \| ChEMBL \| NPASS |

**Table S3: Associations between formulations, constituent plants, phytochemicals, and human molecular targets. The table lists the formulations, plant name and plant part, phytochemical identifiers, associated human target genes (Entrez Gene ID and gene symbol), and the database sources supporting each phytochemical-target interaction.**

| Formulation | Plant name | Part | PubChem CID | Entrez Gene ID | Gene Symbol | Source database(s) |
|---|---|---|---|---|---|---|
| Arogyavardhini | Phyllanthus emblica | Fruit | CID_5280343 | 761 | CA3 | BindingDB \| ChEMBL \| NPASS |
| Arogyavardhini | Phyllanthus emblica | Fruit | CID_5280343 | 762 | CA4 | BindingDB \| ChEMBL \| NPASS |
| Arogyavardhini | Phyllanthus emblica | Fruit | CID_5280343 | 763 | CA5A | BindingDB \| ChEMBL \| NPASS |
| Arogyavardhini | Phyllanthus emblica | Fruit | CID_5280343 | 765 | CA6 | ChEMBL \| NPASS |
| Arogyavardhini | Phyllanthus emblica | Fruit | CID_5280343 | 766 | CA7 | BindingDB \| ChEMBL \| NPASS |
| Arogyavardhini | Phyllanthus emblica | Fruit | CID_5280343 | 768 | CA9 | BindingDB \| ChEMBL \| NPASS |
| Arogyavardhini | Phyllanthus emblica | Fruit | CID_5280343 | 771 | CA12 | BindingDB \| ChEMBL \| NPASS |
| Arogyavardhini | Phyllanthus emblica | Fruit | CID_5280343 | 79915 | ATAD5 | ChEMBL \| NPASS |
| Arogyavardhini | Phyllanthus emblica | Fruit | CID_5280343 | 805 | CALM2 | ChEMBL |
| Arogyavardhini | Phyllanthus emblica | Fruit | CID_5280343 | 814 | CAMK4 | ChEMBL \| NPASS |
| Arogyavardhini | Phyllanthus emblica | Fruit | CID_5280343 | 815 | CAMK2A | ChEMBL \| NPASS |
| Arogyavardhini | Phyllanthus emblica | Fruit | CID_5280343 | 816 | CAMK2B | BindingDB \| ChEMBL \| NPASS |
| Arogyavardhini | Phyllanthus emblica | Fruit | CID_5280343 | 817 | CAMK2D | ChEMBL \| NPASS |
| Arogyavardhini | Phyllanthus emblica | Fruit | CID_5280343 | 818 | CAMK2G | ChEMBL \| NPASS |
| Arogyavardhini | Phyllanthus emblica | Fruit | CID_5280343 | 85417 | CCNB3 | ChEMBL |
| Arogyavardhini | Phyllanthus emblica | Fruit | CID_5280343 | 8550 | MAPKAPK5 | ChEMBL \| NPASS |
| Arogyavardhini | Phyllanthus emblica | Fruit | CID_5280343 | 8576 | STK16 | ChEMBL \| NPASS |
| Arogyavardhini | Phyllanthus emblica | Fruit | CID_5280343 | 8647 | ABCB11 | NPASS |
| Arogyavardhini | Phyllanthus emblica | Fruit | CID_5280343 | 8654 | PDE5A | ChEMBL \| NPASS |
| Arogyavardhini | Phyllanthus emblica | Fruit | CID_5280343 | 8714 | ABCC3 | BindingDB \| NPASS |
| Arogyavardhini | Phyllanthus emblica | Fruit | CID_5280343 | 873 | CBR1 | ChEMBL \| NPASS |
| Arogyavardhini | Phyllanthus emblica | Fruit | CID_5280343 | 874 | CBR3 | ChEMBL \| NPASS |
| Arogyavardhini | Phyllanthus emblica | Fruit | CID_5280343 | 8814 | CDKL1 | ChEMBL \| NPASS |
| Arogyavardhini | Phyllanthus emblica | Fruit | CID_5280343 | 8851 | CDK5R1 | ChEMBL \| NPASS |
| Arogyavardhini | Phyllanthus emblica | Fruit | CID_5280343 | 886 | CCKAR | ChEMBL |
| Arogyavardhini | Phyllanthus emblica | Fruit | CID_5280343 | 887 | CCKBR | ChEMBL |
| Arogyavardhini | Phyllanthus emblica | Fruit | CID_5280343 | 8877 | SPHK1 | ChEMBL |
| Arogyavardhini | Phyllanthus emblica | Fruit | CID_5280343 | 891 | CCNB1 | ChEMBL \| NPASS |
| Arogyavardhini | Phyllanthus emblica | Fruit | CID_5280343 | 8972 | MGAM | ChEMBL \| NPASS |
| Arogyavardhini | Phyllanthus emblica | Fruit | CID_5280343 | 9099 | USP2 | NPASS |
| Arogyavardhini | Phyllanthus emblica | Fruit | CID_5280343 | 9133 | CCNB2 | ChEMBL \| NPASS |
| Arogyavardhini | Phyllanthus emblica | Fruit | CID_5280343 | 9212 | AURKB | BindingDB \| ChEMBL \| NPASS |
| Arogyavardhini | Phyllanthus emblica | Fruit | CID_5280343 | 9252 | RPS6KA5 | ChEMBL \| NPASS |
| Arogyavardhini | Phyllanthus emblica | Fruit | CID_5280343 | 9261 | MAPKAPK2 | ChEMBL \| NPASS |
| Arogyavardhini | Phyllanthus emblica | Fruit | CID_5280343 | 9263 | STK17A | ChEMBL \| NPASS |
| Arogyavardhini | Phyllanthus emblica | Fruit | CID_5280343 | 93432 | MGAM2 | ChEMBL \| NPASS |
| Arogyavardhini | Phyllanthus emblica | Fruit | CID_5280343 | 9429 | ABCG2 | BindingDB \| ChEMBL \| NPASS |
| Arogyavardhini | Phyllanthus emblica | Fruit | CID_5280343 | 952 | CD38 | BindingDB \| ChEMBL \| NPASS |
| Arogyavardhini | Phyllanthus emblica | Fruit | CID_5280343 | 9748 | SLK | ChEMBL \| NPASS |
| Arogyavardhini | Phyllanthus emblica | Fruit | CID_5280343 | 983 | CDK1 | BindingDB \| ChEMBL \| NPASS |
| Arogyavardhini | Phyllanthus emblica | Fruit | CID_5280343 | 9891 | NUAK1 | BindingDB \| ChEMBL \| NPASS |
| Arogyavardhini | Phyllanthus emblica | Fruit | CID_5280343 | 9943 | OXSR1 | ChEMBL \| NPASS |
| Arogyavardhini | Phyllanthus emblica | Fruit | CID_5281855 | 10013 | HDAC6 | ChEMBL |
| Arogyavardhini | Phyllanthus emblica | Fruit | CID_5281855 | 10073 | SNUPN | NPASS |
| Arogyavardhini | Phyllanthus emblica | Fruit | CID_5281855 | 1017 | CDK2 | BindingDB \| ChEMBL |
| Arogyavardhini | Phyllanthus emblica | Fruit | CID_5281855 | 1019 | CDK4 | BindingDB \| ChEMBL |
| Arogyavardhini | Phyllanthus emblica | Fruit | CID_5281855 | 10411 | RAPGEF3 | NPASS |
| Arogyavardhini | Phyllanthus emblica | Fruit | CID_5281855 | 10498 | CARM1 | ChEMBL \| NPASS |
| Arogyavardhini | Phyllanthus emblica | Fruit | CID_5281855 | 10587 | TXNRD2 | ChEMBL |
| Arogyavardhini | Phyllanthus emblica | Fruit | CID_5281855 | 10599 | SLCO1B1 | ChEMBL \| NPASS |
| Arogyavardhini | Phyllanthus emblica | Fruit | CID_5281855 | 10733 | PLK4 | BindingDB \| ChEMBL \| NPASS |
| Arogyavardhini | Phyllanthus emblica | Fruit | CID_5281855 | 10919 | EHMT2 | ChEMBL \| NPASS |
| Arogyavardhini | Phyllanthus emblica | Fruit | CID_5281855 | 10951 | CBX1 | NPASS |
| Arogyavardhini | Phyllanthus emblica | Fruit | CID_5281855 | 11069 | RAPGEF4 | NPASS |
| Arogyavardhini | Phyllanthus emblica | Fruit | CID_5281855 | 11201 | POLI | ChEMBL \| NPASS |
| Arogyavardhini | Phyllanthus emblica | Fruit | CID_5281855 | 11238 | CA5B | BindingDB \| ChEMBL \| NPASS |
| Arogyavardhini | Phyllanthus emblica | Fruit | CID_5281855 | 114112 | TXNRD3 | ChEMBL \| NPASS |
| Arogyavardhini | Phyllanthus emblica | Fruit | CID_5281855 | 1326 | MAP3K8 | BindingDB \| ChEMBL \| NPASS |
| Arogyavardhini | Phyllanthus emblica | Fruit | CID_5281855 | 1457 | CSNK2A1 | BindingDB \| ChEMBL \| NPASS |
| Arogyavardhini | Phyllanthus emblica | Fruit | CID_5281855 | 1459 | CSNK2A2 | ChEMBL \| NPASS |
| Arogyavardhini | Phyllanthus emblica | Fruit | CID_5281855 | 1460 | CSNK2B | ChEMBL |
| Arogyavardhini | Phyllanthus emblica | Fruit | CID_5281855 | 1557 | CYP2C19 | ChEMBL \| NPASS |
| Arogyavardhini | Phyllanthus emblica | Fruit | CID_5281855 | 1559 | CYP2C9 | NPASS |
| Arogyavardhini | Phyllanthus emblica | Fruit | CID_5281855 | 1636 | ACE | BindingDB \| NPASS |
| Arogyavardhini | Phyllanthus emblica | Fruit | CID_5281855 | 1845 | DUSP3 | BindingDB \| NPASS |
| Arogyavardhini | Phyllanthus emblica | Fruit | CID_5281855 | 1859 | DYRK1A | BindingDB \| ChEMBL \| NPASS |
| Arogyavardhini | Phyllanthus emblica | Fruit | CID_5281855 | 1956 | EGFR | BindingDB \| ChEMBL \| NPASS |
| Arogyavardhini | Phyllanthus emblica | Fruit | CID_5281855 | 1981 | EIF4G1 | BindingDB |
| Arogyavardhini | Phyllanthus emblica | Fruit | CID_5281855 | 1995 | ELAVL3 | ChEMBL \| NPASS |
| Arogyavardhini | Phyllanthus emblica | Fruit | CID_5281855 | 2050 | EPHB4 | BindingDB \| ChEMBL \| NPASS |

**Table S3: Associations between formulations, constituent plants, phytochemicals, and human molecular targets. The table lists the formulations, plant name and plant part, phytochemical identifiers, associated human target genes (Entrez Gene ID and gene symbol), and the database sources supporting each phytochemical-target interaction.**

| Formulation | Plant name | Part | PubChem CID | Entrez Gene ID | Gene Symbol | Source database(s) |
|---|---|---|---|---|---|---|
| Arogyavardhini | Phyllanthus emblica | Fruit | CID_5281855 | 2053 | EPHX2 | ChEMBL \| NPASS |
| Arogyavardhini | Phyllanthus emblica | Fruit | CID_5281855 | 2064 | ERBB2 | ChEMBL \| NPASS |
| Arogyavardhini | Phyllanthus emblica | Fruit | CID_5281855 | 207 | AKT1 | ChEMBL \| NPASS |
| Arogyavardhini | Phyllanthus emblica | Fruit | CID_5281855 | 2237 | FEN1 | ChEMBL \| NPASS |
| Arogyavardhini | Phyllanthus emblica | Fruit | CID_5281855 | 231 | AKR1B1 | ChEMBL \| NPASS |
| Arogyavardhini | Phyllanthus emblica | Fruit | CID_5281855 | 23192 | ATG4B | BindingDB \| ChEMBL \| NPASS |
| Arogyavardhini | Phyllanthus emblica | Fruit | CID_5281855 | 2322 | FLT3 | ChEMBL \| NPASS |
| Arogyavardhini | Phyllanthus emblica | Fruit | CID_5281855 | 2324 | FLT4 | BindingDB \| ChEMBL \| NPASS |
| Arogyavardhini | Phyllanthus emblica | Fruit | CID_5281855 | 23405 | DICER1 | ChEMBL |
| Arogyavardhini | Phyllanthus emblica | Fruit | CID_5281855 | 23621 | BACE1 | BindingDB \| ChEMBL \| NPASS |
| Arogyavardhini | Phyllanthus emblica | Fruit | CID_5281855 | 23632 | CA14 | BindingDB \| ChEMBL \| NPASS |
| Arogyavardhini | Phyllanthus emblica | Fruit | CID_5281855 | 238 | ALK | BindingDB \| ChEMBL |
| Arogyavardhini | Phyllanthus emblica | Fruit | CID_5281855 | 246 | ALOX15 | NPASS |
| Arogyavardhini | Phyllanthus emblica | Fruit | CID_5281855 | 247 | ALOX15B | NPASS |
| Arogyavardhini | Phyllanthus emblica | Fruit | CID_5281855 | 2548 | GAA | NPASS |
| Arogyavardhini | Phyllanthus emblica | Fruit | CID_5281855 | 2590 | GALNT2 | ChEMBL \| NPASS |
| Arogyavardhini | Phyllanthus emblica | Fruit | CID_5281855 | 2597 | GAPDH | NPASS |
| Arogyavardhini | Phyllanthus emblica | Fruit | CID_5281855 | 2648 | KAT2A | NPASS |
| Arogyavardhini | Phyllanthus emblica | Fruit | CID_5281855 | 2744 | GLS | ChEMBL \| NPASS |
| Arogyavardhini | Phyllanthus emblica | Fruit | CID_5281855 | 28234 | SLCO1B3 | ChEMBL \| NPASS |
| Arogyavardhini | Phyllanthus emblica | Fruit | CID_5281855 | 2859 | GPR35 | BindingDB \| ChEMBL \| NPASS |
| Arogyavardhini | Phyllanthus emblica | Fruit | CID_5281855 | 2932 | GSK3B | BindingDB \| ChEMBL \| NPASS |
| Arogyavardhini | Phyllanthus emblica | Fruit | CID_5281855 | 2936 | GSR | ChEMBL \| NPASS |
| Arogyavardhini | Phyllanthus emblica | Fruit | CID_5281855 | 2950 | GSTP1 | ChEMBL \| NPASS |
| Arogyavardhini | Phyllanthus emblica | Fruit | CID_5281855 | 29994 | BAZ2B | ChEMBL \| NPASS |
| Arogyavardhini | Phyllanthus emblica | Fruit | CID_5281855 | 3028 | HSD17B10 | ChEMBL \| NPASS |
| Arogyavardhini | Phyllanthus emblica | Fruit | CID_5281855 | 3043 | HBB | ChEMBL \| NPASS |
| Arogyavardhini | Phyllanthus emblica | Fruit | CID_5281855 | 3248 | HPGD | ChEMBL \| NPASS |
| Arogyavardhini | Phyllanthus emblica | Fruit | CID_5281855 | 328 | APEX1 | ChEMBL \| NPASS |
| Arogyavardhini | Phyllanthus emblica | Fruit | CID_5281855 | 3303 | HSPA1A | BindingDB \| ChEMBL \| NPASS |
| Arogyavardhini | Phyllanthus emblica | Fruit | CID_5281855 | 3309 | HSPA5 | NPASS |
| Arogyavardhini | Phyllanthus emblica | Fruit | CID_5281855 | 3312 | HSPA8 | BindingDB \| ChEMBL \| NPASS |
| Arogyavardhini | Phyllanthus emblica | Fruit | CID_5281855 | 3417 | IDH1 | NPASS |
| Arogyavardhini | Phyllanthus emblica | Fruit | CID_5281855 | 3480 | IGF1R | BindingDB \| ChEMBL \| NPASS |
| Arogyavardhini | Phyllanthus emblica | Fruit | CID_5281855 | 351 | APP | BindingDB \| ChEMBL \| NPASS |
| Arogyavardhini | Phyllanthus emblica | Fruit | CID_5281855 | 3551 | IKBKB | BindingDB \| ChEMBL \| NPASS |
| Arogyavardhini | Phyllanthus emblica | Fruit | CID_5281855 | 3643 | INSR | BindingDB \| ChEMBL \| NPASS |
| Arogyavardhini | Phyllanthus emblica | Fruit | CID_5281855 | 377677 | CA13 | ChEMBL \| NPASS |
| Arogyavardhini | Phyllanthus emblica | Fruit | CID_5281855 | 3791 | KDR | BindingDB \| ChEMBL \| NPASS |
| Arogyavardhini | Phyllanthus emblica | Fruit | CID_5281855 | 3837 | KPNB1 | NPASS |
| Arogyavardhini | Phyllanthus emblica | Fruit | CID_5281855 | 390245 | KDM4E | ChEMBL \| NPASS |
| Arogyavardhini | Phyllanthus emblica | Fruit | CID_5281855 | 4000 | LMNA | NPASS |
| Arogyavardhini | Phyllanthus emblica | Fruit | CID_5281855 | 4088 | SMAD3 | BindingDB \| ChEMBL \| NPASS |
| Arogyavardhini | Phyllanthus emblica | Fruit | CID_5281855 | 4128 | MAOA | ChEMBL \| NPASS |
| Arogyavardhini | Phyllanthus emblica | Fruit | CID_5281855 | 4137 | MAPT | ChEMBL \| NPASS |
| Arogyavardhini | Phyllanthus emblica | Fruit | CID_5281855 | 4154 | MBNL1 | NPASS |
| Arogyavardhini | Phyllanthus emblica | Fruit | CID_5281855 | 4170 | MCL1 | BindingDB \| ChEMBL \| NPASS |
| Arogyavardhini | Phyllanthus emblica | Fruit | CID_5281855 | 4233 | MET | BindingDB \| ChEMBL \| NPASS |
| Arogyavardhini | Phyllanthus emblica | Fruit | CID_5281855 | 4282 | MIF | BindingDB \| ChEMBL \| NPASS |
| Arogyavardhini | Phyllanthus emblica | Fruit | CID_5281855 | 4297 | KMT2A | NPASS |
| Arogyavardhini | Phyllanthus emblica | Fruit | CID_5281855 | 43 | ACHE | BindingDB |
| Arogyavardhini | Phyllanthus emblica | Fruit | CID_5281855 | 472 | ATM | NPASS |
| Arogyavardhini | Phyllanthus emblica | Fruit | CID_5281855 | 4780 | NFE2L2 | NPASS |
| Arogyavardhini | Phyllanthus emblica | Fruit | CID_5281855 | 4869 | NPM1 | ChEMBL \| NPASS |
| Arogyavardhini | Phyllanthus emblica | Fruit | CID_5281855 | 4907 | NT5E | ChEMBL \| NPASS |
| Arogyavardhini | Phyllanthus emblica | Fruit | CID_5281855 | 51053 | GMNN | NPASS |
| Arogyavardhini | Phyllanthus emblica | Fruit | CID_5281855 | 51426 | POLK | ChEMBL \| NPASS |
| Arogyavardhini | Phyllanthus emblica | Fruit | CID_5281855 | 5159 | PDGFRB | BindingDB \| ChEMBL \| NPASS |
| Arogyavardhini | Phyllanthus emblica | Fruit | CID_5281855 | 5300 | PIN1 | NPASS |
| Arogyavardhini | Phyllanthus emblica | Fruit | CID_5281855 | 5313 | PKLR | ChEMBL |
| Arogyavardhini | Phyllanthus emblica | Fruit | CID_5281855 | 5347 | PLK1 | BindingDB \| ChEMBL \| NPASS |
| Arogyavardhini | Phyllanthus emblica | Fruit | CID_5281855 | 5423 | POLB | ChEMBL \| NPASS |
| Arogyavardhini | Phyllanthus emblica | Fruit | CID_5281855 | 5429 | POLH | BindingDB \| ChEMBL \| NPASS |
| Arogyavardhini | Phyllanthus emblica | Fruit | CID_5281855 | 5566 | PRKACA | BindingDB \| ChEMBL |
| Arogyavardhini | Phyllanthus emblica | Fruit | CID_5281855 | 5567 | PRKACB | ChEMBL |
| Arogyavardhini | Phyllanthus emblica | Fruit | CID_5281855 | 5568 | PRKACG | ChEMBL \| NPASS |
| Arogyavardhini | Phyllanthus emblica | Fruit | CID_5281855 | 55775 | TDP1 | ChEMBL \| NPASS |
| Arogyavardhini | Phyllanthus emblica | Fruit | CID_5281855 | 55867 | SLC22A11 | ChEMBL \| NPASS |
| Arogyavardhini | Phyllanthus emblica | Fruit | CID_5281855 | 5594 | MAPK1 | ChEMBL \| NPASS |
| Arogyavardhini | Phyllanthus emblica | Fruit | CID_5281855 | 5747 | PTK2 | BindingDB \| ChEMBL \| NPASS |

**Table S3: Associations between formulations, constituent plants, phytochemicals, and human molecular targets. The table lists the formulations, plant name and plant part, phytochemical identifiers, associated human target genes (Entrez Gene ID and gene symbol), and the database sources supporting each phytochemical-target interaction.**

| Formulation | Plant name | Part | PubChem CID | Entrez Gene ID | Gene Symbol | Source database(s) |
|---|---|---|---|---|---|---|
| Arogyavardhini | Phyllanthus emblica | Fruit | CID_5281855 | 5770 | PTPN1 | BindingDB |
| Arogyavardhini | Phyllanthus emblica | Fruit | CID_5281855 | 57787 | MARK4 | ChEMBL |
| Arogyavardhini | Phyllanthus emblica | Fruit | CID_5281855 | 595 | CCND1 | ChEMBL \| NPASS |
| Arogyavardhini | Phyllanthus emblica | Fruit | CID_5281855 | 5965 | RECQL | ChEMBL \| NPASS |
| Arogyavardhini | Phyllanthus emblica | Fruit | CID_5281855 | 5979 | RET | BindingDB \| ChEMBL \| NPASS |
| Arogyavardhini | Phyllanthus emblica | Fruit | CID_5281855 | 5999 | RGS4 | NPASS |
| Arogyavardhini | Phyllanthus emblica | Fruit | CID_5281855 | 641 | BLM | ChEMBL \| NPASS |
| Arogyavardhini | Phyllanthus emblica | Fruit | CID_5281855 | 6714 | SRC | ChEMBL \| NPASS |
| Arogyavardhini | Phyllanthus emblica | Fruit | CID_5281855 | 673 | BRAF | BindingDB \| ChEMBL \| NPASS |
| Arogyavardhini | Phyllanthus emblica | Fruit | CID_5281855 | 6790 | AURKA | BindingDB \| ChEMBL \| NPASS |
| Arogyavardhini | Phyllanthus emblica | Fruit | CID_5281855 | 7010 | TEK | BindingDB \| ChEMBL \| NPASS |
| Arogyavardhini | Phyllanthus emblica | Fruit | CID_5281855 | 7068 | THRB | NPASS |
| Arogyavardhini | Phyllanthus emblica | Fruit | CID_5281855 | 7155 | TOP2B | ChEMBL \| NPASS |
| Arogyavardhini | Phyllanthus emblica | Fruit | CID_5281855 | 7157 | TP53 | NPASS |
| Arogyavardhini | Phyllanthus emblica | Fruit | CID_5281855 | 7257 | TSNAX | ChEMBL \| NPASS |
| Arogyavardhini | Phyllanthus emblica | Fruit | CID_5281855 | 7296 | TXNRD1 | ChEMBL |
| Arogyavardhini | Phyllanthus emblica | Fruit | CID_5281855 | 7398 | USP1 | NPASS |
| Arogyavardhini | Phyllanthus emblica | Fruit | CID_5281855 | 7421 | VDR | ChEMBL \| NPASS |
| Arogyavardhini | Phyllanthus emblica | Fruit | CID_5281855 | 759 | CA1 | BindingDB \| ChEMBL \| NPASS |
| Arogyavardhini | Phyllanthus emblica | Fruit | CID_5281855 | 760 | CA2 | BindingDB \| ChEMBL \| NPASS |
| Arogyavardhini | Phyllanthus emblica | Fruit | CID_5281855 | 761 | CA3 | BindingDB \| ChEMBL \| NPASS |
| Arogyavardhini | Phyllanthus emblica | Fruit | CID_5281855 | 762 | CA4 | BindingDB \| ChEMBL \| NPASS |
| Arogyavardhini | Phyllanthus emblica | Fruit | CID_5281855 | 763 | CA5A | BindingDB \| ChEMBL \| NPASS |
| Arogyavardhini | Phyllanthus emblica | Fruit | CID_5281855 | 765 | CA6 | ChEMBL \| NPASS |
| Arogyavardhini | Phyllanthus emblica | Fruit | CID_5281855 | 766 | CA7 | BindingDB \| ChEMBL \| NPASS |
| Arogyavardhini | Phyllanthus emblica | Fruit | CID_5281855 | 768 | CA9 | BindingDB \| ChEMBL \| NPASS |
| Arogyavardhini | Phyllanthus emblica | Fruit | CID_5281855 | 771 | CA12 | BindingDB \| ChEMBL \| NPASS |
| Arogyavardhini | Phyllanthus emblica | Fruit | CID_5281855 | 861 | RUNX1 | ChEMBL |
| Arogyavardhini | Phyllanthus emblica | Fruit | CID_5281855 | 865 | CBFB | ChEMBL \| NPASS |
| Arogyavardhini | Phyllanthus emblica | Fruit | CID_5281855 | 8877 | SPHK1 | ChEMBL |
| Arogyavardhini | Phyllanthus emblica | Fruit | CID_5281855 | 890 | CCNA2 | ChEMBL |
| Arogyavardhini | Phyllanthus emblica | Fruit | CID_5281855 | 8900 | CCNA1 | ChEMBL \| NPASS |
| Arogyavardhini | Phyllanthus emblica | Fruit | CID_5281855 | 9099 | USP2 | ChEMBL \| NPASS |
| Arogyavardhini | Phyllanthus emblica | Fruit | CID_5281855 | 9212 | AURKB | BindingDB \| ChEMBL \| NPASS |
| Arogyavardhini | Phyllanthus emblica | Fruit | CID_5281855 | 9356 | SLC22A6 | BindingDB \| ChEMBL \| NPASS |
| Arogyavardhini | Phyllanthus emblica | Fruit | CID_5281855 | 9682 | KDM4A | ChEMBL \| NPASS |
| Arogyavardhini | Phyllanthus emblica | Fruit | CID_5281855 | 9891 | NUAK1 | BindingDB \| ChEMBL \| NPASS |
| Arogyavardhini | Phyllanthus emblica | Fruit | CID_54670067 | 10280 | SIGMAR1 | ChEMBL |
| Arogyavardhini | Phyllanthus emblica | Fruit | CID_54670067 | 10599 | SLCO1B1 | ChEMBL \| NPASS |
| Arogyavardhini | Phyllanthus emblica | Fruit | CID_54670067 | 10800 | CYSLTR1 | ChEMBL |
| Arogyavardhini | Phyllanthus emblica | Fruit | CID_54670067 | 10919 | EHMT2 | ChEMBL |
| Arogyavardhini | Phyllanthus emblica | Fruit | CID_54670067 | 112398 | EGLN2 | ChEMBL \| NPASS |
| Arogyavardhini | Phyllanthus emblica | Fruit | CID_54670067 | 112399 | EGLN3 | ChEMBL \| NPASS |
| Arogyavardhini | Phyllanthus emblica | Fruit | CID_54670067 | 11255 | HRH3 | ChEMBL |
| Arogyavardhini | Phyllanthus emblica | Fruit | CID_54670067 | 1128 | CHRM1 | ChEMBL |
| Arogyavardhini | Phyllanthus emblica | Fruit | CID_54670067 | 1129 | CHRM2 | ChEMBL |
| Arogyavardhini | Phyllanthus emblica | Fruit | CID_54670067 | 1131 | CHRM3 | ChEMBL |
| Arogyavardhini | Phyllanthus emblica | Fruit | CID_54670067 | 1132 | CHRM4 | ChEMBL |
| Arogyavardhini | Phyllanthus emblica | Fruit | CID_54670067 | 1133 | CHRM5 | ChEMBL |
| Arogyavardhini | Phyllanthus emblica | Fruit | CID_54670067 | 1233 | CCR4 | ChEMBL |
| Arogyavardhini | Phyllanthus emblica | Fruit | CID_54670067 | 1234 | CCR5 | ChEMBL |
| Arogyavardhini | Phyllanthus emblica | Fruit | CID_54670067 | 1268 | CNR1 | ChEMBL |
| Arogyavardhini | Phyllanthus emblica | Fruit | CID_54670067 | 134 | ADORA1 | ChEMBL |
| Arogyavardhini | Phyllanthus emblica | Fruit | CID_54670067 | 135 | ADORA2A | ChEMBL |
| Arogyavardhini | Phyllanthus emblica | Fruit | CID_54670067 | 140 | ADORA3 | ChEMBL |
| Arogyavardhini | Phyllanthus emblica | Fruit | CID_54670067 | 1432 | MAPK14 | ChEMBL |
| Arogyavardhini | Phyllanthus emblica | Fruit | CID_54670067 | 146 | ADRA1D | ChEMBL |
| Arogyavardhini | Phyllanthus emblica | Fruit | CID_54670067 | 148 | ADRA1A | ChEMBL |
| Arogyavardhini | Phyllanthus emblica | Fruit | CID_54670067 | 150 | ADRA2A | ChEMBL |
| Arogyavardhini | Phyllanthus emblica | Fruit | CID_54670067 | 151 | ADRA2B | ChEMBL \| NPASS |
| Arogyavardhini | Phyllanthus emblica | Fruit | CID_54670067 | 1511 | CTSG | ChEMBL |
| Arogyavardhini | Phyllanthus emblica | Fruit | CID_54670067 | 152 | ADRA2C | ChEMBL |
| Arogyavardhini | Phyllanthus emblica | Fruit | CID_54670067 | 153 | ADRB1 | ChEMBL |
| Arogyavardhini | Phyllanthus emblica | Fruit | CID_54670067 | 154 | ADRB2 | ChEMBL |
| Arogyavardhini | Phyllanthus emblica | Fruit | CID_54670067 | 1544 | CYP1A2 | ChEMBL |
| Arogyavardhini | Phyllanthus emblica | Fruit | CID_54670067 | 1548 | CYP2A6 | ChEMBL |
| Arogyavardhini | Phyllanthus emblica | Fruit | CID_54670067 | 155 | ADRB3 | ChEMBL |
| Arogyavardhini | Phyllanthus emblica | Fruit | CID_54670067 | 1557 | CYP2C19 | ChEMBL |
| Arogyavardhini | Phyllanthus emblica | Fruit | CID_54670067 | 1559 | CYP2C9 | ChEMBL |
| Arogyavardhini | Phyllanthus emblica | Fruit | CID_54670067 | 1565 | CYP2D6 | ChEMBL |

**Table S3: Associations between formulations, constituent plants, phytochemicals, and human molecular targets. The table lists the formulations, plant name and plant part, phytochemical identifiers, associated human target genes (Entrez Gene ID and gene symbol), and the database sources supporting each phytochemical-target interaction.**

| Formulation | Plant name | Part | PubChem CID | Entrez Gene ID | Gene Symbol | Source database(s) |
|---|---|---|---|---|---|---|
| Arogyavardhini | Phyllanthus emblica | Fruit | CID_54670067 | 1571 | CYP2E1 | ChEMBL |
| Arogyavardhini | Phyllanthus emblica | Fruit | CID_54670067 | 1576 | CYP3A4 | ChEMBL |
| Arogyavardhini | Phyllanthus emblica | Fruit | CID_54670067 | 1812 | DRD1 | ChEMBL |
| Arogyavardhini | Phyllanthus emblica | Fruit | CID_54670067 | 1813 | DRD2 | ChEMBL |
| Arogyavardhini | Phyllanthus emblica | Fruit | CID_54670067 | 1814 | DRD3 | ChEMBL |
| Arogyavardhini | Phyllanthus emblica | Fruit | CID_54670067 | 1815 | DRD4 | ChEMBL |
| Arogyavardhini | Phyllanthus emblica | Fruit | CID_54670067 | 186 | AGTR2 | ChEMBL |
| Arogyavardhini | Phyllanthus emblica | Fruit | CID_54670067 | 1909 | EDNRA | ChEMBL |
| Arogyavardhini | Phyllanthus emblica | Fruit | CID_54670067 | 1956 | EGFR | ChEMBL |
| Arogyavardhini | Phyllanthus emblica | Fruit | CID_54670067 | 1991 | ELANE | ChEMBL |
| Arogyavardhini | Phyllanthus emblica | Fruit | CID_54670067 | 2064 | ERBB2 | ChEMBL |
| Arogyavardhini | Phyllanthus emblica | Fruit | CID_54670067 | 2099 | ESR1 | ChEMBL |
| Arogyavardhini | Phyllanthus emblica | Fruit | CID_54670067 | 2100 | ESR2 | ChEMBL |
| Arogyavardhini | Phyllanthus emblica | Fruit | CID_54670067 | 2147 | F2 | ChEMBL |
| Arogyavardhini | Phyllanthus emblica | Fruit | CID_54670067 | 2321 | FLT1 | ChEMBL |
| Arogyavardhini | Phyllanthus emblica | Fruit | CID_54670067 | 239 | ALOX12 | ChEMBL \| NPASS |
| Arogyavardhini | Phyllanthus emblica | Fruit | CID_54670067 | 240 | ALOX5 | ChEMBL \| NPASS |
| Arogyavardhini | Phyllanthus emblica | Fruit | CID_54670067 | 246 | ALOX15 | ChEMBL \| NPASS |
| Arogyavardhini | Phyllanthus emblica | Fruit | CID_54670067 | 247 | ALOX15B | ChEMBL \| NPASS |
| Arogyavardhini | Phyllanthus emblica | Fruit | CID_54670067 | 2534 | FYN | ChEMBL |
| Arogyavardhini | Phyllanthus emblica | Fruit | CID_54670067 | 2554 | GABRA1 | ChEMBL |
| Arogyavardhini | Phyllanthus emblica | Fruit | CID_54670067 | 2555 | GABRA2 | ChEMBL |
| Arogyavardhini | Phyllanthus emblica | Fruit | CID_54670067 | 2561 | GABRB2 | ChEMBL |
| Arogyavardhini | Phyllanthus emblica | Fruit | CID_54670067 | 2566 | GABRG2 | ChEMBL |
| Arogyavardhini | Phyllanthus emblica | Fruit | CID_54670067 | 25939 | SAMHD1 | ChEMBL |
| Arogyavardhini | Phyllanthus emblica | Fruit | CID_54670067 | 276 | AMY1A | ChEMBL |
| Arogyavardhini | Phyllanthus emblica | Fruit | CID_54670067 | 279 | AMY2A | ChEMBL \| NPASS |
| Arogyavardhini | Phyllanthus emblica | Fruit | CID_54670067 | 28234 | SLCO1B3 | ChEMBL \| NPASS |
| Arogyavardhini | Phyllanthus emblica | Fruit | CID_54670067 | 2908 | NR3C1 | ChEMBL |
| Arogyavardhini | Phyllanthus emblica | Fruit | CID_54670067 | 2932 | GSK3B | ChEMBL \| NPASS |
| Arogyavardhini | Phyllanthus emblica | Fruit | CID_54670067 | 3156 | HMGCR | ChEMBL |
| Arogyavardhini | Phyllanthus emblica | Fruit | CID_54670067 | 3269 | HRH1 | ChEMBL |
| Arogyavardhini | Phyllanthus emblica | Fruit | CID_54670067 | 3274 | HRH2 | ChEMBL |
| Arogyavardhini | Phyllanthus emblica | Fruit | CID_54670067 | 3350 | HTR1A | ChEMBL |
| Arogyavardhini | Phyllanthus emblica | Fruit | CID_54670067 | 3356 | HTR2A | ChEMBL |
| Arogyavardhini | Phyllanthus emblica | Fruit | CID_54670067 | 3357 | HTR2B | ChEMBL |
| Arogyavardhini | Phyllanthus emblica | Fruit | CID_54670067 | 3358 | HTR2C | ChEMBL |
| Arogyavardhini | Phyllanthus emblica | Fruit | CID_54670067 | 3362 | HTR6 | ChEMBL |
| Arogyavardhini | Phyllanthus emblica | Fruit | CID_54670067 | 351 | APP | ChEMBL \| NPASS |
| Arogyavardhini | Phyllanthus emblica | Fruit | CID_54670067 | 3577 | CXCR1 | ChEMBL |
| Arogyavardhini | Phyllanthus emblica | Fruit | CID_54670067 | 3579 | CXCR2 | ChEMBL |
| Arogyavardhini | Phyllanthus emblica | Fruit | CID_54670067 | 367 | AR | ChEMBL \| NPASS |
| Arogyavardhini | Phyllanthus emblica | Fruit | CID_54670067 | 3757 | KCNH2 | ChEMBL |
| Arogyavardhini | Phyllanthus emblica | Fruit | CID_54670067 | 3791 | KDR | ChEMBL |
| Arogyavardhini | Phyllanthus emblica | Fruit | CID_54670067 | 3932 | LCK | ChEMBL |
| Arogyavardhini | Phyllanthus emblica | Fruit | CID_54670067 | 4128 | MAOA | ChEMBL |
| Arogyavardhini | Phyllanthus emblica | Fruit | CID_54670067 | 4159 | MC3R | ChEMBL |
| Arogyavardhini | Phyllanthus emblica | Fruit | CID_54670067 | 4160 | MC4R | ChEMBL |
| Arogyavardhini | Phyllanthus emblica | Fruit | CID_54670067 | 4161 | MC5R | ChEMBL |
| Arogyavardhini | Phyllanthus emblica | Fruit | CID_54670067 | 43 | ACHE | ChEMBL |
| Arogyavardhini | Phyllanthus emblica | Fruit | CID_54670067 | 4312 | MMP1 | ChEMBL |
| Arogyavardhini | Phyllanthus emblica | Fruit | CID_54670067 | 4318 | MMP9 | ChEMBL |
| Arogyavardhini | Phyllanthus emblica | Fruit | CID_54670067 | 4780 | NFE2L2 | ChEMBL \| NPASS |
| Arogyavardhini | Phyllanthus emblica | Fruit | CID_54670067 | 4886 | NPY1R | ChEMBL |
| Arogyavardhini | Phyllanthus emblica | Fruit | CID_54670067 | 4887 | NPY2R | ChEMBL |
| Arogyavardhini | Phyllanthus emblica | Fruit | CID_54670067 | 4985 | OPRD1 | ChEMBL |
| Arogyavardhini | Phyllanthus emblica | Fruit | CID_54670067 | 4986 | OPRK1 | ChEMBL |
| Arogyavardhini | Phyllanthus emblica | Fruit | CID_54670067 | 4988 | OPRM1 | ChEMBL |
| Arogyavardhini | Phyllanthus emblica | Fruit | CID_54670067 | 5033 | P4HA1 | ChEMBL \| NPASS |
| Arogyavardhini | Phyllanthus emblica | Fruit | CID_54670067 | 50507 | NOX4 | ChEMBL \| NPASS |
| Arogyavardhini | Phyllanthus emblica | Fruit | CID_54670067 | 5139 | PDE3A | ChEMBL |
| Arogyavardhini | Phyllanthus emblica | Fruit | CID_54670067 | 5141 | PDE4A | ChEMBL |
| Arogyavardhini | Phyllanthus emblica | Fruit | CID_54670067 | 5241 | PGR | ChEMBL |
| Arogyavardhini | Phyllanthus emblica | Fruit | CID_54670067 | 54583 | EGLN1 | ChEMBL \| NPASS |
| Arogyavardhini | Phyllanthus emblica | Fruit | CID_54670067 | 552 | AVPR1A | ChEMBL |
| Arogyavardhini | Phyllanthus emblica | Fruit | CID_54670067 | 5530 | PPP3CA | ChEMBL |
| Arogyavardhini | Phyllanthus emblica | Fruit | CID_54670067 | 55662 | HIF1AN | ChEMBL \| NPASS |
| Arogyavardhini | Phyllanthus emblica | Fruit | CID_54670067 | 5578 | PRKCA | ChEMBL |
| Arogyavardhini | Phyllanthus emblica | Fruit | CID_54670067 | 5594 | MAPK1 | ChEMBL |
| Arogyavardhini | Phyllanthus emblica | Fruit | CID_54670067 | 5595 | MAPK3 | ChEMBL |

**Table S3: Associations between formulations, constituent plants, phytochemicals, and human molecular targets. The table lists the formulations, plant name and plant part, phytochemical identifiers, associated human target genes (Entrez Gene ID and gene symbol), and the database sources supporting each phytochemical-target interaction.**

| Formulation | Plant name | Part | PubChem CID | Entrez Gene ID | Gene Symbol | Source database(s) |
|---|---|---|---|---|---|---|
| Arogyavardhini | Phyllanthus emblica | Fruit | CID_54670067 | 5724 | PTAFR | ChEMBL |
| Arogyavardhini | Phyllanthus emblica | Fruit | CID_54670067 | 5742 | PTGS1 | ChEMBL |
| Arogyavardhini | Phyllanthus emblica | Fruit | CID_54670067 | 5743 | PTGS2 | ChEMBL |
| Arogyavardhini | Phyllanthus emblica | Fruit | CID_54670067 | 5788 | PTPRC | ChEMBL |
| Arogyavardhini | Phyllanthus emblica | Fruit | CID_54670067 | 624 | BDKRB2 | ChEMBL |
| Arogyavardhini | Phyllanthus emblica | Fruit | CID_54670067 | 6530 | SLC6A2 | ChEMBL |
| Arogyavardhini | Phyllanthus emblica | Fruit | CID_54670067 | 6531 | SLC6A3 | ChEMBL |
| Arogyavardhini | Phyllanthus emblica | Fruit | CID_54670067 | 6532 | SLC6A4 | ChEMBL |
| Arogyavardhini | Phyllanthus emblica | Fruit | CID_54670067 | 6865 | TACR2 | ChEMBL |
| Arogyavardhini | Phyllanthus emblica | Fruit | CID_54670067 | 6869 | TACR1 | ChEMBL |
| Arogyavardhini | Phyllanthus emblica | Fruit | CID_54670067 | 6915 | TBXA2R | ChEMBL |
| Arogyavardhini | Phyllanthus emblica | Fruit | CID_54670067 | 6916 | TBXAS1 | ChEMBL |
| Arogyavardhini | Phyllanthus emblica | Fruit | CID_54670067 | 729230 | CCR2 | ChEMBL |
| Arogyavardhini | Phyllanthus emblica | Fruit | CID_54670067 | 7299 | TYR | ChEMBL \| NPASS |
| Arogyavardhini | Phyllanthus emblica | Fruit | CID_54670067 | 7433 | VIPR1 | ChEMBL |
| Arogyavardhini | Phyllanthus emblica | Fruit | CID_54670067 | 760 | CA2 | ChEMBL |
| Arogyavardhini | Phyllanthus emblica | Fruit | CID_54670067 | 799 | CALCR | ChEMBL |
| Arogyavardhini | Phyllanthus emblica | Fruit | CID_54670067 | 834 | CASP1 | ChEMBL |
| Arogyavardhini | Phyllanthus emblica | Fruit | CID_54670067 | 8654 | PDE5A | ChEMBL |
| Arogyavardhini | Phyllanthus emblica | Fruit | CID_54670067 | 886 | CCKAR | ChEMBL |
| Arogyavardhini | Phyllanthus emblica | Fruit | CID_7428 | 10919 | EHMT2 | ChEMBL \| NPASS |
| Arogyavardhini | Phyllanthus emblica | Fruit | CID_7428 | 10951 | CBX1 | NPASS |
| Arogyavardhini | Phyllanthus emblica | Fruit | CID_7428 | 118429 | ANTXR2 | NPASS |
| Arogyavardhini | Phyllanthus emblica | Fruit | CID_7428 | 2237 | FEN1 | NPASS |
| Arogyavardhini | Phyllanthus emblica | Fruit | CID_7428 | 2529 | FUT7 | ChEMBL \| NPASS |
| Arogyavardhini | Phyllanthus emblica | Fruit | CID_7428 | 2548 | GAA | NPASS |
| Arogyavardhini | Phyllanthus emblica | Fruit | CID_7428 | 26013 | L3MBTL1 | NPASS |
| Arogyavardhini | Phyllanthus emblica | Fruit | CID_7428 | 2671 | GFER | ChEMBL \| NPASS |
| Arogyavardhini | Phyllanthus emblica | Fruit | CID_7428 | 28689 | TRAV4 | BindingDB |
| Arogyavardhini | Phyllanthus emblica | Fruit | CID_7428 | 29994 | BAZ2B | ChEMBL \| NPASS |
| Arogyavardhini | Phyllanthus emblica | Fruit | CID_7428 | 328 | APEX1 | ChEMBL \| NPASS |
| Arogyavardhini | Phyllanthus emblica | Fruit | CID_7428 | 390245 | KDM4E | ChEMBL \| NPASS |
| Arogyavardhini | Phyllanthus emblica | Fruit | CID_7428 | 4000 | LMNA | NPASS |
| Arogyavardhini | Phyllanthus emblica | Fruit | CID_7428 | 4790 | NFKB1 | ChEMBL |
| Arogyavardhini | Phyllanthus emblica | Fruit | CID_7428 | 4791 | NFKB2 | ChEMBL |
| Arogyavardhini | Phyllanthus emblica | Fruit | CID_7428 | 5300 | PIN1 | NPASS |
| Arogyavardhini | Phyllanthus emblica | Fruit | CID_7428 | 5423 | POLB | NPASS |
| Arogyavardhini | Phyllanthus emblica | Fruit | CID_7428 | 55775 | TDP1 | NPASS |
| Arogyavardhini | Phyllanthus emblica | Fruit | CID_7428 | 5970 | RELA | ChEMBL |
| Arogyavardhini | Phyllanthus emblica | Fruit | CID_7428 | 7498 | XDH | BindingDB \| ChEMBL \| NPASS |
| Arogyavardhini | Phyllanthus emblica | Fruit | CID_7428 | 9682 | KDM4A | ChEMBL \| NPASS |
| Arogyavardhini | Picrorhiza kurroa | Rhizome | CID_155942 | 5742 | PTGS1 | ChEMBL \| NPASS |
| Arogyavardhini | Picrorhiza kurroa | Rhizome | CID_155942 | 5743 | PTGS2 | ChEMBL \| NPASS |
| Arogyavardhini | Picrorhiza kurroa | Rhizome | CID_2214 | 10013 | HDAC6 | ChEMBL |
| Arogyavardhini | Picrorhiza kurroa | Rhizome | CID_2214 | 10951 | CBX1 | NPASS |
| Arogyavardhini | Picrorhiza kurroa | Rhizome | CID_2214 | 51053 | GMNN | NPASS |
| Arogyavardhini | Picrorhiza kurroa | Rhizome | CID_2214 | 55775 | TDP1 | NPASS |
| Arogyavardhini | Picrorhiza kurroa | Rhizome | CID_8468 | 10013 | HDAC6 | ChEMBL |
| Arogyavardhini | Picrorhiza kurroa | Rhizome | CID_8468 | 10014 | HDAC5 | ChEMBL |
| Arogyavardhini | Picrorhiza kurroa | Rhizome | CID_8468 | 1019 | CDK4 | BindingDB |
| Arogyavardhini | Picrorhiza kurroa | Rhizome | CID_8468 | 1021 | CDK6 | BindingDB |
| Arogyavardhini | Picrorhiza kurroa | Rhizome | CID_8468 | 10461 | MERTK | BindingDB |
| Arogyavardhini | Picrorhiza kurroa | Rhizome | CID_8468 | 11201 | POLI | NPASS |
| Arogyavardhini | Picrorhiza kurroa | Rhizome | CID_8468 | 118429 | ANTXR2 | NPASS |
| Arogyavardhini | Picrorhiza kurroa | Rhizome | CID_8468 | 1956 | EGFR | BindingDB |
| Arogyavardhini | Picrorhiza kurroa | Rhizome | CID_8468 | 1969 | EPHA2 | BindingDB |
| Arogyavardhini | Picrorhiza kurroa | Rhizome | CID_8468 | 2066 | ERBB4 | BindingDB |
| Arogyavardhini | Picrorhiza kurroa | Rhizome | CID_8468 | 2099 | ESR1 | ChEMBL \| NPASS |
| Arogyavardhini | Picrorhiza kurroa | Rhizome | CID_8468 | 2260 | FGFR1 | BindingDB |
| Arogyavardhini | Picrorhiza kurroa | Rhizome | CID_8468 | 2261 | FGFR3 | BindingDB |
| Arogyavardhini | Picrorhiza kurroa | Rhizome | CID_8468 | 2263 | FGFR2 | BindingDB |
| Arogyavardhini | Picrorhiza kurroa | Rhizome | CID_8468 | 231 | AKR1B1 | ChEMBL \| NPASS |
| Arogyavardhini | Picrorhiza kurroa | Rhizome | CID_8468 | 2321 | FLT1 | BindingDB |
| Arogyavardhini | Picrorhiza kurroa | Rhizome | CID_8468 | 238 | ALK | BindingDB |
| Arogyavardhini | Picrorhiza kurroa | Rhizome | CID_8468 | 25 | ABL1 | BindingDB |
| Arogyavardhini | Picrorhiza kurroa | Rhizome | CID_8468 | 3064 | HTT | NPASS |
| Arogyavardhini | Picrorhiza kurroa | Rhizome | CID_8468 | 3065 | HDAC1 | ChEMBL |
| Arogyavardhini | Picrorhiza kurroa | Rhizome | CID_8468 | 3066 | HDAC2 | ChEMBL |
| Arogyavardhini | Picrorhiza kurroa | Rhizome | CID_8468 | 3480 | IGF1R | BindingDB |
| Arogyavardhini | Picrorhiza kurroa | Rhizome | CID_8468 | 3643 | INSR | BindingDB |

**Table S3: Associations between formulations, constituent plants, phytochemicals, and human molecular targets. The table lists the formulations, plant name and plant part, phytochemical identifiers, associated human target genes (Entrez Gene ID and gene symbol), and the database sources supporting each phytochemical-target interaction.**

| Formulation | Plant name | Part | PubChem CID | Entrez Gene ID | Gene Symbol | Source database(s) |
|---|---|---|---|---|---|---|
| Arogyavardhini | Picrorhiza kurroa | Rhizome | CID_8468 | 3791 | KDR | BindingDB |
| Arogyavardhini | Picrorhiza kurroa | Rhizome | CID_8468 | 4058 | LTK | BindingDB |
| Arogyavardhini | Picrorhiza kurroa | Rhizome | CID_8468 | 4233 | MET | BindingDB |
| Arogyavardhini | Picrorhiza kurroa | Rhizome | CID_8468 | 4921 | DDR2 | BindingDB |
| Arogyavardhini | Picrorhiza kurroa | Rhizome | CID_8468 | 5156 | PDGFRA | BindingDB |
| Arogyavardhini | Picrorhiza kurroa | Rhizome | CID_8468 | 51564 | HDAC7 | ChEMBL |
| Arogyavardhini | Picrorhiza kurroa | Rhizome | CID_8468 | 5159 | PDGFRB | BindingDB |
| Arogyavardhini | Picrorhiza kurroa | Rhizome | CID_8468 | 55869 | HDAC8 | ChEMBL |
| Arogyavardhini | Picrorhiza kurroa | Rhizome | CID_8468 | 5594 | MAPK1 | BindingDB |
| Arogyavardhini | Picrorhiza kurroa | Rhizome | CID_8468 | 5979 | RET | BindingDB |
| Arogyavardhini | Picrorhiza kurroa | Rhizome | CID_8468 | 7172 | TPMT | NPASS |
| Arogyavardhini | Picrorhiza kurroa | Rhizome | CID_8468 | 7301 | TYRO3 | BindingDB |
| Arogyavardhini | Picrorhiza kurroa | Rhizome | CID_8468 | 79885 | HDAC11 | ChEMBL |
| Arogyavardhini | Picrorhiza kurroa | Rhizome | CID_8468 | 83933 | HDAC10 | ChEMBL \| NPASS |
| Arogyavardhini | Picrorhiza kurroa | Rhizome | CID_8468 | 8841 | HDAC3 | ChEMBL |
| Arogyavardhini | Picrorhiza kurroa | Rhizome | CID_8468 | 9734 | HDAC9 | ChEMBL |
| Arogyavardhini | Picrorhiza kurroa | Rhizome | CID_8468 | 9759 | HDAC4 | ChEMBL |
| Arogyavardhini | Picrorhiza kurroa | Rhizome | CID_8468 | 983 | CDK1 | BindingDB |
| Arogyavardhini | Picrorhiza kurroa | Rhizome | CID_91458 | 2099 | ESR1 | ChEMBL \| NPASS |
| Arogyavardhini | Picrorhiza kurroa | Rhizome | CID_91458 | 26013 | L3MBTL1 | NPASS |
| Arogyavardhini | Picrorhiza kurroa | Rhizome | CID_91458 | 51053 | GMNN | NPASS |
| Arogyavardhini | Picrorhiza kurroa | Rhizome | CID_91458 | 5742 | PTGS1 | ChEMBL \| NPASS |
| Arogyavardhini | Picrorhiza kurroa | Rhizome | CID_91458 | 5743 | PTGS2 | ChEMBL \| NPASS |
| Arogyavardhini | Picrorhiza kurroa | Rhizome | CID_91520 | 2099 | ESR1 | ChEMBL \| NPASS |
| Arogyavardhini | Picrorhiza kurroa | Rhizome | CID_91520 | 2648 | KAT2A | NPASS |
| Arogyavardhini | Picrorhiza kurroa | Rhizome | CID_91520 | 5742 | PTGS1 | ChEMBL \| NPASS |
| Arogyavardhini | Picrorhiza kurroa | Rhizome | CID_91520 | 5743 | PTGS2 | ChEMBL \| NPASS |
| Arogyavardhini | Picrorhiza kurroa | Rhizome | CID_92123 | 10951 | CBX1 | NPASS |
| Arogyavardhini | Picrorhiza kurroa | Rhizome | CID_92123 | 2139 | EYA2 | NPASS |
| Arogyavardhini | Picrorhiza kurroa | Rhizome | CID_92123 | 51053 | GMNN | NPASS |
| Arogyavardhini | Plumbago zeylanica | Root | CID_10205 | 1543 | CYP1A1 | ChEMBL \| NPASS |
| Arogyavardhini | Plumbago zeylanica | Root | CID_10205 | 1544 | CYP1A2 | ChEMBL \| NPASS |
| Arogyavardhini | Plumbago zeylanica | Root | CID_10205 | 1545 | CYP1B1 | ChEMBL \| NPASS |
| Arogyavardhini | Plumbago zeylanica | Root | CID_10205 | 1557 | CYP2C19 | ChEMBL \| NPASS |
| Arogyavardhini | Plumbago zeylanica | Root | CID_10205 | 1559 | CYP2C9 | ChEMBL \| NPASS |
| Arogyavardhini | Plumbago zeylanica | Root | CID_10205 | 1565 | CYP2D6 | ChEMBL \| NPASS |
| Arogyavardhini | Plumbago zeylanica | Root | CID_10205 | 1576 | CYP3A4 | ChEMBL \| NPASS |
| Arogyavardhini | Plumbago zeylanica | Root | CID_10205 | 1956 | EGFR | ChEMBL |
| Arogyavardhini | Plumbago zeylanica | Root | CID_10205 | 2033 | EP300 | BindingDB \| ChEMBL \| NPASS |
| Arogyavardhini | Plumbago zeylanica | Root | CID_10205 | 2064 | ERBB2 | ChEMBL \| NPASS |
| Arogyavardhini | Plumbago zeylanica | Root | CID_10205 | 207 | AKT1 | ChEMBL |
| Arogyavardhini | Plumbago zeylanica | Root | CID_10205 | 216 | ALDH1A1 | ChEMBL \| NPASS |
| Arogyavardhini | Plumbago zeylanica | Root | CID_10205 | 246 | ALOX15 | NPASS |
| Arogyavardhini | Plumbago zeylanica | Root | CID_10205 | 2648 | KAT2A | NPASS |
| Arogyavardhini | Plumbago zeylanica | Root | CID_10205 | 3028 | HSD17B10 | ChEMBL \| NPASS |
| Arogyavardhini | Plumbago zeylanica | Root | CID_10205 | 3248 | HPGD | ChEMBL \| NPASS |
| Arogyavardhini | Plumbago zeylanica | Root | CID_10205 | 328 | APEX1 | NPASS |
| Arogyavardhini | Plumbago zeylanica | Root | CID_10205 | 3309 | HSPA5 | NPASS |
| Arogyavardhini | Plumbago zeylanica | Root | CID_10205 | 3329 | HSPD1 | ChEMBL \| NPASS |
| Arogyavardhini | Plumbago zeylanica | Root | CID_10205 | 3336 | HSPE1 | ChEMBL \| NPASS |
| Arogyavardhini | Plumbago zeylanica | Root | CID_10205 | 390245 | KDM4E | ChEMBL \| NPASS |
| Arogyavardhini | Plumbago zeylanica | Root | CID_10205 | 4137 | MAPT | ChEMBL \| NPASS |
| Arogyavardhini | Plumbago zeylanica | Root | CID_10205 | 4297 | KMT2A | NPASS |
| Arogyavardhini | Plumbago zeylanica | Root | CID_10205 | 51426 | POLK | ChEMBL \| NPASS |
| Arogyavardhini | Plumbago zeylanica | Root | CID_10205 | 55775 | TDP1 | ChEMBL \| NPASS |
| Arogyavardhini | Plumbago zeylanica | Root | CID_10205 | 5594 | MAPK1 | ChEMBL \| NPASS |
| Arogyavardhini | Plumbago zeylanica | Root | CID_10205 | 5595 | MAPK3 | ChEMBL |
| Arogyavardhini | Plumbago zeylanica | Root | CID_10205 | 5604 | MAP2K1 | NPASS |
| Arogyavardhini | Plumbago zeylanica | Root | CID_10205 | 5609 | MAP2K7 | ChEMBL \| NPASS |
| Arogyavardhini | Plumbago zeylanica | Root | CID_10205 | 5747 | PTK2 | ChEMBL |
| Arogyavardhini | Plumbago zeylanica | Root | CID_10205 | 5894 | RAF1 | NPASS |
| Arogyavardhini | Plumbago zeylanica | Root | CID_10205 | 5965 | RECQL | NPASS |
| Arogyavardhini | Plumbago zeylanica | Root | CID_10205 | 5970 | RELA | ChEMBL |
| Arogyavardhini | Plumbago zeylanica | Root | CID_10205 | 641 | BLM | NPASS |
| Arogyavardhini | Plumbago zeylanica | Root | CID_10205 | 6606 | SMN1 | NPASS |
| Arogyavardhini | Plumbago zeylanica | Root | CID_10205 | 6774 | STAT3 | ChEMBL \| NPASS |
| Arogyavardhini | Plumbago zeylanica | Root | CID_10205 | 7150 | TOP1 | ChEMBL \| NPASS |
| Arogyavardhini | Plumbago zeylanica | Root | CID_10205 | 7157 | TP53 | ChEMBL \| NPASS |
| Arogyavardhini | Plumbago zeylanica | Root | CID_10205 | 7263 | TST | ChEMBL \| NPASS |
| Arogyavardhini | Plumbago zeylanica | Root | CID_10205 | 7398 | USP1 | ChEMBL \| NPASS |

**Table S3: Associations between formulations, constituent plants, phytochemicals, and human molecular targets. The table lists the formulations, plant name and plant part, phytochemical identifiers, associated human target genes (Entrez Gene ID and gene symbol), and the database sources supporting each phytochemical-target interaction.**

| Formulation | Plant name | Part | PubChem CID | Entrez Gene ID | Gene Symbol | Source database(s) |
|---|---|---|---|---|---|---|
| Arogyavardhini | Plumbago zeylanica | Root | CID_10205 | 7421 | VDR | ChEMBL \| NPASS |
| Arogyavardhini | Plumbago zeylanica | Root | CID_10205 | 7514 | XPO1 | ChEMBL |
| Arogyavardhini | Plumbago zeylanica | Root | CID_10205 | 834 | CASP1 | NPASS |
| Arogyavardhini | Plumbago zeylanica | Root | CID_10205 | 840 | CASP7 | NPASS |
| Arogyavardhini | Plumbago zeylanica | Root | CID_10205 | 8850 | KAT2B | BindingDB \| ChEMBL \| NPASS |
| Arogyavardhini | Plumbago zeylanica | Root | CID_10205 | 8989 | TRPA1 | ChEMBL \| NPASS |
| Arogyavardhini | Plumbago zeylanica | Root | CID_10205 | 9020 | MAP3K14 | ChEMBL \| NPASS |
| Arogyavardhini | Plumbago zeylanica | Root | CID_10205 | 993 | CDC25A | BindingDB \| ChEMBL \| NPASS |
| Arogyavardhini | Plumbago zeylanica | Root | CID_10205 | 994 | CDC25B | BindingDB \| ChEMBL \| NPASS |
| Arogyavardhini | Plumbago zeylanica | Root | CID_271296 | 1981 | EIF4G1 | BindingDB |
| Arogyavardhini | Plumbago zeylanica | Root | CID_271296 | 216 | ALDH1A1 | ChEMBL |
| Arogyavardhini | Plumbago zeylanica | Root | CID_271296 | 26013 | L3MBTL1 | ChEMBL |
| Arogyavardhini | Plumbago zeylanica | Root | CID_271296 | 2648 | KAT2A | ChEMBL |
| Arogyavardhini | Plumbago zeylanica | Root | CID_271296 | 2944 | GSTM1 | BindingDB |
| Arogyavardhini | Plumbago zeylanica | Root | CID_271296 | 29994 | BAZ2B | ChEMBL |
| Arogyavardhini | Plumbago zeylanica | Root | CID_271296 | 328 | APEX1 | ChEMBL |
| Arogyavardhini | Plumbago zeylanica | Root | CID_271296 | 4297 | KMT2A | ChEMBL |
| Arogyavardhini | Plumbago zeylanica | Root | CID_271296 | 5423 | POLB | ChEMBL |
| Arogyavardhini | Plumbago zeylanica | Root | CID_271296 | 5965 | RECQL | ChEMBL |
| Arogyavardhini | Plumbago zeylanica | Root | CID_271296 | 597 | BCL2A1 | BindingDB |
| Arogyavardhini | Plumbago zeylanica | Root | CID_271296 | 641 | BLM | ChEMBL |
| Arogyavardhini | Plumbago zeylanica | Root | CID_271296 | 7068 | THRB | ChEMBL |
| Arogyavardhini | Plumbago zeylanica | Root | CID_271296 | 7296 | TXNRD1 | ChEMBL |
| Arogyavardhini | Plumbago zeylanica | Root | CID_5793 | 231 | AKR1B1 | ChEMBL \| NPASS |
| Arogyavardhini | Plumbago zeylanica | Root | CID_5793 | 49 | ACR | ChEMBL \| NPASS |
| Arogyavardhini | Plumbago zeylanica | Root | CID_5793 | 51426 | POLK | NPASS |
| Arogyavardhini | Plumbago zeylanica | Root | CID_5793 | 5834 | PYGB | BindingDB \| ChEMBL \| NPASS |
| Arogyavardhini | Plumbago zeylanica | Root | CID_5793 | 5837 | PYGM | BindingDB \| ChEMBL \| NPASS |
| Arogyavardhini | Plumbago zeylanica | Root | CID_5793 | 6513 | SLC2A1 | ChEMBL |
| Arogyavardhini | Plumbago zeylanica | Root | CID_5793 | 84750 | POFUT3 | ChEMBL \| NPASS |
| Arogyavardhini | Plumbago zeylanica | Root | CID_5793 | 84929 | FIBCD1 | ChEMBL \| NPASS |
| Arogyavardhini | Terminalia bellirica | Fruit | CID_124021 | 231 | AKR1B1 | ChEMBL \| NPASS |
| Arogyavardhini | Terminalia bellirica | Fruit | CID_124021 | 23683 | PRKD3 | ChEMBL |
| Arogyavardhini | Terminalia bellirica | Fruit | CID_124021 | 5578 | PRKCA | BindingDB \| ChEMBL |
| Arogyavardhini | Terminalia bellirica | Fruit | CID_124021 | 5579 | PRKCB | ChEMBL |
| Arogyavardhini | Terminalia bellirica | Fruit | CID_124021 | 5580 | PRKCD | ChEMBL |
| Arogyavardhini | Terminalia bellirica | Fruit | CID_124021 | 5581 | PRKCE | ChEMBL |
| Arogyavardhini | Terminalia bellirica | Fruit | CID_124021 | 5582 | PRKCG | ChEMBL |
| Arogyavardhini | Terminalia bellirica | Fruit | CID_124021 | 5583 | PRKCH | ChEMBL |
| Arogyavardhini | Terminalia bellirica | Fruit | CID_124021 | 5584 | PRKCI | ChEMBL |
| Arogyavardhini | Terminalia bellirica | Fruit | CID_124021 | 5587 | PRKD1 | ChEMBL |
| Arogyavardhini | Terminalia bellirica | Fruit | CID_124021 | 5588 | PRKCQ | ChEMBL \| NPASS |
| Arogyavardhini | Terminalia bellirica | Fruit | CID_124021 | 5590 | PRKCZ | ChEMBL |
| Arogyavardhini | Terminalia bellirica | Fruit | CID_13250 | 2099 | ESR1 | ChEMBL |
| Arogyavardhini | Terminalia bellirica | Fruit | CID_13250 | 22943 | DKK1 | ChEMBL |
| Arogyavardhini | Terminalia bellirica | Fruit | CID_13250 | 4040 | LRP6 | ChEMBL |
| Arogyavardhini | Terminalia bellirica | Fruit | CID_13250 | 2099 | ESR1 | ChEMBL |
| Arogyavardhini | Terminalia bellirica | Fruit | CID_13250 | 22943 | DKK1 | ChEMBL |
| Arogyavardhini | Terminalia bellirica | Fruit | CID_13250 | 4040 | LRP6 | ChEMBL |
| Arogyavardhini | Terminalia bellirica | Fruit | CID_222284 | 10013 | HDAC6 | ChEMBL |
| Arogyavardhini | Terminalia bellirica | Fruit | CID_222284 | 10599 | SLCO1B1 | ChEMBL \| NPASS |
| Arogyavardhini | Terminalia bellirica | Fruit | CID_222284 | 1565 | CYP2D6 | ChEMBL \| NPASS |
| Arogyavardhini | Terminalia bellirica | Fruit | CID_222284 | 1576 | CYP3A4 | BindingDB \| ChEMBL \| NPASS |
| Arogyavardhini | Terminalia bellirica | Fruit | CID_222284 | 1803 | DPP4 | ChEMBL \| NPASS |
| Arogyavardhini | Terminalia bellirica | Fruit | CID_222284 | 2147 | F2 | ChEMBL \| NPASS |
| Arogyavardhini | Terminalia bellirica | Fruit | CID_222284 | 28234 | SLCO1B3 | ChEMBL \| NPASS |
| Arogyavardhini | Terminalia bellirica | Fruit | CID_222284 | 2932 | GSK3B | ChEMBL |
| Arogyavardhini | Terminalia bellirica | Fruit | CID_222284 | 3417 | IDH1 | ChEMBL \| NPASS |
| Arogyavardhini | Terminalia bellirica | Fruit | CID_222284 | 51422 | PRKAG2 | ChEMBL |
| Arogyavardhini | Terminalia bellirica | Fruit | CID_222284 | 5243 | ABCB1 | ChEMBL |
| Arogyavardhini | Terminalia bellirica | Fruit | CID_222284 | 53632 | PRKAG3 | ChEMBL |
| Arogyavardhini | Terminalia bellirica | Fruit | CID_222284 | 5423 | POLB | BindingDB \| ChEMBL \| NPASS |
| Arogyavardhini | Terminalia bellirica | Fruit | CID_222284 | 5562 | PRKAA1 | ChEMBL |
| Arogyavardhini | Terminalia bellirica | Fruit | CID_222284 | 5563 | PRKAA2 | ChEMBL |
| Arogyavardhini | Terminalia bellirica | Fruit | CID_222284 | 5564 | PRKAB1 | ChEMBL |
| Arogyavardhini | Terminalia bellirica | Fruit | CID_222284 | 5565 | PRKAB2 | ChEMBL |
| Arogyavardhini | Terminalia bellirica | Fruit | CID_222284 | 5571 | PRKAG1 | ChEMBL |
| Arogyavardhini | Terminalia bellirica | Fruit | CID_222284 | 7299 | TYR | ChEMBL \| NPASS |
| Arogyavardhini | Terminalia bellirica | Fruit | CID_25310 | 10013 | HDAC6 | ChEMBL |
| Arogyavardhini | Terminalia bellirica | Fruit | CID_370 | 10013 | HDAC6 | ChEMBL |

**Table S3: Associations between formulations, constituent plants, phytochemicals, and human molecular targets. The table lists the formulations, plant name and plant part, phytochemical identifiers, associated human target genes (Entrez Gene ID and gene symbol), and the database sources supporting each phytochemical-target interaction.**

| Formulation | Plant name | Part | PubChem CID | Entrez Gene ID | Gene Symbol | Source database(s) |
|---|---|---|---|---|---|---|
| Arogyavardhini | Terminalia bellirica | Fruit | CID_370 | 10599 | SLCO1B1 | ChEMBL \| NPASS |
| Arogyavardhini | Terminalia bellirica | Fruit | CID_370 | 10919 | EHMT2 | ChEMBL \| NPASS |
| Arogyavardhini | Terminalia bellirica | Fruit | CID_370 | 11238 | CA5B | BindingDB \| ChEMBL \| NPASS |
| Arogyavardhini | Terminalia bellirica | Fruit | CID_370 | 118429 | ANTXR2 | NPASS |
| Arogyavardhini | Terminalia bellirica | Fruit | CID_370 | 1588 | CYP19A1 | ChEMBL \| NPASS |
| Arogyavardhini | Terminalia bellirica | Fruit | CID_370 | 1636 | ACE | BindingDB \| NPASS |
| Arogyavardhini | Terminalia bellirica | Fruit | CID_370 | 1803 | DPP4 | BindingDB \| ChEMBL \| NPASS |
| Arogyavardhini | Terminalia bellirica | Fruit | CID_370 | 2099 | ESR1 | ChEMBL \| NPASS |
| Arogyavardhini | Terminalia bellirica | Fruit | CID_370 | 216 | ALDH1A1 | ChEMBL \| NPASS |
| Arogyavardhini | Terminalia bellirica | Fruit | CID_370 | 231 | AKR1B1 | NPASS |
| Arogyavardhini | Terminalia bellirica | Fruit | CID_370 | 23632 | CA14 | BindingDB \| ChEMBL \| NPASS |
| Arogyavardhini | Terminalia bellirica | Fruit | CID_370 | 2526 | FUT4 | ChEMBL \| NPASS |
| Arogyavardhini | Terminalia bellirica | Fruit | CID_370 | 2529 | FUT7 | ChEMBL \| NPASS |
| Arogyavardhini | Terminalia bellirica | Fruit | CID_370 | 2590 | GALNT2 | ChEMBL \| NPASS |
| Arogyavardhini | Terminalia bellirica | Fruit | CID_370 | 259285 | TAS2R39 | ChEMBL \| NPASS |
| Arogyavardhini | Terminalia bellirica | Fruit | CID_370 | 2597 | GAPDH | ChEMBL \| NPASS |
| Arogyavardhini | Terminalia bellirica | Fruit | CID_370 | 2648 | KAT2A | NPASS |
| Arogyavardhini | Terminalia bellirica | Fruit | CID_370 | 2744 | GLS | ChEMBL \| NPASS |
| Arogyavardhini | Terminalia bellirica | Fruit | CID_370 | 28234 | SLCO1B3 | ChEMBL \| NPASS |
| Arogyavardhini | Terminalia bellirica | Fruit | CID_370 | 2908 | NR3C1 | NPASS |
| Arogyavardhini | Terminalia bellirica | Fruit | CID_370 | 3028 | HSD17B10 | ChEMBL \| NPASS |
| Arogyavardhini | Terminalia bellirica | Fruit | CID_370 | 3248 | HPGD | ChEMBL \| NPASS |
| Arogyavardhini | Terminalia bellirica | Fruit | CID_370 | 328 | APEX1 | NPASS |
| Arogyavardhini | Terminalia bellirica | Fruit | CID_370 | 3315 | HSPB1 | NPASS |
| Arogyavardhini | Terminalia bellirica | Fruit | CID_370 | 3417 | IDH1 | ChEMBL \| NPASS |
| Arogyavardhini | Terminalia bellirica | Fruit | CID_370 | 351 | APP | BindingDB \| ChEMBL |
| Arogyavardhini | Terminalia bellirica | Fruit | CID_370 | 3576 | CXCL8 | NPASS |
| Arogyavardhini | Terminalia bellirica | Fruit | CID_370 | 367 | AR | NPASS |
| Arogyavardhini | Terminalia bellirica | Fruit | CID_370 | 3725 | JUN | NPASS |
| Arogyavardhini | Terminalia bellirica | Fruit | CID_370 | 390245 | KDM4E | ChEMBL \| NPASS |
| Arogyavardhini | Terminalia bellirica | Fruit | CID_370 | 4000 | LMNA | NPASS |
| Arogyavardhini | Terminalia bellirica | Fruit | CID_370 | 4137 | MAPT | ChEMBL \| NPASS |
| Arogyavardhini | Terminalia bellirica | Fruit | CID_370 | 4316 | MMP7 | BindingDB |
| Arogyavardhini | Terminalia bellirica | Fruit | CID_370 | 4780 | NFE2L2 | ChEMBL \| NPASS |
| Arogyavardhini | Terminalia bellirica | Fruit | CID_370 | 5054 | SERPINE1 | BindingDB |
| Arogyavardhini | Terminalia bellirica | Fruit | CID_370 | 51426 | POLK | ChEMBL \| NPASS |
| Arogyavardhini | Terminalia bellirica | Fruit | CID_370 | 54658 | UGT1A1 | ChEMBL \| NPASS |
| Arogyavardhini | Terminalia bellirica | Fruit | CID_370 | 55775 | TDP1 | ChEMBL \| NPASS |
| Arogyavardhini | Terminalia bellirica | Fruit | CID_370 | 5594 | MAPK1 | NPASS |
| Arogyavardhini | Terminalia bellirica | Fruit | CID_370 | 5693 | PSMB5 | BindingDB \| NPASS |
| Arogyavardhini | Terminalia bellirica | Fruit | CID_370 | 5743 | PTGS2 | ChEMBL \| NPASS |
| Arogyavardhini | Terminalia bellirica | Fruit | CID_370 | 6401 | SELE | BindingDB \| NPASS |
| Arogyavardhini | Terminalia bellirica | Fruit | CID_370 | 6402 | SELL | BindingDB \| ChEMBL \| NPASS |
| Arogyavardhini | Terminalia bellirica | Fruit | CID_370 | 6403 | SELP | BindingDB \| ChEMBL \| NPASS |
| Arogyavardhini | Terminalia bellirica | Fruit | CID_370 | 6606 | SMN1 | NPASS |
| Arogyavardhini | Terminalia bellirica | Fruit | CID_370 | 6622 | SNCA | ChEMBL \| NPASS |
| Arogyavardhini | Terminalia bellirica | Fruit | CID_370 | 6868 | ADAM17 | ChEMBL \| NPASS |
| Arogyavardhini | Terminalia bellirica | Fruit | CID_370 | 7157 | TP53 | ChEMBL \| NPASS |
| Arogyavardhini | Terminalia bellirica | Fruit | CID_370 | 7421 | VDR | ChEMBL \| NPASS |
| Arogyavardhini | Terminalia bellirica | Fruit | CID_370 | 759 | CA1 | BindingDB \| ChEMBL \| NPASS |
| Arogyavardhini | Terminalia bellirica | Fruit | CID_370 | 760 | CA2 | BindingDB \| ChEMBL \| NPASS |
| Arogyavardhini | Terminalia bellirica | Fruit | CID_370 | 761 | CA3 | BindingDB \| ChEMBL \| NPASS |
| Arogyavardhini | Terminalia bellirica | Fruit | CID_370 | 762 | CA4 | BindingDB \| ChEMBL \| NPASS |
| Arogyavardhini | Terminalia bellirica | Fruit | CID_370 | 763 | CA5A | BindingDB \| ChEMBL \| NPASS |
| Arogyavardhini | Terminalia bellirica | Fruit | CID_370 | 765 | CA6 | ChEMBL \| NPASS |
| Arogyavardhini | Terminalia bellirica | Fruit | CID_370 | 766 | CA7 | BindingDB \| ChEMBL \| NPASS |
| Arogyavardhini | Terminalia bellirica | Fruit | CID_370 | 768 | CA9 | BindingDB \| ChEMBL \| NPASS |
| Arogyavardhini | Terminalia bellirica | Fruit | CID_370 | 771 | CA12 | BindingDB \| ChEMBL \| NPASS |
| Arogyavardhini | Terminalia bellirica | Fruit | CID_5281855 | 10013 | HDAC6 | ChEMBL |
| Arogyavardhini | Terminalia bellirica | Fruit | CID_5281855 | 10073 | SNUPN | NPASS |
| Arogyavardhini | Terminalia bellirica | Fruit | CID_5281855 | 1017 | CDK2 | BindingDB \| ChEMBL |
| Arogyavardhini | Terminalia bellirica | Fruit | CID_5281855 | 1019 | CDK4 | BindingDB \| ChEMBL |
| Arogyavardhini | Terminalia bellirica | Fruit | CID_5281855 | 10411 | RAPGEF3 | NPASS |
| Arogyavardhini | Terminalia bellirica | Fruit | CID_5281855 | 10498 | CARM1 | ChEMBL \| NPASS |
| Arogyavardhini | Terminalia bellirica | Fruit | CID_5281855 | 10587 | TXNRD2 | ChEMBL |
| Arogyavardhini | Terminalia bellirica | Fruit | CID_5281855 | 10599 | SLCO1B1 | ChEMBL \| NPASS |
| Arogyavardhini | Terminalia bellirica | Fruit | CID_5281855 | 10733 | PLK4 | BindingDB \| ChEMBL \| NPASS |
| Arogyavardhini | Terminalia bellirica | Fruit | CID_5281855 | 10919 | EHMT2 | ChEMBL \| NPASS |
| Arogyavardhini | Terminalia bellirica | Fruit | CID_5281855 | 10951 | CBX1 | NPASS |
| Arogyavardhini | Terminalia bellirica | Fruit | CID_5281855 | 11069 | RAPGEF4 | NPASS |

**Table S3: Associations between formulations, constituent plants, phytochemicals, and human molecular targets. The table lists the formulations, plant name and plant part, phytochemical identifiers, associated human target genes (Entrez Gene ID and gene symbol), and the database sources supporting each phytochemical-target interaction.**

| Formulation | Plant name | Part | PubChem CID | Entrez Gene ID | Gene Symbol | Source database(s) |
|---|---|---|---|---|---|---|
| Arogyavardhini | Terminalia bellirica | Fruit | CID_5281855 | 11201 | POLI | ChEMBL \| NPASS |
| Arogyavardhini | Terminalia bellirica | Fruit | CID_5281855 | 11238 | CA5B | BindingDB \| ChEMBL \| NPASS |
| Arogyavardhini | Terminalia bellirica | Fruit | CID_5281855 | 114112 | TXNRD3 | ChEMBL \| NPASS |
| Arogyavardhini | Terminalia bellirica | Fruit | CID_5281855 | 1326 | MAP3K8 | BindingDB \| ChEMBL \| NPASS |
| Arogyavardhini | Terminalia bellirica | Fruit | CID_5281855 | 1457 | CSNK2A1 | BindingDB \| ChEMBL \| NPASS |
| Arogyavardhini | Terminalia bellirica | Fruit | CID_5281855 | 1459 | CSNK2A2 | ChEMBL \| NPASS |
| Arogyavardhini | Terminalia bellirica | Fruit | CID_5281855 | 1460 | CSNK2B | ChEMBL |
| Arogyavardhini | Terminalia bellirica | Fruit | CID_5281855 | 1557 | CYP2C19 | ChEMBL \| NPASS |
| Arogyavardhini | Terminalia bellirica | Fruit | CID_5281855 | 1559 | CYP2C9 | NPASS |
| Arogyavardhini | Terminalia bellirica | Fruit | CID_5281855 | 1636 | ACE | BindingDB \| NPASS |
| Arogyavardhini | Terminalia bellirica | Fruit | CID_5281855 | 1845 | DUSP3 | BindingDB \| NPASS |
| Arogyavardhini | Terminalia bellirica | Fruit | CID_5281855 | 1859 | DYRK1A | BindingDB \| ChEMBL \| NPASS |
| Arogyavardhini | Terminalia bellirica | Fruit | CID_5281855 | 1956 | EGFR | BindingDB \| ChEMBL \| NPASS |
| Arogyavardhini | Terminalia bellirica | Fruit | CID_5281855 | 1981 | EIF4G1 | BindingDB |
| Arogyavardhini | Terminalia bellirica | Fruit | CID_5281855 | 1995 | ELAVL3 | ChEMBL \| NPASS |
| Arogyavardhini | Terminalia bellirica | Fruit | CID_5281855 | 2050 | EPHB4 | BindingDB \| ChEMBL \| NPASS |
| Arogyavardhini | Terminalia bellirica | Fruit | CID_5281855 | 2053 | EPHX2 | ChEMBL \| NPASS |
| Arogyavardhini | Terminalia bellirica | Fruit | CID_5281855 | 2064 | ERBB2 | ChEMBL \| NPASS |
| Arogyavardhini | Terminalia bellirica | Fruit | CID_5281855 | 207 | AKT1 | ChEMBL \| NPASS |
| Arogyavardhini | Terminalia bellirica | Fruit | CID_5281855 | 2237 | FEN1 | ChEMBL \| NPASS |
| Arogyavardhini | Terminalia bellirica | Fruit | CID_5281855 | 231 | AKR1B1 | ChEMBL \| NPASS |
| Arogyavardhini | Terminalia bellirica | Fruit | CID_5281855 | 23192 | ATG4B | BindingDB \| ChEMBL \| NPASS |
| Arogyavardhini | Terminalia bellirica | Fruit | CID_5281855 | 2322 | FLT3 | ChEMBL \| NPASS |
| Arogyavardhini | Terminalia bellirica | Fruit | CID_5281855 | 2324 | FLT4 | BindingDB \| ChEMBL \| NPASS |
| Arogyavardhini | Terminalia bellirica | Fruit | CID_5281855 | 23405 | DICER1 | ChEMBL |
| Arogyavardhini | Terminalia bellirica | Fruit | CID_5281855 | 23621 | BACE1 | BindingDB \| ChEMBL \| NPASS |
| Arogyavardhini | Terminalia bellirica | Fruit | CID_5281855 | 23632 | CA14 | BindingDB \| ChEMBL \| NPASS |
| Arogyavardhini | Terminalia bellirica | Fruit | CID_5281855 | 238 | ALK | BindingDB \| ChEMBL |
| Arogyavardhini | Terminalia bellirica | Fruit | CID_5281855 | 246 | ALOX15 | NPASS |
| Arogyavardhini | Terminalia bellirica | Fruit | CID_5281855 | 247 | ALOX15B | NPASS |
| Arogyavardhini | Terminalia bellirica | Fruit | CID_5281855 | 2548 | GAA | NPASS |
| Arogyavardhini | Terminalia bellirica | Fruit | CID_5281855 | 2590 | GALNT2 | ChEMBL \| NPASS |
| Arogyavardhini | Terminalia bellirica | Fruit | CID_5281855 | 2597 | GAPDH | NPASS |
| Arogyavardhini | Terminalia bellirica | Fruit | CID_5281855 | 2648 | KAT2A | NPASS |
| Arogyavardhini | Terminalia bellirica | Fruit | CID_5281855 | 2744 | GLS | ChEMBL \| NPASS |
| Arogyavardhini | Terminalia bellirica | Fruit | CID_5281855 | 28234 | SLCO1B3 | ChEMBL \| NPASS |
| Arogyavardhini | Terminalia bellirica | Fruit | CID_5281855 | 2859 | GPR35 | BindingDB \| ChEMBL \| NPASS |
| Arogyavardhini | Terminalia bellirica | Fruit | CID_5281855 | 2932 | GSK3B | BindingDB \| ChEMBL \| NPASS |
| Arogyavardhini | Terminalia bellirica | Fruit | CID_5281855 | 2936 | GSR | ChEMBL \| NPASS |
| Arogyavardhini | Terminalia bellirica | Fruit | CID_5281855 | 2950 | GSTP1 | ChEMBL \| NPASS |
| Arogyavardhini | Terminalia bellirica | Fruit | CID_5281855 | 29994 | BAZ2B | ChEMBL \| NPASS |
| Arogyavardhini | Terminalia bellirica | Fruit | CID_5281855 | 3028 | HSD17B10 | ChEMBL \| NPASS |
| Arogyavardhini | Terminalia bellirica | Fruit | CID_5281855 | 3043 | HBB | ChEMBL \| NPASS |
| Arogyavardhini | Terminalia bellirica | Fruit | CID_5281855 | 3248 | HPGD | ChEMBL \| NPASS |
| Arogyavardhini | Terminalia bellirica | Fruit | CID_5281855 | 328 | APEX1 | ChEMBL \| NPASS |
| Arogyavardhini | Terminalia bellirica | Fruit | CID_5281855 | 3303 | HSPA1A | BindingDB \| ChEMBL \| NPASS |
| Arogyavardhini | Terminalia bellirica | Fruit | CID_5281855 | 3309 | HSPA5 | NPASS |
| Arogyavardhini | Terminalia bellirica | Fruit | CID_5281855 | 3312 | HSPA8 | BindingDB \| ChEMBL \| NPASS |
| Arogyavardhini | Terminalia bellirica | Fruit | CID_5281855 | 3417 | IDH1 | NPASS |
| Arogyavardhini | Terminalia bellirica | Fruit | CID_5281855 | 3480 | IGF1R | BindingDB \| ChEMBL \| NPASS |
| Arogyavardhini | Terminalia bellirica | Fruit | CID_5281855 | 351 | APP | BindingDB \| ChEMBL \| NPASS |
| Arogyavardhini | Terminalia bellirica | Fruit | CID_5281855 | 3551 | IKBKB | BindingDB \| ChEMBL \| NPASS |
| Arogyavardhini | Terminalia bellirica | Fruit | CID_5281855 | 3643 | INSR | BindingDB \| ChEMBL \| NPASS |
| Arogyavardhini | Terminalia bellirica | Fruit | CID_5281855 | 377677 | CA13 | ChEMBL \| NPASS |
| Arogyavardhini | Terminalia bellirica | Fruit | CID_5281855 | 3791 | KDR | BindingDB \| ChEMBL \| NPASS |
| Arogyavardhini | Terminalia bellirica | Fruit | CID_5281855 | 3837 | KPNB1 | NPASS |
| Arogyavardhini | Terminalia bellirica | Fruit | CID_5281855 | 390245 | KDM4E | ChEMBL \| NPASS |
| Arogyavardhini | Terminalia bellirica | Fruit | CID_5281855 | 4000 | LMNA | NPASS |
| Arogyavardhini | Terminalia bellirica | Fruit | CID_5281855 | 4088 | SMAD3 | BindingDB \| ChEMBL \| NPASS |
| Arogyavardhini | Terminalia bellirica | Fruit | CID_5281855 | 4128 | MAOA | ChEMBL \| NPASS |
| Arogyavardhini | Terminalia bellirica | Fruit | CID_5281855 | 4137 | MAPT | ChEMBL \| NPASS |
| Arogyavardhini | Terminalia bellirica | Fruit | CID_5281855 | 4154 | MBNL1 | NPASS |
| Arogyavardhini | Terminalia bellirica | Fruit | CID_5281855 | 4170 | MCL1 | BindingDB \| ChEMBL \| NPASS |
| Arogyavardhini | Terminalia bellirica | Fruit | CID_5281855 | 4233 | MET | BindingDB \| ChEMBL \| NPASS |
| Arogyavardhini | Terminalia bellirica | Fruit | CID_5281855 | 4282 | MIF | BindingDB \| ChEMBL \| NPASS |
| Arogyavardhini | Terminalia bellirica | Fruit | CID_5281855 | 4297 | KMT2A | NPASS |
| Arogyavardhini | Terminalia bellirica | Fruit | CID_5281855 | 43 | ACHE | BindingDB |
| Arogyavardhini | Terminalia bellirica | Fruit | CID_5281855 | 472 | ATM | NPASS |
| Arogyavardhini | Terminalia bellirica | Fruit | CID_5281855 | 4780 | NFE2L2 | NPASS |
| Arogyavardhini | Terminalia bellirica | Fruit | CID_5281855 | 4869 | NPM1 | ChEMBL \| NPASS |

| Table S3: Associations between formulations, constituent plants, phytochemicals, and human molecular targets. The table lists the formulations, plant name and plant part, phytochemical identifiers, associated human target genes (Entrez Gene ID and gene symbol), and the database sources supporting each phytochemical-target interaction. | | | | | | |
|---|---|---|---|---|---|---|
| Formulation | Plant name | Part | PubChem CID | Entrez Gene ID | Gene Symbol | Source database(s) |
| Arogyavardhini | Terminalia bellirica | Fruit | CID_5281855 | 4907 | NT5E | ChEMBL \| NPASS |
| Arogyavardhini | Terminalia bellirica | Fruit | CID_5281855 | 51053 | GMNN | NPASS |
| Arogyavardhini | Terminalia bellirica | Fruit | CID_5281855 | 51426 | POLK | ChEMBL \| NPASS |
| Arogyavardhini | Terminalia bellirica | Fruit | CID_5281855 | 5159 | PDGFRB | BindingDB \| ChEMBL \| NPASS |
| Arogyavardhini | Terminalia bellirica | Fruit | CID_5281855 | 5300 | PIN1 | NPASS |
| Arogyavardhini | Terminalia bellirica | Fruit | CID_5281855 | 5313 | PKLR | ChEMBL |
| Arogyavardhini | Terminalia bellirica | Fruit | CID_5281855 | 5347 | PLK1 | BindingDB \| ChEMBL \| NPASS |
| Arogyavardhini | Terminalia bellirica | Fruit | CID_5281855 | 5423 | POLB | ChEMBL \| NPASS |
| Arogyavardhini | Terminalia bellirica | Fruit | CID_5281855 | 5429 | POLH | BindingDB \| ChEMBL \| NPASS |
| Arogyavardhini | Terminalia bellirica | Fruit | CID_5281855 | 5566 | PRKACA | BindingDB \| ChEMBL |
| Arogyavardhini | Terminalia bellirica | Fruit | CID_5281855 | 5567 | PRKACB | ChEMBL |
| Arogyavardhini | Terminalia bellirica | Fruit | CID_5281855 | 5568 | PRKACG | ChEMBL \| NPASS |
| Arogyavardhini | Terminalia bellirica | Fruit | CID_5281855 | 55775 | TDP1 | ChEMBL \| NPASS |
| Arogyavardhini | Terminalia bellirica | Fruit | CID_5281855 | 55867 | SLC22A11 | ChEMBL \| NPASS |
| Arogyavardhini | Terminalia bellirica | Fruit | CID_5281855 | 5594 | MAPK1 | ChEMBL \| NPASS |
| Arogyavardhini | Terminalia bellirica | Fruit | CID_5281855 | 5747 | PTK2 | BindingDB \| ChEMBL \| NPASS |
| Arogyavardhini | Terminalia bellirica | Fruit | CID_5281855 | 5770 | PTPN1 | BindingDB |
| Arogyavardhini | Terminalia bellirica | Fruit | CID_5281855 | 57787 | MARK4 | ChEMBL |
| Arogyavardhini | Terminalia bellirica | Fruit | CID_5281855 | 595 | CCND1 | ChEMBL \| NPASS |
| Arogyavardhini | Terminalia bellirica | Fruit | CID_5281855 | 5965 | RECQL | ChEMBL \| NPASS |
| Arogyavardhini | Terminalia bellirica | Fruit | CID_5281855 | 5979 | RET | BindingDB \| ChEMBL \| NPASS |
| Arogyavardhini | Terminalia bellirica | Fruit | CID_5281855 | 5999 | RGS4 | NPASS |
| Arogyavardhini | Terminalia bellirica | Fruit | CID_5281855 | 641 | BLM | ChEMBL \| NPASS |
| Arogyavardhini | Terminalia bellirica | Fruit | CID_5281855 | 6714 | SRC | ChEMBL \| NPASS |
| Arogyavardhini | Terminalia bellirica | Fruit | CID_5281855 | 673 | BRAF | BindingDB \| ChEMBL \| NPASS |
| Arogyavardhini | Terminalia bellirica | Fruit | CID_5281855 | 6790 | AURKA | BindingDB \| ChEMBL \| NPASS |
| Arogyavardhini | Terminalia bellirica | Fruit | CID_5281855 | 7010 | TEK | BindingDB \| ChEMBL \| NPASS |
| Arogyavardhini | Terminalia bellirica | Fruit | CID_5281855 | 7068 | THRB | NPASS |
| Arogyavardhini | Terminalia bellirica | Fruit | CID_5281855 | 7155 | TOP2B | ChEMBL \| NPASS |
| Arogyavardhini | Terminalia bellirica | Fruit | CID_5281855 | 7157 | TP53 | NPASS |
| Arogyavardhini | Terminalia bellirica | Fruit | CID_5281855 | 7257 | TSNAX | ChEMBL \| NPASS |
| Arogyavardhini | Terminalia bellirica | Fruit | CID_5281855 | 7296 | TXNRD1 | ChEMBL |
| Arogyavardhini | Terminalia bellirica | Fruit | CID_5281855 | 7398 | USP1 | NPASS |
| Arogyavardhini | Terminalia bellirica | Fruit | CID_5281855 | 7421 | VDR | ChEMBL \| NPASS |
| Arogyavardhini | Terminalia bellirica | Fruit | CID_5281855 | 759 | CA1 | BindingDB \| ChEMBL \| NPASS |
| Arogyavardhini | Terminalia bellirica | Fruit | CID_5281855 | 760 | CA2 | BindingDB \| ChEMBL \| NPASS |
| Arogyavardhini | Terminalia bellirica | Fruit | CID_5281855 | 761 | CA3 | BindingDB \| ChEMBL \| NPASS |
| Arogyavardhini | Terminalia bellirica | Fruit | CID_5281855 | 762 | CA4 | BindingDB \| ChEMBL \| NPASS |
| Arogyavardhini | Terminalia bellirica | Fruit | CID_5281855 | 763 | CA5A | BindingDB \| ChEMBL \| NPASS |
| Arogyavardhini | Terminalia bellirica | Fruit | CID_5281855 | 765 | CA6 | ChEMBL \| NPASS |
| Arogyavardhini | Terminalia bellirica | Fruit | CID_5281855 | 766 | CA7 | BindingDB \| ChEMBL \| NPASS |
| Arogyavardhini | Terminalia bellirica | Fruit | CID_5281855 | 768 | CA9 | BindingDB \| ChEMBL \| NPASS |
| Arogyavardhini | Terminalia bellirica | Fruit | CID_5281855 | 771 | CA12 | BindingDB \| ChEMBL \| NPASS |
| Arogyavardhini | Terminalia bellirica | Fruit | CID_5281855 | 861 | RUNX1 | ChEMBL |
| Arogyavardhini | Terminalia bellirica | Fruit | CID_5281855 | 865 | CBFB | ChEMBL \| NPASS |
| Arogyavardhini | Terminalia bellirica | Fruit | CID_5281855 | 8877 | SPHK1 | ChEMBL |
| Arogyavardhini | Terminalia bellirica | Fruit | CID_5281855 | 890 | CCNA2 | ChEMBL |
| Arogyavardhini | Terminalia bellirica | Fruit | CID_5281855 | 8900 | CCNA1 | ChEMBL \| NPASS |
| Arogyavardhini | Terminalia bellirica | Fruit | CID_5281855 | 9099 | USP2 | ChEMBL \| NPASS |
| Arogyavardhini | Terminalia bellirica | Fruit | CID_5281855 | 9212 | AURKB | BindingDB \| ChEMBL \| NPASS |
| Arogyavardhini | Terminalia bellirica | Fruit | CID_5281855 | 9356 | SLC22A6 | BindingDB \| ChEMBL \| NPASS |
| Arogyavardhini | Terminalia bellirica | Fruit | CID_5281855 | 9682 | KDM4A | ChEMBL \| NPASS |
| Arogyavardhini | Terminalia bellirica | Fruit | CID_5281855 | 9891 | NUAK1 | BindingDB \| ChEMBL \| NPASS |
| Arogyavardhini | Terminalia bellirica | Fruit | CID_5793 | 231 | AKR1B1 | ChEMBL \| NPASS |
| Arogyavardhini | Terminalia bellirica | Fruit | CID_5793 | 49 | ACR | ChEMBL \| NPASS |
| Arogyavardhini | Terminalia bellirica | Fruit | CID_5793 | 51426 | POLK | NPASS |
| Arogyavardhini | Terminalia bellirica | Fruit | CID_5793 | 5834 | PYGB | BindingDB \| ChEMBL \| NPASS |
| Arogyavardhini | Terminalia bellirica | Fruit | CID_5793 | 5837 | PYGM | BindingDB \| ChEMBL \| NPASS |
| Arogyavardhini | Terminalia bellirica | Fruit | CID_5793 | 6513 | SLC2A1 | ChEMBL |
| Arogyavardhini | Terminalia bellirica | Fruit | CID_5793 | 84750 | POFUT3 | ChEMBL \| NPASS |
| Arogyavardhini | Terminalia bellirica | Fruit | CID_5793 | 84929 | FIBCD1 | ChEMBL \| NPASS |
| Arogyavardhini | Terminalia bellirica | Fruit | CID_6036 | 10073 | SNUPN | NPASS |
| Arogyavardhini | Terminalia bellirica | Fruit | CID_6036 | 10951 | CBX1 | NPASS |
| Arogyavardhini | Terminalia bellirica | Fruit | CID_6036 | 2629 | GBA1 | ChEMBL \| NPASS |
| Arogyavardhini | Terminalia bellirica | Fruit | CID_6036 | 2648 | KAT2A | NPASS |
| Arogyavardhini | Terminalia bellirica | Fruit | CID_6036 | 2720 | GLB1 | ChEMBL \| NPASS |
| Arogyavardhini | Terminalia bellirica | Fruit | CID_6036 | 328 | APEX1 | NPASS |
| Arogyavardhini | Terminalia bellirica | Fruit | CID_6036 | 3417 | IDH1 | NPASS |
| Arogyavardhini | Terminalia bellirica | Fruit | CID_6036 | 3837 | KPNB1 | NPASS |
| Arogyavardhini | Terminalia bellirica | Fruit | CID_6036 | 51426 | POLK | NPASS |

| Table S3: Associations between formulations, constituent plants, phytochemicals, and human molecular targets. The table lists the formulations, plant name and plant part, phytochemical identifiers, associated human target genes (Entrez Gene ID and gene symbol), and the database sources supporting each phytochemical-target interaction. | | | | | | |
|---|---|---|---|---|---|---|
| Formulation | Plant name | Part | PubChem CID | Entrez Gene ID | Gene Symbol | Source database(s) |
| Arogyavardhini | Terminalia bellirica | Fruit | CID_6036 | 7398 | USP1 | NPASS |
| Arogyavardhini | Terminalia bellirica | Fruit | CID_6036 | 84750 | POFUT3 | ChEMBL \| NPASS |
| Arogyavardhini | Terminalia bellirica | Fruit | CID_6036 | 84929 | FIBCD1 | ChEMBL \| NPASS |
| Arogyavardhini | Terminalia bellirica | Fruit | CID_6251 | 10280 | SIGMAR1 | ChEMBL |
| Arogyavardhini | Terminalia bellirica | Fruit | CID_6251 | 10800 | CYSLTR1 | ChEMBL |
| Arogyavardhini | Terminalia bellirica | Fruit | CID_6251 | 11255 | HRH3 | ChEMBL |
| Arogyavardhini | Terminalia bellirica | Fruit | CID_6251 | 1128 | CHRM1 | ChEMBL |
| Arogyavardhini | Terminalia bellirica | Fruit | CID_6251 | 1129 | CHRM2 | ChEMBL |
| Arogyavardhini | Terminalia bellirica | Fruit | CID_6251 | 1131 | CHRM3 | ChEMBL |
| Arogyavardhini | Terminalia bellirica | Fruit | CID_6251 | 1132 | CHRM4 | ChEMBL |
| Arogyavardhini | Terminalia bellirica | Fruit | CID_6251 | 1133 | CHRM5 | ChEMBL |
| Arogyavardhini | Terminalia bellirica | Fruit | CID_6251 | 1233 | CCR4 | ChEMBL |
| Arogyavardhini | Terminalia bellirica | Fruit | CID_6251 | 1234 | CCR5 | ChEMBL |
| Arogyavardhini | Terminalia bellirica | Fruit | CID_6251 | 1268 | CNR1 | ChEMBL |
| Arogyavardhini | Terminalia bellirica | Fruit | CID_6251 | 134 | ADORA1 | ChEMBL |
| Arogyavardhini | Terminalia bellirica | Fruit | CID_6251 | 135 | ADORA2A | ChEMBL |
| Arogyavardhini | Terminalia bellirica | Fruit | CID_6251 | 140 | ADORA3 | ChEMBL |
| Arogyavardhini | Terminalia bellirica | Fruit | CID_6251 | 1432 | MAPK14 | ChEMBL |
| Arogyavardhini | Terminalia bellirica | Fruit | CID_6251 | 146 | ADRA1D | ChEMBL |
| Arogyavardhini | Terminalia bellirica | Fruit | CID_6251 | 148 | ADRA1A | ChEMBL |
| Arogyavardhini | Terminalia bellirica | Fruit | CID_6251 | 150 | ADRA2A | ChEMBL |
| Arogyavardhini | Terminalia bellirica | Fruit | CID_6251 | 151 | ADRA2B | ChEMBL |
| Arogyavardhini | Terminalia bellirica | Fruit | CID_6251 | 1511 | CTSG | ChEMBL |
| Arogyavardhini | Terminalia bellirica | Fruit | CID_6251 | 152 | ADRA2C | ChEMBL |
| Arogyavardhini | Terminalia bellirica | Fruit | CID_6251 | 153 | ADRB1 | ChEMBL |
| Arogyavardhini | Terminalia bellirica | Fruit | CID_6251 | 154 | ADRB2 | ChEMBL |
| Arogyavardhini | Terminalia bellirica | Fruit | CID_6251 | 1544 | CYP1A2 | ChEMBL |
| Arogyavardhini | Terminalia bellirica | Fruit | CID_6251 | 1548 | CYP2A6 | ChEMBL |
| Arogyavardhini | Terminalia bellirica | Fruit | CID_6251 | 155 | ADRB3 | ChEMBL |
| Arogyavardhini | Terminalia bellirica | Fruit | CID_6251 | 1557 | CYP2C19 | ChEMBL |
| Arogyavardhini | Terminalia bellirica | Fruit | CID_6251 | 1559 | CYP2C9 | ChEMBL |
| Arogyavardhini | Terminalia bellirica | Fruit | CID_6251 | 1565 | CYP2D6 | ChEMBL |
| Arogyavardhini | Terminalia bellirica | Fruit | CID_6251 | 1571 | CYP2E1 | ChEMBL |
| Arogyavardhini | Terminalia bellirica | Fruit | CID_6251 | 1576 | CYP3A4 | ChEMBL |
| Arogyavardhini | Terminalia bellirica | Fruit | CID_6251 | 1812 | DRD1 | ChEMBL |
| Arogyavardhini | Terminalia bellirica | Fruit | CID_6251 | 1813 | DRD2 | ChEMBL |
| Arogyavardhini | Terminalia bellirica | Fruit | CID_6251 | 1814 | DRD3 | ChEMBL |
| Arogyavardhini | Terminalia bellirica | Fruit | CID_6251 | 1815 | DRD4 | ChEMBL |
| Arogyavardhini | Terminalia bellirica | Fruit | CID_6251 | 186 | AGTR2 | ChEMBL |
| Arogyavardhini | Terminalia bellirica | Fruit | CID_6251 | 1909 | EDNRA | ChEMBL |
| Arogyavardhini | Terminalia bellirica | Fruit | CID_6251 | 1956 | EGFR | ChEMBL |
| Arogyavardhini | Terminalia bellirica | Fruit | CID_6251 | 1991 | ELANE | ChEMBL |
| Arogyavardhini | Terminalia bellirica | Fruit | CID_6251 | 2064 | ERBB2 | ChEMBL |
| Arogyavardhini | Terminalia bellirica | Fruit | CID_6251 | 2099 | ESR1 | ChEMBL \| NPASS |
| Arogyavardhini | Terminalia bellirica | Fruit | CID_6251 | 2100 | ESR2 | ChEMBL |
| Arogyavardhini | Terminalia bellirica | Fruit | CID_6251 | 2147 | F2 | ChEMBL |
| Arogyavardhini | Terminalia bellirica | Fruit | CID_6251 | 2321 | FLT1 | ChEMBL |
| Arogyavardhini | Terminalia bellirica | Fruit | CID_6251 | 23435 | TARDBP | NPASS |
| Arogyavardhini | Terminalia bellirica | Fruit | CID_6251 | 2534 | FYN | ChEMBL |
| Arogyavardhini | Terminalia bellirica | Fruit | CID_6251 | 2554 | GABRA1 | ChEMBL |
| Arogyavardhini | Terminalia bellirica | Fruit | CID_6251 | 2555 | GABRA2 | ChEMBL |
| Arogyavardhini | Terminalia bellirica | Fruit | CID_6251 | 2561 | GABRB2 | ChEMBL |
| Arogyavardhini | Terminalia bellirica | Fruit | CID_6251 | 2566 | GABRG2 | ChEMBL |
| Arogyavardhini | Terminalia bellirica | Fruit | CID_6251 | 26013 | L3MBTL1 | NPASS |
| Arogyavardhini | Terminalia bellirica | Fruit | CID_6251 | 2908 | NR3C1 | ChEMBL \| NPASS |
| Arogyavardhini | Terminalia bellirica | Fruit | CID_6251 | 3156 | HMGCR | ChEMBL |
| Arogyavardhini | Terminalia bellirica | Fruit | CID_6251 | 3269 | HRH1 | ChEMBL |
| Arogyavardhini | Terminalia bellirica | Fruit | CID_6251 | 3274 | HRH2 | ChEMBL |
| Arogyavardhini | Terminalia bellirica | Fruit | CID_6251 | 3350 | HTR1A | ChEMBL |
| Arogyavardhini | Terminalia bellirica | Fruit | CID_6251 | 3356 | HTR2A | ChEMBL |
| Arogyavardhini | Terminalia bellirica | Fruit | CID_6251 | 3357 | HTR2B | ChEMBL |
| Arogyavardhini | Terminalia bellirica | Fruit | CID_6251 | 3358 | HTR2C | ChEMBL |
| Arogyavardhini | Terminalia bellirica | Fruit | CID_6251 | 3362 | HTR6 | ChEMBL |
| Arogyavardhini | Terminalia bellirica | Fruit | CID_6251 | 3577 | CXCR1 | ChEMBL |
| Arogyavardhini | Terminalia bellirica | Fruit | CID_6251 | 3579 | CXCR2 | ChEMBL |
| Arogyavardhini | Terminalia bellirica | Fruit | CID_6251 | 367 | AR | ChEMBL \| NPASS |
| Arogyavardhini | Terminalia bellirica | Fruit | CID_6251 | 3757 | KCNH2 | ChEMBL |
| Arogyavardhini | Terminalia bellirica | Fruit | CID_6251 | 3791 | KDR | ChEMBL |
| Arogyavardhini | Terminalia bellirica | Fruit | CID_6251 | 3932 | LCK | ChEMBL |
| Arogyavardhini | Terminalia bellirica | Fruit | CID_6251 | 4000 | LMNA | NPASS |

| Table S3: Associations between formulations, constituent plants, phytochemicals, and human molecular targets. The table lists the formulations, plant name and plant part, phytochemical identifiers, associated human target genes (Entrez Gene ID and gene symbol), and the database sources supporting each phytochemical-target interaction. | | | | | | |
|---|---|---|---|---|---|---|
| **Formulation** | **Plant name** | **Part** | **PubChem CID** | **Entrez Gene ID** | **Gene Symbol** | **Source database(s)** |
| Arogyavardhini | Terminalia bellirica | Fruit | CID_6251 | 4128 | MAOA | ChEMBL |
| Arogyavardhini | Terminalia bellirica | Fruit | CID_6251 | 4159 | MC3R | ChEMBL |
| Arogyavardhini | Terminalia bellirica | Fruit | CID_6251 | 4160 | MC4R | ChEMBL |
| Arogyavardhini | Terminalia bellirica | Fruit | CID_6251 | 4161 | MC5R | ChEMBL |
| Arogyavardhini | Terminalia bellirica | Fruit | CID_6251 | 43 | ACHE | ChEMBL |
| Arogyavardhini | Terminalia bellirica | Fruit | CID_6251 | 4312 | MMP1 | ChEMBL |
| Arogyavardhini | Terminalia bellirica | Fruit | CID_6251 | 4318 | MMP9 | ChEMBL |
| Arogyavardhini | Terminalia bellirica | Fruit | CID_6251 | 4780 | NFE2L2 | NPASS |
| Arogyavardhini | Terminalia bellirica | Fruit | CID_6251 | 4886 | NPY1R | ChEMBL |
| Arogyavardhini | Terminalia bellirica | Fruit | CID_6251 | 4887 | NPY2R | ChEMBL |
| Arogyavardhini | Terminalia bellirica | Fruit | CID_6251 | 49 | ACR | ChEMBL \| NPASS |
| Arogyavardhini | Terminalia bellirica | Fruit | CID_6251 | 4985 | OPRD1 | ChEMBL |
| Arogyavardhini | Terminalia bellirica | Fruit | CID_6251 | 4986 | OPRK1 | ChEMBL |
| Arogyavardhini | Terminalia bellirica | Fruit | CID_6251 | 4988 | OPRM1 | ChEMBL |
| Arogyavardhini | Terminalia bellirica | Fruit | CID_6251 | 51053 | GMNN | NPASS |
| Arogyavardhini | Terminalia bellirica | Fruit | CID_6251 | 5139 | PDE3A | ChEMBL |
| Arogyavardhini | Terminalia bellirica | Fruit | CID_6251 | 5141 | PDE4A | ChEMBL |
| Arogyavardhini | Terminalia bellirica | Fruit | CID_6251 | 5241 | PGR | ChEMBL |
| Arogyavardhini | Terminalia bellirica | Fruit | CID_6251 | 5243 | ABCB1 | ChEMBL \| NPASS |
| Arogyavardhini | Terminalia bellirica | Fruit | CID_6251 | 552 | AVPR1A | ChEMBL |
| Arogyavardhini | Terminalia bellirica | Fruit | CID_6251 | 5530 | PPP3CA | ChEMBL |
| Arogyavardhini | Terminalia bellirica | Fruit | CID_6251 | 5578 | PRKCA | ChEMBL |
| Arogyavardhini | Terminalia bellirica | Fruit | CID_6251 | 5594 | MAPK1 | ChEMBL |
| Arogyavardhini | Terminalia bellirica | Fruit | CID_6251 | 5595 | MAPK3 | ChEMBL |
| Arogyavardhini | Terminalia bellirica | Fruit | CID_6251 | 5724 | PTAFR | ChEMBL |
| Arogyavardhini | Terminalia bellirica | Fruit | CID_6251 | 5742 | PTGS1 | ChEMBL |
| Arogyavardhini | Terminalia bellirica | Fruit | CID_6251 | 5743 | PTGS2 | ChEMBL |
| Arogyavardhini | Terminalia bellirica | Fruit | CID_6251 | 5788 | PTPRC | ChEMBL |
| Arogyavardhini | Terminalia bellirica | Fruit | CID_6251 | 624 | BDKRB2 | ChEMBL |
| Arogyavardhini | Terminalia bellirica | Fruit | CID_6251 | 6530 | SLC6A2 | ChEMBL |
| Arogyavardhini | Terminalia bellirica | Fruit | CID_6251 | 6531 | SLC6A3 | ChEMBL |
| Arogyavardhini | Terminalia bellirica | Fruit | CID_6251 | 6532 | SLC6A4 | ChEMBL |
| Arogyavardhini | Terminalia bellirica | Fruit | CID_6251 | 6580 | SLC22A1 | ChEMBL \| NPASS |
| Arogyavardhini | Terminalia bellirica | Fruit | CID_6251 | 6865 | TACR2 | ChEMBL |
| Arogyavardhini | Terminalia bellirica | Fruit | CID_6251 | 6869 | TACR1 | ChEMBL |
| Arogyavardhini | Terminalia bellirica | Fruit | CID_6251 | 6915 | TBXA2R | ChEMBL |
| Arogyavardhini | Terminalia bellirica | Fruit | CID_6251 | 6916 | TBXAS1 | ChEMBL |
| Arogyavardhini | Terminalia bellirica | Fruit | CID_6251 | 729230 | CCR2 | ChEMBL |
| Arogyavardhini | Terminalia bellirica | Fruit | CID_6251 | 7398 | USP1 | NPASS |
| Arogyavardhini | Terminalia bellirica | Fruit | CID_6251 | 7433 | VIPR1 | ChEMBL |
| Arogyavardhini | Terminalia bellirica | Fruit | CID_6251 | 760 | CA2 | ChEMBL |
| Arogyavardhini | Terminalia bellirica | Fruit | CID_6251 | 799 | CALCR | ChEMBL |
| Arogyavardhini | Terminalia bellirica | Fruit | CID_6251 | 834 | CASP1 | ChEMBL |
| Arogyavardhini | Terminalia bellirica | Fruit | CID_6251 | 8654 | PDE5A | ChEMBL |
| Arogyavardhini | Terminalia bellirica | Fruit | CID_6251 | 886 | CCKAR | ChEMBL |
| Arogyavardhini | Terminalia bellirica | Fruit | CID_6251 | 9429 | ABCG2 | ChEMBL \| NPASS |
| Arogyavardhini | Terminalia chebula | Fruit | CID_10467 | 5465 | PPARA | ChEMBL \| NPASS |
| Arogyavardhini | Terminalia chebula | Fruit | CID_10467 | 5467 | PPARD | ChEMBL \| NPASS |
| Arogyavardhini | Terminalia chebula | Fruit | CID_10467 | 5468 | PPARG | BindingDB \| ChEMBL \| NPASS |
| Arogyavardhini | Terminalia chebula | Fruit | CID_10467 | 5770 | PTPN1 | BindingDB \| ChEMBL \| NPASS |
| Arogyavardhini | Terminalia chebula | Fruit | CID_10742 | 11238 | CA5B | BindingDB \| ChEMBL \| NPASS |
| Arogyavardhini | Terminalia chebula | Fruit | CID_10742 | 231 | AKR1B1 | NPASS |
| Arogyavardhini | Terminalia chebula | Fruit | CID_10742 | 23632 | CA14 | BindingDB \| ChEMBL \| NPASS |
| Arogyavardhini | Terminalia chebula | Fruit | CID_10742 | 5315 | PKM | ChEMBL |
| Arogyavardhini | Terminalia chebula | Fruit | CID_10742 | 759 | CA1 | BindingDB \| ChEMBL \| NPASS |
| Arogyavardhini | Terminalia chebula | Fruit | CID_10742 | 760 | CA2 | BindingDB \| ChEMBL \| NPASS |
| Arogyavardhini | Terminalia chebula | Fruit | CID_10742 | 761 | CA3 | BindingDB \| ChEMBL \| NPASS |
| Arogyavardhini | Terminalia chebula | Fruit | CID_10742 | 762 | CA4 | BindingDB \| ChEMBL \| NPASS |
| Arogyavardhini | Terminalia chebula | Fruit | CID_10742 | 763 | CA5A | BindingDB \| ChEMBL \| NPASS |
| Arogyavardhini | Terminalia chebula | Fruit | CID_10742 | 765 | CA6 | ChEMBL \| NPASS |
| Arogyavardhini | Terminalia chebula | Fruit | CID_10742 | 766 | CA7 | BindingDB \| ChEMBL \| NPASS |
| Arogyavardhini | Terminalia chebula | Fruit | CID_10742 | 768 | CA9 | BindingDB \| ChEMBL \| NPASS |
| Arogyavardhini | Terminalia chebula | Fruit | CID_10742 | 771 | CA12 | BindingDB \| ChEMBL \| NPASS |
| Arogyavardhini | Terminalia chebula | Fruit | CID_13250 | 2099 | ESR1 | ChEMBL |
| Arogyavardhini | Terminalia chebula | Fruit | CID_13250 | 22943 | DKK1 | ChEMBL |
| Arogyavardhini | Terminalia chebula | Fruit | CID_13250 | 4040 | LRP6 | ChEMBL |
| Arogyavardhini | Terminalia chebula | Fruit | CID_222284 | 10013 | HDAC6 | ChEMBL |
| Arogyavardhini | Terminalia chebula | Fruit | CID_222284 | 10599 | SLCO1B1 | ChEMBL \| NPASS |
| Arogyavardhini | Terminalia chebula | Fruit | CID_222284 | 1565 | CYP2D6 | ChEMBL \| NPASS |
| Arogyavardhini | Terminalia chebula | Fruit | CID_222284 | 1576 | CYP3A4 | BindingDB \| ChEMBL \| NPASS |

| Table S3: Associations between formulations, constituent plants, phytochemicals, and human molecular targets. The table lists the formulations, plant name and plant part, phytochemical identifiers, associated human target genes (Entrez Gene ID and gene symbol), and the database sources supporting each phytochemical-target interaction. | | | | | | |
|---|---|---|---|---|---|---|
| Formulation | Plant name | Part | PubChem CID | Entrez Gene ID | Gene Symbol | Source database(s) |
| Arogyavardhini | Terminalia chebula | Fruit | CID_222284 | 1803 | DPP4 | ChEMBL \| NPASS |
| Arogyavardhini | Terminalia chebula | Fruit | CID_222284 | 2147 | F2 | ChEMBL \| NPASS |
| Arogyavardhini | Terminalia chebula | Fruit | CID_222284 | 28234 | SLCO1B3 | ChEMBL \| NPASS |
| Arogyavardhini | Terminalia chebula | Fruit | CID_222284 | 2932 | GSK3B | ChEMBL |
| Arogyavardhini | Terminalia chebula | Fruit | CID_222284 | 3417 | IDH1 | ChEMBL \| NPASS |
| Arogyavardhini | Terminalia chebula | Fruit | CID_222284 | 51422 | PRKAG2 | ChEMBL |
| Arogyavardhini | Terminalia chebula | Fruit | CID_222284 | 5243 | ABCB1 | ChEMBL |
| Arogyavardhini | Terminalia chebula | Fruit | CID_222284 | 53632 | PRKAG3 | ChEMBL |
| Arogyavardhini | Terminalia chebula | Fruit | CID_222284 | 5423 | POLB | BindingDB \| ChEMBL \| NPASS |
| Arogyavardhini | Terminalia chebula | Fruit | CID_222284 | 5562 | PRKAA1 | ChEMBL |
| Arogyavardhini | Terminalia chebula | Fruit | CID_222284 | 5563 | PRKAA2 | ChEMBL |
| Arogyavardhini | Terminalia chebula | Fruit | CID_222284 | 5564 | PRKAB1 | ChEMBL |
| Arogyavardhini | Terminalia chebula | Fruit | CID_222284 | 5565 | PRKAB2 | ChEMBL |
| Arogyavardhini | Terminalia chebula | Fruit | CID_222284 | 5571 | PRKAG1 | ChEMBL |
| Arogyavardhini | Terminalia chebula | Fruit | CID_222284 | 7299 | TYR | ChEMBL \| NPASS |
| Arogyavardhini | Terminalia chebula | Fruit | CID_370 | 10013 | HDAC6 | ChEMBL |
| Arogyavardhini | Terminalia chebula | Fruit | CID_370 | 10599 | SLCO1B1 | ChEMBL \| NPASS |
| Arogyavardhini | Terminalia chebula | Fruit | CID_370 | 10919 | EHMT2 | ChEMBL \| NPASS |
| Arogyavardhini | Terminalia chebula | Fruit | CID_370 | 11238 | CA5B | BindingDB \| ChEMBL \| NPASS |
| Arogyavardhini | Terminalia chebula | Fruit | CID_370 | 118429 | ANTXR2 | NPASS |
| Arogyavardhini | Terminalia chebula | Fruit | CID_370 | 1588 | CYP19A1 | ChEMBL \| NPASS |
| Arogyavardhini | Terminalia chebula | Fruit | CID_370 | 1636 | ACE | BindingDB \| NPASS |
| Arogyavardhini | Terminalia chebula | Fruit | CID_370 | 1803 | DPP4 | BindingDB \| ChEMBL \| NPASS |
| Arogyavardhini | Terminalia chebula | Fruit | CID_370 | 2099 | ESR1 | ChEMBL \| NPASS |
| Arogyavardhini | Terminalia chebula | Fruit | CID_370 | 216 | ALDH1A1 | ChEMBL \| NPASS |
| Arogyavardhini | Terminalia chebula | Fruit | CID_370 | 231 | AKR1B1 | NPASS |
| Arogyavardhini | Terminalia chebula | Fruit | CID_370 | 23632 | CA14 | BindingDB \| ChEMBL \| NPASS |
| Arogyavardhini | Terminalia chebula | Fruit | CID_370 | 2526 | FUT4 | ChEMBL \| NPASS |
| Arogyavardhini | Terminalia chebula | Fruit | CID_370 | 2529 | FUT7 | ChEMBL \| NPASS |
| Arogyavardhini | Terminalia chebula | Fruit | CID_370 | 2590 | GALNT2 | ChEMBL \| NPASS |
| Arogyavardhini | Terminalia chebula | Fruit | CID_370 | 259285 | TAS2R39 | ChEMBL \| NPASS |
| Arogyavardhini | Terminalia chebula | Fruit | CID_370 | 2597 | GAPDH | ChEMBL \| NPASS |
| Arogyavardhini | Terminalia chebula | Fruit | CID_370 | 2648 | KAT2A | NPASS |
| Arogyavardhini | Terminalia chebula | Fruit | CID_370 | 2744 | GLS | ChEMBL \| NPASS |
| Arogyavardhini | Terminalia chebula | Fruit | CID_370 | 28234 | SLCO1B3 | ChEMBL \| NPASS |
| Arogyavardhini | Terminalia chebula | Fruit | CID_370 | 2908 | NR3C1 | NPASS |
| Arogyavardhini | Terminalia chebula | Fruit | CID_370 | 3028 | HSD17B10 | ChEMBL \| NPASS |
| Arogyavardhini | Terminalia chebula | Fruit | CID_370 | 3248 | HPGD | ChEMBL \| NPASS |
| Arogyavardhini | Terminalia chebula | Fruit | CID_370 | 328 | APEX1 | NPASS |
| Arogyavardhini | Terminalia chebula | Fruit | CID_370 | 3315 | HSPB1 | NPASS |
| Arogyavardhini | Terminalia chebula | Fruit | CID_370 | 3417 | IDH1 | ChEMBL \| NPASS |
| Arogyavardhini | Terminalia chebula | Fruit | CID_370 | 351 | APP | BindingDB \| ChEMBL |
| Arogyavardhini | Terminalia chebula | Fruit | CID_370 | 3576 | CXCL8 | NPASS |
| Arogyavardhini | Terminalia chebula | Fruit | CID_370 | 367 | AR | NPASS |
| Arogyavardhini | Terminalia chebula | Fruit | CID_370 | 3725 | JUN | NPASS |
| Arogyavardhini | Terminalia chebula | Fruit | CID_370 | 390245 | KDM4E | ChEMBL \| NPASS |
| Arogyavardhini | Terminalia chebula | Fruit | CID_370 | 4000 | LMNA | NPASS |
| Arogyavardhini | Terminalia chebula | Fruit | CID_370 | 4137 | MAPT | ChEMBL \| NPASS |
| Arogyavardhini | Terminalia chebula | Fruit | CID_370 | 4316 | MMP7 | BindingDB |
| Arogyavardhini | Terminalia chebula | Fruit | CID_370 | 4780 | NFE2L2 | ChEMBL \| NPASS |
| Arogyavardhini | Terminalia chebula | Fruit | CID_370 | 5054 | SERPINE1 | BindingDB |
| Arogyavardhini | Terminalia chebula | Fruit | CID_370 | 51426 | POLK | ChEMBL \| NPASS |
| Arogyavardhini | Terminalia chebula | Fruit | CID_370 | 54658 | UGT1A1 | ChEMBL \| NPASS |
| Arogyavardhini | Terminalia chebula | Fruit | CID_370 | 55775 | TDP1 | ChEMBL \| NPASS |
| Arogyavardhini | Terminalia chebula | Fruit | CID_370 | 5594 | MAPK1 | NPASS |
| Arogyavardhini | Terminalia chebula | Fruit | CID_370 | 5693 | PSMB5 | BindingDB \| NPASS |
| Arogyavardhini | Terminalia chebula | Fruit | CID_370 | 5743 | PTGS2 | ChEMBL \| NPASS |
| Arogyavardhini | Terminalia chebula | Fruit | CID_370 | 6401 | SELE | BindingDB \| NPASS |
| Arogyavardhini | Terminalia chebula | Fruit | CID_370 | 6402 | SELL | BindingDB \| ChEMBL \| NPASS |
| Arogyavardhini | Terminalia chebula | Fruit | CID_370 | 6403 | SELP | BindingDB \| ChEMBL \| NPASS |
| Arogyavardhini | Terminalia chebula | Fruit | CID_370 | 6606 | SMN1 | NPASS |
| Arogyavardhini | Terminalia chebula | Fruit | CID_370 | 6622 | SNCA | ChEMBL \| NPASS |
| Arogyavardhini | Terminalia chebula | Fruit | CID_370 | 6868 | ADAM17 | ChEMBL \| NPASS |
| Arogyavardhini | Terminalia chebula | Fruit | CID_370 | 7157 | TP53 | ChEMBL \| NPASS |
| Arogyavardhini | Terminalia chebula | Fruit | CID_370 | 7421 | VDR | ChEMBL \| NPASS |
| Arogyavardhini | Terminalia chebula | Fruit | CID_370 | 759 | CA1 | BindingDB \| ChEMBL \| NPASS |
| Arogyavardhini | Terminalia chebula | Fruit | CID_370 | 760 | CA2 | BindingDB \| ChEMBL \| NPASS |
| Arogyavardhini | Terminalia chebula | Fruit | CID_370 | 761 | CA3 | BindingDB \| ChEMBL \| NPASS |
| Arogyavardhini | Terminalia chebula | Fruit | CID_370 | 762 | CA4 | BindingDB \| ChEMBL \| NPASS |
| Arogyavardhini | Terminalia chebula | Fruit | CID_370 | 763 | CA5A | BindingDB \| ChEMBL \| NPASS |

**Table S3: Associations between formulations, constituent plants, phytochemicals, and human molecular targets. The table lists the formulations, plant name and plant part, phytochemical identifiers, associated human target genes (Entrez Gene ID and gene symbol), and the database sources supporting each phytochemical-target interaction.**

| Formulation | Plant name | Part | PubChem CID | Entrez Gene ID | Gene Symbol | Source database(s) |
|---|---|---|---|---|---|---|
| Arogyavardhini | Terminalia chebula | Fruit | CID_370 | 765 | CA6 | ChEMBL \| NPASS |
| Arogyavardhini | Terminalia chebula | Fruit | CID_370 | 766 | CA7 | BindingDB \| ChEMBL \| NPASS |
| Arogyavardhini | Terminalia chebula | Fruit | CID_370 | 768 | CA9 | BindingDB \| ChEMBL \| NPASS |
| Arogyavardhini | Terminalia chebula | Fruit | CID_370 | 771 | CA12 | BindingDB \| ChEMBL \| NPASS |
| Arogyavardhini | Terminalia chebula | Fruit | CID_445639 | 10018 | BCL2L11 | BindingDB |
| Arogyavardhini | Terminalia chebula | Fruit | CID_445639 | 100861540 | CYP3A7-CYP3A51P | ChEMBL |
| Arogyavardhini | Terminalia chebula | Fruit | CID_445639 | 10411 | RAPGEF3 | NPASS |
| Arogyavardhini | Terminalia chebula | Fruit | CID_445639 | 10951 | CBX1 | NPASS |
| Arogyavardhini | Terminalia chebula | Fruit | CID_445639 | 11069 | RAPGEF4 | NPASS |
| Arogyavardhini | Terminalia chebula | Fruit | CID_445639 | 11255 | HRH3 | ChEMBL |
| Arogyavardhini | Terminalia chebula | Fruit | CID_445639 | 1128 | CHRM1 | ChEMBL |
| Arogyavardhini | Terminalia chebula | Fruit | CID_445639 | 11283 | CYP4F8 | ChEMBL |
| Arogyavardhini | Terminalia chebula | Fruit | CID_445639 | 1129 | CHRM2 | ChEMBL |
| Arogyavardhini | Terminalia chebula | Fruit | CID_445639 | 113612 | CYP2U1 | ChEMBL |
| Arogyavardhini | Terminalia chebula | Fruit | CID_445639 | 120227 | CYP2R1 | ChEMBL |
| Arogyavardhini | Terminalia chebula | Fruit | CID_445639 | 122706 | PSMB11 | ChEMBL \| NPASS |
| Arogyavardhini | Terminalia chebula | Fruit | CID_445639 | 140 | ADORA3 | ChEMBL |
| Arogyavardhini | Terminalia chebula | Fruit | CID_445639 | 143471 | PSMA8 | ChEMBL \| NPASS |
| Arogyavardhini | Terminalia chebula | Fruit | CID_445639 | 148 | ADRA1A | ChEMBL |
| Arogyavardhini | Terminalia chebula | Fruit | CID_445639 | 150 | ADRA2A | ChEMBL |
| Arogyavardhini | Terminalia chebula | Fruit | CID_445639 | 1543 | CYP1A1 | ChEMBL |
| Arogyavardhini | Terminalia chebula | Fruit | CID_445639 | 1544 | CYP1A2 | ChEMBL |
| Arogyavardhini | Terminalia chebula | Fruit | CID_445639 | 1545 | CYP1B1 | ChEMBL |
| Arogyavardhini | Terminalia chebula | Fruit | CID_445639 | 1548 | CYP2A6 | ChEMBL |
| Arogyavardhini | Terminalia chebula | Fruit | CID_445639 | 1549 | CYP2A7 | ChEMBL |
| Arogyavardhini | Terminalia chebula | Fruit | CID_445639 | 1553 | CYP2A13 | ChEMBL |
| Arogyavardhini | Terminalia chebula | Fruit | CID_445639 | 1555 | CYP2B6 | ChEMBL |
| Arogyavardhini | Terminalia chebula | Fruit | CID_445639 | 1557 | CYP2C19 | ChEMBL |
| Arogyavardhini | Terminalia chebula | Fruit | CID_445639 | 1558 | CYP2C8 | ChEMBL |
| Arogyavardhini | Terminalia chebula | Fruit | CID_445639 | 1559 | CYP2C9 | ChEMBL |
| Arogyavardhini | Terminalia chebula | Fruit | CID_445639 | 1562 | CYP2C18 | ChEMBL |
| Arogyavardhini | Terminalia chebula | Fruit | CID_445639 | 1565 | CYP2D6 | ChEMBL |
| Arogyavardhini | Terminalia chebula | Fruit | CID_445639 | 1571 | CYP2E1 | ChEMBL |
| Arogyavardhini | Terminalia chebula | Fruit | CID_445639 | 1572 | CYP2F1 | ChEMBL |
| Arogyavardhini | Terminalia chebula | Fruit | CID_445639 | 1573 | CYP2J2 | ChEMBL |
| Arogyavardhini | Terminalia chebula | Fruit | CID_445639 | 1576 | CYP3A4 | ChEMBL |
| Arogyavardhini | Terminalia chebula | Fruit | CID_445639 | 1577 | CYP3A5 | ChEMBL |
| Arogyavardhini | Terminalia chebula | Fruit | CID_445639 | 1579 | CYP4A11 | ChEMBL |
| Arogyavardhini | Terminalia chebula | Fruit | CID_445639 | 1580 | CYP4B1 | ChEMBL |
| Arogyavardhini | Terminalia chebula | Fruit | CID_445639 | 1588 | CYP19A1 | BindingDB \| ChEMBL \| NPASS |
| Arogyavardhini | Terminalia chebula | Fruit | CID_445639 | 1812 | DRD1 | ChEMBL |
| Arogyavardhini | Terminalia chebula | Fruit | CID_445639 | 1814 | DRD3 | ChEMBL |
| Arogyavardhini | Terminalia chebula | Fruit | CID_445639 | 2052 | EPHX1 | ChEMBL |
| Arogyavardhini | Terminalia chebula | Fruit | CID_445639 | 2053 | EPHX2 | ChEMBL |
| Arogyavardhini | Terminalia chebula | Fruit | CID_445639 | 2099 | ESR1 | ChEMBL |
| Arogyavardhini | Terminalia chebula | Fruit | CID_445639 | 2147 | F2 | ChEMBL |
| Arogyavardhini | Terminalia chebula | Fruit | CID_445639 | 2152 | F3 | BindingDB \| ChEMBL \| NPASS |
| Arogyavardhini | Terminalia chebula | Fruit | CID_445639 | 2155 | F7 | ChEMBL |
| Arogyavardhini | Terminalia chebula | Fruit | CID_445639 | 2166 | FAAH | ChEMBL \| NPASS |
| Arogyavardhini | Terminalia chebula | Fruit | CID_445639 | 2167 | FABP4 | ChEMBL \| NPASS |
| Arogyavardhini | Terminalia chebula | Fruit | CID_445639 | 2168 | FABP1 | BindingDB \| ChEMBL \| NPASS |
| Arogyavardhini | Terminalia chebula | Fruit | CID_445639 | 2171 | FABP5 | ChEMBL \| NPASS |
| Arogyavardhini | Terminalia chebula | Fruit | CID_445639 | 2554 | GABRA1 | ChEMBL |
| Arogyavardhini | Terminalia chebula | Fruit | CID_445639 | 2555 | GABRA2 | ChEMBL |
| Arogyavardhini | Terminalia chebula | Fruit | CID_445639 | 2561 | GABRB2 | ChEMBL |
| Arogyavardhini | Terminalia chebula | Fruit | CID_445639 | 2566 | GABRG2 | ChEMBL |
| Arogyavardhini | Terminalia chebula | Fruit | CID_445639 | 2648 | KAT2A | ChEMBL \| NPASS |
| Arogyavardhini | Terminalia chebula | Fruit | CID_445639 | 284541 | CYP4A22 | ChEMBL |
| Arogyavardhini | Terminalia chebula | Fruit | CID_445639 | 29785 | CYP2S1 | ChEMBL |
| Arogyavardhini | Terminalia chebula | Fruit | CID_445639 | 3028 | HSD17B10 | ChEMBL \| NPASS |
| Arogyavardhini | Terminalia chebula | Fruit | CID_445639 | 328 | APEX1 | NPASS |
| Arogyavardhini | Terminalia chebula | Fruit | CID_445639 | 3315 | HSPB1 | NPASS |
| Arogyavardhini | Terminalia chebula | Fruit | CID_445639 | 3350 | HTR1A | ChEMBL |
| Arogyavardhini | Terminalia chebula | Fruit | CID_445639 | 3357 | HTR2B | ChEMBL |
| Arogyavardhini | Terminalia chebula | Fruit | CID_445639 | 367 | AR | ChEMBL |
| Arogyavardhini | Terminalia chebula | Fruit | CID_445639 | 3757 | KCNH2 | ChEMBL \| NPASS |
| Arogyavardhini | Terminalia chebula | Fruit | CID_445639 | 3791 | KDR | ChEMBL |
| Arogyavardhini | Terminalia chebula | Fruit | CID_445639 | 4051 | CYP4F3 | ChEMBL |
| Arogyavardhini | Terminalia chebula | Fruit | CID_445639 | 4128 | MAOA | ChEMBL |
| Arogyavardhini | Terminalia chebula | Fruit | CID_445639 | 4137 | MAPT | NPASS |

**Table S3: Associations between formulations, constituent plants, phytochemicals, and human molecular targets. The table lists the formulations, plant name and plant part, phytochemical identifiers, associated human target genes (Entrez Gene ID and gene symbol), and the database sources supporting each phytochemical-target interaction.**

| Formulation | Plant name | Part | PubChem CID | Entrez Gene ID | Gene Symbol | Source database(s) |
|---|---|---|---|---|---|---|
| Arogyavardhini | Terminalia chebula | Fruit | CID_445639 | 4297 | KMT2A | NPASS |
| Arogyavardhini | Terminalia chebula | Fruit | CID_445639 | 43 | ACHE | ChEMBL |
| Arogyavardhini | Terminalia chebula | Fruit | CID_445639 | 472 | ATM | NPASS |
| Arogyavardhini | Terminalia chebula | Fruit | CID_445639 | 4929 | NR4A2 | BindingDB \| ChEMBL |
| Arogyavardhini | Terminalia chebula | Fruit | CID_445639 | 4988 | OPRM1 | ChEMBL |
| Arogyavardhini | Terminalia chebula | Fruit | CID_445639 | 51053 | GMNN | ChEMBL \| NPASS |
| Arogyavardhini | Terminalia chebula | Fruit | CID_445639 | 5139 | PDE3A | ChEMBL |
| Arogyavardhini | Terminalia chebula | Fruit | CID_445639 | 5141 | PDE4A | ChEMBL |
| Arogyavardhini | Terminalia chebula | Fruit | CID_445639 | 51426 | POLK | NPASS |
| Arogyavardhini | Terminalia chebula | Fruit | CID_445639 | 51548 | SIRT6 | ChEMBL |
| Arogyavardhini | Terminalia chebula | Fruit | CID_445639 | 5241 | PGR | ChEMBL |
| Arogyavardhini | Terminalia chebula | Fruit | CID_445639 | 5347 | PLK1 | ChEMBL \| NPASS |
| Arogyavardhini | Terminalia chebula | Fruit | CID_445639 | 5465 | PPARA | ChEMBL \| NPASS |
| Arogyavardhini | Terminalia chebula | Fruit | CID_445639 | 5467 | PPARD | ChEMBL \| NPASS |
| Arogyavardhini | Terminalia chebula | Fruit | CID_445639 | 5468 | PPARG | BindingDB \| ChEMBL \| NPASS |
| Arogyavardhini | Terminalia chebula | Fruit | CID_445639 | 54905 | CYP2W1 | ChEMBL |
| Arogyavardhini | Terminalia chebula | Fruit | CID_445639 | 5578 | PRKCA | ChEMBL \| NPASS |
| Arogyavardhini | Terminalia chebula | Fruit | CID_445639 | 5682 | PSMA1 | ChEMBL \| NPASS |
| Arogyavardhini | Terminalia chebula | Fruit | CID_445639 | 5683 | PSMA2 | ChEMBL \| NPASS |
| Arogyavardhini | Terminalia chebula | Fruit | CID_445639 | 5684 | PSMA3 | ChEMBL \| NPASS |
| Arogyavardhini | Terminalia chebula | Fruit | CID_445639 | 5685 | PSMA4 | ChEMBL \| NPASS |
| Arogyavardhini | Terminalia chebula | Fruit | CID_445639 | 5686 | PSMA5 | ChEMBL \| NPASS |
| Arogyavardhini | Terminalia chebula | Fruit | CID_445639 | 5687 | PSMA6 | ChEMBL \| NPASS |
| Arogyavardhini | Terminalia chebula | Fruit | CID_445639 | 5688 | PSMA7 | ChEMBL \| NPASS |
| Arogyavardhini | Terminalia chebula | Fruit | CID_445639 | 5689 | PSMB1 | ChEMBL \| NPASS |
| Arogyavardhini | Terminalia chebula | Fruit | CID_445639 | 5690 | PSMB2 | ChEMBL \| NPASS |
| Arogyavardhini | Terminalia chebula | Fruit | CID_445639 | 5691 | PSMB3 | ChEMBL \| NPASS |
| Arogyavardhini | Terminalia chebula | Fruit | CID_445639 | 5692 | PSMB4 | ChEMBL \| NPASS |
| Arogyavardhini | Terminalia chebula | Fruit | CID_445639 | 5693 | PSMB5 | ChEMBL \| NPASS |
| Arogyavardhini | Terminalia chebula | Fruit | CID_445639 | 5694 | PSMB6 | ChEMBL \| NPASS |
| Arogyavardhini | Terminalia chebula | Fruit | CID_445639 | 5695 | PSMB7 | ChEMBL \| NPASS |
| Arogyavardhini | Terminalia chebula | Fruit | CID_445639 | 5696 | PSMB8 | ChEMBL \| NPASS |
| Arogyavardhini | Terminalia chebula | Fruit | CID_445639 | 5698 | PSMB9 | ChEMBL \| NPASS |
| Arogyavardhini | Terminalia chebula | Fruit | CID_445639 | 5699 | PSMB10 | ChEMBL \| NPASS |
| Arogyavardhini | Terminalia chebula | Fruit | CID_445639 | 5742 | PTGS1 | ChEMBL |
| Arogyavardhini | Terminalia chebula | Fruit | CID_445639 | 5745 | PTH1R | NPASS |
| Arogyavardhini | Terminalia chebula | Fruit | CID_445639 | 5770 | PTPN1 | BindingDB \| ChEMBL \| NPASS |
| Arogyavardhini | Terminalia chebula | Fruit | CID_445639 | 6097 | RORC | NPASS |
| Arogyavardhini | Terminalia chebula | Fruit | CID_445639 | 641 | BLM | NPASS |
| Arogyavardhini | Terminalia chebula | Fruit | CID_445639 | 64816 | CYP3A43 | ChEMBL |
| Arogyavardhini | Terminalia chebula | Fruit | CID_445639 | 6530 | SLC6A2 | ChEMBL |
| Arogyavardhini | Terminalia chebula | Fruit | CID_445639 | 6531 | SLC6A3 | ChEMBL |
| Arogyavardhini | Terminalia chebula | Fruit | CID_445639 | 6532 | SLC6A4 | ChEMBL |
| Arogyavardhini | Terminalia chebula | Fruit | CID_445639 | 6646 | SOAT1 | BindingDB \| NPASS |
| Arogyavardhini | Terminalia chebula | Fruit | CID_445639 | 6915 | TBXA2R | ChEMBL |
| Arogyavardhini | Terminalia chebula | Fruit | CID_445639 | 7015 | TERT | ChEMBL \| NPASS |
| Arogyavardhini | Terminalia chebula | Fruit | CID_445639 | 7150 | TOP1 | ChEMBL \| NPASS |
| Arogyavardhini | Terminalia chebula | Fruit | CID_445639 | 7398 | USP1 | ChEMBL \| NPASS |
| Arogyavardhini | Terminalia chebula | Fruit | CID_445639 | 79915 | ATAD5 | NPASS |
| Arogyavardhini | Terminalia chebula | Fruit | CID_445639 | 8529 | CYP4F2 | ChEMBL |
| Arogyavardhini | Terminalia chebula | Fruit | CID_445639 | 9682 | KDM4A | ChEMBL \| NPASS |
| Arogyavardhini | Terminalia chebula | Fruit | CID_445639 | 9971 | NR1H4 | NPASS |
| Arogyavardhini | Terminalia chebula | Fruit | CID_445858 | 10013 | HDAC6 | ChEMBL |
| Arogyavardhini | Terminalia chebula | Fruit | CID_445858 | 10014 | HDAC5 | ChEMBL |
| Arogyavardhini | Terminalia chebula | Fruit | CID_445858 | 10280 | SIGMAR1 | ChEMBL |
| Arogyavardhini | Terminalia chebula | Fruit | CID_445858 | 10599 | SLCO1B1 | ChEMBL \| NPASS |
| Arogyavardhini | Terminalia chebula | Fruit | CID_445858 | 10800 | CYSLTR1 | ChEMBL |
| Arogyavardhini | Terminalia chebula | Fruit | CID_445858 | 11238 | CA5B | BindingDB \| ChEMBL \| NPASS |
| Arogyavardhini | Terminalia chebula | Fruit | CID_445858 | 1128 | CHRM1 | ChEMBL |
| Arogyavardhini | Terminalia chebula | Fruit | CID_445858 | 1129 | CHRM2 | ChEMBL |
| Arogyavardhini | Terminalia chebula | Fruit | CID_445858 | 1131 | CHRM3 | ChEMBL |
| Arogyavardhini | Terminalia chebula | Fruit | CID_445858 | 1132 | CHRM4 | ChEMBL |
| Arogyavardhini | Terminalia chebula | Fruit | CID_445858 | 1133 | CHRM5 | ChEMBL |
| Arogyavardhini | Terminalia chebula | Fruit | CID_445858 | 1233 | CCR4 | ChEMBL |
| Arogyavardhini | Terminalia chebula | Fruit | CID_445858 | 1234 | CCR5 | ChEMBL |
| Arogyavardhini | Terminalia chebula | Fruit | CID_445858 | 1268 | CNR1 | ChEMBL |
| Arogyavardhini | Terminalia chebula | Fruit | CID_445858 | 134 | ADORA1 | ChEMBL |
| Arogyavardhini | Terminalia chebula | Fruit | CID_445858 | 135 | ADORA2A | ChEMBL |
| Arogyavardhini | Terminalia chebula | Fruit | CID_445858 | 140 | ADORA3 | ChEMBL |
| Arogyavardhini | Terminalia chebula | Fruit | CID_445858 | 1432 | MAPK14 | ChEMBL |

**Table S3: Associations between formulations, constituent plants, phytochemicals, and human molecular targets. The table lists the formulations, plant name and plant part, phytochemical identifiers, associated human target genes (Entrez Gene ID and gene symbol), and the database sources supporting each phytochemical-target interaction.**

| Formulation | Plant name | Part | PubChem CID | Entrez Gene ID | Gene Symbol | Source database(s) |
|---|---|---|---|---|---|---|
| Arogyavardhini | Terminalia chebula | Fruit | CID_445858 | 1457 | CSNK2A1 | ChEMBL |
| Arogyavardhini | Terminalia chebula | Fruit | CID_445858 | 1459 | CSNK2A2 | ChEMBL |
| Arogyavardhini | Terminalia chebula | Fruit | CID_445858 | 146 | ADRA1D | ChEMBL |
| Arogyavardhini | Terminalia chebula | Fruit | CID_445858 | 1460 | CSNK2B | ChEMBL |
| Arogyavardhini | Terminalia chebula | Fruit | CID_445858 | 150 | ADRA2A | ChEMBL |
| Arogyavardhini | Terminalia chebula | Fruit | CID_445858 | 151 | ADRA2B | ChEMBL |
| Arogyavardhini | Terminalia chebula | Fruit | CID_445858 | 1511 | CTSG | ChEMBL |
| Arogyavardhini | Terminalia chebula | Fruit | CID_445858 | 152 | ADRA2C | ChEMBL |
| Arogyavardhini | Terminalia chebula | Fruit | CID_445858 | 153 | ADRB1 | ChEMBL |
| Arogyavardhini | Terminalia chebula | Fruit | CID_445858 | 154 | ADRB2 | ChEMBL |
| Arogyavardhini | Terminalia chebula | Fruit | CID_445858 | 1544 | CYP1A2 | ChEMBL |
| Arogyavardhini | Terminalia chebula | Fruit | CID_445858 | 1548 | CYP2A6 | ChEMBL |
| Arogyavardhini | Terminalia chebula | Fruit | CID_445858 | 155 | ADRB3 | ChEMBL |
| Arogyavardhini | Terminalia chebula | Fruit | CID_445858 | 1557 | CYP2C19 | ChEMBL |
| Arogyavardhini | Terminalia chebula | Fruit | CID_445858 | 1559 | CYP2C9 | ChEMBL |
| Arogyavardhini | Terminalia chebula | Fruit | CID_445858 | 1565 | CYP2D6 | ChEMBL |
| Arogyavardhini | Terminalia chebula | Fruit | CID_445858 | 1571 | CYP2E1 | ChEMBL |
| Arogyavardhini | Terminalia chebula | Fruit | CID_445858 | 1576 | CYP3A4 | ChEMBL |
| Arogyavardhini | Terminalia chebula | Fruit | CID_445858 | 1812 | DRD1 | ChEMBL |
| Arogyavardhini | Terminalia chebula | Fruit | CID_445858 | 1813 | DRD2 | ChEMBL |
| Arogyavardhini | Terminalia chebula | Fruit | CID_445858 | 1814 | DRD3 | ChEMBL |
| Arogyavardhini | Terminalia chebula | Fruit | CID_445858 | 1815 | DRD4 | ChEMBL |
| Arogyavardhini | Terminalia chebula | Fruit | CID_445858 | 186 | AGTR2 | ChEMBL |
| Arogyavardhini | Terminalia chebula | Fruit | CID_445858 | 1909 | EDNRA | ChEMBL |
| Arogyavardhini | Terminalia chebula | Fruit | CID_445858 | 1956 | EGFR | ChEMBL |
| Arogyavardhini | Terminalia chebula | Fruit | CID_445858 | 1991 | ELANE | ChEMBL |
| Arogyavardhini | Terminalia chebula | Fruit | CID_445858 | 2064 | ERBB2 | ChEMBL |
| Arogyavardhini | Terminalia chebula | Fruit | CID_445858 | 2099 | ESR1 | ChEMBL \| NPASS |
| Arogyavardhini | Terminalia chebula | Fruit | CID_445858 | 2100 | ESR2 | ChEMBL |
| Arogyavardhini | Terminalia chebula | Fruit | CID_445858 | 2260 | FGFR1 | ChEMBL \| NPASS |
| Arogyavardhini | Terminalia chebula | Fruit | CID_445858 | 2263 | FGFR2 | ChEMBL \| NPASS |
| Arogyavardhini | Terminalia chebula | Fruit | CID_445858 | 2321 | FLT1 | ChEMBL |
| Arogyavardhini | Terminalia chebula | Fruit | CID_445858 | 23632 | CA14 | BindingDB \| ChEMBL \| NPASS |
| Arogyavardhini | Terminalia chebula | Fruit | CID_445858 | 2534 | FYN | ChEMBL |
| Arogyavardhini | Terminalia chebula | Fruit | CID_445858 | 259285 | TAS2R39 | ChEMBL \| NPASS |
| Arogyavardhini | Terminalia chebula | Fruit | CID_445858 | 28234 | SLCO1B3 | ChEMBL \| NPASS |
| Arogyavardhini | Terminalia chebula | Fruit | CID_445858 | 283106 | CSNK2A3 | BindingDB \| ChEMBL |
| Arogyavardhini | Terminalia chebula | Fruit | CID_445858 | 2908 | NR3C1 | ChEMBL |
| Arogyavardhini | Terminalia chebula | Fruit | CID_445858 | 3065 | HDAC1 | ChEMBL |
| Arogyavardhini | Terminalia chebula | Fruit | CID_445858 | 3066 | HDAC2 | ChEMBL |
| Arogyavardhini | Terminalia chebula | Fruit | CID_445858 | 3156 | HMGCR | ChEMBL |
| Arogyavardhini | Terminalia chebula | Fruit | CID_445858 | 3269 | HRH1 | ChEMBL |
| Arogyavardhini | Terminalia chebula | Fruit | CID_445858 | 3274 | HRH2 | ChEMBL |
| Arogyavardhini | Terminalia chebula | Fruit | CID_445858 | 3356 | HTR2A | ChEMBL |
| Arogyavardhini | Terminalia chebula | Fruit | CID_445858 | 3357 | HTR2B | ChEMBL |
| Arogyavardhini | Terminalia chebula | Fruit | CID_445858 | 3358 | HTR2C | ChEMBL |
| Arogyavardhini | Terminalia chebula | Fruit | CID_445858 | 3362 | HTR6 | ChEMBL |
| Arogyavardhini | Terminalia chebula | Fruit | CID_445858 | 3363 | HTR7 | ChEMBL \| NPASS |
| Arogyavardhini | Terminalia chebula | Fruit | CID_445858 | 3373 | HYAL1 | BindingDB \| NPASS |
| Arogyavardhini | Terminalia chebula | Fruit | CID_445858 | 338442 | HCAR2 | ChEMBL \| NPASS |
| Arogyavardhini | Terminalia chebula | Fruit | CID_445858 | 351 | APP | BindingDB \| ChEMBL \| NPASS |
| Arogyavardhini | Terminalia chebula | Fruit | CID_445858 | 3577 | CXCR1 | ChEMBL |
| Arogyavardhini | Terminalia chebula | Fruit | CID_445858 | 3579 | CXCR2 | ChEMBL |
| Arogyavardhini | Terminalia chebula | Fruit | CID_445858 | 3757 | KCNH2 | ChEMBL |
| Arogyavardhini | Terminalia chebula | Fruit | CID_445858 | 390245 | KDM4E | NPASS |
| Arogyavardhini | Terminalia chebula | Fruit | CID_445858 | 3932 | LCK | ChEMBL |
| Arogyavardhini | Terminalia chebula | Fruit | CID_445858 | 4128 | MAOA | ChEMBL |
| Arogyavardhini | Terminalia chebula | Fruit | CID_445858 | 4159 | MC3R | ChEMBL |
| Arogyavardhini | Terminalia chebula | Fruit | CID_445858 | 4160 | MC4R | ChEMBL |
| Arogyavardhini | Terminalia chebula | Fruit | CID_445858 | 4161 | MC5R | ChEMBL |
| Arogyavardhini | Terminalia chebula | Fruit | CID_445858 | 43 | ACHE | BindingDB \| ChEMBL \| NPASS |
| Arogyavardhini | Terminalia chebula | Fruit | CID_445858 | 4312 | MMP1 | ChEMBL |
| Arogyavardhini | Terminalia chebula | Fruit | CID_445858 | 4318 | MMP9 | ChEMBL |
| Arogyavardhini | Terminalia chebula | Fruit | CID_445858 | 4886 | NPY1R | ChEMBL |
| Arogyavardhini | Terminalia chebula | Fruit | CID_445858 | 4887 | NPY2R | ChEMBL |
| Arogyavardhini | Terminalia chebula | Fruit | CID_445858 | 4985 | OPRD1 | ChEMBL |
| Arogyavardhini | Terminalia chebula | Fruit | CID_445858 | 4986 | OPRK1 | ChEMBL |
| Arogyavardhini | Terminalia chebula | Fruit | CID_445858 | 4988 | OPRM1 | ChEMBL |
| Arogyavardhini | Terminalia chebula | Fruit | CID_445858 | 51053 | GMNN | NPASS |
| Arogyavardhini | Terminalia chebula | Fruit | CID_445858 | 51564 | HDAC7 | ChEMBL |

| Table S3: Associations between formulations, constituent plants, phytochemicals, and human molecular targets. The table lists the formulations, plant name and plant part, phytochemical identifiers, associated human target genes (Entrez Gene ID and gene symbol), and the database sources supporting each phytochemical-target interaction. | | | | | | |
|---|---|---|---|---|---|---|
| Formulation | Plant name | Part | PubChem CID | Entrez Gene ID | Gene Symbol | Source database(s) |
| Arogyavardhini | Terminalia chebula | Fruit | CID_445858 | 5315 | PKM | ChEMBL |
| Arogyavardhini | Terminalia chebula | Fruit | CID_445858 | 54657 | UGT1A4 | ChEMBL \| NPASS |
| Arogyavardhini | Terminalia chebula | Fruit | CID_445858 | 54658 | UGT1A1 | ChEMBL \| NPASS |
| Arogyavardhini | Terminalia chebula | Fruit | CID_445858 | 552 | AVPR1A | ChEMBL |
| Arogyavardhini | Terminalia chebula | Fruit | CID_445858 | 5530 | PPP3CA | ChEMBL |
| Arogyavardhini | Terminalia chebula | Fruit | CID_445858 | 55775 | TDP1 | NPASS |
| Arogyavardhini | Terminalia chebula | Fruit | CID_445858 | 5578 | PRKCA | ChEMBL |
| Arogyavardhini | Terminalia chebula | Fruit | CID_445858 | 55869 | HDAC8 | ChEMBL |
| Arogyavardhini | Terminalia chebula | Fruit | CID_445858 | 5594 | MAPK1 | ChEMBL |
| Arogyavardhini | Terminalia chebula | Fruit | CID_445858 | 5595 | MAPK3 | ChEMBL |
| Arogyavardhini | Terminalia chebula | Fruit | CID_445858 | 5724 | PTAFR | ChEMBL |
| Arogyavardhini | Terminalia chebula | Fruit | CID_445858 | 5742 | PTGS1 | ChEMBL |
| Arogyavardhini | Terminalia chebula | Fruit | CID_445858 | 5743 | PTGS2 | ChEMBL |
| Arogyavardhini | Terminalia chebula | Fruit | CID_445858 | 5745 | PTH1R | NPASS |
| Arogyavardhini | Terminalia chebula | Fruit | CID_445858 | 5788 | PTPRC | ChEMBL |
| Arogyavardhini | Terminalia chebula | Fruit | CID_445858 | 590 | BCHE | BindingDB \| ChEMBL \| NPASS |
| Arogyavardhini | Terminalia chebula | Fruit | CID_445858 | 624 | BDKRB2 | ChEMBL |
| Arogyavardhini | Terminalia chebula | Fruit | CID_445858 | 6530 | SLC6A2 | ChEMBL |
| Arogyavardhini | Terminalia chebula | Fruit | CID_445858 | 6531 | SLC6A3 | ChEMBL |
| Arogyavardhini | Terminalia chebula | Fruit | CID_445858 | 6532 | SLC6A4 | ChEMBL |
| Arogyavardhini | Terminalia chebula | Fruit | CID_445858 | 6622 | SNCA | ChEMBL \| NPASS |
| Arogyavardhini | Terminalia chebula | Fruit | CID_445858 | 6865 | TACR2 | ChEMBL |
| Arogyavardhini | Terminalia chebula | Fruit | CID_445858 | 6869 | TACR1 | ChEMBL |
| Arogyavardhini | Terminalia chebula | Fruit | CID_445858 | 6916 | TBXAS1 | ChEMBL |
| Arogyavardhini | Terminalia chebula | Fruit | CID_445858 | 729230 | CCR2 | ChEMBL |
| Arogyavardhini | Terminalia chebula | Fruit | CID_445858 | 7299 | TYR | ChEMBL \| NPASS |
| Arogyavardhini | Terminalia chebula | Fruit | CID_445858 | 7366 | UGT2B15 | ChEMBL \| NPASS |
| Arogyavardhini | Terminalia chebula | Fruit | CID_445858 | 7433 | VIPR1 | ChEMBL |
| Arogyavardhini | Terminalia chebula | Fruit | CID_445858 | 759 | CA1 | BindingDB \| ChEMBL \| NPASS |
| Arogyavardhini | Terminalia chebula | Fruit | CID_445858 | 760 | CA2 | BindingDB \| ChEMBL \| NPASS |
| Arogyavardhini | Terminalia chebula | Fruit | CID_445858 | 761 | CA3 | BindingDB \| ChEMBL \| NPASS |
| Arogyavardhini | Terminalia chebula | Fruit | CID_445858 | 762 | CA4 | BindingDB \| ChEMBL \| NPASS |
| Arogyavardhini | Terminalia chebula | Fruit | CID_445858 | 763 | CA5A | BindingDB \| ChEMBL \| NPASS |
| Arogyavardhini | Terminalia chebula | Fruit | CID_445858 | 765 | CA6 | ChEMBL \| NPASS |
| Arogyavardhini | Terminalia chebula | Fruit | CID_445858 | 766 | CA7 | BindingDB \| ChEMBL \| NPASS |
| Arogyavardhini | Terminalia chebula | Fruit | CID_445858 | 768 | CA9 | BindingDB \| ChEMBL \| NPASS |
| Arogyavardhini | Terminalia chebula | Fruit | CID_445858 | 771 | CA12 | BindingDB \| ChEMBL \| NPASS |
| Arogyavardhini | Terminalia chebula | Fruit | CID_445858 | 79885 | HDAC11 | ChEMBL |
| Arogyavardhini | Terminalia chebula | Fruit | CID_445858 | 799 | CALCR | ChEMBL |
| Arogyavardhini | Terminalia chebula | Fruit | CID_445858 | 834 | CASP1 | ChEMBL |
| Arogyavardhini | Terminalia chebula | Fruit | CID_445858 | 83933 | HDAC10 | ChEMBL \| NPASS |
| Arogyavardhini | Terminalia chebula | Fruit | CID_445858 | 8654 | PDE5A | ChEMBL |
| Arogyavardhini | Terminalia chebula | Fruit | CID_445858 | 8841 | HDAC3 | ChEMBL |
| Arogyavardhini | Terminalia chebula | Fruit | CID_445858 | 886 | CCKAR | ChEMBL |
| Arogyavardhini | Terminalia chebula | Fruit | CID_445858 | 9734 | HDAC9 | ChEMBL |
| Arogyavardhini | Terminalia chebula | Fruit | CID_445858 | 9759 | HDAC4 | ChEMBL |
| Arogyavardhini | Terminalia chebula | Fruit | CID_5280450 | 1576 | CYP3A4 | NPASS |
| Arogyavardhini | Terminalia chebula | Fruit | CID_5280450 | 1588 | CYP19A1 | BindingDB \| ChEMBL \| NPASS |
| Arogyavardhini | Terminalia chebula | Fruit | CID_5280450 | 1845 | DUSP3 | BindingDB \| ChEMBL \| NPASS |
| Arogyavardhini | Terminalia chebula | Fruit | CID_5280450 | 2099 | ESR1 | ChEMBL \| NPASS |
| Arogyavardhini | Terminalia chebula | Fruit | CID_5280450 | 2152 | F3 | BindingDB \| ChEMBL \| NPASS |
| Arogyavardhini | Terminalia chebula | Fruit | CID_5280450 | 2155 | F7 | ChEMBL |
| Arogyavardhini | Terminalia chebula | Fruit | CID_5280450 | 216 | ALDH1A1 | ChEMBL \| NPASS |
| Arogyavardhini | Terminalia chebula | Fruit | CID_5280450 | 2167 | FABP4 | ChEMBL \| NPASS |
| Arogyavardhini | Terminalia chebula | Fruit | CID_5280450 | 2170 | FABP3 | BindingDB \| ChEMBL \| NPASS |
| Arogyavardhini | Terminalia chebula | Fruit | CID_5280450 | 246 | ALOX15 | ChEMBL \| NPASS |
| Arogyavardhini | Terminalia chebula | Fruit | CID_5280450 | 2648 | KAT2A | NPASS |
| Arogyavardhini | Terminalia chebula | Fruit | CID_5280450 | 2769 | GNA15 | ChEMBL |
| Arogyavardhini | Terminalia chebula | Fruit | CID_5280450 | 2864 | FFAR1 | ChEMBL \| NPASS |
| Arogyavardhini | Terminalia chebula | Fruit | CID_5280450 | 3028 | HSD17B10 | NPASS |
| Arogyavardhini | Terminalia chebula | Fruit | CID_5280450 | 3248 | HPGD | ChEMBL \| NPASS |
| Arogyavardhini | Terminalia chebula | Fruit | CID_5280450 | 3315 | HSPB1 | NPASS |
| Arogyavardhini | Terminalia chebula | Fruit | CID_5280450 | 338557 | FFAR4 | BindingDB \| ChEMBL \| NPASS |
| Arogyavardhini | Terminalia chebula | Fruit | CID_5280450 | 367 | AR | NPASS |
| Arogyavardhini | Terminalia chebula | Fruit | CID_5280450 | 3725 | JUN | NPASS |
| Arogyavardhini | Terminalia chebula | Fruit | CID_5280450 | 4000 | LMNA | NPASS |
| Arogyavardhini | Terminalia chebula | Fruit | CID_5280450 | 4780 | NFE2L2 | NPASS |
| Arogyavardhini | Terminalia chebula | Fruit | CID_5280450 | 4790 | NFKB1 | NPASS |
| Arogyavardhini | Terminalia chebula | Fruit | CID_5280450 | 4929 | NR4A2 | BindingDB \| ChEMBL |
| Arogyavardhini | Terminalia chebula | Fruit | CID_5280450 | 51053 | GMNN | NPASS |

**Table S3: Associations between formulations, constituent plants, phytochemicals, and human molecular targets. The table lists the formulations, plant name and plant part, phytochemical identifiers, associated human target genes (Entrez Gene ID and gene symbol), and the database sources supporting each phytochemical-target interaction.**

| Formulation | Plant name | Part | PubChem CID | Entrez Gene ID | Gene Symbol | Source database(s) |
|---|---|---|---|---|---|---|
| Arogyavardhini | Terminalia chebula | Fruit | CID_5280450 | 51426 | POLK | NPASS |
| Arogyavardhini | Terminalia chebula | Fruit | CID_5280450 | 51548 | SIRT6 | ChEMBL |
| Arogyavardhini | Terminalia chebula | Fruit | CID_5280450 | 5465 | PPARA | ChEMBL \| NPASS |
| Arogyavardhini | Terminalia chebula | Fruit | CID_5280450 | 5467 | PPARD | ChEMBL \| NPASS |
| Arogyavardhini | Terminalia chebula | Fruit | CID_5280450 | 5468 | PPARG | BindingDB \| ChEMBL \| NPASS |
| Arogyavardhini | Terminalia chebula | Fruit | CID_5280450 | 55775 | TDP1 | NPASS |
| Arogyavardhini | Terminalia chebula | Fruit | CID_5280450 | 5742 | PTGS1 | ChEMBL \| NPASS |
| Arogyavardhini | Terminalia chebula | Fruit | CID_5280450 | 5778 | PTPN7 | BindingDB \| ChEMBL \| NPASS |
| Arogyavardhini | Terminalia chebula | Fruit | CID_5280450 | 6256 | RXRA | NPASS |
| Arogyavardhini | Terminalia chebula | Fruit | CID_5280450 | 6646 | SOAT1 | BindingDB \| NPASS |
| Arogyavardhini | Terminalia chebula | Fruit | CID_5280450 | 7015 | TERT | ChEMBL \| NPASS |
| Arogyavardhini | Terminalia chebula | Fruit | CID_5280450 | 7253 | TSHR | NPASS |
| Arogyavardhini | Terminalia chebula | Fruit | CID_5280450 | 7398 | USP1 | NPASS |
| Arogyavardhini | Terminalia chebula | Fruit | CID_5280450 | 8644 | AKR1C3 | BindingDB \| ChEMBL \| NPASS |
| Arogyavardhini | Terminalia chebula | Fruit | CID_5280450 | 9971 | NR1H4 | NPASS |
| Arogyavardhini | Terminalia chebula | Fruit | CID_5281 | 11255 | HRH3 | ChEMBL |
| Arogyavardhini | Terminalia chebula | Fruit | CID_5281 | 1128 | CHRM1 | ChEMBL |
| Arogyavardhini | Terminalia chebula | Fruit | CID_5281 | 1129 | CHRM2 | ChEMBL |
| Arogyavardhini | Terminalia chebula | Fruit | CID_5281 | 140 | ADORA3 | ChEMBL |
| Arogyavardhini | Terminalia chebula | Fruit | CID_5281 | 148 | ADRA1A | ChEMBL |
| Arogyavardhini | Terminalia chebula | Fruit | CID_5281 | 150 | ADRA2A | ChEMBL |
| Arogyavardhini | Terminalia chebula | Fruit | CID_5281 | 1588 | CYP19A1 | ChEMBL \| NPASS |
| Arogyavardhini | Terminalia chebula | Fruit | CID_5281 | 1812 | DRD1 | ChEMBL |
| Arogyavardhini | Terminalia chebula | Fruit | CID_5281 | 1814 | DRD3 | ChEMBL |
| Arogyavardhini | Terminalia chebula | Fruit | CID_5281 | 2099 | ESR1 | ChEMBL |
| Arogyavardhini | Terminalia chebula | Fruit | CID_5281 | 2147 | F2 | ChEMBL |
| Arogyavardhini | Terminalia chebula | Fruit | CID_5281 | 2167 | FABP4 | ChEMBL \| NPASS |
| Arogyavardhini | Terminalia chebula | Fruit | CID_5281 | 246 | ALOX15 | NPASS |
| Arogyavardhini | Terminalia chebula | Fruit | CID_5281 | 2554 | GABRA1 | ChEMBL |
| Arogyavardhini | Terminalia chebula | Fruit | CID_5281 | 2555 | GABRA2 | ChEMBL |
| Arogyavardhini | Terminalia chebula | Fruit | CID_5281 | 2561 | GABRB2 | ChEMBL |
| Arogyavardhini | Terminalia chebula | Fruit | CID_5281 | 2566 | GABRG2 | ChEMBL |
| Arogyavardhini | Terminalia chebula | Fruit | CID_5281 | 2908 | NR3C1 | NPASS |
| Arogyavardhini | Terminalia chebula | Fruit | CID_5281 | 3028 | HSD17B10 | ChEMBL \| NPASS |
| Arogyavardhini | Terminalia chebula | Fruit | CID_5281 | 328 | APEX1 | NPASS |
| Arogyavardhini | Terminalia chebula | Fruit | CID_5281 | 3350 | HTR1A | ChEMBL |
| Arogyavardhini | Terminalia chebula | Fruit | CID_5281 | 3357 | HTR2B | ChEMBL |
| Arogyavardhini | Terminalia chebula | Fruit | CID_5281 | 367 | AR | ChEMBL |
| Arogyavardhini | Terminalia chebula | Fruit | CID_5281 | 3757 | KCNH2 | ChEMBL |
| Arogyavardhini | Terminalia chebula | Fruit | CID_5281 | 3791 | KDR | ChEMBL |
| Arogyavardhini | Terminalia chebula | Fruit | CID_5281 | 4128 | MAOA | ChEMBL |
| Arogyavardhini | Terminalia chebula | Fruit | CID_5281 | 4297 | KMT2A | NPASS |
| Arogyavardhini | Terminalia chebula | Fruit | CID_5281 | 43 | ACHE | ChEMBL |
| Arogyavardhini | Terminalia chebula | Fruit | CID_5281 | 4790 | NFKB1 | NPASS |
| Arogyavardhini | Terminalia chebula | Fruit | CID_5281 | 4988 | OPRM1 | ChEMBL |
| Arogyavardhini | Terminalia chebula | Fruit | CID_5281 | 5139 | PDE3A | ChEMBL |
| Arogyavardhini | Terminalia chebula | Fruit | CID_5281 | 5141 | PDE4A | ChEMBL |
| Arogyavardhini | Terminalia chebula | Fruit | CID_5281 | 5241 | PGR | ChEMBL |
| Arogyavardhini | Terminalia chebula | Fruit | CID_5281 | 5465 | PPARA | ChEMBL \| NPASS |
| Arogyavardhini | Terminalia chebula | Fruit | CID_5281 | 5467 | PPARD | ChEMBL \| NPASS |
| Arogyavardhini | Terminalia chebula | Fruit | CID_5281 | 5468 | PPARG | BindingDB \| ChEMBL \| NPASS |
| Arogyavardhini | Terminalia chebula | Fruit | CID_5281 | 55775 | TDP1 | NPASS |
| Arogyavardhini | Terminalia chebula | Fruit | CID_5281 | 5742 | PTGS1 | ChEMBL |
| Arogyavardhini | Terminalia chebula | Fruit | CID_5281 | 5770 | PTPN1 | BindingDB \| ChEMBL \| NPASS |
| Arogyavardhini | Terminalia chebula | Fruit | CID_5281 | 6530 | SLC6A2 | ChEMBL |
| Arogyavardhini | Terminalia chebula | Fruit | CID_5281 | 6531 | SLC6A3 | ChEMBL |
| Arogyavardhini | Terminalia chebula | Fruit | CID_5281 | 6532 | SLC6A4 | ChEMBL |
| Arogyavardhini | Terminalia chebula | Fruit | CID_5281 | 6915 | TBXA2R | ChEMBL |
| Arogyavardhini | Terminalia chebula | Fruit | CID_5281 | 7068 | THRB | NPASS |
| Arogyavardhini | Terminalia chebula | Fruit | CID_5281 | 7253 | TSHR | NPASS |
| Arogyavardhini | Terminalia chebula | Fruit | CID_5281 | 7421 | VDR | NPASS |
| Arogyavardhini | Terminalia chebula | Fruit | CID_5281855 | 10013 | HDAC6 | ChEMBL |
| Arogyavardhini | Terminalia chebula | Fruit | CID_5281855 | 10073 | SNUPN | NPASS |
| Arogyavardhini | Terminalia chebula | Fruit | CID_5281855 | 1017 | CDK2 | BindingDB \| ChEMBL |
| Arogyavardhini | Terminalia chebula | Fruit | CID_5281855 | 1019 | CDK4 | BindingDB \| ChEMBL |
| Arogyavardhini | Terminalia chebula | Fruit | CID_5281855 | 10411 | RAPGEF3 | NPASS |
| Arogyavardhini | Terminalia chebula | Fruit | CID_5281855 | 10498 | CARM1 | ChEMBL \| NPASS |
| Arogyavardhini | Terminalia chebula | Fruit | CID_5281855 | 10587 | TXNRD2 | ChEMBL |
| Arogyavardhini | Terminalia chebula | Fruit | CID_5281855 | 10599 | SLCO1B1 | ChEMBL \| NPASS |
| Arogyavardhini | Terminalia chebula | Fruit | CID_5281855 | 10733 | PLK4 | BindingDB \| ChEMBL \| NPASS |

**Table S3: Associations between formulations, constituent plants, phytochemicals, and human molecular targets. The table lists the formulations, plant name and plant part, phytochemical identifiers, associated human target genes (Entrez Gene ID and gene symbol), and the database sources supporting each phytochemical-target interaction.**

| Formulation | Plant name | Part | PubChem CID | Entrez Gene ID | Gene Symbol | Source database(s) |
|---|---|---|---|---|---|---|
| Arogyavardhini | Terminalia chebula | Fruit | CID_5281855 | 10919 | EHMT2 | ChEMBL \| NPASS |
| Arogyavardhini | Terminalia chebula | Fruit | CID_5281855 | 10951 | CBX1 | NPASS |
| Arogyavardhini | Terminalia chebula | Fruit | CID_5281855 | 11069 | RAPGEF4 | NPASS |
| Arogyavardhini | Terminalia chebula | Fruit | CID_5281855 | 11201 | POLI | ChEMBL \| NPASS |
| Arogyavardhini | Terminalia chebula | Fruit | CID_5281855 | 11238 | CA5B | BindingDB \| ChEMBL \| NPASS |
| Arogyavardhini | Terminalia chebula | Fruit | CID_5281855 | 114112 | TXNRD3 | ChEMBL \| NPASS |
| Arogyavardhini | Terminalia chebula | Fruit | CID_5281855 | 1326 | MAP3K8 | BindingDB \| ChEMBL \| NPASS |
| Arogyavardhini | Terminalia chebula | Fruit | CID_5281855 | 1457 | CSNK2A1 | BindingDB \| ChEMBL \| NPASS |
| Arogyavardhini | Terminalia chebula | Fruit | CID_5281855 | 1459 | CSNK2A2 | ChEMBL \| NPASS |
| Arogyavardhini | Terminalia chebula | Fruit | CID_5281855 | 1460 | CSNK2B | ChEMBL |
| Arogyavardhini | Terminalia chebula | Fruit | CID_5281855 | 1557 | CYP2C19 | ChEMBL \| NPASS |
| Arogyavardhini | Terminalia chebula | Fruit | CID_5281855 | 1559 | CYP2C9 | NPASS |
| Arogyavardhini | Terminalia chebula | Fruit | CID_5281855 | 1636 | ACE | BindingDB \| NPASS |
| Arogyavardhini | Terminalia chebula | Fruit | CID_5281855 | 1845 | DUSP3 | BindingDB \| NPASS |
| Arogyavardhini | Terminalia chebula | Fruit | CID_5281855 | 1859 | DYRK1A | BindingDB \| ChEMBL \| NPASS |
| Arogyavardhini | Terminalia chebula | Fruit | CID_5281855 | 1956 | EGFR | BindingDB \| ChEMBL \| NPASS |
| Arogyavardhini | Terminalia chebula | Fruit | CID_5281855 | 1981 | EIF4G1 | BindingDB |
| Arogyavardhini | Terminalia chebula | Fruit | CID_5281855 | 1995 | ELAVL3 | ChEMBL \| NPASS |
| Arogyavardhini | Terminalia chebula | Fruit | CID_5281855 | 2050 | EPHB4 | BindingDB \| ChEMBL \| NPASS |
| Arogyavardhini | Terminalia chebula | Fruit | CID_5281855 | 2053 | EPHX2 | ChEMBL \| NPASS |
| Arogyavardhini | Terminalia chebula | Fruit | CID_5281855 | 2064 | ERBB2 | ChEMBL \| NPASS |
| Arogyavardhini | Terminalia chebula | Fruit | CID_5281855 | 207 | AKT1 | ChEMBL \| NPASS |
| Arogyavardhini | Terminalia chebula | Fruit | CID_5281855 | 2237 | FEN1 | ChEMBL \| NPASS |
| Arogyavardhini | Terminalia chebula | Fruit | CID_5281855 | 231 | AKR1B1 | ChEMBL \| NPASS |
| Arogyavardhini | Terminalia chebula | Fruit | CID_5281855 | 23192 | ATG4B | BindingDB \| ChEMBL \| NPASS |
| Arogyavardhini | Terminalia chebula | Fruit | CID_5281855 | 2322 | FLT3 | ChEMBL \| NPASS |
| Arogyavardhini | Terminalia chebula | Fruit | CID_5281855 | 2324 | FLT4 | BindingDB \| ChEMBL \| NPASS |
| Arogyavardhini | Terminalia chebula | Fruit | CID_5281855 | 23405 | DICER1 | ChEMBL |
| Arogyavardhini | Terminalia chebula | Fruit | CID_5281855 | 23621 | BACE1 | BindingDB \| ChEMBL \| NPASS |
| Arogyavardhini | Terminalia chebula | Fruit | CID_5281855 | 23632 | CA14 | BindingDB \| ChEMBL \| NPASS |
| Arogyavardhini | Terminalia chebula | Fruit | CID_5281855 | 238 | ALK | BindingDB \| ChEMBL |
| Arogyavardhini | Terminalia chebula | Fruit | CID_5281855 | 246 | ALOX15 | NPASS |
| Arogyavardhini | Terminalia chebula | Fruit | CID_5281855 | 247 | ALOX15B | NPASS |
| Arogyavardhini | Terminalia chebula | Fruit | CID_5281855 | 2548 | GAA | NPASS |
| Arogyavardhini | Terminalia chebula | Fruit | CID_5281855 | 2590 | GALNT2 | ChEMBL \| NPASS |
| Arogyavardhini | Terminalia chebula | Fruit | CID_5281855 | 2597 | GAPDH | NPASS |
| Arogyavardhini | Terminalia chebula | Fruit | CID_5281855 | 2648 | KAT2A | NPASS |
| Arogyavardhini | Terminalia chebula | Fruit | CID_5281855 | 2744 | GLS | ChEMBL \| NPASS |
| Arogyavardhini | Terminalia chebula | Fruit | CID_5281855 | 28234 | SLCO1B3 | ChEMBL \| NPASS |
| Arogyavardhini | Terminalia chebula | Fruit | CID_5281855 | 2859 | GPR35 | BindingDB \| ChEMBL \| NPASS |
| Arogyavardhini | Terminalia chebula | Fruit | CID_5281855 | 2932 | GSK3B | BindingDB \| ChEMBL \| NPASS |
| Arogyavardhini | Terminalia chebula | Fruit | CID_5281855 | 2936 | GSR | ChEMBL \| NPASS |
| Arogyavardhini | Terminalia chebula | Fruit | CID_5281855 | 2950 | GSTP1 | ChEMBL \| NPASS |
| Arogyavardhini | Terminalia chebula | Fruit | CID_5281855 | 29994 | BAZ2B | ChEMBL \| NPASS |
| Arogyavardhini | Terminalia chebula | Fruit | CID_5281855 | 3028 | HSD17B10 | ChEMBL \| NPASS |
| Arogyavardhini | Terminalia chebula | Fruit | CID_5281855 | 3043 | HBB | ChEMBL \| NPASS |
| Arogyavardhini | Terminalia chebula | Fruit | CID_5281855 | 3248 | HPGD | ChEMBL \| NPASS |
| Arogyavardhini | Terminalia chebula | Fruit | CID_5281855 | 328 | APEX1 | ChEMBL \| NPASS |
| Arogyavardhini | Terminalia chebula | Fruit | CID_5281855 | 3303 | HSPA1A | BindingDB \| ChEMBL \| NPASS |
| Arogyavardhini | Terminalia chebula | Fruit | CID_5281855 | 3309 | HSPA5 | NPASS |
| Arogyavardhini | Terminalia chebula | Fruit | CID_5281855 | 3312 | HSPA8 | BindingDB \| ChEMBL \| NPASS |
| Arogyavardhini | Terminalia chebula | Fruit | CID_5281855 | 3417 | IDH1 | NPASS |
| Arogyavardhini | Terminalia chebula | Fruit | CID_5281855 | 3480 | IGF1R | BindingDB \| ChEMBL \| NPASS |
| Arogyavardhini | Terminalia chebula | Fruit | CID_5281855 | 351 | APP | BindingDB \| ChEMBL \| NPASS |
| Arogyavardhini | Terminalia chebula | Fruit | CID_5281855 | 3551 | IKBKB | BindingDB \| ChEMBL \| NPASS |
| Arogyavardhini | Terminalia chebula | Fruit | CID_5281855 | 3643 | INSR | BindingDB \| ChEMBL \| NPASS |
| Arogyavardhini | Terminalia chebula | Fruit | CID_5281855 | 377677 | CA13 | ChEMBL \| NPASS |
| Arogyavardhini | Terminalia chebula | Fruit | CID_5281855 | 3791 | KDR | BindingDB \| ChEMBL \| NPASS |
| Arogyavardhini | Terminalia chebula | Fruit | CID_5281855 | 3837 | KPNB1 | NPASS |
| Arogyavardhini | Terminalia chebula | Fruit | CID_5281855 | 390245 | KDM4E | ChEMBL \| NPASS |
| Arogyavardhini | Terminalia chebula | Fruit | CID_5281855 | 4000 | LMNA | NPASS |
| Arogyavardhini | Terminalia chebula | Fruit | CID_5281855 | 4088 | SMAD3 | BindingDB \| ChEMBL \| NPASS |
| Arogyavardhini | Terminalia chebula | Fruit | CID_5281855 | 4128 | MAOA | ChEMBL \| NPASS |
| Arogyavardhini | Terminalia chebula | Fruit | CID_5281855 | 4137 | MAPT | ChEMBL \| NPASS |
| Arogyavardhini | Terminalia chebula | Fruit | CID_5281855 | 4154 | MBNL1 | NPASS |
| Arogyavardhini | Terminalia chebula | Fruit | CID_5281855 | 4170 | MCL1 | BindingDB \| ChEMBL \| NPASS |
| Arogyavardhini | Terminalia chebula | Fruit | CID_5281855 | 4233 | MET | BindingDB \| ChEMBL \| NPASS |
| Arogyavardhini | Terminalia chebula | Fruit | CID_5281855 | 4282 | MIF | BindingDB \| ChEMBL \| NPASS |
| Arogyavardhini | Terminalia chebula | Fruit | CID_5281855 | 4297 | KMT2A | NPASS |
| Arogyavardhini | Terminalia chebula | Fruit | CID_5281855 | 43 | ACHE | BindingDB |

**Table S3: Associations between formulations, constituent plants, phytochemicals, and human molecular targets. The table lists the formulations, plant name and plant part, phytochemical identifiers, associated human target genes (Entrez Gene ID and gene symbol), and the database sources supporting each phytochemical-target interaction.**

| Formulation | Plant name | Part | PubChem CID | Entrez Gene ID | Gene Symbol | Source database(s) |
|---|---|---|---|---|---|---|
| Arogyavardhini | Terminalia chebula | Fruit | CID_5281855 | 472 | ATM | NPASS |
| Arogyavardhini | Terminalia chebula | Fruit | CID_5281855 | 4780 | NFE2L2 | NPASS |
| Arogyavardhini | Terminalia chebula | Fruit | CID_5281855 | 4869 | NPM1 | ChEMBL \| NPASS |
| Arogyavardhini | Terminalia chebula | Fruit | CID_5281855 | 4907 | NT5E | ChEMBL \| NPASS |
| Arogyavardhini | Terminalia chebula | Fruit | CID_5281855 | 51053 | GMNN | NPASS |
| Arogyavardhini | Terminalia chebula | Fruit | CID_5281855 | 51426 | POLK | ChEMBL \| NPASS |
| Arogyavardhini | Terminalia chebula | Fruit | CID_5281855 | 5159 | PDGFRB | BindingDB \| ChEMBL \| NPASS |
| Arogyavardhini | Terminalia chebula | Fruit | CID_5281855 | 5300 | PIN1 | NPASS |
| Arogyavardhini | Terminalia chebula | Fruit | CID_5281855 | 5313 | PKLR | ChEMBL |
| Arogyavardhini | Terminalia chebula | Fruit | CID_5281855 | 5347 | PLK1 | BindingDB \| ChEMBL \| NPASS |
| Arogyavardhini | Terminalia chebula | Fruit | CID_5281855 | 5423 | POLB | ChEMBL \| NPASS |
| Arogyavardhini | Terminalia chebula | Fruit | CID_5281855 | 5429 | POLH | BindingDB \| ChEMBL \| NPASS |
| Arogyavardhini | Terminalia chebula | Fruit | CID_5281855 | 5566 | PRKACA | BindingDB \| ChEMBL |
| Arogyavardhini | Terminalia chebula | Fruit | CID_5281855 | 5567 | PRKACB | ChEMBL |
| Arogyavardhini | Terminalia chebula | Fruit | CID_5281855 | 5568 | PRKACG | ChEMBL \| NPASS |
| Arogyavardhini | Terminalia chebula | Fruit | CID_5281855 | 55775 | TDP1 | ChEMBL \| NPASS |
| Arogyavardhini | Terminalia chebula | Fruit | CID_5281855 | 55867 | SLC22A11 | ChEMBL \| NPASS |
| Arogyavardhini | Terminalia chebula | Fruit | CID_5281855 | 5594 | MAPK1 | ChEMBL \| NPASS |
| Arogyavardhini | Terminalia chebula | Fruit | CID_5281855 | 5747 | PTK2 | BindingDB \| ChEMBL \| NPASS |
| Arogyavardhini | Terminalia chebula | Fruit | CID_5281855 | 5770 | PTPN1 | BindingDB |
| Arogyavardhini | Terminalia chebula | Fruit | CID_5281855 | 57787 | MARK4 | ChEMBL |
| Arogyavardhini | Terminalia chebula | Fruit | CID_5281855 | 595 | CCND1 | ChEMBL \| NPASS |
| Arogyavardhini | Terminalia chebula | Fruit | CID_5281855 | 5965 | RECQL | ChEMBL \| NPASS |
| Arogyavardhini | Terminalia chebula | Fruit | CID_5281855 | 5979 | RET | BindingDB \| ChEMBL \| NPASS |
| Arogyavardhini | Terminalia chebula | Fruit | CID_5281855 | 5999 | RGS4 | NPASS |
| Arogyavardhini | Terminalia chebula | Fruit | CID_5281855 | 641 | BLM | ChEMBL \| NPASS |
| Arogyavardhini | Terminalia chebula | Fruit | CID_5281855 | 6714 | SRC | ChEMBL \| NPASS |
| Arogyavardhini | Terminalia chebula | Fruit | CID_5281855 | 673 | BRAF | BindingDB \| ChEMBL \| NPASS |
| Arogyavardhini | Terminalia chebula | Fruit | CID_5281855 | 6790 | AURKA | BindingDB \| ChEMBL \| NPASS |
| Arogyavardhini | Terminalia chebula | Fruit | CID_5281855 | 7010 | TEK | BindingDB \| ChEMBL \| NPASS |
| Arogyavardhini | Terminalia chebula | Fruit | CID_5281855 | 7068 | THRB | NPASS |
| Arogyavardhini | Terminalia chebula | Fruit | CID_5281855 | 7155 | TOP2B | ChEMBL \| NPASS |
| Arogyavardhini | Terminalia chebula | Fruit | CID_5281855 | 7157 | TP53 | NPASS |
| Arogyavardhini | Terminalia chebula | Fruit | CID_5281855 | 7257 | TSNAX | ChEMBL \| NPASS |
| Arogyavardhini | Terminalia chebula | Fruit | CID_5281855 | 7296 | TXNRD1 | ChEMBL |
| Arogyavardhini | Terminalia chebula | Fruit | CID_5281855 | 7398 | USP1 | NPASS |
| Arogyavardhini | Terminalia chebula | Fruit | CID_5281855 | 7421 | VDR | ChEMBL \| NPASS |
| Arogyavardhini | Terminalia chebula | Fruit | CID_5281855 | 759 | CA1 | BindingDB \| ChEMBL \| NPASS |
| Arogyavardhini | Terminalia chebula | Fruit | CID_5281855 | 760 | CA2 | BindingDB \| ChEMBL \| NPASS |
| Arogyavardhini | Terminalia chebula | Fruit | CID_5281855 | 761 | CA3 | BindingDB \| ChEMBL \| NPASS |
| Arogyavardhini | Terminalia chebula | Fruit | CID_5281855 | 762 | CA4 | BindingDB \| ChEMBL \| NPASS |
| Arogyavardhini | Terminalia chebula | Fruit | CID_5281855 | 763 | CA5A | BindingDB \| ChEMBL \| NPASS |
| Arogyavardhini | Terminalia chebula | Fruit | CID_5281855 | 765 | CA6 | ChEMBL \| NPASS |
| Arogyavardhini | Terminalia chebula | Fruit | CID_5281855 | 766 | CA7 | BindingDB \| ChEMBL \| NPASS |
| Arogyavardhini | Terminalia chebula | Fruit | CID_5281855 | 768 | CA9 | BindingDB \| ChEMBL \| NPASS |
| Arogyavardhini | Terminalia chebula | Fruit | CID_5281855 | 771 | CA12 | BindingDB \| ChEMBL \| NPASS |
| Arogyavardhini | Terminalia chebula | Fruit | CID_5281855 | 861 | RUNX1 | ChEMBL |
| Arogyavardhini | Terminalia chebula | Fruit | CID_5281855 | 865 | CBFB | ChEMBL \| NPASS |
| Arogyavardhini | Terminalia chebula | Fruit | CID_5281855 | 8877 | SPHK1 | ChEMBL |
| Arogyavardhini | Terminalia chebula | Fruit | CID_5281855 | 890 | CCNA2 | ChEMBL |
| Arogyavardhini | Terminalia chebula | Fruit | CID_5281855 | 8900 | CCNA1 | ChEMBL \| NPASS |
| Arogyavardhini | Terminalia chebula | Fruit | CID_5281855 | 9099 | USP2 | ChEMBL \| NPASS |
| Arogyavardhini | Terminalia chebula | Fruit | CID_5281855 | 9212 | AURKB | BindingDB \| ChEMBL \| NPASS |
| Arogyavardhini | Terminalia chebula | Fruit | CID_5281855 | 9356 | SLC22A6 | BindingDB \| ChEMBL \| NPASS |
| Arogyavardhini | Terminalia chebula | Fruit | CID_5281855 | 9682 | KDM4A | ChEMBL \| NPASS |
| Arogyavardhini | Terminalia chebula | Fruit | CID_5281855 | 9891 | NUAK1 | BindingDB \| ChEMBL \| NPASS |
| Arogyavardhini | Terminalia chebula | Fruit | CID_54670067 | 10280 | SIGMAR1 | ChEMBL |
| Arogyavardhini | Terminalia chebula | Fruit | CID_54670067 | 10599 | SLCO1B1 | ChEMBL \| NPASS |
| Arogyavardhini | Terminalia chebula | Fruit | CID_54670067 | 10800 | CYSLTR1 | ChEMBL |
| Arogyavardhini | Terminalia chebula | Fruit | CID_54670067 | 10919 | EHMT2 | ChEMBL |
| Arogyavardhini | Terminalia chebula | Fruit | CID_54670067 | 112398 | EGLN2 | ChEMBL \| NPASS |
| Arogyavardhini | Terminalia chebula | Fruit | CID_54670067 | 112399 | EGLN3 | ChEMBL \| NPASS |
| Arogyavardhini | Terminalia chebula | Fruit | CID_54670067 | 11255 | HRH3 | ChEMBL |
| Arogyavardhini | Terminalia chebula | Fruit | CID_54670067 | 1128 | CHRM1 | ChEMBL |
| Arogyavardhini | Terminalia chebula | Fruit | CID_54670067 | 1129 | CHRM2 | ChEMBL |
| Arogyavardhini | Terminalia chebula | Fruit | CID_54670067 | 1131 | CHRM3 | ChEMBL |
| Arogyavardhini | Terminalia chebula | Fruit | CID_54670067 | 1132 | CHRM4 | ChEMBL |
| Arogyavardhini | Terminalia chebula | Fruit | CID_54670067 | 1133 | CHRM5 | ChEMBL |
| Arogyavardhini | Terminalia chebula | Fruit | CID_54670067 | 1233 | CCR4 | ChEMBL |
| Arogyavardhini | Terminalia chebula | Fruit | CID_54670067 | 1234 | CCR5 | ChEMBL |

| Table S3: Associations between formulations, constituent plants, phytochemicals, and human molecular targets. The table lists the formulations, plant name and plant part, phytochemical identifiers, associated human target genes (Entrez Gene ID and gene symbol), and the database sources supporting each phytochemical-target interaction. | | | | | | |
|---|---|---|---|---|---|---|
| **Formulation** | **Plant name** | **Part** | **PubChem CID** | **Entrez Gene ID** | **Gene Symbol** | **Source database(s)** |
| Arogyavardhini | Terminalia chebula | Fruit | CID_54670067 | 1268 | CNR1 | ChEMBL |
| Arogyavardhini | Terminalia chebula | Fruit | CID_54670067 | 134 | ADORA1 | ChEMBL |
| Arogyavardhini | Terminalia chebula | Fruit | CID_54670067 | 135 | ADORA2A | ChEMBL |
| Arogyavardhini | Terminalia chebula | Fruit | CID_54670067 | 140 | ADORA3 | ChEMBL |
| Arogyavardhini | Terminalia chebula | Fruit | CID_54670067 | 1432 | MAPK14 | ChEMBL |
| Arogyavardhini | Terminalia chebula | Fruit | CID_54670067 | 146 | ADRA1D | ChEMBL |
| Arogyavardhini | Terminalia chebula | Fruit | CID_54670067 | 148 | ADRA1A | ChEMBL |
| Arogyavardhini | Terminalia chebula | Fruit | CID_54670067 | 150 | ADRA2A | ChEMBL |
| Arogyavardhini | Terminalia chebula | Fruit | CID_54670067 | 151 | ADRA2B | ChEMBL \| NPASS |
| Arogyavardhini | Terminalia chebula | Fruit | CID_54670067 | 1511 | CTSG | ChEMBL |
| Arogyavardhini | Terminalia chebula | Fruit | CID_54670067 | 152 | ADRA2C | ChEMBL |
| Arogyavardhini | Terminalia chebula | Fruit | CID_54670067 | 153 | ADRB1 | ChEMBL |
| Arogyavardhini | Terminalia chebula | Fruit | CID_54670067 | 154 | ADRB2 | ChEMBL |
| Arogyavardhini | Terminalia chebula | Fruit | CID_54670067 | 1544 | CYP1A2 | ChEMBL |
| Arogyavardhini | Terminalia chebula | Fruit | CID_54670067 | 1548 | CYP2A6 | ChEMBL |
| Arogyavardhini | Terminalia chebula | Fruit | CID_54670067 | 155 | ADRB3 | ChEMBL |
| Arogyavardhini | Terminalia chebula | Fruit | CID_54670067 | 1557 | CYP2C19 | ChEMBL |
| Arogyavardhini | Terminalia chebula | Fruit | CID_54670067 | 1559 | CYP2C9 | ChEMBL |
| Arogyavardhini | Terminalia chebula | Fruit | CID_54670067 | 1565 | CYP2D6 | ChEMBL |
| Arogyavardhini | Terminalia chebula | Fruit | CID_54670067 | 1571 | CYP2E1 | ChEMBL |
| Arogyavardhini | Terminalia chebula | Fruit | CID_54670067 | 1576 | CYP3A4 | ChEMBL |
| Arogyavardhini | Terminalia chebula | Fruit | CID_54670067 | 1812 | DRD1 | ChEMBL |
| Arogyavardhini | Terminalia chebula | Fruit | CID_54670067 | 1813 | DRD2 | ChEMBL |
| Arogyavardhini | Terminalia chebula | Fruit | CID_54670067 | 1814 | DRD3 | ChEMBL |
| Arogyavardhini | Terminalia chebula | Fruit | CID_54670067 | 1815 | DRD4 | ChEMBL |
| Arogyavardhini | Terminalia chebula | Fruit | CID_54670067 | 186 | AGTR2 | ChEMBL |
| Arogyavardhini | Terminalia chebula | Fruit | CID_54670067 | 1909 | EDNRA | ChEMBL |
| Arogyavardhini | Terminalia chebula | Fruit | CID_54670067 | 1956 | EGFR | ChEMBL |
| Arogyavardhini | Terminalia chebula | Fruit | CID_54670067 | 1991 | ELANE | ChEMBL |
| Arogyavardhini | Terminalia chebula | Fruit | CID_54670067 | 2064 | ERBB2 | ChEMBL |
| Arogyavardhini | Terminalia chebula | Fruit | CID_54670067 | 2099 | ESR1 | ChEMBL |
| Arogyavardhini | Terminalia chebula | Fruit | CID_54670067 | 2100 | ESR2 | ChEMBL |
| Arogyavardhini | Terminalia chebula | Fruit | CID_54670067 | 2147 | F2 | ChEMBL |
| Arogyavardhini | Terminalia chebula | Fruit | CID_54670067 | 2321 | FLT1 | ChEMBL |
| Arogyavardhini | Terminalia chebula | Fruit | CID_54670067 | 239 | ALOX12 | ChEMBL \| NPASS |
| Arogyavardhini | Terminalia chebula | Fruit | CID_54670067 | 240 | ALOX5 | ChEMBL \| NPASS |
| Arogyavardhini | Terminalia chebula | Fruit | CID_54670067 | 246 | ALOX15 | ChEMBL \| NPASS |
| Arogyavardhini | Terminalia chebula | Fruit | CID_54670067 | 247 | ALOX15B | ChEMBL \| NPASS |
| Arogyavardhini | Terminalia chebula | Fruit | CID_54670067 | 2534 | FYN | ChEMBL |
| Arogyavardhini | Terminalia chebula | Fruit | CID_54670067 | 2554 | GABRA1 | ChEMBL |
| Arogyavardhini | Terminalia chebula | Fruit | CID_54670067 | 2555 | GABRA2 | ChEMBL |
| Arogyavardhini | Terminalia chebula | Fruit | CID_54670067 | 2561 | GABRB2 | ChEMBL |
| Arogyavardhini | Terminalia chebula | Fruit | CID_54670067 | 2566 | GABRG2 | ChEMBL |
| Arogyavardhini | Terminalia chebula | Fruit | CID_54670067 | 25939 | SAMHD1 | ChEMBL |
| Arogyavardhini | Terminalia chebula | Fruit | CID_54670067 | 276 | AMY1A | ChEMBL |
| Arogyavardhini | Terminalia chebula | Fruit | CID_54670067 | 279 | AMY2A | ChEMBL \| NPASS |
| Arogyavardhini | Terminalia chebula | Fruit | CID_54670067 | 28234 | SLCO1B3 | ChEMBL \| NPASS |
| Arogyavardhini | Terminalia chebula | Fruit | CID_54670067 | 2908 | NR3C1 | ChEMBL |
| Arogyavardhini | Terminalia chebula | Fruit | CID_54670067 | 2932 | GSK3B | ChEMBL \| NPASS |
| Arogyavardhini | Terminalia chebula | Fruit | CID_54670067 | 3156 | HMGCR | ChEMBL |
| Arogyavardhini | Terminalia chebula | Fruit | CID_54670067 | 3269 | HRH1 | ChEMBL |
| Arogyavardhini | Terminalia chebula | Fruit | CID_54670067 | 3274 | HRH2 | ChEMBL |
| Arogyavardhini | Terminalia chebula | Fruit | CID_54670067 | 3350 | HTR1A | ChEMBL |
| Arogyavardhini | Terminalia chebula | Fruit | CID_54670067 | 3356 | HTR2A | ChEMBL |
| Arogyavardhini | Terminalia chebula | Fruit | CID_54670067 | 3357 | HTR2B | ChEMBL |
| Arogyavardhini | Terminalia chebula | Fruit | CID_54670067 | 3358 | HTR2C | ChEMBL |
| Arogyavardhini | Terminalia chebula | Fruit | CID_54670067 | 3362 | HTR6 | ChEMBL |
| Arogyavardhini | Terminalia chebula | Fruit | CID_54670067 | 351 | APP | ChEMBL \| NPASS |
| Arogyavardhini | Terminalia chebula | Fruit | CID_54670067 | 3577 | CXCR1 | ChEMBL |
| Arogyavardhini | Terminalia chebula | Fruit | CID_54670067 | 3579 | CXCR2 | ChEMBL |
| Arogyavardhini | Terminalia chebula | Fruit | CID_54670067 | 367 | AR | ChEMBL \| NPASS |
| Arogyavardhini | Terminalia chebula | Fruit | CID_54670067 | 3757 | KCNH2 | ChEMBL |
| Arogyavardhini | Terminalia chebula | Fruit | CID_54670067 | 3791 | KDR | ChEMBL |
| Arogyavardhini | Terminalia chebula | Fruit | CID_54670067 | 3932 | LCK | ChEMBL |
| Arogyavardhini | Terminalia chebula | Fruit | CID_54670067 | 4128 | MAOA | ChEMBL |
| Arogyavardhini | Terminalia chebula | Fruit | CID_54670067 | 4159 | MC3R | ChEMBL |
| Arogyavardhini | Terminalia chebula | Fruit | CID_54670067 | 4160 | MC4R | ChEMBL |
| Arogyavardhini | Terminalia chebula | Fruit | CID_54670067 | 4161 | MC5R | ChEMBL |
| Arogyavardhini | Terminalia chebula | Fruit | CID_54670067 | 43 | ACHE | ChEMBL |
| Arogyavardhini | Terminalia chebula | Fruit | CID_54670067 | 4312 | MMP1 | ChEMBL |

**Table S3: Associations between formulations, constituent plants, phytochemicals, and human molecular targets. The table lists the formulations, plant name and plant part, phytochemical identifiers, associated human target genes (Entrez Gene ID and gene symbol), and the database sources supporting each phytochemical-target interaction.**

| Formulation | Plant name | Part | PubChem CID | Entrez Gene ID | Gene Symbol | Source database(s) |
|---|---|---|---|---|---|---|
| Arogyavardhini | Terminalia chebula | Fruit | CID_54670067 | 4318 | MMP9 | ChEMBL |
| Arogyavardhini | Terminalia chebula | Fruit | CID_54670067 | 4780 | NFE2L2 | ChEMBL \| NPASS |
| Arogyavardhini | Terminalia chebula | Fruit | CID_54670067 | 4886 | NPY1R | ChEMBL |
| Arogyavardhini | Terminalia chebula | Fruit | CID_54670067 | 4887 | NPY2R | ChEMBL |
| Arogyavardhini | Terminalia chebula | Fruit | CID_54670067 | 4985 | OPRD1 | ChEMBL |
| Arogyavardhini | Terminalia chebula | Fruit | CID_54670067 | 4986 | OPRK1 | ChEMBL |
| Arogyavardhini | Terminalia chebula | Fruit | CID_54670067 | 4988 | OPRM1 | ChEMBL |
| Arogyavardhini | Terminalia chebula | Fruit | CID_54670067 | 5033 | P4HA1 | ChEMBL \| NPASS |
| Arogyavardhini | Terminalia chebula | Fruit | CID_54670067 | 50507 | NOX4 | ChEMBL \| NPASS |
| Arogyavardhini | Terminalia chebula | Fruit | CID_54670067 | 5139 | PDE3A | ChEMBL |
| Arogyavardhini | Terminalia chebula | Fruit | CID_54670067 | 5141 | PDE4A | ChEMBL |
| Arogyavardhini | Terminalia chebula | Fruit | CID_54670067 | 5241 | PGR | ChEMBL |
| Arogyavardhini | Terminalia chebula | Fruit | CID_54670067 | 54583 | EGLN1 | ChEMBL \| NPASS |
| Arogyavardhini | Terminalia chebula | Fruit | CID_54670067 | 552 | AVPR1A | ChEMBL |
| Arogyavardhini | Terminalia chebula | Fruit | CID_54670067 | 5530 | PPP3CA | ChEMBL |
| Arogyavardhini | Terminalia chebula | Fruit | CID_54670067 | 55662 | HIF1AN | ChEMBL \| NPASS |
| Arogyavardhini | Terminalia chebula | Fruit | CID_54670067 | 5578 | PRKCA | ChEMBL |
| Arogyavardhini | Terminalia chebula | Fruit | CID_54670067 | 5594 | MAPK1 | ChEMBL |
| Arogyavardhini | Terminalia chebula | Fruit | CID_54670067 | 5595 | MAPK3 | ChEMBL |
| Arogyavardhini | Terminalia chebula | Fruit | CID_54670067 | 5724 | PTAFR | ChEMBL |
| Arogyavardhini | Terminalia chebula | Fruit | CID_54670067 | 5742 | PTGS1 | ChEMBL |
| Arogyavardhini | Terminalia chebula | Fruit | CID_54670067 | 5743 | PTGS2 | ChEMBL |
| Arogyavardhini | Terminalia chebula | Fruit | CID_54670067 | 5788 | PTPRC | ChEMBL |
| Arogyavardhini | Terminalia chebula | Fruit | CID_54670067 | 624 | BDKRB2 | ChEMBL |
| Arogyavardhini | Terminalia chebula | Fruit | CID_54670067 | 6530 | SLC6A2 | ChEMBL |
| Arogyavardhini | Terminalia chebula | Fruit | CID_54670067 | 6531 | SLC6A3 | ChEMBL |
| Arogyavardhini | Terminalia chebula | Fruit | CID_54670067 | 6532 | SLC6A4 | ChEMBL |
| Arogyavardhini | Terminalia chebula | Fruit | CID_54670067 | 6865 | TACR2 | ChEMBL |
| Arogyavardhini | Terminalia chebula | Fruit | CID_54670067 | 6869 | TACR1 | ChEMBL |
| Arogyavardhini | Terminalia chebula | Fruit | CID_54670067 | 6915 | TBXA2R | ChEMBL |
| Arogyavardhini | Terminalia chebula | Fruit | CID_54670067 | 6916 | TBXAS1 | ChEMBL |
| Arogyavardhini | Terminalia chebula | Fruit | CID_54670067 | 729230 | CCR2 | ChEMBL |
| Arogyavardhini | Terminalia chebula | Fruit | CID_54670067 | 7299 | TYR | ChEMBL \| NPASS |
| Arogyavardhini | Terminalia chebula | Fruit | CID_54670067 | 7433 | VIPR1 | ChEMBL |
| Arogyavardhini | Terminalia chebula | Fruit | CID_54670067 | 760 | CA2 | ChEMBL |
| Arogyavardhini | Terminalia chebula | Fruit | CID_54670067 | 799 | CALCR | ChEMBL |
| Arogyavardhini | Terminalia chebula | Fruit | CID_54670067 | 834 | CASP1 | ChEMBL |
| Arogyavardhini | Terminalia chebula | Fruit | CID_54670067 | 8654 | PDE5A | ChEMBL |
| Arogyavardhini | Terminalia chebula | Fruit | CID_54670067 | 886 | CCKAR | ChEMBL |
| Arogyavardhini | Terminalia chebula | Fruit | CID_5742590 | 10013 | HDAC6 | ChEMBL |
| Arogyavardhini | Terminalia chebula | Fruit | CID_5742590 | 2932 | GSK3B | ChEMBL |
| Arogyavardhini | Terminalia chebula | Fruit | CID_5742590 | 3417 | IDH1 | ChEMBL \| NPASS |
| Arogyavardhini | Terminalia chebula | Fruit | CID_5742590 | 51422 | PRKAG2 | ChEMBL |
| Arogyavardhini | Terminalia chebula | Fruit | CID_5742590 | 53632 | PRKAG3 | ChEMBL |
| Arogyavardhini | Terminalia chebula | Fruit | CID_5742590 | 5423 | POLB | BindingDB \| ChEMBL \| NPASS |
| Arogyavardhini | Terminalia chebula | Fruit | CID_5742590 | 5562 | PRKAA1 | ChEMBL |
| Arogyavardhini | Terminalia chebula | Fruit | CID_5742590 | 5563 | PRKAA2 | ChEMBL |
| Arogyavardhini | Terminalia chebula | Fruit | CID_5742590 | 5564 | PRKAB1 | ChEMBL |
| Arogyavardhini | Terminalia chebula | Fruit | CID_5742590 | 5565 | PRKAB2 | ChEMBL |
| Arogyavardhini | Terminalia chebula | Fruit | CID_5742590 | 5571 | PRKAG1 | ChEMBL |
| Arogyavardhini | Terminalia chebula | Fruit | CID_637542 | 10013 | HDAC6 | ChEMBL |
| Arogyavardhini | Terminalia chebula | Fruit | CID_637542 | 1109 | AKR1C4 | ChEMBL \| NPASS |
| Arogyavardhini | Terminalia chebula | Fruit | CID_637542 | 11238 | CA5B | BindingDB \| ChEMBL \| NPASS |
| Arogyavardhini | Terminalia chebula | Fruit | CID_637542 | 1645 | AKR1C1 | ChEMBL \| NPASS |
| Arogyavardhini | Terminalia chebula | Fruit | CID_637542 | 1646 | AKR1C2 | ChEMBL \| NPASS |
| Arogyavardhini | Terminalia chebula | Fruit | CID_637542 | 1803 | DPP4 | ChEMBL \| NPASS |
| Arogyavardhini | Terminalia chebula | Fruit | CID_637542 | 1956 | EGFR | BindingDB \| NPASS |
| Arogyavardhini | Terminalia chebula | Fruit | CID_637542 | 2099 | ESR1 | ChEMBL \| NPASS |
| Arogyavardhini | Terminalia chebula | Fruit | CID_637542 | 2100 | ESR2 | ChEMBL \| NPASS |
| Arogyavardhini | Terminalia chebula | Fruit | CID_637542 | 231 | AKR1B1 | ChEMBL \| NPASS |
| Arogyavardhini | Terminalia chebula | Fruit | CID_637542 | 23632 | CA14 | BindingDB \| ChEMBL \| NPASS |
| Arogyavardhini | Terminalia chebula | Fruit | CID_637542 | 259285 | TAS2R39 | ChEMBL \| NPASS |
| Arogyavardhini | Terminalia chebula | Fruit | CID_637542 | 338442 | HCAR2 | BindingDB \| ChEMBL \| NPASS |
| Arogyavardhini | Terminalia chebula | Fruit | CID_637542 | 5315 | PKM | ChEMBL |
| Arogyavardhini | Terminalia chebula | Fruit | CID_637542 | 57016 | AKR1B10 | ChEMBL \| NPASS |
| Arogyavardhini | Terminalia chebula | Fruit | CID_637542 | 5743 | PTGS2 | ChEMBL \| NPASS |
| Arogyavardhini | Terminalia chebula | Fruit | CID_637542 | 7299 | TYR | BindingDB \| ChEMBL \| NPASS |
| Arogyavardhini | Terminalia chebula | Fruit | CID_637542 | 759 | CA1 | BindingDB \| ChEMBL \| NPASS |
| Arogyavardhini | Terminalia chebula | Fruit | CID_637542 | 760 | CA2 | BindingDB \| ChEMBL \| NPASS |
| Arogyavardhini | Terminalia chebula | Fruit | CID_637542 | 761 | CA3 | BindingDB \| ChEMBL \| NPASS |

| Table S3: Associations between formulations, constituent plants, phytochemicals, and human molecular targets. The table lists the formulations, plant name and plant part, phytochemical identifiers, associated human target genes (Entrez Gene ID and gene symbol), and the database sources supporting each phytochemical-target interaction. | | | | | | |
|---|---|---|---|---|---|---|
| **Formulation** | **Plant name** | **Part** | **PubChem CID** | **Entrez Gene ID** | **Gene Symbol** | **Source database(s)** |
| Arogyavardhini | Terminalia chebula | Fruit | CID_637542 | 762 | CA4 | BindingDB \| ChEMBL \| NPASS |
| Arogyavardhini | Terminalia chebula | Fruit | CID_637542 | 763 | CA5A | BindingDB \| ChEMBL \| NPASS |
| Arogyavardhini | Terminalia chebula | Fruit | CID_637542 | 765 | CA6 | ChEMBL \| NPASS |
| Arogyavardhini | Terminalia chebula | Fruit | CID_637542 | 766 | CA7 | BindingDB \| ChEMBL \| NPASS |
| Arogyavardhini | Terminalia chebula | Fruit | CID_637542 | 768 | CA9 | BindingDB \| ChEMBL \| NPASS |
| Arogyavardhini | Terminalia chebula | Fruit | CID_637542 | 771 | CA12 | BindingDB \| ChEMBL \| NPASS |
| Arogyavardhini | Terminalia chebula | Fruit | CID_637542 | 815 | CAMK2A | BindingDB \| ChEMBL |
| Arogyavardhini | Terminalia chebula | Fruit | CID_637542 | 8644 | AKR1C3 | ChEMBL \| NPASS |
| Arogyavardhini | Terminalia chebula | Fruit | CID_6780 | 10013 | HDAC6 | ChEMBL |
| Arogyavardhini | Terminalia chebula | Fruit | CID_6780 | 10599 | SLCO1B1 | ChEMBL \| NPASS |
| Arogyavardhini | Terminalia chebula | Fruit | CID_6780 | 10919 | EHMT2 | ChEMBL \| NPASS |
| Arogyavardhini | Terminalia chebula | Fruit | CID_6780 | 2099 | ESR1 | ChEMBL \| NPASS |
| Arogyavardhini | Terminalia chebula | Fruit | CID_6780 | 2100 | ESR2 | ChEMBL |
| Arogyavardhini | Terminalia chebula | Fruit | CID_6780 | 28234 | SLCO1B3 | ChEMBL \| NPASS |
| Arogyavardhini | Terminalia chebula | Fruit | CID_6780 | 4129 | MAOB | ChEMBL |
| Arogyavardhini | Terminalia chebula | Fruit | CID_6780 | 4297 | KMT2A | NPASS |
| Arogyavardhini | Terminalia chebula | Fruit | CID_6780 | 4780 | NFE2L2 | NPASS |
| Arogyavardhini | Terminalia chebula | Fruit | CID_6780 | 5166 | PDK4 | ChEMBL \| NPASS |
| Arogyavardhini | Terminalia chebula | Fruit | CID_6780 | 54576 | UGT1A8 | ChEMBL |
| Arogyavardhini | Terminalia chebula | Fruit | CID_6780 | 54658 | UGT1A1 | ChEMBL \| NPASS |
| Arogyavardhini | Terminalia chebula | Fruit | CID_6780 | 6256 | RXRA | ChEMBL \| NPASS |
| Arogyavardhini | Terminalia chebula | Fruit | CID_6780 | 79813 | EHMT1 | BindingDB \| ChEMBL \| NPASS |
| Arogyavardhini | Terminalia chebula | Fruit | CID_6780 | 994 | CDC25B | BindingDB \| ChEMBL \| NPASS |
| Arogyavardhini | Terminalia chebula | Fruit | CID_689043 | 10013 | HDAC6 | ChEMBL |
| Arogyavardhini | Terminalia chebula | Fruit | CID_689043 | 10014 | HDAC5 | ChEMBL |
| Arogyavardhini | Terminalia chebula | Fruit | CID_689043 | 10073 | SNUPN | NPASS |
| Arogyavardhini | Terminalia chebula | Fruit | CID_689043 | 10599 | SLCO1B1 | ChEMBL \| NPASS |
| Arogyavardhini | Terminalia chebula | Fruit | CID_689043 | 10765 | KDM5B | ChEMBL |
| Arogyavardhini | Terminalia chebula | Fruit | CID_689043 | 10919 | EHMT2 | ChEMBL \| NPASS |
| Arogyavardhini | Terminalia chebula | Fruit | CID_689043 | 1109 | AKR1C4 | ChEMBL \| NPASS |
| Arogyavardhini | Terminalia chebula | Fruit | CID_689043 | 11238 | CA5B | BindingDB \| ChEMBL \| NPASS |
| Arogyavardhini | Terminalia chebula | Fruit | CID_689043 | 118429 | ANTXR2 | NPASS |
| Arogyavardhini | Terminalia chebula | Fruit | CID_689043 | 1576 | CYP3A4 | NPASS |
| Arogyavardhini | Terminalia chebula | Fruit | CID_689043 | 1645 | AKR1C1 | ChEMBL \| NPASS |
| Arogyavardhini | Terminalia chebula | Fruit | CID_689043 | 1646 | AKR1C2 | ChEMBL \| NPASS |
| Arogyavardhini | Terminalia chebula | Fruit | CID_689043 | 1803 | DPP4 | BindingDB \| ChEMBL \| NPASS |
| Arogyavardhini | Terminalia chebula | Fruit | CID_689043 | 1845 | DUSP3 | BindingDB \| ChEMBL \| NPASS |
| Arogyavardhini | Terminalia chebula | Fruit | CID_689043 | 185 | AGTR1 | BindingDB \| ChEMBL |
| Arogyavardhini | Terminalia chebula | Fruit | CID_689043 | 1956 | EGFR | BindingDB \| NPASS |
| Arogyavardhini | Terminalia chebula | Fruit | CID_689043 | 2099 | ESR1 | ChEMBL \| NPASS |
| Arogyavardhini | Terminalia chebula | Fruit | CID_689043 | 216 | ALDH1A1 | NPASS |
| Arogyavardhini | Terminalia chebula | Fruit | CID_689043 | 231 | AKR1B1 | ChEMBL \| NPASS |
| Arogyavardhini | Terminalia chebula | Fruit | CID_689043 | 23621 | BACE1 | BindingDB \| ChEMBL |
| Arogyavardhini | Terminalia chebula | Fruit | CID_689043 | 23632 | CA14 | BindingDB \| ChEMBL \| NPASS |
| Arogyavardhini | Terminalia chebula | Fruit | CID_689043 | 240 | ALOX5 | BindingDB \| ChEMBL \| NPASS |
| Arogyavardhini | Terminalia chebula | Fruit | CID_689043 | 2548 | GAA | ChEMBL \| NPASS |
| Arogyavardhini | Terminalia chebula | Fruit | CID_689043 | 259285 | TAS2R39 | ChEMBL \| NPASS |
| Arogyavardhini | Terminalia chebula | Fruit | CID_689043 | 2717 | GLA | NPASS |
| Arogyavardhini | Terminalia chebula | Fruit | CID_689043 | 2744 | GLS | NPASS |
| Arogyavardhini | Terminalia chebula | Fruit | CID_689043 | 28234 | SLCO1B3 | ChEMBL \| NPASS |
| Arogyavardhini | Terminalia chebula | Fruit | CID_689043 | 2875 | GPT | ChEMBL \| NPASS |
| Arogyavardhini | Terminalia chebula | Fruit | CID_689043 | 29994 | BAZ2B | NPASS |
| Arogyavardhini | Terminalia chebula | Fruit | CID_689043 | 3028 | HSD17B10 | ChEMBL \| NPASS |
| Arogyavardhini | Terminalia chebula | Fruit | CID_689043 | 3065 | HDAC1 | BindingDB \| ChEMBL |
| Arogyavardhini | Terminalia chebula | Fruit | CID_689043 | 3066 | HDAC2 | ChEMBL |
| Arogyavardhini | Terminalia chebula | Fruit | CID_689043 | 3248 | HPGD | NPASS |
| Arogyavardhini | Terminalia chebula | Fruit | CID_689043 | 328 | APEX1 | NPASS |
| Arogyavardhini | Terminalia chebula | Fruit | CID_689043 | 3320 | HSP90AA1 | BindingDB |
| Arogyavardhini | Terminalia chebula | Fruit | CID_689043 | 3326 | HSP90AB1 | NPASS |
| Arogyavardhini | Terminalia chebula | Fruit | CID_689043 | 3373 | HYAL1 | BindingDB \| NPASS |
| Arogyavardhini | Terminalia chebula | Fruit | CID_689043 | 338442 | HCAR2 | ChEMBL \| NPASS |
| Arogyavardhini | Terminalia chebula | Fruit | CID_689043 | 351 | APP | BindingDB \| ChEMBL \| NPASS |
| Arogyavardhini | Terminalia chebula | Fruit | CID_689043 | 3757 | KCNH2 | NPASS |
| Arogyavardhini | Terminalia chebula | Fruit | CID_689043 | 3837 | KPNB1 | NPASS |
| Arogyavardhini | Terminalia chebula | Fruit | CID_689043 | 390245 | KDM4E | ChEMBL \| NPASS |
| Arogyavardhini | Terminalia chebula | Fruit | CID_689043 | 4000 | LMNA | NPASS |
| Arogyavardhini | Terminalia chebula | Fruit | CID_689043 | 410 | ARSA | NPASS |
| Arogyavardhini | Terminalia chebula | Fruit | CID_689043 | 4128 | MAOA | ChEMBL |
| Arogyavardhini | Terminalia chebula | Fruit | CID_689043 | 4129 | MAOB | ChEMBL |
| Arogyavardhini | Terminalia chebula | Fruit | CID_689043 | 4137 | MAPT | NPASS |

**Table S3: Associations between formulations, constituent plants, phytochemicals, and human molecular targets. The table lists the formulations, plant name and plant part, phytochemical identifiers, associated human target genes (Entrez Gene ID and gene symbol), and the database sources supporting each phytochemical-target interaction.**

| Formulation | Plant name | Part | PubChem CID | Entrez Gene ID | Gene Symbol | Source database(s) |
|---|---|---|---|---|---|---|
| Arogyavardhini | Terminalia chebula | Fruit | CID_689043 | 4297 | KMT2A | NPASS |
| Arogyavardhini | Terminalia chebula | Fruit | CID_689043 | 43 | ACHE | BindingDB \| ChEMBL \| NPASS |
| Arogyavardhini | Terminalia chebula | Fruit | CID_689043 | 4312 | MMP1 | ChEMBL \| NPASS |
| Arogyavardhini | Terminalia chebula | Fruit | CID_689043 | 4313 | MMP2 | ChEMBL \| NPASS |
| Arogyavardhini | Terminalia chebula | Fruit | CID_689043 | 4318 | MMP9 | BindingDB \| ChEMBL \| NPASS |
| Arogyavardhini | Terminalia chebula | Fruit | CID_689043 | 4350 | MPG | ChEMBL \| NPASS |
| Arogyavardhini | Terminalia chebula | Fruit | CID_689043 | 4790 | NFKB1 | NPASS |
| Arogyavardhini | Terminalia chebula | Fruit | CID_689043 | 51426 | POLK | NPASS |
| Arogyavardhini | Terminalia chebula | Fruit | CID_689043 | 51564 | HDAC7 | ChEMBL |
| Arogyavardhini | Terminalia chebula | Fruit | CID_689043 | 5376 | PMP22 | NPASS |
| Arogyavardhini | Terminalia chebula | Fruit | CID_689043 | 5423 | POLB | NPASS |
| Arogyavardhini | Terminalia chebula | Fruit | CID_689043 | 55775 | TDP1 | NPASS |
| Arogyavardhini | Terminalia chebula | Fruit | CID_689043 | 55869 | HDAC8 | ChEMBL |
| Arogyavardhini | Terminalia chebula | Fruit | CID_689043 | 5594 | MAPK1 | NPASS |
| Arogyavardhini | Terminalia chebula | Fruit | CID_689043 | 57016 | AKR1B10 | ChEMBL \| NPASS |
| Arogyavardhini | Terminalia chebula | Fruit | CID_689043 | 5742 | PTGS1 | ChEMBL \| NPASS |
| Arogyavardhini | Terminalia chebula | Fruit | CID_689043 | 5743 | PTGS2 | BindingDB \| ChEMBL \| NPASS |
| Arogyavardhini | Terminalia chebula | Fruit | CID_689043 | 5770 | PTPN1 | BindingDB \| ChEMBL \| NPASS |
| Arogyavardhini | Terminalia chebula | Fruit | CID_689043 | 5778 | PTPN7 | BindingDB \| ChEMBL \| NPASS |
| Arogyavardhini | Terminalia chebula | Fruit | CID_689043 | 7068 | THRB | NPASS |
| Arogyavardhini | Terminalia chebula | Fruit | CID_689043 | 7276 | TTR | ChEMBL \| NPASS |
| Arogyavardhini | Terminalia chebula | Fruit | CID_689043 | 7299 | TYR | ChEMBL \| NPASS |
| Arogyavardhini | Terminalia chebula | Fruit | CID_689043 | 7498 | XDH | ChEMBL \| NPASS |
| Arogyavardhini | Terminalia chebula | Fruit | CID_689043 | 759 | CA1 | BindingDB \| ChEMBL \| NPASS |
| Arogyavardhini | Terminalia chebula | Fruit | CID_689043 | 760 | CA2 | BindingDB \| ChEMBL \| NPASS |
| Arogyavardhini | Terminalia chebula | Fruit | CID_689043 | 761 | CA3 | BindingDB \| ChEMBL \| NPASS |
| Arogyavardhini | Terminalia chebula | Fruit | CID_689043 | 762 | CA4 | BindingDB \| ChEMBL \| NPASS |
| Arogyavardhini | Terminalia chebula | Fruit | CID_689043 | 763 | CA5A | BindingDB \| ChEMBL \| NPASS |
| Arogyavardhini | Terminalia chebula | Fruit | CID_689043 | 765 | CA6 | ChEMBL \| NPASS |
| Arogyavardhini | Terminalia chebula | Fruit | CID_689043 | 766 | CA7 | BindingDB \| ChEMBL \| NPASS |
| Arogyavardhini | Terminalia chebula | Fruit | CID_689043 | 768 | CA9 | BindingDB \| ChEMBL \| NPASS |
| Arogyavardhini | Terminalia chebula | Fruit | CID_689043 | 771 | CA12 | BindingDB \| ChEMBL \| NPASS |
| Arogyavardhini | Terminalia chebula | Fruit | CID_689043 | 79885 | HDAC11 | ChEMBL |
| Arogyavardhini | Terminalia chebula | Fruit | CID_689043 | 83933 | HDAC10 | ChEMBL \| NPASS |
| Arogyavardhini | Terminalia chebula | Fruit | CID_689043 | 8644 | AKR1C3 | ChEMBL \| NPASS |
| Arogyavardhini | Terminalia chebula | Fruit | CID_689043 | 8841 | HDAC3 | ChEMBL |
| Arogyavardhini | Terminalia chebula | Fruit | CID_689043 | 9682 | KDM4A | NPASS |
| Arogyavardhini | Terminalia chebula | Fruit | CID_689043 | 9734 | HDAC9 | ChEMBL |
| Arogyavardhini | Terminalia chebula | Fruit | CID_689043 | 9759 | HDAC4 | ChEMBL |
| Arogyavardhini | Terminalia chebula | Fruit | CID_73641 | 151306 | GPBAR1 | BindingDB \| ChEMBL \| NPASS |
| Arogyavardhini | Terminalia chebula | Fruit | CID_73641 | 240 | ALOX5 | BindingDB \| ChEMBL \| NPASS |
| Arogyavardhini | Terminalia chebula | Fruit | CID_73641 | 247 | ALOX15B | ChEMBL \| NPASS |
| Arogyavardhini | Terminalia chebula | Fruit | CID_73641 | 5743 | PTGS2 | BindingDB \| ChEMBL \| NPASS |
| Arogyavardhini | Terminalia chebula | Fruit | CID_73641 | 9971 | NR1H4 | NPASS |
| Arogyavardhini | Terminalia chebula | Fruit | CID_73659 | 112398 | EGLN2 | ChEMBL \| NPASS |
| Arogyavardhini | Terminalia chebula | Fruit | CID_73659 | 112399 | EGLN3 | ChEMBL \| NPASS |
| Arogyavardhini | Terminalia chebula | Fruit | CID_73659 | 231 | AKR1B1 | ChEMBL \| NPASS |
| Arogyavardhini | Terminalia chebula | Fruit | CID_73659 | 240 | ALOX5 | BindingDB \| ChEMBL \| NPASS |
| Arogyavardhini | Terminalia chebula | Fruit | CID_73659 | 247 | ALOX15B | ChEMBL \| NPASS |
| Arogyavardhini | Terminalia chebula | Fruit | CID_73659 | 51422 | PRKAG2 | ChEMBL |
| Arogyavardhini | Terminalia chebula | Fruit | CID_73659 | 53632 | PRKAG3 | ChEMBL |
| Arogyavardhini | Terminalia chebula | Fruit | CID_73659 | 54583 | EGLN1 | BindingDB \| ChEMBL \| NPASS |
| Arogyavardhini | Terminalia chebula | Fruit | CID_73659 | 54681 | P4HTM | ChEMBL \| NPASS |
| Arogyavardhini | Terminalia chebula | Fruit | CID_73659 | 5562 | PRKAA1 | ChEMBL |
| Arogyavardhini | Terminalia chebula | Fruit | CID_73659 | 5563 | PRKAA2 | ChEMBL |
| Arogyavardhini | Terminalia chebula | Fruit | CID_73659 | 5564 | PRKAB1 | ChEMBL |
| Arogyavardhini | Terminalia chebula | Fruit | CID_73659 | 5565 | PRKAB2 | ChEMBL |
| Arogyavardhini | Terminalia chebula | Fruit | CID_73659 | 5571 | PRKAG1 | ChEMBL |
| Arogyavardhini | Terminalia chebula | Fruit | CID_73659 | 57016 | AKR1B10 | ChEMBL \| NPASS |
| Arogyavardhini | Terminalia chebula | Fruit | CID_73659 | 5743 | PTGS2 | BindingDB \| ChEMBL \| NPASS |
| Arogyavardhini | Terminalia chebula | Fruit | CID_73659 | 5770 | PTPN1 | BindingDB \| ChEMBL \| NPASS |
| Arogyavardhini | Terminalia chebula | Fruit | CID_73659 | 5771 | PTPN2 | BindingDB \| ChEMBL \| NPASS |
| Arogyavardhini | Terminalia chebula | Fruit | CID_73659 | 5777 | PTPN6 | BindingDB \| ChEMBL \| NPASS |
| Arogyavardhini | Terminalia chebula | Fruit | CID_73659 | 5781 | PTPN11 | BindingDB \| ChEMBL \| NPASS |
| Arogyavardhini | Terminalia chebula | Fruit | CID_73659 | 5792 | PTPRF | BindingDB \| ChEMBL \| NPASS |
| Arogyavardhini | Terminalia chebula | Fruit | CID_73659 | 5837 | PYGM | BindingDB \| ChEMBL \| NPASS |
| Arogyavardhini | Terminalia chebula | Fruit | CID_8215 | 10013 | HDAC6 | ChEMBL |
| Arogyavardhini | Terminalia chebula | Fruit | CID_8215 | 5465 | PPARA | ChEMBL \| NPASS |
| Arogyavardhini | Terminalia chebula | Fruit | CID_8215 | 5467 | PPARD | ChEMBL \| NPASS |
| Arogyavardhini | Terminalia chebula | Fruit | CID_8215 | 5468 | PPARG | BindingDB \| ChEMBL \| NPASS |

Table S3: Associations between formulations, constituent plants, phytochemicals, and human molecular targets. The table lists the formulations, plant name and plant part, phytochemical identifiers, associated human target genes (Entrez Gene ID and gene symbol), and the database sources supporting each phytochemical-target interaction.

| Formulation | Plant name | Part | PubChem CID | Entrez Gene ID | Gene Symbol | Source database(s) |
|---|---|---|---|---|---|---|
| Arogyavardhini | Terminalia chebula | Fruit | CID_8468 | 10013 | HDAC6 | ChEMBL |
| Arogyavardhini | Terminalia chebula | Fruit | CID_8468 | 10014 | HDAC5 | ChEMBL |
| Arogyavardhini | Terminalia chebula | Fruit | CID_8468 | 1019 | CDK4 | BindingDB |
| Arogyavardhini | Terminalia chebula | Fruit | CID_8468 | 1021 | CDK6 | BindingDB |
| Arogyavardhini | Terminalia chebula | Fruit | CID_8468 | 10461 | MERTK | BindingDB |
| Arogyavardhini | Terminalia chebula | Fruit | CID_8468 | 11201 | POLI | NPASS |
| Arogyavardhini | Terminalia chebula | Fruit | CID_8468 | 118429 | ANTXR2 | NPASS |
| Arogyavardhini | Terminalia chebula | Fruit | CID_8468 | 1956 | EGFR | BindingDB |
| Arogyavardhini | Terminalia chebula | Fruit | CID_8468 | 1969 | EPHA2 | BindingDB |
| Arogyavardhini | Terminalia chebula | Fruit | CID_8468 | 2066 | ERBB4 | BindingDB |
| Arogyavardhini | Terminalia chebula | Fruit | CID_8468 | 2099 | ESR1 | ChEMBL \| NPASS |
| Arogyavardhini | Terminalia chebula | Fruit | CID_8468 | 2260 | FGFR1 | BindingDB |
| Arogyavardhini | Terminalia chebula | Fruit | CID_8468 | 2261 | FGFR3 | BindingDB |
| Arogyavardhini | Terminalia chebula | Fruit | CID_8468 | 2263 | FGFR2 | BindingDB |
| Arogyavardhini | Terminalia chebula | Fruit | CID_8468 | 231 | AKR1B1 | ChEMBL \| NPASS |
| Arogyavardhini | Terminalia chebula | Fruit | CID_8468 | 2321 | FLT1 | BindingDB |
| Arogyavardhini | Terminalia chebula | Fruit | CID_8468 | 238 | ALK | BindingDB |
| Arogyavardhini | Terminalia chebula | Fruit | CID_8468 | 25 | ABL1 | BindingDB |
| Arogyavardhini | Terminalia chebula | Fruit | CID_8468 | 3064 | HTT | NPASS |
| Arogyavardhini | Terminalia chebula | Fruit | CID_8468 | 3065 | HDAC1 | ChEMBL |
| Arogyavardhini | Terminalia chebula | Fruit | CID_8468 | 3066 | HDAC2 | ChEMBL |
| Arogyavardhini | Terminalia chebula | Fruit | CID_8468 | 3480 | IGF1R | BindingDB |
| Arogyavardhini | Terminalia chebula | Fruit | CID_8468 | 3643 | INSR | BindingDB |
| Arogyavardhini | Terminalia chebula | Fruit | CID_8468 | 3791 | KDR | BindingDB |
| Arogyavardhini | Terminalia chebula | Fruit | CID_8468 | 4058 | LTK | BindingDB |
| Arogyavardhini | Terminalia chebula | Fruit | CID_8468 | 4233 | MET | BindingDB |
| Arogyavardhini | Terminalia chebula | Fruit | CID_8468 | 4921 | DDR2 | BindingDB |
| Arogyavardhini | Terminalia chebula | Fruit | CID_8468 | 5156 | PDGFRA | BindingDB |
| Arogyavardhini | Terminalia chebula | Fruit | CID_8468 | 51564 | HDAC7 | ChEMBL |
| Arogyavardhini | Terminalia chebula | Fruit | CID_8468 | 5159 | PDGFRB | BindingDB |
| Arogyavardhini | Terminalia chebula | Fruit | CID_8468 | 55869 | HDAC8 | ChEMBL |
| Arogyavardhini | Terminalia chebula | Fruit | CID_8468 | 5594 | MAPK1 | BindingDB |
| Arogyavardhini | Terminalia chebula | Fruit | CID_8468 | 5979 | RET | BindingDB |
| Arogyavardhini | Terminalia chebula | Fruit | CID_8468 | 7172 | TPMT | NPASS |
| Arogyavardhini | Terminalia chebula | Fruit | CID_8468 | 7301 | TYRO3 | BindingDB |
| Arogyavardhini | Terminalia chebula | Fruit | CID_8468 | 79885 | HDAC11 | ChEMBL |
| Arogyavardhini | Terminalia chebula | Fruit | CID_8468 | 83933 | HDAC10 | ChEMBL \| NPASS |
| Arogyavardhini | Terminalia chebula | Fruit | CID_8468 | 8841 | HDAC3 | ChEMBL |
| Arogyavardhini | Terminalia chebula | Fruit | CID_8468 | 9734 | HDAC9 | ChEMBL |
| Arogyavardhini | Terminalia chebula | Fruit | CID_8468 | 9759 | HDAC4 | ChEMBL |
| Arogyavardhini | Terminalia chebula | Fruit | CID_8468 | 983 | CDK1 | BindingDB |
| Arogyavardhini | Terminalia chebula | Fruit | CID_8742 | 10599 | SLCO1B1 | ChEMBL \| NPASS |
| Arogyavardhini | Terminalia chebula | Fruit | CID_8742 | 28234 | SLCO1B3 | ChEMBL \| NPASS |
| Arogyavardhini | Terminalia chebula | Fruit | CID_8742 | 328 | APEX1 | NPASS |
| Arogyavardhini | Terminalia chebula | Fruit | CID_8742 | 51053 | GMNN | NPASS |
| Arogyavardhini | Terminalia chebula | Fruit | CID_8742 | 54578 | UGT1A6 | ChEMBL \| NPASS |
| Arogyavardhini | Terminalia chebula | Fruit | CID_8742 | 54600 | UGT1A9 | ChEMBL \| NPASS |
| Arogyavardhini | Terminalia chebula | Fruit | CID_8742 | 5743 | PTGS2 | ChEMBL \| NPASS |
| Arogyavardhini | Terminalia chebula | Fruit | CID_8742 | 6401 | SELE | ChEMBL \| NPASS |
| Arogyavardhini | Terminalia chebula | Fruit | CID_8742 | 6402 | SELL | ChEMBL \| NPASS |
| Arogyavardhini | Terminalia chebula | Fruit | CID_8742 | 6403 | SELP | ChEMBL \| NPASS |
| Arogyavardhini | Terminalia chebula | Fruit | CID_985 | 1588 | CYP19A1 | ChEMBL \| NPASS |
| Arogyavardhini | Terminalia chebula | Fruit | CID_985 | 2099 | ESR1 | ChEMBL \| NPASS |
| Arogyavardhini | Terminalia chebula | Fruit | CID_985 | 2167 | FABP4 | ChEMBL \| NPASS |
| Arogyavardhini | Terminalia chebula | Fruit | CID_985 | 2169 | FABP2 | ChEMBL \| NPASS |
| Arogyavardhini | Terminalia chebula | Fruit | CID_985 | 2170 | FABP3 | BindingDB \| ChEMBL \| NPASS |
| Arogyavardhini | Terminalia chebula | Fruit | CID_985 | 2171 | FABP5 | ChEMBL \| NPASS |
| Arogyavardhini | Terminalia chebula | Fruit | CID_985 | 2908 | NR3C1 | NPASS |
| Arogyavardhini | Terminalia chebula | Fruit | CID_985 | 367 | AR | NPASS |
| Arogyavardhini | Terminalia chebula | Fruit | CID_985 | 4780 | NFE2L2 | NPASS |
| Arogyavardhini | Terminalia chebula | Fruit | CID_985 | 51053 | GMNN | NPASS |
| Arogyavardhini | Terminalia chebula | Fruit | CID_985 | 5243 | ABCB1 | ChEMBL |
| Arogyavardhini | Terminalia chebula | Fruit | CID_985 | 5465 | PPARA | ChEMBL \| NPASS |
| Arogyavardhini | Terminalia chebula | Fruit | CID_985 | 5467 | PPARD | ChEMBL \| NPASS |
| Arogyavardhini | Terminalia chebula | Fruit | CID_985 | 5468 | PPARG | BindingDB \| ChEMBL \| NPASS |
| Arogyavardhini | Terminalia chebula | Fruit | CID_985 | 5770 | PTPN1 | BindingDB \| ChEMBL \| NPASS |
| Arogyavardhini | Terminalia chebula | Fruit | CID_985 | 7004 | TEAD4 | ChEMBL |
| Arogyavardhini | Terminalia chebula | Fruit | CID_985 | 7097 | TLR2 | ChEMBL \| NPASS |
| Arogyavardhini | Terminalia chebula | Fruit | CID_985 | 7398 | USP1 | NPASS |
| Arogyavardhini | Terminalia chebula | Fruit | CID_985 | 81285 | OR51E2 | ChEMBL \| NPASS |

**Table S3: Associations between formulations, constituent plants, phytochemicals, and human molecular targets. The table lists the formulations, plant name and plant part, phytochemical identifiers, associated human target genes (Entrez Gene ID and gene symbol), and the database sources supporting each phytochemical-target interaction.**

| Formulation | Plant name | Part | PubChem CID | Entrez Gene ID | Gene Symbol | Source database(s) |
|---|---|---|---|---|---|---|
| Arogyavardhini | Terminalia chebula | Fruit | CID_985 | 9971 | NR1H4 | NPASS |
| Nishamalaki | Curcuma longa | Rhizome | CID_10364 | 10941 | UGT2A1 | ChEMBL |
| Nishamalaki | Curcuma longa | Rhizome | CID_10364 | 3725 | JUN | NPASS |
| Nishamalaki | Curcuma longa | Rhizome | CID_10364 | 43 | ACHE | BindingDB \| NPASS |
| Nishamalaki | Curcuma longa | Rhizome | CID_10364 | 54575 | UGT1A10 | ChEMBL \| NPASS |
| Nishamalaki | Curcuma longa | Rhizome | CID_10364 | 54576 | UGT1A8 | ChEMBL \| NPASS |
| Nishamalaki | Curcuma longa | Rhizome | CID_10364 | 54578 | UGT1A6 | ChEMBL \| NPASS |
| Nishamalaki | Curcuma longa | Rhizome | CID_10364 | 54600 | UGT1A9 | ChEMBL \| NPASS |
| Nishamalaki | Curcuma longa | Rhizome | CID_10364 | 54657 | UGT1A4 | ChEMBL \| NPASS |
| Nishamalaki | Curcuma longa | Rhizome | CID_10364 | 54658 | UGT1A1 | ChEMBL \| NPASS |
| Nishamalaki | Curcuma longa | Rhizome | CID_10364 | 5742 | PTGS1 | ChEMBL \| NPASS |
| Nishamalaki | Curcuma longa | Rhizome | CID_10364 | 7366 | UGT2B15 | ChEMBL \| NPASS |
| Nishamalaki | Curcuma longa | Rhizome | CID_10364 | 775 | CACNA1C | ChEMBL \| NPASS |
| Nishamalaki | Curcuma longa | Rhizome | CID_10364 | 8989 | TRPA1 | ChEMBL \| NPASS |
| Nishamalaki | Curcuma longa | Rhizome | CID_11230 | 54657 | UGT1A4 | ChEMBL \| NPASS |
| Nishamalaki | Curcuma longa | Rhizome | CID_1140 | 239 | ALOX12 | NPASS |
| Nishamalaki | Curcuma longa | Rhizome | CID_1140 | 4000 | LMNA | NPASS |
| Nishamalaki | Curcuma longa | Rhizome | CID_1140 | 43 | ACHE | BindingDB \| ChEMBL \| NPASS |
| Nishamalaki | Curcuma longa | Rhizome | CID_1140 | 7068 | THRB | NPASS |
| Nishamalaki | Curcuma longa | Rhizome | CID_1140 | 7253 | TSHR | NPASS |
| Nishamalaki | Curcuma longa | Rhizome | CID_11463 | 5467 | PPARD | NPASS |
| Nishamalaki | Curcuma longa | Rhizome | CID_12310886 | 5770 | PTPN1 | ChEMBL \| NPASS |
| Nishamalaki | Curcuma longa | Rhizome | CID_12388 | 5467 | PPARD | NPASS |
| Nishamalaki | Curcuma longa | Rhizome | CID_14240392 | 10013 | HDAC6 | ChEMBL |
| Nishamalaki | Curcuma longa | Rhizome | CID_153845 | 231 | AKR1B1 | ChEMBL \| NPASS |
| Nishamalaki | Curcuma longa | Rhizome | CID_15448 | 8989 | TRPA1 | ChEMBL \| NPASS |
| Nishamalaki | Curcuma longa | Rhizome | CID_1549026 | 216 | ALDH1A1 | NPASS |
| Nishamalaki | Curcuma longa | Rhizome | CID_1549026 | 246 | ALOX15 | NPASS |
| Nishamalaki | Curcuma longa | Rhizome | CID_1549026 | 367 | AR | NPASS |
| Nishamalaki | Curcuma longa | Rhizome | CID_1549026 | 4780 | NFE2L2 | NPASS |
| Nishamalaki | Curcuma longa | Rhizome | CID_1549026 | 55775 | TDP1 | NPASS |
| Nishamalaki | Curcuma longa | Rhizome | CID_1549026 | 5745 | PTH1R | NPASS |
| Nishamalaki | Curcuma longa | Rhizome | CID_1549026 | 7442 | TRPV1 | BindingDB \| ChEMBL \| NPASS |
| Nishamalaki | Curcuma longa | Rhizome | CID_1549026 | 9971 | NR1H4 | NPASS |
| Nishamalaki | Curcuma longa | Rhizome | CID_160512 | 43 | ACHE | BindingDB \| NPASS |
| Nishamalaki | Curcuma longa | Rhizome | CID_160512 | 6774 | STAT3 | ChEMBL \| NPASS |
| Nishamalaki | Curcuma longa | Rhizome | CID_17100 | 2554 | GABRA1 | ChEMBL |
| Nishamalaki | Curcuma longa | Rhizome | CID_17100 | 2561 | GABRB2 | ChEMBL |
| Nishamalaki | Curcuma longa | Rhizome | CID_17100 | 2566 | GABRG2 | ChEMBL \| NPASS |
| Nishamalaki | Curcuma longa | Rhizome | CID_17100 | 4000 | LMNA | NPASS |
| Nishamalaki | Curcuma longa | Rhizome | CID_1742210 | 7442 | TRPV1 | ChEMBL \| NPASS |
| Nishamalaki | Curcuma longa | Rhizome | CID_1742210 | 84876 | ORAI1 | ChEMBL \| NPASS |
| Nishamalaki | Curcuma longa | Rhizome | CID_189061 | 5770 | PTPN1 | BindingDB \| ChEMBL \| NPASS |
| Nishamalaki | Curcuma longa | Rhizome | CID_189061 | 5771 | PTPN2 | BindingDB \| ChEMBL \| NPASS |
| Nishamalaki | Curcuma longa | Rhizome | CID_222284 | 10013 | HDAC6 | ChEMBL |
| Nishamalaki | Curcuma longa | Rhizome | CID_222284 | 10599 | SLCO1B1 | ChEMBL \| NPASS |
| Nishamalaki | Curcuma longa | Rhizome | CID_222284 | 1565 | CYP2D6 | ChEMBL \| NPASS |
| Nishamalaki | Curcuma longa | Rhizome | CID_222284 | 1576 | CYP3A4 | BindingDB \| ChEMBL \| NPASS |
| Nishamalaki | Curcuma longa | Rhizome | CID_222284 | 1803 | DPP4 | ChEMBL \| NPASS |
| Nishamalaki | Curcuma longa | Rhizome | CID_222284 | 2147 | F2 | ChEMBL \| NPASS |
| Nishamalaki | Curcuma longa | Rhizome | CID_222284 | 28234 | SLCO1B3 | ChEMBL \| NPASS |
| Nishamalaki | Curcuma longa | Rhizome | CID_222284 | 2932 | GSK3B | ChEMBL |
| Nishamalaki | Curcuma longa | Rhizome | CID_222284 | 3417 | IDH1 | ChEMBL \| NPASS |
| Nishamalaki | Curcuma longa | Rhizome | CID_222284 | 51422 | PRKAG2 | ChEMBL |
| Nishamalaki | Curcuma longa | Rhizome | CID_222284 | 5243 | ABCB1 | ChEMBL |
| Nishamalaki | Curcuma longa | Rhizome | CID_222284 | 53632 | PRKAG3 | ChEMBL |
| Nishamalaki | Curcuma longa | Rhizome | CID_222284 | 5423 | POLB | BindingDB \| ChEMBL \| NPASS |
| Nishamalaki | Curcuma longa | Rhizome | CID_222284 | 5562 | PRKAA1 | ChEMBL |
| Nishamalaki | Curcuma longa | Rhizome | CID_222284 | 5563 | PRKAA2 | ChEMBL |
| Nishamalaki | Curcuma longa | Rhizome | CID_222284 | 5564 | PRKAB1 | ChEMBL |
| Nishamalaki | Curcuma longa | Rhizome | CID_222284 | 5565 | PRKAB2 | ChEMBL |
| Nishamalaki | Curcuma longa | Rhizome | CID_222284 | 5571 | PRKAG1 | ChEMBL |
| Nishamalaki | Curcuma longa | Rhizome | CID_222284 | 7299 | TYR | ChEMBL \| NPASS |
| Nishamalaki | Curcuma longa | Rhizome | CID_22311 | 367 | AR | NPASS |
| Nishamalaki | Curcuma longa | Rhizome | CID_22311 | 4000 | LMNA | NPASS |
| Nishamalaki | Curcuma longa | Rhizome | CID_22311 | 4780 | NFE2L2 | ChEMBL \| NPASS |
| Nishamalaki | Curcuma longa | Rhizome | CID_22311 | 4790 | NFKB1 | NPASS |
| Nishamalaki | Curcuma longa | Rhizome | CID_22311 | 5467 | PPARD | NPASS |
| Nishamalaki | Curcuma longa | Rhizome | CID_22311 | 6256 | RXRA | NPASS |
| Nishamalaki | Curcuma longa | Rhizome | CID_22311 | 7068 | THRB | NPASS |

**Table S3: Associations between formulations, constituent plants, phytochemicals, and human molecular targets. The table lists the formulations, plant name and plant part, phytochemical identifiers, associated human target genes (Entrez Gene ID and gene symbol), and the database sources supporting each phytochemical-target interaction.**

| Formulation | Plant name | Part | PubChem CID | Entrez Gene ID | Gene Symbol | Source database(s) |
|---|---|---|---|---|---|---|
| Nishamalaki | Curcuma longa | Rhizome | CID_22311 | 7157 | TP53 | NPASS |
| Nishamalaki | Curcuma longa | Rhizome | CID_24834047 | 5770 | PTPN1 | BindingDB \| ChEMBL \| NPASS |
| Nishamalaki | Curcuma longa | Rhizome | CID_24834047 | 5771 | PTPN2 | BindingDB \| ChEMBL \| NPASS |
| Nishamalaki | Curcuma longa | Rhizome | CID_2537 | 10013 | HDAC6 | ChEMBL |
| Nishamalaki | Curcuma longa | Rhizome | CID_2537 | 162514 | TRPV3 | ChEMBL |
| Nishamalaki | Curcuma longa | Rhizome | CID_2537 | 2099 | ESR1 | ChEMBL |
| Nishamalaki | Curcuma longa | Rhizome | CID_2537 | 216 | ALDH1A1 | ChEMBL |
| Nishamalaki | Curcuma longa | Rhizome | CID_2537 | 4137 | MAPT | NPASS |
| Nishamalaki | Curcuma longa | Rhizome | CID_2537 | 55775 | TDP1 | ChEMBL |
| Nishamalaki | Curcuma longa | Rhizome | CID_26049 | 3315 | HSPB1 | NPASS |
| Nishamalaki | Curcuma longa | Rhizome | CID_26049 | 367 | AR | NPASS |
| Nishamalaki | Curcuma longa | Rhizome | CID_26049 | 4780 | NFE2L2 | NPASS |
| Nishamalaki | Curcuma longa | Rhizome | CID_2758 | 10062 | NR1H3 | ChEMBL \| NPASS |
| Nishamalaki | Curcuma longa | Rhizome | CID_2758 | 10280 | SIGMAR1 | ChEMBL |
| Nishamalaki | Curcuma longa | Rhizome | CID_2758 | 10800 | CYSLTR1 | ChEMBL |
| Nishamalaki | Curcuma longa | Rhizome | CID_2758 | 10951 | CBX1 | NPASS |
| Nishamalaki | Curcuma longa | Rhizome | CID_2758 | 1128 | CHRM1 | ChEMBL |
| Nishamalaki | Curcuma longa | Rhizome | CID_2758 | 1129 | CHRM2 | ChEMBL |
| Nishamalaki | Curcuma longa | Rhizome | CID_2758 | 1131 | CHRM3 | ChEMBL |
| Nishamalaki | Curcuma longa | Rhizome | CID_2758 | 1132 | CHRM4 | ChEMBL |
| Nishamalaki | Curcuma longa | Rhizome | CID_2758 | 1133 | CHRM5 | ChEMBL |
| Nishamalaki | Curcuma longa | Rhizome | CID_2758 | 1233 | CCR4 | ChEMBL |
| Nishamalaki | Curcuma longa | Rhizome | CID_2758 | 1234 | CCR5 | ChEMBL |
| Nishamalaki | Curcuma longa | Rhizome | CID_2758 | 1268 | CNR1 | ChEMBL |
| Nishamalaki | Curcuma longa | Rhizome | CID_2758 | 134 | ADORA1 | ChEMBL |
| Nishamalaki | Curcuma longa | Rhizome | CID_2758 | 135 | ADORA2A | ChEMBL |
| Nishamalaki | Curcuma longa | Rhizome | CID_2758 | 140 | ADORA3 | ChEMBL |
| Nishamalaki | Curcuma longa | Rhizome | CID_2758 | 1432 | MAPK14 | ChEMBL |
| Nishamalaki | Curcuma longa | Rhizome | CID_2758 | 146 | ADRA1D | ChEMBL |
| Nishamalaki | Curcuma longa | Rhizome | CID_2758 | 150 | ADRA2A | ChEMBL |
| Nishamalaki | Curcuma longa | Rhizome | CID_2758 | 151 | ADRA2B | ChEMBL |
| Nishamalaki | Curcuma longa | Rhizome | CID_2758 | 1511 | CTSG | ChEMBL |
| Nishamalaki | Curcuma longa | Rhizome | CID_2758 | 152 | ADRA2C | ChEMBL |
| Nishamalaki | Curcuma longa | Rhizome | CID_2758 | 153 | ADRB1 | ChEMBL |
| Nishamalaki | Curcuma longa | Rhizome | CID_2758 | 154 | ADRB2 | ChEMBL |
| Nishamalaki | Curcuma longa | Rhizome | CID_2758 | 1544 | CYP1A2 | ChEMBL |
| Nishamalaki | Curcuma longa | Rhizome | CID_2758 | 1548 | CYP2A6 | ChEMBL |
| Nishamalaki | Curcuma longa | Rhizome | CID_2758 | 155 | ADRB3 | ChEMBL |
| Nishamalaki | Curcuma longa | Rhizome | CID_2758 | 1557 | CYP2C19 | ChEMBL |
| Nishamalaki | Curcuma longa | Rhizome | CID_2758 | 1559 | CYP2C9 | ChEMBL |
| Nishamalaki | Curcuma longa | Rhizome | CID_2758 | 1565 | CYP2D6 | ChEMBL |
| Nishamalaki | Curcuma longa | Rhizome | CID_2758 | 1571 | CYP2E1 | ChEMBL |
| Nishamalaki | Curcuma longa | Rhizome | CID_2758 | 1576 | CYP3A4 | ChEMBL |
| Nishamalaki | Curcuma longa | Rhizome | CID_2758 | 1812 | DRD1 | ChEMBL |
| Nishamalaki | Curcuma longa | Rhizome | CID_2758 | 1813 | DRD2 | ChEMBL |
| Nishamalaki | Curcuma longa | Rhizome | CID_2758 | 1814 | DRD3 | ChEMBL |
| Nishamalaki | Curcuma longa | Rhizome | CID_2758 | 1815 | DRD4 | ChEMBL |
| Nishamalaki | Curcuma longa | Rhizome | CID_2758 | 186 | AGTR2 | ChEMBL |
| Nishamalaki | Curcuma longa | Rhizome | CID_2758 | 1909 | EDNRA | ChEMBL |
| Nishamalaki | Curcuma longa | Rhizome | CID_2758 | 1956 | EGFR | ChEMBL |
| Nishamalaki | Curcuma longa | Rhizome | CID_2758 | 1991 | ELANE | ChEMBL |
| Nishamalaki | Curcuma longa | Rhizome | CID_2758 | 2064 | ERBB2 | ChEMBL |
| Nishamalaki | Curcuma longa | Rhizome | CID_2758 | 2099 | ESR1 | ChEMBL |
| Nishamalaki | Curcuma longa | Rhizome | CID_2758 | 2100 | ESR2 | ChEMBL |
| Nishamalaki | Curcuma longa | Rhizome | CID_2758 | 2321 | FLT1 | ChEMBL |
| Nishamalaki | Curcuma longa | Rhizome | CID_2758 | 2534 | FYN | ChEMBL |
| Nishamalaki | Curcuma longa | Rhizome | CID_2758 | 25939 | SAMHD1 | ChEMBL |
| Nishamalaki | Curcuma longa | Rhizome | CID_2758 | 2908 | NR3C1 | ChEMBL |
| Nishamalaki | Curcuma longa | Rhizome | CID_2758 | 3156 | HMGCR | ChEMBL |
| Nishamalaki | Curcuma longa | Rhizome | CID_2758 | 3269 | HRH1 | ChEMBL |
| Nishamalaki | Curcuma longa | Rhizome | CID_2758 | 3274 | HRH2 | ChEMBL |
| Nishamalaki | Curcuma longa | Rhizome | CID_2758 | 328 | APEX1 | NPASS |
| Nishamalaki | Curcuma longa | Rhizome | CID_2758 | 3356 | HTR2A | ChEMBL |
| Nishamalaki | Curcuma longa | Rhizome | CID_2758 | 3357 | HTR2B | ChEMBL |
| Nishamalaki | Curcuma longa | Rhizome | CID_2758 | 3358 | HTR2C | ChEMBL |
| Nishamalaki | Curcuma longa | Rhizome | CID_2758 | 3359 | HTR3A | BindingDB \| NPASS |
| Nishamalaki | Curcuma longa | Rhizome | CID_2758 | 3362 | HTR6 | ChEMBL |
| Nishamalaki | Curcuma longa | Rhizome | CID_2758 | 3577 | CXCR1 | ChEMBL |
| Nishamalaki | Curcuma longa | Rhizome | CID_2758 | 3579 | CXCR2 | ChEMBL |
| Nishamalaki | Curcuma longa | Rhizome | CID_2758 | 367 | AR | NPASS |

**Table S3: Associations between formulations, constituent plants, phytochemicals, and human molecular targets. The table lists the formulations, plant name and plant part, phytochemical identifiers, associated human target genes (Entrez Gene ID and gene symbol), and the database sources supporting each phytochemical-target interaction.**

| Formulation | Plant name | Part | PubChem CID | Entrez Gene ID | Gene Symbol | Source database(s) |
|---|---|---|---|---|---|---|
| Nishamalaki | Curcuma longa | Rhizome | CID_2758 | 3725 | JUN | NPASS |
| Nishamalaki | Curcuma longa | Rhizome | CID_2758 | 3757 | KCNH2 | ChEMBL |
| Nishamalaki | Curcuma longa | Rhizome | CID_2758 | 3932 | LCK | ChEMBL |
| Nishamalaki | Curcuma longa | Rhizome | CID_2758 | 4000 | LMNA | NPASS |
| Nishamalaki | Curcuma longa | Rhizome | CID_2758 | 4128 | MAOA | ChEMBL |
| Nishamalaki | Curcuma longa | Rhizome | CID_2758 | 4137 | MAPT | NPASS |
| Nishamalaki | Curcuma longa | Rhizome | CID_2758 | 4159 | MC3R | ChEMBL |
| Nishamalaki | Curcuma longa | Rhizome | CID_2758 | 4160 | MC4R | ChEMBL |
| Nishamalaki | Curcuma longa | Rhizome | CID_2758 | 4161 | MC5R | ChEMBL |
| Nishamalaki | Curcuma longa | Rhizome | CID_2758 | 43 | ACHE | ChEMBL |
| Nishamalaki | Curcuma longa | Rhizome | CID_2758 | 4312 | MMP1 | ChEMBL |
| Nishamalaki | Curcuma longa | Rhizome | CID_2758 | 4318 | MMP9 | ChEMBL |
| Nishamalaki | Curcuma longa | Rhizome | CID_2758 | 4886 | NPY1R | ChEMBL |
| Nishamalaki | Curcuma longa | Rhizome | CID_2758 | 4887 | NPY2R | ChEMBL |
| Nishamalaki | Curcuma longa | Rhizome | CID_2758 | 4985 | OPRD1 | ChEMBL |
| Nishamalaki | Curcuma longa | Rhizome | CID_2758 | 4986 | OPRK1 | ChEMBL |
| Nishamalaki | Curcuma longa | Rhizome | CID_2758 | 4988 | OPRM1 | ChEMBL |
| Nishamalaki | Curcuma longa | Rhizome | CID_2758 | 51053 | GMNN | NPASS |
| Nishamalaki | Curcuma longa | Rhizome | CID_2758 | 5467 | PPARD | NPASS |
| Nishamalaki | Curcuma longa | Rhizome | CID_2758 | 552 | AVPR1A | ChEMBL |
| Nishamalaki | Curcuma longa | Rhizome | CID_2758 | 5530 | PPP3CA | ChEMBL |
| Nishamalaki | Curcuma longa | Rhizome | CID_2758 | 55775 | TDP1 | NPASS |
| Nishamalaki | Curcuma longa | Rhizome | CID_2758 | 5578 | PRKCA | ChEMBL |
| Nishamalaki | Curcuma longa | Rhizome | CID_2758 | 5594 | MAPK1 | ChEMBL \| NPASS |
| Nishamalaki | Curcuma longa | Rhizome | CID_2758 | 5595 | MAPK3 | ChEMBL |
| Nishamalaki | Curcuma longa | Rhizome | CID_2758 | 5724 | PTAFR | ChEMBL |
| Nishamalaki | Curcuma longa | Rhizome | CID_2758 | 5742 | PTGS1 | ChEMBL |
| Nishamalaki | Curcuma longa | Rhizome | CID_2758 | 5743 | PTGS2 | ChEMBL \| NPASS |
| Nishamalaki | Curcuma longa | Rhizome | CID_2758 | 5788 | PTPRC | ChEMBL |
| Nishamalaki | Curcuma longa | Rhizome | CID_2758 | 5965 | RECQL | NPASS |
| Nishamalaki | Curcuma longa | Rhizome | CID_2758 | 624 | BDKRB2 | ChEMBL |
| Nishamalaki | Curcuma longa | Rhizome | CID_2758 | 6530 | SLC6A2 | ChEMBL |
| Nishamalaki | Curcuma longa | Rhizome | CID_2758 | 6531 | SLC6A3 | ChEMBL |
| Nishamalaki | Curcuma longa | Rhizome | CID_2758 | 6532 | SLC6A4 | ChEMBL |
| Nishamalaki | Curcuma longa | Rhizome | CID_2758 | 6865 | TACR2 | ChEMBL |
| Nishamalaki | Curcuma longa | Rhizome | CID_2758 | 6869 | TACR1 | ChEMBL |
| Nishamalaki | Curcuma longa | Rhizome | CID_2758 | 6916 | TBXAS1 | ChEMBL |
| Nishamalaki | Curcuma longa | Rhizome | CID_2758 | 7253 | TSHR | NPASS |
| Nishamalaki | Curcuma longa | Rhizome | CID_2758 | 729230 | CCR2 | ChEMBL |
| Nishamalaki | Curcuma longa | Rhizome | CID_2758 | 7433 | VIPR1 | ChEMBL |
| Nishamalaki | Curcuma longa | Rhizome | CID_2758 | 760 | CA2 | ChEMBL |
| Nishamalaki | Curcuma longa | Rhizome | CID_2758 | 79054 | TRPM8 | ChEMBL \| NPASS |
| Nishamalaki | Curcuma longa | Rhizome | CID_2758 | 799 | CALCR | ChEMBL |
| Nishamalaki | Curcuma longa | Rhizome | CID_2758 | 834 | CASP1 | ChEMBL |
| Nishamalaki | Curcuma longa | Rhizome | CID_2758 | 8654 | PDE5A | ChEMBL |
| Nishamalaki | Curcuma longa | Rhizome | CID_2758 | 886 | CCKAR | ChEMBL |
| Nishamalaki | Curcuma longa | Rhizome | CID_2969 | 2908 | NR3C1 | NPASS |
| Nishamalaki | Curcuma longa | Rhizome | CID_2969 | 3315 | HSPB1 | NPASS |
| Nishamalaki | Curcuma longa | Rhizome | CID_2969 | 367 | AR | NPASS |
| Nishamalaki | Curcuma longa | Rhizome | CID_2969 | 4780 | NFE2L2 | NPASS |
| Nishamalaki | Curcuma longa | Rhizome | CID_2969 | 51053 | GMNN | NPASS |
| Nishamalaki | Curcuma longa | Rhizome | CID_2969 | 53831 | GPR84 | ChEMBL \| NPASS |
| Nishamalaki | Curcuma longa | Rhizome | CID_2969 | 54575 | UGT1A10 | ChEMBL \| NPASS |
| Nishamalaki | Curcuma longa | Rhizome | CID_2969 | 54576 | UGT1A8 | ChEMBL \| NPASS |
| Nishamalaki | Curcuma longa | Rhizome | CID_2969 | 5465 | PPARA | ChEMBL \| NPASS |
| Nishamalaki | Curcuma longa | Rhizome | CID_2969 | 54659 | UGT1A3 | ChEMBL \| NPASS |
| Nishamalaki | Curcuma longa | Rhizome | CID_2969 | 5467 | PPARD | ChEMBL \| NPASS |
| Nishamalaki | Curcuma longa | Rhizome | CID_2969 | 5468 | PPARG | ChEMBL \| NPASS |
| Nishamalaki | Curcuma longa | Rhizome | CID_2969 | 55775 | TDP1 | NPASS |
| Nishamalaki | Curcuma longa | Rhizome | CID_2969 | 6097 | RORC | NPASS |
| Nishamalaki | Curcuma longa | Rhizome | CID_2969 | 7253 | TSHR | ChEMBL \| NPASS |
| Nishamalaki | Curcuma longa | Rhizome | CID_31253 | 2908 | NR3C1 | NPASS |
| Nishamalaki | Curcuma longa | Rhizome | CID_31253 | 3315 | HSPB1 | NPASS |
| Nishamalaki | Curcuma longa | Rhizome | CID_31253 | 3725 | JUN | NPASS |
| Nishamalaki | Curcuma longa | Rhizome | CID_31253 | 4790 | NFKB1 | NPASS |
| Nishamalaki | Curcuma longa | Rhizome | CID_31253 | 5467 | PPARD | NPASS |
| Nishamalaki | Curcuma longa | Rhizome | CID_31253 | 7253 | TSHR | NPASS |
| Nishamalaki | Curcuma longa | Rhizome | CID_31253 | 9971 | NR1H4 | NPASS |
| Nishamalaki | Curcuma longa | Rhizome | CID_325 | 5467 | PPARD | NPASS |
| Nishamalaki | Curcuma longa | Rhizome | CID_3314 | 10280 | SIGMAR1 | ChEMBL |

**Table S3: Associations between formulations, constituent plants, phytochemicals, and human molecular targets. The table lists the formulations, plant name and plant part, phytochemical identifiers, associated human target genes (Entrez Gene ID and gene symbol), and the database sources supporting each phytochemical-target interaction.**

| Formulation | Plant name | Part | PubChem CID | Entrez Gene ID | Gene Symbol | Source database(s) |
|---|---|---|---|---|---|---|
| Nishamalaki | Curcuma longa | Rhizome | CID_3314 | 10599 | SLCO1B1 | ChEMBL \| NPASS |
| Nishamalaki | Curcuma longa | Rhizome | CID_3314 | 10800 | CYSLTR1 | ChEMBL |
| Nishamalaki | Curcuma longa | Rhizome | CID_3314 | 10941 | UGT2A1 | ChEMBL |
| Nishamalaki | Curcuma longa | Rhizome | CID_3314 | 1128 | CHRM1 | ChEMBL |
| Nishamalaki | Curcuma longa | Rhizome | CID_3314 | 1129 | CHRM2 | ChEMBL |
| Nishamalaki | Curcuma longa | Rhizome | CID_3314 | 1131 | CHRM3 | ChEMBL |
| Nishamalaki | Curcuma longa | Rhizome | CID_3314 | 1132 | CHRM4 | ChEMBL |
| Nishamalaki | Curcuma longa | Rhizome | CID_3314 | 1133 | CHRM5 | ChEMBL |
| Nishamalaki | Curcuma longa | Rhizome | CID_3314 | 1233 | CCR4 | ChEMBL |
| Nishamalaki | Curcuma longa | Rhizome | CID_3314 | 1234 | CCR5 | ChEMBL |
| Nishamalaki | Curcuma longa | Rhizome | CID_3314 | 1268 | CNR1 | ChEMBL |
| Nishamalaki | Curcuma longa | Rhizome | CID_3314 | 134 | ADORA1 | ChEMBL |
| Nishamalaki | Curcuma longa | Rhizome | CID_3314 | 135 | ADORA2A | ChEMBL |
| Nishamalaki | Curcuma longa | Rhizome | CID_3314 | 140 | ADORA3 | ChEMBL |
| Nishamalaki | Curcuma longa | Rhizome | CID_3314 | 1432 | MAPK14 | ChEMBL |
| Nishamalaki | Curcuma longa | Rhizome | CID_3314 | 146 | ADRA1D | ChEMBL |
| Nishamalaki | Curcuma longa | Rhizome | CID_3314 | 150 | ADRA2A | ChEMBL |
| Nishamalaki | Curcuma longa | Rhizome | CID_3314 | 151 | ADRA2B | ChEMBL |
| Nishamalaki | Curcuma longa | Rhizome | CID_3314 | 1511 | CTSG | ChEMBL |
| Nishamalaki | Curcuma longa | Rhizome | CID_3314 | 152 | ADRA2C | ChEMBL |
| Nishamalaki | Curcuma longa | Rhizome | CID_3314 | 153 | ADRB1 | ChEMBL |
| Nishamalaki | Curcuma longa | Rhizome | CID_3314 | 154 | ADRB2 | ChEMBL |
| Nishamalaki | Curcuma longa | Rhizome | CID_3314 | 1544 | CYP1A2 | ChEMBL |
| Nishamalaki | Curcuma longa | Rhizome | CID_3314 | 1548 | CYP2A6 | ChEMBL |
| Nishamalaki | Curcuma longa | Rhizome | CID_3314 | 155 | ADRB3 | ChEMBL |
| Nishamalaki | Curcuma longa | Rhizome | CID_3314 | 1557 | CYP2C19 | ChEMBL |
| Nishamalaki | Curcuma longa | Rhizome | CID_3314 | 1559 | CYP2C9 | ChEMBL |
| Nishamalaki | Curcuma longa | Rhizome | CID_3314 | 1565 | CYP2D6 | ChEMBL |
| Nishamalaki | Curcuma longa | Rhizome | CID_3314 | 1571 | CYP2E1 | ChEMBL |
| Nishamalaki | Curcuma longa | Rhizome | CID_3314 | 1576 | CYP3A4 | ChEMBL \| NPASS |
| Nishamalaki | Curcuma longa | Rhizome | CID_3314 | 1812 | DRD1 | ChEMBL |
| Nishamalaki | Curcuma longa | Rhizome | CID_3314 | 1813 | DRD2 | ChEMBL |
| Nishamalaki | Curcuma longa | Rhizome | CID_3314 | 1814 | DRD3 | ChEMBL |
| Nishamalaki | Curcuma longa | Rhizome | CID_3314 | 1815 | DRD4 | ChEMBL |
| Nishamalaki | Curcuma longa | Rhizome | CID_3314 | 186 | AGTR2 | ChEMBL |
| Nishamalaki | Curcuma longa | Rhizome | CID_3314 | 1909 | EDNRA | ChEMBL |
| Nishamalaki | Curcuma longa | Rhizome | CID_3314 | 1956 | EGFR | ChEMBL |
| Nishamalaki | Curcuma longa | Rhizome | CID_3314 | 1991 | ELANE | ChEMBL |
| Nishamalaki | Curcuma longa | Rhizome | CID_3314 | 2064 | ERBB2 | ChEMBL |
| Nishamalaki | Curcuma longa | Rhizome | CID_3314 | 2099 | ESR1 | ChEMBL |
| Nishamalaki | Curcuma longa | Rhizome | CID_3314 | 2100 | ESR2 | ChEMBL |
| Nishamalaki | Curcuma longa | Rhizome | CID_3314 | 216 | ALDH1A1 | ChEMBL \| NPASS |
| Nishamalaki | Curcuma longa | Rhizome | CID_3314 | 2321 | FLT1 | ChEMBL |
| Nishamalaki | Curcuma longa | Rhizome | CID_3314 | 240 | ALOX5 | BindingDB \| ChEMBL \| NPASS |
| Nishamalaki | Curcuma longa | Rhizome | CID_3314 | 2534 | FYN | ChEMBL |
| Nishamalaki | Curcuma longa | Rhizome | CID_3314 | 26339 | OR5K1 | ChEMBL \| NPASS |
| Nishamalaki | Curcuma longa | Rhizome | CID_3314 | 28234 | SLCO1B3 | ChEMBL \| NPASS |
| Nishamalaki | Curcuma longa | Rhizome | CID_3314 | 2908 | NR3C1 | ChEMBL |
| Nishamalaki | Curcuma longa | Rhizome | CID_3314 | 2990 | GUSB | NPASS |
| Nishamalaki | Curcuma longa | Rhizome | CID_3314 | 3156 | HMGCR | ChEMBL |
| Nishamalaki | Curcuma longa | Rhizome | CID_3314 | 3269 | HRH1 | ChEMBL |
| Nishamalaki | Curcuma longa | Rhizome | CID_3314 | 3274 | HRH2 | ChEMBL |
| Nishamalaki | Curcuma longa | Rhizome | CID_3314 | 3356 | HTR2A | ChEMBL |
| Nishamalaki | Curcuma longa | Rhizome | CID_3314 | 3357 | HTR2B | ChEMBL |
| Nishamalaki | Curcuma longa | Rhizome | CID_3314 | 3358 | HTR2C | ChEMBL |
| Nishamalaki | Curcuma longa | Rhizome | CID_3314 | 3362 | HTR6 | ChEMBL |
| Nishamalaki | Curcuma longa | Rhizome | CID_3314 | 3576 | CXCL8 | NPASS |
| Nishamalaki | Curcuma longa | Rhizome | CID_3314 | 3577 | CXCR1 | ChEMBL |
| Nishamalaki | Curcuma longa | Rhizome | CID_3314 | 3579 | CXCR2 | ChEMBL |
| Nishamalaki | Curcuma longa | Rhizome | CID_3314 | 367 | AR | BindingDB \| ChEMBL \| NPASS |
| Nishamalaki | Curcuma longa | Rhizome | CID_3314 | 3725 | JUN | NPASS |
| Nishamalaki | Curcuma longa | Rhizome | CID_3314 | 3757 | KCNH2 | ChEMBL |
| Nishamalaki | Curcuma longa | Rhizome | CID_3314 | 3932 | LCK | ChEMBL |
| Nishamalaki | Curcuma longa | Rhizome | CID_3314 | 4128 | MAOA | ChEMBL |
| Nishamalaki | Curcuma longa | Rhizome | CID_3314 | 4159 | MC3R | ChEMBL |
| Nishamalaki | Curcuma longa | Rhizome | CID_3314 | 4160 | MC4R | ChEMBL |
| Nishamalaki | Curcuma longa | Rhizome | CID_3314 | 4161 | MC5R | ChEMBL |
| Nishamalaki | Curcuma longa | Rhizome | CID_3314 | 43 | ACHE | ChEMBL |
| Nishamalaki | Curcuma longa | Rhizome | CID_3314 | 4312 | MMP1 | ChEMBL |
| Nishamalaki | Curcuma longa | Rhizome | CID_3314 | 4318 | MMP9 | ChEMBL |

**Table S3: Associations between formulations, constituent plants, phytochemicals, and human molecular targets. The table lists the formulations, plant name and plant part, phytochemical identifiers, associated human target genes (Entrez Gene ID and gene symbol), and the database sources supporting each phytochemical-target interaction.**

| Formulation | Plant name | Part | PubChem CID | Entrez Gene ID | Gene Symbol | Source database(s) |
|---|---|---|---|---|---|---|
| Nishamalaki | Curcuma longa | Rhizome | CID_3314 | 4780 | NFE2L2 | NPASS |
| Nishamalaki | Curcuma longa | Rhizome | CID_3314 | 4886 | NPY1R | ChEMBL |
| Nishamalaki | Curcuma longa | Rhizome | CID_3314 | 4887 | NPY2R | ChEMBL |
| Nishamalaki | Curcuma longa | Rhizome | CID_3314 | 4985 | OPRD1 | ChEMBL |
| Nishamalaki | Curcuma longa | Rhizome | CID_3314 | 4986 | OPRK1 | ChEMBL |
| Nishamalaki | Curcuma longa | Rhizome | CID_3314 | 4988 | OPRM1 | ChEMBL |
| Nishamalaki | Curcuma longa | Rhizome | CID_3314 | 51053 | GMNN | NPASS |
| Nishamalaki | Curcuma longa | Rhizome | CID_3314 | 54575 | UGT1A10 | ChEMBL \| NPASS |
| Nishamalaki | Curcuma longa | Rhizome | CID_3314 | 54576 | UGT1A8 | ChEMBL \| NPASS |
| Nishamalaki | Curcuma longa | Rhizome | CID_3314 | 54578 | UGT1A6 | ChEMBL \| NPASS |
| Nishamalaki | Curcuma longa | Rhizome | CID_3314 | 54600 | UGT1A9 | ChEMBL \| NPASS |
| Nishamalaki | Curcuma longa | Rhizome | CID_3314 | 54657 | UGT1A4 | ChEMBL \| NPASS |
| Nishamalaki | Curcuma longa | Rhizome | CID_3314 | 54658 | UGT1A1 | ChEMBL \| NPASS |
| Nishamalaki | Curcuma longa | Rhizome | CID_3314 | 54659 | UGT1A3 | ChEMBL \| NPASS |
| Nishamalaki | Curcuma longa | Rhizome | CID_3314 | 552 | AVPR1A | ChEMBL |
| Nishamalaki | Curcuma longa | Rhizome | CID_3314 | 5530 | PPP3CA | ChEMBL |
| Nishamalaki | Curcuma longa | Rhizome | CID_3314 | 5578 | PRKCA | ChEMBL |
| Nishamalaki | Curcuma longa | Rhizome | CID_3314 | 5594 | MAPK1 | ChEMBL |
| Nishamalaki | Curcuma longa | Rhizome | CID_3314 | 5595 | MAPK3 | ChEMBL |
| Nishamalaki | Curcuma longa | Rhizome | CID_3314 | 5724 | PTAFR | ChEMBL |
| Nishamalaki | Curcuma longa | Rhizome | CID_3314 | 5742 | PTGS1 | ChEMBL \| NPASS |
| Nishamalaki | Curcuma longa | Rhizome | CID_3314 | 5743 | PTGS2 | BindingDB \| ChEMBL \| NPASS |
| Nishamalaki | Curcuma longa | Rhizome | CID_3314 | 5788 | PTPRC | ChEMBL |
| Nishamalaki | Curcuma longa | Rhizome | CID_3314 | 60489 | APOBEC3G | NPASS |
| Nishamalaki | Curcuma longa | Rhizome | CID_3314 | 624 | BDKRB2 | ChEMBL |
| Nishamalaki | Curcuma longa | Rhizome | CID_3314 | 6530 | SLC6A2 | ChEMBL |
| Nishamalaki | Curcuma longa | Rhizome | CID_3314 | 6531 | SLC6A3 | ChEMBL |
| Nishamalaki | Curcuma longa | Rhizome | CID_3314 | 6532 | SLC6A4 | ChEMBL |
| Nishamalaki | Curcuma longa | Rhizome | CID_3314 | 6865 | TACR2 | ChEMBL |
| Nishamalaki | Curcuma longa | Rhizome | CID_3314 | 6869 | TACR1 | ChEMBL |
| Nishamalaki | Curcuma longa | Rhizome | CID_3314 | 6916 | TBXAS1 | ChEMBL |
| Nishamalaki | Curcuma longa | Rhizome | CID_3314 | 729230 | CCR2 | ChEMBL |
| Nishamalaki | Curcuma longa | Rhizome | CID_3314 | 7366 | UGT2B15 | ChEMBL \| NPASS |
| Nishamalaki | Curcuma longa | Rhizome | CID_3314 | 7367 | UGT2B17 | ChEMBL \| NPASS |
| Nishamalaki | Curcuma longa | Rhizome | CID_3314 | 7433 | VIPR1 | ChEMBL |
| Nishamalaki | Curcuma longa | Rhizome | CID_3314 | 760 | CA2 | ChEMBL |
| Nishamalaki | Curcuma longa | Rhizome | CID_3314 | 775 | CACNA1C | NPASS |
| Nishamalaki | Curcuma longa | Rhizome | CID_3314 | 79054 | TRPM8 | ChEMBL |
| Nishamalaki | Curcuma longa | Rhizome | CID_3314 | 799 | CALCR | ChEMBL |
| Nishamalaki | Curcuma longa | Rhizome | CID_3314 | 834 | CASP1 | ChEMBL |
| Nishamalaki | Curcuma longa | Rhizome | CID_3314 | 8654 | PDE5A | ChEMBL |
| Nishamalaki | Curcuma longa | Rhizome | CID_3314 | 886 | CCKAR | ChEMBL |
| Nishamalaki | Curcuma longa | Rhizome | CID_3314 | 8989 | TRPA1 | ChEMBL \| NPASS |
| Nishamalaki | Curcuma longa | Rhizome | CID_442343 | 10013 | HDAC6 | ChEMBL |
| Nishamalaki | Curcuma longa | Rhizome | CID_442343 | 10599 | SLCO1B1 | ChEMBL \| NPASS |
| Nishamalaki | Curcuma longa | Rhizome | CID_442343 | 1139 | CHRNA7 | BindingDB \| ChEMBL \| NPASS |
| Nishamalaki | Curcuma longa | Rhizome | CID_442343 | 28234 | SLCO1B3 | ChEMBL \| NPASS |
| Nishamalaki | Curcuma longa | Rhizome | CID_442343 | 3315 | HSPB1 | NPASS |
| Nishamalaki | Curcuma longa | Rhizome | CID_442343 | 3725 | JUN | NPASS |
| Nishamalaki | Curcuma longa | Rhizome | CID_442343 | 6774 | STAT3 | ChEMBL \| NPASS |
| Nishamalaki | Curcuma longa | Rhizome | CID_442495 | 367 | AR | NPASS |
| Nishamalaki | Curcuma longa | Rhizome | CID_442495 | 4790 | NFKB1 | NPASS |
| Nishamalaki | Curcuma longa | Rhizome | CID_445070 | 10599 | SLCO1B1 | ChEMBL \| NPASS |
| Nishamalaki | Curcuma longa | Rhizome | CID_445070 | 10919 | EHMT2 | NPASS |
| Nishamalaki | Curcuma longa | Rhizome | CID_445070 | 1576 | CYP3A4 | NPASS |
| Nishamalaki | Curcuma longa | Rhizome | CID_445070 | 246 | ALOX15 | NPASS |
| Nishamalaki | Curcuma longa | Rhizome | CID_445070 | 28234 | SLCO1B3 | ChEMBL \| NPASS |
| Nishamalaki | Curcuma longa | Rhizome | CID_445070 | 3315 | HSPB1 | NPASS |
| Nishamalaki | Curcuma longa | Rhizome | CID_445070 | 367 | AR | NPASS |
| Nishamalaki | Curcuma longa | Rhizome | CID_445070 | 3725 | JUN | NPASS |
| Nishamalaki | Curcuma longa | Rhizome | CID_445070 | 4129 | MAOB | BindingDB |
| Nishamalaki | Curcuma longa | Rhizome | CID_445070 | 4137 | MAPT | NPASS |
| Nishamalaki | Curcuma longa | Rhizome | CID_445070 | 4297 | KMT2A | NPASS |
| Nishamalaki | Curcuma longa | Rhizome | CID_445070 | 4780 | NFE2L2 | NPASS |
| Nishamalaki | Curcuma longa | Rhizome | CID_445070 | 54658 | UGT1A1 | ChEMBL \| NPASS |
| Nishamalaki | Curcuma longa | Rhizome | CID_445070 | 5467 | PPARD | NPASS |
| Nishamalaki | Curcuma longa | Rhizome | CID_445070 | 55775 | TDP1 | NPASS |
| Nishamalaki | Curcuma longa | Rhizome | CID_445070 | 7398 | USP1 | NPASS |
| Nishamalaki | Curcuma longa | Rhizome | CID_5144 | 10280 | SIGMAR1 | ChEMBL |
| Nishamalaki | Curcuma longa | Rhizome | CID_5144 | 10599 | SLCO1B1 | ChEMBL \| NPASS |

**Table S3: Associations between formulations, constituent plants, phytochemicals, and human molecular targets. The table lists the formulations, plant name and plant part, phytochemical identifiers, associated human target genes (Entrez Gene ID and gene symbol), and the database sources supporting each phytochemical-target interaction.**

| Formulation | Plant name | Part | PubChem CID | Entrez Gene ID | Gene Symbol | Source database(s) |
|---|---|---|---|---|---|---|
| Nishamalaki | Curcuma longa | Rhizome | CID_5144 | 10800 | CYSLTR1 | ChEMBL |
| Nishamalaki | Curcuma longa | Rhizome | CID_5144 | 1128 | CHRM1 | ChEMBL |
| Nishamalaki | Curcuma longa | Rhizome | CID_5144 | 1129 | CHRM2 | ChEMBL |
| Nishamalaki | Curcuma longa | Rhizome | CID_5144 | 1131 | CHRM3 | ChEMBL |
| Nishamalaki | Curcuma longa | Rhizome | CID_5144 | 1132 | CHRM4 | ChEMBL |
| Nishamalaki | Curcuma longa | Rhizome | CID_5144 | 1133 | CHRM5 | ChEMBL |
| Nishamalaki | Curcuma longa | Rhizome | CID_5144 | 1233 | CCR4 | ChEMBL |
| Nishamalaki | Curcuma longa | Rhizome | CID_5144 | 1234 | CCR5 | ChEMBL |
| Nishamalaki | Curcuma longa | Rhizome | CID_5144 | 1268 | CNR1 | ChEMBL |
| Nishamalaki | Curcuma longa | Rhizome | CID_5144 | 134 | ADORA1 | ChEMBL |
| Nishamalaki | Curcuma longa | Rhizome | CID_5144 | 135 | ADORA2A | ChEMBL |
| Nishamalaki | Curcuma longa | Rhizome | CID_5144 | 140 | ADORA3 | ChEMBL |
| Nishamalaki | Curcuma longa | Rhizome | CID_5144 | 1432 | MAPK14 | ChEMBL |
| Nishamalaki | Curcuma longa | Rhizome | CID_5144 | 146 | ADRA1D | ChEMBL |
| Nishamalaki | Curcuma longa | Rhizome | CID_5144 | 150 | ADRA2A | ChEMBL |
| Nishamalaki | Curcuma longa | Rhizome | CID_5144 | 151 | ADRA2B | ChEMBL |
| Nishamalaki | Curcuma longa | Rhizome | CID_5144 | 1511 | CTSG | ChEMBL |
| Nishamalaki | Curcuma longa | Rhizome | CID_5144 | 152 | ADRA2C | ChEMBL |
| Nishamalaki | Curcuma longa | Rhizome | CID_5144 | 153 | ADRB1 | ChEMBL |
| Nishamalaki | Curcuma longa | Rhizome | CID_5144 | 154 | ADRB2 | ChEMBL |
| Nishamalaki | Curcuma longa | Rhizome | CID_5144 | 1544 | CYP1A2 | ChEMBL \| NPASS |
| Nishamalaki | Curcuma longa | Rhizome | CID_5144 | 1548 | CYP2A6 | ChEMBL |
| Nishamalaki | Curcuma longa | Rhizome | CID_5144 | 155 | ADRB3 | ChEMBL |
| Nishamalaki | Curcuma longa | Rhizome | CID_5144 | 1557 | CYP2C19 | ChEMBL |
| Nishamalaki | Curcuma longa | Rhizome | CID_5144 | 1559 | CYP2C9 | ChEMBL |
| Nishamalaki | Curcuma longa | Rhizome | CID_5144 | 1565 | CYP2D6 | ChEMBL |
| Nishamalaki | Curcuma longa | Rhizome | CID_5144 | 1571 | CYP2E1 | ChEMBL |
| Nishamalaki | Curcuma longa | Rhizome | CID_5144 | 1576 | CYP3A4 | ChEMBL \| NPASS |
| Nishamalaki | Curcuma longa | Rhizome | CID_5144 | 1812 | DRD1 | ChEMBL |
| Nishamalaki | Curcuma longa | Rhizome | CID_5144 | 1813 | DRD2 | ChEMBL |
| Nishamalaki | Curcuma longa | Rhizome | CID_5144 | 1814 | DRD3 | ChEMBL |
| Nishamalaki | Curcuma longa | Rhizome | CID_5144 | 1815 | DRD4 | ChEMBL |
| Nishamalaki | Curcuma longa | Rhizome | CID_5144 | 186 | AGTR2 | ChEMBL |
| Nishamalaki | Curcuma longa | Rhizome | CID_5144 | 1909 | EDNRA | ChEMBL |
| Nishamalaki | Curcuma longa | Rhizome | CID_5144 | 1956 | EGFR | ChEMBL |
| Nishamalaki | Curcuma longa | Rhizome | CID_5144 | 1991 | ELANE | ChEMBL |
| Nishamalaki | Curcuma longa | Rhizome | CID_5144 | 2064 | ERBB2 | ChEMBL |
| Nishamalaki | Curcuma longa | Rhizome | CID_5144 | 2099 | ESR1 | ChEMBL |
| Nishamalaki | Curcuma longa | Rhizome | CID_5144 | 2100 | ESR2 | ChEMBL |
| Nishamalaki | Curcuma longa | Rhizome | CID_5144 | 2321 | FLT1 | ChEMBL |
| Nishamalaki | Curcuma longa | Rhizome | CID_5144 | 2534 | FYN | ChEMBL |
| Nishamalaki | Curcuma longa | Rhizome | CID_5144 | 28234 | SLCO1B3 | ChEMBL \| NPASS |
| Nishamalaki | Curcuma longa | Rhizome | CID_5144 | 2908 | NR3C1 | ChEMBL \| NPASS |
| Nishamalaki | Curcuma longa | Rhizome | CID_5144 | 3156 | HMGCR | ChEMBL |
| Nishamalaki | Curcuma longa | Rhizome | CID_5144 | 3269 | HRH1 | ChEMBL |
| Nishamalaki | Curcuma longa | Rhizome | CID_5144 | 3274 | HRH2 | ChEMBL |
| Nishamalaki | Curcuma longa | Rhizome | CID_5144 | 3356 | HTR2A | ChEMBL |
| Nishamalaki | Curcuma longa | Rhizome | CID_5144 | 3357 | HTR2B | ChEMBL |
| Nishamalaki | Curcuma longa | Rhizome | CID_5144 | 3358 | HTR2C | ChEMBL |
| Nishamalaki | Curcuma longa | Rhizome | CID_5144 | 3362 | HTR6 | ChEMBL |
| Nishamalaki | Curcuma longa | Rhizome | CID_5144 | 3577 | CXCR1 | ChEMBL |
| Nishamalaki | Curcuma longa | Rhizome | CID_5144 | 3579 | CXCR2 | ChEMBL |
| Nishamalaki | Curcuma longa | Rhizome | CID_5144 | 3757 | KCNH2 | ChEMBL |
| Nishamalaki | Curcuma longa | Rhizome | CID_5144 | 3932 | LCK | ChEMBL |
| Nishamalaki | Curcuma longa | Rhizome | CID_5144 | 4128 | MAOA | ChEMBL |
| Nishamalaki | Curcuma longa | Rhizome | CID_5144 | 4137 | MAPT | NPASS |
| Nishamalaki | Curcuma longa | Rhizome | CID_5144 | 4159 | MC3R | ChEMBL |
| Nishamalaki | Curcuma longa | Rhizome | CID_5144 | 4160 | MC4R | ChEMBL |
| Nishamalaki | Curcuma longa | Rhizome | CID_5144 | 4161 | MC5R | ChEMBL |
| Nishamalaki | Curcuma longa | Rhizome | CID_5144 | 43 | ACHE | ChEMBL |
| Nishamalaki | Curcuma longa | Rhizome | CID_5144 | 4312 | MMP1 | ChEMBL |
| Nishamalaki | Curcuma longa | Rhizome | CID_5144 | 4318 | MMP9 | ChEMBL |
| Nishamalaki | Curcuma longa | Rhizome | CID_5144 | 4780 | NFE2L2 | NPASS |
| Nishamalaki | Curcuma longa | Rhizome | CID_5144 | 4886 | NPY1R | ChEMBL |
| Nishamalaki | Curcuma longa | Rhizome | CID_5144 | 4887 | NPY2R | ChEMBL |
| Nishamalaki | Curcuma longa | Rhizome | CID_5144 | 4985 | OPRD1 | ChEMBL |
| Nishamalaki | Curcuma longa | Rhizome | CID_5144 | 4986 | OPRK1 | ChEMBL |
| Nishamalaki | Curcuma longa | Rhizome | CID_5144 | 4988 | OPRM1 | ChEMBL |
| Nishamalaki | Curcuma longa | Rhizome | CID_5144 | 552 | AVPR1A | ChEMBL |
| Nishamalaki | Curcuma longa | Rhizome | CID_5144 | 5530 | PPP3CA | ChEMBL |

**Table S3: Associations between formulations, constituent plants, phytochemicals, and human molecular targets. The table lists the formulations, plant name and plant part, phytochemical identifiers, associated human target genes (Entrez Gene ID and gene symbol), and the database sources supporting each phytochemical-target interaction.**

| Formulation | Plant name | Part | PubChem CID | Entrez Gene ID | Gene Symbol | Source database(s) |
|---|---|---|---|---|---|---|
| Nishamalaki | Curcuma longa | Rhizome | CID_5144 | 5578 | PRKCA | ChEMBL |
| Nishamalaki | Curcuma longa | Rhizome | CID_5144 | 5594 | MAPK1 | ChEMBL |
| Nishamalaki | Curcuma longa | Rhizome | CID_5144 | 5595 | MAPK3 | ChEMBL |
| Nishamalaki | Curcuma longa | Rhizome | CID_5144 | 5724 | PTAFR | ChEMBL |
| Nishamalaki | Curcuma longa | Rhizome | CID_5144 | 5742 | PTGS1 | ChEMBL |
| Nishamalaki | Curcuma longa | Rhizome | CID_5144 | 5743 | PTGS2 | ChEMBL |
| Nishamalaki | Curcuma longa | Rhizome | CID_5144 | 5788 | PTPRC | ChEMBL |
| Nishamalaki | Curcuma longa | Rhizome | CID_5144 | 624 | BDKRB2 | ChEMBL |
| Nishamalaki | Curcuma longa | Rhizome | CID_5144 | 6530 | SLC6A2 | ChEMBL |
| Nishamalaki | Curcuma longa | Rhizome | CID_5144 | 6531 | SLC6A3 | ChEMBL |
| Nishamalaki | Curcuma longa | Rhizome | CID_5144 | 6532 | SLC6A4 | ChEMBL |
| Nishamalaki | Curcuma longa | Rhizome | CID_5144 | 6865 | TACR2 | ChEMBL |
| Nishamalaki | Curcuma longa | Rhizome | CID_5144 | 6869 | TACR1 | ChEMBL |
| Nishamalaki | Curcuma longa | Rhizome | CID_5144 | 6916 | TBXAS1 | ChEMBL |
| Nishamalaki | Curcuma longa | Rhizome | CID_5144 | 729230 | CCR2 | ChEMBL |
| Nishamalaki | Curcuma longa | Rhizome | CID_5144 | 7433 | VIPR1 | ChEMBL |
| Nishamalaki | Curcuma longa | Rhizome | CID_5144 | 760 | CA2 | ChEMBL |
| Nishamalaki | Curcuma longa | Rhizome | CID_5144 | 799 | CALCR | ChEMBL |
| Nishamalaki | Curcuma longa | Rhizome | CID_5144 | 834 | CASP1 | ChEMBL |
| Nishamalaki | Curcuma longa | Rhizome | CID_5144 | 8654 | PDE5A | ChEMBL |
| Nishamalaki | Curcuma longa | Rhizome | CID_5144 | 886 | CCKAR | ChEMBL |
| Nishamalaki | Curcuma longa | Rhizome | CID_5280794 | 10013 | HDAC6 | ChEMBL |
| Nishamalaki | Curcuma longa | Rhizome | CID_5280794 | 10599 | SLCO1B1 | ChEMBL \| NPASS |
| Nishamalaki | Curcuma longa | Rhizome | CID_5280794 | 28234 | SLCO1B3 | ChEMBL \| NPASS |
| Nishamalaki | Curcuma longa | Rhizome | CID_5280794 | 3363 | HTR7 | ChEMBL \| NPASS |
| Nishamalaki | Curcuma longa | Rhizome | CID_5280794 | 3417 | IDH1 | ChEMBL \| NPASS |
| Nishamalaki | Curcuma longa | Rhizome | CID_5280794 | 5423 | POLB | BindingDB \| ChEMBL \| NPASS |
| Nishamalaki | Curcuma longa | Rhizome | CID_5281515 | 10013 | HDAC6 | ChEMBL |
| Nishamalaki | Curcuma longa | Rhizome | CID_5281515 | 10599 | SLCO1B1 | ChEMBL \| NPASS |
| Nishamalaki | Curcuma longa | Rhizome | CID_5281515 | 1269 | CNR2 | BindingDB \| ChEMBL \| NPASS |
| Nishamalaki | Curcuma longa | Rhizome | CID_5281515 | 28234 | SLCO1B3 | ChEMBL \| NPASS |
| Nishamalaki | Curcuma longa | Rhizome | CID_5281515 | 4297 | KMT2A | NPASS |
| Nishamalaki | Curcuma longa | Rhizome | CID_5281515 | 5465 | PPARA | ChEMBL |
| Nishamalaki | Curcuma longa | Rhizome | CID_5281520 | 10599 | SLCO1B1 | ChEMBL \| NPASS |
| Nishamalaki | Curcuma longa | Rhizome | CID_5281520 | 28234 | SLCO1B3 | ChEMBL \| NPASS |
| Nishamalaki | Curcuma longa | Rhizome | CID_5284507 | 10599 | SLCO1B1 | ChEMBL \| NPASS |
| Nishamalaki | Curcuma longa | Rhizome | CID_5284507 | 246 | ALOX15 | ChEMBL \| NPASS |
| Nishamalaki | Curcuma longa | Rhizome | CID_5284507 | 28234 | SLCO1B3 | ChEMBL \| NPASS |
| Nishamalaki | Curcuma longa | Rhizome | CID_5284507 | 7068 | THRB | NPASS |
| Nishamalaki | Curcuma longa | Rhizome | CID_5284507 | 7398 | USP1 | NPASS |
| Nishamalaki | Curcuma longa | Rhizome | CID_5315472 | 1786 | DNMT1 | ChEMBL |
| Nishamalaki | Curcuma longa | Rhizome | CID_5315472 | 231 | AKR1B1 | ChEMBL \| NPASS |
| Nishamalaki | Curcuma longa | Rhizome | CID_5315472 | 2739 | GLO1 | ChEMBL \| NPASS |
| Nishamalaki | Curcuma longa | Rhizome | CID_5315472 | 351 | APP | ChEMBL \| NPASS |
| Nishamalaki | Curcuma longa | Rhizome | CID_5315472 | 4087 | SMAD2 | ChEMBL |
| Nishamalaki | Curcuma longa | Rhizome | CID_5315472 | 4088 | SMAD3 | ChEMBL |
| Nishamalaki | Curcuma longa | Rhizome | CID_5315472 | 4129 | MAOB | ChEMBL |
| Nishamalaki | Curcuma longa | Rhizome | CID_5315472 | 4780 | NFE2L2 | ChEMBL |
| Nishamalaki | Curcuma longa | Rhizome | CID_5315472 | 57016 | AKR1B10 | ChEMBL \| NPASS |
| Nishamalaki | Curcuma longa | Rhizome | CID_5315472 | 5743 | PTGS2 | BindingDB \| ChEMBL \| NPASS |
| Nishamalaki | Curcuma longa | Rhizome | CID_5315472 | 815 | CAMK2A | BindingDB \| ChEMBL \| NPASS |
| Nishamalaki | Curcuma longa | Rhizome | CID_5469424 | 23621 | BACE1 | BindingDB |
| Nishamalaki | Curcuma longa | Rhizome | CID_5469424 | 248 | ALPI | BindingDB |
| Nishamalaki | Curcuma longa | Rhizome | CID_5469424 | 249 | ALPL | BindingDB |
| Nishamalaki | Curcuma longa | Rhizome | CID_5469424 | 251 | ALPG | BindingDB |
| Nishamalaki | Curcuma longa | Rhizome | CID_5469424 | 4128 | MAOA | ChEMBL |
| Nishamalaki | Curcuma longa | Rhizome | CID_5469424 | 4129 | MAOB | ChEMBL |
| Nishamalaki | Curcuma longa | Rhizome | CID_5469424 | 5243 | ABCB1 | ChEMBL |
| Nishamalaki | Curcuma longa | Rhizome | CID_5469424 | 5743 | PTGS2 | BindingDB |
| Nishamalaki | Curcuma longa | Rhizome | CID_5469424 | 65975 | STK33 | BindingDB |
| Nishamalaki | Curcuma longa | Rhizome | CID_5469424 | 815 | CAMK2A | BindingDB |
| Nishamalaki | Curcuma longa | Rhizome | CID_5997 | 10013 | HDAC6 | ChEMBL |
| Nishamalaki | Curcuma longa | Rhizome | CID_5997 | 10062 | NR1H3 | ChEMBL \| NPASS |
| Nishamalaki | Curcuma longa | Rhizome | CID_5997 | 10599 | SLCO1B1 | ChEMBL \| NPASS |
| Nishamalaki | Curcuma longa | Rhizome | CID_5997 | 10858 | CYP46A1 | ChEMBL \| NPASS |
| Nishamalaki | Curcuma longa | Rhizome | CID_5997 | 1595 | CYP51A1 | NPASS |
| Nishamalaki | Curcuma longa | Rhizome | CID_5997 | 23762 | OSBP2 | BindingDB \| ChEMBL |
| Nishamalaki | Curcuma longa | Rhizome | CID_5997 | 28234 | SLCO1B3 | ChEMBL \| NPASS |
| Nishamalaki | Curcuma longa | Rhizome | CID_5997 | 2908 | NR3C1 | NPASS |
| Nishamalaki | Curcuma longa | Rhizome | CID_5997 | 3417 | IDH1 | ChEMBL \| NPASS |

**Table S3: Associations between formulations, constituent plants, phytochemicals, and human molecular targets. The table lists the formulations, plant name and plant part, phytochemical identifiers, associated human target genes (Entrez Gene ID and gene symbol), and the database sources supporting each phytochemical-target interaction.**

| Formulation | Plant name | Part | PubChem CID | Entrez Gene ID | Gene Symbol | Source database(s) |
|---|---|---|---|---|---|---|
| Nishamalaki | Curcuma longa | Rhizome | CID_5997 | 367 | AR | NPASS |
| Nishamalaki | Curcuma longa | Rhizome | CID_5997 | 4000 | LMNA | NPASS |
| Nishamalaki | Curcuma longa | Rhizome | CID_5997 | 4780 | NFE2L2 | ChEMBL \| NPASS |
| Nishamalaki | Curcuma longa | Rhizome | CID_5997 | 5007 | OSBP | ChEMBL |
| Nishamalaki | Curcuma longa | Rhizome | CID_5997 | 5243 | ABCB1 | ChEMBL \| NPASS |
| Nishamalaki | Curcuma longa | Rhizome | CID_5997 | 5422 | POLA1 | BindingDB \| NPASS |
| Nishamalaki | Curcuma longa | Rhizome | CID_5997 | 5467 | PPARD | NPASS |
| Nishamalaki | Curcuma longa | Rhizome | CID_5997 | 6097 | RORC | ChEMBL \| NPASS |
| Nishamalaki | Curcuma longa | Rhizome | CID_5997 | 6323 | SCN1A | ChEMBL |
| Nishamalaki | Curcuma longa | Rhizome | CID_5997 | 6326 | SCN2A | ChEMBL |
| Nishamalaki | Curcuma longa | Rhizome | CID_5997 | 6328 | SCN3A | ChEMBL \| NPASS |
| Nishamalaki | Curcuma longa | Rhizome | CID_5997 | 6622 | SNCA | ChEMBL \| NPASS |
| Nishamalaki | Curcuma longa | Rhizome | CID_5997 | 6820 | SULT2B1 | ChEMBL |
| Nishamalaki | Curcuma longa | Rhizome | CID_5997 | 7155 | TOP2B | NPASS |
| Nishamalaki | Curcuma longa | Rhizome | CID_5997 | 9429 | ABCG2 | ChEMBL |
| Nishamalaki | Curcuma longa | Rhizome | CID_637563 | 10599 | SLCO1B1 | ChEMBL \| NPASS |
| Nishamalaki | Curcuma longa | Rhizome | CID_637563 | 216 | ALDH1A1 | NPASS |
| Nishamalaki | Curcuma longa | Rhizome | CID_637563 | 28234 | SLCO1B3 | ChEMBL \| NPASS |
| Nishamalaki | Curcuma longa | Rhizome | CID_637563 | 367 | AR | NPASS |
| Nishamalaki | Curcuma longa | Rhizome | CID_637563 | 390245 | KDM4E | NPASS |
| Nishamalaki | Curcuma longa | Rhizome | CID_637563 | 4318 | MMP9 | ChEMBL |
| Nishamalaki | Curcuma longa | Rhizome | CID_637563 | 4780 | NFE2L2 | NPASS |
| Nishamalaki | Curcuma longa | Rhizome | CID_637563 | 5467 | PPARD | NPASS |
| Nishamalaki | Curcuma longa | Rhizome | CID_637566 | 10280 | SIGMAR1 | ChEMBL |
| Nishamalaki | Curcuma longa | Rhizome | CID_637566 | 10800 | CYSLTR1 | ChEMBL |
| Nishamalaki | Curcuma longa | Rhizome | CID_637566 | 10941 | UGT2A1 | ChEMBL |
| Nishamalaki | Curcuma longa | Rhizome | CID_637566 | 1128 | CHRM1 | ChEMBL |
| Nishamalaki | Curcuma longa | Rhizome | CID_637566 | 1129 | CHRM2 | ChEMBL |
| Nishamalaki | Curcuma longa | Rhizome | CID_637566 | 1131 | CHRM3 | ChEMBL |
| Nishamalaki | Curcuma longa | Rhizome | CID_637566 | 1132 | CHRM4 | ChEMBL |
| Nishamalaki | Curcuma longa | Rhizome | CID_637566 | 1133 | CHRM5 | ChEMBL |
| Nishamalaki | Curcuma longa | Rhizome | CID_637566 | 1233 | CCR4 | ChEMBL |
| Nishamalaki | Curcuma longa | Rhizome | CID_637566 | 1234 | CCR5 | ChEMBL |
| Nishamalaki | Curcuma longa | Rhizome | CID_637566 | 1268 | CNR1 | ChEMBL |
| Nishamalaki | Curcuma longa | Rhizome | CID_637566 | 134 | ADORA1 | ChEMBL |
| Nishamalaki | Curcuma longa | Rhizome | CID_637566 | 135 | ADORA2A | ChEMBL |
| Nishamalaki | Curcuma longa | Rhizome | CID_637566 | 140 | ADORA3 | ChEMBL |
| Nishamalaki | Curcuma longa | Rhizome | CID_637566 | 1432 | MAPK14 | ChEMBL |
| Nishamalaki | Curcuma longa | Rhizome | CID_637566 | 146 | ADRA1D | ChEMBL |
| Nishamalaki | Curcuma longa | Rhizome | CID_637566 | 150 | ADRA2A | ChEMBL |
| Nishamalaki | Curcuma longa | Rhizome | CID_637566 | 151 | ADRA2B | ChEMBL |
| Nishamalaki | Curcuma longa | Rhizome | CID_637566 | 1511 | CTSG | ChEMBL |
| Nishamalaki | Curcuma longa | Rhizome | CID_637566 | 152 | ADRA2C | ChEMBL |
| Nishamalaki | Curcuma longa | Rhizome | CID_637566 | 153 | ADRB1 | ChEMBL |
| Nishamalaki | Curcuma longa | Rhizome | CID_637566 | 154 | ADRB2 | ChEMBL |
| Nishamalaki | Curcuma longa | Rhizome | CID_637566 | 1544 | CYP1A2 | ChEMBL |
| Nishamalaki | Curcuma longa | Rhizome | CID_637566 | 1548 | CYP2A6 | ChEMBL |
| Nishamalaki | Curcuma longa | Rhizome | CID_637566 | 155 | ADRB3 | ChEMBL |
| Nishamalaki | Curcuma longa | Rhizome | CID_637566 | 1557 | CYP2C19 | ChEMBL |
| Nishamalaki | Curcuma longa | Rhizome | CID_637566 | 1559 | CYP2C9 | ChEMBL |
| Nishamalaki | Curcuma longa | Rhizome | CID_637566 | 1565 | CYP2D6 | ChEMBL |
| Nishamalaki | Curcuma longa | Rhizome | CID_637566 | 1571 | CYP2E1 | ChEMBL |
| Nishamalaki | Curcuma longa | Rhizome | CID_637566 | 1576 | CYP3A4 | ChEMBL \| NPASS |
| Nishamalaki | Curcuma longa | Rhizome | CID_637566 | 1812 | DRD1 | ChEMBL |
| Nishamalaki | Curcuma longa | Rhizome | CID_637566 | 1813 | DRD2 | ChEMBL |
| Nishamalaki | Curcuma longa | Rhizome | CID_637566 | 1814 | DRD3 | ChEMBL |
| Nishamalaki | Curcuma longa | Rhizome | CID_637566 | 1815 | DRD4 | ChEMBL |
| Nishamalaki | Curcuma longa | Rhizome | CID_637566 | 186 | AGTR2 | ChEMBL |
| Nishamalaki | Curcuma longa | Rhizome | CID_637566 | 1909 | EDNRA | ChEMBL |
| Nishamalaki | Curcuma longa | Rhizome | CID_637566 | 1956 | EGFR | ChEMBL |
| Nishamalaki | Curcuma longa | Rhizome | CID_637566 | 1991 | ELANE | ChEMBL |
| Nishamalaki | Curcuma longa | Rhizome | CID_637566 | 2064 | ERBB2 | ChEMBL |
| Nishamalaki | Curcuma longa | Rhizome | CID_637566 | 2099 | ESR1 | ChEMBL |
| Nishamalaki | Curcuma longa | Rhizome | CID_637566 | 2100 | ESR2 | ChEMBL |
| Nishamalaki | Curcuma longa | Rhizome | CID_637566 | 2321 | FLT1 | ChEMBL |
| Nishamalaki | Curcuma longa | Rhizome | CID_637566 | 246 | ALOX15 | ChEMBL \| NPASS |
| Nishamalaki | Curcuma longa | Rhizome | CID_637566 | 2534 | FYN | ChEMBL |
| Nishamalaki | Curcuma longa | Rhizome | CID_637566 | 25939 | SAMHD1 | ChEMBL |
| Nishamalaki | Curcuma longa | Rhizome | CID_637566 | 26339 | OR5K1 | ChEMBL \| NPASS |
| Nishamalaki | Curcuma longa | Rhizome | CID_637566 | 2908 | NR3C1 | ChEMBL \| NPASS |

| Table S3: Associations between formulations, constituent plants, phytochemicals, and human molecular targets. The table lists the formulations, plant name and plant part, phytochemical identifiers, associated human target genes (Entrez Gene ID and gene symbol), and the database sources supporting each phytochemical-target interaction. | | | | | | |
|---|---|---|---|---|---|---|
| Formulation | Plant name | Part | PubChem CID | Entrez Gene ID | Gene Symbol | Source database(s) |
| Nishamalaki | Curcuma longa | Rhizome | CID_637566 | 3156 | HMGCR | ChEMBL |
| Nishamalaki | Curcuma longa | Rhizome | CID_637566 | 3269 | HRH1 | ChEMBL |
| Nishamalaki | Curcuma longa | Rhizome | CID_637566 | 3274 | HRH2 | ChEMBL |
| Nishamalaki | Curcuma longa | Rhizome | CID_637566 | 3356 | HTR2A | ChEMBL |
| Nishamalaki | Curcuma longa | Rhizome | CID_637566 | 3357 | HTR2B | ChEMBL |
| Nishamalaki | Curcuma longa | Rhizome | CID_637566 | 3358 | HTR2C | ChEMBL |
| Nishamalaki | Curcuma longa | Rhizome | CID_637566 | 3362 | HTR6 | ChEMBL |
| Nishamalaki | Curcuma longa | Rhizome | CID_637566 | 3577 | CXCR1 | ChEMBL |
| Nishamalaki | Curcuma longa | Rhizome | CID_637566 | 3579 | CXCR2 | ChEMBL |
| Nishamalaki | Curcuma longa | Rhizome | CID_637566 | 3757 | KCNH2 | ChEMBL |
| Nishamalaki | Curcuma longa | Rhizome | CID_637566 | 3932 | LCK | ChEMBL |
| Nishamalaki | Curcuma longa | Rhizome | CID_637566 | 4128 | MAOA | ChEMBL |
| Nishamalaki | Curcuma longa | Rhizome | CID_637566 | 4159 | MC3R | ChEMBL |
| Nishamalaki | Curcuma longa | Rhizome | CID_637566 | 4160 | MC4R | ChEMBL |
| Nishamalaki | Curcuma longa | Rhizome | CID_637566 | 4161 | MC5R | ChEMBL |
| Nishamalaki | Curcuma longa | Rhizome | CID_637566 | 4297 | KMT2A | NPASS |
| Nishamalaki | Curcuma longa | Rhizome | CID_637566 | 43 | ACHE | ChEMBL |
| Nishamalaki | Curcuma longa | Rhizome | CID_637566 | 4312 | MMP1 | ChEMBL |
| Nishamalaki | Curcuma longa | Rhizome | CID_637566 | 4318 | MMP9 | ChEMBL |
| Nishamalaki | Curcuma longa | Rhizome | CID_637566 | 4790 | NFKB1 | NPASS |
| Nishamalaki | Curcuma longa | Rhizome | CID_637566 | 4886 | NPY1R | ChEMBL |
| Nishamalaki | Curcuma longa | Rhizome | CID_637566 | 4887 | NPY2R | ChEMBL |
| Nishamalaki | Curcuma longa | Rhizome | CID_637566 | 4985 | OPRD1 | ChEMBL |
| Nishamalaki | Curcuma longa | Rhizome | CID_637566 | 4986 | OPRK1 | ChEMBL |
| Nishamalaki | Curcuma longa | Rhizome | CID_637566 | 4988 | OPRM1 | ChEMBL |
| Nishamalaki | Curcuma longa | Rhizome | CID_637566 | 5467 | PPARD | NPASS |
| Nishamalaki | Curcuma longa | Rhizome | CID_637566 | 552 | AVPR1A | ChEMBL |
| Nishamalaki | Curcuma longa | Rhizome | CID_637566 | 5530 | PPP3CA | ChEMBL |
| Nishamalaki | Curcuma longa | Rhizome | CID_637566 | 5578 | PRKCA | ChEMBL |
| Nishamalaki | Curcuma longa | Rhizome | CID_637566 | 5594 | MAPK1 | ChEMBL |
| Nishamalaki | Curcuma longa | Rhizome | CID_637566 | 5595 | MAPK3 | ChEMBL |
| Nishamalaki | Curcuma longa | Rhizome | CID_637566 | 5724 | PTAFR | ChEMBL |
| Nishamalaki | Curcuma longa | Rhizome | CID_637566 | 5742 | PTGS1 | ChEMBL |
| Nishamalaki | Curcuma longa | Rhizome | CID_637566 | 5743 | PTGS2 | ChEMBL |
| Nishamalaki | Curcuma longa | Rhizome | CID_637566 | 5788 | PTPRC | ChEMBL |
| Nishamalaki | Curcuma longa | Rhizome | CID_637566 | 624 | BDKRB2 | ChEMBL |
| Nishamalaki | Curcuma longa | Rhizome | CID_637566 | 6530 | SLC6A2 | ChEMBL |
| Nishamalaki | Curcuma longa | Rhizome | CID_637566 | 6531 | SLC6A3 | ChEMBL |
| Nishamalaki | Curcuma longa | Rhizome | CID_637566 | 6532 | SLC6A4 | ChEMBL |
| Nishamalaki | Curcuma longa | Rhizome | CID_637566 | 6865 | TACR2 | ChEMBL |
| Nishamalaki | Curcuma longa | Rhizome | CID_637566 | 6869 | TACR1 | ChEMBL |
| Nishamalaki | Curcuma longa | Rhizome | CID_637566 | 6916 | TBXAS1 | ChEMBL |
| Nishamalaki | Curcuma longa | Rhizome | CID_637566 | 729230 | CCR2 | ChEMBL |
| Nishamalaki | Curcuma longa | Rhizome | CID_637566 | 7421 | VDR | NPASS |
| Nishamalaki | Curcuma longa | Rhizome | CID_637566 | 7433 | VIPR1 | ChEMBL |
| Nishamalaki | Curcuma longa | Rhizome | CID_637566 | 7442 | TRPV1 | BindingDB \| ChEMBL \| NPASS |
| Nishamalaki | Curcuma longa | Rhizome | CID_637566 | 760 | CA2 | ChEMBL |
| Nishamalaki | Curcuma longa | Rhizome | CID_637566 | 79054 | TRPM8 | ChEMBL \| NPASS |
| Nishamalaki | Curcuma longa | Rhizome | CID_637566 | 799 | CALCR | ChEMBL |
| Nishamalaki | Curcuma longa | Rhizome | CID_637566 | 834 | CASP1 | ChEMBL |
| Nishamalaki | Curcuma longa | Rhizome | CID_637566 | 8654 | PDE5A | ChEMBL |
| Nishamalaki | Curcuma longa | Rhizome | CID_637566 | 886 | CCKAR | ChEMBL |
| Nishamalaki | Curcuma longa | Rhizome | CID_638011 | 216 | ALDH1A1 | NPASS |
| Nishamalaki | Curcuma longa | Rhizome | CID_638011 | 2166 | FAAH | NPASS |
| Nishamalaki | Curcuma longa | Rhizome | CID_638011 | 26339 | OR5K1 | ChEMBL \| NPASS |
| Nishamalaki | Curcuma longa | Rhizome | CID_638011 | 367 | AR | NPASS |
| Nishamalaki | Curcuma longa | Rhizome | CID_638011 | 7068 | THRB | NPASS |
| Nishamalaki | Curcuma longa | Rhizome | CID_638011 | 865 | CBFB | NPASS |
| Nishamalaki | Curcuma longa | Rhizome | CID_638024 | 10013 | HDAC6 | ChEMBL |
| Nishamalaki | Curcuma longa | Rhizome | CID_638024 | 10599 | SLCO1B1 | NPASS |
| Nishamalaki | Curcuma longa | Rhizome | CID_638024 | 11201 | POLI | NPASS |
| Nishamalaki | Curcuma longa | Rhizome | CID_638024 | 1509 | CTSD | BindingDB \| ChEMBL \| NPASS |
| Nishamalaki | Curcuma longa | Rhizome | CID_638024 | 1543 | CYP1A1 | ChEMBL \| NPASS |
| Nishamalaki | Curcuma longa | Rhizome | CID_638024 | 1576 | CYP3A4 | ChEMBL \| NPASS |
| Nishamalaki | Curcuma longa | Rhizome | CID_638024 | 1723 | DHODH | BindingDB \| ChEMBL |
| Nishamalaki | Curcuma longa | Rhizome | CID_638024 | 196 | AHR | ChEMBL \| NPASS |
| Nishamalaki | Curcuma longa | Rhizome | CID_638024 | 216 | ALDH1A1 | NPASS |
| Nishamalaki | Curcuma longa | Rhizome | CID_638024 | 240 | ALOX5 | BindingDB \| ChEMBL \| NPASS |
| Nishamalaki | Curcuma longa | Rhizome | CID_638024 | 2554 | GABRA1 | BindingDB \| ChEMBL |
| Nishamalaki | Curcuma longa | Rhizome | CID_638024 | 2561 | GABRB2 | ChEMBL |

**Table S3: Associations between formulations, constituent plants, phytochemicals, and human molecular targets. The table lists the formulations, plant name and plant part, phytochemical identifiers, associated human target genes (Entrez Gene ID and gene symbol), and the database sources supporting each phytochemical-target interaction.**

| Formulation | Plant name | Part | PubChem CID | Entrez Gene ID | Gene Symbol | Source database(s) |
|---|---|---|---|---|---|---|
| Nishamalaki | Curcuma longa | Rhizome | CID_638024 | 2566 | GABRG2 | ChEMBL \| NPASS |
| Nishamalaki | Curcuma longa | Rhizome | CID_638024 | 25939 | SAMHD1 | ChEMBL |
| Nishamalaki | Curcuma longa | Rhizome | CID_638024 | 28234 | SLCO1B3 | ChEMBL \| NPASS |
| Nishamalaki | Curcuma longa | Rhizome | CID_638024 | 2908 | NR3C1 | NPASS |
| Nishamalaki | Curcuma longa | Rhizome | CID_638024 | 3028 | HSD17B10 | NPASS |
| Nishamalaki | Curcuma longa | Rhizome | CID_638024 | 3248 | HPGD | ChEMBL \| NPASS |
| Nishamalaki | Curcuma longa | Rhizome | CID_638024 | 367 | AR | NPASS |
| Nishamalaki | Curcuma longa | Rhizome | CID_638024 | 390245 | KDM4E | NPASS |
| Nishamalaki | Curcuma longa | Rhizome | CID_638024 | 4128 | MAOA | ChEMBL \| NPASS |
| Nishamalaki | Curcuma longa | Rhizome | CID_638024 | 4129 | MAOB | ChEMBL \| NPASS |
| Nishamalaki | Curcuma longa | Rhizome | CID_638024 | 4780 | NFE2L2 | NPASS |
| Nishamalaki | Curcuma longa | Rhizome | CID_638024 | 4864 | NPC1 | ChEMBL \| NPASS |
| Nishamalaki | Curcuma longa | Rhizome | CID_638024 | 5243 | ABCB1 | ChEMBL \| NPASS |
| Nishamalaki | Curcuma longa | Rhizome | CID_638024 | 5467 | PPARD | NPASS |
| Nishamalaki | Curcuma longa | Rhizome | CID_638024 | 5468 | PPARG | NPASS |
| Nishamalaki | Curcuma longa | Rhizome | CID_638024 | 55775 | TDP1 | NPASS |
| Nishamalaki | Curcuma longa | Rhizome | CID_638024 | 5594 | MAPK1 | NPASS |
| Nishamalaki | Curcuma longa | Rhizome | CID_638024 | 6256 | RXRA | NPASS |
| Nishamalaki | Curcuma longa | Rhizome | CID_638024 | 6311 | ATXN2 | NPASS |
| Nishamalaki | Curcuma longa | Rhizome | CID_638024 | 6606 | SMN1 | ChEMBL \| NPASS |
| Nishamalaki | Curcuma longa | Rhizome | CID_638024 | 6609 | SMPD1 | NPASS |
| Nishamalaki | Curcuma longa | Rhizome | CID_638024 | 672 | BRCA1 | ChEMBL \| NPASS |
| Nishamalaki | Curcuma longa | Rhizome | CID_638024 | 7442 | TRPV1 | BindingDB \| ChEMBL \| NPASS |
| Nishamalaki | Curcuma longa | Rhizome | CID_638024 | 759 | CA1 | BindingDB |
| Nishamalaki | Curcuma longa | Rhizome | CID_638024 | 760 | CA2 | BindingDB |
| Nishamalaki | Curcuma longa | Rhizome | CID_638024 | 79915 | ATAD5 | ChEMBL \| NPASS |
| Nishamalaki | Curcuma longa | Rhizome | CID_638024 | 9367 | RAB9A | ChEMBL \| NPASS |
| Nishamalaki | Curcuma longa | Rhizome | CID_638024 | 9971 | NR1H4 | NPASS |
| Nishamalaki | Curcuma longa | Rhizome | CID_6532 | 4790 | NFKB1 | NPASS |
| Nishamalaki | Curcuma longa | Rhizome | CID_6549 | 10013 | HDAC6 | ChEMBL |
| Nishamalaki | Curcuma longa | Rhizome | CID_6549 | 10073 | SNUPN | NPASS |
| Nishamalaki | Curcuma longa | Rhizome | CID_6549 | 10941 | UGT2A1 | ChEMBL |
| Nishamalaki | Curcuma longa | Rhizome | CID_6549 | 10951 | CBX1 | NPASS |
| Nishamalaki | Curcuma longa | Rhizome | CID_6549 | 3359 | HTR3A | BindingDB \| NPASS |
| Nishamalaki | Curcuma longa | Rhizome | CID_6549 | 3837 | KPNB1 | NPASS |
| Nishamalaki | Curcuma longa | Rhizome | CID_6549 | 54657 | UGT1A4 | ChEMBL \| NPASS |
| Nishamalaki | Curcuma longa | Rhizome | CID_6549 | 54658 | UGT1A1 | ChEMBL \| NPASS |
| Nishamalaki | Curcuma longa | Rhizome | CID_6549 | 7068 | THRB | NPASS |
| Nishamalaki | Curcuma longa | Rhizome | CID_6616 | 5467 | PPARD | NPASS |
| Nishamalaki | Curcuma longa | Rhizome | CID_6616 | 7442 | TRPV1 | ChEMBL \| NPASS |
| Nishamalaki | Curcuma longa | Rhizome | CID_6616 | 84876 | ORAI1 | ChEMBL \| NPASS |
| Nishamalaki | Curcuma longa | Rhizome | CID_6654 | 5594 | MAPK1 | NPASS |
| Nishamalaki | Curcuma longa | Rhizome | CID_6654 | 6097 | RORC | NPASS |
| Nishamalaki | Curcuma longa | Rhizome | CID_6654 | 7253 | TSHR | NPASS |
| Nishamalaki | Curcuma longa | Rhizome | CID_68381 | 8989 | TRPA1 | ChEMBL \| NPASS |
| Nishamalaki | Curcuma longa | Rhizome | CID_6918391 | 10013 | HDAC6 | ChEMBL |
| Nishamalaki | Curcuma longa | Rhizome | CID_6918391 | 5315 | PKM | ChEMBL |
| Nishamalaki | Curcuma longa | Rhizome | CID_6989 | 10013 | HDAC6 | ChEMBL |
| Nishamalaki | Curcuma longa | Rhizome | CID_6989 | 10941 | UGT2A1 | ChEMBL |
| Nishamalaki | Curcuma longa | Rhizome | CID_6989 | 11255 | HRH3 | ChEMBL |
| Nishamalaki | Curcuma longa | Rhizome | CID_6989 | 1128 | CHRM1 | ChEMBL |
| Nishamalaki | Curcuma longa | Rhizome | CID_6989 | 1129 | CHRM2 | ChEMBL |
| Nishamalaki | Curcuma longa | Rhizome | CID_6989 | 140 | ADORA3 | ChEMBL |
| Nishamalaki | Curcuma longa | Rhizome | CID_6989 | 148 | ADRA1A | ChEMBL |
| Nishamalaki | Curcuma longa | Rhizome | CID_6989 | 150 | ADRA2A | ChEMBL |
| Nishamalaki | Curcuma longa | Rhizome | CID_6989 | 1812 | DRD1 | ChEMBL |
| Nishamalaki | Curcuma longa | Rhizome | CID_6989 | 1814 | DRD3 | ChEMBL |
| Nishamalaki | Curcuma longa | Rhizome | CID_6989 | 2099 | ESR1 | ChEMBL |
| Nishamalaki | Curcuma longa | Rhizome | CID_6989 | 2147 | F2 | ChEMBL |
| Nishamalaki | Curcuma longa | Rhizome | CID_6989 | 2554 | GABRA1 | ChEMBL |
| Nishamalaki | Curcuma longa | Rhizome | CID_6989 | 2555 | GABRA2 | ChEMBL |
| Nishamalaki | Curcuma longa | Rhizome | CID_6989 | 2561 | GABRB2 | ChEMBL |
| Nishamalaki | Curcuma longa | Rhizome | CID_6989 | 2566 | GABRG2 | ChEMBL |
| Nishamalaki | Curcuma longa | Rhizome | CID_6989 | 26339 | OR5K1 | ChEMBL \| NPASS |
| Nishamalaki | Curcuma longa | Rhizome | CID_6989 | 3315 | HSPB1 | NPASS |
| Nishamalaki | Curcuma longa | Rhizome | CID_6989 | 3350 | HTR1A | ChEMBL |
| Nishamalaki | Curcuma longa | Rhizome | CID_6989 | 3357 | HTR2B | ChEMBL |
| Nishamalaki | Curcuma longa | Rhizome | CID_6989 | 367 | AR | ChEMBL \| NPASS |
| Nishamalaki | Curcuma longa | Rhizome | CID_6989 | 3757 | KCNH2 | ChEMBL |
| Nishamalaki | Curcuma longa | Rhizome | CID_6989 | 3791 | KDR | ChEMBL |

| Table S3: Associations between formulations, constituent plants, phytochemicals, and human molecular targets. The table lists the formulations, plant name and plant part, phytochemical identifiers, associated human target genes (Entrez Gene ID and gene symbol), and the database sources supporting each phytochemical-target interaction. | | | | | | |
|---|---|---|---|---|---|---|
| **Formulation** | **Plant name** | **Part** | **PubChem CID** | **Entrez Gene ID** | **Gene Symbol** | **Source database(s)** |
| Nishamalaki | Curcuma longa | Rhizome | CID_6989 | 4000 | LMNA | NPASS |
| Nishamalaki | Curcuma longa | Rhizome | CID_6989 | 4128 | MAOA | ChEMBL |
| Nishamalaki | Curcuma longa | Rhizome | CID_6989 | 43 | ACHE | ChEMBL |
| Nishamalaki | Curcuma longa | Rhizome | CID_6989 | 4780 | NFE2L2 | NPASS |
| Nishamalaki | Curcuma longa | Rhizome | CID_6989 | 4988 | OPRM1 | ChEMBL |
| Nishamalaki | Curcuma longa | Rhizome | CID_6989 | 5139 | PDE3A | ChEMBL |
| Nishamalaki | Curcuma longa | Rhizome | CID_6989 | 5141 | PDE4A | ChEMBL |
| Nishamalaki | Curcuma longa | Rhizome | CID_6989 | 5241 | PGR | ChEMBL |
| Nishamalaki | Curcuma longa | Rhizome | CID_6989 | 54578 | UGT1A6 | ChEMBL \| NPASS |
| Nishamalaki | Curcuma longa | Rhizome | CID_6989 | 54600 | UGT1A9 | ChEMBL \| NPASS |
| Nishamalaki | Curcuma longa | Rhizome | CID_6989 | 54657 | UGT1A4 | ChEMBL \| NPASS |
| Nishamalaki | Curcuma longa | Rhizome | CID_6989 | 54658 | UGT1A1 | ChEMBL \| NPASS |
| Nishamalaki | Curcuma longa | Rhizome | CID_6989 | 54659 | UGT1A3 | ChEMBL \| NPASS |
| Nishamalaki | Curcuma longa | Rhizome | CID_6989 | 5742 | PTGS1 | ChEMBL |
| Nishamalaki | Curcuma longa | Rhizome | CID_6989 | 5743 | PTGS2 | ChEMBL \| NPASS |
| Nishamalaki | Curcuma longa | Rhizome | CID_6989 | 6530 | SLC6A2 | ChEMBL |
| Nishamalaki | Curcuma longa | Rhizome | CID_6989 | 6531 | SLC6A3 | ChEMBL |
| Nishamalaki | Curcuma longa | Rhizome | CID_6989 | 6532 | SLC6A4 | ChEMBL |
| Nishamalaki | Curcuma longa | Rhizome | CID_6989 | 6915 | TBXA2R | ChEMBL |
| Nishamalaki | Curcuma longa | Rhizome | CID_6989 | 7366 | UGT2B15 | ChEMBL \| NPASS |
| Nishamalaki | Curcuma longa | Rhizome | CID_6989 | 775 | CACNA1C | NPASS |
| Nishamalaki | Curcuma longa | Rhizome | CID_6989 | 8989 | TRPA1 | ChEMBL \| NPASS |
| Nishamalaki | Curcuma longa | Rhizome | CID_7127 | 216 | ALDH1A1 | NPASS |
| Nishamalaki | Curcuma longa | Rhizome | CID_7127 | 2648 | KAT2A | NPASS |
| Nishamalaki | Curcuma longa | Rhizome | CID_7127 | 2908 | NR3C1 | NPASS |
| Nishamalaki | Curcuma longa | Rhizome | CID_7127 | 367 | AR | NPASS |
| Nishamalaki | Curcuma longa | Rhizome | CID_7127 | 4088 | SMAD3 | NPASS |
| Nishamalaki | Curcuma longa | Rhizome | CID_7127 | 4780 | NFE2L2 | ChEMBL \| NPASS |
| Nishamalaki | Curcuma longa | Rhizome | CID_7127 | 5465 | PPARA | ChEMBL \| NPASS |
| Nishamalaki | Curcuma longa | Rhizome | CID_7127 | 5467 | PPARD | ChEMBL \| NPASS |
| Nishamalaki | Curcuma longa | Rhizome | CID_7127 | 5468 | PPARG | BindingDB \| ChEMBL \| NPASS |
| Nishamalaki | Curcuma longa | Rhizome | CID_7127 | 7068 | THRB | NPASS |
| Nishamalaki | Curcuma longa | Rhizome | CID_7127 | 79915 | ATAD5 | NPASS |
| Nishamalaki | Curcuma longa | Rhizome | CID_7127 | 9971 | NR1H4 | NPASS |
| Nishamalaki | Curcuma longa | Rhizome | CID_7460 | 4780 | NFE2L2 | NPASS |
| Nishamalaki | Curcuma longa | Rhizome | CID_8051 | 3315 | HSPB1 | NPASS |
| Nishamalaki | Curcuma longa | Rhizome | CID_8163 | 8989 | TRPA1 | ChEMBL \| NPASS |
| Nishamalaki | Curcuma longa | Rhizome | CID_8184 | 3315 | HSPB1 | NPASS |
| Nishamalaki | Curcuma longa | Rhizome | CID_8184 | 367 | AR | NPASS |
| Nishamalaki | Curcuma longa | Rhizome | CID_8184 | 3725 | JUN | NPASS |
| Nishamalaki | Curcuma longa | Rhizome | CID_8184 | 4780 | NFE2L2 | NPASS |
| Nishamalaki | Curcuma longa | Rhizome | CID_8184 | 4790 | NFKB1 | NPASS |
| Nishamalaki | Curcuma longa | Rhizome | CID_8184 | 8989 | TRPA1 | ChEMBL \| NPASS |
| Nishamalaki | Curcuma longa | Rhizome | CID_8294 | 2908 | NR3C1 | NPASS |
| Nishamalaki | Curcuma longa | Rhizome | CID_8294 | 4790 | NFKB1 | NPASS |
| Nishamalaki | Curcuma longa | Rhizome | CID_91457 | 8989 | TRPA1 | ChEMBL \| NPASS |
| Nishamalaki | Curcuma longa | Rhizome | CID_969516 | 10013 | HDAC6 | ChEMBL |
| Nishamalaki | Curcuma longa | Rhizome | CID_969516 | 10014 | HDAC5 | ChEMBL |
| Nishamalaki | Curcuma longa | Rhizome | CID_969516 | 10073 | SNUPN | NPASS |
| Nishamalaki | Curcuma longa | Rhizome | CID_969516 | 1020 | CDK5 | ChEMBL |
| Nishamalaki | Curcuma longa | Rhizome | CID_969516 | 10376 | TUBA1B | ChEMBL |
| Nishamalaki | Curcuma longa | Rhizome | CID_969516 | 10381 | TUBB3 | ChEMBL |
| Nishamalaki | Curcuma longa | Rhizome | CID_969516 | 10382 | TUBB4A | ChEMBL |
| Nishamalaki | Curcuma longa | Rhizome | CID_969516 | 10383 | TUBB4B | ChEMBL |
| Nishamalaki | Curcuma longa | Rhizome | CID_969516 | 10524 | KAT5 | BindingDB |
| Nishamalaki | Curcuma longa | Rhizome | CID_969516 | 10587 | TXNRD2 | ChEMBL |
| Nishamalaki | Curcuma longa | Rhizome | CID_969516 | 10919 | EHMT2 | NPASS |
| Nishamalaki | Curcuma longa | Rhizome | CID_969516 | 11201 | POLI | NPASS |
| Nishamalaki | Curcuma longa | Rhizome | CID_969516 | 11238 | CA5B | BindingDB |
| Nishamalaki | Curcuma longa | Rhizome | CID_969516 | 112714 | TUBA3E | ChEMBL |
| Nishamalaki | Curcuma longa | Rhizome | CID_969516 | 114112 | TXNRD3 | ChEMBL \| NPASS |
| Nishamalaki | Curcuma longa | Rhizome | CID_969516 | 114548 | NLRP3 | BindingDB \| ChEMBL \| NPASS |
| Nishamalaki | Curcuma longa | Rhizome | CID_969516 | 1195 | CLK1 | BindingDB \| ChEMBL |
| Nishamalaki | Curcuma longa | Rhizome | CID_969516 | 1196 | CLK2 | BindingDB \| ChEMBL |
| Nishamalaki | Curcuma longa | Rhizome | CID_969516 | 1198 | CLK3 | ChEMBL |
| Nishamalaki | Curcuma longa | Rhizome | CID_969516 | 122706 | PSMB11 | ChEMBL |
| Nishamalaki | Curcuma longa | Rhizome | CID_969516 | 1387 | CREBBP | BindingDB \| ChEMBL \| NPASS |
| Nishamalaki | Curcuma longa | Rhizome | CID_969516 | 143471 | PSMA8 | ChEMBL |
| Nishamalaki | Curcuma longa | Rhizome | CID_969516 | 1454 | CSNK1E | ChEMBL |
| Nishamalaki | Curcuma longa | Rhizome | CID_969516 | 1544 | CYP1A2 | BindingDB |

| Table S3: Associations between formulations, constituent plants, phytochemicals, and human molecular targets. The table lists the formulations, plant name and plant part, phytochemical identifiers, associated human target genes (Entrez Gene ID and gene symbol), and the database sources supporting each phytochemical-target interaction. | | | | | | |
|---|---|---|---|---|---|---|
| Formulation | Plant name | Part | PubChem CID | Entrez Gene ID | Gene Symbol | Source database(s) |
| Nishamalaki | Curcuma longa | Rhizome | CID_969516 | 1559 | CYP2C9 | BindingDB |
| Nishamalaki | Curcuma longa | Rhizome | CID_969516 | 1576 | CYP3A4 | BindingDB \| ChEMBL \| NPASS |
| Nishamalaki | Curcuma longa | Rhizome | CID_969516 | 1859 | DYRK1A | BindingDB \| ChEMBL |
| Nishamalaki | Curcuma longa | Rhizome | CID_969516 | 1956 | EGFR | BindingDB \| ChEMBL \| NPASS |
| Nishamalaki | Curcuma longa | Rhizome | CID_969516 | 196 | AHR | NPASS |
| Nishamalaki | Curcuma longa | Rhizome | CID_969516 | 200316 | APOBEC3F | NPASS |
| Nishamalaki | Curcuma longa | Rhizome | CID_969516 | 203068 | TUBB | ChEMBL |
| Nishamalaki | Curcuma longa | Rhizome | CID_969516 | 2033 | EP300 | BindingDB \| ChEMBL \| NPASS |
| Nishamalaki | Curcuma longa | Rhizome | CID_969516 | 2078 | ERG | NPASS |
| Nishamalaki | Curcuma longa | Rhizome | CID_969516 | 2099 | ESR1 | NPASS |
| Nishamalaki | Curcuma longa | Rhizome | CID_969516 | 2152 | F3 | BindingDB \| ChEMBL \| NPASS |
| Nishamalaki | Curcuma longa | Rhizome | CID_969516 | 216 | ALDH1A1 | ChEMBL \| NPASS |
| Nishamalaki | Curcuma longa | Rhizome | CID_969516 | 2237 | FEN1 | NPASS |
| Nishamalaki | Curcuma longa | Rhizome | CID_969516 | 23028 | KDM1A | BindingDB \| ChEMBL \| NPASS |
| Nishamalaki | Curcuma longa | Rhizome | CID_969516 | 23435 | TARDBP | NPASS |
| Nishamalaki | Curcuma longa | Rhizome | CID_969516 | 23621 | BACE1 | BindingDB \| NPASS |
| Nishamalaki | Curcuma longa | Rhizome | CID_969516 | 23632 | CA14 | BindingDB |
| Nishamalaki | Curcuma longa | Rhizome | CID_969516 | 23643 | LY96 | NPASS |
| Nishamalaki | Curcuma longa | Rhizome | CID_969516 | 239 | ALOX12 | NPASS |
| Nishamalaki | Curcuma longa | Rhizome | CID_969516 | 240 | ALOX5 | BindingDB \| ChEMBL \| NPASS |
| Nishamalaki | Curcuma longa | Rhizome | CID_969516 | 246 | ALOX15 | BindingDB \| NPASS |
| Nishamalaki | Curcuma longa | Rhizome | CID_969516 | 247 | ALOX15B | NPASS |
| Nishamalaki | Curcuma longa | Rhizome | CID_969516 | 248 | ALPI | BindingDB \| NPASS |
| Nishamalaki | Curcuma longa | Rhizome | CID_969516 | 249 | ALPL | BindingDB \| NPASS |
| Nishamalaki | Curcuma longa | Rhizome | CID_969516 | 251 | ALPG | BindingDB \| NPASS |
| Nishamalaki | Curcuma longa | Rhizome | CID_969516 | 2548 | GAA | NPASS |
| Nishamalaki | Curcuma longa | Rhizome | CID_969516 | 2629 | GBA1 | NPASS |
| Nishamalaki | Curcuma longa | Rhizome | CID_969516 | 2739 | GLO1 | ChEMBL \| NPASS |
| Nishamalaki | Curcuma longa | Rhizome | CID_969516 | 2744 | GLS | NPASS |
| Nishamalaki | Curcuma longa | Rhizome | CID_969516 | 2778 | GNAS | NPASS |
| Nishamalaki | Curcuma longa | Rhizome | CID_969516 | 290 | ANPEP | BindingDB \| ChEMBL \| NPASS |
| Nishamalaki | Curcuma longa | Rhizome | CID_969516 | 2908 | NR3C1 | NPASS |
| Nishamalaki | Curcuma longa | Rhizome | CID_969516 | 2932 | GSK3B | BindingDB \| ChEMBL \| NPASS |
| Nishamalaki | Curcuma longa | Rhizome | CID_969516 | 2946 | GSTM2 | BindingDB \| ChEMBL |
| Nishamalaki | Curcuma longa | Rhizome | CID_969516 | 29994 | BAZ2B | ChEMBL \| NPASS |
| Nishamalaki | Curcuma longa | Rhizome | CID_969516 | 3028 | HSD17B10 | ChEMBL \| NPASS |
| Nishamalaki | Curcuma longa | Rhizome | CID_969516 | 3043 | HBB | NPASS |
| Nishamalaki | Curcuma longa | Rhizome | CID_969516 | 3064 | HTT | NPASS |
| Nishamalaki | Curcuma longa | Rhizome | CID_969516 | 3065 | HDAC1 | BindingDB \| ChEMBL \| NPASS |
| Nishamalaki | Curcuma longa | Rhizome | CID_969516 | 3066 | HDAC2 | BindingDB \| ChEMBL \| NPASS |
| Nishamalaki | Curcuma longa | Rhizome | CID_969516 | 3091 | HIF1A | NPASS |
| Nishamalaki | Curcuma longa | Rhizome | CID_969516 | 3248 | HPGD | ChEMBL \| NPASS |
| Nishamalaki | Curcuma longa | Rhizome | CID_969516 | 3290 | HSD11B1 | BindingDB \| ChEMBL |
| Nishamalaki | Curcuma longa | Rhizome | CID_969516 | 3291 | HSD11B2 | BindingDB \| ChEMBL |
| Nishamalaki | Curcuma longa | Rhizome | CID_969516 | 3292 | HSD17B1 | ChEMBL \| NPASS |
| Nishamalaki | Curcuma longa | Rhizome | CID_969516 | 3293 | HSD17B3 | BindingDB |
| Nishamalaki | Curcuma longa | Rhizome | CID_969516 | 3294 | HSD17B2 | BindingDB \| ChEMBL \| NPASS |
| Nishamalaki | Curcuma longa | Rhizome | CID_969516 | 3309 | HSPA5 | NPASS |
| Nishamalaki | Curcuma longa | Rhizome | CID_969516 | 3315 | HSPB1 | NPASS |
| Nishamalaki | Curcuma longa | Rhizome | CID_969516 | 3329 | HSPD1 | ChEMBL \| NPASS |
| Nishamalaki | Curcuma longa | Rhizome | CID_969516 | 3336 | HSPE1 | ChEMBL \| NPASS |
| Nishamalaki | Curcuma longa | Rhizome | CID_969516 | 3417 | IDH1 | NPASS |
| Nishamalaki | Curcuma longa | Rhizome | CID_969516 | 347688 | TUBB8 | ChEMBL |
| Nishamalaki | Curcuma longa | Rhizome | CID_969516 | 347733 | TUBB2B | ChEMBL |
| Nishamalaki | Curcuma longa | Rhizome | CID_969516 | 351 | APP | BindingDB \| ChEMBL \| NPASS |
| Nishamalaki | Curcuma longa | Rhizome | CID_969516 | 3576 | CXCL8 | NPASS |
| Nishamalaki | Curcuma longa | Rhizome | CID_969516 | 367 | AR | NPASS |
| Nishamalaki | Curcuma longa | Rhizome | CID_969516 | 3725 | JUN | NPASS |
| Nishamalaki | Curcuma longa | Rhizome | CID_969516 | 3757 | KCNH2 | ChEMBL \| NPASS |
| Nishamalaki | Curcuma longa | Rhizome | CID_969516 | 3837 | KPNB1 | NPASS |
| Nishamalaki | Curcuma longa | Rhizome | CID_969516 | 387129 | NPSR1 | NPASS |
| Nishamalaki | Curcuma longa | Rhizome | CID_969516 | 390245 | KDM4E | ChEMBL \| NPASS |
| Nishamalaki | Curcuma longa | Rhizome | CID_969516 | 4000 | LMNA | NPASS |
| Nishamalaki | Curcuma longa | Rhizome | CID_969516 | 4088 | SMAD3 | NPASS |
| Nishamalaki | Curcuma longa | Rhizome | CID_969516 | 4128 | MAOA | ChEMBL \| NPASS |
| Nishamalaki | Curcuma longa | Rhizome | CID_969516 | 4129 | MAOB | ChEMBL \| NPASS |
| Nishamalaki | Curcuma longa | Rhizome | CID_969516 | 4137 | MAPT | BindingDB \| ChEMBL \| NPASS |
| Nishamalaki | Curcuma longa | Rhizome | CID_969516 | 4154 | MBNL1 | NPASS |
| Nishamalaki | Curcuma longa | Rhizome | CID_969516 | 4297 | KMT2A | NPASS |
| Nishamalaki | Curcuma longa | Rhizome | CID_969516 | 43 | ACHE | BindingDB \| ChEMBL \| NPASS |

**Table S3: Associations between formulations, constituent plants, phytochemicals, and human molecular targets. The table lists the formulations, plant name and plant part, phytochemical identifiers, associated human target genes (Entrez Gene ID and gene symbol), and the database sources supporting each phytochemical-target interaction.**

| Formulation | Plant name | Part | PubChem CID | Entrez Gene ID | Gene Symbol | Source database(s) |
|---|---|---|---|---|---|---|
| Nishamalaki | Curcuma longa | Rhizome | CID_969516 | 4318 | MMP9 | BindingDB |
| Nishamalaki | Curcuma longa | Rhizome | CID_969516 | 4322 | MMP13 | BindingDB |
| Nishamalaki | Curcuma longa | Rhizome | CID_969516 | 440503 | PLIN5 | BindingDB \| NPASS |
| Nishamalaki | Curcuma longa | Rhizome | CID_969516 | 4780 | NFE2L2 | ChEMBL \| NPASS |
| Nishamalaki | Curcuma longa | Rhizome | CID_969516 | 4790 | NFKB1 | NPASS |
| Nishamalaki | Curcuma longa | Rhizome | CID_969516 | 4843 | NOS2 | BindingDB |
| Nishamalaki | Curcuma longa | Rhizome | CID_969516 | 488 | ATP2A2 | ChEMBL \| NPASS |
| Nishamalaki | Curcuma longa | Rhizome | CID_969516 | 489 | ATP2A3 | ChEMBL \| NPASS |
| Nishamalaki | Curcuma longa | Rhizome | CID_969516 | 5058 | PAK1 | ChEMBL \| NPASS |
| Nishamalaki | Curcuma longa | Rhizome | CID_969516 | 51053 | GMNN | ChEMBL \| NPASS |
| Nishamalaki | Curcuma longa | Rhizome | CID_969516 | 51099 | ABHD5 | ChEMBL \| NPASS |
| Nishamalaki | Curcuma longa | Rhizome | CID_969516 | 51426 | POLK | NPASS |
| Nishamalaki | Curcuma longa | Rhizome | CID_969516 | 51564 | HDAC7 | ChEMBL |
| Nishamalaki | Curcuma longa | Rhizome | CID_969516 | 5300 | PIN1 | NPASS |
| Nishamalaki | Curcuma longa | Rhizome | CID_969516 | 5315 | PKM | ChEMBL \| NPASS |
| Nishamalaki | Curcuma longa | Rhizome | CID_969516 | 5319 | PLA2G1B | BindingDB |
| Nishamalaki | Curcuma longa | Rhizome | CID_969516 | 5346 | PLIN1 | BindingDB \| ChEMBL \| NPASS |
| Nishamalaki | Curcuma longa | Rhizome | CID_969516 | 54106 | TLR9 | ChEMBL \| NPASS |
| Nishamalaki | Curcuma longa | Rhizome | CID_969516 | 5423 | POLB | NPASS |
| Nishamalaki | Curcuma longa | Rhizome | CID_969516 | 5467 | PPARD | NPASS |
| Nishamalaki | Curcuma longa | Rhizome | CID_969516 | 5468 | PPARG | NPASS |
| Nishamalaki | Curcuma longa | Rhizome | CID_969516 | 55775 | TDP1 | ChEMBL \| NPASS |
| Nishamalaki | Curcuma longa | Rhizome | CID_969516 | 5580 | PRKCD | BindingDB |
| Nishamalaki | Curcuma longa | Rhizome | CID_969516 | 5581 | PRKCE | BindingDB |
| Nishamalaki | Curcuma longa | Rhizome | CID_969516 | 55847 | CISD1 | BindingDB \| ChEMBL \| NPASS |
| Nishamalaki | Curcuma longa | Rhizome | CID_969516 | 55869 | HDAC8 | BindingDB \| ChEMBL \| NPASS |
| Nishamalaki | Curcuma longa | Rhizome | CID_969516 | 5588 | PRKCQ | BindingDB |
| Nishamalaki | Curcuma longa | Rhizome | CID_969516 | 5594 | MAPK1 | ChEMBL \| NPASS |
| Nishamalaki | Curcuma longa | Rhizome | CID_969516 | 5599 | MAPK8 | ChEMBL |
| Nishamalaki | Curcuma longa | Rhizome | CID_969516 | 5682 | PSMA1 | ChEMBL |
| Nishamalaki | Curcuma longa | Rhizome | CID_969516 | 5683 | PSMA2 | ChEMBL |
| Nishamalaki | Curcuma longa | Rhizome | CID_969516 | 5684 | PSMA3 | ChEMBL |
| Nishamalaki | Curcuma longa | Rhizome | CID_969516 | 5685 | PSMA4 | ChEMBL |
| Nishamalaki | Curcuma longa | Rhizome | CID_969516 | 5686 | PSMA5 | ChEMBL |
| Nishamalaki | Curcuma longa | Rhizome | CID_969516 | 5687 | PSMA6 | ChEMBL |
| Nishamalaki | Curcuma longa | Rhizome | CID_969516 | 5688 | PSMA7 | ChEMBL |
| Nishamalaki | Curcuma longa | Rhizome | CID_969516 | 5689 | PSMB1 | ChEMBL |
| Nishamalaki | Curcuma longa | Rhizome | CID_969516 | 5690 | PSMB2 | ChEMBL |
| Nishamalaki | Curcuma longa | Rhizome | CID_969516 | 5691 | PSMB3 | ChEMBL |
| Nishamalaki | Curcuma longa | Rhizome | CID_969516 | 5692 | PSMB4 | ChEMBL |
| Nishamalaki | Curcuma longa | Rhizome | CID_969516 | 5693 | PSMB5 | BindingDB \| ChEMBL \| NPASS |
| Nishamalaki | Curcuma longa | Rhizome | CID_969516 | 5694 | PSMB6 | ChEMBL |
| Nishamalaki | Curcuma longa | Rhizome | CID_969516 | 5695 | PSMB7 | ChEMBL |
| Nishamalaki | Curcuma longa | Rhizome | CID_969516 | 5696 | PSMB8 | ChEMBL |
| Nishamalaki | Curcuma longa | Rhizome | CID_969516 | 5698 | PSMB9 | ChEMBL |
| Nishamalaki | Curcuma longa | Rhizome | CID_969516 | 5699 | PSMB10 | ChEMBL |
| Nishamalaki | Curcuma longa | Rhizome | CID_969516 | 57396 | CLK4 | BindingDB \| ChEMBL |
| Nishamalaki | Curcuma longa | Rhizome | CID_969516 | 5743 | PTGS2 | BindingDB \| ChEMBL \| NPASS |
| Nishamalaki | Curcuma longa | Rhizome | CID_969516 | 590 | BCHE | BindingDB \| ChEMBL \| NPASS |
| Nishamalaki | Curcuma longa | Rhizome | CID_969516 | 5901 | RAN | NPASS |
| Nishamalaki | Curcuma longa | Rhizome | CID_969516 | 5970 | RELA | ChEMBL |
| Nishamalaki | Curcuma longa | Rhizome | CID_969516 | 60489 | APOBEC3G | NPASS |
| Nishamalaki | Curcuma longa | Rhizome | CID_969516 | 6097 | RORC | NPASS |
| Nishamalaki | Curcuma longa | Rhizome | CID_969516 | 6256 | RXRA | NPASS |
| Nishamalaki | Curcuma longa | Rhizome | CID_969516 | 6554 | SLC10A1 | BindingDB \| ChEMBL |
| Nishamalaki | Curcuma longa | Rhizome | CID_969516 | 65975 | STK33 | BindingDB |
| Nishamalaki | Curcuma longa | Rhizome | CID_969516 | 6622 | SNCA | BindingDB \| ChEMBL \| NPASS |
| Nishamalaki | Curcuma longa | Rhizome | CID_969516 | 672 | BRCA1 | NPASS |
| Nishamalaki | Curcuma longa | Rhizome | CID_969516 | 7068 | THRB | NPASS |
| Nishamalaki | Curcuma longa | Rhizome | CID_969516 | 7157 | TP53 | ChEMBL \| NPASS |
| Nishamalaki | Curcuma longa | Rhizome | CID_969516 | 7251 | TSG101 | NPASS |
| Nishamalaki | Curcuma longa | Rhizome | CID_969516 | 7253 | TSHR | NPASS |
| Nishamalaki | Curcuma longa | Rhizome | CID_969516 | 7263 | TST | ChEMBL \| NPASS |
| Nishamalaki | Curcuma longa | Rhizome | CID_969516 | 7277 | TUBA4A | ChEMBL |
| Nishamalaki | Curcuma longa | Rhizome | CID_969516 | 7278 | TUBA3C | ChEMBL |
| Nishamalaki | Curcuma longa | Rhizome | CID_969516 | 7280 | TUBB2A | ChEMBL |
| Nishamalaki | Curcuma longa | Rhizome | CID_969516 | 7296 | TXNRD1 | ChEMBL |
| Nishamalaki | Curcuma longa | Rhizome | CID_969516 | 7299 | TYR | BindingDB \| ChEMBL \| NPASS |
| Nishamalaki | Curcuma longa | Rhizome | CID_969516 | 7398 | USP1 | ChEMBL \| NPASS |
| Nishamalaki | Curcuma longa | Rhizome | CID_969516 | 7421 | VDR | BindingDB \| ChEMBL \| NPASS |

**Table S3: Associations between formulations, constituent plants, phytochemicals, and human molecular targets. The table lists the formulations, plant name and plant part, phytochemical identifiers, associated human target genes (Entrez Gene ID and gene symbol), and the database sources supporting each phytochemical-target interaction.**

| Formulation | Plant name | Part | PubChem CID | Entrez Gene ID | Gene Symbol | Source database(s) |
|---|---|---|---|---|---|---|
| Nishamalaki | Curcuma longa | Rhizome | CID_969516 | 759 | CA1 | BindingDB |
| Nishamalaki | Curcuma longa | Rhizome | CID_969516 | 760 | CA2 | BindingDB |
| Nishamalaki | Curcuma longa | Rhizome | CID_969516 | 761 | CA3 | BindingDB |
| Nishamalaki | Curcuma longa | Rhizome | CID_969516 | 762 | CA4 | BindingDB |
| Nishamalaki | Curcuma longa | Rhizome | CID_969516 | 763 | CA5A | BindingDB |
| Nishamalaki | Curcuma longa | Rhizome | CID_969516 | 766 | CA7 | BindingDB |
| Nishamalaki | Curcuma longa | Rhizome | CID_969516 | 768 | CA9 | BindingDB |
| Nishamalaki | Curcuma longa | Rhizome | CID_969516 | 771 | CA12 | BindingDB |
| Nishamalaki | Curcuma longa | Rhizome | CID_969516 | 7846 | TUBA1A | ChEMBL |
| Nishamalaki | Curcuma longa | Rhizome | CID_969516 | 79885 | HDAC11 | ChEMBL |
| Nishamalaki | Curcuma longa | Rhizome | CID_969516 | 79915 | ATAD5 | NPASS |
| Nishamalaki | Curcuma longa | Rhizome | CID_969516 | 81027 | TUBB1 | ChEMBL |
| Nishamalaki | Curcuma longa | Rhizome | CID_969516 | 815 | CAMK2A | BindingDB |
| Nishamalaki | Curcuma longa | Rhizome | CID_969516 | 834 | CASP1 | NPASS |
| Nishamalaki | Curcuma longa | Rhizome | CID_969516 | 83933 | HDAC10 | ChEMBL \| NPASS |
| Nishamalaki | Curcuma longa | Rhizome | CID_969516 | 840 | CASP7 | NPASS |
| Nishamalaki | Curcuma longa | Rhizome | CID_969516 | 8444 | DYRK3 | BindingDB \| ChEMBL |
| Nishamalaki | Curcuma longa | Rhizome | CID_969516 | 8445 | DYRK2 | BindingDB \| ChEMBL |
| Nishamalaki | Curcuma longa | Rhizome | CID_969516 | 84617 | TUBB6 | ChEMBL |
| Nishamalaki | Curcuma longa | Rhizome | CID_969516 | 84790 | TUBA1C | ChEMBL |
| Nishamalaki | Curcuma longa | Rhizome | CID_969516 | 8798 | DYRK4 | BindingDB \| ChEMBL |
| Nishamalaki | Curcuma longa | Rhizome | CID_969516 | 8841 | HDAC3 | ChEMBL |
| Nishamalaki | Curcuma longa | Rhizome | CID_969516 | 8851 | CDK5R1 | ChEMBL |
| Nishamalaki | Curcuma longa | Rhizome | CID_969516 | 9149 | DYRK1B | ChEMBL |
| Nishamalaki | Curcuma longa | Rhizome | CID_969516 | 9429 | ABCG2 | BindingDB \| ChEMBL |
| Nishamalaki | Curcuma longa | Rhizome | CID_969516 | 9536 | PTGES | BindingDB \| ChEMBL \| NPASS |
| Nishamalaki | Curcuma longa | Rhizome | CID_969516 | 9682 | KDM4A | NPASS |
| Nishamalaki | Curcuma longa | Rhizome | CID_969516 | 9734 | HDAC9 | ChEMBL |
| Nishamalaki | Curcuma longa | Rhizome | CID_969516 | 9759 | HDAC4 | ChEMBL |
| Nishamalaki | Curcuma longa | Rhizome | CID_969516 | 9971 | NR1H4 | NPASS |
| Nishamalaki | Phyllanthus emblica | Fruit | CID_1057 | 1017 | CDK2 | BindingDB \| ChEMBL |
| Nishamalaki | Phyllanthus emblica | Fruit | CID_1057 | 10280 | SIGMAR1 | ChEMBL |
| Nishamalaki | Phyllanthus emblica | Fruit | CID_1057 | 10800 | CYSLTR1 | ChEMBL |
| Nishamalaki | Phyllanthus emblica | Fruit | CID_1057 | 10919 | EHMT2 | ChEMBL \| NPASS |
| Nishamalaki | Phyllanthus emblica | Fruit | CID_1057 | 10988 | METAP2 | ChEMBL \| NPASS |
| Nishamalaki | Phyllanthus emblica | Fruit | CID_1057 | 11096 | ADAMTS5 | BindingDB \| ChEMBL |
| Nishamalaki | Phyllanthus emblica | Fruit | CID_1057 | 1128 | CHRM1 | ChEMBL |
| Nishamalaki | Phyllanthus emblica | Fruit | CID_1057 | 1129 | CHRM2 | ChEMBL |
| Nishamalaki | Phyllanthus emblica | Fruit | CID_1057 | 1131 | CHRM3 | ChEMBL |
| Nishamalaki | Phyllanthus emblica | Fruit | CID_1057 | 1132 | CHRM4 | ChEMBL |
| Nishamalaki | Phyllanthus emblica | Fruit | CID_1057 | 1133 | CHRM5 | ChEMBL |
| Nishamalaki | Phyllanthus emblica | Fruit | CID_1057 | 1233 | CCR4 | ChEMBL |
| Nishamalaki | Phyllanthus emblica | Fruit | CID_1057 | 1234 | CCR5 | ChEMBL |
| Nishamalaki | Phyllanthus emblica | Fruit | CID_1057 | 1268 | CNR1 | ChEMBL |
| Nishamalaki | Phyllanthus emblica | Fruit | CID_1057 | 134 | ADORA1 | ChEMBL |
| Nishamalaki | Phyllanthus emblica | Fruit | CID_1057 | 135 | ADORA2A | ChEMBL |
| Nishamalaki | Phyllanthus emblica | Fruit | CID_1057 | 140 | ADORA3 | ChEMBL |
| Nishamalaki | Phyllanthus emblica | Fruit | CID_1057 | 1432 | MAPK14 | ChEMBL |
| Nishamalaki | Phyllanthus emblica | Fruit | CID_1057 | 146 | ADRA1D | ChEMBL |
| Nishamalaki | Phyllanthus emblica | Fruit | CID_1057 | 150 | ADRA2A | ChEMBL |
| Nishamalaki | Phyllanthus emblica | Fruit | CID_1057 | 151 | ADRA2B | ChEMBL |
| Nishamalaki | Phyllanthus emblica | Fruit | CID_1057 | 1511 | CTSG | ChEMBL |
| Nishamalaki | Phyllanthus emblica | Fruit | CID_1057 | 152 | ADRA2C | ChEMBL |
| Nishamalaki | Phyllanthus emblica | Fruit | CID_1057 | 153 | ADRB1 | ChEMBL |
| Nishamalaki | Phyllanthus emblica | Fruit | CID_1057 | 154 | ADRB2 | ChEMBL |
| Nishamalaki | Phyllanthus emblica | Fruit | CID_1057 | 1544 | CYP1A2 | ChEMBL |
| Nishamalaki | Phyllanthus emblica | Fruit | CID_1057 | 1548 | CYP2A6 | ChEMBL |
| Nishamalaki | Phyllanthus emblica | Fruit | CID_1057 | 155 | ADRB3 | ChEMBL |
| Nishamalaki | Phyllanthus emblica | Fruit | CID_1057 | 1557 | CYP2C19 | ChEMBL |
| Nishamalaki | Phyllanthus emblica | Fruit | CID_1057 | 1559 | CYP2C9 | ChEMBL |
| Nishamalaki | Phyllanthus emblica | Fruit | CID_1057 | 1565 | CYP2D6 | ChEMBL |
| Nishamalaki | Phyllanthus emblica | Fruit | CID_1057 | 1571 | CYP2E1 | ChEMBL |
| Nishamalaki | Phyllanthus emblica | Fruit | CID_1057 | 1576 | CYP3A4 | ChEMBL |
| Nishamalaki | Phyllanthus emblica | Fruit | CID_1057 | 1812 | DRD1 | ChEMBL |
| Nishamalaki | Phyllanthus emblica | Fruit | CID_1057 | 1813 | DRD2 | ChEMBL |
| Nishamalaki | Phyllanthus emblica | Fruit | CID_1057 | 1814 | DRD3 | ChEMBL |
| Nishamalaki | Phyllanthus emblica | Fruit | CID_1057 | 1815 | DRD4 | ChEMBL |
| Nishamalaki | Phyllanthus emblica | Fruit | CID_1057 | 186 | AGTR2 | ChEMBL |
| Nishamalaki | Phyllanthus emblica | Fruit | CID_1057 | 1909 | EDNRA | ChEMBL |
| Nishamalaki | Phyllanthus emblica | Fruit | CID_1057 | 1956 | EGFR | ChEMBL \| NPASS |

| Table S3: Associations between formulations, constituent plants, phytochemicals, and human molecular targets. The table lists the formulations, plant name and plant part, phytochemical identifiers, associated human target genes (Entrez Gene ID and gene symbol), and the database sources supporting each phytochemical-target interaction. | | | | | | |
|---|---|---|---|---|---|---|
| Formulation | Plant name | Part | PubChem CID | Entrez Gene ID | Gene Symbol | Source database(s) |
| Nishamalaki | Phyllanthus emblica | Fruit | CID_1057 | 1991 | ELANE | ChEMBL |
| Nishamalaki | Phyllanthus emblica | Fruit | CID_1057 | 2064 | ERBB2 | ChEMBL |
| Nishamalaki | Phyllanthus emblica | Fruit | CID_1057 | 2099 | ESR1 | ChEMBL \| NPASS |
| Nishamalaki | Phyllanthus emblica | Fruit | CID_1057 | 2100 | ESR2 | ChEMBL |
| Nishamalaki | Phyllanthus emblica | Fruit | CID_1057 | 213 | ALB | ChEMBL |
| Nishamalaki | Phyllanthus emblica | Fruit | CID_1057 | 216 | ALDH1A1 | ChEMBL \| NPASS |
| Nishamalaki | Phyllanthus emblica | Fruit | CID_1057 | 2321 | FLT1 | ChEMBL |
| Nishamalaki | Phyllanthus emblica | Fruit | CID_1057 | 246 | ALOX15 | ChEMBL \| NPASS |
| Nishamalaki | Phyllanthus emblica | Fruit | CID_1057 | 2534 | FYN | ChEMBL \| NPASS |
| Nishamalaki | Phyllanthus emblica | Fruit | CID_1057 | 2744 | GLS | NPASS |
| Nishamalaki | Phyllanthus emblica | Fruit | CID_1057 | 2908 | NR3C1 | ChEMBL |
| Nishamalaki | Phyllanthus emblica | Fruit | CID_1057 | 3028 | HSD17B10 | ChEMBL \| NPASS |
| Nishamalaki | Phyllanthus emblica | Fruit | CID_1057 | 3156 | HMGCR | ChEMBL |
| Nishamalaki | Phyllanthus emblica | Fruit | CID_1057 | 3248 | HPGD | ChEMBL \| NPASS |
| Nishamalaki | Phyllanthus emblica | Fruit | CID_1057 | 3269 | HRH1 | ChEMBL |
| Nishamalaki | Phyllanthus emblica | Fruit | CID_1057 | 3274 | HRH2 | ChEMBL |
| Nishamalaki | Phyllanthus emblica | Fruit | CID_1057 | 328 | APEX1 | ChEMBL \| NPASS |
| Nishamalaki | Phyllanthus emblica | Fruit | CID_1057 | 3356 | HTR2A | ChEMBL |
| Nishamalaki | Phyllanthus emblica | Fruit | CID_1057 | 3357 | HTR2B | ChEMBL |
| Nishamalaki | Phyllanthus emblica | Fruit | CID_1057 | 3358 | HTR2C | ChEMBL |
| Nishamalaki | Phyllanthus emblica | Fruit | CID_1057 | 3362 | HTR6 | ChEMBL |
| Nishamalaki | Phyllanthus emblica | Fruit | CID_1057 | 3576 | CXCL8 | ChEMBL \| NPASS |
| Nishamalaki | Phyllanthus emblica | Fruit | CID_1057 | 3577 | CXCR1 | ChEMBL |
| Nishamalaki | Phyllanthus emblica | Fruit | CID_1057 | 3579 | CXCR2 | ChEMBL |
| Nishamalaki | Phyllanthus emblica | Fruit | CID_1057 | 367 | AR | NPASS |
| Nishamalaki | Phyllanthus emblica | Fruit | CID_1057 | 3757 | KCNH2 | ChEMBL \| NPASS |
| Nishamalaki | Phyllanthus emblica | Fruit | CID_1057 | 383 | ARG1 | ChEMBL \| NPASS |
| Nishamalaki | Phyllanthus emblica | Fruit | CID_1057 | 3932 | LCK | ChEMBL |
| Nishamalaki | Phyllanthus emblica | Fruit | CID_1057 | 4000 | LMNA | NPASS |
| Nishamalaki | Phyllanthus emblica | Fruit | CID_1057 | 4128 | MAOA | ChEMBL |
| Nishamalaki | Phyllanthus emblica | Fruit | CID_1057 | 4159 | MC3R | ChEMBL |
| Nishamalaki | Phyllanthus emblica | Fruit | CID_1057 | 4160 | MC4R | ChEMBL |
| Nishamalaki | Phyllanthus emblica | Fruit | CID_1057 | 4161 | MC5R | ChEMBL |
| Nishamalaki | Phyllanthus emblica | Fruit | CID_1057 | 43 | ACHE | ChEMBL |
| Nishamalaki | Phyllanthus emblica | Fruit | CID_1057 | 4312 | MMP1 | ChEMBL |
| Nishamalaki | Phyllanthus emblica | Fruit | CID_1057 | 4313 | MMP2 | ChEMBL |
| Nishamalaki | Phyllanthus emblica | Fruit | CID_1057 | 4317 | MMP8 | BindingDB \| ChEMBL |
| Nishamalaki | Phyllanthus emblica | Fruit | CID_1057 | 4318 | MMP9 | BindingDB \| ChEMBL |
| Nishamalaki | Phyllanthus emblica | Fruit | CID_1057 | 4321 | MMP12 | BindingDB \| ChEMBL |
| Nishamalaki | Phyllanthus emblica | Fruit | CID_1057 | 4780 | NFE2L2 | ChEMBL \| NPASS |
| Nishamalaki | Phyllanthus emblica | Fruit | CID_1057 | 4886 | NPY1R | ChEMBL |
| Nishamalaki | Phyllanthus emblica | Fruit | CID_1057 | 4887 | NPY2R | ChEMBL |
| Nishamalaki | Phyllanthus emblica | Fruit | CID_1057 | 4985 | OPRD1 | ChEMBL |
| Nishamalaki | Phyllanthus emblica | Fruit | CID_1057 | 4986 | OPRK1 | ChEMBL |
| Nishamalaki | Phyllanthus emblica | Fruit | CID_1057 | 4988 | OPRM1 | ChEMBL |
| Nishamalaki | Phyllanthus emblica | Fruit | CID_1057 | 51426 | POLK | ChEMBL \| NPASS |
| Nishamalaki | Phyllanthus emblica | Fruit | CID_1057 | 54658 | UGT1A1 | ChEMBL \| NPASS |
| Nishamalaki | Phyllanthus emblica | Fruit | CID_1057 | 552 | AVPR1A | ChEMBL |
| Nishamalaki | Phyllanthus emblica | Fruit | CID_1057 | 5530 | PPP3CA | ChEMBL |
| Nishamalaki | Phyllanthus emblica | Fruit | CID_1057 | 55775 | TDP1 | ChEMBL \| NPASS |
| Nishamalaki | Phyllanthus emblica | Fruit | CID_1057 | 5578 | PRKCA | ChEMBL |
| Nishamalaki | Phyllanthus emblica | Fruit | CID_1057 | 5594 | MAPK1 | ChEMBL |
| Nishamalaki | Phyllanthus emblica | Fruit | CID_1057 | 5595 | MAPK3 | ChEMBL |
| Nishamalaki | Phyllanthus emblica | Fruit | CID_1057 | 5724 | PTAFR | ChEMBL |
| Nishamalaki | Phyllanthus emblica | Fruit | CID_1057 | 5742 | PTGS1 | ChEMBL |
| Nishamalaki | Phyllanthus emblica | Fruit | CID_1057 | 5743 | PTGS2 | BindingDB \| ChEMBL \| NPASS |
| Nishamalaki | Phyllanthus emblica | Fruit | CID_1057 | 5788 | PTPRC | ChEMBL |
| Nishamalaki | Phyllanthus emblica | Fruit | CID_1057 | 5965 | RECQL | ChEMBL \| NPASS |
| Nishamalaki | Phyllanthus emblica | Fruit | CID_1057 | 624 | BDKRB2 | ChEMBL |
| Nishamalaki | Phyllanthus emblica | Fruit | CID_1057 | 6401 | SELE | ChEMBL \| NPASS |
| Nishamalaki | Phyllanthus emblica | Fruit | CID_1057 | 6402 | SELL | ChEMBL \| NPASS |
| Nishamalaki | Phyllanthus emblica | Fruit | CID_1057 | 6403 | SELP | ChEMBL \| NPASS |
| Nishamalaki | Phyllanthus emblica | Fruit | CID_1057 | 6530 | SLC6A2 | ChEMBL |
| Nishamalaki | Phyllanthus emblica | Fruit | CID_1057 | 6531 | SLC6A3 | ChEMBL |
| Nishamalaki | Phyllanthus emblica | Fruit | CID_1057 | 6532 | SLC6A4 | ChEMBL |
| Nishamalaki | Phyllanthus emblica | Fruit | CID_1057 | 6865 | TACR2 | ChEMBL |
| Nishamalaki | Phyllanthus emblica | Fruit | CID_1057 | 6869 | TACR1 | ChEMBL |
| Nishamalaki | Phyllanthus emblica | Fruit | CID_1057 | 6916 | TBXAS1 | ChEMBL |
| Nishamalaki | Phyllanthus emblica | Fruit | CID_1057 | 729230 | CCR2 | ChEMBL |
| Nishamalaki | Phyllanthus emblica | Fruit | CID_1057 | 7299 | TYR | ChEMBL \| NPASS |

**Table S3: Associations between formulations, constituent plants, phytochemicals, and human molecular targets. The table lists the formulations, plant name and plant part, phytochemical identifiers, associated human target genes (Entrez Gene ID and gene symbol), and the database sources supporting each phytochemical-target interaction.**

| Formulation | Plant name | Part | PubChem CID | Entrez Gene ID | Gene Symbol | Source database(s) |
|---|---|---|---|---|---|---|
| Nishamalaki | Phyllanthus emblica | Fruit | CID_1057 | 7366 | UGT2B15 | ChEMBL \| NPASS |
| Nishamalaki | Phyllanthus emblica | Fruit | CID_1057 | 7421 | VDR | ChEMBL \| NPASS |
| Nishamalaki | Phyllanthus emblica | Fruit | CID_1057 | 7433 | VIPR1 | ChEMBL |
| Nishamalaki | Phyllanthus emblica | Fruit | CID_1057 | 759 | CA1 | BindingDB \| ChEMBL \| NPASS |
| Nishamalaki | Phyllanthus emblica | Fruit | CID_1057 | 760 | CA2 | BindingDB \| ChEMBL \| NPASS |
| Nishamalaki | Phyllanthus emblica | Fruit | CID_1057 | 762 | CA4 | BindingDB |
| Nishamalaki | Phyllanthus emblica | Fruit | CID_1057 | 765 | CA6 | ChEMBL \| NPASS |
| Nishamalaki | Phyllanthus emblica | Fruit | CID_1057 | 799 | CALCR | ChEMBL |
| Nishamalaki | Phyllanthus emblica | Fruit | CID_1057 | 834 | CASP1 | ChEMBL |
| Nishamalaki | Phyllanthus emblica | Fruit | CID_1057 | 861 | RUNX1 | ChEMBL |
| Nishamalaki | Phyllanthus emblica | Fruit | CID_1057 | 865 | CBFB | ChEMBL \| NPASS |
| Nishamalaki | Phyllanthus emblica | Fruit | CID_1057 | 8654 | PDE5A | ChEMBL |
| Nishamalaki | Phyllanthus emblica | Fruit | CID_1057 | 886 | CCKAR | ChEMBL |
| Nishamalaki | Phyllanthus emblica | Fruit | CID_1057 | 9507 | ADAMTS4 | BindingDB \| ChEMBL |
| Nishamalaki | Phyllanthus emblica | Fruit | CID_124021 | 231 | AKR1B1 | ChEMBL \| NPASS |
| Nishamalaki | Phyllanthus emblica | Fruit | CID_124021 | 23683 | PRKD3 | ChEMBL |
| Nishamalaki | Phyllanthus emblica | Fruit | CID_124021 | 5578 | PRKCA | BindingDB \| ChEMBL |
| Nishamalaki | Phyllanthus emblica | Fruit | CID_124021 | 5579 | PRKCB | ChEMBL |
| Nishamalaki | Phyllanthus emblica | Fruit | CID_124021 | 5580 | PRKCD | ChEMBL |
| Nishamalaki | Phyllanthus emblica | Fruit | CID_124021 | 5581 | PRKCE | ChEMBL |
| Nishamalaki | Phyllanthus emblica | Fruit | CID_124021 | 5582 | PRKCG | ChEMBL |
| Nishamalaki | Phyllanthus emblica | Fruit | CID_124021 | 5583 | PRKCH | ChEMBL |
| Nishamalaki | Phyllanthus emblica | Fruit | CID_124021 | 5584 | PRKCI | ChEMBL |
| Nishamalaki | Phyllanthus emblica | Fruit | CID_124021 | 5587 | PRKD1 | ChEMBL |
| Nishamalaki | Phyllanthus emblica | Fruit | CID_124021 | 5588 | PRKCQ | ChEMBL \| NPASS |
| Nishamalaki | Phyllanthus emblica | Fruit | CID_124021 | 5590 | PRKCZ | ChEMBL |
| Nishamalaki | Phyllanthus emblica | Fruit | CID_13250 | 2099 | ESR1 | ChEMBL |
| Nishamalaki | Phyllanthus emblica | Fruit | CID_13250 | 22943 | DKK1 | ChEMBL |
| Nishamalaki | Phyllanthus emblica | Fruit | CID_13250 | 4040 | LRP6 | ChEMBL |
| Nishamalaki | Phyllanthus emblica | Fruit | CID_289 | 10280 | SIGMAR1 | ChEMBL |
| Nishamalaki | Phyllanthus emblica | Fruit | CID_289 | 10800 | CYSLTR1 | ChEMBL |
| Nishamalaki | Phyllanthus emblica | Fruit | CID_289 | 10919 | EHMT2 | ChEMBL \| NPASS |
| Nishamalaki | Phyllanthus emblica | Fruit | CID_289 | 10951 | CBX1 | NPASS |
| Nishamalaki | Phyllanthus emblica | Fruit | CID_289 | 11201 | POLI | ChEMBL \| NPASS |
| Nishamalaki | Phyllanthus emblica | Fruit | CID_289 | 11238 | CA5B | BindingDB \| ChEMBL \| NPASS |
| Nishamalaki | Phyllanthus emblica | Fruit | CID_289 | 1128 | CHRM1 | ChEMBL |
| Nishamalaki | Phyllanthus emblica | Fruit | CID_289 | 1129 | CHRM2 | ChEMBL |
| Nishamalaki | Phyllanthus emblica | Fruit | CID_289 | 1131 | CHRM3 | ChEMBL |
| Nishamalaki | Phyllanthus emblica | Fruit | CID_289 | 1132 | CHRM4 | ChEMBL |
| Nishamalaki | Phyllanthus emblica | Fruit | CID_289 | 1133 | CHRM5 | ChEMBL |
| Nishamalaki | Phyllanthus emblica | Fruit | CID_289 | 1233 | CCR4 | ChEMBL |
| Nishamalaki | Phyllanthus emblica | Fruit | CID_289 | 1234 | CCR5 | ChEMBL |
| Nishamalaki | Phyllanthus emblica | Fruit | CID_289 | 1268 | CNR1 | ChEMBL |
| Nishamalaki | Phyllanthus emblica | Fruit | CID_289 | 134 | ADORA1 | ChEMBL |
| Nishamalaki | Phyllanthus emblica | Fruit | CID_289 | 135 | ADORA2A | ChEMBL |
| Nishamalaki | Phyllanthus emblica | Fruit | CID_289 | 140 | ADORA3 | ChEMBL |
| Nishamalaki | Phyllanthus emblica | Fruit | CID_289 | 1432 | MAPK14 | ChEMBL |
| Nishamalaki | Phyllanthus emblica | Fruit | CID_289 | 146 | ADRA1D | ChEMBL |
| Nishamalaki | Phyllanthus emblica | Fruit | CID_289 | 150 | ADRA2A | ChEMBL |
| Nishamalaki | Phyllanthus emblica | Fruit | CID_289 | 151 | ADRA2B | ChEMBL |
| Nishamalaki | Phyllanthus emblica | Fruit | CID_289 | 1511 | CTSG | ChEMBL |
| Nishamalaki | Phyllanthus emblica | Fruit | CID_289 | 152 | ADRA2C | ChEMBL |
| Nishamalaki | Phyllanthus emblica | Fruit | CID_289 | 153 | ADRB1 | ChEMBL |
| Nishamalaki | Phyllanthus emblica | Fruit | CID_289 | 154 | ADRB2 | ChEMBL |
| Nishamalaki | Phyllanthus emblica | Fruit | CID_289 | 1544 | CYP1A2 | ChEMBL |
| Nishamalaki | Phyllanthus emblica | Fruit | CID_289 | 1548 | CYP2A6 | ChEMBL |
| Nishamalaki | Phyllanthus emblica | Fruit | CID_289 | 155 | ADRB3 | ChEMBL |
| Nishamalaki | Phyllanthus emblica | Fruit | CID_289 | 1557 | CYP2C19 | ChEMBL |
| Nishamalaki | Phyllanthus emblica | Fruit | CID_289 | 1559 | CYP2C9 | ChEMBL |
| Nishamalaki | Phyllanthus emblica | Fruit | CID_289 | 1565 | CYP2D6 | ChEMBL |
| Nishamalaki | Phyllanthus emblica | Fruit | CID_289 | 1571 | CYP2E1 | ChEMBL |
| Nishamalaki | Phyllanthus emblica | Fruit | CID_289 | 1576 | CYP3A4 | ChEMBL \| NPASS |
| Nishamalaki | Phyllanthus emblica | Fruit | CID_289 | 1812 | DRD1 | ChEMBL |
| Nishamalaki | Phyllanthus emblica | Fruit | CID_289 | 1813 | DRD2 | ChEMBL |
| Nishamalaki | Phyllanthus emblica | Fruit | CID_289 | 1814 | DRD3 | ChEMBL |
| Nishamalaki | Phyllanthus emblica | Fruit | CID_289 | 1815 | DRD4 | ChEMBL |
| Nishamalaki | Phyllanthus emblica | Fruit | CID_289 | 186 | AGTR2 | ChEMBL |
| Nishamalaki | Phyllanthus emblica | Fruit | CID_289 | 1909 | EDNRA | ChEMBL |
| Nishamalaki | Phyllanthus emblica | Fruit | CID_289 | 1956 | EGFR | ChEMBL \| NPASS |
| Nishamalaki | Phyllanthus emblica | Fruit | CID_289 | 1991 | ELANE | ChEMBL |

**Table S3: Associations between formulations, constituent plants, phytochemicals, and human molecular targets. The table lists the formulations, plant name and plant part, phytochemical identifiers, associated human target genes (Entrez Gene ID and gene symbol), and the database sources supporting each phytochemical-target interaction.**

| Formulation | Plant name | Part | PubChem CID | Entrez Gene ID | Gene Symbol | Source database(s) |
|---|---|---|---|---|---|---|
| Nishamalaki | Phyllanthus emblica | Fruit | CID_289 | 2064 | ERBB2 | ChEMBL |
| Nishamalaki | Phyllanthus emblica | Fruit | CID_289 | 2099 | ESR1 | ChEMBL \| NPASS |
| Nishamalaki | Phyllanthus emblica | Fruit | CID_289 | 2100 | ESR2 | ChEMBL |
| Nishamalaki | Phyllanthus emblica | Fruit | CID_289 | 216 | ALDH1A1 | ChEMBL \| NPASS |
| Nishamalaki | Phyllanthus emblica | Fruit | CID_289 | 2237 | FEN1 | ChEMBL \| NPASS |
| Nishamalaki | Phyllanthus emblica | Fruit | CID_289 | 2321 | FLT1 | ChEMBL |
| Nishamalaki | Phyllanthus emblica | Fruit | CID_289 | 23632 | CA14 | BindingDB \| ChEMBL \| NPASS |
| Nishamalaki | Phyllanthus emblica | Fruit | CID_289 | 240 | ALOX5 | ChEMBL \| NPASS |
| Nishamalaki | Phyllanthus emblica | Fruit | CID_289 | 246 | ALOX15 | ChEMBL \| NPASS |
| Nishamalaki | Phyllanthus emblica | Fruit | CID_289 | 247 | ALOX15B | NPASS |
| Nishamalaki | Phyllanthus emblica | Fruit | CID_289 | 2475 | MTOR | ChEMBL \| NPASS |
| Nishamalaki | Phyllanthus emblica | Fruit | CID_289 | 2534 | FYN | ChEMBL \| NPASS |
| Nishamalaki | Phyllanthus emblica | Fruit | CID_289 | 25797 | QPCT | BindingDB \| ChEMBL |
| Nishamalaki | Phyllanthus emblica | Fruit | CID_289 | 26013 | L3MBTL1 | ChEMBL \| NPASS |
| Nishamalaki | Phyllanthus emblica | Fruit | CID_289 | 2744 | GLS | NPASS |
| Nishamalaki | Phyllanthus emblica | Fruit | CID_289 | 2908 | NR3C1 | ChEMBL \| NPASS |
| Nishamalaki | Phyllanthus emblica | Fruit | CID_289 | 3028 | HSD17B10 | ChEMBL \| NPASS |
| Nishamalaki | Phyllanthus emblica | Fruit | CID_289 | 3091 | HIF1A | ChEMBL \| NPASS |
| Nishamalaki | Phyllanthus emblica | Fruit | CID_289 | 3156 | HMGCR | ChEMBL |
| Nishamalaki | Phyllanthus emblica | Fruit | CID_289 | 3248 | HPGD | ChEMBL \| NPASS |
| Nishamalaki | Phyllanthus emblica | Fruit | CID_289 | 3269 | HRH1 | ChEMBL |
| Nishamalaki | Phyllanthus emblica | Fruit | CID_289 | 3274 | HRH2 | ChEMBL |
| Nishamalaki | Phyllanthus emblica | Fruit | CID_289 | 328 | APEX1 | ChEMBL \| NPASS |
| Nishamalaki | Phyllanthus emblica | Fruit | CID_289 | 3309 | HSPA5 | NPASS |
| Nishamalaki | Phyllanthus emblica | Fruit | CID_289 | 3315 | HSPB1 | NPASS |
| Nishamalaki | Phyllanthus emblica | Fruit | CID_289 | 3356 | HTR2A | ChEMBL |
| Nishamalaki | Phyllanthus emblica | Fruit | CID_289 | 3357 | HTR2B | ChEMBL |
| Nishamalaki | Phyllanthus emblica | Fruit | CID_289 | 3358 | HTR2C | ChEMBL |
| Nishamalaki | Phyllanthus emblica | Fruit | CID_289 | 3362 | HTR6 | ChEMBL |
| Nishamalaki | Phyllanthus emblica | Fruit | CID_289 | 3417 | IDH1 | NPASS |
| Nishamalaki | Phyllanthus emblica | Fruit | CID_289 | 3576 | CXCL8 | NPASS |
| Nishamalaki | Phyllanthus emblica | Fruit | CID_289 | 3577 | CXCR1 | ChEMBL |
| Nishamalaki | Phyllanthus emblica | Fruit | CID_289 | 3579 | CXCR2 | ChEMBL |
| Nishamalaki | Phyllanthus emblica | Fruit | CID_289 | 3620 | IDO1 | ChEMBL \| NPASS |
| Nishamalaki | Phyllanthus emblica | Fruit | CID_289 | 367 | AR | NPASS |
| Nishamalaki | Phyllanthus emblica | Fruit | CID_289 | 3725 | JUN | NPASS |
| Nishamalaki | Phyllanthus emblica | Fruit | CID_289 | 3757 | KCNH2 | ChEMBL |
| Nishamalaki | Phyllanthus emblica | Fruit | CID_289 | 390245 | KDM4E | ChEMBL \| NPASS |
| Nishamalaki | Phyllanthus emblica | Fruit | CID_289 | 3932 | LCK | ChEMBL |
| Nishamalaki | Phyllanthus emblica | Fruit | CID_289 | 4000 | LMNA | NPASS |
| Nishamalaki | Phyllanthus emblica | Fruit | CID_289 | 4128 | MAOA | ChEMBL |
| Nishamalaki | Phyllanthus emblica | Fruit | CID_289 | 4137 | MAPT | NPASS |
| Nishamalaki | Phyllanthus emblica | Fruit | CID_289 | 4159 | MC3R | ChEMBL |
| Nishamalaki | Phyllanthus emblica | Fruit | CID_289 | 4160 | MC4R | ChEMBL |
| Nishamalaki | Phyllanthus emblica | Fruit | CID_289 | 4161 | MC5R | ChEMBL |
| Nishamalaki | Phyllanthus emblica | Fruit | CID_289 | 4193 | MDM2 | BindingDB \| ChEMBL \| NPASS |
| Nishamalaki | Phyllanthus emblica | Fruit | CID_289 | 43 | ACHE | ChEMBL |
| Nishamalaki | Phyllanthus emblica | Fruit | CID_289 | 4312 | MMP1 | ChEMBL \| NPASS |
| Nishamalaki | Phyllanthus emblica | Fruit | CID_289 | 4313 | MMP2 | ChEMBL \| NPASS |
| Nishamalaki | Phyllanthus emblica | Fruit | CID_289 | 4314 | MMP3 | ChEMBL \| NPASS |
| Nishamalaki | Phyllanthus emblica | Fruit | CID_289 | 4317 | MMP8 | ChEMBL \| NPASS |
| Nishamalaki | Phyllanthus emblica | Fruit | CID_289 | 4318 | MMP9 | ChEMBL \| NPASS |
| Nishamalaki | Phyllanthus emblica | Fruit | CID_289 | 4780 | NFE2L2 | ChEMBL \| NPASS |
| Nishamalaki | Phyllanthus emblica | Fruit | CID_289 | 4790 | NFKB1 | NPASS |
| Nishamalaki | Phyllanthus emblica | Fruit | CID_289 | 4864 | NPC1 | NPASS |
| Nishamalaki | Phyllanthus emblica | Fruit | CID_289 | 4886 | NPY1R | ChEMBL |
| Nishamalaki | Phyllanthus emblica | Fruit | CID_289 | 4887 | NPY2R | ChEMBL |
| Nishamalaki | Phyllanthus emblica | Fruit | CID_289 | 4985 | OPRD1 | ChEMBL |
| Nishamalaki | Phyllanthus emblica | Fruit | CID_289 | 4986 | OPRK1 | ChEMBL |
| Nishamalaki | Phyllanthus emblica | Fruit | CID_289 | 4988 | OPRM1 | ChEMBL |
| Nishamalaki | Phyllanthus emblica | Fruit | CID_289 | 51053 | GMNN | NPASS |
| Nishamalaki | Phyllanthus emblica | Fruit | CID_289 | 51426 | POLK | ChEMBL \| NPASS |
| Nishamalaki | Phyllanthus emblica | Fruit | CID_289 | 5290 | PIK3CA | BindingDB \| ChEMBL \| NPASS |
| Nishamalaki | Phyllanthus emblica | Fruit | CID_289 | 5300 | PIN1 | NPASS |
| Nishamalaki | Phyllanthus emblica | Fruit | CID_289 | 5376 | PMP22 | NPASS |
| Nishamalaki | Phyllanthus emblica | Fruit | CID_289 | 5423 | POLB | ChEMBL \| NPASS |
| Nishamalaki | Phyllanthus emblica | Fruit | CID_289 | 5429 | POLH | NPASS |
| Nishamalaki | Phyllanthus emblica | Fruit | CID_289 | 5467 | PPARD | NPASS |
| Nishamalaki | Phyllanthus emblica | Fruit | CID_289 | 5468 | PPARG | NPASS |
| Nishamalaki | Phyllanthus emblica | Fruit | CID_289 | 54737 | MPHOSPH8 | ChEMBL \| NPASS |

**Table S3: Associations between formulations, constituent plants, phytochemicals, and human molecular targets. The table lists the formulations, plant name and plant part, phytochemical identifiers, associated human target genes (Entrez Gene ID and gene symbol), and the database sources supporting each phytochemical-target interaction.**

| Formulation | Plant name | Part | PubChem CID | Entrez Gene ID | Gene Symbol | Source database(s) |
|---|---|---|---|---|---|---|
| Nishamalaki | Phyllanthus emblica | Fruit | CID_289 | 552 | AVPR1A | ChEMBL |
| Nishamalaki | Phyllanthus emblica | Fruit | CID_289 | 5530 | PPP3CA | ChEMBL |
| Nishamalaki | Phyllanthus emblica | Fruit | CID_289 | 55775 | TDP1 | NPASS |
| Nishamalaki | Phyllanthus emblica | Fruit | CID_289 | 5578 | PRKCA | ChEMBL |
| Nishamalaki | Phyllanthus emblica | Fruit | CID_289 | 5594 | MAPK1 | ChEMBL |
| Nishamalaki | Phyllanthus emblica | Fruit | CID_289 | 5595 | MAPK3 | ChEMBL |
| Nishamalaki | Phyllanthus emblica | Fruit | CID_289 | 5724 | PTAFR | ChEMBL |
| Nishamalaki | Phyllanthus emblica | Fruit | CID_289 | 5742 | PTGS1 | ChEMBL |
| Nishamalaki | Phyllanthus emblica | Fruit | CID_289 | 5743 | PTGS2 | ChEMBL |
| Nishamalaki | Phyllanthus emblica | Fruit | CID_289 | 5745 | PTH1R | NPASS |
| Nishamalaki | Phyllanthus emblica | Fruit | CID_289 | 5788 | PTPRC | ChEMBL |
| Nishamalaki | Phyllanthus emblica | Fruit | CID_289 | 5965 | RECQL | ChEMBL \| NPASS |
| Nishamalaki | Phyllanthus emblica | Fruit | CID_289 | 5999 | RGS4 | NPASS |
| Nishamalaki | Phyllanthus emblica | Fruit | CID_289 | 60489 | APOBEC3G | NPASS |
| Nishamalaki | Phyllanthus emblica | Fruit | CID_289 | 6097 | RORC | ChEMBL \| NPASS |
| Nishamalaki | Phyllanthus emblica | Fruit | CID_289 | 624 | BDKRB2 | ChEMBL |
| Nishamalaki | Phyllanthus emblica | Fruit | CID_289 | 6311 | ATXN2 | NPASS |
| Nishamalaki | Phyllanthus emblica | Fruit | CID_289 | 641 | BLM | NPASS |
| Nishamalaki | Phyllanthus emblica | Fruit | CID_289 | 6530 | SLC6A2 | ChEMBL |
| Nishamalaki | Phyllanthus emblica | Fruit | CID_289 | 6531 | SLC6A3 | ChEMBL |
| Nishamalaki | Phyllanthus emblica | Fruit | CID_289 | 6532 | SLC6A4 | ChEMBL |
| Nishamalaki | Phyllanthus emblica | Fruit | CID_289 | 6606 | SMN1 | NPASS |
| Nishamalaki | Phyllanthus emblica | Fruit | CID_289 | 6865 | TACR2 | ChEMBL |
| Nishamalaki | Phyllanthus emblica | Fruit | CID_289 | 6869 | TACR1 | ChEMBL |
| Nishamalaki | Phyllanthus emblica | Fruit | CID_289 | 6916 | TBXAS1 | ChEMBL |
| Nishamalaki | Phyllanthus emblica | Fruit | CID_289 | 7066 | THPO | NPASS |
| Nishamalaki | Phyllanthus emblica | Fruit | CID_289 | 729230 | CCR2 | ChEMBL |
| Nishamalaki | Phyllanthus emblica | Fruit | CID_289 | 7433 | VIPR1 | ChEMBL |
| Nishamalaki | Phyllanthus emblica | Fruit | CID_289 | 7498 | XDH | ChEMBL \| NPASS |
| Nishamalaki | Phyllanthus emblica | Fruit | CID_289 | 759 | CA1 | BindingDB \| ChEMBL \| NPASS |
| Nishamalaki | Phyllanthus emblica | Fruit | CID_289 | 760 | CA2 | BindingDB \| ChEMBL \| NPASS |
| Nishamalaki | Phyllanthus emblica | Fruit | CID_289 | 761 | CA3 | BindingDB \| ChEMBL \| NPASS |
| Nishamalaki | Phyllanthus emblica | Fruit | CID_289 | 762 | CA4 | BindingDB \| ChEMBL \| NPASS |
| Nishamalaki | Phyllanthus emblica | Fruit | CID_289 | 763 | CA5A | BindingDB \| ChEMBL \| NPASS |
| Nishamalaki | Phyllanthus emblica | Fruit | CID_289 | 765 | CA6 | NPASS |
| Nishamalaki | Phyllanthus emblica | Fruit | CID_289 | 766 | CA7 | BindingDB \| NPASS |
| Nishamalaki | Phyllanthus emblica | Fruit | CID_289 | 768 | CA9 | BindingDB \| NPASS |
| Nishamalaki | Phyllanthus emblica | Fruit | CID_289 | 771 | CA12 | BindingDB \| ChEMBL \| NPASS |
| Nishamalaki | Phyllanthus emblica | Fruit | CID_289 | 799 | CALCR | ChEMBL |
| Nishamalaki | Phyllanthus emblica | Fruit | CID_289 | 834 | CASP1 | ChEMBL |
| Nishamalaki | Phyllanthus emblica | Fruit | CID_289 | 8654 | PDE5A | ChEMBL |
| Nishamalaki | Phyllanthus emblica | Fruit | CID_289 | 886 | CCKAR | ChEMBL |
| Nishamalaki | Phyllanthus emblica | Fruit | CID_289 | 9367 | RAB9A | NPASS |
| Nishamalaki | Phyllanthus emblica | Fruit | CID_289 | 9682 | KDM4A | ChEMBL \| NPASS |
| Nishamalaki | Phyllanthus emblica | Fruit | CID_289 | 9971 | NR1H4 | NPASS |
| Nishamalaki | Phyllanthus emblica | Fruit | CID_359 | 10013 | HDAC6 | ChEMBL |
| Nishamalaki | Phyllanthus emblica | Fruit | CID_359 | 23621 | BACE1 | BindingDB \| ChEMBL \| NPASS |
| Nishamalaki | Phyllanthus emblica | Fruit | CID_359 | 51426 | POLK | NPASS |
| Nishamalaki | Phyllanthus emblica | Fruit | CID_359 | 7299 | TYR | ChEMBL \| NPASS |
| Nishamalaki | Phyllanthus emblica | Fruit | CID_370 | 10013 | HDAC6 | ChEMBL |
| Nishamalaki | Phyllanthus emblica | Fruit | CID_370 | 10599 | SLCO1B1 | ChEMBL \| NPASS |
| Nishamalaki | Phyllanthus emblica | Fruit | CID_370 | 10919 | EHMT2 | ChEMBL \| NPASS |
| Nishamalaki | Phyllanthus emblica | Fruit | CID_370 | 11238 | CA5B | BindingDB \| ChEMBL \| NPASS |
| Nishamalaki | Phyllanthus emblica | Fruit | CID_370 | 118429 | ANTXR2 | NPASS |
| Nishamalaki | Phyllanthus emblica | Fruit | CID_370 | 1588 | CYP19A1 | ChEMBL \| NPASS |
| Nishamalaki | Phyllanthus emblica | Fruit | CID_370 | 1636 | ACE | BindingDB \| NPASS |
| Nishamalaki | Phyllanthus emblica | Fruit | CID_370 | 1803 | DPP4 | BindingDB \| ChEMBL \| NPASS |
| Nishamalaki | Phyllanthus emblica | Fruit | CID_370 | 2099 | ESR1 | ChEMBL \| NPASS |
| Nishamalaki | Phyllanthus emblica | Fruit | CID_370 | 216 | ALDH1A1 | ChEMBL \| NPASS |
| Nishamalaki | Phyllanthus emblica | Fruit | CID_370 | 231 | AKR1B1 | NPASS |
| Nishamalaki | Phyllanthus emblica | Fruit | CID_370 | 23632 | CA14 | BindingDB \| ChEMBL \| NPASS |
| Nishamalaki | Phyllanthus emblica | Fruit | CID_370 | 2526 | FUT4 | ChEMBL \| NPASS |
| Nishamalaki | Phyllanthus emblica | Fruit | CID_370 | 2529 | FUT7 | ChEMBL \| NPASS |
| Nishamalaki | Phyllanthus emblica | Fruit | CID_370 | 2590 | GALNT2 | ChEMBL \| NPASS |
| Nishamalaki | Phyllanthus emblica | Fruit | CID_370 | 259285 | TAS2R39 | ChEMBL \| NPASS |
| Nishamalaki | Phyllanthus emblica | Fruit | CID_370 | 2597 | GAPDH | ChEMBL \| NPASS |
| Nishamalaki | Phyllanthus emblica | Fruit | CID_370 | 2648 | KAT2A | NPASS |
| Nishamalaki | Phyllanthus emblica | Fruit | CID_370 | 2744 | GLS | ChEMBL \| NPASS |
| Nishamalaki | Phyllanthus emblica | Fruit | CID_370 | 28234 | SLCO1B3 | ChEMBL \| NPASS |
| Nishamalaki | Phyllanthus emblica | Fruit | CID_370 | 2908 | NR3C1 | NPASS |

**Table S3: Associations between formulations, constituent plants, phytochemicals, and human molecular targets. The table lists the formulations, plant name and plant part, phytochemical identifiers, associated human target genes (Entrez Gene ID and gene symbol), and the database sources supporting each phytochemical-target interaction.**

| Formulation | Plant name | Part | PubChem CID | Entrez Gene ID | Gene Symbol | Source database(s) |
|---|---|---|---|---|---|---|
| Nishamalaki | Phyllanthus emblica | Fruit | CID_370 | 3028 | HSD17B10 | ChEMBL \| NPASS |
| Nishamalaki | Phyllanthus emblica | Fruit | CID_370 | 3248 | HPGD | ChEMBL \| NPASS |
| Nishamalaki | Phyllanthus emblica | Fruit | CID_370 | 328 | APEX1 | NPASS |
| Nishamalaki | Phyllanthus emblica | Fruit | CID_370 | 3315 | HSPB1 | NPASS |
| Nishamalaki | Phyllanthus emblica | Fruit | CID_370 | 3417 | IDH1 | ChEMBL \| NPASS |
| Nishamalaki | Phyllanthus emblica | Fruit | CID_370 | 351 | APP | BindingDB \| ChEMBL |
| Nishamalaki | Phyllanthus emblica | Fruit | CID_370 | 3576 | CXCL8 | NPASS |
| Nishamalaki | Phyllanthus emblica | Fruit | CID_370 | 367 | AR | NPASS |
| Nishamalaki | Phyllanthus emblica | Fruit | CID_370 | 3725 | JUN | NPASS |
| Nishamalaki | Phyllanthus emblica | Fruit | CID_370 | 390245 | KDM4E | ChEMBL \| NPASS |
| Nishamalaki | Phyllanthus emblica | Fruit | CID_370 | 4000 | LMNA | NPASS |
| Nishamalaki | Phyllanthus emblica | Fruit | CID_370 | 4137 | MAPT | ChEMBL \| NPASS |
| Nishamalaki | Phyllanthus emblica | Fruit | CID_370 | 4316 | MMP7 | BindingDB |
| Nishamalaki | Phyllanthus emblica | Fruit | CID_370 | 4780 | NFE2L2 | ChEMBL \| NPASS |
| Nishamalaki | Phyllanthus emblica | Fruit | CID_370 | 5054 | SERPINE1 | BindingDB |
| Nishamalaki | Phyllanthus emblica | Fruit | CID_370 | 51426 | POLK | ChEMBL \| NPASS |
| Nishamalaki | Phyllanthus emblica | Fruit | CID_370 | 54658 | UGT1A1 | ChEMBL \| NPASS |
| Nishamalaki | Phyllanthus emblica | Fruit | CID_370 | 55775 | TDP1 | ChEMBL \| NPASS |
| Nishamalaki | Phyllanthus emblica | Fruit | CID_370 | 5594 | MAPK1 | NPASS |
| Nishamalaki | Phyllanthus emblica | Fruit | CID_370 | 5693 | PSMB5 | BindingDB \| NPASS |
| Nishamalaki | Phyllanthus emblica | Fruit | CID_370 | 5743 | PTGS2 | ChEMBL \| NPASS |
| Nishamalaki | Phyllanthus emblica | Fruit | CID_370 | 6401 | SELE | BindingDB \| NPASS |
| Nishamalaki | Phyllanthus emblica | Fruit | CID_370 | 6402 | SELL | BindingDB \| ChEMBL \| NPASS |
| Nishamalaki | Phyllanthus emblica | Fruit | CID_370 | 6403 | SELP | BindingDB \| ChEMBL \| NPASS |
| Nishamalaki | Phyllanthus emblica | Fruit | CID_370 | 6606 | SMN1 | NPASS |
| Nishamalaki | Phyllanthus emblica | Fruit | CID_370 | 6622 | SNCA | ChEMBL \| NPASS |
| Nishamalaki | Phyllanthus emblica | Fruit | CID_370 | 6868 | ADAM17 | ChEMBL \| NPASS |
| Nishamalaki | Phyllanthus emblica | Fruit | CID_370 | 7157 | TP53 | ChEMBL \| NPASS |
| Nishamalaki | Phyllanthus emblica | Fruit | CID_370 | 7421 | VDR | ChEMBL \| NPASS |
| Nishamalaki | Phyllanthus emblica | Fruit | CID_370 | 759 | CA1 | BindingDB \| ChEMBL \| NPASS |
| Nishamalaki | Phyllanthus emblica | Fruit | CID_370 | 760 | CA2 | BindingDB \| ChEMBL \| NPASS |
| Nishamalaki | Phyllanthus emblica | Fruit | CID_370 | 761 | CA3 | BindingDB \| ChEMBL \| NPASS |
| Nishamalaki | Phyllanthus emblica | Fruit | CID_370 | 762 | CA4 | BindingDB \| ChEMBL \| NPASS |
| Nishamalaki | Phyllanthus emblica | Fruit | CID_370 | 763 | CA5A | BindingDB \| ChEMBL \| NPASS |
| Nishamalaki | Phyllanthus emblica | Fruit | CID_370 | 765 | CA6 | ChEMBL \| NPASS |
| Nishamalaki | Phyllanthus emblica | Fruit | CID_370 | 766 | CA7 | BindingDB \| ChEMBL \| NPASS |
| Nishamalaki | Phyllanthus emblica | Fruit | CID_370 | 768 | CA9 | BindingDB \| ChEMBL \| NPASS |
| Nishamalaki | Phyllanthus emblica | Fruit | CID_370 | 771 | CA12 | BindingDB \| ChEMBL \| NPASS |
| Nishamalaki | Phyllanthus emblica | Fruit | CID_449093 | 10013 | HDAC6 | ChEMBL |
| Nishamalaki | Phyllanthus emblica | Fruit | CID_449093 | 328 | APEX1 | NPASS |
| Nishamalaki | Phyllanthus emblica | Fruit | CID_449093 | 5745 | PTH1R | NPASS |
| Nishamalaki | Phyllanthus emblica | Fruit | CID_449093 | 5965 | RECQL | NPASS |
| Nishamalaki | Phyllanthus emblica | Fruit | CID_449093 | 7253 | TSHR | NPASS |
| Nishamalaki | Phyllanthus emblica | Fruit | CID_493570 | 10013 | HDAC6 | ChEMBL |
| Nishamalaki | Phyllanthus emblica | Fruit | CID_493570 | 10599 | SLCO1B1 | ChEMBL \| NPASS |
| Nishamalaki | Phyllanthus emblica | Fruit | CID_493570 | 10919 | EHMT2 | ChEMBL |
| Nishamalaki | Phyllanthus emblica | Fruit | CID_493570 | 11330 | CTRC | ChEMBL \| NPASS |
| Nishamalaki | Phyllanthus emblica | Fruit | CID_493570 | 122706 | PSMB11 | ChEMBL |
| Nishamalaki | Phyllanthus emblica | Fruit | CID_493570 | 143471 | PSMA8 | ChEMBL |
| Nishamalaki | Phyllanthus emblica | Fruit | CID_493570 | 1544 | CYP1A2 | ChEMBL |
| Nishamalaki | Phyllanthus emblica | Fruit | CID_493570 | 1576 | CYP3A4 | NPASS |
| Nishamalaki | Phyllanthus emblica | Fruit | CID_493570 | 216 | ALDH1A1 | NPASS |
| Nishamalaki | Phyllanthus emblica | Fruit | CID_493570 | 246 | ALOX15 | NPASS |
| Nishamalaki | Phyllanthus emblica | Fruit | CID_493570 | 28234 | SLCO1B3 | ChEMBL \| NPASS |
| Nishamalaki | Phyllanthus emblica | Fruit | CID_493570 | 3028 | HSD17B10 | ChEMBL \| NPASS |
| Nishamalaki | Phyllanthus emblica | Fruit | CID_493570 | 3248 | HPGD | NPASS |
| Nishamalaki | Phyllanthus emblica | Fruit | CID_493570 | 3315 | HSPB1 | NPASS |
| Nishamalaki | Phyllanthus emblica | Fruit | CID_493570 | 3329 | HSPD1 | ChEMBL \| NPASS |
| Nishamalaki | Phyllanthus emblica | Fruit | CID_493570 | 3336 | HSPE1 | ChEMBL \| NPASS |
| Nishamalaki | Phyllanthus emblica | Fruit | CID_493570 | 367 | AR | NPASS |
| Nishamalaki | Phyllanthus emblica | Fruit | CID_493570 | 3725 | JUN | NPASS |
| Nishamalaki | Phyllanthus emblica | Fruit | CID_493570 | 4780 | NFE2L2 | NPASS |
| Nishamalaki | Phyllanthus emblica | Fruit | CID_493570 | 4790 | NFKB1 | NPASS |
| Nishamalaki | Phyllanthus emblica | Fruit | CID_493570 | 5467 | PPARD | NPASS |
| Nishamalaki | Phyllanthus emblica | Fruit | CID_493570 | 5468 | PPARG | NPASS |
| Nishamalaki | Phyllanthus emblica | Fruit | CID_493570 | 55775 | TDP1 | NPASS |
| Nishamalaki | Phyllanthus emblica | Fruit | CID_493570 | 5594 | MAPK1 | NPASS |
| Nishamalaki | Phyllanthus emblica | Fruit | CID_493570 | 5682 | PSMA1 | ChEMBL |
| Nishamalaki | Phyllanthus emblica | Fruit | CID_493570 | 5683 | PSMA2 | ChEMBL |
| Nishamalaki | Phyllanthus emblica | Fruit | CID_493570 | 5684 | PSMA3 | ChEMBL |

**Table S3: Associations between formulations, constituent plants, phytochemicals, and human molecular targets. The table lists the formulations, plant name and plant part, phytochemical identifiers, associated human target genes (Entrez Gene ID and gene symbol), and the database sources supporting each phytochemical-target interaction.**

| Formulation | Plant name | Part | PubChem CID | Entrez Gene ID | Gene Symbol | Source database(s) |
|---|---|---|---|---|---|---|
| Nishamalaki | Phyllanthus emblica | Fruit | CID_493570 | 5685 | PSMA4 | ChEMBL |
| Nishamalaki | Phyllanthus emblica | Fruit | CID_493570 | 5686 | PSMA5 | ChEMBL |
| Nishamalaki | Phyllanthus emblica | Fruit | CID_493570 | 5687 | PSMA6 | ChEMBL |
| Nishamalaki | Phyllanthus emblica | Fruit | CID_493570 | 5688 | PSMA7 | ChEMBL |
| Nishamalaki | Phyllanthus emblica | Fruit | CID_493570 | 5689 | PSMB1 | ChEMBL |
| Nishamalaki | Phyllanthus emblica | Fruit | CID_493570 | 5690 | PSMB2 | ChEMBL |
| Nishamalaki | Phyllanthus emblica | Fruit | CID_493570 | 5691 | PSMB3 | ChEMBL |
| Nishamalaki | Phyllanthus emblica | Fruit | CID_493570 | 5692 | PSMB4 | ChEMBL |
| Nishamalaki | Phyllanthus emblica | Fruit | CID_493570 | 5693 | PSMB5 | ChEMBL |
| Nishamalaki | Phyllanthus emblica | Fruit | CID_493570 | 5694 | PSMB6 | ChEMBL |
| Nishamalaki | Phyllanthus emblica | Fruit | CID_493570 | 5695 | PSMB7 | ChEMBL |
| Nishamalaki | Phyllanthus emblica | Fruit | CID_493570 | 5696 | PSMB8 | ChEMBL |
| Nishamalaki | Phyllanthus emblica | Fruit | CID_493570 | 5698 | PSMB9 | ChEMBL |
| Nishamalaki | Phyllanthus emblica | Fruit | CID_493570 | 5699 | PSMB10 | ChEMBL |
| Nishamalaki | Phyllanthus emblica | Fruit | CID_493570 | 5931 | RBBP7 | ChEMBL |
| Nishamalaki | Phyllanthus emblica | Fruit | CID_493570 | 5965 | RECQL | NPASS |
| Nishamalaki | Phyllanthus emblica | Fruit | CID_493570 | 6256 | RXRA | NPASS |
| Nishamalaki | Phyllanthus emblica | Fruit | CID_493570 | 672 | BRCA1 | ChEMBL \| NPASS |
| Nishamalaki | Phyllanthus emblica | Fruit | CID_493570 | 7068 | THRB | NPASS |
| Nishamalaki | Phyllanthus emblica | Fruit | CID_493570 | 7124 | TNF | NPASS |
| Nishamalaki | Phyllanthus emblica | Fruit | CID_493570 | 7263 | TST | ChEMBL \| NPASS |
| Nishamalaki | Phyllanthus emblica | Fruit | CID_493570 | 7398 | USP1 | ChEMBL |
| Nishamalaki | Phyllanthus emblica | Fruit | CID_493570 | 7421 | VDR | NPASS |
| Nishamalaki | Phyllanthus emblica | Fruit | CID_493570 | 834 | CASP1 | NPASS |
| Nishamalaki | Phyllanthus emblica | Fruit | CID_493570 | 840 | CASP7 | NPASS |
| Nishamalaki | Phyllanthus emblica | Fruit | CID_493570 | 8520 | HAT1 | ChEMBL |
| Nishamalaki | Phyllanthus emblica | Fruit | CID_493570 | 9971 | NR1H4 | NPASS |
| Nishamalaki | Phyllanthus emblica | Fruit | CID_5280343 | 10013 | HDAC6 | ChEMBL |
| Nishamalaki | Phyllanthus emblica | Fruit | CID_5280343 | 10072 | DPP3 | BindingDB |
| Nishamalaki | Phyllanthus emblica | Fruit | CID_5280343 | 1017 | CDK2 | BindingDB \| ChEMBL \| NPASS |
| Nishamalaki | Phyllanthus emblica | Fruit | CID_5280343 | 1019 | CDK4 | BindingDB \| ChEMBL |
| Nishamalaki | Phyllanthus emblica | Fruit | CID_5280343 | 1020 | CDK5 | ChEMBL |
| Nishamalaki | Phyllanthus emblica | Fruit | CID_5280343 | 1021 | CDK6 | BindingDB \| ChEMBL \| NPASS |
| Nishamalaki | Phyllanthus emblica | Fruit | CID_5280343 | 10257 | ABCC4 | BindingDB \| ChEMBL \| NPASS |
| Nishamalaki | Phyllanthus emblica | Fruit | CID_5280343 | 10298 | PAK4 | ChEMBL \| NPASS |
| Nishamalaki | Phyllanthus emblica | Fruit | CID_5280343 | 10494 | STK25 | ChEMBL \| NPASS |
| Nishamalaki | Phyllanthus emblica | Fruit | CID_5280343 | 10599 | SLCO1B1 | ChEMBL \| NPASS |
| Nishamalaki | Phyllanthus emblica | Fruit | CID_5280343 | 10645 | CAMKK2 | ChEMBL \| NPASS |
| Nishamalaki | Phyllanthus emblica | Fruit | CID_5280343 | 106821730 | BUB1B-PAK6 | ChEMBL |
| Nishamalaki | Phyllanthus emblica | Fruit | CID_5280343 | 10733 | PLK4 | ChEMBL \| NPASS |
| Nishamalaki | Phyllanthus emblica | Fruit | CID_5280343 | 10783 | NEK6 | BindingDB \| ChEMBL \| NPASS |
| Nishamalaki | Phyllanthus emblica | Fruit | CID_5280343 | 1080 | CFTR | ChEMBL |
| Nishamalaki | Phyllanthus emblica | Fruit | CID_5280343 | 10919 | EHMT2 | ChEMBL \| NPASS |
| Nishamalaki | Phyllanthus emblica | Fruit | CID_5280343 | 10951 | CBX1 | NPASS |
| Nishamalaki | Phyllanthus emblica | Fruit | CID_5280343 | 11040 | PIM2 | ChEMBL \| NPASS |
| Nishamalaki | Phyllanthus emblica | Fruit | CID_5280343 | 1111 | CHEK1 | ChEMBL \| NPASS |
| Nishamalaki | Phyllanthus emblica | Fruit | CID_5280343 | 11200 | CHEK2 | ChEMBL \| NPASS |
| Nishamalaki | Phyllanthus emblica | Fruit | CID_5280343 | 11201 | POLI | ChEMBL \| NPASS |
| Nishamalaki | Phyllanthus emblica | Fruit | CID_5280343 | 11238 | CA5B | BindingDB \| ChEMBL \| NPASS |
| Nishamalaki | Phyllanthus emblica | Fruit | CID_5280343 | 11255 | HRH3 | ChEMBL |
| Nishamalaki | Phyllanthus emblica | Fruit | CID_5280343 | 1128 | CHRM1 | ChEMBL |
| Nishamalaki | Phyllanthus emblica | Fruit | CID_5280343 | 1129 | CHRM2 | ChEMBL |
| Nishamalaki | Phyllanthus emblica | Fruit | CID_5280343 | 11309 | SLCO2B1 | ChEMBL \| NPASS |
| Nishamalaki | Phyllanthus emblica | Fruit | CID_5280343 | 1131 | CHRM3 | BindingDB \| ChEMBL |
| Nishamalaki | Phyllanthus emblica | Fruit | CID_5280343 | 11329 | STK38 | ChEMBL \| NPASS |
| Nishamalaki | Phyllanthus emblica | Fruit | CID_5280343 | 1133 | CHRM5 | BindingDB |
| Nishamalaki | Phyllanthus emblica | Fruit | CID_5280343 | 11330 | CTRC | BindingDB \| ChEMBL \| NPASS |
| Nishamalaki | Phyllanthus emblica | Fruit | CID_5280343 | 1137 | CHRNA4 | ChEMBL |
| Nishamalaki | Phyllanthus emblica | Fruit | CID_5280343 | 118429 | ANTXR2 | NPASS |
| Nishamalaki | Phyllanthus emblica | Fruit | CID_5280343 | 1195 | CLK1 | ChEMBL \| NPASS |
| Nishamalaki | Phyllanthus emblica | Fruit | CID_5280343 | 1196 | CLK2 | ChEMBL \| NPASS |
| Nishamalaki | Phyllanthus emblica | Fruit | CID_5280343 | 1198 | CLK3 | ChEMBL \| NPASS |
| Nishamalaki | Phyllanthus emblica | Fruit | CID_5280343 | 1233 | CCR4 | ChEMBL \| NPASS |
| Nishamalaki | Phyllanthus emblica | Fruit | CID_5280343 | 1244 | ABCC2 | ChEMBL \| NPASS |
| Nishamalaki | Phyllanthus emblica | Fruit | CID_5280343 | 1268 | CNR1 | ChEMBL |
| Nishamalaki | Phyllanthus emblica | Fruit | CID_5280343 | 134 | ADORA1 | ChEMBL |
| Nishamalaki | Phyllanthus emblica | Fruit | CID_5280343 | 135 | ADORA2A | ChEMBL |
| Nishamalaki | Phyllanthus emblica | Fruit | CID_5280343 | 1394 | CRHR1 | ChEMBL |
| Nishamalaki | Phyllanthus emblica | Fruit | CID_5280343 | 1395 | CRHR2 | ChEMBL |
| Nishamalaki | Phyllanthus emblica | Fruit | CID_5280343 | 140 | ADORA3 | BindingDB \| ChEMBL \| NPASS |

**Table S3: Associations between formulations, constituent plants, phytochemicals, and human molecular targets. The table lists the formulations, plant name and plant part, phytochemical identifiers, associated human target genes (Entrez Gene ID and gene symbol), and the database sources supporting each phytochemical-target interaction.**

| Formulation | Plant name | Part | PubChem CID | Entrez Gene ID | Gene Symbol | Source database(s) |
|---|---|---|---|---|---|---|
| Nishamalaki | Phyllanthus emblica | Fruit | CID_5280343 | 1432 | MAPK14 | BindingDB \| ChEMBL \| NPASS |
| Nishamalaki | Phyllanthus emblica | Fruit | CID_5280343 | 1455 | CSNK1G2 | ChEMBL \| NPASS |
| Nishamalaki | Phyllanthus emblica | Fruit | CID_5280343 | 1456 | CSNK1G3 | ChEMBL \| NPASS |
| Nishamalaki | Phyllanthus emblica | Fruit | CID_5280343 | 1457 | CSNK2A1 | BindingDB \| ChEMBL \| NPASS |
| Nishamalaki | Phyllanthus emblica | Fruit | CID_5280343 | 1459 | CSNK2A2 | ChEMBL \| NPASS |
| Nishamalaki | Phyllanthus emblica | Fruit | CID_5280343 | 1460 | CSNK2B | ChEMBL \| NPASS |
| Nishamalaki | Phyllanthus emblica | Fruit | CID_5280343 | 148 | ADRA1A | BindingDB \| ChEMBL |
| Nishamalaki | Phyllanthus emblica | Fruit | CID_5280343 | 1499 | CTNNB1 | BindingDB \| ChEMBL |
| Nishamalaki | Phyllanthus emblica | Fruit | CID_5280343 | 150 | ADRA2A | ChEMBL |
| Nishamalaki | Phyllanthus emblica | Fruit | CID_5280343 | 1504 | CTRB1 | BindingDB \| ChEMBL \| NPASS |
| Nishamalaki | Phyllanthus emblica | Fruit | CID_5280343 | 151 | ADRA2B | ChEMBL |
| Nishamalaki | Phyllanthus emblica | Fruit | CID_5280343 | 1514 | CTSL | BindingDB |
| Nishamalaki | Phyllanthus emblica | Fruit | CID_5280343 | 152 | ADRA2C | ChEMBL |
| Nishamalaki | Phyllanthus emblica | Fruit | CID_5280343 | 153 | ADRB1 | ChEMBL |
| Nishamalaki | Phyllanthus emblica | Fruit | CID_5280343 | 154 | ADRB2 | ChEMBL |
| Nishamalaki | Phyllanthus emblica | Fruit | CID_5280343 | 1543 | CYP1A1 | BindingDB \| ChEMBL \| NPASS |
| Nishamalaki | Phyllanthus emblica | Fruit | CID_5280343 | 1544 | CYP1A2 | BindingDB \| ChEMBL \| NPASS |
| Nishamalaki | Phyllanthus emblica | Fruit | CID_5280343 | 1545 | CYP1B1 | BindingDB \| ChEMBL \| NPASS |
| Nishamalaki | Phyllanthus emblica | Fruit | CID_5280343 | 155 | ADRB3 | ChEMBL |
| Nishamalaki | Phyllanthus emblica | Fruit | CID_5280343 | 1557 | CYP2C19 | BindingDB \| ChEMBL \| NPASS |
| Nishamalaki | Phyllanthus emblica | Fruit | CID_5280343 | 1558 | CYP2C8 | ChEMBL \| NPASS |
| Nishamalaki | Phyllanthus emblica | Fruit | CID_5280343 | 1559 | CYP2C9 | BindingDB \| ChEMBL \| NPASS |
| Nishamalaki | Phyllanthus emblica | Fruit | CID_5280343 | 1565 | CYP2D6 | ChEMBL \| NPASS |
| Nishamalaki | Phyllanthus emblica | Fruit | CID_5280343 | 1573 | CYP2J2 | ChEMBL \| NPASS |
| Nishamalaki | Phyllanthus emblica | Fruit | CID_5280343 | 1576 | CYP3A4 | BindingDB \| ChEMBL \| NPASS |
| Nishamalaki | Phyllanthus emblica | Fruit | CID_5280343 | 1577 | CYP3A5 | ChEMBL |
| Nishamalaki | Phyllanthus emblica | Fruit | CID_5280343 | 1588 | CYP19A1 | BindingDB \| ChEMBL \| NPASS |
| Nishamalaki | Phyllanthus emblica | Fruit | CID_5280343 | 1612 | DAPK1 | BindingDB \| ChEMBL \| NPASS |
| Nishamalaki | Phyllanthus emblica | Fruit | CID_5280343 | 1613 | DAPK3 | ChEMBL \| NPASS |
| Nishamalaki | Phyllanthus emblica | Fruit | CID_5280343 | 1760 | DMPK | ChEMBL \| NPASS |
| Nishamalaki | Phyllanthus emblica | Fruit | CID_5280343 | 1803 | DPP4 | BindingDB \| ChEMBL \| NPASS |
| Nishamalaki | Phyllanthus emblica | Fruit | CID_5280343 | 1812 | DRD1 | ChEMBL \| NPASS |
| Nishamalaki | Phyllanthus emblica | Fruit | CID_5280343 | 1813 | DRD2 | ChEMBL |
| Nishamalaki | Phyllanthus emblica | Fruit | CID_5280343 | 1814 | DRD3 | ChEMBL |
| Nishamalaki | Phyllanthus emblica | Fruit | CID_5280343 | 1815 | DRD4 | BindingDB \| ChEMBL \| NPASS |
| Nishamalaki | Phyllanthus emblica | Fruit | CID_5280343 | 185 | AGTR1 | ChEMBL |
| Nishamalaki | Phyllanthus emblica | Fruit | CID_5280343 | 1909 | EDNRA | ChEMBL |
| Nishamalaki | Phyllanthus emblica | Fruit | CID_5280343 | 1910 | EDNRB | ChEMBL |
| Nishamalaki | Phyllanthus emblica | Fruit | CID_5280343 | 1956 | EGFR | BindingDB \| ChEMBL \| NPASS |
| Nishamalaki | Phyllanthus emblica | Fruit | CID_5280343 | 1991 | ELANE | BindingDB \| ChEMBL \| NPASS |
| Nishamalaki | Phyllanthus emblica | Fruit | CID_5280343 | 1994 | ELAVL1 | BindingDB \| ChEMBL \| NPASS |
| Nishamalaki | Phyllanthus emblica | Fruit | CID_5280343 | 1995 | ELAVL3 | ChEMBL \| NPASS |
| Nishamalaki | Phyllanthus emblica | Fruit | CID_5280343 | 2030 | SLC29A1 | ChEMBL |
| Nishamalaki | Phyllanthus emblica | Fruit | CID_5280343 | 2053 | EPHX2 | ChEMBL \| NPASS |
| Nishamalaki | Phyllanthus emblica | Fruit | CID_5280343 | 207 | AKT1 | ChEMBL \| NPASS |
| Nishamalaki | Phyllanthus emblica | Fruit | CID_5280343 | 2099 | ESR1 | ChEMBL \| NPASS |
| Nishamalaki | Phyllanthus emblica | Fruit | CID_5280343 | 2100 | ESR2 | ChEMBL |
| Nishamalaki | Phyllanthus emblica | Fruit | CID_5280343 | 2147 | F2 | ChEMBL \| NPASS |
| Nishamalaki | Phyllanthus emblica | Fruit | CID_5280343 | 216 | ALDH1A1 | ChEMBL \| NPASS |
| Nishamalaki | Phyllanthus emblica | Fruit | CID_5280343 | 2167 | FABP4 | ChEMBL |
| Nishamalaki | Phyllanthus emblica | Fruit | CID_5280343 | 2203 | FBP1 | ChEMBL \| NPASS |
| Nishamalaki | Phyllanthus emblica | Fruit | CID_5280343 | 2237 | FEN1 | ChEMBL \| NPASS |
| Nishamalaki | Phyllanthus emblica | Fruit | CID_5280343 | 22933 | SIRT2 | ChEMBL |
| Nishamalaki | Phyllanthus emblica | Fruit | CID_5280343 | 23028 | KDM1A | BindingDB \| ChEMBL \| NPASS |
| Nishamalaki | Phyllanthus emblica | Fruit | CID_5280343 | 23043 | TNIK | ChEMBL \| NPASS |
| Nishamalaki | Phyllanthus emblica | Fruit | CID_5280343 | 231 | AKR1B1 | ChEMBL \| NPASS |
| Nishamalaki | Phyllanthus emblica | Fruit | CID_5280343 | 2322 | FLT3 | ChEMBL \| NPASS |
| Nishamalaki | Phyllanthus emblica | Fruit | CID_5280343 | 23408 | SIRT5 | ChEMBL |
| Nishamalaki | Phyllanthus emblica | Fruit | CID_5280343 | 23410 | SIRT3 | ChEMBL |
| Nishamalaki | Phyllanthus emblica | Fruit | CID_5280343 | 23411 | SIRT1 | ChEMBL |
| Nishamalaki | Phyllanthus emblica | Fruit | CID_5280343 | 23621 | BACE1 | BindingDB \| ChEMBL \| NPASS |
| Nishamalaki | Phyllanthus emblica | Fruit | CID_5280343 | 23632 | CA14 | BindingDB \| ChEMBL \| NPASS |
| Nishamalaki | Phyllanthus emblica | Fruit | CID_5280343 | 238 | ALK | BindingDB \| ChEMBL \| NPASS |
| Nishamalaki | Phyllanthus emblica | Fruit | CID_5280343 | 239 | ALOX12 | BindingDB \| ChEMBL \| NPASS |
| Nishamalaki | Phyllanthus emblica | Fruit | CID_5280343 | 240 | ALOX5 | BindingDB \| ChEMBL \| NPASS |
| Nishamalaki | Phyllanthus emblica | Fruit | CID_5280343 | 246 | ALOX15 | BindingDB \| ChEMBL \| NPASS |
| Nishamalaki | Phyllanthus emblica | Fruit | CID_5280343 | 247 | ALOX15B | BindingDB \| ChEMBL \| NPASS |
| Nishamalaki | Phyllanthus emblica | Fruit | CID_5280343 | 253430 | IPMK | ChEMBL \| NPASS |
| Nishamalaki | Phyllanthus emblica | Fruit | CID_5280343 | 2548 | GAA | ChEMBL \| NPASS |
| Nishamalaki | Phyllanthus emblica | Fruit | CID_5280343 | 2584 | GALK1 | ChEMBL \| NPASS |

**Table S3: Associations between formulations, constituent plants, phytochemicals, and human molecular targets. The table lists the formulations, plant name and plant part, phytochemical identifiers, associated human target genes (Entrez Gene ID and gene symbol), and the database sources supporting each phytochemical-target interaction.**

| Formulation | Plant name | Part | PubChem CID | Entrez Gene ID | Gene Symbol | Source database(s) |
|---|---|---|---|---|---|---|
| Nishamalaki | Phyllanthus emblica | Fruit | CID_5280343 | 2629 | GBA1 | NPASS |
| Nishamalaki | Phyllanthus emblica | Fruit | CID_5280343 | 2642 | GCGR | ChEMBL |
| Nishamalaki | Phyllanthus emblica | Fruit | CID_5280343 | 2693 | GHSR | ChEMBL |
| Nishamalaki | Phyllanthus emblica | Fruit | CID_5280343 | 2696 | GIPR | ChEMBL |
| Nishamalaki | Phyllanthus emblica | Fruit | CID_5280343 | 2739 | GLO1 | ChEMBL \| NPASS |
| Nishamalaki | Phyllanthus emblica | Fruit | CID_5280343 | 276 | AMY1A | ChEMBL |
| Nishamalaki | Phyllanthus emblica | Fruit | CID_5280343 | 277 | AMY1B | BindingDB |
| Nishamalaki | Phyllanthus emblica | Fruit | CID_5280343 | 28234 | SLCO1B3 | ChEMBL \| NPASS |
| Nishamalaki | Phyllanthus emblica | Fruit | CID_5280343 | 283106 | CSNK2A3 | ChEMBL \| NPASS |
| Nishamalaki | Phyllanthus emblica | Fruit | CID_5280343 | 2859 | GPR35 | BindingDB \| ChEMBL \| NPASS |
| Nishamalaki | Phyllanthus emblica | Fruit | CID_5280343 | 2862 | MLNR | ChEMBL |
| Nishamalaki | Phyllanthus emblica | Fruit | CID_5280343 | 2908 | NR3C1 | ChEMBL |
| Nishamalaki | Phyllanthus emblica | Fruit | CID_5280343 | 2931 | GSK3A | ChEMBL |
| Nishamalaki | Phyllanthus emblica | Fruit | CID_5280343 | 2932 | GSK3B | BindingDB \| ChEMBL \| NPASS |
| Nishamalaki | Phyllanthus emblica | Fruit | CID_5280343 | 2936 | GSR | NPASS |
| Nishamalaki | Phyllanthus emblica | Fruit | CID_5280343 | 3028 | HSD17B10 | ChEMBL \| NPASS |
| Nishamalaki | Phyllanthus emblica | Fruit | CID_5280343 | 3091 | HIF1A | ChEMBL \| NPASS |
| Nishamalaki | Phyllanthus emblica | Fruit | CID_5280343 | 3248 | HPGD | ChEMBL \| NPASS |
| Nishamalaki | Phyllanthus emblica | Fruit | CID_5280343 | 3269 | HRH1 | ChEMBL |
| Nishamalaki | Phyllanthus emblica | Fruit | CID_5280343 | 3274 | HRH2 | ChEMBL |
| Nishamalaki | Phyllanthus emblica | Fruit | CID_5280343 | 328 | APEX1 | ChEMBL \| NPASS |
| Nishamalaki | Phyllanthus emblica | Fruit | CID_5280343 | 3290 | HSD11B1 | ChEMBL \| NPASS |
| Nishamalaki | Phyllanthus emblica | Fruit | CID_5280343 | 3292 | HSD17B1 | ChEMBL \| NPASS |
| Nishamalaki | Phyllanthus emblica | Fruit | CID_5280343 | 3294 | HSD17B2 | BindingDB \| ChEMBL \| NPASS |
| Nishamalaki | Phyllanthus emblica | Fruit | CID_5280343 | 3297 | HSF1 | BindingDB \| ChEMBL \| NPASS |
| Nishamalaki | Phyllanthus emblica | Fruit | CID_5280343 | 3309 | HSPA5 | NPASS |
| Nishamalaki | Phyllanthus emblica | Fruit | CID_5280343 | 3315 | HSPB1 | ChEMBL |
| Nishamalaki | Phyllanthus emblica | Fruit | CID_5280343 | 3350 | HTR1A | ChEMBL \| NPASS |
| Nishamalaki | Phyllanthus emblica | Fruit | CID_5280343 | 3352 | HTR1D | BindingDB |
| Nishamalaki | Phyllanthus emblica | Fruit | CID_5280343 | 3356 | HTR2A | ChEMBL |
| Nishamalaki | Phyllanthus emblica | Fruit | CID_5280343 | 3357 | HTR2B | ChEMBL |
| Nishamalaki | Phyllanthus emblica | Fruit | CID_5280343 | 3358 | HTR2C | ChEMBL |
| Nishamalaki | Phyllanthus emblica | Fruit | CID_5280343 | 3359 | HTR3A | BindingDB \| ChEMBL |
| Nishamalaki | Phyllanthus emblica | Fruit | CID_5280343 | 3362 | HTR6 | BindingDB |
| Nishamalaki | Phyllanthus emblica | Fruit | CID_5280343 | 3363 | HTR7 | BindingDB |
| Nishamalaki | Phyllanthus emblica | Fruit | CID_5280343 | 3480 | IGF1R | BindingDB \| ChEMBL \| NPASS |
| Nishamalaki | Phyllanthus emblica | Fruit | CID_5280343 | 351 | APP | BindingDB \| ChEMBL \| NPASS |
| Nishamalaki | Phyllanthus emblica | Fruit | CID_5280343 | 3577 | CXCR1 | ChEMBL \| NPASS |
| Nishamalaki | Phyllanthus emblica | Fruit | CID_5280343 | 367 | AR | NPASS |
| Nishamalaki | Phyllanthus emblica | Fruit | CID_5280343 | 3716 | JAK1 | ChEMBL \| NPASS |
| Nishamalaki | Phyllanthus emblica | Fruit | CID_5280343 | 3725 | JUN | NPASS |
| Nishamalaki | Phyllanthus emblica | Fruit | CID_5280343 | 3757 | KCNH2 | ChEMBL |
| Nishamalaki | Phyllanthus emblica | Fruit | CID_5280343 | 3791 | KDR | BindingDB \| ChEMBL \| NPASS |
| Nishamalaki | Phyllanthus emblica | Fruit | CID_5280343 | 387129 | NPSR1 | NPASS |
| Nishamalaki | Phyllanthus emblica | Fruit | CID_5280343 | 390245 | KDM4E | ChEMBL \| NPASS |
| Nishamalaki | Phyllanthus emblica | Fruit | CID_5280343 | 3932 | LCK | ChEMBL \| NPASS |
| Nishamalaki | Phyllanthus emblica | Fruit | CID_5280343 | 4000 | LMNA | NPASS |
| Nishamalaki | Phyllanthus emblica | Fruit | CID_5280343 | 4088 | SMAD3 | NPASS |
| Nishamalaki | Phyllanthus emblica | Fruit | CID_5280343 | 410 | ARSA | NPASS |
| Nishamalaki | Phyllanthus emblica | Fruit | CID_5280343 | 4128 | MAOA | ChEMBL \| NPASS |
| Nishamalaki | Phyllanthus emblica | Fruit | CID_5280343 | 4129 | MAOB | ChEMBL \| NPASS |
| Nishamalaki | Phyllanthus emblica | Fruit | CID_5280343 | 4137 | MAPT | ChEMBL \| NPASS |
| Nishamalaki | Phyllanthus emblica | Fruit | CID_5280343 | 415116 | PIM3 | ChEMBL \| NPASS |
| Nishamalaki | Phyllanthus emblica | Fruit | CID_5280343 | 4154 | MBNL1 | NPASS |
| Nishamalaki | Phyllanthus emblica | Fruit | CID_5280343 | 4159 | MC3R | ChEMBL |
| Nishamalaki | Phyllanthus emblica | Fruit | CID_5280343 | 4160 | MC4R | ChEMBL |
| Nishamalaki | Phyllanthus emblica | Fruit | CID_5280343 | 4217 | MAP3K5 | ChEMBL \| NPASS |
| Nishamalaki | Phyllanthus emblica | Fruit | CID_5280343 | 4233 | MET | BindingDB \| ChEMBL \| NPASS |
| Nishamalaki | Phyllanthus emblica | Fruit | CID_5280343 | 4297 | KMT2A | NPASS |
| Nishamalaki | Phyllanthus emblica | Fruit | CID_5280343 | 43 | ACHE | BindingDB \| ChEMBL \| NPASS |
| Nishamalaki | Phyllanthus emblica | Fruit | CID_5280343 | 4312 | MMP1 | ChEMBL |
| Nishamalaki | Phyllanthus emblica | Fruit | CID_5280343 | 4313 | MMP2 | ChEMBL \| NPASS |
| Nishamalaki | Phyllanthus emblica | Fruit | CID_5280343 | 4314 | MMP3 | ChEMBL \| NPASS |
| Nishamalaki | Phyllanthus emblica | Fruit | CID_5280343 | 4318 | MMP9 | BindingDB \| ChEMBL \| NPASS |
| Nishamalaki | Phyllanthus emblica | Fruit | CID_5280343 | 4321 | MMP12 | BindingDB \| ChEMBL \| NPASS |
| Nishamalaki | Phyllanthus emblica | Fruit | CID_5280343 | 4322 | MMP13 | BindingDB \| ChEMBL \| NPASS |
| Nishamalaki | Phyllanthus emblica | Fruit | CID_5280343 | 4350 | MPG | ChEMBL \| NPASS |
| Nishamalaki | Phyllanthus emblica | Fruit | CID_5280343 | 4353 | MPO | BindingDB \| ChEMBL \| NPASS |
| Nishamalaki | Phyllanthus emblica | Fruit | CID_5280343 | 4363 | ABCC1 | ChEMBL \| NPASS |
| Nishamalaki | Phyllanthus emblica | Fruit | CID_5280343 | 4751 | NEK2 | ChEMBL \| NPASS |

| Table S3: Associations between formulations, constituent plants, phytochemicals, and human molecular targets. The table lists the formulations, plant name and plant part, phytochemical identifiers, associated human target genes (Entrez Gene ID and gene symbol), and the database sources supporting each phytochemical-target interaction. | | | | | | |
|---|---|---|---|---|---|---|
| Formulation | Plant name | Part | PubChem CID | Entrez Gene ID | Gene Symbol | Source database(s) |
| Nishamalaki | Phyllanthus emblica | Fruit | CID_5280343 | 4759 | NEU2 | BindingDB \| NPASS |
| Nishamalaki | Phyllanthus emblica | Fruit | CID_5280343 | 4790 | NFKB1 | ChEMBL \| NPASS |
| Nishamalaki | Phyllanthus emblica | Fruit | CID_5280343 | 4791 | NFKB2 | ChEMBL |
| Nishamalaki | Phyllanthus emblica | Fruit | CID_5280343 | 4886 | NPY1R | ChEMBL |
| Nishamalaki | Phyllanthus emblica | Fruit | CID_5280343 | 4887 | NPY2R | ChEMBL |
| Nishamalaki | Phyllanthus emblica | Fruit | CID_5280343 | 4907 | NT5E | BindingDB \| ChEMBL |
| Nishamalaki | Phyllanthus emblica | Fruit | CID_5280343 | 4923 | NTSR1 | ChEMBL |
| Nishamalaki | Phyllanthus emblica | Fruit | CID_5280343 | 4985 | OPRD1 | ChEMBL |
| Nishamalaki | Phyllanthus emblica | Fruit | CID_5280343 | 4986 | OPRK1 | BindingDB \| ChEMBL |
| Nishamalaki | Phyllanthus emblica | Fruit | CID_5280343 | 4988 | OPRM1 | BindingDB \| ChEMBL |
| Nishamalaki | Phyllanthus emblica | Fruit | CID_5280343 | 50507 | NOX4 | BindingDB \| ChEMBL \| NPASS |
| Nishamalaki | Phyllanthus emblica | Fruit | CID_5280343 | 51231 | VRK3 | ChEMBL \| NPASS |
| Nishamalaki | Phyllanthus emblica | Fruit | CID_5280343 | 5127 | CDK16 | ChEMBL \| NPASS |
| Nishamalaki | Phyllanthus emblica | Fruit | CID_5280343 | 5139 | PDE3A | ChEMBL |
| Nishamalaki | Phyllanthus emblica | Fruit | CID_5280343 | 51426 | POLK | ChEMBL \| NPASS |
| Nishamalaki | Phyllanthus emblica | Fruit | CID_5280343 | 5144 | PDE4D | ChEMBL |
| Nishamalaki | Phyllanthus emblica | Fruit | CID_5280343 | 51447 | IP6K2 | ChEMBL \| NPASS |
| Nishamalaki | Phyllanthus emblica | Fruit | CID_5280343 | 51548 | SIRT6 | ChEMBL |
| Nishamalaki | Phyllanthus emblica | Fruit | CID_5280343 | 5163 | PDK1 | ChEMBL \| NPASS |
| Nishamalaki | Phyllanthus emblica | Fruit | CID_5280343 | 5170 | PDPK1 | ChEMBL \| NPASS |
| Nishamalaki | Phyllanthus emblica | Fruit | CID_5280343 | 51765 | STK26 | ChEMBL \| NPASS |
| Nishamalaki | Phyllanthus emblica | Fruit | CID_5280343 | 5241 | PGR | ChEMBL |
| Nishamalaki | Phyllanthus emblica | Fruit | CID_5280343 | 5243 | ABCB1 | ChEMBL \| NPASS |
| Nishamalaki | Phyllanthus emblica | Fruit | CID_5280343 | 5290 | PIK3CA | ChEMBL |
| Nishamalaki | Phyllanthus emblica | Fruit | CID_5280343 | 5292 | PIM1 | ChEMBL \| NPASS |
| Nishamalaki | Phyllanthus emblica | Fruit | CID_5280343 | 5294 | PIK3CG | BindingDB \| ChEMBL \| NPASS |
| Nishamalaki | Phyllanthus emblica | Fruit | CID_5280343 | 5295 | PIK3R1 | ChEMBL \| NPASS |
| Nishamalaki | Phyllanthus emblica | Fruit | CID_5280343 | 5300 | PIN1 | NPASS |
| Nishamalaki | Phyllanthus emblica | Fruit | CID_5280343 | 5319 | PLA2G1B | BindingDB \| ChEMBL \| NPASS |
| Nishamalaki | Phyllanthus emblica | Fruit | CID_5280343 | 5328 | PLAU | BindingDB \| ChEMBL \| NPASS |
| Nishamalaki | Phyllanthus emblica | Fruit | CID_5280343 | 5340 | PLG | ChEMBL \| NPASS |
| Nishamalaki | Phyllanthus emblica | Fruit | CID_5280343 | 5347 | PLK1 | BindingDB \| ChEMBL \| NPASS |
| Nishamalaki | Phyllanthus emblica | Fruit | CID_5280343 | 5376 | PMP22 | NPASS |
| Nishamalaki | Phyllanthus emblica | Fruit | CID_5280343 | 53944 | CSNK1G1 | ChEMBL \| NPASS |
| Nishamalaki | Phyllanthus emblica | Fruit | CID_5280343 | 5406 | PNLIP | BindingDB \| ChEMBL |
| Nishamalaki | Phyllanthus emblica | Fruit | CID_5280343 | 5423 | POLB | ChEMBL \| NPASS |
| Nishamalaki | Phyllanthus emblica | Fruit | CID_5280343 | 5429 | POLH | BindingDB \| ChEMBL \| NPASS |
| Nishamalaki | Phyllanthus emblica | Fruit | CID_5280343 | 5444 | PON1 | ChEMBL \| NPASS |
| Nishamalaki | Phyllanthus emblica | Fruit | CID_5280343 | 54575 | UGT1A10 | ChEMBL \| NPASS |
| Nishamalaki | Phyllanthus emblica | Fruit | CID_5280343 | 54576 | UGT1A8 | ChEMBL \| NPASS |
| Nishamalaki | Phyllanthus emblica | Fruit | CID_5280343 | 54578 | UGT1A6 | ChEMBL \| NPASS |
| Nishamalaki | Phyllanthus emblica | Fruit | CID_5280343 | 54600 | UGT1A9 | ChEMBL \| NPASS |
| Nishamalaki | Phyllanthus emblica | Fruit | CID_5280343 | 54657 | UGT1A4 | ChEMBL \| NPASS |
| Nishamalaki | Phyllanthus emblica | Fruit | CID_5280343 | 54658 | UGT1A1 | ChEMBL \| NPASS |
| Nishamalaki | Phyllanthus emblica | Fruit | CID_5280343 | 54659 | UGT1A3 | ChEMBL \| NPASS |
| Nishamalaki | Phyllanthus emblica | Fruit | CID_5280343 | 5467 | PPARD | NPASS |
| Nishamalaki | Phyllanthus emblica | Fruit | CID_5280343 | 5468 | PPARG | BindingDB \| ChEMBL \| NPASS |
| Nishamalaki | Phyllanthus emblica | Fruit | CID_5280343 | 54737 | MPHOSPH8 | ChEMBL \| NPASS |
| Nishamalaki | Phyllanthus emblica | Fruit | CID_5280343 | 552 | AVPR1A | ChEMBL |
| Nishamalaki | Phyllanthus emblica | Fruit | CID_5280343 | 554 | AVPR2 | ChEMBL \| NPASS |
| Nishamalaki | Phyllanthus emblica | Fruit | CID_5280343 | 5550 | PREP | ChEMBL \| NPASS |
| Nishamalaki | Phyllanthus emblica | Fruit | CID_5280343 | 5563 | PRKAA2 | ChEMBL \| NPASS |
| Nishamalaki | Phyllanthus emblica | Fruit | CID_5280343 | 5566 | PRKACA | ChEMBL \| NPASS |
| Nishamalaki | Phyllanthus emblica | Fruit | CID_5280343 | 55662 | HIF1AN | ChEMBL \| NPASS |
| Nishamalaki | Phyllanthus emblica | Fruit | CID_5280343 | 55775 | TDP1 | ChEMBL \| NPASS |
| Nishamalaki | Phyllanthus emblica | Fruit | CID_5280343 | 5578 | PRKCA | ChEMBL \| NPASS |
| Nishamalaki | Phyllanthus emblica | Fruit | CID_5280343 | 55781 | RIOK2 | ChEMBL \| NPASS |
| Nishamalaki | Phyllanthus emblica | Fruit | CID_5280343 | 558 | AXL | ChEMBL \| NPASS |
| Nishamalaki | Phyllanthus emblica | Fruit | CID_5280343 | 5585 | PKN1 | BindingDB \| ChEMBL \| NPASS |
| Nishamalaki | Phyllanthus emblica | Fruit | CID_5280343 | 55872 | PBK | ChEMBL \| NPASS |
| Nishamalaki | Phyllanthus emblica | Fruit | CID_5280343 | 5594 | MAPK1 | ChEMBL \| NPASS |
| Nishamalaki | Phyllanthus emblica | Fruit | CID_5280343 | 5595 | MAPK3 | ChEMBL \| NPASS |
| Nishamalaki | Phyllanthus emblica | Fruit | CID_5280343 | 5597 | MAPK6 | ChEMBL \| NPASS |
| Nishamalaki | Phyllanthus emblica | Fruit | CID_5280343 | 5599 | MAPK8 | ChEMBL \| NPASS |
| Nishamalaki | Phyllanthus emblica | Fruit | CID_5280343 | 5600 | MAPK11 | ChEMBL \| NPASS |
| Nishamalaki | Phyllanthus emblica | Fruit | CID_5280343 | 5601 | MAPK9 | ChEMBL |
| Nishamalaki | Phyllanthus emblica | Fruit | CID_5280343 | 5602 | MAPK10 | BindingDB |
| Nishamalaki | Phyllanthus emblica | Fruit | CID_5280343 | 5603 | MAPK13 | ChEMBL \| NPASS |
| Nishamalaki | Phyllanthus emblica | Fruit | CID_5280343 | 5604 | MAP2K1 | ChEMBL \| NPASS |
| Nishamalaki | Phyllanthus emblica | Fruit | CID_5280343 | 5605 | MAP2K2 | ChEMBL \| NPASS |

**Table S3: Associations between formulations, constituent plants, phytochemicals, and human molecular targets. The table lists the formulations, plant name and plant part, phytochemical identifiers, associated human target genes (Entrez Gene ID and gene symbol), and the database sources supporting each phytochemical-target interaction.**

| Formulation | Plant name | Part | PubChem CID | Entrez Gene ID | Gene Symbol | Source database(s) |
|---|---|---|---|---|---|---|
| Nishamalaki | Phyllanthus emblica | Fruit | CID_5280343 | 5608 | MAP2K6 | ChEMBL \| NPASS |
| Nishamalaki | Phyllanthus emblica | Fruit | CID_5280343 | 56339 | METTL3 | ChEMBL |
| Nishamalaki | Phyllanthus emblica | Fruit | CID_5280343 | 5644 | PRSS1 | BindingDB \| ChEMBL |
| Nishamalaki | Phyllanthus emblica | Fruit | CID_5280343 | 5645 | PRSS2 | ChEMBL |
| Nishamalaki | Phyllanthus emblica | Fruit | CID_5280343 | 5646 | PRSS3 | ChEMBL \| NPASS |
| Nishamalaki | Phyllanthus emblica | Fruit | CID_5280343 | 56924 | PAK6 | NPASS |
| Nishamalaki | Phyllanthus emblica | Fruit | CID_5280343 | 5693 | PSMB5 | BindingDB \| ChEMBL \| NPASS |
| Nishamalaki | Phyllanthus emblica | Fruit | CID_5280343 | 57016 | AKR1B10 | ChEMBL \| NPASS |
| Nishamalaki | Phyllanthus emblica | Fruit | CID_5280343 | 57118 | CAMK1D | ChEMBL \| NPASS |
| Nishamalaki | Phyllanthus emblica | Fruit | CID_5280343 | 57144 | PAK5 | ChEMBL \| NPASS |
| Nishamalaki | Phyllanthus emblica | Fruit | CID_5280343 | 57172 | CAMK1G | ChEMBL \| NPASS |
| Nishamalaki | Phyllanthus emblica | Fruit | CID_5280343 | 5742 | PTGS1 | ChEMBL \| NPASS |
| Nishamalaki | Phyllanthus emblica | Fruit | CID_5280343 | 5743 | PTGS2 | BindingDB \| ChEMBL \| NPASS |
| Nishamalaki | Phyllanthus emblica | Fruit | CID_5280343 | 5747 | PTK2 | BindingDB \| ChEMBL \| NPASS |
| Nishamalaki | Phyllanthus emblica | Fruit | CID_5280343 | 5770 | PTPN1 | BindingDB \| ChEMBL \| NPASS |
| Nishamalaki | Phyllanthus emblica | Fruit | CID_5280343 | 57721 | METTL14 | ChEMBL |
| Nishamalaki | Phyllanthus emblica | Fruit | CID_5280343 | 5836 | PYGL | BindingDB \| ChEMBL \| NPASS |
| Nishamalaki | Phyllanthus emblica | Fruit | CID_5280343 | 590 | BCHE | BindingDB \| ChEMBL \| NPASS |
| Nishamalaki | Phyllanthus emblica | Fruit | CID_5280343 | 595 | CCND1 | ChEMBL \| NPASS |
| Nishamalaki | Phyllanthus emblica | Fruit | CID_5280343 | 5965 | RECQL | ChEMBL \| NPASS |
| Nishamalaki | Phyllanthus emblica | Fruit | CID_5280343 | 5970 | RELA | ChEMBL |
| Nishamalaki | Phyllanthus emblica | Fruit | CID_5280343 | 5979 | RET | BindingDB \| ChEMBL \| NPASS |
| Nishamalaki | Phyllanthus emblica | Fruit | CID_5280343 | 5999 | RGS4 | NPASS |
| Nishamalaki | Phyllanthus emblica | Fruit | CID_5280343 | 607 | BCL9 | BindingDB \| ChEMBL |
| Nishamalaki | Phyllanthus emblica | Fruit | CID_5280343 | 6197 | RPS6KA3 | ChEMBL \| NPASS |
| Nishamalaki | Phyllanthus emblica | Fruit | CID_5280343 | 6198 | RPS6KB1 | ChEMBL \| NPASS |
| Nishamalaki | Phyllanthus emblica | Fruit | CID_5280343 | 623 | BDKRB1 | ChEMBL |
| Nishamalaki | Phyllanthus emblica | Fruit | CID_5280343 | 624 | BDKRB2 | ChEMBL |
| Nishamalaki | Phyllanthus emblica | Fruit | CID_5280343 | 6300 | MAPK12 | ChEMBL \| NPASS |
| Nishamalaki | Phyllanthus emblica | Fruit | CID_5280343 | 641 | BLM | ChEMBL \| NPASS |
| Nishamalaki | Phyllanthus emblica | Fruit | CID_5280343 | 6446 | SGK1 | ChEMBL \| NPASS |
| Nishamalaki | Phyllanthus emblica | Fruit | CID_5280343 | 6476 | SI | ChEMBL \| NPASS |
| Nishamalaki | Phyllanthus emblica | Fruit | CID_5280343 | 6514 | SLC2A2 | ChEMBL \| NPASS |
| Nishamalaki | Phyllanthus emblica | Fruit | CID_5280343 | 6530 | SLC6A2 | BindingDB \| ChEMBL |
| Nishamalaki | Phyllanthus emblica | Fruit | CID_5280343 | 6531 | SLC6A3 | ChEMBL |
| Nishamalaki | Phyllanthus emblica | Fruit | CID_5280343 | 6532 | SLC6A4 | BindingDB \| ChEMBL |
| Nishamalaki | Phyllanthus emblica | Fruit | CID_5280343 | 65975 | STK33 | BindingDB |
| Nishamalaki | Phyllanthus emblica | Fruit | CID_5280343 | 6606 | SMN1 | NPASS |
| Nishamalaki | Phyllanthus emblica | Fruit | CID_5280343 | 6622 | SNCA | ChEMBL \| NPASS |
| Nishamalaki | Phyllanthus emblica | Fruit | CID_5280343 | 6652 | SORD | BindingDB \| ChEMBL \| NPASS |
| Nishamalaki | Phyllanthus emblica | Fruit | CID_5280343 | 6714 | SRC | ChEMBL \| NPASS |
| Nishamalaki | Phyllanthus emblica | Fruit | CID_5280343 | 673 | BRAF | ChEMBL \| NPASS |
| Nishamalaki | Phyllanthus emblica | Fruit | CID_5280343 | 6778 | STAT6 | NPASS |
| Nishamalaki | Phyllanthus emblica | Fruit | CID_5280343 | 6789 | STK4 | ChEMBL \| NPASS |
| Nishamalaki | Phyllanthus emblica | Fruit | CID_5280343 | 6793 | STK10 | ChEMBL \| NPASS |
| Nishamalaki | Phyllanthus emblica | Fruit | CID_5280343 | 6850 | SYK | BindingDB \| ChEMBL \| NPASS |
| Nishamalaki | Phyllanthus emblica | Fruit | CID_5280343 | 6915 | TBXA2R | ChEMBL |
| Nishamalaki | Phyllanthus emblica | Fruit | CID_5280343 | 7068 | THRB | NPASS |
| Nishamalaki | Phyllanthus emblica | Fruit | CID_5280343 | 7150 | TOP1 | ChEMBL |
| Nishamalaki | Phyllanthus emblica | Fruit | CID_5280343 | 7153 | TOP2A | ChEMBL \| NPASS |
| Nishamalaki | Phyllanthus emblica | Fruit | CID_5280343 | 7155 | TOP2B | ChEMBL \| NPASS |
| Nishamalaki | Phyllanthus emblica | Fruit | CID_5280343 | 7157 | TP53 | NPASS |
| Nishamalaki | Phyllanthus emblica | Fruit | CID_5280343 | 7224 | TRPC5 | ChEMBL \| NPASS |
| Nishamalaki | Phyllanthus emblica | Fruit | CID_5280343 | 7253 | TSHR | NPASS |
| Nishamalaki | Phyllanthus emblica | Fruit | CID_5280343 | 7257 | TSNAX | ChEMBL \| NPASS |
| Nishamalaki | Phyllanthus emblica | Fruit | CID_5280343 | 7299 | TYR | BindingDB \| ChEMBL \| NPASS |
| Nishamalaki | Phyllanthus emblica | Fruit | CID_5280343 | 7366 | UGT2B15 | ChEMBL \| NPASS |
| Nishamalaki | Phyllanthus emblica | Fruit | CID_5280343 | 7398 | USP1 | NPASS |
| Nishamalaki | Phyllanthus emblica | Fruit | CID_5280343 | 7443 | VRK1 | ChEMBL \| NPASS |
| Nishamalaki | Phyllanthus emblica | Fruit | CID_5280343 | 7444 | VRK2 | ChEMBL \| NPASS |
| Nishamalaki | Phyllanthus emblica | Fruit | CID_5280343 | 7498 | XDH | BindingDB \| ChEMBL \| NPASS |
| Nishamalaki | Phyllanthus emblica | Fruit | CID_5280343 | 7525 | YES1 | BindingDB \| ChEMBL \| NPASS |
| Nishamalaki | Phyllanthus emblica | Fruit | CID_5280343 | 759 | CA1 | BindingDB \| ChEMBL \| NPASS |
| Nishamalaki | Phyllanthus emblica | Fruit | CID_5280343 | 760 | CA2 | BindingDB \| ChEMBL \| NPASS |
| Nishamalaki | Phyllanthus emblica | Fruit | CID_5280343 | 761 | CA3 | BindingDB \| ChEMBL \| NPASS |
| Nishamalaki | Phyllanthus emblica | Fruit | CID_5280343 | 762 | CA4 | BindingDB \| ChEMBL \| NPASS |
| Nishamalaki | Phyllanthus emblica | Fruit | CID_5280343 | 763 | CA5A | BindingDB \| ChEMBL \| NPASS |
| Nishamalaki | Phyllanthus emblica | Fruit | CID_5280343 | 765 | CA6 | ChEMBL \| NPASS |
| Nishamalaki | Phyllanthus emblica | Fruit | CID_5280343 | 766 | CA7 | BindingDB \| ChEMBL \| NPASS |
| Nishamalaki | Phyllanthus emblica | Fruit | CID_5280343 | 768 | CA9 | BindingDB \| ChEMBL \| NPASS |

**Table S3: Associations between formulations, constituent plants, phytochemicals, and human molecular targets. The table lists the formulations, plant name and plant part, phytochemical identifiers, associated human target genes (Entrez Gene ID and gene symbol), and the database sources supporting each phytochemical-target interaction.**

| Formulation | Plant name | Part | PubChem CID | Entrez Gene ID | Gene Symbol | Source database(s) |
|---|---|---|---|---|---|---|
| Nishamalaki | Phyllanthus emblica | Fruit | CID_5280343 | 771 | CA12 | BindingDB \| ChEMBL \| NPASS |
| Nishamalaki | Phyllanthus emblica | Fruit | CID_5280343 | 79915 | ATAD5 | ChEMBL \| NPASS |
| Nishamalaki | Phyllanthus emblica | Fruit | CID_5280343 | 805 | CALM2 | ChEMBL |
| Nishamalaki | Phyllanthus emblica | Fruit | CID_5280343 | 814 | CAMK4 | ChEMBL \| NPASS |
| Nishamalaki | Phyllanthus emblica | Fruit | CID_5280343 | 815 | CAMK2A | ChEMBL \| NPASS |
| Nishamalaki | Phyllanthus emblica | Fruit | CID_5280343 | 816 | CAMK2B | BindingDB \| ChEMBL \| NPASS |
| Nishamalaki | Phyllanthus emblica | Fruit | CID_5280343 | 817 | CAMK2D | ChEMBL \| NPASS |
| Nishamalaki | Phyllanthus emblica | Fruit | CID_5280343 | 818 | CAMK2G | ChEMBL \| NPASS |
| Nishamalaki | Phyllanthus emblica | Fruit | CID_5280343 | 85417 | CCNB3 | ChEMBL |
| Nishamalaki | Phyllanthus emblica | Fruit | CID_5280343 | 8550 | MAPKAPK5 | ChEMBL \| NPASS |
| Nishamalaki | Phyllanthus emblica | Fruit | CID_5280343 | 8576 | STK16 | ChEMBL \| NPASS |
| Nishamalaki | Phyllanthus emblica | Fruit | CID_5280343 | 8647 | ABCB11 | NPASS |
| Nishamalaki | Phyllanthus emblica | Fruit | CID_5280343 | 8654 | PDE5A | ChEMBL \| NPASS |
| Nishamalaki | Phyllanthus emblica | Fruit | CID_5280343 | 8714 | ABCC3 | BindingDB \| NPASS |
| Nishamalaki | Phyllanthus emblica | Fruit | CID_5280343 | 873 | CBR1 | ChEMBL \| NPASS |
| Nishamalaki | Phyllanthus emblica | Fruit | CID_5280343 | 874 | CBR3 | ChEMBL \| NPASS |
| Nishamalaki | Phyllanthus emblica | Fruit | CID_5280343 | 8814 | CDKL1 | ChEMBL \| NPASS |
| Nishamalaki | Phyllanthus emblica | Fruit | CID_5280343 | 8851 | CDK5R1 | ChEMBL \| NPASS |
| Nishamalaki | Phyllanthus emblica | Fruit | CID_5280343 | 886 | CCKAR | ChEMBL |
| Nishamalaki | Phyllanthus emblica | Fruit | CID_5280343 | 887 | CCKBR | ChEMBL |
| Nishamalaki | Phyllanthus emblica | Fruit | CID_5280343 | 8877 | SPHK1 | ChEMBL |
| Nishamalaki | Phyllanthus emblica | Fruit | CID_5280343 | 891 | CCNB1 | ChEMBL \| NPASS |
| Nishamalaki | Phyllanthus emblica | Fruit | CID_5280343 | 8972 | MGAM | ChEMBL \| NPASS |
| Nishamalaki | Phyllanthus emblica | Fruit | CID_5280343 | 9099 | USP2 | NPASS |
| Nishamalaki | Phyllanthus emblica | Fruit | CID_5280343 | 9133 | CCNB2 | ChEMBL \| NPASS |
| Nishamalaki | Phyllanthus emblica | Fruit | CID_5280343 | 9212 | AURKB | BindingDB \| ChEMBL \| NPASS |
| Nishamalaki | Phyllanthus emblica | Fruit | CID_5280343 | 9252 | RPS6KA5 | ChEMBL \| NPASS |
| Nishamalaki | Phyllanthus emblica | Fruit | CID_5280343 | 9261 | MAPKAPK2 | ChEMBL \| NPASS |
| Nishamalaki | Phyllanthus emblica | Fruit | CID_5280343 | 9263 | STK17A | ChEMBL \| NPASS |
| Nishamalaki | Phyllanthus emblica | Fruit | CID_5280343 | 93432 | MGAM2 | ChEMBL \| NPASS |
| Nishamalaki | Phyllanthus emblica | Fruit | CID_5280343 | 9429 | ABCG2 | BindingDB \| ChEMBL \| NPASS |
| Nishamalaki | Phyllanthus emblica | Fruit | CID_5280343 | 952 | CD38 | BindingDB \| ChEMBL \| NPASS |
| Nishamalaki | Phyllanthus emblica | Fruit | CID_5280343 | 9748 | SLK | ChEMBL \| NPASS |
| Nishamalaki | Phyllanthus emblica | Fruit | CID_5280343 | 983 | CDK1 | BindingDB \| ChEMBL \| NPASS |
| Nishamalaki | Phyllanthus emblica | Fruit | CID_5280343 | 9891 | NUAK1 | BindingDB \| ChEMBL \| NPASS |
| Nishamalaki | Phyllanthus emblica | Fruit | CID_5280343 | 9943 | OXSR1 | ChEMBL \| NPASS |
| Nishamalaki | Phyllanthus emblica | Fruit | CID_5281855 | 10013 | HDAC6 | ChEMBL |
| Nishamalaki | Phyllanthus emblica | Fruit | CID_5281855 | 10073 | SNUPN | NPASS |
| Nishamalaki | Phyllanthus emblica | Fruit | CID_5281855 | 1017 | CDK2 | BindingDB \| ChEMBL |
| Nishamalaki | Phyllanthus emblica | Fruit | CID_5281855 | 1019 | CDK4 | BindingDB \| ChEMBL |
| Nishamalaki | Phyllanthus emblica | Fruit | CID_5281855 | 10411 | RAPGEF3 | NPASS |
| Nishamalaki | Phyllanthus emblica | Fruit | CID_5281855 | 10498 | CARM1 | ChEMBL \| NPASS |
| Nishamalaki | Phyllanthus emblica | Fruit | CID_5281855 | 10587 | TXNRD2 | ChEMBL |
| Nishamalaki | Phyllanthus emblica | Fruit | CID_5281855 | 10599 | SLCO1B1 | ChEMBL \| NPASS |
| Nishamalaki | Phyllanthus emblica | Fruit | CID_5281855 | 10733 | PLK4 | BindingDB \| ChEMBL \| NPASS |
| Nishamalaki | Phyllanthus emblica | Fruit | CID_5281855 | 10919 | EHMT2 | ChEMBL \| NPASS |
| Nishamalaki | Phyllanthus emblica | Fruit | CID_5281855 | 10951 | CBX1 | NPASS |
| Nishamalaki | Phyllanthus emblica | Fruit | CID_5281855 | 11069 | RAPGEF4 | NPASS |
| Nishamalaki | Phyllanthus emblica | Fruit | CID_5281855 | 11201 | POLI | ChEMBL \| NPASS |
| Nishamalaki | Phyllanthus emblica | Fruit | CID_5281855 | 11238 | CA5B | BindingDB \| ChEMBL \| NPASS |
| Nishamalaki | Phyllanthus emblica | Fruit | CID_5281855 | 114112 | TXNRD3 | ChEMBL \| NPASS |
| Nishamalaki | Phyllanthus emblica | Fruit | CID_5281855 | 1326 | MAP3K8 | BindingDB \| ChEMBL \| NPASS |
| Nishamalaki | Phyllanthus emblica | Fruit | CID_5281855 | 1457 | CSNK2A1 | BindingDB \| ChEMBL \| NPASS |
| Nishamalaki | Phyllanthus emblica | Fruit | CID_5281855 | 1459 | CSNK2A2 | ChEMBL \| NPASS |
| Nishamalaki | Phyllanthus emblica | Fruit | CID_5281855 | 1460 | CSNK2B | ChEMBL |
| Nishamalaki | Phyllanthus emblica | Fruit | CID_5281855 | 1557 | CYP2C19 | ChEMBL \| NPASS |
| Nishamalaki | Phyllanthus emblica | Fruit | CID_5281855 | 1559 | CYP2C9 | NPASS |
| Nishamalaki | Phyllanthus emblica | Fruit | CID_5281855 | 1636 | ACE | BindingDB \| NPASS |
| Nishamalaki | Phyllanthus emblica | Fruit | CID_5281855 | 1845 | DUSP3 | BindingDB \| NPASS |
| Nishamalaki | Phyllanthus emblica | Fruit | CID_5281855 | 1859 | DYRK1A | BindingDB \| ChEMBL \| NPASS |
| Nishamalaki | Phyllanthus emblica | Fruit | CID_5281855 | 1956 | EGFR | BindingDB \| ChEMBL \| NPASS |
| Nishamalaki | Phyllanthus emblica | Fruit | CID_5281855 | 1981 | EIF4G1 | BindingDB |
| Nishamalaki | Phyllanthus emblica | Fruit | CID_5281855 | 1995 | ELAVL3 | ChEMBL \| NPASS |
| Nishamalaki | Phyllanthus emblica | Fruit | CID_5281855 | 2050 | EPHB4 | BindingDB \| ChEMBL \| NPASS |
| Nishamalaki | Phyllanthus emblica | Fruit | CID_5281855 | 2053 | EPHX2 | ChEMBL \| NPASS |
| Nishamalaki | Phyllanthus emblica | Fruit | CID_5281855 | 2064 | ERBB2 | ChEMBL \| NPASS |
| Nishamalaki | Phyllanthus emblica | Fruit | CID_5281855 | 207 | AKT1 | ChEMBL \| NPASS |
| Nishamalaki | Phyllanthus emblica | Fruit | CID_5281855 | 2237 | FEN1 | ChEMBL \| NPASS |
| Nishamalaki | Phyllanthus emblica | Fruit | CID_5281855 | 231 | AKR1B1 | ChEMBL \| NPASS |
| Nishamalaki | Phyllanthus emblica | Fruit | CID_5281855 | 23192 | ATG4B | BindingDB \| ChEMBL \| NPASS |

**Table S3: Associations between formulations, constituent plants, phytochemicals, and human molecular targets. The table lists the formulations, plant name and plant part, phytochemical identifiers, associated human target genes (Entrez Gene ID and gene symbol), and the database sources supporting each phytochemical-target interaction.**

| Formulation | Plant name | Part | PubChem CID | Entrez Gene ID | Gene Symbol | Source database(s) |
|---|---|---|---|---|---|---|
| Nishamalaki | Phyllanthus emblica | Fruit | CID_5281855 | 2322 | FLT3 | ChEMBL \| NPASS |
| Nishamalaki | Phyllanthus emblica | Fruit | CID_5281855 | 2324 | FLT4 | BindingDB \| ChEMBL \| NPASS |
| Nishamalaki | Phyllanthus emblica | Fruit | CID_5281855 | 23405 | DICER1 | ChEMBL |
| Nishamalaki | Phyllanthus emblica | Fruit | CID_5281855 | 23621 | BACE1 | BindingDB \| ChEMBL \| NPASS |
| Nishamalaki | Phyllanthus emblica | Fruit | CID_5281855 | 23632 | CA14 | BindingDB \| ChEMBL \| NPASS |
| Nishamalaki | Phyllanthus emblica | Fruit | CID_5281855 | 238 | ALK | BindingDB \| ChEMBL |
| Nishamalaki | Phyllanthus emblica | Fruit | CID_5281855 | 246 | ALOX15 | NPASS |
| Nishamalaki | Phyllanthus emblica | Fruit | CID_5281855 | 247 | ALOX15B | NPASS |
| Nishamalaki | Phyllanthus emblica | Fruit | CID_5281855 | 2548 | GAA | NPASS |
| Nishamalaki | Phyllanthus emblica | Fruit | CID_5281855 | 2590 | GALNT2 | ChEMBL \| NPASS |
| Nishamalaki | Phyllanthus emblica | Fruit | CID_5281855 | 2597 | GAPDH | NPASS |
| Nishamalaki | Phyllanthus emblica | Fruit | CID_5281855 | 2648 | KAT2A | NPASS |
| Nishamalaki | Phyllanthus emblica | Fruit | CID_5281855 | 2744 | GLS | ChEMBL \| NPASS |
| Nishamalaki | Phyllanthus emblica | Fruit | CID_5281855 | 28234 | SLCO1B3 | ChEMBL \| NPASS |
| Nishamalaki | Phyllanthus emblica | Fruit | CID_5281855 | 2859 | GPR35 | BindingDB \| ChEMBL \| NPASS |
| Nishamalaki | Phyllanthus emblica | Fruit | CID_5281855 | 2932 | GSK3B | BindingDB \| ChEMBL \| NPASS |
| Nishamalaki | Phyllanthus emblica | Fruit | CID_5281855 | 2936 | GSR | ChEMBL \| NPASS |
| Nishamalaki | Phyllanthus emblica | Fruit | CID_5281855 | 2950 | GSTP1 | ChEMBL \| NPASS |
| Nishamalaki | Phyllanthus emblica | Fruit | CID_5281855 | 29994 | BAZ2B | ChEMBL \| NPASS |
| Nishamalaki | Phyllanthus emblica | Fruit | CID_5281855 | 3028 | HSD17B10 | ChEMBL \| NPASS |
| Nishamalaki | Phyllanthus emblica | Fruit | CID_5281855 | 3043 | HBB | ChEMBL \| NPASS |
| Nishamalaki | Phyllanthus emblica | Fruit | CID_5281855 | 3248 | HPGD | ChEMBL \| NPASS |
| Nishamalaki | Phyllanthus emblica | Fruit | CID_5281855 | 328 | APEX1 | ChEMBL \| NPASS |
| Nishamalaki | Phyllanthus emblica | Fruit | CID_5281855 | 3303 | HSPA1A | BindingDB \| ChEMBL \| NPASS |
| Nishamalaki | Phyllanthus emblica | Fruit | CID_5281855 | 3309 | HSPA5 | NPASS |
| Nishamalaki | Phyllanthus emblica | Fruit | CID_5281855 | 3312 | HSPA8 | BindingDB \| ChEMBL \| NPASS |
| Nishamalaki | Phyllanthus emblica | Fruit | CID_5281855 | 3417 | IDH1 | NPASS |
| Nishamalaki | Phyllanthus emblica | Fruit | CID_5281855 | 3480 | IGF1R | BindingDB \| ChEMBL \| NPASS |
| Nishamalaki | Phyllanthus emblica | Fruit | CID_5281855 | 351 | APP | BindingDB \| ChEMBL \| NPASS |
| Nishamalaki | Phyllanthus emblica | Fruit | CID_5281855 | 3551 | IKBKB | BindingDB \| ChEMBL \| NPASS |
| Nishamalaki | Phyllanthus emblica | Fruit | CID_5281855 | 3643 | INSR | BindingDB \| ChEMBL \| NPASS |
| Nishamalaki | Phyllanthus emblica | Fruit | CID_5281855 | 377677 | CA13 | ChEMBL \| NPASS |
| Nishamalaki | Phyllanthus emblica | Fruit | CID_5281855 | 3791 | KDR | BindingDB \| ChEMBL \| NPASS |
| Nishamalaki | Phyllanthus emblica | Fruit | CID_5281855 | 3837 | KPNB1 | NPASS |
| Nishamalaki | Phyllanthus emblica | Fruit | CID_5281855 | 390245 | KDM4E | ChEMBL \| NPASS |
| Nishamalaki | Phyllanthus emblica | Fruit | CID_5281855 | 4000 | LMNA | NPASS |
| Nishamalaki | Phyllanthus emblica | Fruit | CID_5281855 | 4088 | SMAD3 | BindingDB \| ChEMBL \| NPASS |
| Nishamalaki | Phyllanthus emblica | Fruit | CID_5281855 | 4128 | MAOA | ChEMBL \| NPASS |
| Nishamalaki | Phyllanthus emblica | Fruit | CID_5281855 | 4137 | MAPT | ChEMBL \| NPASS |
| Nishamalaki | Phyllanthus emblica | Fruit | CID_5281855 | 4154 | MBNL1 | NPASS |
| Nishamalaki | Phyllanthus emblica | Fruit | CID_5281855 | 4170 | MCL1 | BindingDB \| ChEMBL \| NPASS |
| Nishamalaki | Phyllanthus emblica | Fruit | CID_5281855 | 4233 | MET | BindingDB \| ChEMBL \| NPASS |
| Nishamalaki | Phyllanthus emblica | Fruit | CID_5281855 | 4282 | MIF | BindingDB \| ChEMBL \| NPASS |
| Nishamalaki | Phyllanthus emblica | Fruit | CID_5281855 | 4297 | KMT2A | NPASS |
| Nishamalaki | Phyllanthus emblica | Fruit | CID_5281855 | 43 | ACHE | BindingDB |
| Nishamalaki | Phyllanthus emblica | Fruit | CID_5281855 | 472 | ATM | NPASS |
| Nishamalaki | Phyllanthus emblica | Fruit | CID_5281855 | 4780 | NFE2L2 | NPASS |
| Nishamalaki | Phyllanthus emblica | Fruit | CID_5281855 | 4869 | NPM1 | ChEMBL \| NPASS |
| Nishamalaki | Phyllanthus emblica | Fruit | CID_5281855 | 4907 | NT5E | ChEMBL \| NPASS |
| Nishamalaki | Phyllanthus emblica | Fruit | CID_5281855 | 51053 | GMNN | NPASS |
| Nishamalaki | Phyllanthus emblica | Fruit | CID_5281855 | 51426 | POLK | ChEMBL \| NPASS |
| Nishamalaki | Phyllanthus emblica | Fruit | CID_5281855 | 5159 | PDGFRB | BindingDB \| ChEMBL \| NPASS |
| Nishamalaki | Phyllanthus emblica | Fruit | CID_5281855 | 5300 | PIN1 | NPASS |
| Nishamalaki | Phyllanthus emblica | Fruit | CID_5281855 | 5313 | PKLR | ChEMBL |
| Nishamalaki | Phyllanthus emblica | Fruit | CID_5281855 | 5347 | PLK1 | BindingDB \| ChEMBL \| NPASS |
| Nishamalaki | Phyllanthus emblica | Fruit | CID_5281855 | 5423 | POLB | ChEMBL \| NPASS |
| Nishamalaki | Phyllanthus emblica | Fruit | CID_5281855 | 5429 | POLH | BindingDB \| ChEMBL \| NPASS |
| Nishamalaki | Phyllanthus emblica | Fruit | CID_5281855 | 5566 | PRKACA | BindingDB \| ChEMBL |
| Nishamalaki | Phyllanthus emblica | Fruit | CID_5281855 | 5567 | PRKACB | ChEMBL |
| Nishamalaki | Phyllanthus emblica | Fruit | CID_5281855 | 5568 | PRKACG | ChEMBL \| NPASS |
| Nishamalaki | Phyllanthus emblica | Fruit | CID_5281855 | 55775 | TDP1 | ChEMBL \| NPASS |
| Nishamalaki | Phyllanthus emblica | Fruit | CID_5281855 | 55867 | SLC22A11 | ChEMBL \| NPASS |
| Nishamalaki | Phyllanthus emblica | Fruit | CID_5281855 | 5594 | MAPK1 | ChEMBL \| NPASS |
| Nishamalaki | Phyllanthus emblica | Fruit | CID_5281855 | 5747 | PTK2 | BindingDB \| ChEMBL \| NPASS |
| Nishamalaki | Phyllanthus emblica | Fruit | CID_5281855 | 5770 | PTPN1 | BindingDB |
| Nishamalaki | Phyllanthus emblica | Fruit | CID_5281855 | 57787 | MARK4 | ChEMBL |
| Nishamalaki | Phyllanthus emblica | Fruit | CID_5281855 | 595 | CCND1 | ChEMBL \| NPASS |
| Nishamalaki | Phyllanthus emblica | Fruit | CID_5281855 | 5965 | RECQL | ChEMBL \| NPASS |
| Nishamalaki | Phyllanthus emblica | Fruit | CID_5281855 | 5979 | RET | BindingDB \| ChEMBL \| NPASS |
| Nishamalaki | Phyllanthus emblica | Fruit | CID_5281855 | 5999 | RGS4 | NPASS |

**Table S3: Associations between formulations, constituent plants, phytochemicals, and human molecular targets. The table lists the formulations, plant name and plant part, phytochemical identifiers, associated human target genes (Entrez Gene ID and gene symbol), and the database sources supporting each phytochemical-target interaction.**

| Formulation | Plant name | Part | PubChem CID | Entrez Gene ID | Gene Symbol | Source database(s) |
|---|---|---|---|---|---|---|
| Nishamalaki | Phyllanthus emblica | Fruit | CID_5281855 | 641 | BLM | ChEMBL \| NPASS |
| Nishamalaki | Phyllanthus emblica | Fruit | CID_5281855 | 6714 | SRC | ChEMBL \| NPASS |
| Nishamalaki | Phyllanthus emblica | Fruit | CID_5281855 | 673 | BRAF | BindingDB \| ChEMBL \| NPASS |
| Nishamalaki | Phyllanthus emblica | Fruit | CID_5281855 | 6790 | AURKA | BindingDB \| ChEMBL \| NPASS |
| Nishamalaki | Phyllanthus emblica | Fruit | CID_5281855 | 7010 | TEK | BindingDB \| ChEMBL \| NPASS |
| Nishamalaki | Phyllanthus emblica | Fruit | CID_5281855 | 7068 | THRB | NPASS |
| Nishamalaki | Phyllanthus emblica | Fruit | CID_5281855 | 7155 | TOP2B | ChEMBL \| NPASS |
| Nishamalaki | Phyllanthus emblica | Fruit | CID_5281855 | 7157 | TP53 | NPASS |
| Nishamalaki | Phyllanthus emblica | Fruit | CID_5281855 | 7257 | TSNAX | ChEMBL \| NPASS |
| Nishamalaki | Phyllanthus emblica | Fruit | CID_5281855 | 7296 | TXNRD1 | ChEMBL |
| Nishamalaki | Phyllanthus emblica | Fruit | CID_5281855 | 7398 | USP1 | NPASS |
| Nishamalaki | Phyllanthus emblica | Fruit | CID_5281855 | 7421 | VDR | ChEMBL \| NPASS |
| Nishamalaki | Phyllanthus emblica | Fruit | CID_5281855 | 759 | CA1 | BindingDB \| ChEMBL \| NPASS |
| Nishamalaki | Phyllanthus emblica | Fruit | CID_5281855 | 760 | CA2 | BindingDB \| ChEMBL \| NPASS |
| Nishamalaki | Phyllanthus emblica | Fruit | CID_5281855 | 761 | CA3 | BindingDB \| ChEMBL \| NPASS |
| Nishamalaki | Phyllanthus emblica | Fruit | CID_5281855 | 762 | CA4 | BindingDB \| ChEMBL \| NPASS |
| Nishamalaki | Phyllanthus emblica | Fruit | CID_5281855 | 763 | CA5A | BindingDB \| ChEMBL \| NPASS |
| Nishamalaki | Phyllanthus emblica | Fruit | CID_5281855 | 765 | CA6 | ChEMBL \| NPASS |
| Nishamalaki | Phyllanthus emblica | Fruit | CID_5281855 | 766 | CA7 | BindingDB \| ChEMBL \| NPASS |
| Nishamalaki | Phyllanthus emblica | Fruit | CID_5281855 | 768 | CA9 | BindingDB \| ChEMBL \| NPASS |
| Nishamalaki | Phyllanthus emblica | Fruit | CID_5281855 | 771 | CA12 | BindingDB \| ChEMBL \| NPASS |
| Nishamalaki | Phyllanthus emblica | Fruit | CID_5281855 | 861 | RUNX1 | ChEMBL |
| Nishamalaki | Phyllanthus emblica | Fruit | CID_5281855 | 865 | CBFB | ChEMBL \| NPASS |
| Nishamalaki | Phyllanthus emblica | Fruit | CID_5281855 | 8877 | SPHK1 | ChEMBL |
| Nishamalaki | Phyllanthus emblica | Fruit | CID_5281855 | 890 | CCNA2 | ChEMBL |
| Nishamalaki | Phyllanthus emblica | Fruit | CID_5281855 | 8900 | CCNA1 | ChEMBL \| NPASS |
| Nishamalaki | Phyllanthus emblica | Fruit | CID_5281855 | 9099 | USP2 | ChEMBL \| NPASS |
| Nishamalaki | Phyllanthus emblica | Fruit | CID_5281855 | 9212 | AURKB | BindingDB \| ChEMBL \| NPASS |
| Nishamalaki | Phyllanthus emblica | Fruit | CID_5281855 | 9356 | SLC22A6 | BindingDB \| ChEMBL \| NPASS |
| Nishamalaki | Phyllanthus emblica | Fruit | CID_5281855 | 9682 | KDM4A | ChEMBL \| NPASS |
| Nishamalaki | Phyllanthus emblica | Fruit | CID_5281855 | 9891 | NUAK1 | BindingDB \| ChEMBL \| NPASS |
| Nishamalaki | Phyllanthus emblica | Fruit | CID_54670067 | 10280 | SIGMAR1 | ChEMBL |
| Nishamalaki | Phyllanthus emblica | Fruit | CID_54670067 | 10599 | SLCO1B1 | ChEMBL \| NPASS |
| Nishamalaki | Phyllanthus emblica | Fruit | CID_54670067 | 10800 | CYSLTR1 | ChEMBL |
| Nishamalaki | Phyllanthus emblica | Fruit | CID_54670067 | 10919 | EHMT2 | ChEMBL |
| Nishamalaki | Phyllanthus emblica | Fruit | CID_54670067 | 112398 | EGLN2 | ChEMBL \| NPASS |
| Nishamalaki | Phyllanthus emblica | Fruit | CID_54670067 | 112399 | EGLN3 | ChEMBL \| NPASS |
| Nishamalaki | Phyllanthus emblica | Fruit | CID_54670067 | 11255 | HRH3 | ChEMBL |
| Nishamalaki | Phyllanthus emblica | Fruit | CID_54670067 | 1128 | CHRM1 | ChEMBL |
| Nishamalaki | Phyllanthus emblica | Fruit | CID_54670067 | 1129 | CHRM2 | ChEMBL |
| Nishamalaki | Phyllanthus emblica | Fruit | CID_54670067 | 1131 | CHRM3 | ChEMBL |
| Nishamalaki | Phyllanthus emblica | Fruit | CID_54670067 | 1132 | CHRM4 | ChEMBL |
| Nishamalaki | Phyllanthus emblica | Fruit | CID_54670067 | 1133 | CHRM5 | ChEMBL |
| Nishamalaki | Phyllanthus emblica | Fruit | CID_54670067 | 1233 | CCR4 | ChEMBL |
| Nishamalaki | Phyllanthus emblica | Fruit | CID_54670067 | 1234 | CCR5 | ChEMBL |
| Nishamalaki | Phyllanthus emblica | Fruit | CID_54670067 | 1268 | CNR1 | ChEMBL |
| Nishamalaki | Phyllanthus emblica | Fruit | CID_54670067 | 134 | ADORA1 | ChEMBL |
| Nishamalaki | Phyllanthus emblica | Fruit | CID_54670067 | 135 | ADORA2A | ChEMBL |
| Nishamalaki | Phyllanthus emblica | Fruit | CID_54670067 | 140 | ADORA3 | ChEMBL |
| Nishamalaki | Phyllanthus emblica | Fruit | CID_54670067 | 1432 | MAPK14 | ChEMBL |
| Nishamalaki | Phyllanthus emblica | Fruit | CID_54670067 | 146 | ADRA1D | ChEMBL |
| Nishamalaki | Phyllanthus emblica | Fruit | CID_54670067 | 148 | ADRA1A | ChEMBL |
| Nishamalaki | Phyllanthus emblica | Fruit | CID_54670067 | 150 | ADRA2A | ChEMBL |
| Nishamalaki | Phyllanthus emblica | Fruit | CID_54670067 | 151 | ADRA2B | ChEMBL \| NPASS |
| Nishamalaki | Phyllanthus emblica | Fruit | CID_54670067 | 1511 | CTSG | ChEMBL |
| Nishamalaki | Phyllanthus emblica | Fruit | CID_54670067 | 152 | ADRA2C | ChEMBL |
| Nishamalaki | Phyllanthus emblica | Fruit | CID_54670067 | 153 | ADRB1 | ChEMBL |
| Nishamalaki | Phyllanthus emblica | Fruit | CID_54670067 | 154 | ADRB2 | ChEMBL |
| Nishamalaki | Phyllanthus emblica | Fruit | CID_54670067 | 1544 | CYP1A2 | ChEMBL |
| Nishamalaki | Phyllanthus emblica | Fruit | CID_54670067 | 1548 | CYP2A6 | ChEMBL |
| Nishamalaki | Phyllanthus emblica | Fruit | CID_54670067 | 155 | ADRB3 | ChEMBL |
| Nishamalaki | Phyllanthus emblica | Fruit | CID_54670067 | 1557 | CYP2C19 | ChEMBL |
| Nishamalaki | Phyllanthus emblica | Fruit | CID_54670067 | 1559 | CYP2C9 | ChEMBL |
| Nishamalaki | Phyllanthus emblica | Fruit | CID_54670067 | 1565 | CYP2D6 | ChEMBL |
| Nishamalaki | Phyllanthus emblica | Fruit | CID_54670067 | 1571 | CYP2E1 | ChEMBL |
| Nishamalaki | Phyllanthus emblica | Fruit | CID_54670067 | 1576 | CYP3A4 | ChEMBL |
| Nishamalaki | Phyllanthus emblica | Fruit | CID_54670067 | 1812 | DRD1 | ChEMBL |
| Nishamalaki | Phyllanthus emblica | Fruit | CID_54670067 | 1813 | DRD2 | ChEMBL |
| Nishamalaki | Phyllanthus emblica | Fruit | CID_54670067 | 1814 | DRD3 | ChEMBL |
| Nishamalaki | Phyllanthus emblica | Fruit | CID_54670067 | 1815 | DRD4 | ChEMBL |

**Table S3: Associations between formulations, constituent plants, phytochemicals, and human molecular targets. The table lists the formulations, plant name and plant part, phytochemical identifiers, associated human target genes (Entrez Gene ID and gene symbol), and the database sources supporting each phytochemical-target interaction.**

| Formulation | Plant name | Part | PubChem CID | Entrez Gene ID | Gene Symbol | Source database(s) |
|---|---|---|---|---|---|---|
| Nishamalaki | Phyllanthus emblica | Fruit | CID_54670067 | 186 | AGTR2 | ChEMBL |
| Nishamalaki | Phyllanthus emblica | Fruit | CID_54670067 | 1909 | EDNRA | ChEMBL |
| Nishamalaki | Phyllanthus emblica | Fruit | CID_54670067 | 1956 | EGFR | ChEMBL |
| Nishamalaki | Phyllanthus emblica | Fruit | CID_54670067 | 1991 | ELANE | ChEMBL |
| Nishamalaki | Phyllanthus emblica | Fruit | CID_54670067 | 2064 | ERBB2 | ChEMBL |
| Nishamalaki | Phyllanthus emblica | Fruit | CID_54670067 | 2099 | ESR1 | ChEMBL |
| Nishamalaki | Phyllanthus emblica | Fruit | CID_54670067 | 2100 | ESR2 | ChEMBL |
| Nishamalaki | Phyllanthus emblica | Fruit | CID_54670067 | 2147 | F2 | ChEMBL |
| Nishamalaki | Phyllanthus emblica | Fruit | CID_54670067 | 2321 | FLT1 | ChEMBL |
| Nishamalaki | Phyllanthus emblica | Fruit | CID_54670067 | 239 | ALOX12 | ChEMBL \| NPASS |
| Nishamalaki | Phyllanthus emblica | Fruit | CID_54670067 | 240 | ALOX5 | ChEMBL \| NPASS |
| Nishamalaki | Phyllanthus emblica | Fruit | CID_54670067 | 246 | ALOX15 | ChEMBL \| NPASS |
| Nishamalaki | Phyllanthus emblica | Fruit | CID_54670067 | 247 | ALOX15B | ChEMBL \| NPASS |
| Nishamalaki | Phyllanthus emblica | Fruit | CID_54670067 | 2534 | FYN | ChEMBL |
| Nishamalaki | Phyllanthus emblica | Fruit | CID_54670067 | 2554 | GABRA1 | ChEMBL |
| Nishamalaki | Phyllanthus emblica | Fruit | CID_54670067 | 2555 | GABRA2 | ChEMBL |
| Nishamalaki | Phyllanthus emblica | Fruit | CID_54670067 | 2561 | GABRB2 | ChEMBL |
| Nishamalaki | Phyllanthus emblica | Fruit | CID_54670067 | 2566 | GABRG2 | ChEMBL |
| Nishamalaki | Phyllanthus emblica | Fruit | CID_54670067 | 25939 | SAMHD1 | ChEMBL |
| Nishamalaki | Phyllanthus emblica | Fruit | CID_54670067 | 276 | AMY1A | ChEMBL |
| Nishamalaki | Phyllanthus emblica | Fruit | CID_54670067 | 279 | AMY2A | ChEMBL \| NPASS |
| Nishamalaki | Phyllanthus emblica | Fruit | CID_54670067 | 28234 | SLCO1B3 | ChEMBL \| NPASS |
| Nishamalaki | Phyllanthus emblica | Fruit | CID_54670067 | 2908 | NR3C1 | ChEMBL |
| Nishamalaki | Phyllanthus emblica | Fruit | CID_54670067 | 2932 | GSK3B | ChEMBL \| NPASS |
| Nishamalaki | Phyllanthus emblica | Fruit | CID_54670067 | 3156 | HMGCR | ChEMBL |
| Nishamalaki | Phyllanthus emblica | Fruit | CID_54670067 | 3269 | HRH1 | ChEMBL |
| Nishamalaki | Phyllanthus emblica | Fruit | CID_54670067 | 3274 | HRH2 | ChEMBL |
| Nishamalaki | Phyllanthus emblica | Fruit | CID_54670067 | 3350 | HTR1A | ChEMBL |
| Nishamalaki | Phyllanthus emblica | Fruit | CID_54670067 | 3356 | HTR2A | ChEMBL |
| Nishamalaki | Phyllanthus emblica | Fruit | CID_54670067 | 3357 | HTR2B | ChEMBL |
| Nishamalaki | Phyllanthus emblica | Fruit | CID_54670067 | 3358 | HTR2C | ChEMBL |
| Nishamalaki | Phyllanthus emblica | Fruit | CID_54670067 | 3362 | HTR6 | ChEMBL |
| Nishamalaki | Phyllanthus emblica | Fruit | CID_54670067 | 351 | APP | ChEMBL \| NPASS |
| Nishamalaki | Phyllanthus emblica | Fruit | CID_54670067 | 3577 | CXCR1 | ChEMBL |
| Nishamalaki | Phyllanthus emblica | Fruit | CID_54670067 | 3579 | CXCR2 | ChEMBL |
| Nishamalaki | Phyllanthus emblica | Fruit | CID_54670067 | 367 | AR | ChEMBL \| NPASS |
| Nishamalaki | Phyllanthus emblica | Fruit | CID_54670067 | 3757 | KCNH2 | ChEMBL |
| Nishamalaki | Phyllanthus emblica | Fruit | CID_54670067 | 3791 | KDR | ChEMBL |
| Nishamalaki | Phyllanthus emblica | Fruit | CID_54670067 | 3932 | LCK | ChEMBL |
| Nishamalaki | Phyllanthus emblica | Fruit | CID_54670067 | 4128 | MAOA | ChEMBL |
| Nishamalaki | Phyllanthus emblica | Fruit | CID_54670067 | 4159 | MC3R | ChEMBL |
| Nishamalaki | Phyllanthus emblica | Fruit | CID_54670067 | 4160 | MC4R | ChEMBL |
| Nishamalaki | Phyllanthus emblica | Fruit | CID_54670067 | 4161 | MC5R | ChEMBL |
| Nishamalaki | Phyllanthus emblica | Fruit | CID_54670067 | 43 | ACHE | ChEMBL |
| Nishamalaki | Phyllanthus emblica | Fruit | CID_54670067 | 4312 | MMP1 | ChEMBL |
| Nishamalaki | Phyllanthus emblica | Fruit | CID_54670067 | 4318 | MMP9 | ChEMBL |
| Nishamalaki | Phyllanthus emblica | Fruit | CID_54670067 | 4780 | NFE2L2 | ChEMBL \| NPASS |
| Nishamalaki | Phyllanthus emblica | Fruit | CID_54670067 | 4886 | NPY1R | ChEMBL |
| Nishamalaki | Phyllanthus emblica | Fruit | CID_54670067 | 4887 | NPY2R | ChEMBL |
| Nishamalaki | Phyllanthus emblica | Fruit | CID_54670067 | 4985 | OPRD1 | ChEMBL |
| Nishamalaki | Phyllanthus emblica | Fruit | CID_54670067 | 4986 | OPRK1 | ChEMBL |
| Nishamalaki | Phyllanthus emblica | Fruit | CID_54670067 | 4988 | OPRM1 | ChEMBL |
| Nishamalaki | Phyllanthus emblica | Fruit | CID_54670067 | 5033 | P4HA1 | ChEMBL \| NPASS |
| Nishamalaki | Phyllanthus emblica | Fruit | CID_54670067 | 50507 | NOX4 | ChEMBL \| NPASS |
| Nishamalaki | Phyllanthus emblica | Fruit | CID_54670067 | 5139 | PDE3A | ChEMBL |
| Nishamalaki | Phyllanthus emblica | Fruit | CID_54670067 | 5141 | PDE4A | ChEMBL |
| Nishamalaki | Phyllanthus emblica | Fruit | CID_54670067 | 5241 | PGR | ChEMBL |
| Nishamalaki | Phyllanthus emblica | Fruit | CID_54670067 | 54583 | EGLN1 | ChEMBL \| NPASS |
| Nishamalaki | Phyllanthus emblica | Fruit | CID_54670067 | 552 | AVPR1A | ChEMBL |
| Nishamalaki | Phyllanthus emblica | Fruit | CID_54670067 | 5530 | PPP3CA | ChEMBL |
| Nishamalaki | Phyllanthus emblica | Fruit | CID_54670067 | 55662 | HIF1AN | ChEMBL \| NPASS |
| Nishamalaki | Phyllanthus emblica | Fruit | CID_54670067 | 5578 | PRKCA | ChEMBL |
| Nishamalaki | Phyllanthus emblica | Fruit | CID_54670067 | 5594 | MAPK1 | ChEMBL |
| Nishamalaki | Phyllanthus emblica | Fruit | CID_54670067 | 5595 | MAPK3 | ChEMBL |
| Nishamalaki | Phyllanthus emblica | Fruit | CID_54670067 | 5724 | PTAFR | ChEMBL |
| Nishamalaki | Phyllanthus emblica | Fruit | CID_54670067 | 5742 | PTGS1 | ChEMBL |
| Nishamalaki | Phyllanthus emblica | Fruit | CID_54670067 | 5743 | PTGS2 | ChEMBL |
| Nishamalaki | Phyllanthus emblica | Fruit | CID_54670067 | 5788 | PTPRC | ChEMBL |
| Nishamalaki | Phyllanthus emblica | Fruit | CID_54670067 | 624 | BDKRB2 | ChEMBL |
| Nishamalaki | Phyllanthus emblica | Fruit | CID_54670067 | 6530 | SLC6A2 | ChEMBL |

**Table S3: Associations between formulations, constituent plants, phytochemicals, and human molecular targets. The table lists the formulations, plant name and plant part, phytochemical identifiers, associated human target genes (Entrez Gene ID and gene symbol), and the database sources supporting each phytochemical-target interaction.**

| Formulation | Plant name | Part | PubChem CID | Entrez Gene ID | Gene Symbol | Source database(s) |
|---|---|---|---|---|---|---|
| Nishamalaki | Phyllanthus emblica | Fruit | CID_54670067 | 6531 | SLC6A3 | ChEMBL |
| Nishamalaki | Phyllanthus emblica | Fruit | CID_54670067 | 6532 | SLC6A4 | ChEMBL |
| Nishamalaki | Phyllanthus emblica | Fruit | CID_54670067 | 6865 | TACR2 | ChEMBL |
| Nishamalaki | Phyllanthus emblica | Fruit | CID_54670067 | 6869 | TACR1 | ChEMBL |
| Nishamalaki | Phyllanthus emblica | Fruit | CID_54670067 | 6915 | TBXA2R | ChEMBL |
| Nishamalaki | Phyllanthus emblica | Fruit | CID_54670067 | 6916 | TBXAS1 | ChEMBL |
| Nishamalaki | Phyllanthus emblica | Fruit | CID_54670067 | 729230 | CCR2 | ChEMBL |
| Nishamalaki | Phyllanthus emblica | Fruit | CID_54670067 | 7299 | TYR | ChEMBL \| NPASS |
| Nishamalaki | Phyllanthus emblica | Fruit | CID_54670067 | 7433 | VIPR1 | ChEMBL |
| Nishamalaki | Phyllanthus emblica | Fruit | CID_54670067 | 760 | CA2 | ChEMBL |
| Nishamalaki | Phyllanthus emblica | Fruit | CID_54670067 | 799 | CALCR | ChEMBL |
| Nishamalaki | Phyllanthus emblica | Fruit | CID_54670067 | 834 | CASP1 | ChEMBL |
| Nishamalaki | Phyllanthus emblica | Fruit | CID_54670067 | 8654 | PDE5A | ChEMBL |
| Nishamalaki | Phyllanthus emblica | Fruit | CID_54670067 | 886 | CCKAR | ChEMBL |
| Nishamalaki | Phyllanthus emblica | Fruit | CID_7428 | 10919 | EHMT2 | ChEMBL \| NPASS |
| Nishamalaki | Phyllanthus emblica | Fruit | CID_7428 | 10951 | CBX1 | NPASS |
| Nishamalaki | Phyllanthus emblica | Fruit | CID_7428 | 118429 | ANTXR2 | NPASS |
| Nishamalaki | Phyllanthus emblica | Fruit | CID_7428 | 2237 | FEN1 | NPASS |
| Nishamalaki | Phyllanthus emblica | Fruit | CID_7428 | 2529 | FUT7 | ChEMBL \| NPASS |
| Nishamalaki | Phyllanthus emblica | Fruit | CID_7428 | 2548 | GAA | NPASS |
| Nishamalaki | Phyllanthus emblica | Fruit | CID_7428 | 26013 | L3MBTL1 | NPASS |
| Nishamalaki | Phyllanthus emblica | Fruit | CID_7428 | 2671 | GFER | ChEMBL \| NPASS |
| Nishamalaki | Phyllanthus emblica | Fruit | CID_7428 | 28689 | TRAV4 | BindingDB |
| Nishamalaki | Phyllanthus emblica | Fruit | CID_7428 | 29994 | BAZ2B | ChEMBL \| NPASS |
| Nishamalaki | Phyllanthus emblica | Fruit | CID_7428 | 328 | APEX1 | ChEMBL \| NPASS |
| Nishamalaki | Phyllanthus emblica | Fruit | CID_7428 | 390245 | KDM4E | ChEMBL \| NPASS |
| Nishamalaki | Phyllanthus emblica | Fruit | CID_7428 | 4000 | LMNA | NPASS |
| Nishamalaki | Phyllanthus emblica | Fruit | CID_7428 | 4790 | NFKB1 | ChEMBL |
| Nishamalaki | Phyllanthus emblica | Fruit | CID_7428 | 4791 | NFKB2 | ChEMBL |
| Nishamalaki | Phyllanthus emblica | Fruit | CID_7428 | 5300 | PIN1 | NPASS |
| Nishamalaki | Phyllanthus emblica | Fruit | CID_7428 | 5423 | POLB | NPASS |
| Nishamalaki | Phyllanthus emblica | Fruit | CID_7428 | 55775 | TDP1 | NPASS |
| Nishamalaki | Phyllanthus emblica | Fruit | CID_7428 | 5970 | RELA | ChEMBL |
| Nishamalaki | Phyllanthus emblica | Fruit | CID_7428 | 7498 | XDH | BindingDB \| ChEMBL \| NPASS |
| Nishamalaki | Phyllanthus emblica | Fruit | CID_7428 | 9682 | KDM4A | ChEMBL \| NPASS |

| Table S4: Curated gene set associated with Type 2 diabetes mellitus. Genes were compiled from multiple resources. The table includes genes that are either known targets of FDA-approved antidiabetic drugs or supported by at least two of the four considered databases. For each gene, the Entrez gene ID, gene symbol, evidence source(s), and approved drug information (where applicable) are reported. | | | | |
|---|---|---|---|---|
| **Entrez Gene ID** | **Gene symbol** | **Approved drug target** | **Drug names** | **Evidence source** |
| 2740 | GLP1R | Yes | Pramlintide \| Semaglutide \| Tirzepatide | DrugBank (TTD) \| DisGeNET \| DISEASES \| GeneCards \| CTD |
| 3767 | KCNJ11 | Yes | Glimepiride \| Glipizide \| Gliquidone \| Glyburide \| Nateglinide \| Repaglinide \| Tolazamide | DrugBank (Drug Central \| Santos et al., Drug Central) \| DisGeNET \| DISEASES \| GeneCards \| CTD |
| 5468 | PPARG | Yes | Pioglitazone | DrugBank (Santos et al., Drug Central \| TTD) \| DisGeNET \| DISEASES \| GeneCards \| CTD |
| 1803 | DPP4 | Yes | Alogliptin \| Linagliptin \| Saxagliptin \| Sitagliptin | DrugBank (Drug Central \| Santos et al., Drug Central \| TTD) \| DisGeNET \| DISEASES \| GeneCards |
| 3643 | INSR | Yes | Metformin | DrugBank (TTD) \| DisGeNET \| DISEASES \| GeneCards |
| 6833 | ABCC8 | Yes | Glimepiride \| Glipizide \| Gliquidone \| Glyburide \| Nateglinide \| Repaglinide \| Tolazamide | DrugBank (Drug Central \| Santos et al., Drug Central) \| DisGeNET \| GeneCards \| CTD |
| 4535 | MT-ND1\|ND1 | Yes | Metformin | DrugBank (Santos et al., Drug Central) \| DisGeNET \| GeneCards |
| 6524 | SLC5A2 | Yes | Bexagliflozin \| Canagliflozin \| Dapagliflozin \| Empagliflozin \| Ertugliflozin \| Sotagliflozin | DrugBank (Drug Central \| Santos et al., Drug Central \| TTD) \| DISEASES \| GeneCards |
| 6523 | SLC5A1 | Yes | Sotagliflozin | DrugBank (Drug Central) \| DISEASES |
| 8972 | MGAM | Yes | Miglitol | DrugBank (Santos et al., Drug Central) \| DISEASES |
| 10060 | ABCC9 | Yes | Glimepiride | DrugBank (Santos et al., Drug Central) |
| 126328 | NDUFA11 | Yes | Metformin | DrugBank (Santos et al., Drug Central) |
| 2548 | GAA | Yes | Miglitol | DrugBank (Santos et al., Drug Central) |
| 25915 | NDUFAF3 | Yes | Metformin | DrugBank (Santos et al., Drug Central) |
| 29078 | NDUFAF4 | Yes | Metformin | DrugBank (Santos et al., Drug Central) |
| 32 | ACACB | Yes | Metformin | DrugBank (TTD) |
| 374291 | NDUFS7 | Yes | Metformin | DrugBank (Santos et al., Drug Central) |
| 4536 | MT-ND2 | Yes | Metformin | DrugBank (Santos et al., Drug Central) |
| 4537 | MT-ND3 | Yes | Metformin | DrugBank (Santos et al., Drug Central) |
| 4538 | MT-ND4 | Yes | Metformin | DrugBank (Santos et al., Drug Central) |
| 4539 | MT-ND4L | Yes | Metformin | DrugBank (Santos et al., Drug Central) |
| 4540 | MT-ND5 | Yes | Metformin | DrugBank (Santos et al., Drug Central) |
| 4541 | MT-ND6 | Yes | Metformin | DrugBank (Santos et al., Drug Central) |
| 4694 | NDUFA1 | Yes | Metformin | DrugBank (Santos et al., Drug Central) |
| 4695 | NDUFA2 | Yes | Metformin | DrugBank (Santos et al., Drug Central) |
| 4696 | NDUFA3 | Yes | Metformin | DrugBank (Santos et al., Drug Central) |
| 4697 | COXFA4 | Yes | Metformin | DrugBank (Santos et al., Drug Central) |
| 4698 | NDUFA5 | Yes | Metformin | DrugBank (Santos et al., Drug Central) |
| 4700 | NDUFA6 | Yes | Metformin | DrugBank (Santos et al., Drug Central) |
| 4701 | NDUFA7 | Yes | Metformin | DrugBank (Santos et al., Drug Central) |
| 4702 | NDUFA8 | Yes | Metformin | DrugBank (Santos et al., Drug Central) |
| 4704 | NDUFA9 | Yes | Metformin | DrugBank (Santos et al., Drug Central) |
| 4705 | NDUFA10 | Yes | Metformin | DrugBank (Santos et al., Drug Central) |
| 4706 | NDUFAB1 | Yes | Metformin | DrugBank (Santos et al., Drug Central) |
| 4707 | NDUFB1 | Yes | Metformin | DrugBank (Santos et al., Drug Central) |
| 4708 | NDUFB2 | Yes | Metformin | DrugBank (Santos et al., Drug Central) |
| 4709 | NDUFB3 | Yes | Metformin | DrugBank (Santos et al., Drug Central) |
| 4710 | NDUFB4 | Yes | Metformin | DrugBank (Santos et al., Drug Central) |
| 4711 | NDUFB5 | Yes | Metformin | DrugBank (Santos et al., Drug Central) |
| 4712 | NDUFB6 | Yes | Metformin | DrugBank (Santos et al., Drug Central) |
| 4713 | NDUFB7 | Yes | Metformin | DrugBank (Santos et al., Drug Central) |
| 4714 | NDUFB8 | Yes | Metformin | DrugBank (Santos et al., Drug Central) |
| 4715 | NDUFB9 | Yes | Metformin | DrugBank (Santos et al., Drug Central) |
| 4716 | NDUFB10 | Yes | Metformin | DrugBank (Santos et al., Drug Central) |
| 4717 | NDUFC1 | Yes | Metformin | DrugBank (Santos et al., Drug Central) |
| 4718 | NDUFC2 | Yes | Metformin | DrugBank (Santos et al., Drug Central) |
| 4719 | NDUFS1 | Yes | Metformin | DrugBank (Santos et al., Drug Central) |
| 4720 | NDUFS2 | Yes | Metformin | DrugBank (Santos et al., Drug Central) |
| 4722 | NDUFS3 | Yes | Metformin | DrugBank (Santos et al., Drug Central) |
| 4723 | NDUFV1 | Yes | Metformin | DrugBank (Santos et al., Drug Central) |
| 4724 | NDUFS4 | Yes | Metformin | DrugBank (Santos et al., Drug Central) |
| 4725 | NDUFS5 | Yes | Metformin | DrugBank (Santos et al., Drug Central) |
| 4726 | NDUFS6 | Yes | Metformin | DrugBank (Santos et al., Drug Central) |
| 4728 | NDUFS8 | Yes | Metformin | DrugBank (Santos et al., Drug Central) |
| 4729 | NDUFV2 | Yes | Metformin | DrugBank (Santos et al., Drug Central) |
| 4731 | NDUFV3 | Yes | Metformin | DrugBank (Santos et al., Drug Central) |
| 51079 | NDUFA13 | Yes | Metformin | DrugBank (Santos et al., Drug Central) |
| 51103 | NDUFAF1 | Yes | Metformin | DrugBank (Santos et al., Drug Central) |
| 54539 | NDUFB11 | Yes | Metformin | DrugBank (Santos et al., Drug Central) |
| 55244 | SLC47A1 | Yes | Metformin | DrugBank (TTD) |
| 55967 | NDUFA12 | Yes | Metformin | DrugBank (Santos et al., Drug Central) |
| 56901 | COXFA4L2 | Yes | Metformin | DrugBank (Santos et al., Drug Central) |
| 91942 | NDUFAF2 | Yes | Metformin | DrugBank (Santos et al., Drug Central) |
| 10891 | PPARGC1A | No | - | DisGeNET \| DISEASES \| GeneCards \| CTD |
| 169026 | SLC30A8 | No | - | DisGeNET \| DISEASES \| GeneCards \| CTD |
| 23411 | SIRT1 | No | - | DisGeNET \| DISEASES \| GeneCards \| CTD |
| 26291 | FGF21 | No | - | DisGeNET \| DISEASES \| GeneCards \| CTD |
| 2641 | GCG | No | - | DisGeNET \| DISEASES \| GeneCards \| CTD |
| 2645 | GCK | No | - | DisGeNET \| DISEASES \| GeneCards \| CTD |

| Table S4: Curated gene set associated with Type 2 diabetes mellitus. Genes were compiled from multiple resources. The table includes genes that are either known targets of FDA-approved antidiabetic drugs or supported by at least two of the four considered databases. For each gene, the Entrez gene ID, gene symbol, evidence source(s), and approved drug information (where applicable) are reported. | | | | |
|---|---|---|---|---|
| **Entrez Gene ID** | **Gene symbol** | **Approved drug target** | **Drug names** | **Evidence source** |
| 3375 | IAPP | No | - | DisGeNET \| DISEASES \| GeneCards \| CTD |
| 3383 | ICAM1 | No | - | DisGeNET \| DISEASES \| GeneCards \| CTD |
| 3569 | IL6 | No | - | DisGeNET \| DISEASES \| GeneCards \| CTD |
| 3630 | INS | No | - | DisGeNET \| DISEASES \| GeneCards \| CTD |
| 3667 | IRS1 | No | - | DisGeNET \| DISEASES \| GeneCards \| CTD |
| 3952 | LEP | No | - | DisGeNET \| DISEASES \| GeneCards \| CTD |
| 3953 | LEPR | No | - | DisGeNET \| DISEASES \| GeneCards \| CTD |
| 4846 | NOS3 | No | - | DisGeNET \| DISEASES \| GeneCards \| CTD |
| 5465 | PPARA | No | - | DisGeNET \| DISEASES \| GeneCards \| CTD |
| 54901 | CDKAL1 | No | - | DisGeNET \| DISEASES \| GeneCards \| CTD |
| 56729 | RETN | No | - | DisGeNET \| DISEASES \| GeneCards \| CTD |
| 5770 | PTPN1 | No | - | DisGeNET \| DISEASES \| GeneCards \| CTD |
| 6462 | SHBG | No | - | DisGeNET \| DISEASES \| GeneCards \| CTD |
| 6514 | SLC2A2 | No | - | DisGeNET \| DISEASES \| GeneCards \| CTD |
| 6517 | SLC2A4 | No | - | DisGeNET \| DISEASES \| GeneCards \| CTD |
| 6934 | TCF7L2 | No | - | DisGeNET \| DISEASES \| GeneCards \| CTD |
| 7124 | TNF | No | - | DisGeNET \| DISEASES \| GeneCards \| CTD |
| 7351 | UCP2 | No | - | DisGeNET \| DISEASES \| GeneCards \| CTD |
| 79068 | FTO | No | - | DisGeNET \| DISEASES \| GeneCards \| CTD |
| 8660 | IRS2 | No | - | DisGeNET \| DISEASES \| GeneCards \| CTD |
| 9370 | ADIPOQ | No | - | DisGeNET \| DISEASES \| GeneCards \| CTD |
| 948 | CD36 | No | - | DisGeNET \| DISEASES \| GeneCards \| CTD |
| 10644 | IGF2BP2 | No | - | DisGeNET \| GeneCards \| CTD |
| 11132 | CAPN10 | No | - | DisGeNET \| GeneCards \| CTD |
| 1636 | ACE | No | - | DisGeNET \| DISEASES \| GeneCards |
| 1906 | EDN1 | No | - | DisGeNET \| DISEASES \| CTD |
| 207 | AKT1 | No | - | DisGeNET \| DISEASES \| CTD |
| 208 | AKT2 | No | - | DisGeNET \| GeneCards \| CTD |
| 221895 | JAZF1 | No | - | DisGeNET \| GeneCards \| CTD |
| 2646 | GCKR | No | - | DisGeNET \| GeneCards \| CTD |
| 2820 | GPD2 | No | - | DisGeNET \| GeneCards \| CTD |
| 3087 | HHEX | No | - | DisGeNET \| GeneCards \| CTD |
| 3159 | HMGA1 | No | - | DisGeNET \| GeneCards \| CTD |
| 3172 | HNF4A | No | - | DisGeNET \| GeneCards \| CTD |
| 338 | APOB | No | - | DisGeNET \| DISEASES \| GeneCards |
| 3479 | IGF1 | No | - | DisGeNET \| DISEASES \| GeneCards |
| 3553 | IL1B | No | - | DisGeNET \| DISEASES \| GeneCards |
| 3586 | IL10 | No | - | DisGeNET \| DISEASES \| GeneCards |
| 3651 | PDX1 | No | - | DisGeNET \| GeneCards \| CTD |
| 3784 | KCNQ1 | No | - | DisGeNET \| GeneCards \| CTD |
| 3990 | LIPC | No | - | DisGeNET \| GeneCards \| CTD |
| 4023 | LPL | No | - | DisGeNET \| DISEASES \| GeneCards |
| 4544 | MTNR1B | No | - | DisGeNET \| GeneCards \| CTD |
| 4760 | NEUROD1 | No | - | DisGeNET \| GeneCards \| CTD |
| 4790 | NFKB1 | No | - | DisGeNET \| DISEASES \| CTD |
| 4853 | NOTCH2 | No | - | DisGeNET \| GeneCards \| CTD |
| 5078 | PAX4 | No | - | DisGeNET \| GeneCards \| CTD |
| 5167 | ENPP1 | No | - | DisGeNET \| GeneCards \| CTD |
| 51738 | GHRL | No | - | DisGeNET \| DISEASES \| GeneCards |
| 5294 | PIK3CG | No | - | DisGeNET \| DISEASES \| CTD |
| 5506 | PPP1R3A | No | - | DisGeNET \| GeneCards \| CTD |
| 55600 | ITLN1 | No | - | DisGeNET \| GeneCards \| CTD |
| 56999 | ADAMTS9 | No | - | DisGeNET \| GeneCards \| CTD |
| 5950 | RBP4 | No | - | DisGeNET \| DISEASES \| GeneCards |
| 63892 | THADA | No | - | DisGeNET \| GeneCards \| CTD |
| 6648 | SOD2\|SOD2-2 | No | - | DisGeNET \| DISEASES \| CTD |
| 6720 | SREBF1 | No | - | DisGeNET \| DISEASES \| GeneCards |
| 6774 | STAT3 | No | - | DisGeNET \| DISEASES \| GeneCards |
| 6927 | HNF1A | No | - | DisGeNET \| GeneCards \| CTD |
| 6928 | HNF1B | No | - | DisGeNET \| GeneCards \| CTD |
| 7099 | TLR4 | No | - | DisGeNET \| DISEASES \| GeneCards |
| 7350 | UCP1 | No | - | DisGeNET \| DISEASES \| GeneCards |
| 7466 | WFS1 | No | - | DisGeNET \| GeneCards \| CTD |
| 79602 | ADIPOR2 | No | - | DisGeNET \| DISEASES \| GeneCards |
| 836 | CASP3 | No | - | DisGeNET \| DISEASES \| CTD |
| 847 | CAT | No | - | DisGeNET \| GeneCards \| CTD |
| 9479 | MAPK8IP1 | No | - | DisGeNET \| GeneCards \| CTD |
| 10018 | BCL2L11 | No | - | DisGeNET \| CTD |
| 10239 | AP3S2 | No | - | DisGeNET \| CTD |
| 10296 | MAEA | No | - | DisGeNET \| CTD |

**Table S4: Curated gene set associated with Type 2 diabetes mellitus. Genes were compiled from multiple resources. The table includes genes that are either known targets of FDA-approved antidiabetic drugs or supported by at least two of the four considered databases. For each gene, the Entrez gene ID, gene symbol, evidence source(s), and approved drug information (where applicable) are reported.**

| Entrez Gene ID | Gene symbol | Approved drug target | Drug names | Evidence source |
|---|---|---|---|---|
| 1036 | CDO1 | No | - | DisGeNET \| CTD |
| 10363 | HMG20A | No | - | DisGeNET \| CTD |
| 111 | ADCY5 | No | - | DisGeNET \| CTD |
| 11317 | RBPJL | No | - | DisGeNET \| GeneCards |
| 116150 | NUS1 | No | - | DisGeNET \| CTD |
| 129787 | TMEM18 | No | - | DisGeNET \| CTD |
| 132332 | SMIM43 | No | - | DisGeNET \| CTD |
| 136259 | KLF14 | No | - | DisGeNET \| CTD |
| 1363 | CPE | No | - | DisGeNET \| GeneCards |
| 1374 | CPT1A | No | - | DisGeNET \| CTD |
| 1401 | CRP | No | - | DISEASES \| GeneCards |
| 145741 | C2CD4A | No | - | DisGeNET \| CTD |
| 150 | ADRA2A | No | - | DisGeNET \| GeneCards |
| 1535 | CYBA | No | - | DisGeNET \| CTD |
| 1544 | CYP1A2 | No | - | DisGeNET \| CTD |
| 155 | ADRB3 | No | - | DisGeNET \| GeneCards |
| 157855 | KCNU1 | No | - | DisGeNET \| CTD |
| 169792 | GLIS3 | No | - | DisGeNET \| CTD |
| 177 | AGER | No | - | DisGeNET \| GeneCards |
| 181 | AGRP | No | - | DisGeNET \| DISEASES |
| 1889 | ECE1 | No | - | DisGeNET \| CTD |
| 19 | ABCA1 | No | - | DisGeNET \| GeneCards |
| 1909 | EDNRA | No | - | DisGeNET \| CTD |
| 1910 | EDNRB | No | - | DisGeNET \| CTD |
| 1956 | EGFR | No | - | DisGeNET \| CTD |
| 197 | AHSG | No | - | DisGeNET \| GeneCards |
| 2113 | ETS1 | No | - | DisGeNET \| CTD |
| 213 | ALB | No | - | DISEASES \| GeneCards |
| 2167 | FABP4 | No | - | DISEASES \| GeneCards |
| 22808 | MRAS | No | - | DisGeNET \| CTD |
| 22866 | CNKSR2 | No | - | DisGeNET \| CTD |
| 23338 | JADE2 | No | - | DisGeNET \| CTD |
| 2475 | MTOR | No | - | DisGeNET \| DISEASES |
| 252995 | FNDC5 | No | - | DISEASES \| GeneCards |
| 26053 | AUTS2 | No | - | DisGeNET \| CTD |
| 26122 | EPC2 | No | - | DisGeNET \| CTD |
| 2642 | GCGR | No | - | DisGeNET \| GeneCards |
| 2695 | GIP | No | - | DisGeNET \| DISEASES |
| 2729 | GCLC | No | - | DisGeNET \| CTD |
| 2730 | GCLM | No | - | DisGeNET \| CTD |
| 2784 | GNB3 | No | - | DisGeNET \| CTD |
| 2813 | GP2 | No | - | DisGeNET \| CTD |
| 283450 | HECTD4 | No | - | DisGeNET \| CTD |
| 283455 | KSR2 | No | - | DisGeNET \| CTD |
| 2864 | FFAR1 | No | - | DisGeNET \| GeneCards |
| 2876 | GPX1 | No | - | DisGeNET \| CTD |
| 2878 | GPX3 | No | - | DisGeNET \| DISEASES |
| 2888 | GRB14 | No | - | DisGeNET \| CTD |
| 2944 | GSTM1 | No | - | DisGeNET \| CTD |
| 3039 | HBA1 | No | - | DisGeNET \| CTD |
| 3077 | HFE | No | - | DisGeNET \| GeneCards |
| 3098 | HK1 | No | - | DisGeNET \| CTD |
| 3156 | HMGCR | No | - | DisGeNET \| DISEASES |
| 3162 | HMOX1 | No | - | DisGeNET \| CTD |
| 3170 | FOXA2 | No | - | DisGeNET \| GeneCards |
| 3240 | HP | No | - | DisGeNET \| CTD |
| 3263 | HPX | No | - | DisGeNET \| CTD |
| 335 | APOA1 | No | - | DISEASES \| GeneCards |
| 3397 | ID1 | No | - | DisGeNET \| CTD |
| 3416 | IDE | No | - | DisGeNET \| GeneCards |
| 345 | APOC3 | No | - | DISEASES \| GeneCards |
| 3458 | IFNG | No | - | DisGeNET \| DISEASES |
| 348 | APOE | No | - | DISEASES \| GeneCards |
| 3516 | RBPJ | No | - | DisGeNET \| GeneCards |
| 355 | FAS | No | - | DisGeNET \| CTD |
| 3597 | IL13RA1 | No | - | DisGeNET \| CTD |
| 3636 | INPPL1 | No | - | DisGeNET \| CTD |
| 3672 | ITGA1 | No | - | DisGeNET \| CTD |
| 388125 | C2CD4B | No | - | DisGeNET \| CTD |
| 4000 | LMNA | No | - | DisGeNET \| GeneCards |

**Table S4: Curated gene set associated with Type 2 diabetes mellitus. Genes were compiled from multiple resources. The table includes genes that are either known targets of FDA-approved antidiabetic drugs or supported by at least two of the four considered databases. For each gene, the Entrez gene ID, gene symbol, evidence source(s), and approved drug information (where applicable) are reported.**

| Entrez Gene ID | Gene symbol | Approved drug target | Drug names | Evidence source |
|---|---|---|---|---|
| 4090 | SMAD5 | No | - | DisGeNET \| CTD |
| 4143 | MAT1A | No | - | DisGeNET \| CTD |
| 4160 | MC4R | No | - | DISEASES \| GeneCards |
| 4524 | MTHFR | No | - | DisGeNET \| GeneCards |
| 467 | ATF3 | No | - | DisGeNET \| CTD |
| 4773 | NFATC2 | No | - | DisGeNET \| CTD |
| 4825 | NKX6-1 | No | - | DisGeNET \| CTD |
| 4843 | NOS2 | No | - | DisGeNET \| CTD |
| 4852 | NPY | No | - | DisGeNET \| DISEASES |
| 488 | ATP2A2 | No | - | DisGeNET \| CTD |
| 489 | ATP2A3 | No | - | DisGeNET \| CTD |
| 493856 | CISD2 | No | - | DisGeNET \| CTD |
| 4968 | OGG1 | No | - | DisGeNET \| CTD |
| 5054 | SERPINE1 | No | - | DISEASES \| GeneCards |
| 5066 | PAM | No | - | DisGeNET \| CTD |
| 50674 | NEUROG3 | No | - | DisGeNET \| GeneCards |
| 5080 | PAX6 | No | - | DisGeNET \| CTD |
| 51094 | ADIPOR1 | No | - | DISEASES \| GeneCards |
| 5126 | PCSK2 | No | - | DisGeNET \| CTD |
| 51530 | ZC3HC1 | No | - | DisGeNET \| CTD |
| 5184 | PEPD | No | - | DisGeNET \| CTD |
| 5346 | PLIN1 | No | - | DISEASES \| GeneCards |
| 5444 | PON1 | No | - | DisGeNET \| GeneCards |
| 5467 | PPARD | No | - | DisGeNET \| DISEASES |
| 5563 | PRKAA2 | No | - | DisGeNET \| GeneCards |
| 5579 | PRKCB | No | - | DisGeNET \| CTD |
| 5629 | PROX1 | No | - | DisGeNET \| CTD |
| 57818 | G6PC2 | No | - | DISEASES \| GeneCards |
| 581 | BAX | No | - | DisGeNET \| CTD |
| 59338 | PLEKHA1 | No | - | DisGeNET \| CTD |
| 596 | BCL2 | No | - | DisGeNET \| CTD |
| 598 | BCL2L1 | No | - | DisGeNET \| CTD |
| 6049 | RNF6 | No | - | DisGeNET \| CTD |
| 60685 | ZFAND3 | No | - | DisGeNET \| CTD |
| 6277 | S100A6 | No | - | DisGeNET \| CTD |
| 6344 | SCTR | No | - | DisGeNET \| CTD |
| 6347 | CCL2 | No | - | DISEASES \| GeneCards |
| 635 | BHMT | No | - | DisGeNET \| CTD |
| 6424 | SFRP4 | No | - | DisGeNET \| CTD |
| 6480 | ST6GAL1 | No | - | DisGeNET \| CTD |
| 64897 | C12orf43 | No | - | DisGeNET \| GeneCards |
| 6506 | SLC1A2 | No | - | DisGeNET \| CTD |
| 6513 | SLC2A1 | No | - | DisGeNET \| CTD |
| 6581 | SLC22A3 | No | - | DisGeNET \| CTD |
| 6616 | SNAP25 | No | - | DisGeNET \| CTD |
| 6647 | SOD1 | No | - | DisGeNET \| CTD |
| 673 | BRAF | No | - | DisGeNET \| CTD |
| 7040 | TGFB1 | No | - | DISEASES \| GeneCards |
| 7076 | TIMP1 | No | - | DisGeNET \| CTD |
| 7132 | TNFRSF1A | No | - | DisGeNET \| CTD |
| 7133 | TNFRSF1B | No | - | DisGeNET \| CTD |
| 718 | C3 | No | - | DisGeNET \| CTD |
| 723961 | INS-IGF2 | No | - | DisGeNET \| GeneCards |
| 7325 | UBE2E2 | No | - | DisGeNET \| CTD |
| 7352 | UCP3 | No | - | DisGeNET \| GeneCards |
| 7486 | WRN | No | - | DisGeNET \| GeneCards |
| 80212 | CCDC92 | No | - | DisGeNET \| CTD |
| 80339 | PNPLA3 | No | - | DisGeNET \| DISEASES |
| 80790 | CMIP | No | - | DisGeNET \| CTD |
| 83795 | KCNK16 | No | - | DisGeNET \| CTD |
| 841 | CASP8 | No | - | DisGeNET \| CTD |
| 84196 | USP48 | No | - | DisGeNET \| CTD |
| 8462 | KLF11 | No | - | DisGeNET \| GeneCards |
| 8527 | DGKD | No | - | DisGeNET \| CTD |
| 875 | CBS | No | - | DisGeNET \| CTD |
| 894 | CCND2 | No | - | DisGeNET \| CTD |
| 9365 | KL | No | - | DisGeNET \| CTD |
| 9559 | VPS26A | No | - | DisGeNET \| CTD |
| 9861 | PSMD6 | No | - | DisGeNET \| CTD |
| 9882 | TBC1D4 | No | - | DISEASES \| GeneCards |

**Table S5: Curated gene set associated with fatty liver disease. Genes were compiled from multiple resources. The table includes genes that are either known targets of FDA-approved drugs for fattty liver disease or supported by at least two of the four considered databases. For each gene, the Entrez gene ID, gene symbol, evidence source(s), and approved drug information (where applicable) are reported.**

| Entrez Gene ID | Gene symbol | Approved drug target | Drug Names | Evidence source |
|---|---|---|---|---|
| 3156 | HMGCR | Yes | Pravastatin \| Lovastatin \| Simvastatin \| Atorvastatin \| Pitavastatin | DrugBank (Santos et al., Drug Central \| TTD) |
| 8694 | DGAT1 | Yes | Icosapent ethyl | DrugBank (Santos et al., Drug Central) \| DISEASES |
| 5465 | PPARA | Yes | Fenofibrate \| Fenofibric acid | DrugBank (Santos et al., Drug Central \| TTD) \| DisGeNET \| DISEASES \| GeneCards \| CTD |
| 4615 | MYD88 | Yes | Atorvastatin | DrugBank (TTD) |
| 123 | PLIN2 | No | - | DisGeNET \| CTD |
| 127 | ADH4 | No | - | DisGeNET \| CTD |
| 191 | AHCY | No | - | DisGeNET \| CTD |
| 196 | AHR | No | - | DisGeNET \| CTD |
| 207 | AKT1 | No | - | DisGeNET \| DISEASES \| CTD |
| 216 | ALDH1A1 | No | - | DisGeNET \| CTD |
| 217 | ALDH2 | No | - | DisGeNET \| CTD |
| 219 | ALDH1B1 | No | - | DisGeNET \| CTD |
| 330 | BIRC3 | No | - | DisGeNET \| CTD |
| 338 | APOB | No | - | DisGeNET \| DISEASES \| GeneCards \| CTD |
| 348 | APOE | No | - | DisGeNET \| CTD |
| 350 | APOH | No | - | DisGeNET \| CTD |
| 355 | FAS | No | - | DisGeNET \| CTD |
| 412 | STS | No | - | DisGeNET \| CTD |
| 468 | ATF4 | No | - | DisGeNET \| CTD |
| 506 | ATP5F1B | No | - | DisGeNET \| CTD |
| 540 | ATP7B | No | - | DisGeNET \| CTD |
| 714 | C1QC | No | - | DisGeNET \| CTD |
| 847 | CAT | No | - | DisGeNET \| CTD |
| 866 | SERPINA6 | No | - | DisGeNET \| CTD |
| 929 | CD14 | No | - | DisGeNET \| CTD |
| 949 | SCARB1 | No | - | DisGeNET \| DISEASES \| CTD |
| 1051 | CEBPB | No | - | DisGeNET \| CTD |
| 1244 | ABCC2 | No | - | DisGeNET \| CTD |
| 1281 | COL3A1 | No | - | DisGeNET \| CTD |
| 1374 | CPT1A | No | - | DisGeNET \| DISEASES \| CTD |
| 1437 | CSF2 | No | - | DisGeNET \| CTD |
| 1544 | CYP1A2 | No | - | DisGeNET \| CTD |
| 1545 | CYP1B1 | No | - | DisGeNET \| CTD |
| 1571 | CYP2E1 | No | - | DisGeNET \| DISEASES \| CTD |
| 1586 | CYP17A1 | No | - | DisGeNET \| CTD |
| 1588 | CYP19A1 | No | - | DisGeNET \| CTD |
| 1636 | ACE | No | - | DisGeNET \| CTD |
| 1649 | DDIT3 | No | - | DisGeNET \| CTD |
| 1728 | NQO1 | No | - | DisGeNET \| CTD |
| 2099 | ESR1 | No | - | DisGeNET \| CTD |
| 2147 | F2 | No | - | DisGeNET \| CTD |
| 2149 | F2R | No | - | DisGeNET \| CTD |
| 2350 | FOLR2 | No | - | DisGeNET \| CTD |
| 2494 | NR5A2 | No | - | DisGeNET \| CTD |
| 2875 | GPT | No | - | DisGeNET \| DISEASES \| GeneCards \| CTD |
| 2876 | GPX1 | No | - | DisGeNET \| CTD |
| 2879 | GPX4 | No | - | DisGeNET \| CTD |
| 2938 | GSTA1 | No | - | DisGeNET \| CTD |
| 2944 | GSTM1 | No | - | DisGeNET \| CTD |
| 2950 | GSTP1 | No | - | DisGeNET \| CTD |
| 2952 | GSTT1 | No | - | DisGeNET \| CTD |
| 3032 | HADHB | No | - | DisGeNET \| CTD |
| 3038 | HAS3 | No | - | DisGeNET \| CTD |
| 3077 | HFE | No | - | DisGeNET \| GeneCards |
| 3087 | HHEX | No | - | DisGeNET \| CTD |
| 3169 | FOXA1 | No | - | DisGeNET \| CTD |
| 3552 | IL1A | No | - | DisGeNET \| CTD |
| 3553 | IL1B | No | - | DisGeNET \| DISEASES \| CTD |
| 3557 | IL1RN | No | - | DisGeNET \| CTD |
| 3562 | IL3 | No | - | DisGeNET \| CTD |
| 3565 | IL4 | No | - | DisGeNET \| CTD |
| 3569 | IL6 | No | - | DisGeNET \| DISEASES \| GeneCards |
| 3630 | INS | No | - | DisGeNET \| DISEASES \| GeneCards \| CTD |
| 3717 | JAK2 | No | - | DisGeNET \| CTD |
| 3949 | LDLR | No | - | DisGeNET \| CTD |
| 3952 | LEP | No | - | DisGeNET \| DISEASES \| GeneCards \| CTD |
| 3953 | LEPR | No | - | DisGeNET \| CTD |
| 3976 | LIF | No | - | DisGeNET \| CTD |
| 4143 | MAT1A | No | - | DisGeNET \| CTD |
| 4233 | MET | No | - | DisGeNET \| CTD |

**Table S5: Curated gene set associated with fatty liver disease. Genes were compiled from multiple resources. The table includes genes that are either known targets of FDA-approved drugs for fattty liver disease or supported by at least two of the four considered databases. For each gene, the Entrez gene ID, gene symbol, evidence source(s), and approved drug information (where applicable) are reported.**

| Entrez Gene ID | Gene symbol | Approved drug target | Drug Names | Evidence source |
|---|---|---|---|---|
| 4312 | MMP1 | No | - | DisGeNET \| CTD |
| 4547 | MTTP | No | - | DisGeNET \| DISEASES \| CTD |
| 4609 | MYC | No | - | DisGeNET \| CTD |
| 4780 | NFE2L2 | No | - | DisGeNET \| CTD |
| 5054 | SERPINE1 | No | - | DisGeNET \| CTD |
| 5055 | SERPINB2 | No | - | DisGeNET \| CTD |
| 5166 | PDK4 | No | - | DisGeNET \| CTD |
| 5244 | ABCB4 | No | - | DisGeNET \| CTD |
| 5265 | SERPINA1 | No | - | DisGeNET \| CTD |
| 5443 | POMC | No | - | DisGeNET \| CTD |
| 5444 | PON1 | No | - | DisGeNET \| CTD |
| 5467 | PPARD | No | - | DisGeNET \| CTD |
| 5468 | PPARG | No | - | DisGeNET \| DISEASES \| GeneCards \| CTD |
| 5551 | PRF1 | No | - | DisGeNET \| CTD |
| 5566 | PRKACA | No | - | DisGeNET \| CTD |
| 5578 | PRKCA | No | - | DisGeNET \| CTD |
| 5580 | PRKCD | No | - | DisGeNET \| CTD |
| 5581 | PRKCE | No | - | DisGeNET \| CTD |
| 5686 | PSMA5 | No | - | DisGeNET \| CTD |
| 5728 | PTEN | No | - | DisGeNET \| CTD |
| 5757 | PTMA | No | - | DisGeNET \| CTD |
| 5897 | RAG2 | No | - | DisGeNET \| CTD |
| 5962 | RDX | No | - | DisGeNET \| CTD |
| 6319 | SCD | No | - | DisGeNET \| DISEASES |
| 6347 | CCL2 | No | - | DisGeNET \| DISEASES \| CTD |
| 6514 | SLC2A2 | No | - | DisGeNET \| CTD |
| 6647 | SOD1 | No | - | DisGeNET \| CTD |
| 6648 | SOD2 | No | - | DisGeNET \| CTD |
| 6720 | SREBF1 | No | - | DisGeNET \| DISEASES \| GeneCards \| CTD |
| 7018 | TF | No | - | DisGeNET \| CTD |
| 7040 | TGFB1 | No | - | DisGeNET \| DISEASES \| CTD |
| 7099 | TLR4 | No | - | DisGeNET \| DISEASES \| CTD |
| 7124 | TNF | No | - | DisGeNET \| DISEASES \| GeneCards \| CTD |
| 7133 | TNFRSF1B | No | - | DisGeNET \| CTD |
| 7351 | UCP2 | No | - | DisGeNET \| CTD |
| 7436 | VLDLR | No | - | DisGeNET \| CTD |
| 7494 | XBP1 | No | - | DisGeNET \| CTD |
| 8431 | NR0B2 | No | - | DisGeNET \| CTD |
| 8517 | IKBKG | No | - | DisGeNET \| CTD |
| 8614 | STC2 | No | - | DisGeNET \| CTD |
| 8659 | ALDH4A1 | No | - | DisGeNET \| CTD |
| 8856 | NR1I2 | No | - | DisGeNET \| CTD |
| 9315 | NREP | No | - | DisGeNET \| CTD |
| 9370 | ADIPOQ | No | - | DisGeNET \| DISEASES \| GeneCards \| CTD |
| 9376 | SLC22A8 | No | - | DisGeNET \| CTD |
| 9572 | NR1D1 | No | - | DisGeNET \| CTD |
| 9971 | NR1H4 | No | - | DisGeNET \| DISEASES \| CTD |
| 10062 | NR1H3 | No | - | DisGeNET \| CTD |
| 10221 | TRIB1 | No | - | DisGeNET \| DISEASES \| CTD |
| 10400 | PEMT | No | - | DisGeNET \| CTD |
| 10437 | IFI30 | No | - | DisGeNET \| CTD |
| 10891 | PPARGC1A | No | - | DisGeNET \| DISEASES |
| 10935 | PRDX3 | No | - | DisGeNET \| CTD |
| 22837 | COBLL1 | No | - | DisGeNET \| CTD |
| 23171 | GPD1L | No | - | DisGeNET \| CTD |
| 23411 | SIRT1 | No | - | DisGeNET \| DISEASES \| CTD |
| 26291 | FGF21 | No | - | DisGeNET \| DISEASES \| GeneCards \| CTD |
| 27087 | B3GAT1 | No | - | DisGeNET \| CTD |
| 27102 | EIF2AK1 | No | - | DisGeNET \| CTD |
| 27232 | GNMT | No | - | DisGeNET \| CTD |
| 51129 | ANGPTL4 | No | - | DisGeNET \| CTD |
| 51547 | SIRT7 | No | - | DisGeNET \| CTD |
| 51548 | SIRT6 | No | - | DisGeNET \| CTD |
| 51635 | DHRS7 | No | - | DisGeNET \| CTD |
| 53345 | TM6SF2 | No | - | DisGeNET \| DISEASES \| GeneCards \| CTD |
| 54205 | CYCS | No | - | DisGeNET \| CTD |
| 55748 | CNDP2 | No | - | DisGeNET \| CTD |
| 57591 | MRTFA | No | - | DisGeNET \| CTD |
| 57761 | TRIB3 | No | - | DisGeNET \| CTD |
| 79068 | FTO | No | - | DisGeNET \| CTD |

| Table S5: Curated gene set associated with fatty liver disease. Genes were compiled from multiple resources. The table includes genes that are either known targets of FDA-approved drugs for fattty liver disease or supported by at least two of the four considered databases. For each gene, the Entrez gene ID, gene symbol, evidence source(s), and approved drug information (where applicable) are reported. | | | | |
|---|---|---|---|---|
| **Entrez Gene ID** | **Gene symbol** | **Approved drug target** | **Drug Names** | **Evidence source** |
| 79661 | NEIL1 | No | - | DisGeNET \| CTD |
| 80339 | PNPLA3 | No | - | DisGeNET \| DISEASES \| GeneCards \| CTD |
| 84649 | DGAT2 | No | - | DisGeNET \| DISEASES |
| 93974 | ATP5IF1 | No | - | DisGeNET \| CTD |
| 114548 | NLRP3 | No | - | DisGeNET \| CTD |
| 116150 | NUS1 | No | - | DisGeNET \| CTD |
| 152831 | KLB | No | - | DisGeNET \| CTD |
| 284111 | SLC13A5 | No | - | DisGeNET \| CTD |
| 284217 | LAMA1 | No | - | DisGeNET \| CTD |

**Table S6: Curated gene set associated with obesity. Genes were compiled from multiple resources. The table includes genes that are either known targets of FDA-approved antiobesity drugs or supported by at least two of the four considered databases. For each gene, the Entrez gene ID, gene symbol, evidence source(s), and approved drug information (where applicable) are reported.**

| Entrez Gene ID | Gene symbol | Approved drug target | Drug Names | Evidence source |
|---|---|---|---|---|
| 4160 | MC4R | Yes | Setmelanotide | DrugBank (TTD) \| DisGeNET \| DISEASES \| GeneCards \| CTD |
| 2194 | FASN | Yes | Orlistat | DrugBank (Santos et al., Drug Central) \| DisGeNET \| DISEASES \| CTD |
| 5406 | PNLIP | Yes | Orlistat | DrugBank (Santos et al., Drug Central \| TTD) \| DISEASES \| GeneCards |
| 6531 | SLC6A3 | Yes | Amphetamine \| Diethylpropion \| Metamfetamine \| Phendimetrazine | DrugBank (Drug Central \| Santos et al., Drug Central) \| DisGeNET |
| 134864 | TAAR1 | Yes | Amphetamine | DrugBank (Santos et al., Drug Central) |
| 147 | ADRA1B | Yes | Phendimetrazine | DrugBank (TTD) |
| 56144 | PCDHA4 | Yes | Orlistat | DrugBank (TTD) |
| 6530 | SLC6A2 | Yes | Amphetamine \| Benzphetamine \| Diethylpropion \| Phendimetrazine \| Phentermine | DrugBank (Santos et al., Drug Central \| TTD) |
| 6571 | SLC18A2 | Yes | Benzphetamine | DrugBank (Santos et al., Drug Central) |
| 8513 | LIPF | Yes | Orlistat | DrugBank (Santos et al., Drug Central) |
| 5468 | PPARG | No | - | DisGeNET \| DISEASES \| GeneCards \| CTD |
| 2740 | GLP1R | No | - | DisGeNET \| DISEASES \| GeneCards |
| 32 | ACACB | No | - | DisGeNET \| CTD |
| 3643 | INSR | No | - | DISEASES \| GeneCards |
| 10135 | NAMPT | No | - | DisGeNET \| DISEASES \| GeneCards \| CTD |
| 109 | ADCY3 | No | - | DisGeNET \| DISEASES \| GeneCards \| CTD |
| 1268 | CNR1 | No | - | DisGeNET \| DISEASES \| GeneCards \| CTD |
| 1363 | CPE | No | - | DisGeNET \| DISEASES \| GeneCards \| CTD |
| 1401 | CRP | No | - | DisGeNET \| DISEASES \| GeneCards \| CTD |
| 155 | ADRB3 | No | - | DisGeNET \| DISEASES \| GeneCards \| CTD |
| 181 | AGRP | No | - | DisGeNET \| DISEASES \| GeneCards \| CTD |
| 2641 | GCG | No | - | DisGeNET \| DISEASES \| GeneCards \| CTD |
| 27125 | AFF4 | No | - | DisGeNET \| DISEASES \| GeneCards \| CTD |
| 2778 | GNAS | No | - | DisGeNET \| DISEASES \| GeneCards \| CTD |
| 348 | APOE | No | - | DisGeNET \| DISEASES \| GeneCards \| CTD |
| 3569 | IL6 | No | - | DisGeNET \| DISEASES \| GeneCards \| CTD |
| 3630 | INS | No | - | DisGeNET \| DISEASES \| GeneCards \| CTD |
| 3667 | IRS1 | No | - | DisGeNET \| DISEASES \| GeneCards \| CTD |
| 3952 | LEP | No | - | DisGeNET \| DISEASES \| GeneCards \| CTD |
| 3953 | LEPR | No | - | DisGeNET \| DISEASES \| GeneCards \| CTD |
| 4023 | LPL | No | - | DisGeNET \| DISEASES \| GeneCards \| CTD |
| 4915 | NTRK2 | No | - | DisGeNET \| DISEASES \| GeneCards \| CTD |
| 5054 | SERPINE1 | No | - | DisGeNET \| DISEASES \| GeneCards \| CTD |
| 5122 | PCSK1 | No | - | DisGeNET \| DISEASES \| GeneCards \| CTD |
| 5167 | ENPP1 | No | - | DisGeNET \| DISEASES \| GeneCards \| CTD |
| 51738 | GHRL | No | - | DisGeNET \| DISEASES \| GeneCards \| CTD |
| 5346 | PLIN1 | No | - | DisGeNET \| DISEASES \| GeneCards \| CTD |
| 5443 | POMC | No | - | DisGeNET \| DISEASES \| GeneCards \| CTD |
| 56623 | INPP5E | No | - | DisGeNET \| DISEASES \| GeneCards \| CTD |
| 5697 | PYY | No | - | DisGeNET \| DISEASES \| GeneCards \| CTD |
| 5770 | PTPN1 | No | - | DisGeNET \| DISEASES \| GeneCards \| CTD |
| 6347 | CCL2 | No | - | DisGeNET \| DISEASES \| GeneCards \| CTD |
| 6720 | SREBF1 | No | - | DisGeNET \| DISEASES \| GeneCards \| CTD |
| 7124 | TNF | No | - | DisGeNET \| DISEASES \| GeneCards \| CTD |
| 7350 | UCP1 | No | - | DisGeNET \| DISEASES \| GeneCards \| CTD |
| 7351 | UCP2 | No | - | DisGeNET \| DISEASES \| GeneCards \| CTD |
| 7352 | UCP3 | No | - | DisGeNET \| DISEASES \| GeneCards \| CTD |
| 79068 | FTO | No | - | DisGeNET \| DISEASES \| GeneCards \| CTD |
| 8431 | NR0B2 | No | - | DisGeNET \| DISEASES \| GeneCards \| CTD |
| 9370 | ADIPOQ | No | - | DisGeNET \| DISEASES \| GeneCards \| CTD |
| 9607 | CARTPT | No | - | DisGeNET \| DISEASES \| GeneCards \| CTD |
| 1050 | CEBPA | No | - | DisGeNET \| DISEASES \| CTD |
| 11254 | SLC6A14 | No | - | DisGeNET \| DISEASES \| GeneCards |
| 1149 | CIDEA | No | - | DisGeNET \| DISEASES \| CTD |
| 133522 | PPARGC1B | No | - | DisGeNET \| GeneCards \| CTD |
| 154 | ADRB2 | No | - | DisGeNET \| GeneCards \| CTD |
| 207 | AKT1 | No | - | DisGeNET \| DISEASES \| CTD |
| 2099 | ESR1 | No | - | DisGeNET \| DISEASES \| CTD |
| 23411 | SIRT1 | No | - | DisGeNET \| DISEASES \| CTD |
| 25970 | SH2B1 | No | - | DisGeNET \| GeneCards \| CTD |
| 26291 | FGF21 | No | - | DisGeNET \| DISEASES \| CTD |
| 2688 | GH1 | No | - | DisGeNET \| GeneCards \| CTD |
| 3060 | HCRT | No | - | DisGeNET \| DISEASES \| CTD |
| 3290 | HSD11B1 | No | - | DisGeNET \| GeneCards \| CTD |
| 3383 | ICAM1 | No | - | DisGeNET \| DISEASES \| CTD |
| 364 | AQP7 | No | - | DisGeNET \| GeneCards \| CTD |
| 3949 | LDLR | No | - | DisGeNET \| DISEASES \| CTD |
| 5465 | PPARA | No | - | DisGeNET \| DISEASES \| CTD |
| 5467 | PPARD | No | - | DisGeNET \| DISEASES \| CTD |
| 5573 | PRKAR1A | No | - | DisGeNET \| GeneCards \| CTD |

**Table S6: Curated gene set associated with obesity. Genes were compiled from multiple resources. The table includes genes that are either known targets of FDA-approved antiobesity drugs or supported by at least two of the four considered databases. For each gene, the Entrez gene ID, gene symbol, evidence source(s), and approved drug information (where applicable) are reported.**

| Entrez Gene ID | Gene symbol | Approved drug target | Drug Names | Evidence source |
|---|---|---|---|---|
| 56729 | RETN | No | - | DisGeNET \| DISEASES \| GeneCards |
| 583 | BBS2 | No | - | DisGeNET \| DISEASES \| GeneCards |
| 585 | BBS4 | No | - | DisGeNET \| DISEASES \| GeneCards |
| 627 | BDNF | No | - | DisGeNET \| DISEASES \| GeneCards |
| 6648 | SOD2 | No | - | DisGeNET \| DISEASES \| CTD |
| 6774 | STAT3 | No | - | DisGeNET \| DISEASES \| GeneCards |
| 8195 | MKKS | No | - | DisGeNET \| DISEASES \| GeneCards |
| 9672 | SDC3 | No | - | DisGeNET \| GeneCards \| CTD |
| 10059 | DNM1L | No | - | DisGeNET \| GeneCards |
| 10411 | RAPGEF3 | No | - | DisGeNET \| CTD |
| 1066 | CES1 | No | - | DisGeNET \| CTD |
| 10726 | NUDC | No | - | DisGeNET \| GeneCards |
| 10743 | RAI1 | No | - | DisGeNET \| CTD |
| 10806 | SDCCAG8 | No | - | DISEASES \| GeneCards |
| 10840 | ALDH1L1 | No | - | DisGeNET \| CTD |
| 10891 | PPARGC1A | No | - | DISEASES \| GeneCards |
| 112609 | MRAP2 | No | - | DISEASES \| GeneCards |
| 114294 | LACTB | No | - | DisGeNET \| CTD |
| 123016 | TTC8 | No | - | DISEASES \| GeneCards |
| 129787 | TMEM18 | No | - | DisGeNET \| CTD |
| 129880 | BBS5 | No | - | DISEASES \| GeneCards |
| 1312 | COMT | No | - | DisGeNET \| GeneCards |
| 132789 | GNPDA2 | No | - | DisGeNET \| CTD |
| 1350 | COX7C | No | - | DisGeNET \| CTD |
| 1361 | CPB2 | No | - | DisGeNET \| CTD |
| 142 | PARP1 | No | - | DisGeNET \| CTD |
| 1431 | CS | No | - | DisGeNET \| CTD |
| 1489 | CTF1 | No | - | DisGeNET \| CTD |
| 151742 | PPM1L | No | - | DisGeNET \| CTD |
| 1520 | CTSS | No | - | DisGeNET \| CTD |
| 1528 | CYB5A | No | - | DisGeNET \| CTD |
| 153 | ADRB1 | No | - | DisGeNET \| CTD |
| 1545 | CYP1B1 | No | - | DisGeNET \| CTD |
| 1571 | CYP2E1 | No | - | DisGeNET \| CTD |
| 157657 | C8orf37\|CFAP418 | No | - | DISEASES \| GeneCards |
| 157680 | VPS13B | No | - | DISEASES \| GeneCards |
| 1636 | ACE | No | - | DISEASES \| GeneCards |
| 1649 | DDIT3 | No | - | DisGeNET \| CTD |
| 166379 | BBS12 | No | - | DISEASES \| GeneCards |
| 169841 | ZNF169 | No | - | DisGeNET \| CTD |
| 1728 | NQO1 | No | - | DisGeNET \| CTD |
| 1788 | DNMT3A | No | - | DisGeNET \| GeneCards |
| 1806 | DPYD | No | - | DisGeNET \| CTD |
| 183 | AGT | No | - | DisGeNET \| DISEASES |
| 1892 | ECHS1 | No | - | DisGeNET \| CTD |
| 1947 | EFNB1 | No | - | DisGeNET \| CTD |
| 196 | AHR | No | - | DisGeNET \| CTD |
| 2033 | EP300 | No | - | DisGeNET \| CTD |
| 2110 | ETFDH | No | - | DisGeNET \| CTD |
| 2147 | F2 | No | - | DisGeNET \| CTD |
| 2166 | FAAH | No | - | DisGeNET \| CTD |
| 2180 | ACSL1 | No | - | DisGeNET \| CTD |
| 23011 | RAB21 | No | - | DisGeNET \| CTD |
| 2309 | FOXO3 | No | - | DisGeNET \| CTD |
| 23216 | TBC1D1 | No | - | DisGeNET \| CTD |
| 23217 | ZFR2 | No | - | DisGeNET \| CTD |
| 23410 | SIRT3 | No | - | DisGeNET \| CTD |
| 2353 | FOS | No | - | DisGeNET \| CTD |
| 23788 | MTCH2 | No | - | DisGeNET \| CTD |
| 2512 | FTL | No | - | DisGeNET \| CTD |
| 252995 | FNDC5 | No | - | DISEASES \| GeneCards |
| 257194 | NEGR1 | No | - | DisGeNET \| CTD |
| 2673 | GFPT1 | No | - | DisGeNET \| CTD |
| 2695 | GIP | No | - | DisGeNET \| DISEASES |
| 27241 | BBS9 | No | - | DISEASES \| GeneCards |
| 2731 | GLDC | No | - | DisGeNET \| GeneCards |
| 2752 | GLUL | No | - | DisGeNET \| CTD |
| 2784 | GNB3 | No | - | DisGeNET \| CTD |
| 283455 | KSR2 | No | - | DisGeNET \| GeneCards |
| 2840 | GPR17 | No | - | DisGeNET \| CTD |

**Table S6: Curated gene set associated with obesity. Genes were compiled from multiple resources. The table includes genes that are either known targets of FDA-approved antiobesity drugs or supported by at least two of the four considered databases. For each gene, the Entrez gene ID, gene symbol, evidence source(s), and approved drug information (where applicable) are reported.**

| Entrez Gene ID | Gene symbol | Approved drug target | Drug Names | Evidence source |
|---|---|---|---|---|
| 2876 | GPX1 | No | - | DisGeNET \| CTD |
| 2878 | GPX3 | No | - | DisGeNET \| CTD |
| 2908 | NR3C1 | No | - | DisGeNET \| GeneCards |
| 3033 | HADH | No | - | DisGeNET \| CTD |
| 3099 | HK2 | No | - | DisGeNET \| CTD |
| 3148 | HMGB2 | No | - | DisGeNET \| CTD |
| 3162 | HMOX1 | No | - | DisGeNET \| CTD |
| 3215 | HOXB5 | No | - | DisGeNET \| CTD |
| 3291 | HSD11B2 | No | - | DisGeNET \| GeneCards |
| 3309 | HSPA5 | No | - | DisGeNET \| CTD |
| 3356 | HTR2A | No | - | DisGeNET \| CTD |
| 3358 | HTR2C | No | - | DisGeNET \| CTD |
| 338 | APOB | No | - | DISEASES \| GeneCards |
| 34 | ACADM | No | - | DisGeNET \| CTD |
| 3479 | IGF1 | No | - | DISEASES \| GeneCards |
| 3481 | IGF2 | No | - | DisGeNET \| CTD |
| 3485 | IGFBP2 | No | - | DisGeNET \| CTD |
| 3553 | IL1B | No | - | DISEASES \| GeneCards |
| 3574 | IL7 | No | - | DisGeNET \| CTD |
| 3620 | IDO1 | No | - | DisGeNET \| CTD |
| 3636 | INPPL1 | No | - | DisGeNET \| GeneCards |
| 3684 | ITGAM | No | - | DisGeNET \| CTD |
| 3778 | KCNMA1 | No | - | DisGeNET \| CTD |
| 3929 | LBP | No | - | DisGeNET \| CTD |
| 412 | STS | No | - | DisGeNET \| CTD |
| 4159 | MC3R | No | - | DisGeNET \| GeneCards |
| 4199 | ME1 | No | - | DisGeNET \| CTD |
| 43 | ACHE | No | - | DisGeNET \| CTD |
| 4318 | MMP9 | No | - | DisGeNET \| CTD |
| 4329 | ALDH6A1 | No | - | DisGeNET \| CTD |
| 434 | ASIP | No | - | DisGeNET \| DISEASES |
| 4360 | MRC1 | No | - | DisGeNET \| CTD |
| 4547 | MTTP | No | - | DisGeNET \| GeneCards |
| 4594 | MMUT | No | - | DisGeNET \| CTD |
| 463 | ZFHX3 | No | - | DisGeNET \| CTD |
| 47 | ACLY | No | - | DisGeNET \| CTD |
| 4808 | NHLH2 | No | - | DisGeNET \| CTD |
| 4852 | NPY | No | - | DISEASES \| GeneCards |
| 4864 | NPC1 | No | - | DisGeNET \| CTD |
| 4886 | NPY1R | No | - | DisGeNET \| CTD |
| 4889 | NPY5R | No | - | DisGeNET \| CTD |
| 4968 | OGG1 | No | - | DisGeNET \| CTD |
| 51043 | ZBTB7B | No | - | DisGeNET \| CTD |
| 5105 | PCK1 | No | - | DisGeNET \| CTD |
| 51094 | ADIPOR1 | No | - | DISEASES \| GeneCards |
| 51129 | ANGPTL4 | No | - | DisGeNET \| CTD |
| 51181 | DCXR | No | - | DisGeNET \| CTD |
| 5209 | PFKFB3 | No | - | DisGeNET \| CTD |
| 5290 | PIK3CA | No | - | DISEASES \| GeneCards |
| 5367 | PMCH | No | - | DisGeNET \| CTD |
| 54 | ACP5 | No | - | DisGeNET \| CTD |
| 54205 | CYCS | No | - | DisGeNET \| CTD |
| 54496 | PRMT7 | No | - | DisGeNET \| GeneCards |
| 54551 | MAGEL2 | No | - | DISEASES \| GeneCards |
| 54585 | LZTFL1 | No | - | DISEASES \| GeneCards |
| 54901 | CDKAL1 | No | - | DisGeNET \| GeneCards |
| 55212 | BBS7 | No | - | DISEASES \| GeneCards |
| 55277 | FGGY | No | - | DisGeNET \| CTD |
| 55690 | PACS1 | No | - | DisGeNET \| CTD |
| 5577 | PRKAR2B | No | - | DisGeNET \| CTD |
| 5583 | PRKCH | No | - | DisGeNET \| CTD |
| 56603 | CYP26B1 | No | - | DisGeNET \| CTD |
| 57104 | PNPLA2 | No | - | DISEASES \| GeneCards |
| 5743 | PTGS2 | No | - | DisGeNET \| CTD |
| 57498 | KIDINS220 | No | - | DISEASES \| GeneCards |
| 582 | BBS1 | No | - | DISEASES \| GeneCards |
| 590 | BCHE | No | - | DisGeNET \| CTD |
| 5919 | RARRES2 | No | - | DISEASES \| GeneCards |
| 6303 | SAT1 | No | - | DisGeNET \| CTD |
| 6319 | SCD | No | - | DisGeNET \| DISEASES |

**Table S6: Curated gene set associated with obesity. Genes were compiled from multiple resources. The table includes genes that are either known targets of FDA-approved antiobesity drugs or supported by at least two of the four considered databases. For each gene, the Entrez gene ID, gene symbol, evidence source(s), and approved drug information (where applicable) are reported.**

| Entrez Gene ID | Gene symbol | Approved drug target | Drug Names | Evidence source |
|---|---|---|---|---|
| 6462 | SHBG | No | - | DISEASES \| GeneCards |
| 64756 | ATPAF1 | No | - | DisGeNET \| CTD |
| 6492 | SIM1 | No | - | DisGeNET \| GeneCards |
| 6517 | SLC2A4 | No | - | DISEASES \| GeneCards |
| 6580 | SLC22A1 | No | - | DisGeNET \| CTD |
| 6581 | SLC22A3 | No | - | DisGeNET \| CTD |
| 6582 | SLC22A2 | No | - | DisGeNET \| CTD |
| 6647 | SOD1 | No | - | DisGeNET \| CTD |
| 7018 | TF | No | - | DisGeNET \| CTD |
| 7037 | TFRC | No | - | DisGeNET \| CTD |
| 7133 | TNFRSF1B | No | - | DisGeNET \| CTD |
| 7275 | TUB | No | - | DISEASES \| GeneCards |
| 7385 | UQCRC2 | No | - | DisGeNET \| CTD |
| 7425 | VGF | No | - | DisGeNET \| CTD |
| 7436 | VLDLR | No | - | DisGeNET \| CTD |
| 7442 | TRPV1 | No | - | DisGeNET \| CTD |
| 761 | CA3 | No | - | DisGeNET \| CTD |
| 7840 | ALMS1 | No | - | DISEASES \| GeneCards |
| 79047 | KCTD15 | No | - | DisGeNET \| CTD |
| 79661 | NEIL1 | No | - | DisGeNET \| CTD |
| 79738 | BBS10 | No | - | DISEASES \| GeneCards |
| 80184 | CEP290 | No | - | DISEASES \| GeneCards |
| 8165 | AKAP1 | No | - | DisGeNET \| CTD |
| 834 | CASP1 | No | - | DisGeNET \| CTD |
| 84100 | ARL 6.00\|ARL6 | No | - | DISEASES \| GeneCards |
| 84872 | ZC3H10 | No | - | DisGeNET \| CTD |
| 84984 | CEP19 | No | - | DISEASES \| GeneCards |
| 8522 | GAS7 | No | - | DisGeNET \| CTD |
| 8694 | DGAT1 | No | - | DISEASES \| GeneCards |
| 8800 | PEX11A | No | - | DisGeNET \| CTD |
| 8801 | SUCLG2 | No | - | DisGeNET \| CTD |
| 8803 | SUCLA2 | No | - | DisGeNET \| CTD |
| 8856 | NR1I2 | No | - | DisGeNET \| CTD |
| 89780 | WNT3A | No | - | DisGeNET \| CTD |
| 91147 | TMEM67 | No | - | DISEASES \| GeneCards |
| 9149 | DYRK1B | No | - | DISEASES \| GeneCards |
| 9194 | SLC16A7 | No | - | DisGeNET \| CTD |
| 92482 | BBIP1 | No | - | DISEASES \| GeneCards |
| 9332 | CD163 | No | - | DisGeNET \| CTD |
| 955 | ENTPD6 | No | - | DisGeNET \| CTD |
| 958 | CD40 | No | - | DisGeNET \| CTD |
| 968 | CD68 | No | - | DisGeNET \| CTD |
| 9970 | NR1I3 | No | - | DisGeNET \| CTD |

**Table S7: Network proximity of formulation-associated targets to type 2 diabetes-, fatty liver-, and obesity-associated gene modules. The table reports the average closest shortest path length (d_c), corresponding Z-scores, and one-tailed p-values for each formulation-disease pair. Negative Z-scores indicate that formulation-associated targets are positioned closer to the corresponding disease-associated gene module than expected by chance in the human interactome .**

| | Type 2 diabetes | | | Fatty liver | | | Obesity | | |
|---|---|---|---|---|---|---|---|---|---|
| Formulation | Average shortest path length (d_c) | Z-score | One-tailed p-value | Average shortest path length (dc) | Z-score | One-tailed p-value | Average shortest path length (dc) | Z-score | One-tailed p-value |
| Arogyavardhini | 1.174382716 | -5.584496817 | 0 | 1.25154321 | -4.045398035 | 0 | 1.160493827 | -5.62843838 | 0 |
| Nishamalaki | 1.158238173 | -5.77617662 | 0 | 1.239804241 | -3.721061507 | 0.003 | 1.145187602 | -5.240616468 | 0 |

**Table S8: Top 20 enriched KEGG pathways identified from pathway enrichment analyses of formulation-associated and disease-associated target gene sets. Enrichment analysis was performed separately for each formulation and disease condition. Pathways were ranked based on statistical significance, and the top 20 pathways satisfying the predefined filtering criteria (adjusted p-value < 0.01 and minimum pathway size ≥ 50 genes). The table reports pathway-level statistics, including overlap ratio, p-value, adjusted p-value, odds ratio, combined score, overlapping genes, total pathway gene count, and KEGG subclass annotation.**

| Pathways associated with Arogyavardhini | | | | | | | | | | Pathways associated with Type 2 diabetes | | | | | | | | | | | |
|---|---|---|---|---|---|---|---|---|---|---|---|---|---|---|---|---|---|---|---|---|---|
| Term | Overlap | P-value | Adjusted P-value | Odds Ratio | Combined Score | Genes | Overlap gene number | Total genes in pathway | Subclass | Term | Overlap | P-value | Adjusted P-value | Old P-value | Old Adjusted P-value | Odds Ratio | Combined Score | Genes | Overlap gene number | Total genes in pathway | Subclass |
| NEUROACTIVE LIGAND-RECEPTOR INTERACTION | 78/366 | 4.697505294e-41 | 7.351595785e-39 | 8.913176454 | 827.6682711 | CHRM2;GLP1R;GABRB2;CHRM3;VIPR1;CHRM1;THRB;THRA;CHRM4;CHRM5;NPY2R;MLNR;HTR2B;HTR2C;ADRA1D;PTH1R;HTR2A;NR3C1;ADRA1A;MC4R;HTR6;EDNRA;CYSLTR1;EDNRB;HTR7;CCKAR;ADORA3;ADORA1;BDKRB2;BDKRB1;CTSG;PRSS3;PRSS2;GHSR;GPR35;AVPR2;TACR2;NPY1R;AVPR1A;TACR1;OPRM1;ADRA2C;ADRA2B;GABRG2;ADRA2A;MC3R;CCKBR;ADRB3;ADORA2A;AGTR1;AGTR2;NPSR1;OPRD1;GIPR;CHRNA4;PTAFR;ADRB1;PLG;ADRB2;CRHR1;CRHR2;CALCR;HRH3;TBXA2R;HRH2;DRD1;DRD2;DRD3;NTSR1;DRD4;GABRA2;GABRA1;GCGR;HTR1D;HTR1A;OPRK1;TSHR;MC5R | 78 | 366 | Signaling molecules and interaction | NON-ALCOHOLIC Fatty liver DISEASE | 63/154 | 3.053122614e-77 | 8.273962285e-75 | 0 | 0 | 62.91666667 | 11084.83999 | NDUFA13;NDUFA11;IRS1;NDUFA12;NDUFA10;IRS2;COXFA4L2;TNF;CASP8;AKT2;CASP3;AKT1;SREBF1;ADIPOQ;NDUFC2;NDUFC1;TNFRSF1A;NDUFS8;NDUFS7;NDUFS6;NDUFS5;IL1B;NDUFS4;NDUFS3;NDUFS2;COXFA4;PPARG;NDUFS1;PPARA;NDUFB9;NDUFB8;PRKAA2;NDUFB7;NDUFB10;NDUFB6;NDUFB11;NDUFB5;NDUFB4;NDUFB3;NDUFB2;NDUFB1;ADIPOR1;ADIPOR2;INS;BCL2L11;NDUFV3;NDUFV2;NDUFV1;NDUFA9;NDUFA8;NDUFA7;TGFB1;NDUFA6;NDUFA5;INSR;NDUFA3;NDUFA2;NDUFA1;NFKB1;IL6;LEP;NDUFAB1;BAX | 63 | 154 | Endocrine and metabolic disease |
| CALCIUM SIGNALING PATHWAY | 62/250 | 8.157610181e-37 | 8.511106622e-35 | 10.61693188 | 882.2319734 | RET;CHRM2;CHRM3;CHRM1;FLT1;FLT4;CHRM5;HTR2B;HTR2C;ADRA1D;HTR2A;ADRA1A;PPP3CA;HTR6;EDNRA;CYSLTR1;HTR7;EDNRB;CCKAR;PRKACG;KDR;BDKRB2;CD38;BDKRB1;PRKACB;PRKCG;PDGFRB;PDGFRA;CAMK1D;PRKCB;SPHK1;TACR2;PRKCA;TACR1;AVPR1A;CCKBR;ADRB3;ADORA2A;AGTR1;CAMK1G;CAMK2B;CAMK2D;PTAFR;CAMK2A;ADRB1;ADRB2;STK4;EGFR;GNA15;HRH2;TBXA2R;ERBB4;ERBB2;DRD1;CAMK2G;NTSR1;CAMK4;GNAS;CALM2;FGFR3;FGFR2;FGFR1 | 62 | 250 | Signal transduction | DIABETIC CARDIOMYOPATHY | 58/203 | 3.094734312e-60 | 4.193364992e-58 | 0 | 0 | 35.43167421 | 4855.039456 | NDUFA13;NDUFA11;IRS1;NDUFA12;NDUFA10;SLC2A1;ATP2A3;ATP2A2;SLC2A4;COXFA4L2;AKT2;AKT1;CD36;PRKCB;NDUFC2;CYBA;NDUFC1;NDUFS8;NDUFS7;TBC1D4;NDUFS6;NDUFS5;NDUFS4;NDUFS3;NDUFS2;COXFA4;NDUFS1;PPARA;NDUFB9;NDUFB8;NDUFB7;NDUFB10;NDUFB6;NDUFB11;NDUFB5;NDUFB4;NDUFB3;NDUFB2;NDUFB1;AGER;INS;NDUFV3;NDUFV2;NDUFV1;NDUFA9;NDUFA8;NDUFA7;ACE;TGFB1;NDUFA6;NDUFA5;INSR;NDUFA3;NDUFA2;NDUFA1;NFKB1;MTOR;NDUFAB1 | 58 | 203 | Cardiovascular disease |
| LIPID AND ATHEROSCLEROSIS | 57/214 | 9.847922775e-36 | 6.164799657e-34 | 11.60936306 | 935.7820292 | GSK3B;CXCL8;HSP90AB1;TNF;IKBKB;PPP3CA;CASP7;CASP1;AKT1;MAP3K5;HSP90AA1;MMP1;PDPK1;MMP3;PRKCA;MMP9;CYP2A7;CYP2A6;PIK3CA;PPARG;TP53;TLR2;CAMK2B;CAMK2D;SRC;CAMK2A;PIK3R1;RELA;HSPD1;MAPK9;MAPK8;CYP2B6;RXRA;MAPK1;MAP2K7;CAMK2G;MAP2K6;MAPK3;CYP2J2;HSPA8;JUN;HSPA5;STAT3;MAPK14;SELE;PTK2;NFKB1;MAPK12;MAPK13;MAPK10;CYP2C9;MAPK11;CYP2C8;CYP1A1;CALM2;HSPA1A;NFE2L2 | 57 | 214 | Cardiovascular disease | RETROGRADE ENDOCANNABINOID SIGNALING | 42/146 | 1.14322436e-43 | 5.163563361e-42 | 0 | 0 | 33.42721519 | 3305.192996 | NDUFB9;NDUFA13;NDUFB8;NDUFB7;NDUFA11;NDUFB10;NDUFB6;NDUFA12;NDUFB11;NDUFB5;NDUFB4;NDUFA10;NDUFB3;NDUFB2;NDUFB1;COXFA4L2;ADCY5;NDUFV3;NDUFV2;NDUFV1;NDUFA9;NDUFA8;NDUFA7;NDUFA6;NDUFA5;PRKCB;NDUFA3;NDUFA2;NDUFA1;NDUFC2;NDUFC1;NDUFS8;NDUFS7;NDUFS6;NDUFS5;NDUFS4;NDUFAB1;NDUFS3;GNB3;NDUFS2;COXFA4;NDUFS1 | 42 | 146 | Nervous system |
| INSULIN RESISTANCE | 41/107 | 3.905654467e-33 | 2.037449747e-31 | 19.44010724 | 1450.676842 | PYGB;GSK3B;PRKAA1;PRKAA2;SLC2A1;PRKAG1;PRKAG2;SLC2A2;PYGM;PIK3R1;PYGL;TNF;PRKCZ;PRKAG3;PTPRF;RELA;IKBKB;RPS6KA3;MAPK9;MAPK8;AKT1;PTPN1;PRKAB2;PRKCB;PDPK1;PRKCE;INSR;NR1H2;PRKCD;STAT3;NR1H3;PTPN11;PRKAB1;NFKB1;MTOR;MAPK10;PIK3CA;RPS6KB1;PTPRA;PRKCQ;PPARA | 41 | 107 | Endocrine and metabolic disease | OXIDATIVE PHOSPHORYLATION | 39/134 | 8.52251483e-41 | 2.887001899e-39 | 0 | 0 | 33.57078947 | 3097.35106 | NDUFB9;NDUFA13;NDUFB8;NDUFB7;NDUFA11;NDUFB10;NDUFB6;NDUFA12;NDUFB11;NDUFB5;NDUFB4;NDUFA10;NDUFB3;NDUFB2;NDUFB1;COXFA4L2;NDUFV3;NDUFV2;NDUFV1;NDUFA9;NDUFA8;NDUFA7;NDUFA6;NDUFA5;NDUFA3;NDUFA2;NDUFA1;NDUFC2;NDUFC1;NDUFS8;NDUFS7;NDUFS6;NDUFS5;NDUFS4;NDUFAB1;NDUFS3;NDUFS2;COXFA4;NDUFS1 | 39 | 134 | Energy metabolism |
| FOXO SIGNALING PATHWAY | 44/130 | 9.301623727e-33 | 4.15915461e-31 | 16.07246102 | 1185.426277 | PRKAA1;PRKAA2;PRKAG1;PRKAG2;PIK3R1;STK4;PRKAG3;EGFR;IGF1R;CCNB3;IKBKB;CCNB2;MAPK9;CCNB1;MAPK8;BCL2L11;CCND1;AKT1;EP300;MAPK1;MAPK3;PLK4;MAP2K1;PRKAB2;USP7;MAP2K2;SMAD3;PDPK1;INSR;PLK1;STAT3;BRAF;MAPK14;PRKAB1;SIRT1;MAPK12;MAPK13;MAPK10;MAPK11;PIK3CA;CDK2;MDM2;ATM;SGK1 | 44 | 130 | Signal transduction | ADIPOCYTOKINE SIGNALING PATHWAY | 25/70 | 5.105868944e-29 | 1.064377295e-27 | 0 | 0 | 43.03587052 | 2803.553583 | PRKAA2;IRS1;SLC2A1;IRS2;SLC2A4;ADIPOR1;TNF;ADIPOR2;ACACB;AKT2;NPY;AKT1;CD36;PPARGC1A;CPT1A;ADIPOQ;STAT3;TNFRSF1B;MTOR;NFKB1;TNFRSF1A;G6PC2;LEP;AGRP;PPARA | 25 | 70 | Endocrine system |
| CAMP SIGNALING PATHWAY | 54/222 | 1.06727923e-31 | 4.175729988e-30 | 10.22121535 | 728.9262309 | CHRM2;GLP1R;CHRM1;HTR6;EDNRA;PRKACG;ADORA1;AKT1;EP300;PDE4A;PRKACB;GHSR;ABCC4;MAP2K1;MAP2K2;PDE4D;NPY1R;PRKCA;ADORA2A;PIK3CA;PPARA;CFTR;RAPGEF3;RAPGEF4;CAMK2B;CAMK2D;GIPR;CAMK2A;ADRB1;PIK3R1;ADRB2;CRHR1;RELA;CRHR2;HCAR2;MAPK9;MAPK8;MAPK1;DRD1;DRD2;CAMK2G;MAPK3;JUN;HTR1D;HTR1A;BRAF;NFKB1;TSHR;MAPK10;CAMK4;PDE3A;GNAS;PKN1;CALM2 | 54 | 222 | Signal transduction | INSULIN RESISTANCE | 27/107 | 1.309093204e-26 | 2.534030417e-25 | 0 | 0 | 26.30491071 | 1567.716852 | PRKAA2;IRS1;SLC2A1;SLC2A2;IRS2;PPP1R3A;SLC2A4;TNF;ACACB;INS;AKT2;AKT1;CD36;PPARGC1A;SREBF1;PTPN1;CPT1A;PRKCB;INSR;STAT3;NFKB1;MTOR;TNFRSF1A;G6PC2;IL6;TBC1D4;PPARA | 27 | 107 | Endocrine and metabolic disease |
| NEUROACTIVE LIGAND SIGNALING | 51/195 | 1.220912248e-31 | 4.246061485e-30 | 11.22053768 | 798.6839261 | CHRM2;ACHE;GABRB2;CHRM3;CHRM1;CHRM4;CHRM5;HTR2B;HTR2C;ADRA1D;HTR2A;SLC6A2;ADRA1A;SLC6A3;SLC6A4;MC4R;HTR6;HTR7;PRKACG;ADORA3;ADORA1;PRKACB;PRKCG;PRKCB;TACR2;PRKCA;HTR3A;TACR1;OPRM1;ADRA2C;ADRA2B;GABRG2;ADRA2A;MC3R;ADORA2A;OPRD1;CHRNA4;ADRB1;ADRB2;HRH3;HRH2;DRD1;DRD2;DRD3;DRD4;GABRA2;GABRA1;HTR1D;HTR1A;OPRK1;GNAS | 51 | 195 | Signaling molecules and interaction | AMPK SIGNALING PATHWAY | 24/121 | 4.418713298e-21 | 7.983142025e-20 | 0 | 0 | 19.04087326 | 892.4159955 | SREBF1;CPT1A;PRKAA2;IRS1;INSR;ADIPOQ;IRS2;HNF1A;IGF1;SLC2A4;HMGCR;ADIPOR1;ADIPOR2;ACACB;SIRT1;MTOR;INS;G6PC2;AKT2;LEP;AKT1;PPARG;CD36;PPARGC1A | 24 | 121 | Signal transduction |
| MAPK SIGNALING PATHWAY | 61/296 | 2.462996668e-31 | 7.709179571e-30 | 8.322061973 | 586.5286012 | RET;FLT1;FLT3;FLT4;HSPB1;TNF;IGF1R;IKBKB;RPS6KA3;PPP3CA;RPS6KA5;PRKACG;KDR;AKT1;MAP3K8;PRKACB;MAP3K5;PRKCG;PDGFRB;PDGFRA;MAP2K1;DUSP3;MAP2K2;PRKCB;PRKCA;CDC25B;MAPKAPK2;MAPKAPK5;MAPT;TP53;EPHA2;STK4;EGFR;RELA;MAPK9;MAPK8;ERBB4;ERBB2;MAPK1;MAP2K7;MAP2K6;MAPK3;HSPA8;JUN;INSR;BRAF;MAPK14;NFKB1;MAPK12;NFKB2;MAPK13;MAPK10;MAPK11;PKN1;PTPN7;TEK;MAP3K14;FGFR3;FGFR2;HSPA1A;FGFR1 | 61 | 296 | Signal transduction | LONGEVITY REGULATING PATHWAY | 20/87 | 4.040129907e-19 | 6.842970029e-18 | 0 | 0 | 22.65199101 | 959.3761491 | KL;PRKAA2;IRS1;INSR;ADIPOQ;IRS2;IGF1;ADIPOR1;SOD2;SIRT1;ADIPOR2;MTOR;NFKB1;INS;ADCY5;AKT2;AKT1;BAX;PPARG;PPARGC1A | 20 | 87 | Aging |
| HORMONE SIGNALING | 53/218 | 4.082114656e-31 | 1.16154717e-29 | 10.19901666 | 713.6611239 | GLP1R;VIPR1;THRB;THRA;MLNR;ADRA1D;PTH1R;ADRA1A;IGF1R;MC4R;CCKAR;PRKACG;PRKACB;GHSR;AVPR2;PRKCA;AVPR1A;OPRM1;ADRA2C;ADRA2B;ADRA2A;MC3R;CCKBR;ADRB3;PIK3CA;AGTR1;AGTR2;RAPGEF3;RAPGEF4;OPRD1;GIPR;SRC;ADRB1;PIK3R1;ADRB2;CRHR1;CRHR2;GNA15;CALCR;THPO;DRD1;DRD2;DRD3;DRD4;INSR;GCGR;STAT3;OPRK1;ESR1;ESR2;TSHR;MC5R;GNAS | 53 | 218 | Signaling molecules and interaction | ALCOHOLIC LIVER DISEASE | 23/140 | 2.597264466e-18 | 4.14034511e-17 | 0 | 0 | 15.05381944 | 609.5603546 | SREBF1;TCF7L2;CPT1A;PRKAA2;ADIPOQ;NDUFA2;ADIPOR1;TNF;ADIPOR2;ACACB;SIRT1;NFKB1;TNFRSF1A;C3;IL6;CASP8;AKT2;CASP3;IL1B;AKT1;PPARA;PPARGC1A;TLR4 | 23 | 140 | Endocrine and metabolic disease |
| ERBB SIGNALING PATHWAY | 32/83 | 3.222851108e-26 | 6.30470248e-25 | 19.3676549 | 1136.822194 | CAMK2B;GSK3B;CAMK2D;SRC;CAMK2A;PIK3R1;EGFR;MAPK9;MAPK8;ERBB4;ERBB2;ABL1;AKT1;MAPK1;MAP2K7;CAMK2G;PAK4;MAPK3;PRKCG;MAP2K1;JUN;MAP2K2;PRKCB;PRKCA;BRAF;MTOR;PTK2;BUB1B-PAK6;MAPK10;PIK3CA;RPS6KB1;PKN1 | 32 | 83 | Signal transduction | LIPID AND ATHEROSCLEROSIS | 26/214 | 3.439267302e-17 | 5.178007993e-16 | 0 | 0 | 10.67736103 | 404.7647493 | TNF;AGER;ICAM1;CASP8;AKT2;CASP3;AKT1;CCL2;CD36;APOB;ABCA1;STAT3;NDUFA2;NFATC2;APOA1;CYBA;SOD2;NFKB1;TNFRSF1A;IL6;IL1B;BCL2;BAX;PPARG;TLR4;BCL2L1 | 26 | 214 | Cardiovascular disease |
| INFLAMMATORY MEDIATOR REGULATION OF TRP CHANNELS | 34/97 | 3.709836161e-26 | 6.830463049e-25 | 16.70157201 | 977.9810038 | CAMK2B;CAMK2D;SRC;CAMK2A;HTR2B;HTR2C;ALOX12;PIK3R1;HTR2A;MAPK9;MAPK8;PRKACG;BDKRB2;BDKRB1;PRKACB;CAMK2G;MAP2K6;PRKCG;CYP2J2;PRKCH;TRPA1;PRKCB;PRKCE;PRKCD;PRKCA;MAPK14;MAPK12;MAPK13;MAPK10;MAPK11;PIK3CA;GNAS;PRKCQ;CALM2 | 34 | 97 | Sensory system | HIF-1 SIGNALING PATHWAY | 19/110 | 9.354595078e-16 | 1.267547633e-14 | 0 | 0 | 15.76373626 | 545.5118777 | EDN1;PRKCB;INSR;SERPINE1;STAT3;SLC2A1;IGF1;NFKB1;EGFR;MTOR;INS;IL6;IFNG;AKT2;BCL2;AKT1;HMOX1;TIMP1;TLR4 | 19 | 110 | Signal transduction |
| SEROTONERGIC SYNAPSE | 35/111 | 4.276727398e-25 | 6.084616708e-24 | 14.26520985 | 800.4414556 | APP;GABRB2;MAOB;MAOA;ALOX15;HTR2B;HTR2C;ALOX12;HTR2A;CYP2C19;PTGS2;CYP2C18;SLC6A4;PTGS1;HTR6;HTR7;PRKACG;ALOX5;MAPK1;PRKACB;MAPK3;PRKCG;CYP2J2;MAP2K1;PRKCB;HTR1D;HTR1A;PRKCA;BRAF;HTR3A;ALOX15B;CYP2C9;CYP2C8;GNAS;RAPGEF3 | 35 | 111 | Nervous system | FOXO SIGNALING PATHWAY | 20/130 | 1.667546743e-15 | 2.151929369e-14 | 0 | 0 | 13.76693577 | 468.4533449 | IL10;TGFB1;PRKAA2;IRS1;INSR;STAT3;IRS2;BRAF;IGF1;SLC2A4;SOD2;SIRT1;EGFR;INS;G6PC2;IL6;CCND2;BCL2L11;AKT2;AKT1 | 20 | 130 | Signal transduction |
| NEUROTROPHIN SIGNALING PATHWAY | 36/119 | 4.51614803e-25 | 6.145888406e-24 | 13.45206915 | 754.0822083 | CAMK2B;GSK3B;CAMK2D;CAMK2A;PIK3R1;RELA;IKBKB;RPS6KA3;MAPK9;MAPK8;RPS6KA5;ABL1;AKT1;MAPK1;MAP2K7;CAMK2G;MAP3K5;MAPK3;MAP2K1;JUN;MAP2K2;PDPK1;PRKCD;BRAF;PTPN11;MAPK14;NFKB1;MAPK12;MAPK13;MAPK10;MAPK11;PIK3CA;CAMK4;MAPKAPK2;TP53;CALM2 | 36 | 119 | Nervous system | AGE-RAGE SIGNALING PATHWAY IN DIABETIC COMPLICATIONS | 17/100 | 4.150244715e-14 | 5.112346899e-13 | 0 | 0 | 15.35206475 | 473.0435395 | EDN1;TGFB1;PRKCB;SERPINE1;STAT3;TNF;AGER;NFKB1;ICAM1;IL6;AKT2;CASP3;IL1B;BCL2;AKT1;CCL2;BAX | 17 | 100 | Endocrine and metabolic disease |
| METABOLISM OF XENOBIOTICS BY CYTOCHROME P450 | 30/76 | 5.098937103e-25 | 6.649863805e-24 | 20.07177033 | 1122.726421 | UGT1A10;GSTP1;CYP3A4;CYP3A5;HSD11B1;CYP2B6;CYP2A13;CYP1B1;UGT1A4;UGT1A3;UGT2A1;UGT1A9;CYP2F1;UGT1A8;UGT1A6;CBR3;CBR1;GSTM1;UGT1A1;AKR1C1;UGT2B15;EPHX1;UGT2B17;CYP2A7;CYP2C9;CYP2A6;CYP2S1;CYP1A2;CYP1A1;CYP2E1 | 30 | 76 | Xenobiotics biodegradation and metabolism | LONGEVITY REGULATING PATHWAY - MULTIPLE SPECIES | 14/60 | 7.452718257e-14 | 8.781246295e-13 | 0 | 0 | 22.59639048 | 683.0349348 | PRKAA2;IRS1;INSR;IRS2;IGF1;SOD2;SIRT1;MTOR;INS;ADCY5;SOD1;AKT2;AKT1;FOXA2 | 14 | 60 | Aging |
| PROGESTERONE-MEDIATED OOCYTE MATURATION | 34/108 | 2.18186712e-24 | 2.626632341e-23 | 14.21079346 | 774.2304774 | HSP90AB1;PIK3R1;IGF1R;AURKA;CCNB3;RPS6KA3;STK10;CCNB2;MAPK9;CCNB1;MAPK8;PRKACG;AKT1;MAPK1;PRKACB;MAPK3;MAP2K1;HSP90AA1;PLK1;PRKCA;BRAF;MAPK14;CDC25A;MAPK12;CDC25B;MAPK13;CCNA2;MAPK10;CCNA1;MAPK11;PIK3CA;CDK2;CDK1;PGR | 34 | 108 | Endocrine system | HORMONE SIGNALING | 21/218 | 4.065634074e-12 | 4.237641669e-11 | 0 | 0 | 8.066816196 | 211.5800966 | GLP1R;IRS1;INSR;GCGR;STAT3;ADIPOQ;GCG;IRS2;IGF1;ADIPOR1;ADIPOR2;SMAD5;ADRA2A;ADCY5;INS;MC4R;ADRB3;LEP;GNB3;GHRL;IAPP | 21 | 218 | Signaling molecules and interaction |
| HIF-1 SIGNALING PATHWAY | 34/110 | 4.315281347e-24 | 5.002529858e-23 | 13.83538903 | 744.3422658 | CAMK2B;CAMK2D;FLT1;SERPINE1;CAMK2A;SLC2A1;PIK3R1;HIF1A;EGFR;RELA;IGF1R;ERBB2;AKT1;EP300;MAPK1;CAMK2G;MAPK3;PRKCG;EGLN1;MAP2K1;EGLN3;MAP2K2;EGLN2;PRKCB;PDPK1;INSR;STAT3;PRKCA;NFKB1;MTOR;PIK3CA;RPS6KB1;TEK;GAPDH | 34 | 110 | Signal transduction | INSULIN SIGNALING PATHWAY | 17/137 | 8.290750676e-12 | 8.32145716e-11 | 0 | 0 | 10.59850509 | 270.4301904 | SREBF1;PTPN1;PRKAA2;IRS1;INSR;INPPL1;IRS2;PPP1R3A;BRAF;SLC2A4;ACACB;MTOR;INS;G6PC2;AKT2;AKT1;PPARGC1A | 17 | 137 | Endocrine system |
| BILE SECRETION | 31/88 | 4.91580819e-24 | 5.495171298e-23 | 16.75539488 | 899.2551073 | UGT1A10;ABCB1;SLC22A1;SLC2A1;HMGCR;ABCB11;CYP3A4;SLCO1B1;RXRA;CA2;PRKACG;SLCO1B3;UGT1A4;UGT1A3;UGT2A1;UGT1A9;PRKACB;UGT1A8;UGT1A6;ABCC3;ABCC4;ABCC2;UGT1A1;UGT2B15;EPHX1;NR1H4;UGT2B17;PRKCA;GNAS;CFTR;ABCG2 | 31 | 88 | Digestive system | REGULATION OF LIPOLYSIS IN ADIPOCYTES | 11/58 | 4.072746254e-10 | 3.560368499e-09 | 0 | 0 | 17.18116862 | 371.4832113 | FABP4;ADRB3;IRS1;AKT2;NPY;INSR;AKT1;IRS2;PLIN1;INS;ADCY5 | 11 | 58 | Endocrine system |
| T CELL RECEPTOR SIGNALING PATHWAY | 35/119 | 6.01347705e-24 | 6.274061056e-23 | 12.90125938 | 689.805041 | GSK3B;PIK3R1;TNF;RELA;IKBKB;PPP3CA;MAPK9;MAPK8;AKT1;MAPK1;MAP3K8;FYN;MAP2K7;PAK4;MAPK3;MAP2K1;JUN;MAP2K2;PDPK1;PTPN11;MAPK14;NFKB1;MAPK12;BUB1B-PAK6;MAPK13;MAPK10;MAPK11;PTPRC;PIK3CA;CDK4;LCK;PRKCQ;PTPN6;PKN1;MAP3K14 | 35 | 119 | Immune system | PPAR SIGNALING PATHWAY | 12/76 | 6.014544749e-10 | 4.939217051e-09 | 0 | 0 | 13.80407303 | 293.083527 | CPT1A;FABP4;ADIPOQ;UCP1;APOA1;LPL;APOC3;PPARG;CD36;PLIN1;PPARA;PPARD | 12 | 76 | Endocrine system |
| STEROID HORMONE BIOSYNTHESIS | 27/63 | 8.41432592e-24 | 8.38453167e-23 | 22.98452381 | 1221.216167 | UGT1A10;CYP3A4;CYP19A1;CYP3A5;CYP17A1;HSD11B1;HSD17B1;HSD17B2;CYP1B1;UGT1A4;UGT1A3;UGT2A1;UGT1A9;UGT1A8;UGT1A6;CYP3A7-CYP3A51P;UGT1A1;AKR1C1;UGT2B15;AKR1C3;UGT2B17;AKR1C2;AKR1C4;SULT2B1;CYP1A2;CYP1A1;CYP2E1 | 27 | 63 | Lipid metabolism | FLUID SHEAR STRESS AND ATHEROSCLEROSIS | 15/138 | 9.506256212e-10 | 7.577045393e-09 | 0 | 0 | 9.053030303 | 188.0667534 | EDN1;GSTM1;PRKAA2;CYBA;TNF;NFKB1;ICAM1;TNFRSF1A;IFNG;AKT2;IL1B;BCL2;AKT1;HMOX1;CCL2 | 15 | 138 | Cardiovascular disease |
| INSULIN SIGNALING PATHWAY | 37/137 | 8.572045158e-24 | 8.38453167e-23 | 11.48372581 | 609.9412827 | PYGB;GSK3B;PRKAA1;PRKAA2;PRKAG1;PRKAG2;PYGM;PIK3R1;PYGL;PRKCZ;PRKAG3;PTPRF;IKBKB;MAPK9;MAPK8;PRKACG;AKT1;MAPK1;PRKACB;MAPK3;PTPN1;MAP2K1;PRKAB2;PRKCI;MAP2K2;PKLR;PDPK1;INSR;PRKCA;BRAF;PRKAB1;MTOR;MAPK10;PIK3CA;RPS6KB1;CALM2;FBP1 | 37 | 137 | Endocrine system | CGMP-PKG SIGNALING PATHWAY | 16/161 | 9.872380602e-10 | 7.644043266e-09 | 0 | 0 | 8.213347319 | 170.3128727 | IRS1;INSR;ATP2A3;NFATC2;ATP2A2;IRS2;PIK3CG;ADRA2A;INS;ADCY5;EDNRA;KCNU1;EDNRB;ADRB3;AKT2;AKT1 | 16 | 161 | Signal transduction |
| Pathways associated with Nishamalaki | | | | | | | | | | Pathways associated with Fatty liver | | | | | | | | | | | |
| Term | Overlap | P-value | Adjusted P-value | Odds Ratio | Combined Score | Genes | Overlap gene number | Total genes in pathway | Subclass | Term | Overlap | P-value | Adjusted P-value | Old P-value | Old Adjusted P-value | Odds Ratio | Combined Score | Genes | Overlap gene number | Total genes in pathway | Subclass |
| NEUROACTIVE LIGAND-RECEPTOR INTERACTION | 78/366 | 5.156230647e-43 | 1.582962808e-40 | 9.576466049 | 932.4696273 | CHRM2;GABRB2;CHRM3;VIPR1;CHRM1;THRB;CHRM4;CHRM5;NPY2R;MLNR;HTR2B;HTR2C;ADRA1D;PTH1R;HTR2A;NR3C1;ADRA1A;MC4R;HTR6;EDNRA;CYSLTR1;EDNRB;HTR7;CCKAR;ADORA3;ADORA1;BDKRB2;BDKRB1;CTSG;PRSS3;PRSS2;GHSR;GPR35;AVPR2;TACR2;NPY1R;AVPR1A;TACR1;OPRM1;ADRA2C;ADRA2B;GABRG2;ADRA2A;MC3R;CCKBR;ADRB3;ADORA2A;AGTR1;AGTR2;NPSR1;OPRD1;GIPR;CHRNA4;PTAFR;ADRB1;PLG;ADRB2;CRHR1;CRHR2;CALCR;HRH3;TBXA2R;HRH2;NLRP3;DRD1;DRD2;DRD3;NTSR1;DRD4;GABRA2;GABRA1;GCGR;HTR1D;HTR1A;OPRK1;TRPV1;TSHR;MC5R | 78 | 366 | Signaling molecules and interaction | LIPID AND ATHEROSCLEROSIS | 22/214 | 5.087534552e-19 | 1.338021587e-16 | 0 | 0 | 17.73695866 | 747.1219093 | XBP1;MMP1;PRKCA;VLDLR;SOD2;TNF;IL6;DDIT3;IL1B;AKT1;NLRP3;CYCS;CCL2;PPARG;IKBKG;APOB;JAK2;TLR4;LDLR;MYD88;ATF4;NFE2L2 | 22 | 214 | Cardiovascular disease |
| CALCIUM SIGNALING PATHWAY | 60/250 | 2.754472214e-36 | 1.600255819e-34 | 10.86134692 | 889.3253205 | RET;CHRM2;CHRM3;CHRM1;FLT1;FLT4;CHRM5;HTR2B;ATP2A3;ATP2A2;HTR2C;ADRA1D;HTR2A;ADRA1A;PPP3CA;HTR6;EDNRA;CYSLTR1;HTR7;EDNRB;CCKAR;PRKACG;KDR;BDKRB2;CD38;BDKRB1;PRKACB;PRKCG;PDGFRB;CAMK1D;PRKCB;SPHK1;TACR2;PRKCA;TACR1;AVPR1A;CCKBR;ADRB3;ADORA2A;AGTR1;ORAI1;CAMK1G;CAMK2B;CAMK2D;PTAFR;CAMK2A;ADRB1;CACNA1C;ADRB2;STK4;EGFR;HRH2;TBXA2R;ERBB2;DRD1;CAMK2G;NTSR1;CAMK4;GNAS;CALM2 | 60 | 250 | Signal transduction | ALCOHOLIC LIVER DISEASE | 18/140 | 1.717282567e-17 | 2.258226576e-15 | 0 | 0 | 22.22012264 | 857.7679147 | SREBF1;CPT1A;ADIPOQ;TNF;SIRT1;IL6;ALDH2;ALDH1B1;SCD;IL1B;AKT1;CYP2E1;IKBKG;PPARA;PPARGC1A;TLR4;MYD88;C1QC | 18 | 140 | Endocrine and metabolic disease |
| CAMP SIGNALING PATHWAY | 57/222 | 3.127535803e-36 | 1.600255819e-34 | 11.83351159 | 967.4229183 | CHRM2;CHRM1;ATP2A3;ATP2A2;HTR6;EDNRA;PRKACG;ADORA1;AKT1;EP300;PDE4A;PRKACB;GHSR;ABCC4;MAP2K1;MAP2K2;PDE4D;NPY1R;PRKCA;ADORA2A;PIK3CA;ORAI1;PPARA;CFTR;RAPGEF3;RAPGEF4;CAMK2B;CAMK2D;GIPR;CAMK2A;ADRB1;PIK3R1;CACNA1C;ADRB2;CRHR1;RELA;CRHR2;MAPK9;MAPK8;MAPK1;DRD1;DRD2;CAMK2G;MAPK3;CREBBP;JUN;HTR1D;HTR1A;BRAF;NFKB1;TSHR;MAPK10;CAMK4;PDE3A;GNAS;PKN1;CALM2 | 57 | 222 | Signal transduction | NON-ALCOHOLIC Fatty liver DISEASE | 18/154 | 9.746123108e-17 | 8.544101258e-15 | 0 | 0 | 19.91861248 | 734.3410202 | SREBF1;XBP1;TGFB1;ADIPOQ;NR1H3;TNF;INS;IL1A;IL6;DDIT3;IL1B;LEP;AKT1;CYCS;PPARG;CYP2E1;PPARA;ATF4 | 18 | 154 | Endocrine and metabolic disease |
| FOXO SIGNALING PATHWAY | 44/130 | 6.781116357e-34 | 2.602253402e-32 | 17.19925452 | 1313.571589 | PRKAA1;PRKAA2;PRKAG1;PRKAG2;PIK3R1;STK4;PRKAG3;EGFR;IGF1R;CCNB3;IKBKB;CCNB2;MAPK9;CCNB1;MAPK8;CCND1;AKT1;EP300;MAPK1;MAPK3;PLK4;MAP2K1;PRKAB2;CREBBP;MAP2K2;SMAD3;PDPK1;INSR;PLK1;STAT3;BRAF;CSNK1E;MAPK14;PRKAB1;SIRT1;MAPK12;MAPK13;MAPK10;MAPK11;PIK3CA;CDK2;MDM2;ATM;SGK1 | 44 | 130 | Signal transduction | INSULIN RESISTANCE | 14/107 | 5.85741462e-14 | 3.08100009e-12 | 0 | 0 | 22.03201912 | 671.2821996 | SREBF1;CPT1A;PRKCE;PRKCD;PTEN;SLC2A2;NR1H3;TNF;INS;IL6;AKT1;TRIB3;PPARA;PPARGC1A | 14 | 107 | Endocrine and metabolic disease |
| NEUROACTIVE LIGAND SIGNALING | 51/195 | 6.290667457e-33 | 2.145816566e-31 | 12.01668136 | 890.9917511 | CHRM2;ACHE;GABRB2;CHRM3;CHRM1;CHRM4;CHRM5;HTR2B;HTR2C;ADRA1D;HTR2A;SLC6A2;ADRA1A;SLC6A3;SLC6A4;MC4R;HTR6;HTR7;PRKACG;ADORA3;ADORA1;PRKACB;PRKCG;PRKCB;TACR2;PRKCA;HTR3A;TACR1;OPRM1;ADRA2C;ADRA2B;GABRG2;ADRA2A;MC3R;ADORA2A;OPRD1;CHRNA4;ADRB1;ADRB2;HRH3;HRH2;DRD1;DRD2;DRD3;DRD4;GABRA2;GABRA1;HTR1D;HTR1A;OPRK1;GNAS | 51 | 195 | Signaling molecules and interaction | AGE-RAGE SIGNALING PATHWAY IN DIABETIC COMPLICATIONS | 13/100 | 5.222129684e-13 | 2.289033511e-11 | 0 | 0 | 21.71501014 | 614.115707 | TGFB1;PRKCE;SERPINE1;PRKCD;PRKCA;TNF;IL1A;IL6;COL3A1;IL1B;AKT1;CCL2;JAK2 | 13 | 100 | Endocrine and metabolic disease |

**Table S8: Top 20 enriched KEGG pathways identified from pathway enrichment analyses of formulation-associated and disease-associated target gene sets. Enrichment analysis was performed separately for each formulation and disease condition. Pathways were ranked based on statistical significance, and the top 20 pathways satisfying the predefined filtering criteria (adjusted p-value < 0.01 and minimum pathway size ≥ 50 genes). The table reports pathway-level statistics, including overlap ratio, p-value, adjusted p-value, odds ratio, combined score, overlapping genes, total pathway gene count, and KEGG subclass annotation.**

| **Pathways associated with Arogyavardhini** | | | | | | | | | |
|---|---|---|---|---|---|---|---|---|---|
| Term | Overlap | P-value | Adjusted P-value | Odds Ratio | Combined Score | Genes | Overlap gene number | Total genes in pathway | Subclass |
| LIPID AND ATHEROSCLEROSIS | 53/214 | 6.994450776e-33 | 2.147296388e-31 | 11.19895564 | 829.1728145 | GSK3B;CXCL8;TNF;IKBKB;PPP3CA;CASP7;CASP1;AKT1;MAP3K5;MMP1;PDPK1;MMP3;PRKCA;MMP9;CYP2A6;PIK3CA;PPARG;TP53;CAMK2B;CAMK2D;SRC;CAMK2A;LY96;PIK3R1;RELA;HSPD1;MAPK9;MAPK8;RXRA;NLRP3;MAPK1;CAMK2G;MAP2K6;MAPK3;CYP2J2;HSPA8;JUN;HSPA5;STAT3;MAPK14;SELE;PTK2;NFKB1;MAPK12;MAPK13;MAPK10;CYP2C9;MAPK11;CYP2C8;CYP1A1;CALM2;HSPA1A;NFE2L2 | 53 | 214 | Cardiovascular disease |
| INFLAMMATORY MEDIATOR REGULATION OF TRP CHANNELS | 37/97 | 6.901218479e-31 | 1.765561728e-29 | 20.50814687 | 1424.258805 | CAMK2B;CAMK2D;SRC;CAMK2A;HTR2B;HTR2C;ALOX12;PIK3R1;HTR2A;MAPK9;MAPK8;PRKACG;BDKRB2;BDKRB1;TRPM8;PRKACB;CAMK2G;MAP2K6;PRKCG;CYP2J2;PRKCH;TRPA1;PRKCB;PRKCE;PRKCD;TRPV3;PRKCA;TRPV1;MAPK14;MAPK12;MAPK13;MAPK10;MAPK11;PIK3CA;GNAS;PRKCQ;CALM2 | 37 | 97 | Sensory system |
| HORMONE SIGNALING | 51/218 | 2.208803541e-30 | 5.216174515e-29 | 10.3493014 | 706.7030995 | VIPR1;THRB;MLNR;ADRA1D;PTH1R;ADRA1A;IGF1R;MC4R;CCKAR;PRKACG;PRKACB;GHSR;AVPR2;PRKCA;AVPR1A;OPRM1;ADRA2C;ADRA2B;ADRA2A;MC3R;CCKBR;ADRB3;PIK3CA;AGTR1;AGTR2;RAPGEF3;RAPGEF4;OPRD1;GIPR;SRC;ADRB1;PIK3R1;ADRB2;CRHR1;CRHR2;CALCR;THPO;NLRP3;DRD1;DRD2;DRD3;DRD4;INSR;GCGR;STAT3;OPRK1;ESR1;ESR2;TSHR;MC5R;GNAS | 51 | 218 | Signaling molecules and interaction |
| GAP JUNCTION | 34/85 | 2.620956367e-29 | 5.364224031e-28 | 22.0673516 | 1452.28393 | SRC;HTR2B;HTR2C;ADRB1;HTR2A;EGFR;TUBB8;TUBA1C;TUBA3E;TUBB6;TUBA1B;TUBA1A;TUBA3C;TUBB3;PRKACG;MAPK1;DRD1;DRD2;PRKACB;MAPK3;PRKCG;PDGFRB;MAP2K1;MAP2K2;PRKCB;TUBB;PRKCA;TUBB4B;TUBB4A;TUBA4A;TUBB2B;TUBB2A;GNAS;CDK1 | 34 | 85 | Cellular community - eukaryotes |
| SEROTONERGIC SYNAPSE | 35/111 | 5.491526708e-26 | 9.917051173e-25 | 15.2502934 | 887.0181586 | APP;GABRB2;MAOB;MAOA;ALOX15;HTR2B;HTR2C;ALOX12;CACNA1C;HTR2A;CYP2C19;PTGS2;SLC6A4;PTGS1;HTR6;HTR7;PRKACG;ALOX5;MAPK1;PRKACB;MAPK3;PRKCG;CYP2J2;MAP2K1;PRKCB;HTR1D;HTR1A;PRKCA;BRAF;HTR3A;ALOX15B;CYP2C9;CYP2C8;GNAS;RAPGEF3 | 35 | 111 | Nervous system |
| INSULIN RESISTANCE | 34/107 | 2.099271838e-25 | 3.391981338e-24 | 15.39937136 | 875.0390455 | GSK3B;PRKAA1;PRKAA2;PRKAG1;PRKAG2;SLC2A2;PIK3R1;PYGL;TNF;PRKCZ;PRKAG3;RELA;IKBKB;RPS6KA3;MAPK9;MAPK8;AKT1;PTPN1;PRKAB2;PRKCB;PDPK1;PRKCE;INSR;PRKCD;STAT3;NR1H3;PRKAB1;NFKB1;MTOR;MAPK10;PIK3CA;RPS6KB1;PRKCQ;PPARA | 34 | 107 | Endocrine and metabolic disease |
| AGE-RAGE SIGNALING PATHWAY IN DIABETIC COMPLICATIONS | 33/100 | 2.697815803e-25 | 4.141147258e-24 | 16.26215078 | 919.9854001 | CXCL8;SERPINE1;PIK3R1;TNF;PRKCZ;RELA;MAPK9;MAPK8;CCND1;AKT1;MAPK1;MAPK3;SMAD2;JUN;SMAD3;PRKCB;PRKCE;MMP2;PRKCD;STAT3;PRKCA;MAPK14;SELE;F3;NFKB1;MAPK12;MAPK13;MAPK10;MAPK11;PIK3CA;CDK4;AGTR1;NOX4 | 33 | 100 | Endocrine and metabolic disease |
| HIF-1 SIGNALING PATHWAY | 34/110 | 5.910269154e-25 | 8.64025062e-24 | 14.78920332 | 825.0591274 | CAMK2B;CAMK2D;FLT1;SERPINE1;CAMK2A;PIK3R1;HIF1A;EGFR;RELA;IGF1R;ERBB2;AKT1;EP300;MAPK1;CAMK2G;MAPK3;PRKCG;EGLN1;MAP2K1;CREBBP;EGLN3;MAP2K2;EGLN2;PRKCB;PDPK1;INSR;STAT3;PRKCA;NFKB1;MTOR;PIK3CA;RPS6KB1;TEK;GAPDH | 34 | 110 | Signal transduction |
| MAPK SIGNALING PATHWAY | 52/296 | 1.181587168e-24 | 1.533755504e-23 | 7.205989689 | 397.0153252 | RET;FLT1;FLT3;FLT4;HSPB1;TNF;IGF1R;IKBKB;RPS6KA3;PPP3CA;RPS6KA5;PRKACG;KDR;AKT1;MAP3K8;PRKACB;MAP3K5;PRKCG;PDGFRB;MAP2K1;DUSP3;MAP2K2;PRKCB;PRKCA;MAPKAPK2;MAPKAPK5;MAPT;TP53;CACNA1C;STK4;EGFR;RELA;MAPK9;MAPK8;ERBB2;MAPK1;MAP2K6;MAPK3;HSPA8;JUN;INSR;BRAF;MAPK14;NFKB1;MAPK12;NFKB2;MAPK13;MAPK10;MAPK11;PKN1;TEK;HSPA1A | 52 | 296 | Signal transduction |
| CELLULAR SENESCENCE | 38/156 | 1.376416829e-23 | 1.637048334e-22 | 10.69596727 | 563.0354339 | CXCL8;SERPINE1;PIK3R1;RELA;CCNB3;CCNB2;PPP3CA;CCNB1;CCND1;CHEK2;CHEK1;AKT1;MAPK1;MAP2K6;MAPK3;SMAD2;MAP2K1;MAP2K2;SMAD3;MAPK14;SIRT1;NFKB1;MTOR;MAPK12;MAPK13;CCNA2;CCNA1;MAPK11;CDK6;PIK3CA;CDK4;MAPKAPK2;CDK2;MDM2;CDK1;ATM;TP53;CALM2 | 38 | 156 | Cell growth and death |
| ERBB SIGNALING PATHWAY | 29/83 | 3.426990532e-23 | 3.889383412e-22 | 17.62283846 | 911.5902166 | CAMK2B;GSK3B;CAMK2D;SRC;CAMK2A;PIK3R1;EGFR;MAPK9;MAPK8;ERBB2;AKT1;MAPK1;CAMK2G;PAK4;MAPK3;PRKCG;MAP2K1;JUN;MAP2K2;PRKCB;PRKCA;BRAF;MTOR;PTK2;BUB1B-PAK6;MAPK10;PIK3CA;RPS6KB1;PKN1 | 29 | 83 | Signal transduction |
| SPHINGOLIPID SIGNALING PATHWAY | 34/123 | 3.547320376e-23 | 3.889383412e-22 | 12.62047868 | 652.3937445 | OPRD1;PIK3R1;TNF;PRKCZ;RELA;MAPK9;MAPK8;ADORA3;SMPD1;ADORA1;BDKRB2;AKT1;MAPK1;FYN;CTSD;MAP3K5;MAPK3;PRKCG;MAP2K1;ABCC1;MAP2K2;PRKCB;PDPK1;PRKCE;SPHK1;PRKCA;MAPK14;NFKB1;MAPK12;MAPK13;MAPK10;MAPK11;PIK3CA;TP53 | 34 | 123 | Signal transduction |
| NEUROTROPHIN SIGNALING PATHWAY | 33/119 | 1.401401912e-22 | 1.43410129e-21 | 12.6568873 | 636.8869614 | CAMK2B;GSK3B;CAMK2D;CAMK2A;PIK3R1;RELA;IKBKB;RPS6KA3;MAPK9;MAPK8;RPS6KA5;AKT1;MAPK1;CAMK2G;MAP3K5;MAPK3;MAP2K1;JUN;MAP2K2;PDPK1;PRKCD;BRAF;MAPK14;NFKB1;MAPK12;MAPK13;MAPK10;MAPK11;PIK3CA;CAMK4;MAPKAPK2;TP53;CALM2 | 33 | 119 | Nervous system |
| BILE SECRETION | 29/88 | 2.320537369e-22 | 2.226265538e-21 | 16.12520503 | 803.2782699 | UGT1A10;ABCB1;HMGCR;ABCB11;CYP3A4;SLCO1B1;RXRA;CA2;PRKACG;SLCO1B3;UGT1A4;UGT1A3;UGT2A1;UGT1A9;PRKACB;UGT1A8;UGT1A6;ABCC3;ABCC4;ABCC2;UGT1A1;SLC10A1;UGT2B15;NR1H4;UGT2B17;PRKCA;GNAS;CFTR;ABCG2 | 29 | 88 | Digestive system |
| NON-ALCOHOLIC Fatty liver DISEASE | 36/154 | 9.377452247e-22 | 8.467287764e-21 | 10.09820024 | 488.9403542 | GSK3B;GSK3A;PRKAA1;CXCL8;PRKAA2;PRKAG1;PRKAG2;PIK3R1;PTGS2;TNF;PRKAG3;RELA;PTGS1;IKBKB;MAPK9;CASP7;MAPK8;RXRA;AKT1;MAP3K5;PRKAB2;JUN;PKLR;INSR;NR1H3;MAPK14;PRKAB1;NFKB1;MAPK12;MAPK13;MAPK10;MAPK11;PIK3CA;PPARG;CYP2E1;PPARA | 36 | 154 | Endocrine and metabolic disease |

| **Pathways associated with Type 2 diabetes** | | | | | | | | | | | |
|---|---|---|---|---|---|---|---|---|---|---|---|
| Term | Overlap | P-value | Adjusted P-value | Old P-value | Old Adjusted P-value | Odds Ratio | Combined Score | Genes | Overlap gene number | Total genes in pathway | Subclass |
| ADIPOCYTOKINE SIGNALING PATHWAY | 11/70 | 4.009636278e-12 | 1.50647763e-10 | 0 | 0 | 26.73937607 | 701.7032791 | POMC;CPT1A;LEP;ADIPOQ;AKT1;IKBKG;JAK2;TNFRSF1B;PPARA;TNF;PPARGC1A | 11 | 70 | Endocrine system |
| TNF SIGNALING PATHWAY | 12/118 | 8.098693299e-11 | 2.129956338e-09 | 0 | 0 | 16.31593444 | 379.1290171 | IL6;CEBPB;CSF2;IL1B;LIF;AKT1;CCL2;IKBKG;TNFRSF1B;TNF;ATF4;BIRC3 | 12 | 118 | Signal transduction |
| PPAR SIGNALING PATHWAY | 10/76 | 2.434813543e-10 | 5.821417834e-09 | 0 | 0 | 21.56638326 | 477.3930446 | CPT1A;MMP1;SCD;ADIPOQ;NR1H3;PLIN2;PPARG;ANGPTL4;PPARA;PPARD | 10 | 76 | Endocrine system |
| FLUID SHEAR STRESS AND ATHEROSCLEROSIS | 12/138 | 5.0898822e-10 | 1.029722322e-08 | 0 | 0 | 13.71220021 | 293.4218359 | IL1A;NQO1;GSTM1;IL1B;GSTP1;GSTA1;AKT1;CCL2;GSTT1;IKBKG;TNF;NFE2L2 | 12 | 138 | Cardiovascular disease |
| AMPK SIGNALING PATHWAY | 11/121 | 1.711716576e-09 | 3.00120973e-08 | 0 | 0 | 14.30507246 | 288.7588901 | SREBF1;CPT1A;SCD;LEP;ADIPOQ;AKT1;PPARG;HMGCR;PPARGC1A;SIRT1;INS | 11 | 121 | Signal transduction |
| LONGEVITY REGULATING PATHWAY | 9/87 | 1.734476956e-08 | 2.524697084e-07 | 0 | 0 | 16.29642857 | 291.2167686 | ADIPOQ;AKT1;PRKCA;PPARG;SOD2;SIRT1;PPARGC1A;ATF4;INS | 9 | 87 | Aging |
| BILE SECRETION | 9/88 | 1.919921737e-08 | 2.524697084e-07 | 0 | 0 | 16.08933092 | 285.8816084 | SCARB1;ABCC2;ABCB4;NR1H4;PRKCA;HMGCR;SLC22A8;LDLR;NR0B2 | 9 | 88 | Digestive system |
| IL-17 SIGNALING PATHWAY | 9/93 | 3.131697814e-08 | 3.743802387e-07 | 0 | 0 | 15.12780612 | 261.3949573 | IL4;CEBPB;IL6;CSF2;MMP1;IL1B;CCL2;IKBKG;TNF | 9 | 93 | Immune system |
| PI3K-AKT SIGNALING PATHWAY | 15/355 | 6.926134672e-08 | 7.286293675e-07 | 0 | 0 | 6.423726953 | 105.8975725 | LAMA1;F2R;PTEN;PRKCA;INS;IL4;IL3;IL6;MYC;AKT1;IKBKG;JAK2;TLR4;ATF4;FGF21 | 15 | 355 | Signal transduction |
| OVARIAN STEROIDOGENESIS | 7/51 | 9.727689946e-08 | 9.1370802e-07 | 0 | 0 | 22.1909411 | 358.288373 | SCARB1;CYP1B1;PRKCA;CYP19A1;LDLR;INS;CYP17A1 | 7 | 51 | Endocrine system |
| CYTOKINE-CYTOKINE RECEPTOR INTERACTION | 13/296 | 3.482199896e-07 | 2.616624493e-06 | 0 | 0 | 6.609436708 | 98.28517517 | IL1RN;CSF2;TGFB1;LIF;TNFRSF1B;TNF;IL4;IL3;IL1A;IL6;IL1B;LEP;CCL2 | 13 | 296 | Signaling molecules and interaction |
| MAPK SIGNALING PATHWAY | 13/296 | 3.482199896e-07 | 2.616624493e-06 | 0 | 0 | 6.609436708 | 98.28517517 | TGFB1;PRKCA;TNF;INS;IL1A;MYC;DDIT3;IL1B;AKT1;IKBKG;MYD88;ATF4;FGF21 | 13 | 296 | Signal transduction |
| DIABETIC CARDIOMYOPATHY | 11/203 | 3.655087071e-07 | 2.670244165e-06 | 0 | 0 | 8.161571558 | 120.9706156 | COL3A1;ACE;TGFB1;ATP5F1B;PRKCD;PDK4;PTEN;AKT1;PRKCA;PPARA;INS | 11 | 203 | Cardiovascular disease |
| GLUCAGON SIGNALING PATHWAY | 8/104 | 1.106541134e-06 | 6.929055199e-06 | 0 | 0 | 11.67553191 | 160.1214146 | CPT1A;AKT1;SLC2A2;PRKCA;PPARA;PPARGC1A;SIRT1;ATF4 | 8 | 104 | Endocrine system |
| NOD-LIKE RECEPTOR SIGNALING PATHWAY | 10/185 | 1.292425469e-06 | 7.725179506e-06 | 0 | 0 | 8.088797533 | 109.675924 | IL6;IL1B;PRKCD;NLRP3;CCL2;IKBKG;TNF;TLR4;MYD88;BIRC3 | 10 | 185 | Immune system |

| Pathways associated with Obesity | | | | | | | | | | | |
|---|---|---|---|---|---|---|---|---|---|---|---|
| Term | Overlap | P-value | Adjusted P-value | Old P-value | Old Adjusted P-value | Odds Ratio | Combined Score | Genes | Overlap gene number | Total genes in pathway | Subclass |
| AMPK SIGNALING PATHWAY | 22/121 | 1.782268471e-19 | 4.919060979e-17 | 0 | 0 | 18.9013949 | 815.9964627 | SREBF1;PFKFB3;IRS1;INSR;ADIPOQ;CIDEA;IGF1;SLC2A4;FOXO3;ADIPOR1;ACACB;SIRT1;INS;TBC1D1;PIK3CA;SCD;FASN;LEP;AKT1;PPARG;PCK1;PPARGC1A | 22 | 121 | Signal transduction |
| HORMONE SIGNALING | 27/218 | 3.961677794e-19 | 5.467115355e-17 | 0 | 0 | 12.23212714 | 518.3051851 | GLP1R;IRS1;PMCH;ADRB1;ADRB2;ADIPOR1;ADRA1B;INS;MC4R;HCRT;INSR;STAT3;ADIPOQ;IGF2;GCG;IGF1;ESR1;POMC;MC3R;GH1;ADRB3;PIK3CA;LEP;GNAS;GNB3;GHRL;RAPGEF3 | 27 | 218 | Signaling molecules and interaction |
| NON-ALCOHOLIC Fatty liver DISEASE | 23/154 | 2.645842253e-18 | 2.434174873e-16 | 0 | 0 | 14.9740458 | 606.0526751 | SREBF1;CEBPA;IRS1;INSR;ADIPOQ;FOS;ADIPOR1;PTGS2;TNF;COX7C;INS;IL6;PIK3CA;DDIT3;FASN;IL1B;LEP;AKT1;CYCS;PPARG;UQCRC2;CYP2E1;PPARA | 23 | 154 | Endocrine and metabolic disease |
| ADIPOCYTOKINE SIGNALING PATHWAY | 17/70 | 1.372872059e-17 | 9.472817206e-16 | 0 | 0 | 26.76670931 | 1039.272137 | IRS1;ACSL1;ADIPOQ;STAT3;SLC2A4;TNFRSF1B;ADIPOR1;TNF;ACACB;POMC;NPY;LEP;AKT1;AGRP;PPARA;PCK1;PPARGC1A | 17 | 70 | Endocrine system |
| REGULATION OF LIPOLYSIS IN ADIPOCYTES | 15/58 | 4.10618511e-16 | 2.266614181e-14 | 0 | 0 | 28.88020324 | 1023.192882 | IRS1;INSR;NPY1R;AQP7;ADRB1;ADRB2;PTGS2;INS;PIK3CA;ADRB3;NPY;GNAS;AKT1;PLIN1;PNPLA2 | 15 | 58 | Endocrine system |
| INSULIN RESISTANCE | 17/107 | 2.670662968e-14 | 1.228504965e-12 | 0 | 0 | 15.73300377 | 491.7171688 | SREBF1;PTPN1;IRS1;INSR;GFPT1;STAT3;SLC2A4;TNF;ACACB;INS;IL6;PIK3CA;AKT1;PCK1;PPARA;PPARGC1A;PPARGC1B | 17 | 107 | Endocrine and metabolic disease |
| LIPID AND ATHEROSCLEROSIS | 22/214 | 4.424983127e-14 | 1.744707633e-12 | 0 | 0 | 9.699900794 | 298.2615204 | CD40;HSPA5;STAT3;VLDLR;FOS;SOD2;TNF;MMP9;ICAM1;IL6;PIK3CA;DDIT3;FASN;IL1B;CASP1;AKT1;CYCS;CCL2;PPARG;LBP;APOB;LDLR | 22 | 214 | Cardiovascular disease |
| ALCOHOLIC LIVER DISEASE | 17/140 | 2.475563621e-12 | 8.540694492e-11 | 0 | 0 | 11.49262781 | 307.1353412 | SREBF1;ADIPOQ;FOXO3;ADIPOR1;TNF;ACACB;SIRT1;IL6;SCD;FASN;IL1B;AKT1;CYP2E1;ACADM;LBP;PPARA;PPARGC1A | 17 | 140 | Endocrine and metabolic disease |
| NEUROACTIVE LIGAND-RECEPTOR INTERACTION | 25/366 | 8.233715088e-12 | 2.52500596e-10 | 0 | 0 | 6.240031898 | 159.262985 | GLP1R;PMCH;HTR2C;ADRB1;ADRB2;HTR2A;NR3C1;ADRA1B;MC4R;NPY;PCDHA4;HCRT;NPY5R;NPY1R;GCG;TRPV1;TAAR1;POMC;MC3R;PYY;GH1;ADRB3;VGF;LEP;GHRL | 25 | 366 | Signaling molecules and interaction |
| PPAR SIGNALING PATHWAY | 13/76 | 1.174433815e-11 | 3.241437329e-10 | 0 | 0 | 16.92407407 | 425.9391704 | ACSL1;ADIPOQ;UCP1;AQP7;LPL;SCD;PPARG;ACADM;PLIN1;ANGPTL4;PPARA;PCK1;PPARD | 13 | 76 | Endocrine system |
| LONGEVITY REGULATING PATHWAY | 13/87 | 6.956873193e-11 | 1.745542728e-09 | 0 | 0 | 14.40028153 | 336.8039497 | IRS1;INSR;ADIPOQ;IGF1;FOXO3;ADIPOR1;SOD2;SIRT1;INS;PIK3CA;AKT1;PPARG;PPARGC1A | 13 | 87 | Aging |
| INSULIN SIGNALING PATHWAY | 15/137 | 2.20756449e-10 | 5.077398328e-09 | 0 | 0 | 10.13827662 | 225.4140477 | SREBF1;PTPN1;IRS1;INSR;INPPL1;SLC2A4;ACACB;INS;PIK3CA;PRKAR1A;PRKAR2B;FASN;AKT1;PCK1;PPARGC1A | 15 | 137 | Endocrine system |
| FOXO SIGNALING PATHWAY | 14/130 | 1.113127822e-09 | 2.363255991e-08 | 0 | 0 | 9.913215986 | 204.3717721 | IRS1;INSR;STAT3;IGF1;SLC2A4;FOXO3;SOD2;SIRT1;INS;IL6;PIK3CA;AKT1;EP300;PCK1 | 14 | 130 | Signal transduction |
| HIF-1 SIGNALING PATHWAY | 13/110 | 1.393440416e-09 | 2.747068249e-08 | 0 | 0 | 10.97293814 | 223.7545588 | PFKFB3;TFRC;INSR;SERPINE1;STAT3;IGF1;INS;TF;IL6;PIK3CA;AKT1;EP300;HMOX1 | 13 | 110 | Signal transduction |
| TNF SIGNALING PATHWAY | 13/118 | 3.342764817e-09 | 6.150687263e-08 | 0 | 0 | 10.13277778 | 197.756029 | FOS;TNFRSF1B;PTGS2;TNF;MMP9;ICAM1;IL6;PIK3CA;FASN;IL1B;AKT1;CCL2;DNM1L | 13 | 118 | Signal transduction |
| LONGEVITY REGULATING PATHWAY - MULTIPLE SPECIES | 10/60 | 3.801025186e-09 | 6.556768446e-08 | 0 | 0 | 16.21152263 | 314.3089201 | PIK3CA;IRS1;INSR;AKT1;IGF1;FOXO3;SOD2;SIRT1;INS;SOD1 | 10 | 60 | Aging |
| CAMP SIGNALING PATHWAY | 16/222 | 2.521242841e-08 | 4.093311907e-07 | 0 | 0 | 6.403998198 | 112.0438963 | GLP1R;BDNF;NPY1R;GCG;ADRB1;FOS;ADRB2;POMC;PIK3CA;NPY;GNAS;AKT1;EP300;GHRL;PPARA;RAPGEF3 | 16 | 222 | Signal transduction |
| NEUROACTIVE LIGAND SIGNALING | 15/195 | 2.926405784e-08 | 4.487155535e-07 | 0 | 0 | 6.851190476 | 118.8469556 | ACHE;HTR2C;ADRB1;ADRB2;HTR2A;ADRA1B;SLC6A2;SLC6A3;MC4R;POMC;MC3R;GNAS;GNB3;AGRP;SLC18A2 | 15 | 195 | Signaling molecules and interaction |
| GROWTH HORMONE SYNTHESIS, SECRETION AND ACTION | 11/118 | 3.102178654e-07 | 3.891824129e-06 | 0 | 0 | 8.34324554 | 125.0318025 | GH1;PIK3CA;IRS1;STAT3;GNAS;AKT1;EP300;GHRL;IGF1;FOS;MRAP2 | 11 | 118 | Endocrine system |
| DIABETIC CARDIOMYOPATHY | 14/203 | 3.288293041e-07 | 3.94595165e-06 | 0 | 0 | 6.06167674 | 90.48705585 | ACE;PARP1;IRS1;INSR;GFPT1;SLC2A4;PTGS2;COX7C;MMP9;INS;PIK3CA;AKT1;UQCRC2;PPARA | 14 | 203 | Cardiovascular disease |

**Table S9: Complete node and relationship data used to construct the knowledge graph. This table summarizes interactions among FDA-approved drugs, formulation, herbs, diseases (type 2 diabetes, fatty liver, obesity), phytochemicals, and their target genes/proteins. Each entry includes source and target node details, edge type, and node annotations (Entrez Gene ID and gene symbol, where applicable).**

| Source Node Label | Source Node ID | Source Node Name | Source Node Properties | Edge Type | Edge Properties | Target Node Label | Target Node ID | Target Node Name | Target Node Properties |
|---|---|---|---|---|---|---|---|---|---|
| FDA approved drug | - | Alogliptin | {"name": "Alogliptin"} | TREATS | - | Disease | - | Type 2 Diabetes | {"name": "Type 2 Diabetes"} |
| FDA approved drug | - | Alogliptin | {"name": "Alogliptin"} | AFFECTS | - | Target | 1803 | DPP4 | {"name": "DPP4", "id": "1803"} |
| FDA approved drug | - | Amphetamine | {"name": "Amphetamine"} | TREATS | - | Disease | - | Obesity | {"name": "Obesity"} |
| FDA approved drug | - | Amphetamine | {"name": "Amphetamine"} | AFFECTS | - | Target | 6531 | SLC6A3 | {"name": "SLC6A3", "id": "6531"} |
| FDA approved drug | - | Amphetamine | {"name": "Amphetamine"} | AFFECTS | - | Target | 134864 | TAAR1 | {"name": "TAAR1", "id": "134864"} |
| FDA approved drug | - | Amphetamine | {"name": "Amphetamine"} | AFFECTS | - | Target | 6530 | SLC6A2 | {"name": "SLC6A2", "id": "6530"} |
| FDA approved drug | - | Atorvastatin | {"name": "Atorvastatin"} | TREATS | - | Disease | - | Fatty Liver | {"name": "Fatty Liver"} |
| FDA approved drug | - | Atorvastatin | {"name": "Atorvastatin"} | AFFECTS | - | Target | 3156 | HMGCR | {"name": "HMGCR", "id": "3156"} |
| FDA approved drug | - | Atorvastatin | {"name": "Atorvastatin"} | AFFECTS | - | Target | 4615 | MYD88 | {"name": "MYD88", "id": "4615"} |
| FDA approved drug | - | Benzphetamine | {"name": "Benzphetamine"} | TREATS | - | Disease | - | Obesity | {"name": "Obesity"} |
| FDA approved drug | - | Benzphetamine | {"name": "Benzphetamine"} | AFFECTS | - | Target | 6530 | SLC6A2 | {"name": "SLC6A2", "id": "6530"} |
| FDA approved drug | - | Benzphetamine | {"name": "Benzphetamine"} | AFFECTS | - | Target | 6571 | SLC18A2 | {"name": "SLC18A2", "id": "6571"} |
| FDA approved drug | - | Bexagliflozin | {"name": "Bexagliflozin"} | TREATS | - | Disease | - | Type 2 Diabetes | {"name": "Type 2 Diabetes"} |
| FDA approved drug | - | Bexagliflozin | {"name": "Bexagliflozin"} | AFFECTS | - | Target | 6524 | SLC5A2 | {"name": "SLC5A2", "id": "6524"} |
| FDA approved drug | - | Canagliflozin | {"name": "Canagliflozin"} | TREATS | - | Disease | - | Type 2 Diabetes | {"name": "Type 2 Diabetes"} |
| FDA approved drug | - | Canagliflozin | {"name": "Canagliflozin"} | AFFECTS | - | Target | 6524 | SLC5A2 | {"name": "SLC5A2", "id": "6524"} |
| FDA approved drug | - | Dapagliflozin | {"name": "Dapagliflozin"} | TREATS | - | Disease | - | Type 2 Diabetes | {"name": "Type 2 Diabetes"} |
| FDA approved drug | - | Dapagliflozin | {"name": "Dapagliflozin"} | AFFECTS | - | Target | 6524 | SLC5A2 | {"name": "SLC5A2", "id": "6524"} |
| FDA approved drug | - | Diethylpropion | {"name": "Diethylpropion"} | TREATS | - | Disease | - | Obesity | {"name": "Obesity"} |
| FDA approved drug | - | Diethylpropion | {"name": "Diethylpropion"} | AFFECTS | - | Target | 6531 | SLC6A3 | {"name": "SLC6A3", "id": "6531"} |
| FDA approved drug | - | Diethylpropion | {"name": "Diethylpropion"} | AFFECTS | - | Target | 6530 | SLC6A2 | {"name": "SLC6A2", "id": "6530"} |
| FDA approved drug | - | Empagliflozin | {"name": "Empagliflozin"} | TREATS | - | Disease | - | Type 2 Diabetes | {"name": "Type 2 Diabetes"} |
| FDA approved drug | - | Empagliflozin | {"name": "Empagliflozin"} | AFFECTS | - | Target | 6524 | SLC5A2 | {"name": "SLC5A2", "id": "6524"} |
| FDA approved drug | - | Ertugliflozin | {"name": "Ertugliflozin"} | TREATS | - | Disease | - | Type 2 Diabetes | {"name": "Type 2 Diabetes"} |
| FDA approved drug | - | Ertugliflozin | {"name": "Ertugliflozin"} | AFFECTS | - | Target | 6524 | SLC5A2 | {"name": "SLC5A2", "id": "6524"} |
| FDA approved drug | - | Fenofibrate | {"name": "Fenofibrate"} | TREATS | - | Disease | - | Fatty Liver | {"name": "Fatty Liver"} |
| FDA approved drug | - | Fenofibrate | {"name": "Fenofibrate"} | AFFECTS | - | Target | 5465 | PPARA | {"name": "PPARA", "id": "5465"} |
| FDA approved drug | - | Fenofibric acid | {"name": "Fenofibric acid"} | TREATS | - | Disease | - | Fatty Liver | {"name": "Fatty Liver"} |
| FDA approved drug | - | Fenofibric acid | {"name": "Fenofibric acid"} | AFFECTS | - | Target | 5465 | PPARA | {"name": "PPARA", "id": "5465"} |
| FDA approved drug | - | Glimepiride | {"name": "Glimepiride"} | TREATS | - | Disease | - | Type 2 Diabetes | {"name": "Type 2 Diabetes"} |
| FDA approved drug | - | Glimepiride | {"name": "Glimepiride"} | AFFECTS | - | Target | 3767 | KCNJ11 | {"name": "KCNJ11", "id": "3767"} |
| FDA approved drug | - | Glimepiride | {"name": "Glimepiride"} | AFFECTS | - | Target | 6833 | ABCC8 | {"name": "ABCC8", "id": "6833"} |
| FDA approved drug | - | Glimepiride | {"name": "Glimepiride"} | AFFECTS | - | Target | 10060 | ABCC9 | {"name": "ABCC9", "id": "10060"} |
| FDA approved drug | - | Glipizide | {"name": "Glipizide"} | TREATS | - | Disease | - | Type 2 Diabetes | {"name": "Type 2 Diabetes"} |
| FDA approved drug | - | Glipizide | {"name": "Glipizide"} | AFFECTS | - | Target | 3767 | KCNJ11 | {"name": "KCNJ11", "id": "3767"} |
| FDA approved drug | - | Glipizide | {"name": "Glipizide"} | AFFECTS | - | Target | 6833 | ABCC8 | {"name": "ABCC8", "id": "6833"} |
| FDA approved drug | - | Gliquidone | {"name": "Gliquidone"} | TREATS | - | Disease | - | Type 2 Diabetes | {"name": "Type 2 Diabetes"} |
| FDA approved drug | - | Gliquidone | {"name": "Gliquidone"} | AFFECTS | - | Target | 3767 | KCNJ11 | {"name": "KCNJ11", "id": "3767"} |
| FDA approved drug | - | Gliquidone | {"name": "Gliquidone"} | AFFECTS | - | Target | 6833 | ABCC8 | {"name": "ABCC8", "id": "6833"} |
| FDA approved drug | - | Glyburide | {"name": "Glyburide"} | TREATS | - | Disease | - | Type 2 Diabetes | {"name": "Type 2 Diabetes"} |
| FDA approved drug | - | Glyburide | {"name": "Glyburide"} | AFFECTS | - | Target | 3767 | KCNJ11 | {"name": "KCNJ11", "id": "3767"} |
| FDA approved drug | - | Glyburide | {"name": "Glyburide"} | AFFECTS | - | Target | 6833 | ABCC8 | {"name": "ABCC8", "id": "6833"} |
| FDA approved drug | - | Icosapent ethyl | {"name": "Icosapent ethyl"} | TREATS | - | Disease | - | Fatty Liver | {"name": "Fatty Liver"} |
| FDA approved drug | - | Icosapent ethyl | {"name": "Icosapent ethyl"} | AFFECTS | - | Target | 8694 | DGAT1 | {"name": "DGAT1", "id": "8694"} |
| FDA approved drug | - | Linagliptin | {"name": "Linagliptin"} | TREATS | - | Disease | - | Type 2 Diabetes | {"name": "Type 2 Diabetes"} |
| FDA approved drug | - | Linagliptin | {"name": "Linagliptin"} | AFFECTS | - | Target | 1803 | DPP4 | {"name": "DPP4", "id": "1803"} |
| FDA approved drug | - | Lovastatin | {"name": "Lovastatin"} | TREATS | - | Disease | - | Fatty Liver | {"name": "Fatty Liver"} |
| FDA approved drug | - | Lovastatin | {"name": "Lovastatin"} | AFFECTS | - | Target | 3156 | HMGCR | {"name": "HMGCR", "id": "3156"} |
| FDA approved drug | - | Metamfetamine | {"name": "Metamfetamine"} | TREATS | - | Disease | - | Obesity | {"name": "Obesity"} |
| FDA approved drug | - | Metamfetamine | {"name": "Metamfetamine"} | AFFECTS | - | Target | 6531 | SLC6A3 | {"name": "SLC6A3", "id": "6531"} |
| FDA approved drug | - | Metformin | {"name": "Metformin"} | TREATS | - | Disease | - | Type 2 Diabetes | {"name": "Type 2 Diabetes"} |
| FDA approved drug | - | Metformin | {"name": "Metformin"} | AFFECTS | - | Target | 3643 | INSR | {"name": "INSR", "id": "3643"} |
| FDA approved drug | - | Metformin | {"name": "Metformin"} | AFFECTS | - | Target | 4535 | MT-ND1\|ND1 | {"name": "MT-ND1\|ND1", "id": "4535"} |
| FDA approved drug | - | Metformin | {"name": "Metformin"} | AFFECTS | - | Target | 126328 | NDUFA11 | {"name": "NDUFA11", "id": "126328"} |
| FDA approved drug | - | Metformin | {"name": "Metformin"} | AFFECTS | - | Target | 25915 | NDUFAF3 | {"name": "NDUFAF3", "id": "25915"} |
| FDA approved drug | - | Metformin | {"name": "Metformin"} | AFFECTS | - | Target | 29078 | NDUFAF4 | {"name": "NDUFAF4", "id": "29078"} |
| FDA approved drug | - | Metformin | {"name": "Metformin"} | AFFECTS | - | Target | 32 | ACACB | {"name": "ACACB", "id": "32"} |
| FDA approved drug | - | Metformin | {"name": "Metformin"} | AFFECTS | - | Target | 374291 | NDUFS7 | {"name": "NDUFS7", "id": "374291"} |
| FDA approved drug | - | Metformin | {"name": "Metformin"} | AFFECTS | - | Target | 4536 | MT-ND2 | {"name": "MT-ND2", "id": "4536"} |
| FDA approved drug | - | Metformin | {"name": "Metformin"} | AFFECTS | - | Target | 4537 | MT-ND3 | {"name": "MT-ND3", "id": "4537"} |
| FDA approved drug | - | Metformin | {"name": "Metformin"} | AFFECTS | - | Target | 4538 | MT-ND4 | {"name": "MT-ND4", "id": "4538"} |
| FDA approved drug | - | Metformin | {"name": "Metformin"} | AFFECTS | - | Target | 4539 | MT-ND4L | {"name": "MT-ND4L", "id": "4539"} |
| FDA approved drug | - | Metformin | {"name": "Metformin"} | AFFECTS | - | Target | 4540 | MT-ND5 | {"name": "MT-ND5", "id": "4540"} |
| FDA approved drug | - | Metformin | {"name": "Metformin"} | AFFECTS | - | Target | 4541 | MT-ND6 | {"name": "MT-ND6", "id": "4541"} |
| FDA approved drug | - | Metformin | {"name": "Metformin"} | AFFECTS | - | Target | 4694 | NDUFA1 | {"name": "NDUFA1", "id": "4694"} |
| FDA approved drug | - | Metformin | {"name": "Metformin"} | AFFECTS | - | Target | 4695 | NDUFA2 | {"name": "NDUFA2", "id": "4695"} |
| FDA approved drug | - | Metformin | {"name": "Metformin"} | AFFECTS | - | Target | 4696 | NDUFA3 | {"name": "NDUFA3", "id": "4696"} |
| FDA approved drug | - | Metformin | {"name": "Metformin"} | AFFECTS | - | Target | 4697 | COXFA4 | {"name": "COXFA4", "id": "4697"} |
| FDA approved drug | - | Metformin | {"name": "Metformin"} | AFFECTS | - | Target | 4698 | NDUFA5 | {"name": "NDUFA5", "id": "4698"} |
| FDA approved drug | - | Metformin | {"name": "Metformin"} | AFFECTS | - | Target | 4700 | NDUFA6 | {"name": "NDUFA6", "id": "4700"} |
| FDA approved drug | - | Metformin | {"name": "Metformin"} | AFFECTS | - | Target | 4701 | NDUFA7 | {"name": "NDUFA7", "id": "4701"} |
| FDA approved drug | - | Metformin | {"name": "Metformin"} | AFFECTS | - | Target | 4702 | NDUFA8 | {"name": "NDUFA8", "id": "4702"} |
| FDA approved drug | - | Metformin | {"name": "Metformin"} | AFFECTS | - | Target | 4704 | NDUFA9 | {"name": "NDUFA9", "id": "4704"} |
| FDA approved drug | - | Metformin | {"name": "Metformin"} | AFFECTS | - | Target | 4705 | NDUFA10 | {"name": "NDUFA10", "id": "4705"} |
| FDA approved drug | - | Metformin | {"name": "Metformin"} | AFFECTS | - | Target | 4706 | NDUFAB1 | {"name": "NDUFAB1", "id": "4706"} |
| FDA approved drug | - | Metformin | {"name": "Metformin"} | AFFECTS | - | Target | 4707 | NDUFB1 | {"name": "NDUFB1", "id": "4707"} |

**Table S9: Complete node and relationship data used to construct the knowledge graph. This table summarizes interactions among FDA-approved drugs, formulation, herbs, diseases (type 2 diabetes, fatty liver, obesity), phytochemicals, and their target genes/proteins. Each entry includes source and target node details, edge type, and node annotations (Entrez Gene ID and gene symbol, where applicable).**

| Source Node Label | Source Node ID | Source Node Name | Source Node Properties | Edge Type | Edge Properties | Target Node Label | Target Node ID | Target Node Name | Target Node Properties |
|---|---|---|---|---|---|---|---|---|---|
| FDA approved drug | - | Metformin | {"name": "Metformin"} | AFFECTS | - | Target | 4708 | NDUFB2 | {"name": "NDUFB2", "id": "4708"} |
| FDA approved drug | - | Metformin | {"name": "Metformin"} | AFFECTS | - | Target | 4709 | NDUFB3 | {"name": "NDUFB3", "id": "4709"} |
| FDA approved drug | - | Metformin | {"name": "Metformin"} | AFFECTS | - | Target | 4710 | NDUFB4 | {"name": "NDUFB4", "id": "4710"} |
| FDA approved drug | - | Metformin | {"name": "Metformin"} | AFFECTS | - | Target | 4711 | NDUFB5 | {"name": "NDUFB5", "id": "4711"} |
| FDA approved drug | - | Metformin | {"name": "Metformin"} | AFFECTS | - | Target | 4712 | NDUFB6 | {"name": "NDUFB6", "id": "4712"} |
| FDA approved drug | - | Metformin | {"name": "Metformin"} | AFFECTS | - | Target | 4713 | NDUFB7 | {"name": "NDUFB7", "id": "4713"} |
| FDA approved drug | - | Metformin | {"name": "Metformin"} | AFFECTS | - | Target | 4714 | NDUFB8 | {"name": "NDUFB8", "id": "4714"} |
| FDA approved drug | - | Metformin | {"name": "Metformin"} | AFFECTS | - | Target | 4715 | NDUFB9 | {"name": "NDUFB9", "id": "4715"} |
| FDA approved drug | - | Metformin | {"name": "Metformin"} | AFFECTS | - | Target | 4716 | NDUFB10 | {"name": "NDUFB10", "id": "4716"} |
| FDA approved drug | - | Metformin | {"name": "Metformin"} | AFFECTS | - | Target | 4717 | NDUFC1 | {"name": "NDUFC1", "id": "4717"} |
| FDA approved drug | - | Metformin | {"name": "Metformin"} | AFFECTS | - | Target | 4718 | NDUFC2 | {"name": "NDUFC2", "id": "4718"} |
| FDA approved drug | - | Metformin | {"name": "Metformin"} | AFFECTS | - | Target | 4719 | NDUFS1 | {"name": "NDUFS1", "id": "4719"} |
| FDA approved drug | - | Metformin | {"name": "Metformin"} | AFFECTS | - | Target | 4720 | NDUFS2 | {"name": "NDUFS2", "id": "4720"} |
| FDA approved drug | - | Metformin | {"name": "Metformin"} | AFFECTS | - | Target | 4722 | NDUFS3 | {"name": "NDUFS3", "id": "4722"} |
| FDA approved drug | - | Metformin | {"name": "Metformin"} | AFFECTS | - | Target | 4723 | NDUFV1 | {"name": "NDUFV1", "id": "4723"} |
| FDA approved drug | - | Metformin | {"name": "Metformin"} | AFFECTS | - | Target | 4724 | NDUFS4 | {"name": "NDUFS4", "id": "4724"} |
| FDA approved drug | - | Metformin | {"name": "Metformin"} | AFFECTS | - | Target | 4725 | NDUFS5 | {"name": "NDUFS5", "id": "4725"} |
| FDA approved drug | - | Metformin | {"name": "Metformin"} | AFFECTS | - | Target | 4726 | NDUFS6 | {"name": "NDUFS6", "id": "4726"} |
| FDA approved drug | - | Metformin | {"name": "Metformin"} | AFFECTS | - | Target | 4728 | NDUFS8 | {"name": "NDUFS8", "id": "4728"} |
| FDA approved drug | - | Metformin | {"name": "Metformin"} | AFFECTS | - | Target | 4729 | NDUFV2 | {"name": "NDUFV2", "id": "4729"} |
| FDA approved drug | - | Metformin | {"name": "Metformin"} | AFFECTS | - | Target | 4731 | NDUFV3 | {"name": "NDUFV3", "id": "4731"} |
| FDA approved drug | - | Metformin | {"name": "Metformin"} | AFFECTS | - | Target | 51079 | NDUFA13 | {"name": "NDUFA13", "id": "51079"} |
| FDA approved drug | - | Metformin | {"name": "Metformin"} | AFFECTS | - | Target | 51103 | NDUFAF1 | {"name": "NDUFAF1", "id": "51103"} |
| FDA approved drug | - | Metformin | {"name": "Metformin"} | AFFECTS | - | Target | 54539 | NDUFB11 | {"name": "NDUFB11", "id": "54539"} |
| FDA approved drug | - | Metformin | {"name": "Metformin"} | AFFECTS | - | Target | 55244 | SLC47A1 | {"name": "SLC47A1", "id": "55244"} |
| FDA approved drug | - | Metformin | {"name": "Metformin"} | AFFECTS | - | Target | 55967 | NDUFA12 | {"name": "NDUFA12", "id": "55967"} |
| FDA approved drug | - | Metformin | {"name": "Metformin"} | AFFECTS | - | Target | 56901 | COXFA4L2 | {"name": "COXFA4L2", "id": "56901"} |
| FDA approved drug | - | Metformin | {"name": "Metformin"} | AFFECTS | - | Target | 91942 | NDUFAF2 | {"name": "NDUFAF2", "id": "91942"} |
| FDA approved drug | - | Miglitol | {"name": "Miglitol"} | TREATS | - | Disease | - | Type 2 Diabetes | {"name": "Type 2 Diabetes"} |
| FDA approved drug | - | Miglitol | {"name": "Miglitol"} | AFFECTS | - | Target | 8972 | MGAM | {"name": "MGAM", "id": "8972"} |
| FDA approved drug | - | Miglitol | {"name": "Miglitol"} | AFFECTS | - | Target | 2548 | GAA | {"name": "GAA", "id": "2548"} |
| FDA approved drug | - | Nateglinide | {"name": "Nateglinide"} | TREATS | - | Disease | - | Type 2 Diabetes | {"name": "Type 2 Diabetes"} |
| FDA approved drug | - | Nateglinide | {"name": "Nateglinide"} | AFFECTS | - | Target | 3767 | KCNJ11 | {"name": "KCNJ11", "id": "3767"} |
| FDA approved drug | - | Nateglinide | {"name": "Nateglinide"} | AFFECTS | - | Target | 6833 | ABCC8 | {"name": "ABCC8", "id": "6833"} |
| FDA approved drug | - | Orlistat | {"name": "Orlistat"} | TREATS | - | Disease | - | Obesity | {"name": "Obesity"} |
| FDA approved drug | - | Orlistat | {"name": "Orlistat"} | AFFECTS | - | Target | 2194 | FASN | {"name": "FASN", "id": "2194"} |
| FDA approved drug | - | Orlistat | {"name": "Orlistat"} | AFFECTS | - | Target | 5406 | PNLIP | {"name": "PNLIP", "id": "5406"} |
| FDA approved drug | - | Orlistat | {"name": "Orlistat"} | AFFECTS | - | Target | 56144 | PCDHA4 | {"name": "PCDHA4", "id": "56144"} |
| FDA approved drug | - | Orlistat | {"name": "Orlistat"} | AFFECTS | - | Target | 8513 | LIPF | {"name": "LIPF", "id": "8513"} |
| FDA approved drug | - | Phendimetrazine | {"name": "Phendimetrazine"} | TREATS | - | Disease | - | Obesity | {"name": "Obesity"} |
| FDA approved drug | - | Phendimetrazine | {"name": "Phendimetrazine"} | AFFECTS | - | Target | 6531 | SLC6A3 | {"name": "SLC6A3", "id": "6531"} |
| FDA approved drug | - | Phendimetrazine | {"name": "Phendimetrazine"} | AFFECTS | - | Target | 147 | ADRA1B | {"name": "ADRA1B", "id": "147"} |
| FDA approved drug | - | Phendimetrazine | {"name": "Phendimetrazine"} | AFFECTS | - | Target | 6530 | SLC6A2 | {"name": "SLC6A2", "id": "6530"} |
| FDA approved drug | - | Phentermine | {"name": "Phentermine"} | TREATS | - | Disease | - | Obesity | {"name": "Obesity"} |
| FDA approved drug | - | Phentermine | {"name": "Phentermine"} | AFFECTS | - | Target | 6530 | SLC6A2 | {"name": "SLC6A2", "id": "6530"} |
| FDA approved drug | - | Pioglitazone | {"name": "Pioglitazone"} | TREATS | - | Disease | - | Type 2 Diabetes | {"name": "Type 2 Diabetes"} |
| FDA approved drug | - | Pioglitazone | {"name": "Pioglitazone"} | AFFECTS | - | Target | 5468 | PPARG | {"name": "PPARG", "id": "5468"} |
| FDA approved drug | - | Pitavastatin | {"name": "Pitavastatin"} | TREATS | - | Disease | - | Fatty Liver | {"name": "Fatty Liver"} |
| FDA approved drug | - | Pitavastatin | {"name": "Pitavastatin"} | AFFECTS | - | Target | 3156 | HMGCR | {"name": "HMGCR", "id": "3156"} |
| FDA approved drug | - | Pramlintide | {"name": "Pramlintide"} | TREATS | - | Disease | - | Type 2 Diabetes | {"name": "Type 2 Diabetes"} |
| FDA approved drug | - | Pramlintide | {"name": "Pramlintide"} | AFFECTS | - | Target | 2740 | GLP1R | {"name": "GLP1R", "id": "2740"} |
| FDA approved drug | - | Pravastatin | {"name": "Pravastatin"} | TREATS | - | Disease | - | Fatty Liver | {"name": "Fatty Liver"} |
| FDA approved drug | - | Pravastatin | {"name": "Pravastatin"} | AFFECTS | - | Target | 3156 | HMGCR | {"name": "HMGCR", "id": "3156"} |
| FDA approved drug | - | Repaglinide | {"name": "Repaglinide"} | TREATS | - | Disease | - | Type 2 Diabetes | {"name": "Type 2 Diabetes"} |
| FDA approved drug | - | Repaglinide | {"name": "Repaglinide"} | AFFECTS | - | Target | 3767 | KCNJ11 | {"name": "KCNJ11", "id": "3767"} |
| FDA approved drug | - | Repaglinide | {"name": "Repaglinide"} | AFFECTS | - | Target | 6833 | ABCC8 | {"name": "ABCC8", "id": "6833"} |
| FDA approved drug | - | Saxagliptin | {"name": "Saxagliptin"} | TREATS | - | Disease | - | Type 2 Diabetes | {"name": "Type 2 Diabetes"} |
| FDA approved drug | - | Saxagliptin | {"name": "Saxagliptin"} | AFFECTS | - | Target | 1803 | DPP4 | {"name": "DPP4", "id": "1803"} |
| FDA approved drug | - | Semaglutide | {"name": "Semaglutide"} | TREATS | - | Disease | - | Type 2 Diabetes | {"name": "Type 2 Diabetes"} |
| FDA approved drug | - | Semaglutide | {"name": "Semaglutide"} | AFFECTS | - | Target | 2740 | GLP1R | {"name": "GLP1R", "id": "2740"} |
| FDA approved drug | - | Setmelanotide | {"name": "Setmelanotide"} | TREATS | - | Disease | - | Obesity | {"name": "Obesity"} |
| FDA approved drug | - | Setmelanotide | {"name": "Setmelanotide"} | AFFECTS | - | Target | 4160 | MC4R | {"name": "MC4R", "id": "4160"} |
| FDA approved drug | - | Simvastatin | {"name": "Simvastatin"} | TREATS | - | Disease | - | Fatty Liver | {"name": "Fatty Liver"} |
| FDA approved drug | - | Simvastatin | {"name": "Simvastatin"} | AFFECTS | - | Target | 3156 | HMGCR | {"name": "HMGCR", "id": "3156"} |
| FDA approved drug | - | Sitagliptin | {"name": "Sitagliptin"} | TREATS | - | Disease | - | Type 2 Diabetes | {"name": "Type 2 Diabetes"} |
| FDA approved drug | - | Sitagliptin | {"name": "Sitagliptin"} | AFFECTS | - | Target | 1803 | DPP4 | {"name": "DPP4", "id": "1803"} |
| FDA approved drug | - | Sotagliflozin | {"name": "Sotagliflozin"} | TREATS | - | Disease | - | Type 2 Diabetes | {"name": "Type 2 Diabetes"} |
| FDA approved drug | - | Sotagliflozin | {"name": "Sotagliflozin"} | AFFECTS | - | Target | 6524 | SLC5A2 | {"name": "SLC5A2", "id": "6524"} |
| FDA approved drug | - | Sotagliflozin | {"name": "Sotagliflozin"} | AFFECTS | - | Target | 6523 | SLC5A1 | {"name": "SLC5A1", "id": "6523"} |
| FDA approved drug | - | Tirzepatide | {"name": "Tirzepatide"} | TREATS | - | Disease | - | Type 2 Diabetes | {"name": "Type 2 Diabetes"} |
| FDA approved drug | - | Tirzepatide | {"name": "Tirzepatide"} | AFFECTS | - | Target | 2740 | GLP1R | {"name": "GLP1R", "id": "2740"} |
| FDA approved drug | - | Tolazamide | {"name": "Tolazamide"} | TREATS | - | Disease | - | Type 2 Diabetes | {"name": "Type 2 Diabetes"} |
| FDA approved drug | - | Tolazamide | {"name": "Tolazamide"} | AFFECTS | - | Target | 3767 | KCNJ11 | {"name": "KCNJ11", "id": "3767"} |
| FDA approved drug | - | Tolazamide | {"name": "Tolazamide"} | AFFECTS | - | Target | 6833 | ABCC8 | {"name": "ABCC8", "id": "6833"} |
| Formulation | - | Arogyavardhini | {"name": "Arogyavardhini"} | CONTAINS_HERB | - | Herb | - | azadirachta indica | {"name": "azadirachta indica", "part": "leaf"} |

**Table S9: Complete node and relationship data used to construct the knowledge graph. This table summarizes interactions among FDA-approved drugs, formulation, herbs, diseases (type 2 diabetes, fatty liver, obesity), phytochemicals, and their target genes/proteins. Each entry includes source and target node details, edge type, and node annotations (Entrez Gene ID and gene symbol, where applicable).**

| Source Node Label | Source Node ID | Source Node Name | Source Node Properties | Edge Type | Edge Properties | Target Node Label | Target Node ID | Target Node Name | Target Node Properties |
|---|---|---|---|---|---|---|---|---|---|
| Formulation | - | Arogyavardhini | {"name": "Arogyavardhini"} | CONTAINS_HERB | - | Herb | - | commiphora wightii | {"name": "commiphora wightii", "part": "plant exudate"} |
| Formulation | - | Arogyavardhini | {"name": "Arogyavardhini"} | CONTAINS_HERB | - | Herb | - | phyllanthus emblica | {"name": "phyllanthus emblica", "part": "fruit"} |
| Formulation | - | Arogyavardhini | {"name": "Arogyavardhini"} | CONTAINS_HERB | - | Herb | - | picrorhiza kurroa | {"name": "picrorhiza kurroa", "part": "rhizome"} |
| Formulation | - | Arogyavardhini | {"name": "Arogyavardhini"} | CONTAINS_HERB | - | Herb | - | plumbago zeylanica | {"name": "plumbago zeylanica", "part": "root"} |
| Formulation | - | Arogyavardhini | {"name": "Arogyavardhini"} | CONTAINS_HERB | - | Herb | - | terminalia bellirica | {"name": "terminalia bellirica", "part": "fruit"} |
| Formulation | - | Arogyavardhini | {"name": "Arogyavardhini"} | CONTAINS_HERB | - | Herb | - | terminalia chebula | {"name": "terminalia chebula", "part": "fruit"} |
| Formulation | - | Nishamalaki | {"name": "Nishamalaki"} | CONTAINS_HERB | - | Herb | - | curcuma longa | {"name": "curcuma longa", "part": "rhizome"} |
| Formulation | - | Nishamalaki | {"name": "Nishamalaki"} | CONTAINS_HERB | - | Herb | - | phyllanthus emblica | {"name": "phyllanthus emblica", "part": "fruit"} |
| Herb | - | azadirachta indica | {"name": "azadirachta indica", "part": "leaf"} | CONTAINS_CHEMICAL | - | Phytochemical | CAS_105377-74-0 | Isomargosinolide | {"name": "Isomargosinolide", "id": "CAS_105377-74-0"} |
| Herb | - | azadirachta indica | {"name": "azadirachta indica", "part": "leaf"} | CONTAINS_CHEMICAL | - | Phytochemical | CHEMSPIDER_10298763 | Margosinolide | {"name": "Margosinolide", "id": "CHEMSPIDER_10298763"} |
| Herb | - | azadirachta indica | {"name": "azadirachta indica", "part": "leaf"} | CONTAINS_CHEMICAL | - | Phytochemical | CID_100017 | Methyl 2-[6-(furan-3-yl)-7,9,11,15-tetramethyl-12,16-dioxo-3,17-dioxapentacyclo[9.6.1.02,9.04,8.015,18]octadeca-7,13-dien-10-yl]acetate | {"name": "Methyl 2-[6-(furan-3-yl)-7,9,11,15-tetramethyl-12,16-dioxo-3,17-dioxapentacyclo[9.6.1.02,9.04,8.015,18]octadeca-7,13-dien-10-yl]acetate", "id": "CID_100017"} |
| Herb | - | azadirachta indica | {"name": "azadirachta indica", "part": "leaf"} | CONTAINS_CHEMICAL | - | Phytochemical | CID_101289833 | Meldenin | {"name": "Meldenin", "id": "CID_101289833"} |
| Herb | - | azadirachta indica | {"name": "azadirachta indica", "part": "leaf"} | CONTAINS_CHEMICAL | - | Phytochemical | CID_101425842 | Isoazadirolide | {"name": "Isoazadirolide", "id": "CID_101425842"} |
| Herb | - | azadirachta indica | {"name": "azadirachta indica", "part": "leaf"} | CONTAINS_CHEMICAL | - | Phytochemical | CID_101650343 | (1S)-1-[(2R,4S,5R)-5-hydroxy-4-[(3S,5R,9R,10R,13S,14S,17S)-3-hydroxy-4,4,10,13,14-pentamethyl-2,3,5,6,9,11,12,15,16,17-decahydro-1H-cyclopenta[a]phenanthren-17-yl]oxolan-2-yl]-2-methylpropane-1,2-diol | {"name": "(1S)-1-[(2R,4S,5R)-5-hydroxy-4-[(3S,5R,9R,10R,13S,14S,17S)-3-hydroxy-4,4,10,13,14-pentamethyl-2,3,5,6,9,11,12,15,16,17-decahydro-1H-cyclopenta[a]phenanthren-17-yl]oxolan-2-yl]-2-methylpropane-1,2-diol", "id": "CID_101650343"} |
| Herb | - | azadirachta indica | {"name": "azadirachta indica", "part": "leaf"} | CONTAINS_CHEMICAL | - | Phytochemical | CID_102090424 | Vilasinin | {"name": "Vilasinin", "id": "CID_102090424"} |
| Herb | - | azadirachta indica | {"name": "azadirachta indica", "part": "leaf"} | CONTAINS_CHEMICAL | - | Phytochemical | CID_10505484 | 6-Deacetylnimbin | {"name": "6-Deacetylnimbin", "id": "CID_10505484"} |
| Herb | - | azadirachta indica | {"name": "azadirachta indica", "part": "leaf"} | CONTAINS_CHEMICAL | - | Phytochemical | CID_1107 | Sterol | {"name": "Sterol", "id": "CID_1107"} |
| Herb | - | azadirachta indica | {"name": "azadirachta indica", "part": "leaf"} | CONTAINS_CHEMICAL | - | Phytochemical | CID_12309055 | beta-Sitosterol-beta-D-glucoside | {"name": "beta-Sitosterol-beta-D-glucoside", "id": "CID_12309055"} |
| Herb | - | azadirachta indica | {"name": "azadirachta indica", "part": "leaf"} | CONTAINS_CHEMICAL | - | Phytochemical | CID_12309449 | delta-Elemene | {"name": "delta-Elemene", "id": "CID_12309449"} |
| Herb | - | azadirachta indica | {"name": "azadirachta indica", "part": "leaf"} | CONTAINS_CHEMICAL | - | Phytochemical | CID_12313020 | gamma-Muurolene | {"name": "gamma-Muurolene", "id": "CID_12313020"} |
| Herb | - | azadirachta indica | {"name": "azadirachta indica", "part": "leaf"} | CONTAINS_CHEMICAL | - | Phytochemical | CID_12409 | Nonacosane | {"name": "Nonacosane", "id": "CID_12409"} |
| Herb | - | azadirachta indica | {"name": "azadirachta indica", "part": "leaf"} | CONTAINS_CHEMICAL | - | Phytochemical | CID_13875774 | Nimbocinolide | {"name": "Nimbocinolide", "id": "CID_13875774"} |
| Herb | - | azadirachta indica | {"name": "azadirachta indica", "part": "leaf"} | CONTAINS_CHEMICAL | - | Phytochemical | CID_14458886 | 3-Deacetylsalannin | {"name": "3-Deacetylsalannin", "id": "CID_14458886"} |
| Herb | - | azadirachta indica | {"name": "azadirachta indica", "part": "leaf"} | CONTAINS_CHEMICAL | - | Phytochemical | CID_14467538 | 28-Deoxonimbolide | {"name": "28-Deoxonimbolide", "id": "CID_14467538"} |
| Herb | - | azadirachta indica | {"name": "azadirachta indica", "part": "leaf"} | CONTAINS_CHEMICAL | - | Phytochemical | CID_14492795 | 5,7-Dihydroxy-4'-methoxy-8,3'-di-C-prenylflavanone | {"name": "5,7-Dihydroxy-4'-methoxy-8,3'-di-C-prenylflavanone", "id": "CID_14492795"} |
| Herb | - | azadirachta indica | {"name": "azadirachta indica", "part": "leaf"} | CONTAINS_CHEMICAL | - | Phytochemical | CID_157277 | 2H-Cyclopenta(b)naphtho(2,3-d)furan-10-carboxylic acid,2-(3-furanyl)-3,3a,4a,5,5a,6,9,9a,10,10a-decahydro-5,6-dihydroxy-1,6,9a,10a-tetramethyl-9-oxo-, methyl ester, (2R,3aS,4aS,5R,5aS,6R,9aR,10S,10aR)- | {"name": "2H-Cyclopenta(b)naphtho(2,3-d)furan-10-carboxylic acid,2-(3-furanyl)-3,3a,4a,5,5a,6,9,9a,10,10a-decahydro-5,6-dihydroxy-1,6,9a,10a-tetramethyl-9-oxo-, methyl ester, (2R,3aS,4aS,5R,5aS,6R,9aR,10S,10aR)-", "id": "CID_157277"} |
| Herb | - | azadirachta indica | {"name": "azadirachta indica", "part": "leaf"} | CONTAINS_CHEMICAL | - | Phytochemical | CID_178770 | Nimocinol | {"name": "Nimocinol", "id": "CID_178770"} |
| Herb | - | azadirachta indica | {"name": "azadirachta indica", "part": "leaf"} | CONTAINS_CHEMICAL | - | Phytochemical | CID_184310 | [(5R,6R,7S,9R,10R,13S,17R)-6-hydroxy-17-(2-hydroxy-5-oxo-2H-furan-3-yl)-4,4,8,10,13-pentamethyl-3-oxo-5,6,7,9,11,12,16,17-octahydrocyclopenta[a]phenanthren-7-yl] acetate | {"name": "[(5R,6R,7S,9R,10R,13S,17R)-6-hydroxy-17-(2-hydroxy-5-oxo-2H-furan-3-yl)-4,4,8,10,13-pentamethyl-3-oxo-5,6,7,9,11,12,16,17-octahydrocyclopenta[a]phenanthren-7-yl] acetate", "id": "CID_184310"} |
| Herb | - | azadirachta indica | {"name": "azadirachta indica", "part": "leaf"} | CONTAINS_CHEMICAL | - | Phytochemical | CID_184503 | (9R,10R,13S,14S,17R)-17-[(2R)-2-[(1R,2S)-1,3-dihydroxy-2-methylpropyl]-2,3-dihydrofuran-4-yl]-4,4,10,13,14-pentamethyl-1,2,5,6,9,11,12,15,16,17-decahydrocyclopenta[a]phenanthren-3-one | {"name": "(9R,10R,13S,14S,17R)-17-[(2R)-2-[(1R,2S)-1,3-dihydroxy-2-methylpropyl]-2,3-dihydrofuran-4-yl]-4,4,10,13,14-pentamethyl-1,2,5,6,9,11,12,15,16,17-decahydrocyclopenta[a]phenanthren-3-one", "id": "CID_184503"} |
| Herb | - | azadirachta indica | {"name": "azadirachta indica", "part": "leaf"} | CONTAINS_CHEMICAL | - | Phytochemical | CID_222284 | beta-Sitosterol | {"name": "beta-Sitosterol", "id": "CID_222284"} |
| Herb | - | azadirachta indica | {"name": "azadirachta indica", "part": "leaf"} | CONTAINS_CHEMICAL | - | Phytochemical | CID_243696 | 1-Nonacosanol | {"name": "1-Nonacosanol", "id": "CID_243696"} |
| Herb | - | azadirachta indica | {"name": "azadirachta indica", "part": "leaf"} | CONTAINS_CHEMICAL | - | Phytochemical | CID_31291 | Tetradecanal | {"name": "Tetradecanal", "id": "CID_31291"} |
| Herb | - | azadirachta indica | {"name": "azadirachta indica", "part": "leaf"} | CONTAINS_CHEMICAL | - | Phytochemical | CID_42608158 | Allo-Aromadendrene | {"name": "Allo-Aromadendrene", "id": "CID_42608158"} |
| Herb | - | azadirachta indica | {"name": "azadirachta indica", "part": "leaf"} | CONTAINS_CHEMICAL | - | Phytochemical | CID_44715635 | Nimbinene | {"name": "Nimbinene", "id": "CID_44715635"} |
| Herb | - | azadirachta indica | {"name": "azadirachta indica", "part": "leaf"} | CONTAINS_CHEMICAL | - | Phytochemical | CID_521710 | alpha-Patchoulene | {"name": "alpha-Patchoulene", "id": "CID_521710"} |
| Herb | - | azadirachta indica | {"name": "azadirachta indica", "part": "leaf"} | CONTAINS_CHEMICAL | - | Phytochemical | CID_5280343 | Quercetin | {"name": "Quercetin", "id": "CID_5280343"} |
| Herb | - | azadirachta indica | {"name": "azadirachta indica", "part": "leaf"} | CONTAINS_CHEMICAL | - | Phytochemical | CID_5280460 | Scopoletin | {"name": "Scopoletin", "id": "CID_5280460"} |
| Herb | - | azadirachta indica | {"name": "azadirachta indica", "part": "leaf"} | CONTAINS_CHEMICAL | - | Phytochemical | CID_5280489 | beta-Carotene | {"name": "beta-Carotene", "id": "CID_5280489"} |
| Herb | - | azadirachta indica | {"name": "azadirachta indica", "part": "leaf"} | CONTAINS_CHEMICAL | - | Phytochemical | CID_5280794 | Stigmasterol | {"name": "Stigmasterol", "id": "CID_5280794"} |
| Herb | - | azadirachta indica | {"name": "azadirachta indica", "part": "leaf"} | CONTAINS_CHEMICAL | - | Phytochemical | CID_5281243 | Lutein | {"name": "Lutein", "id": "CID_5281243"} |
| Herb | - | azadirachta indica | {"name": "azadirachta indica", "part": "leaf"} | CONTAINS_CHEMICAL | - | Phytochemical | CID_5281515 | beta-Caryophyllene | {"name": "beta-Caryophyllene", "id": "CID_5281515"} |
| Herb | - | azadirachta indica | {"name": "azadirachta indica", "part": "leaf"} | CONTAINS_CHEMICAL | - | Phytochemical | CID_5281519 | Germacrene B | {"name": "Germacrene B", "id": "CID_5281519"} |
| Herb | - | azadirachta indica | {"name": "azadirachta indica", "part": "leaf"} | CONTAINS_CHEMICAL | - | Phytochemical | CID_5281520 | Humulene | {"name": "Humulene", "id": "CID_5281520"} |

**Table S9: Complete node and relationship data used to construct the knowledge graph. This table summarizes interactions among FDA-approved drugs, formulation, herbs, diseases (type 2 diabetes, fatty liver, obesity), phytochemicals, and their target genes/proteins. Each entry includes source and target node details, edge type, and node annotations (Entrez Gene ID and gene symbol, where applicable).**

| Source Node Label | Source Node ID | Source Node Name | Source Node Properties | Edge Type | Edge Properties | Target Node Label | Target Node ID | Target Node Name | Target Node Properties |
|---|---|---|---|---|---|---|---|---|---|
| Herb | - | azadirachta indica | {"name": "azadirachta indica", "part": "leaf"} | CONTAINS_CHEMICAL | - | Phytochemical | CID_5281654 | Isorhamnetin | {"name": "Isorhamnetin", "id": "CID_5281654"} |
| Herb | - | azadirachta indica | {"name": "azadirachta indica", "part": "leaf"} | CONTAINS_CHEMICAL | - | Phytochemical | CID_5742590 | Daucosterol | {"name": "Daucosterol", "id": "CID_5742590"} |
| Herb | - | azadirachta indica | {"name": "azadirachta indica", "part": "leaf"} | CONTAINS_CHEMICAL | - | Phytochemical | CID_6432312 | gamma-Elemene | {"name": "gamma-Elemene", "id": "CID_6432312"} |
| Herb | - | azadirachta indica | {"name": "azadirachta indica", "part": "leaf"} | CONTAINS_CHEMICAL | - | Phytochemical | CID_6442906 | [(4S,5R,6R,11R,12R,15S,16R,17Z)-17-(2,2-dihydroxyethylidene)-5-hydroxy-3,7,7,11,15-pentamethyl-8,18-dioxo-19-oxapentacyclo[14.3.1.02,15.03,12.06,11]icosa-1,9-dien-4-yl] acetate | {"name": "[(4S,5R,6R,11R,12R,15S,16R,17Z)-17-(2,2-dihydroxyethylidene)-5-hydroxy-3,7,7,11,15-pentamethyl-8,18-dioxo-19-oxapentacyclo[14.3.1.02,15.03,12.06,11]icosa-1,9-dien-4-yl] acetate", "id": "CID_6442906"} |
| Herb | - | azadirachta indica | {"name": "azadirachta indica", "part": "leaf"} | CONTAINS_CHEMICAL | - | Phytochemical | CID_68171 | 1-Hexacosanol | {"name": "1-Hexacosanol", "id": "CID_68171"} |
| Herb | - | azadirachta indica | {"name": "azadirachta indica", "part": "leaf"} | CONTAINS_CHEMICAL | - | Phytochemical | CID_6918391 | beta-Elemene | {"name": "beta-Elemene", "id": "CID_6918391"} |
| Herb | - | azadirachta indica | {"name": "azadirachta indica", "part": "leaf"} | CONTAINS_CHEMICAL | - | Phytochemical | CID_6918774 | Corosolic acid | {"name": "Corosolic acid", "id": "CID_6918774"} |
| Herb | - | azadirachta indica | {"name": "azadirachta indica", "part": "leaf"} | CONTAINS_CHEMICAL | - | Phytochemical | CID_70678558 | alpha-Copaene | {"name": "alpha-Copaene", "id": "CID_70678558"} |
| Herb | - | azadirachta indica | {"name": "azadirachta indica", "part": "leaf"} | CONTAINS_CHEMICAL | - | Phytochemical | CID_71597391 | Triterpenoids | {"name": "Triterpenoids", "id": "CID_71597391"} |
| Herb | - | azadirachta indica | {"name": "azadirachta indica", "part": "leaf"} | CONTAINS_CHEMICAL | - | Phytochemical | CID_76316558 | Isomeldenin | {"name": "Isomeldenin", "id": "CID_76316558"} |
| Herb | - | azadirachta indica | {"name": "azadirachta indica", "part": "leaf"} | CONTAINS_CHEMICAL | - | Phytochemical | CID_91723653 | (S,1Z,6Z)-8-Isopropyl-1-methyl-5-methylenecyclodeca-1,6-diene | {"name": "(S,1Z,6Z)-8-Isopropyl-1-methyl-5-methylenecyclodeca-1,6-diene", "id": "CID_91723653"} |
| Herb | - | azadirachta indica | {"name": "azadirachta indica", "part": "leaf"} | CONTAINS_CHEMICAL | - | Phytochemical | CID_91886694 | 2',3'-Dehydrosalannol | {"name": "2',3'-Dehydrosalannol", "id": "CID_91886694"} |
| Herb | - | azadirachta indica | {"name": "azadirachta indica", "part": "leaf"} | CONTAINS_CHEMICAL | - | Phytochemical | CID_984 | Hexadecanal | {"name": "Hexadecanal", "id": "CID_984"} |
| Herb | - | commiphora wightii | {"name": "commiphora wightii", "part": "plant exudate"} | CONTAINS_CHEMICAL | - | Phytochemical | CID_101297673 | Guggulsterol I | {"name": "Guggulsterol I", "id": "CID_101297673"} |
| Herb | - | commiphora wightii | {"name": "commiphora wightii", "part": "plant exudate"} | CONTAINS_CHEMICAL | - | Phytochemical | CID_101297674 | Guggulsterol II | {"name": "Guggulsterol II", "id": "CID_101297674"} |
| Herb | - | commiphora wightii | {"name": "commiphora wightii", "part": "plant exudate"} | CONTAINS_CHEMICAL | - | Phytochemical | CID_101297675 | Guggulsterol III | {"name": "Guggulsterol III", "id": "CID_101297675"} |
| Herb | - | commiphora wightii | {"name": "commiphora wightii", "part": "plant exudate"} | CONTAINS_CHEMICAL | - | Phytochemical | CID_101750 | alpha-Camphorene | {"name": "alpha-Camphorene", "id": "CID_101750"} |
| Herb | - | commiphora wightii | {"name": "commiphora wightii", "part": "plant exudate"} | CONTAINS_CHEMICAL | - | Phytochemical | CID_11747713 | Thunbergene | {"name": "Thunbergene", "id": "CID_11747713"} |
| Herb | - | commiphora wightii | {"name": "commiphora wightii", "part": "plant exudate"} | CONTAINS_CHEMICAL | - | Phytochemical | CID_122173119 | (8S,9R,10R,13R,14S,17Z)-17-ethylidene-10,13-dimethyl-1,2,6,7,8,9,11,12,14,15-decahydrocyclopenta[a]phenanthrene-3,16-dione | {"name": "(8S,9R,10R,13R,14S,17Z)-17-ethylidene-10,13-dimethyl-1,2,6,7,8,9,11,12,14,15-decahydrocyclopenta[a]phenanthrene-3,16-dione", "id": "CID_122173119"} |
| Herb | - | commiphora wightii | {"name": "commiphora wightii", "part": "plant exudate"} | CONTAINS_CHEMICAL | - | Phytochemical | CID_13255923 | Mukulol | {"name": "Mukulol", "id": "CID_13255923"} |
| Herb | - | commiphora wightii | {"name": "commiphora wightii", "part": "plant exudate"} | CONTAINS_CHEMICAL | - | Phytochemical | CID_161109 | 20-Hydroxypregn-4-en-3-one | {"name": "20-Hydroxypregn-4-en-3-one", "id": "CID_161109"} |
| Herb | - | commiphora wightii | {"name": "commiphora wightii", "part": "plant exudate"} | CONTAINS_CHEMICAL | - | Phytochemical | CID_222284 | beta-Sitosterol | {"name": "beta-Sitosterol", "id": "CID_222284"} |
| Herb | - | commiphora wightii | {"name": "commiphora wightii", "part": "plant exudate"} | CONTAINS_CHEMICAL | - | Phytochemical | CID_31253 | Myrcene | {"name": "Myrcene", "id": "CID_31253"} |
| Herb | - | commiphora wightii | {"name": "commiphora wightii", "part": "plant exudate"} | CONTAINS_CHEMICAL | - | Phytochemical | CID_445858 | Ferulic acid | {"name": "Ferulic acid", "id": "CID_445858"} |
| Herb | - | commiphora wightii | {"name": "commiphora wightii", "part": "plant exudate"} | CONTAINS_CHEMICAL | - | Phytochemical | CID_5204 | 5-[3-(1,3-Benzodioxol-5-yl)-1,3,3a,4,6,6a-hexahydrofuro[3,4-c]furan-6-yl]-1,3-benzodioxole | {"name": "5-[3-(1,3-Benzodioxol-5-yl)-1,3,3a,4,6,6a-hexahydrofuro[3,4-c]furan-6-yl]-1,3-benzodioxole", "id": "CID_5204"} |
| Herb | - | commiphora wightii | {"name": "commiphora wightii", "part": "plant exudate"} | CONTAINS_CHEMICAL | - | Phytochemical | CID_5280794 | Stigmasterol | {"name": "Stigmasterol", "id": "CID_5280794"} |
| Herb | - | commiphora wightii | {"name": "commiphora wightii", "part": "plant exudate"} | CONTAINS_CHEMICAL | - | Phytochemical | CID_5281384 | Neocembrene | {"name": "Neocembrene", "id": "CID_5281384"} |
| Herb | - | commiphora wightii | {"name": "commiphora wightii", "part": "plant exudate"} | CONTAINS_CHEMICAL | - | Phytochemical | CID_5281515 | beta-Caryophyllene | {"name": "beta-Caryophyllene", "id": "CID_5281515"} |
| Herb | - | commiphora wightii | {"name": "commiphora wightii", "part": "plant exudate"} | CONTAINS_CHEMICAL | - | Phytochemical | CID_5368823 | (2E,6Z,10E)-3,7,11-trimethyl-14-propan-2-ylcyclotetradeca-2,6,10-trien-1-ol | {"name": "(2E,6Z,10E)-3,7,11-trimethyl-14-propan-2-ylcyclotetradeca-2,6,10-trien-1-ol", "id": "CID_5368823"} |
| Herb | - | commiphora wightii | {"name": "commiphora wightii", "part": "plant exudate"} | CONTAINS_CHEMICAL | - | Phytochemical | CID_5997 | Cholesterol | {"name": "Cholesterol", "id": "CID_5997"} |
| Herb | - | commiphora wightii | {"name": "commiphora wightii", "part": "plant exudate"} | CONTAINS_CHEMICAL | - | Phytochemical | CID_6439929 | E-Guggulsterone | {"name": "E-Guggulsterone", "id": "CID_6439929"} |
| Herb | - | commiphora wightii | {"name": "commiphora wightii", "part": "plant exudate"} | CONTAINS_CHEMICAL | - | Phytochemical | CID_68972 | 1-Triacontanol | {"name": "1-Triacontanol", "id": "CID_68972"} |
| Herb | - | commiphora wightii | {"name": "commiphora wightii", "part": "plant exudate"} | CONTAINS_CHEMICAL | - | Phytochemical | CID_70695727 | Pluviatilol | {"name": "Pluviatilol", "id": "CID_70695727"} |
| Herb | - | commiphora wightii | {"name": "commiphora wightii", "part": "plant exudate"} | CONTAINS_CHEMICAL | - | Phytochemical | CID_92747 | (20R)-20-Hydroxypregn-4-en-3-one | {"name": "(20R)-20-Hydroxypregn-4-en-3-one", "id": "CID_92747"} |
| Herb | - | curcuma longa | {"name": "curcuma longa", "part": "rhizome"} | CONTAINS_CHEMICAL | - | Phytochemical | CHEMSPIDER_16737047 | Ar-dihydro-turmerone | {"name": "Ar-dihydro-turmerone", "id": "CHEMSPIDER_16737047"} |
| Herb | - | curcuma longa | {"name": "curcuma longa", "part": "rhizome"} | CONTAINS_CHEMICAL | - | Phytochemical | CID_10104370 | beta-Bisabolene | {"name": "beta-Bisabolene", "id": "CID_10104370"} |
| Herb | - | curcuma longa | {"name": "curcuma longa", "part": "rhizome"} | CONTAINS_CHEMICAL | - | Phytochemical | CID_101297706 | Humuladienone | {"name": "Humuladienone", "id": "CID_101297706"} |
| Herb | - | curcuma longa | {"name": "curcuma longa", "part": "rhizome"} | CONTAINS_CHEMICAL | - | Phytochemical | CID_101680 | Camphene hydrate | {"name": "Camphene hydrate", "id": "CID_101680"} |
| Herb | - | curcuma longa | {"name": "curcuma longa", "part": "rhizome"} | CONTAINS_CHEMICAL | - | Phytochemical | CID_101926857 | Bisabola-3,10-diene-2-one | {"name": "Bisabola-3,10-diene-2-one", "id": "CID_101926857"} |
| Herb | - | curcuma longa | {"name": "curcuma longa", "part": "rhizome"} | CONTAINS_CHEMICAL | - | Phytochemical | CID_102024 | 2-Cyclohexen-1-ol, 2-methyl-5-(1-methylethenyl)-, acetate, cis- | {"name": "2-Cyclohexen-1-ol, 2-methyl-5-(1-methylethenyl)-, acetate, cis-", "id": "CID_102024"} |
| Herb | - | curcuma longa | {"name": "curcuma longa", "part": "rhizome"} | CONTAINS_CHEMICAL | - | Phytochemical | CID_10248 | Elemicin | {"name": "Elemicin", "id": "CID_10248"} |
| Herb | - | curcuma longa | {"name": "curcuma longa", "part": "rhizome"} | CONTAINS_CHEMICAL | - | Phytochemical | CID_10263440 | Epiprocurcumenol | {"name": "Epiprocurcumenol", "id": "CID_10263440"} |
| Herb | - | curcuma longa | {"name": "curcuma longa", "part": "rhizome"} | CONTAINS_CHEMICAL | - | Phytochemical | CID_10364 | Carvacrol | {"name": "Carvacrol", "id": "CID_10364"} |
| Herb | - | curcuma longa | {"name": "curcuma longa", "part": "rhizome"} | CONTAINS_CHEMICAL | - | Phytochemical | CID_10421034 | 1-Bisabolone | {"name": "1-Bisabolone", "id": "CID_10421034"} |

**Table S9: Complete node and relationship data used to construct the knowledge graph. This table summarizes interactions among FDA-approved drugs, formulation, herbs, diseases (type 2 diabetes, fatty liver, obesity), phytochemicals, and their target genes/proteins. Each entry includes source and target node details, edge type, and node annotations (Entrez Gene ID and gene symbol, where applicable).**

| Source Node Label | Source Node ID | Source Node Name | Source Node Properties | Edge Type | Edge Properties | Target Node Label | Target Node ID | Target Node Name | Target Node Properties |
|---|---|---|---|---|---|---|---|---|---|
| Herb | - | curcuma longa | {"name": "curcuma longa", "part": "rhizome"} | CONTAINS_CHEMICAL | - | Phytochemical | CID_10429233 | (1E)-1,7-bis(4-hydroxy-3-methoxyphenyl)hept-1-ene-3,5-dione | {"name": "(1E)-1,7-bis(4-hydroxy-3-methoxyphenyl)hept-1-ene-3,5-dione", "id": "CID_10429233"} |
| Herb | - | curcuma longa | {"name": "curcuma longa", "part": "rhizome"} | CONTAINS_CHEMICAL | - | Phytochemical | CID_10469828 | (1E,4E)-1-(4-Hydroxy-3-methoxyphenyl)-5-(4-hydroxyphenyl)-1,4-pentadiene-3-one | {"name": "(1E,4E)-1-(4-Hydroxy-3-methoxyphenyl)-5-(4-hydroxyphenyl)-1,4-pentadiene-3-one", "id": "CID_10469828"} |
| Herb | - | curcuma longa | {"name": "curcuma longa", "part": "rhizome"} | CONTAINS_CHEMICAL | - | Phytochemical | CID_10582 | Myrtenol | {"name": "Myrtenol", "id": "CID_10582"} |
| Herb | - | curcuma longa | {"name": "curcuma longa", "part": "rhizome"} | CONTAINS_CHEMICAL | - | Phytochemical | CID_10703 | O-Cymene | {"name": "O-Cymene", "id": "CID_10703"} |
| Herb | - | curcuma longa | {"name": "curcuma longa", "part": "rhizome"} | CONTAINS_CHEMICAL | - | Phytochemical | CID_10704181 | Humulene epoxide II | {"name": "Humulene epoxide II", "id": "CID_10704181"} |
| Herb | - | curcuma longa | {"name": "curcuma longa", "part": "rhizome"} | CONTAINS_CHEMICAL | - | Phytochemical | CID_10856614 | alpha-Selinene | {"name": "alpha-Selinene", "id": "CID_10856614"} |
| Herb | - | curcuma longa | {"name": "curcuma longa", "part": "rhizome"} | CONTAINS_CHEMICAL | - | Phytochemical | CID_10955433 | 2-Hepten-4-one, 6-(2-hydroxy-4-methylphenyl)-2-methyl- | {"name": "2-Hepten-4-one, 6-(2-hydroxy-4-methylphenyl)-2-methyl-", "id": "CID_10955433"} |
| Herb | - | curcuma longa | {"name": "curcuma longa", "part": "rhizome"} | CONTAINS_CHEMICAL | - | Phytochemical | CID_10976 | 2-Heptanol | {"name": "2-Heptanol", "id": "CID_10976"} |
| Herb | - | curcuma longa | {"name": "curcuma longa", "part": "rhizome"} | CONTAINS_CHEMICAL | - | Phytochemical | CID_11116492 | 2-Furanmethanol, 5-ethenyltetrahydro-alpha,alpha,5-trimethyl-, cis- | {"name": "2-Furanmethanol, 5-ethenyltetrahydro-alpha,alpha,5-trimethyl-, cis-", "id": "CID_11116492"} |
| Herb | - | curcuma longa | {"name": "curcuma longa", "part": "rhizome"} | CONTAINS_CHEMICAL | - | Phytochemical | CID_11230 | 4-Carvomenthenol | {"name": "4-Carvomenthenol", "id": "CID_11230"} |
| Herb | - | curcuma longa | {"name": "curcuma longa", "part": "rhizome"} | CONTAINS_CHEMICAL | - | Phytochemical | CID_1140 | Toluene | {"name": "Toluene", "id": "CID_1140"} |
| Herb | - | curcuma longa | {"name": "curcuma longa", "part": "rhizome"} | CONTAINS_CHEMICAL | - | Phytochemical | CID_11463 | Terpinolene | {"name": "Terpinolene", "id": "CID_11463"} |
| Herb | - | curcuma longa | {"name": "curcuma longa", "part": "rhizome"} | CONTAINS_CHEMICAL | - | Phytochemical | CID_11467 | gamma-Terpineol | {"name": "gamma-Terpineol", "id": "CID_11467"} |
| Herb | - | curcuma longa | {"name": "curcuma longa", "part": "rhizome"} | CONTAINS_CHEMICAL | - | Phytochemical | CID_11586487 | beta-Himachalene | {"name": "beta-Himachalene", "id": "CID_11586487"} |
| Herb | - | curcuma longa | {"name": "curcuma longa", "part": "rhizome"} | CONTAINS_CHEMICAL | - | Phytochemical | CID_11716 | 3-Buten-2-OL | {"name": "3-Buten-2-OL", "id": "CID_11716"} |
| Herb | - | curcuma longa | {"name": "curcuma longa", "part": "rhizome"} | CONTAINS_CHEMICAL | - | Phytochemical | CID_11996452 | Viridiflorol | {"name": "Viridiflorol", "id": "CID_11996452"} |
| Herb | - | curcuma longa | {"name": "curcuma longa", "part": "rhizome"} | CONTAINS_CHEMICAL | - | Phytochemical | CID_122130015 | Ethene;2-methoxyphenol;phenol;propanedial | {"name": "Ethene;2-methoxyphenol;phenol;propanedial", "id": "CID_122130015"} |
| Herb | - | curcuma longa | {"name": "curcuma longa", "part": "rhizome"} | CONTAINS_CHEMICAL | - | Phytochemical | CID_12233170 | cis-Carvotanacetol | {"name": "cis-Carvotanacetol", "id": "CID_12233170"} |
| Herb | - | curcuma longa | {"name": "curcuma longa", "part": "rhizome"} | CONTAINS_CHEMICAL | - | Phytochemical | CID_122484 | 2-Cyclohexen-1-ol, 1-methyl-4-(1-methylethyl)-, trans- | {"name": "2-Cyclohexen-1-ol, 1-methyl-4-(1-methylethyl)-, trans-", "id": "CID_122484"} |
| Herb | - | curcuma longa | {"name": "curcuma longa", "part": "rhizome"} | CONTAINS_CHEMICAL | - | Phytochemical | CID_12299867 | alpha-Atlantone | {"name": "alpha-Atlantone", "id": "CID_12299867"} |
| Herb | - | curcuma longa | {"name": "curcuma longa", "part": "rhizome"} | CONTAINS_CHEMICAL | - | Phytochemical | CID_12300148 | 6-Epi-beta-bisabolol | {"name": "6-Epi-beta-bisabolol", "id": "CID_12300148"} |
| Herb | - | curcuma longa | {"name": "curcuma longa", "part": "rhizome"} | CONTAINS_CHEMICAL | - | Phytochemical | CID_12302222 | tau-Cadinol | {"name": "tau-Cadinol", "id": "CID_12302222"} |
| Herb | - | curcuma longa | {"name": "curcuma longa", "part": "rhizome"} | CONTAINS_CHEMICAL | - | Phytochemical | CID_12306047 | alpha-Muurolene | {"name": "alpha-Muurolene", "id": "CID_12306047"} |
| Herb | - | curcuma longa | {"name": "curcuma longa", "part": "rhizome"} | CONTAINS_CHEMICAL | - | Phytochemical | CID_12306048 | (+)-alpha-Cadinene | {"name": "(+)-alpha-Cadinene", "id": "CID_12306048"} |
| Herb | - | curcuma longa | {"name": "curcuma longa", "part": "rhizome"} | CONTAINS_CHEMICAL | - | Phytochemical | CID_12309449 | delta-Elemene | {"name": "delta-Elemene", "id": "CID_12309449"} |
| Herb | - | curcuma longa | {"name": "curcuma longa", "part": "rhizome"} | CONTAINS_CHEMICAL | - | Phytochemical | CID_12310886 | (1S,2S,5S,8S)-2-methyl-6-methylidene-9-propan-2-ylidene-11-oxatricyclo[6.2.1.01,5]undecan-8-ol | {"name": "(1S,2S,5S,8S)-2-methyl-6-methylidene-9-propan-2-ylidene-11-oxatricyclo[6.2.1.01,5]undecan-8-ol", "id": "CID_12310886"} |
| Herb | - | curcuma longa | {"name": "curcuma longa", "part": "rhizome"} | CONTAINS_CHEMICAL | - | Phytochemical | CID_12367 | 2-Nonanol | {"name": "2-Nonanol", "id": "CID_12367"} |
| Herb | - | curcuma longa | {"name": "curcuma longa", "part": "rhizome"} | CONTAINS_CHEMICAL | - | Phytochemical | CID_12388 | Tridecane | {"name": "Tridecane", "id": "CID_12388"} |
| Herb | - | curcuma longa | {"name": "curcuma longa", "part": "rhizome"} | CONTAINS_CHEMICAL | - | Phytochemical | CID_12389 | Tetradecane | {"name": "Tetradecane", "id": "CID_12389"} |
| Herb | - | curcuma longa | {"name": "curcuma longa", "part": "rhizome"} | CONTAINS_CHEMICAL | - | Phytochemical | CID_14014430 | (-)-beta-Curcumene | {"name": "(-)-beta-Curcumene", "id": "CID_14014430"} |
| Herb | - | curcuma longa | {"name": "curcuma longa", "part": "rhizome"} | CONTAINS_CHEMICAL | - | Phytochemical | CID_14191392 | (6Z)-6,10-dimethyl-3-propan-2-ylidenecyclodec-6-ene-1,4-dione | {"name": "(6Z)-6,10-dimethyl-3-propan-2-ylidenecyclodec-6-ene-1,4-dione", "id": "CID_14191392"} |
| Herb | - | curcuma longa | {"name": "curcuma longa", "part": "rhizome"} | CONTAINS_CHEMICAL | - | Phytochemical | CID_14240392 | Curcumol | {"name": "Curcumol", "id": "CID_14240392"} |
| Herb | - | curcuma longa | {"name": "curcuma longa", "part": "rhizome"} | CONTAINS_CHEMICAL | - | Phytochemical | CID_14254 | 2-Decanol | {"name": "2-Decanol", "id": "CID_14254"} |
| Herb | - | curcuma longa | {"name": "curcuma longa", "part": "rhizome"} | CONTAINS_CHEMICAL | - | Phytochemical | CID_14287397 | Bisacurone | {"name": "Bisacurone", "id": "CID_14287397"} |
| Herb | - | curcuma longa | {"name": "curcuma longa", "part": "rhizome"} | CONTAINS_CHEMICAL | - | Phytochemical | CID_14529 | 2-(4-Methylphenyl)propan-2-ol | {"name": "2-(4-Methylphenyl)propan-2-ol", "id": "CID_14529"} |
| Herb | - | curcuma longa | {"name": "curcuma longa", "part": "rhizome"} | CONTAINS_CHEMICAL | - | Phytochemical | CID_14543198 | Isoprocurcumenol | {"name": "Isoprocurcumenol", "id": "CID_14543198"} |
| Herb | - | curcuma longa | {"name": "curcuma longa", "part": "rhizome"} | CONTAINS_CHEMICAL | - | Phytochemical | CID_14632996 | alpha-Turmerone | {"name": "alpha-Turmerone", "id": "CID_14632996"} |
| Herb | - | curcuma longa | {"name": "curcuma longa", "part": "rhizome"} | CONTAINS_CHEMICAL | - | Phytochemical | CID_14633011 | Procurcumadiol | {"name": "Procurcumadiol", "id": "CID_14633011"} |
| Herb | - | curcuma longa | {"name": "curcuma longa", "part": "rhizome"} | CONTAINS_CHEMICAL | - | Phytochemical | CID_14896 | beta-Pinene | {"name": "beta-Pinene", "id": "CID_14896"} |
| Herb | - | curcuma longa | {"name": "curcuma longa", "part": "rhizome"} | CONTAINS_CHEMICAL | - | Phytochemical | CID_153845 | Bicyclo(4.1.0)heptan-3-one, 1-methyl-4-(1-methylethylidene)-7-(3-oxobutyl)-, (1S,6R,7R)- | {"name": "Bicyclo(4.1.0)heptan-3-one, 1-methyl-4-(1-methylethylidene)-7-(3-oxobutyl)-, (1S,6R,7R)-", "id": "CID_153845"} |
| Herb | - | curcuma longa | {"name": "curcuma longa", "part": "rhizome"} | CONTAINS_CHEMICAL | - | Phytochemical | CID_15448 | 2-Undecanol | {"name": "2-Undecanol", "id": "CID_15448"} |
| Herb | - | curcuma longa | {"name": "curcuma longa", "part": "rhizome"} | CONTAINS_CHEMICAL | - | Phytochemical | CID_1549026 | Geranyl acetate | {"name": "Geranyl acetate", "id": "CID_1549026"} |
| Herb | - | curcuma longa | {"name": "curcuma longa", "part": "rhizome"} | CONTAINS_CHEMICAL | - | Phytochemical | CID_1550890 | Cinnamyl cinnamate | {"name": "Cinnamyl cinnamate", "id": "CID_1550890"} |
| Herb | - | curcuma longa | {"name": "curcuma longa", "part": "rhizome"} | CONTAINS_CHEMICAL | - | Phytochemical | CID_15858385 | Turmeronol A | {"name": "Turmeronol A", "id": "CID_15858385"} |
| Herb | - | curcuma longa | {"name": "curcuma longa", "part": "rhizome"} | CONTAINS_CHEMICAL | - | Phytochemical | CID_160512 | ar-Turmerone | {"name": "ar-Turmerone", "id": "CID_160512"} |

**Table S9: Complete node and relationship data used to construct the knowledge graph. This table summarizes interactions among FDA-approved drugs, formulation, herbs, diseases (type 2 diabetes, fatty liver, obesity), phytochemicals, and their target genes/proteins. Each entry includes source and target node details, edge type, and node annotations (Entrez Gene ID and gene symbol, where applicable).**

| Source Node Label | Source Node ID | Source Node Name | Source Node Properties | Edge Type | Edge Properties | Target Node Label | Target Node ID | Target Node Name | Target Node Properties |
|---|---|---|---|---|---|---|---|---|---|
| Herb | - | curcuma longa | {"name": "curcuma longa", "part": "rhizome"} | CONTAINS_CHEMICAL | - | Phytochemical | CID_167812 | (+)-Curcumenol | {"name": "(+)-Curcumenol", "id": "CID_167812"} |
| Herb | - | curcuma longa | {"name": "curcuma longa", "part": "rhizome"} | CONTAINS_CHEMICAL | - | Phytochemical | CID_17100 | alpha-Terpineol | {"name": "alpha-Terpineol", "id": "CID_17100"} |
| Herb | - | curcuma longa | {"name": "curcuma longa", "part": "rhizome"} | CONTAINS_CHEMICAL | - | Phytochemical | CID_173183 | Campesterol | {"name": "Campesterol", "id": "CID_173183"} |
| Herb | - | curcuma longa | {"name": "curcuma longa", "part": "rhizome"} | CONTAINS_CHEMICAL | - | Phytochemical | CID_1742210 | Caryophyllene oxide | {"name": "Caryophyllene oxide", "id": "CID_1742210"} |
| Herb | - | curcuma longa | {"name": "curcuma longa", "part": "rhizome"} | CONTAINS_CHEMICAL | - | Phytochemical | CID_176983 | p-Mentha-1,3,8-triene | {"name": "p-Mentha-1,3,8-triene", "id": "CID_176983"} |
| Herb | - | curcuma longa | {"name": "curcuma longa", "part": "rhizome"} | CONTAINS_CHEMICAL | - | Phytochemical | CID_18818 | Sabinene | {"name": "Sabinene", "id": "CID_18818"} |
| Herb | - | curcuma longa | {"name": "curcuma longa", "part": "rhizome"} | CONTAINS_CHEMICAL | - | Phytochemical | CID_189061 | Procurcumenol | {"name": "Procurcumenol", "id": "CID_189061"} |
| Herb | - | curcuma longa | {"name": "curcuma longa", "part": "rhizome"} | CONTAINS_CHEMICAL | - | Phytochemical | CID_196216 | Curlone | {"name": "Curlone", "id": "CID_196216"} |
| Herb | - | curcuma longa | {"name": "curcuma longa", "part": "rhizome"} | CONTAINS_CHEMICAL | - | Phytochemical | CID_20055538 | Turmerol | {"name": "Turmerol", "id": "CID_20055538"} |
| Herb | - | curcuma longa | {"name": "curcuma longa", "part": "rhizome"} | CONTAINS_CHEMICAL | - | Phytochemical | CID_20083 | 2-Octanol | {"name": "2-Octanol", "id": "CID_20083"} |
| Herb | - | curcuma longa | {"name": "curcuma longa", "part": "rhizome"} | CONTAINS_CHEMICAL | - | Phytochemical | CID_20928 | 1-Hexen-3-OL | {"name": "1-Hexen-3-OL", "id": "CID_20928"} |
| Herb | - | curcuma longa | {"name": "curcuma longa", "part": "rhizome"} | CONTAINS_CHEMICAL | - | Phytochemical | CID_222284 | beta-Sitosterol | {"name": "beta-Sitosterol", "id": "CID_222284"} |
| Herb | - | curcuma longa | {"name": "curcuma longa", "part": "rhizome"} | CONTAINS_CHEMICAL | - | Phytochemical | CID_22311 | Limonene | {"name": "Limonene", "id": "CID_22311"} |
| Herb | - | curcuma longa | {"name": "curcuma longa", "part": "rhizome"} | CONTAINS_CHEMICAL | - | Phytochemical | CID_247573 | Isobornyl acetate | {"name": "Isobornyl acetate", "id": "CID_247573"} |
| Herb | - | curcuma longa | {"name": "curcuma longa", "part": "rhizome"} | CONTAINS_CHEMICAL | - | Phytochemical | CID_24834047 | 4,10-Epizedoarondiol | {"name": "4,10-Epizedoarondiol", "id": "CID_24834047"} |
| Herb | - | curcuma longa | {"name": "curcuma longa", "part": "rhizome"} | CONTAINS_CHEMICAL | - | Phytochemical | CID_25202482 | Sesquisabinene | {"name": "Sesquisabinene", "id": "CID_25202482"} |
| Herb | - | curcuma longa | {"name": "curcuma longa", "part": "rhizome"} | CONTAINS_CHEMICAL | - | Phytochemical | CID_2537 | Camphor | {"name": "Camphor", "id": "CID_2537"} |
| Herb | - | curcuma longa | {"name": "curcuma longa", "part": "rhizome"} | CONTAINS_CHEMICAL | - | Phytochemical | CID_26049 | 3-Carene | {"name": "3-Carene", "id": "CID_26049"} |
| Herb | - | curcuma longa | {"name": "curcuma longa", "part": "rhizome"} | CONTAINS_CHEMICAL | - | Phytochemical | CID_2758 | Eucalyptol | {"name": "Eucalyptol", "id": "CID_2758"} |
| Herb | - | curcuma longa | {"name": "curcuma longa", "part": "rhizome"} | CONTAINS_CHEMICAL | - | Phytochemical | CID_28930 | alpha-Fenchene | {"name": "alpha-Fenchene", "id": "CID_28930"} |
| Herb | - | curcuma longa | {"name": "curcuma longa", "part": "rhizome"} | CONTAINS_CHEMICAL | - | Phytochemical | CID_2969 | Decanoic acid | {"name": "Decanoic acid", "id": "CID_2969"} |
| Herb | - | curcuma longa | {"name": "curcuma longa", "part": "rhizome"} | CONTAINS_CHEMICAL | - | Phytochemical | CID_3081930 | Curzerenone | {"name": "Curzerenone", "id": "CID_3081930"} |
| Herb | - | curcuma longa | {"name": "curcuma longa", "part": "rhizome"} | CONTAINS_CHEMICAL | - | Phytochemical | CID_31253 | Myrcene | {"name": "Myrcene", "id": "CID_31253"} |
| Herb | - | curcuma longa | {"name": "curcuma longa", "part": "rhizome"} | CONTAINS_CHEMICAL | - | Phytochemical | CID_31291 | Tetradecanal | {"name": "Tetradecanal", "id": "CID_31291"} |
| Herb | - | curcuma longa | {"name": "curcuma longa", "part": "rhizome"} | CONTAINS_CHEMICAL | - | Phytochemical | CID_325 | 4-Isopropylbenzyl alcohol | {"name": "4-Isopropylbenzyl alcohol", "id": "CID_325"} |
| Herb | - | curcuma longa | {"name": "curcuma longa", "part": "rhizome"} | CONTAINS_CHEMICAL | - | Phytochemical | CID_330573 | (-)-cis-Carveol | {"name": "(-)-cis-Carveol", "id": "CID_330573"} |
| Herb | - | curcuma longa | {"name": "curcuma longa", "part": "rhizome"} | CONTAINS_CHEMICAL | - | Phytochemical | CID_3314 | Eugenol | {"name": "Eugenol", "id": "CID_3314"} |
| Herb | - | curcuma longa | {"name": "curcuma longa", "part": "rhizome"} | CONTAINS_CHEMICAL | - | Phytochemical | CID_360253 | Curcuphenol | {"name": "Curcuphenol", "id": "CID_360253"} |
| Herb | - | curcuma longa | {"name": "curcuma longa", "part": "rhizome"} | CONTAINS_CHEMICAL | - | Phytochemical | CID_42608158 | Allo-Aromadendrene | {"name": "Allo-Aromadendrene", "id": "CID_42608158"} |
| Herb | - | curcuma longa | {"name": "curcuma longa", "part": "rhizome"} | CONTAINS_CHEMICAL | - | Phytochemical | CID_439711 | alpha-Fenchol | {"name": "alpha-Fenchol", "id": "CID_439711"} |
| Herb | - | curcuma longa | {"name": "curcuma longa", "part": "rhizome"} | CONTAINS_CHEMICAL | - | Phytochemical | CID_441005 | (+)-delta-Cadinene | {"name": "(+)-delta-Cadinene", "id": "CID_441005"} |
| Herb | - | curcuma longa | {"name": "curcuma longa", "part": "rhizome"} | CONTAINS_CHEMICAL | - | Phytochemical | CID_442343 | Levomenol | {"name": "Levomenol", "id": "CID_442343"} |
| Herb | - | curcuma longa | {"name": "curcuma longa", "part": "rhizome"} | CONTAINS_CHEMICAL | - | Phytochemical | CID_442484 | (+)-beta-Phellandrene | {"name": "(+)-beta-Phellandrene", "id": "CID_442484"} |
| Herb | - | curcuma longa | {"name": "curcuma longa", "part": "rhizome"} | CONTAINS_CHEMICAL | - | Phytochemical | CID_442495 | Pulegone | {"name": "Pulegone", "id": "CID_442495"} |
| Herb | - | curcuma longa | {"name": "curcuma longa", "part": "rhizome"} | CONTAINS_CHEMICAL | - | Phytochemical | CID_445070 | Farnesol | {"name": "Farnesol", "id": "CID_445070"} |
| Herb | - | curcuma longa | {"name": "curcuma longa", "part": "rhizome"} | CONTAINS_CHEMICAL | - | Phytochemical | CID_5144 | Safrole | {"name": "Safrole", "id": "CID_5144"} |
| Herb | - | curcuma longa | {"name": "curcuma longa", "part": "rhizome"} | CONTAINS_CHEMICAL | - | Phytochemical | CID_519764 | 3-(1,5-Dimethyl-4-hexenyl)-6-methylene-1-cyclohexene | {"name": "3-(1,5-Dimethyl-4-hexenyl)-6-methylene-1-cyclohexene", "id": "CID_519764"} |
| Herb | - | curcuma longa | {"name": "curcuma longa", "part": "rhizome"} | CONTAINS_CHEMICAL | - | Phytochemical | CID_520453 | Bisabolane | {"name": "Bisabolane", "id": "CID_520453"} |
| Herb | - | curcuma longa | {"name": "curcuma longa", "part": "rhizome"} | CONTAINS_CHEMICAL | - | Phytochemical | CID_522296 | 3,7(11)-Eudesmadiene | {"name": "3,7(11)-Eudesmadiene", "id": "CID_522296"} |
| Herb | - | curcuma longa | {"name": "curcuma longa", "part": "rhizome"} | CONTAINS_CHEMICAL | - | Phytochemical | CID_5280794 | Stigmasterol | {"name": "Stigmasterol", "id": "CID_5280794"} |
| Herb | - | curcuma longa | {"name": "curcuma longa", "part": "rhizome"} | CONTAINS_CHEMICAL | - | Phytochemical | CID_5281167 | cis-3-Hexen-1-ol | {"name": "cis-3-Hexen-1-ol", "id": "CID_5281167"} |
| Herb | - | curcuma longa | {"name": "curcuma longa", "part": "rhizome"} | CONTAINS_CHEMICAL | - | Phytochemical | CID_5281515 | beta-Caryophyllene | {"name": "beta-Caryophyllene", "id": "CID_5281515"} |
| Herb | - | curcuma longa | {"name": "curcuma longa", "part": "rhizome"} | CONTAINS_CHEMICAL | - | Phytochemical | CID_5281517 | beta-Farnesene | {"name": "beta-Farnesene", "id": "CID_5281517"} |
| Herb | - | curcuma longa | {"name": "curcuma longa", "part": "rhizome"} | CONTAINS_CHEMICAL | - | Phytochemical | CID_5281519 | Germacrene B | {"name": "Germacrene B", "id": "CID_5281519"} |
| Herb | - | curcuma longa | {"name": "curcuma longa", "part": "rhizome"} | CONTAINS_CHEMICAL | - | Phytochemical | CID_5281520 | Humulene | {"name": "Humulene", "id": "CID_5281520"} |
| Herb | - | curcuma longa | {"name": "curcuma longa", "part": "rhizome"} | CONTAINS_CHEMICAL | - | Phytochemical | CID_5281553 | (E)-beta-ocimene | {"name": "(E)-beta-ocimene", "id": "CID_5281553"} |
| Herb | - | curcuma longa | {"name": "curcuma longa", "part": "rhizome"} | CONTAINS_CHEMICAL | - | Phytochemical | CID_5282109 | Geranyl formate | {"name": "Geranyl formate", "id": "CID_5282109"} |
| Herb | - | curcuma longa | {"name": "curcuma longa", "part": "rhizome"} | CONTAINS_CHEMICAL | - | Phytochemical | CID_5284507 | Nerolidol | {"name": "Nerolidol", "id": "CID_5284507"} |
| Herb | - | curcuma longa | {"name": "curcuma longa", "part": "rhizome"} | CONTAINS_CHEMICAL | - | Phytochemical | CID_530426 | Isoledene | {"name": "Isoledene", "id": "CID_530426"} |

**Table S9: Complete node and relationship data used to construct the knowledge graph. This table summarizes interactions among FDA-approved drugs, formulation, herbs, diseases (type 2 diabetes, fatty liver, obesity), phytochemicals, and their target genes/proteins. Each entry includes source and target node details, edge type, and node annotations (Entrez Gene ID and gene symbol, where applicable).**

| Source Node Label | Source Node ID | Source Node Name | Source Node Properties | Edge Type | Edge Properties | Target Node Label | Target Node ID | Target Node Name | Target Node Properties |
|---|---|---|---|---|---|---|---|---|---|
| Herb | - | curcuma longa | {"name": "curcuma longa", "part": "rhizome"} | CONTAINS_CHEMICAL | - | Phytochemical | CID_5315469 | Bisacumol | {"name": "Bisacumol", "id": "CID_5315469"} |
| Herb | - | curcuma longa | {"name": "curcuma longa", "part": "rhizome"} | CONTAINS_CHEMICAL | - | Phytochemical | CID_5315472 | Bisdemethoxycurcumin | {"name": "Bisdemethoxycurcumin", "id": "CID_5315472"} |
| Herb | - | curcuma longa | {"name": "curcuma longa", "part": "rhizome"} | CONTAINS_CHEMICAL | - | Phytochemical | CID_5317319 | cis-beta-Farnesene | {"name": "cis-beta-Farnesene", "id": "CID_5317319"} |
| Herb | - | curcuma longa | {"name": "curcuma longa", "part": "rhizome"} | CONTAINS_CHEMICAL | - | Phytochemical | CID_5317844 | alpha-Guaiene | {"name": "alpha-Guaiene", "id": "CID_5317844"} |
| Herb | - | curcuma longa | {"name": "curcuma longa", "part": "rhizome"} | CONTAINS_CHEMICAL | - | Phytochemical | CID_5320250 | (Z)-beta-Ocimene | {"name": "(Z)-beta-Ocimene", "id": "CID_5320250"} |
| Herb | - | curcuma longa | {"name": "curcuma longa", "part": "rhizome"} | CONTAINS_CHEMICAL | - | Phytochemical | CID_5355856 | Geranyl butyrate | {"name": "Geranyl butyrate", "id": "CID_5355856"} |
| Herb | - | curcuma longa | {"name": "curcuma longa", "part": "rhizome"} | CONTAINS_CHEMICAL | - | Phytochemical | CID_5365992 | Geranyl hexanoate | {"name": "Geranyl hexanoate", "id": "CID_5365992"} |
| Herb | - | curcuma longa | {"name": "curcuma longa", "part": "rhizome"} | CONTAINS_CHEMICAL | - | Phytochemical | CID_5469424 | Demethoxycurcumin | {"name": "Demethoxycurcumin", "id": "CID_5469424"} |
| Herb | - | curcuma longa | {"name": "curcuma longa", "part": "rhizome"} | CONTAINS_CHEMICAL | - | Phytochemical | CID_572766 | Curzerene | {"name": "Curzerene", "id": "CID_572766"} |
| Herb | - | curcuma longa | {"name": "curcuma longa", "part": "rhizome"} | CONTAINS_CHEMICAL | - | Phytochemical | CID_57386731 | delta-Curcumene | {"name": "delta-Curcumene", "id": "CID_57386731"} |
| Herb | - | curcuma longa | {"name": "curcuma longa", "part": "rhizome"} | CONTAINS_CHEMICAL | - | Phytochemical | CID_5997 | Cholesterol | {"name": "Cholesterol", "id": "CID_5997"} |
| Herb | - | curcuma longa | {"name": "curcuma longa", "part": "rhizome"} | CONTAINS_CHEMICAL | - | Phytochemical | CID_61060 | d-Borneol | {"name": "d-Borneol", "id": "CID_61060"} |
| Herb | - | curcuma longa | {"name": "curcuma longa", "part": "rhizome"} | CONTAINS_CHEMICAL | - | Phytochemical | CID_61098 | Linalyl propionate | {"name": "Linalyl propionate", "id": "CID_61098"} |
| Herb | - | curcuma longa | {"name": "curcuma longa", "part": "rhizome"} | CONTAINS_CHEMICAL | - | Phytochemical | CID_61130 | Myrtenal | {"name": "Myrtenal", "id": "CID_61130"} |
| Herb | - | curcuma longa | {"name": "curcuma longa", "part": "rhizome"} | CONTAINS_CHEMICAL | - | Phytochemical | CID_62385 | 1-Methyl-4-(prop-1-en-2-yl)benzene | {"name": "1-Methyl-4-(prop-1-en-2-yl)benzene", "id": "CID_62385"} |
| Herb | - | curcuma longa | {"name": "curcuma longa", "part": "rhizome"} | CONTAINS_CHEMICAL | - | Phytochemical | CID_637563 | Anethole | {"name": "Anethole", "id": "CID_637563"} |
| Herb | - | curcuma longa | {"name": "curcuma longa", "part": "rhizome"} | CONTAINS_CHEMICAL | - | Phytochemical | CID_637566 | Geraniol | {"name": "Geraniol", "id": "CID_637566"} |
| Herb | - | curcuma longa | {"name": "curcuma longa", "part": "rhizome"} | CONTAINS_CHEMICAL | - | Phytochemical | CID_638011 | Citral | {"name": "Citral", "id": "CID_638011"} |
| Herb | - | curcuma longa | {"name": "curcuma longa", "part": "rhizome"} | CONTAINS_CHEMICAL | - | Phytochemical | CID_638024 | Piperine | {"name": "Piperine", "id": "CID_638024"} |
| Herb | - | curcuma longa | {"name": "curcuma longa", "part": "rhizome"} | CONTAINS_CHEMICAL | - | Phytochemical | CID_6428435 | 7-epi-cis-Sesquisabinene hydrate | {"name": "7-epi-cis-Sesquisabinene hydrate", "id": "CID_6428435"} |
| Herb | - | curcuma longa | {"name": "curcuma longa", "part": "rhizome"} | CONTAINS_CHEMICAL | - | Phytochemical | CID_6428444 | trans-Sesquisabinene hydrate | {"name": "trans-Sesquisabinene hydrate", "id": "CID_6428444"} |
| Herb | - | curcuma longa | {"name": "curcuma longa", "part": "rhizome"} | CONTAINS_CHEMICAL | - | Phytochemical | CID_6429302 | trans-alpha-Bergamotene | {"name": "trans-alpha-Bergamotene", "id": "CID_6429302"} |
| Herb | - | curcuma longa | {"name": "curcuma longa", "part": "rhizome"} | CONTAINS_CHEMICAL | - | Phytochemical | CID_6430754 | 10-epi-gamma-Eudesmol | {"name": "10-epi-gamma-Eudesmol", "id": "CID_6430754"} |
| Herb | - | curcuma longa | {"name": "curcuma longa", "part": "rhizome"} | CONTAINS_CHEMICAL | - | Phytochemical | CID_6432254 | trans-Linalool oxide | {"name": "trans-Linalool oxide", "id": "CID_6432254"} |
| Herb | - | curcuma longa | {"name": "curcuma longa", "part": "rhizome"} | CONTAINS_CHEMICAL | - | Phytochemical | CID_6432312 | gamma-Elemene | {"name": "gamma-Elemene", "id": "CID_6432312"} |
| Herb | - | curcuma longa | {"name": "curcuma longa", "part": "rhizome"} | CONTAINS_CHEMICAL | - | Phytochemical | CID_6436348 | Germacrone | {"name": "Germacrone", "id": "CID_6436348"} |
| Herb | - | curcuma longa | {"name": "curcuma longa", "part": "rhizome"} | CONTAINS_CHEMICAL | - | Phytochemical | CID_643779 | Neral | {"name": "Neral", "id": "CID_643779"} |
| Herb | - | curcuma longa | {"name": "curcuma longa", "part": "rhizome"} | CONTAINS_CHEMICAL | - | Phytochemical | CID_6441391 | Germacr-1(10)-ene-5,8-dione | {"name": "Germacr-1(10)-ene-5,8-dione", "id": "CID_6441391"} |
| Herb | - | curcuma longa | {"name": "curcuma longa", "part": "rhizome"} | CONTAINS_CHEMICAL | - | Phytochemical | CID_6442374 | Furanodienon | {"name": "Furanodienon", "id": "CID_6442374"} |
| Herb | - | curcuma longa | {"name": "curcuma longa", "part": "rhizome"} | CONTAINS_CHEMICAL | - | Phytochemical | CID_6451618 | (1R)-2-methyl-5-propan-2-ylbicyclo[3.1.0]hex-2-ene | {"name": "(1R)-2-methyl-5-propan-2-ylbicyclo[3.1.0]hex-2-ene", "id": "CID_6451618"} |
| Herb | - | curcuma longa | {"name": "curcuma longa", "part": "rhizome"} | CONTAINS_CHEMICAL | - | Phytochemical | CID_6532 | Linalyl isobutyrate | {"name": "Linalyl isobutyrate", "id": "CID_6532"} |
| Herb | - | curcuma longa | {"name": "curcuma longa", "part": "rhizome"} | CONTAINS_CHEMICAL | - | Phytochemical | CID_6549 | Linalool | {"name": "Linalool", "id": "CID_6549"} |
| Herb | - | curcuma longa | {"name": "curcuma longa", "part": "rhizome"} | CONTAINS_CHEMICAL | - | Phytochemical | CID_6616 | Camphene | {"name": "Camphene", "id": "CID_6616"} |
| Herb | - | curcuma longa | {"name": "curcuma longa", "part": "rhizome"} | CONTAINS_CHEMICAL | - | Phytochemical | CID_6654 | alpha-Pinene | {"name": "alpha-Pinene", "id": "CID_6654"} |
| Herb | - | curcuma longa | {"name": "curcuma longa", "part": "rhizome"} | CONTAINS_CHEMICAL | - | Phytochemical | CID_68252 | Thymol acetate | {"name": "Thymol acetate", "id": "CID_68252"} |
| Herb | - | curcuma longa | {"name": "curcuma longa", "part": "rhizome"} | CONTAINS_CHEMICAL | - | Phytochemical | CID_68381 | Perilla ketone | {"name": "Perilla ketone", "id": "CID_68381"} |
| Herb | - | curcuma longa | {"name": "curcuma longa", "part": "rhizome"} | CONTAINS_CHEMICAL | - | Phytochemical | CID_6918391 | beta-Elemene | {"name": "beta-Elemene", "id": "CID_6918391"} |
| Herb | - | curcuma longa | {"name": "curcuma longa", "part": "rhizome"} | CONTAINS_CHEMICAL | - | Phytochemical | CID_69879809 | Cyclocurcumin | {"name": "Cyclocurcumin", "id": "CID_69879809"} |
| Herb | - | curcuma longa | {"name": "curcuma longa", "part": "rhizome"} | CONTAINS_CHEMICAL | - | Phytochemical | CID_6989 | Thymol | {"name": "Thymol", "id": "CID_6989"} |
| Herb | - | curcuma longa | {"name": "curcuma longa", "part": "rhizome"} | CONTAINS_CHEMICAL | - | Phytochemical | CID_7127 | Methyleugenol | {"name": "Methyleugenol", "id": "CID_7127"} |
| Herb | - | curcuma longa | {"name": "curcuma longa", "part": "rhizome"} | CONTAINS_CHEMICAL | - | Phytochemical | CID_71346280 | 1,7-Bis(4-hydroxyphenyl)-1,4,6-heptatrien-3-one | {"name": "1,7-Bis(4-hydroxyphenyl)-1,4,6-heptatrien-3-one", "id": "CID_71346280"} |
| Herb | - | curcuma longa | {"name": "curcuma longa", "part": "rhizome"} | CONTAINS_CHEMICAL | - | Phytochemical | CID_7439 | Carvone | {"name": "Carvone", "id": "CID_7439"} |
| Herb | - | curcuma longa | {"name": "curcuma longa", "part": "rhizome"} | CONTAINS_CHEMICAL | - | Phytochemical | CID_7460 | alpha-Phellandrene | {"name": "alpha-Phellandrene", "id": "CID_7460"} |
| Herb | - | curcuma longa | {"name": "curcuma longa", "part": "rhizome"} | CONTAINS_CHEMICAL | - | Phytochemical | CID_7461 | gamma-Terpinene | {"name": "gamma-Terpinene", "id": "CID_7461"} |
| Herb | - | curcuma longa | {"name": "curcuma longa", "part": "rhizome"} | CONTAINS_CHEMICAL | - | Phytochemical | CID_7462 | alpha-Terpinene | {"name": "alpha-Terpinene", "id": "CID_7462"} |
| Herb | - | curcuma longa | {"name": "curcuma longa", "part": "rhizome"} | CONTAINS_CHEMICAL | - | Phytochemical | CID_7463 | p-Cymene | {"name": "p-Cymene", "id": "CID_7463"} |
| Herb | - | curcuma longa | {"name": "curcuma longa", "part": "rhizome"} | CONTAINS_CHEMICAL | - | Phytochemical | CID_79035 | Tricyclene | {"name": "Tricyclene", "id": "CID_79035"} |
| Herb | - | curcuma longa | {"name": "curcuma longa", "part": "rhizome"} | CONTAINS_CHEMICAL | - | Phytochemical | CID_80420 | Heptyl salicylate | {"name": "Heptyl salicylate", "id": "CID_80420"} |
| Herb | - | curcuma longa | {"name": "curcuma longa", "part": "rhizome"} | CONTAINS_CHEMICAL | - | Phytochemical | CID_8051 | 2-Heptanone | {"name": "2-Heptanone", "id": "CID_8051"} |

**Table S9: Complete node and relationship data used to construct the knowledge graph. This table summarizes interactions among FDA-approved drugs, formulation, herbs, diseases (type 2 diabetes, fatty liver, obesity), phytochemicals, and their target genes/proteins. Each entry includes source and target node details, edge type, and node annotations (Entrez Gene ID and gene symbol, where applicable).**

| Source Node Label | Source Node ID | Source Node Name | Source Node Properties | Edge Type | Edge Properties | Target Node Label | Target Node ID | Target Node Name | Target Node Properties |
|---|---|---|---|---|---|---|---|---|---|
| Herb | - | curcuma longa | {"name": "curcuma longa", "part": "rhizome"} | CONTAINS_CHEMICAL | - | Phytochemical | CID_8163 | 2-Undecanone | {"name": "2-Undecanone", "id": "CID_8163"} |
| Herb | - | curcuma longa | {"name": "curcuma longa", "part": "rhizome"} | CONTAINS_CHEMICAL | - | Phytochemical | CID_8184 | 1-Undecanol | {"name": "1-Undecanol", "id": "CID_8184"} |
| Herb | - | curcuma longa | {"name": "curcuma longa", "part": "rhizome"} | CONTAINS_CHEMICAL | - | Phytochemical | CID_8294 | Linalyl acetate | {"name": "Linalyl acetate", "id": "CID_8294"} |
| Herb | - | curcuma longa | {"name": "curcuma longa", "part": "rhizome"} | CONTAINS_CHEMICAL | - | Phytochemical | CID_8500 | 4'-Methylacetophenone | {"name": "4'-Methylacetophenone", "id": "CID_8500"} |
| Herb | - | curcuma longa | {"name": "curcuma longa", "part": "rhizome"} | CONTAINS_CHEMICAL | - | Phytochemical | CID_87014 | 2-(Hydroxymethyl)anthraquinone | {"name": "2-(Hydroxymethyl)anthraquinone", "id": "CID_87014"} |
| Herb | - | curcuma longa | {"name": "curcuma longa", "part": "rhizome"} | CONTAINS_CHEMICAL | - | Phytochemical | CID_91457 | beta-Eudesmol | {"name": "beta-Eudesmol", "id": "CID_91457"} |
| Herb | - | curcuma longa | {"name": "curcuma longa", "part": "rhizome"} | CONTAINS_CHEMICAL | - | Phytochemical | CID_91723653 | (S,1Z,6Z)-8-Isopropyl-1-methyl-5-methylenecyclodeca-1,6-diene | {"name": "(S,1Z,6Z)-8-Isopropyl-1-methyl-5-methylenecyclodeca-1,6-diene", "id": "CID_91723653"} |
| Herb | - | curcuma longa | {"name": "curcuma longa", "part": "rhizome"} | CONTAINS_CHEMICAL | - | Phytochemical | CID_91753231 | (4S,5S)-Germacrone-4,5-epoxide | {"name": "(4S,5S)-Germacrone-4,5-epoxide", "id": "CID_91753231"} |
| Herb | - | curcuma longa | {"name": "curcuma longa", "part": "rhizome"} | CONTAINS_CHEMICAL | - | Phytochemical | CID_91753502 | cis-alpha-Bergamotene | {"name": "cis-alpha-Bergamotene", "id": "CID_91753502"} |
| Herb | - | curcuma longa | {"name": "curcuma longa", "part": "rhizome"} | CONTAINS_CHEMICAL | - | Phytochemical | CID_92139 | alpha-Curcumene | {"name": "alpha-Curcumene", "id": "CID_92139"} |
| Herb | - | curcuma longa | {"name": "curcuma longa", "part": "rhizome"} | CONTAINS_CHEMICAL | - | Phytochemical | CID_92776 | Zingiberene | {"name": "Zingiberene", "id": "CID_92776"} |
| Herb | - | curcuma longa | {"name": "curcuma longa", "part": "rhizome"} | CONTAINS_CHEMICAL | - | Phytochemical | CID_93135 | Xanthorrhizol | {"name": "Xanthorrhizol", "id": "CID_93135"} |
| Herb | - | curcuma longa | {"name": "curcuma longa", "part": "rhizome"} | CONTAINS_CHEMICAL | - | Phytochemical | CID_94147 | (+)-cis-Sabinol | {"name": "(+)-cis-Sabinol", "id": "CID_94147"} |
| Herb | - | curcuma longa | {"name": "curcuma longa", "part": "rhizome"} | CONTAINS_CHEMICAL | - | Phytochemical | CID_94266 | Sabinyl acetate | {"name": "Sabinyl acetate", "id": "CID_94266"} |
| Herb | - | curcuma longa | {"name": "curcuma longa", "part": "rhizome"} | CONTAINS_CHEMICAL | - | Phytochemical | CID_969516 | Curcumin | {"name": "Curcumin", "id": "CID_969516"} |
| Herb | - | phyllanthus emblica | {"name": "phyllanthus emblica", "part": "fruit"} | CONTAINS_CHEMICAL | - | Phytochemical | CID_1057 | Pyrogallol | {"name": "Pyrogallol", "id": "CID_1057"} |
| Herb | - | phyllanthus emblica | {"name": "phyllanthus emblica", "part": "fruit"} | CONTAINS_CHEMICAL | - | Phytochemical | CID_12314211 | Phyllantidine | {"name": "Phyllantidine", "id": "CID_12314211"} |
| Herb | - | phyllanthus emblica | {"name": "phyllanthus emblica", "part": "fruit"} | CONTAINS_CHEMICAL | - | Phytochemical | CID_124021 | beta-Glucogallin | {"name": "beta-Glucogallin", "id": "CID_124021"} |
| Herb | - | phyllanthus emblica | {"name": "phyllanthus emblica", "part": "fruit"} | CONTAINS_CHEMICAL | - | Phytochemical | CID_13250 | Ethyl gallate | {"name": "Ethyl gallate", "id": "CID_13250"} |
| Herb | - | phyllanthus emblica | {"name": "phyllanthus emblica", "part": "fruit"} | CONTAINS_CHEMICAL | - | Phytochemical | CID_193552 | Phyllanthine | {"name": "Phyllanthine", "id": "CID_193552"} |
| Herb | - | phyllanthus emblica | {"name": "phyllanthus emblica", "part": "fruit"} | CONTAINS_CHEMICAL | - | Phytochemical | CID_289 | Catechol | {"name": "Catechol", "id": "CID_289"} |
| Herb | - | phyllanthus emblica | {"name": "phyllanthus emblica", "part": "fruit"} | CONTAINS_CHEMICAL | - | Phytochemical | CID_359 | Phloroglucinol | {"name": "Phloroglucinol", "id": "CID_359"} |
| Herb | - | phyllanthus emblica | {"name": "phyllanthus emblica", "part": "fruit"} | CONTAINS_CHEMICAL | - | Phytochemical | CID_370 | Gallic acid | {"name": "Gallic acid", "id": "CID_370"} |
| Herb | - | phyllanthus emblica | {"name": "phyllanthus emblica", "part": "fruit"} | CONTAINS_CHEMICAL | - | Phytochemical | CID_449093 | trans-Zeatin | {"name": "trans-Zeatin", "id": "CID_449093"} |
| Herb | - | phyllanthus emblica | {"name": "phyllanthus emblica", "part": "fruit"} | CONTAINS_CHEMICAL | - | Phytochemical | CID_493570 | Riboflavin | {"name": "Riboflavin", "id": "CID_493570"} |
| Herb | - | phyllanthus emblica | {"name": "phyllanthus emblica", "part": "fruit"} | CONTAINS_CHEMICAL | - | Phytochemical | CID_5280343 | Quercetin | {"name": "Quercetin", "id": "CID_5280343"} |
| Herb | - | phyllanthus emblica | {"name": "phyllanthus emblica", "part": "fruit"} | CONTAINS_CHEMICAL | - | Phytochemical | CID_5281855 | Ellagic acid | {"name": "Ellagic acid", "id": "CID_5281855"} |
| Herb | - | phyllanthus emblica | {"name": "phyllanthus emblica", "part": "fruit"} | CONTAINS_CHEMICAL | - | Phytochemical | CID_54670067 | Ascorbic acid | {"name": "Ascorbic acid", "id": "CID_54670067"} |
| Herb | - | phyllanthus emblica | {"name": "phyllanthus emblica", "part": "fruit"} | CONTAINS_CHEMICAL | - | Phytochemical | CID_7428 | Methyl gallate | {"name": "Methyl gallate", "id": "CID_7428"} |
| Herb | - | picrorhiza kurroa | {"name": "picrorhiza kurroa", "part": "rhizome"} | CONTAINS_CHEMICAL | - | Phytochemical | CID_8468 | Vanillic acid | {"name": "Vanillic acid", "id": "CID_8468"} |
| Herb | - | picrorhiza kurroa | {"name": "picrorhiza kurroa", "part": "rhizome"} | CONTAINS_CHEMICAL | - | Phytochemical | CID_91458 | Aucubin | {"name": "Aucubin", "id": "CID_91458"} |
| Herb | - | picrorhiza kurroa | {"name": "picrorhiza kurroa", "part": "rhizome"} | CONTAINS_CHEMICAL | - | Phytochemical | CID_91520 | Catalpol | {"name": "Catalpol", "id": "CID_91520"} |
| Herb | - | picrorhiza kurroa | {"name": "picrorhiza kurroa", "part": "rhizome"} | CONTAINS_CHEMICAL | - | Phytochemical | CID_92123 | Picein | {"name": "Picein", "id": "CID_92123"} |
| Herb | - | picrorhiza kurroa | {"name": "picrorhiza kurroa", "part": "rhizome"} | CONTAINS_CHEMICAL | - | Phytochemical | CID_ 119287 | cucurbitacins | {"name": "cucurbitacins", "id": "CID_ 119287"} |
| Herb | - | picrorhiza kurroa | {"name": "picrorhiza kurroa", "part": "rhizome"} | CONTAINS_CHEMICAL | - | Phytochemical | CID_14081907 | Bartsioside | {"name": "Bartsioside", "id": "CID_14081907"} |
| Herb | - | picrorhiza kurroa | {"name": "picrorhiza kurroa", "part": "rhizome"} | CONTAINS_CHEMICAL | - | Phytochemical | CID_155942 | 8-Epiiridotrial glucoside | {"name": "8-Epiiridotrial glucoside", "id": "CID_155942"} |
| Herb | - | picrorhiza kurroa | {"name": "picrorhiza kurroa", "part": "rhizome"} | CONTAINS_CHEMICAL | - | Phytochemical | CID_164648 | Androsin | {"name": "Androsin", "id": "CID_164648"} |
| Herb | - | picrorhiza kurroa | {"name": "picrorhiza kurroa", "part": "rhizome"} | CONTAINS_CHEMICAL | - | Phytochemical | CID_2214 | Acetovanillone | {"name": "Acetovanillone", "id": "CID_2214"} |
| Herb | - | picrorhiza kurroa | {"name": "picrorhiza kurroa", "part": "rhizome"} | CONTAINS_CHEMICAL | - | Phytochemical | CID_443353 | 10-Deoxygeniposidic acid | {"name": "10-Deoxygeniposidic acid", "id": "CID_443353"} |
| Herb | - | picrorhiza kurroa | {"name": "picrorhiza kurroa", "part": "rhizome"} | CONTAINS_CHEMICAL | - | Phytochemical | CID_6440892 | Picroside I | {"name": "Picroside I", "id": "CID_6440892"} |
| Herb | - | plumbago zeylanica | {"name": "plumbago zeylanica", "part": "root"} | CONTAINS_CHEMICAL | - | Phytochemical | CID_100947536 | Isozeylanone | {"name": "Isozeylanone", "id": "CID_100947536"} |
| Herb | - | plumbago zeylanica | {"name": "plumbago zeylanica", "part": "root"} | CONTAINS_CHEMICAL | - | Phytochemical | CID_100947539 | Plumbazeylanone | {"name": "Plumbazeylanone", "id": "CID_100947539"} |
| Herb | - | plumbago zeylanica | {"name": "plumbago zeylanica", "part": "root"} | CONTAINS_CHEMICAL | - | Phytochemical | CID_10205 | Plumbagin | {"name": "Plumbagin", "id": "CID_10205"} |
| Herb | - | plumbago zeylanica | {"name": "plumbago zeylanica", "part": "root"} | CONTAINS_CHEMICAL | - | Phytochemical | CID_146680 | (2,2'-Binaphthalene)-5,5',8,8'-tetrone, 1,1'-dihydroxy-6,6'-dimethyl- | {"name": "(2,2'-Binaphthalene)-5,5',8,8'-tetrone, 1,1'-dihydroxy-6,6'-dimethyl-", "id": "CID_146680"} |
| Herb | - | plumbago zeylanica | {"name": "plumbago zeylanica", "part": "root"} | CONTAINS_CHEMICAL | - | Phytochemical | CID_271296 | Methylnaphthazarin | {"name": "Methylnaphthazarin", "id": "CID_271296"} |
| Herb | - | plumbago zeylanica | {"name": "plumbago zeylanica", "part": "root"} | CONTAINS_CHEMICAL | - | Phytochemical | CID_2723872 | D-Fructose | {"name": "D-Fructose", "id": "CID_2723872"} |
| Herb | - | plumbago zeylanica | {"name": "plumbago zeylanica", "part": "root"} | CONTAINS_CHEMICAL | - | Phytochemical | CID_338719 | 3-Chloro-5-hydroxy-2-methylnaphthalene-1,4-dione | {"name": "3-Chloro-5-hydroxy-2-methylnaphthalene-1,4-dione", "id": "CID_338719"} |
| Herb | - | plumbago zeylanica | {"name": "plumbago zeylanica", "part": "root"} | CONTAINS_CHEMICAL | - | Phytochemical | CID_442739 | Droserone | {"name": "Droserone", "id": "CID_442739"} |

**Table S9: Complete node and relationship data used to construct the knowledge graph. This table summarizes interactions among FDA-approved drugs, formulation, herbs, diseases (type 2 diabetes, fatty liver, obesity), phytochemicals, and their target genes/proteins. Each entry includes source and target node details, edge type, and node annotations (Entrez Gene ID and gene symbol, where applicable).**

| Source Node Label | Source Node ID | Source Node Name | Source Node Properties | Edge Type | Edge Properties | Target Node Label | Target Node ID | Target Node Name | Target Node Properties |
|---|---|---|---|---|---|---|---|---|---|
| Herb | - | plumbago zeylanica | {"name": "plumbago zeylanica", "part": "root"} | CONTAINS_CHEMICAL | - | Phytochemical | CID_5276618 | Zeylanone | {"name": "Zeylanone", "id": "CID_5276618"} |
| Herb | - | plumbago zeylanica | {"name": "plumbago zeylanica", "part": "root"} | CONTAINS_CHEMICAL | - | Phytochemical | CID_5793 | D-Glucose | {"name": "D-Glucose", "id": "CID_5793"} |
| Herb | - | plumbago zeylanica | {"name": "plumbago zeylanica", "part": "root"} | CONTAINS_CHEMICAL | - | Phytochemical | CID_633024 | 5-Hydroxy-8-(4-hydroxy-7-methyl-5,8-dioxonaphthalen-1-yl)-2-methylnaphthalene-1,4-dione | {"name": "5-Hydroxy-8-(4-hydroxy-7-methyl-5,8-dioxonaphthalen-1-yl)-2-methylnaphthalene-1,4-dione", "id": "CID_633024"} |
| Herb | - | plumbago zeylanica | {"name": "plumbago zeylanica", "part": "root"} | CONTAINS_CHEMICAL | - | Phytochemical | CID_633072 | Chitranone | {"name": "Chitranone", "id": "CID_633072"} |
| Herb | - | terminalia bellirica | {"name": "terminalia bellirica", "part": "fruit"} | CONTAINS_CHEMICAL | - | Phytochemical | CID_124021 | beta-Glucogallin | {"name": "beta-Glucogallin", "id": "CID_124021"} |
| Herb | - | terminalia bellirica | {"name": "terminalia bellirica", "part": "fruit"} | CONTAINS_CHEMICAL | - | Phytochemical | CID_12411 | Tritriacontane | {"name": "Tritriacontane", "id": "CID_12411"} |
| Herb | - | terminalia bellirica | {"name": "terminalia bellirica", "part": "fruit"} | CONTAINS_CHEMICAL | - | Phytochemical | CID_12444386 | Arjungenin | {"name": "Arjungenin", "id": "CID_12444386"} |
| Herb | - | terminalia bellirica | {"name": "terminalia bellirica", "part": "fruit"} | CONTAINS_CHEMICAL | - | Phytochemical | CID_129688132 | Tritriacontan-9-one | {"name": "Tritriacontan-9-one", "id": "CID_129688132"} |
| Herb | - | terminalia bellirica | {"name": "terminalia bellirica", "part": "fruit"} | CONTAINS_CHEMICAL | - | Phytochemical | CID_13250 | Ethyl gallate | {"name": "Ethyl gallate", "id": "CID_13250"} |
| Herb | - | terminalia bellirica | {"name": "terminalia bellirica", "part": "fruit"} | CONTAINS_CHEMICAL | - | Phytochemical | CID_14105933 | (4aS,6aR,6aS,6bR,8aR,10R,11R,12aR,14bS)-10,11-dihydroxy-9,9-bis(hydroxymethyl)-2,2,6a,6b,12a-pentamethyl-1,3,4,5,6,6a,7,8,8a,10,11,12,13,14b-tetradecahydropicene-4a-carboxylic acid | {"name": "(4aS,6aR,6aS,6bR,8aR,10R,11R,12aR,14bS)-10,11-dihydroxy-9,9-bis(hydroxymethyl)-2,2,6a,6b,12a-pentamethyl-1,3,4,5,6,6a,7,8,8a,10,11,12,13,14b-tetradecahydropicene-4a-carboxylic acid", "id": "CID_14105933"} |
| Herb | - | terminalia bellirica | {"name": "terminalia bellirica", "part": "fruit"} | CONTAINS_CHEMICAL | - | Phytochemical | CID_21726781 | (4aS,6aR,6aR,6bR,7R,8aR,9R,10R,11R,12aR,14bS)-7,10,11-trihydroxy-9-(hydroxymethyl)-2,2,6a,6b,9,12a-hexamethyl-1,3,4,5,6,6a,7,8,8a,10,11,12,13,14b-tetradecahydropicene-4a-carboxylic acid | {"name": "(4aS,6aR,6aR,6bR,7R,8aR,9R,10R,11R,12aR,14bS)-7,10,11-trihydroxy-9-(hydroxymethyl)-2,2,6a,6b,9,12a-hexamethyl-1,3,4,5,6,6a,7,8,8a,10,11,12,13,14b-tetradecahydropicene-4a-carboxylic acid", "id": "CID_21726781"} |
| Herb | - | terminalia bellirica | {"name": "terminalia bellirica", "part": "fruit"} | CONTAINS_CHEMICAL | - | Phytochemical | CID_222284 | beta-Sitosterol | {"name": "beta-Sitosterol", "id": "CID_222284"} |
| Herb | - | terminalia bellirica | {"name": "terminalia bellirica", "part": "fruit"} | CONTAINS_CHEMICAL | - | Phytochemical | CID_25310 | L-Rhamnose | {"name": "L-Rhamnose", "id": "CID_25310"} |
| Herb | - | terminalia bellirica | {"name": "terminalia bellirica", "part": "fruit"} | CONTAINS_CHEMICAL | - | Phytochemical | CID_26519 | Tetratriacontane | {"name": "Tetratriacontane", "id": "CID_26519"} |
| Herb | - | terminalia bellirica | {"name": "terminalia bellirica", "part": "fruit"} | CONTAINS_CHEMICAL | - | Phytochemical | CID_2723872 | D-Fructose | {"name": "D-Fructose", "id": "CID_2723872"} |
| Herb | - | terminalia bellirica | {"name": "terminalia bellirica", "part": "fruit"} | CONTAINS_CHEMICAL | - | Phytochemical | CID_370 | Gallic acid | {"name": "Gallic acid", "id": "CID_370"} |
| Herb | - | terminalia bellirica | {"name": "terminalia bellirica", "part": "fruit"} | CONTAINS_CHEMICAL | - | Phytochemical | CID_466076 | Termilignan | {"name": "Termilignan", "id": "CID_466076"} |
| Herb | - | terminalia bellirica | {"name": "terminalia bellirica", "part": "fruit"} | CONTAINS_CHEMICAL | - | Phytochemical | CID_466078 | 7-Hydroxy-3',4'-methylenedioxyflavan | {"name": "7-Hydroxy-3',4'-methylenedioxyflavan", "id": "CID_466078"} |
| Herb | - | terminalia bellirica | {"name": "terminalia bellirica", "part": "fruit"} | CONTAINS_CHEMICAL | - | Phytochemical | CID_5281855 | Ellagic acid | {"name": "Ellagic acid", "id": "CID_5281855"} |
| Herb | - | terminalia bellirica | {"name": "terminalia bellirica", "part": "fruit"} | CONTAINS_CHEMICAL | - | Phytochemical | CID_5793 | D-Glucose | {"name": "D-Glucose", "id": "CID_5793"} |
| Herb | - | terminalia bellirica | {"name": "terminalia bellirica", "part": "fruit"} | CONTAINS_CHEMICAL | - | Phytochemical | CID_6036 | D-Galactose | {"name": "D-Galactose", "id": "CID_6036"} |
| Herb | - | terminalia bellirica | {"name": "terminalia bellirica", "part": "fruit"} | CONTAINS_CHEMICAL | - | Phytochemical | CID_6251 | Mannitol | {"name": "Mannitol", "id": "CID_6251"} |
| Herb | - | terminalia bellirica | {"name": "terminalia bellirica", "part": "fruit"} | CONTAINS_CHEMICAL | - | Phytochemical | CID_72388 | Anolignan B | {"name": "Anolignan B", "id": "CID_72388"} |
| Herb | - | terminalia chebula | {"name": "terminalia chebula", "part": "fruit"} | CONTAINS_CHEMICAL | - | Phytochemical | CID_10467 | Arachidic acid | {"name": "Arachidic acid", "id": "CID_10467"} |
| Herb | - | terminalia chebula | {"name": "terminalia chebula", "part": "fruit"} | CONTAINS_CHEMICAL | - | Phytochemical | CID_10471 | Triacontanoic acid | {"name": "Triacontanoic acid", "id": "CID_10471"} |
| Herb | - | terminalia chebula | {"name": "terminalia chebula", "part": "fruit"} | CONTAINS_CHEMICAL | - | Phytochemical | CID_10742 | Syringic acid | {"name": "Syringic acid", "id": "CID_10742"} |
| Herb | - | terminalia chebula | {"name": "terminalia chebula", "part": "fruit"} | CONTAINS_CHEMICAL | - | Phytochemical | CID_12444386 | Arjungenin | {"name": "Arjungenin", "id": "CID_12444386"} |
| Herb | - | terminalia chebula | {"name": "terminalia chebula", "part": "fruit"} | CONTAINS_CHEMICAL | - | Phytochemical | CID_13250 | Ethyl gallate | {"name": "Ethyl gallate", "id": "CID_13250"} |
| Herb | - | terminalia chebula | {"name": "terminalia chebula", "part": "fruit"} | CONTAINS_CHEMICAL | - | Phytochemical | CID_222284 | beta-Sitosterol | {"name": "beta-Sitosterol", "id": "CID_222284"} |
| Herb | - | terminalia chebula | {"name": "terminalia chebula", "part": "fruit"} | CONTAINS_CHEMICAL | - | Phytochemical | CID_370 | Gallic acid | {"name": "Gallic acid", "id": "CID_370"} |
| Herb | - | terminalia chebula | {"name": "terminalia chebula", "part": "fruit"} | CONTAINS_CHEMICAL | - | Phytochemical | CID_445639 | Oleic acid | {"name": "Oleic acid", "id": "CID_445639"} |
| Herb | - | terminalia chebula | {"name": "terminalia chebula", "part": "fruit"} | CONTAINS_CHEMICAL | - | Phytochemical | CID_445858 | Ferulic acid | {"name": "Ferulic acid", "id": "CID_445858"} |
| Herb | - | terminalia chebula | {"name": "terminalia chebula", "part": "fruit"} | CONTAINS_CHEMICAL | - | Phytochemical | CID_5280450 | Linoleic acid | {"name": "Linoleic acid", "id": "CID_5280450"} |
| Herb | - | terminalia chebula | {"name": "terminalia chebula", "part": "fruit"} | CONTAINS_CHEMICAL | - | Phytochemical | CID_5281 | Stearic acid | {"name": "Stearic acid", "id": "CID_5281"} |
| Herb | - | terminalia chebula | {"name": "terminalia chebula", "part": "fruit"} | CONTAINS_CHEMICAL | - | Phytochemical | CID_5281855 | Ellagic acid | {"name": "Ellagic acid", "id": "CID_5281855"} |
| Herb | - | terminalia chebula | {"name": "terminalia chebula", "part": "fruit"} | CONTAINS_CHEMICAL | - | Phytochemical | CID_54670067 | Ascorbic acid | {"name": "Ascorbic acid", "id": "CID_54670067"} |
| Herb | - | terminalia chebula | {"name": "terminalia chebula", "part": "fruit"} | CONTAINS_CHEMICAL | - | Phytochemical | CID_5742590 | Daucosterol | {"name": "Daucosterol", "id": "CID_5742590"} |
| Herb | - | terminalia chebula | {"name": "terminalia chebula", "part": "fruit"} | CONTAINS_CHEMICAL | - | Phytochemical | CID_637542 | 4-Hydroxycinnamic acid | {"name": "4-Hydroxycinnamic acid", "id": "CID_637542"} |
| Herb | - | terminalia chebula | {"name": "terminalia chebula", "part": "fruit"} | CONTAINS_CHEMICAL | - | Phytochemical | CID_6780 | Anthraquinone | {"name": "Anthraquinone", "id": "CID_6780"} |
| Herb | - | terminalia chebula | {"name": "terminalia chebula", "part": "fruit"} | CONTAINS_CHEMICAL | - | Phytochemical | CID_689043 | Caffeic acid | {"name": "Caffeic acid", "id": "CID_689043"} |
| Herb | - | terminalia chebula | {"name": "terminalia chebula", "part": "fruit"} | CONTAINS_CHEMICAL | - | Phytochemical | CID_73641 | Arjunolic acid | {"name": "Arjunolic acid", "id": "CID_73641"} |
| Herb | - | terminalia chebula | {"name": "terminalia chebula", "part": "fruit"} | CONTAINS_CHEMICAL | - | Phytochemical | CID_73659 | Maslinic acid | {"name": "Maslinic acid", "id": "CID_73659"} |
| Herb | - | terminalia chebula | {"name": "terminalia chebula", "part": "fruit"} | CONTAINS_CHEMICAL | - | Phytochemical | CID_8215 | Docosanoic acid | {"name": "Docosanoic acid", "id": "CID_8215"} |
| Herb | - | terminalia chebula | {"name": "terminalia chebula", "part": "fruit"} | CONTAINS_CHEMICAL | - | Phytochemical | CID_8468 | Vanillic acid | {"name": "Vanillic acid", "id": "CID_8468"} |
| Herb | - | terminalia chebula | {"name": "terminalia chebula", "part": "fruit"} | CONTAINS_CHEMICAL | - | Phytochemical | CID_8742 | Shikimic acid | {"name": "Shikimic acid", "id": "CID_8742"} |

**Table S9: Complete node and relationship data used to construct the knowledge graph. This table summarizes interactions among FDA-approved drugs, formulation, herbs, diseases (type 2 diabetes, fatty liver, obesity), phytochemicals, and their target genes/proteins. Each entry includes source and target node details, edge type, and node annotations (Entrez Gene ID and gene symbol, where applicable).**

| Source Node Label | Source Node ID | Source Node Name | Source Node Properties | Edge Type | Edge Properties | Target Node Label | Target Node ID | Target Node Name | Target Node Properties |
|---|---|---|---|---|---|---|---|---|---|
| Herb | - | terminalia chebula | {"name": "terminalia chebula", "part": "fruit"} | CONTAINS_CHEMICAL | - | Phytochemical | CID_985 | Palmitic acid | {"name": "Palmitic acid", "id": "CID_985"} |
| Phytochemical | CID_12310886 | (1S,2S,5S,8S)-2-methyl-6-methylidene-9-propan-2-ylidene-11-oxatricyclo[6.2.1.01,5]undecan-8-ol | {"name": "(1S,2S,5S,8S)-2-methyl-6-methylidene-9-propan-2-ylidene-11-oxatricyclo[6.2.1.01,5]undecan-8-ol", "id": "CID_12310886"} | INTERACTS_WITH | - | Target | 5770 | PTPN1 | {"name": "PTPN1", "id": "5770"} |
| Phytochemical | CID_92747 | (20R)-20-Hydroxypregn-4-en-3-one | {"name": "(20R)-20-Hydroxypregn-4-en-3-one", "id": "CID_92747"} | INTERACTS_WITH | - | Target | 1586 | CYP17A1 | {"name": "CYP17A1", "id": "1586"} |
| Phytochemical | CID_92747 | (20R)-20-Hydroxypregn-4-en-3-one | {"name": "(20R)-20-Hydroxypregn-4-en-3-one", "id": "CID_92747"} | INTERACTS_WITH | - | Target | 5241 | PGR | {"name": "PGR", "id": "5241"} |
| Phytochemical | CID_8184 | 1-Undecanol | {"name": "1-Undecanol", "id": "CID_8184"} | INTERACTS_WITH | - | Target | 3315 | HSPB1 | {"name": "HSPB1", "id": "3315"} |
| Phytochemical | CID_8184 | 1-Undecanol | {"name": "1-Undecanol", "id": "CID_8184"} | INTERACTS_WITH | - | Target | 367 | AR | {"name": "AR", "id": "367"} |
| Phytochemical | CID_8184 | 1-Undecanol | {"name": "1-Undecanol", "id": "CID_8184"} | INTERACTS_WITH | - | Target | 3725 | JUN | {"name": "JUN", "id": "3725"} |
| Phytochemical | CID_8184 | 1-Undecanol | {"name": "1-Undecanol", "id": "CID_8184"} | INTERACTS_WITH | - | Target | 4780 | NFE2L2 | {"name": "NFE2L2", "id": "4780"} |
| Phytochemical | CID_8184 | 1-Undecanol | {"name": "1-Undecanol", "id": "CID_8184"} | INTERACTS_WITH | - | Target | 4790 | NFKB1 | {"name": "NFKB1", "id": "4790"} |
| Phytochemical | CID_8184 | 1-Undecanol | {"name": "1-Undecanol", "id": "CID_8184"} | INTERACTS_WITH | - | Target | 8989 | TRPA1 | {"name": "TRPA1", "id": "8989"} |
| Phytochemical | CID_8051 | 2-Heptanone | {"name": "2-Heptanone", "id": "CID_8051"} | INTERACTS_WITH | - | Target | 3315 | HSPB1 | {"name": "HSPB1", "id": "3315"} |
| Phytochemical | CID_15448 | 2-Undecanol | {"name": "2-Undecanol", "id": "CID_15448"} | INTERACTS_WITH | - | Target | 8989 | TRPA1 | {"name": "TRPA1", "id": "8989"} |
| Phytochemical | CID_8163 | 2-Undecanone | {"name": "2-Undecanone", "id": "CID_8163"} | INTERACTS_WITH | - | Target | 8989 | TRPA1 | {"name": "TRPA1", "id": "8989"} |
| Phytochemical | CID_26049 | 3-Carene | {"name": "3-Carene", "id": "CID_26049"} | INTERACTS_WITH | - | Target | 3315 | HSPB1 | {"name": "HSPB1", "id": "3315"} |
| Phytochemical | CID_26049 | 3-Carene | {"name": "3-Carene", "id": "CID_26049"} | INTERACTS_WITH | - | Target | 367 | AR | {"name": "AR", "id": "367"} |
| Phytochemical | CID_26049 | 3-Carene | {"name": "3-Carene", "id": "CID_26049"} | INTERACTS_WITH | - | Target | 4780 | NFE2L2 | {"name": "NFE2L2", "id": "4780"} |
| Phytochemical | CID_14458886 | 3-Deacetylsalannin | {"name": "3-Deacetylsalannin", "id": "CID_14458886"} | INTERACTS_WITH | - | Target | 3320 | HSP90AA1 | {"name": "HSP90AA1", "id": "3320"} |
| Phytochemical | CID_24834047 | 4,10-Epizedoarondiol | {"name": "4,10-Epizedoarondiol", "id": "CID_24834047"} | INTERACTS_WITH | - | Target | 5770 | PTPN1 | {"name": "PTPN1", "id": "5770"} |
| Phytochemical | CID_24834047 | 4,10-Epizedoarondiol | {"name": "4,10-Epizedoarondiol", "id": "CID_24834047"} | INTERACTS_WITH | - | Target | 5771 | PTPN2 | {"name": "PTPN2", "id": "5771"} |
| Phytochemical | CID_11230 | 4-Carvomenthenol | {"name": "4-Carvomenthenol", "id": "CID_11230"} | INTERACTS_WITH | - | Target | 54657 | UGT1A4 | {"name": "UGT1A4", "id": "54657"} |
| Phytochemical | CID_637542 | 4-Hydroxycinnamic acid | {"name": "4-Hydroxycinnamic acid", "id": "CID_637542"} | INTERACTS_WITH | - | Target | 10013 | HDAC6 | {"name": "HDAC6", "id": "10013"} |
| Phytochemical | CID_637542 | 4-Hydroxycinnamic acid | {"name": "4-Hydroxycinnamic acid", "id": "CID_637542"} | INTERACTS_WITH | - | Target | 1109 | AKR1C4 | {"name": "AKR1C4", "id": "1109"} |
| Phytochemical | CID_637542 | 4-Hydroxycinnamic acid | {"name": "4-Hydroxycinnamic acid", "id": "CID_637542"} | INTERACTS_WITH | - | Target | 11238 | CA5B | {"name": "CA5B", "id": "11238"} |
| Phytochemical | CID_637542 | 4-Hydroxycinnamic acid | {"name": "4-Hydroxycinnamic acid", "id": "CID_637542"} | INTERACTS_WITH | - | Target | 1645 | AKR1C1 | {"name": "AKR1C1", "id": "1645"} |
| Phytochemical | CID_637542 | 4-Hydroxycinnamic acid | {"name": "4-Hydroxycinnamic acid", "id": "CID_637542"} | INTERACTS_WITH | - | Target | 1646 | AKR1C2 | {"name": "AKR1C2", "id": "1646"} |
| Phytochemical | CID_637542 | 4-Hydroxycinnamic acid | {"name": "4-Hydroxycinnamic acid", "id": "CID_637542"} | INTERACTS_WITH | - | Target | 1803 | DPP4 | {"name": "DPP4", "id": "1803"} |
| Phytochemical | CID_637542 | 4-Hydroxycinnamic acid | {"name": "4-Hydroxycinnamic acid", "id": "CID_637542"} | INTERACTS_WITH | - | Target | 1956 | EGFR | {"name": "EGFR", "id": "1956"} |
| Phytochemical | CID_637542 | 4-Hydroxycinnamic acid | {"name": "4-Hydroxycinnamic acid", "id": "CID_637542"} | INTERACTS_WITH | - | Target | 2099 | ESR1 | {"name": "ESR1", "id": "2099"} |
| Phytochemical | CID_637542 | 4-Hydroxycinnamic acid | {"name": "4-Hydroxycinnamic acid", "id": "CID_637542"} | INTERACTS_WITH | - | Target | 2100 | ESR2 | {"name": "ESR2", "id": "2100"} |
| Phytochemical | CID_637542 | 4-Hydroxycinnamic acid | {"name": "4-Hydroxycinnamic acid", "id": "CID_637542"} | INTERACTS_WITH | - | Target | 231 | AKR1B1 | {"name": "AKR1B1", "id": "231"} |
| Phytochemical | CID_637542 | 4-Hydroxycinnamic acid | {"name": "4-Hydroxycinnamic acid", "id": "CID_637542"} | INTERACTS_WITH | - | Target | 23632 | CA14 | {"name": "CA14", "id": "23632"} |
| Phytochemical | CID_637542 | 4-Hydroxycinnamic acid | {"name": "4-Hydroxycinnamic acid", "id": "CID_637542"} | INTERACTS_WITH | - | Target | 259285 | TAS2R39 | {"name": "TAS2R39", "id": "259285"} |
| Phytochemical | CID_637542 | 4-Hydroxycinnamic acid | {"name": "4-Hydroxycinnamic acid", "id": "CID_637542"} | INTERACTS_WITH | - | Target | 338442 | HCAR2 | {"name": "HCAR2", "id": "338442"} |
| Phytochemical | CID_637542 | 4-Hydroxycinnamic acid | {"name": "4-Hydroxycinnamic acid", "id": "CID_637542"} | INTERACTS_WITH | - | Target | 5315 | PKM | {"name": "PKM", "id": "5315"} |
| Phytochemical | CID_637542 | 4-Hydroxycinnamic acid | {"name": "4-Hydroxycinnamic acid", "id": "CID_637542"} | INTERACTS_WITH | - | Target | 57016 | AKR1B10 | {"name": "AKR1B10", "id": "57016"} |
| Phytochemical | CID_637542 | 4-Hydroxycinnamic acid | {"name": "4-Hydroxycinnamic acid", "id": "CID_637542"} | INTERACTS_WITH | - | Target | 5743 | PTGS2 | {"name": "PTGS2", "id": "5743"} |
| Phytochemical | CID_637542 | 4-Hydroxycinnamic acid | {"name": "4-Hydroxycinnamic acid", "id": "CID_637542"} | INTERACTS_WITH | - | Target | 7299 | TYR | {"name": "TYR", "id": "7299"} |
| Phytochemical | CID_637542 | 4-Hydroxycinnamic acid | {"name": "4-Hydroxycinnamic acid", "id": "CID_637542"} | INTERACTS_WITH | - | Target | 759 | CA1 | {"name": "CA1", "id": "759"} |
| Phytochemical | CID_637542 | 4-Hydroxycinnamic acid | {"name": "4-Hydroxycinnamic acid", "id": "CID_637542"} | INTERACTS_WITH | - | Target | 760 | CA2 | {"name": "CA2", "id": "760"} |
| Phytochemical | CID_637542 | 4-Hydroxycinnamic acid | {"name": "4-Hydroxycinnamic acid", "id": "CID_637542"} | INTERACTS_WITH | - | Target | 761 | CA3 | {"name": "CA3", "id": "761"} |
| Phytochemical | CID_637542 | 4-Hydroxycinnamic acid | {"name": "4-Hydroxycinnamic acid", "id": "CID_637542"} | INTERACTS_WITH | - | Target | 762 | CA4 | {"name": "CA4", "id": "762"} |
| Phytochemical | CID_637542 | 4-Hydroxycinnamic acid | {"name": "4-Hydroxycinnamic acid", "id": "CID_637542"} | INTERACTS_WITH | - | Target | 763 | CA5A | {"name": "CA5A", "id": "763"} |
| Phytochemical | CID_637542 | 4-Hydroxycinnamic acid | {"name": "4-Hydroxycinnamic acid", "id": "CID_637542"} | INTERACTS_WITH | - | Target | 765 | CA6 | {"name": "CA6", "id": "765"} |
| Phytochemical | CID_637542 | 4-Hydroxycinnamic acid | {"name": "4-Hydroxycinnamic acid", "id": "CID_637542"} | INTERACTS_WITH | - | Target | 766 | CA7 | {"name": "CA7", "id": "766"} |
| Phytochemical | CID_637542 | 4-Hydroxycinnamic acid | {"name": "4-Hydroxycinnamic acid", "id": "CID_637542"} | INTERACTS_WITH | - | Target | 768 | CA9 | {"name": "CA9", "id": "768"} |
| Phytochemical | CID_637542 | 4-Hydroxycinnamic acid | {"name": "4-Hydroxycinnamic acid", "id": "CID_637542"} | INTERACTS_WITH | - | Target | 771 | CA12 | {"name": "CA12", "id": "771"} |
| Phytochemical | CID_637542 | 4-Hydroxycinnamic acid | {"name": "4-Hydroxycinnamic acid", "id": "CID_637542"} | INTERACTS_WITH | - | Target | 815 | CAMK2A | {"name": "CAMK2A", "id": "815"} |
| Phytochemical | CID_637542 | 4-Hydroxycinnamic acid | {"name": "4-Hydroxycinnamic acid", "id": "CID_637542"} | INTERACTS_WITH | - | Target | 8644 | AKR1C3 | {"name": "AKR1C3", "id": "8644"} |
| Phytochemical | CID_325 | 4-Isopropylbenzyl alcohol | {"name": "4-Isopropylbenzyl alcohol", "id": "CID_325"} | INTERACTS_WITH | - | Target | 5467 | PPARD | {"name": "PPARD", "id": "5467"} |

**Table S9: Complete node and relationship data used to construct the knowledge graph. This table summarizes interactions among FDA-approved drugs, formulation, herbs, diseases (type 2 diabetes, fatty liver, obesity), phytochemicals, and their target genes/proteins. Each entry includes source and target node details, edge type, and node annotations (Entrez Gene ID and gene symbol, where applicable).**

| Source Node Label | Source Node ID | Source Node Name | Source Node Properties | Edge Type | Edge Properties | Target Node Label | Target Node ID | Target Node Name | Target Node Properties |
|---|---|---|---|---|---|---|---|---|---|
| Phytochemical | CID_5204 | 5-[3-(1,3-Benzodioxol-5-yl)-1,3,3a,4,6,6a-hexahydrofuro[3,4-c]furan-6-yl]-1,3-benzodioxole | {"name": "5-[3-(1,3-Benzodioxol-5-yl)-1,3,3a,4,6,6a-hexahydrofuro[3,4-c]furan-6-yl]-1,3-benzodioxole", "id": "CID_5204"} | INTERACTS_WITH | - | Target | 10951 | CBX1 | {"name": "CBX1", "id": "10951"} |
| Phytochemical | CID_5204 | 5-[3-(1,3-Benzodioxol-5-yl)-1,3,3a,4,6,6a-hexahydrofuro[3,4-c]furan-6-yl]-1,3-benzodioxole | {"name": "5-[3-(1,3-Benzodioxol-5-yl)-1,3,3a,4,6,6a-hexahydrofuro[3,4-c]furan-6-yl]-1,3-benzodioxole", "id": "CID_5204"} | INTERACTS_WITH | - | Target | 11201 | POLI | {"name": "POLI", "id": "11201"} |
| Phytochemical | CID_5204 | 5-[3-(1,3-Benzodioxol-5-yl)-1,3,3a,4,6,6a-hexahydrofuro[3,4-c]furan-6-yl]-1,3-benzodioxole | {"name": "5-[3-(1,3-Benzodioxol-5-yl)-1,3,3a,4,6,6a-hexahydrofuro[3,4-c]furan-6-yl]-1,3-benzodioxole", "id": "CID_5204"} | INTERACTS_WITH | - | Target | 23435 | TARDBP | {"name": "TARDBP", "id": "23435"} |
| Phytochemical | CID_5204 | 5-[3-(1,3-Benzodioxol-5-yl)-1,3,3a,4,6,6a-hexahydrofuro[3,4-c]furan-6-yl]-1,3-benzodioxole | {"name": "5-[3-(1,3-Benzodioxol-5-yl)-1,3,3a,4,6,6a-hexahydrofuro[3,4-c]furan-6-yl]-1,3-benzodioxole", "id": "CID_5204"} | INTERACTS_WITH | - | Target | 2740 | GLP1R | {"name": "GLP1R", "id": "2740"} |
| Phytochemical | CID_5204 | 5-[3-(1,3-Benzodioxol-5-yl)-1,3,3a,4,6,6a-hexahydrofuro[3,4-c]furan-6-yl]-1,3-benzodioxole | {"name": "5-[3-(1,3-Benzodioxol-5-yl)-1,3,3a,4,6,6a-hexahydrofuro[3,4-c]furan-6-yl]-1,3-benzodioxole", "id": "CID_5204"} | INTERACTS_WITH | - | Target | 51053 | GMNN | {"name": "GMNN", "id": "51053"} |
| Phytochemical | CID_5204 | 5-[3-(1,3-Benzodioxol-5-yl)-1,3,3a,4,6,6a-hexahydrofuro[3,4-c]furan-6-yl]-1,3-benzodioxole | {"name": "5-[3-(1,3-Benzodioxol-5-yl)-1,3,3a,4,6,6a-hexahydrofuro[3,4-c]furan-6-yl]-1,3-benzodioxole", "id": "CID_5204"} | INTERACTS_WITH | - | Target | 55775 | TDP1 | {"name": "TDP1", "id": "55775"} |
| Phytochemical | CID_5204 | 5-[3-(1,3-Benzodioxol-5-yl)-1,3,3a,4,6,6a-hexahydrofuro[3,4-c]furan-6-yl]-1,3-benzodioxole | {"name": "5-[3-(1,3-Benzodioxol-5-yl)-1,3,3a,4,6,6a-hexahydrofuro[3,4-c]furan-6-yl]-1,3-benzodioxole", "id": "CID_5204"} | INTERACTS_WITH | - | Target | 5745 | PTH1R | {"name": "PTH1R", "id": "5745"} |
| Phytochemical | CID_155942 | 8-Epiiridotrial glucoside | {"name": "8-Epiiridotrial glucoside", "id": "CID_155942"} | INTERACTS_WITH | - | Target | 5742 | PTGS1 | {"name": "PTGS1", "id": "5742"} |
| Phytochemical | CID_155942 | 8-Epiiridotrial glucoside | {"name": "8-Epiiridotrial glucoside", "id": "CID_155942"} | INTERACTS_WITH | - | Target | 5743 | PTGS2 | {"name": "PTGS2", "id": "5743"} |
| Phytochemical | CID_2214 | Acetovanillone | {"name": "Acetovanillone", "id": "CID_2214"} | INTERACTS_WITH | - | Target | 10013 | HDAC6 | {"name": "HDAC6", "id": "10013"} |
| Phytochemical | CID_2214 | Acetovanillone | {"name": "Acetovanillone", "id": "CID_2214"} | INTERACTS_WITH | - | Target | 10951 | CBX1 | {"name": "CBX1", "id": "10951"} |
| Phytochemical | CID_2214 | Acetovanillone | {"name": "Acetovanillone", "id": "CID_2214"} | INTERACTS_WITH | - | Target | 51053 | GMNN | {"name": "GMNN", "id": "51053"} |
| Phytochemical | CID_2214 | Acetovanillone | {"name": "Acetovanillone", "id": "CID_2214"} | INTERACTS_WITH | - | Target | 55775 | TDP1 | {"name": "TDP1", "id": "55775"} |
| Phytochemical | CID_637563 | Anethole | {"name": "Anethole", "id": "CID_637563"} | INTERACTS_WITH | - | Target | 10599 | SLCO1B1 | {"name": "SLCO1B1", "id": "10599"} |
| Phytochemical | CID_637563 | Anethole | {"name": "Anethole", "id": "CID_637563"} | INTERACTS_WITH | - | Target | 216 | ALDH1A1 | {"name": "ALDH1A1", "id": "216"} |
| Phytochemical | CID_637563 | Anethole | {"name": "Anethole", "id": "CID_637563"} | INTERACTS_WITH | - | Target | 28234 | SLCO1B3 | {"name": "SLCO1B3", "id": "28234"} |
| Phytochemical | CID_637563 | Anethole | {"name": "Anethole", "id": "CID_637563"} | INTERACTS_WITH | - | Target | 367 | AR | {"name": "AR", "id": "367"} |
| Phytochemical | CID_637563 | Anethole | {"name": "Anethole", "id": "CID_637563"} | INTERACTS_WITH | - | Target | 390245 | KDM4E | {"name": "KDM4E", "id": "390245"} |
| Phytochemical | CID_637563 | Anethole | {"name": "Anethole", "id": "CID_637563"} | INTERACTS_WITH | - | Target | 4318 | MMP9 | {"name": "MMP9", "id": "4318"} |
| Phytochemical | CID_637563 | Anethole | {"name": "Anethole", "id": "CID_637563"} | INTERACTS_WITH | - | Target | 4780 | NFE2L2 | {"name": "NFE2L2", "id": "4780"} |
| Phytochemical | CID_637563 | Anethole | {"name": "Anethole", "id": "CID_637563"} | INTERACTS_WITH | - | Target | 5467 | PPARD | {"name": "PPARD", "id": "5467"} |
| Phytochemical | CID_6780 | Anthraquinone | {"name": "Anthraquinone", "id": "CID_6780"} | INTERACTS_WITH | - | Target | 10013 | HDAC6 | {"name": "HDAC6", "id": "10013"} |
| Phytochemical | CID_6780 | Anthraquinone | {"name": "Anthraquinone", "id": "CID_6780"} | INTERACTS_WITH | - | Target | 10599 | SLCO1B1 | {"name": "SLCO1B1", "id": "10599"} |
| Phytochemical | CID_6780 | Anthraquinone | {"name": "Anthraquinone", "id": "CID_6780"} | INTERACTS_WITH | - | Target | 10919 | EHMT2 | {"name": "EHMT2", "id": "10919"} |
| Phytochemical | CID_6780 | Anthraquinone | {"name": "Anthraquinone", "id": "CID_6780"} | INTERACTS_WITH | - | Target | 2099 | ESR1 | {"name": "ESR1", "id": "2099"} |
| Phytochemical | CID_6780 | Anthraquinone | {"name": "Anthraquinone", "id": "CID_6780"} | INTERACTS_WITH | - | Target | 2100 | ESR2 | {"name": "ESR2", "id": "2100"} |
| Phytochemical | CID_6780 | Anthraquinone | {"name": "Anthraquinone", "id": "CID_6780"} | INTERACTS_WITH | - | Target | 28234 | SLCO1B3 | {"name": "SLCO1B3", "id": "28234"} |
| Phytochemical | CID_6780 | Anthraquinone | {"name": "Anthraquinone", "id": "CID_6780"} | INTERACTS_WITH | - | Target | 4129 | MAOB | {"name": "MAOB", "id": "4129"} |
| Phytochemical | CID_6780 | Anthraquinone | {"name": "Anthraquinone", "id": "CID_6780"} | INTERACTS_WITH | - | Target | 4297 | KMT2A | {"name": "KMT2A", "id": "4297"} |
| Phytochemical | CID_6780 | Anthraquinone | {"name": "Anthraquinone", "id": "CID_6780"} | INTERACTS_WITH | - | Target | 4780 | NFE2L2 | {"name": "NFE2L2", "id": "4780"} |
| Phytochemical | CID_6780 | Anthraquinone | {"name": "Anthraquinone", "id": "CID_6780"} | INTERACTS_WITH | - | Target | 5166 | PDK4 | {"name": "PDK4", "id": "5166"} |
| Phytochemical | CID_6780 | Anthraquinone | {"name": "Anthraquinone", "id": "CID_6780"} | INTERACTS_WITH | - | Target | 54576 | UGT1A8 | {"name": "UGT1A8", "id": "54576"} |
| Phytochemical | CID_6780 | Anthraquinone | {"name": "Anthraquinone", "id": "CID_6780"} | INTERACTS_WITH | - | Target | 54658 | UGT1A1 | {"name": "UGT1A1", "id": "54658"} |
| Phytochemical | CID_6780 | Anthraquinone | {"name": "Anthraquinone", "id": "CID_6780"} | INTERACTS_WITH | - | Target | 6256 | RXRA | {"name": "RXRA", "id": "6256"} |
| Phytochemical | CID_6780 | Anthraquinone | {"name": "Anthraquinone", "id": "CID_6780"} | INTERACTS_WITH | - | Target | 79813 | EHMT1 | {"name": "EHMT1", "id": "79813"} |
| Phytochemical | CID_6780 | Anthraquinone | {"name": "Anthraquinone", "id": "CID_6780"} | INTERACTS_WITH | - | Target | 994 | CDC25B | {"name": "CDC25B", "id": "994"} |
| Phytochemical | CID_10467 | Arachidic acid | {"name": "Arachidic acid", "id": "CID_10467"} | INTERACTS_WITH | - | Target | 5465 | PPARA | {"name": "PPARA", "id": "5465"} |
| Phytochemical | CID_10467 | Arachidic acid | {"name": "Arachidic acid", "id": "CID_10467"} | INTERACTS_WITH | - | Target | 5467 | PPARD | {"name": "PPARD", "id": "5467"} |
| Phytochemical | CID_10467 | Arachidic acid | {"name": "Arachidic acid", "id": "CID_10467"} | INTERACTS_WITH | - | Target | 5468 | PPARG | {"name": "PPARG", "id": "5468"} |
| Phytochemical | CID_10467 | Arachidic acid | {"name": "Arachidic acid", "id": "CID_10467"} | INTERACTS_WITH | - | Target | 5770 | PTPN1 | {"name": "PTPN1", "id": "5770"} |
| Phytochemical | CID_73641 | Arjunolic acid | {"name": "Arjunolic acid", "id": "CID_73641"} | INTERACTS_WITH | - | Target | 151306 | GPBAR1 | {"name": "GPBAR1", "id": "151306"} |
| Phytochemical | CID_73641 | Arjunolic acid | {"name": "Arjunolic acid", "id": "CID_73641"} | INTERACTS_WITH | - | Target | 240 | ALOX5 | {"name": "ALOX5", "id": "240"} |
| Phytochemical | CID_73641 | Arjunolic acid | {"name": "Arjunolic acid", "id": "CID_73641"} | INTERACTS_WITH | - | Target | 247 | ALOX15B | {"name": "ALOX15B", "id": "247"} |
| Phytochemical | CID_73641 | Arjunolic acid | {"name": "Arjunolic acid", "id": "CID_73641"} | INTERACTS_WITH | - | Target | 5743 | PTGS2 | {"name": "PTGS2", "id": "5743"} |
| Phytochemical | CID_73641 | Arjunolic acid | {"name": "Arjunolic acid", "id": "CID_73641"} | INTERACTS_WITH | - | Target | 9971 | NR1H4 | {"name": "NR1H4", "id": "9971"} |
| Phytochemical | CID_54670067 | Ascorbic acid | {"name": "Ascorbic acid", "id": "CID_54670067"} | INTERACTS_WITH | - | Target | 10280 | SIGMAR1 | {"name": "SIGMAR1", "id": "10280"} |
| Phytochemical | CID_54670067 | Ascorbic acid | {"name": "Ascorbic acid", "id": "CID_54670067"} | INTERACTS_WITH | - | Target | 10599 | SLCO1B1 | {"name": "SLCO1B1", "id": "10599"} |
| Phytochemical | CID_54670067 | Ascorbic acid | {"name": "Ascorbic acid", "id": "CID_54670067"} | INTERACTS_WITH | - | Target | 10800 | CYSLTR1 | {"name": "CYSLTR1", "id": "10800"} |
| Phytochemical | CID_54670067 | Ascorbic acid | {"name": "Ascorbic acid", "id": "CID_54670067"} | INTERACTS_WITH | - | Target | 10919 | EHMT2 | {"name": "EHMT2", "id": "10919"} |

**Table S9: Complete node and relationship data used to construct the knowledge graph. This table summarizes interactions among FDA-approved drugs, formulation, herbs, diseases (type 2 diabetes, fatty liver, obesity), phytochemicals, and their target genes/proteins. Each entry includes source and target node details, edge type, and node annotations (Entrez Gene ID and gene symbol, where applicable).**

| Source Node Label | Source Node ID | Source Node Name | Source Node Properties | Edge Type | Edge Properties | Target Node Label | Target Node ID | Target Node Name | Target Node Properties |
|---|---|---|---|---|---|---|---|---|---|
| Phytochemical | CID_54670067 | Ascorbic acid | {"name": "Ascorbic acid", "id": "CID_54670067"} | INTERACTS_WITH | - | Target | 112398 | EGLN2 | {"name": "EGLN2", "id": "112398"} |
| Phytochemical | CID_54670067 | Ascorbic acid | {"name": "Ascorbic acid", "id": "CID_54670067"} | INTERACTS_WITH | - | Target | 112399 | EGLN3 | {"name": "EGLN3", "id": "112399"} |
| Phytochemical | CID_54670067 | Ascorbic acid | {"name": "Ascorbic acid", "id": "CID_54670067"} | INTERACTS_WITH | - | Target | 11255 | HRH3 | {"name": "HRH3", "id": "11255"} |
| Phytochemical | CID_54670067 | Ascorbic acid | {"name": "Ascorbic acid", "id": "CID_54670067"} | INTERACTS_WITH | - | Target | 1128 | CHRM1 | {"name": "CHRM1", "id": "1128"} |
| Phytochemical | CID_54670067 | Ascorbic acid | {"name": "Ascorbic acid", "id": "CID_54670067"} | INTERACTS_WITH | - | Target | 1129 | CHRM2 | {"name": "CHRM2", "id": "1129"} |
| Phytochemical | CID_54670067 | Ascorbic acid | {"name": "Ascorbic acid", "id": "CID_54670067"} | INTERACTS_WITH | - | Target | 1131 | CHRM3 | {"name": "CHRM3", "id": "1131"} |
| Phytochemical | CID_54670067 | Ascorbic acid | {"name": "Ascorbic acid", "id": "CID_54670067"} | INTERACTS_WITH | - | Target | 1132 | CHRM4 | {"name": "CHRM4", "id": "1132"} |
| Phytochemical | CID_54670067 | Ascorbic acid | {"name": "Ascorbic acid", "id": "CID_54670067"} | INTERACTS_WITH | - | Target | 1133 | CHRM5 | {"name": "CHRM5", "id": "1133"} |
| Phytochemical | CID_54670067 | Ascorbic acid | {"name": "Ascorbic acid", "id": "CID_54670067"} | INTERACTS_WITH | - | Target | 1233 | CCR4 | {"name": "CCR4", "id": "1233"} |
| Phytochemical | CID_54670067 | Ascorbic acid | {"name": "Ascorbic acid", "id": "CID_54670067"} | INTERACTS_WITH | - | Target | 1234 | CCR5 | {"name": "CCR5", "id": "1234"} |
| Phytochemical | CID_54670067 | Ascorbic acid | {"name": "Ascorbic acid", "id": "CID_54670067"} | INTERACTS_WITH | - | Target | 1268 | CNR1 | {"name": "CNR1", "id": "1268"} |
| Phytochemical | CID_54670067 | Ascorbic acid | {"name": "Ascorbic acid", "id": "CID_54670067"} | INTERACTS_WITH | - | Target | 134 | ADORA1 | {"name": "ADORA1", "id": "134"} |
| Phytochemical | CID_54670067 | Ascorbic acid | {"name": "Ascorbic acid", "id": "CID_54670067"} | INTERACTS_WITH | - | Target | 135 | ADORA2A | {"name": "ADORA2A", "id": "135"} |
| Phytochemical | CID_54670067 | Ascorbic acid | {"name": "Ascorbic acid", "id": "CID_54670067"} | INTERACTS_WITH | - | Target | 140 | ADORA3 | {"name": "ADORA3", "id": "140"} |
| Phytochemical | CID_54670067 | Ascorbic acid | {"name": "Ascorbic acid", "id": "CID_54670067"} | INTERACTS_WITH | - | Target | 1432 | MAPK14 | {"name": "MAPK14", "id": "1432"} |
| Phytochemical | CID_54670067 | Ascorbic acid | {"name": "Ascorbic acid", "id": "CID_54670067"} | INTERACTS_WITH | - | Target | 146 | ADRA1D | {"name": "ADRA1D", "id": "146"} |
| Phytochemical | CID_54670067 | Ascorbic acid | {"name": "Ascorbic acid", "id": "CID_54670067"} | INTERACTS_WITH | - | Target | 148 | ADRA1A | {"name": "ADRA1A", "id": "148"} |
| Phytochemical | CID_54670067 | Ascorbic acid | {"name": "Ascorbic acid", "id": "CID_54670067"} | INTERACTS_WITH | - | Target | 150 | ADRA2A | {"name": "ADRA2A", "id": "150"} |
| Phytochemical | CID_54670067 | Ascorbic acid | {"name": "Ascorbic acid", "id": "CID_54670067"} | INTERACTS_WITH | - | Target | 151 | ADRA2B | {"name": "ADRA2B", "id": "151"} |
| Phytochemical | CID_54670067 | Ascorbic acid | {"name": "Ascorbic acid", "id": "CID_54670067"} | INTERACTS_WITH | - | Target | 1511 | CTSG | {"name": "CTSG", "id": "1511"} |
| Phytochemical | CID_54670067 | Ascorbic acid | {"name": "Ascorbic acid", "id": "CID_54670067"} | INTERACTS_WITH | - | Target | 152 | ADRA2C | {"name": "ADRA2C", "id": "152"} |
| Phytochemical | CID_54670067 | Ascorbic acid | {"name": "Ascorbic acid", "id": "CID_54670067"} | INTERACTS_WITH | - | Target | 153 | ADRB1 | {"name": "ADRB1", "id": "153"} |
| Phytochemical | CID_54670067 | Ascorbic acid | {"name": "Ascorbic acid", "id": "CID_54670067"} | INTERACTS_WITH | - | Target | 154 | ADRB2 | {"name": "ADRB2", "id": "154"} |
| Phytochemical | CID_54670067 | Ascorbic acid | {"name": "Ascorbic acid", "id": "CID_54670067"} | INTERACTS_WITH | - | Target | 1544 | CYP1A2 | {"name": "CYP1A2", "id": "1544"} |
| Phytochemical | CID_54670067 | Ascorbic acid | {"name": "Ascorbic acid", "id": "CID_54670067"} | INTERACTS_WITH | - | Target | 1548 | CYP2A6 | {"name": "CYP2A6", "id": "1548"} |
| Phytochemical | CID_54670067 | Ascorbic acid | {"name": "Ascorbic acid", "id": "CID_54670067"} | INTERACTS_WITH | - | Target | 155 | ADRB3 | {"name": "ADRB3", "id": "155"} |
| Phytochemical | CID_54670067 | Ascorbic acid | {"name": "Ascorbic acid", "id": "CID_54670067"} | INTERACTS_WITH | - | Target | 1557 | CYP2C19 | {"name": "CYP2C19", "id": "1557"} |
| Phytochemical | CID_54670067 | Ascorbic acid | {"name": "Ascorbic acid", "id": "CID_54670067"} | INTERACTS_WITH | - | Target | 1559 | CYP2C9 | {"name": "CYP2C9", "id": "1559"} |
| Phytochemical | CID_54670067 | Ascorbic acid | {"name": "Ascorbic acid", "id": "CID_54670067"} | INTERACTS_WITH | - | Target | 1565 | CYP2D6 | {"name": "CYP2D6", "id": "1565"} |
| Phytochemical | CID_54670067 | Ascorbic acid | {"name": "Ascorbic acid", "id": "CID_54670067"} | INTERACTS_WITH | - | Target | 1571 | CYP2E1 | {"name": "CYP2E1", "id": "1571"} |
| Phytochemical | CID_54670067 | Ascorbic acid | {"name": "Ascorbic acid", "id": "CID_54670067"} | INTERACTS_WITH | - | Target | 1576 | CYP3A4 | {"name": "CYP3A4", "id": "1576"} |
| Phytochemical | CID_54670067 | Ascorbic acid | {"name": "Ascorbic acid", "id": "CID_54670067"} | INTERACTS_WITH | - | Target | 1812 | DRD1 | {"name": "DRD1", "id": "1812"} |
| Phytochemical | CID_54670067 | Ascorbic acid | {"name": "Ascorbic acid", "id": "CID_54670067"} | INTERACTS_WITH | - | Target | 1813 | DRD2 | {"name": "DRD2", "id": "1813"} |
| Phytochemical | CID_54670067 | Ascorbic acid | {"name": "Ascorbic acid", "id": "CID_54670067"} | INTERACTS_WITH | - | Target | 1814 | DRD3 | {"name": "DRD3", "id": "1814"} |
| Phytochemical | CID_54670067 | Ascorbic acid | {"name": "Ascorbic acid", "id": "CID_54670067"} | INTERACTS_WITH | - | Target | 1815 | DRD4 | {"name": "DRD4", "id": "1815"} |
| Phytochemical | CID_54670067 | Ascorbic acid | {"name": "Ascorbic acid", "id": "CID_54670067"} | INTERACTS_WITH | - | Target | 186 | AGTR2 | {"name": "AGTR2", "id": "186"} |
| Phytochemical | CID_54670067 | Ascorbic acid | {"name": "Ascorbic acid", "id": "CID_54670067"} | INTERACTS_WITH | - | Target | 1909 | EDNRA | {"name": "EDNRA", "id": "1909"} |
| Phytochemical | CID_54670067 | Ascorbic acid | {"name": "Ascorbic acid", "id": "CID_54670067"} | INTERACTS_WITH | - | Target | 1956 | EGFR | {"name": "EGFR", "id": "1956"} |
| Phytochemical | CID_54670067 | Ascorbic acid | {"name": "Ascorbic acid", "id": "CID_54670067"} | INTERACTS_WITH | - | Target | 1991 | ELANE | {"name": "ELANE", "id": "1991"} |
| Phytochemical | CID_54670067 | Ascorbic acid | {"name": "Ascorbic acid", "id": "CID_54670067"} | INTERACTS_WITH | - | Target | 2064 | ERBB2 | {"name": "ERBB2", "id": "2064"} |
| Phytochemical | CID_54670067 | Ascorbic acid | {"name": "Ascorbic acid", "id": "CID_54670067"} | INTERACTS_WITH | - | Target | 2099 | ESR1 | {"name": "ESR1", "id": "2099"} |
| Phytochemical | CID_54670067 | Ascorbic acid | {"name": "Ascorbic acid", "id": "CID_54670067"} | INTERACTS_WITH | - | Target | 2100 | ESR2 | {"name": "ESR2", "id": "2100"} |
| Phytochemical | CID_54670067 | Ascorbic acid | {"name": "Ascorbic acid", "id": "CID_54670067"} | INTERACTS_WITH | - | Target | 2147 | F2 | {"name": "F2", "id": "2147"} |
| Phytochemical | CID_54670067 | Ascorbic acid | {"name": "Ascorbic acid", "id": "CID_54670067"} | INTERACTS_WITH | - | Target | 2321 | FLT1 | {"name": "FLT1", "id": "2321"} |
| Phytochemical | CID_54670067 | Ascorbic acid | {"name": "Ascorbic acid", "id": "CID_54670067"} | INTERACTS_WITH | - | Target | 239 | ALOX12 | {"name": "ALOX12", "id": "239"} |
| Phytochemical | CID_54670067 | Ascorbic acid | {"name": "Ascorbic acid", "id": "CID_54670067"} | INTERACTS_WITH | - | Target | 240 | ALOX5 | {"name": "ALOX5", "id": "240"} |
| Phytochemical | CID_54670067 | Ascorbic acid | {"name": "Ascorbic acid", "id": "CID_54670067"} | INTERACTS_WITH | - | Target | 246 | ALOX15 | {"name": "ALOX15", "id": "246"} |
| Phytochemical | CID_54670067 | Ascorbic acid | {"name": "Ascorbic acid", "id": "CID_54670067"} | INTERACTS_WITH | - | Target | 247 | ALOX15B | {"name": "ALOX15B", "id": "247"} |
| Phytochemical | CID_54670067 | Ascorbic acid | {"name": "Ascorbic acid", "id": "CID_54670067"} | INTERACTS_WITH | - | Target | 2534 | FYN | {"name": "FYN", "id": "2534"} |

**Table S9: Complete node and relationship data used to construct the knowledge graph. This table summarizes interactions among FDA-approved drugs, formulation, herbs, diseases (type 2 diabetes, fatty liver, obesity), phytochemicals, and their target genes/proteins. Each entry includes source and target node details, edge type, and node annotations (Entrez Gene ID and gene symbol, where applicable).**

| Source Node Label | Source Node ID | Source Node Name | Source Node Properties | Edge Type | Edge Properties | Target Node Label | Target Node ID | Target Node Name | Target Node Properties |
|---|---|---|---|---|---|---|---|---|---|
| Phytochemical | CID_54670067 | Ascorbic acid | {"name": "Ascorbic acid", "id": "CID_54670067"} | INTERACTS_WITH | - | Target | 2554 | GABRA1 | {"name": "GABRA1", "id": "2554"} |
| Phytochemical | CID_54670067 | Ascorbic acid | {"name": "Ascorbic acid", "id": "CID_54670067"} | INTERACTS_WITH | - | Target | 2555 | GABRA2 | {"name": "GABRA2", "id": "2555"} |
| Phytochemical | CID_54670067 | Ascorbic acid | {"name": "Ascorbic acid", "id": "CID_54670067"} | INTERACTS_WITH | - | Target | 2561 | GABRB2 | {"name": "GABRB2", "id": "2561"} |
| Phytochemical | CID_54670067 | Ascorbic acid | {"name": "Ascorbic acid", "id": "CID_54670067"} | INTERACTS_WITH | - | Target | 2566 | GABRG2 | {"name": "GABRG2", "id": "2566"} |
| Phytochemical | CID_54670067 | Ascorbic acid | {"name": "Ascorbic acid", "id": "CID_54670067"} | INTERACTS_WITH | - | Target | 25939 | SAMHD1 | {"name": "SAMHD1", "id": "25939"} |
| Phytochemical | CID_54670067 | Ascorbic acid | {"name": "Ascorbic acid", "id": "CID_54670067"} | INTERACTS_WITH | - | Target | 276 | AMY1A | {"name": "AMY1A", "id": "276"} |
| Phytochemical | CID_54670067 | Ascorbic acid | {"name": "Ascorbic acid", "id": "CID_54670067"} | INTERACTS_WITH | - | Target | 279 | AMY2A | {"name": "AMY2A", "id": "279"} |
| Phytochemical | CID_54670067 | Ascorbic acid | {"name": "Ascorbic acid", "id": "CID_54670067"} | INTERACTS_WITH | - | Target | 28234 | SLCO1B3 | {"name": "SLCO1B3", "id": "28234"} |
| Phytochemical | CID_54670067 | Ascorbic acid | {"name": "Ascorbic acid", "id": "CID_54670067"} | INTERACTS_WITH | - | Target | 2908 | NR3C1 | {"name": "NR3C1", "id": "2908"} |
| Phytochemical | CID_54670067 | Ascorbic acid | {"name": "Ascorbic acid", "id": "CID_54670067"} | INTERACTS_WITH | - | Target | 2932 | GSK3B | {"name": "GSK3B", "id": "2932"} |
| Phytochemical | CID_54670067 | Ascorbic acid | {"name": "Ascorbic acid", "id": "CID_54670067"} | INTERACTS_WITH | - | Target | 3156 | HMGCR | {"name": "HMGCR", "id": "3156"} |
| Phytochemical | CID_54670067 | Ascorbic acid | {"name": "Ascorbic acid", "id": "CID_54670067"} | INTERACTS_WITH | - | Target | 3269 | HRH1 | {"name": "HRH1", "id": "3269"} |
| Phytochemical | CID_54670067 | Ascorbic acid | {"name": "Ascorbic acid", "id": "CID_54670067"} | INTERACTS_WITH | - | Target | 3274 | HRH2 | {"name": "HRH2", "id": "3274"} |
| Phytochemical | CID_54670067 | Ascorbic acid | {"name": "Ascorbic acid", "id": "CID_54670067"} | INTERACTS_WITH | - | Target | 3350 | HTR1A | {"name": "HTR1A", "id": "3350"} |
| Phytochemical | CID_54670067 | Ascorbic acid | {"name": "Ascorbic acid", "id": "CID_54670067"} | INTERACTS_WITH | - | Target | 3356 | HTR2A | {"name": "HTR2A", "id": "3356"} |
| Phytochemical | CID_54670067 | Ascorbic acid | {"name": "Ascorbic acid", "id": "CID_54670067"} | INTERACTS_WITH | - | Target | 3357 | HTR2B | {"name": "HTR2B", "id": "3357"} |
| Phytochemical | CID_54670067 | Ascorbic acid | {"name": "Ascorbic acid", "id": "CID_54670067"} | INTERACTS_WITH | - | Target | 3358 | HTR2C | {"name": "HTR2C", "id": "3358"} |
| Phytochemical | CID_54670067 | Ascorbic acid | {"name": "Ascorbic acid", "id": "CID_54670067"} | INTERACTS_WITH | - | Target | 3362 | HTR6 | {"name": "HTR6", "id": "3362"} |
| Phytochemical | CID_54670067 | Ascorbic acid | {"name": "Ascorbic acid", "id": "CID_54670067"} | INTERACTS_WITH | - | Target | 351 | APP | {"name": "APP", "id": "351"} |
| Phytochemical | CID_54670067 | Ascorbic acid | {"name": "Ascorbic acid", "id": "CID_54670067"} | INTERACTS_WITH | - | Target | 3577 | CXCR1 | {"name": "CXCR1", "id": "3577"} |
| Phytochemical | CID_54670067 | Ascorbic acid | {"name": "Ascorbic acid", "id": "CID_54670067"} | INTERACTS_WITH | - | Target | 3579 | CXCR2 | {"name": "CXCR2", "id": "3579"} |
| Phytochemical | CID_54670067 | Ascorbic acid | {"name": "Ascorbic acid", "id": "CID_54670067"} | INTERACTS_WITH | - | Target | 367 | AR | {"name": "AR", "id": "367"} |
| Phytochemical | CID_54670067 | Ascorbic acid | {"name": "Ascorbic acid", "id": "CID_54670067"} | INTERACTS_WITH | - | Target | 3757 | KCNH2 | {"name": "KCNH2", "id": "3757"} |
| Phytochemical | CID_54670067 | Ascorbic acid | {"name": "Ascorbic acid", "id": "CID_54670067"} | INTERACTS_WITH | - | Target | 3791 | KDR | {"name": "KDR", "id": "3791"} |
| Phytochemical | CID_54670067 | Ascorbic acid | {"name": "Ascorbic acid", "id": "CID_54670067"} | INTERACTS_WITH | - | Target | 3932 | LCK | {"name": "LCK", "id": "3932"} |
| Phytochemical | CID_54670067 | Ascorbic acid | {"name": "Ascorbic acid", "id": "CID_54670067"} | INTERACTS_WITH | - | Target | 4128 | MAOA | {"name": "MAOA", "id": "4128"} |
| Phytochemical | CID_54670067 | Ascorbic acid | {"name": "Ascorbic acid", "id": "CID_54670067"} | INTERACTS_WITH | - | Target | 4159 | MC3R | {"name": "MC3R", "id": "4159"} |
| Phytochemical | CID_54670067 | Ascorbic acid | {"name": "Ascorbic acid", "id": "CID_54670067"} | INTERACTS_WITH | - | Target | 4160 | MC4R | {"name": "MC4R", "id": "4160"} |
| Phytochemical | CID_54670067 | Ascorbic acid | {"name": "Ascorbic acid", "id": "CID_54670067"} | INTERACTS_WITH | - | Target | 4161 | MC5R | {"name": "MC5R", "id": "4161"} |
| Phytochemical | CID_54670067 | Ascorbic acid | {"name": "Ascorbic acid", "id": "CID_54670067"} | INTERACTS_WITH | - | Target | 43 | ACHE | {"name": "ACHE", "id": "43"} |
| Phytochemical | CID_54670067 | Ascorbic acid | {"name": "Ascorbic acid", "id": "CID_54670067"} | INTERACTS_WITH | - | Target | 4312 | MMP1 | {"name": "MMP1", "id": "4312"} |
| Phytochemical | CID_54670067 | Ascorbic acid | {"name": "Ascorbic acid", "id": "CID_54670067"} | INTERACTS_WITH | - | Target | 4318 | MMP9 | {"name": "MMP9", "id": "4318"} |
| Phytochemical | CID_54670067 | Ascorbic acid | {"name": "Ascorbic acid", "id": "CID_54670067"} | INTERACTS_WITH | - | Target | 4780 | NFE2L2 | {"name": "NFE2L2", "id": "4780"} |
| Phytochemical | CID_54670067 | Ascorbic acid | {"name": "Ascorbic acid", "id": "CID_54670067"} | INTERACTS_WITH | - | Target | 4886 | NPY1R | {"name": "NPY1R", "id": "4886"} |
| Phytochemical | CID_54670067 | Ascorbic acid | {"name": "Ascorbic acid", "id": "CID_54670067"} | INTERACTS_WITH | - | Target | 4887 | NPY2R | {"name": "NPY2R", "id": "4887"} |
| Phytochemical | CID_54670067 | Ascorbic acid | {"name": "Ascorbic acid", "id": "CID_54670067"} | INTERACTS_WITH | - | Target | 4985 | OPRD1 | {"name": "OPRD1", "id": "4985"} |
| Phytochemical | CID_54670067 | Ascorbic acid | {"name": "Ascorbic acid", "id": "CID_54670067"} | INTERACTS_WITH | - | Target | 4986 | OPRK1 | {"name": "OPRK1", "id": "4986"} |
| Phytochemical | CID_54670067 | Ascorbic acid | {"name": "Ascorbic acid", "id": "CID_54670067"} | INTERACTS_WITH | - | Target | 4988 | OPRM1 | {"name": "OPRM1", "id": "4988"} |
| Phytochemical | CID_54670067 | Ascorbic acid | {"name": "Ascorbic acid", "id": "CID_54670067"} | INTERACTS_WITH | - | Target | 5033 | P4HA1 | {"name": "P4HA1", "id": "5033"} |
| Phytochemical | CID_54670067 | Ascorbic acid | {"name": "Ascorbic acid", "id": "CID_54670067"} | INTERACTS_WITH | - | Target | 50507 | NOX4 | {"name": "NOX4", "id": "50507"} |
| Phytochemical | CID_54670067 | Ascorbic acid | {"name": "Ascorbic acid", "id": "CID_54670067"} | INTERACTS_WITH | - | Target | 5139 | PDE3A | {"name": "PDE3A", "id": "5139"} |
| Phytochemical | CID_54670067 | Ascorbic acid | {"name": "Ascorbic acid", "id": "CID_54670067"} | INTERACTS_WITH | - | Target | 5141 | PDE4A | {"name": "PDE4A", "id": "5141"} |
| Phytochemical | CID_54670067 | Ascorbic acid | {"name": "Ascorbic acid", "id": "CID_54670067"} | INTERACTS_WITH | - | Target | 5241 | PGR | {"name": "PGR", "id": "5241"} |
| Phytochemical | CID_54670067 | Ascorbic acid | {"name": "Ascorbic acid", "id": "CID_54670067"} | INTERACTS_WITH | - | Target | 54583 | EGLN1 | {"name": "EGLN1", "id": "54583"} |
| Phytochemical | CID_54670067 | Ascorbic acid | {"name": "Ascorbic acid", "id": "CID_54670067"} | INTERACTS_WITH | - | Target | 552 | AVPR1A | {"name": "AVPR1A", "id": "552"} |
| Phytochemical | CID_54670067 | Ascorbic acid | {"name": "Ascorbic acid", "id": "CID_54670067"} | INTERACTS_WITH | - | Target | 5530 | PPP3CA | {"name": "PPP3CA", "id": "5530"} |
| Phytochemical | CID_54670067 | Ascorbic acid | {"name": "Ascorbic acid", "id": "CID_54670067"} | INTERACTS_WITH | - | Target | 55662 | HIF1AN | {"name": "HIF1AN", "id": "55662"} |
| Phytochemical | CID_54670067 | Ascorbic acid | {"name": "Ascorbic acid", "id": "CID_54670067"} | INTERACTS_WITH | - | Target | 5578 | PRKCA | {"name": "PRKCA", "id": "5578"} |
| Phytochemical | CID_54670067 | Ascorbic acid | {"name": "Ascorbic acid", "id": "CID_54670067"} | INTERACTS_WITH | - | Target | 5594 | MAPK1 | {"name": "MAPK1", "id": "5594"} |

**Table S9: Complete node and relationship data used to construct the knowledge graph. This table summarizes interactions among FDA-approved drugs, formulation, herbs, diseases (type 2 diabetes, fatty liver, obesity), phytochemicals, and their target genes/proteins. Each entry includes source and target node details, edge type, and node annotations (Entrez Gene ID and gene symbol, where applicable).**

| Source Node Label | Source Node ID | Source Node Name | Source Node Properties | Edge Type | Edge Properties | Target Node Label | Target Node ID | Target Node Name | Target Node Properties |
|---|---|---|---|---|---|---|---|---|---|
| Phytochemical | CID_54670067 | Ascorbic acid | {"name": "Ascorbic acid", "id": "CID_54670067"} | INTERACTS_WITH | - | Target | 5595 | MAPK3 | {"name": "MAPK3", "id": "5595"} |
| Phytochemical | CID_54670067 | Ascorbic acid | {"name": "Ascorbic acid", "id": "CID_54670067"} | INTERACTS_WITH | - | Target | 5724 | PTAFR | {"name": "PTAFR", "id": "5724"} |
| Phytochemical | CID_54670067 | Ascorbic acid | {"name": "Ascorbic acid", "id": "CID_54670067"} | INTERACTS_WITH | - | Target | 5742 | PTGS1 | {"name": "PTGS1", "id": "5742"} |
| Phytochemical | CID_54670067 | Ascorbic acid | {"name": "Ascorbic acid", "id": "CID_54670067"} | INTERACTS_WITH | - | Target | 5743 | PTGS2 | {"name": "PTGS2", "id": "5743"} |
| Phytochemical | CID_54670067 | Ascorbic acid | {"name": "Ascorbic acid", "id": "CID_54670067"} | INTERACTS_WITH | - | Target | 5788 | PTPRC | {"name": "PTPRC", "id": "5788"} |
| Phytochemical | CID_54670067 | Ascorbic acid | {"name": "Ascorbic acid", "id": "CID_54670067"} | INTERACTS_WITH | - | Target | 624 | BDKRB2 | {"name": "BDKRB2", "id": "624"} |
| Phytochemical | CID_54670067 | Ascorbic acid | {"name": "Ascorbic acid", "id": "CID_54670067"} | INTERACTS_WITH | - | Target | 6530 | SLC6A2 | {"name": "SLC6A2", "id": "6530"} |
| Phytochemical | CID_54670067 | Ascorbic acid | {"name": "Ascorbic acid", "id": "CID_54670067"} | INTERACTS_WITH | - | Target | 6531 | SLC6A3 | {"name": "SLC6A3", "id": "6531"} |
| Phytochemical | CID_54670067 | Ascorbic acid | {"name": "Ascorbic acid", "id": "CID_54670067"} | INTERACTS_WITH | - | Target | 6532 | SLC6A4 | {"name": "SLC6A4", "id": "6532"} |
| Phytochemical | CID_54670067 | Ascorbic acid | {"name": "Ascorbic acid", "id": "CID_54670067"} | INTERACTS_WITH | - | Target | 6865 | TACR2 | {"name": "TACR2", "id": "6865"} |
| Phytochemical | CID_54670067 | Ascorbic acid | {"name": "Ascorbic acid", "id": "CID_54670067"} | INTERACTS_WITH | - | Target | 6869 | TACR1 | {"name": "TACR1", "id": "6869"} |
| Phytochemical | CID_54670067 | Ascorbic acid | {"name": "Ascorbic acid", "id": "CID_54670067"} | INTERACTS_WITH | - | Target | 6915 | TBXA2R | {"name": "TBXA2R", "id": "6915"} |
| Phytochemical | CID_54670067 | Ascorbic acid | {"name": "Ascorbic acid", "id": "CID_54670067"} | INTERACTS_WITH | - | Target | 6916 | TBXAS1 | {"name": "TBXAS1", "id": "6916"} |
| Phytochemical | CID_54670067 | Ascorbic acid | {"name": "Ascorbic acid", "id": "CID_54670067"} | INTERACTS_WITH | - | Target | 729230 | CCR2 | {"name": "CCR2", "id": "729230"} |
| Phytochemical | CID_54670067 | Ascorbic acid | {"name": "Ascorbic acid", "id": "CID_54670067"} | INTERACTS_WITH | - | Target | 7299 | TYR | {"name": "TYR", "id": "7299"} |
| Phytochemical | CID_54670067 | Ascorbic acid | {"name": "Ascorbic acid", "id": "CID_54670067"} | INTERACTS_WITH | - | Target | 7433 | VIPR1 | {"name": "VIPR1", "id": "7433"} |
| Phytochemical | CID_54670067 | Ascorbic acid | {"name": "Ascorbic acid", "id": "CID_54670067"} | INTERACTS_WITH | - | Target | 760 | CA2 | {"name": "CA2", "id": "760"} |
| Phytochemical | CID_54670067 | Ascorbic acid | {"name": "Ascorbic acid", "id": "CID_54670067"} | INTERACTS_WITH | - | Target | 799 | CALCR | {"name": "CALCR", "id": "799"} |
| Phytochemical | CID_54670067 | Ascorbic acid | {"name": "Ascorbic acid", "id": "CID_54670067"} | INTERACTS_WITH | - | Target | 834 | CASP1 | {"name": "CASP1", "id": "834"} |
| Phytochemical | CID_54670067 | Ascorbic acid | {"name": "Ascorbic acid", "id": "CID_54670067"} | INTERACTS_WITH | - | Target | 8654 | PDE5A | {"name": "PDE5A", "id": "8654"} |
| Phytochemical | CID_54670067 | Ascorbic acid | {"name": "Ascorbic acid", "id": "CID_54670067"} | INTERACTS_WITH | - | Target | 886 | CCKAR | {"name": "CCKAR", "id": "886"} |
| Phytochemical | CID_91458 | Aucubin | {"name": "Aucubin", "id": "CID_91458"} | INTERACTS_WITH | - | Target | 2099 | ESR1 | {"name": "ESR1", "id": "2099"} |
| Phytochemical | CID_91458 | Aucubin | {"name": "Aucubin", "id": "CID_91458"} | INTERACTS_WITH | - | Target | 26013 | L3MBTL1 | {"name": "L3MBTL1", "id": "26013"} |
| Phytochemical | CID_91458 | Aucubin | {"name": "Aucubin", "id": "CID_91458"} | INTERACTS_WITH | - | Target | 51053 | GMNN | {"name": "GMNN", "id": "51053"} |
| Phytochemical | CID_91458 | Aucubin | {"name": "Aucubin", "id": "CID_91458"} | INTERACTS_WITH | - | Target | 5742 | PTGS1 | {"name": "PTGS1", "id": "5742"} |
| Phytochemical | CID_91458 | Aucubin | {"name": "Aucubin", "id": "CID_91458"} | INTERACTS_WITH | - | Target | 5743 | PTGS2 | {"name": "PTGS2", "id": "5743"} |
| Phytochemical | CID_153845 | Bicyclo(4.1.0)heptan-3-one, 1-methyl-4-(1-methylethylidene)-7-(3-oxobutyl)-, (1S,6R,7R)- | {"name": "Bicyclo(4.1.0)heptan-3-one, 1-methyl-4-(1-methylethylidene)-7-(3-oxobutyl)-, (1S,6R,7R)-", "id": "CID_153845"} | INTERACTS_WITH | - | Target | 231 | AKR1B1 | {"name": "AKR1B1", "id": "231"} |
| Phytochemical | CID_5315472 | Bisdemethoxycurcumin | {"name": "Bisdemethoxycurcumin", "id": "CID_5315472"} | INTERACTS_WITH | - | Target | 1786 | DNMT1 | {"name": "DNMT1", "id": "1786"} |
| Phytochemical | CID_5315472 | Bisdemethoxycurcumin | {"name": "Bisdemethoxycurcumin", "id": "CID_5315472"} | INTERACTS_WITH | - | Target | 231 | AKR1B1 | {"name": "AKR1B1", "id": "231"} |
| Phytochemical | CID_5315472 | Bisdemethoxycurcumin | {"name": "Bisdemethoxycurcumin", "id": "CID_5315472"} | INTERACTS_WITH | - | Target | 2739 | GLO1 | {"name": "GLO1", "id": "2739"} |
| Phytochemical | CID_5315472 | Bisdemethoxycurcumin | {"name": "Bisdemethoxycurcumin", "id": "CID_5315472"} | INTERACTS_WITH | - | Target | 351 | APP | {"name": "APP", "id": "351"} |
| Phytochemical | CID_5315472 | Bisdemethoxycurcumin | {"name": "Bisdemethoxycurcumin", "id": "CID_5315472"} | INTERACTS_WITH | - | Target | 4087 | SMAD2 | {"name": "SMAD2", "id": "4087"} |
| Phytochemical | CID_5315472 | Bisdemethoxycurcumin | {"name": "Bisdemethoxycurcumin", "id": "CID_5315472"} | INTERACTS_WITH | - | Target | 4088 | SMAD3 | {"name": "SMAD3", "id": "4088"} |
| Phytochemical | CID_5315472 | Bisdemethoxycurcumin | {"name": "Bisdemethoxycurcumin", "id": "CID_5315472"} | INTERACTS_WITH | - | Target | 4129 | MAOB | {"name": "MAOB", "id": "4129"} |
| Phytochemical | CID_5315472 | Bisdemethoxycurcumin | {"name": "Bisdemethoxycurcumin", "id": "CID_5315472"} | INTERACTS_WITH | - | Target | 4780 | NFE2L2 | {"name": "NFE2L2", "id": "4780"} |
| Phytochemical | CID_5315472 | Bisdemethoxycurcumin | {"name": "Bisdemethoxycurcumin", "id": "CID_5315472"} | INTERACTS_WITH | - | Target | 57016 | AKR1B10 | {"name": "AKR1B10", "id": "57016"} |
| Phytochemical | CID_5315472 | Bisdemethoxycurcumin | {"name": "Bisdemethoxycurcumin", "id": "CID_5315472"} | INTERACTS_WITH | - | Target | 5743 | PTGS2 | {"name": "PTGS2", "id": "5743"} |
| Phytochemical | CID_5315472 | Bisdemethoxycurcumin | {"name": "Bisdemethoxycurcumin", "id": "CID_5315472"} | INTERACTS_WITH | - | Target | 815 | CAMK2A | {"name": "CAMK2A", "id": "815"} |
| Phytochemical | CID_689043 | Caffeic acid | {"name": "Caffeic acid", "id": "CID_689043"} | INTERACTS_WITH | - | Target | 10013 | HDAC6 | {"name": "HDAC6", "id": "10013"} |
| Phytochemical | CID_689043 | Caffeic acid | {"name": "Caffeic acid", "id": "CID_689043"} | INTERACTS_WITH | - | Target | 10014 | HDAC5 | {"name": "HDAC5", "id": "10014"} |
| Phytochemical | CID_689043 | Caffeic acid | {"name": "Caffeic acid", "id": "CID_689043"} | INTERACTS_WITH | - | Target | 10073 | SNUPN | {"name": "SNUPN", "id": "10073"} |
| Phytochemical | CID_689043 | Caffeic acid | {"name": "Caffeic acid", "id": "CID_689043"} | INTERACTS_WITH | - | Target | 10599 | SLCO1B1 | {"name": "SLCO1B1", "id": "10599"} |
| Phytochemical | CID_689043 | Caffeic acid | {"name": "Caffeic acid", "id": "CID_689043"} | INTERACTS_WITH | - | Target | 10765 | KDM5B | {"name": "KDM5B", "id": "10765"} |
| Phytochemical | CID_689043 | Caffeic acid | {"name": "Caffeic acid", "id": "CID_689043"} | INTERACTS_WITH | - | Target | 10919 | EHMT2 | {"name": "EHMT2", "id": "10919"} |
| Phytochemical | CID_689043 | Caffeic acid | {"name": "Caffeic acid", "id": "CID_689043"} | INTERACTS_WITH | - | Target | 1109 | AKR1C4 | {"name": "AKR1C4", "id": "1109"} |
| Phytochemical | CID_689043 | Caffeic acid | {"name": "Caffeic acid", "id": "CID_689043"} | INTERACTS_WITH | - | Target | 11238 | CA5B | {"name": "CA5B", "id": "11238"} |
| Phytochemical | CID_689043 | Caffeic acid | {"name": "Caffeic acid", "id": "CID_689043"} | INTERACTS_WITH | - | Target | 118429 | ANTXR2 | {"name": "ANTXR2", "id": "118429"} |
| Phytochemical | CID_689043 | Caffeic acid | {"name": "Caffeic acid", "id": "CID_689043"} | INTERACTS_WITH | - | Target | 1576 | CYP3A4 | {"name": "CYP3A4", "id": "1576"} |
| Phytochemical | CID_689043 | Caffeic acid | {"name": "Caffeic acid", "id": "CID_689043"} | INTERACTS_WITH | - | Target | 1645 | AKR1C1 | {"name": "AKR1C1", "id": "1645"} |
| Phytochemical | CID_689043 | Caffeic acid | {"name": "Caffeic acid", "id": "CID_689043"} | INTERACTS_WITH | - | Target | 1646 | AKR1C2 | {"name": "AKR1C2", "id": "1646"} |

**Table S9: Complete node and relationship data used to construct the knowledge graph. This table summarizes interactions among FDA-approved drugs, formulation, herbs, diseases (type 2 diabetes, fatty liver, obesity), phytochemicals, and their target genes/proteins. Each entry includes source and target node details, edge type, and node annotations (Entrez Gene ID and gene symbol, where applicable).**

| Source Node Label | Source Node ID | Source Node Name | Source Node Properties | Edge Type | Edge Properties | Target Node Label | Target Node ID | Target Node Name | Target Node Properties |
|---|---|---|---|---|---|---|---|---|---|
| Phytochemical | CID_689043 | Caffeic acid | {"name": "Caffeic acid", "id": "CID_689043"} | INTERACTS_WITH | - | Target | 1803 | DPP4 | {"name": "DPP4", "id": "1803"} |
| Phytochemical | CID_689043 | Caffeic acid | {"name": "Caffeic acid", "id": "CID_689043"} | INTERACTS_WITH | - | Target | 1845 | DUSP3 | {"name": "DUSP3", "id": "1845"} |
| Phytochemical | CID_689043 | Caffeic acid | {"name": "Caffeic acid", "id": "CID_689043"} | INTERACTS_WITH | - | Target | 185 | AGTR1 | {"name": "AGTR1", "id": "185"} |
| Phytochemical | CID_689043 | Caffeic acid | {"name": "Caffeic acid", "id": "CID_689043"} | INTERACTS_WITH | - | Target | 1956 | EGFR | {"name": "EGFR", "id": "1956"} |
| Phytochemical | CID_689043 | Caffeic acid | {"name": "Caffeic acid", "id": "CID_689043"} | INTERACTS_WITH | - | Target | 2099 | ESR1 | {"name": "ESR1", "id": "2099"} |
| Phytochemical | CID_689043 | Caffeic acid | {"name": "Caffeic acid", "id": "CID_689043"} | INTERACTS_WITH | - | Target | 216 | ALDH1A1 | {"name": "ALDH1A1", "id": "216"} |
| Phytochemical | CID_689043 | Caffeic acid | {"name": "Caffeic acid", "id": "CID_689043"} | INTERACTS_WITH | - | Target | 231 | AKR1B1 | {"name": "AKR1B1", "id": "231"} |
| Phytochemical | CID_689043 | Caffeic acid | {"name": "Caffeic acid", "id": "CID_689043"} | INTERACTS_WITH | - | Target | 23621 | BACE1 | {"name": "BACE1", "id": "23621"} |
| Phytochemical | CID_689043 | Caffeic acid | {"name": "Caffeic acid", "id": "CID_689043"} | INTERACTS_WITH | - | Target | 23632 | CA14 | {"name": "CA14", "id": "23632"} |
| Phytochemical | CID_689043 | Caffeic acid | {"name": "Caffeic acid", "id": "CID_689043"} | INTERACTS_WITH | - | Target | 240 | ALOX5 | {"name": "ALOX5", "id": "240"} |
| Phytochemical | CID_689043 | Caffeic acid | {"name": "Caffeic acid", "id": "CID_689043"} | INTERACTS_WITH | - | Target | 2548 | GAA | {"name": "GAA", "id": "2548"} |
| Phytochemical | CID_689043 | Caffeic acid | {"name": "Caffeic acid", "id": "CID_689043"} | INTERACTS_WITH | - | Target | 259285 | TAS2R39 | {"name": "TAS2R39", "id": "259285"} |
| Phytochemical | CID_689043 | Caffeic acid | {"name": "Caffeic acid", "id": "CID_689043"} | INTERACTS_WITH | - | Target | 2717 | GLA | {"name": "GLA", "id": "2717"} |
| Phytochemical | CID_689043 | Caffeic acid | {"name": "Caffeic acid", "id": "CID_689043"} | INTERACTS_WITH | - | Target | 2744 | GLS | {"name": "GLS", "id": "2744"} |
| Phytochemical | CID_689043 | Caffeic acid | {"name": "Caffeic acid", "id": "CID_689043"} | INTERACTS_WITH | - | Target | 28234 | SLCO1B3 | {"name": "SLCO1B3", "id": "28234"} |
| Phytochemical | CID_689043 | Caffeic acid | {"name": "Caffeic acid", "id": "CID_689043"} | INTERACTS_WITH | - | Target | 2875 | GPT | {"name": "GPT", "id": "2875"} |
| Phytochemical | CID_689043 | Caffeic acid | {"name": "Caffeic acid", "id": "CID_689043"} | INTERACTS_WITH | - | Target | 29994 | BAZ2B | {"name": "BAZ2B", "id": "29994"} |
| Phytochemical | CID_689043 | Caffeic acid | {"name": "Caffeic acid", "id": "CID_689043"} | INTERACTS_WITH | - | Target | 3028 | HSD17B10 | {"name": "HSD17B10", "id": "3028"} |
| Phytochemical | CID_689043 | Caffeic acid | {"name": "Caffeic acid", "id": "CID_689043"} | INTERACTS_WITH | - | Target | 3065 | HDAC1 | {"name": "HDAC1", "id": "3065"} |
| Phytochemical | CID_689043 | Caffeic acid | {"name": "Caffeic acid", "id": "CID_689043"} | INTERACTS_WITH | - | Target | 3066 | HDAC2 | {"name": "HDAC2", "id": "3066"} |
| Phytochemical | CID_689043 | Caffeic acid | {"name": "Caffeic acid", "id": "CID_689043"} | INTERACTS_WITH | - | Target | 3248 | HPGD | {"name": "HPGD", "id": "3248"} |
| Phytochemical | CID_689043 | Caffeic acid | {"name": "Caffeic acid", "id": "CID_689043"} | INTERACTS_WITH | - | Target | 328 | APEX1 | {"name": "APEX1", "id": "328"} |
| Phytochemical | CID_689043 | Caffeic acid | {"name": "Caffeic acid", "id": "CID_689043"} | INTERACTS_WITH | - | Target | 3320 | HSP90AA1 | {"name": "HSP90AA1", "id": "3320"} |
| Phytochemical | CID_689043 | Caffeic acid | {"name": "Caffeic acid", "id": "CID_689043"} | INTERACTS_WITH | - | Target | 3326 | HSP90AB1 | {"name": "HSP90AB1", "id": "3326"} |
| Phytochemical | CID_689043 | Caffeic acid | {"name": "Caffeic acid", "id": "CID_689043"} | INTERACTS_WITH | - | Target | 3373 | HYAL1 | {"name": "HYAL1", "id": "3373"} |
| Phytochemical | CID_689043 | Caffeic acid | {"name": "Caffeic acid", "id": "CID_689043"} | INTERACTS_WITH | - | Target | 338442 | HCAR2 | {"name": "HCAR2", "id": "338442"} |
| Phytochemical | CID_689043 | Caffeic acid | {"name": "Caffeic acid", "id": "CID_689043"} | INTERACTS_WITH | - | Target | 351 | APP | {"name": "APP", "id": "351"} |
| Phytochemical | CID_689043 | Caffeic acid | {"name": "Caffeic acid", "id": "CID_689043"} | INTERACTS_WITH | - | Target | 3757 | KCNH2 | {"name": "KCNH2", "id": "3757"} |
| Phytochemical | CID_689043 | Caffeic acid | {"name": "Caffeic acid", "id": "CID_689043"} | INTERACTS_WITH | - | Target | 3837 | KPNB1 | {"name": "KPNB1", "id": "3837"} |
| Phytochemical | CID_689043 | Caffeic acid | {"name": "Caffeic acid", "id": "CID_689043"} | INTERACTS_WITH | - | Target | 390245 | KDM4E | {"name": "KDM4E", "id": "390245"} |
| Phytochemical | CID_689043 | Caffeic acid | {"name": "Caffeic acid", "id": "CID_689043"} | INTERACTS_WITH | - | Target | 4000 | LMNA | {"name": "LMNA", "id": "4000"} |
| Phytochemical | CID_689043 | Caffeic acid | {"name": "Caffeic acid", "id": "CID_689043"} | INTERACTS_WITH | - | Target | 410 | ARSA | {"name": "ARSA", "id": "410"} |
| Phytochemical | CID_689043 | Caffeic acid | {"name": "Caffeic acid", "id": "CID_689043"} | INTERACTS_WITH | - | Target | 4128 | MAOA | {"name": "MAOA", "id": "4128"} |
| Phytochemical | CID_689043 | Caffeic acid | {"name": "Caffeic acid", "id": "CID_689043"} | INTERACTS_WITH | - | Target | 4129 | MAOB | {"name": "MAOB", "id": "4129"} |
| Phytochemical | CID_689043 | Caffeic acid | {"name": "Caffeic acid", "id": "CID_689043"} | INTERACTS_WITH | - | Target | 4137 | MAPT | {"name": "MAPT", "id": "4137"} |
| Phytochemical | CID_689043 | Caffeic acid | {"name": "Caffeic acid", "id": "CID_689043"} | INTERACTS_WITH | - | Target | 4297 | KMT2A | {"name": "KMT2A", "id": "4297"} |
| Phytochemical | CID_689043 | Caffeic acid | {"name": "Caffeic acid", "id": "CID_689043"} | INTERACTS_WITH | - | Target | 43 | ACHE | {"name": "ACHE", "id": "43"} |
| Phytochemical | CID_689043 | Caffeic acid | {"name": "Caffeic acid", "id": "CID_689043"} | INTERACTS_WITH | - | Target | 4312 | MMP1 | {"name": "MMP1", "id": "4312"} |
| Phytochemical | CID_689043 | Caffeic acid | {"name": "Caffeic acid", "id": "CID_689043"} | INTERACTS_WITH | - | Target | 4313 | MMP2 | {"name": "MMP2", "id": "4313"} |
| Phytochemical | CID_689043 | Caffeic acid | {"name": "Caffeic acid", "id": "CID_689043"} | INTERACTS_WITH | - | Target | 4318 | MMP9 | {"name": "MMP9", "id": "4318"} |
| Phytochemical | CID_689043 | Caffeic acid | {"name": "Caffeic acid", "id": "CID_689043"} | INTERACTS_WITH | - | Target | 4350 | MPG | {"name": "MPG", "id": "4350"} |
| Phytochemical | CID_689043 | Caffeic acid | {"name": "Caffeic acid", "id": "CID_689043"} | INTERACTS_WITH | - | Target | 4790 | NFKB1 | {"name": "NFKB1", "id": "4790"} |
| Phytochemical | CID_689043 | Caffeic acid | {"name": "Caffeic acid", "id": "CID_689043"} | INTERACTS_WITH | - | Target | 51426 | POLK | {"name": "POLK", "id": "51426"} |
| Phytochemical | CID_689043 | Caffeic acid | {"name": "Caffeic acid", "id": "CID_689043"} | INTERACTS_WITH | - | Target | 51564 | HDAC7 | {"name": "HDAC7", "id": "51564"} |
| Phytochemical | CID_689043 | Caffeic acid | {"name": "Caffeic acid", "id": "CID_689043"} | INTERACTS_WITH | - | Target | 5376 | PMP22 | {"name": "PMP22", "id": "5376"} |
| Phytochemical | CID_689043 | Caffeic acid | {"name": "Caffeic acid", "id": "CID_689043"} | INTERACTS_WITH | - | Target | 5423 | POLB | {"name": "POLB", "id": "5423"} |
| Phytochemical | CID_689043 | Caffeic acid | {"name": "Caffeic acid", "id": "CID_689043"} | INTERACTS_WITH | - | Target | 55775 | TDP1 | {"name": "TDP1", "id": "55775"} |
| Phytochemical | CID_689043 | Caffeic acid | {"name": "Caffeic acid", "id": "CID_689043"} | INTERACTS_WITH | - | Target | 55869 | HDAC8 | {"name": "HDAC8", "id": "55869"} |
| Phytochemical | CID_689043 | Caffeic acid | {"name": "Caffeic acid", "id": "CID_689043"} | INTERACTS_WITH | - | Target | 5594 | MAPK1 | {"name": "MAPK1", "id": "5594"} |

**Table S9: Complete node and relationship data used to construct the knowledge graph. This table summarizes interactions among FDA-approved drugs, formulation, herbs, diseases (type 2 diabetes, fatty liver, obesity), phytochemicals, and their target genes/proteins. Each entry includes source and target node details, edge type, and node annotations (Entrez Gene ID and gene symbol, where applicable).**

| Source Node Label | Source Node ID | Source Node Name | Source Node Properties | Edge Type | Edge Properties | Target Node Label | Target Node ID | Target Node Name | Target Node Properties |
|---|---|---|---|---|---|---|---|---|---|
| Phytochemical | CID_689043 | Caffeic acid | {"name": "Caffeic acid", "id": "CID_689043"} | INTERACTS_WITH | - | Target | 57016 | AKR1B10 | {"name": "AKR1B10", "id": "57016"} |
| Phytochemical | CID_689043 | Caffeic acid | {"name": "Caffeic acid", "id": "CID_689043"} | INTERACTS_WITH | - | Target | 5742 | PTGS1 | {"name": "PTGS1", "id": "5742"} |
| Phytochemical | CID_689043 | Caffeic acid | {"name": "Caffeic acid", "id": "CID_689043"} | INTERACTS_WITH | - | Target | 5743 | PTGS2 | {"name": "PTGS2", "id": "5743"} |
| Phytochemical | CID_689043 | Caffeic acid | {"name": "Caffeic acid", "id": "CID_689043"} | INTERACTS_WITH | - | Target | 5770 | PTPN1 | {"name": "PTPN1", "id": "5770"} |
| Phytochemical | CID_689043 | Caffeic acid | {"name": "Caffeic acid", "id": "CID_689043"} | INTERACTS_WITH | - | Target | 5778 | PTPN7 | {"name": "PTPN7", "id": "5778"} |
| Phytochemical | CID_689043 | Caffeic acid | {"name": "Caffeic acid", "id": "CID_689043"} | INTERACTS_WITH | - | Target | 7068 | THRB | {"name": "THRB", "id": "7068"} |
| Phytochemical | CID_689043 | Caffeic acid | {"name": "Caffeic acid", "id": "CID_689043"} | INTERACTS_WITH | - | Target | 7276 | TTR | {"name": "TTR", "id": "7276"} |
| Phytochemical | CID_689043 | Caffeic acid | {"name": "Caffeic acid", "id": "CID_689043"} | INTERACTS_WITH | - | Target | 7299 | TYR | {"name": "TYR", "id": "7299"} |
| Phytochemical | CID_689043 | Caffeic acid | {"name": "Caffeic acid", "id": "CID_689043"} | INTERACTS_WITH | - | Target | 7498 | XDH | {"name": "XDH", "id": "7498"} |
| Phytochemical | CID_689043 | Caffeic acid | {"name": "Caffeic acid", "id": "CID_689043"} | INTERACTS_WITH | - | Target | 759 | CA1 | {"name": "CA1", "id": "759"} |
| Phytochemical | CID_689043 | Caffeic acid | {"name": "Caffeic acid", "id": "CID_689043"} | INTERACTS_WITH | - | Target | 760 | CA2 | {"name": "CA2", "id": "760"} |
| Phytochemical | CID_689043 | Caffeic acid | {"name": "Caffeic acid", "id": "CID_689043"} | INTERACTS_WITH | - | Target | 761 | CA3 | {"name": "CA3", "id": "761"} |
| Phytochemical | CID_689043 | Caffeic acid | {"name": "Caffeic acid", "id": "CID_689043"} | INTERACTS_WITH | - | Target | 762 | CA4 | {"name": "CA4", "id": "762"} |
| Phytochemical | CID_689043 | Caffeic acid | {"name": "Caffeic acid", "id": "CID_689043"} | INTERACTS_WITH | - | Target | 763 | CA5A | {"name": "CA5A", "id": "763"} |
| Phytochemical | CID_689043 | Caffeic acid | {"name": "Caffeic acid", "id": "CID_689043"} | INTERACTS_WITH | - | Target | 765 | CA6 | {"name": "CA6", "id": "765"} |
| Phytochemical | CID_689043 | Caffeic acid | {"name": "Caffeic acid", "id": "CID_689043"} | INTERACTS_WITH | - | Target | 766 | CA7 | {"name": "CA7", "id": "766"} |
| Phytochemical | CID_689043 | Caffeic acid | {"name": "Caffeic acid", "id": "CID_689043"} | INTERACTS_WITH | - | Target | 768 | CA9 | {"name": "CA9", "id": "768"} |
| Phytochemical | CID_689043 | Caffeic acid | {"name": "Caffeic acid", "id": "CID_689043"} | INTERACTS_WITH | - | Target | 771 | CA12 | {"name": "CA12", "id": "771"} |
| Phytochemical | CID_689043 | Caffeic acid | {"name": "Caffeic acid", "id": "CID_689043"} | INTERACTS_WITH | - | Target | 79885 | HDAC11 | {"name": "HDAC11", "id": "79885"} |
| Phytochemical | CID_689043 | Caffeic acid | {"name": "Caffeic acid", "id": "CID_689043"} | INTERACTS_WITH | - | Target | 83933 | HDAC10 | {"name": "HDAC10", "id": "83933"} |
| Phytochemical | CID_689043 | Caffeic acid | {"name": "Caffeic acid", "id": "CID_689043"} | INTERACTS_WITH | - | Target | 8644 | AKR1C3 | {"name": "AKR1C3", "id": "8644"} |
| Phytochemical | CID_689043 | Caffeic acid | {"name": "Caffeic acid", "id": "CID_689043"} | INTERACTS_WITH | - | Target | 8841 | HDAC3 | {"name": "HDAC3", "id": "8841"} |
| Phytochemical | CID_689043 | Caffeic acid | {"name": "Caffeic acid", "id": "CID_689043"} | INTERACTS_WITH | - | Target | 9682 | KDM4A | {"name": "KDM4A", "id": "9682"} |
| Phytochemical | CID_689043 | Caffeic acid | {"name": "Caffeic acid", "id": "CID_689043"} | INTERACTS_WITH | - | Target | 9734 | HDAC9 | {"name": "HDAC9", "id": "9734"} |
| Phytochemical | CID_689043 | Caffeic acid | {"name": "Caffeic acid", "id": "CID_689043"} | INTERACTS_WITH | - | Target | 9759 | HDAC4 | {"name": "HDAC4", "id": "9759"} |
| Phytochemical | CID_6616 | Camphene | {"name": "Camphene", "id": "CID_6616"} | INTERACTS_WITH | - | Target | 5467 | PPARD | {"name": "PPARD", "id": "5467"} |
| Phytochemical | CID_6616 | Camphene | {"name": "Camphene", "id": "CID_6616"} | INTERACTS_WITH | - | Target | 7442 | TRPV1 | {"name": "TRPV1", "id": "7442"} |
| Phytochemical | CID_6616 | Camphene | {"name": "Camphene", "id": "CID_6616"} | INTERACTS_WITH | - | Target | 84876 | ORAI1 | {"name": "ORAI1", "id": "84876"} |
| Phytochemical | CID_2537 | Camphor | {"name": "Camphor", "id": "CID_2537"} | INTERACTS_WITH | - | Target | 10013 | HDAC6 | {"name": "HDAC6", "id": "10013"} |
| Phytochemical | CID_2537 | Camphor | {"name": "Camphor", "id": "CID_2537"} | INTERACTS_WITH | - | Target | 162514 | TRPV3 | {"name": "TRPV3", "id": "162514"} |
| Phytochemical | CID_2537 | Camphor | {"name": "Camphor", "id": "CID_2537"} | INTERACTS_WITH | - | Target | 2099 | ESR1 | {"name": "ESR1", "id": "2099"} |
| Phytochemical | CID_2537 | Camphor | {"name": "Camphor", "id": "CID_2537"} | INTERACTS_WITH | - | Target | 216 | ALDH1A1 | {"name": "ALDH1A1", "id": "216"} |
| Phytochemical | CID_2537 | Camphor | {"name": "Camphor", "id": "CID_2537"} | INTERACTS_WITH | - | Target | 4137 | MAPT | {"name": "MAPT", "id": "4137"} |
| Phytochemical | CID_2537 | Camphor | {"name": "Camphor", "id": "CID_2537"} | INTERACTS_WITH | - | Target | 55775 | TDP1 | {"name": "TDP1", "id": "55775"} |
| Phytochemical | CID_10364 | Carvacrol | {"name": "Carvacrol", "id": "CID_10364"} | INTERACTS_WITH | - | Target | 10941 | UGT2A1 | {"name": "UGT2A1", "id": "10941"} |
| Phytochemical | CID_10364 | Carvacrol | {"name": "Carvacrol", "id": "CID_10364"} | INTERACTS_WITH | - | Target | 3725 | JUN | {"name": "JUN", "id": "3725"} |
| Phytochemical | CID_10364 | Carvacrol | {"name": "Carvacrol", "id": "CID_10364"} | INTERACTS_WITH | - | Target | 43 | ACHE | {"name": "ACHE", "id": "43"} |
| Phytochemical | CID_10364 | Carvacrol | {"name": "Carvacrol", "id": "CID_10364"} | INTERACTS_WITH | - | Target | 54575 | UGT1A10 | {"name": "UGT1A10", "id": "54575"} |
| Phytochemical | CID_10364 | Carvacrol | {"name": "Carvacrol", "id": "CID_10364"} | INTERACTS_WITH | - | Target | 54576 | UGT1A8 | {"name": "UGT1A8", "id": "54576"} |
| Phytochemical | CID_10364 | Carvacrol | {"name": "Carvacrol", "id": "CID_10364"} | INTERACTS_WITH | - | Target | 54578 | UGT1A6 | {"name": "UGT1A6", "id": "54578"} |
| Phytochemical | CID_10364 | Carvacrol | {"name": "Carvacrol", "id": "CID_10364"} | INTERACTS_WITH | - | Target | 54600 | UGT1A9 | {"name": "UGT1A9", "id": "54600"} |
| Phytochemical | CID_10364 | Carvacrol | {"name": "Carvacrol", "id": "CID_10364"} | INTERACTS_WITH | - | Target | 54657 | UGT1A4 | {"name": "UGT1A4", "id": "54657"} |
| Phytochemical | CID_10364 | Carvacrol | {"name": "Carvacrol", "id": "CID_10364"} | INTERACTS_WITH | - | Target | 54658 | UGT1A1 | {"name": "UGT1A1", "id": "54658"} |
| Phytochemical | CID_10364 | Carvacrol | {"name": "Carvacrol", "id": "CID_10364"} | INTERACTS_WITH | - | Target | 5742 | PTGS1 | {"name": "PTGS1", "id": "5742"} |
| Phytochemical | CID_10364 | Carvacrol | {"name": "Carvacrol", "id": "CID_10364"} | INTERACTS_WITH | - | Target | 7366 | UGT2B15 | {"name": "UGT2B15", "id": "7366"} |
| Phytochemical | CID_10364 | Carvacrol | {"name": "Carvacrol", "id": "CID_10364"} | INTERACTS_WITH | - | Target | 775 | CACNA1C | {"name": "CACNA1C", "id": "775"} |
| Phytochemical | CID_10364 | Carvacrol | {"name": "Carvacrol", "id": "CID_10364"} | INTERACTS_WITH | - | Target | 8989 | TRPA1 | {"name": "TRPA1", "id": "8989"} |
| Phytochemical | CID_1742210 | Caryophyllene oxide | {"name": "Caryophyllene oxide", "id": "CID_1742210"} | INTERACTS_WITH | - | Target | 7442 | TRPV1 | {"name": "TRPV1", "id": "7442"} |
| Phytochemical | CID_1742210 | Caryophyllene oxide | {"name": "Caryophyllene oxide", "id": "CID_1742210"} | INTERACTS_WITH | - | Target | 84876 | ORAI1 | {"name": "ORAI1", "id": "84876"} |
| Phytochemical | CID_91520 | Catalpol | {"name": "Catalpol", "id": "CID_91520"} | INTERACTS_WITH | - | Target | 2099 | ESR1 | {"name": "ESR1", "id": "2099"} |
| Phytochemical | CID_91520 | Catalpol | {"name": "Catalpol", "id": "CID_91520"} | INTERACTS_WITH | - | Target | 2648 | KAT2A | {"name": "KAT2A", "id": "2648"} |
| Phytochemical | CID_91520 | Catalpol | {"name": "Catalpol", "id": "CID_91520"} | INTERACTS_WITH | - | Target | 5742 | PTGS1 | {"name": "PTGS1", "id": "5742"} |
| Phytochemical | CID_91520 | Catalpol | {"name": "Catalpol", "id": "CID_91520"} | INTERACTS_WITH | - | Target | 5743 | PTGS2 | {"name": "PTGS2", "id": "5743"} |
| Phytochemical | CID_289 | Catechol | {"name": "Catechol", "id": "CID_289"} | INTERACTS_WITH | - | Target | 10280 | SIGMAR1 | {"name": "SIGMAR1", "id": "10280"} |
| Phytochemical | CID_289 | Catechol | {"name": "Catechol", "id": "CID_289"} | INTERACTS_WITH | - | Target | 10800 | CYSLTR1 | {"name": "CYSLTR1", "id": "10800"} |
| Phytochemical | CID_289 | Catechol | {"name": "Catechol", "id": "CID_289"} | INTERACTS_WITH | - | Target | 10919 | EHMT2 | {"name": "EHMT2", "id": "10919"} |
| Phytochemical | CID_289 | Catechol | {"name": "Catechol", "id": "CID_289"} | INTERACTS_WITH | - | Target | 10951 | CBX1 | {"name": "CBX1", "id": "10951"} |
| Phytochemical | CID_289 | Catechol | {"name": "Catechol", "id": "CID_289"} | INTERACTS_WITH | - | Target | 11201 | POLI | {"name": "POLI", "id": "11201"} |
| Phytochemical | CID_289 | Catechol | {"name": "Catechol", "id": "CID_289"} | INTERACTS_WITH | - | Target | 11238 | CA5B | {"name": "CA5B", "id": "11238"} |
| Phytochemical | CID_289 | Catechol | {"name": "Catechol", "id": "CID_289"} | INTERACTS_WITH | - | Target | 1128 | CHRM1 | {"name": "CHRM1", "id": "1128"} |
| Phytochemical | CID_289 | Catechol | {"name": "Catechol", "id": "CID_289"} | INTERACTS_WITH | - | Target | 1129 | CHRM2 | {"name": "CHRM2", "id": "1129"} |

**Table S9: Complete node and relationship data used to construct the knowledge graph. This table summarizes interactions among FDA-approved drugs, formulation, herbs, diseases (type 2 diabetes, fatty liver, obesity), phytochemicals, and their target genes/proteins. Each entry includes source and target node details, edge type, and node annotations (Entrez Gene ID and gene symbol, where applicable).**

| Source Node Label | Source Node ID | Source Node Name | Source Node Properties | Edge Type | Edge Properties | Target Node Label | Target Node ID | Target Node Name | Target Node Properties |
|---|---|---|---|---|---|---|---|---|---|
| Phytochemical | CID_289 | Catechol | {"name": "Catechol", "id": "CID_289"} | INTERACTS_WITH | - | Target | 1131 | CHRM3 | {"name": "CHRM3", "id": "1131"} |
| Phytochemical | CID_289 | Catechol | {"name": "Catechol", "id": "CID_289"} | INTERACTS_WITH | - | Target | 1132 | CHRM4 | {"name": "CHRM4", "id": "1132"} |
| Phytochemical | CID_289 | Catechol | {"name": "Catechol", "id": "CID_289"} | INTERACTS_WITH | - | Target | 1133 | CHRM5 | {"name": "CHRM5", "id": "1133"} |
| Phytochemical | CID_289 | Catechol | {"name": "Catechol", "id": "CID_289"} | INTERACTS_WITH | - | Target | 1233 | CCR4 | {"name": "CCR4", "id": "1233"} |
| Phytochemical | CID_289 | Catechol | {"name": "Catechol", "id": "CID_289"} | INTERACTS_WITH | - | Target | 1234 | CCR5 | {"name": "CCR5", "id": "1234"} |
| Phytochemical | CID_289 | Catechol | {"name": "Catechol", "id": "CID_289"} | INTERACTS_WITH | - | Target | 1268 | CNR1 | {"name": "CNR1", "id": "1268"} |
| Phytochemical | CID_289 | Catechol | {"name": "Catechol", "id": "CID_289"} | INTERACTS_WITH | - | Target | 134 | ADORA1 | {"name": "ADORA1", "id": "134"} |
| Phytochemical | CID_289 | Catechol | {"name": "Catechol", "id": "CID_289"} | INTERACTS_WITH | - | Target | 135 | ADORA2A | {"name": "ADORA2A", "id": "135"} |
| Phytochemical | CID_289 | Catechol | {"name": "Catechol", "id": "CID_289"} | INTERACTS_WITH | - | Target | 140 | ADORA3 | {"name": "ADORA3", "id": "140"} |
| Phytochemical | CID_289 | Catechol | {"name": "Catechol", "id": "CID_289"} | INTERACTS_WITH | - | Target | 1432 | MAPK14 | {"name": "MAPK14", "id": "1432"} |
| Phytochemical | CID_289 | Catechol | {"name": "Catechol", "id": "CID_289"} | INTERACTS_WITH | - | Target | 146 | ADRA1D | {"name": "ADRA1D", "id": "146"} |
| Phytochemical | CID_289 | Catechol | {"name": "Catechol", "id": "CID_289"} | INTERACTS_WITH | - | Target | 150 | ADRA2A | {"name": "ADRA2A", "id": "150"} |
| Phytochemical | CID_289 | Catechol | {"name": "Catechol", "id": "CID_289"} | INTERACTS_WITH | - | Target | 151 | ADRA2B | {"name": "ADRA2B", "id": "151"} |
| Phytochemical | CID_289 | Catechol | {"name": "Catechol", "id": "CID_289"} | INTERACTS_WITH | - | Target | 1511 | CTSG | {"name": "CTSG", "id": "1511"} |
| Phytochemical | CID_289 | Catechol | {"name": "Catechol", "id": "CID_289"} | INTERACTS_WITH | - | Target | 152 | ADRA2C | {"name": "ADRA2C", "id": "152"} |
| Phytochemical | CID_289 | Catechol | {"name": "Catechol", "id": "CID_289"} | INTERACTS_WITH | - | Target | 153 | ADRB1 | {"name": "ADRB1", "id": "153"} |
| Phytochemical | CID_289 | Catechol | {"name": "Catechol", "id": "CID_289"} | INTERACTS_WITH | - | Target | 154 | ADRB2 | {"name": "ADRB2", "id": "154"} |
| Phytochemical | CID_289 | Catechol | {"name": "Catechol", "id": "CID_289"} | INTERACTS_WITH | - | Target | 1544 | CYP1A2 | {"name": "CYP1A2", "id": "1544"} |
| Phytochemical | CID_289 | Catechol | {"name": "Catechol", "id": "CID_289"} | INTERACTS_WITH | - | Target | 1548 | CYP2A6 | {"name": "CYP2A6", "id": "1548"} |
| Phytochemical | CID_289 | Catechol | {"name": "Catechol", "id": "CID_289"} | INTERACTS_WITH | - | Target | 155 | ADRB3 | {"name": "ADRB3", "id": "155"} |
| Phytochemical | CID_289 | Catechol | {"name": "Catechol", "id": "CID_289"} | INTERACTS_WITH | - | Target | 1557 | CYP2C19 | {"name": "CYP2C19", "id": "1557"} |
| Phytochemical | CID_289 | Catechol | {"name": "Catechol", "id": "CID_289"} | INTERACTS_WITH | - | Target | 1559 | CYP2C9 | {"name": "CYP2C9", "id": "1559"} |
| Phytochemical | CID_289 | Catechol | {"name": "Catechol", "id": "CID_289"} | INTERACTS_WITH | - | Target | 1565 | CYP2D6 | {"name": "CYP2D6", "id": "1565"} |
| Phytochemical | CID_289 | Catechol | {"name": "Catechol", "id": "CID_289"} | INTERACTS_WITH | - | Target | 1571 | CYP2E1 | {"name": "CYP2E1", "id": "1571"} |
| Phytochemical | CID_289 | Catechol | {"name": "Catechol", "id": "CID_289"} | INTERACTS_WITH | - | Target | 1576 | CYP3A4 | {"name": "CYP3A4", "id": "1576"} |
| Phytochemical | CID_289 | Catechol | {"name": "Catechol", "id": "CID_289"} | INTERACTS_WITH | - | Target | 1812 | DRD1 | {"name": "DRD1", "id": "1812"} |
| Phytochemical | CID_289 | Catechol | {"name": "Catechol", "id": "CID_289"} | INTERACTS_WITH | - | Target | 1813 | DRD2 | {"name": "DRD2", "id": "1813"} |
| Phytochemical | CID_289 | Catechol | {"name": "Catechol", "id": "CID_289"} | INTERACTS_WITH | - | Target | 1814 | DRD3 | {"name": "DRD3", "id": "1814"} |
| Phytochemical | CID_289 | Catechol | {"name": "Catechol", "id": "CID_289"} | INTERACTS_WITH | - | Target | 1815 | DRD4 | {"name": "DRD4", "id": "1815"} |
| Phytochemical | CID_289 | Catechol | {"name": "Catechol", "id": "CID_289"} | INTERACTS_WITH | - | Target | 186 | AGTR2 | {"name": "AGTR2", "id": "186"} |
| Phytochemical | CID_289 | Catechol | {"name": "Catechol", "id": "CID_289"} | INTERACTS_WITH | - | Target | 1909 | EDNRA | {"name": "EDNRA", "id": "1909"} |
| Phytochemical | CID_289 | Catechol | {"name": "Catechol", "id": "CID_289"} | INTERACTS_WITH | - | Target | 1956 | EGFR | {"name": "EGFR", "id": "1956"} |
| Phytochemical | CID_289 | Catechol | {"name": "Catechol", "id": "CID_289"} | INTERACTS_WITH | - | Target | 1991 | ELANE | {"name": "ELANE", "id": "1991"} |
| Phytochemical | CID_289 | Catechol | {"name": "Catechol", "id": "CID_289"} | INTERACTS_WITH | - | Target | 2064 | ERBB2 | {"name": "ERBB2", "id": "2064"} |
| Phytochemical | CID_289 | Catechol | {"name": "Catechol", "id": "CID_289"} | INTERACTS_WITH | - | Target | 2099 | ESR1 | {"name": "ESR1", "id": "2099"} |
| Phytochemical | CID_289 | Catechol | {"name": "Catechol", "id": "CID_289"} | INTERACTS_WITH | - | Target | 2100 | ESR2 | {"name": "ESR2", "id": "2100"} |
| Phytochemical | CID_289 | Catechol | {"name": "Catechol", "id": "CID_289"} | INTERACTS_WITH | - | Target | 216 | ALDH1A1 | {"name": "ALDH1A1", "id": "216"} |
| Phytochemical | CID_289 | Catechol | {"name": "Catechol", "id": "CID_289"} | INTERACTS_WITH | - | Target | 2237 | FEN1 | {"name": "FEN1", "id": "2237"} |
| Phytochemical | CID_289 | Catechol | {"name": "Catechol", "id": "CID_289"} | INTERACTS_WITH | - | Target | 2321 | FLT1 | {"name": "FLT1", "id": "2321"} |
| Phytochemical | CID_289 | Catechol | {"name": "Catechol", "id": "CID_289"} | INTERACTS_WITH | - | Target | 23632 | CA14 | {"name": "CA14", "id": "23632"} |
| Phytochemical | CID_289 | Catechol | {"name": "Catechol", "id": "CID_289"} | INTERACTS_WITH | - | Target | 240 | ALOX5 | {"name": "ALOX5", "id": "240"} |
| Phytochemical | CID_289 | Catechol | {"name": "Catechol", "id": "CID_289"} | INTERACTS_WITH | - | Target | 246 | ALOX15 | {"name": "ALOX15", "id": "246"} |
| Phytochemical | CID_289 | Catechol | {"name": "Catechol", "id": "CID_289"} | INTERACTS_WITH | - | Target | 247 | ALOX15B | {"name": "ALOX15B", "id": "247"} |
| Phytochemical | CID_289 | Catechol | {"name": "Catechol", "id": "CID_289"} | INTERACTS_WITH | - | Target | 2475 | MTOR | {"name": "MTOR", "id": "2475"} |
| Phytochemical | CID_289 | Catechol | {"name": "Catechol", "id": "CID_289"} | INTERACTS_WITH | - | Target | 2534 | FYN | {"name": "FYN", "id": "2534"} |
| Phytochemical | CID_289 | Catechol | {"name": "Catechol", "id": "CID_289"} | INTERACTS_WITH | - | Target | 25797 | QPCT | {"name": "QPCT", "id": "25797"} |
| Phytochemical | CID_289 | Catechol | {"name": "Catechol", "id": "CID_289"} | INTERACTS_WITH | - | Target | 26013 | L3MBTL1 | {"name": "L3MBTL1", "id": "26013"} |
| Phytochemical | CID_289 | Catechol | {"name": "Catechol", "id": "CID_289"} | INTERACTS_WITH | - | Target | 2744 | GLS | {"name": "GLS", "id": "2744"} |
| Phytochemical | CID_289 | Catechol | {"name": "Catechol", "id": "CID_289"} | INTERACTS_WITH | - | Target | 2908 | NR3C1 | {"name": "NR3C1", "id": "2908"} |
| Phytochemical | CID_289 | Catechol | {"name": "Catechol", "id": "CID_289"} | INTERACTS_WITH | - | Target | 3028 | HSD17B10 | {"name": "HSD17B10", "id": "3028"} |
| Phytochemical | CID_289 | Catechol | {"name": "Catechol", "id": "CID_289"} | INTERACTS_WITH | - | Target | 3091 | HIF1A | {"name": "HIF1A", "id": "3091"} |
| Phytochemical | CID_289 | Catechol | {"name": "Catechol", "id": "CID_289"} | INTERACTS_WITH | - | Target | 3156 | HMGCR | {"name": "HMGCR", "id": "3156"} |
| Phytochemical | CID_289 | Catechol | {"name": "Catechol", "id": "CID_289"} | INTERACTS_WITH | - | Target | 3248 | HPGD | {"name": "HPGD", "id": "3248"} |
| Phytochemical | CID_289 | Catechol | {"name": "Catechol", "id": "CID_289"} | INTERACTS_WITH | - | Target | 3269 | HRH1 | {"name": "HRH1", "id": "3269"} |
| Phytochemical | CID_289 | Catechol | {"name": "Catechol", "id": "CID_289"} | INTERACTS_WITH | - | Target | 3274 | HRH2 | {"name": "HRH2", "id": "3274"} |
| Phytochemical | CID_289 | Catechol | {"name": "Catechol", "id": "CID_289"} | INTERACTS_WITH | - | Target | 328 | APEX1 | {"name": "APEX1", "id": "328"} |
| Phytochemical | CID_289 | Catechol | {"name": "Catechol", "id": "CID_289"} | INTERACTS_WITH | - | Target | 3309 | HSPA5 | {"name": "HSPA5", "id": "3309"} |
| Phytochemical | CID_289 | Catechol | {"name": "Catechol", "id": "CID_289"} | INTERACTS_WITH | - | Target | 3315 | HSPB1 | {"name": "HSPB1", "id": "3315"} |
| Phytochemical | CID_289 | Catechol | {"name": "Catechol", "id": "CID_289"} | INTERACTS_WITH | - | Target | 3356 | HTR2A | {"name": "HTR2A", "id": "3356"} |
| Phytochemical | CID_289 | Catechol | {"name": "Catechol", "id": "CID_289"} | INTERACTS_WITH | - | Target | 3357 | HTR2B | {"name": "HTR2B", "id": "3357"} |
| Phytochemical | CID_289 | Catechol | {"name": "Catechol", "id": "CID_289"} | INTERACTS_WITH | - | Target | 3358 | HTR2C | {"name": "HTR2C", "id": "3358"} |
| Phytochemical | CID_289 | Catechol | {"name": "Catechol", "id": "CID_289"} | INTERACTS_WITH | - | Target | 3362 | HTR6 | {"name": "HTR6", "id": "3362"} |
| Phytochemical | CID_289 | Catechol | {"name": "Catechol", "id": "CID_289"} | INTERACTS_WITH | - | Target | 3417 | IDH1 | {"name": "IDH1", "id": "3417"} |
| Phytochemical | CID_289 | Catechol | {"name": "Catechol", "id": "CID_289"} | INTERACTS_WITH | - | Target | 3576 | CXCL8 | {"name": "CXCL8", "id": "3576"} |
| Phytochemical | CID_289 | Catechol | {"name": "Catechol", "id": "CID_289"} | INTERACTS_WITH | - | Target | 3577 | CXCR1 | {"name": "CXCR1", "id": "3577"} |
| Phytochemical | CID_289 | Catechol | {"name": "Catechol", "id": "CID_289"} | INTERACTS_WITH | - | Target | 3579 | CXCR2 | {"name": "CXCR2", "id": "3579"} |
| Phytochemical | CID_289 | Catechol | {"name": "Catechol", "id": "CID_289"} | INTERACTS_WITH | - | Target | 3620 | IDO1 | {"name": "IDO1", "id": "3620"} |
| Phytochemical | CID_289 | Catechol | {"name": "Catechol", "id": "CID_289"} | INTERACTS_WITH | - | Target | 367 | AR | {"name": "AR", "id": "367"} |
| Phytochemical | CID_289 | Catechol | {"name": "Catechol", "id": "CID_289"} | INTERACTS_WITH | - | Target | 3725 | JUN | {"name": "JUN", "id": "3725"} |
| Phytochemical | CID_289 | Catechol | {"name": "Catechol", "id": "CID_289"} | INTERACTS_WITH | - | Target | 3757 | KCNH2 | {"name": "KCNH2", "id": "3757"} |
| Phytochemical | CID_289 | Catechol | {"name": "Catechol", "id": "CID_289"} | INTERACTS_WITH | - | Target | 390245 | KDM4E | {"name": "KDM4E", "id": "390245"} |
| Phytochemical | CID_289 | Catechol | {"name": "Catechol", "id": "CID_289"} | INTERACTS_WITH | - | Target | 3932 | LCK | {"name": "LCK", "id": "3932"} |
| Phytochemical | CID_289 | Catechol | {"name": "Catechol", "id": "CID_289"} | INTERACTS_WITH | - | Target | 4000 | LMNA | {"name": "LMNA", "id": "4000"} |
| Phytochemical | CID_289 | Catechol | {"name": "Catechol", "id": "CID_289"} | INTERACTS_WITH | - | Target | 4128 | MAOA | {"name": "MAOA", "id": "4128"} |
| Phytochemical | CID_289 | Catechol | {"name": "Catechol", "id": "CID_289"} | INTERACTS_WITH | - | Target | 4137 | MAPT | {"name": "MAPT", "id": "4137"} |
| Phytochemical | CID_289 | Catechol | {"name": "Catechol", "id": "CID_289"} | INTERACTS_WITH | - | Target | 4159 | MC3R | {"name": "MC3R", "id": "4159"} |

**Table S9: Complete node and relationship data used to construct the knowledge graph. This table summarizes interactions among FDA-approved drugs, formulation, herbs, diseases (type 2 diabetes, fatty liver, obesity), phytochemicals, and their target genes/proteins. Each entry includes source and target node details, edge type, and node annotations (Entrez Gene ID and gene symbol, where applicable).**

| Source Node Label | Source Node ID | Source Node Name | Source Node Properties | Edge Type | Edge Properties | Target Node Label | Target Node ID | Target Node Name | Target Node Properties |
|---|---|---|---|---|---|---|---|---|---|
| Phytochemical | CID_289 | Catechol | {"name": "Catechol", "id": "CID_289"} | INTERACTS_WITH | - | Target | 4160 | MC4R | {"name": "MC4R", "id": "4160"} |
| Phytochemical | CID_289 | Catechol | {"name": "Catechol", "id": "CID_289"} | INTERACTS_WITH | - | Target | 4161 | MC5R | {"name": "MC5R", "id": "4161"} |
| Phytochemical | CID_289 | Catechol | {"name": "Catechol", "id": "CID_289"} | INTERACTS_WITH | - | Target | 4193 | MDM2 | {"name": "MDM2", "id": "4193"} |
| Phytochemical | CID_289 | Catechol | {"name": "Catechol", "id": "CID_289"} | INTERACTS_WITH | - | Target | 43 | ACHE | {"name": "ACHE", "id": "43"} |
| Phytochemical | CID_289 | Catechol | {"name": "Catechol", "id": "CID_289"} | INTERACTS_WITH | - | Target | 4312 | MMP1 | {"name": "MMP1", "id": "4312"} |
| Phytochemical | CID_289 | Catechol | {"name": "Catechol", "id": "CID_289"} | INTERACTS_WITH | - | Target | 4313 | MMP2 | {"name": "MMP2", "id": "4313"} |
| Phytochemical | CID_289 | Catechol | {"name": "Catechol", "id": "CID_289"} | INTERACTS_WITH | - | Target | 4314 | MMP3 | {"name": "MMP3", "id": "4314"} |
| Phytochemical | CID_289 | Catechol | {"name": "Catechol", "id": "CID_289"} | INTERACTS_WITH | - | Target | 4317 | MMP8 | {"name": "MMP8", "id": "4317"} |
| Phytochemical | CID_289 | Catechol | {"name": "Catechol", "id": "CID_289"} | INTERACTS_WITH | - | Target | 4318 | MMP9 | {"name": "MMP9", "id": "4318"} |
| Phytochemical | CID_289 | Catechol | {"name": "Catechol", "id": "CID_289"} | INTERACTS_WITH | - | Target | 4780 | NFE2L2 | {"name": "NFE2L2", "id": "4780"} |
| Phytochemical | CID_289 | Catechol | {"name": "Catechol", "id": "CID_289"} | INTERACTS_WITH | - | Target | 4790 | NFKB1 | {"name": "NFKB1", "id": "4790"} |
| Phytochemical | CID_289 | Catechol | {"name": "Catechol", "id": "CID_289"} | INTERACTS_WITH | - | Target | 4864 | NPC1 | {"name": "NPC1", "id": "4864"} |
| Phytochemical | CID_289 | Catechol | {"name": "Catechol", "id": "CID_289"} | INTERACTS_WITH | - | Target | 4886 | NPY1R | {"name": "NPY1R", "id": "4886"} |
| Phytochemical | CID_289 | Catechol | {"name": "Catechol", "id": "CID_289"} | INTERACTS_WITH | - | Target | 4887 | NPY2R | {"name": "NPY2R", "id": "4887"} |
| Phytochemical | CID_289 | Catechol | {"name": "Catechol", "id": "CID_289"} | INTERACTS_WITH | - | Target | 4985 | OPRD1 | {"name": "OPRD1", "id": "4985"} |
| Phytochemical | CID_289 | Catechol | {"name": "Catechol", "id": "CID_289"} | INTERACTS_WITH | - | Target | 4986 | OPRK1 | {"name": "OPRK1", "id": "4986"} |
| Phytochemical | CID_289 | Catechol | {"name": "Catechol", "id": "CID_289"} | INTERACTS_WITH | - | Target | 4988 | OPRM1 | {"name": "OPRM1", "id": "4988"} |
| Phytochemical | CID_289 | Catechol | {"name": "Catechol", "id": "CID_289"} | INTERACTS_WITH | - | Target | 51053 | GMNN | {"name": "GMNN", "id": "51053"} |
| Phytochemical | CID_289 | Catechol | {"name": "Catechol", "id": "CID_289"} | INTERACTS_WITH | - | Target | 51426 | POLK | {"name": "POLK", "id": "51426"} |
| Phytochemical | CID_289 | Catechol | {"name": "Catechol", "id": "CID_289"} | INTERACTS_WITH | - | Target | 5290 | PIK3CA | {"name": "PIK3CA", "id": "5290"} |
| Phytochemical | CID_289 | Catechol | {"name": "Catechol", "id": "CID_289"} | INTERACTS_WITH | - | Target | 5300 | PIN1 | {"name": "PIN1", "id": "5300"} |
| Phytochemical | CID_289 | Catechol | {"name": "Catechol", "id": "CID_289"} | INTERACTS_WITH | - | Target | 5376 | PMP22 | {"name": "PMP22", "id": "5376"} |
| Phytochemical | CID_289 | Catechol | {"name": "Catechol", "id": "CID_289"} | INTERACTS_WITH | - | Target | 5423 | POLB | {"name": "POLB", "id": "5423"} |
| Phytochemical | CID_289 | Catechol | {"name": "Catechol", "id": "CID_289"} | INTERACTS_WITH | - | Target | 5429 | POLH | {"name": "POLH", "id": "5429"} |
| Phytochemical | CID_289 | Catechol | {"name": "Catechol", "id": "CID_289"} | INTERACTS_WITH | - | Target | 5467 | PPARD | {"name": "PPARD", "id": "5467"} |
| Phytochemical | CID_289 | Catechol | {"name": "Catechol", "id": "CID_289"} | INTERACTS_WITH | - | Target | 5468 | PPARG | {"name": "PPARG", "id": "5468"} |
| Phytochemical | CID_289 | Catechol | {"name": "Catechol", "id": "CID_289"} | INTERACTS_WITH | - | Target | 54737 | MPHOSPH8 | {"name": "MPHOSPH8", "id": "54737"} |
| Phytochemical | CID_289 | Catechol | {"name": "Catechol", "id": "CID_289"} | INTERACTS_WITH | - | Target | 552 | AVPR1A | {"name": "AVPR1A", "id": "552"} |
| Phytochemical | CID_289 | Catechol | {"name": "Catechol", "id": "CID_289"} | INTERACTS_WITH | - | Target | 5530 | PPP3CA | {"name": "PPP3CA", "id": "5530"} |
| Phytochemical | CID_289 | Catechol | {"name": "Catechol", "id": "CID_289"} | INTERACTS_WITH | - | Target | 55775 | TDP1 | {"name": "TDP1", "id": "55775"} |
| Phytochemical | CID_289 | Catechol | {"name": "Catechol", "id": "CID_289"} | INTERACTS_WITH | - | Target | 5578 | PRKCA | {"name": "PRKCA", "id": "5578"} |
| Phytochemical | CID_289 | Catechol | {"name": "Catechol", "id": "CID_289"} | INTERACTS_WITH | - | Target | 5594 | MAPK1 | {"name": "MAPK1", "id": "5594"} |
| Phytochemical | CID_289 | Catechol | {"name": "Catechol", "id": "CID_289"} | INTERACTS_WITH | - | Target | 5595 | MAPK3 | {"name": "MAPK3", "id": "5595"} |
| Phytochemical | CID_289 | Catechol | {"name": "Catechol", "id": "CID_289"} | INTERACTS_WITH | - | Target | 5724 | PTAFR | {"name": "PTAFR", "id": "5724"} |
| Phytochemical | CID_289 | Catechol | {"name": "Catechol", "id": "CID_289"} | INTERACTS_WITH | - | Target | 5742 | PTGS1 | {"name": "PTGS1", "id": "5742"} |
| Phytochemical | CID_289 | Catechol | {"name": "Catechol", "id": "CID_289"} | INTERACTS_WITH | - | Target | 5743 | PTGS2 | {"name": "PTGS2", "id": "5743"} |
| Phytochemical | CID_289 | Catechol | {"name": "Catechol", "id": "CID_289"} | INTERACTS_WITH | - | Target | 5745 | PTH1R | {"name": "PTH1R", "id": "5745"} |
| Phytochemical | CID_289 | Catechol | {"name": "Catechol", "id": "CID_289"} | INTERACTS_WITH | - | Target | 5788 | PTPRC | {"name": "PTPRC", "id": "5788"} |
| Phytochemical | CID_289 | Catechol | {"name": "Catechol", "id": "CID_289"} | INTERACTS_WITH | - | Target | 5965 | RECQL | {"name": "RECQL", "id": "5965"} |
| Phytochemical | CID_289 | Catechol | {"name": "Catechol", "id": "CID_289"} | INTERACTS_WITH | - | Target | 5999 | RGS4 | {"name": "RGS4", "id": "5999"} |
| Phytochemical | CID_289 | Catechol | {"name": "Catechol", "id": "CID_289"} | INTERACTS_WITH | - | Target | 60489 | APOBEC3G | {"name": "APOBEC3G", "id": "60489"} |
| Phytochemical | CID_289 | Catechol | {"name": "Catechol", "id": "CID_289"} | INTERACTS_WITH | - | Target | 6097 | RORC | {"name": "RORC", "id": "6097"} |
| Phytochemical | CID_289 | Catechol | {"name": "Catechol", "id": "CID_289"} | INTERACTS_WITH | - | Target | 624 | BDKRB2 | {"name": "BDKRB2", "id": "624"} |
| Phytochemical | CID_289 | Catechol | {"name": "Catechol", "id": "CID_289"} | INTERACTS_WITH | - | Target | 6311 | ATXN2 | {"name": "ATXN2", "id": "6311"} |
| Phytochemical | CID_289 | Catechol | {"name": "Catechol", "id": "CID_289"} | INTERACTS_WITH | - | Target | 641 | BLM | {"name": "BLM", "id": "641"} |
| Phytochemical | CID_289 | Catechol | {"name": "Catechol", "id": "CID_289"} | INTERACTS_WITH | - | Target | 6530 | SLC6A2 | {"name": "SLC6A2", "id": "6530"} |
| Phytochemical | CID_289 | Catechol | {"name": "Catechol", "id": "CID_289"} | INTERACTS_WITH | - | Target | 6531 | SLC6A3 | {"name": "SLC6A3", "id": "6531"} |
| Phytochemical | CID_289 | Catechol | {"name": "Catechol", "id": "CID_289"} | INTERACTS_WITH | - | Target | 6532 | SLC6A4 | {"name": "SLC6A4", "id": "6532"} |
| Phytochemical | CID_289 | Catechol | {"name": "Catechol", "id": "CID_289"} | INTERACTS_WITH | - | Target | 6606 | SMN1 | {"name": "SMN1", "id": "6606"} |
| Phytochemical | CID_289 | Catechol | {"name": "Catechol", "id": "CID_289"} | INTERACTS_WITH | - | Target | 6865 | TACR2 | {"name": "TACR2", "id": "6865"} |
| Phytochemical | CID_289 | Catechol | {"name": "Catechol", "id": "CID_289"} | INTERACTS_WITH | - | Target | 6869 | TACR1 | {"name": "TACR1", "id": "6869"} |
| Phytochemical | CID_289 | Catechol | {"name": "Catechol", "id": "CID_289"} | INTERACTS_WITH | - | Target | 6916 | TBXAS1 | {"name": "TBXAS1", "id": "6916"} |
| Phytochemical | CID_289 | Catechol | {"name": "Catechol", "id": "CID_289"} | INTERACTS_WITH | - | Target | 7066 | THPO | {"name": "THPO", "id": "7066"} |
| Phytochemical | CID_289 | Catechol | {"name": "Catechol", "id": "CID_289"} | INTERACTS_WITH | - | Target | 729230 | CCR2 | {"name": "CCR2", "id": "729230"} |
| Phytochemical | CID_289 | Catechol | {"name": "Catechol", "id": "CID_289"} | INTERACTS_WITH | - | Target | 7433 | VIPR1 | {"name": "VIPR1", "id": "7433"} |
| Phytochemical | CID_289 | Catechol | {"name": "Catechol", "id": "CID_289"} | INTERACTS_WITH | - | Target | 7498 | XDH | {"name": "XDH", "id": "7498"} |
| Phytochemical | CID_289 | Catechol | {"name": "Catechol", "id": "CID_289"} | INTERACTS_WITH | - | Target | 759 | CA1 | {"name": "CA1", "id": "759"} |
| Phytochemical | CID_289 | Catechol | {"name": "Catechol", "id": "CID_289"} | INTERACTS_WITH | - | Target | 760 | CA2 | {"name": "CA2", "id": "760"} |
| Phytochemical | CID_289 | Catechol | {"name": "Catechol", "id": "CID_289"} | INTERACTS_WITH | - | Target | 761 | CA3 | {"name": "CA3", "id": "761"} |
| Phytochemical | CID_289 | Catechol | {"name": "Catechol", "id": "CID_289"} | INTERACTS_WITH | - | Target | 762 | CA4 | {"name": "CA4", "id": "762"} |
| Phytochemical | CID_289 | Catechol | {"name": "Catechol", "id": "CID_289"} | INTERACTS_WITH | - | Target | 763 | CA5A | {"name": "CA5A", "id": "763"} |
| Phytochemical | CID_289 | Catechol | {"name": "Catechol", "id": "CID_289"} | INTERACTS_WITH | - | Target | 765 | CA6 | {"name": "CA6", "id": "765"} |
| Phytochemical | CID_289 | Catechol | {"name": "Catechol", "id": "CID_289"} | INTERACTS_WITH | - | Target | 766 | CA7 | {"name": "CA7", "id": "766"} |
| Phytochemical | CID_289 | Catechol | {"name": "Catechol", "id": "CID_289"} | INTERACTS_WITH | - | Target | 768 | CA9 | {"name": "CA9", "id": "768"} |
| Phytochemical | CID_289 | Catechol | {"name": "Catechol", "id": "CID_289"} | INTERACTS_WITH | - | Target | 771 | CA12 | {"name": "CA12", "id": "771"} |
| Phytochemical | CID_289 | Catechol | {"name": "Catechol", "id": "CID_289"} | INTERACTS_WITH | - | Target | 799 | CALCR | {"name": "CALCR", "id": "799"} |
| Phytochemical | CID_289 | Catechol | {"name": "Catechol", "id": "CID_289"} | INTERACTS_WITH | - | Target | 834 | CASP1 | {"name": "CASP1", "id": "834"} |
| Phytochemical | CID_289 | Catechol | {"name": "Catechol", "id": "CID_289"} | INTERACTS_WITH | - | Target | 8654 | PDE5A | {"name": "PDE5A", "id": "8654"} |
| Phytochemical | CID_289 | Catechol | {"name": "Catechol", "id": "CID_289"} | INTERACTS_WITH | - | Target | 886 | CCKAR | {"name": "CCKAR", "id": "886"} |
| Phytochemical | CID_289 | Catechol | {"name": "Catechol", "id": "CID_289"} | INTERACTS_WITH | - | Target | 9367 | RAB9A | {"name": "RAB9A", "id": "9367"} |
| Phytochemical | CID_289 | Catechol | {"name": "Catechol", "id": "CID_289"} | INTERACTS_WITH | - | Target | 9682 | KDM4A | {"name": "KDM4A", "id": "9682"} |
| Phytochemical | CID_289 | Catechol | {"name": "Catechol", "id": "CID_289"} | INTERACTS_WITH | - | Target | 9971 | NR1H4 | {"name": "NR1H4", "id": "9971"} |
| Phytochemical | CID_5997 | Cholesterol | {"name": "Cholesterol", "id": "CID_5997"} | INTERACTS_WITH | - | Target | 10013 | HDAC6 | {"name": "HDAC6", "id": "10013"} |
| Phytochemical | CID_5997 | Cholesterol | {"name": "Cholesterol", "id": "CID_5997"} | INTERACTS_WITH | - | Target | 10062 | NR1H3 | {"name": "NR1H3", "id": "10062"} |
| Phytochemical | CID_5997 | Cholesterol | {"name": "Cholesterol", "id": "CID_5997"} | INTERACTS_WITH | - | Target | 10599 | SLCO1B1 | {"name": "SLCO1B1", "id": "10599"} |
| Phytochemical | CID_5997 | Cholesterol | {"name": "Cholesterol", "id": "CID_5997"} | INTERACTS_WITH | - | Target | 10858 | CYP46A1 | {"name": "CYP46A1", "id": "10858"} |

**Table S9: Complete node and relationship data used to construct the knowledge graph. This table summarizes interactions among FDA-approved drugs, formulation, herbs, diseases (type 2 diabetes, fatty liver, obesity), phytochemicals, and their target genes/proteins. Each entry includes source and target node details, edge type, and node annotations (Entrez Gene ID and gene symbol, where applicable).**

| Source Node Label | Source Node ID | Source Node Name | Source Node Properties | Edge Type | Edge Properties | Target Node Label | Target Node ID | Target Node Name | Target Node Properties |
|---|---|---|---|---|---|---|---|---|---|
| Phytochemical | CID_5997 | Cholesterol | {"name": "Cholesterol", "id": "CID_5997"} | INTERACTS_WITH | - | Target | 1595 | CYP51A1 | {"name": "CYP51A1", "id": "1595"} |
| Phytochemical | CID_5997 | Cholesterol | {"name": "Cholesterol", "id": "CID_5997"} | INTERACTS_WITH | - | Target | 23762 | OSBP2 | {"name": "OSBP2", "id": "23762"} |
| Phytochemical | CID_5997 | Cholesterol | {"name": "Cholesterol", "id": "CID_5997"} | INTERACTS_WITH | - | Target | 28234 | SLCO1B3 | {"name": "SLCO1B3", "id": "28234"} |
| Phytochemical | CID_5997 | Cholesterol | {"name": "Cholesterol", "id": "CID_5997"} | INTERACTS_WITH | - | Target | 2908 | NR3C1 | {"name": "NR3C1", "id": "2908"} |
| Phytochemical | CID_5997 | Cholesterol | {"name": "Cholesterol", "id": "CID_5997"} | INTERACTS_WITH | - | Target | 3417 | IDH1 | {"name": "IDH1", "id": "3417"} |
| Phytochemical | CID_5997 | Cholesterol | {"name": "Cholesterol", "id": "CID_5997"} | INTERACTS_WITH | - | Target | 367 | AR | {"name": "AR", "id": "367"} |
| Phytochemical | CID_5997 | Cholesterol | {"name": "Cholesterol", "id": "CID_5997"} | INTERACTS_WITH | - | Target | 4000 | LMNA | {"name": "LMNA", "id": "4000"} |
| Phytochemical | CID_5997 | Cholesterol | {"name": "Cholesterol", "id": "CID_5997"} | INTERACTS_WITH | - | Target | 4780 | NFE2L2 | {"name": "NFE2L2", "id": "4780"} |
| Phytochemical | CID_5997 | Cholesterol | {"name": "Cholesterol", "id": "CID_5997"} | INTERACTS_WITH | - | Target | 5007 | OSBP | {"name": "OSBP", "id": "5007"} |
| Phytochemical | CID_5997 | Cholesterol | {"name": "Cholesterol", "id": "CID_5997"} | INTERACTS_WITH | - | Target | 5243 | ABCB1 | {"name": "ABCB1", "id": "5243"} |
| Phytochemical | CID_5997 | Cholesterol | {"name": "Cholesterol", "id": "CID_5997"} | INTERACTS_WITH | - | Target | 5422 | POLA1 | {"name": "POLA1", "id": "5422"} |
| Phytochemical | CID_5997 | Cholesterol | {"name": "Cholesterol", "id": "CID_5997"} | INTERACTS_WITH | - | Target | 5467 | PPARD | {"name": "PPARD", "id": "5467"} |
| Phytochemical | CID_5997 | Cholesterol | {"name": "Cholesterol", "id": "CID_5997"} | INTERACTS_WITH | - | Target | 6097 | RORC | {"name": "RORC", "id": "6097"} |
| Phytochemical | CID_5997 | Cholesterol | {"name": "Cholesterol", "id": "CID_5997"} | INTERACTS_WITH | - | Target | 6323 | SCN1A | {"name": "SCN1A", "id": "6323"} |
| Phytochemical | CID_5997 | Cholesterol | {"name": "Cholesterol", "id": "CID_5997"} | INTERACTS_WITH | - | Target | 6326 | SCN2A | {"name": "SCN2A", "id": "6326"} |
| Phytochemical | CID_5997 | Cholesterol | {"name": "Cholesterol", "id": "CID_5997"} | INTERACTS_WITH | - | Target | 6328 | SCN3A | {"name": "SCN3A", "id": "6328"} |
| Phytochemical | CID_5997 | Cholesterol | {"name": "Cholesterol", "id": "CID_5997"} | INTERACTS_WITH | - | Target | 6622 | SNCA | {"name": "SNCA", "id": "6622"} |
| Phytochemical | CID_5997 | Cholesterol | {"name": "Cholesterol", "id": "CID_5997"} | INTERACTS_WITH | - | Target | 6820 | SULT2B1 | {"name": "SULT2B1", "id": "6820"} |
| Phytochemical | CID_5997 | Cholesterol | {"name": "Cholesterol", "id": "CID_5997"} | INTERACTS_WITH | - | Target | 7155 | TOP2B | {"name": "TOP2B", "id": "7155"} |
| Phytochemical | CID_5997 | Cholesterol | {"name": "Cholesterol", "id": "CID_5997"} | INTERACTS_WITH | - | Target | 9429 | ABCG2 | {"name": "ABCG2", "id": "9429"} |
| Phytochemical | CID_638011 | Citral | {"name": "Citral", "id": "CID_638011"} | INTERACTS_WITH | - | Target | 216 | ALDH1A1 | {"name": "ALDH1A1", "id": "216"} |
| Phytochemical | CID_638011 | Citral | {"name": "Citral", "id": "CID_638011"} | INTERACTS_WITH | - | Target | 2166 | FAAH | {"name": "FAAH", "id": "2166"} |
| Phytochemical | CID_638011 | Citral | {"name": "Citral", "id": "CID_638011"} | INTERACTS_WITH | - | Target | 26339 | OR5K1 | {"name": "OR5K1", "id": "26339"} |
| Phytochemical | CID_638011 | Citral | {"name": "Citral", "id": "CID_638011"} | INTERACTS_WITH | - | Target | 367 | AR | {"name": "AR", "id": "367"} |
| Phytochemical | CID_638011 | Citral | {"name": "Citral", "id": "CID_638011"} | INTERACTS_WITH | - | Target | 7068 | THRB | {"name": "THRB", "id": "7068"} |
| Phytochemical | CID_638011 | Citral | {"name": "Citral", "id": "CID_638011"} | INTERACTS_WITH | - | Target | 865 | CBFB | {"name": "CBFB", "id": "865"} |
| Phytochemical | CID_6918774 | Corosolic acid | {"name": "Corosolic acid", "id": "CID_6918774"} | INTERACTS_WITH | - | Target | 10013 | HDAC6 | {"name": "HDAC6", "id": "10013"} |
| Phytochemical | CID_6918774 | Corosolic acid | {"name": "Corosolic acid", "id": "CID_6918774"} | INTERACTS_WITH | - | Target | 10951 | CBX1 | {"name": "CBX1", "id": "10951"} |
| Phytochemical | CID_6918774 | Corosolic acid | {"name": "Corosolic acid", "id": "CID_6918774"} | INTERACTS_WITH | - | Target | 11069 | RAPGEF4 | {"name": "RAPGEF4", "id": "11069"} |
| Phytochemical | CID_6918774 | Corosolic acid | {"name": "Corosolic acid", "id": "CID_6918774"} | INTERACTS_WITH | - | Target | 240 | ALOX5 | {"name": "ALOX5", "id": "240"} |
| Phytochemical | CID_6918774 | Corosolic acid | {"name": "Corosolic acid", "id": "CID_6918774"} | INTERACTS_WITH | - | Target | 247 | ALOX15B | {"name": "ALOX15B", "id": "247"} |
| Phytochemical | CID_6918774 | Corosolic acid | {"name": "Corosolic acid", "id": "CID_6918774"} | INTERACTS_WITH | - | Target | 29843 | SENP1 | {"name": "SENP1", "id": "29843"} |
| Phytochemical | CID_6918774 | Corosolic acid | {"name": "Corosolic acid", "id": "CID_6918774"} | INTERACTS_WITH | - | Target | 3064 | HTT | {"name": "HTT", "id": "3064"} |
| Phytochemical | CID_6918774 | Corosolic acid | {"name": "Corosolic acid", "id": "CID_6918774"} | INTERACTS_WITH | - | Target | 3290 | HSD11B1 | {"name": "HSD11B1", "id": "3290"} |
| Phytochemical | CID_6918774 | Corosolic acid | {"name": "Corosolic acid", "id": "CID_6918774"} | INTERACTS_WITH | - | Target | 4088 | SMAD3 | {"name": "SMAD3", "id": "4088"} |
| Phytochemical | CID_6918774 | Corosolic acid | {"name": "Corosolic acid", "id": "CID_6918774"} | INTERACTS_WITH | - | Target | 4780 | NFE2L2 | {"name": "NFE2L2", "id": "4780"} |
| Phytochemical | CID_6918774 | Corosolic acid | {"name": "Corosolic acid", "id": "CID_6918774"} | INTERACTS_WITH | - | Target | 51053 | GMNN | {"name": "GMNN", "id": "51053"} |
| Phytochemical | CID_6918774 | Corosolic acid | {"name": "Corosolic acid", "id": "CID_6918774"} | INTERACTS_WITH | - | Target | 51422 | PRKAG2 | {"name": "PRKAG2", "id": "51422"} |
| Phytochemical | CID_6918774 | Corosolic acid | {"name": "Corosolic acid", "id": "CID_6918774"} | INTERACTS_WITH | - | Target | 53632 | PRKAG3 | {"name": "PRKAG3", "id": "53632"} |
| Phytochemical | CID_6918774 | Corosolic acid | {"name": "Corosolic acid", "id": "CID_6918774"} | INTERACTS_WITH | - | Target | 5423 | POLB | {"name": "POLB", "id": "5423"} |
| Phytochemical | CID_6918774 | Corosolic acid | {"name": "Corosolic acid", "id": "CID_6918774"} | INTERACTS_WITH | - | Target | 5562 | PRKAA1 | {"name": "PRKAA1", "id": "5562"} |
| Phytochemical | CID_6918774 | Corosolic acid | {"name": "Corosolic acid", "id": "CID_6918774"} | INTERACTS_WITH | - | Target | 5563 | PRKAA2 | {"name": "PRKAA2", "id": "5563"} |
| Phytochemical | CID_6918774 | Corosolic acid | {"name": "Corosolic acid", "id": "CID_6918774"} | INTERACTS_WITH | - | Target | 5564 | PRKAB1 | {"name": "PRKAB1", "id": "5564"} |
| Phytochemical | CID_6918774 | Corosolic acid | {"name": "Corosolic acid", "id": "CID_6918774"} | INTERACTS_WITH | - | Target | 5565 | PRKAB2 | {"name": "PRKAB2", "id": "5565"} |
| Phytochemical | CID_6918774 | Corosolic acid | {"name": "Corosolic acid", "id": "CID_6918774"} | INTERACTS_WITH | - | Target | 5571 | PRKAG1 | {"name": "PRKAG1", "id": "5571"} |
| Phytochemical | CID_6918774 | Corosolic acid | {"name": "Corosolic acid", "id": "CID_6918774"} | INTERACTS_WITH | - | Target | 55775 | TDP1 | {"name": "TDP1", "id": "55775"} |
| Phytochemical | CID_6918774 | Corosolic acid | {"name": "Corosolic acid", "id": "CID_6918774"} | INTERACTS_WITH | - | Target | 5743 | PTGS2 | {"name": "PTGS2", "id": "5743"} |
| Phytochemical | CID_6918774 | Corosolic acid | {"name": "Corosolic acid", "id": "CID_6918774"} | INTERACTS_WITH | - | Target | 5770 | PTPN1 | {"name": "PTPN1", "id": "5770"} |
| Phytochemical | CID_6918774 | Corosolic acid | {"name": "Corosolic acid", "id": "CID_6918774"} | INTERACTS_WITH | - | Target | 5771 | PTPN2 | {"name": "PTPN2", "id": "5771"} |
| Phytochemical | CID_6918774 | Corosolic acid | {"name": "Corosolic acid", "id": "CID_6918774"} | INTERACTS_WITH | - | Target | 5777 | PTPN6 | {"name": "PTPN6", "id": "5777"} |
| Phytochemical | CID_6918774 | Corosolic acid | {"name": "Corosolic acid", "id": "CID_6918774"} | INTERACTS_WITH | - | Target | 5786 | PTPRA | {"name": "PTPRA", "id": "5786"} |
| Phytochemical | CID_6918774 | Corosolic acid | {"name": "Corosolic acid", "id": "CID_6918774"} | INTERACTS_WITH | - | Target | 5791 | PTPRE | {"name": "PTPRE", "id": "5791"} |
| Phytochemical | CID_6918774 | Corosolic acid | {"name": "Corosolic acid", "id": "CID_6918774"} | INTERACTS_WITH | - | Target | 5792 | PTPRF | {"name": "PTPRF", "id": "5792"} |
| Phytochemical | CID_6918774 | Corosolic acid | {"name": "Corosolic acid", "id": "CID_6918774"} | INTERACTS_WITH | - | Target | 5837 | PYGM | {"name": "PYGM", "id": "5837"} |
| Phytochemical | CID_6918774 | Corosolic acid | {"name": "Corosolic acid", "id": "CID_6918774"} | INTERACTS_WITH | - | Target | 6311 | ATXN2 | {"name": "ATXN2", "id": "6311"} |
| Phytochemical | CID_6918774 | Corosolic acid | {"name": "Corosolic acid", "id": "CID_6918774"} | INTERACTS_WITH | - | Target | 7874 | USP7 | {"name": "USP7", "id": "7874"} |
| Phytochemical | CID_6918774 | Corosolic acid | {"name": "Corosolic acid", "id": "CID_6918774"} | INTERACTS_WITH | - | Target | 79915 | ATAD5 | {"name": "ATAD5", "id": "79915"} |
| Phytochemical | CID_969516 | Curcumin | {"name": "Curcumin", "id": "CID_969516"} | INTERACTS_WITH | - | Target | 10013 | HDAC6 | {"name": "HDAC6", "id": "10013"} |

**Table S9: Complete node and relationship data used to construct the knowledge graph. This table summarizes interactions among FDA-approved drugs, formulation, herbs, diseases (type 2 diabetes, fatty liver, obesity), phytochemicals, and their target genes/proteins. Each entry includes source and target node details, edge type, and node annotations (Entrez Gene ID and gene symbol, where applicable).**

| Source Node Label | Source Node ID | Source Node Name | Source Node Properties | Edge Type | Edge Properties | Target Node Label | Target Node ID | Target Node Name | Target Node Properties |
|---|---|---|---|---|---|---|---|---|---|
| Phytochemical | CID_969516 | Curcumin | {"name": "Curcumin", "id": "CID_969516"} | INTERACTS_WITH | - | Target | 10014 | HDAC5 | {"name": "HDAC5", "id": "10014"} |
| Phytochemical | CID_969516 | Curcumin | {"name": "Curcumin", "id": "CID_969516"} | INTERACTS_WITH | - | Target | 10073 | SNUPN | {"name": "SNUPN", "id": "10073"} |
| Phytochemical | CID_969516 | Curcumin | {"name": "Curcumin", "id": "CID_969516"} | INTERACTS_WITH | - | Target | 1020 | CDK5 | {"name": "CDK5", "id": "1020"} |
| Phytochemical | CID_969516 | Curcumin | {"name": "Curcumin", "id": "CID_969516"} | INTERACTS_WITH | - | Target | 10376 | TUBA1B | {"name": "TUBA1B", "id": "10376"} |
| Phytochemical | CID_969516 | Curcumin | {"name": "Curcumin", "id": "CID_969516"} | INTERACTS_WITH | - | Target | 10381 | TUBB3 | {"name": "TUBB3", "id": "10381"} |
| Phytochemical | CID_969516 | Curcumin | {"name": "Curcumin", "id": "CID_969516"} | INTERACTS_WITH | - | Target | 10382 | TUBB4A | {"name": "TUBB4A", "id": "10382"} |
| Phytochemical | CID_969516 | Curcumin | {"name": "Curcumin", "id": "CID_969516"} | INTERACTS_WITH | - | Target | 10383 | TUBB4B | {"name": "TUBB4B", "id": "10383"} |
| Phytochemical | CID_969516 | Curcumin | {"name": "Curcumin", "id": "CID_969516"} | INTERACTS_WITH | - | Target | 10524 | KAT5 | {"name": "KAT5", "id": "10524"} |
| Phytochemical | CID_969516 | Curcumin | {"name": "Curcumin", "id": "CID_969516"} | INTERACTS_WITH | - | Target | 10587 | TXNRD2 | {"name": "TXNRD2", "id": "10587"} |
| Phytochemical | CID_969516 | Curcumin | {"name": "Curcumin", "id": "CID_969516"} | INTERACTS_WITH | - | Target | 10919 | EHMT2 | {"name": "EHMT2", "id": "10919"} |
| Phytochemical | CID_969516 | Curcumin | {"name": "Curcumin", "id": "CID_969516"} | INTERACTS_WITH | - | Target | 11201 | POLI | {"name": "POLI", "id": "11201"} |
| Phytochemical | CID_969516 | Curcumin | {"name": "Curcumin", "id": "CID_969516"} | INTERACTS_WITH | - | Target | 11238 | CA5B | {"name": "CA5B", "id": "11238"} |
| Phytochemical | CID_969516 | Curcumin | {"name": "Curcumin", "id": "CID_969516"} | INTERACTS_WITH | - | Target | 112714 | TUBA3E | {"name": "TUBA3E", "id": "112714"} |
| Phytochemical | CID_969516 | Curcumin | {"name": "Curcumin", "id": "CID_969516"} | INTERACTS_WITH | - | Target | 114112 | TXNRD3 | {"name": "TXNRD3", "id": "114112"} |
| Phytochemical | CID_969516 | Curcumin | {"name": "Curcumin", "id": "CID_969516"} | INTERACTS_WITH | - | Target | 114548 | NLRP3 | {"name": "NLRP3", "id": "114548"} |
| Phytochemical | CID_969516 | Curcumin | {"name": "Curcumin", "id": "CID_969516"} | INTERACTS_WITH | - | Target | 1195 | CLK1 | {"name": "CLK1", "id": "1195"} |
| Phytochemical | CID_969516 | Curcumin | {"name": "Curcumin", "id": "CID_969516"} | INTERACTS_WITH | - | Target | 1196 | CLK2 | {"name": "CLK2", "id": "1196"} |
| Phytochemical | CID_969516 | Curcumin | {"name": "Curcumin", "id": "CID_969516"} | INTERACTS_WITH | - | Target | 1198 | CLK3 | {"name": "CLK3", "id": "1198"} |
| Phytochemical | CID_969516 | Curcumin | {"name": "Curcumin", "id": "CID_969516"} | INTERACTS_WITH | - | Target | 122706 | PSMB11 | {"name": "PSMB11", "id": "122706"} |
| Phytochemical | CID_969516 | Curcumin | {"name": "Curcumin", "id": "CID_969516"} | INTERACTS_WITH | - | Target | 1387 | CREBBP | {"name": "CREBBP", "id": "1387"} |
| Phytochemical | CID_969516 | Curcumin | {"name": "Curcumin", "id": "CID_969516"} | INTERACTS_WITH | - | Target | 143471 | PSMA8 | {"name": "PSMA8", "id": "143471"} |
| Phytochemical | CID_969516 | Curcumin | {"name": "Curcumin", "id": "CID_969516"} | INTERACTS_WITH | - | Target | 1454 | CSNK1E | {"name": "CSNK1E", "id": "1454"} |
| Phytochemical | CID_969516 | Curcumin | {"name": "Curcumin", "id": "CID_969516"} | INTERACTS_WITH | - | Target | 1544 | CYP1A2 | {"name": "CYP1A2", "id": "1544"} |
| Phytochemical | CID_969516 | Curcumin | {"name": "Curcumin", "id": "CID_969516"} | INTERACTS_WITH | - | Target | 1559 | CYP2C9 | {"name": "CYP2C9", "id": "1559"} |
| Phytochemical | CID_969516 | Curcumin | {"name": "Curcumin", "id": "CID_969516"} | INTERACTS_WITH | - | Target | 1576 | CYP3A4 | {"name": "CYP3A4", "id": "1576"} |
| Phytochemical | CID_969516 | Curcumin | {"name": "Curcumin", "id": "CID_969516"} | INTERACTS_WITH | - | Target | 1859 | DYRK1A | {"name": "DYRK1A", "id": "1859"} |
| Phytochemical | CID_969516 | Curcumin | {"name": "Curcumin", "id": "CID_969516"} | INTERACTS_WITH | - | Target | 1956 | EGFR | {"name": "EGFR", "id": "1956"} |
| Phytochemical | CID_969516 | Curcumin | {"name": "Curcumin", "id": "CID_969516"} | INTERACTS_WITH | - | Target | 196 | AHR | {"name": "AHR", "id": "196"} |
| Phytochemical | CID_969516 | Curcumin | {"name": "Curcumin", "id": "CID_969516"} | INTERACTS_WITH | - | Target | 200316 | APOBEC3F | {"name": "APOBEC3F", "id": "200316"} |
| Phytochemical | CID_969516 | Curcumin | {"name": "Curcumin", "id": "CID_969516"} | INTERACTS_WITH | - | Target | 203068 | TUBB | {"name": "TUBB", "id": "203068"} |
| Phytochemical | CID_969516 | Curcumin | {"name": "Curcumin", "id": "CID_969516"} | INTERACTS_WITH | - | Target | 2033 | EP300 | {"name": "EP300", "id": "2033"} |
| Phytochemical | CID_969516 | Curcumin | {"name": "Curcumin", "id": "CID_969516"} | INTERACTS_WITH | - | Target | 2078 | ERG | {"name": "ERG", "id": "2078"} |
| Phytochemical | CID_969516 | Curcumin | {"name": "Curcumin", "id": "CID_969516"} | INTERACTS_WITH | - | Target | 2099 | ESR1 | {"name": "ESR1", "id": "2099"} |
| Phytochemical | CID_969516 | Curcumin | {"name": "Curcumin", "id": "CID_969516"} | INTERACTS_WITH | - | Target | 2152 | F3 | {"name": "F3", "id": "2152"} |
| Phytochemical | CID_969516 | Curcumin | {"name": "Curcumin", "id": "CID_969516"} | INTERACTS_WITH | - | Target | 216 | ALDH1A1 | {"name": "ALDH1A1", "id": "216"} |
| Phytochemical | CID_969516 | Curcumin | {"name": "Curcumin", "id": "CID_969516"} | INTERACTS_WITH | - | Target | 2237 | FEN1 | {"name": "FEN1", "id": "2237"} |
| Phytochemical | CID_969516 | Curcumin | {"name": "Curcumin", "id": "CID_969516"} | INTERACTS_WITH | - | Target | 23028 | KDM1A | {"name": "KDM1A", "id": "23028"} |
| Phytochemical | CID_969516 | Curcumin | {"name": "Curcumin", "id": "CID_969516"} | INTERACTS_WITH | - | Target | 23435 | TARDBP | {"name": "TARDBP", "id": "23435"} |
| Phytochemical | CID_969516 | Curcumin | {"name": "Curcumin", "id": "CID_969516"} | INTERACTS_WITH | - | Target | 23621 | BACE1 | {"name": "BACE1", "id": "23621"} |
| Phytochemical | CID_969516 | Curcumin | {"name": "Curcumin", "id": "CID_969516"} | INTERACTS_WITH | - | Target | 23632 | CA14 | {"name": "CA14", "id": "23632"} |
| Phytochemical | CID_969516 | Curcumin | {"name": "Curcumin", "id": "CID_969516"} | INTERACTS_WITH | - | Target | 23643 | LY96 | {"name": "LY96", "id": "23643"} |
| Phytochemical | CID_969516 | Curcumin | {"name": "Curcumin", "id": "CID_969516"} | INTERACTS_WITH | - | Target | 239 | ALOX12 | {"name": "ALOX12", "id": "239"} |
| Phytochemical | CID_969516 | Curcumin | {"name": "Curcumin", "id": "CID_969516"} | INTERACTS_WITH | - | Target | 240 | ALOX5 | {"name": "ALOX5", "id": "240"} |
| Phytochemical | CID_969516 | Curcumin | {"name": "Curcumin", "id": "CID_969516"} | INTERACTS_WITH | - | Target | 246 | ALOX15 | {"name": "ALOX15", "id": "246"} |
| Phytochemical | CID_969516 | Curcumin | {"name": "Curcumin", "id": "CID_969516"} | INTERACTS_WITH | - | Target | 247 | ALOX15B | {"name": "ALOX15B", "id": "247"} |
| Phytochemical | CID_969516 | Curcumin | {"name": "Curcumin", "id": "CID_969516"} | INTERACTS_WITH | - | Target | 248 | ALPI | {"name": "ALPI", "id": "248"} |
| Phytochemical | CID_969516 | Curcumin | {"name": "Curcumin", "id": "CID_969516"} | INTERACTS_WITH | - | Target | 249 | ALPL | {"name": "ALPL", "id": "249"} |
| Phytochemical | CID_969516 | Curcumin | {"name": "Curcumin", "id": "CID_969516"} | INTERACTS_WITH | - | Target | 251 | ALPG | {"name": "ALPG", "id": "251"} |
| Phytochemical | CID_969516 | Curcumin | {"name": "Curcumin", "id": "CID_969516"} | INTERACTS_WITH | - | Target | 2548 | GAA | {"name": "GAA", "id": "2548"} |

**Table S9: Complete node and relationship data used to construct the knowledge graph. This table summarizes interactions among FDA-approved drugs, formulation, herbs, diseases (type 2 diabetes, fatty liver, obesity), phytochemicals, and their target genes/proteins. Each entry includes source and target node details, edge type, and node annotations (Entrez Gene ID and gene symbol, where applicable).**

| Source Node Label | Source Node ID | Source Node Name | Source Node Properties | Edge Type | Edge Properties | Target Node Label | Target Node ID | Target Node Name | Target Node Properties |
|---|---|---|---|---|---|---|---|---|---|
| Phytochemical | CID_969516 | Curcumin | {"name": "Curcumin", "id": "CID_969516"} | INTERACTS_WITH | - | Target | 2629 | GBA1 | {"name": "GBA1", "id": "2629"} |
| Phytochemical | CID_969516 | Curcumin | {"name": "Curcumin", "id": "CID_969516"} | INTERACTS_WITH | - | Target | 2739 | GLO1 | {"name": "GLO1", "id": "2739"} |
| Phytochemical | CID_969516 | Curcumin | {"name": "Curcumin", "id": "CID_969516"} | INTERACTS_WITH | - | Target | 2744 | GLS | {"name": "GLS", "id": "2744"} |
| Phytochemical | CID_969516 | Curcumin | {"name": "Curcumin", "id": "CID_969516"} | INTERACTS_WITH | - | Target | 2778 | GNAS | {"name": "GNAS", "id": "2778"} |
| Phytochemical | CID_969516 | Curcumin | {"name": "Curcumin", "id": "CID_969516"} | INTERACTS_WITH | - | Target | 290 | ANPEP | {"name": "ANPEP", "id": "290"} |
| Phytochemical | CID_969516 | Curcumin | {"name": "Curcumin", "id": "CID_969516"} | INTERACTS_WITH | - | Target | 2908 | NR3C1 | {"name": "NR3C1", "id": "2908"} |
| Phytochemical | CID_969516 | Curcumin | {"name": "Curcumin", "id": "CID_969516"} | INTERACTS_WITH | - | Target | 2932 | GSK3B | {"name": "GSK3B", "id": "2932"} |
| Phytochemical | CID_969516 | Curcumin | {"name": "Curcumin", "id": "CID_969516"} | INTERACTS_WITH | - | Target | 2946 | GSTM2 | {"name": "GSTM2", "id": "2946"} |
| Phytochemical | CID_969516 | Curcumin | {"name": "Curcumin", "id": "CID_969516"} | INTERACTS_WITH | - | Target | 29994 | BAZ2B | {"name": "BAZ2B", "id": "29994"} |
| Phytochemical | CID_969516 | Curcumin | {"name": "Curcumin", "id": "CID_969516"} | INTERACTS_WITH | - | Target | 3028 | HSD17B10 | {"name": "HSD17B10", "id": "3028"} |
| Phytochemical | CID_969516 | Curcumin | {"name": "Curcumin", "id": "CID_969516"} | INTERACTS_WITH | - | Target | 3043 | HBB | {"name": "HBB", "id": "3043"} |
| Phytochemical | CID_969516 | Curcumin | {"name": "Curcumin", "id": "CID_969516"} | INTERACTS_WITH | - | Target | 3064 | HTT | {"name": "HTT", "id": "3064"} |
| Phytochemical | CID_969516 | Curcumin | {"name": "Curcumin", "id": "CID_969516"} | INTERACTS_WITH | - | Target | 3065 | HDAC1 | {"name": "HDAC1", "id": "3065"} |
| Phytochemical | CID_969516 | Curcumin | {"name": "Curcumin", "id": "CID_969516"} | INTERACTS_WITH | - | Target | 3066 | HDAC2 | {"name": "HDAC2", "id": "3066"} |
| Phytochemical | CID_969516 | Curcumin | {"name": "Curcumin", "id": "CID_969516"} | INTERACTS_WITH | - | Target | 3091 | HIF1A | {"name": "HIF1A", "id": "3091"} |
| Phytochemical | CID_969516 | Curcumin | {"name": "Curcumin", "id": "CID_969516"} | INTERACTS_WITH | - | Target | 3248 | HPGD | {"name": "HPGD", "id": "3248"} |
| Phytochemical | CID_969516 | Curcumin | {"name": "Curcumin", "id": "CID_969516"} | INTERACTS_WITH | - | Target | 3290 | HSD11B1 | {"name": "HSD11B1", "id": "3290"} |
| Phytochemical | CID_969516 | Curcumin | {"name": "Curcumin", "id": "CID_969516"} | INTERACTS_WITH | - | Target | 3291 | HSD11B2 | {"name": "HSD11B2", "id": "3291"} |
| Phytochemical | CID_969516 | Curcumin | {"name": "Curcumin", "id": "CID_969516"} | INTERACTS_WITH | - | Target | 3292 | HSD17B1 | {"name": "HSD17B1", "id": "3292"} |
| Phytochemical | CID_969516 | Curcumin | {"name": "Curcumin", "id": "CID_969516"} | INTERACTS_WITH | - | Target | 3293 | HSD17B3 | {"name": "HSD17B3", "id": "3293"} |
| Phytochemical | CID_969516 | Curcumin | {"name": "Curcumin", "id": "CID_969516"} | INTERACTS_WITH | - | Target | 3294 | HSD17B2 | {"name": "HSD17B2", "id": "3294"} |
| Phytochemical | CID_969516 | Curcumin | {"name": "Curcumin", "id": "CID_969516"} | INTERACTS_WITH | - | Target | 3309 | HSPA5 | {"name": "HSPA5", "id": "3309"} |
| Phytochemical | CID_969516 | Curcumin | {"name": "Curcumin", "id": "CID_969516"} | INTERACTS_WITH | - | Target | 3315 | HSPB1 | {"name": "HSPB1", "id": "3315"} |
| Phytochemical | CID_969516 | Curcumin | {"name": "Curcumin", "id": "CID_969516"} | INTERACTS_WITH | - | Target | 3329 | HSPD1 | {"name": "HSPD1", "id": "3329"} |
| Phytochemical | CID_969516 | Curcumin | {"name": "Curcumin", "id": "CID_969516"} | INTERACTS_WITH | - | Target | 3336 | HSPE1 | {"name": "HSPE1", "id": "3336"} |
| Phytochemical | CID_969516 | Curcumin | {"name": "Curcumin", "id": "CID_969516"} | INTERACTS_WITH | - | Target | 3417 | IDH1 | {"name": "IDH1", "id": "3417"} |
| Phytochemical | CID_969516 | Curcumin | {"name": "Curcumin", "id": "CID_969516"} | INTERACTS_WITH | - | Target | 347688 | TUBB8 | {"name": "TUBB8", "id": "347688"} |
| Phytochemical | CID_969516 | Curcumin | {"name": "Curcumin", "id": "CID_969516"} | INTERACTS_WITH | - | Target | 347733 | TUBB2B | {"name": "TUBB2B", "id": "347733"} |
| Phytochemical | CID_969516 | Curcumin | {"name": "Curcumin", "id": "CID_969516"} | INTERACTS_WITH | - | Target | 351 | APP | {"name": "APP", "id": "351"} |
| Phytochemical | CID_969516 | Curcumin | {"name": "Curcumin", "id": "CID_969516"} | INTERACTS_WITH | - | Target | 3576 | CXCL8 | {"name": "CXCL8", "id": "3576"} |
| Phytochemical | CID_969516 | Curcumin | {"name": "Curcumin", "id": "CID_969516"} | INTERACTS_WITH | - | Target | 367 | AR | {"name": "AR", "id": "367"} |
| Phytochemical | CID_969516 | Curcumin | {"name": "Curcumin", "id": "CID_969516"} | INTERACTS_WITH | - | Target | 3725 | JUN | {"name": "JUN", "id": "3725"} |
| Phytochemical | CID_969516 | Curcumin | {"name": "Curcumin", "id": "CID_969516"} | INTERACTS_WITH | - | Target | 3757 | KCNH2 | {"name": "KCNH2", "id": "3757"} |
| Phytochemical | CID_969516 | Curcumin | {"name": "Curcumin", "id": "CID_969516"} | INTERACTS_WITH | - | Target | 3837 | KPNB1 | {"name": "KPNB1", "id": "3837"} |
| Phytochemical | CID_969516 | Curcumin | {"name": "Curcumin", "id": "CID_969516"} | INTERACTS_WITH | - | Target | 387129 | NPSR1 | {"name": "NPSR1", "id": "387129"} |
| Phytochemical | CID_969516 | Curcumin | {"name": "Curcumin", "id": "CID_969516"} | INTERACTS_WITH | - | Target | 390245 | KDM4E | {"name": "KDM4E", "id": "390245"} |
| Phytochemical | CID_969516 | Curcumin | {"name": "Curcumin", "id": "CID_969516"} | INTERACTS_WITH | - | Target | 4000 | LMNA | {"name": "LMNA", "id": "4000"} |
| Phytochemical | CID_969516 | Curcumin | {"name": "Curcumin", "id": "CID_969516"} | INTERACTS_WITH | - | Target | 4088 | SMAD3 | {"name": "SMAD3", "id": "4088"} |
| Phytochemical | CID_969516 | Curcumin | {"name": "Curcumin", "id": "CID_969516"} | INTERACTS_WITH | - | Target | 4128 | MAOA | {"name": "MAOA", "id": "4128"} |
| Phytochemical | CID_969516 | Curcumin | {"name": "Curcumin", "id": "CID_969516"} | INTERACTS_WITH | - | Target | 4129 | MAOB | {"name": "MAOB", "id": "4129"} |
| Phytochemical | CID_969516 | Curcumin | {"name": "Curcumin", "id": "CID_969516"} | INTERACTS_WITH | - | Target | 4137 | MAPT | {"name": "MAPT", "id": "4137"} |
| Phytochemical | CID_969516 | Curcumin | {"name": "Curcumin", "id": "CID_969516"} | INTERACTS_WITH | - | Target | 4154 | MBNL1 | {"name": "MBNL1", "id": "4154"} |
| Phytochemical | CID_969516 | Curcumin | {"name": "Curcumin", "id": "CID_969516"} | INTERACTS_WITH | - | Target | 4297 | KMT2A | {"name": "KMT2A", "id": "4297"} |
| Phytochemical | CID_969516 | Curcumin | {"name": "Curcumin", "id": "CID_969516"} | INTERACTS_WITH | - | Target | 43 | ACHE | {"name": "ACHE", "id": "43"} |
| Phytochemical | CID_969516 | Curcumin | {"name": "Curcumin", "id": "CID_969516"} | INTERACTS_WITH | - | Target | 4318 | MMP9 | {"name": "MMP9", "id": "4318"} |
| Phytochemical | CID_969516 | Curcumin | {"name": "Curcumin", "id": "CID_969516"} | INTERACTS_WITH | - | Target | 4322 | MMP13 | {"name": "MMP13", "id": "4322"} |
| Phytochemical | CID_969516 | Curcumin | {"name": "Curcumin", "id": "CID_969516"} | INTERACTS_WITH | - | Target | 440503 | PLIN5 | {"name": "PLIN5", "id": "440503"} |
| Phytochemical | CID_969516 | Curcumin | {"name": "Curcumin", "id": "CID_969516"} | INTERACTS_WITH | - | Target | 4780 | NFE2L2 | {"name": "NFE2L2", "id": "4780"} |
| Phytochemical | CID_969516 | Curcumin | {"name": "Curcumin", "id": "CID_969516"} | INTERACTS_WITH | - | Target | 4790 | NFKB1 | {"name": "NFKB1", "id": "4790"} |

**Table S9: Complete node and relationship data used to construct the knowledge graph. This table summarizes interactions among FDA-approved drugs, formulation, herbs, diseases (type 2 diabetes, fatty liver, obesity), phytochemicals, and their target genes/proteins. Each entry includes source and target node details, edge type, and node annotations (Entrez Gene ID and gene symbol, where applicable).**

| Source Node Label | Source Node ID | Source Node Name | Source Node Properties | Edge Type | Edge Properties | Target Node Label | Target Node ID | Target Node Name | Target Node Properties |
|---|---|---|---|---|---|---|---|---|---|
| Phytochemical | CID_969516 | Curcumin | {"name": "Curcumin", "id": " CID_969516"} | INTERACTS_WITH | - | Target | 4843 | NOS2 | {"name": "NOS2", "id": "4843"} |
| Phytochemical | CID_969516 | Curcumin | {"name": "Curcumin", "id": " CID_969516"} | INTERACTS_WITH | - | Target | 488 | ATP2A2 | {"name": "ATP2A2", "id": "488"} |
| Phytochemical | CID_969516 | Curcumin | {"name": "Curcumin", "id": " CID_969516"} | INTERACTS_WITH | - | Target | 489 | ATP2A3 | {"name": "ATP2A3", "id": "489"} |
| Phytochemical | CID_969516 | Curcumin | {"name": "Curcumin", "id": " CID_969516"} | INTERACTS_WITH | - | Target | 5058 | PAK1 | {"name": "PAK1", "id": "5058"} |
| Phytochemical | CID_969516 | Curcumin | {"name": "Curcumin", "id": " CID_969516"} | INTERACTS_WITH | - | Target | 51053 | GMNN | {"name": "GMNN", "id": "51053"} |
| Phytochemical | CID_969516 | Curcumin | {"name": "Curcumin", "id": " CID_969516"} | INTERACTS_WITH | - | Target | 51099 | ABHD5 | {"name": "ABHD5", "id": "51099"} |
| Phytochemical | CID_969516 | Curcumin | {"name": "Curcumin", "id": " CID_969516"} | INTERACTS_WITH | - | Target | 51426 | POLK | {"name": "POLK", "id": "51426"} |
| Phytochemical | CID_969516 | Curcumin | {"name": "Curcumin", "id": " CID_969516"} | INTERACTS_WITH | - | Target | 51564 | HDAC7 | {"name": "HDAC7", "id": "51564"} |
| Phytochemical | CID_969516 | Curcumin | {"name": "Curcumin", "id": " CID_969516"} | INTERACTS_WITH | - | Target | 5300 | PIN1 | {"name": "PIN1", "id": "5300"} |
| Phytochemical | CID_969516 | Curcumin | {"name": "Curcumin", "id": " CID_969516"} | INTERACTS_WITH | - | Target | 5315 | PKM | {"name": "PKM", "id": "5315"} |
| Phytochemical | CID_969516 | Curcumin | {"name": "Curcumin", "id": " CID_969516"} | INTERACTS_WITH | - | Target | 5319 | PLA2G1B | {"name": "PLA2G1B", "id": "5319"} |
| Phytochemical | CID_969516 | Curcumin | {"name": "Curcumin", "id": " CID_969516"} | INTERACTS_WITH | - | Target | 5346 | PLIN1 | {"name": "PLIN1", "id": "5346"} |
| Phytochemical | CID_969516 | Curcumin | {"name": "Curcumin", "id": " CID_969516"} | INTERACTS_WITH | - | Target | 54106 | TLR9 | {"name": "TLR9", "id": "54106"} |
| Phytochemical | CID_969516 | Curcumin | {"name": "Curcumin", "id": " CID_969516"} | INTERACTS_WITH | - | Target | 5423 | POLB | {"name": "POLB", "id": "5423"} |
| Phytochemical | CID_969516 | Curcumin | {"name": "Curcumin", "id": " CID_969516"} | INTERACTS_WITH | - | Target | 5467 | PPARD | {"name": "PPARD", "id": "5467"} |
| Phytochemical | CID_969516 | Curcumin | {"name": "Curcumin", "id": " CID_969516"} | INTERACTS_WITH | - | Target | 5468 | PPARG | {"name": "PPARG", "id": "5468"} |
| Phytochemical | CID_969516 | Curcumin | {"name": "Curcumin", "id": " CID_969516"} | INTERACTS_WITH | - | Target | 55775 | TDP1 | {"name": "TDP1", "id": "55775"} |
| Phytochemical | CID_969516 | Curcumin | {"name": "Curcumin", "id": " CID_969516"} | INTERACTS_WITH | - | Target | 5580 | PRKCD | {"name": "PRKCD", "id": "5580"} |
| Phytochemical | CID_969516 | Curcumin | {"name": "Curcumin", "id": " CID_969516"} | INTERACTS_WITH | - | Target | 5581 | PRKCE | {"name": "PRKCE", "id": "5581"} |
| Phytochemical | CID_969516 | Curcumin | {"name": "Curcumin", "id": " CID_969516"} | INTERACTS_WITH | - | Target | 55847 | CISD1 | {"name": "CISD1", "id": "55847"} |
| Phytochemical | CID_969516 | Curcumin | {"name": "Curcumin", "id": " CID_969516"} | INTERACTS_WITH | - | Target | 55869 | HDAC8 | {"name": "HDAC8", "id": "55869"} |
| Phytochemical | CID_969516 | Curcumin | {"name": "Curcumin", "id": " CID_969516"} | INTERACTS_WITH | - | Target | 5588 | PRKCQ | {"name": "PRKCQ", "id": "5588"} |
| Phytochemical | CID_969516 | Curcumin | {"name": "Curcumin", "id": " CID_969516"} | INTERACTS_WITH | - | Target | 5594 | MAPK1 | {"name": "MAPK1", "id": "5594"} |
| Phytochemical | CID_969516 | Curcumin | {"name": "Curcumin", "id": " CID_969516"} | INTERACTS_WITH | - | Target | 5599 | MAPK8 | {"name": "MAPK8", "id": "5599"} |
| Phytochemical | CID_969516 | Curcumin | {"name": "Curcumin", "id": " CID_969516"} | INTERACTS_WITH | - | Target | 5682 | PSMA1 | {"name": "PSMA1", "id": "5682"} |
| Phytochemical | CID_969516 | Curcumin | {"name": "Curcumin", "id": " CID_969516"} | INTERACTS_WITH | - | Target | 5683 | PSMA2 | {"name": "PSMA2", "id": "5683"} |
| Phytochemical | CID_969516 | Curcumin | {"name": "Curcumin", "id": " CID_969516"} | INTERACTS_WITH | - | Target | 5684 | PSMA3 | {"name": "PSMA3", "id": "5684"} |
| Phytochemical | CID_969516 | Curcumin | {"name": "Curcumin", "id": " CID_969516"} | INTERACTS_WITH | - | Target | 5685 | PSMA4 | {"name": "PSMA4", "id": "5685"} |
| Phytochemical | CID_969516 | Curcumin | {"name": "Curcumin", "id": " CID_969516"} | INTERACTS_WITH | - | Target | 5686 | PSMA5 | {"name": "PSMA5", "id": "5686"} |
| Phytochemical | CID_969516 | Curcumin | {"name": "Curcumin", "id": " CID_969516"} | INTERACTS_WITH | - | Target | 5687 | PSMA6 | {"name": "PSMA6", "id": "5687"} |
| Phytochemical | CID_969516 | Curcumin | {"name": "Curcumin", "id": " CID_969516"} | INTERACTS_WITH | - | Target | 5688 | PSMA7 | {"name": "PSMA7", "id": "5688"} |
| Phytochemical | CID_969516 | Curcumin | {"name": "Curcumin", "id": " CID_969516"} | INTERACTS_WITH | - | Target | 5689 | PSMB1 | {"name": "PSMB1", "id": "5689"} |
| Phytochemical | CID_969516 | Curcumin | {"name": "Curcumin", "id": " CID_969516"} | INTERACTS_WITH | - | Target | 5690 | PSMB2 | {"name": "PSMB2", "id": "5690"} |
| Phytochemical | CID_969516 | Curcumin | {"name": "Curcumin", "id": " CID_969516"} | INTERACTS_WITH | - | Target | 5691 | PSMB3 | {"name": "PSMB3", "id": "5691"} |
| Phytochemical | CID_969516 | Curcumin | {"name": "Curcumin", "id": " CID_969516"} | INTERACTS_WITH | - | Target | 5692 | PSMB4 | {"name": "PSMB4", "id": "5692"} |
| Phytochemical | CID_969516 | Curcumin | {"name": "Curcumin", "id": " CID_969516"} | INTERACTS_WITH | - | Target | 5693 | PSMB5 | {"name": "PSMB5", "id": "5693"} |
| Phytochemical | CID_969516 | Curcumin | {"name": "Curcumin", "id": " CID_969516"} | INTERACTS_WITH | - | Target | 5694 | PSMB6 | {"name": "PSMB6", "id": "5694"} |
| Phytochemical | CID_969516 | Curcumin | {"name": "Curcumin", "id": " CID_969516"} | INTERACTS_WITH | - | Target | 5695 | PSMB7 | {"name": "PSMB7", "id": "5695"} |
| Phytochemical | CID_969516 | Curcumin | {"name": "Curcumin", "id": " CID_969516"} | INTERACTS_WITH | - | Target | 5696 | PSMB8 | {"name": "PSMB8", "id": "5696"} |
| Phytochemical | CID_969516 | Curcumin | {"name": "Curcumin", "id": " CID_969516"} | INTERACTS_WITH | - | Target | 5698 | PSMB9 | {"name": "PSMB9", "id": "5698"} |
| Phytochemical | CID_969516 | Curcumin | {"name": "Curcumin", "id": " CID_969516"} | INTERACTS_WITH | - | Target | 5699 | PSMB10 | {"name": "PSMB10", "id": "5699"} |
| Phytochemical | CID_969516 | Curcumin | {"name": "Curcumin", "id": " CID_969516"} | INTERACTS_WITH | - | Target | 57396 | CLK4 | {"name": "CLK4", "id": "57396"} |
| Phytochemical | CID_969516 | Curcumin | {"name": "Curcumin", "id": " CID_969516"} | INTERACTS_WITH | - | Target | 5743 | PTGS2 | {"name": "PTGS2", "id": "5743"} |
| Phytochemical | CID_969516 | Curcumin | {"name": "Curcumin", "id": " CID_969516"} | INTERACTS_WITH | - | Target | 590 | BCHE | {"name": "BCHE", "id": "590"} |
| Phytochemical | CID_969516 | Curcumin | {"name": "Curcumin", "id": " CID_969516"} | INTERACTS_WITH | - | Target | 5901 | RAN | {"name": "RAN", "id": "5901"} |
| Phytochemical | CID_969516 | Curcumin | {"name": "Curcumin", "id": " CID_969516"} | INTERACTS_WITH | - | Target | 5970 | RELA | {"name": "RELA", "id": "5970"} |
| Phytochemical | CID_969516 | Curcumin | {"name": "Curcumin", "id": " CID_969516"} | INTERACTS_WITH | - | Target | 60489 | APOBEC3G | {"name": "APOBEC3G", "id": "60489"} |
| Phytochemical | CID_969516 | Curcumin | {"name": "Curcumin", "id": " CID_969516"} | INTERACTS_WITH | - | Target | 6097 | RORC | {"name": "RORC", "id": "6097"} |
| Phytochemical | CID_969516 | Curcumin | {"name": "Curcumin", "id": " CID_969516"} | INTERACTS_WITH | - | Target | 6256 | RXRA | {"name": "RXRA", "id": "6256"} |

**Table S9: Complete node and relationship data used to construct the knowledge graph. This table summarizes interactions among FDA-approved drugs, formulation, herbs, diseases (type 2 diabetes, fatty liver, obesity), phytochemicals, and their target genes/proteins. Each entry includes source and target node details, edge type, and node annotations (Entrez Gene ID and gene symbol, where applicable).**

| Source Node Label | Source Node ID | Source Node Name | Source Node Properties | Edge Type | Edge Properties | Target Node Label | Target Node ID | Target Node Name | Target Node Properties |
|---|---|---|---|---|---|---|---|---|---|
| Phytochemical | CID_969516 | Curcumin | {"name": "Curcumin", "id": " CID_969516"} | INTERACTS_WITH | - | Target | 6554 | SLC10A1 | {"name": "SLC10A1", "id": "6554"} |
| Phytochemical | CID_969516 | Curcumin | {"name": "Curcumin", "id": " CID_969516"} | INTERACTS_WITH | - | Target | 65975 | STK33 | {"name": "STK33", "id": "65975"} |
| Phytochemical | CID_969516 | Curcumin | {"name": "Curcumin", "id": " CID_969516"} | INTERACTS_WITH | - | Target | 6622 | SNCA | {"name": "SNCA", "id": "6622"} |
| Phytochemical | CID_969516 | Curcumin | {"name": "Curcumin", "id": " CID_969516"} | INTERACTS_WITH | - | Target | 672 | BRCA1 | {"name": "BRCA1", "id": "672"} |
| Phytochemical | CID_969516 | Curcumin | {"name": "Curcumin", "id": " CID_969516"} | INTERACTS_WITH | - | Target | 7068 | THRB | {"name": "THRB", "id": "7068"} |
| Phytochemical | CID_969516 | Curcumin | {"name": "Curcumin", "id": " CID_969516"} | INTERACTS_WITH | - | Target | 7157 | TP53 | {"name": "TP53", "id": "7157"} |
| Phytochemical | CID_969516 | Curcumin | {"name": "Curcumin", "id": " CID_969516"} | INTERACTS_WITH | - | Target | 7251 | TSG101 | {"name": "TSG101", "id": "7251"} |
| Phytochemical | CID_969516 | Curcumin | {"name": "Curcumin", "id": " CID_969516"} | INTERACTS_WITH | - | Target | 7253 | TSHR | {"name": "TSHR", "id": "7253"} |
| Phytochemical | CID_969516 | Curcumin | {"name": "Curcumin", "id": " CID_969516"} | INTERACTS_WITH | - | Target | 7263 | TST | {"name": "TST", "id": "7263"} |
| Phytochemical | CID_969516 | Curcumin | {"name": "Curcumin", "id": " CID_969516"} | INTERACTS_WITH | - | Target | 7277 | TUBA4A | {"name": "TUBA4A", "id": "7277"} |
| Phytochemical | CID_969516 | Curcumin | {"name": "Curcumin", "id": " CID_969516"} | INTERACTS_WITH | - | Target | 7278 | TUBA3C | {"name": "TUBA3C", "id": "7278"} |
| Phytochemical | CID_969516 | Curcumin | {"name": "Curcumin", "id": " CID_969516"} | INTERACTS_WITH | - | Target | 7280 | TUBB2A | {"name": "TUBB2A", "id": "7280"} |
| Phytochemical | CID_969516 | Curcumin | {"name": "Curcumin", "id": " CID_969516"} | INTERACTS_WITH | - | Target | 7296 | TXNRD1 | {"name": "TXNRD1", "id": "7296"} |
| Phytochemical | CID_969516 | Curcumin | {"name": "Curcumin", "id": " CID_969516"} | INTERACTS_WITH | - | Target | 7299 | TYR | {"name": "TYR", "id": "7299"} |
| Phytochemical | CID_969516 | Curcumin | {"name": "Curcumin", "id": " CID_969516"} | INTERACTS_WITH | - | Target | 7398 | USP1 | {"name": "USP1", "id": "7398"} |
| Phytochemical | CID_969516 | Curcumin | {"name": "Curcumin", "id": " CID_969516"} | INTERACTS_WITH | - | Target | 7421 | VDR | {"name": "VDR", "id": "7421"} |
| Phytochemical | CID_969516 | Curcumin | {"name": "Curcumin", "id": " CID_969516"} | INTERACTS_WITH | - | Target | 759 | CA1 | {"name": "CA1", "id": "759"} |
| Phytochemical | CID_969516 | Curcumin | {"name": "Curcumin", "id": " CID_969516"} | INTERACTS_WITH | - | Target | 760 | CA2 | {"name": "CA2", "id": "760"} |
| Phytochemical | CID_969516 | Curcumin | {"name": "Curcumin", "id": " CID_969516"} | INTERACTS_WITH | - | Target | 761 | CA3 | {"name": "CA3", "id": "761"} |
| Phytochemical | CID_969516 | Curcumin | {"name": "Curcumin", "id": " CID_969516"} | INTERACTS_WITH | - | Target | 762 | CA4 | {"name": "CA4", "id": "762"} |
| Phytochemical | CID_969516 | Curcumin | {"name": "Curcumin", "id": " CID_969516"} | INTERACTS_WITH | - | Target | 763 | CA5A | {"name": "CA5A", "id": "763"} |
| Phytochemical | CID_969516 | Curcumin | {"name": "Curcumin", "id": " CID_969516"} | INTERACTS_WITH | - | Target | 766 | CA7 | {"name": "CA7", "id": "766"} |
| Phytochemical | CID_969516 | Curcumin | {"name": "Curcumin", "id": " CID_969516"} | INTERACTS_WITH | - | Target | 768 | CA9 | {"name": "CA9", "id": "768"} |
| Phytochemical | CID_969516 | Curcumin | {"name": "Curcumin", "id": " CID_969516"} | INTERACTS_WITH | - | Target | 771 | CA12 | {"name": "CA12", "id": "771"} |
| Phytochemical | CID_969516 | Curcumin | {"name": "Curcumin", "id": " CID_969516"} | INTERACTS_WITH | - | Target | 7846 | TUBA1A | {"name": "TUBA1A", "id": "7846"} |
| Phytochemical | CID_969516 | Curcumin | {"name": "Curcumin", "id": " CID_969516"} | INTERACTS_WITH | - | Target | 79885 | HDAC11 | {"name": "HDAC11", "id": "79885"} |
| Phytochemical | CID_969516 | Curcumin | {"name": "Curcumin", "id": " CID_969516"} | INTERACTS_WITH | - | Target | 79915 | ATAD5 | {"name": "ATAD5", "id": "79915"} |
| Phytochemical | CID_969516 | Curcumin | {"name": "Curcumin", "id": " CID_969516"} | INTERACTS_WITH | - | Target | 81027 | TUBB1 | {"name": "TUBB1", "id": "81027"} |
| Phytochemical | CID_969516 | Curcumin | {"name": "Curcumin", "id": " CID_969516"} | INTERACTS_WITH | - | Target | 815 | CAMK2A | {"name": "CAMK2A", "id": "815"} |
| Phytochemical | CID_969516 | Curcumin | {"name": "Curcumin", "id": " CID_969516"} | INTERACTS_WITH | - | Target | 834 | CASP1 | {"name": "CASP1", "id": "834"} |
| Phytochemical | CID_969516 | Curcumin | {"name": "Curcumin", "id": " CID_969516"} | INTERACTS_WITH | - | Target | 83933 | HDAC10 | {"name": "HDAC10", "id": "83933"} |
| Phytochemical | CID_969516 | Curcumin | {"name": "Curcumin", "id": " CID_969516"} | INTERACTS_WITH | - | Target | 840 | CASP7 | {"name": "CASP7", "id": "840"} |
| Phytochemical | CID_969516 | Curcumin | {"name": "Curcumin", "id": " CID_969516"} | INTERACTS_WITH | - | Target | 8444 | DYRK3 | {"name": "DYRK3", "id": "8444"} |
| Phytochemical | CID_969516 | Curcumin | {"name": "Curcumin", "id": " CID_969516"} | INTERACTS_WITH | - | Target | 8445 | DYRK2 | {"name": "DYRK2", "id": "8445"} |
| Phytochemical | CID_969516 | Curcumin | {"name": "Curcumin", "id": " CID_969516"} | INTERACTS_WITH | - | Target | 84617 | TUBB6 | {"name": "TUBB6", "id": "84617"} |
| Phytochemical | CID_969516 | Curcumin | {"name": "Curcumin", "id": " CID_969516"} | INTERACTS_WITH | - | Target | 84790 | TUBA1C | {"name": "TUBA1C", "id": "84790"} |
| Phytochemical | CID_969516 | Curcumin | {"name": "Curcumin", "id": " CID_969516"} | INTERACTS_WITH | - | Target | 8798 | DYRK4 | {"name": "DYRK4", "id": "8798"} |
| Phytochemical | CID_969516 | Curcumin | {"name": "Curcumin", "id": " CID_969516"} | INTERACTS_WITH | - | Target | 8841 | HDAC3 | {"name": "HDAC3", "id": "8841"} |
| Phytochemical | CID_969516 | Curcumin | {"name": "Curcumin", "id": " CID_969516"} | INTERACTS_WITH | - | Target | 8851 | CDK5R1 | {"name": "CDK5R1", "id": "8851"} |
| Phytochemical | CID_969516 | Curcumin | {"name": "Curcumin", "id": " CID_969516"} | INTERACTS_WITH | - | Target | 9149 | DYRK1B | {"name": "DYRK1B", "id": "9149"} |
| Phytochemical | CID_969516 | Curcumin | {"name": "Curcumin", "id": " CID_969516"} | INTERACTS_WITH | - | Target | 9429 | ABCG2 | {"name": "ABCG2", "id": "9429"} |
| Phytochemical | CID_969516 | Curcumin | {"name": "Curcumin", "id": " CID_969516"} | INTERACTS_WITH | - | Target | 9536 | PTGES | {"name": "PTGES", "id": "9536"} |
| Phytochemical | CID_969516 | Curcumin | {"name": "Curcumin", "id": " CID_969516"} | INTERACTS_WITH | - | Target | 9682 | KDM4A | {"name": "KDM4A", "id": "9682"} |
| Phytochemical | CID_969516 | Curcumin | {"name": "Curcumin", "id": " CID_969516"} | INTERACTS_WITH | - | Target | 9734 | HDAC9 | {"name": "HDAC9", "id": "9734"} |
| Phytochemical | CID_969516 | Curcumin | {"name": "Curcumin", "id": " CID_969516"} | INTERACTS_WITH | - | Target | 9759 | HDAC4 | {"name": "HDAC4", "id": "9759"} |
| Phytochemical | CID_969516 | Curcumin | {"name": "Curcumin", "id": " CID_969516"} | INTERACTS_WITH | - | Target | 9971 | NR1H4 | {"name": "NR1H4", "id": "9971"} |
| Phytochemical | CID_14240392 | Curcumol | {"name": "Curcumol", "id": " CID_14240392"} | INTERACTS_WITH | - | Target | 10013 | HDAC6 | {"name": "HDAC6", "id": "10013"} |
| Phytochemical | CID_6036 | D-Galactose | {"name": "D-Galactose", "id": " CID_6036"} | INTERACTS_WITH | - | Target | 10073 | SNUPN | {"name": "SNUPN", "id": "10073"} |
| Phytochemical | CID_6036 | D-Galactose | {"name": "D-Galactose", "id": " CID_6036"} | INTERACTS_WITH | - | Target | 10951 | CBX1 | {"name": "CBX1", "id": "10951"} |

**Table S9: Complete node and relationship data used to construct the knowledge graph. This table summarizes interactions among FDA-approved drugs, formulation, herbs, diseases (type 2 diabetes, fatty liver, obesity), phytochemicals, and their target genes/proteins. Each entry includes source and target node details, edge type, and node annotations (Entrez Gene ID and gene symbol, where applicable).**

| Source Node Label | Source Node ID | Source Node Name | Source Node Properties | Edge Type | Edge Properties | Target Node Label | Target Node ID | Target Node Name | Target Node Properties |
|---|---|---|---|---|---|---|---|---|---|
| Phytochemical | CID_6036 | D-Galactose | {"name": "D-Galactose", "id": "CID_6036"} | INTERACTS_WITH | - | Target | 2629 | GBA1 | {"name": "GBA1", "id": "2629"} |
| Phytochemical | CID_6036 | D-Galactose | {"name": "D-Galactose", "id": "CID_6036"} | INTERACTS_WITH | - | Target | 2648 | KAT2A | {"name": "KAT2A", "id": "2648"} |
| Phytochemical | CID_6036 | D-Galactose | {"name": "D-Galactose", "id": "CID_6036"} | INTERACTS_WITH | - | Target | 2720 | GLB1 | {"name": "GLB1", "id": "2720"} |
| Phytochemical | CID_6036 | D-Galactose | {"name": "D-Galactose", "id": "CID_6036"} | INTERACTS_WITH | - | Target | 328 | APEX1 | {"name": "APEX1", "id": "328"} |
| Phytochemical | CID_6036 | D-Galactose | {"name": "D-Galactose", "id": "CID_6036"} | INTERACTS_WITH | - | Target | 3417 | IDH1 | {"name": "IDH1", "id": "3417"} |
| Phytochemical | CID_6036 | D-Galactose | {"name": "D-Galactose", "id": "CID_6036"} | INTERACTS_WITH | - | Target | 3837 | KPNB1 | {"name": "KPNB1", "id": "3837"} |
| Phytochemical | CID_6036 | D-Galactose | {"name": "D-Galactose", "id": "CID_6036"} | INTERACTS_WITH | - | Target | 51426 | POLK | {"name": "POLK", "id": "51426"} |
| Phytochemical | CID_6036 | D-Galactose | {"name": "D-Galactose", "id": "CID_6036"} | INTERACTS_WITH | - | Target | 7398 | USP1 | {"name": "USP1", "id": "7398"} |
| Phytochemical | CID_6036 | D-Galactose | {"name": "D-Galactose", "id": "CID_6036"} | INTERACTS_WITH | - | Target | 84750 | POFUT3 | {"name": "POFUT3", "id": "84750"} |
| Phytochemical | CID_6036 | D-Galactose | {"name": "D-Galactose", "id": "CID_6036"} | INTERACTS_WITH | - | Target | 84929 | FIBCD1 | {"name": "FIBCD1", "id": "84929"} |
| Phytochemical | CID_5793 | D-Glucose | {"name": "D-Glucose", "id": "CID_5793"} | INTERACTS_WITH | - | Target | 231 | AKR1B1 | {"name": "AKR1B1", "id": "231"} |
| Phytochemical | CID_5793 | D-Glucose | {"name": "D-Glucose", "id": "CID_5793"} | INTERACTS_WITH | - | Target | 49 | ACR | {"name": "ACR", "id": "49"} |
| Phytochemical | CID_5793 | D-Glucose | {"name": "D-Glucose", "id": "CID_5793"} | INTERACTS_WITH | - | Target | 51426 | POLK | {"name": "POLK", "id": "51426"} |
| Phytochemical | CID_5793 | D-Glucose | {"name": "D-Glucose", "id": "CID_5793"} | INTERACTS_WITH | - | Target | 5834 | PYGB | {"name": "PYGB", "id": "5834"} |
| Phytochemical | CID_5793 | D-Glucose | {"name": "D-Glucose", "id": "CID_5793"} | INTERACTS_WITH | - | Target | 5837 | PYGM | {"name": "PYGM", "id": "5837"} |
| Phytochemical | CID_5793 | D-Glucose | {"name": "D-Glucose", "id": "CID_5793"} | INTERACTS_WITH | - | Target | 6513 | SLC2A1 | {"name": "SLC2A1", "id": "6513"} |
| Phytochemical | CID_5793 | D-Glucose | {"name": "D-Glucose", "id": "CID_5793"} | INTERACTS_WITH | - | Target | 84750 | POFUT3 | {"name": "POFUT3", "id": "84750"} |
| Phytochemical | CID_5793 | D-Glucose | {"name": "D-Glucose", "id": "CID_5793"} | INTERACTS_WITH | - | Target | 84929 | FIBCD1 | {"name": "FIBCD1", "id": "84929"} |
| Phytochemical | CID_5742590 | Daucosterol | {"name": "Daucosterol", "id": "CID_5742590"} | INTERACTS_WITH | - | Target | 10013 | HDAC6 | {"name": "HDAC6", "id": "10013"} |
| Phytochemical | CID_5742590 | Daucosterol | {"name": "Daucosterol", "id": "CID_5742590"} | INTERACTS_WITH | - | Target | 2932 | GSK3B | {"name": "GSK3B", "id": "2932"} |
| Phytochemical | CID_5742590 | Daucosterol | {"name": "Daucosterol", "id": "CID_5742590"} | INTERACTS_WITH | - | Target | 3417 | IDH1 | {"name": "IDH1", "id": "3417"} |
| Phytochemical | CID_5742590 | Daucosterol | {"name": "Daucosterol", "id": "CID_5742590"} | INTERACTS_WITH | - | Target | 51422 | PRKAG2 | {"name": "PRKAG2", "id": "51422"} |
| Phytochemical | CID_5742590 | Daucosterol | {"name": "Daucosterol", "id": "CID_5742590"} | INTERACTS_WITH | - | Target | 53632 | PRKAG3 | {"name": "PRKAG3", "id": "53632"} |
| Phytochemical | CID_5742590 | Daucosterol | {"name": "Daucosterol", "id": "CID_5742590"} | INTERACTS_WITH | - | Target | 5423 | POLB | {"name": "POLB", "id": "5423"} |
| Phytochemical | CID_5742590 | Daucosterol | {"name": "Daucosterol", "id": "CID_5742590"} | INTERACTS_WITH | - | Target | 5562 | PRKAA1 | {"name": "PRKAA1", "id": "5562"} |
| Phytochemical | CID_5742590 | Daucosterol | {"name": "Daucosterol", "id": "CID_5742590"} | INTERACTS_WITH | - | Target | 5563 | PRKAA2 | {"name": "PRKAA2", "id": "5563"} |
| Phytochemical | CID_5742590 | Daucosterol | {"name": "Daucosterol", "id": "CID_5742590"} | INTERACTS_WITH | - | Target | 5564 | PRKAB1 | {"name": "PRKAB1", "id": "5564"} |
| Phytochemical | CID_5742590 | Daucosterol | {"name": "Daucosterol", "id": "CID_5742590"} | INTERACTS_WITH | - | Target | 5565 | PRKAB2 | {"name": "PRKAB2", "id": "5565"} |
| Phytochemical | CID_5742590 | Daucosterol | {"name": "Daucosterol", "id": "CID_5742590"} | INTERACTS_WITH | - | Target | 5571 | PRKAG1 | {"name": "PRKAG1", "id": "5571"} |
| Phytochemical | CID_2969 | Decanoic acid | {"name": "Decanoic acid", "id": "CID_2969"} | INTERACTS_WITH | - | Target | 2908 | NR3C1 | {"name": "NR3C1", "id": "2908"} |
| Phytochemical | CID_2969 | Decanoic acid | {"name": "Decanoic acid", "id": "CID_2969"} | INTERACTS_WITH | - | Target | 3315 | HSPB1 | {"name": "HSPB1", "id": "3315"} |
| Phytochemical | CID_2969 | Decanoic acid | {"name": "Decanoic acid", "id": "CID_2969"} | INTERACTS_WITH | - | Target | 367 | AR | {"name": "AR", "id": "367"} |
| Phytochemical | CID_2969 | Decanoic acid | {"name": "Decanoic acid", "id": "CID_2969"} | INTERACTS_WITH | - | Target | 4780 | NFE2L2 | {"name": "NFE2L2", "id": "4780"} |
| Phytochemical | CID_2969 | Decanoic acid | {"name": "Decanoic acid", "id": "CID_2969"} | INTERACTS_WITH | - | Target | 51053 | GMNN | {"name": "GMNN", "id": "51053"} |
| Phytochemical | CID_2969 | Decanoic acid | {"name": "Decanoic acid", "id": "CID_2969"} | INTERACTS_WITH | - | Target | 53831 | GPR84 | {"name": "GPR84", "id": "53831"} |
| Phytochemical | CID_2969 | Decanoic acid | {"name": "Decanoic acid", "id": "CID_2969"} | INTERACTS_WITH | - | Target | 54575 | UGT1A10 | {"name": "UGT1A10", "id": "54575"} |
| Phytochemical | CID_2969 | Decanoic acid | {"name": "Decanoic acid", "id": "CID_2969"} | INTERACTS_WITH | - | Target | 54576 | UGT1A8 | {"name": "UGT1A8", "id": "54576"} |
| Phytochemical | CID_2969 | Decanoic acid | {"name": "Decanoic acid", "id": "CID_2969"} | INTERACTS_WITH | - | Target | 5465 | PPARA | {"name": "PPARA", "id": "5465"} |
| Phytochemical | CID_2969 | Decanoic acid | {"name": "Decanoic acid", "id": "CID_2969"} | INTERACTS_WITH | - | Target | 54659 | UGT1A3 | {"name": "UGT1A3", "id": "54659"} |
| Phytochemical | CID_2969 | Decanoic acid | {"name": "Decanoic acid", "id": "CID_2969"} | INTERACTS_WITH | - | Target | 5467 | PPARD | {"name": "PPARD", "id": "5467"} |
| Phytochemical | CID_2969 | Decanoic acid | {"name": "Decanoic acid", "id": "CID_2969"} | INTERACTS_WITH | - | Target | 5468 | PPARG | {"name": "PPARG", "id": "5468"} |
| Phytochemical | CID_2969 | Decanoic acid | {"name": "Decanoic acid", "id": "CID_2969"} | INTERACTS_WITH | - | Target | 55775 | TDP1 | {"name": "TDP1", "id": "55775"} |
| Phytochemical | CID_2969 | Decanoic acid | {"name": "Decanoic acid", "id": "CID_2969"} | INTERACTS_WITH | - | Target | 6097 | RORC | {"name": "RORC", "id": "6097"} |
| Phytochemical | CID_2969 | Decanoic acid | {"name": "Decanoic acid", "id": "CID_2969"} | INTERACTS_WITH | - | Target | 7253 | TSHR | {"name": "TSHR", "id": "7253"} |
| Phytochemical | CID_5469424 | Demethoxycurcumin | {"name": "Demethoxycurcumin", "id": "CID_5469424"} | INTERACTS_WITH | - | Target | 23621 | BACE1 | {"name": "BACE1", "id": "23621"} |
| Phytochemical | CID_5469424 | Demethoxycurcumin | {"name": "Demethoxycurcumin", "id": "CID_5469424"} | INTERACTS_WITH | - | Target | 248 | ALPI | {"name": "ALPI", "id": "248"} |
| Phytochemical | CID_5469424 | Demethoxycurcumin | {"name": "Demethoxycurcumin", "id": "CID_5469424"} | INTERACTS_WITH | - | Target | 249 | ALPL | {"name": "ALPL", "id": "249"} |
| Phytochemical | CID_5469424 | Demethoxycurcumin | {"name": "Demethoxycurcumin", "id": "CID_5469424"} | INTERACTS_WITH | - | Target | 251 | ALPG | {"name": "ALPG", "id": "251"} |
| Phytochemical | CID_5469424 | Demethoxycurcumin | {"name": "Demethoxycurcumin", "id": "CID_5469424"} | INTERACTS_WITH | - | Target | 4128 | MAOA | {"name": "MAOA", "id": "4128"} |
| Phytochemical | CID_5469424 | Demethoxycurcumin | {"name": "Demethoxycurcumin", "id": "CID_5469424"} | INTERACTS_WITH | - | Target | 4129 | MAOB | {"name": "MAOB", "id": "4129"} |
| Phytochemical | CID_5469424 | Demethoxycurcumin | {"name": "Demethoxycurcumin", "id": "CID_5469424"} | INTERACTS_WITH | - | Target | 5243 | ABCB1 | {"name": "ABCB1", "id": "5243"} |

**Table S9: Complete node and relationship data used to construct the knowledge graph. This table summarizes interactions among FDA-approved drugs, formulation, herbs, diseases (type 2 diabetes, fatty liver, obesity), phytochemicals, and their target genes/proteins. Each entry includes source and target node details, edge type, and node annotations (Entrez Gene ID and gene symbol, where applicable).**

| Source Node Label | Source Node ID | Source Node Name | Source Node Properties | Edge Type | Edge Properties | Target Node Label | Target Node ID | Target Node Name | Target Node Properties |
|---|---|---|---|---|---|---|---|---|---|
| Phytochemical | CID_5469424 | Demethoxycurcumin | {"name": "Demethoxycurcumin", "id": "CID_5469424"} | INTERACTS_WITH | - | Target | 5743 | PTGS2 | {"name": "PTGS2", "id": "5743"} |
| Phytochemical | CID_5469424 | Demethoxycurcumin | {"name": "Demethoxycurcumin", "id": "CID_5469424"} | INTERACTS_WITH | - | Target | 65975 | STK33 | {"name": "STK33", "id": "65975"} |
| Phytochemical | CID_5469424 | Demethoxycurcumin | {"name": "Demethoxycurcumin", "id": "CID_5469424"} | INTERACTS_WITH | - | Target | 815 | CAMK2A | {"name": "CAMK2A", "id": "815"} |
| Phytochemical | CID_8215 | Docosanoic acid | {"name": "Docosanoic acid", "id": "CID_8215"} | INTERACTS_WITH | - | Target | 10013 | HDAC6 | {"name": "HDAC6", "id": "10013"} |
| Phytochemical | CID_8215 | Docosanoic acid | {"name": "Docosanoic acid", "id": "CID_8215"} | INTERACTS_WITH | - | Target | 5465 | PPARA | {"name": "PPARA", "id": "5465"} |
| Phytochemical | CID_8215 | Docosanoic acid | {"name": "Docosanoic acid", "id": "CID_8215"} | INTERACTS_WITH | - | Target | 5467 | PPARD | {"name": "PPARD", "id": "5467"} |
| Phytochemical | CID_8215 | Docosanoic acid | {"name": "Docosanoic acid", "id": "CID_8215"} | INTERACTS_WITH | - | Target | 5468 | PPARG | {"name": "PPARG", "id": "5468"} |
| Phytochemical | CID_6439929 | E-Guggulsterone | {"name": "E-Guggulsterone", "id": "CID_6439929"} | INTERACTS_WITH | - | Target | 10062 | NR1H3 | {"name": "NR1H3", "id": "10062"} |
| Phytochemical | CID_6439929 | E-Guggulsterone | {"name": "E-Guggulsterone", "id": "CID_6439929"} | INTERACTS_WITH | - | Target | 1576 | CYP3A4 | {"name": "CYP3A4", "id": "1576"} |
| Phytochemical | CID_6439929 | E-Guggulsterone | {"name": "E-Guggulsterone", "id": "CID_6439929"} | INTERACTS_WITH | - | Target | 2099 | ESR1 | {"name": "ESR1", "id": "2099"} |
| Phytochemical | CID_6439929 | E-Guggulsterone | {"name": "E-Guggulsterone", "id": "CID_6439929"} | INTERACTS_WITH | - | Target | 2100 | ESR2 | {"name": "ESR2", "id": "2100"} |
| Phytochemical | CID_6439929 | E-Guggulsterone | {"name": "E-Guggulsterone", "id": "CID_6439929"} | INTERACTS_WITH | - | Target | 216 | ALDH1A1 | {"name": "ALDH1A1", "id": "216"} |
| Phytochemical | CID_6439929 | E-Guggulsterone | {"name": "E-Guggulsterone", "id": "CID_6439929"} | INTERACTS_WITH | - | Target | 2908 | NR3C1 | {"name": "NR3C1", "id": "2908"} |
| Phytochemical | CID_6439929 | E-Guggulsterone | {"name": "E-Guggulsterone", "id": "CID_6439929"} | INTERACTS_WITH | - | Target | 3315 | HSPB1 | {"name": "HSPB1", "id": "3315"} |
| Phytochemical | CID_6439929 | E-Guggulsterone | {"name": "E-Guggulsterone", "id": "CID_6439929"} | INTERACTS_WITH | - | Target | 3576 | CXCL8 | {"name": "CXCL8", "id": "3576"} |
| Phytochemical | CID_6439929 | E-Guggulsterone | {"name": "E-Guggulsterone", "id": "CID_6439929"} | INTERACTS_WITH | - | Target | 367 | AR | {"name": "AR", "id": "367"} |
| Phytochemical | CID_6439929 | E-Guggulsterone | {"name": "E-Guggulsterone", "id": "CID_6439929"} | INTERACTS_WITH | - | Target | 3757 | KCNH2 | {"name": "KCNH2", "id": "3757"} |
| Phytochemical | CID_6439929 | E-Guggulsterone | {"name": "E-Guggulsterone", "id": "CID_6439929"} | INTERACTS_WITH | - | Target | 4306 | NR3C2 | {"name": "NR3C2", "id": "4306"} |
| Phytochemical | CID_6439929 | E-Guggulsterone | {"name": "E-Guggulsterone", "id": "CID_6439929"} | INTERACTS_WITH | - | Target | 4780 | NFE2L2 | {"name": "NFE2L2", "id": "4780"} |
| Phytochemical | CID_6439929 | E-Guggulsterone | {"name": "E-Guggulsterone", "id": "CID_6439929"} | INTERACTS_WITH | - | Target | 4790 | NFKB1 | {"name": "NFKB1", "id": "4790"} |
| Phytochemical | CID_6439929 | E-Guggulsterone | {"name": "E-Guggulsterone", "id": "CID_6439929"} | INTERACTS_WITH | - | Target | 5241 | PGR | {"name": "PGR", "id": "5241"} |
| Phytochemical | CID_6439929 | E-Guggulsterone | {"name": "E-Guggulsterone", "id": "CID_6439929"} | INTERACTS_WITH | - | Target | 5465 | PPARA | {"name": "PPARA", "id": "5465"} |
| Phytochemical | CID_6439929 | E-Guggulsterone | {"name": "E-Guggulsterone", "id": "CID_6439929"} | INTERACTS_WITH | - | Target | 5467 | PPARD | {"name": "PPARD", "id": "5467"} |
| Phytochemical | CID_6439929 | E-Guggulsterone | {"name": "E-Guggulsterone", "id": "CID_6439929"} | INTERACTS_WITH | - | Target | 5468 | PPARG | {"name": "PPARG", "id": "5468"} |
| Phytochemical | CID_6439929 | E-Guggulsterone | {"name": "E-Guggulsterone", "id": "CID_6439929"} | INTERACTS_WITH | - | Target | 5594 | MAPK1 | {"name": "MAPK1", "id": "5594"} |
| Phytochemical | CID_6439929 | E-Guggulsterone | {"name": "E-Guggulsterone", "id": "CID_6439929"} | INTERACTS_WITH | - | Target | 5914 | RARA | {"name": "RARA", "id": "5914"} |
| Phytochemical | CID_6439929 | E-Guggulsterone | {"name": "E-Guggulsterone", "id": "CID_6439929"} | INTERACTS_WITH | - | Target | 5915 | RARB | {"name": "RARB", "id": "5915"} |
| Phytochemical | CID_6439929 | E-Guggulsterone | {"name": "E-Guggulsterone", "id": "CID_6439929"} | INTERACTS_WITH | - | Target | 5916 | RARG | {"name": "RARG", "id": "5916"} |
| Phytochemical | CID_6439929 | E-Guggulsterone | {"name": "E-Guggulsterone", "id": "CID_6439929"} | INTERACTS_WITH | - | Target | 6097 | RORC | {"name": "RORC", "id": "6097"} |
| Phytochemical | CID_6439929 | E-Guggulsterone | {"name": "E-Guggulsterone", "id": "CID_6439929"} | INTERACTS_WITH | - | Target | 6256 | RXRA | {"name": "RXRA", "id": "6256"} |
| Phytochemical | CID_6439929 | E-Guggulsterone | {"name": "E-Guggulsterone", "id": "CID_6439929"} | INTERACTS_WITH | - | Target | 6606 | SMN1 | {"name": "SMN1", "id": "6606"} |
| Phytochemical | CID_6439929 | E-Guggulsterone | {"name": "E-Guggulsterone", "id": "CID_6439929"} | INTERACTS_WITH | - | Target | 7067 | THRA | {"name": "THRA", "id": "7067"} |
| Phytochemical | CID_6439929 | E-Guggulsterone | {"name": "E-Guggulsterone", "id": "CID_6439929"} | INTERACTS_WITH | - | Target | 7068 | THRB | {"name": "THRB", "id": "7068"} |
| Phytochemical | CID_6439929 | E-Guggulsterone | {"name": "E-Guggulsterone", "id": "CID_6439929"} | INTERACTS_WITH | - | Target | 7157 | TP53 | {"name": "TP53", "id": "7157"} |
| Phytochemical | CID_6439929 | E-Guggulsterone | {"name": "E-Guggulsterone", "id": "CID_6439929"} | INTERACTS_WITH | - | Target | 7253 | TSHR | {"name": "TSHR", "id": "7253"} |
| Phytochemical | CID_6439929 | E-Guggulsterone | {"name": "E-Guggulsterone", "id": "CID_6439929"} | INTERACTS_WITH | - | Target | 7376 | NR1H2 | {"name": "NR1H2", "id": "7376"} |
| Phytochemical | CID_6439929 | E-Guggulsterone | {"name": "E-Guggulsterone", "id": "CID_6439929"} | INTERACTS_WITH | - | Target | 7421 | VDR | {"name": "VDR", "id": "7421"} |
| Phytochemical | CID_6439929 | E-Guggulsterone | {"name": "E-Guggulsterone", "id": "CID_6439929"} | INTERACTS_WITH | - | Target | 865 | CBFB | {"name": "CBFB", "id": "865"} |
| Phytochemical | CID_6439929 | E-Guggulsterone | {"name": "E-Guggulsterone", "id": "CID_6439929"} | INTERACTS_WITH | - | Target | 8856 | NR1I2 | {"name": "NR1I2", "id": "8856"} |
| Phytochemical | CID_6439929 | E-Guggulsterone | {"name": "E-Guggulsterone", "id": "CID_6439929"} | INTERACTS_WITH | - | Target | 9971 | NR1H4 | {"name": "NR1H4", "id": "9971"} |
| Phytochemical | CID_5281855 | Ellagic acid | {"name": "Ellagic acid", "id": "CID_5281855"} | INTERACTS_WITH | - | Target | 10013 | HDAC6 | {"name": "HDAC6", "id": "10013"} |
| Phytochemical | CID_5281855 | Ellagic acid | {"name": "Ellagic acid", "id": "CID_5281855"} | INTERACTS_WITH | - | Target | 10073 | SNUPN | {"name": "SNUPN", "id": "10073"} |
| Phytochemical | CID_5281855 | Ellagic acid | {"name": "Ellagic acid", "id": "CID_5281855"} | INTERACTS_WITH | - | Target | 1017 | CDK2 | {"name": "CDK2", "id": "1017"} |
| Phytochemical | CID_5281855 | Ellagic acid | {"name": "Ellagic acid", "id": "CID_5281855"} | INTERACTS_WITH | - | Target | 1019 | CDK4 | {"name": "CDK4", "id": "1019"} |
| Phytochemical | CID_5281855 | Ellagic acid | {"name": "Ellagic acid", "id": "CID_5281855"} | INTERACTS_WITH | - | Target | 10411 | RAPGEF3 | {"name": "RAPGEF3", "id": "10411"} |
| Phytochemical | CID_5281855 | Ellagic acid | {"name": "Ellagic acid", "id": "CID_5281855"} | INTERACTS_WITH | - | Target | 10498 | CARM1 | {"name": "CARM1", "id": "10498"} |
| Phytochemical | CID_5281855 | Ellagic acid | {"name": "Ellagic acid", "id": "CID_5281855"} | INTERACTS_WITH | - | Target | 10587 | TXNRD2 | {"name": "TXNRD2", "id": "10587"} |
| Phytochemical | CID_5281855 | Ellagic acid | {"name": "Ellagic acid", "id": "CID_5281855"} | INTERACTS_WITH | - | Target | 10599 | SLCO1B1 | {"name": "SLCO1B1", "id": "10599"} |
| Phytochemical | CID_5281855 | Ellagic acid | {"name": "Ellagic acid", "id": "CID_5281855"} | INTERACTS_WITH | - | Target | 10733 | PLK4 | {"name": "PLK4", "id": "10733"} |

**Table S9: Complete node and relationship data used to construct the knowledge graph. This table summarizes interactions among FDA-approved drugs, formulation, herbs, diseases (type 2 diabetes, fatty liver, obesity), phytochemicals, and their target genes/proteins. Each entry includes source and target node details, edge type, and node annotations (Entrez Gene ID and gene symbol, where applicable).**

| Source Node Label | Source Node ID | Source Node Name | Source Node Properties | Edge Type | Edge Properties | Target Node Label | Target Node ID | Target Node Name | Target Node Properties |
|---|---|---|---|---|---|---|---|---|---|
| Phytochemical | CID_5281855 | Ellagic acid | {"name": "Ellagic acid", "id": "CID_5281855"} | INTERACTS_WITH | - | Target | 10919 | EHMT2 | {"name": "EHMT2", "id": "10919"} |
| Phytochemical | CID_5281855 | Ellagic acid | {"name": "Ellagic acid", "id": "CID_5281855"} | INTERACTS_WITH | - | Target | 10951 | CBX1 | {"name": "CBX1", "id": "10951"} |
| Phytochemical | CID_5281855 | Ellagic acid | {"name": "Ellagic acid", "id": "CID_5281855"} | INTERACTS_WITH | - | Target | 11069 | RAPGEF4 | {"name": "RAPGEF4", "id": "11069"} |
| Phytochemical | CID_5281855 | Ellagic acid | {"name": "Ellagic acid", "id": "CID_5281855"} | INTERACTS_WITH | - | Target | 11201 | POLI | {"name": "POLI", "id": "11201"} |
| Phytochemical | CID_5281855 | Ellagic acid | {"name": "Ellagic acid", "id": "CID_5281855"} | INTERACTS_WITH | - | Target | 11238 | CA5B | {"name": "CA5B", "id": "11238"} |
| Phytochemical | CID_5281855 | Ellagic acid | {"name": "Ellagic acid", "id": "CID_5281855"} | INTERACTS_WITH | - | Target | 114112 | TXNRD3 | {"name": "TXNRD3", "id": "114112"} |
| Phytochemical | CID_5281855 | Ellagic acid | {"name": "Ellagic acid", "id": "CID_5281855"} | INTERACTS_WITH | - | Target | 1326 | MAP3K8 | {"name": "MAP3K8", "id": "1326"} |
| Phytochemical | CID_5281855 | Ellagic acid | {"name": "Ellagic acid", "id": "CID_5281855"} | INTERACTS_WITH | - | Target | 1457 | CSNK2A1 | {"name": "CSNK2A1", "id": "1457"} |
| Phytochemical | CID_5281855 | Ellagic acid | {"name": "Ellagic acid", "id": "CID_5281855"} | INTERACTS_WITH | - | Target | 1459 | CSNK2A2 | {"name": "CSNK2A2", "id": "1459"} |
| Phytochemical | CID_5281855 | Ellagic acid | {"name": "Ellagic acid", "id": "CID_5281855"} | INTERACTS_WITH | - | Target | 1460 | CSNK2B | {"name": "CSNK2B", "id": "1460"} |
| Phytochemical | CID_5281855 | Ellagic acid | {"name": "Ellagic acid", "id": "CID_5281855"} | INTERACTS_WITH | - | Target | 1557 | CYP2C19 | {"name": "CYP2C19", "id": "1557"} |
| Phytochemical | CID_5281855 | Ellagic acid | {"name": "Ellagic acid", "id": "CID_5281855"} | INTERACTS_WITH | - | Target | 1559 | CYP2C9 | {"name": "CYP2C9", "id": "1559"} |
| Phytochemical | CID_5281855 | Ellagic acid | {"name": "Ellagic acid", "id": "CID_5281855"} | INTERACTS_WITH | - | Target | 1636 | ACE | {"name": "ACE", "id": "1636"} |
| Phytochemical | CID_5281855 | Ellagic acid | {"name": "Ellagic acid", "id": "CID_5281855"} | INTERACTS_WITH | - | Target | 1845 | DUSP3 | {"name": "DUSP3", "id": "1845"} |
| Phytochemical | CID_5281855 | Ellagic acid | {"name": "Ellagic acid", "id": "CID_5281855"} | INTERACTS_WITH | - | Target | 1859 | DYRK1A | {"name": "DYRK1A", "id": "1859"} |
| Phytochemical | CID_5281855 | Ellagic acid | {"name": "Ellagic acid", "id": "CID_5281855"} | INTERACTS_WITH | - | Target | 1956 | EGFR | {"name": "EGFR", "id": "1956"} |
| Phytochemical | CID_5281855 | Ellagic acid | {"name": "Ellagic acid", "id": "CID_5281855"} | INTERACTS_WITH | - | Target | 1981 | EIF4G1 | {"name": "EIF4G1", "id": "1981"} |
| Phytochemical | CID_5281855 | Ellagic acid | {"name": "Ellagic acid", "id": "CID_5281855"} | INTERACTS_WITH | - | Target | 1995 | ELAVL3 | {"name": "ELAVL3", "id": "1995"} |
| Phytochemical | CID_5281855 | Ellagic acid | {"name": "Ellagic acid", "id": "CID_5281855"} | INTERACTS_WITH | - | Target | 2050 | EPHB4 | {"name": "EPHB4", "id": "2050"} |
| Phytochemical | CID_5281855 | Ellagic acid | {"name": "Ellagic acid", "id": "CID_5281855"} | INTERACTS_WITH | - | Target | 2053 | EPHX2 | {"name": "EPHX2", "id": "2053"} |
| Phytochemical | CID_5281855 | Ellagic acid | {"name": "Ellagic acid", "id": "CID_5281855"} | INTERACTS_WITH | - | Target | 2064 | ERBB2 | {"name": "ERBB2", "id": "2064"} |
| Phytochemical | CID_5281855 | Ellagic acid | {"name": "Ellagic acid", "id": "CID_5281855"} | INTERACTS_WITH | - | Target | 207 | AKT1 | {"name": "AKT1", "id": "207"} |
| Phytochemical | CID_5281855 | Ellagic acid | {"name": "Ellagic acid", "id": "CID_5281855"} | INTERACTS_WITH | - | Target | 2237 | FEN1 | {"name": "FEN1", "id": "2237"} |
| Phytochemical | CID_5281855 | Ellagic acid | {"name": "Ellagic acid", "id": "CID_5281855"} | INTERACTS_WITH | - | Target | 231 | AKR1B1 | {"name": "AKR1B1", "id": "231"} |
| Phytochemical | CID_5281855 | Ellagic acid | {"name": "Ellagic acid", "id": "CID_5281855"} | INTERACTS_WITH | - | Target | 23192 | ATG4B | {"name": "ATG4B", "id": "23192"} |
| Phytochemical | CID_5281855 | Ellagic acid | {"name": "Ellagic acid", "id": "CID_5281855"} | INTERACTS_WITH | - | Target | 2322 | FLT3 | {"name": "FLT3", "id": "2322"} |
| Phytochemical | CID_5281855 | Ellagic acid | {"name": "Ellagic acid", "id": "CID_5281855"} | INTERACTS_WITH | - | Target | 2324 | FLT4 | {"name": "FLT4", "id": "2324"} |
| Phytochemical | CID_5281855 | Ellagic acid | {"name": "Ellagic acid", "id": "CID_5281855"} | INTERACTS_WITH | - | Target | 23405 | DICER1 | {"name": "DICER1", "id": "23405"} |
| Phytochemical | CID_5281855 | Ellagic acid | {"name": "Ellagic acid", "id": "CID_5281855"} | INTERACTS_WITH | - | Target | 23621 | BACE1 | {"name": "BACE1", "id": "23621"} |
| Phytochemical | CID_5281855 | Ellagic acid | {"name": "Ellagic acid", "id": "CID_5281855"} | INTERACTS_WITH | - | Target | 23632 | CA14 | {"name": "CA14", "id": "23632"} |
| Phytochemical | CID_5281855 | Ellagic acid | {"name": "Ellagic acid", "id": "CID_5281855"} | INTERACTS_WITH | - | Target | 238 | ALK | {"name": "ALK", "id": "238"} |
| Phytochemical | CID_5281855 | Ellagic acid | {"name": "Ellagic acid", "id": "CID_5281855"} | INTERACTS_WITH | - | Target | 246 | ALOX15 | {"name": "ALOX15", "id": "246"} |
| Phytochemical | CID_5281855 | Ellagic acid | {"name": "Ellagic acid", "id": "CID_5281855"} | INTERACTS_WITH | - | Target | 247 | ALOX15B | {"name": "ALOX15B", "id": "247"} |
| Phytochemical | CID_5281855 | Ellagic acid | {"name": "Ellagic acid", "id": "CID_5281855"} | INTERACTS_WITH | - | Target | 2548 | GAA | {"name": "GAA", "id": "2548"} |
| Phytochemical | CID_5281855 | Ellagic acid | {"name": "Ellagic acid", "id": "CID_5281855"} | INTERACTS_WITH | - | Target | 2590 | GALNT2 | {"name": "GALNT2", "id": "2590"} |
| Phytochemical | CID_5281855 | Ellagic acid | {"name": "Ellagic acid", "id": "CID_5281855"} | INTERACTS_WITH | - | Target | 2597 | GAPDH | {"name": "GAPDH", "id": "2597"} |
| Phytochemical | CID_5281855 | Ellagic acid | {"name": "Ellagic acid", "id": "CID_5281855"} | INTERACTS_WITH | - | Target | 2648 | KAT2A | {"name": "KAT2A", "id": "2648"} |
| Phytochemical | CID_5281855 | Ellagic acid | {"name": "Ellagic acid", "id": "CID_5281855"} | INTERACTS_WITH | - | Target | 2744 | GLS | {"name": "GLS", "id": "2744"} |
| Phytochemical | CID_5281855 | Ellagic acid | {"name": "Ellagic acid", "id": "CID_5281855"} | INTERACTS_WITH | - | Target | 28234 | SLCO1B3 | {"name": "SLCO1B3", "id": "28234"} |
| Phytochemical | CID_5281855 | Ellagic acid | {"name": "Ellagic acid", "id": "CID_5281855"} | INTERACTS_WITH | - | Target | 2859 | GPR35 | {"name": "GPR35", "id": "2859"} |
| Phytochemical | CID_5281855 | Ellagic acid | {"name": "Ellagic acid", "id": "CID_5281855"} | INTERACTS_WITH | - | Target | 2932 | GSK3B | {"name": "GSK3B", "id": "2932"} |
| Phytochemical | CID_5281855 | Ellagic acid | {"name": "Ellagic acid", "id": "CID_5281855"} | INTERACTS_WITH | - | Target | 2936 | GSR | {"name": "GSR", "id": "2936"} |
| Phytochemical | CID_5281855 | Ellagic acid | {"name": "Ellagic acid", "id": "CID_5281855"} | INTERACTS_WITH | - | Target | 2950 | GSTP1 | {"name": "GSTP1", "id": "2950"} |
| Phytochemical | CID_5281855 | Ellagic acid | {"name": "Ellagic acid", "id": "CID_5281855"} | INTERACTS_WITH | - | Target | 29994 | BAZ2B | {"name": "BAZ2B", "id": "29994"} |
| Phytochemical | CID_5281855 | Ellagic acid | {"name": "Ellagic acid", "id": "CID_5281855"} | INTERACTS_WITH | - | Target | 3028 | HSD17B10 | {"name": "HSD17B10", "id": "3028"} |
| Phytochemical | CID_5281855 | Ellagic acid | {"name": "Ellagic acid", "id": "CID_5281855"} | INTERACTS_WITH | - | Target | 3043 | HBB | {"name": "HBB", "id": "3043"} |
| Phytochemical | CID_5281855 | Ellagic acid | {"name": "Ellagic acid", "id": "CID_5281855"} | INTERACTS_WITH | - | Target | 3248 | HPGD | {"name": "HPGD", "id": "3248"} |
| Phytochemical | CID_5281855 | Ellagic acid | {"name": "Ellagic acid", "id": "CID_5281855"} | INTERACTS_WITH | - | Target | 328 | APEX1 | {"name": "APEX1", "id": "328"} |
| Phytochemical | CID_5281855 | Ellagic acid | {"name": "Ellagic acid", "id": "CID_5281855"} | INTERACTS_WITH | - | Target | 3303 | HSPA1A | {"name": "HSPA1A", "id": "3303"} |

**Table S9: Complete node and relationship data used to construct the knowledge graph. This table summarizes interactions among FDA-approved drugs, formulation, herbs, diseases (type 2 diabetes, fatty liver, obesity), phytochemicals, and their target genes/proteins. Each entry includes source and target node details, edge type, and node annotations (Entrez Gene ID and gene symbol, where applicable).**

| Source Node Label | Source Node ID | Source Node Name | Source Node Properties | Edge Type | Edge Properties | Target Node Label | Target Node ID | Target Node Name | Target Node Properties |
|---|---|---|---|---|---|---|---|---|---|
| Phytochemical | CID_5281855 | Ellagic acid | {"name": "Ellagic acid", "id": "CID_5281855"} | INTERACTS_WITH | - | Target | 3309 | HSPA5 | {"name": "HSPA5", "id": "3309"} |
| Phytochemical | CID_5281855 | Ellagic acid | {"name": "Ellagic acid", "id": "CID_5281855"} | INTERACTS_WITH | - | Target | 3312 | HSPA8 | {"name": "HSPA8", "id": "3312"} |
| Phytochemical | CID_5281855 | Ellagic acid | {"name": "Ellagic acid", "id": "CID_5281855"} | INTERACTS_WITH | - | Target | 3417 | IDH1 | {"name": "IDH1", "id": "3417"} |
| Phytochemical | CID_5281855 | Ellagic acid | {"name": "Ellagic acid", "id": "CID_5281855"} | INTERACTS_WITH | - | Target | 3480 | IGF1R | {"name": "IGF1R", "id": "3480"} |
| Phytochemical | CID_5281855 | Ellagic acid | {"name": "Ellagic acid", "id": "CID_5281855"} | INTERACTS_WITH | - | Target | 351 | APP | {"name": "APP", "id": "351"} |
| Phytochemical | CID_5281855 | Ellagic acid | {"name": "Ellagic acid", "id": "CID_5281855"} | INTERACTS_WITH | - | Target | 3551 | IKBKB | {"name": "IKBKB", "id": "3551"} |
| Phytochemical | CID_5281855 | Ellagic acid | {"name": "Ellagic acid", "id": "CID_5281855"} | INTERACTS_WITH | - | Target | 3643 | INSR | {"name": "INSR", "id": "3643"} |
| Phytochemical | CID_5281855 | Ellagic acid | {"name": "Ellagic acid", "id": "CID_5281855"} | INTERACTS_WITH | - | Target | 377677 | CA13 | {"name": "CA13", "id": "377677"} |
| Phytochemical | CID_5281855 | Ellagic acid | {"name": "Ellagic acid", "id": "CID_5281855"} | INTERACTS_WITH | - | Target | 3791 | KDR | {"name": "KDR", "id": "3791"} |
| Phytochemical | CID_5281855 | Ellagic acid | {"name": "Ellagic acid", "id": "CID_5281855"} | INTERACTS_WITH | - | Target | 3837 | KPNB1 | {"name": "KPNB1", "id": "3837"} |
| Phytochemical | CID_5281855 | Ellagic acid | {"name": "Ellagic acid", "id": "CID_5281855"} | INTERACTS_WITH | - | Target | 390245 | KDM4E | {"name": "KDM4E", "id": "390245"} |
| Phytochemical | CID_5281855 | Ellagic acid | {"name": "Ellagic acid", "id": "CID_5281855"} | INTERACTS_WITH | - | Target | 4000 | LMNA | {"name": "LMNA", "id": "4000"} |
| Phytochemical | CID_5281855 | Ellagic acid | {"name": "Ellagic acid", "id": "CID_5281855"} | INTERACTS_WITH | - | Target | 4088 | SMAD3 | {"name": "SMAD3", "id": "4088"} |
| Phytochemical | CID_5281855 | Ellagic acid | {"name": "Ellagic acid", "id": "CID_5281855"} | INTERACTS_WITH | - | Target | 4128 | MAOA | {"name": "MAOA", "id": "4128"} |
| Phytochemical | CID_5281855 | Ellagic acid | {"name": "Ellagic acid", "id": "CID_5281855"} | INTERACTS_WITH | - | Target | 4137 | MAPT | {"name": "MAPT", "id": "4137"} |
| Phytochemical | CID_5281855 | Ellagic acid | {"name": "Ellagic acid", "id": "CID_5281855"} | INTERACTS_WITH | - | Target | 4154 | MBNL1 | {"name": "MBNL1", "id": "4154"} |
| Phytochemical | CID_5281855 | Ellagic acid | {"name": "Ellagic acid", "id": "CID_5281855"} | INTERACTS_WITH | - | Target | 4170 | MCL1 | {"name": "MCL1", "id": "4170"} |
| Phytochemical | CID_5281855 | Ellagic acid | {"name": "Ellagic acid", "id": "CID_5281855"} | INTERACTS_WITH | - | Target | 4233 | MET | {"name": "MET", "id": "4233"} |
| Phytochemical | CID_5281855 | Ellagic acid | {"name": "Ellagic acid", "id": "CID_5281855"} | INTERACTS_WITH | - | Target | 4282 | MIF | {"name": "MIF", "id": "4282"} |
| Phytochemical | CID_5281855 | Ellagic acid | {"name": "Ellagic acid", "id": "CID_5281855"} | INTERACTS_WITH | - | Target | 4297 | KMT2A | {"name": "KMT2A", "id": "4297"} |
| Phytochemical | CID_5281855 | Ellagic acid | {"name": "Ellagic acid", "id": "CID_5281855"} | INTERACTS_WITH | - | Target | 43 | ACHE | {"name": "ACHE", "id": "43"} |
| Phytochemical | CID_5281855 | Ellagic acid | {"name": "Ellagic acid", "id": "CID_5281855"} | INTERACTS_WITH | - | Target | 472 | ATM | {"name": "ATM", "id": "472"} |
| Phytochemical | CID_5281855 | Ellagic acid | {"name": "Ellagic acid", "id": "CID_5281855"} | INTERACTS_WITH | - | Target | 4780 | NFE2L2 | {"name": "NFE2L2", "id": "4780"} |
| Phytochemical | CID_5281855 | Ellagic acid | {"name": "Ellagic acid", "id": "CID_5281855"} | INTERACTS_WITH | - | Target | 4869 | NPM1 | {"name": "NPM1", "id": "4869"} |
| Phytochemical | CID_5281855 | Ellagic acid | {"name": "Ellagic acid", "id": "CID_5281855"} | INTERACTS_WITH | - | Target | 4907 | NT5E | {"name": "NT5E", "id": "4907"} |
| Phytochemical | CID_5281855 | Ellagic acid | {"name": "Ellagic acid", "id": "CID_5281855"} | INTERACTS_WITH | - | Target | 51053 | GMNN | {"name": "GMNN", "id": "51053"} |
| Phytochemical | CID_5281855 | Ellagic acid | {"name": "Ellagic acid", "id": "CID_5281855"} | INTERACTS_WITH | - | Target | 51426 | POLK | {"name": "POLK", "id": "51426"} |
| Phytochemical | CID_5281855 | Ellagic acid | {"name": "Ellagic acid", "id": "CID_5281855"} | INTERACTS_WITH | - | Target | 5159 | PDGFRB | {"name": "PDGFRB", "id": "5159"} |
| Phytochemical | CID_5281855 | Ellagic acid | {"name": "Ellagic acid", "id": "CID_5281855"} | INTERACTS_WITH | - | Target | 5300 | PIN1 | {"name": "PIN1", "id": "5300"} |
| Phytochemical | CID_5281855 | Ellagic acid | {"name": "Ellagic acid", "id": "CID_5281855"} | INTERACTS_WITH | - | Target | 5313 | PKLR | {"name": "PKLR", "id": "5313"} |
| Phytochemical | CID_5281855 | Ellagic acid | {"name": "Ellagic acid", "id": "CID_5281855"} | INTERACTS_WITH | - | Target | 5347 | PLK1 | {"name": "PLK1", "id": "5347"} |
| Phytochemical | CID_5281855 | Ellagic acid | {"name": "Ellagic acid", "id": "CID_5281855"} | INTERACTS_WITH | - | Target | 5423 | POLB | {"name": "POLB", "id": "5423"} |
| Phytochemical | CID_5281855 | Ellagic acid | {"name": "Ellagic acid", "id": "CID_5281855"} | INTERACTS_WITH | - | Target | 5429 | POLH | {"name": "POLH", "id": "5429"} |
| Phytochemical | CID_5281855 | Ellagic acid | {"name": "Ellagic acid", "id": "CID_5281855"} | INTERACTS_WITH | - | Target | 5566 | PRKACA | {"name": "PRKACA", "id": "5566"} |
| Phytochemical | CID_5281855 | Ellagic acid | {"name": "Ellagic acid", "id": "CID_5281855"} | INTERACTS_WITH | - | Target | 5567 | PRKACB | {"name": "PRKACB", "id": "5567"} |
| Phytochemical | CID_5281855 | Ellagic acid | {"name": "Ellagic acid", "id": "CID_5281855"} | INTERACTS_WITH | - | Target | 5568 | PRKACG | {"name": "PRKACG", "id": "5568"} |
| Phytochemical | CID_5281855 | Ellagic acid | {"name": "Ellagic acid", "id": "CID_5281855"} | INTERACTS_WITH | - | Target | 55775 | TDP1 | {"name": "TDP1", "id": "55775"} |
| Phytochemical | CID_5281855 | Ellagic acid | {"name": "Ellagic acid", "id": "CID_5281855"} | INTERACTS_WITH | - | Target | 55867 | SLC22A11 | {"name": "SLC22A11", "id": "55867"} |
| Phytochemical | CID_5281855 | Ellagic acid | {"name": "Ellagic acid", "id": "CID_5281855"} | INTERACTS_WITH | - | Target | 5594 | MAPK1 | {"name": "MAPK1", "id": "5594"} |
| Phytochemical | CID_5281855 | Ellagic acid | {"name": "Ellagic acid", "id": "CID_5281855"} | INTERACTS_WITH | - | Target | 5747 | PTK2 | {"name": "PTK2", "id": "5747"} |
| Phytochemical | CID_5281855 | Ellagic acid | {"name": "Ellagic acid", "id": "CID_5281855"} | INTERACTS_WITH | - | Target | 5770 | PTPN1 | {"name": "PTPN1", "id": "5770"} |
| Phytochemical | CID_5281855 | Ellagic acid | {"name": "Ellagic acid", "id": "CID_5281855"} | INTERACTS_WITH | - | Target | 57787 | MARK4 | {"name": "MARK4", "id": "57787"} |
| Phytochemical | CID_5281855 | Ellagic acid | {"name": "Ellagic acid", "id": "CID_5281855"} | INTERACTS_WITH | - | Target | 595 | CCND1 | {"name": "CCND1", "id": "595"} |
| Phytochemical | CID_5281855 | Ellagic acid | {"name": "Ellagic acid", "id": "CID_5281855"} | INTERACTS_WITH | - | Target | 5965 | RECQL | {"name": "RECQL", "id": "5965"} |
| Phytochemical | CID_5281855 | Ellagic acid | {"name": "Ellagic acid", "id": "CID_5281855"} | INTERACTS_WITH | - | Target | 5979 | RET | {"name": "RET", "id": "5979"} |
| Phytochemical | CID_5281855 | Ellagic acid | {"name": "Ellagic acid", "id": "CID_5281855"} | INTERACTS_WITH | - | Target | 5999 | RGS4 | {"name": "RGS4", "id": "5999"} |
| Phytochemical | CID_5281855 | Ellagic acid | {"name": "Ellagic acid", "id": "CID_5281855"} | INTERACTS_WITH | - | Target | 641 | BLM | {"name": "BLM", "id": "641"} |
| Phytochemical | CID_5281855 | Ellagic acid | {"name": "Ellagic acid", "id": "CID_5281855"} | INTERACTS_WITH | - | Target | 6714 | SRC | {"name": "SRC", "id": "6714"} |
| Phytochemical | CID_5281855 | Ellagic acid | {"name": "Ellagic acid", "id": "CID_5281855"} | INTERACTS_WITH | - | Target | 673 | BRAF | {"name": "BRAF", "id": "673"} |

**Table S9: Complete node and relationship data used to construct the knowledge graph. This table summarizes interactions among FDA-approved drugs, formulation, herbs, diseases (type 2 diabetes, fatty liver, obesity), phytochemicals, and their target genes/proteins. Each entry includes source and target node details, edge type, and node annotations (Entrez Gene ID and gene symbol, where applicable).**

| Source Node Label | Source Node ID | Source Node Name | Source Node Properties | Edge Type | Edge Properties | Target Node Label | Target Node ID | Target Node Name | Target Node Properties |
|---|---|---|---|---|---|---|---|---|---|
| Phytochemical | CID_5281855 | Ellagic acid | {"name": "Ellagic acid", "id": "CID_5281855"} | INTERACTS_WITH | - | Target | 6790 | AURKA | {"name": "AURKA", "id": "6790"} |
| Phytochemical | CID_5281855 | Ellagic acid | {"name": "Ellagic acid", "id": "CID_5281855"} | INTERACTS_WITH | - | Target | 7010 | TEK | {"name": "TEK", "id": "7010"} |
| Phytochemical | CID_5281855 | Ellagic acid | {"name": "Ellagic acid", "id": "CID_5281855"} | INTERACTS_WITH | - | Target | 7068 | THRB | {"name": "THRB", "id": "7068"} |
| Phytochemical | CID_5281855 | Ellagic acid | {"name": "Ellagic acid", "id": "CID_5281855"} | INTERACTS_WITH | - | Target | 7155 | TOP2B | {"name": "TOP2B", "id": "7155"} |
| Phytochemical | CID_5281855 | Ellagic acid | {"name": "Ellagic acid", "id": "CID_5281855"} | INTERACTS_WITH | - | Target | 7157 | TP53 | {"name": "TP53", "id": "7157"} |
| Phytochemical | CID_5281855 | Ellagic acid | {"name": "Ellagic acid", "id": "CID_5281855"} | INTERACTS_WITH | - | Target | 7257 | TSNAX | {"name": "TSNAX", "id": "7257"} |
| Phytochemical | CID_5281855 | Ellagic acid | {"name": "Ellagic acid", "id": "CID_5281855"} | INTERACTS_WITH | - | Target | 7296 | TXNRD1 | {"name": "TXNRD1", "id": "7296"} |
| Phytochemical | CID_5281855 | Ellagic acid | {"name": "Ellagic acid", "id": "CID_5281855"} | INTERACTS_WITH | - | Target | 7398 | USP1 | {"name": "USP1", "id": "7398"} |
| Phytochemical | CID_5281855 | Ellagic acid | {"name": "Ellagic acid", "id": "CID_5281855"} | INTERACTS_WITH | - | Target | 7421 | VDR | {"name": "VDR", "id": "7421"} |
| Phytochemical | CID_5281855 | Ellagic acid | {"name": "Ellagic acid", "id": "CID_5281855"} | INTERACTS_WITH | - | Target | 759 | CA1 | {"name": "CA1", "id": "759"} |
| Phytochemical | CID_5281855 | Ellagic acid | {"name": "Ellagic acid", "id": "CID_5281855"} | INTERACTS_WITH | - | Target | 760 | CA2 | {"name": "CA2", "id": "760"} |
| Phytochemical | CID_5281855 | Ellagic acid | {"name": "Ellagic acid", "id": "CID_5281855"} | INTERACTS_WITH | - | Target | 761 | CA3 | {"name": "CA3", "id": "761"} |
| Phytochemical | CID_5281855 | Ellagic acid | {"name": "Ellagic acid", "id": "CID_5281855"} | INTERACTS_WITH | - | Target | 762 | CA4 | {"name": "CA4", "id": "762"} |
| Phytochemical | CID_5281855 | Ellagic acid | {"name": "Ellagic acid", "id": "CID_5281855"} | INTERACTS_WITH | - | Target | 763 | CA5A | {"name": "CA5A", "id": "763"} |
| Phytochemical | CID_5281855 | Ellagic acid | {"name": "Ellagic acid", "id": "CID_5281855"} | INTERACTS_WITH | - | Target | 765 | CA6 | {"name": "CA6", "id": "765"} |
| Phytochemical | CID_5281855 | Ellagic acid | {"name": "Ellagic acid", "id": "CID_5281855"} | INTERACTS_WITH | - | Target | 766 | CA7 | {"name": "CA7", "id": "766"} |
| Phytochemical | CID_5281855 | Ellagic acid | {"name": "Ellagic acid", "id": "CID_5281855"} | INTERACTS_WITH | - | Target | 768 | CA9 | {"name": "CA9", "id": "768"} |
| Phytochemical | CID_5281855 | Ellagic acid | {"name": "Ellagic acid", "id": "CID_5281855"} | INTERACTS_WITH | - | Target | 771 | CA12 | {"name": "CA12", "id": "771"} |
| Phytochemical | CID_5281855 | Ellagic acid | {"name": "Ellagic acid", "id": "CID_5281855"} | INTERACTS_WITH | - | Target | 861 | RUNX1 | {"name": "RUNX1", "id": "861"} |
| Phytochemical | CID_5281855 | Ellagic acid | {"name": "Ellagic acid", "id": "CID_5281855"} | INTERACTS_WITH | - | Target | 865 | CBFB | {"name": "CBFB", "id": "865"} |
| Phytochemical | CID_5281855 | Ellagic acid | {"name": "Ellagic acid", "id": "CID_5281855"} | INTERACTS_WITH | - | Target | 8877 | SPHK1 | {"name": "SPHK1", "id": "8877"} |
| Phytochemical | CID_5281855 | Ellagic acid | {"name": "Ellagic acid", "id": "CID_5281855"} | INTERACTS_WITH | - | Target | 890 | CCNA2 | {"name": "CCNA2", "id": "890"} |
| Phytochemical | CID_5281855 | Ellagic acid | {"name": "Ellagic acid", "id": "CID_5281855"} | INTERACTS_WITH | - | Target | 8900 | CCNA1 | {"name": "CCNA1", "id": "8900"} |
| Phytochemical | CID_5281855 | Ellagic acid | {"name": "Ellagic acid", "id": "CID_5281855"} | INTERACTS_WITH | - | Target | 9099 | USP2 | {"name": "USP2", "id": "9099"} |
| Phytochemical | CID_5281855 | Ellagic acid | {"name": "Ellagic acid", "id": "CID_5281855"} | INTERACTS_WITH | - | Target | 9212 | AURKB | {"name": "AURKB", "id": "9212"} |
| Phytochemical | CID_5281855 | Ellagic acid | {"name": "Ellagic acid", "id": "CID_5281855"} | INTERACTS_WITH | - | Target | 9356 | SLC22A6 | {"name": "SLC22A6", "id": "9356"} |
| Phytochemical | CID_5281855 | Ellagic acid | {"name": "Ellagic acid", "id": "CID_5281855"} | INTERACTS_WITH | - | Target | 9682 | KDM4A | {"name": "KDM4A", "id": "9682"} |
| Phytochemical | CID_5281855 | Ellagic acid | {"name": "Ellagic acid", "id": "CID_5281855"} | INTERACTS_WITH | - | Target | 9891 | NUAK1 | {"name": "NUAK1", "id": "9891"} |
| Phytochemical | CID_13250 | Ethyl gallate | {"name": "Ethyl gallate", "id": "CID_13250"} | INTERACTS_WITH | - | Target | 2099 | ESR1 | {"name": "ESR1", "id": "2099"} |
| Phytochemical | CID_13250 | Ethyl gallate | {"name": "Ethyl gallate", "id": "CID_13250"} | INTERACTS_WITH | - | Target | 22943 | DKK1 | {"name": "DKK1", "id": "22943"} |
| Phytochemical | CID_13250 | Ethyl gallate | {"name": "Ethyl gallate", "id": "CID_13250"} | INTERACTS_WITH | - | Target | 4040 | LRP6 | {"name": "LRP6", "id": "4040"} |
| Phytochemical | CID_2758 | Eucalyptol | {"name": "Eucalyptol", "id": "CID_2758"} | INTERACTS_WITH | - | Target | 10062 | NR1H3 | {"name": "NR1H3", "id": "10062"} |
| Phytochemical | CID_2758 | Eucalyptol | {"name": "Eucalyptol", "id": "CID_2758"} | INTERACTS_WITH | - | Target | 10280 | SIGMAR1 | {"name": "SIGMAR1", "id": "10280"} |
| Phytochemical | CID_2758 | Eucalyptol | {"name": "Eucalyptol", "id": "CID_2758"} | INTERACTS_WITH | - | Target | 10800 | CYSLTR1 | {"name": "CYSLTR1", "id": "10800"} |
| Phytochemical | CID_2758 | Eucalyptol | {"name": "Eucalyptol", "id": "CID_2758"} | INTERACTS_WITH | - | Target | 10951 | CBX1 | {"name": "CBX1", "id": "10951"} |
| Phytochemical | CID_2758 | Eucalyptol | {"name": "Eucalyptol", "id": "CID_2758"} | INTERACTS_WITH | - | Target | 1128 | CHRM1 | {"name": "CHRM1", "id": "1128"} |
| Phytochemical | CID_2758 | Eucalyptol | {"name": "Eucalyptol", "id": "CID_2758"} | INTERACTS_WITH | - | Target | 1129 | CHRM2 | {"name": "CHRM2", "id": "1129"} |
| Phytochemical | CID_2758 | Eucalyptol | {"name": "Eucalyptol", "id": "CID_2758"} | INTERACTS_WITH | - | Target | 1131 | CHRM3 | {"name": "CHRM3", "id": "1131"} |
| Phytochemical | CID_2758 | Eucalyptol | {"name": "Eucalyptol", "id": "CID_2758"} | INTERACTS_WITH | - | Target | 1132 | CHRM4 | {"name": "CHRM4", "id": "1132"} |
| Phytochemical | CID_2758 | Eucalyptol | {"name": "Eucalyptol", "id": "CID_2758"} | INTERACTS_WITH | - | Target | 1133 | CHRM5 | {"name": "CHRM5", "id": "1133"} |
| Phytochemical | CID_2758 | Eucalyptol | {"name": "Eucalyptol", "id": "CID_2758"} | INTERACTS_WITH | - | Target | 1233 | CCR4 | {"name": "CCR4", "id": "1233"} |
| Phytochemical | CID_2758 | Eucalyptol | {"name": "Eucalyptol", "id": "CID_2758"} | INTERACTS_WITH | - | Target | 1234 | CCR5 | {"name": "CCR5", "id": "1234"} |
| Phytochemical | CID_2758 | Eucalyptol | {"name": "Eucalyptol", "id": "CID_2758"} | INTERACTS_WITH | - | Target | 1268 | CNR1 | {"name": "CNR1", "id": "1268"} |
| Phytochemical | CID_2758 | Eucalyptol | {"name": "Eucalyptol", "id": "CID_2758"} | INTERACTS_WITH | - | Target | 134 | ADORA1 | {"name": "ADORA1", "id": "134"} |
| Phytochemical | CID_2758 | Eucalyptol | {"name": "Eucalyptol", "id": "CID_2758"} | INTERACTS_WITH | - | Target | 135 | ADORA2A | {"name": "ADORA2A", "id": "135"} |
| Phytochemical | CID_2758 | Eucalyptol | {"name": "Eucalyptol", "id": "CID_2758"} | INTERACTS_WITH | - | Target | 140 | ADORA3 | {"name": "ADORA3", "id": "140"} |
| Phytochemical | CID_2758 | Eucalyptol | {"name": "Eucalyptol", "id": "CID_2758"} | INTERACTS_WITH | - | Target | 1432 | MAPK14 | {"name": "MAPK14", "id": "1432"} |
| Phytochemical | CID_2758 | Eucalyptol | {"name": "Eucalyptol", "id": "CID_2758"} | INTERACTS_WITH | - | Target | 146 | ADRA1D | {"name": "ADRA1D", "id": "146"} |
| Phytochemical | CID_2758 | Eucalyptol | {"name": "Eucalyptol", "id": "CID_2758"} | INTERACTS_WITH | - | Target | 150 | ADRA2A | {"name": "ADRA2A", "id": "150"} |
| Phytochemical | CID_2758 | Eucalyptol | {"name": "Eucalyptol", "id": "CID_2758"} | INTERACTS_WITH | - | Target | 151 | ADRA2B | {"name": "ADRA2B", "id": "151"} |
| Phytochemical | CID_2758 | Eucalyptol | {"name": "Eucalyptol", "id": "CID_2758"} | INTERACTS_WITH | - | Target | 1511 | CTSG | {"name": "CTSG", "id": "1511"} |
| Phytochemical | CID_2758 | Eucalyptol | {"name": "Eucalyptol", "id": "CID_2758"} | INTERACTS_WITH | - | Target | 152 | ADRA2C | {"name": "ADRA2C", "id": "152"} |
| Phytochemical | CID_2758 | Eucalyptol | {"name": "Eucalyptol", "id": "CID_2758"} | INTERACTS_WITH | - | Target | 153 | ADRB1 | {"name": "ADRB1", "id": "153"} |
| Phytochemical | CID_2758 | Eucalyptol | {"name": "Eucalyptol", "id": "CID_2758"} | INTERACTS_WITH | - | Target | 154 | ADRB2 | {"name": "ADRB2", "id": "154"} |
| Phytochemical | CID_2758 | Eucalyptol | {"name": "Eucalyptol", "id": "CID_2758"} | INTERACTS_WITH | - | Target | 1544 | CYP1A2 | {"name": "CYP1A2", "id": "1544"} |
| Phytochemical | CID_2758 | Eucalyptol | {"name": "Eucalyptol", "id": "CID_2758"} | INTERACTS_WITH | - | Target | 1548 | CYP2A6 | {"name": "CYP2A6", "id": "1548"} |
| Phytochemical | CID_2758 | Eucalyptol | {"name": "Eucalyptol", "id": "CID_2758"} | INTERACTS_WITH | - | Target | 155 | ADRB3 | {"name": "ADRB3", "id": "155"} |
| Phytochemical | CID_2758 | Eucalyptol | {"name": "Eucalyptol", "id": "CID_2758"} | INTERACTS_WITH | - | Target | 1557 | CYP2C19 | {"name": "CYP2C19", "id": "1557"} |
| Phytochemical | CID_2758 | Eucalyptol | {"name": "Eucalyptol", "id": "CID_2758"} | INTERACTS_WITH | - | Target | 1559 | CYP2C9 | {"name": "CYP2C9", "id": "1559"} |

**Table S9: Complete node and relationship data used to construct the knowledge graph. This table summarizes interactions among FDA-approved drugs, formulation, herbs, diseases (type 2 diabetes, fatty liver, obesity), phytochemicals, and their target genes/proteins. Each entry includes source and target node details, edge type, and node annotations (Entrez Gene ID and gene symbol, where applicable).**

| Source Node Label | Source Node ID | Source Node Name | Source Node Properties | Edge Type | Edge Properties | Target Node Label | Target Node ID | Target Node Name | Target Node Properties |
|---|---|---|---|---|---|---|---|---|---|
| Phytochemical | CID_2758 | Eucalyptol | {"name": "Eucalyptol", "id": "CID_2758"} | INTERACTS_WITH | - | Target | 1565 | CYP2D6 | {"name": "CYP2D6", "id": "1565"} |
| Phytochemical | CID_2758 | Eucalyptol | {"name": "Eucalyptol", "id": "CID_2758"} | INTERACTS_WITH | - | Target | 1571 | CYP2E1 | {"name": "CYP2E1", "id": "1571"} |
| Phytochemical | CID_2758 | Eucalyptol | {"name": "Eucalyptol", "id": "CID_2758"} | INTERACTS_WITH | - | Target | 1576 | CYP3A4 | {"name": "CYP3A4", "id": "1576"} |
| Phytochemical | CID_2758 | Eucalyptol | {"name": "Eucalyptol", "id": "CID_2758"} | INTERACTS_WITH | - | Target | 1812 | DRD1 | {"name": "DRD1", "id": "1812"} |
| Phytochemical | CID_2758 | Eucalyptol | {"name": "Eucalyptol", "id": "CID_2758"} | INTERACTS_WITH | - | Target | 1813 | DRD2 | {"name": "DRD2", "id": "1813"} |
| Phytochemical | CID_2758 | Eucalyptol | {"name": "Eucalyptol", "id": "CID_2758"} | INTERACTS_WITH | - | Target | 1814 | DRD3 | {"name": "DRD3", "id": "1814"} |
| Phytochemical | CID_2758 | Eucalyptol | {"name": "Eucalyptol", "id": "CID_2758"} | INTERACTS_WITH | - | Target | 1815 | DRD4 | {"name": "DRD4", "id": "1815"} |
| Phytochemical | CID_2758 | Eucalyptol | {"name": "Eucalyptol", "id": "CID_2758"} | INTERACTS_WITH | - | Target | 186 | AGTR2 | {"name": "AGTR2", "id": "186"} |
| Phytochemical | CID_2758 | Eucalyptol | {"name": "Eucalyptol", "id": "CID_2758"} | INTERACTS_WITH | - | Target | 1909 | EDNRA | {"name": "EDNRA", "id": "1909"} |
| Phytochemical | CID_2758 | Eucalyptol | {"name": "Eucalyptol", "id": "CID_2758"} | INTERACTS_WITH | - | Target | 1956 | EGFR | {"name": "EGFR", "id": "1956"} |
| Phytochemical | CID_2758 | Eucalyptol | {"name": "Eucalyptol", "id": "CID_2758"} | INTERACTS_WITH | - | Target | 1991 | ELANE | {"name": "ELANE", "id": "1991"} |
| Phytochemical | CID_2758 | Eucalyptol | {"name": "Eucalyptol", "id": "CID_2758"} | INTERACTS_WITH | - | Target | 2064 | ERBB2 | {"name": "ERBB2", "id": "2064"} |
| Phytochemical | CID_2758 | Eucalyptol | {"name": "Eucalyptol", "id": "CID_2758"} | INTERACTS_WITH | - | Target | 2099 | ESR1 | {"name": "ESR1", "id": "2099"} |
| Phytochemical | CID_2758 | Eucalyptol | {"name": "Eucalyptol", "id": "CID_2758"} | INTERACTS_WITH | - | Target | 2100 | ESR2 | {"name": "ESR2", "id": "2100"} |
| Phytochemical | CID_2758 | Eucalyptol | {"name": "Eucalyptol", "id": "CID_2758"} | INTERACTS_WITH | - | Target | 2321 | FLT1 | {"name": "FLT1", "id": "2321"} |
| Phytochemical | CID_2758 | Eucalyptol | {"name": "Eucalyptol", "id": "CID_2758"} | INTERACTS_WITH | - | Target | 2534 | FYN | {"name": "FYN", "id": "2534"} |
| Phytochemical | CID_2758 | Eucalyptol | {"name": "Eucalyptol", "id": "CID_2758"} | INTERACTS_WITH | - | Target | 25939 | SAMHD1 | {"name": "SAMHD1", "id": "25939"} |
| Phytochemical | CID_2758 | Eucalyptol | {"name": "Eucalyptol", "id": "CID_2758"} | INTERACTS_WITH | - | Target | 2908 | NR3C1 | {"name": "NR3C1", "id": "2908"} |
| Phytochemical | CID_2758 | Eucalyptol | {"name": "Eucalyptol", "id": "CID_2758"} | INTERACTS_WITH | - | Target | 3156 | HMGCR | {"name": "HMGCR", "id": "3156"} |
| Phytochemical | CID_2758 | Eucalyptol | {"name": "Eucalyptol", "id": "CID_2758"} | INTERACTS_WITH | - | Target | 3269 | HRH1 | {"name": "HRH1", "id": "3269"} |
| Phytochemical | CID_2758 | Eucalyptol | {"name": "Eucalyptol", "id": "CID_2758"} | INTERACTS_WITH | - | Target | 3274 | HRH2 | {"name": "HRH2", "id": "3274"} |
| Phytochemical | CID_2758 | Eucalyptol | {"name": "Eucalyptol", "id": "CID_2758"} | INTERACTS_WITH | - | Target | 328 | APEX1 | {"name": "APEX1", "id": "328"} |
| Phytochemical | CID_2758 | Eucalyptol | {"name": "Eucalyptol", "id": "CID_2758"} | INTERACTS_WITH | - | Target | 3356 | HTR2A | {"name": "HTR2A", "id": "3356"} |
| Phytochemical | CID_2758 | Eucalyptol | {"name": "Eucalyptol", "id": "CID_2758"} | INTERACTS_WITH | - | Target | 3357 | HTR2B | {"name": "HTR2B", "id": "3357"} |
| Phytochemical | CID_2758 | Eucalyptol | {"name": "Eucalyptol", "id": "CID_2758"} | INTERACTS_WITH | - | Target | 3358 | HTR2C | {"name": "HTR2C", "id": "3358"} |
| Phytochemical | CID_2758 | Eucalyptol | {"name": "Eucalyptol", "id": "CID_2758"} | INTERACTS_WITH | - | Target | 3359 | HTR3A | {"name": "HTR3A", "id": "3359"} |
| Phytochemical | CID_2758 | Eucalyptol | {"name": "Eucalyptol", "id": "CID_2758"} | INTERACTS_WITH | - | Target | 3362 | HTR6 | {"name": "HTR6", "id": "3362"} |
| Phytochemical | CID_2758 | Eucalyptol | {"name": "Eucalyptol", "id": "CID_2758"} | INTERACTS_WITH | - | Target | 3577 | CXCR1 | {"name": "CXCR1", "id": "3577"} |
| Phytochemical | CID_2758 | Eucalyptol | {"name": "Eucalyptol", "id": "CID_2758"} | INTERACTS_WITH | - | Target | 3579 | CXCR2 | {"name": "CXCR2", "id": "3579"} |
| Phytochemical | CID_2758 | Eucalyptol | {"name": "Eucalyptol", "id": "CID_2758"} | INTERACTS_WITH | - | Target | 367 | AR | {"name": "AR", "id": "367"} |
| Phytochemical | CID_2758 | Eucalyptol | {"name": "Eucalyptol", "id": "CID_2758"} | INTERACTS_WITH | - | Target | 3725 | JUN | {"name": "JUN", "id": "3725"} |
| Phytochemical | CID_2758 | Eucalyptol | {"name": "Eucalyptol", "id": "CID_2758"} | INTERACTS_WITH | - | Target | 3757 | KCNH2 | {"name": "KCNH2", "id": "3757"} |
| Phytochemical | CID_2758 | Eucalyptol | {"name": "Eucalyptol", "id": "CID_2758"} | INTERACTS_WITH | - | Target | 3932 | LCK | {"name": "LCK", "id": "3932"} |
| Phytochemical | CID_2758 | Eucalyptol | {"name": "Eucalyptol", "id": "CID_2758"} | INTERACTS_WITH | - | Target | 4000 | LMNA | {"name": "LMNA", "id": "4000"} |
| Phytochemical | CID_2758 | Eucalyptol | {"name": "Eucalyptol", "id": "CID_2758"} | INTERACTS_WITH | - | Target | 4128 | MAOA | {"name": "MAOA", "id": "4128"} |
| Phytochemical | CID_2758 | Eucalyptol | {"name": "Eucalyptol", "id": "CID_2758"} | INTERACTS_WITH | - | Target | 4137 | MAPT | {"name": "MAPT", "id": "4137"} |
| Phytochemical | CID_2758 | Eucalyptol | {"name": "Eucalyptol", "id": "CID_2758"} | INTERACTS_WITH | - | Target | 4159 | MC3R | {"name": "MC3R", "id": "4159"} |
| Phytochemical | CID_2758 | Eucalyptol | {"name": "Eucalyptol", "id": "CID_2758"} | INTERACTS_WITH | - | Target | 4160 | MC4R | {"name": "MC4R", "id": "4160"} |
| Phytochemical | CID_2758 | Eucalyptol | {"name": "Eucalyptol", "id": "CID_2758"} | INTERACTS_WITH | - | Target | 4161 | MC5R | {"name": "MC5R", "id": "4161"} |
| Phytochemical | CID_2758 | Eucalyptol | {"name": "Eucalyptol", "id": "CID_2758"} | INTERACTS_WITH | - | Target | 43 | ACHE | {"name": "ACHE", "id": "43"} |
| Phytochemical | CID_2758 | Eucalyptol | {"name": "Eucalyptol", "id": "CID_2758"} | INTERACTS_WITH | - | Target | 4312 | MMP1 | {"name": "MMP1", "id": "4312"} |
| Phytochemical | CID_2758 | Eucalyptol | {"name": "Eucalyptol", "id": "CID_2758"} | INTERACTS_WITH | - | Target | 4318 | MMP9 | {"name": "MMP9", "id": "4318"} |
| Phytochemical | CID_2758 | Eucalyptol | {"name": "Eucalyptol", "id": "CID_2758"} | INTERACTS_WITH | - | Target | 4886 | NPY1R | {"name": "NPY1R", "id": "4886"} |
| Phytochemical | CID_2758 | Eucalyptol | {"name": "Eucalyptol", "id": "CID_2758"} | INTERACTS_WITH | - | Target | 4887 | NPY2R | {"name": "NPY2R", "id": "4887"} |
| Phytochemical | CID_2758 | Eucalyptol | {"name": "Eucalyptol", "id": "CID_2758"} | INTERACTS_WITH | - | Target | 4985 | OPRD1 | {"name": "OPRD1", "id": "4985"} |
| Phytochemical | CID_2758 | Eucalyptol | {"name": "Eucalyptol", "id": "CID_2758"} | INTERACTS_WITH | - | Target | 4986 | OPRK1 | {"name": "OPRK1", "id": "4986"} |
| Phytochemical | CID_2758 | Eucalyptol | {"name": "Eucalyptol", "id": "CID_2758"} | INTERACTS_WITH | - | Target | 4988 | OPRM1 | {"name": "OPRM1", "id": "4988"} |
| Phytochemical | CID_2758 | Eucalyptol | {"name": "Eucalyptol", "id": "CID_2758"} | INTERACTS_WITH | - | Target | 51053 | GMNN | {"name": "GMNN", "id": "51053"} |
| Phytochemical | CID_2758 | Eucalyptol | {"name": "Eucalyptol", "id": "CID_2758"} | INTERACTS_WITH | - | Target | 5467 | PPARD | {"name": "PPARD", "id": "5467"} |
| Phytochemical | CID_2758 | Eucalyptol | {"name": "Eucalyptol", "id": "CID_2758"} | INTERACTS_WITH | - | Target | 552 | AVPR1A | {"name": "AVPR1A", "id": "552"} |
| Phytochemical | CID_2758 | Eucalyptol | {"name": "Eucalyptol", "id": "CID_2758"} | INTERACTS_WITH | - | Target | 5530 | PPP3CA | {"name": "PPP3CA", "id": "5530"} |
| Phytochemical | CID_2758 | Eucalyptol | {"name": "Eucalyptol", "id": "CID_2758"} | INTERACTS_WITH | - | Target | 55775 | TDP1 | {"name": "TDP1", "id": "55775"} |
| Phytochemical | CID_2758 | Eucalyptol | {"name": "Eucalyptol", "id": "CID_2758"} | INTERACTS_WITH | - | Target | 5578 | PRKCA | {"name": "PRKCA", "id": "5578"} |
| Phytochemical | CID_2758 | Eucalyptol | {"name": "Eucalyptol", "id": "CID_2758"} | INTERACTS_WITH | - | Target | 5594 | MAPK1 | {"name": "MAPK1", "id": "5594"} |
| Phytochemical | CID_2758 | Eucalyptol | {"name": "Eucalyptol", "id": "CID_2758"} | INTERACTS_WITH | - | Target | 5595 | MAPK3 | {"name": "MAPK3", "id": "5595"} |
| Phytochemical | CID_2758 | Eucalyptol | {"name": "Eucalyptol", "id": "CID_2758"} | INTERACTS_WITH | - | Target | 5724 | PTAFR | {"name": "PTAFR", "id": "5724"} |
| Phytochemical | CID_2758 | Eucalyptol | {"name": "Eucalyptol", "id": "CID_2758"} | INTERACTS_WITH | - | Target | 5742 | PTGS1 | {"name": "PTGS1", "id": "5742"} |
| Phytochemical | CID_2758 | Eucalyptol | {"name": "Eucalyptol", "id": "CID_2758"} | INTERACTS_WITH | - | Target | 5743 | PTGS2 | {"name": "PTGS2", "id": "5743"} |
| Phytochemical | CID_2758 | Eucalyptol | {"name": "Eucalyptol", "id": "CID_2758"} | INTERACTS_WITH | - | Target | 5788 | PTPRC | {"name": "PTPRC", "id": "5788"} |
| Phytochemical | CID_2758 | Eucalyptol | {"name": "Eucalyptol", "id": "CID_2758"} | INTERACTS_WITH | - | Target | 5965 | RECQL | {"name": "RECQL", "id": "5965"} |
| Phytochemical | CID_2758 | Eucalyptol | {"name": "Eucalyptol", "id": "CID_2758"} | INTERACTS_WITH | - | Target | 624 | BDKRB2 | {"name": "BDKRB2", "id": "624"} |
| Phytochemical | CID_2758 | Eucalyptol | {"name": "Eucalyptol", "id": "CID_2758"} | INTERACTS_WITH | - | Target | 6530 | SLC6A2 | {"name": "SLC6A2", "id": "6530"} |
| Phytochemical | CID_2758 | Eucalyptol | {"name": "Eucalyptol", "id": "CID_2758"} | INTERACTS_WITH | - | Target | 6531 | SLC6A3 | {"name": "SLC6A3", "id": "6531"} |
| Phytochemical | CID_2758 | Eucalyptol | {"name": "Eucalyptol", "id": "CID_2758"} | INTERACTS_WITH | - | Target | 6532 | SLC6A4 | {"name": "SLC6A4", "id": "6532"} |
| Phytochemical | CID_2758 | Eucalyptol | {"name": "Eucalyptol", "id": "CID_2758"} | INTERACTS_WITH | - | Target | 6865 | TACR2 | {"name": "TACR2", "id": "6865"} |
| Phytochemical | CID_2758 | Eucalyptol | {"name": "Eucalyptol", "id": "CID_2758"} | INTERACTS_WITH | - | Target | 6869 | TACR1 | {"name": "TACR1", "id": "6869"} |
| Phytochemical | CID_2758 | Eucalyptol | {"name": "Eucalyptol", "id": "CID_2758"} | INTERACTS_WITH | - | Target | 6916 | TBXAS1 | {"name": "TBXAS1", "id": "6916"} |
| Phytochemical | CID_2758 | Eucalyptol | {"name": "Eucalyptol", "id": "CID_2758"} | INTERACTS_WITH | - | Target | 7253 | TSHR | {"name": "TSHR", "id": "7253"} |
| Phytochemical | CID_2758 | Eucalyptol | {"name": "Eucalyptol", "id": "CID_2758"} | INTERACTS_WITH | - | Target | 729230 | CCR2 | {"name": "CCR2", "id": "729230"} |
| Phytochemical | CID_2758 | Eucalyptol | {"name": "Eucalyptol", "id": "CID_2758"} | INTERACTS_WITH | - | Target | 7433 | VIPR1 | {"name": "VIPR1", "id": "7433"} |
| Phytochemical | CID_2758 | Eucalyptol | {"name": "Eucalyptol", "id": "CID_2758"} | INTERACTS_WITH | - | Target | 760 | CA2 | {"name": "CA2", "id": "760"} |
| Phytochemical | CID_2758 | Eucalyptol | {"name": "Eucalyptol", "id": "CID_2758"} | INTERACTS_WITH | - | Target | 79054 | TRPM8 | {"name": "TRPM8", "id": "79054"} |
| Phytochemical | CID_2758 | Eucalyptol | {"name": "Eucalyptol", "id": "CID_2758"} | INTERACTS_WITH | - | Target | 799 | CALCR | {"name": "CALCR", "id": "799"} |
| Phytochemical | CID_2758 | Eucalyptol | {"name": "Eucalyptol", "id": "CID_2758"} | INTERACTS_WITH | - | Target | 834 | CASP1 | {"name": "CASP1", "id": "834"} |
| Phytochemical | CID_2758 | Eucalyptol | {"name": "Eucalyptol", "id": "CID_2758"} | INTERACTS_WITH | - | Target | 8654 | PDE5A | {"name": "PDE5A", "id": "8654"} |
| Phytochemical | CID_2758 | Eucalyptol | {"name": "Eucalyptol", "id": "CID_2758"} | INTERACTS_WITH | - | Target | 886 | CCKAR | {"name": "CCKAR", "id": "886"} |

**Table S9: Complete node and relationship data used to construct the knowledge graph. This table summarizes interactions among FDA-approved drugs, formulation, herbs, diseases (type 2 diabetes, fatty liver, obesity), phytochemicals, and their target genes/proteins. Each entry includes source and target node details, edge type, and node annotations (Entrez Gene ID and gene symbol, where applicable).**

| Source Node Label | Source Node ID | Source Node Name | Source Node Properties | Edge Type | Edge Properties | Target Node Label | Target Node ID | Target Node Name | Target Node Properties |
|---|---|---|---|---|---|---|---|---|---|
| Phytochemical | CID_3314 | Eugenol | {"name": "Eugenol", "id": "CID_3314"} | INTERACTS_WITH | - | Target | 10280 | SIGMAR1 | {"name": "SIGMAR1", "id": "10280"} |
| Phytochemical | CID_3314 | Eugenol | {"name": "Eugenol", "id": "CID_3314"} | INTERACTS_WITH | - | Target | 10599 | SLCO1B1 | {"name": "SLCO1B1", "id": "10599"} |
| Phytochemical | CID_3314 | Eugenol | {"name": "Eugenol", "id": "CID_3314"} | INTERACTS_WITH | - | Target | 10800 | CYSLTR1 | {"name": "CYSLTR1", "id": "10800"} |
| Phytochemical | CID_3314 | Eugenol | {"name": "Eugenol", "id": "CID_3314"} | INTERACTS_WITH | - | Target | 10941 | UGT2A1 | {"name": "UGT2A1", "id": "10941"} |
| Phytochemical | CID_3314 | Eugenol | {"name": "Eugenol", "id": "CID_3314"} | INTERACTS_WITH | - | Target | 1128 | CHRM1 | {"name": "CHRM1", "id": "1128"} |
| Phytochemical | CID_3314 | Eugenol | {"name": "Eugenol", "id": "CID_3314"} | INTERACTS_WITH | - | Target | 1129 | CHRM2 | {"name": "CHRM2", "id": "1129"} |
| Phytochemical | CID_3314 | Eugenol | {"name": "Eugenol", "id": "CID_3314"} | INTERACTS_WITH | - | Target | 1131 | CHRM3 | {"name": "CHRM3", "id": "1131"} |
| Phytochemical | CID_3314 | Eugenol | {"name": "Eugenol", "id": "CID_3314"} | INTERACTS_WITH | - | Target | 1132 | CHRM4 | {"name": "CHRM4", "id": "1132"} |
| Phytochemical | CID_3314 | Eugenol | {"name": "Eugenol", "id": "CID_3314"} | INTERACTS_WITH | - | Target | 1133 | CHRM5 | {"name": "CHRM5", "id": "1133"} |
| Phytochemical | CID_3314 | Eugenol | {"name": "Eugenol", "id": "CID_3314"} | INTERACTS_WITH | - | Target | 1233 | CCR4 | {"name": "CCR4", "id": "1233"} |
| Phytochemical | CID_3314 | Eugenol | {"name": "Eugenol", "id": "CID_3314"} | INTERACTS_WITH | - | Target | 1234 | CCR5 | {"name": "CCR5", "id": "1234"} |
| Phytochemical | CID_3314 | Eugenol | {"name": "Eugenol", "id": "CID_3314"} | INTERACTS_WITH | - | Target | 1268 | CNR1 | {"name": "CNR1", "id": "1268"} |
| Phytochemical | CID_3314 | Eugenol | {"name": "Eugenol", "id": "CID_3314"} | INTERACTS_WITH | - | Target | 134 | ADORA1 | {"name": "ADORA1", "id": "134"} |
| Phytochemical | CID_3314 | Eugenol | {"name": "Eugenol", "id": "CID_3314"} | INTERACTS_WITH | - | Target | 135 | ADORA2A | {"name": "ADORA2A", "id": "135"} |
| Phytochemical | CID_3314 | Eugenol | {"name": "Eugenol", "id": "CID_3314"} | INTERACTS_WITH | - | Target | 140 | ADORA3 | {"name": "ADORA3", "id": "140"} |
| Phytochemical | CID_3314 | Eugenol | {"name": "Eugenol", "id": "CID_3314"} | INTERACTS_WITH | - | Target | 1432 | MAPK14 | {"name": "MAPK14", "id": "1432"} |
| Phytochemical | CID_3314 | Eugenol | {"name": "Eugenol", "id": "CID_3314"} | INTERACTS_WITH | - | Target | 146 | ADRA1D | {"name": "ADRA1D", "id": "146"} |
| Phytochemical | CID_3314 | Eugenol | {"name": "Eugenol", "id": "CID_3314"} | INTERACTS_WITH | - | Target | 150 | ADRA2A | {"name": "ADRA2A", "id": "150"} |
| Phytochemical | CID_3314 | Eugenol | {"name": "Eugenol", "id": "CID_3314"} | INTERACTS_WITH | - | Target | 151 | ADRA2B | {"name": "ADRA2B", "id": "151"} |
| Phytochemical | CID_3314 | Eugenol | {"name": "Eugenol", "id": "CID_3314"} | INTERACTS_WITH | - | Target | 1511 | CTSG | {"name": "CTSG", "id": "1511"} |
| Phytochemical | CID_3314 | Eugenol | {"name": "Eugenol", "id": "CID_3314"} | INTERACTS_WITH | - | Target | 152 | ADRA2C | {"name": "ADRA2C", "id": "152"} |
| Phytochemical | CID_3314 | Eugenol | {"name": "Eugenol", "id": "CID_3314"} | INTERACTS_WITH | - | Target | 153 | ADRB1 | {"name": "ADRB1", "id": "153"} |
| Phytochemical | CID_3314 | Eugenol | {"name": "Eugenol", "id": "CID_3314"} | INTERACTS_WITH | - | Target | 154 | ADRB2 | {"name": "ADRB2", "id": "154"} |
| Phytochemical | CID_3314 | Eugenol | {"name": "Eugenol", "id": "CID_3314"} | INTERACTS_WITH | - | Target | 1544 | CYP1A2 | {"name": "CYP1A2", "id": "1544"} |
| Phytochemical | CID_3314 | Eugenol | {"name": "Eugenol", "id": "CID_3314"} | INTERACTS_WITH | - | Target | 1548 | CYP2A6 | {"name": "CYP2A6", "id": "1548"} |
| Phytochemical | CID_3314 | Eugenol | {"name": "Eugenol", "id": "CID_3314"} | INTERACTS_WITH | - | Target | 155 | ADRB3 | {"name": "ADRB3", "id": "155"} |
| Phytochemical | CID_3314 | Eugenol | {"name": "Eugenol", "id": "CID_3314"} | INTERACTS_WITH | - | Target | 1557 | CYP2C19 | {"name": "CYP2C19", "id": "1557"} |
| Phytochemical | CID_3314 | Eugenol | {"name": "Eugenol", "id": "CID_3314"} | INTERACTS_WITH | - | Target | 1559 | CYP2C9 | {"name": "CYP2C9", "id": "1559"} |
| Phytochemical | CID_3314 | Eugenol | {"name": "Eugenol", "id": "CID_3314"} | INTERACTS_WITH | - | Target | 1565 | CYP2D6 | {"name": "CYP2D6", "id": "1565"} |
| Phytochemical | CID_3314 | Eugenol | {"name": "Eugenol", "id": "CID_3314"} | INTERACTS_WITH | - | Target | 1571 | CYP2E1 | {"name": "CYP2E1", "id": "1571"} |
| Phytochemical | CID_3314 | Eugenol | {"name": "Eugenol", "id": "CID_3314"} | INTERACTS_WITH | - | Target | 1576 | CYP3A4 | {"name": "CYP3A4", "id": "1576"} |
| Phytochemical | CID_3314 | Eugenol | {"name": "Eugenol", "id": "CID_3314"} | INTERACTS_WITH | - | Target | 1812 | DRD1 | {"name": "DRD1", "id": "1812"} |
| Phytochemical | CID_3314 | Eugenol | {"name": "Eugenol", "id": "CID_3314"} | INTERACTS_WITH | - | Target | 1813 | DRD2 | {"name": "DRD2", "id": "1813"} |
| Phytochemical | CID_3314 | Eugenol | {"name": "Eugenol", "id": "CID_3314"} | INTERACTS_WITH | - | Target | 1814 | DRD3 | {"name": "DRD3", "id": "1814"} |
| Phytochemical | CID_3314 | Eugenol | {"name": "Eugenol", "id": "CID_3314"} | INTERACTS_WITH | - | Target | 1815 | DRD4 | {"name": "DRD4", "id": "1815"} |
| Phytochemical | CID_3314 | Eugenol | {"name": "Eugenol", "id": "CID_3314"} | INTERACTS_WITH | - | Target | 186 | AGTR2 | {"name": "AGTR2", "id": "186"} |
| Phytochemical | CID_3314 | Eugenol | {"name": "Eugenol", "id": "CID_3314"} | INTERACTS_WITH | - | Target | 1909 | EDNRA | {"name": "EDNRA", "id": "1909"} |
| Phytochemical | CID_3314 | Eugenol | {"name": "Eugenol", "id": "CID_3314"} | INTERACTS_WITH | - | Target | 1956 | EGFR | {"name": "EGFR", "id": "1956"} |
| Phytochemical | CID_3314 | Eugenol | {"name": "Eugenol", "id": "CID_3314"} | INTERACTS_WITH | - | Target | 1991 | ELANE | {"name": "ELANE", "id": "1991"} |
| Phytochemical | CID_3314 | Eugenol | {"name": "Eugenol", "id": "CID_3314"} | INTERACTS_WITH | - | Target | 2064 | ERBB2 | {"name": "ERBB2", "id": "2064"} |
| Phytochemical | CID_3314 | Eugenol | {"name": "Eugenol", "id": "CID_3314"} | INTERACTS_WITH | - | Target | 2099 | ESR1 | {"name": "ESR1", "id": "2099"} |
| Phytochemical | CID_3314 | Eugenol | {"name": "Eugenol", "id": "CID_3314"} | INTERACTS_WITH | - | Target | 2100 | ESR2 | {"name": "ESR2", "id": "2100"} |
| Phytochemical | CID_3314 | Eugenol | {"name": "Eugenol", "id": "CID_3314"} | INTERACTS_WITH | - | Target | 216 | ALDH1A1 | {"name": "ALDH1A1", "id": "216"} |
| Phytochemical | CID_3314 | Eugenol | {"name": "Eugenol", "id": "CID_3314"} | INTERACTS_WITH | - | Target | 2321 | FLT1 | {"name": "FLT1", "id": "2321"} |
| Phytochemical | CID_3314 | Eugenol | {"name": "Eugenol", "id": "CID_3314"} | INTERACTS_WITH | - | Target | 240 | ALOX5 | {"name": "ALOX5", "id": "240"} |
| Phytochemical | CID_3314 | Eugenol | {"name": "Eugenol", "id": "CID_3314"} | INTERACTS_WITH | - | Target | 2534 | FYN | {"name": "FYN", "id": "2534"} |
| Phytochemical | CID_3314 | Eugenol | {"name": "Eugenol", "id": "CID_3314"} | INTERACTS_WITH | - | Target | 26339 | OR5K1 | {"name": "OR5K1", "id": "26339"} |
| Phytochemical | CID_3314 | Eugenol | {"name": "Eugenol", "id": "CID_3314"} | INTERACTS_WITH | - | Target | 28234 | SLCO1B3 | {"name": "SLCO1B3", "id": "28234"} |
| Phytochemical | CID_3314 | Eugenol | {"name": "Eugenol", "id": "CID_3314"} | INTERACTS_WITH | - | Target | 2908 | NR3C1 | {"name": "NR3C1", "id": "2908"} |
| Phytochemical | CID_3314 | Eugenol | {"name": "Eugenol", "id": "CID_3314"} | INTERACTS_WITH | - | Target | 2990 | GUSB | {"name": "GUSB", "id": "2990"} |
| Phytochemical | CID_3314 | Eugenol | {"name": "Eugenol", "id": "CID_3314"} | INTERACTS_WITH | - | Target | 3156 | HMGCR | {"name": "HMGCR", "id": "3156"} |
| Phytochemical | CID_3314 | Eugenol | {"name": "Eugenol", "id": "CID_3314"} | INTERACTS_WITH | - | Target | 3269 | HRH1 | {"name": "HRH1", "id": "3269"} |
| Phytochemical | CID_3314 | Eugenol | {"name": "Eugenol", "id": "CID_3314"} | INTERACTS_WITH | - | Target | 3274 | HRH2 | {"name": "HRH2", "id": "3274"} |
| Phytochemical | CID_3314 | Eugenol | {"name": "Eugenol", "id": "CID_3314"} | INTERACTS_WITH | - | Target | 3356 | HTR2A | {"name": "HTR2A", "id": "3356"} |
| Phytochemical | CID_3314 | Eugenol | {"name": "Eugenol", "id": "CID_3314"} | INTERACTS_WITH | - | Target | 3357 | HTR2B | {"name": "HTR2B", "id": "3357"} |
| Phytochemical | CID_3314 | Eugenol | {"name": "Eugenol", "id": "CID_3314"} | INTERACTS_WITH | - | Target | 3358 | HTR2C | {"name": "HTR2C", "id": "3358"} |
| Phytochemical | CID_3314 | Eugenol | {"name": "Eugenol", "id": "CID_3314"} | INTERACTS_WITH | - | Target | 3362 | HTR6 | {"name": "HTR6", "id": "3362"} |
| Phytochemical | CID_3314 | Eugenol | {"name": "Eugenol", "id": "CID_3314"} | INTERACTS_WITH | - | Target | 3576 | CXCL8 | {"name": "CXCL8", "id": "3576"} |
| Phytochemical | CID_3314 | Eugenol | {"name": "Eugenol", "id": "CID_3314"} | INTERACTS_WITH | - | Target | 3577 | CXCR1 | {"name": "CXCR1", "id": "3577"} |
| Phytochemical | CID_3314 | Eugenol | {"name": "Eugenol", "id": "CID_3314"} | INTERACTS_WITH | - | Target | 3579 | CXCR2 | {"name": "CXCR2", "id": "3579"} |
| Phytochemical | CID_3314 | Eugenol | {"name": "Eugenol", "id": "CID_3314"} | INTERACTS_WITH | - | Target | 367 | AR | {"name": "AR", "id": "367"} |
| Phytochemical | CID_3314 | Eugenol | {"name": "Eugenol", "id": "CID_3314"} | INTERACTS_WITH | - | Target | 3725 | JUN | {"name": "JUN", "id": "3725"} |
| Phytochemical | CID_3314 | Eugenol | {"name": "Eugenol", "id": "CID_3314"} | INTERACTS_WITH | - | Target | 3757 | KCNH2 | {"name": "KCNH2", "id": "3757"} |
| Phytochemical | CID_3314 | Eugenol | {"name": "Eugenol", "id": "CID_3314"} | INTERACTS_WITH | - | Target | 3932 | LCK | {"name": "LCK", "id": "3932"} |
| Phytochemical | CID_3314 | Eugenol | {"name": "Eugenol", "id": "CID_3314"} | INTERACTS_WITH | - | Target | 4128 | MAOA | {"name": "MAOA", "id": "4128"} |
| Phytochemical | CID_3314 | Eugenol | {"name": "Eugenol", "id": "CID_3314"} | INTERACTS_WITH | - | Target | 4159 | MC3R | {"name": "MC3R", "id": "4159"} |
| Phytochemical | CID_3314 | Eugenol | {"name": "Eugenol", "id": "CID_3314"} | INTERACTS_WITH | - | Target | 4160 | MC4R | {"name": "MC4R", "id": "4160"} |
| Phytochemical | CID_3314 | Eugenol | {"name": "Eugenol", "id": "CID_3314"} | INTERACTS_WITH | - | Target | 4161 | MC5R | {"name": "MC5R", "id": "4161"} |
| Phytochemical | CID_3314 | Eugenol | {"name": "Eugenol", "id": "CID_3314"} | INTERACTS_WITH | - | Target | 43 | ACHE | {"name": "ACHE", "id": "43"} |
| Phytochemical | CID_3314 | Eugenol | {"name": "Eugenol", "id": "CID_3314"} | INTERACTS_WITH | - | Target | 4312 | MMP1 | {"name": "MMP1", "id": "4312"} |
| Phytochemical | CID_3314 | Eugenol | {"name": "Eugenol", "id": "CID_3314"} | INTERACTS_WITH | - | Target | 4318 | MMP9 | {"name": "MMP9", "id": "4318"} |
| Phytochemical | CID_3314 | Eugenol | {"name": "Eugenol", "id": "CID_3314"} | INTERACTS_WITH | - | Target | 4780 | NFE2L2 | {"name": "NFE2L2", "id": "4780"} |
| Phytochemical | CID_3314 | Eugenol | {"name": "Eugenol", "id": "CID_3314"} | INTERACTS_WITH | - | Target | 4886 | NPY1R | {"name": "NPY1R", "id": "4886"} |
| Phytochemical | CID_3314 | Eugenol | {"name": "Eugenol", "id": "CID_3314"} | INTERACTS_WITH | - | Target | 4887 | NPY2R | {"name": "NPY2R", "id": "4887"} |
| Phytochemical | CID_3314 | Eugenol | {"name": "Eugenol", "id": "CID_3314"} | INTERACTS_WITH | - | Target | 4985 | OPRD1 | {"name": "OPRD1", "id": "4985"} |
| Phytochemical | CID_3314 | Eugenol | {"name": "Eugenol", "id": "CID_3314"} | INTERACTS_WITH | - | Target | 4986 | OPRK1 | {"name": "OPRK1", "id": "4986"} |

**Table S9: Complete node and relationship data used to construct the knowledge graph. This table summarizes interactions among FDA-approved drugs, formulation, herbs, diseases (type 2 diabetes, fatty liver, obesity), phytochemicals, and their target genes/proteins. Each entry includes source and target node details, edge type, and node annotations (Entrez Gene ID and gene symbol, where applicable).**

| Source Node Label | Source Node ID | Source Node Name | Source Node Properties | Edge Type | Edge Properties | Target Node Label | Target Node ID | Target Node Name | Target Node Properties |
|---|---|---|---|---|---|---|---|---|---|
| Phytochemical | CID_3314 | Eugenol | {"name": "Eugenol", "id": "CID_3314"} | INTERACTS_WITH | - | Target | 4988 | OPRM1 | {"name": "OPRM1", "id": "4988"} |
| Phytochemical | CID_3314 | Eugenol | {"name": "Eugenol", "id": "CID_3314"} | INTERACTS_WITH | - | Target | 51053 | GMNN | {"name": "GMNN", "id": "51053"} |
| Phytochemical | CID_3314 | Eugenol | {"name": "Eugenol", "id": "CID_3314"} | INTERACTS_WITH | - | Target | 54575 | UGT1A10 | {"name": "UGT1A10", "id": "54575"} |
| Phytochemical | CID_3314 | Eugenol | {"name": "Eugenol", "id": "CID_3314"} | INTERACTS_WITH | - | Target | 54576 | UGT1A8 | {"name": "UGT1A8", "id": "54576"} |
| Phytochemical | CID_3314 | Eugenol | {"name": "Eugenol", "id": "CID_3314"} | INTERACTS_WITH | - | Target | 54578 | UGT1A6 | {"name": "UGT1A6", "id": "54578"} |
| Phytochemical | CID_3314 | Eugenol | {"name": "Eugenol", "id": "CID_3314"} | INTERACTS_WITH | - | Target | 54600 | UGT1A9 | {"name": "UGT1A9", "id": "54600"} |
| Phytochemical | CID_3314 | Eugenol | {"name": "Eugenol", "id": "CID_3314"} | INTERACTS_WITH | - | Target | 54657 | UGT1A4 | {"name": "UGT1A4", "id": "54657"} |
| Phytochemical | CID_3314 | Eugenol | {"name": "Eugenol", "id": "CID_3314"} | INTERACTS_WITH | - | Target | 54658 | UGT1A1 | {"name": "UGT1A1", "id": "54658"} |
| Phytochemical | CID_3314 | Eugenol | {"name": "Eugenol", "id": "CID_3314"} | INTERACTS_WITH | - | Target | 54659 | UGT1A3 | {"name": "UGT1A3", "id": "54659"} |
| Phytochemical | CID_3314 | Eugenol | {"name": "Eugenol", "id": "CID_3314"} | INTERACTS_WITH | - | Target | 552 | AVPR1A | {"name": "AVPR1A", "id": "552"} |
| Phytochemical | CID_3314 | Eugenol | {"name": "Eugenol", "id": "CID_3314"} | INTERACTS_WITH | - | Target | 5530 | PPP3CA | {"name": "PPP3CA", "id": "5530"} |
| Phytochemical | CID_3314 | Eugenol | {"name": "Eugenol", "id": "CID_3314"} | INTERACTS_WITH | - | Target | 5578 | PRKCA | {"name": "PRKCA", "id": "5578"} |
| Phytochemical | CID_3314 | Eugenol | {"name": "Eugenol", "id": "CID_3314"} | INTERACTS_WITH | - | Target | 5594 | MAPK1 | {"name": "MAPK1", "id": "5594"} |
| Phytochemical | CID_3314 | Eugenol | {"name": "Eugenol", "id": "CID_3314"} | INTERACTS_WITH | - | Target | 5595 | MAPK3 | {"name": "MAPK3", "id": "5595"} |
| Phytochemical | CID_3314 | Eugenol | {"name": "Eugenol", "id": "CID_3314"} | INTERACTS_WITH | - | Target | 5724 | PTAFR | {"name": "PTAFR", "id": "5724"} |
| Phytochemical | CID_3314 | Eugenol | {"name": "Eugenol", "id": "CID_3314"} | INTERACTS_WITH | - | Target | 5742 | PTGS1 | {"name": "PTGS1", "id": "5742"} |
| Phytochemical | CID_3314 | Eugenol | {"name": "Eugenol", "id": "CID_3314"} | INTERACTS_WITH | - | Target | 5743 | PTGS2 | {"name": "PTGS2", "id": "5743"} |
| Phytochemical | CID_3314 | Eugenol | {"name": "Eugenol", "id": "CID_3314"} | INTERACTS_WITH | - | Target | 5788 | PTPRC | {"name": "PTPRC", "id": "5788"} |
| Phytochemical | CID_3314 | Eugenol | {"name": "Eugenol", "id": "CID_3314"} | INTERACTS_WITH | - | Target | 60489 | APOBEC3G | {"name": "APOBEC3G", "id": "60489"} |
| Phytochemical | CID_3314 | Eugenol | {"name": "Eugenol", "id": "CID_3314"} | INTERACTS_WITH | - | Target | 624 | BDKRB2 | {"name": "BDKRB2", "id": "624"} |
| Phytochemical | CID_3314 | Eugenol | {"name": "Eugenol", "id": "CID_3314"} | INTERACTS_WITH | - | Target | 6530 | SLC6A2 | {"name": "SLC6A2", "id": "6530"} |
| Phytochemical | CID_3314 | Eugenol | {"name": "Eugenol", "id": "CID_3314"} | INTERACTS_WITH | - | Target | 6531 | SLC6A3 | {"name": "SLC6A3", "id": "6531"} |
| Phytochemical | CID_3314 | Eugenol | {"name": "Eugenol", "id": "CID_3314"} | INTERACTS_WITH | - | Target | 6532 | SLC6A4 | {"name": "SLC6A4", "id": "6532"} |
| Phytochemical | CID_3314 | Eugenol | {"name": "Eugenol", "id": "CID_3314"} | INTERACTS_WITH | - | Target | 6865 | TACR2 | {"name": "TACR2", "id": "6865"} |
| Phytochemical | CID_3314 | Eugenol | {"name": "Eugenol", "id": "CID_3314"} | INTERACTS_WITH | - | Target | 6869 | TACR1 | {"name": "TACR1", "id": "6869"} |
| Phytochemical | CID_3314 | Eugenol | {"name": "Eugenol", "id": "CID_3314"} | INTERACTS_WITH | - | Target | 6916 | TBXAS1 | {"name": "TBXAS1", "id": "6916"} |
| Phytochemical | CID_3314 | Eugenol | {"name": "Eugenol", "id": "CID_3314"} | INTERACTS_WITH | - | Target | 729230 | CCR2 | {"name": "CCR2", "id": "729230"} |
| Phytochemical | CID_3314 | Eugenol | {"name": "Eugenol", "id": "CID_3314"} | INTERACTS_WITH | - | Target | 7366 | UGT2B15 | {"name": "UGT2B15", "id": "7366"} |
| Phytochemical | CID_3314 | Eugenol | {"name": "Eugenol", "id": "CID_3314"} | INTERACTS_WITH | - | Target | 7367 | UGT2B17 | {"name": "UGT2B17", "id": "7367"} |
| Phytochemical | CID_3314 | Eugenol | {"name": "Eugenol", "id": "CID_3314"} | INTERACTS_WITH | - | Target | 7433 | VIPR1 | {"name": "VIPR1", "id": "7433"} |
| Phytochemical | CID_3314 | Eugenol | {"name": "Eugenol", "id": "CID_3314"} | INTERACTS_WITH | - | Target | 760 | CA2 | {"name": "CA2", "id": "760"} |
| Phytochemical | CID_3314 | Eugenol | {"name": "Eugenol", "id": "CID_3314"} | INTERACTS_WITH | - | Target | 775 | CACNA1C | {"name": "CACNA1C", "id": "775"} |
| Phytochemical | CID_3314 | Eugenol | {"name": "Eugenol", "id": "CID_3314"} | INTERACTS_WITH | - | Target | 79054 | TRPM8 | {"name": "TRPM8", "id": "79054"} |
| Phytochemical | CID_3314 | Eugenol | {"name": "Eugenol", "id": "CID_3314"} | INTERACTS_WITH | - | Target | 799 | CALCR | {"name": "CALCR", "id": "799"} |
| Phytochemical | CID_3314 | Eugenol | {"name": "Eugenol", "id": "CID_3314"} | INTERACTS_WITH | - | Target | 834 | CASP1 | {"name": "CASP1", "id": "834"} |
| Phytochemical | CID_3314 | Eugenol | {"name": "Eugenol", "id": "CID_3314"} | INTERACTS_WITH | - | Target | 8654 | PDE5A | {"name": "PDE5A", "id": "8654"} |
| Phytochemical | CID_3314 | Eugenol | {"name": "Eugenol", "id": "CID_3314"} | INTERACTS_WITH | - | Target | 886 | CCKAR | {"name": "CCKAR", "id": "886"} |
| Phytochemical | CID_3314 | Eugenol | {"name": "Eugenol", "id": "CID_3314"} | INTERACTS_WITH | - | Target | 8989 | TRPA1 | {"name": "TRPA1", "id": "8989"} |
| Phytochemical | CID_445070 | Farnesol | {"name": "Farnesol", "id": "CID_445070"} | INTERACTS_WITH | - | Target | 10599 | SLCO1B1 | {"name": "SLCO1B1", "id": "10599"} |
| Phytochemical | CID_445070 | Farnesol | {"name": "Farnesol", "id": "CID_445070"} | INTERACTS_WITH | - | Target | 10919 | EHMT2 | {"name": "EHMT2", "id": "10919"} |
| Phytochemical | CID_445070 | Farnesol | {"name": "Farnesol", "id": "CID_445070"} | INTERACTS_WITH | - | Target | 1576 | CYP3A4 | {"name": "CYP3A4", "id": "1576"} |
| Phytochemical | CID_445070 | Farnesol | {"name": "Farnesol", "id": "CID_445070"} | INTERACTS_WITH | - | Target | 246 | ALOX15 | {"name": "ALOX15", "id": "246"} |
| Phytochemical | CID_445070 | Farnesol | {"name": "Farnesol", "id": "CID_445070"} | INTERACTS_WITH | - | Target | 28234 | SLCO1B3 | {"name": "SLCO1B3", "id": "28234"} |
| Phytochemical | CID_445070 | Farnesol | {"name": "Farnesol", "id": "CID_445070"} | INTERACTS_WITH | - | Target | 3315 | HSPB1 | {"name": "HSPB1", "id": "3315"} |
| Phytochemical | CID_445070 | Farnesol | {"name": "Farnesol", "id": "CID_445070"} | INTERACTS_WITH | - | Target | 367 | AR | {"name": "AR", "id": "367"} |
| Phytochemical | CID_445070 | Farnesol | {"name": "Farnesol", "id": "CID_445070"} | INTERACTS_WITH | - | Target | 3725 | JUN | {"name": "JUN", "id": "3725"} |
| Phytochemical | CID_445070 | Farnesol | {"name": "Farnesol", "id": "CID_445070"} | INTERACTS_WITH | - | Target | 4129 | MAOB | {"name": "MAOB", "id": "4129"} |
| Phytochemical | CID_445070 | Farnesol | {"name": "Farnesol", "id": "CID_445070"} | INTERACTS_WITH | - | Target | 4137 | MAPT | {"name": "MAPT", "id": "4137"} |
| Phytochemical | CID_445070 | Farnesol | {"name": "Farnesol", "id": "CID_445070"} | INTERACTS_WITH | - | Target | 4297 | KMT2A | {"name": "KMT2A", "id": "4297"} |
| Phytochemical | CID_445070 | Farnesol | {"name": "Farnesol", "id": "CID_445070"} | INTERACTS_WITH | - | Target | 4780 | NFE2L2 | {"name": "NFE2L2", "id": "4780"} |
| Phytochemical | CID_445070 | Farnesol | {"name": "Farnesol", "id": "CID_445070"} | INTERACTS_WITH | - | Target | 54658 | UGT1A1 | {"name": "UGT1A1", "id": "54658"} |
| Phytochemical | CID_445070 | Farnesol | {"name": "Farnesol", "id": "CID_445070"} | INTERACTS_WITH | - | Target | 5467 | PPARD | {"name": "PPARD", "id": "5467"} |
| Phytochemical | CID_445070 | Farnesol | {"name": "Farnesol", "id": "CID_445070"} | INTERACTS_WITH | - | Target | 55775 | TDP1 | {"name": "TDP1", "id": "55775"} |
| Phytochemical | CID_445070 | Farnesol | {"name": "Farnesol", "id": "CID_445070"} | INTERACTS_WITH | - | Target | 7398 | USP1 | {"name": "USP1", "id": "7398"} |
| Phytochemical | CID_445858 | Ferulic acid | {"name": "Ferulic acid", "id": "CID_445858"} | INTERACTS_WITH | - | Target | 10013 | HDAC6 | {"name": "HDAC6", "id": "10013"} |
| Phytochemical | CID_445858 | Ferulic acid | {"name": "Ferulic acid", "id": "CID_445858"} | INTERACTS_WITH | - | Target | 10014 | HDAC5 | {"name": "HDAC5", "id": "10014"} |
| Phytochemical | CID_445858 | Ferulic acid | {"name": "Ferulic acid", "id": "CID_445858"} | INTERACTS_WITH | - | Target | 10280 | SIGMAR1 | {"name": "SIGMAR1", "id": "10280"} |
| Phytochemical | CID_445858 | Ferulic acid | {"name": "Ferulic acid", "id": "CID_445858"} | INTERACTS_WITH | - | Target | 10599 | SLCO1B1 | {"name": "SLCO1B1", "id": "10599"} |
| Phytochemical | CID_445858 | Ferulic acid | {"name": "Ferulic acid", "id": "CID_445858"} | INTERACTS_WITH | - | Target | 10800 | CYSLTR1 | {"name": "CYSLTR1", "id": "10800"} |
| Phytochemical | CID_445858 | Ferulic acid | {"name": "Ferulic acid", "id": "CID_445858"} | INTERACTS_WITH | - | Target | 11238 | CA5B | {"name": "CA5B", "id": "11238"} |
| Phytochemical | CID_445858 | Ferulic acid | {"name": "Ferulic acid", "id": "CID_445858"} | INTERACTS_WITH | - | Target | 1128 | CHRM1 | {"name": "CHRM1", "id": "1128"} |
| Phytochemical | CID_445858 | Ferulic acid | {"name": "Ferulic acid", "id": "CID_445858"} | INTERACTS_WITH | - | Target | 1129 | CHRM2 | {"name": "CHRM2", "id": "1129"} |
| Phytochemical | CID_445858 | Ferulic acid | {"name": "Ferulic acid", "id": "CID_445858"} | INTERACTS_WITH | - | Target | 1131 | CHRM3 | {"name": "CHRM3", "id": "1131"} |
| Phytochemical | CID_445858 | Ferulic acid | {"name": "Ferulic acid", "id": "CID_445858"} | INTERACTS_WITH | - | Target | 1132 | CHRM4 | {"name": "CHRM4", "id": "1132"} |
| Phytochemical | CID_445858 | Ferulic acid | {"name": "Ferulic acid", "id": "CID_445858"} | INTERACTS_WITH | - | Target | 1133 | CHRM5 | {"name": "CHRM5", "id": "1133"} |
| Phytochemical | CID_445858 | Ferulic acid | {"name": "Ferulic acid", "id": "CID_445858"} | INTERACTS_WITH | - | Target | 1233 | CCR4 | {"name": "CCR4", "id": "1233"} |
| Phytochemical | CID_445858 | Ferulic acid | {"name": "Ferulic acid", "id": "CID_445858"} | INTERACTS_WITH | - | Target | 1234 | CCR5 | {"name": "CCR5", "id": "1234"} |
| Phytochemical | CID_445858 | Ferulic acid | {"name": "Ferulic acid", "id": "CID_445858"} | INTERACTS_WITH | - | Target | 1268 | CNR1 | {"name": "CNR1", "id": "1268"} |

**Table S9: Complete node and relationship data used to construct the knowledge graph. This table summarizes interactions among FDA-approved drugs, formulation, herbs, diseases (type 2 diabetes, fatty liver, obesity), phytochemicals, and their target genes/proteins. Each entry includes source and target node details, edge type, and node annotations (Entrez Gene ID and gene symbol, where applicable).**

| Source Node Label | Source Node ID | Source Node Name | Source Node Properties | Edge Type | Edge Properties | Target Node Label | Target Node ID | Target Node Name | Target Node Properties |
|---|---|---|---|---|---|---|---|---|---|
| Phytochemical | CID_445858 | Ferulic acid | {"name": "Ferulic acid", "id": " CID_445858"} | INTERACTS_WITH | - | Target | 134 | ADORA1 | {"name": "ADORA1", "id": "134"} |
| Phytochemical | CID_445858 | Ferulic acid | {"name": "Ferulic acid", "id": " CID_445858"} | INTERACTS_WITH | - | Target | 135 | ADORA2A | {"name": "ADORA2A", "id": "135"} |
| Phytochemical | CID_445858 | Ferulic acid | {"name": "Ferulic acid", "id": " CID_445858"} | INTERACTS_WITH | - | Target | 140 | ADORA3 | {"name": "ADORA3", "id": "140"} |
| Phytochemical | CID_445858 | Ferulic acid | {"name": "Ferulic acid", "id": " CID_445858"} | INTERACTS_WITH | - | Target | 1432 | MAPK14 | {"name": "MAPK14", "id": "1432"} |
| Phytochemical | CID_445858 | Ferulic acid | {"name": "Ferulic acid", "id": " CID_445858"} | INTERACTS_WITH | - | Target | 1457 | CSNK2A1 | {"name": "CSNK2A1", "id": "1457"} |
| Phytochemical | CID_445858 | Ferulic acid | {"name": "Ferulic acid", "id": " CID_445858"} | INTERACTS_WITH | - | Target | 1459 | CSNK2A2 | {"name": "CSNK2A2", "id": "1459"} |
| Phytochemical | CID_445858 | Ferulic acid | {"name": "Ferulic acid", "id": " CID_445858"} | INTERACTS_WITH | - | Target | 146 | ADRA1D | {"name": "ADRA1D", "id": "146"} |
| Phytochemical | CID_445858 | Ferulic acid | {"name": "Ferulic acid", "id": " CID_445858"} | INTERACTS_WITH | - | Target | 1460 | CSNK2B | {"name": "CSNK2B", "id": "1460"} |
| Phytochemical | CID_445858 | Ferulic acid | {"name": "Ferulic acid", "id": " CID_445858"} | INTERACTS_WITH | - | Target | 150 | ADRA2A | {"name": "ADRA2A", "id": "150"} |
| Phytochemical | CID_445858 | Ferulic acid | {"name": "Ferulic acid", "id": " CID_445858"} | INTERACTS_WITH | - | Target | 151 | ADRA2B | {"name": "ADRA2B", "id": "151"} |
| Phytochemical | CID_445858 | Ferulic acid | {"name": "Ferulic acid", "id": " CID_445858"} | INTERACTS_WITH | - | Target | 1511 | CTSG | {"name": "CTSG", "id": "1511"} |
| Phytochemical | CID_445858 | Ferulic acid | {"name": "Ferulic acid", "id": " CID_445858"} | INTERACTS_WITH | - | Target | 152 | ADRA2C | {"name": "ADRA2C", "id": "152"} |
| Phytochemical | CID_445858 | Ferulic acid | {"name": "Ferulic acid", "id": " CID_445858"} | INTERACTS_WITH | - | Target | 153 | ADRB1 | {"name": "ADRB1", "id": "153"} |
| Phytochemical | CID_445858 | Ferulic acid | {"name": "Ferulic acid", "id": " CID_445858"} | INTERACTS_WITH | - | Target | 154 | ADRB2 | {"name": "ADRB2", "id": "154"} |
| Phytochemical | CID_445858 | Ferulic acid | {"name": "Ferulic acid", "id": " CID_445858"} | INTERACTS_WITH | - | Target | 1544 | CYP1A2 | {"name": "CYP1A2", "id": "1544"} |
| Phytochemical | CID_445858 | Ferulic acid | {"name": "Ferulic acid", "id": " CID_445858"} | INTERACTS_WITH | - | Target | 1548 | CYP2A6 | {"name": "CYP2A6", "id": "1548"} |
| Phytochemical | CID_445858 | Ferulic acid | {"name": "Ferulic acid", "id": " CID_445858"} | INTERACTS_WITH | - | Target | 155 | ADRB3 | {"name": "ADRB3", "id": "155"} |
| Phytochemical | CID_445858 | Ferulic acid | {"name": "Ferulic acid", "id": " CID_445858"} | INTERACTS_WITH | - | Target | 1557 | CYP2C19 | {"name": "CYP2C19", "id": "1557"} |
| Phytochemical | CID_445858 | Ferulic acid | {"name": "Ferulic acid", "id": " CID_445858"} | INTERACTS_WITH | - | Target | 1559 | CYP2C9 | {"name": "CYP2C9", "id": "1559"} |
| Phytochemical | CID_445858 | Ferulic acid | {"name": "Ferulic acid", "id": " CID_445858"} | INTERACTS_WITH | - | Target | 1565 | CYP2D6 | {"name": "CYP2D6", "id": "1565"} |
| Phytochemical | CID_445858 | Ferulic acid | {"name": "Ferulic acid", "id": " CID_445858"} | INTERACTS_WITH | - | Target | 1571 | CYP2E1 | {"name": "CYP2E1", "id": "1571"} |
| Phytochemical | CID_445858 | Ferulic acid | {"name": "Ferulic acid", "id": " CID_445858"} | INTERACTS_WITH | - | Target | 1576 | CYP3A4 | {"name": "CYP3A4", "id": "1576"} |
| Phytochemical | CID_445858 | Ferulic acid | {"name": "Ferulic acid", "id": " CID_445858"} | INTERACTS_WITH | - | Target | 1812 | DRD1 | {"name": "DRD1", "id": "1812"} |
| Phytochemical | CID_445858 | Ferulic acid | {"name": "Ferulic acid", "id": " CID_445858"} | INTERACTS_WITH | - | Target | 1813 | DRD2 | {"name": "DRD2", "id": "1813"} |
| Phytochemical | CID_445858 | Ferulic acid | {"name": "Ferulic acid", "id": " CID_445858"} | INTERACTS_WITH | - | Target | 1814 | DRD3 | {"name": "DRD3", "id": "1814"} |
| Phytochemical | CID_445858 | Ferulic acid | {"name": "Ferulic acid", "id": " CID_445858"} | INTERACTS_WITH | - | Target | 1815 | DRD4 | {"name": "DRD4", "id": "1815"} |
| Phytochemical | CID_445858 | Ferulic acid | {"name": "Ferulic acid", "id": " CID_445858"} | INTERACTS_WITH | - | Target | 186 | AGTR2 | {"name": "AGTR2", "id": "186"} |
| Phytochemical | CID_445858 | Ferulic acid | {"name": "Ferulic acid", "id": " CID_445858"} | INTERACTS_WITH | - | Target | 1909 | EDNRA | {"name": "EDNRA", "id": "1909"} |
| Phytochemical | CID_445858 | Ferulic acid | {"name": "Ferulic acid", "id": " CID_445858"} | INTERACTS_WITH | - | Target | 1956 | EGFR | {"name": "EGFR", "id": "1956"} |
| Phytochemical | CID_445858 | Ferulic acid | {"name": "Ferulic acid", "id": " CID_445858"} | INTERACTS_WITH | - | Target | 1991 | ELANE | {"name": "ELANE", "id": "1991"} |
| Phytochemical | CID_445858 | Ferulic acid | {"name": "Ferulic acid", "id": " CID_445858"} | INTERACTS_WITH | - | Target | 2064 | ERBB2 | {"name": "ERBB2", "id": "2064"} |
| Phytochemical | CID_445858 | Ferulic acid | {"name": "Ferulic acid", "id": " CID_445858"} | INTERACTS_WITH | - | Target | 2099 | ESR1 | {"name": "ESR1", "id": "2099"} |
| Phytochemical | CID_445858 | Ferulic acid | {"name": "Ferulic acid", "id": " CID_445858"} | INTERACTS_WITH | - | Target | 2100 | ESR2 | {"name": "ESR2", "id": "2100"} |
| Phytochemical | CID_445858 | Ferulic acid | {"name": "Ferulic acid", "id": " CID_445858"} | INTERACTS_WITH | - | Target | 2260 | FGFR1 | {"name": "FGFR1", "id": "2260"} |
| Phytochemical | CID_445858 | Ferulic acid | {"name": "Ferulic acid", "id": " CID_445858"} | INTERACTS_WITH | - | Target | 2263 | FGFR2 | {"name": "FGFR2", "id": "2263"} |
| Phytochemical | CID_445858 | Ferulic acid | {"name": "Ferulic acid", "id": " CID_445858"} | INTERACTS_WITH | - | Target | 2321 | FLT1 | {"name": "FLT1", "id": "2321"} |
| Phytochemical | CID_445858 | Ferulic acid | {"name": "Ferulic acid", "id": " CID_445858"} | INTERACTS_WITH | - | Target | 23632 | CA14 | {"name": "CA14", "id": "23632"} |
| Phytochemical | CID_445858 | Ferulic acid | {"name": "Ferulic acid", "id": " CID_445858"} | INTERACTS_WITH | - | Target | 2534 | FYN | {"name": "FYN", "id": "2534"} |
| Phytochemical | CID_445858 | Ferulic acid | {"name": "Ferulic acid", "id": " CID_445858"} | INTERACTS_WITH | - | Target | 259285 | TAS2R39 | {"name": "TAS2R39", "id": "259285"} |
| Phytochemical | CID_445858 | Ferulic acid | {"name": "Ferulic acid", "id": " CID_445858"} | INTERACTS_WITH | - | Target | 28234 | SLCO1B3 | {"name": "SLCO1B3", "id": "28234"} |
| Phytochemical | CID_445858 | Ferulic acid | {"name": "Ferulic acid", "id": " CID_445858"} | INTERACTS_WITH | - | Target | 283106 | CSNK2A3 | {"name": "CSNK2A3", "id": "283106"} |
| Phytochemical | CID_445858 | Ferulic acid | {"name": "Ferulic acid", "id": " CID_445858"} | INTERACTS_WITH | - | Target | 2908 | NR3C1 | {"name": "NR3C1", "id": "2908"} |
| Phytochemical | CID_445858 | Ferulic acid | {"name": "Ferulic acid", "id": " CID_445858"} | INTERACTS_WITH | - | Target | 3065 | HDAC1 | {"name": "HDAC1", "id": "3065"} |
| Phytochemical | CID_445858 | Ferulic acid | {"name": "Ferulic acid", "id": " CID_445858"} | INTERACTS_WITH | - | Target | 3066 | HDAC2 | {"name": "HDAC2", "id": "3066"} |
| Phytochemical | CID_445858 | Ferulic acid | {"name": "Ferulic acid", "id": " CID_445858"} | INTERACTS_WITH | - | Target | 3156 | HMGCR | {"name": "HMGCR", "id": "3156"} |
| Phytochemical | CID_445858 | Ferulic acid | {"name": "Ferulic acid", "id": " CID_445858"} | INTERACTS_WITH | - | Target | 3269 | HRH1 | {"name": "HRH1", "id": "3269"} |
| Phytochemical | CID_445858 | Ferulic acid | {"name": "Ferulic acid", "id": " CID_445858"} | INTERACTS_WITH | - | Target | 3274 | HRH2 | {"name": "HRH2", "id": "3274"} |
| Phytochemical | CID_445858 | Ferulic acid | {"name": "Ferulic acid", "id": " CID_445858"} | INTERACTS_WITH | - | Target | 3356 | HTR2A | {"name": "HTR2A", "id": "3356"} |
| Phytochemical | CID_445858 | Ferulic acid | {"name": "Ferulic acid", "id": " CID_445858"} | INTERACTS_WITH | - | Target | 3357 | HTR2B | {"name": "HTR2B", "id": "3357"} |

Table S9: Complete node and relationship data used to construct the knowledge graph. This table summarizes interactions among FDA-approved drugs, formulation, herbs, diseases (type 2 diabetes, fatty liver, obesity), phytochemicals, and their target genes/proteins. Each entry includes source and target node details, edge type, and node annotations (Entrez Gene ID and gene symbol, where applicable).

| Source Node Label | Source Node ID | Source Node Name | Source Node Properties | Edge Type | Edge Properties | Target Node Label | Target Node ID | Target Node Name | Target Node Properties |
|---|---|---|---|---|---|---|---|---|---|
| Phytochemical | CID_445858 | Ferulic acid | {"name": "Ferulic acid", "id": " CID_445858"} | INTERACTS_WITH | - | Target | 3358 | HTR2C | {"name": "HTR2C", "id": "3358"} |
| Phytochemical | CID_445858 | Ferulic acid | {"name": "Ferulic acid", "id": " CID_445858"} | INTERACTS_WITH | - | Target | 3362 | HTR6 | {"name": "HTR6", "id": "3362"} |
| Phytochemical | CID_445858 | Ferulic acid | {"name": "Ferulic acid", "id": " CID_445858"} | INTERACTS_WITH | - | Target | 3363 | HTR7 | {"name": "HTR7", "id": "3363"} |
| Phytochemical | CID_445858 | Ferulic acid | {"name": "Ferulic acid", "id": " CID_445858"} | INTERACTS_WITH | - | Target | 3373 | HYAL1 | {"name": "HYAL1", "id": "3373"} |
| Phytochemical | CID_445858 | Ferulic acid | {"name": "Ferulic acid", "id": " CID_445858"} | INTERACTS_WITH | - | Target | 338442 | HCAR2 | {"name": "HCAR2", "id": "338442"} |
| Phytochemical | CID_445858 | Ferulic acid | {"name": "Ferulic acid", "id": " CID_445858"} | INTERACTS_WITH | - | Target | 351 | APP | {"name": "APP", "id": "351"} |
| Phytochemical | CID_445858 | Ferulic acid | {"name": "Ferulic acid", "id": " CID_445858"} | INTERACTS_WITH | - | Target | 3577 | CXCR1 | {"name": "CXCR1", "id": "3577"} |
| Phytochemical | CID_445858 | Ferulic acid | {"name": "Ferulic acid", "id": " CID_445858"} | INTERACTS_WITH | - | Target | 3579 | CXCR2 | {"name": "CXCR2", "id": "3579"} |
| Phytochemical | CID_445858 | Ferulic acid | {"name": "Ferulic acid", "id": " CID_445858"} | INTERACTS_WITH | - | Target | 3757 | KCNH2 | {"name": "KCNH2", "id": "3757"} |
| Phytochemical | CID_445858 | Ferulic acid | {"name": "Ferulic acid", "id": " CID_445858"} | INTERACTS_WITH | - | Target | 390245 | KDM4E | {"name": "KDM4E", "id": "390245"} |
| Phytochemical | CID_445858 | Ferulic acid | {"name": "Ferulic acid", "id": " CID_445858"} | INTERACTS_WITH | - | Target | 3932 | LCK | {"name": "LCK", "id": "3932"} |
| Phytochemical | CID_445858 | Ferulic acid | {"name": "Ferulic acid", "id": " CID_445858"} | INTERACTS_WITH | - | Target | 4128 | MAOA | {"name": "MAOA", "id": "4128"} |
| Phytochemical | CID_445858 | Ferulic acid | {"name": "Ferulic acid", "id": " CID_445858"} | INTERACTS_WITH | - | Target | 4159 | MC3R | {"name": "MC3R", "id": "4159"} |
| Phytochemical | CID_445858 | Ferulic acid | {"name": "Ferulic acid", "id": " CID_445858"} | INTERACTS_WITH | - | Target | 4160 | MC4R | {"name": "MC4R", "id": "4160"} |
| Phytochemical | CID_445858 | Ferulic acid | {"name": "Ferulic acid", "id": " CID_445858"} | INTERACTS_WITH | - | Target | 4161 | MC5R | {"name": "MC5R", "id": "4161"} |
| Phytochemical | CID_445858 | Ferulic acid | {"name": "Ferulic acid", "id": " CID_445858"} | INTERACTS_WITH | - | Target | 43 | ACHE | {"name": "ACHE", "id": "43"} |
| Phytochemical | CID_445858 | Ferulic acid | {"name": "Ferulic acid", "id": " CID_445858"} | INTERACTS_WITH | - | Target | 4312 | MMP1 | {"name": "MMP1", "id": "4312"} |
| Phytochemical | CID_445858 | Ferulic acid | {"name": "Ferulic acid", "id": " CID_445858"} | INTERACTS_WITH | - | Target | 4318 | MMP9 | {"name": "MMP9", "id": "4318"} |
| Phytochemical | CID_445858 | Ferulic acid | {"name": "Ferulic acid", "id": " CID_445858"} | INTERACTS_WITH | - | Target | 4886 | NPY1R | {"name": "NPY1R", "id": "4886"} |
| Phytochemical | CID_445858 | Ferulic acid | {"name": "Ferulic acid", "id": " CID_445858"} | INTERACTS_WITH | - | Target | 4887 | NPY2R | {"name": "NPY2R", "id": "4887"} |
| Phytochemical | CID_445858 | Ferulic acid | {"name": "Ferulic acid", "id": " CID_445858"} | INTERACTS_WITH | - | Target | 4985 | OPRD1 | {"name": "OPRD1", "id": "4985"} |
| Phytochemical | CID_445858 | Ferulic acid | {"name": "Ferulic acid", "id": " CID_445858"} | INTERACTS_WITH | - | Target | 4986 | OPRK1 | {"name": "OPRK1", "id": "4986"} |
| Phytochemical | CID_445858 | Ferulic acid | {"name": "Ferulic acid", "id": " CID_445858"} | INTERACTS_WITH | - | Target | 4988 | OPRM1 | {"name": "OPRM1", "id": "4988"} |
| Phytochemical | CID_445858 | Ferulic acid | {"name": "Ferulic acid", "id": " CID_445858"} | INTERACTS_WITH | - | Target | 51053 | GMNN | {"name": "GMNN", "id": "51053"} |
| Phytochemical | CID_445858 | Ferulic acid | {"name": "Ferulic acid", "id": " CID_445858"} | INTERACTS_WITH | - | Target | 51564 | HDAC7 | {"name": "HDAC7", "id": "51564"} |
| Phytochemical | CID_445858 | Ferulic acid | {"name": "Ferulic acid", "id": " CID_445858"} | INTERACTS_WITH | - | Target | 5315 | PKM | {"name": "PKM", "id": "5315"} |
| Phytochemical | CID_445858 | Ferulic acid | {"name": "Ferulic acid", "id": " CID_445858"} | INTERACTS_WITH | - | Target | 54657 | UGT1A4 | {"name": "UGT1A4", "id": "54657"} |
| Phytochemical | CID_445858 | Ferulic acid | {"name": "Ferulic acid", "id": " CID_445858"} | INTERACTS_WITH | - | Target | 54658 | UGT1A1 | {"name": "UGT1A1", "id": "54658"} |
| Phytochemical | CID_445858 | Ferulic acid | {"name": "Ferulic acid", "id": " CID_445858"} | INTERACTS_WITH | - | Target | 552 | AVPR1A | {"name": "AVPR1A", "id": "552"} |
| Phytochemical | CID_445858 | Ferulic acid | {"name": "Ferulic acid", "id": " CID_445858"} | INTERACTS_WITH | - | Target | 5530 | PPP3CA | {"name": "PPP3CA", "id": "5530"} |
| Phytochemical | CID_445858 | Ferulic acid | {"name": "Ferulic acid", "id": " CID_445858"} | INTERACTS_WITH | - | Target | 55775 | TDP1 | {"name": "TDP1", "id": "55775"} |
| Phytochemical | CID_445858 | Ferulic acid | {"name": "Ferulic acid", "id": " CID_445858"} | INTERACTS_WITH | - | Target | 5578 | PRKCA | {"name": "PRKCA", "id": "5578"} |
| Phytochemical | CID_445858 | Ferulic acid | {"name": "Ferulic acid", "id": " CID_445858"} | INTERACTS_WITH | - | Target | 55869 | HDAC8 | {"name": "HDAC8", "id": "55869"} |
| Phytochemical | CID_445858 | Ferulic acid | {"name": "Ferulic acid", "id": " CID_445858"} | INTERACTS_WITH | - | Target | 5594 | MAPK1 | {"name": "MAPK1", "id": "5594"} |
| Phytochemical | CID_445858 | Ferulic acid | {"name": "Ferulic acid", "id": " CID_445858"} | INTERACTS_WITH | - | Target | 5595 | MAPK3 | {"name": "MAPK3", "id": "5595"} |
| Phytochemical | CID_445858 | Ferulic acid | {"name": "Ferulic acid", "id": " CID_445858"} | INTERACTS_WITH | - | Target | 5724 | PTAFR | {"name": "PTAFR", "id": "5724"} |
| Phytochemical | CID_445858 | Ferulic acid | {"name": "Ferulic acid", "id": " CID_445858"} | INTERACTS_WITH | - | Target | 5742 | PTGS1 | {"name": "PTGS1", "id": "5742"} |
| Phytochemical | CID_445858 | Ferulic acid | {"name": "Ferulic acid", "id": " CID_445858"} | INTERACTS_WITH | - | Target | 5743 | PTGS2 | {"name": "PTGS2", "id": "5743"} |
| Phytochemical | CID_445858 | Ferulic acid | {"name": "Ferulic acid", "id": " CID_445858"} | INTERACTS_WITH | - | Target | 5745 | PTH1R | {"name": "PTH1R", "id": "5745"} |
| Phytochemical | CID_445858 | Ferulic acid | {"name": "Ferulic acid", "id": " CID_445858"} | INTERACTS_WITH | - | Target | 5788 | PTPRC | {"name": "PTPRC", "id": "5788"} |
| Phytochemical | CID_445858 | Ferulic acid | {"name": "Ferulic acid", "id": " CID_445858"} | INTERACTS_WITH | - | Target | 590 | BCHE | {"name": "BCHE", "id": "590"} |
| Phytochemical | CID_445858 | Ferulic acid | {"name": "Ferulic acid", "id": " CID_445858"} | INTERACTS_WITH | - | Target | 624 | BDKRB2 | {"name": "BDKRB2", "id": "624"} |
| Phytochemical | CID_445858 | Ferulic acid | {"name": "Ferulic acid", "id": " CID_445858"} | INTERACTS_WITH | - | Target | 6530 | SLC6A2 | {"name": "SLC6A2", "id": "6530"} |
| Phytochemical | CID_445858 | Ferulic acid | {"name": "Ferulic acid", "id": " CID_445858"} | INTERACTS_WITH | - | Target | 6531 | SLC6A3 | {"name": "SLC6A3", "id": "6531"} |
| Phytochemical | CID_445858 | Ferulic acid | {"name": "Ferulic acid", "id": " CID_445858"} | INTERACTS_WITH | - | Target | 6532 | SLC6A4 | {"name": "SLC6A4", "id": "6532"} |
| Phytochemical | CID_445858 | Ferulic acid | {"name": "Ferulic acid", "id": " CID_445858"} | INTERACTS_WITH | - | Target | 6622 | SNCA | {"name": "SNCA", "id": "6622"} |
| Phytochemical | CID_445858 | Ferulic acid | {"name": "Ferulic acid", "id": " CID_445858"} | INTERACTS_WITH | - | Target | 6865 | TACR2 | {"name": "TACR2", "id": "6865"} |
| Phytochemical | CID_445858 | Ferulic acid | {"name": "Ferulic acid", "id": " CID_445858"} | INTERACTS_WITH | - | Target | 6869 | TACR1 | {"name": "TACR1", "id": "6869"} |
| Phytochemical | CID_445858 | Ferulic acid | {"name": "Ferulic acid", "id": " CID_445858"} | INTERACTS_WITH | - | Target | 6916 | TBXAS1 | {"name": "TBXAS1", "id": "6916"} |

**Table S9: Complete node and relationship data used to construct the knowledge graph. This table summarizes interactions among FDA-approved drugs, formulation, herbs, diseases (type 2 diabetes, fatty liver, obesity), phytochemicals, and their target genes/proteins. Each entry includes source and target node details, edge type, and node annotations (Entrez Gene ID and gene symbol, where applicable).**

| Source Node Label | Source Node ID | Source Node Name | Source Node Properties | Edge Type | Edge Properties | Target Node Label | Target Node ID | Target Node Name | Target Node Properties |
|---|---|---|---|---|---|---|---|---|---|
| Phytochemical | CID_445858 | Ferulic acid | {"name": "Ferulic acid", "id": "CID_445858"} | INTERACTS_WITH | - | Target | 729230 | CCR2 | {"name": "CCR2", "id": "729230"} |
| Phytochemical | CID_445858 | Ferulic acid | {"name": "Ferulic acid", "id": "CID_445858"} | INTERACTS_WITH | - | Target | 7299 | TYR | {"name": "TYR", "id": "7299"} |
| Phytochemical | CID_445858 | Ferulic acid | {"name": "Ferulic acid", "id": "CID_445858"} | INTERACTS_WITH | - | Target | 7366 | UGT2B15 | {"name": "UGT2B15", "id": "7366"} |
| Phytochemical | CID_445858 | Ferulic acid | {"name": "Ferulic acid", "id": "CID_445858"} | INTERACTS_WITH | - | Target | 7433 | VIPR1 | {"name": "VIPR1", "id": "7433"} |
| Phytochemical | CID_445858 | Ferulic acid | {"name": "Ferulic acid", "id": "CID_445858"} | INTERACTS_WITH | - | Target | 759 | CA1 | {"name": "CA1", "id": "759"} |
| Phytochemical | CID_445858 | Ferulic acid | {"name": "Ferulic acid", "id": "CID_445858"} | INTERACTS_WITH | - | Target | 760 | CA2 | {"name": "CA2", "id": "760"} |
| Phytochemical | CID_445858 | Ferulic acid | {"name": "Ferulic acid", "id": "CID_445858"} | INTERACTS_WITH | - | Target | 761 | CA3 | {"name": "CA3", "id": "761"} |
| Phytochemical | CID_445858 | Ferulic acid | {"name": "Ferulic acid", "id": "CID_445858"} | INTERACTS_WITH | - | Target | 762 | CA4 | {"name": "CA4", "id": "762"} |
| Phytochemical | CID_445858 | Ferulic acid | {"name": "Ferulic acid", "id": "CID_445858"} | INTERACTS_WITH | - | Target | 763 | CA5A | {"name": "CA5A", "id": "763"} |
| Phytochemical | CID_445858 | Ferulic acid | {"name": "Ferulic acid", "id": "CID_445858"} | INTERACTS_WITH | - | Target | 765 | CA6 | {"name": "CA6", "id": "765"} |
| Phytochemical | CID_445858 | Ferulic acid | {"name": "Ferulic acid", "id": "CID_445858"} | INTERACTS_WITH | - | Target | 766 | CA7 | {"name": "CA7", "id": "766"} |
| Phytochemical | CID_445858 | Ferulic acid | {"name": "Ferulic acid", "id": "CID_445858"} | INTERACTS_WITH | - | Target | 768 | CA9 | {"name": "CA9", "id": "768"} |
| Phytochemical | CID_445858 | Ferulic acid | {"name": "Ferulic acid", "id": "CID_445858"} | INTERACTS_WITH | - | Target | 771 | CA12 | {"name": "CA12", "id": "771"} |
| Phytochemical | CID_445858 | Ferulic acid | {"name": "Ferulic acid", "id": "CID_445858"} | INTERACTS_WITH | - | Target | 79885 | HDAC11 | {"name": "HDAC11", "id": "79885"} |
| Phytochemical | CID_445858 | Ferulic acid | {"name": "Ferulic acid", "id": "CID_445858"} | INTERACTS_WITH | - | Target | 799 | CALCR | {"name": "CALCR", "id": "799"} |
| Phytochemical | CID_445858 | Ferulic acid | {"name": "Ferulic acid", "id": "CID_445858"} | INTERACTS_WITH | - | Target | 834 | CASP1 | {"name": "CASP1", "id": "834"} |
| Phytochemical | CID_445858 | Ferulic acid | {"name": "Ferulic acid", "id": "CID_445858"} | INTERACTS_WITH | - | Target | 83933 | HDAC10 | {"name": "HDAC10", "id": "83933"} |
| Phytochemical | CID_445858 | Ferulic acid | {"name": "Ferulic acid", "id": "CID_445858"} | INTERACTS_WITH | - | Target | 8654 | PDE5A | {"name": "PDE5A", "id": "8654"} |
| Phytochemical | CID_445858 | Ferulic acid | {"name": "Ferulic acid", "id": "CID_445858"} | INTERACTS_WITH | - | Target | 8841 | HDAC3 | {"name": "HDAC3", "id": "8841"} |
| Phytochemical | CID_445858 | Ferulic acid | {"name": "Ferulic acid", "id": "CID_445858"} | INTERACTS_WITH | - | Target | 886 | CCKAR | {"name": "CCKAR", "id": "886"} |
| Phytochemical | CID_445858 | Ferulic acid | {"name": "Ferulic acid", "id": "CID_445858"} | INTERACTS_WITH | - | Target | 9734 | HDAC9 | {"name": "HDAC9", "id": "9734"} |
| Phytochemical | CID_445858 | Ferulic acid | {"name": "Ferulic acid", "id": "CID_445858"} | INTERACTS_WITH | - | Target | 9759 | HDAC4 | {"name": "HDAC4", "id": "9759"} |
| Phytochemical | CID_370 | Gallic acid | {"name": "Gallic acid", "id": "CID_370"} | INTERACTS_WITH | - | Target | 10013 | HDAC6 | {"name": "HDAC6", "id": "10013"} |
| Phytochemical | CID_370 | Gallic acid | {"name": "Gallic acid", "id": "CID_370"} | INTERACTS_WITH | - | Target | 10599 | SLCO1B1 | {"name": "SLCO1B1", "id": "10599"} |
| Phytochemical | CID_370 | Gallic acid | {"name": "Gallic acid", "id": "CID_370"} | INTERACTS_WITH | - | Target | 10919 | EHMT2 | {"name": "EHMT2", "id": "10919"} |
| Phytochemical | CID_370 | Gallic acid | {"name": "Gallic acid", "id": "CID_370"} | INTERACTS_WITH | - | Target | 11238 | CA5B | {"name": "CA5B", "id": "11238"} |
| Phytochemical | CID_370 | Gallic acid | {"name": "Gallic acid", "id": "CID_370"} | INTERACTS_WITH | - | Target | 118429 | ANTXR2 | {"name": "ANTXR2", "id": "118429"} |
| Phytochemical | CID_370 | Gallic acid | {"name": "Gallic acid", "id": "CID_370"} | INTERACTS_WITH | - | Target | 1588 | CYP19A1 | {"name": "CYP19A1", "id": "1588"} |
| Phytochemical | CID_370 | Gallic acid | {"name": "Gallic acid", "id": "CID_370"} | INTERACTS_WITH | - | Target | 1636 | ACE | {"name": "ACE", "id": "1636"} |
| Phytochemical | CID_370 | Gallic acid | {"name": "Gallic acid", "id": "CID_370"} | INTERACTS_WITH | - | Target | 1803 | DPP4 | {"name": "DPP4", "id": "1803"} |
| Phytochemical | CID_370 | Gallic acid | {"name": "Gallic acid", "id": "CID_370"} | INTERACTS_WITH | - | Target | 2099 | ESR1 | {"name": "ESR1", "id": "2099"} |
| Phytochemical | CID_370 | Gallic acid | {"name": "Gallic acid", "id": "CID_370"} | INTERACTS_WITH | - | Target | 216 | ALDH1A1 | {"name": "ALDH1A1", "id": "216"} |
| Phytochemical | CID_370 | Gallic acid | {"name": "Gallic acid", "id": "CID_370"} | INTERACTS_WITH | - | Target | 231 | AKR1B1 | {"name": "AKR1B1", "id": "231"} |
| Phytochemical | CID_370 | Gallic acid | {"name": "Gallic acid", "id": "CID_370"} | INTERACTS_WITH | - | Target | 23632 | CA14 | {"name": "CA14", "id": "23632"} |
| Phytochemical | CID_370 | Gallic acid | {"name": "Gallic acid", "id": "CID_370"} | INTERACTS_WITH | - | Target | 2526 | FUT4 | {"name": "FUT4", "id": "2526"} |
| Phytochemical | CID_370 | Gallic acid | {"name": "Gallic acid", "id": "CID_370"} | INTERACTS_WITH | - | Target | 2529 | FUT7 | {"name": "FUT7", "id": "2529"} |
| Phytochemical | CID_370 | Gallic acid | {"name": "Gallic acid", "id": "CID_370"} | INTERACTS_WITH | - | Target | 2590 | GALNT2 | {"name": "GALNT2", "id": "2590"} |
| Phytochemical | CID_370 | Gallic acid | {"name": "Gallic acid", "id": "CID_370"} | INTERACTS_WITH | - | Target | 259285 | TAS2R39 | {"name": "TAS2R39", "id": "259285"} |
| Phytochemical | CID_370 | Gallic acid | {"name": "Gallic acid", "id": "CID_370"} | INTERACTS_WITH | - | Target | 2597 | GAPDH | {"name": "GAPDH", "id": "2597"} |
| Phytochemical | CID_370 | Gallic acid | {"name": "Gallic acid", "id": "CID_370"} | INTERACTS_WITH | - | Target | 2648 | KAT2A | {"name": "KAT2A", "id": "2648"} |
| Phytochemical | CID_370 | Gallic acid | {"name": "Gallic acid", "id": "CID_370"} | INTERACTS_WITH | - | Target | 2744 | GLS | {"name": "GLS", "id": "2744"} |
| Phytochemical | CID_370 | Gallic acid | {"name": "Gallic acid", "id": "CID_370"} | INTERACTS_WITH | - | Target | 28234 | SLCO1B3 | {"name": "SLCO1B3", "id": "28234"} |
| Phytochemical | CID_370 | Gallic acid | {"name": "Gallic acid", "id": "CID_370"} | INTERACTS_WITH | - | Target | 2908 | NR3C1 | {"name": "NR3C1", "id": "2908"} |
| Phytochemical | CID_370 | Gallic acid | {"name": "Gallic acid", "id": "CID_370"} | INTERACTS_WITH | - | Target | 3028 | HSD17B10 | {"name": "HSD17B10", "id": "3028"} |
| Phytochemical | CID_370 | Gallic acid | {"name": "Gallic acid", "id": "CID_370"} | INTERACTS_WITH | - | Target | 3248 | HPGD | {"name": "HPGD", "id": "3248"} |
| Phytochemical | CID_370 | Gallic acid | {"name": "Gallic acid", "id": "CID_370"} | INTERACTS_WITH | - | Target | 328 | APEX1 | {"name": "APEX1", "id": "328"} |
| Phytochemical | CID_370 | Gallic acid | {"name": "Gallic acid", "id": "CID_370"} | INTERACTS_WITH | - | Target | 3315 | HSPB1 | {"name": "HSPB1", "id": "3315"} |
| Phytochemical | CID_370 | Gallic acid | {"name": "Gallic acid", "id": "CID_370"} | INTERACTS_WITH | - | Target | 3417 | IDH1 | {"name": "IDH1", "id": "3417"} |
| Phytochemical | CID_370 | Gallic acid | {"name": "Gallic acid", "id": "CID_370"} | INTERACTS_WITH | - | Target | 351 | APP | {"name": "APP", "id": "351"} |
| Phytochemical | CID_370 | Gallic acid | {"name": "Gallic acid", "id": "CID_370"} | INTERACTS_WITH | - | Target | 3576 | CXCL8 | {"name": "CXCL8", "id": "3576"} |
| Phytochemical | CID_370 | Gallic acid | {"name": "Gallic acid", "id": "CID_370"} | INTERACTS_WITH | - | Target | 367 | AR | {"name": "AR", "id": "367"} |
| Phytochemical | CID_370 | Gallic acid | {"name": "Gallic acid", "id": "CID_370"} | INTERACTS_WITH | - | Target | 3725 | JUN | {"name": "JUN", "id": "3725"} |
| Phytochemical | CID_370 | Gallic acid | {"name": "Gallic acid", "id": "CID_370"} | INTERACTS_WITH | - | Target | 390245 | KDM4E | {"name": "KDM4E", "id": "390245"} |
| Phytochemical | CID_370 | Gallic acid | {"name": "Gallic acid", "id": "CID_370"} | INTERACTS_WITH | - | Target | 4000 | LMNA | {"name": "LMNA", "id": "4000"} |
| Phytochemical | CID_370 | Gallic acid | {"name": "Gallic acid", "id": "CID_370"} | INTERACTS_WITH | - | Target | 4137 | MAPT | {"name": "MAPT", "id": "4137"} |
| Phytochemical | CID_370 | Gallic acid | {"name": "Gallic acid", "id": "CID_370"} | INTERACTS_WITH | - | Target | 4316 | MMP7 | {"name": "MMP7", "id": "4316"} |
| Phytochemical | CID_370 | Gallic acid | {"name": "Gallic acid", "id": "CID_370"} | INTERACTS_WITH | - | Target | 4780 | NFE2L2 | {"name": "NFE2L2", "id": "4780"} |
| Phytochemical | CID_370 | Gallic acid | {"name": "Gallic acid", "id": "CID_370"} | INTERACTS_WITH | - | Target | 5054 | SERPINE1 | {"name": "SERPINE1", "id": "5054"} |
| Phytochemical | CID_370 | Gallic acid | {"name": "Gallic acid", "id": "CID_370"} | INTERACTS_WITH | - | Target | 51426 | POLK | {"name": "POLK", "id": "51426"} |
| Phytochemical | CID_370 | Gallic acid | {"name": "Gallic acid", "id": "CID_370"} | INTERACTS_WITH | - | Target | 54658 | UGT1A1 | {"name": "UGT1A1", "id": "54658"} |
| Phytochemical | CID_370 | Gallic acid | {"name": "Gallic acid", "id": "CID_370"} | INTERACTS_WITH | - | Target | 55775 | TDP1 | {"name": "TDP1", "id": "55775"} |
| Phytochemical | CID_370 | Gallic acid | {"name": "Gallic acid", "id": "CID_370"} | INTERACTS_WITH | - | Target | 5594 | MAPK1 | {"name": "MAPK1", "id": "5594"} |
| Phytochemical | CID_370 | Gallic acid | {"name": "Gallic acid", "id": "CID_370"} | INTERACTS_WITH | - | Target | 5693 | PSMB5 | {"name": "PSMB5", "id": "5693"} |
| Phytochemical | CID_370 | Gallic acid | {"name": "Gallic acid", "id": "CID_370"} | INTERACTS_WITH | - | Target | 5743 | PTGS2 | {"name": "PTGS2", "id": "5743"} |

**Table S9: Complete node and relationship data used to construct the knowledge graph. This table summarizes interactions among FDA-approved drugs, formulation, herbs, diseases (type 2 diabetes, fatty liver, obesity), phytochemicals, and their target genes/proteins. Each entry includes source and target node details, edge type, and node annotations (Entrez Gene ID and gene symbol, where applicable).**

| Source Node Label | Source Node ID | Source Node Name | Source Node Properties | Edge Type | Edge Properties | Target Node Label | Target Node ID | Target Node Name | Target Node Properties |
|---|---|---|---|---|---|---|---|---|---|
| Phytochemical | CID_370 | Gallic acid | {"name": "Gallic acid", "id": "CID_370"} | INTERACTS_WITH | - | Target | 6401 | SELE | {"name": "SELE", "id": "6401"} |
| Phytochemical | CID_370 | Gallic acid | {"name": "Gallic acid", "id": "CID_370"} | INTERACTS_WITH | - | Target | 6402 | SELL | {"name": "SELL", "id": "6402"} |
| Phytochemical | CID_370 | Gallic acid | {"name": "Gallic acid", "id": "CID_370"} | INTERACTS_WITH | - | Target | 6403 | SELP | {"name": "SELP", "id": "6403"} |
| Phytochemical | CID_370 | Gallic acid | {"name": "Gallic acid", "id": "CID_370"} | INTERACTS_WITH | - | Target | 6606 | SMN1 | {"name": "SMN1", "id": "6606"} |
| Phytochemical | CID_370 | Gallic acid | {"name": "Gallic acid", "id": "CID_370"} | INTERACTS_WITH | - | Target | 6622 | SNCA | {"name": "SNCA", "id": "6622"} |
| Phytochemical | CID_370 | Gallic acid | {"name": "Gallic acid", "id": "CID_370"} | INTERACTS_WITH | - | Target | 6868 | ADAM17 | {"name": "ADAM17", "id": "6868"} |
| Phytochemical | CID_370 | Gallic acid | {"name": "Gallic acid", "id": "CID_370"} | INTERACTS_WITH | - | Target | 7157 | TP53 | {"name": "TP53", "id": "7157"} |
| Phytochemical | CID_370 | Gallic acid | {"name": "Gallic acid", "id": "CID_370"} | INTERACTS_WITH | - | Target | 7421 | VDR | {"name": "VDR", "id": "7421"} |
| Phytochemical | CID_370 | Gallic acid | {"name": "Gallic acid", "id": "CID_370"} | INTERACTS_WITH | - | Target | 759 | CA1 | {"name": "CA1", "id": "759"} |
| Phytochemical | CID_370 | Gallic acid | {"name": "Gallic acid", "id": "CID_370"} | INTERACTS_WITH | - | Target | 760 | CA2 | {"name": "CA2", "id": "760"} |
| Phytochemical | CID_370 | Gallic acid | {"name": "Gallic acid", "id": "CID_370"} | INTERACTS_WITH | - | Target | 761 | CA3 | {"name": "CA3", "id": "761"} |
| Phytochemical | CID_370 | Gallic acid | {"name": "Gallic acid", "id": "CID_370"} | INTERACTS_WITH | - | Target | 762 | CA4 | {"name": "CA4", "id": "762"} |
| Phytochemical | CID_370 | Gallic acid | {"name": "Gallic acid", "id": "CID_370"} | INTERACTS_WITH | - | Target | 763 | CA5A | {"name": "CA5A", "id": "763"} |
| Phytochemical | CID_370 | Gallic acid | {"name": "Gallic acid", "id": "CID_370"} | INTERACTS_WITH | - | Target | 765 | CA6 | {"name": "CA6", "id": "765"} |
| Phytochemical | CID_370 | Gallic acid | {"name": "Gallic acid", "id": "CID_370"} | INTERACTS_WITH | - | Target | 766 | CA7 | {"name": "CA7", "id": "766"} |
| Phytochemical | CID_370 | Gallic acid | {"name": "Gallic acid", "id": "CID_370"} | INTERACTS_WITH | - | Target | 768 | CA9 | {"name": "CA9", "id": "768"} |
| Phytochemical | CID_370 | Gallic acid | {"name": "Gallic acid", "id": "CID_370"} | INTERACTS_WITH | - | Target | 771 | CA12 | {"name": "CA12", "id": "771"} |
| Phytochemical | CID_637566 | Geraniol | {"name": "Geraniol", "id": "CID_637566"} | INTERACTS_WITH | - | Target | 10280 | SIGMAR1 | {"name": "SIGMAR1", "id": "10280"} |
| Phytochemical | CID_637566 | Geraniol | {"name": "Geraniol", "id": "CID_637566"} | INTERACTS_WITH | - | Target | 10800 | CYSLTR1 | {"name": "CYSLTR1", "id": "10800"} |
| Phytochemical | CID_637566 | Geraniol | {"name": "Geraniol", "id": "CID_637566"} | INTERACTS_WITH | - | Target | 10941 | UGT2A1 | {"name": "UGT2A1", "id": "10941"} |
| Phytochemical | CID_637566 | Geraniol | {"name": "Geraniol", "id": "CID_637566"} | INTERACTS_WITH | - | Target | 1128 | CHRM1 | {"name": "CHRM1", "id": "1128"} |
| Phytochemical | CID_637566 | Geraniol | {"name": "Geraniol", "id": "CID_637566"} | INTERACTS_WITH | - | Target | 1129 | CHRM2 | {"name": "CHRM2", "id": "1129"} |
| Phytochemical | CID_637566 | Geraniol | {"name": "Geraniol", "id": "CID_637566"} | INTERACTS_WITH | - | Target | 1131 | CHRM3 | {"name": "CHRM3", "id": "1131"} |
| Phytochemical | CID_637566 | Geraniol | {"name": "Geraniol", "id": "CID_637566"} | INTERACTS_WITH | - | Target | 1132 | CHRM4 | {"name": "CHRM4", "id": "1132"} |
| Phytochemical | CID_637566 | Geraniol | {"name": "Geraniol", "id": "CID_637566"} | INTERACTS_WITH | - | Target | 1133 | CHRM5 | {"name": "CHRM5", "id": "1133"} |
| Phytochemical | CID_637566 | Geraniol | {"name": "Geraniol", "id": "CID_637566"} | INTERACTS_WITH | - | Target | 1233 | CCR4 | {"name": "CCR4", "id": "1233"} |
| Phytochemical | CID_637566 | Geraniol | {"name": "Geraniol", "id": "CID_637566"} | INTERACTS_WITH | - | Target | 1234 | CCR5 | {"name": "CCR5", "id": "1234"} |
| Phytochemical | CID_637566 | Geraniol | {"name": "Geraniol", "id": "CID_637566"} | INTERACTS_WITH | - | Target | 1268 | CNR1 | {"name": "CNR1", "id": "1268"} |
| Phytochemical | CID_637566 | Geraniol | {"name": "Geraniol", "id": "CID_637566"} | INTERACTS_WITH | - | Target | 134 | ADORA1 | {"name": "ADORA1", "id": "134"} |
| Phytochemical | CID_637566 | Geraniol | {"name": "Geraniol", "id": "CID_637566"} | INTERACTS_WITH | - | Target | 135 | ADORA2A | {"name": "ADORA2A", "id": "135"} |
| Phytochemical | CID_637566 | Geraniol | {"name": "Geraniol", "id": "CID_637566"} | INTERACTS_WITH | - | Target | 140 | ADORA3 | {"name": "ADORA3", "id": "140"} |
| Phytochemical | CID_637566 | Geraniol | {"name": "Geraniol", "id": "CID_637566"} | INTERACTS_WITH | - | Target | 1432 | MAPK14 | {"name": "MAPK14", "id": "1432"} |
| Phytochemical | CID_637566 | Geraniol | {"name": "Geraniol", "id": "CID_637566"} | INTERACTS_WITH | - | Target | 146 | ADRA1D | {"name": "ADRA1D", "id": "146"} |
| Phytochemical | CID_637566 | Geraniol | {"name": "Geraniol", "id": "CID_637566"} | INTERACTS_WITH | - | Target | 150 | ADRA2A | {"name": "ADRA2A", "id": "150"} |
| Phytochemical | CID_637566 | Geraniol | {"name": "Geraniol", "id": "CID_637566"} | INTERACTS_WITH | - | Target | 151 | ADRA2B | {"name": "ADRA2B", "id": "151"} |
| Phytochemical | CID_637566 | Geraniol | {"name": "Geraniol", "id": "CID_637566"} | INTERACTS_WITH | - | Target | 1511 | CTSG | {"name": "CTSG", "id": "1511"} |
| Phytochemical | CID_637566 | Geraniol | {"name": "Geraniol", "id": "CID_637566"} | INTERACTS_WITH | - | Target | 152 | ADRA2C | {"name": "ADRA2C", "id": "152"} |
| Phytochemical | CID_637566 | Geraniol | {"name": "Geraniol", "id": "CID_637566"} | INTERACTS_WITH | - | Target | 153 | ADRB1 | {"name": "ADRB1", "id": "153"} |
| Phytochemical | CID_637566 | Geraniol | {"name": "Geraniol", "id": "CID_637566"} | INTERACTS_WITH | - | Target | 154 | ADRB2 | {"name": "ADRB2", "id": "154"} |
| Phytochemical | CID_637566 | Geraniol | {"name": "Geraniol", "id": "CID_637566"} | INTERACTS_WITH | - | Target | 1544 | CYP1A2 | {"name": "CYP1A2", "id": "1544"} |
| Phytochemical | CID_637566 | Geraniol | {"name": "Geraniol", "id": "CID_637566"} | INTERACTS_WITH | - | Target | 1548 | CYP2A6 | {"name": "CYP2A6", "id": "1548"} |
| Phytochemical | CID_637566 | Geraniol | {"name": "Geraniol", "id": "CID_637566"} | INTERACTS_WITH | - | Target | 155 | ADRB3 | {"name": "ADRB3", "id": "155"} |
| Phytochemical | CID_637566 | Geraniol | {"name": "Geraniol", "id": "CID_637566"} | INTERACTS_WITH | - | Target | 1557 | CYP2C19 | {"name": "CYP2C19", "id": "1557"} |
| Phytochemical | CID_637566 | Geraniol | {"name": "Geraniol", "id": "CID_637566"} | INTERACTS_WITH | - | Target | 1559 | CYP2C9 | {"name": "CYP2C9", "id": "1559"} |
| Phytochemical | CID_637566 | Geraniol | {"name": "Geraniol", "id": "CID_637566"} | INTERACTS_WITH | - | Target | 1565 | CYP2D6 | {"name": "CYP2D6", "id": "1565"} |
| Phytochemical | CID_637566 | Geraniol | {"name": "Geraniol", "id": "CID_637566"} | INTERACTS_WITH | - | Target | 1571 | CYP2E1 | {"name": "CYP2E1", "id": "1571"} |
| Phytochemical | CID_637566 | Geraniol | {"name": "Geraniol", "id": "CID_637566"} | INTERACTS_WITH | - | Target | 1576 | CYP3A4 | {"name": "CYP3A4", "id": "1576"} |
| Phytochemical | CID_637566 | Geraniol | {"name": "Geraniol", "id": "CID_637566"} | INTERACTS_WITH | - | Target | 1812 | DRD1 | {"name": "DRD1", "id": "1812"} |
| Phytochemical | CID_637566 | Geraniol | {"name": "Geraniol", "id": "CID_637566"} | INTERACTS_WITH | - | Target | 1813 | DRD2 | {"name": "DRD2", "id": "1813"} |
| Phytochemical | CID_637566 | Geraniol | {"name": "Geraniol", "id": "CID_637566"} | INTERACTS_WITH | - | Target | 1814 | DRD3 | {"name": "DRD3", "id": "1814"} |
| Phytochemical | CID_637566 | Geraniol | {"name": "Geraniol", "id": "CID_637566"} | INTERACTS_WITH | - | Target | 1815 | DRD4 | {"name": "DRD4", "id": "1815"} |
| Phytochemical | CID_637566 | Geraniol | {"name": "Geraniol", "id": "CID_637566"} | INTERACTS_WITH | - | Target | 186 | AGTR2 | {"name": "AGTR2", "id": "186"} |
| Phytochemical | CID_637566 | Geraniol | {"name": "Geraniol", "id": "CID_637566"} | INTERACTS_WITH | - | Target | 1909 | EDNRA | {"name": "EDNRA", "id": "1909"} |
| Phytochemical | CID_637566 | Geraniol | {"name": "Geraniol", "id": "CID_637566"} | INTERACTS_WITH | - | Target | 1956 | EGFR | {"name": "EGFR", "id": "1956"} |
| Phytochemical | CID_637566 | Geraniol | {"name": "Geraniol", "id": "CID_637566"} | INTERACTS_WITH | - | Target | 1991 | ELANE | {"name": "ELANE", "id": "1991"} |
| Phytochemical | CID_637566 | Geraniol | {"name": "Geraniol", "id": "CID_637566"} | INTERACTS_WITH | - | Target | 2064 | ERBB2 | {"name": "ERBB2", "id": "2064"} |
| Phytochemical | CID_637566 | Geraniol | {"name": "Geraniol", "id": "CID_637566"} | INTERACTS_WITH | - | Target | 2099 | ESR1 | {"name": "ESR1", "id": "2099"} |
| Phytochemical | CID_637566 | Geraniol | {"name": "Geraniol", "id": "CID_637566"} | INTERACTS_WITH | - | Target | 2100 | ESR2 | {"name": "ESR2", "id": "2100"} |
| Phytochemical | CID_637566 | Geraniol | {"name": "Geraniol", "id": "CID_637566"} | INTERACTS_WITH | - | Target | 2321 | FLT1 | {"name": "FLT1", "id": "2321"} |
| Phytochemical | CID_637566 | Geraniol | {"name": "Geraniol", "id": "CID_637566"} | INTERACTS_WITH | - | Target | 246 | ALOX15 | {"name": "ALOX15", "id": "246"} |
| Phytochemical | CID_637566 | Geraniol | {"name": "Geraniol", "id": "CID_637566"} | INTERACTS_WITH | - | Target | 2534 | FYN | {"name": "FYN", "id": "2534"} |
| Phytochemical | CID_637566 | Geraniol | {"name": "Geraniol", "id": "CID_637566"} | INTERACTS_WITH | - | Target | 25939 | SAMHD1 | {"name": "SAMHD1", "id": "25939"} |
| Phytochemical | CID_637566 | Geraniol | {"name": "Geraniol", "id": "CID_637566"} | INTERACTS_WITH | - | Target | 26339 | OR5K1 | {"name": "OR5K1", "id": "26339"} |
| Phytochemical | CID_637566 | Geraniol | {"name": "Geraniol", "id": "CID_637566"} | INTERACTS_WITH | - | Target | 2908 | NR3C1 | {"name": "NR3C1", "id": "2908"} |
| Phytochemical | CID_637566 | Geraniol | {"name": "Geraniol", "id": "CID_637566"} | INTERACTS_WITH | - | Target | 3156 | HMGCR | {"name": "HMGCR", "id": "3156"} |
| Phytochemical | CID_637566 | Geraniol | {"name": "Geraniol", "id": "CID_637566"} | INTERACTS_WITH | - | Target | 3269 | HRH1 | {"name": "HRH1", "id": "3269"} |
| Phytochemical | CID_637566 | Geraniol | {"name": "Geraniol", "id": "CID_637566"} | INTERACTS_WITH | - | Target | 3274 | HRH2 | {"name": "HRH2", "id": "3274"} |
| Phytochemical | CID_637566 | Geraniol | {"name": "Geraniol", "id": "CID_637566"} | INTERACTS_WITH | - | Target | 3356 | HTR2A | {"name": "HTR2A", "id": "3356"} |
| Phytochemical | CID_637566 | Geraniol | {"name": "Geraniol", "id": "CID_637566"} | INTERACTS_WITH | - | Target | 3357 | HTR2B | {"name": "HTR2B", "id": "3357"} |
| Phytochemical | CID_637566 | Geraniol | {"name": "Geraniol", "id": "CID_637566"} | INTERACTS_WITH | - | Target | 3358 | HTR2C | {"name": "HTR2C", "id": "3358"} |
| Phytochemical | CID_637566 | Geraniol | {"name": "Geraniol", "id": "CID_637566"} | INTERACTS_WITH | - | Target | 3362 | HTR6 | {"name": "HTR6", "id": "3362"} |
| Phytochemical | CID_637566 | Geraniol | {"name": "Geraniol", "id": "CID_637566"} | INTERACTS_WITH | - | Target | 3577 | CXCR1 | {"name": "CXCR1", "id": "3577"} |
| Phytochemical | CID_637566 | Geraniol | {"name": "Geraniol", "id": "CID_637566"} | INTERACTS_WITH | - | Target | 3579 | CXCR2 | {"name": "CXCR2", "id": "3579"} |
| Phytochemical | CID_637566 | Geraniol | {"name": "Geraniol", "id": "CID_637566"} | INTERACTS_WITH | - | Target | 3757 | KCNH2 | {"name": "KCNH2", "id": "3757"} |
| Phytochemical | CID_637566 | Geraniol | {"name": "Geraniol", "id": "CID_637566"} | INTERACTS_WITH | - | Target | 3932 | LCK | {"name": "LCK", "id": "3932"} |
| Phytochemical | CID_637566 | Geraniol | {"name": "Geraniol", "id": "CID_637566"} | INTERACTS_WITH | - | Target | 4128 | MAOA | {"name": "MAOA", "id": "4128"} |

**Table S9: Complete node and relationship data used to construct the knowledge graph. This table summarizes interactions among FDA-approved drugs, formulation, herbs, diseases (type 2 diabetes, fatty liver, obesity), phytochemicals, and their target genes/proteins. Each entry includes source and target node details, edge type, and node annotations (Entrez Gene ID and gene symbol, where applicable).**

| Source Node Label | Source Node ID | Source Node Name | Source Node Properties | Edge Type | Edge Properties | Target Node Label | Target Node ID | Target Node Name | Target Node Properties |
|---|---|---|---|---|---|---|---|---|---|
| Phytochemical | CID_637566 | Geraniol | {"name": "Geraniol", "id": "CID_637566"} | INTERACTS_WITH | - | Target | 4159 | MC3R | {"name": "MC3R", "id": "4159"} |
| Phytochemical | CID_637566 | Geraniol | {"name": "Geraniol", "id": "CID_637566"} | INTERACTS_WITH | - | Target | 4160 | MC4R | {"name": "MC4R", "id": "4160"} |
| Phytochemical | CID_637566 | Geraniol | {"name": "Geraniol", "id": "CID_637566"} | INTERACTS_WITH | - | Target | 4161 | MC5R | {"name": "MC5R", "id": "4161"} |
| Phytochemical | CID_637566 | Geraniol | {"name": "Geraniol", "id": "CID_637566"} | INTERACTS_WITH | - | Target | 4297 | KMT2A | {"name": "KMT2A", "id": "4297"} |
| Phytochemical | CID_637566 | Geraniol | {"name": "Geraniol", "id": "CID_637566"} | INTERACTS_WITH | - | Target | 43 | ACHE | {"name": "ACHE", "id": "43"} |
| Phytochemical | CID_637566 | Geraniol | {"name": "Geraniol", "id": "CID_637566"} | INTERACTS_WITH | - | Target | 4312 | MMP1 | {"name": "MMP1", "id": "4312"} |
| Phytochemical | CID_637566 | Geraniol | {"name": "Geraniol", "id": "CID_637566"} | INTERACTS_WITH | - | Target | 4318 | MMP9 | {"name": "MMP9", "id": "4318"} |
| Phytochemical | CID_637566 | Geraniol | {"name": "Geraniol", "id": "CID_637566"} | INTERACTS_WITH | - | Target | 4790 | NFKB1 | {"name": "NFKB1", "id": "4790"} |
| Phytochemical | CID_637566 | Geraniol | {"name": "Geraniol", "id": "CID_637566"} | INTERACTS_WITH | - | Target | 4886 | NPY1R | {"name": "NPY1R", "id": "4886"} |
| Phytochemical | CID_637566 | Geraniol | {"name": "Geraniol", "id": "CID_637566"} | INTERACTS_WITH | - | Target | 4887 | NPY2R | {"name": "NPY2R", "id": "4887"} |
| Phytochemical | CID_637566 | Geraniol | {"name": "Geraniol", "id": "CID_637566"} | INTERACTS_WITH | - | Target | 4985 | OPRD1 | {"name": "OPRD1", "id": "4985"} |
| Phytochemical | CID_637566 | Geraniol | {"name": "Geraniol", "id": "CID_637566"} | INTERACTS_WITH | - | Target | 4986 | OPRK1 | {"name": "OPRK1", "id": "4986"} |
| Phytochemical | CID_637566 | Geraniol | {"name": "Geraniol", "id": "CID_637566"} | INTERACTS_WITH | - | Target | 4988 | OPRM1 | {"name": "OPRM1", "id": "4988"} |
| Phytochemical | CID_637566 | Geraniol | {"name": "Geraniol", "id": "CID_637566"} | INTERACTS_WITH | - | Target | 5467 | PPARD | {"name": "PPARD", "id": "5467"} |
| Phytochemical | CID_637566 | Geraniol | {"name": "Geraniol", "id": "CID_637566"} | INTERACTS_WITH | - | Target | 552 | AVPR1A | {"name": "AVPR1A", "id": "552"} |
| Phytochemical | CID_637566 | Geraniol | {"name": "Geraniol", "id": "CID_637566"} | INTERACTS_WITH | - | Target | 5530 | PPP3CA | {"name": "PPP3CA", "id": "5530"} |
| Phytochemical | CID_637566 | Geraniol | {"name": "Geraniol", "id": "CID_637566"} | INTERACTS_WITH | - | Target | 5578 | PRKCA | {"name": "PRKCA", "id": "5578"} |
| Phytochemical | CID_637566 | Geraniol | {"name": "Geraniol", "id": "CID_637566"} | INTERACTS_WITH | - | Target | 5594 | MAPK1 | {"name": "MAPK1", "id": "5594"} |
| Phytochemical | CID_637566 | Geraniol | {"name": "Geraniol", "id": "CID_637566"} | INTERACTS_WITH | - | Target | 5595 | MAPK3 | {"name": "MAPK3", "id": "5595"} |
| Phytochemical | CID_637566 | Geraniol | {"name": "Geraniol", "id": "CID_637566"} | INTERACTS_WITH | - | Target | 5724 | PTAFR | {"name": "PTAFR", "id": "5724"} |
| Phytochemical | CID_637566 | Geraniol | {"name": "Geraniol", "id": "CID_637566"} | INTERACTS_WITH | - | Target | 5742 | PTGS1 | {"name": "PTGS1", "id": "5742"} |
| Phytochemical | CID_637566 | Geraniol | {"name": "Geraniol", "id": "CID_637566"} | INTERACTS_WITH | - | Target | 5743 | PTGS2 | {"name": "PTGS2", "id": "5743"} |
| Phytochemical | CID_637566 | Geraniol | {"name": "Geraniol", "id": "CID_637566"} | INTERACTS_WITH | - | Target | 5788 | PTPRC | {"name": "PTPRC", "id": "5788"} |
| Phytochemical | CID_637566 | Geraniol | {"name": "Geraniol", "id": "CID_637566"} | INTERACTS_WITH | - | Target | 624 | BDKRB2 | {"name": "BDKRB2", "id": "624"} |
| Phytochemical | CID_637566 | Geraniol | {"name": "Geraniol", "id": "CID_637566"} | INTERACTS_WITH | - | Target | 6530 | SLC6A2 | {"name": "SLC6A2", "id": "6530"} |
| Phytochemical | CID_637566 | Geraniol | {"name": "Geraniol", "id": "CID_637566"} | INTERACTS_WITH | - | Target | 6531 | SLC6A3 | {"name": "SLC6A3", "id": "6531"} |
| Phytochemical | CID_637566 | Geraniol | {"name": "Geraniol", "id": "CID_637566"} | INTERACTS_WITH | - | Target | 6532 | SLC6A4 | {"name": "SLC6A4", "id": "6532"} |
| Phytochemical | CID_637566 | Geraniol | {"name": "Geraniol", "id": "CID_637566"} | INTERACTS_WITH | - | Target | 6865 | TACR2 | {"name": "TACR2", "id": "6865"} |
| Phytochemical | CID_637566 | Geraniol | {"name": "Geraniol", "id": "CID_637566"} | INTERACTS_WITH | - | Target | 6869 | TACR1 | {"name": "TACR1", "id": "6869"} |
| Phytochemical | CID_637566 | Geraniol | {"name": "Geraniol", "id": "CID_637566"} | INTERACTS_WITH | - | Target | 6916 | TBXAS1 | {"name": "TBXAS1", "id": "6916"} |
| Phytochemical | CID_637566 | Geraniol | {"name": "Geraniol", "id": "CID_637566"} | INTERACTS_WITH | - | Target | 729230 | CCR2 | {"name": "CCR2", "id": "729230"} |
| Phytochemical | CID_637566 | Geraniol | {"name": "Geraniol", "id": "CID_637566"} | INTERACTS_WITH | - | Target | 7421 | VDR | {"name": "VDR", "id": "7421"} |
| Phytochemical | CID_637566 | Geraniol | {"name": "Geraniol", "id": "CID_637566"} | INTERACTS_WITH | - | Target | 7433 | VIPR1 | {"name": "VIPR1", "id": "7433"} |
| Phytochemical | CID_637566 | Geraniol | {"name": "Geraniol", "id": "CID_637566"} | INTERACTS_WITH | - | Target | 7442 | TRPV1 | {"name": "TRPV1", "id": "7442"} |
| Phytochemical | CID_637566 | Geraniol | {"name": "Geraniol", "id": "CID_637566"} | INTERACTS_WITH | - | Target | 760 | CA2 | {"name": "CA2", "id": "760"} |
| Phytochemical | CID_637566 | Geraniol | {"name": "Geraniol", "id": "CID_637566"} | INTERACTS_WITH | - | Target | 79054 | TRPM8 | {"name": "TRPM8", "id": "79054"} |
| Phytochemical | CID_637566 | Geraniol | {"name": "Geraniol", "id": "CID_637566"} | INTERACTS_WITH | - | Target | 799 | CALCR | {"name": "CALCR", "id": "799"} |
| Phytochemical | CID_637566 | Geraniol | {"name": "Geraniol", "id": "CID_637566"} | INTERACTS_WITH | - | Target | 834 | CASP1 | {"name": "CASP1", "id": "834"} |
| Phytochemical | CID_637566 | Geraniol | {"name": "Geraniol", "id": "CID_637566"} | INTERACTS_WITH | - | Target | 8654 | PDE5A | {"name": "PDE5A", "id": "8654"} |
| Phytochemical | CID_637566 | Geraniol | {"name": "Geraniol", "id": "CID_637566"} | INTERACTS_WITH | - | Target | 886 | CCKAR | {"name": "CCKAR", "id": "886"} |
| Phytochemical | CID_1549026 | Geranyl acetate | {"name": "Geranyl acetate", "id": "CID_1549026"} | INTERACTS_WITH | - | Target | 216 | ALDH1A1 | {"name": "ALDH1A1", "id": "216"} |
| Phytochemical | CID_1549026 | Geranyl acetate | {"name": "Geranyl acetate", "id": "CID_1549026"} | INTERACTS_WITH | - | Target | 246 | ALOX15 | {"name": "ALOX15", "id": "246"} |
| Phytochemical | CID_1549026 | Geranyl acetate | {"name": "Geranyl acetate", "id": "CID_1549026"} | INTERACTS_WITH | - | Target | 367 | AR | {"name": "AR", "id": "367"} |
| Phytochemical | CID_1549026 | Geranyl acetate | {"name": "Geranyl acetate", "id": "CID_1549026"} | INTERACTS_WITH | - | Target | 4780 | NFE2L2 | {"name": "NFE2L2", "id": "4780"} |
| Phytochemical | CID_1549026 | Geranyl acetate | {"name": "Geranyl acetate", "id": "CID_1549026"} | INTERACTS_WITH | - | Target | 55775 | TDP1 | {"name": "TDP1", "id": "55775"} |
| Phytochemical | CID_1549026 | Geranyl acetate | {"name": "Geranyl acetate", "id": "CID_1549026"} | INTERACTS_WITH | - | Target | 5745 | PTH1R | {"name": "PTH1R", "id": "5745"} |
| Phytochemical | CID_1549026 | Geranyl acetate | {"name": "Geranyl acetate", "id": "CID_1549026"} | INTERACTS_WITH | - | Target | 7442 | TRPV1 | {"name": "TRPV1", "id": "7442"} |
| Phytochemical | CID_1549026 | Geranyl acetate | {"name": "Geranyl acetate", "id": "CID_1549026"} | INTERACTS_WITH | - | Target | 9971 | NR1H4 | {"name": "NR1H4", "id": "9971"} |
| Phytochemical | CID_984 | Hexadecanal | {"name": "Hexadecanal", "id": "CID_984"} | INTERACTS_WITH | - | Target | 10013 | HDAC6 | {"name": "HDAC6", "id": "10013"} |
| Phytochemical | CID_984 | Hexadecanal | {"name": "Hexadecanal", "id": "CID_984"} | INTERACTS_WITH | - | Target | 4023 | LPL | {"name": "LPL", "id": "4023"} |
| Phytochemical | CID_984 | Hexadecanal | {"name": "Hexadecanal", "id": "CID_984"} | INTERACTS_WITH | - | Target | 9388 | LIPG | {"name": "LIPG", "id": "9388"} |
| Phytochemical | CID_5281520 | Humulene | {"name": "Humulene", "id": "CID_5281520"} | INTERACTS_WITH | - | Target | 10599 | SLCO1B1 | {"name": "SLCO1B1", "id": "10599"} |
| Phytochemical | CID_5281520 | Humulene | {"name": "Humulene", "id": "CID_5281520"} | INTERACTS_WITH | - | Target | 28234 | SLCO1B3 | {"name": "SLCO1B3", "id": "28234"} |
| Phytochemical | CID_5281654 | Isorhamnetin | {"name": "Isorhamnetin", "id": "CID_5281654"} | INTERACTS_WITH | - | Target | 1432 | MAPK14 | {"name": "MAPK14", "id": "1432"} |
| Phytochemical | CID_5281654 | Isorhamnetin | {"name": "Isorhamnetin", "id": "CID_5281654"} | INTERACTS_WITH | - | Target | 1543 | CYP1A1 | {"name": "CYP1A1", "id": "1543"} |
| Phytochemical | CID_5281654 | Isorhamnetin | {"name": "Isorhamnetin", "id": "CID_5281654"} | INTERACTS_WITH | - | Target | 1544 | CYP1A2 | {"name": "CYP1A2", "id": "1544"} |
| Phytochemical | CID_5281654 | Isorhamnetin | {"name": "Isorhamnetin", "id": "CID_5281654"} | INTERACTS_WITH | - | Target | 1545 | CYP1B1 | {"name": "CYP1B1", "id": "1545"} |
| Phytochemical | CID_5281654 | Isorhamnetin | {"name": "Isorhamnetin", "id": "CID_5281654"} | INTERACTS_WITH | - | Target | 2237 | FEN1 | {"name": "FEN1", "id": "2237"} |
| Phytochemical | CID_5281654 | Isorhamnetin | {"name": "Isorhamnetin", "id": "CID_5281654"} | INTERACTS_WITH | - | Target | 253430 | IPMK | {"name": "IPMK", "id": "253430"} |
| Phytochemical | CID_5281654 | Isorhamnetin | {"name": "Isorhamnetin", "id": "CID_5281654"} | INTERACTS_WITH | - | Target | 276 | AMY1A | {"name": "AMY1A", "id": "276"} |
| Phytochemical | CID_5281654 | Isorhamnetin | {"name": "Isorhamnetin", "id": "CID_5281654"} | INTERACTS_WITH | - | Target | 277 | AMY1B | {"name": "AMY1B", "id": "277"} |
| Phytochemical | CID_5281654 | Isorhamnetin | {"name": "Isorhamnetin", "id": "CID_5281654"} | INTERACTS_WITH | - | Target | 328 | APEX1 | {"name": "APEX1", "id": "328"} |
| Phytochemical | CID_5281654 | Isorhamnetin | {"name": "Isorhamnetin", "id": "CID_5281654"} | INTERACTS_WITH | - | Target | 4297 | KMT2A | {"name": "KMT2A", "id": "4297"} |
| Phytochemical | CID_5281654 | Isorhamnetin | {"name": "Isorhamnetin", "id": "CID_5281654"} | INTERACTS_WITH | - | Target | 4790 | NFKB1 | {"name": "NFKB1", "id": "4790"} |

**Table S9: Complete node and relationship data used to construct the knowledge graph. This table summarizes interactions among FDA-approved drugs, formulation, herbs, diseases (type 2 diabetes, fatty liver, obesity), phytochemicals, and their target genes/proteins. Each entry includes source and target node details, edge type, and node annotations (Entrez Gene ID and gene symbol, where applicable).**

| Source Node Label | Source Node ID | Source Node Name | Source Node Properties | Edge Type | Edge Properties | Target Node Label | Target Node ID | Target Node Name | Target Node Properties |
|---|---|---|---|---|---|---|---|---|---|
| Phytochemical | CID_5281654 | Isorhamnetin | {"name": "Isorhamnetin", "id": "CID_5281654"} | INTERACTS_WITH | - | Target | 4791 | NFKB2 | {"name": "NFKB2", "id": "4791"} |
| Phytochemical | CID_5281654 | Isorhamnetin | {"name": "Isorhamnetin", "id": "CID_5281654"} | INTERACTS_WITH | - | Target | 51426 | POLK | {"name": "POLK", "id": "51426"} |
| Phytochemical | CID_5281654 | Isorhamnetin | {"name": "Isorhamnetin", "id": "CID_5281654"} | INTERACTS_WITH | - | Target | 51447 | IP6K2 | {"name": "IP6K2", "id": "51447"} |
| Phytochemical | CID_5281654 | Isorhamnetin | {"name": "Isorhamnetin", "id": "CID_5281654"} | INTERACTS_WITH | - | Target | 5468 | PPARG | {"name": "PPARG", "id": "5468"} |
| Phytochemical | CID_5281654 | Isorhamnetin | {"name": "Isorhamnetin", "id": "CID_5281654"} | INTERACTS_WITH | - | Target | 5594 | MAPK1 | {"name": "MAPK1", "id": "5594"} |
| Phytochemical | CID_5281654 | Isorhamnetin | {"name": "Isorhamnetin", "id": "CID_5281654"} | INTERACTS_WITH | - | Target | 5595 | MAPK3 | {"name": "MAPK3", "id": "5595"} |
| Phytochemical | CID_5281654 | Isorhamnetin | {"name": "Isorhamnetin", "id": "CID_5281654"} | INTERACTS_WITH | - | Target | 5600 | MAPK11 | {"name": "MAPK11", "id": "5600"} |
| Phytochemical | CID_5281654 | Isorhamnetin | {"name": "Isorhamnetin", "id": "CID_5281654"} | INTERACTS_WITH | - | Target | 5602 | MAPK10 | {"name": "MAPK10", "id": "5602"} |
| Phytochemical | CID_5281654 | Isorhamnetin | {"name": "Isorhamnetin", "id": "CID_5281654"} | INTERACTS_WITH | - | Target | 5603 | MAPK13 | {"name": "MAPK13", "id": "5603"} |
| Phytochemical | CID_5281654 | Isorhamnetin | {"name": "Isorhamnetin", "id": "CID_5281654"} | INTERACTS_WITH | - | Target | 5970 | RELA | {"name": "RELA", "id": "5970"} |
| Phytochemical | CID_5281654 | Isorhamnetin | {"name": "Isorhamnetin", "id": "CID_5281654"} | INTERACTS_WITH | - | Target | 6300 | MAPK12 | {"name": "MAPK12", "id": "6300"} |
| Phytochemical | CID_5281654 | Isorhamnetin | {"name": "Isorhamnetin", "id": "CID_5281654"} | INTERACTS_WITH | - | Target | 7498 | XDH | {"name": "XDH", "id": "7498"} |
| Phytochemical | CID_5281654 | Isorhamnetin | {"name": "Isorhamnetin", "id": "CID_5281654"} | INTERACTS_WITH | - | Target | 759 | CA1 | {"name": "CA1", "id": "759"} |
| Phytochemical | CID_5281654 | Isorhamnetin | {"name": "Isorhamnetin", "id": "CID_5281654"} | INTERACTS_WITH | - | Target | 760 | CA2 | {"name": "CA2", "id": "760"} |
| Phytochemical | CID_5281654 | Isorhamnetin | {"name": "Isorhamnetin", "id": "CID_5281654"} | INTERACTS_WITH | - | Target | 762 | CA4 | {"name": "CA4", "id": "762"} |
| Phytochemical | CID_5281654 | Isorhamnetin | {"name": "Isorhamnetin", "id": "CID_5281654"} | INTERACTS_WITH | - | Target | 766 | CA7 | {"name": "CA7", "id": "766"} |
| Phytochemical | CID_5281654 | Isorhamnetin | {"name": "Isorhamnetin", "id": "CID_5281654"} | INTERACTS_WITH | - | Target | 771 | CA12 | {"name": "CA12", "id": "771"} |
| Phytochemical | CID_5281654 | Isorhamnetin | {"name": "Isorhamnetin", "id": "CID_5281654"} | INTERACTS_WITH | - | Target | 9212 | AURKB | {"name": "AURKB", "id": "9212"} |
| Phytochemical | CID_25310 | L-Rhamnose | {"name": "L-Rhamnose", "id": "CID_25310"} | INTERACTS_WITH | - | Target | 10013 | HDAC6 | {"name": "HDAC6", "id": "10013"} |
| Phytochemical | CID_442343 | Levomenol | {"name": "Levomenol", "id": "CID_442343"} | INTERACTS_WITH | - | Target | 10013 | HDAC6 | {"name": "HDAC6", "id": "10013"} |
| Phytochemical | CID_442343 | Levomenol | {"name": "Levomenol", "id": "CID_442343"} | INTERACTS_WITH | - | Target | 10599 | SLCO1B1 | {"name": "SLCO1B1", "id": "10599"} |
| Phytochemical | CID_442343 | Levomenol | {"name": "Levomenol", "id": "CID_442343"} | INTERACTS_WITH | - | Target | 1139 | CHRNA7 | {"name": "CHRNA7", "id": "1139"} |
| Phytochemical | CID_442343 | Levomenol | {"name": "Levomenol", "id": "CID_442343"} | INTERACTS_WITH | - | Target | 28234 | SLCO1B3 | {"name": "SLCO1B3", "id": "28234"} |
| Phytochemical | CID_442343 | Levomenol | {"name": "Levomenol", "id": "CID_442343"} | INTERACTS_WITH | - | Target | 3315 | HSPB1 | {"name": "HSPB1", "id": "3315"} |
| Phytochemical | CID_442343 | Levomenol | {"name": "Levomenol", "id": "CID_442343"} | INTERACTS_WITH | - | Target | 3725 | JUN | {"name": "JUN", "id": "3725"} |
| Phytochemical | CID_442343 | Levomenol | {"name": "Levomenol", "id": "CID_442343"} | INTERACTS_WITH | - | Target | 6774 | STAT3 | {"name": "STAT3", "id": "6774"} |
| Phytochemical | CID_22311 | Limonene | {"name": "Limonene", "id": "CID_22311"} | INTERACTS_WITH | - | Target | 367 | AR | {"name": "AR", "id": "367"} |
| Phytochemical | CID_22311 | Limonene | {"name": "Limonene", "id": "CID_22311"} | INTERACTS_WITH | - | Target | 4000 | LMNA | {"name": "LMNA", "id": "4000"} |
| Phytochemical | CID_22311 | Limonene | {"name": "Limonene", "id": "CID_22311"} | INTERACTS_WITH | - | Target | 4780 | NFE2L2 | {"name": "NFE2L2", "id": "4780"} |
| Phytochemical | CID_22311 | Limonene | {"name": "Limonene", "id": "CID_22311"} | INTERACTS_WITH | - | Target | 4790 | NFKB1 | {"name": "NFKB1", "id": "4790"} |
| Phytochemical | CID_22311 | Limonene | {"name": "Limonene", "id": "CID_22311"} | INTERACTS_WITH | - | Target | 5467 | PPARD | {"name": "PPARD", "id": "5467"} |
| Phytochemical | CID_22311 | Limonene | {"name": "Limonene", "id": "CID_22311"} | INTERACTS_WITH | - | Target | 6256 | RXRA | {"name": "RXRA", "id": "6256"} |
| Phytochemical | CID_22311 | Limonene | {"name": "Limonene", "id": "CID_22311"} | INTERACTS_WITH | - | Target | 7068 | THRB | {"name": "THRB", "id": "7068"} |
| Phytochemical | CID_22311 | Limonene | {"name": "Limonene", "id": "CID_22311"} | INTERACTS_WITH | - | Target | 7157 | TP53 | {"name": "TP53", "id": "7157"} |
| Phytochemical | CID_6549 | Linalool | {"name": "Linalool", "id": "CID_6549"} | INTERACTS_WITH | - | Target | 10013 | HDAC6 | {"name": "HDAC6", "id": "10013"} |
| Phytochemical | CID_6549 | Linalool | {"name": "Linalool", "id": "CID_6549"} | INTERACTS_WITH | - | Target | 10073 | SNUPN | {"name": "SNUPN", "id": "10073"} |
| Phytochemical | CID_6549 | Linalool | {"name": "Linalool", "id": "CID_6549"} | INTERACTS_WITH | - | Target | 10941 | UGT2A1 | {"name": "UGT2A1", "id": "10941"} |
| Phytochemical | CID_6549 | Linalool | {"name": "Linalool", "id": "CID_6549"} | INTERACTS_WITH | - | Target | 10951 | CBX1 | {"name": "CBX1", "id": "10951"} |
| Phytochemical | CID_6549 | Linalool | {"name": "Linalool", "id": "CID_6549"} | INTERACTS_WITH | - | Target | 3359 | HTR3A | {"name": "HTR3A", "id": "3359"} |
| Phytochemical | CID_6549 | Linalool | {"name": "Linalool", "id": "CID_6549"} | INTERACTS_WITH | - | Target | 3837 | KPNB1 | {"name": "KPNB1", "id": "3837"} |
| Phytochemical | CID_6549 | Linalool | {"name": "Linalool", "id": "CID_6549"} | INTERACTS_WITH | - | Target | 54657 | UGT1A4 | {"name": "UGT1A4", "id": "54657"} |
| Phytochemical | CID_6549 | Linalool | {"name": "Linalool", "id": "CID_6549"} | INTERACTS_WITH | - | Target | 54658 | UGT1A1 | {"name": "UGT1A1", "id": "54658"} |
| Phytochemical | CID_6549 | Linalool | {"name": "Linalool", "id": "CID_6549"} | INTERACTS_WITH | - | Target | 7068 | THRB | {"name": "THRB", "id": "7068"} |
| Phytochemical | CID_8294 | Linalyl acetate | {"name": "Linalyl acetate", "id": "CID_8294"} | INTERACTS_WITH | - | Target | 2908 | NR3C1 | {"name": "NR3C1", "id": "2908"} |
| Phytochemical | CID_8294 | Linalyl acetate | {"name": "Linalyl acetate", "id": "CID_8294"} | INTERACTS_WITH | - | Target | 4790 | NFKB1 | {"name": "NFKB1", "id": "4790"} |
| Phytochemical | CID_6532 | Linalyl isobutyrate | {"name": "Linalyl isobutyrate", "id": "CID_6532"} | INTERACTS_WITH | - | Target | 4790 | NFKB1 | {"name": "NFKB1", "id": "4790"} |
| Phytochemical | CID_5280450 | Linoleic acid | {"name": "Linoleic acid", "id": "CID_5280450"} | INTERACTS_WITH | - | Target | 1576 | CYP3A4 | {"name": "CYP3A4", "id": "1576"} |
| Phytochemical | CID_5280450 | Linoleic acid | {"name": "Linoleic acid", "id": "CID_5280450"} | INTERACTS_WITH | - | Target | 1588 | CYP19A1 | {"name": "CYP19A1", "id": "1588"} |
| Phytochemical | CID_5280450 | Linoleic acid | {"name": "Linoleic acid", "id": "CID_5280450"} | INTERACTS_WITH | - | Target | 1845 | DUSP3 | {"name": "DUSP3", "id": "1845"} |
| Phytochemical | CID_5280450 | Linoleic acid | {"name": "Linoleic acid", "id": "CID_5280450"} | INTERACTS_WITH | - | Target | 2099 | ESR1 | {"name": "ESR1", "id": "2099"} |
| Phytochemical | CID_5280450 | Linoleic acid | {"name": "Linoleic acid", "id": "CID_5280450"} | INTERACTS_WITH | - | Target | 2152 | F3 | {"name": "F3", "id": "2152"} |
| Phytochemical | CID_5280450 | Linoleic acid | {"name": "Linoleic acid", "id": "CID_5280450"} | INTERACTS_WITH | - | Target | 2155 | F7 | {"name": "F7", "id": "2155"} |
| Phytochemical | CID_5280450 | Linoleic acid | {"name": "Linoleic acid", "id": "CID_5280450"} | INTERACTS_WITH | - | Target | 216 | ALDH1A1 | {"name": "ALDH1A1", "id": "216"} |
| Phytochemical | CID_5280450 | Linoleic acid | {"name": "Linoleic acid", "id": "CID_5280450"} | INTERACTS_WITH | - | Target | 2167 | FABP4 | {"name": "FABP4", "id": "2167"} |
| Phytochemical | CID_5280450 | Linoleic acid | {"name": "Linoleic acid", "id": "CID_5280450"} | INTERACTS_WITH | - | Target | 2170 | FABP3 | {"name": "FABP3", "id": "2170"} |

**Table S9: Complete node and relationship data used to construct the knowledge graph. This table summarizes interactions among FDA-approved drugs, formulation, herbs, diseases (type 2 diabetes, fatty liver, obesity), phytochemicals, and their target genes/proteins. Each entry includes source and target node details, edge type, and node annotations (Entrez Gene ID and gene symbol, where applicable).**

| Source Node Label | Source Node ID | Source Node Name | Source Node Properties | Edge Type | Edge Properties | Target Node Label | Target Node ID | Target Node Name | Target Node Properties |
|---|---|---|---|---|---|---|---|---|---|
| Phytochemical | CID_5280450 | Linoleic acid | {"name": "Linoleic acid", "id": "CID_5280450"} | INTERACTS_WITH | - | Target | 246 | ALOX15 | {"name": "ALOX15", "id": "246"} |
| Phytochemical | CID_5280450 | Linoleic acid | {"name": "Linoleic acid", "id": "CID_5280450"} | INTERACTS_WITH | - | Target | 2648 | KAT2A | {"name": "KAT2A", "id": "2648"} |
| Phytochemical | CID_5280450 | Linoleic acid | {"name": "Linoleic acid", "id": "CID_5280450"} | INTERACTS_WITH | - | Target | 2769 | GNA15 | {"name": "GNA15", "id": "2769"} |
| Phytochemical | CID_5280450 | Linoleic acid | {"name": "Linoleic acid", "id": "CID_5280450"} | INTERACTS_WITH | - | Target | 2864 | FFAR1 | {"name": "FFAR1", "id": "2864"} |
| Phytochemical | CID_5280450 | Linoleic acid | {"name": "Linoleic acid", "id": "CID_5280450"} | INTERACTS_WITH | - | Target | 3028 | HSD17B10 | {"name": "HSD17B10", "id": "3028"} |
| Phytochemical | CID_5280450 | Linoleic acid | {"name": "Linoleic acid", "id": "CID_5280450"} | INTERACTS_WITH | - | Target | 3248 | HPGD | {"name": "HPGD", "id": "3248"} |
| Phytochemical | CID_5280450 | Linoleic acid | {"name": "Linoleic acid", "id": "CID_5280450"} | INTERACTS_WITH | - | Target | 3315 | HSPB1 | {"name": "HSPB1", "id": "3315"} |
| Phytochemical | CID_5280450 | Linoleic acid | {"name": "Linoleic acid", "id": "CID_5280450"} | INTERACTS_WITH | - | Target | 338557 | FFAR4 | {"name": "FFAR4", "id": "338557"} |
| Phytochemical | CID_5280450 | Linoleic acid | {"name": "Linoleic acid", "id": "CID_5280450"} | INTERACTS_WITH | - | Target | 367 | AR | {"name": "AR", "id": "367"} |
| Phytochemical | CID_5280450 | Linoleic acid | {"name": "Linoleic acid", "id": "CID_5280450"} | INTERACTS_WITH | - | Target | 3725 | JUN | {"name": "JUN", "id": "3725"} |
| Phytochemical | CID_5280450 | Linoleic acid | {"name": "Linoleic acid", "id": "CID_5280450"} | INTERACTS_WITH | - | Target | 4000 | LMNA | {"name": "LMNA", "id": "4000"} |
| Phytochemical | CID_5280450 | Linoleic acid | {"name": "Linoleic acid", "id": "CID_5280450"} | INTERACTS_WITH | - | Target | 4780 | NFE2L2 | {"name": "NFE2L2", "id": "4780"} |
| Phytochemical | CID_5280450 | Linoleic acid | {"name": "Linoleic acid", "id": "CID_5280450"} | INTERACTS_WITH | - | Target | 4790 | NFKB1 | {"name": "NFKB1", "id": "4790"} |
| Phytochemical | CID_5280450 | Linoleic acid | {"name": "Linoleic acid", "id": "CID_5280450"} | INTERACTS_WITH | - | Target | 4929 | NR4A2 | {"name": "NR4A2", "id": "4929"} |
| Phytochemical | CID_5280450 | Linoleic acid | {"name": "Linoleic acid", "id": "CID_5280450"} | INTERACTS_WITH | - | Target | 51053 | GMNN | {"name": "GMNN", "id": "51053"} |
| Phytochemical | CID_5280450 | Linoleic acid | {"name": "Linoleic acid", "id": "CID_5280450"} | INTERACTS_WITH | - | Target | 51426 | POLK | {"name": "POLK", "id": "51426"} |
| Phytochemical | CID_5280450 | Linoleic acid | {"name": "Linoleic acid", "id": "CID_5280450"} | INTERACTS_WITH | - | Target | 51548 | SIRT6 | {"name": "SIRT6", "id": "51548"} |
| Phytochemical | CID_5280450 | Linoleic acid | {"name": "Linoleic acid", "id": "CID_5280450"} | INTERACTS_WITH | - | Target | 5465 | PPARA | {"name": "PPARA", "id": "5465"} |
| Phytochemical | CID_5280450 | Linoleic acid | {"name": "Linoleic acid", "id": "CID_5280450"} | INTERACTS_WITH | - | Target | 5467 | PPARD | {"name": "PPARD", "id": "5467"} |
| Phytochemical | CID_5280450 | Linoleic acid | {"name": "Linoleic acid", "id": "CID_5280450"} | INTERACTS_WITH | - | Target | 5468 | PPARG | {"name": "PPARG", "id": "5468"} |
| Phytochemical | CID_5280450 | Linoleic acid | {"name": "Linoleic acid", "id": "CID_5280450"} | INTERACTS_WITH | - | Target | 55775 | TDP1 | {"name": "TDP1", "id": "55775"} |
| Phytochemical | CID_5280450 | Linoleic acid | {"name": "Linoleic acid", "id": "CID_5280450"} | INTERACTS_WITH | - | Target | 5742 | PTGS1 | {"name": "PTGS1", "id": "5742"} |
| Phytochemical | CID_5280450 | Linoleic acid | {"name": "Linoleic acid", "id": "CID_5280450"} | INTERACTS_WITH | - | Target | 5778 | PTPN7 | {"name": "PTPN7", "id": "5778"} |
| Phytochemical | CID_5280450 | Linoleic acid | {"name": "Linoleic acid", "id": "CID_5280450"} | INTERACTS_WITH | - | Target | 6256 | RXRA | {"name": "RXRA", "id": "6256"} |
| Phytochemical | CID_5280450 | Linoleic acid | {"name": "Linoleic acid", "id": "CID_5280450"} | INTERACTS_WITH | - | Target | 6646 | SOAT1 | {"name": "SOAT1", "id": "6646"} |
| Phytochemical | CID_5280450 | Linoleic acid | {"name": "Linoleic acid", "id": "CID_5280450"} | INTERACTS_WITH | - | Target | 7015 | TERT | {"name": "TERT", "id": "7015"} |
| Phytochemical | CID_5280450 | Linoleic acid | {"name": "Linoleic acid", "id": "CID_5280450"} | INTERACTS_WITH | - | Target | 7253 | TSHR | {"name": "TSHR", "id": "7253"} |
| Phytochemical | CID_5280450 | Linoleic acid | {"name": "Linoleic acid", "id": "CID_5280450"} | INTERACTS_WITH | - | Target | 7398 | USP1 | {"name": "USP1", "id": "7398"} |
| Phytochemical | CID_5280450 | Linoleic acid | {"name": "Linoleic acid", "id": "CID_5280450"} | INTERACTS_WITH | - | Target | 8644 | AKR1C3 | {"name": "AKR1C3", "id": "8644"} |
| Phytochemical | CID_5280450 | Linoleic acid | {"name": "Linoleic acid", "id": "CID_5280450"} | INTERACTS_WITH | - | Target | 9971 | NR1H4 | {"name": "NR1H4", "id": "9971"} |
| Phytochemical | CID_5281243 | Lutein | {"name": "Lutein", "id": "CID_5281243"} | INTERACTS_WITH | - | Target | 10013 | HDAC6 | {"name": "HDAC6", "id": "10013"} |
| Phytochemical | CID_5281243 | Lutein | {"name": "Lutein", "id": "CID_5281243"} | INTERACTS_WITH | - | Target | 367 | AR | {"name": "AR", "id": "367"} |
| Phytochemical | CID_5281243 | Lutein | {"name": "Lutein", "id": "CID_5281243"} | INTERACTS_WITH | - | Target | 55775 | TDP1 | {"name": "TDP1", "id": "55775"} |
| Phytochemical | CID_6251 | Mannitol | {"name": "Mannitol", "id": "CID_6251"} | INTERACTS_WITH | - | Target | 10280 | SIGMAR1 | {"name": "SIGMAR1", "id": "10280"} |
| Phytochemical | CID_6251 | Mannitol | {"name": "Mannitol", "id": "CID_6251"} | INTERACTS_WITH | - | Target | 10800 | CYSLTR1 | {"name": "CYSLTR1", "id": "10800"} |
| Phytochemical | CID_6251 | Mannitol | {"name": "Mannitol", "id": "CID_6251"} | INTERACTS_WITH | - | Target | 11255 | HRH3 | {"name": "HRH3", "id": "11255"} |
| Phytochemical | CID_6251 | Mannitol | {"name": "Mannitol", "id": "CID_6251"} | INTERACTS_WITH | - | Target | 1128 | CHRM1 | {"name": "CHRM1", "id": "1128"} |
| Phytochemical | CID_6251 | Mannitol | {"name": "Mannitol", "id": "CID_6251"} | INTERACTS_WITH | - | Target | 1129 | CHRM2 | {"name": "CHRM2", "id": "1129"} |
| Phytochemical | CID_6251 | Mannitol | {"name": "Mannitol", "id": "CID_6251"} | INTERACTS_WITH | - | Target | 1131 | CHRM3 | {"name": "CHRM3", "id": "1131"} |
| Phytochemical | CID_6251 | Mannitol | {"name": "Mannitol", "id": "CID_6251"} | INTERACTS_WITH | - | Target | 1132 | CHRM4 | {"name": "CHRM4", "id": "1132"} |
| Phytochemical | CID_6251 | Mannitol | {"name": "Mannitol", "id": "CID_6251"} | INTERACTS_WITH | - | Target | 1133 | CHRM5 | {"name": "CHRM5", "id": "1133"} |
| Phytochemical | CID_6251 | Mannitol | {"name": "Mannitol", "id": "CID_6251"} | INTERACTS_WITH | - | Target | 1233 | CCR4 | {"name": "CCR4", "id": "1233"} |
| Phytochemical | CID_6251 | Mannitol | {"name": "Mannitol", "id": "CID_6251"} | INTERACTS_WITH | - | Target | 1234 | CCR5 | {"name": "CCR5", "id": "1234"} |
| Phytochemical | CID_6251 | Mannitol | {"name": "Mannitol", "id": "CID_6251"} | INTERACTS_WITH | - | Target | 1268 | CNR1 | {"name": "CNR1", "id": "1268"} |
| Phytochemical | CID_6251 | Mannitol | {"name": "Mannitol", "id": "CID_6251"} | INTERACTS_WITH | - | Target | 134 | ADORA1 | {"name": "ADORA1", "id": "134"} |
| Phytochemical | CID_6251 | Mannitol | {"name": "Mannitol", "id": "CID_6251"} | INTERACTS_WITH | - | Target | 135 | ADORA2A | {"name": "ADORA2A", "id": "135"} |
| Phytochemical | CID_6251 | Mannitol | {"name": "Mannitol", "id": "CID_6251"} | INTERACTS_WITH | - | Target | 140 | ADORA3 | {"name": "ADORA3", "id": "140"} |
| Phytochemical | CID_6251 | Mannitol | {"name": "Mannitol", "id": "CID_6251"} | INTERACTS_WITH | - | Target | 1432 | MAPK14 | {"name": "MAPK14", "id": "1432"} |
| Phytochemical | CID_6251 | Mannitol | {"name": "Mannitol", "id": "CID_6251"} | INTERACTS_WITH | - | Target | 146 | ADRA1D | {"name": "ADRA1D", "id": "146"} |
| Phytochemical | CID_6251 | Mannitol | {"name": "Mannitol", "id": "CID_6251"} | INTERACTS_WITH | - | Target | 148 | ADRA1A | {"name": "ADRA1A", "id": "148"} |
| Phytochemical | CID_6251 | Mannitol | {"name": "Mannitol", "id": "CID_6251"} | INTERACTS_WITH | - | Target | 150 | ADRA2A | {"name": "ADRA2A", "id": "150"} |
| Phytochemical | CID_6251 | Mannitol | {"name": "Mannitol", "id": "CID_6251"} | INTERACTS_WITH | - | Target | 151 | ADRA2B | {"name": "ADRA2B", "id": "151"} |
| Phytochemical | CID_6251 | Mannitol | {"name": "Mannitol", "id": "CID_6251"} | INTERACTS_WITH | - | Target | 1511 | CTSG | {"name": "CTSG", "id": "1511"} |
| Phytochemical | CID_6251 | Mannitol | {"name": "Mannitol", "id": "CID_6251"} | INTERACTS_WITH | - | Target | 152 | ADRA2C | {"name": "ADRA2C", "id": "152"} |
| Phytochemical | CID_6251 | Mannitol | {"name": "Mannitol", "id": "CID_6251"} | INTERACTS_WITH | - | Target | 153 | ADRB1 | {"name": "ADRB1", "id": "153"} |
| Phytochemical | CID_6251 | Mannitol | {"name": "Mannitol", "id": "CID_6251"} | INTERACTS_WITH | - | Target | 154 | ADRB2 | {"name": "ADRB2", "id": "154"} |
| Phytochemical | CID_6251 | Mannitol | {"name": "Mannitol", "id": "CID_6251"} | INTERACTS_WITH | - | Target | 1544 | CYP1A2 | {"name": "CYP1A2", "id": "1544"} |
| Phytochemical | CID_6251 | Mannitol | {"name": "Mannitol", "id": "CID_6251"} | INTERACTS_WITH | - | Target | 1548 | CYP2A6 | {"name": "CYP2A6", "id": "1548"} |
| Phytochemical | CID_6251 | Mannitol | {"name": "Mannitol", "id": "CID_6251"} | INTERACTS_WITH | - | Target | 155 | ADRB3 | {"name": "ADRB3", "id": "155"} |

**Table S9: Complete node and relationship data used to construct the knowledge graph. This table summarizes interactions among FDA-approved drugs, formulation, herbs, diseases (type 2 diabetes, fatty liver, obesity), phytochemicals, and their target genes/proteins. Each entry includes source and target node details, edge type, and node annotations (Entrez Gene ID and gene symbol, where applicable).**

| Source Node Label | Source Node ID | Source Node Name | Source Node Properties | Edge Type | Edge Properties | Target Node Label | Target Node ID | Target Node Name | Target Node Properties |
|---|---|---|---|---|---|---|---|---|---|
| Phytochemical | CID_6251 | Mannitol | {"name": "Mannitol", "id": "CID_6251"} | INTERACTS_WITH | - | Target | 1557 | CYP2C19 | {"name": "CYP2C19", "id": "1557"} |
| Phytochemical | CID_6251 | Mannitol | {"name": "Mannitol", "id": "CID_6251"} | INTERACTS_WITH | - | Target | 1559 | CYP2C9 | {"name": "CYP2C9", "id": "1559"} |
| Phytochemical | CID_6251 | Mannitol | {"name": "Mannitol", "id": "CID_6251"} | INTERACTS_WITH | - | Target | 1565 | CYP2D6 | {"name": "CYP2D6", "id": "1565"} |
| Phytochemical | CID_6251 | Mannitol | {"name": "Mannitol", "id": "CID_6251"} | INTERACTS_WITH | - | Target | 1571 | CYP2E1 | {"name": "CYP2E1", "id": "1571"} |
| Phytochemical | CID_6251 | Mannitol | {"name": "Mannitol", "id": "CID_6251"} | INTERACTS_WITH | - | Target | 1576 | CYP3A4 | {"name": "CYP3A4", "id": "1576"} |
| Phytochemical | CID_6251 | Mannitol | {"name": "Mannitol", "id": "CID_6251"} | INTERACTS_WITH | - | Target | 1812 | DRD1 | {"name": "DRD1", "id": "1812"} |
| Phytochemical | CID_6251 | Mannitol | {"name": "Mannitol", "id": "CID_6251"} | INTERACTS_WITH | - | Target | 1813 | DRD2 | {"name": "DRD2", "id": "1813"} |
| Phytochemical | CID_6251 | Mannitol | {"name": "Mannitol", "id": "CID_6251"} | INTERACTS_WITH | - | Target | 1814 | DRD3 | {"name": "DRD3", "id": "1814"} |
| Phytochemical | CID_6251 | Mannitol | {"name": "Mannitol", "id": "CID_6251"} | INTERACTS_WITH | - | Target | 1815 | DRD4 | {"name": "DRD4", "id": "1815"} |
| Phytochemical | CID_6251 | Mannitol | {"name": "Mannitol", "id": "CID_6251"} | INTERACTS_WITH | - | Target | 186 | AGTR2 | {"name": "AGTR2", "id": "186"} |
| Phytochemical | CID_6251 | Mannitol | {"name": "Mannitol", "id": "CID_6251"} | INTERACTS_WITH | - | Target | 1909 | EDNRA | {"name": "EDNRA", "id": "1909"} |
| Phytochemical | CID_6251 | Mannitol | {"name": "Mannitol", "id": "CID_6251"} | INTERACTS_WITH | - | Target | 1956 | EGFR | {"name": "EGFR", "id": "1956"} |
| Phytochemical | CID_6251 | Mannitol | {"name": "Mannitol", "id": "CID_6251"} | INTERACTS_WITH | - | Target | 1991 | ELANE | {"name": "ELANE", "id": "1991"} |
| Phytochemical | CID_6251 | Mannitol | {"name": "Mannitol", "id": "CID_6251"} | INTERACTS_WITH | - | Target | 2064 | ERBB2 | {"name": "ERBB2", "id": "2064"} |
| Phytochemical | CID_6251 | Mannitol | {"name": "Mannitol", "id": "CID_6251"} | INTERACTS_WITH | - | Target | 2099 | ESR1 | {"name": "ESR1", "id": "2099"} |
| Phytochemical | CID_6251 | Mannitol | {"name": "Mannitol", "id": "CID_6251"} | INTERACTS_WITH | - | Target | 2100 | ESR2 | {"name": "ESR2", "id": "2100"} |
| Phytochemical | CID_6251 | Mannitol | {"name": "Mannitol", "id": "CID_6251"} | INTERACTS_WITH | - | Target | 2147 | F2 | {"name": "F2", "id": "2147"} |
| Phytochemical | CID_6251 | Mannitol | {"name": "Mannitol", "id": "CID_6251"} | INTERACTS_WITH | - | Target | 2321 | FLT1 | {"name": "FLT1", "id": "2321"} |
| Phytochemical | CID_6251 | Mannitol | {"name": "Mannitol", "id": "CID_6251"} | INTERACTS_WITH | - | Target | 23435 | TARDBP | {"name": "TARDBP", "id": "23435"} |
| Phytochemical | CID_6251 | Mannitol | {"name": "Mannitol", "id": "CID_6251"} | INTERACTS_WITH | - | Target | 2534 | FYN | {"name": "FYN", "id": "2534"} |
| Phytochemical | CID_6251 | Mannitol | {"name": "Mannitol", "id": "CID_6251"} | INTERACTS_WITH | - | Target | 2554 | GABRA1 | {"name": "GABRA1", "id": "2554"} |
| Phytochemical | CID_6251 | Mannitol | {"name": "Mannitol", "id": "CID_6251"} | INTERACTS_WITH | - | Target | 2555 | GABRA2 | {"name": "GABRA2", "id": "2555"} |
| Phytochemical | CID_6251 | Mannitol | {"name": "Mannitol", "id": "CID_6251"} | INTERACTS_WITH | - | Target | 2561 | GABRB2 | {"name": "GABRB2", "id": "2561"} |
| Phytochemical | CID_6251 | Mannitol | {"name": "Mannitol", "id": "CID_6251"} | INTERACTS_WITH | - | Target | 2566 | GABRG2 | {"name": "GABRG2", "id": "2566"} |
| Phytochemical | CID_6251 | Mannitol | {"name": "Mannitol", "id": "CID_6251"} | INTERACTS_WITH | - | Target | 26013 | L3MBTL1 | {"name": "L3MBTL1", "id": "26013"} |
| Phytochemical | CID_6251 | Mannitol | {"name": "Mannitol", "id": "CID_6251"} | INTERACTS_WITH | - | Target | 2908 | NR3C1 | {"name": "NR3C1", "id": "2908"} |
| Phytochemical | CID_6251 | Mannitol | {"name": "Mannitol", "id": "CID_6251"} | INTERACTS_WITH | - | Target | 3156 | HMGCR | {"name": "HMGCR", "id": "3156"} |
| Phytochemical | CID_6251 | Mannitol | {"name": "Mannitol", "id": "CID_6251"} | INTERACTS_WITH | - | Target | 3269 | HRH1 | {"name": "HRH1", "id": "3269"} |
| Phytochemical | CID_6251 | Mannitol | {"name": "Mannitol", "id": "CID_6251"} | INTERACTS_WITH | - | Target | 3274 | HRH2 | {"name": "HRH2", "id": "3274"} |
| Phytochemical | CID_6251 | Mannitol | {"name": "Mannitol", "id": "CID_6251"} | INTERACTS_WITH | - | Target | 3350 | HTR1A | {"name": "HTR1A", "id": "3350"} |
| Phytochemical | CID_6251 | Mannitol | {"name": "Mannitol", "id": "CID_6251"} | INTERACTS_WITH | - | Target | 3356 | HTR2A | {"name": "HTR2A", "id": "3356"} |
| Phytochemical | CID_6251 | Mannitol | {"name": "Mannitol", "id": "CID_6251"} | INTERACTS_WITH | - | Target | 3357 | HTR2B | {"name": "HTR2B", "id": "3357"} |
| Phytochemical | CID_6251 | Mannitol | {"name": "Mannitol", "id": "CID_6251"} | INTERACTS_WITH | - | Target | 3358 | HTR2C | {"name": "HTR2C", "id": "3358"} |
| Phytochemical | CID_6251 | Mannitol | {"name": "Mannitol", "id": "CID_6251"} | INTERACTS_WITH | - | Target | 3362 | HTR6 | {"name": "HTR6", "id": "3362"} |
| Phytochemical | CID_6251 | Mannitol | {"name": "Mannitol", "id": "CID_6251"} | INTERACTS_WITH | - | Target | 3577 | CXCR1 | {"name": "CXCR1", "id": "3577"} |
| Phytochemical | CID_6251 | Mannitol | {"name": "Mannitol", "id": "CID_6251"} | INTERACTS_WITH | - | Target | 3579 | CXCR2 | {"name": "CXCR2", "id": "3579"} |
| Phytochemical | CID_6251 | Mannitol | {"name": "Mannitol", "id": "CID_6251"} | INTERACTS_WITH | - | Target | 367 | AR | {"name": "AR", "id": "367"} |
| Phytochemical | CID_6251 | Mannitol | {"name": "Mannitol", "id": "CID_6251"} | INTERACTS_WITH | - | Target | 3757 | KCNH2 | {"name": "KCNH2", "id": "3757"} |
| Phytochemical | CID_6251 | Mannitol | {"name": "Mannitol", "id": "CID_6251"} | INTERACTS_WITH | - | Target | 3791 | KDR | {"name": "KDR", "id": "3791"} |
| Phytochemical | CID_6251 | Mannitol | {"name": "Mannitol", "id": "CID_6251"} | INTERACTS_WITH | - | Target | 3932 | LCK | {"name": "LCK", "id": "3932"} |
| Phytochemical | CID_6251 | Mannitol | {"name": "Mannitol", "id": "CID_6251"} | INTERACTS_WITH | - | Target | 4000 | LMNA | {"name": "LMNA", "id": "4000"} |
| Phytochemical | CID_6251 | Mannitol | {"name": "Mannitol", "id": "CID_6251"} | INTERACTS_WITH | - | Target | 4128 | MAOA | {"name": "MAOA", "id": "4128"} |
| Phytochemical | CID_6251 | Mannitol | {"name": "Mannitol", "id": "CID_6251"} | INTERACTS_WITH | - | Target | 4159 | MC3R | {"name": "MC3R", "id": "4159"} |
| Phytochemical | CID_6251 | Mannitol | {"name": "Mannitol", "id": "CID_6251"} | INTERACTS_WITH | - | Target | 4160 | MC4R | {"name": "MC4R", "id": "4160"} |
| Phytochemical | CID_6251 | Mannitol | {"name": "Mannitol", "id": "CID_6251"} | INTERACTS_WITH | - | Target | 4161 | MC5R | {"name": "MC5R", "id": "4161"} |
| Phytochemical | CID_6251 | Mannitol | {"name": "Mannitol", "id": "CID_6251"} | INTERACTS_WITH | - | Target | 43 | ACHE | {"name": "ACHE", "id": "43"} |
| Phytochemical | CID_6251 | Mannitol | {"name": "Mannitol", "id": "CID_6251"} | INTERACTS_WITH | - | Target | 4312 | MMP1 | {"name": "MMP1", "id": "4312"} |
| Phytochemical | CID_6251 | Mannitol | {"name": "Mannitol", "id": "CID_6251"} | INTERACTS_WITH | - | Target | 4318 | MMP9 | {"name": "MMP9", "id": "4318"} |
| Phytochemical | CID_6251 | Mannitol | {"name": "Mannitol", "id": "CID_6251"} | INTERACTS_WITH | - | Target | 4780 | NFE2L2 | {"name": "NFE2L2", "id": "4780"} |
| Phytochemical | CID_6251 | Mannitol | {"name": "Mannitol", "id": "CID_6251"} | INTERACTS_WITH | - | Target | 4886 | NPY1R | {"name": "NPY1R", "id": "4886"} |
| Phytochemical | CID_6251 | Mannitol | {"name": "Mannitol", "id": "CID_6251"} | INTERACTS_WITH | - | Target | 4887 | NPY2R | {"name": "NPY2R", "id": "4887"} |
| Phytochemical | CID_6251 | Mannitol | {"name": "Mannitol", "id": "CID_6251"} | INTERACTS_WITH | - | Target | 49 | ACR | {"name": "ACR", "id": "49"} |
| Phytochemical | CID_6251 | Mannitol | {"name": "Mannitol", "id": "CID_6251"} | INTERACTS_WITH | - | Target | 4985 | OPRD1 | {"name": "OPRD1", "id": "4985"} |
| Phytochemical | CID_6251 | Mannitol | {"name": "Mannitol", "id": "CID_6251"} | INTERACTS_WITH | - | Target | 4986 | OPRK1 | {"name": "OPRK1", "id": "4986"} |
| Phytochemical | CID_6251 | Mannitol | {"name": "Mannitol", "id": "CID_6251"} | INTERACTS_WITH | - | Target | 4988 | OPRM1 | {"name": "OPRM1", "id": "4988"} |
| Phytochemical | CID_6251 | Mannitol | {"name": "Mannitol", "id": "CID_6251"} | INTERACTS_WITH | - | Target | 51053 | GMNN | {"name": "GMNN", "id": "51053"} |
| Phytochemical | CID_6251 | Mannitol | {"name": "Mannitol", "id": "CID_6251"} | INTERACTS_WITH | - | Target | 5139 | PDE3A | {"name": "PDE3A", "id": "5139"} |
| Phytochemical | CID_6251 | Mannitol | {"name": "Mannitol", "id": "CID_6251"} | INTERACTS_WITH | - | Target | 5141 | PDE4A | {"name": "PDE4A", "id": "5141"} |
| Phytochemical | CID_6251 | Mannitol | {"name": "Mannitol", "id": "CID_6251"} | INTERACTS_WITH | - | Target | 5241 | PGR | {"name": "PGR", "id": "5241"} |
| Phytochemical | CID_6251 | Mannitol | {"name": "Mannitol", "id": "CID_6251"} | INTERACTS_WITH | - | Target | 5243 | ABCB1 | {"name": "ABCB1", "id": "5243"} |
| Phytochemical | CID_6251 | Mannitol | {"name": "Mannitol", "id": "CID_6251"} | INTERACTS_WITH | - | Target | 552 | AVPR1A | {"name": "AVPR1A", "id": "552"} |
| Phytochemical | CID_6251 | Mannitol | {"name": "Mannitol", "id": "CID_6251"} | INTERACTS_WITH | - | Target | 5530 | PPP3CA | {"name": "PPP3CA", "id": "5530"} |
| Phytochemical | CID_6251 | Mannitol | {"name": "Mannitol", "id": "CID_6251"} | INTERACTS_WITH | - | Target | 5578 | PRKCA | {"name": "PRKCA", "id": "5578"} |
| Phytochemical | CID_6251 | Mannitol | {"name": "Mannitol", "id": "CID_6251"} | INTERACTS_WITH | - | Target | 5594 | MAPK1 | {"name": "MAPK1", "id": "5594"} |
| Phytochemical | CID_6251 | Mannitol | {"name": "Mannitol", "id": "CID_6251"} | INTERACTS_WITH | - | Target | 5595 | MAPK3 | {"name": "MAPK3", "id": "5595"} |
| Phytochemical | CID_6251 | Mannitol | {"name": "Mannitol", "id": "CID_6251"} | INTERACTS_WITH | - | Target | 5724 | PTAFR | {"name": "PTAFR", "id": "5724"} |
| Phytochemical | CID_6251 | Mannitol | {"name": "Mannitol", "id": "CID_6251"} | INTERACTS_WITH | - | Target | 5742 | PTGS1 | {"name": "PTGS1", "id": "5742"} |
| Phytochemical | CID_6251 | Mannitol | {"name": "Mannitol", "id": "CID_6251"} | INTERACTS_WITH | - | Target | 5743 | PTGS2 | {"name": "PTGS2", "id": "5743"} |
| Phytochemical | CID_6251 | Mannitol | {"name": "Mannitol", "id": "CID_6251"} | INTERACTS_WITH | - | Target | 5788 | PTPRC | {"name": "PTPRC", "id": "5788"} |
| Phytochemical | CID_6251 | Mannitol | {"name": "Mannitol", "id": "CID_6251"} | INTERACTS_WITH | - | Target | 624 | BDKRB2 | {"name": "BDKRB2", "id": "624"} |
| Phytochemical | CID_6251 | Mannitol | {"name": "Mannitol", "id": "CID_6251"} | INTERACTS_WITH | - | Target | 6530 | SLC6A2 | {"name": "SLC6A2", "id": "6530"} |
| Phytochemical | CID_6251 | Mannitol | {"name": "Mannitol", "id": "CID_6251"} | INTERACTS_WITH | - | Target | 6531 | SLC6A3 | {"name": "SLC6A3", "id": "6531"} |
| Phytochemical | CID_6251 | Mannitol | {"name": "Mannitol", "id": "CID_6251"} | INTERACTS_WITH | - | Target | 6532 | SLC6A4 | {"name": "SLC6A4", "id": "6532"} |
| Phytochemical | CID_6251 | Mannitol | {"name": "Mannitol", "id": "CID_6251"} | INTERACTS_WITH | - | Target | 6580 | SLC22A1 | {"name": "SLC22A1", "id": "6580"} |
| Phytochemical | CID_6251 | Mannitol | {"name": "Mannitol", "id": "CID_6251"} | INTERACTS_WITH | - | Target | 6865 | TACR2 | {"name": "TACR2", "id": "6865"} |
| Phytochemical | CID_6251 | Mannitol | {"name": "Mannitol", "id": "CID_6251"} | INTERACTS_WITH | - | Target | 6869 | TACR1 | {"name": "TACR1", "id": "6869"} |

**Table S9: Complete node and relationship data used to construct the knowledge graph. This table summarizes interactions among FDA-approved drugs, formulation, herbs, diseases (type 2 diabetes, fatty liver, obesity), phytochemicals, and their target genes/proteins. Each entry includes source and target node details, edge type, and node annotations (Entrez Gene ID and gene symbol, where applicable).**

| Source Node Label | Source Node ID | Source Node Name | Source Node Properties | Edge Type | Edge Properties | Target Node Label | Target Node ID | Target Node Name | Target Node Properties |
|---|---|---|---|---|---|---|---|---|---|
| Phytochemical | CID_6251 | Mannitol | {"name": "Mannitol", "id": "CID_6251"} | INTERACTS_WITH | - | Target | 6915 | TBXA2R | {"name": "TBXA2R", "id": "6915"} |
| Phytochemical | CID_6251 | Mannitol | {"name": "Mannitol", "id": "CID_6251"} | INTERACTS_WITH | - | Target | 6916 | TBXAS1 | {"name": "TBXAS1", "id": "6916"} |
| Phytochemical | CID_6251 | Mannitol | {"name": "Mannitol", "id": "CID_6251"} | INTERACTS_WITH | - | Target | 729230 | CCR2 | {"name": "CCR2", "id": "729230"} |
| Phytochemical | CID_6251 | Mannitol | {"name": "Mannitol", "id": "CID_6251"} | INTERACTS_WITH | - | Target | 7398 | USP1 | {"name": "USP1", "id": "7398"} |
| Phytochemical | CID_6251 | Mannitol | {"name": "Mannitol", "id": "CID_6251"} | INTERACTS_WITH | - | Target | 7433 | VIPR1 | {"name": "VIPR1", "id": "7433"} |
| Phytochemical | CID_6251 | Mannitol | {"name": "Mannitol", "id": "CID_6251"} | INTERACTS_WITH | - | Target | 760 | CA2 | {"name": "CA2", "id": "760"} |
| Phytochemical | CID_6251 | Mannitol | {"name": "Mannitol", "id": "CID_6251"} | INTERACTS_WITH | - | Target | 799 | CALCR | {"name": "CALCR", "id": "799"} |
| Phytochemical | CID_6251 | Mannitol | {"name": "Mannitol", "id": "CID_6251"} | INTERACTS_WITH | - | Target | 834 | CASP1 | {"name": "CASP1", "id": "834"} |
| Phytochemical | CID_6251 | Mannitol | {"name": "Mannitol", "id": "CID_6251"} | INTERACTS_WITH | - | Target | 8654 | PDE5A | {"name": "PDE5A", "id": "8654"} |
| Phytochemical | CID_6251 | Mannitol | {"name": "Mannitol", "id": "CID_6251"} | INTERACTS_WITH | - | Target | 886 | CCKAR | {"name": "CCKAR", "id": "886"} |
| Phytochemical | CID_6251 | Mannitol | {"name": "Mannitol", "id": "CID_6251"} | INTERACTS_WITH | - | Target | 9429 | ABCG2 | {"name": "ABCG2", "id": "9429"} |
| Phytochemical | CID_73659 | Maslinic acid | {"name": "Maslinic acid", "id": "CID_73659"} | INTERACTS_WITH | - | Target | 112398 | EGLN2 | {"name": "EGLN2", "id": "112398"} |
| Phytochemical | CID_73659 | Maslinic acid | {"name": "Maslinic acid", "id": "CID_73659"} | INTERACTS_WITH | - | Target | 112399 | EGLN3 | {"name": "EGLN3", "id": "112399"} |
| Phytochemical | CID_73659 | Maslinic acid | {"name": "Maslinic acid", "id": "CID_73659"} | INTERACTS_WITH | - | Target | 231 | AKR1B1 | {"name": "AKR1B1", "id": "231"} |
| Phytochemical | CID_73659 | Maslinic acid | {"name": "Maslinic acid", "id": "CID_73659"} | INTERACTS_WITH | - | Target | 240 | ALOX5 | {"name": "ALOX5", "id": "240"} |
| Phytochemical | CID_73659 | Maslinic acid | {"name": "Maslinic acid", "id": "CID_73659"} | INTERACTS_WITH | - | Target | 247 | ALOX15B | {"name": "ALOX15B", "id": "247"} |
| Phytochemical | CID_73659 | Maslinic acid | {"name": "Maslinic acid", "id": "CID_73659"} | INTERACTS_WITH | - | Target | 51422 | PRKAG2 | {"name": "PRKAG2", "id": "51422"} |
| Phytochemical | CID_73659 | Maslinic acid | {"name": "Maslinic acid", "id": "CID_73659"} | INTERACTS_WITH | - | Target | 53632 | PRKAG3 | {"name": "PRKAG3", "id": "53632"} |
| Phytochemical | CID_73659 | Maslinic acid | {"name": "Maslinic acid", "id": "CID_73659"} | INTERACTS_WITH | - | Target | 54583 | EGLN1 | {"name": "EGLN1", "id": "54583"} |
| Phytochemical | CID_73659 | Maslinic acid | {"name": "Maslinic acid", "id": "CID_73659"} | INTERACTS_WITH | - | Target | 54681 | P4HTM | {"name": "P4HTM", "id": "54681"} |
| Phytochemical | CID_73659 | Maslinic acid | {"name": "Maslinic acid", "id": "CID_73659"} | INTERACTS_WITH | - | Target | 5562 | PRKAA1 | {"name": "PRKAA1", "id": "5562"} |
| Phytochemical | CID_73659 | Maslinic acid | {"name": "Maslinic acid", "id": "CID_73659"} | INTERACTS_WITH | - | Target | 5563 | PRKAA2 | {"name": "PRKAA2", "id": "5563"} |
| Phytochemical | CID_73659 | Maslinic acid | {"name": "Maslinic acid", "id": "CID_73659"} | INTERACTS_WITH | - | Target | 5564 | PRKAB1 | {"name": "PRKAB1", "id": "5564"} |
| Phytochemical | CID_73659 | Maslinic acid | {"name": "Maslinic acid", "id": "CID_73659"} | INTERACTS_WITH | - | Target | 5565 | PRKAB2 | {"name": "PRKAB2", "id": "5565"} |
| Phytochemical | CID_73659 | Maslinic acid | {"name": "Maslinic acid", "id": "CID_73659"} | INTERACTS_WITH | - | Target | 5571 | PRKAG1 | {"name": "PRKAG1", "id": "5571"} |
| Phytochemical | CID_73659 | Maslinic acid | {"name": "Maslinic acid", "id": "CID_73659"} | INTERACTS_WITH | - | Target | 57016 | AKR1B10 | {"name": "AKR1B10", "id": "57016"} |
| Phytochemical | CID_73659 | Maslinic acid | {"name": "Maslinic acid", "id": "CID_73659"} | INTERACTS_WITH | - | Target | 5743 | PTGS2 | {"name": "PTGS2", "id": "5743"} |
| Phytochemical | CID_73659 | Maslinic acid | {"name": "Maslinic acid", "id": "CID_73659"} | INTERACTS_WITH | - | Target | 5770 | PTPN1 | {"name": "PTPN1", "id": "5770"} |
| Phytochemical | CID_73659 | Maslinic acid | {"name": "Maslinic acid", "id": "CID_73659"} | INTERACTS_WITH | - | Target | 5771 | PTPN2 | {"name": "PTPN2", "id": "5771"} |
| Phytochemical | CID_73659 | Maslinic acid | {"name": "Maslinic acid", "id": "CID_73659"} | INTERACTS_WITH | - | Target | 5777 | PTPN6 | {"name": "PTPN6", "id": "5777"} |
| Phytochemical | CID_73659 | Maslinic acid | {"name": "Maslinic acid", "id": "CID_73659"} | INTERACTS_WITH | - | Target | 5781 | PTPN11 | {"name": "PTPN11", "id": "5781"} |
| Phytochemical | CID_73659 | Maslinic acid | {"name": "Maslinic acid", "id": "CID_73659"} | INTERACTS_WITH | - | Target | 5792 | PTPRF | {"name": "PTPRF", "id": "5792"} |
| Phytochemical | CID_73659 | Maslinic acid | {"name": "Maslinic acid", "id": "CID_73659"} | INTERACTS_WITH | - | Target | 5837 | PYGM | {"name": "PYGM", "id": "5837"} |
| Phytochemical | CID_7428 | Methyl gallate | {"name": "Methyl gallate", "id": "CID_7428"} | INTERACTS_WITH | - | Target | 10919 | EHMT2 | {"name": "EHMT2", "id": "10919"} |
| Phytochemical | CID_7428 | Methyl gallate | {"name": "Methyl gallate", "id": "CID_7428"} | INTERACTS_WITH | - | Target | 10951 | CBX1 | {"name": "CBX1", "id": "10951"} |
| Phytochemical | CID_7428 | Methyl gallate | {"name": "Methyl gallate", "id": "CID_7428"} | INTERACTS_WITH | - | Target | 118429 | ANTXR2 | {"name": "ANTXR2", "id": "118429"} |
| Phytochemical | CID_7428 | Methyl gallate | {"name": "Methyl gallate", "id": "CID_7428"} | INTERACTS_WITH | - | Target | 2237 | FEN1 | {"name": "FEN1", "id": "2237"} |
| Phytochemical | CID_7428 | Methyl gallate | {"name": "Methyl gallate", "id": "CID_7428"} | INTERACTS_WITH | - | Target | 2529 | FUT7 | {"name": "FUT7", "id": "2529"} |
| Phytochemical | CID_7428 | Methyl gallate | {"name": "Methyl gallate", "id": "CID_7428"} | INTERACTS_WITH | - | Target | 2548 | GAA | {"name": "GAA", "id": "2548"} |
| Phytochemical | CID_7428 | Methyl gallate | {"name": "Methyl gallate", "id": "CID_7428"} | INTERACTS_WITH | - | Target | 26013 | L3MBTL1 | {"name": "L3MBTL1", "id": "26013"} |
| Phytochemical | CID_7428 | Methyl gallate | {"name": "Methyl gallate", "id": "CID_7428"} | INTERACTS_WITH | - | Target | 2671 | GFER | {"name": "GFER", "id": "2671"} |
| Phytochemical | CID_7428 | Methyl gallate | {"name": "Methyl gallate", "id": "CID_7428"} | INTERACTS_WITH | - | Target | 28689 | TRAV4 | {"name": "TRAV4", "id": "28689"} |
| Phytochemical | CID_7428 | Methyl gallate | {"name": "Methyl gallate", "id": "CID_7428"} | INTERACTS_WITH | - | Target | 29994 | BAZ2B | {"name": "BAZ2B", "id": "29994"} |
| Phytochemical | CID_7428 | Methyl gallate | {"name": "Methyl gallate", "id": "CID_7428"} | INTERACTS_WITH | - | Target | 328 | APEX1 | {"name": "APEX1", "id": "328"} |
| Phytochemical | CID_7428 | Methyl gallate | {"name": "Methyl gallate", "id": "CID_7428"} | INTERACTS_WITH | - | Target | 390245 | KDM4E | {"name": "KDM4E", "id": "390245"} |
| Phytochemical | CID_7428 | Methyl gallate | {"name": "Methyl gallate", "id": "CID_7428"} | INTERACTS_WITH | - | Target | 4000 | LMNA | {"name": "LMNA", "id": "4000"} |
| Phytochemical | CID_7428 | Methyl gallate | {"name": "Methyl gallate", "id": "CID_7428"} | INTERACTS_WITH | - | Target | 4790 | NFKB1 | {"name": "NFKB1", "id": "4790"} |
| Phytochemical | CID_7428 | Methyl gallate | {"name": "Methyl gallate", "id": "CID_7428"} | INTERACTS_WITH | - | Target | 4791 | NFKB2 | {"name": "NFKB2", "id": "4791"} |
| Phytochemical | CID_7428 | Methyl gallate | {"name": "Methyl gallate", "id": "CID_7428"} | INTERACTS_WITH | - | Target | 5300 | PIN1 | {"name": "PIN1", "id": "5300"} |
| Phytochemical | CID_7428 | Methyl gallate | {"name": "Methyl gallate", "id": "CID_7428"} | INTERACTS_WITH | - | Target | 5423 | POLB | {"name": "POLB", "id": "5423"} |
| Phytochemical | CID_7428 | Methyl gallate | {"name": "Methyl gallate", "id": "CID_7428"} | INTERACTS_WITH | - | Target | 55775 | TDP1 | {"name": "TDP1", "id": "55775"} |
| Phytochemical | CID_7428 | Methyl gallate | {"name": "Methyl gallate", "id": "CID_7428"} | INTERACTS_WITH | - | Target | 5970 | RELA | {"name": "RELA", "id": "5970"} |
| Phytochemical | CID_7428 | Methyl gallate | {"name": "Methyl gallate", "id": "CID_7428"} | INTERACTS_WITH | - | Target | 7498 | XDH | {"name": "XDH", "id": "7498"} |

**Table S9: Complete node and relationship data used to construct the knowledge graph. This table summarizes interactions among FDA-approved drugs, formulation, herbs, diseases (type 2 diabetes, fatty liver, obesity), phytochemicals, and their target genes/proteins. Each entry includes source and target node details, edge type, and node annotations (Entrez Gene ID and gene symbol, where applicable).**

| Source Node Label | Source Node ID | Source Node Name | Source Node Properties | Edge Type | Edge Properties | Target Node Label | Target Node ID | Target Node Name | Target Node Properties |
|---|---|---|---|---|---|---|---|---|---|
| Phytochemical | CID_7428 | Methyl gallate | {"name": "Methyl gallate", "id": "CID_7428"} | INTERACTS_WITH | - | Target | 9682 | KDM4A | {"name": "KDM4A", "id": "9682"} |
| Phytochemical | CID_7127 | Methyleugenol | {"name": "Methyleugenol", "id": "CID_7127"} | INTERACTS_WITH | - | Target | 216 | ALDH1A1 | {"name": "ALDH1A1", "id": "216"} |
| Phytochemical | CID_7127 | Methyleugenol | {"name": "Methyleugenol", "id": "CID_7127"} | INTERACTS_WITH | - | Target | 2648 | KAT2A | {"name": "KAT2A", "id": "2648"} |
| Phytochemical | CID_7127 | Methyleugenol | {"name": "Methyleugenol", "id": "CID_7127"} | INTERACTS_WITH | - | Target | 2908 | NR3C1 | {"name": "NR3C1", "id": "2908"} |
| Phytochemical | CID_7127 | Methyleugenol | {"name": "Methyleugenol", "id": "CID_7127"} | INTERACTS_WITH | - | Target | 367 | AR | {"name": "AR", "id": "367"} |
| Phytochemical | CID_7127 | Methyleugenol | {"name": "Methyleugenol", "id": "CID_7127"} | INTERACTS_WITH | - | Target | 4088 | SMAD3 | {"name": "SMAD3", "id": "4088"} |
| Phytochemical | CID_7127 | Methyleugenol | {"name": "Methyleugenol", "id": "CID_7127"} | INTERACTS_WITH | - | Target | 4780 | NFE2L2 | {"name": "NFE2L2", "id": "4780"} |
| Phytochemical | CID_7127 | Methyleugenol | {"name": "Methyleugenol", "id": "CID_7127"} | INTERACTS_WITH | - | Target | 5465 | PPARA | {"name": "PPARA", "id": "5465"} |
| Phytochemical | CID_7127 | Methyleugenol | {"name": "Methyleugenol", "id": "CID_7127"} | INTERACTS_WITH | - | Target | 5467 | PPARD | {"name": "PPARD", "id": "5467"} |
| Phytochemical | CID_7127 | Methyleugenol | {"name": "Methyleugenol", "id": "CID_7127"} | INTERACTS_WITH | - | Target | 5468 | PPARG | {"name": "PPARG", "id": "5468"} |
| Phytochemical | CID_7127 | Methyleugenol | {"name": "Methyleugenol", "id": "CID_7127"} | INTERACTS_WITH | - | Target | 7068 | THRB | {"name": "THRB", "id": "7068"} |
| Phytochemical | CID_7127 | Methyleugenol | {"name": "Methyleugenol", "id": "CID_7127"} | INTERACTS_WITH | - | Target | 79915 | ATAD5 | {"name": "ATAD5", "id": "79915"} |
| Phytochemical | CID_7127 | Methyleugenol | {"name": "Methyleugenol", "id": "CID_7127"} | INTERACTS_WITH | - | Target | 9971 | NR1H4 | {"name": "NR1H4", "id": "9971"} |
| Phytochemical | CID_271296 | Methylnaphthazarin | {"name": "Methylnaphthazarin", "id": "CID_271296"} | INTERACTS_WITH | - | Target | 1981 | EIF4G1 | {"name": "EIF4G1", "id": "1981"} |
| Phytochemical | CID_271296 | Methylnaphthazarin | {"name": "Methylnaphthazarin", "id": "CID_271296"} | INTERACTS_WITH | - | Target | 216 | ALDH1A1 | {"name": "ALDH1A1", "id": "216"} |
| Phytochemical | CID_271296 | Methylnaphthazarin | {"name": "Methylnaphthazarin", "id": "CID_271296"} | INTERACTS_WITH | - | Target | 26013 | L3MBTL1 | {"name": "L3MBTL1", "id": "26013"} |
| Phytochemical | CID_271296 | Methylnaphthazarin | {"name": "Methylnaphthazarin", "id": "CID_271296"} | INTERACTS_WITH | - | Target | 2648 | KAT2A | {"name": "KAT2A", "id": "2648"} |
| Phytochemical | CID_271296 | Methylnaphthazarin | {"name": "Methylnaphthazarin", "id": "CID_271296"} | INTERACTS_WITH | - | Target | 2944 | GSTM1 | {"name": "GSTM1", "id": "2944"} |
| Phytochemical | CID_271296 | Methylnaphthazarin | {"name": "Methylnaphthazarin", "id": "CID_271296"} | INTERACTS_WITH | - | Target | 29994 | BAZ2B | {"name": "BAZ2B", "id": "29994"} |
| Phytochemical | CID_271296 | Methylnaphthazarin | {"name": "Methylnaphthazarin", "id": "CID_271296"} | INTERACTS_WITH | - | Target | 328 | APEX1 | {"name": "APEX1", "id": "328"} |
| Phytochemical | CID_271296 | Methylnaphthazarin | {"name": "Methylnaphthazarin", "id": "CID_271296"} | INTERACTS_WITH | - | Target | 4297 | KMT2A | {"name": "KMT2A", "id": "4297"} |
| Phytochemical | CID_271296 | Methylnaphthazarin | {"name": "Methylnaphthazarin", "id": "CID_271296"} | INTERACTS_WITH | - | Target | 5423 | POLB | {"name": "POLB", "id": "5423"} |
| Phytochemical | CID_271296 | Methylnaphthazarin | {"name": "Methylnaphthazarin", "id": "CID_271296"} | INTERACTS_WITH | - | Target | 5965 | RECQL | {"name": "RECQL", "id": "5965"} |
| Phytochemical | CID_271296 | Methylnaphthazarin | {"name": "Methylnaphthazarin", "id": "CID_271296"} | INTERACTS_WITH | - | Target | 597 | BCL2A1 | {"name": "BCL2A1", "id": "597"} |
| Phytochemical | CID_271296 | Methylnaphthazarin | {"name": "Methylnaphthazarin", "id": "CID_271296"} | INTERACTS_WITH | - | Target | 641 | BLM | {"name": "BLM", "id": "641"} |
| Phytochemical | CID_271296 | Methylnaphthazarin | {"name": "Methylnaphthazarin", "id": "CID_271296"} | INTERACTS_WITH | - | Target | 7068 | THRB | {"name": "THRB", "id": "7068"} |
| Phytochemical | CID_271296 | Methylnaphthazarin | {"name": "Methylnaphthazarin", "id": "CID_271296"} | INTERACTS_WITH | - | Target | 7296 | TXNRD1 | {"name": "TXNRD1", "id": "7296"} |
| Phytochemical | CID_31253 | Myrcene | {"name": "Myrcene", "id": "CID_31253"} | INTERACTS_WITH | - | Target | 2908 | NR3C1 | {"name": "NR3C1", "id": "2908"} |
| Phytochemical | CID_31253 | Myrcene | {"name": "Myrcene", "id": "CID_31253"} | INTERACTS_WITH | - | Target | 3315 | HSPB1 | {"name": "HSPB1", "id": "3315"} |
| Phytochemical | CID_31253 | Myrcene | {"name": "Myrcene", "id": "CID_31253"} | INTERACTS_WITH | - | Target | 3725 | JUN | {"name": "JUN", "id": "3725"} |
| Phytochemical | CID_31253 | Myrcene | {"name": "Myrcene", "id": "CID_31253"} | INTERACTS_WITH | - | Target | 4790 | NFKB1 | {"name": "NFKB1", "id": "4790"} |
| Phytochemical | CID_31253 | Myrcene | {"name": "Myrcene", "id": "CID_31253"} | INTERACTS_WITH | - | Target | 5467 | PPARD | {"name": "PPARD", "id": "5467"} |
| Phytochemical | CID_31253 | Myrcene | {"name": "Myrcene", "id": "CID_31253"} | INTERACTS_WITH | - | Target | 7253 | TSHR | {"name": "TSHR", "id": "7253"} |
| Phytochemical | CID_31253 | Myrcene | {"name": "Myrcene", "id": "CID_31253"} | INTERACTS_WITH | - | Target | 9971 | NR1H4 | {"name": "NR1H4", "id": "9971"} |
| Phytochemical | CID_5281384 | Neocembrene | {"name": "Neocembrene", "id": "CID_5281384"} | INTERACTS_WITH | - | Target | 5319 | PLA2G1B | {"name": "PLA2G1B", "id": "5319"} |
| Phytochemical | CID_5284507 | Nerolidol | {"name": "Nerolidol", "id": "CID_5284507"} | INTERACTS_WITH | - | Target | 10599 | SLCO1B1 | {"name": "SLCO1B1", "id": "10599"} |
| Phytochemical | CID_5284507 | Nerolidol | {"name": "Nerolidol", "id": "CID_5284507"} | INTERACTS_WITH | - | Target | 246 | ALOX15 | {"name": "ALOX15", "id": "246"} |
| Phytochemical | CID_5284507 | Nerolidol | {"name": "Nerolidol", "id": "CID_5284507"} | INTERACTS_WITH | - | Target | 28234 | SLCO1B3 | {"name": "SLCO1B3", "id": "28234"} |
| Phytochemical | CID_5284507 | Nerolidol | {"name": "Nerolidol", "id": "CID_5284507"} | INTERACTS_WITH | - | Target | 7068 | THRB | {"name": "THRB", "id": "7068"} |
| Phytochemical | CID_5284507 | Nerolidol | {"name": "Nerolidol", "id": "CID_5284507"} | INTERACTS_WITH | - | Target | 7398 | USP1 | {"name": "USP1", "id": "7398"} |
| Phytochemical | CID_445639 | Oleic acid | {"name": "Oleic acid", "id": "CID_445639"} | INTERACTS_WITH | - | Target | 10018 | BCL2L11 | {"name": "BCL2L11", "id": "10018"} |
| Phytochemical | CID_445639 | Oleic acid | {"name": "Oleic acid", "id": "CID_445639"} | INTERACTS_WITH | - | Target | 100861540 | CYP3A7-CYP3A51P | {"name": "CYP3A7-CYP3A51P", "id": "100861540"} |
| Phytochemical | CID_445639 | Oleic acid | {"name": "Oleic acid", "id": "CID_445639"} | INTERACTS_WITH | - | Target | 10411 | RAPGEF3 | {"name": "RAPGEF3", "id": "10411"} |
| Phytochemical | CID_445639 | Oleic acid | {"name": "Oleic acid", "id": "CID_445639"} | INTERACTS_WITH | - | Target | 10951 | CBX1 | {"name": "CBX1", "id": "10951"} |
| Phytochemical | CID_445639 | Oleic acid | {"name": "Oleic acid", "id": "CID_445639"} | INTERACTS_WITH | - | Target | 11069 | RAPGEF4 | {"name": "RAPGEF4", "id": "11069"} |
| Phytochemical | CID_445639 | Oleic acid | {"name": "Oleic acid", "id": "CID_445639"} | INTERACTS_WITH | - | Target | 11255 | HRH3 | {"name": "HRH3", "id": "11255"} |
| Phytochemical | CID_445639 | Oleic acid | {"name": "Oleic acid", "id": "CID_445639"} | INTERACTS_WITH | - | Target | 1128 | CHRM1 | {"name": "CHRM1", "id": "1128"} |
| Phytochemical | CID_445639 | Oleic acid | {"name": "Oleic acid", "id": "CID_445639"} | INTERACTS_WITH | - | Target | 11283 | CYP4F8 | {"name": "CYP4F8", "id": "11283"} |
| Phytochemical | CID_445639 | Oleic acid | {"name": "Oleic acid", "id": "CID_445639"} | INTERACTS_WITH | - | Target | 1129 | CHRM2 | {"name": "CHRM2", "id": "1129"} |
| Phytochemical | CID_445639 | Oleic acid | {"name": "Oleic acid", "id": "CID_445639"} | INTERACTS_WITH | - | Target | 113612 | CYP2U1 | {"name": "CYP2U1", "id": "113612"} |
| Phytochemical | CID_445639 | Oleic acid | {"name": "Oleic acid", "id": "CID_445639"} | INTERACTS_WITH | - | Target | 120227 | CYP2R1 | {"name": "CYP2R1", "id": "120227"} |

**Table S9: Complete node and relationship data used to construct the knowledge graph. This table summarizes interactions among FDA-approved drugs, formulation, herbs, diseases (type 2 diabetes, fatty liver, obesity), phytochemicals, and their target genes/proteins. Each entry includes source and target node details, edge type, and node annotations (Entrez Gene ID and gene symbol, where applicable).**

| Source Node Label | Source Node ID | Source Node Name | Source Node Properties | Edge Type | Edge Properties | Target Node Label | Target Node ID | Target Node Name | Target Node Properties |
|---|---|---|---|---|---|---|---|---|---|
| Phytochemical | CID_445639 | Oleic acid | {"name": "Oleic acid", "id": "CID_445639"} | INTERACTS_WITH | - | Target | 122706 | PSMB11 | {"name": "PSMB11", "id": "122706"} |
| Phytochemical | CID_445639 | Oleic acid | {"name": "Oleic acid", "id": "CID_445639"} | INTERACTS_WITH | - | Target | 140 | ADORA3 | {"name": "ADORA3", "id": "140"} |
| Phytochemical | CID_445639 | Oleic acid | {"name": "Oleic acid", "id": "CID_445639"} | INTERACTS_WITH | - | Target | 143471 | PSMA8 | {"name": "PSMA8", "id": "143471"} |
| Phytochemical | CID_445639 | Oleic acid | {"name": "Oleic acid", "id": "CID_445639"} | INTERACTS_WITH | - | Target | 148 | ADRA1A | {"name": "ADRA1A", "id": "148"} |
| Phytochemical | CID_445639 | Oleic acid | {"name": "Oleic acid", "id": "CID_445639"} | INTERACTS_WITH | - | Target | 150 | ADRA2A | {"name": "ADRA2A", "id": "150"} |
| Phytochemical | CID_445639 | Oleic acid | {"name": "Oleic acid", "id": "CID_445639"} | INTERACTS_WITH | - | Target | 1543 | CYP1A1 | {"name": "CYP1A1", "id": "1543"} |
| Phytochemical | CID_445639 | Oleic acid | {"name": "Oleic acid", "id": "CID_445639"} | INTERACTS_WITH | - | Target | 1544 | CYP1A2 | {"name": "CYP1A2", "id": "1544"} |
| Phytochemical | CID_445639 | Oleic acid | {"name": "Oleic acid", "id": "CID_445639"} | INTERACTS_WITH | - | Target | 1545 | CYP1B1 | {"name": "CYP1B1", "id": "1545"} |
| Phytochemical | CID_445639 | Oleic acid | {"name": "Oleic acid", "id": "CID_445639"} | INTERACTS_WITH | - | Target | 1548 | CYP2A6 | {"name": "CYP2A6", "id": "1548"} |
| Phytochemical | CID_445639 | Oleic acid | {"name": "Oleic acid", "id": "CID_445639"} | INTERACTS_WITH | - | Target | 1549 | CYP2A7 | {"name": "CYP2A7", "id": "1549"} |
| Phytochemical | CID_445639 | Oleic acid | {"name": "Oleic acid", "id": "CID_445639"} | INTERACTS_WITH | - | Target | 1553 | CYP2A13 | {"name": "CYP2A13", "id": "1553"} |
| Phytochemical | CID_445639 | Oleic acid | {"name": "Oleic acid", "id": "CID_445639"} | INTERACTS_WITH | - | Target | 1555 | CYP2B6 | {"name": "CYP2B6", "id": "1555"} |
| Phytochemical | CID_445639 | Oleic acid | {"name": "Oleic acid", "id": "CID_445639"} | INTERACTS_WITH | - | Target | 1557 | CYP2C19 | {"name": "CYP2C19", "id": "1557"} |
| Phytochemical | CID_445639 | Oleic acid | {"name": "Oleic acid", "id": "CID_445639"} | INTERACTS_WITH | - | Target | 1558 | CYP2C8 | {"name": "CYP2C8", "id": "1558"} |
| Phytochemical | CID_445639 | Oleic acid | {"name": "Oleic acid", "id": "CID_445639"} | INTERACTS_WITH | - | Target | 1559 | CYP2C9 | {"name": "CYP2C9", "id": "1559"} |
| Phytochemical | CID_445639 | Oleic acid | {"name": "Oleic acid", "id": "CID_445639"} | INTERACTS_WITH | - | Target | 1562 | CYP2C18 | {"name": "CYP2C18", "id": "1562"} |
| Phytochemical | CID_445639 | Oleic acid | {"name": "Oleic acid", "id": "CID_445639"} | INTERACTS_WITH | - | Target | 1565 | CYP2D6 | {"name": "CYP2D6", "id": "1565"} |
| Phytochemical | CID_445639 | Oleic acid | {"name": "Oleic acid", "id": "CID_445639"} | INTERACTS_WITH | - | Target | 1571 | CYP2E1 | {"name": "CYP2E1", "id": "1571"} |
| Phytochemical | CID_445639 | Oleic acid | {"name": "Oleic acid", "id": "CID_445639"} | INTERACTS_WITH | - | Target | 1572 | CYP2F1 | {"name": "CYP2F1", "id": "1572"} |
| Phytochemical | CID_445639 | Oleic acid | {"name": "Oleic acid", "id": "CID_445639"} | INTERACTS_WITH | - | Target | 1573 | CYP2J2 | {"name": "CYP2J2", "id": "1573"} |
| Phytochemical | CID_445639 | Oleic acid | {"name": "Oleic acid", "id": "CID_445639"} | INTERACTS_WITH | - | Target | 1576 | CYP3A4 | {"name": "CYP3A4", "id": "1576"} |
| Phytochemical | CID_445639 | Oleic acid | {"name": "Oleic acid", "id": "CID_445639"} | INTERACTS_WITH | - | Target | 1577 | CYP3A5 | {"name": "CYP3A5", "id": "1577"} |
| Phytochemical | CID_445639 | Oleic acid | {"name": "Oleic acid", "id": "CID_445639"} | INTERACTS_WITH | - | Target | 1579 | CYP4A11 | {"name": "CYP4A11", "id": "1579"} |
| Phytochemical | CID_445639 | Oleic acid | {"name": "Oleic acid", "id": "CID_445639"} | INTERACTS_WITH | - | Target | 1580 | CYP4B1 | {"name": "CYP4B1", "id": "1580"} |
| Phytochemical | CID_445639 | Oleic acid | {"name": "Oleic acid", "id": "CID_445639"} | INTERACTS_WITH | - | Target | 1588 | CYP19A1 | {"name": "CYP19A1", "id": "1588"} |
| Phytochemical | CID_445639 | Oleic acid | {"name": "Oleic acid", "id": "CID_445639"} | INTERACTS_WITH | - | Target | 1812 | DRD1 | {"name": "DRD1", "id": "1812"} |
| Phytochemical | CID_445639 | Oleic acid | {"name": "Oleic acid", "id": "CID_445639"} | INTERACTS_WITH | - | Target | 1814 | DRD3 | {"name": "DRD3", "id": "1814"} |
| Phytochemical | CID_445639 | Oleic acid | {"name": "Oleic acid", "id": "CID_445639"} | INTERACTS_WITH | - | Target | 2052 | EPHX1 | {"name": "EPHX1", "id": "2052"} |
| Phytochemical | CID_445639 | Oleic acid | {"name": "Oleic acid", "id": "CID_445639"} | INTERACTS_WITH | - | Target | 2053 | EPHX2 | {"name": "EPHX2", "id": "2053"} |
| Phytochemical | CID_445639 | Oleic acid | {"name": "Oleic acid", "id": "CID_445639"} | INTERACTS_WITH | - | Target | 2099 | ESR1 | {"name": "ESR1", "id": "2099"} |
| Phytochemical | CID_445639 | Oleic acid | {"name": "Oleic acid", "id": "CID_445639"} | INTERACTS_WITH | - | Target | 2147 | F2 | {"name": "F2", "id": "2147"} |
| Phytochemical | CID_445639 | Oleic acid | {"name": "Oleic acid", "id": "CID_445639"} | INTERACTS_WITH | - | Target | 2152 | F3 | {"name": "F3", "id": "2152"} |
| Phytochemical | CID_445639 | Oleic acid | {"name": "Oleic acid", "id": "CID_445639"} | INTERACTS_WITH | - | Target | 2155 | F7 | {"name": "F7", "id": "2155"} |
| Phytochemical | CID_445639 | Oleic acid | {"name": "Oleic acid", "id": "CID_445639"} | INTERACTS_WITH | - | Target | 2166 | FAAH | {"name": "FAAH", "id": "2166"} |
| Phytochemical | CID_445639 | Oleic acid | {"name": "Oleic acid", "id": "CID_445639"} | INTERACTS_WITH | - | Target | 2167 | FABP4 | {"name": "FABP4", "id": "2167"} |
| Phytochemical | CID_445639 | Oleic acid | {"name": "Oleic acid", "id": "CID_445639"} | INTERACTS_WITH | - | Target | 2168 | FABP1 | {"name": "FABP1", "id": "2168"} |
| Phytochemical | CID_445639 | Oleic acid | {"name": "Oleic acid", "id": "CID_445639"} | INTERACTS_WITH | - | Target | 2171 | FABP5 | {"name": "FABP5", "id": "2171"} |
| Phytochemical | CID_445639 | Oleic acid | {"name": "Oleic acid", "id": "CID_445639"} | INTERACTS_WITH | - | Target | 2554 | GABRA1 | {"name": "GABRA1", "id": "2554"} |
| Phytochemical | CID_445639 | Oleic acid | {"name": "Oleic acid", "id": "CID_445639"} | INTERACTS_WITH | - | Target | 2555 | GABRA2 | {"name": "GABRA2", "id": "2555"} |
| Phytochemical | CID_445639 | Oleic acid | {"name": "Oleic acid", "id": "CID_445639"} | INTERACTS_WITH | - | Target | 2561 | GABRB2 | {"name": "GABRB2", "id": "2561"} |
| Phytochemical | CID_445639 | Oleic acid | {"name": "Oleic acid", "id": "CID_445639"} | INTERACTS_WITH | - | Target | 2566 | GABRG2 | {"name": "GABRG2", "id": "2566"} |
| Phytochemical | CID_445639 | Oleic acid | {"name": "Oleic acid", "id": "CID_445639"} | INTERACTS_WITH | - | Target | 2648 | KAT2A | {"name": "KAT2A", "id": "2648"} |
| Phytochemical | CID_445639 | Oleic acid | {"name": "Oleic acid", "id": "CID_445639"} | INTERACTS_WITH | - | Target | 284541 | CYP4A22 | {"name": "CYP4A22", "id": "284541"} |
| Phytochemical | CID_445639 | Oleic acid | {"name": "Oleic acid", "id": "CID_445639"} | INTERACTS_WITH | - | Target | 29785 | CYP2S1 | {"name": "CYP2S1", "id": "29785"} |
| Phytochemical | CID_445639 | Oleic acid | {"name": "Oleic acid", "id": "CID_445639"} | INTERACTS_WITH | - | Target | 3028 | HSD17B10 | {"name": "HSD17B10", "id": "3028"} |
| Phytochemical | CID_445639 | Oleic acid | {"name": "Oleic acid", "id": "CID_445639"} | INTERACTS_WITH | - | Target | 328 | APEX1 | {"name": "APEX1", "id": "328"} |
| Phytochemical | CID_445639 | Oleic acid | {"name": "Oleic acid", "id": "CID_445639"} | INTERACTS_WITH | - | Target | 3315 | HSPB1 | {"name": "HSPB1", "id": "3315"} |
| Phytochemical | CID_445639 | Oleic acid | {"name": "Oleic acid", "id": "CID_445639"} | INTERACTS_WITH | - | Target | 3350 | HTR1A | {"name": "HTR1A", "id": "3350"} |
| Phytochemical | CID_445639 | Oleic acid | {"name": "Oleic acid", "id": "CID_445639"} | INTERACTS_WITH | - | Target | 3357 | HTR2B | {"name": "HTR2B", "id": "3357"} |

**Table S9: Complete node and relationship data used to construct the knowledge graph. This table summarizes interactions among FDA-approved drugs, formulation, herbs, diseases (type 2 diabetes, fatty liver, obesity), phytochemicals, and their target genes/proteins. Each entry includes source and target node details, edge type, and node annotations (Entrez Gene ID and gene symbol, where applicable).**

| Source Node Label | Source Node ID | Source Node Name | Source Node Properties | Edge Type | Edge Properties | Target Node Label | Target Node ID | Target Node Name | Target Node Properties |
|---|---|---|---|---|---|---|---|---|---|
| Phytochemical | CID_445639 | Oleic acid | {"name": "Oleic acid", "id": "CID_445639"} | INTERACTS_WITH | - | Target | 367 | AR | {"name": "AR", "id": "367"} |
| Phytochemical | CID_445639 | Oleic acid | {"name": "Oleic acid", "id": "CID_445639"} | INTERACTS_WITH | - | Target | 3757 | KCNH2 | {"name": "KCNH2", "id": "3757"} |
| Phytochemical | CID_445639 | Oleic acid | {"name": "Oleic acid", "id": "CID_445639"} | INTERACTS_WITH | - | Target | 3791 | KDR | {"name": "KDR", "id": "3791"} |
| Phytochemical | CID_445639 | Oleic acid | {"name": "Oleic acid", "id": "CID_445639"} | INTERACTS_WITH | - | Target | 4051 | CYP4F3 | {"name": "CYP4F3", "id": "4051"} |
| Phytochemical | CID_445639 | Oleic acid | {"name": "Oleic acid", "id": "CID_445639"} | INTERACTS_WITH | - | Target | 4128 | MAOA | {"name": "MAOA", "id": "4128"} |
| Phytochemical | CID_445639 | Oleic acid | {"name": "Oleic acid", "id": "CID_445639"} | INTERACTS_WITH | - | Target | 4137 | MAPT | {"name": "MAPT", "id": "4137"} |
| Phytochemical | CID_445639 | Oleic acid | {"name": "Oleic acid", "id": "CID_445639"} | INTERACTS_WITH | - | Target | 4297 | KMT2A | {"name": "KMT2A", "id": "4297"} |
| Phytochemical | CID_445639 | Oleic acid | {"name": "Oleic acid", "id": "CID_445639"} | INTERACTS_WITH | - | Target | 43 | ACHE | {"name": "ACHE", "id": "43"} |
| Phytochemical | CID_445639 | Oleic acid | {"name": "Oleic acid", "id": "CID_445639"} | INTERACTS_WITH | - | Target | 472 | ATM | {"name": "ATM", "id": "472"} |
| Phytochemical | CID_445639 | Oleic acid | {"name": "Oleic acid", "id": "CID_445639"} | INTERACTS_WITH | - | Target | 4929 | NR4A2 | {"name": "NR4A2", "id": "4929"} |
| Phytochemical | CID_445639 | Oleic acid | {"name": "Oleic acid", "id": "CID_445639"} | INTERACTS_WITH | - | Target | 4988 | OPRM1 | {"name": "OPRM1", "id": "4988"} |
| Phytochemical | CID_445639 | Oleic acid | {"name": "Oleic acid", "id": "CID_445639"} | INTERACTS_WITH | - | Target | 51053 | GMNN | {"name": "GMNN", "id": "51053"} |
| Phytochemical | CID_445639 | Oleic acid | {"name": "Oleic acid", "id": "CID_445639"} | INTERACTS_WITH | - | Target | 5139 | PDE3A | {"name": "PDE3A", "id": "5139"} |
| Phytochemical | CID_445639 | Oleic acid | {"name": "Oleic acid", "id": "CID_445639"} | INTERACTS_WITH | - | Target | 5141 | PDE4A | {"name": "PDE4A", "id": "5141"} |
| Phytochemical | CID_445639 | Oleic acid | {"name": "Oleic acid", "id": "CID_445639"} | INTERACTS_WITH | - | Target | 51426 | POLK | {"name": "POLK", "id": "51426"} |
| Phytochemical | CID_445639 | Oleic acid | {"name": "Oleic acid", "id": "CID_445639"} | INTERACTS_WITH | - | Target | 51548 | SIRT6 | {"name": "SIRT6", "id": "51548"} |
| Phytochemical | CID_445639 | Oleic acid | {"name": "Oleic acid", "id": "CID_445639"} | INTERACTS_WITH | - | Target | 5241 | PGR | {"name": "PGR", "id": "5241"} |
| Phytochemical | CID_445639 | Oleic acid | {"name": "Oleic acid", "id": "CID_445639"} | INTERACTS_WITH | - | Target | 5347 | PLK1 | {"name": "PLK1", "id": "5347"} |
| Phytochemical | CID_445639 | Oleic acid | {"name": "Oleic acid", "id": "CID_445639"} | INTERACTS_WITH | - | Target | 5465 | PPARA | {"name": "PPARA", "id": "5465"} |
| Phytochemical | CID_445639 | Oleic acid | {"name": "Oleic acid", "id": "CID_445639"} | INTERACTS_WITH | - | Target | 5467 | PPARD | {"name": "PPARD", "id": "5467"} |
| Phytochemical | CID_445639 | Oleic acid | {"name": "Oleic acid", "id": "CID_445639"} | INTERACTS_WITH | - | Target | 5468 | PPARG | {"name": "PPARG", "id": "5468"} |
| Phytochemical | CID_445639 | Oleic acid | {"name": "Oleic acid", "id": "CID_445639"} | INTERACTS_WITH | - | Target | 54905 | CYP2W1 | {"name": "CYP2W1", "id": "54905"} |
| Phytochemical | CID_445639 | Oleic acid | {"name": "Oleic acid", "id": "CID_445639"} | INTERACTS_WITH | - | Target | 5578 | PRKCA | {"name": "PRKCA", "id": "5578"} |
| Phytochemical | CID_445639 | Oleic acid | {"name": "Oleic acid", "id": "CID_445639"} | INTERACTS_WITH | - | Target | 5682 | PSMA1 | {"name": "PSMA1", "id": "5682"} |
| Phytochemical | CID_445639 | Oleic acid | {"name": "Oleic acid", "id": "CID_445639"} | INTERACTS_WITH | - | Target | 5683 | PSMA2 | {"name": "PSMA2", "id": "5683"} |
| Phytochemical | CID_445639 | Oleic acid | {"name": "Oleic acid", "id": "CID_445639"} | INTERACTS_WITH | - | Target | 5684 | PSMA3 | {"name": "PSMA3", "id": "5684"} |
| Phytochemical | CID_445639 | Oleic acid | {"name": "Oleic acid", "id": "CID_445639"} | INTERACTS_WITH | - | Target | 5685 | PSMA4 | {"name": "PSMA4", "id": "5685"} |
| Phytochemical | CID_445639 | Oleic acid | {"name": "Oleic acid", "id": "CID_445639"} | INTERACTS_WITH | - | Target | 5686 | PSMA5 | {"name": "PSMA5", "id": "5686"} |
| Phytochemical | CID_445639 | Oleic acid | {"name": "Oleic acid", "id": "CID_445639"} | INTERACTS_WITH | - | Target | 5687 | PSMA6 | {"name": "PSMA6", "id": "5687"} |
| Phytochemical | CID_445639 | Oleic acid | {"name": "Oleic acid", "id": "CID_445639"} | INTERACTS_WITH | - | Target | 5688 | PSMA7 | {"name": "PSMA7", "id": "5688"} |
| Phytochemical | CID_445639 | Oleic acid | {"name": "Oleic acid", "id": "CID_445639"} | INTERACTS_WITH | - | Target | 5689 | PSMB1 | {"name": "PSMB1", "id": "5689"} |
| Phytochemical | CID_445639 | Oleic acid | {"name": "Oleic acid", "id": "CID_445639"} | INTERACTS_WITH | - | Target | 5690 | PSMB2 | {"name": "PSMB2", "id": "5690"} |
| Phytochemical | CID_445639 | Oleic acid | {"name": "Oleic acid", "id": "CID_445639"} | INTERACTS_WITH | - | Target | 5691 | PSMB3 | {"name": "PSMB3", "id": "5691"} |
| Phytochemical | CID_445639 | Oleic acid | {"name": "Oleic acid", "id": "CID_445639"} | INTERACTS_WITH | - | Target | 5692 | PSMB4 | {"name": "PSMB4", "id": "5692"} |
| Phytochemical | CID_445639 | Oleic acid | {"name": "Oleic acid", "id": "CID_445639"} | INTERACTS_WITH | - | Target | 5693 | PSMB5 | {"name": "PSMB5", "id": "5693"} |
| Phytochemical | CID_445639 | Oleic acid | {"name": "Oleic acid", "id": "CID_445639"} | INTERACTS_WITH | - | Target | 5694 | PSMB6 | {"name": "PSMB6", "id": "5694"} |
| Phytochemical | CID_445639 | Oleic acid | {"name": "Oleic acid", "id": "CID_445639"} | INTERACTS_WITH | - | Target | 5695 | PSMB7 | {"name": "PSMB7", "id": "5695"} |
| Phytochemical | CID_445639 | Oleic acid | {"name": "Oleic acid", "id": "CID_445639"} | INTERACTS_WITH | - | Target | 5696 | PSMB8 | {"name": "PSMB8", "id": "5696"} |
| Phytochemical | CID_445639 | Oleic acid | {"name": "Oleic acid", "id": "CID_445639"} | INTERACTS_WITH | - | Target | 5698 | PSMB9 | {"name": "PSMB9", "id": "5698"} |
| Phytochemical | CID_445639 | Oleic acid | {"name": "Oleic acid", "id": "CID_445639"} | INTERACTS_WITH | - | Target | 5699 | PSMB10 | {"name": "PSMB10", "id": "5699"} |
| Phytochemical | CID_445639 | Oleic acid | {"name": "Oleic acid", "id": "CID_445639"} | INTERACTS_WITH | - | Target | 5742 | PTGS1 | {"name": "PTGS1", "id": "5742"} |
| Phytochemical | CID_445639 | Oleic acid | {"name": "Oleic acid", "id": "CID_445639"} | INTERACTS_WITH | - | Target | 5745 | PTH1R | {"name": "PTH1R", "id": "5745"} |
| Phytochemical | CID_445639 | Oleic acid | {"name": "Oleic acid", "id": "CID_445639"} | INTERACTS_WITH | - | Target | 5770 | PTPN1 | {"name": "PTPN1", "id": "5770"} |
| Phytochemical | CID_445639 | Oleic acid | {"name": "Oleic acid", "id": "CID_445639"} | INTERACTS_WITH | - | Target | 6097 | RORC | {"name": "RORC", "id": "6097"} |
| Phytochemical | CID_445639 | Oleic acid | {"name": "Oleic acid", "id": "CID_445639"} | INTERACTS_WITH | - | Target | 641 | BLM | {"name": "BLM", "id": "641"} |
| Phytochemical | CID_445639 | Oleic acid | {"name": "Oleic acid", "id": "CID_445639"} | INTERACTS_WITH | - | Target | 64816 | CYP3A43 | {"name": "CYP3A43", "id": "64816"} |
| Phytochemical | CID_445639 | Oleic acid | {"name": "Oleic acid", "id": "CID_445639"} | INTERACTS_WITH | - | Target | 6530 | SLC6A2 | {"name": "SLC6A2", "id": "6530"} |
| Phytochemical | CID_445639 | Oleic acid | {"name": "Oleic acid", "id": "CID_445639"} | INTERACTS_WITH | - | Target | 6531 | SLC6A3 | {"name": "SLC6A3", "id": "6531"} |
| Phytochemical | CID_445639 | Oleic acid | {"name": "Oleic acid", "id": "CID_445639"} | INTERACTS_WITH | - | Target | 6532 | SLC6A4 | {"name": "SLC6A4", "id": "6532"} |

**Table S9: Complete node and relationship data used to construct the knowledge graph. This table summarizes interactions among FDA-approved drugs, formulation, herbs, diseases (type 2 diabetes, fatty liver, obesity), phytochemicals, and their target genes/proteins. Each entry includes source and target node details, edge type, and node annotations (Entrez Gene ID and gene symbol, where applicable).**

| Source Node Label | Source Node ID | Source Node Name | Source Node Properties | Edge Type | Edge Properties | Target Node Label | Target Node ID | Target Node Name | Target Node Properties |
|---|---|---|---|---|---|---|---|---|---|
| Phytochemical | CID_445639 | Oleic acid | {"name": "Oleic acid", "id": "CID_445639"} | INTERACTS_WITH | - | Target | 6646 | SOAT1 | {"name": "SOAT1", "id": "6646"} |
| Phytochemical | CID_445639 | Oleic acid | {"name": "Oleic acid", "id": "CID_445639"} | INTERACTS_WITH | - | Target | 6915 | TBXA2R | {"name": "TBXA2R", "id": "6915"} |
| Phytochemical | CID_445639 | Oleic acid | {"name": "Oleic acid", "id": "CID_445639"} | INTERACTS_WITH | - | Target | 7015 | TERT | {"name": "TERT", "id": "7015"} |
| Phytochemical | CID_445639 | Oleic acid | {"name": "Oleic acid", "id": "CID_445639"} | INTERACTS_WITH | - | Target | 7150 | TOP1 | {"name": "TOP1", "id": "7150"} |
| Phytochemical | CID_445639 | Oleic acid | {"name": "Oleic acid", "id": "CID_445639"} | INTERACTS_WITH | - | Target | 7398 | USP1 | {"name": "USP1", "id": "7398"} |
| Phytochemical | CID_445639 | Oleic acid | {"name": "Oleic acid", "id": "CID_445639"} | INTERACTS_WITH | - | Target | 79915 | ATAD5 | {"name": "ATAD5", "id": "79915"} |
| Phytochemical | CID_445639 | Oleic acid | {"name": "Oleic acid", "id": "CID_445639"} | INTERACTS_WITH | - | Target | 8529 | CYP4F2 | {"name": "CYP4F2", "id": "8529"} |
| Phytochemical | CID_445639 | Oleic acid | {"name": "Oleic acid", "id": "CID_445639"} | INTERACTS_WITH | - | Target | 9682 | KDM4A | {"name": "KDM4A", "id": "9682"} |
| Phytochemical | CID_445639 | Oleic acid | {"name": "Oleic acid", "id": "CID_445639"} | INTERACTS_WITH | - | Target | 9971 | NR1H4 | {"name": "NR1H4", "id": "9971"} |
| Phytochemical | CID_985 | Palmitic acid | {"name": "Palmitic acid", "id": "CID_985"} | INTERACTS_WITH | - | Target | 1588 | CYP19A1 | {"name": "CYP19A1", "id": "1588"} |
| Phytochemical | CID_985 | Palmitic acid | {"name": "Palmitic acid", "id": "CID_985"} | INTERACTS_WITH | - | Target | 2099 | ESR1 | {"name": "ESR1", "id": "2099"} |
| Phytochemical | CID_985 | Palmitic acid | {"name": "Palmitic acid", "id": "CID_985"} | INTERACTS_WITH | - | Target | 2167 | FABP4 | {"name": "FABP4", "id": "2167"} |
| Phytochemical | CID_985 | Palmitic acid | {"name": "Palmitic acid", "id": "CID_985"} | INTERACTS_WITH | - | Target | 2169 | FABP2 | {"name": "FABP2", "id": "2169"} |
| Phytochemical | CID_985 | Palmitic acid | {"name": "Palmitic acid", "id": "CID_985"} | INTERACTS_WITH | - | Target | 2170 | FABP3 | {"name": "FABP3", "id": "2170"} |
| Phytochemical | CID_985 | Palmitic acid | {"name": "Palmitic acid", "id": "CID_985"} | INTERACTS_WITH | - | Target | 2171 | FABP5 | {"name": "FABP5", "id": "2171"} |
| Phytochemical | CID_985 | Palmitic acid | {"name": "Palmitic acid", "id": "CID_985"} | INTERACTS_WITH | - | Target | 2908 | NR3C1 | {"name": "NR3C1", "id": "2908"} |
| Phytochemical | CID_985 | Palmitic acid | {"name": "Palmitic acid", "id": "CID_985"} | INTERACTS_WITH | - | Target | 367 | AR | {"name": "AR", "id": "367"} |
| Phytochemical | CID_985 | Palmitic acid | {"name": "Palmitic acid", "id": "CID_985"} | INTERACTS_WITH | - | Target | 4780 | NFE2L2 | {"name": "NFE2L2", "id": "4780"} |
| Phytochemical | CID_985 | Palmitic acid | {"name": "Palmitic acid", "id": "CID_985"} | INTERACTS_WITH | - | Target | 51053 | GMNN | {"name": "GMNN", "id": "51053"} |
| Phytochemical | CID_985 | Palmitic acid | {"name": "Palmitic acid", "id": "CID_985"} | INTERACTS_WITH | - | Target | 5243 | ABCB1 | {"name": "ABCB1", "id": "5243"} |
| Phytochemical | CID_985 | Palmitic acid | {"name": "Palmitic acid", "id": "CID_985"} | INTERACTS_WITH | - | Target | 5465 | PPARA | {"name": "PPARA", "id": "5465"} |
| Phytochemical | CID_985 | Palmitic acid | {"name": "Palmitic acid", "id": "CID_985"} | INTERACTS_WITH | - | Target | 5467 | PPARD | {"name": "PPARD", "id": "5467"} |
| Phytochemical | CID_985 | Palmitic acid | {"name": "Palmitic acid", "id": "CID_985"} | INTERACTS_WITH | - | Target | 5468 | PPARG | {"name": "PPARG", "id": "5468"} |
| Phytochemical | CID_985 | Palmitic acid | {"name": "Palmitic acid", "id": "CID_985"} | INTERACTS_WITH | - | Target | 5770 | PTPN1 | {"name": "PTPN1", "id": "5770"} |
| Phytochemical | CID_985 | Palmitic acid | {"name": "Palmitic acid", "id": "CID_985"} | INTERACTS_WITH | - | Target | 7004 | TEAD4 | {"name": "TEAD4", "id": "7004"} |
| Phytochemical | CID_985 | Palmitic acid | {"name": "Palmitic acid", "id": "CID_985"} | INTERACTS_WITH | - | Target | 7097 | TLR2 | {"name": "TLR2", "id": "7097"} |
| Phytochemical | CID_985 | Palmitic acid | {"name": "Palmitic acid", "id": "CID_985"} | INTERACTS_WITH | - | Target | 7398 | USP1 | {"name": "USP1", "id": "7398"} |
| Phytochemical | CID_985 | Palmitic acid | {"name": "Palmitic acid", "id": "CID_985"} | INTERACTS_WITH | - | Target | 81285 | OR51E2 | {"name": "OR51E2", "id": "81285"} |
| Phytochemical | CID_985 | Palmitic acid | {"name": "Palmitic acid", "id": "CID_985"} | INTERACTS_WITH | - | Target | 9971 | NR1H4 | {"name": "NR1H4", "id": "9971"} |
| Phytochemical | CID_68381 | Perilla ketone | {"name": "Perilla ketone", "id": "CID_68381"} | INTERACTS_WITH | - | Target | 8989 | TRPA1 | {"name": "TRPA1", "id": "8989"} |
| Phytochemical | CID_359 | Phloroglucinol | {"name": "Phloroglucinol", "id": "CID_359"} | INTERACTS_WITH | - | Target | 10013 | HDAC6 | {"name": "HDAC6", "id": "10013"} |
| Phytochemical | CID_359 | Phloroglucinol | {"name": "Phloroglucinol", "id": "CID_359"} | INTERACTS_WITH | - | Target | 23621 | BACE1 | {"name": "BACE1", "id": "23621"} |
| Phytochemical | CID_359 | Phloroglucinol | {"name": "Phloroglucinol", "id": "CID_359"} | INTERACTS_WITH | - | Target | 51426 | POLK | {"name": "POLK", "id": "51426"} |
| Phytochemical | CID_359 | Phloroglucinol | {"name": "Phloroglucinol", "id": "CID_359"} | INTERACTS_WITH | - | Target | 7299 | TYR | {"name": "TYR", "id": "7299"} |
| Phytochemical | CID_92123 | Picein | {"name": "Picein", "id": "CID_92123"} | INTERACTS_WITH | - | Target | 10951 | CBX1 | {"name": "CBX1", "id": "10951"} |
| Phytochemical | CID_92123 | Picein | {"name": "Picein", "id": "CID_92123"} | INTERACTS_WITH | - | Target | 2139 | EYA2 | {"name": "EYA2", "id": "2139"} |
| Phytochemical | CID_92123 | Picein | {"name": "Picein", "id": "CID_92123"} | INTERACTS_WITH | - | Target | 51053 | GMNN | {"name": "GMNN", "id": "51053"} |
| Phytochemical | CID_638024 | Piperine | {"name": "Piperine", "id": "CID_638024"} | INTERACTS_WITH | - | Target | 10013 | HDAC6 | {"name": "HDAC6", "id": "10013"} |
| Phytochemical | CID_638024 | Piperine | {"name": "Piperine", "id": "CID_638024"} | INTERACTS_WITH | - | Target | 10599 | SLCO1B1 | {"name": "SLCO1B1", "id": "10599"} |
| Phytochemical | CID_638024 | Piperine | {"name": "Piperine", "id": "CID_638024"} | INTERACTS_WITH | - | Target | 11201 | POLI | {"name": "POLI", "id": "11201"} |
| Phytochemical | CID_638024 | Piperine | {"name": "Piperine", "id": "CID_638024"} | INTERACTS_WITH | - | Target | 1509 | CTSD | {"name": "CTSD", "id": "1509"} |
| Phytochemical | CID_638024 | Piperine | {"name": "Piperine", "id": "CID_638024"} | INTERACTS_WITH | - | Target | 1543 | CYP1A1 | {"name": "CYP1A1", "id": "1543"} |
| Phytochemical | CID_638024 | Piperine | {"name": "Piperine", "id": "CID_638024"} | INTERACTS_WITH | - | Target | 1576 | CYP3A4 | {"name": "CYP3A4", "id": "1576"} |
| Phytochemical | CID_638024 | Piperine | {"name": "Piperine", "id": "CID_638024"} | INTERACTS_WITH | - | Target | 1723 | DHODH | {"name": "DHODH", "id": "1723"} |
| Phytochemical | CID_638024 | Piperine | {"name": "Piperine", "id": "CID_638024"} | INTERACTS_WITH | - | Target | 196 | AHR | {"name": "AHR", "id": "196"} |
| Phytochemical | CID_638024 | Piperine | {"name": "Piperine", "id": "CID_638024"} | INTERACTS_WITH | - | Target | 216 | ALDH1A1 | {"name": "ALDH1A1", "id": "216"} |
| Phytochemical | CID_638024 | Piperine | {"name": "Piperine", "id": "CID_638024"} | INTERACTS_WITH | - | Target | 240 | ALOX5 | {"name": "ALOX5", "id": "240"} |
| Phytochemical | CID_638024 | Piperine | {"name": "Piperine", "id": "CID_638024"} | INTERACTS_WITH | - | Target | 2554 | GABRA1 | {"name": "GABRA1", "id": "2554"} |
| Phytochemical | CID_638024 | Piperine | {"name": "Piperine", "id": "CID_638024"} | INTERACTS_WITH | - | Target | 2561 | GABRB2 | {"name": "GABRB2", "id": "2561"} |
| Phytochemical | CID_638024 | Piperine | {"name": "Piperine", "id": "CID_638024"} | INTERACTS_WITH | - | Target | 2566 | GABRG2 | {"name": "GABRG2", "id": "2566"} |
| Phytochemical | CID_638024 | Piperine | {"name": "Piperine", "id": "CID_638024"} | INTERACTS_WITH | - | Target | 25939 | SAMHD1 | {"name": "SAMHD1", "id": "25939"} |
| Phytochemical | CID_638024 | Piperine | {"name": "Piperine", "id": "CID_638024"} | INTERACTS_WITH | - | Target | 28234 | SLCO1B3 | {"name": "SLCO1B3", "id": "28234"} |
| Phytochemical | CID_638024 | Piperine | {"name": "Piperine", "id": "CID_638024"} | INTERACTS_WITH | - | Target | 2908 | NR3C1 | {"name": "NR3C1", "id": "2908"} |
| Phytochemical | CID_638024 | Piperine | {"name": "Piperine", "id": "CID_638024"} | INTERACTS_WITH | - | Target | 3028 | HSD17B10 | {"name": "HSD17B10", "id": "3028"} |
| Phytochemical | CID_638024 | Piperine | {"name": "Piperine", "id": "CID_638024"} | INTERACTS_WITH | - | Target | 3248 | HPGD | {"name": "HPGD", "id": "3248"} |
| Phytochemical | CID_638024 | Piperine | {"name": "Piperine", "id": "CID_638024"} | INTERACTS_WITH | - | Target | 367 | AR | {"name": "AR", "id": "367"} |
| Phytochemical | CID_638024 | Piperine | {"name": "Piperine", "id": "CID_638024"} | INTERACTS_WITH | - | Target | 390245 | KDM4E | {"name": "KDM4E", "id": "390245"} |
| Phytochemical | CID_638024 | Piperine | {"name": "Piperine", "id": "CID_638024"} | INTERACTS_WITH | - | Target | 4128 | MAOA | {"name": "MAOA", "id": "4128"} |
| Phytochemical | CID_638024 | Piperine | {"name": "Piperine", "id": "CID_638024"} | INTERACTS_WITH | - | Target | 4129 | MAOB | {"name": "MAOB", "id": "4129"} |
| Phytochemical | CID_638024 | Piperine | {"name": "Piperine", "id": "CID_638024"} | INTERACTS_WITH | - | Target | 4780 | NFE2L2 | {"name": "NFE2L2", "id": "4780"} |
| Phytochemical | CID_638024 | Piperine | {"name": "Piperine", "id": "CID_638024"} | INTERACTS_WITH | - | Target | 4864 | NPC1 | {"name": "NPC1", "id": "4864"} |
| Phytochemical | CID_638024 | Piperine | {"name": "Piperine", "id": "CID_638024"} | INTERACTS_WITH | - | Target | 5243 | ABCB1 | {"name": "ABCB1", "id": "5243"} |
| Phytochemical | CID_638024 | Piperine | {"name": "Piperine", "id": "CID_638024"} | INTERACTS_WITH | - | Target | 5467 | PPARD | {"name": "PPARD", "id": "5467"} |
| Phytochemical | CID_638024 | Piperine | {"name": "Piperine", "id": "CID_638024"} | INTERACTS_WITH | - | Target | 5468 | PPARG | {"name": "PPARG", "id": "5468"} |
| Phytochemical | CID_638024 | Piperine | {"name": "Piperine", "id": "CID_638024"} | INTERACTS_WITH | - | Target | 55775 | TDP1 | {"name": "TDP1", "id": "55775"} |
| Phytochemical | CID_638024 | Piperine | {"name": "Piperine", "id": "CID_638024"} | INTERACTS_WITH | - | Target | 5594 | MAPK1 | {"name": "MAPK1", "id": "5594"} |
| Phytochemical | CID_638024 | Piperine | {"name": "Piperine", "id": "CID_638024"} | INTERACTS_WITH | - | Target | 6256 | RXRA | {"name": "RXRA", "id": "6256"} |
| Phytochemical | CID_638024 | Piperine | {"name": "Piperine", "id": "CID_638024"} | INTERACTS_WITH | - | Target | 6311 | ATXN2 | {"name": "ATXN2", "id": "6311"} |

**Table S9: Complete node and relationship data used to construct the knowledge graph. This table summarizes interactions among FDA-approved drugs, formulation, herbs, diseases (type 2 diabetes, fatty liver, obesity), phytochemicals, and their target genes/proteins. Each entry includes source and target node details, edge type, and node annotations (Entrez Gene ID and gene symbol, where applicable).**

| Source Node Label | Source Node ID | Source Node Name | Source Node Properties | Edge Type | Edge Properties | Target Node Label | Target Node ID | Target Node Name | Target Node Properties |
|---|---|---|---|---|---|---|---|---|---|
| Phytochemical | CID_638024 | Piperine | {"name": "Piperine", "id": "CID_638024"} | INTERACTS_WITH | - | Target | 6606 | SMN1 | {"name": "SMN1", "id": "6606"} |
| Phytochemical | CID_638024 | Piperine | {"name": "Piperine", "id": "CID_638024"} | INTERACTS_WITH | - | Target | 6609 | SMPD1 | {"name": "SMPD1", "id": "6609"} |
| Phytochemical | CID_638024 | Piperine | {"name": "Piperine", "id": "CID_638024"} | INTERACTS_WITH | - | Target | 672 | BRCA1 | {"name": "BRCA1", "id": "672"} |
| Phytochemical | CID_638024 | Piperine | {"name": "Piperine", "id": "CID_638024"} | INTERACTS_WITH | - | Target | 7442 | TRPV1 | {"name": "TRPV1", "id": "7442"} |
| Phytochemical | CID_638024 | Piperine | {"name": "Piperine", "id": "CID_638024"} | INTERACTS_WITH | - | Target | 759 | CA1 | {"name": "CA1", "id": "759"} |
| Phytochemical | CID_638024 | Piperine | {"name": "Piperine", "id": "CID_638024"} | INTERACTS_WITH | - | Target | 760 | CA2 | {"name": "CA2", "id": "760"} |
| Phytochemical | CID_638024 | Piperine | {"name": "Piperine", "id": "CID_638024"} | INTERACTS_WITH | - | Target | 79915 | ATAD5 | {"name": "ATAD5", "id": "79915"} |
| Phytochemical | CID_638024 | Piperine | {"name": "Piperine", "id": "CID_638024"} | INTERACTS_WITH | - | Target | 9367 | RAB9A | {"name": "RAB9A", "id": "9367"} |
| Phytochemical | CID_638024 | Piperine | {"name": "Piperine", "id": "CID_638024"} | INTERACTS_WITH | - | Target | 9971 | NR1H4 | {"name": "NR1H4", "id": "9971"} |
| Phytochemical | CID_10205 | Plumbagin | {"name": "Plumbagin", "id": "CID_10205"} | INTERACTS_WITH | - | Target | 1543 | CYP1A1 | {"name": "CYP1A1", "id": "1543"} |
| Phytochemical | CID_10205 | Plumbagin | {"name": "Plumbagin", "id": "CID_10205"} | INTERACTS_WITH | - | Target | 1544 | CYP1A2 | {"name": "CYP1A2", "id": "1544"} |
| Phytochemical | CID_10205 | Plumbagin | {"name": "Plumbagin", "id": "CID_10205"} | INTERACTS_WITH | - | Target | 1545 | CYP1B1 | {"name": "CYP1B1", "id": "1545"} |
| Phytochemical | CID_10205 | Plumbagin | {"name": "Plumbagin", "id": "CID_10205"} | INTERACTS_WITH | - | Target | 1557 | CYP2C19 | {"name": "CYP2C19", "id": "1557"} |
| Phytochemical | CID_10205 | Plumbagin | {"name": "Plumbagin", "id": "CID_10205"} | INTERACTS_WITH | - | Target | 1559 | CYP2C9 | {"name": "CYP2C9", "id": "1559"} |
| Phytochemical | CID_10205 | Plumbagin | {"name": "Plumbagin", "id": "CID_10205"} | INTERACTS_WITH | - | Target | 1565 | CYP2D6 | {"name": "CYP2D6", "id": "1565"} |
| Phytochemical | CID_10205 | Plumbagin | {"name": "Plumbagin", "id": "CID_10205"} | INTERACTS_WITH | - | Target | 1576 | CYP3A4 | {"name": "CYP3A4", "id": "1576"} |
| Phytochemical | CID_10205 | Plumbagin | {"name": "Plumbagin", "id": "CID_10205"} | INTERACTS_WITH | - | Target | 1956 | EGFR | {"name": "EGFR", "id": "1956"} |
| Phytochemical | CID_10205 | Plumbagin | {"name": "Plumbagin", "id": "CID_10205"} | INTERACTS_WITH | - | Target | 2033 | EP300 | {"name": "EP300", "id": "2033"} |
| Phytochemical | CID_10205 | Plumbagin | {"name": "Plumbagin", "id": "CID_10205"} | INTERACTS_WITH | - | Target | 2064 | ERBB2 | {"name": "ERBB2", "id": "2064"} |
| Phytochemical | CID_10205 | Plumbagin | {"name": "Plumbagin", "id": "CID_10205"} | INTERACTS_WITH | - | Target | 207 | AKT1 | {"name": "AKT1", "id": "207"} |
| Phytochemical | CID_10205 | Plumbagin | {"name": "Plumbagin", "id": "CID_10205"} | INTERACTS_WITH | - | Target | 216 | ALDH1A1 | {"name": "ALDH1A1", "id": "216"} |
| Phytochemical | CID_10205 | Plumbagin | {"name": "Plumbagin", "id": "CID_10205"} | INTERACTS_WITH | - | Target | 246 | ALOX15 | {"name": "ALOX15", "id": "246"} |
| Phytochemical | CID_10205 | Plumbagin | {"name": "Plumbagin", "id": "CID_10205"} | INTERACTS_WITH | - | Target | 2648 | KAT2A | {"name": "KAT2A", "id": "2648"} |
| Phytochemical | CID_10205 | Plumbagin | {"name": "Plumbagin", "id": "CID_10205"} | INTERACTS_WITH | - | Target | 3028 | HSD17B10 | {"name": "HSD17B10", "id": "3028"} |
| Phytochemical | CID_10205 | Plumbagin | {"name": "Plumbagin", "id": "CID_10205"} | INTERACTS_WITH | - | Target | 3248 | HPGD | {"name": "HPGD", "id": "3248"} |
| Phytochemical | CID_10205 | Plumbagin | {"name": "Plumbagin", "id": "CID_10205"} | INTERACTS_WITH | - | Target | 328 | APEX1 | {"name": "APEX1", "id": "328"} |
| Phytochemical | CID_10205 | Plumbagin | {"name": "Plumbagin", "id": "CID_10205"} | INTERACTS_WITH | - | Target | 3309 | HSPA5 | {"name": "HSPA5", "id": "3309"} |
| Phytochemical | CID_10205 | Plumbagin | {"name": "Plumbagin", "id": "CID_10205"} | INTERACTS_WITH | - | Target | 3329 | HSPD1 | {"name": "HSPD1", "id": "3329"} |
| Phytochemical | CID_10205 | Plumbagin | {"name": "Plumbagin", "id": "CID_10205"} | INTERACTS_WITH | - | Target | 3336 | HSPE1 | {"name": "HSPE1", "id": "3336"} |
| Phytochemical | CID_10205 | Plumbagin | {"name": "Plumbagin", "id": "CID_10205"} | INTERACTS_WITH | - | Target | 390245 | KDM4E | {"name": "KDM4E", "id": "390245"} |
| Phytochemical | CID_10205 | Plumbagin | {"name": "Plumbagin", "id": "CID_10205"} | INTERACTS_WITH | - | Target | 4137 | MAPT | {"name": "MAPT", "id": "4137"} |
| Phytochemical | CID_10205 | Plumbagin | {"name": "Plumbagin", "id": "CID_10205"} | INTERACTS_WITH | - | Target | 4297 | KMT2A | {"name": "KMT2A", "id": "4297"} |
| Phytochemical | CID_10205 | Plumbagin | {"name": "Plumbagin", "id": "CID_10205"} | INTERACTS_WITH | - | Target | 51426 | POLK | {"name": "POLK", "id": "51426"} |
| Phytochemical | CID_10205 | Plumbagin | {"name": "Plumbagin", "id": "CID_10205"} | INTERACTS_WITH | - | Target | 55775 | TDP1 | {"name": "TDP1", "id": "55775"} |
| Phytochemical | CID_10205 | Plumbagin | {"name": "Plumbagin", "id": "CID_10205"} | INTERACTS_WITH | - | Target | 5594 | MAPK1 | {"name": "MAPK1", "id": "5594"} |
| Phytochemical | CID_10205 | Plumbagin | {"name": "Plumbagin", "id": "CID_10205"} | INTERACTS_WITH | - | Target | 5595 | MAPK3 | {"name": "MAPK3", "id": "5595"} |
| Phytochemical | CID_10205 | Plumbagin | {"name": "Plumbagin", "id": "CID_10205"} | INTERACTS_WITH | - | Target | 5604 | MAP2K1 | {"name": "MAP2K1", "id": "5604"} |
| Phytochemical | CID_10205 | Plumbagin | {"name": "Plumbagin", "id": "CID_10205"} | INTERACTS_WITH | - | Target | 5609 | MAP2K7 | {"name": "MAP2K7", "id": "5609"} |
| Phytochemical | CID_10205 | Plumbagin | {"name": "Plumbagin", "id": "CID_10205"} | INTERACTS_WITH | - | Target | 5747 | PTK2 | {"name": "PTK2", "id": "5747"} |
| Phytochemical | CID_10205 | Plumbagin | {"name": "Plumbagin", "id": "CID_10205"} | INTERACTS_WITH | - | Target | 5894 | RAF1 | {"name": "RAF1", "id": "5894"} |
| Phytochemical | CID_10205 | Plumbagin | {"name": "Plumbagin", "id": "CID_10205"} | INTERACTS_WITH | - | Target | 5965 | RECQL | {"name": "RECQL", "id": "5965"} |
| Phytochemical | CID_10205 | Plumbagin | {"name": "Plumbagin", "id": "CID_10205"} | INTERACTS_WITH | - | Target | 5970 | RELA | {"name": "RELA", "id": "5970"} |
| Phytochemical | CID_10205 | Plumbagin | {"name": "Plumbagin", "id": "CID_10205"} | INTERACTS_WITH | - | Target | 641 | BLM | {"name": "BLM", "id": "641"} |
| Phytochemical | CID_10205 | Plumbagin | {"name": "Plumbagin", "id": "CID_10205"} | INTERACTS_WITH | - | Target | 6606 | SMN1 | {"name": "SMN1", "id": "6606"} |
| Phytochemical | CID_10205 | Plumbagin | {"name": "Plumbagin", "id": "CID_10205"} | INTERACTS_WITH | - | Target | 6774 | STAT3 | {"name": "STAT3", "id": "6774"} |
| Phytochemical | CID_10205 | Plumbagin | {"name": "Plumbagin", "id": "CID_10205"} | INTERACTS_WITH | - | Target | 7150 | TOP1 | {"name": "TOP1", "id": "7150"} |
| Phytochemical | CID_10205 | Plumbagin | {"name": "Plumbagin", "id": "CID_10205"} | INTERACTS_WITH | - | Target | 7157 | TP53 | {"name": "TP53", "id": "7157"} |
| Phytochemical | CID_10205 | Plumbagin | {"name": "Plumbagin", "id": "CID_10205"} | INTERACTS_WITH | - | Target | 7263 | TST | {"name": "TST", "id": "7263"} |
| Phytochemical | CID_10205 | Plumbagin | {"name": "Plumbagin", "id": "CID_10205"} | INTERACTS_WITH | - | Target | 7398 | USP1 | {"name": "USP1", "id": "7398"} |
| Phytochemical | CID_10205 | Plumbagin | {"name": "Plumbagin", "id": "CID_10205"} | INTERACTS_WITH | - | Target | 7421 | VDR | {"name": "VDR", "id": "7421"} |
| Phytochemical | CID_10205 | Plumbagin | {"name": "Plumbagin", "id": "CID_10205"} | INTERACTS_WITH | - | Target | 7514 | XPO1 | {"name": "XPO1", "id": "7514"} |
| Phytochemical | CID_10205 | Plumbagin | {"name": "Plumbagin", "id": "CID_10205"} | INTERACTS_WITH | - | Target | 834 | CASP1 | {"name": "CASP1", "id": "834"} |
| Phytochemical | CID_10205 | Plumbagin | {"name": "Plumbagin", "id": "CID_10205"} | INTERACTS_WITH | - | Target | 840 | CASP7 | {"name": "CASP7", "id": "840"} |
| Phytochemical | CID_10205 | Plumbagin | {"name": "Plumbagin", "id": "CID_10205"} | INTERACTS_WITH | - | Target | 8850 | KAT2B | {"name": "KAT2B", "id": "8850"} |
| Phytochemical | CID_10205 | Plumbagin | {"name": "Plumbagin", "id": "CID_10205"} | INTERACTS_WITH | - | Target | 8989 | TRPA1 | {"name": "TRPA1", "id": "8989"} |
| Phytochemical | CID_10205 | Plumbagin | {"name": "Plumbagin", "id": "CID_10205"} | INTERACTS_WITH | - | Target | 9020 | MAP3K14 | {"name": "MAP3K14", "id": "9020"} |
| Phytochemical | CID_10205 | Plumbagin | {"name": "Plumbagin", "id": "CID_10205"} | INTERACTS_WITH | - | Target | 993 | CDC25A | {"name": "CDC25A", "id": "993"} |
| Phytochemical | CID_10205 | Plumbagin | {"name": "Plumbagin", "id": "CID_10205"} | INTERACTS_WITH | - | Target | 994 | CDC25B | {"name": "CDC25B", "id": "994"} |
| Phytochemical | CID_70695727 | Pluviatilol | {"name": "Pluviatilol", "id": " CID_70695727"} | INTERACTS_WITH | - | Target | 5465 | PPARA | {"name": "PPARA", "id": "5465"} |
| Phytochemical | CID_70695727 | Pluviatilol | {"name": "Pluviatilol", "id": " CID_70695727"} | INTERACTS_WITH | - | Target | 5467 | PPARD | {"name": "PPARD", "id": "5467"} |
| Phytochemical | CID_70695727 | Pluviatilol | {"name": "Pluviatilol", "id": " CID_70695727"} | INTERACTS_WITH | - | Target | 5468 | PPARG | {"name": "PPARG", "id": "5468"} |
| Phytochemical | CID_189061 | Procurcumenol | {"name": "Procurcumenol", "id": " CID_189061"} | INTERACTS_WITH | - | Target | 5770 | PTPN1 | {"name": "PTPN1", "id": "5770"} |
| Phytochemical | CID_189061 | Procurcumenol | {"name": "Procurcumenol", "id": " CID_189061"} | INTERACTS_WITH | - | Target | 5771 | PTPN2 | {"name": "PTPN2", "id": "5771"} |
| Phytochemical | CID_442495 | Pulegone | {"name": "Pulegone", "id": "CID_442495"} | INTERACTS_WITH | - | Target | 367 | AR | {"name": "AR", "id": "367"} |
| Phytochemical | CID_442495 | Pulegone | {"name": "Pulegone", "id": "CID_442495"} | INTERACTS_WITH | - | Target | 4790 | NFKB1 | {"name": "NFKB1", "id": "4790"} |
| Phytochemical | CID_1057 | Pyrogallol | {"name": "Pyrogallol", "id": "CID_1057"} | INTERACTS_WITH | - | Target | 1017 | CDK2 | {"name": "CDK2", "id": "1017"} |
| Phytochemical | CID_1057 | Pyrogallol | {"name": "Pyrogallol", "id": "CID_1057"} | INTERACTS_WITH | - | Target | 10280 | SIGMAR1 | {"name": "SIGMAR1", "id": "10280"} |
| Phytochemical | CID_1057 | Pyrogallol | {"name": "Pyrogallol", "id": "CID_1057"} | INTERACTS_WITH | - | Target | 10800 | CYSLTR1 | {"name": "CYSLTR1", "id": "10800"} |
| Phytochemical | CID_1057 | Pyrogallol | {"name": "Pyrogallol", "id": "CID_1057"} | INTERACTS_WITH | - | Target | 10919 | EHMT2 | {"name": "EHMT2", "id": "10919"} |
| Phytochemical | CID_1057 | Pyrogallol | {"name": "Pyrogallol", "id": "CID_1057"} | INTERACTS_WITH | - | Target | 10988 | METAP2 | {"name": "METAP2", "id": "10988"} |
| Phytochemical | CID_1057 | Pyrogallol | {"name": "Pyrogallol", "id": "CID_1057"} | INTERACTS_WITH | - | Target | 11096 | ADAMTS5 | {"name": "ADAMTS5", "id": "11096"} |
| Phytochemical | CID_1057 | Pyrogallol | {"name": "Pyrogallol", "id": "CID_1057"} | INTERACTS_WITH | - | Target | 1128 | CHRM1 | {"name": "CHRM1", "id": "1128"} |
| Phytochemical | CID_1057 | Pyrogallol | {"name": "Pyrogallol", "id": "CID_1057"} | INTERACTS_WITH | - | Target | 1129 | CHRM2 | {"name": "CHRM2", "id": "1129"} |

**Table S9: Complete node and relationship data used to construct the knowledge graph. This table summarizes interactions among FDA-approved drugs, formulation, herbs, diseases (type 2 diabetes, fatty liver, obesity), phytochemicals, and their target genes/proteins. Each entry includes source and target node details, edge type, and node annotations (Entrez Gene ID and gene symbol, where applicable).**

| Source Node Label | Source Node ID | Source Node Name | Source Node Properties | Edge Type | Edge Properties | Target Node Label | Target Node ID | Target Node Name | Target Node Properties |
|---|---|---|---|---|---|---|---|---|---|
| Phytochemical | CID_1057 | Pyrogallol | {"name": "Pyrogallol", "id": "CID_1057"} | INTERACTS_WITH | - | Target | 1131 | CHRM3 | {"name": "CHRM3", "id": "1131"} |
| Phytochemical | CID_1057 | Pyrogallol | {"name": "Pyrogallol", "id": "CID_1057"} | INTERACTS_WITH | - | Target | 1132 | CHRM4 | {"name": "CHRM4", "id": "1132"} |
| Phytochemical | CID_1057 | Pyrogallol | {"name": "Pyrogallol", "id": "CID_1057"} | INTERACTS_WITH | - | Target | 1133 | CHRM5 | {"name": "CHRM5", "id": "1133"} |
| Phytochemical | CID_1057 | Pyrogallol | {"name": "Pyrogallol", "id": "CID_1057"} | INTERACTS_WITH | - | Target | 1233 | CCR4 | {"name": "CCR4", "id": "1233"} |
| Phytochemical | CID_1057 | Pyrogallol | {"name": "Pyrogallol", "id": "CID_1057"} | INTERACTS_WITH | - | Target | 1234 | CCR5 | {"name": "CCR5", "id": "1234"} |
| Phytochemical | CID_1057 | Pyrogallol | {"name": "Pyrogallol", "id": "CID_1057"} | INTERACTS_WITH | - | Target | 1268 | CNR1 | {"name": "CNR1", "id": "1268"} |
| Phytochemical | CID_1057 | Pyrogallol | {"name": "Pyrogallol", "id": "CID_1057"} | INTERACTS_WITH | - | Target | 134 | ADORA1 | {"name": "ADORA1", "id": "134"} |
| Phytochemical | CID_1057 | Pyrogallol | {"name": "Pyrogallol", "id": "CID_1057"} | INTERACTS_WITH | - | Target | 135 | ADORA2A | {"name": "ADORA2A", "id": "135"} |
| Phytochemical | CID_1057 | Pyrogallol | {"name": "Pyrogallol", "id": "CID_1057"} | INTERACTS_WITH | - | Target | 140 | ADORA3 | {"name": "ADORA3", "id": "140"} |
| Phytochemical | CID_1057 | Pyrogallol | {"name": "Pyrogallol", "id": "CID_1057"} | INTERACTS_WITH | - | Target | 1432 | MAPK14 | {"name": "MAPK14", "id": "1432"} |
| Phytochemical | CID_1057 | Pyrogallol | {"name": "Pyrogallol", "id": "CID_1057"} | INTERACTS_WITH | - | Target | 146 | ADRA1D | {"name": "ADRA1D", "id": "146"} |
| Phytochemical | CID_1057 | Pyrogallol | {"name": "Pyrogallol", "id": "CID_1057"} | INTERACTS_WITH | - | Target | 150 | ADRA2A | {"name": "ADRA2A", "id": "150"} |
| Phytochemical | CID_1057 | Pyrogallol | {"name": "Pyrogallol", "id": "CID_1057"} | INTERACTS_WITH | - | Target | 151 | ADRA2B | {"name": "ADRA2B", "id": "151"} |
| Phytochemical | CID_1057 | Pyrogallol | {"name": "Pyrogallol", "id": "CID_1057"} | INTERACTS_WITH | - | Target | 1511 | CTSG | {"name": "CTSG", "id": "1511"} |
| Phytochemical | CID_1057 | Pyrogallol | {"name": "Pyrogallol", "id": "CID_1057"} | INTERACTS_WITH | - | Target | 152 | ADRA2C | {"name": "ADRA2C", "id": "152"} |
| Phytochemical | CID_1057 | Pyrogallol | {"name": "Pyrogallol", "id": "CID_1057"} | INTERACTS_WITH | - | Target | 153 | ADRB1 | {"name": "ADRB1", "id": "153"} |
| Phytochemical | CID_1057 | Pyrogallol | {"name": "Pyrogallol", "id": "CID_1057"} | INTERACTS_WITH | - | Target | 154 | ADRB2 | {"name": "ADRB2", "id": "154"} |
| Phytochemical | CID_1057 | Pyrogallol | {"name": "Pyrogallol", "id": "CID_1057"} | INTERACTS_WITH | - | Target | 1544 | CYP1A2 | {"name": "CYP1A2", "id": "1544"} |
| Phytochemical | CID_1057 | Pyrogallol | {"name": "Pyrogallol", "id": "CID_1057"} | INTERACTS_WITH | - | Target | 1548 | CYP2A6 | {"name": "CYP2A6", "id": "1548"} |
| Phytochemical | CID_1057 | Pyrogallol | {"name": "Pyrogallol", "id": "CID_1057"} | INTERACTS_WITH | - | Target | 155 | ADRB3 | {"name": "ADRB3", "id": "155"} |
| Phytochemical | CID_1057 | Pyrogallol | {"name": "Pyrogallol", "id": "CID_1057"} | INTERACTS_WITH | - | Target | 1557 | CYP2C19 | {"name": "CYP2C19", "id": "1557"} |
| Phytochemical | CID_1057 | Pyrogallol | {"name": "Pyrogallol", "id": "CID_1057"} | INTERACTS_WITH | - | Target | 1559 | CYP2C9 | {"name": "CYP2C9", "id": "1559"} |
| Phytochemical | CID_1057 | Pyrogallol | {"name": "Pyrogallol", "id": "CID_1057"} | INTERACTS_WITH | - | Target | 1565 | CYP2D6 | {"name": "CYP2D6", "id": "1565"} |
| Phytochemical | CID_1057 | Pyrogallol | {"name": "Pyrogallol", "id": "CID_1057"} | INTERACTS_WITH | - | Target | 1571 | CYP2E1 | {"name": "CYP2E1", "id": "1571"} |
| Phytochemical | CID_1057 | Pyrogallol | {"name": "Pyrogallol", "id": "CID_1057"} | INTERACTS_WITH | - | Target | 1576 | CYP3A4 | {"name": "CYP3A4", "id": "1576"} |
| Phytochemical | CID_1057 | Pyrogallol | {"name": "Pyrogallol", "id": "CID_1057"} | INTERACTS_WITH | - | Target | 1812 | DRD1 | {"name": "DRD1", "id": "1812"} |
| Phytochemical | CID_1057 | Pyrogallol | {"name": "Pyrogallol", "id": "CID_1057"} | INTERACTS_WITH | - | Target | 1813 | DRD2 | {"name": "DRD2", "id": "1813"} |
| Phytochemical | CID_1057 | Pyrogallol | {"name": "Pyrogallol", "id": "CID_1057"} | INTERACTS_WITH | - | Target | 1814 | DRD3 | {"name": "DRD3", "id": "1814"} |
| Phytochemical | CID_1057 | Pyrogallol | {"name": "Pyrogallol", "id": "CID_1057"} | INTERACTS_WITH | - | Target | 1815 | DRD4 | {"name": "DRD4", "id": "1815"} |
| Phytochemical | CID_1057 | Pyrogallol | {"name": "Pyrogallol", "id": "CID_1057"} | INTERACTS_WITH | - | Target | 186 | AGTR2 | {"name": "AGTR2", "id": "186"} |
| Phytochemical | CID_1057 | Pyrogallol | {"name": "Pyrogallol", "id": "CID_1057"} | INTERACTS_WITH | - | Target | 1909 | EDNRA | {"name": "EDNRA", "id": "1909"} |
| Phytochemical | CID_1057 | Pyrogallol | {"name": "Pyrogallol", "id": "CID_1057"} | INTERACTS_WITH | - | Target | 1956 | EGFR | {"name": "EGFR", "id": "1956"} |
| Phytochemical | CID_1057 | Pyrogallol | {"name": "Pyrogallol", "id": "CID_1057"} | INTERACTS_WITH | - | Target | 1991 | ELANE | {"name": "ELANE", "id": "1991"} |
| Phytochemical | CID_1057 | Pyrogallol | {"name": "Pyrogallol", "id": "CID_1057"} | INTERACTS_WITH | - | Target | 2064 | ERBB2 | {"name": "ERBB2", "id": "2064"} |
| Phytochemical | CID_1057 | Pyrogallol | {"name": "Pyrogallol", "id": "CID_1057"} | INTERACTS_WITH | - | Target | 2099 | ESR1 | {"name": "ESR1", "id": "2099"} |
| Phytochemical | CID_1057 | Pyrogallol | {"name": "Pyrogallol", "id": "CID_1057"} | INTERACTS_WITH | - | Target | 2100 | ESR2 | {"name": "ESR2", "id": "2100"} |
| Phytochemical | CID_1057 | Pyrogallol | {"name": "Pyrogallol", "id": "CID_1057"} | INTERACTS_WITH | - | Target | 213 | ALB | {"name": "ALB", "id": "213"} |
| Phytochemical | CID_1057 | Pyrogallol | {"name": "Pyrogallol", "id": "CID_1057"} | INTERACTS_WITH | - | Target | 216 | ALDH1A1 | {"name": "ALDH1A1", "id": "216"} |
| Phytochemical | CID_1057 | Pyrogallol | {"name": "Pyrogallol", "id": "CID_1057"} | INTERACTS_WITH | - | Target | 2321 | FLT1 | {"name": "FLT1", "id": "2321"} |
| Phytochemical | CID_1057 | Pyrogallol | {"name": "Pyrogallol", "id": "CID_1057"} | INTERACTS_WITH | - | Target | 246 | ALOX15 | {"name": "ALOX15", "id": "246"} |
| Phytochemical | CID_1057 | Pyrogallol | {"name": "Pyrogallol", "id": "CID_1057"} | INTERACTS_WITH | - | Target | 2534 | FYN | {"name": "FYN", "id": "2534"} |
| Phytochemical | CID_1057 | Pyrogallol | {"name": "Pyrogallol", "id": "CID_1057"} | INTERACTS_WITH | - | Target | 2744 | GLS | {"name": "GLS", "id": "2744"} |
| Phytochemical | CID_1057 | Pyrogallol | {"name": "Pyrogallol", "id": "CID_1057"} | INTERACTS_WITH | - | Target | 2908 | NR3C1 | {"name": "NR3C1", "id": "2908"} |
| Phytochemical | CID_1057 | Pyrogallol | {"name": "Pyrogallol", "id": "CID_1057"} | INTERACTS_WITH | - | Target | 3028 | HSD17B10 | {"name": "HSD17B10", "id": "3028"} |
| Phytochemical | CID_1057 | Pyrogallol | {"name": "Pyrogallol", "id": "CID_1057"} | INTERACTS_WITH | - | Target | 3156 | HMGCR | {"name": "HMGCR", "id": "3156"} |
| Phytochemical | CID_1057 | Pyrogallol | {"name": "Pyrogallol", "id": "CID_1057"} | INTERACTS_WITH | - | Target | 3248 | HPGD | {"name": "HPGD", "id": "3248"} |
| Phytochemical | CID_1057 | Pyrogallol | {"name": "Pyrogallol", "id": "CID_1057"} | INTERACTS_WITH | - | Target | 3269 | HRH1 | {"name": "HRH1", "id": "3269"} |
| Phytochemical | CID_1057 | Pyrogallol | {"name": "Pyrogallol", "id": "CID_1057"} | INTERACTS_WITH | - | Target | 3274 | HRH2 | {"name": "HRH2", "id": "3274"} |
| Phytochemical | CID_1057 | Pyrogallol | {"name": "Pyrogallol", "id": "CID_1057"} | INTERACTS_WITH | - | Target | 328 | APEX1 | {"name": "APEX1", "id": "328"} |
| Phytochemical | CID_1057 | Pyrogallol | {"name": "Pyrogallol", "id": "CID_1057"} | INTERACTS_WITH | - | Target | 3356 | HTR2A | {"name": "HTR2A", "id": "3356"} |
| Phytochemical | CID_1057 | Pyrogallol | {"name": "Pyrogallol", "id": "CID_1057"} | INTERACTS_WITH | - | Target | 3357 | HTR2B | {"name": "HTR2B", "id": "3357"} |
| Phytochemical | CID_1057 | Pyrogallol | {"name": "Pyrogallol", "id": "CID_1057"} | INTERACTS_WITH | - | Target | 3358 | HTR2C | {"name": "HTR2C", "id": "3358"} |
| Phytochemical | CID_1057 | Pyrogallol | {"name": "Pyrogallol", "id": "CID_1057"} | INTERACTS_WITH | - | Target | 3362 | HTR6 | {"name": "HTR6", "id": "3362"} |
| Phytochemical | CID_1057 | Pyrogallol | {"name": "Pyrogallol", "id": "CID_1057"} | INTERACTS_WITH | - | Target | 3576 | CXCL8 | {"name": "CXCL8", "id": "3576"} |
| Phytochemical | CID_1057 | Pyrogallol | {"name": "Pyrogallol", "id": "CID_1057"} | INTERACTS_WITH | - | Target | 3577 | CXCR1 | {"name": "CXCR1", "id": "3577"} |
| Phytochemical | CID_1057 | Pyrogallol | {"name": "Pyrogallol", "id": "CID_1057"} | INTERACTS_WITH | - | Target | 3579 | CXCR2 | {"name": "CXCR2", "id": "3579"} |
| Phytochemical | CID_1057 | Pyrogallol | {"name": "Pyrogallol", "id": "CID_1057"} | INTERACTS_WITH | - | Target | 367 | AR | {"name": "AR", "id": "367"} |
| Phytochemical | CID_1057 | Pyrogallol | {"name": "Pyrogallol", "id": "CID_1057"} | INTERACTS_WITH | - | Target | 3757 | KCNH2 | {"name": "KCNH2", "id": "3757"} |
| Phytochemical | CID_1057 | Pyrogallol | {"name": "Pyrogallol", "id": "CID_1057"} | INTERACTS_WITH | - | Target | 383 | ARG1 | {"name": "ARG1", "id": "383"} |
| Phytochemical | CID_1057 | Pyrogallol | {"name": "Pyrogallol", "id": "CID_1057"} | INTERACTS_WITH | - | Target | 3932 | LCK | {"name": "LCK", "id": "3932"} |
| Phytochemical | CID_1057 | Pyrogallol | {"name": "Pyrogallol", "id": "CID_1057"} | INTERACTS_WITH | - | Target | 4000 | LMNA | {"name": "LMNA", "id": "4000"} |
| Phytochemical | CID_1057 | Pyrogallol | {"name": "Pyrogallol", "id": "CID_1057"} | INTERACTS_WITH | - | Target | 4128 | MAOA | {"name": "MAOA", "id": "4128"} |
| Phytochemical | CID_1057 | Pyrogallol | {"name": "Pyrogallol", "id": "CID_1057"} | INTERACTS_WITH | - | Target | 4159 | MC3R | {"name": "MC3R", "id": "4159"} |
| Phytochemical | CID_1057 | Pyrogallol | {"name": "Pyrogallol", "id": "CID_1057"} | INTERACTS_WITH | - | Target | 4160 | MC4R | {"name": "MC4R", "id": "4160"} |
| Phytochemical | CID_1057 | Pyrogallol | {"name": "Pyrogallol", "id": "CID_1057"} | INTERACTS_WITH | - | Target | 4161 | MC5R | {"name": "MC5R", "id": "4161"} |
| Phytochemical | CID_1057 | Pyrogallol | {"name": "Pyrogallol", "id": "CID_1057"} | INTERACTS_WITH | - | Target | 43 | ACHE | {"name": "ACHE", "id": "43"} |
| Phytochemical | CID_1057 | Pyrogallol | {"name": "Pyrogallol", "id": "CID_1057"} | INTERACTS_WITH | - | Target | 4312 | MMP1 | {"name": "MMP1", "id": "4312"} |
| Phytochemical | CID_1057 | Pyrogallol | {"name": "Pyrogallol", "id": "CID_1057"} | INTERACTS_WITH | - | Target | 4313 | MMP2 | {"name": "MMP2", "id": "4313"} |
| Phytochemical | CID_1057 | Pyrogallol | {"name": "Pyrogallol", "id": "CID_1057"} | INTERACTS_WITH | - | Target | 4317 | MMP8 | {"name": "MMP8", "id": "4317"} |
| Phytochemical | CID_1057 | Pyrogallol | {"name": "Pyrogallol", "id": "CID_1057"} | INTERACTS_WITH | - | Target | 4318 | MMP9 | {"name": "MMP9", "id": "4318"} |
| Phytochemical | CID_1057 | Pyrogallol | {"name": "Pyrogallol", "id": "CID_1057"} | INTERACTS_WITH | - | Target | 4321 | MMP12 | {"name": "MMP12", "id": "4321"} |
| Phytochemical | CID_1057 | Pyrogallol | {"name": "Pyrogallol", "id": "CID_1057"} | INTERACTS_WITH | - | Target | 4780 | NFE2L2 | {"name": "NFE2L2", "id": "4780"} |
| Phytochemical | CID_1057 | Pyrogallol | {"name": "Pyrogallol", "id": "CID_1057"} | INTERACTS_WITH | - | Target | 4886 | NPY1R | {"name": "NPY1R", "id": "4886"} |
| Phytochemical | CID_1057 | Pyrogallol | {"name": "Pyrogallol", "id": "CID_1057"} | INTERACTS_WITH | - | Target | 4887 | NPY2R | {"name": "NPY2R", "id": "4887"} |
| Phytochemical | CID_1057 | Pyrogallol | {"name": "Pyrogallol", "id": "CID_1057"} | INTERACTS_WITH | - | Target | 4985 | OPRD1 | {"name": "OPRD1", "id": "4985"} |
| Phytochemical | CID_1057 | Pyrogallol | {"name": "Pyrogallol", "id": "CID_1057"} | INTERACTS_WITH | - | Target | 4986 | OPRK1 | {"name": "OPRK1", "id": "4986"} |

**Table S9: Complete node and relationship data used to construct the knowledge graph. This table summarizes interactions among FDA-approved drugs, formulation, herbs, diseases (type 2 diabetes, fatty liver, obesity), phytochemicals, and their target genes/proteins. Each entry includes source and target node details, edge type, and node annotations (Entrez Gene ID and gene symbol, where applicable).**

| Source Node Label | Source Node ID | Source Node Name | Source Node Properties | Edge Type | Edge Properties | Target Node Label | Target Node ID | Target Node Name | Target Node Properties |
|---|---|---|---|---|---|---|---|---|---|
| Phytochemical | CID_1057 | Pyrogallol | {"name": "Pyrogallol", "id": "CID_1057"} | INTERACTS_WITH | - | Target | 4988 | OPRM1 | {"name": "OPRM1", "id": "4988"} |
| Phytochemical | CID_1057 | Pyrogallol | {"name": "Pyrogallol", "id": "CID_1057"} | INTERACTS_WITH | - | Target | 51426 | POLK | {"name": "POLK", "id": "51426"} |
| Phytochemical | CID_1057 | Pyrogallol | {"name": "Pyrogallol", "id": "CID_1057"} | INTERACTS_WITH | - | Target | 54658 | UGT1A1 | {"name": "UGT1A1", "id": "54658"} |
| Phytochemical | CID_1057 | Pyrogallol | {"name": "Pyrogallol", "id": "CID_1057"} | INTERACTS_WITH | - | Target | 552 | AVPR1A | {"name": "AVPR1A", "id": "552"} |
| Phytochemical | CID_1057 | Pyrogallol | {"name": "Pyrogallol", "id": "CID_1057"} | INTERACTS_WITH | - | Target | 5530 | PPP3CA | {"name": "PPP3CA", "id": "5530"} |
| Phytochemical | CID_1057 | Pyrogallol | {"name": "Pyrogallol", "id": "CID_1057"} | INTERACTS_WITH | - | Target | 55775 | TDP1 | {"name": "TDP1", "id": "55775"} |
| Phytochemical | CID_1057 | Pyrogallol | {"name": "Pyrogallol", "id": "CID_1057"} | INTERACTS_WITH | - | Target | 5578 | PRKCA | {"name": "PRKCA", "id": "5578"} |
| Phytochemical | CID_1057 | Pyrogallol | {"name": "Pyrogallol", "id": "CID_1057"} | INTERACTS_WITH | - | Target | 5594 | MAPK1 | {"name": "MAPK1", "id": "5594"} |
| Phytochemical | CID_1057 | Pyrogallol | {"name": "Pyrogallol", "id": "CID_1057"} | INTERACTS_WITH | - | Target | 5595 | MAPK3 | {"name": "MAPK3", "id": "5595"} |
| Phytochemical | CID_1057 | Pyrogallol | {"name": "Pyrogallol", "id": "CID_1057"} | INTERACTS_WITH | - | Target | 5724 | PTAFR | {"name": "PTAFR", "id": "5724"} |
| Phytochemical | CID_1057 | Pyrogallol | {"name": "Pyrogallol", "id": "CID_1057"} | INTERACTS_WITH | - | Target | 5742 | PTGS1 | {"name": "PTGS1", "id": "5742"} |
| Phytochemical | CID_1057 | Pyrogallol | {"name": "Pyrogallol", "id": "CID_1057"} | INTERACTS_WITH | - | Target | 5743 | PTGS2 | {"name": "PTGS2", "id": "5743"} |
| Phytochemical | CID_1057 | Pyrogallol | {"name": "Pyrogallol", "id": "CID_1057"} | INTERACTS_WITH | - | Target | 5788 | PTPRC | {"name": "PTPRC", "id": "5788"} |
| Phytochemical | CID_1057 | Pyrogallol | {"name": "Pyrogallol", "id": "CID_1057"} | INTERACTS_WITH | - | Target | 5965 | RECQL | {"name": "RECQL", "id": "5965"} |
| Phytochemical | CID_1057 | Pyrogallol | {"name": "Pyrogallol", "id": "CID_1057"} | INTERACTS_WITH | - | Target | 624 | BDKRB2 | {"name": "BDKRB2", "id": "624"} |
| Phytochemical | CID_1057 | Pyrogallol | {"name": "Pyrogallol", "id": "CID_1057"} | INTERACTS_WITH | - | Target | 6401 | SELE | {"name": "SELE", "id": "6401"} |
| Phytochemical | CID_1057 | Pyrogallol | {"name": "Pyrogallol", "id": "CID_1057"} | INTERACTS_WITH | - | Target | 6402 | SELL | {"name": "SELL", "id": "6402"} |
| Phytochemical | CID_1057 | Pyrogallol | {"name": "Pyrogallol", "id": "CID_1057"} | INTERACTS_WITH | - | Target | 6403 | SELP | {"name": "SELP", "id": "6403"} |
| Phytochemical | CID_1057 | Pyrogallol | {"name": "Pyrogallol", "id": "CID_1057"} | INTERACTS_WITH | - | Target | 6530 | SLC6A2 | {"name": "SLC6A2", "id": "6530"} |
| Phytochemical | CID_1057 | Pyrogallol | {"name": "Pyrogallol", "id": "CID_1057"} | INTERACTS_WITH | - | Target | 6531 | SLC6A3 | {"name": "SLC6A3", "id": "6531"} |
| Phytochemical | CID_1057 | Pyrogallol | {"name": "Pyrogallol", "id": "CID_1057"} | INTERACTS_WITH | - | Target | 6532 | SLC6A4 | {"name": "SLC6A4", "id": "6532"} |
| Phytochemical | CID_1057 | Pyrogallol | {"name": "Pyrogallol", "id": "CID_1057"} | INTERACTS_WITH | - | Target | 6865 | TACR2 | {"name": "TACR2", "id": "6865"} |
| Phytochemical | CID_1057 | Pyrogallol | {"name": "Pyrogallol", "id": "CID_1057"} | INTERACTS_WITH | - | Target | 6869 | TACR1 | {"name": "TACR1", "id": "6869"} |
| Phytochemical | CID_1057 | Pyrogallol | {"name": "Pyrogallol", "id": "CID_1057"} | INTERACTS_WITH | - | Target | 6916 | TBXAS1 | {"name": "TBXAS1", "id": "6916"} |
| Phytochemical | CID_1057 | Pyrogallol | {"name": "Pyrogallol", "id": "CID_1057"} | INTERACTS_WITH | - | Target | 729230 | CCR2 | {"name": "CCR2", "id": "729230"} |
| Phytochemical | CID_1057 | Pyrogallol | {"name": "Pyrogallol", "id": "CID_1057"} | INTERACTS_WITH | - | Target | 7299 | TYR | {"name": "TYR", "id": "7299"} |
| Phytochemical | CID_1057 | Pyrogallol | {"name": "Pyrogallol", "id": "CID_1057"} | INTERACTS_WITH | - | Target | 7366 | UGT2B15 | {"name": "UGT2B15", "id": "7366"} |
| Phytochemical | CID_1057 | Pyrogallol | {"name": "Pyrogallol", "id": "CID_1057"} | INTERACTS_WITH | - | Target | 7421 | VDR | {"name": "VDR", "id": "7421"} |
| Phytochemical | CID_1057 | Pyrogallol | {"name": "Pyrogallol", "id": "CID_1057"} | INTERACTS_WITH | - | Target | 7433 | VIPR1 | {"name": "VIPR1", "id": "7433"} |
| Phytochemical | CID_1057 | Pyrogallol | {"name": "Pyrogallol", "id": "CID_1057"} | INTERACTS_WITH | - | Target | 759 | CA1 | {"name": "CA1", "id": "759"} |
| Phytochemical | CID_1057 | Pyrogallol | {"name": "Pyrogallol", "id": "CID_1057"} | INTERACTS_WITH | - | Target | 760 | CA2 | {"name": "CA2", "id": "760"} |
| Phytochemical | CID_1057 | Pyrogallol | {"name": "Pyrogallol", "id": "CID_1057"} | INTERACTS_WITH | - | Target | 762 | CA4 | {"name": "CA4", "id": "762"} |
| Phytochemical | CID_1057 | Pyrogallol | {"name": "Pyrogallol", "id": "CID_1057"} | INTERACTS_WITH | - | Target | 765 | CA6 | {"name": "CA6", "id": "765"} |
| Phytochemical | CID_1057 | Pyrogallol | {"name": "Pyrogallol", "id": "CID_1057"} | INTERACTS_WITH | - | Target | 799 | CALCR | {"name": "CALCR", "id": "799"} |
| Phytochemical | CID_1057 | Pyrogallol | {"name": "Pyrogallol", "id": "CID_1057"} | INTERACTS_WITH | - | Target | 834 | CASP1 | {"name": "CASP1", "id": "834"} |
| Phytochemical | CID_1057 | Pyrogallol | {"name": "Pyrogallol", "id": "CID_1057"} | INTERACTS_WITH | - | Target | 861 | RUNX1 | {"name": "RUNX1", "id": "861"} |
| Phytochemical | CID_1057 | Pyrogallol | {"name": "Pyrogallol", "id": "CID_1057"} | INTERACTS_WITH | - | Target | 865 | CBFB | {"name": "CBFB", "id": "865"} |
| Phytochemical | CID_1057 | Pyrogallol | {"name": "Pyrogallol", "id": "CID_1057"} | INTERACTS_WITH | - | Target | 8654 | PDE5A | {"name": "PDE5A", "id": "8654"} |
| Phytochemical | CID_1057 | Pyrogallol | {"name": "Pyrogallol", "id": "CID_1057"} | INTERACTS_WITH | - | Target | 886 | CCKAR | {"name": "CCKAR", "id": "886"} |
| Phytochemical | CID_1057 | Pyrogallol | {"name": "Pyrogallol", "id": "CID_1057"} | INTERACTS_WITH | - | Target | 9507 | ADAMTS4 | {"name": "ADAMTS4", "id": "9507"} |
| Phytochemical | CID_5280343 | Quercetin | {"name": "Quercetin", "id": " CID_5280343"} | INTERACTS_WITH | - | Target | 10013 | HDAC6 | {"name": "HDAC6", "id": "10013"} |
| Phytochemical | CID_5280343 | Quercetin | {"name": "Quercetin", "id": " CID_5280343"} | INTERACTS_WITH | - | Target | 10072 | DPP3 | {"name": "DPP3", "id": "10072"} |
| Phytochemical | CID_5280343 | Quercetin | {"name": "Quercetin", "id": " CID_5280343"} | INTERACTS_WITH | - | Target | 1017 | CDK2 | {"name": "CDK2", "id": "1017"} |
| Phytochemical | CID_5280343 | Quercetin | {"name": "Quercetin", "id": " CID_5280343"} | INTERACTS_WITH | - | Target | 1019 | CDK4 | {"name": "CDK4", "id": "1019"} |
| Phytochemical | CID_5280343 | Quercetin | {"name": "Quercetin", "id": " CID_5280343"} | INTERACTS_WITH | - | Target | 1020 | CDK5 | {"name": "CDK5", "id": "1020"} |
| Phytochemical | CID_5280343 | Quercetin | {"name": "Quercetin", "id": " CID_5280343"} | INTERACTS_WITH | - | Target | 1021 | CDK6 | {"name": "CDK6", "id": "1021"} |
| Phytochemical | CID_5280343 | Quercetin | {"name": "Quercetin", "id": " CID_5280343"} | INTERACTS_WITH | - | Target | 10257 | ABCC4 | {"name": "ABCC4", "id": "10257"} |
| Phytochemical | CID_5280343 | Quercetin | {"name": "Quercetin", "id": " CID_5280343"} | INTERACTS_WITH | - | Target | 10298 | PAK4 | {"name": "PAK4", "id": "10298"} |
| Phytochemical | CID_5280343 | Quercetin | {"name": "Quercetin", "id": " CID_5280343"} | INTERACTS_WITH | - | Target | 10494 | STK25 | {"name": "STK25", "id": "10494"} |
| Phytochemical | CID_5280343 | Quercetin | {"name": "Quercetin", "id": " CID_5280343"} | INTERACTS_WITH | - | Target | 10599 | SLCO1B1 | {"name": "SLCO1B1", "id": "10599"} |
| Phytochemical | CID_5280343 | Quercetin | {"name": "Quercetin", "id": " CID_5280343"} | INTERACTS_WITH | - | Target | 10645 | CAMKK2 | {"name": "CAMKK2", "id": "10645"} |
| Phytochemical | CID_5280343 | Quercetin | {"name": "Quercetin", "id": " CID_5280343"} | INTERACTS_WITH | - | Target | 106821730 | BUB1B-PAK6 | {"name": "BUB1B-PAK6", "id": "106821730"} |
| Phytochemical | CID_5280343 | Quercetin | {"name": "Quercetin", "id": " CID_5280343"} | INTERACTS_WITH | - | Target | 10733 | PLK4 | {"name": "PLK4", "id": "10733"} |
| Phytochemical | CID_5280343 | Quercetin | {"name": "Quercetin", "id": " CID_5280343"} | INTERACTS_WITH | - | Target | 10783 | NEK6 | {"name": "NEK6", "id": "10783"} |
| Phytochemical | CID_5280343 | Quercetin | {"name": "Quercetin", "id": " CID_5280343"} | INTERACTS_WITH | - | Target | 1080 | CFTR | {"name": "CFTR", "id": "1080"} |
| Phytochemical | CID_5280343 | Quercetin | {"name": "Quercetin", "id": " CID_5280343"} | INTERACTS_WITH | - | Target | 10919 | EHMT2 | {"name": "EHMT2", "id": "10919"} |
| Phytochemical | CID_5280343 | Quercetin | {"name": "Quercetin", "id": " CID_5280343"} | INTERACTS_WITH | - | Target | 10951 | CBX1 | {"name": "CBX1", "id": "10951"} |
| Phytochemical | CID_5280343 | Quercetin | {"name": "Quercetin", "id": " CID_5280343"} | INTERACTS_WITH | - | Target | 11040 | PIM2 | {"name": "PIM2", "id": "11040"} |
| Phytochemical | CID_5280343 | Quercetin | {"name": "Quercetin", "id": " CID_5280343"} | INTERACTS_WITH | - | Target | 1111 | CHEK1 | {"name": "CHEK1", "id": "1111"} |
| Phytochemical | CID_5280343 | Quercetin | {"name": "Quercetin", "id": " CID_5280343"} | INTERACTS_WITH | - | Target | 11200 | CHEK2 | {"name": "CHEK2", "id": "11200"} |
| Phytochemical | CID_5280343 | Quercetin | {"name": "Quercetin", "id": " CID_5280343"} | INTERACTS_WITH | - | Target | 11201 | POLI | {"name": "POLI", "id": "11201"} |
| Phytochemical | CID_5280343 | Quercetin | {"name": "Quercetin", "id": " CID_5280343"} | INTERACTS_WITH | - | Target | 11238 | CA5B | {"name": "CA5B", "id": "11238"} |
| Phytochemical | CID_5280343 | Quercetin | {"name": "Quercetin", "id": " CID_5280343"} | INTERACTS_WITH | - | Target | 11255 | HRH3 | {"name": "HRH3", "id": "11255"} |

**Table S9: Complete node and relationship data used to construct the knowledge graph. This table summarizes interactions among FDA-approved drugs, formulation, herbs, diseases (type 2 diabetes, fatty liver, obesity), phytochemicals, and their target genes/proteins. Each entry includes source and target node details, edge type, and node annotations (Entrez Gene ID and gene symbol, where applicable).**

| Source Node Label | Source Node ID | Source Node Name | Source Node Properties | Edge Type | Edge Properties | Target Node Label | Target Node ID | Target Node Name | Target Node Properties |
|---|---|---|---|---|---|---|---|---|---|
| Phytochemical | CID_5280343 | Quercetin | {"name": "Quercetin", "id": "CID_5280343"} | INTERACTS_WITH | - | Target | 1128 | CHRM1 | {"name": "CHRM1", "id": "1128"} |
| Phytochemical | CID_5280343 | Quercetin | {"name": "Quercetin", "id": "CID_5280343"} | INTERACTS_WITH | - | Target | 1129 | CHRM2 | {"name": "CHRM2", "id": "1129"} |
| Phytochemical | CID_5280343 | Quercetin | {"name": "Quercetin", "id": "CID_5280343"} | INTERACTS_WITH | - | Target | 11309 | SLCO2B1 | {"name": "SLCO2B1", "id": "11309"} |
| Phytochemical | CID_5280343 | Quercetin | {"name": "Quercetin", "id": "CID_5280343"} | INTERACTS_WITH | - | Target | 1131 | CHRM3 | {"name": "CHRM3", "id": "1131"} |
| Phytochemical | CID_5280343 | Quercetin | {"name": "Quercetin", "id": "CID_5280343"} | INTERACTS_WITH | - | Target | 11329 | STK38 | {"name": "STK38", "id": "11329"} |
| Phytochemical | CID_5280343 | Quercetin | {"name": "Quercetin", "id": "CID_5280343"} | INTERACTS_WITH | - | Target | 1133 | CHRM5 | {"name": "CHRM5", "id": "1133"} |
| Phytochemical | CID_5280343 | Quercetin | {"name": "Quercetin", "id": "CID_5280343"} | INTERACTS_WITH | - | Target | 11330 | CTRC | {"name": "CTRC", "id": "11330"} |
| Phytochemical | CID_5280343 | Quercetin | {"name": "Quercetin", "id": "CID_5280343"} | INTERACTS_WITH | - | Target | 1137 | CHRNA4 | {"name": "CHRNA4", "id": "1137"} |
| Phytochemical | CID_5280343 | Quercetin | {"name": "Quercetin", "id": "CID_5280343"} | INTERACTS_WITH | - | Target | 118429 | ANTXR2 | {"name": "ANTXR2", "id": "118429"} |
| Phytochemical | CID_5280343 | Quercetin | {"name": "Quercetin", "id": "CID_5280343"} | INTERACTS_WITH | - | Target | 1195 | CLK1 | {"name": "CLK1", "id": "1195"} |
| Phytochemical | CID_5280343 | Quercetin | {"name": "Quercetin", "id": "CID_5280343"} | INTERACTS_WITH | - | Target | 1196 | CLK2 | {"name": "CLK2", "id": "1196"} |
| Phytochemical | CID_5280343 | Quercetin | {"name": "Quercetin", "id": "CID_5280343"} | INTERACTS_WITH | - | Target | 1198 | CLK3 | {"name": "CLK3", "id": "1198"} |
| Phytochemical | CID_5280343 | Quercetin | {"name": "Quercetin", "id": "CID_5280343"} | INTERACTS_WITH | - | Target | 1233 | CCR4 | {"name": "CCR4", "id": "1233"} |
| Phytochemical | CID_5280343 | Quercetin | {"name": "Quercetin", "id": "CID_5280343"} | INTERACTS_WITH | - | Target | 1244 | ABCC2 | {"name": "ABCC2", "id": "1244"} |
| Phytochemical | CID_5280343 | Quercetin | {"name": "Quercetin", "id": "CID_5280343"} | INTERACTS_WITH | - | Target | 1268 | CNR1 | {"name": "CNR1", "id": "1268"} |
| Phytochemical | CID_5280343 | Quercetin | {"name": "Quercetin", "id": "CID_5280343"} | INTERACTS_WITH | - | Target | 134 | ADORA1 | {"name": "ADORA1", "id": "134"} |
| Phytochemical | CID_5280343 | Quercetin | {"name": "Quercetin", "id": "CID_5280343"} | INTERACTS_WITH | - | Target | 135 | ADORA2A | {"name": "ADORA2A", "id": "135"} |
| Phytochemical | CID_5280343 | Quercetin | {"name": "Quercetin", "id": "CID_5280343"} | INTERACTS_WITH | - | Target | 1394 | CRHR1 | {"name": "CRHR1", "id": "1394"} |
| Phytochemical | CID_5280343 | Quercetin | {"name": "Quercetin", "id": "CID_5280343"} | INTERACTS_WITH | - | Target | 1395 | CRHR2 | {"name": "CRHR2", "id": "1395"} |
| Phytochemical | CID_5280343 | Quercetin | {"name": "Quercetin", "id": "CID_5280343"} | INTERACTS_WITH | - | Target | 140 | ADORA3 | {"name": "ADORA3", "id": "140"} |
| Phytochemical | CID_5280343 | Quercetin | {"name": "Quercetin", "id": "CID_5280343"} | INTERACTS_WITH | - | Target | 1432 | MAPK14 | {"name": "MAPK14", "id": "1432"} |
| Phytochemical | CID_5280343 | Quercetin | {"name": "Quercetin", "id": "CID_5280343"} | INTERACTS_WITH | - | Target | 1455 | CSNK1G2 | {"name": "CSNK1G2", "id": "1455"} |
| Phytochemical | CID_5280343 | Quercetin | {"name": "Quercetin", "id": "CID_5280343"} | INTERACTS_WITH | - | Target | 1456 | CSNK1G3 | {"name": "CSNK1G3", "id": "1456"} |
| Phytochemical | CID_5280343 | Quercetin | {"name": "Quercetin", "id": "CID_5280343"} | INTERACTS_WITH | - | Target | 1457 | CSNK2A1 | {"name": "CSNK2A1", "id": "1457"} |
| Phytochemical | CID_5280343 | Quercetin | {"name": "Quercetin", "id": "CID_5280343"} | INTERACTS_WITH | - | Target | 1459 | CSNK2A2 | {"name": "CSNK2A2", "id": "1459"} |
| Phytochemical | CID_5280343 | Quercetin | {"name": "Quercetin", "id": "CID_5280343"} | INTERACTS_WITH | - | Target | 1460 | CSNK2B | {"name": "CSNK2B", "id": "1460"} |
| Phytochemical | CID_5280343 | Quercetin | {"name": "Quercetin", "id": "CID_5280343"} | INTERACTS_WITH | - | Target | 148 | ADRA1A | {"name": "ADRA1A", "id": "148"} |
| Phytochemical | CID_5280343 | Quercetin | {"name": "Quercetin", "id": "CID_5280343"} | INTERACTS_WITH | - | Target | 1499 | CTNNB1 | {"name": "CTNNB1", "id": "1499"} |
| Phytochemical | CID_5280343 | Quercetin | {"name": "Quercetin", "id": "CID_5280343"} | INTERACTS_WITH | - | Target | 150 | ADRA2A | {"name": "ADRA2A", "id": "150"} |
| Phytochemical | CID_5280343 | Quercetin | {"name": "Quercetin", "id": "CID_5280343"} | INTERACTS_WITH | - | Target | 1504 | CTRB1 | {"name": "CTRB1", "id": "1504"} |
| Phytochemical | CID_5280343 | Quercetin | {"name": "Quercetin", "id": "CID_5280343"} | INTERACTS_WITH | - | Target | 151 | ADRA2B | {"name": "ADRA2B", "id": "151"} |
| Phytochemical | CID_5280343 | Quercetin | {"name": "Quercetin", "id": "CID_5280343"} | INTERACTS_WITH | - | Target | 1514 | CTSL | {"name": "CTSL", "id": "1514"} |
| Phytochemical | CID_5280343 | Quercetin | {"name": "Quercetin", "id": "CID_5280343"} | INTERACTS_WITH | - | Target | 152 | ADRA2C | {"name": "ADRA2C", "id": "152"} |
| Phytochemical | CID_5280343 | Quercetin | {"name": "Quercetin", "id": "CID_5280343"} | INTERACTS_WITH | - | Target | 153 | ADRB1 | {"name": "ADRB1", "id": "153"} |
| Phytochemical | CID_5280343 | Quercetin | {"name": "Quercetin", "id": "CID_5280343"} | INTERACTS_WITH | - | Target | 154 | ADRB2 | {"name": "ADRB2", "id": "154"} |
| Phytochemical | CID_5280343 | Quercetin | {"name": "Quercetin", "id": "CID_5280343"} | INTERACTS_WITH | - | Target | 1543 | CYP1A1 | {"name": "CYP1A1", "id": "1543"} |
| Phytochemical | CID_5280343 | Quercetin | {"name": "Quercetin", "id": "CID_5280343"} | INTERACTS_WITH | - | Target | 1544 | CYP1A2 | {"name": "CYP1A2", "id": "1544"} |
| Phytochemical | CID_5280343 | Quercetin | {"name": "Quercetin", "id": "CID_5280343"} | INTERACTS_WITH | - | Target | 1545 | CYP1B1 | {"name": "CYP1B1", "id": "1545"} |
| Phytochemical | CID_5280343 | Quercetin | {"name": "Quercetin", "id": "CID_5280343"} | INTERACTS_WITH | - | Target | 155 | ADRB3 | {"name": "ADRB3", "id": "155"} |
| Phytochemical | CID_5280343 | Quercetin | {"name": "Quercetin", "id": "CID_5280343"} | INTERACTS_WITH | - | Target | 1557 | CYP2C19 | {"name": "CYP2C19", "id": "1557"} |
| Phytochemical | CID_5280343 | Quercetin | {"name": "Quercetin", "id": "CID_5280343"} | INTERACTS_WITH | - | Target | 1558 | CYP2C8 | {"name": "CYP2C8", "id": "1558"} |
| Phytochemical | CID_5280343 | Quercetin | {"name": "Quercetin", "id": "CID_5280343"} | INTERACTS_WITH | - | Target | 1559 | CYP2C9 | {"name": "CYP2C9", "id": "1559"} |
| Phytochemical | CID_5280343 | Quercetin | {"name": "Quercetin", "id": "CID_5280343"} | INTERACTS_WITH | - | Target | 1565 | CYP2D6 | {"name": "CYP2D6", "id": "1565"} |
| Phytochemical | CID_5280343 | Quercetin | {"name": "Quercetin", "id": "CID_5280343"} | INTERACTS_WITH | - | Target | 1573 | CYP2J2 | {"name": "CYP2J2", "id": "1573"} |
| Phytochemical | CID_5280343 | Quercetin | {"name": "Quercetin", "id": "CID_5280343"} | INTERACTS_WITH | - | Target | 1576 | CYP3A4 | {"name": "CYP3A4", "id": "1576"} |
| Phytochemical | CID_5280343 | Quercetin | {"name": "Quercetin", "id": "CID_5280343"} | INTERACTS_WITH | - | Target | 1577 | CYP3A5 | {"name": "CYP3A5", "id": "1577"} |
| Phytochemical | CID_5280343 | Quercetin | {"name": "Quercetin", "id": "CID_5280343"} | INTERACTS_WITH | - | Target | 1588 | CYP19A1 | {"name": "CYP19A1", "id": "1588"} |
| Phytochemical | CID_5280343 | Quercetin | {"name": "Quercetin", "id": "CID_5280343"} | INTERACTS_WITH | - | Target | 1612 | DAPK1 | {"name": "DAPK1", "id": "1612"} |
| Phytochemical | CID_5280343 | Quercetin | {"name": "Quercetin", "id": "CID_5280343"} | INTERACTS_WITH | - | Target | 1613 | DAPK3 | {"name": "DAPK3", "id": "1613"} |

**Table S9: Complete node and relationship data used to construct the knowledge graph. This table summarizes interactions among FDA-approved drugs, formulation, herbs, diseases (type 2 diabetes, fatty liver, obesity), phytochemicals, and their target genes/proteins. Each entry includes source and target node details, edge type, and node annotations (Entrez Gene ID and gene symbol, where applicable).**

| Source Node Label | Source Node ID | Source Node Name | Source Node Properties | Edge Type | Edge Properties | Target Node Label | Target Node ID | Target Node Name | Target Node Properties |
|---|---|---|---|---|---|---|---|---|---|
| Phytochemical | CID_5280343 | Quercetin | {"name": "Quercetin", "id": "CID_5280343"} | INTERACTS_WITH | - | Target | 1760 | DMPK | {"name": "DMPK", "id": "1760"} |
| Phytochemical | CID_5280343 | Quercetin | {"name": "Quercetin", "id": "CID_5280343"} | INTERACTS_WITH | - | Target | 1803 | DPP4 | {"name": "DPP4", "id": "1803"} |
| Phytochemical | CID_5280343 | Quercetin | {"name": "Quercetin", "id": "CID_5280343"} | INTERACTS_WITH | - | Target | 1812 | DRD1 | {"name": "DRD1", "id": "1812"} |
| Phytochemical | CID_5280343 | Quercetin | {"name": "Quercetin", "id": "CID_5280343"} | INTERACTS_WITH | - | Target | 1813 | DRD2 | {"name": "DRD2", "id": "1813"} |
| Phytochemical | CID_5280343 | Quercetin | {"name": "Quercetin", "id": "CID_5280343"} | INTERACTS_WITH | - | Target | 1814 | DRD3 | {"name": "DRD3", "id": "1814"} |
| Phytochemical | CID_5280343 | Quercetin | {"name": "Quercetin", "id": "CID_5280343"} | INTERACTS_WITH | - | Target | 1815 | DRD4 | {"name": "DRD4", "id": "1815"} |
| Phytochemical | CID_5280343 | Quercetin | {"name": "Quercetin", "id": "CID_5280343"} | INTERACTS_WITH | - | Target | 185 | AGTR1 | {"name": "AGTR1", "id": "185"} |
| Phytochemical | CID_5280343 | Quercetin | {"name": "Quercetin", "id": "CID_5280343"} | INTERACTS_WITH | - | Target | 1909 | EDNRA | {"name": "EDNRA", "id": "1909"} |
| Phytochemical | CID_5280343 | Quercetin | {"name": "Quercetin", "id": "CID_5280343"} | INTERACTS_WITH | - | Target | 1910 | EDNRB | {"name": "EDNRB", "id": "1910"} |
| Phytochemical | CID_5280343 | Quercetin | {"name": "Quercetin", "id": "CID_5280343"} | INTERACTS_WITH | - | Target | 1956 | EGFR | {"name": "EGFR", "id": "1956"} |
| Phytochemical | CID_5280343 | Quercetin | {"name": "Quercetin", "id": "CID_5280343"} | INTERACTS_WITH | - | Target | 1991 | ELANE | {"name": "ELANE", "id": "1991"} |
| Phytochemical | CID_5280343 | Quercetin | {"name": "Quercetin", "id": "CID_5280343"} | INTERACTS_WITH | - | Target | 1994 | ELAVL1 | {"name": "ELAVL1", "id": "1994"} |
| Phytochemical | CID_5280343 | Quercetin | {"name": "Quercetin", "id": "CID_5280343"} | INTERACTS_WITH | - | Target | 1995 | ELAVL3 | {"name": "ELAVL3", "id": "1995"} |
| Phytochemical | CID_5280343 | Quercetin | {"name": "Quercetin", "id": "CID_5280343"} | INTERACTS_WITH | - | Target | 2030 | SLC29A1 | {"name": "SLC29A1", "id": "2030"} |
| Phytochemical | CID_5280343 | Quercetin | {"name": "Quercetin", "id": "CID_5280343"} | INTERACTS_WITH | - | Target | 2053 | EPHX2 | {"name": "EPHX2", "id": "2053"} |
| Phytochemical | CID_5280343 | Quercetin | {"name": "Quercetin", "id": "CID_5280343"} | INTERACTS_WITH | - | Target | 207 | AKT1 | {"name": "AKT1", "id": "207"} |
| Phytochemical | CID_5280343 | Quercetin | {"name": "Quercetin", "id": "CID_5280343"} | INTERACTS_WITH | - | Target | 2099 | ESR1 | {"name": "ESR1", "id": "2099"} |
| Phytochemical | CID_5280343 | Quercetin | {"name": "Quercetin", "id": "CID_5280343"} | INTERACTS_WITH | - | Target | 2100 | ESR2 | {"name": "ESR2", "id": "2100"} |
| Phytochemical | CID_5280343 | Quercetin | {"name": "Quercetin", "id": "CID_5280343"} | INTERACTS_WITH | - | Target | 2147 | F2 | {"name": "F2", "id": "2147"} |
| Phytochemical | CID_5280343 | Quercetin | {"name": "Quercetin", "id": "CID_5280343"} | INTERACTS_WITH | - | Target | 216 | ALDH1A1 | {"name": "ALDH1A1", "id": "216"} |
| Phytochemical | CID_5280343 | Quercetin | {"name": "Quercetin", "id": "CID_5280343"} | INTERACTS_WITH | - | Target | 2167 | FABP4 | {"name": "FABP4", "id": "2167"} |
| Phytochemical | CID_5280343 | Quercetin | {"name": "Quercetin", "id": "CID_5280343"} | INTERACTS_WITH | - | Target | 2203 | FBP1 | {"name": "FBP1", "id": "2203"} |
| Phytochemical | CID_5280343 | Quercetin | {"name": "Quercetin", "id": "CID_5280343"} | INTERACTS_WITH | - | Target | 2237 | FEN1 | {"name": "FEN1", "id": "2237"} |
| Phytochemical | CID_5280343 | Quercetin | {"name": "Quercetin", "id": "CID_5280343"} | INTERACTS_WITH | - | Target | 22933 | SIRT2 | {"name": "SIRT2", "id": "22933"} |
| Phytochemical | CID_5280343 | Quercetin | {"name": "Quercetin", "id": "CID_5280343"} | INTERACTS_WITH | - | Target | 23028 | KDM1A | {"name": "KDM1A", "id": "23028"} |
| Phytochemical | CID_5280343 | Quercetin | {"name": "Quercetin", "id": "CID_5280343"} | INTERACTS_WITH | - | Target | 23043 | TNIK | {"name": "TNIK", "id": "23043"} |
| Phytochemical | CID_5280343 | Quercetin | {"name": "Quercetin", "id": "CID_5280343"} | INTERACTS_WITH | - | Target | 231 | AKR1B1 | {"name": "AKR1B1", "id": "231"} |
| Phytochemical | CID_5280343 | Quercetin | {"name": "Quercetin", "id": "CID_5280343"} | INTERACTS_WITH | - | Target | 2322 | FLT3 | {"name": "FLT3", "id": "2322"} |
| Phytochemical | CID_5280343 | Quercetin | {"name": "Quercetin", "id": "CID_5280343"} | INTERACTS_WITH | - | Target | 23408 | SIRT5 | {"name": "SIRT5", "id": "23408"} |
| Phytochemical | CID_5280343 | Quercetin | {"name": "Quercetin", "id": "CID_5280343"} | INTERACTS_WITH | - | Target | 23410 | SIRT3 | {"name": "SIRT3", "id": "23410"} |
| Phytochemical | CID_5280343 | Quercetin | {"name": "Quercetin", "id": "CID_5280343"} | INTERACTS_WITH | - | Target | 23411 | SIRT1 | {"name": "SIRT1", "id": "23411"} |
| Phytochemical | CID_5280343 | Quercetin | {"name": "Quercetin", "id": "CID_5280343"} | INTERACTS_WITH | - | Target | 23621 | BACE1 | {"name": "BACE1", "id": "23621"} |
| Phytochemical | CID_5280343 | Quercetin | {"name": "Quercetin", "id": "CID_5280343"} | INTERACTS_WITH | - | Target | 23632 | CA14 | {"name": "CA14", "id": "23632"} |
| Phytochemical | CID_5280343 | Quercetin | {"name": "Quercetin", "id": "CID_5280343"} | INTERACTS_WITH | - | Target | 238 | ALK | {"name": "ALK", "id": "238"} |
| Phytochemical | CID_5280343 | Quercetin | {"name": "Quercetin", "id": "CID_5280343"} | INTERACTS_WITH | - | Target | 239 | ALOX12 | {"name": "ALOX12", "id": "239"} |
| Phytochemical | CID_5280343 | Quercetin | {"name": "Quercetin", "id": "CID_5280343"} | INTERACTS_WITH | - | Target | 240 | ALOX5 | {"name": "ALOX5", "id": "240"} |
| Phytochemical | CID_5280343 | Quercetin | {"name": "Quercetin", "id": "CID_5280343"} | INTERACTS_WITH | - | Target | 246 | ALOX15 | {"name": "ALOX15", "id": "246"} |
| Phytochemical | CID_5280343 | Quercetin | {"name": "Quercetin", "id": "CID_5280343"} | INTERACTS_WITH | - | Target | 247 | ALOX15B | {"name": "ALOX15B", "id": "247"} |
| Phytochemical | CID_5280343 | Quercetin | {"name": "Quercetin", "id": "CID_5280343"} | INTERACTS_WITH | - | Target | 253430 | IPMK | {"name": "IPMK", "id": "253430"} |
| Phytochemical | CID_5280343 | Quercetin | {"name": "Quercetin", "id": "CID_5280343"} | INTERACTS_WITH | - | Target | 2548 | GAA | {"name": "GAA", "id": "2548"} |
| Phytochemical | CID_5280343 | Quercetin | {"name": "Quercetin", "id": "CID_5280343"} | INTERACTS_WITH | - | Target | 2584 | GALK1 | {"name": "GALK1", "id": "2584"} |
| Phytochemical | CID_5280343 | Quercetin | {"name": "Quercetin", "id": "CID_5280343"} | INTERACTS_WITH | - | Target | 2629 | GBA1 | {"name": "GBA1", "id": "2629"} |
| Phytochemical | CID_5280343 | Quercetin | {"name": "Quercetin", "id": "CID_5280343"} | INTERACTS_WITH | - | Target | 2642 | GCGR | {"name": "GCGR", "id": "2642"} |
| Phytochemical | CID_5280343 | Quercetin | {"name": "Quercetin", "id": "CID_5280343"} | INTERACTS_WITH | - | Target | 2693 | GHSR | {"name": "GHSR", "id": "2693"} |
| Phytochemical | CID_5280343 | Quercetin | {"name": "Quercetin", "id": "CID_5280343"} | INTERACTS_WITH | - | Target | 2696 | GIPR | {"name": "GIPR", "id": "2696"} |
| Phytochemical | CID_5280343 | Quercetin | {"name": "Quercetin", "id": "CID_5280343"} | INTERACTS_WITH | - | Target | 2739 | GLO1 | {"name": "GLO1", "id": "2739"} |
| Phytochemical | CID_5280343 | Quercetin | {"name": "Quercetin", "id": "CID_5280343"} | INTERACTS_WITH | - | Target | 276 | AMY1A | {"name": "AMY1A", "id": "276"} |
| Phytochemical | CID_5280343 | Quercetin | {"name": "Quercetin", "id": "CID_5280343"} | INTERACTS_WITH | - | Target | 277 | AMY1B | {"name": "AMY1B", "id": "277"} |
| Phytochemical | CID_5280343 | Quercetin | {"name": "Quercetin", "id": "CID_5280343"} | INTERACTS_WITH | - | Target | 28234 | SLCO1B3 | {"name": "SLCO1B3", "id": "28234"} |

**Table S9: Complete node and relationship data used to construct the knowledge graph. This table summarizes interactions among FDA-approved drugs, formulation, herbs, diseases (type 2 diabetes, fatty liver, obesity), phytochemicals, and their target genes/proteins. Each entry includes source and target node details, edge type, and node annotations (Entrez Gene ID and gene symbol, where applicable).**

| Source Node Label | Source Node ID | Source Node Name | Source Node Properties | Edge Type | Edge Properties | Target Node Label | Target Node ID | Target Node Name | Target Node Properties |
|---|---|---|---|---|---|---|---|---|---|
| Phytochemical | CID_5280343 | Quercetin | {"name": "Quercetin", "id": "CID_5280343"} | INTERACTS_WITH | - | Target | 283106 | CSNK2A3 | {"name": "CSNK2A3", "id": "283106"} |
| Phytochemical | CID_5280343 | Quercetin | {"name": "Quercetin", "id": "CID_5280343"} | INTERACTS_WITH | - | Target | 2859 | GPR35 | {"name": "GPR35", "id": "2859"} |
| Phytochemical | CID_5280343 | Quercetin | {"name": "Quercetin", "id": "CID_5280343"} | INTERACTS_WITH | - | Target | 2862 | MLNR | {"name": "MLNR", "id": "2862"} |
| Phytochemical | CID_5280343 | Quercetin | {"name": "Quercetin", "id": "CID_5280343"} | INTERACTS_WITH | - | Target | 2908 | NR3C1 | {"name": "NR3C1", "id": "2908"} |
| Phytochemical | CID_5280343 | Quercetin | {"name": "Quercetin", "id": "CID_5280343"} | INTERACTS_WITH | - | Target | 2931 | GSK3A | {"name": "GSK3A", "id": "2931"} |
| Phytochemical | CID_5280343 | Quercetin | {"name": "Quercetin", "id": "CID_5280343"} | INTERACTS_WITH | - | Target | 2932 | GSK3B | {"name": "GSK3B", "id": "2932"} |
| Phytochemical | CID_5280343 | Quercetin | {"name": "Quercetin", "id": "CID_5280343"} | INTERACTS_WITH | - | Target | 2936 | GSR | {"name": "GSR", "id": "2936"} |
| Phytochemical | CID_5280343 | Quercetin | {"name": "Quercetin", "id": "CID_5280343"} | INTERACTS_WITH | - | Target | 3028 | HSD17B10 | {"name": "HSD17B10", "id": "3028"} |
| Phytochemical | CID_5280343 | Quercetin | {"name": "Quercetin", "id": "CID_5280343"} | INTERACTS_WITH | - | Target | 3091 | HIF1A | {"name": "HIF1A", "id": "3091"} |
| Phytochemical | CID_5280343 | Quercetin | {"name": "Quercetin", "id": "CID_5280343"} | INTERACTS_WITH | - | Target | 3248 | HPGD | {"name": "HPGD", "id": "3248"} |
| Phytochemical | CID_5280343 | Quercetin | {"name": "Quercetin", "id": "CID_5280343"} | INTERACTS_WITH | - | Target | 3269 | HRH1 | {"name": "HRH1", "id": "3269"} |
| Phytochemical | CID_5280343 | Quercetin | {"name": "Quercetin", "id": "CID_5280343"} | INTERACTS_WITH | - | Target | 3274 | HRH2 | {"name": "HRH2", "id": "3274"} |
| Phytochemical | CID_5280343 | Quercetin | {"name": "Quercetin", "id": "CID_5280343"} | INTERACTS_WITH | - | Target | 328 | APEX1 | {"name": "APEX1", "id": "328"} |
| Phytochemical | CID_5280343 | Quercetin | {"name": "Quercetin", "id": "CID_5280343"} | INTERACTS_WITH | - | Target | 3290 | HSD11B1 | {"name": "HSD11B1", "id": "3290"} |
| Phytochemical | CID_5280343 | Quercetin | {"name": "Quercetin", "id": "CID_5280343"} | INTERACTS_WITH | - | Target | 3292 | HSD17B1 | {"name": "HSD17B1", "id": "3292"} |
| Phytochemical | CID_5280343 | Quercetin | {"name": "Quercetin", "id": "CID_5280343"} | INTERACTS_WITH | - | Target | 3294 | HSD17B2 | {"name": "HSD17B2", "id": "3294"} |
| Phytochemical | CID_5280343 | Quercetin | {"name": "Quercetin", "id": "CID_5280343"} | INTERACTS_WITH | - | Target | 3297 | HSF1 | {"name": "HSF1", "id": "3297"} |
| Phytochemical | CID_5280343 | Quercetin | {"name": "Quercetin", "id": "CID_5280343"} | INTERACTS_WITH | - | Target | 3309 | HSPA5 | {"name": "HSPA5", "id": "3309"} |
| Phytochemical | CID_5280343 | Quercetin | {"name": "Quercetin", "id": "CID_5280343"} | INTERACTS_WITH | - | Target | 3315 | HSPB1 | {"name": "HSPB1", "id": "3315"} |
| Phytochemical | CID_5280343 | Quercetin | {"name": "Quercetin", "id": "CID_5280343"} | INTERACTS_WITH | - | Target | 3350 | HTR1A | {"name": "HTR1A", "id": "3350"} |
| Phytochemical | CID_5280343 | Quercetin | {"name": "Quercetin", "id": "CID_5280343"} | INTERACTS_WITH | - | Target | 3352 | HTR1D | {"name": "HTR1D", "id": "3352"} |
| Phytochemical | CID_5280343 | Quercetin | {"name": "Quercetin", "id": "CID_5280343"} | INTERACTS_WITH | - | Target | 3356 | HTR2A | {"name": "HTR2A", "id": "3356"} |
| Phytochemical | CID_5280343 | Quercetin | {"name": "Quercetin", "id": "CID_5280343"} | INTERACTS_WITH | - | Target | 3357 | HTR2B | {"name": "HTR2B", "id": "3357"} |
| Phytochemical | CID_5280343 | Quercetin | {"name": "Quercetin", "id": "CID_5280343"} | INTERACTS_WITH | - | Target | 3358 | HTR2C | {"name": "HTR2C", "id": "3358"} |
| Phytochemical | CID_5280343 | Quercetin | {"name": "Quercetin", "id": "CID_5280343"} | INTERACTS_WITH | - | Target | 3359 | HTR3A | {"name": "HTR3A", "id": "3359"} |
| Phytochemical | CID_5280343 | Quercetin | {"name": "Quercetin", "id": "CID_5280343"} | INTERACTS_WITH | - | Target | 3362 | HTR6 | {"name": "HTR6", "id": "3362"} |
| Phytochemical | CID_5280343 | Quercetin | {"name": "Quercetin", "id": "CID_5280343"} | INTERACTS_WITH | - | Target | 3363 | HTR7 | {"name": "HTR7", "id": "3363"} |
| Phytochemical | CID_5280343 | Quercetin | {"name": "Quercetin", "id": "CID_5280343"} | INTERACTS_WITH | - | Target | 3480 | IGF1R | {"name": "IGF1R", "id": "3480"} |
| Phytochemical | CID_5280343 | Quercetin | {"name": "Quercetin", "id": "CID_5280343"} | INTERACTS_WITH | - | Target | 351 | APP | {"name": "APP", "id": "351"} |
| Phytochemical | CID_5280343 | Quercetin | {"name": "Quercetin", "id": "CID_5280343"} | INTERACTS_WITH | - | Target | 3577 | CXCR1 | {"name": "CXCR1", "id": "3577"} |
| Phytochemical | CID_5280343 | Quercetin | {"name": "Quercetin", "id": "CID_5280343"} | INTERACTS_WITH | - | Target | 367 | AR | {"name": "AR", "id": "367"} |
| Phytochemical | CID_5280343 | Quercetin | {"name": "Quercetin", "id": "CID_5280343"} | INTERACTS_WITH | - | Target | 3716 | JAK1 | {"name": "JAK1", "id": "3716"} |
| Phytochemical | CID_5280343 | Quercetin | {"name": "Quercetin", "id": "CID_5280343"} | INTERACTS_WITH | - | Target | 3725 | JUN | {"name": "JUN", "id": "3725"} |
| Phytochemical | CID_5280343 | Quercetin | {"name": "Quercetin", "id": "CID_5280343"} | INTERACTS_WITH | - | Target | 3757 | KCNH2 | {"name": "KCNH2", "id": "3757"} |
| Phytochemical | CID_5280343 | Quercetin | {"name": "Quercetin", "id": "CID_5280343"} | INTERACTS_WITH | - | Target | 3791 | KDR | {"name": "KDR", "id": "3791"} |
| Phytochemical | CID_5280343 | Quercetin | {"name": "Quercetin", "id": "CID_5280343"} | INTERACTS_WITH | - | Target | 387129 | NPSR1 | {"name": "NPSR1", "id": "387129"} |
| Phytochemical | CID_5280343 | Quercetin | {"name": "Quercetin", "id": "CID_5280343"} | INTERACTS_WITH | - | Target | 390245 | KDM4E | {"name": "KDM4E", "id": "390245"} |
| Phytochemical | CID_5280343 | Quercetin | {"name": "Quercetin", "id": "CID_5280343"} | INTERACTS_WITH | - | Target | 3932 | LCK | {"name": "LCK", "id": "3932"} |
| Phytochemical | CID_5280343 | Quercetin | {"name": "Quercetin", "id": "CID_5280343"} | INTERACTS_WITH | - | Target | 4000 | LMNA | {"name": "LMNA", "id": "4000"} |
| Phytochemical | CID_5280343 | Quercetin | {"name": "Quercetin", "id": "CID_5280343"} | INTERACTS_WITH | - | Target | 4088 | SMAD3 | {"name": "SMAD3", "id": "4088"} |
| Phytochemical | CID_5280343 | Quercetin | {"name": "Quercetin", "id": "CID_5280343"} | INTERACTS_WITH | - | Target | 410 | ARSA | {"name": "ARSA", "id": "410"} |
| Phytochemical | CID_5280343 | Quercetin | {"name": "Quercetin", "id": "CID_5280343"} | INTERACTS_WITH | - | Target | 4128 | MAOA | {"name": "MAOA", "id": "4128"} |
| Phytochemical | CID_5280343 | Quercetin | {"name": "Quercetin", "id": "CID_5280343"} | INTERACTS_WITH | - | Target | 4129 | MAOB | {"name": "MAOB", "id": "4129"} |
| Phytochemical | CID_5280343 | Quercetin | {"name": "Quercetin", "id": "CID_5280343"} | INTERACTS_WITH | - | Target | 4137 | MAPT | {"name": "MAPT", "id": "4137"} |
| Phytochemical | CID_5280343 | Quercetin | {"name": "Quercetin", "id": "CID_5280343"} | INTERACTS_WITH | - | Target | 415116 | PIM3 | {"name": "PIM3", "id": "415116"} |
| Phytochemical | CID_5280343 | Quercetin | {"name": "Quercetin", "id": "CID_5280343"} | INTERACTS_WITH | - | Target | 4154 | MBNL1 | {"name": "MBNL1", "id": "4154"} |
| Phytochemical | CID_5280343 | Quercetin | {"name": "Quercetin", "id": "CID_5280343"} | INTERACTS_WITH | - | Target | 4159 | MC3R | {"name": "MC3R", "id": "4159"} |
| Phytochemical | CID_5280343 | Quercetin | {"name": "Quercetin", "id": "CID_5280343"} | INTERACTS_WITH | - | Target | 4160 | MC4R | {"name": "MC4R", "id": "4160"} |
| Phytochemical | CID_5280343 | Quercetin | {"name": "Quercetin", "id": "CID_5280343"} | INTERACTS_WITH | - | Target | 4217 | MAP3K5 | {"name": "MAP3K5", "id": "4217"} |

| Table S9: Complete node and relationship data used to construct the knowledge graph. This table summarizes interactions among FDA-approved drugs, formulation, herbs, diseases (type 2 diabetes, fatty liver, obesity), phytochemicals, and their target genes/proteins. Each entry includes source and target node details, edge type, and node annotations (Entrez Gene ID and gene symbol, where applicable). | | | | | | | | | |
|---|---|---|---|---|---|---|---|---|---|
| Source Node Label | Source Node ID | Source Node Name | Source Node Properties | Edge Type | Edge Properties | Target Node Label | Target Node ID | Target Node Name | Target Node Properties |
| Phytochemical | CID_5280343 | Quercetin | {"name": "Quercetin", "id": " CID_5280343"} | INTERACTS_WITH | - | Target | 4233 | MET | {"name": "MET", "id": "4233"} |
| Phytochemical | CID_5280343 | Quercetin | {"name": "Quercetin", "id": " CID_5280343"} | INTERACTS_WITH | - | Target | 4297 | KMT2A | {"name": "KMT2A", "id": "4297"} |
| Phytochemical | CID_5280343 | Quercetin | {"name": "Quercetin", "id": " CID_5280343"} | INTERACTS_WITH | - | Target | 43 | ACHE | {"name": "ACHE", "id": "43"} |
| Phytochemical | CID_5280343 | Quercetin | {"name": "Quercetin", "id": " CID_5280343"} | INTERACTS_WITH | - | Target | 4312 | MMP1 | {"name": "MMP1", "id": "4312"} |
| Phytochemical | CID_5280343 | Quercetin | {"name": "Quercetin", "id": " CID_5280343"} | INTERACTS_WITH | - | Target | 4313 | MMP2 | {"name": "MMP2", "id": "4313"} |
| Phytochemical | CID_5280343 | Quercetin | {"name": "Quercetin", "id": " CID_5280343"} | INTERACTS_WITH | - | Target | 4314 | MMP3 | {"name": "MMP3", "id": "4314"} |
| Phytochemical | CID_5280343 | Quercetin | {"name": "Quercetin", "id": " CID_5280343"} | INTERACTS_WITH | - | Target | 4318 | MMP9 | {"name": "MMP9", "id": "4318"} |
| Phytochemical | CID_5280343 | Quercetin | {"name": "Quercetin", "id": " CID_5280343"} | INTERACTS_WITH | - | Target | 4321 | MMP12 | {"name": "MMP12", "id": "4321"} |
| Phytochemical | CID_5280343 | Quercetin | {"name": "Quercetin", "id": " CID_5280343"} | INTERACTS_WITH | - | Target | 4322 | MMP13 | {"name": "MMP13", "id": "4322"} |
| Phytochemical | CID_5280343 | Quercetin | {"name": "Quercetin", "id": " CID_5280343"} | INTERACTS_WITH | - | Target | 4350 | MPG | {"name": "MPG", "id": "4350"} |
| Phytochemical | CID_5280343 | Quercetin | {"name": "Quercetin", "id": " CID_5280343"} | INTERACTS_WITH | - | Target | 4353 | MPO | {"name": "MPO", "id": "4353"} |
| Phytochemical | CID_5280343 | Quercetin | {"name": "Quercetin", "id": " CID_5280343"} | INTERACTS_WITH | - | Target | 4363 | ABCC1 | {"name": "ABCC1", "id": "4363"} |
| Phytochemical | CID_5280343 | Quercetin | {"name": "Quercetin", "id": " CID_5280343"} | INTERACTS_WITH | - | Target | 4751 | NEK2 | {"name": "NEK2", "id": "4751"} |
| Phytochemical | CID_5280343 | Quercetin | {"name": "Quercetin", "id": " CID_5280343"} | INTERACTS_WITH | - | Target | 4759 | NEU2 | {"name": "NEU2", "id": "4759"} |
| Phytochemical | CID_5280343 | Quercetin | {"name": "Quercetin", "id": " CID_5280343"} | INTERACTS_WITH | - | Target | 4790 | NFKB1 | {"name": "NFKB1", "id": "4790"} |
| Phytochemical | CID_5280343 | Quercetin | {"name": "Quercetin", "id": " CID_5280343"} | INTERACTS_WITH | - | Target | 4791 | NFKB2 | {"name": "NFKB2", "id": "4791"} |
| Phytochemical | CID_5280343 | Quercetin | {"name": "Quercetin", "id": " CID_5280343"} | INTERACTS_WITH | - | Target | 4886 | NPY1R | {"name": "NPY1R", "id": "4886"} |
| Phytochemical | CID_5280343 | Quercetin | {"name": "Quercetin", "id": " CID_5280343"} | INTERACTS_WITH | - | Target | 4887 | NPY2R | {"name": "NPY2R", "id": "4887"} |
| Phytochemical | CID_5280343 | Quercetin | {"name": "Quercetin", "id": " CID_5280343"} | INTERACTS_WITH | - | Target | 4907 | NT5E | {"name": "NT5E", "id": "4907"} |
| Phytochemical | CID_5280343 | Quercetin | {"name": "Quercetin", "id": " CID_5280343"} | INTERACTS_WITH | - | Target | 4923 | NTSR1 | {"name": "NTSR1", "id": "4923"} |
| Phytochemical | CID_5280343 | Quercetin | {"name": "Quercetin", "id": " CID_5280343"} | INTERACTS_WITH | - | Target | 4985 | OPRD1 | {"name": "OPRD1", "id": "4985"} |
| Phytochemical | CID_5280343 | Quercetin | {"name": "Quercetin", "id": " CID_5280343"} | INTERACTS_WITH | - | Target | 4986 | OPRK1 | {"name": "OPRK1", "id": "4986"} |
| Phytochemical | CID_5280343 | Quercetin | {"name": "Quercetin", "id": " CID_5280343"} | INTERACTS_WITH | - | Target | 4988 | OPRM1 | {"name": "OPRM1", "id": "4988"} |
| Phytochemical | CID_5280343 | Quercetin | {"name": "Quercetin", "id": " CID_5280343"} | INTERACTS_WITH | - | Target | 50507 | NOX4 | {"name": "NOX4", "id": "50507"} |
| Phytochemical | CID_5280343 | Quercetin | {"name": "Quercetin", "id": " CID_5280343"} | INTERACTS_WITH | - | Target | 51231 | VRK3 | {"name": "VRK3", "id": "51231"} |
| Phytochemical | CID_5280343 | Quercetin | {"name": "Quercetin", "id": " CID_5280343"} | INTERACTS_WITH | - | Target | 5127 | CDK16 | {"name": "CDK16", "id": "5127"} |
| Phytochemical | CID_5280343 | Quercetin | {"name": "Quercetin", "id": " CID_5280343"} | INTERACTS_WITH | - | Target | 5139 | PDE3A | {"name": "PDE3A", "id": "5139"} |
| Phytochemical | CID_5280343 | Quercetin | {"name": "Quercetin", "id": " CID_5280343"} | INTERACTS_WITH | - | Target | 51426 | POLK | {"name": "POLK", "id": "51426"} |
| Phytochemical | CID_5280343 | Quercetin | {"name": "Quercetin", "id": " CID_5280343"} | INTERACTS_WITH | - | Target | 5144 | PDE4D | {"name": "PDE4D", "id": "5144"} |
| Phytochemical | CID_5280343 | Quercetin | {"name": "Quercetin", "id": " CID_5280343"} | INTERACTS_WITH | - | Target | 51447 | IP6K2 | {"name": "IP6K2", "id": "51447"} |
| Phytochemical | CID_5280343 | Quercetin | {"name": "Quercetin", "id": " CID_5280343"} | INTERACTS_WITH | - | Target | 51548 | SIRT6 | {"name": "SIRT6", "id": "51548"} |
| Phytochemical | CID_5280343 | Quercetin | {"name": "Quercetin", "id": " CID_5280343"} | INTERACTS_WITH | - | Target | 5163 | PDK1 | {"name": "PDK1", "id": "5163"} |
| Phytochemical | CID_5280343 | Quercetin | {"name": "Quercetin", "id": " CID_5280343"} | INTERACTS_WITH | - | Target | 5170 | PDPK1 | {"name": "PDPK1", "id": "5170"} |
| Phytochemical | CID_5280343 | Quercetin | {"name": "Quercetin", "id": " CID_5280343"} | INTERACTS_WITH | - | Target | 51765 | STK26 | {"name": "STK26", "id": "51765"} |
| Phytochemical | CID_5280343 | Quercetin | {"name": "Quercetin", "id": " CID_5280343"} | INTERACTS_WITH | - | Target | 5241 | PGR | {"name": "PGR", "id": "5241"} |
| Phytochemical | CID_5280343 | Quercetin | {"name": "Quercetin", "id": " CID_5280343"} | INTERACTS_WITH | - | Target | 5243 | ABCB1 | {"name": "ABCB1", "id": "5243"} |
| Phytochemical | CID_5280343 | Quercetin | {"name": "Quercetin", "id": " CID_5280343"} | INTERACTS_WITH | - | Target | 5290 | PIK3CA | {"name": "PIK3CA", "id": "5290"} |
| Phytochemical | CID_5280343 | Quercetin | {"name": "Quercetin", "id": " CID_5280343"} | INTERACTS_WITH | - | Target | 5292 | PIM1 | {"name": "PIM1", "id": "5292"} |
| Phytochemical | CID_5280343 | Quercetin | {"name": "Quercetin", "id": " CID_5280343"} | INTERACTS_WITH | - | Target | 5294 | PIK3CG | {"name": "PIK3CG", "id": "5294"} |
| Phytochemical | CID_5280343 | Quercetin | {"name": "Quercetin", "id": " CID_5280343"} | INTERACTS_WITH | - | Target | 5295 | PIK3R1 | {"name": "PIK3R1", "id": "5295"} |
| Phytochemical | CID_5280343 | Quercetin | {"name": "Quercetin", "id": " CID_5280343"} | INTERACTS_WITH | - | Target | 5300 | PIN1 | {"name": "PIN1", "id": "5300"} |
| Phytochemical | CID_5280343 | Quercetin | {"name": "Quercetin", "id": " CID_5280343"} | INTERACTS_WITH | - | Target | 5319 | PLA2G1B | {"name": "PLA2G1B", "id": "5319"} |
| Phytochemical | CID_5280343 | Quercetin | {"name": "Quercetin", "id": " CID_5280343"} | INTERACTS_WITH | - | Target | 5328 | PLAU | {"name": "PLAU", "id": "5328"} |
| Phytochemical | CID_5280343 | Quercetin | {"name": "Quercetin", "id": " CID_5280343"} | INTERACTS_WITH | - | Target | 5340 | PLG | {"name": "PLG", "id": "5340"} |
| Phytochemical | CID_5280343 | Quercetin | {"name": "Quercetin", "id": " CID_5280343"} | INTERACTS_WITH | - | Target | 5347 | PLK1 | {"name": "PLK1", "id": "5347"} |
| Phytochemical | CID_5280343 | Quercetin | {"name": "Quercetin", "id": " CID_5280343"} | INTERACTS_WITH | - | Target | 5376 | PMP22 | {"name": "PMP22", "id": "5376"} |
| Phytochemical | CID_5280343 | Quercetin | {"name": "Quercetin", "id": " CID_5280343"} | INTERACTS_WITH | - | Target | 53944 | CSNK1G1 | {"name": "CSNK1G1", "id": "53944"} |
| Phytochemical | CID_5280343 | Quercetin | {"name": "Quercetin", "id": " CID_5280343"} | INTERACTS_WITH | - | Target | 5406 | PNLIP | {"name": "PNLIP", "id": "5406"} |
| Phytochemical | CID_5280343 | Quercetin | {"name": "Quercetin", "id": " CID_5280343"} | INTERACTS_WITH | - | Target | 5423 | POLB | {"name": "POLB", "id": "5423"} |

**Table S9: Complete node and relationship data used to construct the knowledge graph. This table summarizes interactions among FDA-approved drugs, formulation, herbs, diseases (type 2 diabetes, fatty liver, obesity), phytochemicals, and their target genes/proteins. Each entry includes source and target node details, edge type, and node annotations (Entrez Gene ID and gene symbol, where applicable).**

| Source Node Label | Source Node ID | Source Node Name | Source Node Properties | Edge Type | Edge Properties | Target Node Label | Target Node ID | Target Node Name | Target Node Properties |
|---|---|---|---|---|---|---|---|---|---|
| Phytochemical | CID_5280343 | Quercetin | {"name": "Quercetin", "id": " CID_5280343"} | INTERACTS_WITH | - | Target | 5429 | POLH | {"name": "POLH", "id": "5429"} |
| Phytochemical | CID_5280343 | Quercetin | {"name": "Quercetin", "id": " CID_5280343"} | INTERACTS_WITH | - | Target | 5444 | PON1 | {"name": "PON1", "id": "5444"} |
| Phytochemical | CID_5280343 | Quercetin | {"name": "Quercetin", "id": " CID_5280343"} | INTERACTS_WITH | - | Target | 54575 | UGT1A10 | {"name": "UGT1A10", "id": "54575"} |
| Phytochemical | CID_5280343 | Quercetin | {"name": "Quercetin", "id": " CID_5280343"} | INTERACTS_WITH | - | Target | 54576 | UGT1A8 | {"name": "UGT1A8", "id": "54576"} |
| Phytochemical | CID_5280343 | Quercetin | {"name": "Quercetin", "id": " CID_5280343"} | INTERACTS_WITH | - | Target | 54578 | UGT1A6 | {"name": "UGT1A6", "id": "54578"} |
| Phytochemical | CID_5280343 | Quercetin | {"name": "Quercetin", "id": " CID_5280343"} | INTERACTS_WITH | - | Target | 54600 | UGT1A9 | {"name": "UGT1A9", "id": "54600"} |
| Phytochemical | CID_5280343 | Quercetin | {"name": "Quercetin", "id": " CID_5280343"} | INTERACTS_WITH | - | Target | 54657 | UGT1A4 | {"name": "UGT1A4", "id": "54657"} |
| Phytochemical | CID_5280343 | Quercetin | {"name": "Quercetin", "id": " CID_5280343"} | INTERACTS_WITH | - | Target | 54658 | UGT1A1 | {"name": "UGT1A1", "id": "54658"} |
| Phytochemical | CID_5280343 | Quercetin | {"name": "Quercetin", "id": " CID_5280343"} | INTERACTS_WITH | - | Target | 54659 | UGT1A3 | {"name": "UGT1A3", "id": "54659"} |
| Phytochemical | CID_5280343 | Quercetin | {"name": "Quercetin", "id": " CID_5280343"} | INTERACTS_WITH | - | Target | 5467 | PPARD | {"name": "PPARD", "id": "5467"} |
| Phytochemical | CID_5280343 | Quercetin | {"name": "Quercetin", "id": " CID_5280343"} | INTERACTS_WITH | - | Target | 5468 | PPARG | {"name": "PPARG", "id": "5468"} |
| Phytochemical | CID_5280343 | Quercetin | {"name": "Quercetin", "id": " CID_5280343"} | INTERACTS_WITH | - | Target | 54737 | MPHOSPH8 | {"name": "MPHOSPH8", "id": "54737"} |
| Phytochemical | CID_5280343 | Quercetin | {"name": "Quercetin", "id": " CID_5280343"} | INTERACTS_WITH | - | Target | 552 | AVPR1A | {"name": "AVPR1A", "id": "552"} |
| Phytochemical | CID_5280343 | Quercetin | {"name": "Quercetin", "id": " CID_5280343"} | INTERACTS_WITH | - | Target | 554 | AVPR2 | {"name": "AVPR2", "id": "554"} |
| Phytochemical | CID_5280343 | Quercetin | {"name": "Quercetin", "id": " CID_5280343"} | INTERACTS_WITH | - | Target | 5550 | PREP | {"name": "PREP", "id": "5550"} |
| Phytochemical | CID_5280343 | Quercetin | {"name": "Quercetin", "id": " CID_5280343"} | INTERACTS_WITH | - | Target | 5563 | PRKAA2 | {"name": "PRKAA2", "id": "5563"} |
| Phytochemical | CID_5280343 | Quercetin | {"name": "Quercetin", "id": " CID_5280343"} | INTERACTS_WITH | - | Target | 5566 | PRKACA | {"name": "PRKACA", "id": "5566"} |
| Phytochemical | CID_5280343 | Quercetin | {"name": "Quercetin", "id": " CID_5280343"} | INTERACTS_WITH | - | Target | 55662 | HIF1AN | {"name": "HIF1AN", "id": "55662"} |
| Phytochemical | CID_5280343 | Quercetin | {"name": "Quercetin", "id": " CID_5280343"} | INTERACTS_WITH | - | Target | 55775 | TDP1 | {"name": "TDP1", "id": "55775"} |
| Phytochemical | CID_5280343 | Quercetin | {"name": "Quercetin", "id": " CID_5280343"} | INTERACTS_WITH | - | Target | 5578 | PRKCA | {"name": "PRKCA", "id": "5578"} |
| Phytochemical | CID_5280343 | Quercetin | {"name": "Quercetin", "id": " CID_5280343"} | INTERACTS_WITH | - | Target | 55781 | RIOK2 | {"name": "RIOK2", "id": "55781"} |
| Phytochemical | CID_5280343 | Quercetin | {"name": "Quercetin", "id": " CID_5280343"} | INTERACTS_WITH | - | Target | 558 | AXL | {"name": "AXL", "id": "558"} |
| Phytochemical | CID_5280343 | Quercetin | {"name": "Quercetin", "id": " CID_5280343"} | INTERACTS_WITH | - | Target | 5585 | PKN1 | {"name": "PKN1", "id": "5585"} |
| Phytochemical | CID_5280343 | Quercetin | {"name": "Quercetin", "id": " CID_5280343"} | INTERACTS_WITH | - | Target | 55872 | PBK | {"name": "PBK", "id": "55872"} |
| Phytochemical | CID_5280343 | Quercetin | {"name": "Quercetin", "id": " CID_5280343"} | INTERACTS_WITH | - | Target | 5594 | MAPK1 | {"name": "MAPK1", "id": "5594"} |
| Phytochemical | CID_5280343 | Quercetin | {"name": "Quercetin", "id": " CID_5280343"} | INTERACTS_WITH | - | Target | 5595 | MAPK3 | {"name": "MAPK3", "id": "5595"} |
| Phytochemical | CID_5280343 | Quercetin | {"name": "Quercetin", "id": " CID_5280343"} | INTERACTS_WITH | - | Target | 5597 | MAPK6 | {"name": "MAPK6", "id": "5597"} |
| Phytochemical | CID_5280343 | Quercetin | {"name": "Quercetin", "id": " CID_5280343"} | INTERACTS_WITH | - | Target | 5599 | MAPK8 | {"name": "MAPK8", "id": "5599"} |
| Phytochemical | CID_5280343 | Quercetin | {"name": "Quercetin", "id": " CID_5280343"} | INTERACTS_WITH | - | Target | 5600 | MAPK11 | {"name": "MAPK11", "id": "5600"} |
| Phytochemical | CID_5280343 | Quercetin | {"name": "Quercetin", "id": " CID_5280343"} | INTERACTS_WITH | - | Target | 5601 | MAPK9 | {"name": "MAPK9", "id": "5601"} |
| Phytochemical | CID_5280343 | Quercetin | {"name": "Quercetin", "id": " CID_5280343"} | INTERACTS_WITH | - | Target | 5602 | MAPK10 | {"name": "MAPK10", "id": "5602"} |
| Phytochemical | CID_5280343 | Quercetin | {"name": "Quercetin", "id": " CID_5280343"} | INTERACTS_WITH | - | Target | 5603 | MAPK13 | {"name": "MAPK13", "id": "5603"} |
| Phytochemical | CID_5280343 | Quercetin | {"name": "Quercetin", "id": " CID_5280343"} | INTERACTS_WITH | - | Target | 5604 | MAP2K1 | {"name": "MAP2K1", "id": "5604"} |
| Phytochemical | CID_5280343 | Quercetin | {"name": "Quercetin", "id": " CID_5280343"} | INTERACTS_WITH | - | Target | 5605 | MAP2K2 | {"name": "MAP2K2", "id": "5605"} |
| Phytochemical | CID_5280343 | Quercetin | {"name": "Quercetin", "id": " CID_5280343"} | INTERACTS_WITH | - | Target | 5608 | MAP2K6 | {"name": "MAP2K6", "id": "5608"} |
| Phytochemical | CID_5280343 | Quercetin | {"name": "Quercetin", "id": " CID_5280343"} | INTERACTS_WITH | - | Target | 56339 | METTL3 | {"name": "METTL3", "id": "56339"} |
| Phytochemical | CID_5280343 | Quercetin | {"name": "Quercetin", "id": " CID_5280343"} | INTERACTS_WITH | - | Target | 5644 | PRSS1 | {"name": "PRSS1", "id": "5644"} |
| Phytochemical | CID_5280343 | Quercetin | {"name": "Quercetin", "id": " CID_5280343"} | INTERACTS_WITH | - | Target | 5645 | PRSS2 | {"name": "PRSS2", "id": "5645"} |
| Phytochemical | CID_5280343 | Quercetin | {"name": "Quercetin", "id": " CID_5280343"} | INTERACTS_WITH | - | Target | 5646 | PRSS3 | {"name": "PRSS3", "id": "5646"} |
| Phytochemical | CID_5280343 | Quercetin | {"name": "Quercetin", "id": " CID_5280343"} | INTERACTS_WITH | - | Target | 56924 | PAK6 | {"name": "PAK6", "id": "56924"} |
| Phytochemical | CID_5280343 | Quercetin | {"name": "Quercetin", "id": " CID_5280343"} | INTERACTS_WITH | - | Target | 5693 | PSMB5 | {"name": "PSMB5", "id": "5693"} |
| Phytochemical | CID_5280343 | Quercetin | {"name": "Quercetin", "id": " CID_5280343"} | INTERACTS_WITH | - | Target | 57016 | AKR1B10 | {"name": "AKR1B10", "id": "57016"} |
| Phytochemical | CID_5280343 | Quercetin | {"name": "Quercetin", "id": " CID_5280343"} | INTERACTS_WITH | - | Target | 57118 | CAMK1D | {"name": "CAMK1D", "id": "57118"} |
| Phytochemical | CID_5280343 | Quercetin | {"name": "Quercetin", "id": " CID_5280343"} | INTERACTS_WITH | - | Target | 57144 | PAK5 | {"name": "PAK5", "id": "57144"} |
| Phytochemical | CID_5280343 | Quercetin | {"name": "Quercetin", "id": " CID_5280343"} | INTERACTS_WITH | - | Target | 57172 | CAMK1G | {"name": "CAMK1G", "id": "57172"} |
| Phytochemical | CID_5280343 | Quercetin | {"name": "Quercetin", "id": " CID_5280343"} | INTERACTS_WITH | - | Target | 5742 | PTGS1 | {"name": "PTGS1", "id": "5742"} |
| Phytochemical | CID_5280343 | Quercetin | {"name": "Quercetin", "id": " CID_5280343"} | INTERACTS_WITH | - | Target | 5743 | PTGS2 | {"name": "PTGS2", "id": "5743"} |
| Phytochemical | CID_5280343 | Quercetin | {"name": "Quercetin", "id": " CID_5280343"} | INTERACTS_WITH | - | Target | 5747 | PTK2 | {"name": "PTK2", "id": "5747"} |
| Phytochemical | CID_5280343 | Quercetin | {"name": "Quercetin", "id": " CID_5280343"} | INTERACTS_WITH | - | Target | 5770 | PTPN1 | {"name": "PTPN1", "id": "5770"} |

**Table S9: Complete node and relationship data used to construct the knowledge graph. This table summarizes interactions among FDA-approved drugs, formulation, herbs, diseases (type 2 diabetes, fatty liver, obesity), phytochemicals, and their target genes/proteins. Each entry includes source and target node details, edge type, and node annotations (Entrez Gene ID and gene symbol, where applicable).**

| Source Node Label | Source Node ID | Source Node Name | Source Node Properties | Edge Type | Edge Properties | Target Node Label | Target Node ID | Target Node Name | Target Node Properties |
|---|---|---|---|---|---|---|---|---|---|
| Phytochemical | CID_5280343 | Quercetin | {"name": "Quercetin", "id": "CID_5280343"} | INTERACTS_WITH | - | Target | 57721 | METTL14 | {"name": "METTL14", "id": "57721"} |
| Phytochemical | CID_5280343 | Quercetin | {"name": "Quercetin", "id": "CID_5280343"} | INTERACTS_WITH | - | Target | 5836 | PYGL | {"name": "PYGL", "id": "5836"} |
| Phytochemical | CID_5280343 | Quercetin | {"name": "Quercetin", "id": "CID_5280343"} | INTERACTS_WITH | - | Target | 590 | BCHE | {"name": "BCHE", "id": "590"} |
| Phytochemical | CID_5280343 | Quercetin | {"name": "Quercetin", "id": "CID_5280343"} | INTERACTS_WITH | - | Target | 595 | CCND1 | {"name": "CCND1", "id": "595"} |
| Phytochemical | CID_5280343 | Quercetin | {"name": "Quercetin", "id": "CID_5280343"} | INTERACTS_WITH | - | Target | 5965 | RECQL | {"name": "RECQL", "id": "5965"} |
| Phytochemical | CID_5280343 | Quercetin | {"name": "Quercetin", "id": "CID_5280343"} | INTERACTS_WITH | - | Target | 5970 | RELA | {"name": "RELA", "id": "5970"} |
| Phytochemical | CID_5280343 | Quercetin | {"name": "Quercetin", "id": "CID_5280343"} | INTERACTS_WITH | - | Target | 5979 | RET | {"name": "RET", "id": "5979"} |
| Phytochemical | CID_5280343 | Quercetin | {"name": "Quercetin", "id": "CID_5280343"} | INTERACTS_WITH | - | Target | 5999 | RGS4 | {"name": "RGS4", "id": "5999"} |
| Phytochemical | CID_5280343 | Quercetin | {"name": "Quercetin", "id": "CID_5280343"} | INTERACTS_WITH | - | Target | 607 | BCL9 | {"name": "BCL9", "id": "607"} |
| Phytochemical | CID_5280343 | Quercetin | {"name": "Quercetin", "id": "CID_5280343"} | INTERACTS_WITH | - | Target | 6197 | RPS6KA3 | {"name": "RPS6KA3", "id": "6197"} |
| Phytochemical | CID_5280343 | Quercetin | {"name": "Quercetin", "id": "CID_5280343"} | INTERACTS_WITH | - | Target | 6198 | RPS6KB1 | {"name": "RPS6KB1", "id": "6198"} |
| Phytochemical | CID_5280343 | Quercetin | {"name": "Quercetin", "id": "CID_5280343"} | INTERACTS_WITH | - | Target | 623 | BDKRB1 | {"name": "BDKRB1", "id": "623"} |
| Phytochemical | CID_5280343 | Quercetin | {"name": "Quercetin", "id": "CID_5280343"} | INTERACTS_WITH | - | Target | 624 | BDKRB2 | {"name": "BDKRB2", "id": "624"} |
| Phytochemical | CID_5280343 | Quercetin | {"name": "Quercetin", "id": "CID_5280343"} | INTERACTS_WITH | - | Target | 6300 | MAPK12 | {"name": "MAPK12", "id": "6300"} |
| Phytochemical | CID_5280343 | Quercetin | {"name": "Quercetin", "id": "CID_5280343"} | INTERACTS_WITH | - | Target | 641 | BLM | {"name": "BLM", "id": "641"} |
| Phytochemical | CID_5280343 | Quercetin | {"name": "Quercetin", "id": "CID_5280343"} | INTERACTS_WITH | - | Target | 6446 | SGK1 | {"name": "SGK1", "id": "6446"} |
| Phytochemical | CID_5280343 | Quercetin | {"name": "Quercetin", "id": "CID_5280343"} | INTERACTS_WITH | - | Target | 6476 | SI | {"name": "SI", "id": "6476"} |
| Phytochemical | CID_5280343 | Quercetin | {"name": "Quercetin", "id": "CID_5280343"} | INTERACTS_WITH | - | Target | 6514 | SLC2A2 | {"name": "SLC2A2", "id": "6514"} |
| Phytochemical | CID_5280343 | Quercetin | {"name": "Quercetin", "id": "CID_5280343"} | INTERACTS_WITH | - | Target | 6530 | SLC6A2 | {"name": "SLC6A2", "id": "6530"} |
| Phytochemical | CID_5280343 | Quercetin | {"name": "Quercetin", "id": "CID_5280343"} | INTERACTS_WITH | - | Target | 6531 | SLC6A3 | {"name": "SLC6A3", "id": "6531"} |
| Phytochemical | CID_5280343 | Quercetin | {"name": "Quercetin", "id": "CID_5280343"} | INTERACTS_WITH | - | Target | 6532 | SLC6A4 | {"name": "SLC6A4", "id": "6532"} |
| Phytochemical | CID_5280343 | Quercetin | {"name": "Quercetin", "id": "CID_5280343"} | INTERACTS_WITH | - | Target | 65975 | STK33 | {"name": "STK33", "id": "65975"} |
| Phytochemical | CID_5280343 | Quercetin | {"name": "Quercetin", "id": "CID_5280343"} | INTERACTS_WITH | - | Target | 6606 | SMN1 | {"name": "SMN1", "id": "6606"} |
| Phytochemical | CID_5280343 | Quercetin | {"name": "Quercetin", "id": "CID_5280343"} | INTERACTS_WITH | - | Target | 6622 | SNCA | {"name": "SNCA", "id": "6622"} |
| Phytochemical | CID_5280343 | Quercetin | {"name": "Quercetin", "id": "CID_5280343"} | INTERACTS_WITH | - | Target | 6652 | SORD | {"name": "SORD", "id": "6652"} |
| Phytochemical | CID_5280343 | Quercetin | {"name": "Quercetin", "id": "CID_5280343"} | INTERACTS_WITH | - | Target | 6714 | SRC | {"name": "SRC", "id": "6714"} |
| Phytochemical | CID_5280343 | Quercetin | {"name": "Quercetin", "id": "CID_5280343"} | INTERACTS_WITH | - | Target | 673 | BRAF | {"name": "BRAF", "id": "673"} |
| Phytochemical | CID_5280343 | Quercetin | {"name": "Quercetin", "id": "CID_5280343"} | INTERACTS_WITH | - | Target | 6778 | STAT6 | {"name": "STAT6", "id": "6778"} |
| Phytochemical | CID_5280343 | Quercetin | {"name": "Quercetin", "id": "CID_5280343"} | INTERACTS_WITH | - | Target | 6789 | STK4 | {"name": "STK4", "id": "6789"} |
| Phytochemical | CID_5280343 | Quercetin | {"name": "Quercetin", "id": "CID_5280343"} | INTERACTS_WITH | - | Target | 6793 | STK10 | {"name": "STK10", "id": "6793"} |
| Phytochemical | CID_5280343 | Quercetin | {"name": "Quercetin", "id": "CID_5280343"} | INTERACTS_WITH | - | Target | 6850 | SYK | {"name": "SYK", "id": "6850"} |
| Phytochemical | CID_5280343 | Quercetin | {"name": "Quercetin", "id": "CID_5280343"} | INTERACTS_WITH | - | Target | 6915 | TBXA2R | {"name": "TBXA2R", "id": "6915"} |
| Phytochemical | CID_5280343 | Quercetin | {"name": "Quercetin", "id": "CID_5280343"} | INTERACTS_WITH | - | Target | 7068 | THRB | {"name": "THRB", "id": "7068"} |
| Phytochemical | CID_5280343 | Quercetin | {"name": "Quercetin", "id": "CID_5280343"} | INTERACTS_WITH | - | Target | 7150 | TOP1 | {"name": "TOP1", "id": "7150"} |
| Phytochemical | CID_5280343 | Quercetin | {"name": "Quercetin", "id": "CID_5280343"} | INTERACTS_WITH | - | Target | 7153 | TOP2A | {"name": "TOP2A", "id": "7153"} |
| Phytochemical | CID_5280343 | Quercetin | {"name": "Quercetin", "id": "CID_5280343"} | INTERACTS_WITH | - | Target | 7155 | TOP2B | {"name": "TOP2B", "id": "7155"} |
| Phytochemical | CID_5280343 | Quercetin | {"name": "Quercetin", "id": "CID_5280343"} | INTERACTS_WITH | - | Target | 7157 | TP53 | {"name": "TP53", "id": "7157"} |
| Phytochemical | CID_5280343 | Quercetin | {"name": "Quercetin", "id": "CID_5280343"} | INTERACTS_WITH | - | Target | 7224 | TRPC5 | {"name": "TRPC5", "id": "7224"} |
| Phytochemical | CID_5280343 | Quercetin | {"name": "Quercetin", "id": "CID_5280343"} | INTERACTS_WITH | - | Target | 7253 | TSHR | {"name": "TSHR", "id": "7253"} |
| Phytochemical | CID_5280343 | Quercetin | {"name": "Quercetin", "id": "CID_5280343"} | INTERACTS_WITH | - | Target | 7257 | TSNAX | {"name": "TSNAX", "id": "7257"} |
| Phytochemical | CID_5280343 | Quercetin | {"name": "Quercetin", "id": "CID_5280343"} | INTERACTS_WITH | - | Target | 7299 | TYR | {"name": "TYR", "id": "7299"} |
| Phytochemical | CID_5280343 | Quercetin | {"name": "Quercetin", "id": "CID_5280343"} | INTERACTS_WITH | - | Target | 7366 | UGT2B15 | {"name": "UGT2B15", "id": "7366"} |
| Phytochemical | CID_5280343 | Quercetin | {"name": "Quercetin", "id": "CID_5280343"} | INTERACTS_WITH | - | Target | 7398 | USP1 | {"name": "USP1", "id": "7398"} |
| Phytochemical | CID_5280343 | Quercetin | {"name": "Quercetin", "id": "CID_5280343"} | INTERACTS_WITH | - | Target | 7443 | VRK1 | {"name": "VRK1", "id": "7443"} |
| Phytochemical | CID_5280343 | Quercetin | {"name": "Quercetin", "id": "CID_5280343"} | INTERACTS_WITH | - | Target | 7444 | VRK2 | {"name": "VRK2", "id": "7444"} |
| Phytochemical | CID_5280343 | Quercetin | {"name": "Quercetin", "id": "CID_5280343"} | INTERACTS_WITH | - | Target | 7498 | XDH | {"name": "XDH", "id": "7498"} |
| Phytochemical | CID_5280343 | Quercetin | {"name": "Quercetin", "id": "CID_5280343"} | INTERACTS_WITH | - | Target | 7525 | YES1 | {"name": "YES1", "id": "7525"} |
| Phytochemical | CID_5280343 | Quercetin | {"name": "Quercetin", "id": "CID_5280343"} | INTERACTS_WITH | - | Target | 759 | CA1 | {"name": "CA1", "id": "759"} |
| Phytochemical | CID_5280343 | Quercetin | {"name": "Quercetin", "id": "CID_5280343"} | INTERACTS_WITH | - | Target | 760 | CA2 | {"name": "CA2", "id": "760"} |

**Table S9: Complete node and relationship data used to construct the knowledge graph. This table summarizes interactions among FDA-approved drugs, formulation, herbs, diseases (type 2 diabetes, fatty liver, obesity), phytochemicals, and their target genes/proteins. Each entry includes source and target node details, edge type, and node annotations (Entrez Gene ID and gene symbol, where applicable).**

| Source Node Label | Source Node ID | Source Node Name | Source Node Properties | Edge Type | Edge Properties | Target Node Label | Target Node ID | Target Node Name | Target Node Properties |
|---|---|---|---|---|---|---|---|---|---|
| Phytochemical | CID_5280343 | Quercetin | {"name": "Quercetin", "id": "CID_5280343"} | INTERACTS_WITH | - | Target | 761 | CA3 | {"name": "CA3", "id": "761"} |
| Phytochemical | CID_5280343 | Quercetin | {"name": "Quercetin", "id": "CID_5280343"} | INTERACTS_WITH | - | Target | 762 | CA4 | {"name": "CA4", "id": "762"} |
| Phytochemical | CID_5280343 | Quercetin | {"name": "Quercetin", "id": "CID_5280343"} | INTERACTS_WITH | - | Target | 763 | CA5A | {"name": "CA5A", "id": "763"} |
| Phytochemical | CID_5280343 | Quercetin | {"name": "Quercetin", "id": "CID_5280343"} | INTERACTS_WITH | - | Target | 765 | CA6 | {"name": "CA6", "id": "765"} |
| Phytochemical | CID_5280343 | Quercetin | {"name": "Quercetin", "id": "CID_5280343"} | INTERACTS_WITH | - | Target | 766 | CA7 | {"name": "CA7", "id": "766"} |
| Phytochemical | CID_5280343 | Quercetin | {"name": "Quercetin", "id": "CID_5280343"} | INTERACTS_WITH | - | Target | 768 | CA9 | {"name": "CA9", "id": "768"} |
| Phytochemical | CID_5280343 | Quercetin | {"name": "Quercetin", "id": "CID_5280343"} | INTERACTS_WITH | - | Target | 771 | CA12 | {"name": "CA12", "id": "771"} |
| Phytochemical | CID_5280343 | Quercetin | {"name": "Quercetin", "id": "CID_5280343"} | INTERACTS_WITH | - | Target | 79915 | ATAD5 | {"name": "ATAD5", "id": "79915"} |
| Phytochemical | CID_5280343 | Quercetin | {"name": "Quercetin", "id": "CID_5280343"} | INTERACTS_WITH | - | Target | 805 | CALM2 | {"name": "CALM2", "id": "805"} |
| Phytochemical | CID_5280343 | Quercetin | {"name": "Quercetin", "id": "CID_5280343"} | INTERACTS_WITH | - | Target | 814 | CAMK4 | {"name": "CAMK4", "id": "814"} |
| Phytochemical | CID_5280343 | Quercetin | {"name": "Quercetin", "id": "CID_5280343"} | INTERACTS_WITH | - | Target | 815 | CAMK2A | {"name": "CAMK2A", "id": "815"} |
| Phytochemical | CID_5280343 | Quercetin | {"name": "Quercetin", "id": "CID_5280343"} | INTERACTS_WITH | - | Target | 816 | CAMK2B | {"name": "CAMK2B", "id": "816"} |
| Phytochemical | CID_5280343 | Quercetin | {"name": "Quercetin", "id": "CID_5280343"} | INTERACTS_WITH | - | Target | 817 | CAMK2D | {"name": "CAMK2D", "id": "817"} |
| Phytochemical | CID_5280343 | Quercetin | {"name": "Quercetin", "id": "CID_5280343"} | INTERACTS_WITH | - | Target | 818 | CAMK2G | {"name": "CAMK2G", "id": "818"} |
| Phytochemical | CID_5280343 | Quercetin | {"name": "Quercetin", "id": "CID_5280343"} | INTERACTS_WITH | - | Target | 85417 | CCNB3 | {"name": "CCNB3", "id": "85417"} |
| Phytochemical | CID_5280343 | Quercetin | {"name": "Quercetin", "id": "CID_5280343"} | INTERACTS_WITH | - | Target | 8550 | MAPKAPK5 | {"name": "MAPKAPK5", "id": "8550"} |
| Phytochemical | CID_5280343 | Quercetin | {"name": "Quercetin", "id": "CID_5280343"} | INTERACTS_WITH | - | Target | 8576 | STK16 | {"name": "STK16", "id": "8576"} |
| Phytochemical | CID_5280343 | Quercetin | {"name": "Quercetin", "id": "CID_5280343"} | INTERACTS_WITH | - | Target | 8647 | ABCB11 | {"name": "ABCB11", "id": "8647"} |
| Phytochemical | CID_5280343 | Quercetin | {"name": "Quercetin", "id": "CID_5280343"} | INTERACTS_WITH | - | Target | 8654 | PDE5A | {"name": "PDE5A", "id": "8654"} |
| Phytochemical | CID_5280343 | Quercetin | {"name": "Quercetin", "id": "CID_5280343"} | INTERACTS_WITH | - | Target | 8714 | ABCC3 | {"name": "ABCC3", "id": "8714"} |
| Phytochemical | CID_5280343 | Quercetin | {"name": "Quercetin", "id": "CID_5280343"} | INTERACTS_WITH | - | Target | 873 | CBR1 | {"name": "CBR1", "id": "873"} |
| Phytochemical | CID_5280343 | Quercetin | {"name": "Quercetin", "id": "CID_5280343"} | INTERACTS_WITH | - | Target | 874 | CBR3 | {"name": "CBR3", "id": "874"} |
| Phytochemical | CID_5280343 | Quercetin | {"name": "Quercetin", "id": "CID_5280343"} | INTERACTS_WITH | - | Target | 8814 | CDKL1 | {"name": "CDKL1", "id": "8814"} |
| Phytochemical | CID_5280343 | Quercetin | {"name": "Quercetin", "id": "CID_5280343"} | INTERACTS_WITH | - | Target | 8851 | CDK5R1 | {"name": "CDK5R1", "id": "8851"} |
| Phytochemical | CID_5280343 | Quercetin | {"name": "Quercetin", "id": "CID_5280343"} | INTERACTS_WITH | - | Target | 886 | CCKAR | {"name": "CCKAR", "id": "886"} |
| Phytochemical | CID_5280343 | Quercetin | {"name": "Quercetin", "id": "CID_5280343"} | INTERACTS_WITH | - | Target | 887 | CCKBR | {"name": "CCKBR", "id": "887"} |
| Phytochemical | CID_5280343 | Quercetin | {"name": "Quercetin", "id": "CID_5280343"} | INTERACTS_WITH | - | Target | 8877 | SPHK1 | {"name": "SPHK1", "id": "8877"} |
| Phytochemical | CID_5280343 | Quercetin | {"name": "Quercetin", "id": "CID_5280343"} | INTERACTS_WITH | - | Target | 891 | CCNB1 | {"name": "CCNB1", "id": "891"} |
| Phytochemical | CID_5280343 | Quercetin | {"name": "Quercetin", "id": "CID_5280343"} | INTERACTS_WITH | - | Target | 8972 | MGAM | {"name": "MGAM", "id": "8972"} |
| Phytochemical | CID_5280343 | Quercetin | {"name": "Quercetin", "id": "CID_5280343"} | INTERACTS_WITH | - | Target | 9099 | USP2 | {"name": "USP2", "id": "9099"} |
| Phytochemical | CID_5280343 | Quercetin | {"name": "Quercetin", "id": "CID_5280343"} | INTERACTS_WITH | - | Target | 9133 | CCNB2 | {"name": "CCNB2", "id": "9133"} |
| Phytochemical | CID_5280343 | Quercetin | {"name": "Quercetin", "id": "CID_5280343"} | INTERACTS_WITH | - | Target | 9212 | AURKB | {"name": "AURKB", "id": "9212"} |
| Phytochemical | CID_5280343 | Quercetin | {"name": "Quercetin", "id": "CID_5280343"} | INTERACTS_WITH | - | Target | 9252 | RPS6KA5 | {"name": "RPS6KA5", "id": "9252"} |
| Phytochemical | CID_5280343 | Quercetin | {"name": "Quercetin", "id": "CID_5280343"} | INTERACTS_WITH | - | Target | 9261 | MAPKAPK2 | {"name": "MAPKAPK2", "id": "9261"} |
| Phytochemical | CID_5280343 | Quercetin | {"name": "Quercetin", "id": "CID_5280343"} | INTERACTS_WITH | - | Target | 9263 | STK17A | {"name": "STK17A", "id": "9263"} |
| Phytochemical | CID_5280343 | Quercetin | {"name": "Quercetin", "id": "CID_5280343"} | INTERACTS_WITH | - | Target | 93432 | MGAM2 | {"name": "MGAM2", "id": "93432"} |
| Phytochemical | CID_5280343 | Quercetin | {"name": "Quercetin", "id": "CID_5280343"} | INTERACTS_WITH | - | Target | 9429 | ABCG2 | {"name": "ABCG2", "id": "9429"} |
| Phytochemical | CID_5280343 | Quercetin | {"name": "Quercetin", "id": "CID_5280343"} | INTERACTS_WITH | - | Target | 952 | CD38 | {"name": "CD38", "id": "952"} |
| Phytochemical | CID_5280343 | Quercetin | {"name": "Quercetin", "id": "CID_5280343"} | INTERACTS_WITH | - | Target | 9748 | SLK | {"name": "SLK", "id": "9748"} |
| Phytochemical | CID_5280343 | Quercetin | {"name": "Quercetin", "id": "CID_5280343"} | INTERACTS_WITH | - | Target | 983 | CDK1 | {"name": "CDK1", "id": "983"} |
| Phytochemical | CID_5280343 | Quercetin | {"name": "Quercetin", "id": "CID_5280343"} | INTERACTS_WITH | - | Target | 9891 | NUAK1 | {"name": "NUAK1", "id": "9891"} |
| Phytochemical | CID_5280343 | Quercetin | {"name": "Quercetin", "id": "CID_5280343"} | INTERACTS_WITH | - | Target | 9943 | OXSR1 | {"name": "OXSR1", "id": "9943"} |
| Phytochemical | CID_493570 | Riboflavin | {"name": "Riboflavin", "id": "CID_493570"} | INTERACTS_WITH | - | Target | 10013 | HDAC6 | {"name": "HDAC6", "id": "10013"} |
| Phytochemical | CID_493570 | Riboflavin | {"name": "Riboflavin", "id": "CID_493570"} | INTERACTS_WITH | - | Target | 10599 | SLCO1B1 | {"name": "SLCO1B1", "id": "10599"} |
| Phytochemical | CID_493570 | Riboflavin | {"name": "Riboflavin", "id": "CID_493570"} | INTERACTS_WITH | - | Target | 10919 | EHMT2 | {"name": "EHMT2", "id": "10919"} |
| Phytochemical | CID_493570 | Riboflavin | {"name": "Riboflavin", "id": "CID_493570"} | INTERACTS_WITH | - | Target | 11330 | CTRC | {"name": "CTRC", "id": "11330"} |
| Phytochemical | CID_493570 | Riboflavin | {"name": "Riboflavin", "id": "CID_493570"} | INTERACTS_WITH | - | Target | 122706 | PSMB11 | {"name": "PSMB11", "id": "122706"} |
| Phytochemical | CID_493570 | Riboflavin | {"name": "Riboflavin", "id": "CID_493570"} | INTERACTS_WITH | - | Target | 143471 | PSMA8 | {"name": "PSMA8", "id": "143471"} |
| Phytochemical | CID_493570 | Riboflavin | {"name": "Riboflavin", "id": "CID_493570"} | INTERACTS_WITH | - | Target | 1544 | CYP1A2 | {"name": "CYP1A2", "id": "1544"} |

**Table S9: Complete node and relationship data used to construct the knowledge graph. This table summarizes interactions among FDA-approved drugs, formulation, herbs, diseases (type 2 diabetes, fatty liver, obesity), phytochemicals, and their target genes/proteins. Each entry includes source and target node details, edge type, and node annotations (Entrez Gene ID and gene symbol, where applicable).**

| Source Node Label | Source Node ID | Source Node Name | Source Node Properties | Edge Type | Edge Properties | Target Node Label | Target Node ID | Target Node Name | Target Node Properties |
|---|---|---|---|---|---|---|---|---|---|
| Phytochemical | CID_493570 | Riboflavin | {"name": "Riboflavin", "id": "CID_493570"} | INTERACTS_WITH | - | Target | 1576 | CYP3A4 | {"name": "CYP3A4", "id": "1576"} |
| Phytochemical | CID_493570 | Riboflavin | {"name": "Riboflavin", "id": "CID_493570"} | INTERACTS_WITH | - | Target | 216 | ALDH1A1 | {"name": "ALDH1A1", "id": "216"} |
| Phytochemical | CID_493570 | Riboflavin | {"name": "Riboflavin", "id": "CID_493570"} | INTERACTS_WITH | - | Target | 246 | ALOX15 | {"name": "ALOX15", "id": "246"} |
| Phytochemical | CID_493570 | Riboflavin | {"name": "Riboflavin", "id": "CID_493570"} | INTERACTS_WITH | - | Target | 28234 | SLCO1B3 | {"name": "SLCO1B3", "id": "28234"} |
| Phytochemical | CID_493570 | Riboflavin | {"name": "Riboflavin", "id": "CID_493570"} | INTERACTS_WITH | - | Target | 3028 | HSD17B10 | {"name": "HSD17B10", "id": "3028"} |
| Phytochemical | CID_493570 | Riboflavin | {"name": "Riboflavin", "id": "CID_493570"} | INTERACTS_WITH | - | Target | 3248 | HPGD | {"name": "HPGD", "id": "3248"} |
| Phytochemical | CID_493570 | Riboflavin | {"name": "Riboflavin", "id": "CID_493570"} | INTERACTS_WITH | - | Target | 3315 | HSPB1 | {"name": "HSPB1", "id": "3315"} |
| Phytochemical | CID_493570 | Riboflavin | {"name": "Riboflavin", "id": "CID_493570"} | INTERACTS_WITH | - | Target | 3329 | HSPD1 | {"name": "HSPD1", "id": "3329"} |
| Phytochemical | CID_493570 | Riboflavin | {"name": "Riboflavin", "id": "CID_493570"} | INTERACTS_WITH | - | Target | 3336 | HSPE1 | {"name": "HSPE1", "id": "3336"} |
| Phytochemical | CID_493570 | Riboflavin | {"name": "Riboflavin", "id": "CID_493570"} | INTERACTS_WITH | - | Target | 367 | AR | {"name": "AR", "id": "367"} |
| Phytochemical | CID_493570 | Riboflavin | {"name": "Riboflavin", "id": "CID_493570"} | INTERACTS_WITH | - | Target | 3725 | JUN | {"name": "JUN", "id": "3725"} |
| Phytochemical | CID_493570 | Riboflavin | {"name": "Riboflavin", "id": "CID_493570"} | INTERACTS_WITH | - | Target | 4780 | NFE2L2 | {"name": "NFE2L2", "id": "4780"} |
| Phytochemical | CID_493570 | Riboflavin | {"name": "Riboflavin", "id": "CID_493570"} | INTERACTS_WITH | - | Target | 4790 | NFKB1 | {"name": "NFKB1", "id": "4790"} |
| Phytochemical | CID_493570 | Riboflavin | {"name": "Riboflavin", "id": "CID_493570"} | INTERACTS_WITH | - | Target | 5467 | PPARD | {"name": "PPARD", "id": "5467"} |
| Phytochemical | CID_493570 | Riboflavin | {"name": "Riboflavin", "id": "CID_493570"} | INTERACTS_WITH | - | Target | 5468 | PPARG | {"name": "PPARG", "id": "5468"} |
| Phytochemical | CID_493570 | Riboflavin | {"name": "Riboflavin", "id": "CID_493570"} | INTERACTS_WITH | - | Target | 55775 | TDP1 | {"name": "TDP1", "id": "55775"} |
| Phytochemical | CID_493570 | Riboflavin | {"name": "Riboflavin", "id": "CID_493570"} | INTERACTS_WITH | - | Target | 5594 | MAPK1 | {"name": "MAPK1", "id": "5594"} |
| Phytochemical | CID_493570 | Riboflavin | {"name": "Riboflavin", "id": "CID_493570"} | INTERACTS_WITH | - | Target | 5682 | PSMA1 | {"name": "PSMA1", "id": "5682"} |
| Phytochemical | CID_493570 | Riboflavin | {"name": "Riboflavin", "id": "CID_493570"} | INTERACTS_WITH | - | Target | 5683 | PSMA2 | {"name": "PSMA2", "id": "5683"} |
| Phytochemical | CID_493570 | Riboflavin | {"name": "Riboflavin", "id": "CID_493570"} | INTERACTS_WITH | - | Target | 5684 | PSMA3 | {"name": "PSMA3", "id": "5684"} |
| Phytochemical | CID_493570 | Riboflavin | {"name": "Riboflavin", "id": "CID_493570"} | INTERACTS_WITH | - | Target | 5685 | PSMA4 | {"name": "PSMA4", "id": "5685"} |
| Phytochemical | CID_493570 | Riboflavin | {"name": "Riboflavin", "id": "CID_493570"} | INTERACTS_WITH | - | Target | 5686 | PSMA5 | {"name": "PSMA5", "id": "5686"} |
| Phytochemical | CID_493570 | Riboflavin | {"name": "Riboflavin", "id": "CID_493570"} | INTERACTS_WITH | - | Target | 5687 | PSMA6 | {"name": "PSMA6", "id": "5687"} |
| Phytochemical | CID_493570 | Riboflavin | {"name": "Riboflavin", "id": "CID_493570"} | INTERACTS_WITH | - | Target | 5688 | PSMA7 | {"name": "PSMA7", "id": "5688"} |
| Phytochemical | CID_493570 | Riboflavin | {"name": "Riboflavin", "id": "CID_493570"} | INTERACTS_WITH | - | Target | 5689 | PSMB1 | {"name": "PSMB1", "id": "5689"} |
| Phytochemical | CID_493570 | Riboflavin | {"name": "Riboflavin", "id": "CID_493570"} | INTERACTS_WITH | - | Target | 5690 | PSMB2 | {"name": "PSMB2", "id": "5690"} |
| Phytochemical | CID_493570 | Riboflavin | {"name": "Riboflavin", "id": "CID_493570"} | INTERACTS_WITH | - | Target | 5691 | PSMB3 | {"name": "PSMB3", "id": "5691"} |
| Phytochemical | CID_493570 | Riboflavin | {"name": "Riboflavin", "id": "CID_493570"} | INTERACTS_WITH | - | Target | 5692 | PSMB4 | {"name": "PSMB4", "id": "5692"} |
| Phytochemical | CID_493570 | Riboflavin | {"name": "Riboflavin", "id": "CID_493570"} | INTERACTS_WITH | - | Target | 5693 | PSMB5 | {"name": "PSMB5", "id": "5693"} |
| Phytochemical | CID_493570 | Riboflavin | {"name": "Riboflavin", "id": "CID_493570"} | INTERACTS_WITH | - | Target | 5694 | PSMB6 | {"name": "PSMB6", "id": "5694"} |
| Phytochemical | CID_493570 | Riboflavin | {"name": "Riboflavin", "id": "CID_493570"} | INTERACTS_WITH | - | Target | 5695 | PSMB7 | {"name": "PSMB7", "id": "5695"} |
| Phytochemical | CID_493570 | Riboflavin | {"name": "Riboflavin", "id": "CID_493570"} | INTERACTS_WITH | - | Target | 5696 | PSMB8 | {"name": "PSMB8", "id": "5696"} |
| Phytochemical | CID_493570 | Riboflavin | {"name": "Riboflavin", "id": "CID_493570"} | INTERACTS_WITH | - | Target | 5698 | PSMB9 | {"name": "PSMB9", "id": "5698"} |
| Phytochemical | CID_493570 | Riboflavin | {"name": "Riboflavin", "id": "CID_493570"} | INTERACTS_WITH | - | Target | 5699 | PSMB10 | {"name": "PSMB10", "id": "5699"} |
| Phytochemical | CID_493570 | Riboflavin | {"name": "Riboflavin", "id": "CID_493570"} | INTERACTS_WITH | - | Target | 5931 | RBBP7 | {"name": "RBBP7", "id": "5931"} |
| Phytochemical | CID_493570 | Riboflavin | {"name": "Riboflavin", "id": "CID_493570"} | INTERACTS_WITH | - | Target | 5965 | RECQL | {"name": "RECQL", "id": "5965"} |
| Phytochemical | CID_493570 | Riboflavin | {"name": "Riboflavin", "id": "CID_493570"} | INTERACTS_WITH | - | Target | 6256 | RXRA | {"name": "RXRA", "id": "6256"} |
| Phytochemical | CID_493570 | Riboflavin | {"name": "Riboflavin", "id": "CID_493570"} | INTERACTS_WITH | - | Target | 672 | BRCA1 | {"name": "BRCA1", "id": "672"} |
| Phytochemical | CID_493570 | Riboflavin | {"name": "Riboflavin", "id": "CID_493570"} | INTERACTS_WITH | - | Target | 7068 | THRB | {"name": "THRB", "id": "7068"} |
| Phytochemical | CID_493570 | Riboflavin | {"name": "Riboflavin", "id": "CID_493570"} | INTERACTS_WITH | - | Target | 7124 | TNF | {"name": "TNF", "id": "7124"} |
| Phytochemical | CID_493570 | Riboflavin | {"name": "Riboflavin", "id": "CID_493570"} | INTERACTS_WITH | - | Target | 7263 | TST | {"name": "TST", "id": "7263"} |
| Phytochemical | CID_493570 | Riboflavin | {"name": "Riboflavin", "id": "CID_493570"} | INTERACTS_WITH | - | Target | 7398 | USP1 | {"name": "USP1", "id": "7398"} |
| Phytochemical | CID_493570 | Riboflavin | {"name": "Riboflavin", "id": "CID_493570"} | INTERACTS_WITH | - | Target | 7421 | VDR | {"name": "VDR", "id": "7421"} |
| Phytochemical | CID_493570 | Riboflavin | {"name": "Riboflavin", "id": "CID_493570"} | INTERACTS_WITH | - | Target | 834 | CASP1 | {"name": "CASP1", "id": "834"} |
| Phytochemical | CID_493570 | Riboflavin | {"name": "Riboflavin", "id": "CID_493570"} | INTERACTS_WITH | - | Target | 840 | CASP7 | {"name": "CASP7", "id": "840"} |
| Phytochemical | CID_493570 | Riboflavin | {"name": "Riboflavin", "id": "CID_493570"} | INTERACTS_WITH | - | Target | 8520 | HAT1 | {"name": "HAT1", "id": "8520"} |
| Phytochemical | CID_493570 | Riboflavin | {"name": "Riboflavin", "id": "CID_493570"} | INTERACTS_WITH | - | Target | 9971 | NR1H4 | {"name": "NR1H4", "id": "9971"} |
| Phytochemical | CID_5144 | Safrole | {"name": "Safrole", "id": "CID_5144"} | INTERACTS_WITH | - | Target | 10280 | SIGMAR1 | {"name": "SIGMAR1", "id": "10280"} |
| Phytochemical | CID_5144 | Safrole | {"name": "Safrole", "id": "CID_5144"} | INTERACTS_WITH | - | Target | 10599 | SLCO1B1 | {"name": "SLCO1B1", "id": "10599"} |
| Phytochemical | CID_5144 | Safrole | {"name": "Safrole", "id": "CID_5144"} | INTERACTS_WITH | - | Target | 10800 | CYSLTR1 | {"name": "CYSLTR1", "id": "10800"} |

Table S9: Complete node and relationship data used to construct the knowledge graph. This table summarizes interactions among FDA-approved drugs, formulation, herbs, diseases (type 2 diabetes, fatty liver, obesity), phytochemicals, and their target genes/proteins. Each entry includes source and target node details, edge type, and node annotations (Entrez Gene ID and gene symbol, where applicable).

| Source Node Label | Source Node ID | Source Node Name | Source Node Properties | Edge Type | Edge Properties | Target Node Label | Target Node ID | Target Node Name | Target Node Properties |
|---|---|---|---|---|---|---|---|---|---|
| Phytochemical | CID_5144 | Safrole | {"name": "Safrole", "id": "CID_5144"} | INTERACTS_WITH | - | Target | 1128 | CHRM1 | {"name": "CHRM1", "id": "1128"} |
| Phytochemical | CID_5144 | Safrole | {"name": "Safrole", "id": "CID_5144"} | INTERACTS_WITH | - | Target | 1129 | CHRM2 | {"name": "CHRM2", "id": "1129"} |
| Phytochemical | CID_5144 | Safrole | {"name": "Safrole", "id": "CID_5144"} | INTERACTS_WITH | - | Target | 1131 | CHRM3 | {"name": "CHRM3", "id": "1131"} |
| Phytochemical | CID_5144 | Safrole | {"name": "Safrole", "id": "CID_5144"} | INTERACTS_WITH | - | Target | 1132 | CHRM4 | {"name": "CHRM4", "id": "1132"} |
| Phytochemical | CID_5144 | Safrole | {"name": "Safrole", "id": "CID_5144"} | INTERACTS_WITH | - | Target | 1133 | CHRM5 | {"name": "CHRM5", "id": "1133"} |
| Phytochemical | CID_5144 | Safrole | {"name": "Safrole", "id": "CID_5144"} | INTERACTS_WITH | - | Target | 1233 | CCR4 | {"name": "CCR4", "id": "1233"} |
| Phytochemical | CID_5144 | Safrole | {"name": "Safrole", "id": "CID_5144"} | INTERACTS_WITH | - | Target | 1234 | CCR5 | {"name": "CCR5", "id": "1234"} |
| Phytochemical | CID_5144 | Safrole | {"name": "Safrole", "id": "CID_5144"} | INTERACTS_WITH | - | Target | 1268 | CNR1 | {"name": "CNR1", "id": "1268"} |
| Phytochemical | CID_5144 | Safrole | {"name": "Safrole", "id": "CID_5144"} | INTERACTS_WITH | - | Target | 134 | ADORA1 | {"name": "ADORA1", "id": "134"} |
| Phytochemical | CID_5144 | Safrole | {"name": "Safrole", "id": "CID_5144"} | INTERACTS_WITH | - | Target | 135 | ADORA2A | {"name": "ADORA2A", "id": "135"} |
| Phytochemical | CID_5144 | Safrole | {"name": "Safrole", "id": "CID_5144"} | INTERACTS_WITH | - | Target | 140 | ADORA3 | {"name": "ADORA3", "id": "140"} |
| Phytochemical | CID_5144 | Safrole | {"name": "Safrole", "id": "CID_5144"} | INTERACTS_WITH | - | Target | 1432 | MAPK14 | {"name": "MAPK14", "id": "1432"} |
| Phytochemical | CID_5144 | Safrole | {"name": "Safrole", "id": "CID_5144"} | INTERACTS_WITH | - | Target | 146 | ADRA1D | {"name": "ADRA1D", "id": "146"} |
| Phytochemical | CID_5144 | Safrole | {"name": "Safrole", "id": "CID_5144"} | INTERACTS_WITH | - | Target | 150 | ADRA2A | {"name": "ADRA2A", "id": "150"} |
| Phytochemical | CID_5144 | Safrole | {"name": "Safrole", "id": "CID_5144"} | INTERACTS_WITH | - | Target | 151 | ADRA2B | {"name": "ADRA2B", "id": "151"} |
| Phytochemical | CID_5144 | Safrole | {"name": "Safrole", "id": "CID_5144"} | INTERACTS_WITH | - | Target | 1511 | CTSG | {"name": "CTSG", "id": "1511"} |
| Phytochemical | CID_5144 | Safrole | {"name": "Safrole", "id": "CID_5144"} | INTERACTS_WITH | - | Target | 152 | ADRA2C | {"name": "ADRA2C", "id": "152"} |
| Phytochemical | CID_5144 | Safrole | {"name": "Safrole", "id": "CID_5144"} | INTERACTS_WITH | - | Target | 153 | ADRB1 | {"name": "ADRB1", "id": "153"} |
| Phytochemical | CID_5144 | Safrole | {"name": "Safrole", "id": "CID_5144"} | INTERACTS_WITH | - | Target | 154 | ADRB2 | {"name": "ADRB2", "id": "154"} |
| Phytochemical | CID_5144 | Safrole | {"name": "Safrole", "id": "CID_5144"} | INTERACTS_WITH | - | Target | 1544 | CYP1A2 | {"name": "CYP1A2", "id": "1544"} |
| Phytochemical | CID_5144 | Safrole | {"name": "Safrole", "id": "CID_5144"} | INTERACTS_WITH | - | Target | 1548 | CYP2A6 | {"name": "CYP2A6", "id": "1548"} |
| Phytochemical | CID_5144 | Safrole | {"name": "Safrole", "id": "CID_5144"} | INTERACTS_WITH | - | Target | 155 | ADRB3 | {"name": "ADRB3", "id": "155"} |
| Phytochemical | CID_5144 | Safrole | {"name": "Safrole", "id": "CID_5144"} | INTERACTS_WITH | - | Target | 1557 | CYP2C19 | {"name": "CYP2C19", "id": "1557"} |
| Phytochemical | CID_5144 | Safrole | {"name": "Safrole", "id": "CID_5144"} | INTERACTS_WITH | - | Target | 1559 | CYP2C9 | {"name": "CYP2C9", "id": "1559"} |
| Phytochemical | CID_5144 | Safrole | {"name": "Safrole", "id": "CID_5144"} | INTERACTS_WITH | - | Target | 1565 | CYP2D6 | {"name": "CYP2D6", "id": "1565"} |
| Phytochemical | CID_5144 | Safrole | {"name": "Safrole", "id": "CID_5144"} | INTERACTS_WITH | - | Target | 1571 | CYP2E1 | {"name": "CYP2E1", "id": "1571"} |
| Phytochemical | CID_5144 | Safrole | {"name": "Safrole", "id": "CID_5144"} | INTERACTS_WITH | - | Target | 1576 | CYP3A4 | {"name": "CYP3A4", "id": "1576"} |
| Phytochemical | CID_5144 | Safrole | {"name": "Safrole", "id": "CID_5144"} | INTERACTS_WITH | - | Target | 1812 | DRD1 | {"name": "DRD1", "id": "1812"} |
| Phytochemical | CID_5144 | Safrole | {"name": "Safrole", "id": "CID_5144"} | INTERACTS_WITH | - | Target | 1813 | DRD2 | {"name": "DRD2", "id": "1813"} |
| Phytochemical | CID_5144 | Safrole | {"name": "Safrole", "id": "CID_5144"} | INTERACTS_WITH | - | Target | 1814 | DRD3 | {"name": "DRD3", "id": "1814"} |
| Phytochemical | CID_5144 | Safrole | {"name": "Safrole", "id": "CID_5144"} | INTERACTS_WITH | - | Target | 1815 | DRD4 | {"name": "DRD4", "id": "1815"} |
| Phytochemical | CID_5144 | Safrole | {"name": "Safrole", "id": "CID_5144"} | INTERACTS_WITH | - | Target | 186 | AGTR2 | {"name": "AGTR2", "id": "186"} |
| Phytochemical | CID_5144 | Safrole | {"name": "Safrole", "id": "CID_5144"} | INTERACTS_WITH | - | Target | 1909 | EDNRA | {"name": "EDNRA", "id": "1909"} |
| Phytochemical | CID_5144 | Safrole | {"name": "Safrole", "id": "CID_5144"} | INTERACTS_WITH | - | Target | 1956 | EGFR | {"name": "EGFR", "id": "1956"} |
| Phytochemical | CID_5144 | Safrole | {"name": "Safrole", "id": "CID_5144"} | INTERACTS_WITH | - | Target | 1991 | ELANE | {"name": "ELANE", "id": "1991"} |
| Phytochemical | CID_5144 | Safrole | {"name": "Safrole", "id": "CID_5144"} | INTERACTS_WITH | - | Target | 2064 | ERBB2 | {"name": "ERBB2", "id": "2064"} |
| Phytochemical | CID_5144 | Safrole | {"name": "Safrole", "id": "CID_5144"} | INTERACTS_WITH | - | Target | 2099 | ESR1 | {"name": "ESR1", "id": "2099"} |
| Phytochemical | CID_5144 | Safrole | {"name": "Safrole", "id": "CID_5144"} | INTERACTS_WITH | - | Target | 2100 | ESR2 | {"name": "ESR2", "id": "2100"} |
| Phytochemical | CID_5144 | Safrole | {"name": "Safrole", "id": "CID_5144"} | INTERACTS_WITH | - | Target | 2321 | FLT1 | {"name": "FLT1", "id": "2321"} |
| Phytochemical | CID_5144 | Safrole | {"name": "Safrole", "id": "CID_5144"} | INTERACTS_WITH | - | Target | 2534 | FYN | {"name": "FYN", "id": "2534"} |
| Phytochemical | CID_5144 | Safrole | {"name": "Safrole", "id": "CID_5144"} | INTERACTS_WITH | - | Target | 28234 | SLCO1B3 | {"name": "SLCO1B3", "id": "28234"} |
| Phytochemical | CID_5144 | Safrole | {"name": "Safrole", "id": "CID_5144"} | INTERACTS_WITH | - | Target | 2908 | NR3C1 | {"name": "NR3C1", "id": "2908"} |
| Phytochemical | CID_5144 | Safrole | {"name": "Safrole", "id": "CID_5144"} | INTERACTS_WITH | - | Target | 3156 | HMGCR | {"name": "HMGCR", "id": "3156"} |
| Phytochemical | CID_5144 | Safrole | {"name": "Safrole", "id": "CID_5144"} | INTERACTS_WITH | - | Target | 3269 | HRH1 | {"name": "HRH1", "id": "3269"} |
| Phytochemical | CID_5144 | Safrole | {"name": "Safrole", "id": "CID_5144"} | INTERACTS_WITH | - | Target | 3274 | HRH2 | {"name": "HRH2", "id": "3274"} |
| Phytochemical | CID_5144 | Safrole | {"name": "Safrole", "id": "CID_5144"} | INTERACTS_WITH | - | Target | 3356 | HTR2A | {"name": "HTR2A", "id": "3356"} |
| Phytochemical | CID_5144 | Safrole | {"name": "Safrole", "id": "CID_5144"} | INTERACTS_WITH | - | Target | 3357 | HTR2B | {"name": "HTR2B", "id": "3357"} |
| Phytochemical | CID_5144 | Safrole | {"name": "Safrole", "id": "CID_5144"} | INTERACTS_WITH | - | Target | 3358 | HTR2C | {"name": "HTR2C", "id": "3358"} |
| Phytochemical | CID_5144 | Safrole | {"name": "Safrole", "id": "CID_5144"} | INTERACTS_WITH | - | Target | 3362 | HTR6 | {"name": "HTR6", "id": "3362"} |
| Phytochemical | CID_5144 | Safrole | {"name": "Safrole", "id": "CID_5144"} | INTERACTS_WITH | - | Target | 3577 | CXCR1 | {"name": "CXCR1", "id": "3577"} |
| Phytochemical | CID_5144 | Safrole | {"name": "Safrole", "id": "CID_5144"} | INTERACTS_WITH | - | Target | 3579 | CXCR2 | {"name": "CXCR2", "id": "3579"} |
| Phytochemical | CID_5144 | Safrole | {"name": "Safrole", "id": "CID_5144"} | INTERACTS_WITH | - | Target | 3757 | KCNH2 | {"name": "KCNH2", "id": "3757"} |
| Phytochemical | CID_5144 | Safrole | {"name": "Safrole", "id": "CID_5144"} | INTERACTS_WITH | - | Target | 3932 | LCK | {"name": "LCK", "id": "3932"} |
| Phytochemical | CID_5144 | Safrole | {"name": "Safrole", "id": "CID_5144"} | INTERACTS_WITH | - | Target | 4128 | MAOA | {"name": "MAOA", "id": "4128"} |
| Phytochemical | CID_5144 | Safrole | {"name": "Safrole", "id": "CID_5144"} | INTERACTS_WITH | - | Target | 4137 | MAPT | {"name": "MAPT", "id": "4137"} |
| Phytochemical | CID_5144 | Safrole | {"name": "Safrole", "id": "CID_5144"} | INTERACTS_WITH | - | Target | 4159 | MC3R | {"name": "MC3R", "id": "4159"} |
| Phytochemical | CID_5144 | Safrole | {"name": "Safrole", "id": "CID_5144"} | INTERACTS_WITH | - | Target | 4160 | MC4R | {"name": "MC4R", "id": "4160"} |
| Phytochemical | CID_5144 | Safrole | {"name": "Safrole", "id": "CID_5144"} | INTERACTS_WITH | - | Target | 4161 | MC5R | {"name": "MC5R", "id": "4161"} |
| Phytochemical | CID_5144 | Safrole | {"name": "Safrole", "id": "CID_5144"} | INTERACTS_WITH | - | Target | 43 | ACHE | {"name": "ACHE", "id": "43"} |
| Phytochemical | CID_5144 | Safrole | {"name": "Safrole", "id": "CID_5144"} | INTERACTS_WITH | - | Target | 4312 | MMP1 | {"name": "MMP1", "id": "4312"} |
| Phytochemical | CID_5144 | Safrole | {"name": "Safrole", "id": "CID_5144"} | INTERACTS_WITH | - | Target | 4318 | MMP9 | {"name": "MMP9", "id": "4318"} |
| Phytochemical | CID_5144 | Safrole | {"name": "Safrole", "id": "CID_5144"} | INTERACTS_WITH | - | Target | 4780 | NFE2L2 | {"name": "NFE2L2", "id": "4780"} |
| Phytochemical | CID_5144 | Safrole | {"name": "Safrole", "id": "CID_5144"} | INTERACTS_WITH | - | Target | 4886 | NPY1R | {"name": "NPY1R", "id": "4886"} |
| Phytochemical | CID_5144 | Safrole | {"name": "Safrole", "id": "CID_5144"} | INTERACTS_WITH | - | Target | 4887 | NPY2R | {"name": "NPY2R", "id": "4887"} |
| Phytochemical | CID_5144 | Safrole | {"name": "Safrole", "id": "CID_5144"} | INTERACTS_WITH | - | Target | 4985 | OPRD1 | {"name": "OPRD1", "id": "4985"} |
| Phytochemical | CID_5144 | Safrole | {"name": "Safrole", "id": "CID_5144"} | INTERACTS_WITH | - | Target | 4986 | OPRK1 | {"name": "OPRK1", "id": "4986"} |
| Phytochemical | CID_5144 | Safrole | {"name": "Safrole", "id": "CID_5144"} | INTERACTS_WITH | - | Target | 4988 | OPRM1 | {"name": "OPRM1", "id": "4988"} |
| Phytochemical | CID_5144 | Safrole | {"name": "Safrole", "id": "CID_5144"} | INTERACTS_WITH | - | Target | 552 | AVPR1A | {"name": "AVPR1A", "id": "552"} |
| Phytochemical | CID_5144 | Safrole | {"name": "Safrole", "id": "CID_5144"} | INTERACTS_WITH | - | Target | 5530 | PPP3CA | {"name": "PPP3CA", "id": "5530"} |
| Phytochemical | CID_5144 | Safrole | {"name": "Safrole", "id": "CID_5144"} | INTERACTS_WITH | - | Target | 5578 | PRKCA | {"name": "PRKCA", "id": "5578"} |
| Phytochemical | CID_5144 | Safrole | {"name": "Safrole", "id": "CID_5144"} | INTERACTS_WITH | - | Target | 5594 | MAPK1 | {"name": "MAPK1", "id": "5594"} |
| Phytochemical | CID_5144 | Safrole | {"name": "Safrole", "id": "CID_5144"} | INTERACTS_WITH | - | Target | 5595 | MAPK3 | {"name": "MAPK3", "id": "5595"} |
| Phytochemical | CID_5144 | Safrole | {"name": "Safrole", "id": "CID_5144"} | INTERACTS_WITH | - | Target | 5724 | PTAFR | {"name": "PTAFR", "id": "5724"} |
| Phytochemical | CID_5144 | Safrole | {"name": "Safrole", "id": "CID_5144"} | INTERACTS_WITH | - | Target | 5742 | PTGS1 | {"name": "PTGS1", "id": "5742"} |
| Phytochemical | CID_5144 | Safrole | {"name": "Safrole", "id": "CID_5144"} | INTERACTS_WITH | - | Target | 5743 | PTGS2 | {"name": "PTGS2", "id": "5743"} |
| Phytochemical | CID_5144 | Safrole | {"name": "Safrole", "id": "CID_5144"} | INTERACTS_WITH | - | Target | 5788 | PTPRC | {"name": "PTPRC", "id": "5788"} |

**Table S9: Complete node and relationship data used to construct the knowledge graph. This table summarizes interactions among FDA-approved drugs, formulation, herbs, diseases (type 2 diabetes, fatty liver, obesity), phytochemicals, and their target genes/proteins. Each entry includes source and target node details, edge type, and node annotations (Entrez Gene ID and gene symbol, where applicable).**

| Source Node Label | Source Node ID | Source Node Name | Source Node Properties | Edge Type | Edge Properties | Target Node Label | Target Node ID | Target Node Name | Target Node Properties |
|---|---|---|---|---|---|---|---|---|---|
| Phytochemical | CID_5144 | Safrole | {"name": "Safrole", "id": "CID_5144"} | INTERACTS_WITH | - | Target | 624 | BDKRB2 | {"name": "BDKRB2", "id": "624"} |
| Phytochemical | CID_5144 | Safrole | {"name": "Safrole", "id": "CID_5144"} | INTERACTS_WITH | - | Target | 6530 | SLC6A2 | {"name": "SLC6A2", "id": "6530"} |
| Phytochemical | CID_5144 | Safrole | {"name": "Safrole", "id": "CID_5144"} | INTERACTS_WITH | - | Target | 6531 | SLC6A3 | {"name": "SLC6A3", "id": "6531"} |
| Phytochemical | CID_5144 | Safrole | {"name": "Safrole", "id": "CID_5144"} | INTERACTS_WITH | - | Target | 6532 | SLC6A4 | {"name": "SLC6A4", "id": "6532"} |
| Phytochemical | CID_5144 | Safrole | {"name": "Safrole", "id": "CID_5144"} | INTERACTS_WITH | - | Target | 6865 | TACR2 | {"name": "TACR2", "id": "6865"} |
| Phytochemical | CID_5144 | Safrole | {"name": "Safrole", "id": "CID_5144"} | INTERACTS_WITH | - | Target | 6869 | TACR1 | {"name": "TACR1", "id": "6869"} |
| Phytochemical | CID_5144 | Safrole | {"name": "Safrole", "id": "CID_5144"} | INTERACTS_WITH | - | Target | 6916 | TBXAS1 | {"name": "TBXAS1", "id": "6916"} |
| Phytochemical | CID_5144 | Safrole | {"name": "Safrole", "id": "CID_5144"} | INTERACTS_WITH | - | Target | 729230 | CCR2 | {"name": "CCR2", "id": "729230"} |
| Phytochemical | CID_5144 | Safrole | {"name": "Safrole", "id": "CID_5144"} | INTERACTS_WITH | - | Target | 7433 | VIPR1 | {"name": "VIPR1", "id": "7433"} |
| Phytochemical | CID_5144 | Safrole | {"name": "Safrole", "id": "CID_5144"} | INTERACTS_WITH | - | Target | 760 | CA2 | {"name": "CA2", "id": "760"} |
| Phytochemical | CID_5144 | Safrole | {"name": "Safrole", "id": "CID_5144"} | INTERACTS_WITH | - | Target | 799 | CALCR | {"name": "CALCR", "id": "799"} |
| Phytochemical | CID_5144 | Safrole | {"name": "Safrole", "id": "CID_5144"} | INTERACTS_WITH | - | Target | 834 | CASP1 | {"name": "CASP1", "id": "834"} |
| Phytochemical | CID_5144 | Safrole | {"name": "Safrole", "id": "CID_5144"} | INTERACTS_WITH | - | Target | 8654 | PDE5A | {"name": "PDE5A", "id": "8654"} |
| Phytochemical | CID_5144 | Safrole | {"name": "Safrole", "id": "CID_5144"} | INTERACTS_WITH | - | Target | 886 | CCKAR | {"name": "CCKAR", "id": "886"} |
| Phytochemical | CID_5280460 | Scopoletin | {"name": "Scopoletin", "id": "CID_5280460"} | INTERACTS_WITH | - | Target | 10073 | SNUPN | {"name": "SNUPN", "id": "10073"} |
| Phytochemical | CID_5280460 | Scopoletin | {"name": "Scopoletin", "id": "CID_5280460"} | INTERACTS_WITH | - | Target | 10599 | SLCO1B1 | {"name": "SLCO1B1", "id": "10599"} |
| Phytochemical | CID_5280460 | Scopoletin | {"name": "Scopoletin", "id": "CID_5280460"} | INTERACTS_WITH | - | Target | 10919 | EHMT2 | {"name": "EHMT2", "id": "10919"} |
| Phytochemical | CID_5280460 | Scopoletin | {"name": "Scopoletin", "id": "CID_5280460"} | INTERACTS_WITH | - | Target | 10941 | UGT2A1 | {"name": "UGT2A1", "id": "10941"} |
| Phytochemical | CID_5280460 | Scopoletin | {"name": "Scopoletin", "id": "CID_5280460"} | INTERACTS_WITH | - | Target | 1544 | CYP1A2 | {"name": "CYP1A2", "id": "1544"} |
| Phytochemical | CID_5280460 | Scopoletin | {"name": "Scopoletin", "id": "CID_5280460"} | INTERACTS_WITH | - | Target | 216 | ALDH1A1 | {"name": "ALDH1A1", "id": "216"} |
| Phytochemical | CID_5280460 | Scopoletin | {"name": "Scopoletin", "id": "CID_5280460"} | INTERACTS_WITH | - | Target | 231 | AKR1B1 | {"name": "AKR1B1", "id": "231"} |
| Phytochemical | CID_5280460 | Scopoletin | {"name": "Scopoletin", "id": "CID_5280460"} | INTERACTS_WITH | - | Target | 240 | ALOX5 | {"name": "ALOX5", "id": "240"} |
| Phytochemical | CID_5280460 | Scopoletin | {"name": "Scopoletin", "id": "CID_5280460"} | INTERACTS_WITH | - | Target | 2548 | GAA | {"name": "GAA", "id": "2548"} |
| Phytochemical | CID_5280460 | Scopoletin | {"name": "Scopoletin", "id": "CID_5280460"} | INTERACTS_WITH | - | Target | 2597 | GAPDH | {"name": "GAPDH", "id": "2597"} |
| Phytochemical | CID_5280460 | Scopoletin | {"name": "Scopoletin", "id": "CID_5280460"} | INTERACTS_WITH | - | Target | 2629 | GBA1 | {"name": "GBA1", "id": "2629"} |
| Phytochemical | CID_5280460 | Scopoletin | {"name": "Scopoletin", "id": "CID_5280460"} | INTERACTS_WITH | - | Target | 2648 | KAT2A | {"name": "KAT2A", "id": "2648"} |
| Phytochemical | CID_5280460 | Scopoletin | {"name": "Scopoletin", "id": "CID_5280460"} | INTERACTS_WITH | - | Target | 2671 | GFER | {"name": "GFER", "id": "2671"} |
| Phytochemical | CID_5280460 | Scopoletin | {"name": "Scopoletin", "id": "CID_5280460"} | INTERACTS_WITH | - | Target | 2717 | GLA | {"name": "GLA", "id": "2717"} |
| Phytochemical | CID_5280460 | Scopoletin | {"name": "Scopoletin", "id": "CID_5280460"} | INTERACTS_WITH | - | Target | 2778 | GNAS | {"name": "GNAS", "id": "2778"} |
| Phytochemical | CID_5280460 | Scopoletin | {"name": "Scopoletin", "id": "CID_5280460"} | INTERACTS_WITH | - | Target | 28234 | SLCO1B3 | {"name": "SLCO1B3", "id": "28234"} |
| Phytochemical | CID_5280460 | Scopoletin | {"name": "Scopoletin", "id": "CID_5280460"} | INTERACTS_WITH | - | Target | 3028 | HSD17B10 | {"name": "HSD17B10", "id": "3028"} |
| Phytochemical | CID_5280460 | Scopoletin | {"name": "Scopoletin", "id": "CID_5280460"} | INTERACTS_WITH | - | Target | 3091 | HIF1A | {"name": "HIF1A", "id": "3091"} |
| Phytochemical | CID_5280460 | Scopoletin | {"name": "Scopoletin", "id": "CID_5280460"} | INTERACTS_WITH | - | Target | 3248 | HPGD | {"name": "HPGD", "id": "3248"} |
| Phytochemical | CID_5280460 | Scopoletin | {"name": "Scopoletin", "id": "CID_5280460"} | INTERACTS_WITH | - | Target | 328 | APEX1 | {"name": "APEX1", "id": "328"} |
| Phytochemical | CID_5280460 | Scopoletin | {"name": "Scopoletin", "id": "CID_5280460"} | INTERACTS_WITH | - | Target | 377677 | CA13 | {"name": "CA13", "id": "377677"} |
| Phytochemical | CID_5280460 | Scopoletin | {"name": "Scopoletin", "id": "CID_5280460"} | INTERACTS_WITH | - | Target | 3837 | KPNB1 | {"name": "KPNB1", "id": "3837"} |
| Phytochemical | CID_5280460 | Scopoletin | {"name": "Scopoletin", "id": "CID_5280460"} | INTERACTS_WITH | - | Target | 390245 | KDM4E | {"name": "KDM4E", "id": "390245"} |
| Phytochemical | CID_5280460 | Scopoletin | {"name": "Scopoletin", "id": "CID_5280460"} | INTERACTS_WITH | - | Target | 4128 | MAOA | {"name": "MAOA", "id": "4128"} |
| Phytochemical | CID_5280460 | Scopoletin | {"name": "Scopoletin", "id": "CID_5280460"} | INTERACTS_WITH | - | Target | 4129 | MAOB | {"name": "MAOB", "id": "4129"} |
| Phytochemical | CID_5280460 | Scopoletin | {"name": "Scopoletin", "id": "CID_5280460"} | INTERACTS_WITH | - | Target | 4170 | MCL1 | {"name": "MCL1", "id": "4170"} |
| Phytochemical | CID_5280460 | Scopoletin | {"name": "Scopoletin", "id": "CID_5280460"} | INTERACTS_WITH | - | Target | 4297 | KMT2A | {"name": "KMT2A", "id": "4297"} |
| Phytochemical | CID_5280460 | Scopoletin | {"name": "Scopoletin", "id": "CID_5280460"} | INTERACTS_WITH | - | Target | 43 | ACHE | {"name": "ACHE", "id": "43"} |
| Phytochemical | CID_5280460 | Scopoletin | {"name": "Scopoletin", "id": "CID_5280460"} | INTERACTS_WITH | - | Target | 51053 | GMNN | {"name": "GMNN", "id": "51053"} |
| Phytochemical | CID_5280460 | Scopoletin | {"name": "Scopoletin", "id": "CID_5280460"} | INTERACTS_WITH | - | Target | 51752 | ERAP1 | {"name": "ERAP1", "id": "51752"} |
| Phytochemical | CID_5280460 | Scopoletin | {"name": "Scopoletin", "id": "CID_5280460"} | INTERACTS_WITH | - | Target | 5300 | PIN1 | {"name": "PIN1", "id": "5300"} |
| Phytochemical | CID_5280460 | Scopoletin | {"name": "Scopoletin", "id": "CID_5280460"} | INTERACTS_WITH | - | Target | 54575 | UGT1A10 | {"name": "UGT1A10", "id": "54575"} |
| Phytochemical | CID_5280460 | Scopoletin | {"name": "Scopoletin", "id": "CID_5280460"} | INTERACTS_WITH | - | Target | 54576 | UGT1A8 | {"name": "UGT1A8", "id": "54576"} |
| Phytochemical | CID_5280460 | Scopoletin | {"name": "Scopoletin", "id": "CID_5280460"} | INTERACTS_WITH | - | Target | 54600 | UGT1A9 | {"name": "UGT1A9", "id": "54600"} |
| Phytochemical | CID_5280460 | Scopoletin | {"name": "Scopoletin", "id": "CID_5280460"} | INTERACTS_WITH | - | Target | 54657 | UGT1A4 | {"name": "UGT1A4", "id": "54657"} |
| Phytochemical | CID_5280460 | Scopoletin | {"name": "Scopoletin", "id": "CID_5280460"} | INTERACTS_WITH | - | Target | 54658 | UGT1A1 | {"name": "UGT1A1", "id": "54658"} |
| Phytochemical | CID_5280460 | Scopoletin | {"name": "Scopoletin", "id": "CID_5280460"} | INTERACTS_WITH | - | Target | 54659 | UGT1A3 | {"name": "UGT1A3", "id": "54659"} |
| Phytochemical | CID_5280460 | Scopoletin | {"name": "Scopoletin", "id": "CID_5280460"} | INTERACTS_WITH | - | Target | 55775 | TDP1 | {"name": "TDP1", "id": "55775"} |
| Phytochemical | CID_5280460 | Scopoletin | {"name": "Scopoletin", "id": "CID_5280460"} | INTERACTS_WITH | - | Target | 5743 | PTGS2 | {"name": "PTGS2", "id": "5743"} |
| Phytochemical | CID_5280460 | Scopoletin | {"name": "Scopoletin", "id": "CID_5280460"} | INTERACTS_WITH | - | Target | 5901 | RAN | {"name": "RAN", "id": "5901"} |

**Table S9: Complete node and relationship data used to construct the knowledge graph. This table summarizes interactions among FDA-approved drugs, formulation, herbs, diseases (type 2 diabetes, fatty liver, obesity), phytochemicals, and their target genes/proteins. Each entry includes source and target node details, edge type, and node annotations (Entrez Gene ID and gene symbol, where applicable).**

| Source Node Label | Source Node ID | Source Node Name | Source Node Properties | Edge Type | Edge Properties | Target Node Label | Target Node ID | Target Node Name | Target Node Properties |
|---|---|---|---|---|---|---|---|---|---|
| Phytochemical | CID_5280460 | Scopoletin | {"name": "Scopoletin", "id": "CID_5280460"} | INTERACTS_WITH | - | Target | 5965 | RECQL | {"name": "RECQL", "id": "5965"} |
| Phytochemical | CID_5280460 | Scopoletin | {"name": "Scopoletin", "id": "CID_5280460"} | INTERACTS_WITH | - | Target | 5970 | RELA | {"name": "RELA", "id": "5970"} |
| Phytochemical | CID_5280460 | Scopoletin | {"name": "Scopoletin", "id": "CID_5280460"} | INTERACTS_WITH | - | Target | 6311 | ATXN2 | {"name": "ATXN2", "id": "6311"} |
| Phytochemical | CID_5280460 | Scopoletin | {"name": "Scopoletin", "id": "CID_5280460"} | INTERACTS_WITH | - | Target | 64167 | ERAP2 | {"name": "ERAP2", "id": "64167"} |
| Phytochemical | CID_5280460 | Scopoletin | {"name": "Scopoletin", "id": "CID_5280460"} | INTERACTS_WITH | - | Target | 6652 | SORD | {"name": "SORD", "id": "6652"} |
| Phytochemical | CID_5280460 | Scopoletin | {"name": "Scopoletin", "id": "CID_5280460"} | INTERACTS_WITH | - | Target | 7366 | UGT2B15 | {"name": "UGT2B15", "id": "7366"} |
| Phytochemical | CID_5280460 | Scopoletin | {"name": "Scopoletin", "id": "CID_5280460"} | INTERACTS_WITH | - | Target | 7367 | UGT2B17 | {"name": "UGT2B17", "id": "7367"} |
| Phytochemical | CID_5280460 | Scopoletin | {"name": "Scopoletin", "id": "CID_5280460"} | INTERACTS_WITH | - | Target | 759 | CA1 | {"name": "CA1", "id": "759"} |
| Phytochemical | CID_5280460 | Scopoletin | {"name": "Scopoletin", "id": "CID_5280460"} | INTERACTS_WITH | - | Target | 760 | CA2 | {"name": "CA2", "id": "760"} |
| Phytochemical | CID_5280460 | Scopoletin | {"name": "Scopoletin", "id": "CID_5280460"} | INTERACTS_WITH | - | Target | 766 | CA7 | {"name": "CA7", "id": "766"} |
| Phytochemical | CID_5280460 | Scopoletin | {"name": "Scopoletin", "id": "CID_5280460"} | INTERACTS_WITH | - | Target | 768 | CA9 | {"name": "CA9", "id": "768"} |
| Phytochemical | CID_5280460 | Scopoletin | {"name": "Scopoletin", "id": "CID_5280460"} | INTERACTS_WITH | - | Target | 771 | CA12 | {"name": "CA12", "id": "771"} |
| Phytochemical | CID_5280460 | Scopoletin | {"name": "Scopoletin", "id": "CID_5280460"} | INTERACTS_WITH | - | Target | 834 | CASP1 | {"name": "CASP1", "id": "834"} |
| Phytochemical | CID_5280460 | Scopoletin | {"name": "Scopoletin", "id": "CID_5280460"} | INTERACTS_WITH | - | Target | 840 | CASP7 | {"name": "CASP7", "id": "840"} |
| Phytochemical | CID_8742 | Shikimic acid | {"name": "Shikimic acid", "id": "CID_8742"} | INTERACTS_WITH | - | Target | 10599 | SLCO1B1 | {"name": "SLCO1B1", "id": "10599"} |
| Phytochemical | CID_8742 | Shikimic acid | {"name": "Shikimic acid", "id": "CID_8742"} | INTERACTS_WITH | - | Target | 28234 | SLCO1B3 | {"name": "SLCO1B3", "id": "28234"} |
| Phytochemical | CID_8742 | Shikimic acid | {"name": "Shikimic acid", "id": "CID_8742"} | INTERACTS_WITH | - | Target | 328 | APEX1 | {"name": "APEX1", "id": "328"} |
| Phytochemical | CID_8742 | Shikimic acid | {"name": "Shikimic acid", "id": "CID_8742"} | INTERACTS_WITH | - | Target | 51053 | GMNN | {"name": "GMNN", "id": "51053"} |
| Phytochemical | CID_8742 | Shikimic acid | {"name": "Shikimic acid", "id": "CID_8742"} | INTERACTS_WITH | - | Target | 54578 | UGT1A6 | {"name": "UGT1A6", "id": "54578"} |
| Phytochemical | CID_8742 | Shikimic acid | {"name": "Shikimic acid", "id": "CID_8742"} | INTERACTS_WITH | - | Target | 54600 | UGT1A9 | {"name": "UGT1A9", "id": "54600"} |
| Phytochemical | CID_8742 | Shikimic acid | {"name": "Shikimic acid", "id": "CID_8742"} | INTERACTS_WITH | - | Target | 5743 | PTGS2 | {"name": "PTGS2", "id": "5743"} |
| Phytochemical | CID_8742 | Shikimic acid | {"name": "Shikimic acid", "id": "CID_8742"} | INTERACTS_WITH | - | Target | 6401 | SELE | {"name": "SELE", "id": "6401"} |
| Phytochemical | CID_8742 | Shikimic acid | {"name": "Shikimic acid", "id": "CID_8742"} | INTERACTS_WITH | - | Target | 6402 | SELL | {"name": "SELL", "id": "6402"} |
| Phytochemical | CID_8742 | Shikimic acid | {"name": "Shikimic acid", "id": "CID_8742"} | INTERACTS_WITH | - | Target | 6403 | SELP | {"name": "SELP", "id": "6403"} |
| Phytochemical | CID_5281 | Stearic acid | {"name": "Stearic acid", "id": "CID_5281"} | INTERACTS_WITH | - | Target | 11255 | HRH3 | {"name": "HRH3", "id": "11255"} |
| Phytochemical | CID_5281 | Stearic acid | {"name": "Stearic acid", "id": "CID_5281"} | INTERACTS_WITH | - | Target | 1128 | CHRM1 | {"name": "CHRM1", "id": "1128"} |
| Phytochemical | CID_5281 | Stearic acid | {"name": "Stearic acid", "id": "CID_5281"} | INTERACTS_WITH | - | Target | 1129 | CHRM2 | {"name": "CHRM2", "id": "1129"} |
| Phytochemical | CID_5281 | Stearic acid | {"name": "Stearic acid", "id": "CID_5281"} | INTERACTS_WITH | - | Target | 140 | ADORA3 | {"name": "ADORA3", "id": "140"} |
| Phytochemical | CID_5281 | Stearic acid | {"name": "Stearic acid", "id": "CID_5281"} | INTERACTS_WITH | - | Target | 148 | ADRA1A | {"name": "ADRA1A", "id": "148"} |
| Phytochemical | CID_5281 | Stearic acid | {"name": "Stearic acid", "id": "CID_5281"} | INTERACTS_WITH | - | Target | 150 | ADRA2A | {"name": "ADRA2A", "id": "150"} |
| Phytochemical | CID_5281 | Stearic acid | {"name": "Stearic acid", "id": "CID_5281"} | INTERACTS_WITH | - | Target | 1588 | CYP19A1 | {"name": "CYP19A1", "id": "1588"} |
| Phytochemical | CID_5281 | Stearic acid | {"name": "Stearic acid", "id": "CID_5281"} | INTERACTS_WITH | - | Target | 1812 | DRD1 | {"name": "DRD1", "id": "1812"} |
| Phytochemical | CID_5281 | Stearic acid | {"name": "Stearic acid", "id": "CID_5281"} | INTERACTS_WITH | - | Target | 1814 | DRD3 | {"name": "DRD3", "id": "1814"} |
| Phytochemical | CID_5281 | Stearic acid | {"name": "Stearic acid", "id": "CID_5281"} | INTERACTS_WITH | - | Target | 2099 | ESR1 | {"name": "ESR1", "id": "2099"} |
| Phytochemical | CID_5281 | Stearic acid | {"name": "Stearic acid", "id": "CID_5281"} | INTERACTS_WITH | - | Target | 2147 | F2 | {"name": "F2", "id": "2147"} |
| Phytochemical | CID_5281 | Stearic acid | {"name": "Stearic acid", "id": "CID_5281"} | INTERACTS_WITH | - | Target | 2167 | FABP4 | {"name": "FABP4", "id": "2167"} |
| Phytochemical | CID_5281 | Stearic acid | {"name": "Stearic acid", "id": "CID_5281"} | INTERACTS_WITH | - | Target | 246 | ALOX15 | {"name": "ALOX15", "id": "246"} |
| Phytochemical | CID_5281 | Stearic acid | {"name": "Stearic acid", "id": "CID_5281"} | INTERACTS_WITH | - | Target | 2554 | GABRA1 | {"name": "GABRA1", "id": "2554"} |
| Phytochemical | CID_5281 | Stearic acid | {"name": "Stearic acid", "id": "CID_5281"} | INTERACTS_WITH | - | Target | 2555 | GABRA2 | {"name": "GABRA2", "id": "2555"} |
| Phytochemical | CID_5281 | Stearic acid | {"name": "Stearic acid", "id": "CID_5281"} | INTERACTS_WITH | - | Target | 2561 | GABRB2 | {"name": "GABRB2", "id": "2561"} |
| Phytochemical | CID_5281 | Stearic acid | {"name": "Stearic acid", "id": "CID_5281"} | INTERACTS_WITH | - | Target | 2566 | GABRG2 | {"name": "GABRG2", "id": "2566"} |
| Phytochemical | CID_5281 | Stearic acid | {"name": "Stearic acid", "id": "CID_5281"} | INTERACTS_WITH | - | Target | 2908 | NR3C1 | {"name": "NR3C1", "id": "2908"} |
| Phytochemical | CID_5281 | Stearic acid | {"name": "Stearic acid", "id": "CID_5281"} | INTERACTS_WITH | - | Target | 3028 | HSD17B10 | {"name": "HSD17B10", "id": "3028"} |
| Phytochemical | CID_5281 | Stearic acid | {"name": "Stearic acid", "id": "CID_5281"} | INTERACTS_WITH | - | Target | 328 | APEX1 | {"name": "APEX1", "id": "328"} |
| Phytochemical | CID_5281 | Stearic acid | {"name": "Stearic acid", "id": "CID_5281"} | INTERACTS_WITH | - | Target | 3350 | HTR1A | {"name": "HTR1A", "id": "3350"} |
| Phytochemical | CID_5281 | Stearic acid | {"name": "Stearic acid", "id": "CID_5281"} | INTERACTS_WITH | - | Target | 3357 | HTR2B | {"name": "HTR2B", "id": "3357"} |
| Phytochemical | CID_5281 | Stearic acid | {"name": "Stearic acid", "id": "CID_5281"} | INTERACTS_WITH | - | Target | 367 | AR | {"name": "AR", "id": "367"} |
| Phytochemical | CID_5281 | Stearic acid | {"name": "Stearic acid", "id": "CID_5281"} | INTERACTS_WITH | - | Target | 3757 | KCNH2 | {"name": "KCNH2", "id": "3757"} |
| Phytochemical | CID_5281 | Stearic acid | {"name": "Stearic acid", "id": "CID_5281"} | INTERACTS_WITH | - | Target | 3791 | KDR | {"name": "KDR", "id": "3791"} |
| Phytochemical | CID_5281 | Stearic acid | {"name": "Stearic acid", "id": "CID_5281"} | INTERACTS_WITH | - | Target | 4128 | MAOA | {"name": "MAOA", "id": "4128"} |
| Phytochemical | CID_5281 | Stearic acid | {"name": "Stearic acid", "id": "CID_5281"} | INTERACTS_WITH | - | Target | 4297 | KMT2A | {"name": "KMT2A", "id": "4297"} |
| Phytochemical | CID_5281 | Stearic acid | {"name": "Stearic acid", "id": "CID_5281"} | INTERACTS_WITH | - | Target | 43 | ACHE | {"name": "ACHE", "id": "43"} |
| Phytochemical | CID_5281 | Stearic acid | {"name": "Stearic acid", "id": "CID_5281"} | INTERACTS_WITH | - | Target | 4790 | NFKB1 | {"name": "NFKB1", "id": "4790"} |
| Phytochemical | CID_5281 | Stearic acid | {"name": "Stearic acid", "id": "CID_5281"} | INTERACTS_WITH | - | Target | 4988 | OPRM1 | {"name": "OPRM1", "id": "4988"} |
| Phytochemical | CID_5281 | Stearic acid | {"name": "Stearic acid", "id": "CID_5281"} | INTERACTS_WITH | - | Target | 5139 | PDE3A | {"name": "PDE3A", "id": "5139"} |
| Phytochemical | CID_5281 | Stearic acid | {"name": "Stearic acid", "id": "CID_5281"} | INTERACTS_WITH | - | Target | 5141 | PDE4A | {"name": "PDE4A", "id": "5141"} |
| Phytochemical | CID_5281 | Stearic acid | {"name": "Stearic acid", "id": "CID_5281"} | INTERACTS_WITH | - | Target | 5241 | PGR | {"name": "PGR", "id": "5241"} |
| Phytochemical | CID_5281 | Stearic acid | {"name": "Stearic acid", "id": "CID_5281"} | INTERACTS_WITH | - | Target | 5465 | PPARA | {"name": "PPARA", "id": "5465"} |
| Phytochemical | CID_5281 | Stearic acid | {"name": "Stearic acid", "id": "CID_5281"} | INTERACTS_WITH | - | Target | 5467 | PPARD | {"name": "PPARD", "id": "5467"} |
| Phytochemical | CID_5281 | Stearic acid | {"name": "Stearic acid", "id": "CID_5281"} | INTERACTS_WITH | - | Target | 5468 | PPARG | {"name": "PPARG", "id": "5468"} |
| Phytochemical | CID_5281 | Stearic acid | {"name": "Stearic acid", "id": "CID_5281"} | INTERACTS_WITH | - | Target | 55775 | TDP1 | {"name": "TDP1", "id": "55775"} |
| Phytochemical | CID_5281 | Stearic acid | {"name": "Stearic acid", "id": "CID_5281"} | INTERACTS_WITH | - | Target | 5742 | PTGS1 | {"name": "PTGS1", "id": "5742"} |

**Table S9: Complete node and relationship data used to construct the knowledge graph. This table summarizes interactions among FDA-approved drugs, formulation, herbs, diseases (type 2 diabetes, fatty liver, obesity), phytochemicals, and their target genes/proteins. Each entry includes source and target node details, edge type, and node annotations (Entrez Gene ID and gene symbol, where applicable).**

| Source Node Label | Source Node ID | Source Node Name | Source Node Properties | Edge Type | Edge Properties | Target Node Label | Target Node ID | Target Node Name | Target Node Properties |
|---|---|---|---|---|---|---|---|---|---|
| Phytochemical | CID_5281 | Stearic acid | {"name": "Stearic acid", "id": "CID_5281"} | INTERACTS_WITH | - | Target | 5770 | PTPN1 | {"name": "PTPN1", "id": "5770"} |
| Phytochemical | CID_5281 | Stearic acid | {"name": "Stearic acid", "id": "CID_5281"} | INTERACTS_WITH | - | Target | 6530 | SLC6A2 | {"name": "SLC6A2", "id": "6530"} |
| Phytochemical | CID_5281 | Stearic acid | {"name": "Stearic acid", "id": "CID_5281"} | INTERACTS_WITH | - | Target | 6531 | SLC6A3 | {"name": "SLC6A3", "id": "6531"} |
| Phytochemical | CID_5281 | Stearic acid | {"name": "Stearic acid", "id": "CID_5281"} | INTERACTS_WITH | - | Target | 6532 | SLC6A4 | {"name": "SLC6A4", "id": "6532"} |
| Phytochemical | CID_5281 | Stearic acid | {"name": "Stearic acid", "id": "CID_5281"} | INTERACTS_WITH | - | Target | 6915 | TBXA2R | {"name": "TBXA2R", "id": "6915"} |
| Phytochemical | CID_5281 | Stearic acid | {"name": "Stearic acid", "id": "CID_5281"} | INTERACTS_WITH | - | Target | 7068 | THRB | {"name": "THRB", "id": "7068"} |
| Phytochemical | CID_5281 | Stearic acid | {"name": "Stearic acid", "id": "CID_5281"} | INTERACTS_WITH | - | Target | 7253 | TSHR | {"name": "TSHR", "id": "7253"} |
| Phytochemical | CID_5281 | Stearic acid | {"name": "Stearic acid", "id": "CID_5281"} | INTERACTS_WITH | - | Target | 7421 | VDR | {"name": "VDR", "id": "7421"} |
| Phytochemical | CID_5280794 | Stigmasterol | {"name": "Stigmasterol", "id": " CID_5280794"} | INTERACTS_WITH | - | Target | 10013 | HDAC6 | {"name": "HDAC6", "id": "10013"} |
| Phytochemical | CID_5280794 | Stigmasterol | {"name": "Stigmasterol", "id": " CID_5280794"} | INTERACTS_WITH | - | Target | 10599 | SLCO1B1 | {"name": "SLCO1B1", "id": "10599"} |
| Phytochemical | CID_5280794 | Stigmasterol | {"name": "Stigmasterol", "id": " CID_5280794"} | INTERACTS_WITH | - | Target | 28234 | SLCO1B3 | {"name": "SLCO1B3", "id": "28234"} |
| Phytochemical | CID_5280794 | Stigmasterol | {"name": "Stigmasterol", "id": " CID_5280794"} | INTERACTS_WITH | - | Target | 3363 | HTR7 | {"name": "HTR7", "id": "3363"} |
| Phytochemical | CID_5280794 | Stigmasterol | {"name": "Stigmasterol", "id": " CID_5280794"} | INTERACTS_WITH | - | Target | 3417 | IDH1 | {"name": "IDH1", "id": "3417"} |
| Phytochemical | CID_5280794 | Stigmasterol | {"name": "Stigmasterol", "id": " CID_5280794"} | INTERACTS_WITH | - | Target | 5423 | POLB | {"name": "POLB", "id": "5423"} |
| Phytochemical | CID_10742 | Syringic acid | {"name": "Syringic acid", "id": " CID_10742"} | INTERACTS_WITH | - | Target | 11238 | CA5B | {"name": "CA5B", "id": "11238"} |
| Phytochemical | CID_10742 | Syringic acid | {"name": "Syringic acid", "id": " CID_10742"} | INTERACTS_WITH | - | Target | 231 | AKR1B1 | {"name": "AKR1B1", "id": "231"} |
| Phytochemical | CID_10742 | Syringic acid | {"name": "Syringic acid", "id": " CID_10742"} | INTERACTS_WITH | - | Target | 23632 | CA14 | {"name": "CA14", "id": "23632"} |
| Phytochemical | CID_10742 | Syringic acid | {"name": "Syringic acid", "id": " CID_10742"} | INTERACTS_WITH | - | Target | 5315 | PKM | {"name": "PKM", "id": "5315"} |
| Phytochemical | CID_10742 | Syringic acid | {"name": "Syringic acid", "id": " CID_10742"} | INTERACTS_WITH | - | Target | 759 | CA1 | {"name": "CA1", "id": "759"} |
| Phytochemical | CID_10742 | Syringic acid | {"name": "Syringic acid", "id": " CID_10742"} | INTERACTS_WITH | - | Target | 760 | CA2 | {"name": "CA2", "id": "760"} |
| Phytochemical | CID_10742 | Syringic acid | {"name": "Syringic acid", "id": " CID_10742"} | INTERACTS_WITH | - | Target | 761 | CA3 | {"name": "CA3", "id": "761"} |
| Phytochemical | CID_10742 | Syringic acid | {"name": "Syringic acid", "id": " CID_10742"} | INTERACTS_WITH | - | Target | 762 | CA4 | {"name": "CA4", "id": "762"} |
| Phytochemical | CID_10742 | Syringic acid | {"name": "Syringic acid", "id": " CID_10742"} | INTERACTS_WITH | - | Target | 763 | CA5A | {"name": "CA5A", "id": "763"} |
| Phytochemical | CID_10742 | Syringic acid | {"name": "Syringic acid", "id": " CID_10742"} | INTERACTS_WITH | - | Target | 765 | CA6 | {"name": "CA6", "id": "765"} |
| Phytochemical | CID_10742 | Syringic acid | {"name": "Syringic acid", "id": " CID_10742"} | INTERACTS_WITH | - | Target | 766 | CA7 | {"name": "CA7", "id": "766"} |
| Phytochemical | CID_10742 | Syringic acid | {"name": "Syringic acid", "id": " CID_10742"} | INTERACTS_WITH | - | Target | 768 | CA9 | {"name": "CA9", "id": "768"} |
| Phytochemical | CID_10742 | Syringic acid | {"name": "Syringic acid", "id": " CID_10742"} | INTERACTS_WITH | - | Target | 771 | CA12 | {"name": "CA12", "id": "771"} |
| Phytochemical | CID_11463 | Terpinolene | {"name": "Terpinolene", "id": " CID_11463"} | INTERACTS_WITH | - | Target | 5467 | PPARD | {"name": "PPARD", "id": "5467"} |
| Phytochemical | CID_6989 | Thymol | {"name": "Thymol", "id": "CID_6989"} | INTERACTS_WITH | - | Target | 10013 | HDAC6 | {"name": "HDAC6", "id": "10013"} |
| Phytochemical | CID_6989 | Thymol | {"name": "Thymol", "id": "CID_6989"} | INTERACTS_WITH | - | Target | 10941 | UGT2A1 | {"name": "UGT2A1", "id": "10941"} |
| Phytochemical | CID_6989 | Thymol | {"name": "Thymol", "id": "CID_6989"} | INTERACTS_WITH | - | Target | 11255 | HRH3 | {"name": "HRH3", "id": "11255"} |
| Phytochemical | CID_6989 | Thymol | {"name": "Thymol", "id": "CID_6989"} | INTERACTS_WITH | - | Target | 1128 | CHRM1 | {"name": "CHRM1", "id": "1128"} |
| Phytochemical | CID_6989 | Thymol | {"name": "Thymol", "id": "CID_6989"} | INTERACTS_WITH | - | Target | 1129 | CHRM2 | {"name": "CHRM2", "id": "1129"} |
| Phytochemical | CID_6989 | Thymol | {"name": "Thymol", "id": "CID_6989"} | INTERACTS_WITH | - | Target | 140 | ADORA3 | {"name": "ADORA3", "id": "140"} |
| Phytochemical | CID_6989 | Thymol | {"name": "Thymol", "id": "CID_6989"} | INTERACTS_WITH | - | Target | 148 | ADRA1A | {"name": "ADRA1A", "id": "148"} |
| Phytochemical | CID_6989 | Thymol | {"name": "Thymol", "id": "CID_6989"} | INTERACTS_WITH | - | Target | 150 | ADRA2A | {"name": "ADRA2A", "id": "150"} |
| Phytochemical | CID_6989 | Thymol | {"name": "Thymol", "id": "CID_6989"} | INTERACTS_WITH | - | Target | 1812 | DRD1 | {"name": "DRD1", "id": "1812"} |
| Phytochemical | CID_6989 | Thymol | {"name": "Thymol", "id": "CID_6989"} | INTERACTS_WITH | - | Target | 1814 | DRD3 | {"name": "DRD3", "id": "1814"} |
| Phytochemical | CID_6989 | Thymol | {"name": "Thymol", "id": "CID_6989"} | INTERACTS_WITH | - | Target | 2099 | ESR1 | {"name": "ESR1", "id": "2099"} |
| Phytochemical | CID_6989 | Thymol | {"name": "Thymol", "id": "CID_6989"} | INTERACTS_WITH | - | Target | 2147 | F2 | {"name": "F2", "id": "2147"} |
| Phytochemical | CID_6989 | Thymol | {"name": "Thymol", "id": "CID_6989"} | INTERACTS_WITH | - | Target | 2554 | GABRA1 | {"name": "GABRA1", "id": "2554"} |
| Phytochemical | CID_6989 | Thymol | {"name": "Thymol", "id": "CID_6989"} | INTERACTS_WITH | - | Target | 2555 | GABRA2 | {"name": "GABRA2", "id": "2555"} |
| Phytochemical | CID_6989 | Thymol | {"name": "Thymol", "id": "CID_6989"} | INTERACTS_WITH | - | Target | 2561 | GABRB2 | {"name": "GABRB2", "id": "2561"} |
| Phytochemical | CID_6989 | Thymol | {"name": "Thymol", "id": "CID_6989"} | INTERACTS_WITH | - | Target | 2566 | GABRG2 | {"name": "GABRG2", "id": "2566"} |
| Phytochemical | CID_6989 | Thymol | {"name": "Thymol", "id": "CID_6989"} | INTERACTS_WITH | - | Target | 26339 | OR5K1 | {"name": "OR5K1", "id": "26339"} |
| Phytochemical | CID_6989 | Thymol | {"name": "Thymol", "id": "CID_6989"} | INTERACTS_WITH | - | Target | 3315 | HSPB1 | {"name": "HSPB1", "id": "3315"} |
| Phytochemical | CID_6989 | Thymol | {"name": "Thymol", "id": "CID_6989"} | INTERACTS_WITH | - | Target | 3350 | HTR1A | {"name": "HTR1A", "id": "3350"} |
| Phytochemical | CID_6989 | Thymol | {"name": "Thymol", "id": "CID_6989"} | INTERACTS_WITH | - | Target | 3357 | HTR2B | {"name": "HTR2B", "id": "3357"} |
| Phytochemical | CID_6989 | Thymol | {"name": "Thymol", "id": "CID_6989"} | INTERACTS_WITH | - | Target | 367 | AR | {"name": "AR", "id": "367"} |
| Phytochemical | CID_6989 | Thymol | {"name": "Thymol", "id": "CID_6989"} | INTERACTS_WITH | - | Target | 3757 | KCNH2 | {"name": "KCNH2", "id": "3757"} |
| Phytochemical | CID_6989 | Thymol | {"name": "Thymol", "id": "CID_6989"} | INTERACTS_WITH | - | Target | 3791 | KDR | {"name": "KDR", "id": "3791"} |
| Phytochemical | CID_6989 | Thymol | {"name": "Thymol", "id": "CID_6989"} | INTERACTS_WITH | - | Target | 4000 | LMNA | {"name": "LMNA", "id": "4000"} |
| Phytochemical | CID_6989 | Thymol | {"name": "Thymol", "id": "CID_6989"} | INTERACTS_WITH | - | Target | 4128 | MAOA | {"name": "MAOA", "id": "4128"} |
| Phytochemical | CID_6989 | Thymol | {"name": "Thymol", "id": "CID_6989"} | INTERACTS_WITH | - | Target | 43 | ACHE | {"name": "ACHE", "id": "43"} |
| Phytochemical | CID_6989 | Thymol | {"name": "Thymol", "id": "CID_6989"} | INTERACTS_WITH | - | Target | 4780 | NFE2L2 | {"name": "NFE2L2", "id": "4780"} |
| Phytochemical | CID_6989 | Thymol | {"name": "Thymol", "id": "CID_6989"} | INTERACTS_WITH | - | Target | 4988 | OPRM1 | {"name": "OPRM1", "id": "4988"} |
| Phytochemical | CID_6989 | Thymol | {"name": "Thymol", "id": "CID_6989"} | INTERACTS_WITH | - | Target | 5139 | PDE3A | {"name": "PDE3A", "id": "5139"} |
| Phytochemical | CID_6989 | Thymol | {"name": "Thymol", "id": "CID_6989"} | INTERACTS_WITH | - | Target | 5141 | PDE4A | {"name": "PDE4A", "id": "5141"} |
| Phytochemical | CID_6989 | Thymol | {"name": "Thymol", "id": "CID_6989"} | INTERACTS_WITH | - | Target | 5241 | PGR | {"name": "PGR", "id": "5241"} |
| Phytochemical | CID_6989 | Thymol | {"name": "Thymol", "id": "CID_6989"} | INTERACTS_WITH | - | Target | 54578 | UGT1A6 | {"name": "UGT1A6", "id": "54578"} |
| Phytochemical | CID_6989 | Thymol | {"name": "Thymol", "id": "CID_6989"} | INTERACTS_WITH | - | Target | 54600 | UGT1A9 | {"name": "UGT1A9", "id": "54600"} |
| Phytochemical | CID_6989 | Thymol | {"name": "Thymol", "id": "CID_6989"} | INTERACTS_WITH | - | Target | 54657 | UGT1A4 | {"name": "UGT1A4", "id": "54657"} |
| Phytochemical | CID_6989 | Thymol | {"name": "Thymol", "id": "CID_6989"} | INTERACTS_WITH | - | Target | 54658 | UGT1A1 | {"name": "UGT1A1", "id": "54658"} |
| Phytochemical | CID_6989 | Thymol | {"name": "Thymol", "id": "CID_6989"} | INTERACTS_WITH | - | Target | 54659 | UGT1A3 | {"name": "UGT1A3", "id": "54659"} |
| Phytochemical | CID_6989 | Thymol | {"name": "Thymol", "id": "CID_6989"} | INTERACTS_WITH | - | Target | 5742 | PTGS1 | {"name": "PTGS1", "id": "5742"} |

**Table S9: Complete node and relationship data used to construct the knowledge graph. This table summarizes interactions among FDA-approved drugs, formulation, herbs, diseases (type 2 diabetes, fatty liver, obesity), phytochemicals, and their target genes/proteins. Each entry includes source and target node details, edge type, and node annotations (Entrez Gene ID and gene symbol, where applicable).**

| Source Node Label | Source Node ID | Source Node Name | Source Node Properties | Edge Type | Edge Properties | Target Node Label | Target Node ID | Target Node Name | Target Node Properties |
|---|---|---|---|---|---|---|---|---|---|
| Phytochemical | CID_6989 | Thymol | {"name": "Thymol", "id": "CID_6989"} | INTERACTS_WITH | - | Target | 5743 | PTGS2 | {"name": "PTGS2", "id": "5743"} |
| Phytochemical | CID_6989 | Thymol | {"name": "Thymol", "id": "CID_6989"} | INTERACTS_WITH | - | Target | 6530 | SLC6A2 | {"name": "SLC6A2", "id": "6530"} |
| Phytochemical | CID_6989 | Thymol | {"name": "Thymol", "id": "CID_6989"} | INTERACTS_WITH | - | Target | 6531 | SLC6A3 | {"name": "SLC6A3", "id": "6531"} |
| Phytochemical | CID_6989 | Thymol | {"name": "Thymol", "id": "CID_6989"} | INTERACTS_WITH | - | Target | 6532 | SLC6A4 | {"name": "SLC6A4", "id": "6532"} |
| Phytochemical | CID_6989 | Thymol | {"name": "Thymol", "id": "CID_6989"} | INTERACTS_WITH | - | Target | 6915 | TBXA2R | {"name": "TBXA2R", "id": "6915"} |
| Phytochemical | CID_6989 | Thymol | {"name": "Thymol", "id": "CID_6989"} | INTERACTS_WITH | - | Target | 7366 | UGT2B15 | {"name": "UGT2B15", "id": "7366"} |
| Phytochemical | CID_6989 | Thymol | {"name": "Thymol", "id": "CID_6989"} | INTERACTS_WITH | - | Target | 775 | CACNA1C | {"name": "CACNA1C", "id": "775"} |
| Phytochemical | CID_6989 | Thymol | {"name": "Thymol", "id": "CID_6989"} | INTERACTS_WITH | - | Target | 8989 | TRPA1 | {"name": "TRPA1", "id": "8989"} |
| Phytochemical | CID_1140 | Toluene | {"name": "Toluene", "id": "CID_1140"} | INTERACTS_WITH | - | Target | 239 | ALOX12 | {"name": "ALOX12", "id": "239"} |
| Phytochemical | CID_1140 | Toluene | {"name": "Toluene", "id": "CID_1140"} | INTERACTS_WITH | - | Target | 4000 | LMNA | {"name": "LMNA", "id": "4000"} |
| Phytochemical | CID_1140 | Toluene | {"name": "Toluene", "id": "CID_1140"} | INTERACTS_WITH | - | Target | 43 | ACHE | {"name": "ACHE", "id": "43"} |
| Phytochemical | CID_1140 | Toluene | {"name": "Toluene", "id": "CID_1140"} | INTERACTS_WITH | - | Target | 7068 | THRB | {"name": "THRB", "id": "7068"} |
| Phytochemical | CID_1140 | Toluene | {"name": "Toluene", "id": "CID_1140"} | INTERACTS_WITH | - | Target | 7253 | TSHR | {"name": "TSHR", "id": "7253"} |
| Phytochemical | CID_12388 | Tridecane | {"name": "Tridecane", "id": "CID_12388"} | INTERACTS_WITH | - | Target | 5467 | PPARD | {"name": "PPARD", "id": "5467"} |
| Phytochemical | CID_8468 | Vanillic acid | {"name": "Vanillic acid", "id": "CID_8468"} | INTERACTS_WITH | - | Target | 10013 | HDAC6 | {"name": "HDAC6", "id": "10013"} |
| Phytochemical | CID_8468 | Vanillic acid | {"name": "Vanillic acid", "id": "CID_8468"} | INTERACTS_WITH | - | Target | 10014 | HDAC5 | {"name": "HDAC5", "id": "10014"} |
| Phytochemical | CID_8468 | Vanillic acid | {"name": "Vanillic acid", "id": "CID_8468"} | INTERACTS_WITH | - | Target | 1019 | CDK4 | {"name": "CDK4", "id": "1019"} |
| Phytochemical | CID_8468 | Vanillic acid | {"name": "Vanillic acid", "id": "CID_8468"} | INTERACTS_WITH | - | Target | 1021 | CDK6 | {"name": "CDK6", "id": "1021"} |
| Phytochemical | CID_8468 | Vanillic acid | {"name": "Vanillic acid", "id": "CID_8468"} | INTERACTS_WITH | - | Target | 10461 | MERTK | {"name": "MERTK", "id": "10461"} |
| Phytochemical | CID_8468 | Vanillic acid | {"name": "Vanillic acid", "id": "CID_8468"} | INTERACTS_WITH | - | Target | 11201 | POLI | {"name": "POLI", "id": "11201"} |
| Phytochemical | CID_8468 | Vanillic acid | {"name": "Vanillic acid", "id": "CID_8468"} | INTERACTS_WITH | - | Target | 118429 | ANTXR2 | {"name": "ANTXR2", "id": "118429"} |
| Phytochemical | CID_8468 | Vanillic acid | {"name": "Vanillic acid", "id": "CID_8468"} | INTERACTS_WITH | - | Target | 1956 | EGFR | {"name": "EGFR", "id": "1956"} |
| Phytochemical | CID_8468 | Vanillic acid | {"name": "Vanillic acid", "id": "CID_8468"} | INTERACTS_WITH | - | Target | 1969 | EPHA2 | {"name": "EPHA2", "id": "1969"} |
| Phytochemical | CID_8468 | Vanillic acid | {"name": "Vanillic acid", "id": "CID_8468"} | INTERACTS_WITH | - | Target | 2066 | ERBB4 | {"name": "ERBB4", "id": "2066"} |
| Phytochemical | CID_8468 | Vanillic acid | {"name": "Vanillic acid", "id": "CID_8468"} | INTERACTS_WITH | - | Target | 2099 | ESR1 | {"name": "ESR1", "id": "2099"} |
| Phytochemical | CID_8468 | Vanillic acid | {"name": "Vanillic acid", "id": "CID_8468"} | INTERACTS_WITH | - | Target | 2260 | FGFR1 | {"name": "FGFR1", "id": "2260"} |
| Phytochemical | CID_8468 | Vanillic acid | {"name": "Vanillic acid", "id": "CID_8468"} | INTERACTS_WITH | - | Target | 2261 | FGFR3 | {"name": "FGFR3", "id": "2261"} |
| Phytochemical | CID_8468 | Vanillic acid | {"name": "Vanillic acid", "id": "CID_8468"} | INTERACTS_WITH | - | Target | 2263 | FGFR2 | {"name": "FGFR2", "id": "2263"} |
| Phytochemical | CID_8468 | Vanillic acid | {"name": "Vanillic acid", "id": "CID_8468"} | INTERACTS_WITH | - | Target | 231 | AKR1B1 | {"name": "AKR1B1", "id": "231"} |
| Phytochemical | CID_8468 | Vanillic acid | {"name": "Vanillic acid", "id": "CID_8468"} | INTERACTS_WITH | - | Target | 2321 | FLT1 | {"name": "FLT1", "id": "2321"} |
| Phytochemical | CID_8468 | Vanillic acid | {"name": "Vanillic acid", "id": "CID_8468"} | INTERACTS_WITH | - | Target | 238 | ALK | {"name": "ALK", "id": "238"} |
| Phytochemical | CID_8468 | Vanillic acid | {"name": "Vanillic acid", "id": "CID_8468"} | INTERACTS_WITH | - | Target | 25 | ABL1 | {"name": "ABL1", "id": "25"} |
| Phytochemical | CID_8468 | Vanillic acid | {"name": "Vanillic acid", "id": "CID_8468"} | INTERACTS_WITH | - | Target | 3064 | HTT | {"name": "HTT", "id": "3064"} |
| Phytochemical | CID_8468 | Vanillic acid | {"name": "Vanillic acid", "id": "CID_8468"} | INTERACTS_WITH | - | Target | 3065 | HDAC1 | {"name": "HDAC1", "id": "3065"} |
| Phytochemical | CID_8468 | Vanillic acid | {"name": "Vanillic acid", "id": "CID_8468"} | INTERACTS_WITH | - | Target | 3066 | HDAC2 | {"name": "HDAC2", "id": "3066"} |
| Phytochemical | CID_8468 | Vanillic acid | {"name": "Vanillic acid", "id": "CID_8468"} | INTERACTS_WITH | - | Target | 3480 | IGF1R | {"name": "IGF1R", "id": "3480"} |
| Phytochemical | CID_8468 | Vanillic acid | {"name": "Vanillic acid", "id": "CID_8468"} | INTERACTS_WITH | - | Target | 3643 | INSR | {"name": "INSR", "id": "3643"} |
| Phytochemical | CID_8468 | Vanillic acid | {"name": "Vanillic acid", "id": "CID_8468"} | INTERACTS_WITH | - | Target | 3791 | KDR | {"name": "KDR", "id": "3791"} |
| Phytochemical | CID_8468 | Vanillic acid | {"name": "Vanillic acid", "id": "CID_8468"} | INTERACTS_WITH | - | Target | 4058 | LTK | {"name": "LTK", "id": "4058"} |
| Phytochemical | CID_8468 | Vanillic acid | {"name": "Vanillic acid", "id": "CID_8468"} | INTERACTS_WITH | - | Target | 4233 | MET | {"name": "MET", "id": "4233"} |
| Phytochemical | CID_8468 | Vanillic acid | {"name": "Vanillic acid", "id": "CID_8468"} | INTERACTS_WITH | - | Target | 4921 | DDR2 | {"name": "DDR2", "id": "4921"} |
| Phytochemical | CID_8468 | Vanillic acid | {"name": "Vanillic acid", "id": "CID_8468"} | INTERACTS_WITH | - | Target | 5156 | PDGFRA | {"name": "PDGFRA", "id": "5156"} |
| Phytochemical | CID_8468 | Vanillic acid | {"name": "Vanillic acid", "id": "CID_8468"} | INTERACTS_WITH | - | Target | 51564 | HDAC7 | {"name": "HDAC7", "id": "51564"} |
| Phytochemical | CID_8468 | Vanillic acid | {"name": "Vanillic acid", "id": "CID_8468"} | INTERACTS_WITH | - | Target | 5159 | PDGFRB | {"name": "PDGFRB", "id": "5159"} |
| Phytochemical | CID_8468 | Vanillic acid | {"name": "Vanillic acid", "id": "CID_8468"} | INTERACTS_WITH | - | Target | 55869 | HDAC8 | {"name": "HDAC8", "id": "55869"} |
| Phytochemical | CID_8468 | Vanillic acid | {"name": "Vanillic acid", "id": "CID_8468"} | INTERACTS_WITH | - | Target | 5594 | MAPK1 | {"name": "MAPK1", "id": "5594"} |
| Phytochemical | CID_8468 | Vanillic acid | {"name": "Vanillic acid", "id": "CID_8468"} | INTERACTS_WITH | - | Target | 5979 | RET | {"name": "RET", "id": "5979"} |
| Phytochemical | CID_8468 | Vanillic acid | {"name": "Vanillic acid", "id": "CID_8468"} | INTERACTS_WITH | - | Target | 7172 | TPMT | {"name": "TPMT", "id": "7172"} |
| Phytochemical | CID_8468 | Vanillic acid | {"name": "Vanillic acid", "id": "CID_8468"} | INTERACTS_WITH | - | Target | 7301 | TYRO3 | {"name": "TYRO3", "id": "7301"} |
| Phytochemical | CID_8468 | Vanillic acid | {"name": "Vanillic acid", "id": "CID_8468"} | INTERACTS_WITH | - | Target | 79885 | HDAC11 | {"name": "HDAC11", "id": "79885"} |
| Phytochemical | CID_8468 | Vanillic acid | {"name": "Vanillic acid", "id": "CID_8468"} | INTERACTS_WITH | - | Target | 83933 | HDAC10 | {"name": "HDAC10", "id": "83933"} |
| Phytochemical | CID_8468 | Vanillic acid | {"name": "Vanillic acid", "id": "CID_8468"} | INTERACTS_WITH | - | Target | 8841 | HDAC3 | {"name": "HDAC3", "id": "8841"} |
| Phytochemical | CID_8468 | Vanillic acid | {"name": "Vanillic acid", "id": "CID_8468"} | INTERACTS_WITH | - | Target | 9734 | HDAC9 | {"name": "HDAC9", "id": "9734"} |
| Phytochemical | CID_8468 | Vanillic acid | {"name": "Vanillic acid", "id": "CID_8468"} | INTERACTS_WITH | - | Target | 9759 | HDAC4 | {"name": "HDAC4", "id": "9759"} |

**Table S9: Complete node and relationship data used to construct the knowledge graph. This table summarizes interactions among FDA-approved drugs, formulation, herbs, diseases (type 2 diabetes, fatty liver, obesity), phytochemicals, and their target genes/proteins. Each entry includes source and target node details, edge type, and node annotations (Entrez Gene ID and gene symbol, where applicable).**

| Source Node Label | Source Node ID | Source Node Name | Source Node Properties | Edge Type | Edge Properties | Target Node Label | Target Node ID | Target Node Name | Target Node Properties |
|---|---|---|---|---|---|---|---|---|---|
| Phytochemical | CID_8468 | Vanillic acid | {"name": "Vanillic acid", "id": "CID_8468"} | INTERACTS_WITH | - | Target | 983 | CDK1 | {"name": "CDK1", "id": "983"} |
| Phytochemical | CID_7460 | alpha-Phellandrene | {"name": "alpha-Phellandrene", "id": "CID_7460"} | INTERACTS_WITH | - | Target | 4780 | NFE2L2 | {"name": "NFE2L2", "id": "4780"} |
| Phytochemical | CID_6654 | alpha-Pinene | {"name": "alpha-Pinene", "id": "CID_6654"} | INTERACTS_WITH | - | Target | 5594 | MAPK1 | {"name": "MAPK1", "id": "5594"} |
| Phytochemical | CID_6654 | alpha-Pinene | {"name": "alpha-Pinene", "id": "CID_6654"} | INTERACTS_WITH | - | Target | 6097 | RORC | {"name": "RORC", "id": "6097"} |
| Phytochemical | CID_6654 | alpha-Pinene | {"name": "alpha-Pinene", "id": "CID_6654"} | INTERACTS_WITH | - | Target | 7253 | TSHR | {"name": "TSHR", "id": "7253"} |
| Phytochemical | CID_17100 | alpha-Terpineol | {"name": "alpha-Terpineol", "id": "CID_17100"} | INTERACTS_WITH | - | Target | 2554 | GABRA1 | {"name": "GABRA1", "id": "2554"} |
| Phytochemical | CID_17100 | alpha-Terpineol | {"name": "alpha-Terpineol", "id": "CID_17100"} | INTERACTS_WITH | - | Target | 2561 | GABRB2 | {"name": "GABRB2", "id": "2561"} |
| Phytochemical | CID_17100 | alpha-Terpineol | {"name": "alpha-Terpineol", "id": "CID_17100"} | INTERACTS_WITH | - | Target | 2566 | GABRG2 | {"name": "GABRG2", "id": "2566"} |
| Phytochemical | CID_17100 | alpha-Terpineol | {"name": "alpha-Terpineol", "id": "CID_17100"} | INTERACTS_WITH | - | Target | 4000 | LMNA | {"name": "LMNA", "id": "4000"} |
| Phytochemical | CID_160512 | ar-Turmerone | {"name": "ar-Turmerone", "id": "CID_160512"} | INTERACTS_WITH | - | Target | 43 | ACHE | {"name": "ACHE", "id": "43"} |
| Phytochemical | CID_160512 | ar-Turmerone | {"name": "ar-Turmerone", "id": "CID_160512"} | INTERACTS_WITH | - | Target | 6774 | STAT3 | {"name": "STAT3", "id": "6774"} |
| Phytochemical | CID_5280489 | beta-Carotene | {"name": "beta-Carotene", "id": "CID_5280489"} | INTERACTS_WITH | - | Target | 10013 | HDAC6 | {"name": "HDAC6", "id": "10013"} |
| Phytochemical | CID_5280489 | beta-Carotene | {"name": "beta-Carotene", "id": "CID_5280489"} | INTERACTS_WITH | - | Target | 10599 | SLCO1B1 | {"name": "SLCO1B1", "id": "10599"} |
| Phytochemical | CID_5280489 | beta-Carotene | {"name": "beta-Carotene", "id": "CID_5280489"} | INTERACTS_WITH | - | Target | 11255 | HRH3 | {"name": "HRH3", "id": "11255"} |
| Phytochemical | CID_5280489 | beta-Carotene | {"name": "beta-Carotene", "id": "CID_5280489"} | INTERACTS_WITH | - | Target | 1128 | CHRM1 | {"name": "CHRM1", "id": "1128"} |
| Phytochemical | CID_5280489 | beta-Carotene | {"name": "beta-Carotene", "id": "CID_5280489"} | INTERACTS_WITH | - | Target | 1129 | CHRM2 | {"name": "CHRM2", "id": "1129"} |
| Phytochemical | CID_5280489 | beta-Carotene | {"name": "beta-Carotene", "id": "CID_5280489"} | INTERACTS_WITH | - | Target | 140 | ADORA3 | {"name": "ADORA3", "id": "140"} |
| Phytochemical | CID_5280489 | beta-Carotene | {"name": "beta-Carotene", "id": "CID_5280489"} | INTERACTS_WITH | - | Target | 148 | ADRA1A | {"name": "ADRA1A", "id": "148"} |
| Phytochemical | CID_5280489 | beta-Carotene | {"name": "beta-Carotene", "id": "CID_5280489"} | INTERACTS_WITH | - | Target | 150 | ADRA2A | {"name": "ADRA2A", "id": "150"} |
| Phytochemical | CID_5280489 | beta-Carotene | {"name": "beta-Carotene", "id": "CID_5280489"} | INTERACTS_WITH | - | Target | 153 | ADRB1 | {"name": "ADRB1", "id": "153"} |
| Phytochemical | CID_5280489 | beta-Carotene | {"name": "beta-Carotene", "id": "CID_5280489"} | INTERACTS_WITH | - | Target | 154 | ADRB2 | {"name": "ADRB2", "id": "154"} |
| Phytochemical | CID_5280489 | beta-Carotene | {"name": "beta-Carotene", "id": "CID_5280489"} | INTERACTS_WITH | - | Target | 155 | ADRB3 | {"name": "ADRB3", "id": "155"} |
| Phytochemical | CID_5280489 | beta-Carotene | {"name": "beta-Carotene", "id": "CID_5280489"} | INTERACTS_WITH | - | Target | 1812 | DRD1 | {"name": "DRD1", "id": "1812"} |
| Phytochemical | CID_5280489 | beta-Carotene | {"name": "beta-Carotene", "id": "CID_5280489"} | INTERACTS_WITH | - | Target | 1814 | DRD3 | {"name": "DRD3", "id": "1814"} |
| Phytochemical | CID_5280489 | beta-Carotene | {"name": "beta-Carotene", "id": "CID_5280489"} | INTERACTS_WITH | - | Target | 2099 | ESR1 | {"name": "ESR1", "id": "2099"} |
| Phytochemical | CID_5280489 | beta-Carotene | {"name": "beta-Carotene", "id": "CID_5280489"} | INTERACTS_WITH | - | Target | 2147 | F2 | {"name": "F2", "id": "2147"} |
| Phytochemical | CID_5280489 | beta-Carotene | {"name": "beta-Carotene", "id": "CID_5280489"} | INTERACTS_WITH | - | Target | 216 | ALDH1A1 | {"name": "ALDH1A1", "id": "216"} |
| Phytochemical | CID_5280489 | beta-Carotene | {"name": "beta-Carotene", "id": "CID_5280489"} | INTERACTS_WITH | - | Target | 2648 | KAT2A | {"name": "KAT2A", "id": "2648"} |
| Phytochemical | CID_5280489 | beta-Carotene | {"name": "beta-Carotene", "id": "CID_5280489"} | INTERACTS_WITH | - | Target | 28234 | SLCO1B3 | {"name": "SLCO1B3", "id": "28234"} |
| Phytochemical | CID_5280489 | beta-Carotene | {"name": "beta-Carotene", "id": "CID_5280489"} | INTERACTS_WITH | - | Target | 3350 | HTR1A | {"name": "HTR1A", "id": "3350"} |
| Phytochemical | CID_5280489 | beta-Carotene | {"name": "beta-Carotene", "id": "CID_5280489"} | INTERACTS_WITH | - | Target | 3357 | HTR2B | {"name": "HTR2B", "id": "3357"} |
| Phytochemical | CID_5280489 | beta-Carotene | {"name": "beta-Carotene", "id": "CID_5280489"} | INTERACTS_WITH | - | Target | 367 | AR | {"name": "AR", "id": "367"} |
| Phytochemical | CID_5280489 | beta-Carotene | {"name": "beta-Carotene", "id": "CID_5280489"} | INTERACTS_WITH | - | Target | 3757 | KCNH2 | {"name": "KCNH2", "id": "3757"} |
| Phytochemical | CID_5280489 | beta-Carotene | {"name": "beta-Carotene", "id": "CID_5280489"} | INTERACTS_WITH | - | Target | 3791 | KDR | {"name": "KDR", "id": "3791"} |
| Phytochemical | CID_5280489 | beta-Carotene | {"name": "beta-Carotene", "id": "CID_5280489"} | INTERACTS_WITH | - | Target | 4000 | LMNA | {"name": "LMNA", "id": "4000"} |
| Phytochemical | CID_5280489 | beta-Carotene | {"name": "beta-Carotene", "id": "CID_5280489"} | INTERACTS_WITH | - | Target | 4128 | MAOA | {"name": "MAOA", "id": "4128"} |
| Phytochemical | CID_5280489 | beta-Carotene | {"name": "beta-Carotene", "id": "CID_5280489"} | INTERACTS_WITH | - | Target | 43 | ACHE | {"name": "ACHE", "id": "43"} |
| Phytochemical | CID_5280489 | beta-Carotene | {"name": "beta-Carotene", "id": "CID_5280489"} | INTERACTS_WITH | - | Target | 4988 | OPRM1 | {"name": "OPRM1", "id": "4988"} |
| Phytochemical | CID_5280489 | beta-Carotene | {"name": "beta-Carotene", "id": "CID_5280489"} | INTERACTS_WITH | - | Target | 51053 | GMNN | {"name": "GMNN", "id": "51053"} |
| Phytochemical | CID_5280489 | beta-Carotene | {"name": "beta-Carotene", "id": "CID_5280489"} | INTERACTS_WITH | - | Target | 5139 | PDE3A | {"name": "PDE3A", "id": "5139"} |
| Phytochemical | CID_5280489 | beta-Carotene | {"name": "beta-Carotene", "id": "CID_5280489"} | INTERACTS_WITH | - | Target | 5141 | PDE4A | {"name": "PDE4A", "id": "5141"} |
| Phytochemical | CID_5280489 | beta-Carotene | {"name": "beta-Carotene", "id": "CID_5280489"} | INTERACTS_WITH | - | Target | 5241 | PGR | {"name": "PGR", "id": "5241"} |
| Phytochemical | CID_5280489 | beta-Carotene | {"name": "beta-Carotene", "id": "CID_5280489"} | INTERACTS_WITH | - | Target | 55775 | TDP1 | {"name": "TDP1", "id": "55775"} |
| Phytochemical | CID_5280489 | beta-Carotene | {"name": "beta-Carotene", "id": "CID_5280489"} | INTERACTS_WITH | - | Target | 5742 | PTGS1 | {"name": "PTGS1", "id": "5742"} |
| Phytochemical | CID_5280489 | beta-Carotene | {"name": "beta-Carotene", "id": "CID_5280489"} | INTERACTS_WITH | - | Target | 6530 | SLC6A2 | {"name": "SLC6A2", "id": "6530"} |
| Phytochemical | CID_5280489 | beta-Carotene | {"name": "beta-Carotene", "id": "CID_5280489"} | INTERACTS_WITH | - | Target | 6531 | SLC6A3 | {"name": "SLC6A3", "id": "6531"} |
| Phytochemical | CID_5280489 | beta-Carotene | {"name": "beta-Carotene", "id": "CID_5280489"} | INTERACTS_WITH | - | Target | 6532 | SLC6A4 | {"name": "SLC6A4", "id": "6532"} |
| Phytochemical | CID_5280489 | beta-Carotene | {"name": "beta-Carotene", "id": "CID_5280489"} | INTERACTS_WITH | - | Target | 6915 | TBXA2R | {"name": "TBXA2R", "id": "6915"} |
| Phytochemical | CID_5280489 | beta-Carotene | {"name": "beta-Carotene", "id": "CID_5280489"} | INTERACTS_WITH | - | Target | 8647 | ABCB11 | {"name": "ABCB11", "id": "8647"} |

**Table S9: Complete node and relationship data used to construct the knowledge graph. This table summarizes interactions among FDA-approved drugs, formulation, herbs, diseases (type 2 diabetes, fatty liver, obesity), phytochemicals, and their target genes/proteins. Each entry includes source and target node details, edge type, and node annotations (Entrez Gene ID and gene symbol, where applicable).**

| Source Node Label | Source Node ID | Source Node Name | Source Node Properties | Edge Type | Edge Properties | Target Node Label | Target Node ID | Target Node Name | Target Node Properties |
|---|---|---|---|---|---|---|---|---|---|
| Phytochemical | CID_5281515 | beta-Caryophyllene | {"name": "beta-Caryophyllene", "id": "CID_5281515"} | INTERACTS_WITH | - | Target | 10013 | HDAC6 | {"name": "HDAC6", "id": "10013"} |
| Phytochemical | CID_5281515 | beta-Caryophyllene | {"name": "beta-Caryophyllene", "id": "CID_5281515"} | INTERACTS_WITH | - | Target | 10599 | SLCO1B1 | {"name": "SLCO1B1", "id": "10599"} |
| Phytochemical | CID_5281515 | beta-Caryophyllene | {"name": "beta-Caryophyllene", "id": "CID_5281515"} | INTERACTS_WITH | - | Target | 1269 | CNR2 | {"name": "CNR2", "id": "1269"} |
| Phytochemical | CID_5281515 | beta-Caryophyllene | {"name": "beta-Caryophyllene", "id": "CID_5281515"} | INTERACTS_WITH | - | Target | 28234 | SLCO1B3 | {"name": "SLCO1B3", "id": "28234"} |
| Phytochemical | CID_5281515 | beta-Caryophyllene | {"name": "beta-Caryophyllene", "id": "CID_5281515"} | INTERACTS_WITH | - | Target | 4297 | KMT2A | {"name": "KMT2A", "id": "4297"} |
| Phytochemical | CID_5281515 | beta-Caryophyllene | {"name": "beta-Caryophyllene", "id": "CID_5281515"} | INTERACTS_WITH | - | Target | 5465 | PPARA | {"name": "PPARA", "id": "5465"} |
| Phytochemical | CID_6918391 | beta-Elemene | {"name": "beta-Elemene", "id": "CID_6918391"} | INTERACTS_WITH | - | Target | 10013 | HDAC6 | {"name": "HDAC6", "id": "10013"} |
| Phytochemical | CID_6918391 | beta-Elemene | {"name": "beta-Elemene", "id": "CID_6918391"} | INTERACTS_WITH | - | Target | 5315 | PKM | {"name": "PKM", "id": "5315"} |
| Phytochemical | CID_91457 | beta-Eudesmol | {"name": "beta-Eudesmol", "id": "CID_91457"} | INTERACTS_WITH | - | Target | 8989 | TRPA1 | {"name": "TRPA1", "id": "8989"} |
| Phytochemical | CID_124021 | beta-Glucogallin | {"name": "beta-Glucogallin", "id": "CID_124021"} | INTERACTS_WITH | - | Target | 231 | AKR1B1 | {"name": "AKR1B1", "id": "231"} |
| Phytochemical | CID_124021 | beta-Glucogallin | {"name": "beta-Glucogallin", "id": "CID_124021"} | INTERACTS_WITH | - | Target | 23683 | PRKD3 | {"name": "PRKD3", "id": "23683"} |
| Phytochemical | CID_124021 | beta-Glucogallin | {"name": "beta-Glucogallin", "id": "CID_124021"} | INTERACTS_WITH | - | Target | 5578 | PRKCA | {"name": "PRKCA", "id": "5578"} |
| Phytochemical | CID_124021 | beta-Glucogallin | {"name": "beta-Glucogallin", "id": "CID_124021"} | INTERACTS_WITH | - | Target | 5579 | PRKCB | {"name": "PRKCB", "id": "5579"} |
| Phytochemical | CID_124021 | beta-Glucogallin | {"name": "beta-Glucogallin", "id": "CID_124021"} | INTERACTS_WITH | - | Target | 5580 | PRKCD | {"name": "PRKCD", "id": "5580"} |
| Phytochemical | CID_124021 | beta-Glucogallin | {"name": "beta-Glucogallin", "id": "CID_124021"} | INTERACTS_WITH | - | Target | 5581 | PRKCE | {"name": "PRKCE", "id": "5581"} |
| Phytochemical | CID_124021 | beta-Glucogallin | {"name": "beta-Glucogallin", "id": "CID_124021"} | INTERACTS_WITH | - | Target | 5582 | PRKCG | {"name": "PRKCG", "id": "5582"} |
| Phytochemical | CID_124021 | beta-Glucogallin | {"name": "beta-Glucogallin", "id": "CID_124021"} | INTERACTS_WITH | - | Target | 5583 | PRKCH | {"name": "PRKCH", "id": "5583"} |
| Phytochemical | CID_124021 | beta-Glucogallin | {"name": "beta-Glucogallin", "id": "CID_124021"} | INTERACTS_WITH | - | Target | 5584 | PRKCI | {"name": "PRKCI", "id": "5584"} |
| Phytochemical | CID_124021 | beta-Glucogallin | {"name": "beta-Glucogallin", "id": "CID_124021"} | INTERACTS_WITH | - | Target | 5587 | PRKD1 | {"name": "PRKD1", "id": "5587"} |
| Phytochemical | CID_124021 | beta-Glucogallin | {"name": "beta-Glucogallin", "id": "CID_124021"} | INTERACTS_WITH | - | Target | 5588 | PRKCQ | {"name": "PRKCQ", "id": "5588"} |
| Phytochemical | CID_124021 | beta-Glucogallin | {"name": "beta-Glucogallin", "id": "CID_124021"} | INTERACTS_WITH | - | Target | 5590 | PRKCZ | {"name": "PRKCZ", "id": "5590"} |
| Phytochemical | CID_222284 | beta-Sitosterol | {"name": "beta-Sitosterol", "id": "CID_222284"} | INTERACTS_WITH | - | Target | 10013 | HDAC6 | {"name": "HDAC6", "id": "10013"} |
| Phytochemical | CID_222284 | beta-Sitosterol | {"name": "beta-Sitosterol", "id": "CID_222284"} | INTERACTS_WITH | - | Target | 10599 | SLCO1B1 | {"name": "SLCO1B1", "id": "10599"} |
| Phytochemical | CID_222284 | beta-Sitosterol | {"name": "beta-Sitosterol", "id": "CID_222284"} | INTERACTS_WITH | - | Target | 1565 | CYP2D6 | {"name": "CYP2D6", "id": "1565"} |
| Phytochemical | CID_222284 | beta-Sitosterol | {"name": "beta-Sitosterol", "id": "CID_222284"} | INTERACTS_WITH | - | Target | 1576 | CYP3A4 | {"name": "CYP3A4", "id": "1576"} |
| Phytochemical | CID_222284 | beta-Sitosterol | {"name": "beta-Sitosterol", "id": "CID_222284"} | INTERACTS_WITH | - | Target | 1803 | DPP4 | {"name": "DPP4", "id": "1803"} |
| Phytochemical | CID_222284 | beta-Sitosterol | {"name": "beta-Sitosterol", "id": "CID_222284"} | INTERACTS_WITH | - | Target | 2147 | F2 | {"name": "F2", "id": "2147"} |
| Phytochemical | CID_222284 | beta-Sitosterol | {"name": "beta-Sitosterol", "id": "CID_222284"} | INTERACTS_WITH | - | Target | 28234 | SLCO1B3 | {"name": "SLCO1B3", "id": "28234"} |
| Phytochemical | CID_222284 | beta-Sitosterol | {"name": "beta-Sitosterol", "id": "CID_222284"} | INTERACTS_WITH | - | Target | 2932 | GSK3B | {"name": "GSK3B", "id": "2932"} |
| Phytochemical | CID_222284 | beta-Sitosterol | {"name": "beta-Sitosterol", "id": "CID_222284"} | INTERACTS_WITH | - | Target | 3417 | IDH1 | {"name": "IDH1", "id": "3417"} |
| Phytochemical | CID_222284 | beta-Sitosterol | {"name": "beta-Sitosterol", "id": "CID_222284"} | INTERACTS_WITH | - | Target | 51422 | PRKAG2 | {"name": "PRKAG2", "id": "51422"} |
| Phytochemical | CID_222284 | beta-Sitosterol | {"name": "beta-Sitosterol", "id": "CID_222284"} | INTERACTS_WITH | - | Target | 5243 | ABCB1 | {"name": "ABCB1", "id": "5243"} |
| Phytochemical | CID_222284 | beta-Sitosterol | {"name": "beta-Sitosterol", "id": "CID_222284"} | INTERACTS_WITH | - | Target | 53632 | PRKAG3 | {"name": "PRKAG3", "id": "53632"} |
| Phytochemical | CID_222284 | beta-Sitosterol | {"name": "beta-Sitosterol", "id": "CID_222284"} | INTERACTS_WITH | - | Target | 5423 | POLB | {"name": "POLB", "id": "5423"} |
| Phytochemical | CID_222284 | beta-Sitosterol | {"name": "beta-Sitosterol", "id": "CID_222284"} | INTERACTS_WITH | - | Target | 5562 | PRKAA1 | {"name": "PRKAA1", "id": "5562"} |
| Phytochemical | CID_222284 | beta-Sitosterol | {"name": "beta-Sitosterol", "id": "CID_222284"} | INTERACTS_WITH | - | Target | 5563 | PRKAA2 | {"name": "PRKAA2", "id": "5563"} |
| Phytochemical | CID_222284 | beta-Sitosterol | {"name": "beta-Sitosterol", "id": "CID_222284"} | INTERACTS_WITH | - | Target | 5564 | PRKAB1 | {"name": "PRKAB1", "id": "5564"} |
| Phytochemical | CID_222284 | beta-Sitosterol | {"name": "beta-Sitosterol", "id": "CID_222284"} | INTERACTS_WITH | - | Target | 5565 | PRKAB2 | {"name": "PRKAB2", "id": "5565"} |
| Phytochemical | CID_222284 | beta-Sitosterol | {"name": "beta-Sitosterol", "id": "CID_222284"} | INTERACTS_WITH | - | Target | 5571 | PRKAG1 | {"name": "PRKAG1", "id": "5571"} |
| Phytochemical | CID_222284 | beta-Sitosterol | {"name": "beta-Sitosterol", "id": "CID_222284"} | INTERACTS_WITH | - | Target | 7299 | TYR | {"name": "TYR", "id": "7299"} |
| Phytochemical | CID_449093 | trans-Zeatin | {"name": "trans-Zeatin", "id": "CID_449093"} | INTERACTS_WITH | - | Target | 10013 | HDAC6 | {"name": "HDAC6", "id": "10013"} |
| Phytochemical | CID_449093 | trans-Zeatin | {"name": "trans-Zeatin", "id": "CID_449093"} | INTERACTS_WITH | - | Target | 328 | APEX1 | {"name": "APEX1", "id": "328"} |
| Phytochemical | CID_449093 | trans-Zeatin | {"name": "trans-Zeatin", "id": "CID_449093"} | INTERACTS_WITH | - | Target | 5745 | PTH1R | {"name": "PTH1R", "id": "5745"} |
| Phytochemical | CID_449093 | trans-Zeatin | {"name": "trans-Zeatin", "id": "CID_449093"} | INTERACTS_WITH | - | Target | 5965 | RECQL | {"name": "RECQL", "id": "5965"} |
| Phytochemical | CID_449093 | trans-Zeatin | {"name": "trans-Zeatin", "id": "CID_449093"} | INTERACTS_WITH | - | Target | 7253 | TSHR | {"name": "TSHR", "id": "7253"} |
| Target | 19 | ABCA1 | {"name": "ABCA1", "id": "19"} | CORRESPONDS_TO | - | Disease | - | Type 2 Diabetes | {"name": "Type 2 Diabetes"} |
| Target | 5244 | ABCB4 | {"name": "ABCB4", "id": "5244"} | CORRESPONDS_TO | - | Disease | - | Fatty Liver | {"name": "Fatty Liver"} |
| Target | 1244 | ABCC2 | {"name": "ABCC2", "id": "1244"} | CORRESPONDS_TO | - | Disease | - | Fatty Liver | {"name": "Fatty Liver"} |
| Target | 6833 | ABCC8 | {"name": "ABCC8", "id": "6833"} | CORRESPONDS_TO | - | Disease | - | Type 2 Diabetes | {"name": "Type 2 Diabetes"} |
| Target | 10060 | ABCC9 | {"name": "ABCC9", "id": "10060"} | CORRESPONDS_TO | - | Disease | - | Type 2 Diabetes | {"name": "Type 2 Diabetes"} |
| Target | 32 | ACACB | {"name": "ACACB", "id": "32"} | CORRESPONDS_TO | - | Disease | - | Type 2 Diabetes | {"name": "Type 2 Diabetes"} |

**Table S9: Complete node and relationship data used to construct the knowledge graph. This table summarizes interactions among FDA-approved drugs, formulation, herbs, diseases (type 2 diabetes, fatty liver, obesity), phytochemicals, and their target genes/proteins. Each entry includes source and target node details, edge type, and node annotations (Entrez Gene ID and gene symbol, where applicable).**

| Source Node Label | Source Node ID | Source Node Name | Source Node Properties | Edge Type | Edge Properties | Target Node Label | Target Node ID | Target Node Name | Target Node Properties |
|---|---|---|---|---|---|---|---|---|---|
| Target | 32 | ACACB | {"name": "ACACB", "id": "32"} | CORRESPONDS_TO | - | Disease | - | Obesity | {"name": "Obesity"} |
| Target | 34 | ACADM | {"name": "ACADM", "id": "34"} | CORRESPONDS_TO | - | Disease | - | Obesity | {"name": "Obesity"} |
| Target | 1636 | ACE | {"name": "ACE", "id": "1636"} | CORRESPONDS_TO | - | Disease | - | Type 2 Diabetes | {"name": "Type 2 Diabetes"} |
| Target | 1636 | ACE | {"name": "ACE", "id": "1636"} | CORRESPONDS_TO | - | Disease | - | Fatty Liver | {"name": "Fatty Liver"} |
| Target | 1636 | ACE | {"name": "ACE", "id": "1636"} | CORRESPONDS_TO | - | Disease | - | Obesity | {"name": "Obesity"} |
| Target | 43 | ACHE | {"name": "ACHE", "id": "43"} | CORRESPONDS_TO | - | Disease | - | Obesity | {"name": "Obesity"} |
| Target | 47 | ACLY | {"name": "ACLY", "id": "47"} | CORRESPONDS_TO | - | Disease | - | Obesity | {"name": "Obesity"} |
| Target | 54 | ACP5 | {"name": "ACP5", "id": "54"} | CORRESPONDS_TO | - | Disease | - | Obesity | {"name": "Obesity"} |
| Target | 2180 | ACSL1 | {"name": "ACSL1", "id": "2180"} | CORRESPONDS_TO | - | Disease | - | Obesity | {"name": "Obesity"} |
| Target | 56999 | ADAMTS9 | {"name": "ADAMTS9", "id": "56999"} | CORRESPONDS_TO | - | Disease | - | Type 2 Diabetes | {"name": "Type 2 Diabetes"} |
| Target | 109 | ADCY3 | {"name": "ADCY3", "id": "109"} | CORRESPONDS_TO | - | Disease | - | Obesity | {"name": "Obesity"} |
| Target | 111 | ADCY5 | {"name": "ADCY5", "id": "111"} | CORRESPONDS_TO | - | Disease | - | Type 2 Diabetes | {"name": "Type 2 Diabetes"} |
| Target | 127 | ADH4 | {"name": "ADH4", "id": "127"} | CORRESPONDS_TO | - | Disease | - | Fatty Liver | {"name": "Fatty Liver"} |
| Target | 9370 | ADIPOQ | {"name": "ADIPOQ", "id": "9370"} | CORRESPONDS_TO | - | Disease | - | Type 2 Diabetes | {"name": "Type 2 Diabetes"} |
| Target | 9370 | ADIPOQ | {"name": "ADIPOQ", "id": "9370"} | CORRESPONDS_TO | - | Disease | - | Fatty Liver | {"name": "Fatty Liver"} |
| Target | 9370 | ADIPOQ | {"name": "ADIPOQ", "id": "9370"} | CORRESPONDS_TO | - | Disease | - | Obesity | {"name": "Obesity"} |
| Target | 51094 | ADIPOR1 | {"name": "ADIPOR1", "id": "51094"} | CORRESPONDS_TO | - | Disease | - | Type 2 Diabetes | {"name": "Type 2 Diabetes"} |
| Target | 51094 | ADIPOR1 | {"name": "ADIPOR1", "id": "51094"} | CORRESPONDS_TO | - | Disease | - | Obesity | {"name": "Obesity"} |
| Target | 79602 | ADIPOR2 | {"name": "ADIPOR2", "id": "79602"} | CORRESPONDS_TO | - | Disease | - | Type 2 Diabetes | {"name": "Type 2 Diabetes"} |
| Target | 147 | ADRA1B | {"name": "ADRA1B", "id": "147"} | CORRESPONDS_TO | - | Disease | - | Obesity | {"name": "Obesity"} |
| Target | 150 | ADRA2A | {"name": "ADRA2A", "id": "150"} | CORRESPONDS_TO | - | Disease | - | Type 2 Diabetes | {"name": "Type 2 Diabetes"} |
| Target | 153 | ADRB1 | {"name": "ADRB1", "id": "153"} | CORRESPONDS_TO | - | Disease | - | Obesity | {"name": "Obesity"} |
| Target | 154 | ADRB2 | {"name": "ADRB2", "id": "154"} | CORRESPONDS_TO | - | Disease | - | Obesity | {"name": "Obesity"} |
| Target | 155 | ADRB3 | {"name": "ADRB3", "id": "155"} | CORRESPONDS_TO | - | Disease | - | Type 2 Diabetes | {"name": "Type 2 Diabetes"} |
| Target | 155 | ADRB3 | {"name": "ADRB3", "id": "155"} | CORRESPONDS_TO | - | Disease | - | Obesity | {"name": "Obesity"} |
| Target | 27125 | AFF4 | {"name": "AFF4", "id": "27125"} | CORRESPONDS_TO | - | Disease | - | Obesity | {"name": "Obesity"} |
| Target | 177 | AGER | {"name": "AGER", "id": "177"} | CORRESPONDS_TO | - | Disease | - | Type 2 Diabetes | {"name": "Type 2 Diabetes"} |
| Target | 181 | AGRP | {"name": "AGRP", "id": "181"} | CORRESPONDS_TO | - | Disease | - | Type 2 Diabetes | {"name": "Type 2 Diabetes"} |
| Target | 181 | AGRP | {"name": "AGRP", "id": "181"} | CORRESPONDS_TO | - | Disease | - | Obesity | {"name": "Obesity"} |
| Target | 183 | AGT | {"name": "AGT", "id": "183"} | CORRESPONDS_TO | - | Disease | - | Obesity | {"name": "Obesity"} |
| Target | 191 | AHCY | {"name": "AHCY", "id": "191"} | CORRESPONDS_TO | - | Disease | - | Fatty Liver | {"name": "Fatty Liver"} |
| Target | 196 | AHR | {"name": "AHR", "id": "196"} | CORRESPONDS_TO | - | Disease | - | Fatty Liver | {"name": "Fatty Liver"} |
| Target | 196 | AHR | {"name": "AHR", "id": "196"} | CORRESPONDS_TO | - | Disease | - | Obesity | {"name": "Obesity"} |
| Target | 197 | AHSG | {"name": "AHSG", "id": "197"} | CORRESPONDS_TO | - | Disease | - | Type 2 Diabetes | {"name": "Type 2 Diabetes"} |
| Target | 8165 | AKAP1 | {"name": "AKAP1", "id": "8165"} | CORRESPONDS_TO | - | Disease | - | Obesity | {"name": "Obesity"} |
| Target | 207 | AKT1 | {"name": "AKT1", "id": "207"} | CORRESPONDS_TO | - | Disease | - | Type 2 Diabetes | {"name": "Type 2 Diabetes"} |
| Target | 207 | AKT1 | {"name": "AKT1", "id": "207"} | CORRESPONDS_TO | - | Disease | - | Fatty Liver | {"name": "Fatty Liver"} |
| Target | 207 | AKT1 | {"name": "AKT1", "id": "207"} | CORRESPONDS_TO | - | Disease | - | Obesity | {"name": "Obesity"} |
| Target | 208 | AKT2 | {"name": "AKT2", "id": "208"} | CORRESPONDS_TO | - | Disease | - | Type 2 Diabetes | {"name": "Type 2 Diabetes"} |
| Target | 213 | ALB | {"name": "ALB", "id": "213"} | CORRESPONDS_TO | - | Disease | - | Type 2 Diabetes | {"name": "Type 2 Diabetes"} |
| Target | 216 | ALDH1A1 | {"name": "ALDH1A1", "id": "216"} | CORRESPONDS_TO | - | Disease | - | Fatty Liver | {"name": "Fatty Liver"} |
| Target | 219 | ALDH1B1 | {"name": "ALDH1B1", "id": "219"} | CORRESPONDS_TO | - | Disease | - | Fatty Liver | {"name": "Fatty Liver"} |
| Target | 10840 | ALDH1L1 | {"name": "ALDH1L1", "id": "10840"} | CORRESPONDS_TO | - | Disease | - | Obesity | {"name": "Obesity"} |
| Target | 217 | ALDH2 | {"name": "ALDH2", "id": "217"} | CORRESPONDS_TO | - | Disease | - | Fatty Liver | {"name": "Fatty Liver"} |
| Target | 8659 | ALDH4A1 | {"name": "ALDH4A1", "id": "8659"} | CORRESPONDS_TO | - | Disease | - | Fatty Liver | {"name": "Fatty Liver"} |
| Target | 4329 | ALDH6A1 | {"name": "ALDH6A1", "id": "4329"} | CORRESPONDS_TO | - | Disease | - | Obesity | {"name": "Obesity"} |
| Target | 7840 | ALMS1 | {"name": "ALMS1", "id": "7840"} | CORRESPONDS_TO | - | Disease | - | Obesity | {"name": "Obesity"} |
| Target | 51129 | ANGPTL4 | {"name": "ANGPTL4", "id": "51129"} | CORRESPONDS_TO | - | Disease | - | Fatty Liver | {"name": "Fatty Liver"} |
| Target | 51129 | ANGPTL4 | {"name": "ANGPTL4", "id": "51129"} | CORRESPONDS_TO | - | Disease | - | Obesity | {"name": "Obesity"} |
| Target | 10239 | AP3S2 | {"name": "AP3S2", "id": "10239"} | CORRESPONDS_TO | - | Disease | - | Type 2 Diabetes | {"name": "Type 2 Diabetes"} |
| Target | 335 | APOA1 | {"name": "APOA1", "id": "335"} | CORRESPONDS_TO | - | Disease | - | Type 2 Diabetes | {"name": "Type 2 Diabetes"} |
| Target | 338 | APOB | {"name": "APOB", "id": "338"} | CORRESPONDS_TO | - | Disease | - | Type 2 Diabetes | {"name": "Type 2 Diabetes"} |
| Target | 338 | APOB | {"name": "APOB", "id": "338"} | CORRESPONDS_TO | - | Disease | - | Fatty Liver | {"name": "Fatty Liver"} |
| Target | 338 | APOB | {"name": "APOB", "id": "338"} | CORRESPONDS_TO | - | Disease | - | Obesity | {"name": "Obesity"} |
| Target | 345 | APOC3 | {"name": "APOC3", "id": "345"} | CORRESPONDS_TO | - | Disease | - | Type 2 Diabetes | {"name": "Type 2 Diabetes"} |
| Target | 348 | APOE | {"name": "APOE", "id": "348"} | CORRESPONDS_TO | - | Disease | - | Type 2 Diabetes | {"name": "Type 2 Diabetes"} |
| Target | 348 | APOE | {"name": "APOE", "id": "348"} | CORRESPONDS_TO | - | Disease | - | Fatty Liver | {"name": "Fatty Liver"} |
| Target | 348 | APOE | {"name": "APOE", "id": "348"} | CORRESPONDS_TO | - | Disease | - | Obesity | {"name": "Obesity"} |
| Target | 350 | APOH | {"name": "APOH", "id": "350"} | CORRESPONDS_TO | - | Disease | - | Fatty Liver | {"name": "Fatty Liver"} |
| Target | 364 | AQP7 | {"name": "AQP7", "id": "364"} | CORRESPONDS_TO | - | Disease | - | Obesity | {"name": "Obesity"} |
| Target | 84100 | ARL 6.00\|ARL6 | {"name": "ARL 6.00\|ARL6", "id": "84100"} | CORRESPONDS_TO | - | Disease | - | Obesity | {"name": "Obesity"} |
| Target | 434 | ASIP | {"name": "ASIP", "id": "434"} | CORRESPONDS_TO | - | Disease | - | Obesity | {"name": "Obesity"} |
| Target | 467 | ATF3 | {"name": "ATF3", "id": "467"} | CORRESPONDS_TO | - | Disease | - | Type 2 Diabetes | {"name": "Type 2 Diabetes"} |
| Target | 468 | ATF4 | {"name": "ATF4", "id": "468"} | CORRESPONDS_TO | - | Disease | - | Fatty Liver | {"name": "Fatty Liver"} |
| Target | 488 | ATP2A2 | {"name": "ATP2A2", "id": "488"} | CORRESPONDS_TO | - | Disease | - | Type 2 Diabetes | {"name": "Type 2 Diabetes"} |
| Target | 489 | ATP2A3 | {"name": "ATP2A3", "id": "489"} | CORRESPONDS_TO | - | Disease | - | Type 2 Diabetes | {"name": "Type 2 Diabetes"} |
| Target | 506 | ATP5F1B | {"name": "ATP5F1B", "id": "506"} | CORRESPONDS_TO | - | Disease | - | Fatty Liver | {"name": "Fatty Liver"} |
| Target | 93974 | ATP5IF1 | {"name": "ATP5IF1", "id": "93974"} | CORRESPONDS_TO | - | Disease | - | Fatty Liver | {"name": "Fatty Liver"} |
| Target | 540 | ATP7B | {"name": "ATP7B", "id": "540"} | CORRESPONDS_TO | - | Disease | - | Fatty Liver | {"name": "Fatty Liver"} |
| Target | 64756 | ATPAF1 | {"name": "ATPAF1", "id": "64756"} | CORRESPONDS_TO | - | Disease | - | Obesity | {"name": "Obesity"} |
| Target | 26053 | AUTS2 | {"name": "AUTS2", "id": "26053"} | CORRESPONDS_TO | - | Disease | - | Type 2 Diabetes | {"name": "Type 2 Diabetes"} |
| Target | 27087 | B3GAT1 | {"name": "B3GAT1", "id": "27087"} | CORRESPONDS_TO | - | Disease | - | Fatty Liver | {"name": "Fatty Liver"} |
| Target | 581 | BAX | {"name": "BAX", "id": "581"} | CORRESPONDS_TO | - | Disease | - | Type 2 Diabetes | {"name": "Type 2 Diabetes"} |
| Target | 92482 | BBIP1 | {"name": "BBIP1", "id": "92482"} | CORRESPONDS_TO | - | Disease | - | Obesity | {"name": "Obesity"} |
| Target | 582 | BBS1 | {"name": "BBS1", "id": "582"} | CORRESPONDS_TO | - | Disease | - | Obesity | {"name": "Obesity"} |
| Target | 79738 | BBS10 | {"name": "BBS10", "id": "79738"} | CORRESPONDS_TO | - | Disease | - | Obesity | {"name": "Obesity"} |

**Table S9: Complete node and relationship data used to construct the knowledge graph. This table summarizes interactions among FDA-approved drugs, formulation, herbs, diseases (type 2 diabetes, fatty liver, obesity), phytochemicals, and their target genes/proteins. Each entry includes source and target node details, edge type, and node annotations (Entrez Gene ID and gene symbol, where applicable).**

| Source Node Label | Source Node ID | Source Node Name | Source Node Properties | Edge Type | Edge Properties | Target Node Label | Target Node ID | Target Node Name | Target Node Properties |
|---|---|---|---|---|---|---|---|---|---|
| Target | 166379 | BBS12 | {"name": "BBS12", "id": "166379"} | CORRESPONDS_TO | - | Disease | - | Obesity | {"name": "Obesity"} |
| Target | 583 | BBS2 | {"name": "BBS2", "id": "583"} | CORRESPONDS_TO | - | Disease | - | Obesity | {"name": "Obesity"} |
| Target | 585 | BBS4 | {"name": "BBS4", "id": "585"} | CORRESPONDS_TO | - | Disease | - | Obesity | {"name": "Obesity"} |
| Target | 129880 | BBS5 | {"name": "BBS5", "id": "129880"} | CORRESPONDS_TO | - | Disease | - | Obesity | {"name": "Obesity"} |
| Target | 55212 | BBS7 | {"name": "BBS7", "id": "55212"} | CORRESPONDS_TO | - | Disease | - | Obesity | {"name": "Obesity"} |
| Target | 27241 | BBS9 | {"name": "BBS9", "id": "27241"} | CORRESPONDS_TO | - | Disease | - | Obesity | {"name": "Obesity"} |
| Target | 590 | BCHE | {"name": "BCHE", "id": "590"} | CORRESPONDS_TO | - | Disease | - | Obesity | {"name": "Obesity"} |
| Target | 596 | BCL2 | {"name": "BCL2", "id": "596"} | CORRESPONDS_TO | - | Disease | - | Type 2 Diabetes | {"name": "Type 2 Diabetes"} |
| Target | 598 | BCL2L1 | {"name": "BCL2L1", "id": "598"} | CORRESPONDS_TO | - | Disease | - | Type 2 Diabetes | {"name": "Type 2 Diabetes"} |
| Target | 10018 | BCL2L11 | {"name": "BCL2L11", "id": "10018"} | CORRESPONDS_TO | - | Disease | - | Type 2 Diabetes | {"name": "Type 2 Diabetes"} |
| Target | 627 | BDNF | {"name": "BDNF", "id": "627"} | CORRESPONDS_TO | - | Disease | - | Obesity | {"name": "Obesity"} |
| Target | 635 | BHMT | {"name": "BHMT", "id": "635"} | CORRESPONDS_TO | - | Disease | - | Type 2 Diabetes | {"name": "Type 2 Diabetes"} |
| Target | 330 | BIRC3 | {"name": "BIRC3", "id": "330"} | CORRESPONDS_TO | - | Disease | - | Fatty Liver | {"name": "Fatty Liver"} |
| Target | 673 | BRAF | {"name": "BRAF", "id": "673"} | CORRESPONDS_TO | - | Disease | - | Type 2 Diabetes | {"name": "Type 2 Diabetes"} |
| Target | 64897 | C12orf43 | {"name": "C12orf43", "id": "64897"} | CORRESPONDS_TO | - | Disease | - | Type 2 Diabetes | {"name": "Type 2 Diabetes"} |
| Target | 714 | C1QC | {"name": "C1QC", "id": "714"} | CORRESPONDS_TO | - | Disease | - | Fatty Liver | {"name": "Fatty Liver"} |
| Target | 145741 | C2CD4A | {"name": "C2CD4A", "id": "145741"} | CORRESPONDS_TO | - | Disease | - | Type 2 Diabetes | {"name": "Type 2 Diabetes"} |
| Target | 388125 | C2CD4B | {"name": "C2CD4B", "id": "388125"} | CORRESPONDS_TO | - | Disease | - | Type 2 Diabetes | {"name": "Type 2 Diabetes"} |
| Target | 718 | C3 | {"name": "C3", "id": "718"} | CORRESPONDS_TO | - | Disease | - | Type 2 Diabetes | {"name": "Type 2 Diabetes"} |
| Target | 157657 | C8orf37\|CFAP418 | {"name": "C8orf37\|CFAP418", "id": " 157657"} | CORRESPONDS_TO | - | Disease | - | Obesity | {"name": "Obesity"} |
| Target | 761 | CA3 | {"name": "CA3", "id": "761"} | CORRESPONDS_TO | - | Disease | - | Obesity | {"name": "Obesity"} |
| Target | 11132 | CAPN10 | {"name": "CAPN10", "id": "11132"} | CORRESPONDS_TO | - | Disease | - | Type 2 Diabetes | {"name": "Type 2 Diabetes"} |
| Target | 9607 | CARTPT | {"name": "CARTPT", "id": "9607"} | CORRESPONDS_TO | - | Disease | - | Obesity | {"name": "Obesity"} |
| Target | 834 | CASP1 | {"name": "CASP1", "id": "834"} | CORRESPONDS_TO | - | Disease | - | Obesity | {"name": "Obesity"} |
| Target | 836 | CASP3 | {"name": "CASP3", "id": "836"} | CORRESPONDS_TO | - | Disease | - | Type 2 Diabetes | {"name": "Type 2 Diabetes"} |
| Target | 841 | CASP8 | {"name": "CASP8", "id": "841"} | CORRESPONDS_TO | - | Disease | - | Type 2 Diabetes | {"name": "Type 2 Diabetes"} |
| Target | 847 | CAT | {"name": "CAT", "id": "847"} | CORRESPONDS_TO | - | Disease | - | Type 2 Diabetes | {"name": "Type 2 Diabetes"} |
| Target | 847 | CAT | {"name": "CAT", "id": "847"} | CORRESPONDS_TO | - | Disease | - | Fatty Liver | {"name": "Fatty Liver"} |
| Target | 875 | CBS | {"name": "CBS", "id": "875"} | CORRESPONDS_TO | - | Disease | - | Type 2 Diabetes | {"name": "Type 2 Diabetes"} |
| Target | 80212 | CCDC92 | {"name": "CCDC92", "id": "80212"} | CORRESPONDS_TO | - | Disease | - | Type 2 Diabetes | {"name": "Type 2 Diabetes"} |
| Target | 6347 | CCL2 | {"name": "CCL2", "id": "6347"} | CORRESPONDS_TO | - | Disease | - | Type 2 Diabetes | {"name": "Type 2 Diabetes"} |
| Target | 6347 | CCL2 | {"name": "CCL2", "id": "6347"} | CORRESPONDS_TO | - | Disease | - | Fatty Liver | {"name": "Fatty Liver"} |
| Target | 6347 | CCL2 | {"name": "CCL2", "id": "6347"} | CORRESPONDS_TO | - | Disease | - | Obesity | {"name": "Obesity"} |
| Target | 894 | CCND2 | {"name": "CCND2", "id": "894"} | CORRESPONDS_TO | - | Disease | - | Type 2 Diabetes | {"name": "Type 2 Diabetes"} |
| Target | 929 | CD14 | {"name": "CD14", "id": "929"} | CORRESPONDS_TO | - | Disease | - | Fatty Liver | {"name": "Fatty Liver"} |
| Target | 9332 | CD163 | {"name": "CD163", "id": "9332"} | CORRESPONDS_TO | - | Disease | - | Obesity | {"name": "Obesity"} |
| Target | 948 | CD36 | {"name": "CD36", "id": "948"} | CORRESPONDS_TO | - | Disease | - | Type 2 Diabetes | {"name": "Type 2 Diabetes"} |
| Target | 958 | CD40 | {"name": "CD40", "id": "958"} | CORRESPONDS_TO | - | Disease | - | Obesity | {"name": "Obesity"} |
| Target | 968 | CD68 | {"name": "CD68", "id": "968"} | CORRESPONDS_TO | - | Disease | - | Obesity | {"name": "Obesity"} |
| Target | 54901 | CDKAL1 | {"name": "CDKAL1", "id": "54901"} | CORRESPONDS_TO | - | Disease | - | Type 2 Diabetes | {"name": "Type 2 Diabetes"} |
| Target | 54901 | CDKAL1 | {"name": "CDKAL1", "id": "54901"} | CORRESPONDS_TO | - | Disease | - | Obesity | {"name": "Obesity"} |
| Target | 1036 | CDO1 | {"name": "CDO1", "id": "1036"} | CORRESPONDS_TO | - | Disease | - | Type 2 Diabetes | {"name": "Type 2 Diabetes"} |
| Target | 1050 | CEBPA | {"name": "CEBPA", "id": "1050"} | CORRESPONDS_TO | - | Disease | - | Obesity | {"name": "Obesity"} |
| Target | 1051 | CEBPB | {"name": "CEBPB", "id": "1051"} | CORRESPONDS_TO | - | Disease | - | Fatty Liver | {"name": "Fatty Liver"} |
| Target | 84984 | CEP19 | {"name": "CEP19", "id": "84984"} | CORRESPONDS_TO | - | Disease | - | Obesity | {"name": "Obesity"} |
| Target | 80184 | CEP290 | {"name": "CEP290", "id": "80184"} | CORRESPONDS_TO | - | Disease | - | Obesity | {"name": "Obesity"} |
| Target | 1066 | CES1 | {"name": "CES1", "id": "1066"} | CORRESPONDS_TO | - | Disease | - | Obesity | {"name": "Obesity"} |
| Target | 1149 | CIDEA | {"name": "CIDEA", "id": "1149"} | CORRESPONDS_TO | - | Disease | - | Obesity | {"name": "Obesity"} |
| Target | 493856 | CISD2 | {"name": "CISD2", "id": "493856"} | CORRESPONDS_TO | - | Disease | - | Type 2 Diabetes | {"name": "Type 2 Diabetes"} |
| Target | 80790 | CMIP | {"name": "CMIP", "id": "80790"} | CORRESPONDS_TO | - | Disease | - | Type 2 Diabetes | {"name": "Type 2 Diabetes"} |
| Target | 55748 | CNDP2 | {"name": "CNDP2", "id": "55748"} | CORRESPONDS_TO | - | Disease | - | Fatty Liver | {"name": "Fatty Liver"} |
| Target | 22866 | CNKSR2 | {"name": "CNKSR2", "id": "22866"} | CORRESPONDS_TO | - | Disease | - | Type 2 Diabetes | {"name": "Type 2 Diabetes"} |
| Target | 1268 | CNR1 | {"name": "CNR1", "id": "1268"} | CORRESPONDS_TO | - | Disease | - | Obesity | {"name": "Obesity"} |
| Target | 22837 | COBLL1 | {"name": "COBLL1", "id": "22837"} | CORRESPONDS_TO | - | Disease | - | Fatty Liver | {"name": "Fatty Liver"} |
| Target | 1281 | COL3A1 | {"name": "COL3A1", "id": "1281"} | CORRESPONDS_TO | - | Disease | - | Fatty Liver | {"name": "Fatty Liver"} |
| Target | 1312 | COMT | {"name": "COMT", "id": "1312"} | CORRESPONDS_TO | - | Disease | - | Obesity | {"name": "Obesity"} |
| Target | 1350 | COX7C | {"name": "COX7C", "id": "1350"} | CORRESPONDS_TO | - | Disease | - | Obesity | {"name": "Obesity"} |
| Target | 4697 | COXFA4 | {"name": "COXFA4", "id": "4697"} | CORRESPONDS_TO | - | Disease | - | Type 2 Diabetes | {"name": "Type 2 Diabetes"} |
| Target | 56901 | COXFA4L2 | {"name": "COXFA4L2", "id": "56901"} | CORRESPONDS_TO | - | Disease | - | Type 2 Diabetes | {"name": "Type 2 Diabetes"} |
| Target | 1361 | CPB2 | {"name": "CPB2", "id": "1361"} | CORRESPONDS_TO | - | Disease | - | Obesity | {"name": "Obesity"} |
| Target | 1363 | CPE | {"name": "CPE", "id": "1363"} | CORRESPONDS_TO | - | Disease | - | Type 2 Diabetes | {"name": "Type 2 Diabetes"} |
| Target | 1363 | CPE | {"name": "CPE", "id": "1363"} | CORRESPONDS_TO | - | Disease | - | Obesity | {"name": "Obesity"} |
| Target | 1374 | CPT1A | {"name": "CPT1A", "id": "1374"} | CORRESPONDS_TO | - | Disease | - | Type 2 Diabetes | {"name": "Type 2 Diabetes"} |
| Target | 1374 | CPT1A | {"name": "CPT1A", "id": "1374"} | CORRESPONDS_TO | - | Disease | - | Fatty Liver | {"name": "Fatty Liver"} |
| Target | 1401 | CRP | {"name": "CRP", "id": "1401"} | CORRESPONDS_TO | - | Disease | - | Type 2 Diabetes | {"name": "Type 2 Diabetes"} |
| Target | 1401 | CRP | {"name": "CRP", "id": "1401"} | CORRESPONDS_TO | - | Disease | - | Obesity | {"name": "Obesity"} |
| Target | 1431 | CS | {"name": "CS", "id": "1431"} | CORRESPONDS_TO | - | Disease | - | Obesity | {"name": "Obesity"} |
| Target | 1437 | CSF2 | {"name": "CSF2", "id": "1437"} | CORRESPONDS_TO | - | Disease | - | Fatty Liver | {"name": "Fatty Liver"} |
| Target | 1489 | CTF1 | {"name": "CTF1", "id": "1489"} | CORRESPONDS_TO | - | Disease | - | Obesity | {"name": "Obesity"} |
| Target | 1520 | CTSS | {"name": "CTSS", "id": "1520"} | CORRESPONDS_TO | - | Disease | - | Obesity | {"name": "Obesity"} |
| Target | 1528 | CYB5A | {"name": "CYB5A", "id": "1528"} | CORRESPONDS_TO | - | Disease | - | Obesity | {"name": "Obesity"} |
| Target | 1535 | CYBA | {"name": "CYBA", "id": "1535"} | CORRESPONDS_TO | - | Disease | - | Type 2 Diabetes | {"name": "Type 2 Diabetes"} |
| Target | 54205 | CYCS | {"name": "CYCS", "id": "54205"} | CORRESPONDS_TO | - | Disease | - | Fatty Liver | {"name": "Fatty Liver"} |
| Target | 54205 | CYCS | {"name": "CYCS", "id": "54205"} | CORRESPONDS_TO | - | Disease | - | Obesity | {"name": "Obesity"} |
| Target | 1586 | CYP17A1 | {"name": "CYP17A1", "id": "1586"} | CORRESPONDS_TO | - | Disease | - | Fatty Liver | {"name": "Fatty Liver"} |

| Table S9: Complete node and relationship data used to construct the knowledge graph. This table summarizes interactions among FDA-approved drugs, formulation, herbs, diseases (type 2 diabetes, fatty liver, obesity), phytochemicals, and their target genes/proteins. Each entry includes source and target node details, edge type, and node annotations (Entrez Gene ID and gene symbol, where applicable). | | | | | | | | | |
|---|---|---|---|---|---|---|---|---|---|
| **Source Node Label** | **Source Node ID** | **Source Node Name** | **Source Node Properties** | **Edge Type** | **Edge Properties** | **Target Node Label** | **Target Node ID** | **Target Node Name** | **Target Node Properties** |
| Target | 1588 | CYP19A1 | {"name": "CYP19A1", "id": "1588"} | CORRESPONDS_TO | - | Disease | - | Fatty Liver | {"name": "Fatty Liver"} |
| Target | 1544 | CYP1A2 | {"name": "CYP1A2", "id": "1544"} | CORRESPONDS_TO | - | Disease | - | Type 2 Diabetes | {"name": "Type 2 Diabetes"} |
| Target | 1544 | CYP1A2 | {"name": "CYP1A2", "id": "1544"} | CORRESPONDS_TO | - | Disease | - | Fatty Liver | {"name": "Fatty Liver"} |
| Target | 1545 | CYP1B1 | {"name": "CYP1B1", "id": "1545"} | CORRESPONDS_TO | - | Disease | - | Fatty Liver | {"name": "Fatty Liver"} |
| Target | 1545 | CYP1B1 | {"name": "CYP1B1", "id": "1545"} | CORRESPONDS_TO | - | Disease | - | Obesity | {"name": "Obesity"} |
| Target | 56603 | CYP26B1 | {"name": "CYP26B1", "id": "56603"} | CORRESPONDS_TO | - | Disease | - | Obesity | {"name": "Obesity"} |
| Target | 1571 | CYP2E1 | {"name": "CYP2E1", "id": "1571"} | CORRESPONDS_TO | - | Disease | - | Fatty Liver | {"name": "Fatty Liver"} |
| Target | 1571 | CYP2E1 | {"name": "CYP2E1", "id": "1571"} | CORRESPONDS_TO | - | Disease | - | Obesity | {"name": "Obesity"} |
| Target | 51181 | DCXR | {"name": "DCXR", "id": "51181"} | CORRESPONDS_TO | - | Disease | - | Obesity | {"name": "Obesity"} |
| Target | 1649 | DDIT3 | {"name": "DDIT3", "id": "1649"} | CORRESPONDS_TO | - | Disease | - | Fatty Liver | {"name": "Fatty Liver"} |
| Target | 1649 | DDIT3 | {"name": "DDIT3", "id": "1649"} | CORRESPONDS_TO | - | Disease | - | Obesity | {"name": "Obesity"} |
| Target | 8694 | DGAT1 | {"name": "DGAT1", "id": "8694"} | CORRESPONDS_TO | - | Disease | - | Fatty Liver | {"name": "Fatty Liver"} |
| Target | 8694 | DGAT1 | {"name": "DGAT1", "id": "8694"} | CORRESPONDS_TO | - | Disease | - | Obesity | {"name": "Obesity"} |
| Target | 84649 | DGAT2 | {"name": "DGAT2", "id": "84649"} | CORRESPONDS_TO | - | Disease | - | Fatty Liver | {"name": "Fatty Liver"} |
| Target | 8527 | DGKD | {"name": "DGKD", "id": "8527"} | CORRESPONDS_TO | - | Disease | - | Type 2 Diabetes | {"name": "Type 2 Diabetes"} |
| Target | 51635 | DHRS7 | {"name": "DHRS7", "id": "51635"} | CORRESPONDS_TO | - | Disease | - | Fatty Liver | {"name": "Fatty Liver"} |
| Target | 10059 | DNM1L | {"name": "DNM1L", "id": "10059"} | CORRESPONDS_TO | - | Disease | - | Obesity | {"name": "Obesity"} |
| Target | 1788 | DNMT3A | {"name": "DNMT3A", "id": "1788"} | CORRESPONDS_TO | - | Disease | - | Obesity | {"name": "Obesity"} |
| Target | 1803 | DPP4 | {"name": "DPP4", "id": "1803"} | CORRESPONDS_TO | - | Disease | - | Type 2 Diabetes | {"name": "Type 2 Diabetes"} |
| Target | 1806 | DPYD | {"name": "DPYD", "id": "1806"} | CORRESPONDS_TO | - | Disease | - | Obesity | {"name": "Obesity"} |
| Target | 9149 | DYRK1B | {"name": "DYRK1B", "id": "9149"} | CORRESPONDS_TO | - | Disease | - | Obesity | {"name": "Obesity"} |
| Target | 1889 | ECE1 | {"name": "ECE1", "id": "1889"} | CORRESPONDS_TO | - | Disease | - | Type 2 Diabetes | {"name": "Type 2 Diabetes"} |
| Target | 1892 | ECHS1 | {"name": "ECHS1", "id": "1892"} | CORRESPONDS_TO | - | Disease | - | Obesity | {"name": "Obesity"} |
| Target | 1906 | EDN1 | {"name": "EDN1", "id": "1906"} | CORRESPONDS_TO | - | Disease | - | Type 2 Diabetes | {"name": "Type 2 Diabetes"} |
| Target | 1909 | EDNRA | {"name": "EDNRA", "id": "1909"} | CORRESPONDS_TO | - | Disease | - | Type 2 Diabetes | {"name": "Type 2 Diabetes"} |
| Target | 1910 | EDNRB | {"name": "EDNRB", "id": "1910"} | CORRESPONDS_TO | - | Disease | - | Type 2 Diabetes | {"name": "Type 2 Diabetes"} |
| Target | 1947 | EFNB1 | {"name": "EFNB1", "id": "1947"} | CORRESPONDS_TO | - | Disease | - | Obesity | {"name": "Obesity"} |
| Target | 1956 | EGFR | {"name": "EGFR", "id": "1956"} | CORRESPONDS_TO | - | Disease | - | Type 2 Diabetes | {"name": "Type 2 Diabetes"} |
| Target | 27102 | EIF2AK1 | {"name": "EIF2AK1", "id": "27102"} | CORRESPONDS_TO | - | Disease | - | Fatty Liver | {"name": "Fatty Liver"} |
| Target | 5167 | ENPP1 | {"name": "ENPP1", "id": "5167"} | CORRESPONDS_TO | - | Disease | - | Type 2 Diabetes | {"name": "Type 2 Diabetes"} |
| Target | 5167 | ENPP1 | {"name": "ENPP1", "id": "5167"} | CORRESPONDS_TO | - | Disease | - | Obesity | {"name": "Obesity"} |
| Target | 955 | ENTPD6 | {"name": "ENTPD6", "id": "955"} | CORRESPONDS_TO | - | Disease | - | Obesity | {"name": "Obesity"} |
| Target | 2033 | EP300 | {"name": "EP300", "id": "2033"} | CORRESPONDS_TO | - | Disease | - | Obesity | {"name": "Obesity"} |
| Target | 26122 | EPC2 | {"name": "EPC2", "id": "26122"} | CORRESPONDS_TO | - | Disease | - | Type 2 Diabetes | {"name": "Type 2 Diabetes"} |
| Target | 2099 | ESR1 | {"name": "ESR1", "id": "2099"} | CORRESPONDS_TO | - | Disease | - | Fatty Liver | {"name": "Fatty Liver"} |
| Target | 2099 | ESR1 | {"name": "ESR1", "id": "2099"} | CORRESPONDS_TO | - | Disease | - | Obesity | {"name": "Obesity"} |
| Target | 2110 | ETFDH | {"name": "ETFDH", "id": "2110"} | CORRESPONDS_TO | - | Disease | - | Obesity | {"name": "Obesity"} |
| Target | 2113 | ETS1 | {"name": "ETS1", "id": "2113"} | CORRESPONDS_TO | - | Disease | - | Type 2 Diabetes | {"name": "Type 2 Diabetes"} |
| Target | 2147 | F2 | {"name": "F2", "id": "2147"} | CORRESPONDS_TO | - | Disease | - | Fatty Liver | {"name": "Fatty Liver"} |
| Target | 2147 | F2 | {"name": "F2", "id": "2147"} | CORRESPONDS_TO | - | Disease | - | Obesity | {"name": "Obesity"} |
| Target | 2149 | F2R | {"name": "F2R", "id": "2149"} | CORRESPONDS_TO | - | Disease | - | Fatty Liver | {"name": "Fatty Liver"} |
| Target | 2166 | FAAH | {"name": "FAAH", "id": "2166"} | CORRESPONDS_TO | - | Disease | - | Obesity | {"name": "Obesity"} |
| Target | 2167 | FABP4 | {"name": "FABP4", "id": "2167"} | CORRESPONDS_TO | - | Disease | - | Type 2 Diabetes | {"name": "Type 2 Diabetes"} |
| Target | 355 | FAS | {"name": "FAS", "id": "355"} | CORRESPONDS_TO | - | Disease | - | Type 2 Diabetes | {"name": "Type 2 Diabetes"} |
| Target | 355 | FAS | {"name": "FAS", "id": "355"} | CORRESPONDS_TO | - | Disease | - | Fatty Liver | {"name": "Fatty Liver"} |
| Target | 2194 | FASN | {"name": "FASN", "id": "2194"} | CORRESPONDS_TO | - | Disease | - | Obesity | {"name": "Obesity"} |
| Target | 2864 | FFAR1 | {"name": "FFAR1", "id": "2864"} | CORRESPONDS_TO | - | Disease | - | Type 2 Diabetes | {"name": "Type 2 Diabetes"} |
| Target | 26291 | FGF21 | {"name": "FGF21", "id": "26291"} | CORRESPONDS_TO | - | Disease | - | Type 2 Diabetes | {"name": "Type 2 Diabetes"} |
| Target | 26291 | FGF21 | {"name": "FGF21", "id": "26291"} | CORRESPONDS_TO | - | Disease | - | Fatty Liver | {"name": "Fatty Liver"} |
| Target | 26291 | FGF21 | {"name": "FGF21", "id": "26291"} | CORRESPONDS_TO | - | Disease | - | Obesity | {"name": "Obesity"} |
| Target | 55277 | FGGY | {"name": "FGGY", "id": "55277"} | CORRESPONDS_TO | - | Disease | - | Obesity | {"name": "Obesity"} |
| Target | 252995 | FNDC5 | {"name": "FNDC5", "id": "252995"} | CORRESPONDS_TO | - | Disease | - | Type 2 Diabetes | {"name": "Type 2 Diabetes"} |
| Target | 252995 | FNDC5 | {"name": "FNDC5", "id": "252995"} | CORRESPONDS_TO | - | Disease | - | Obesity | {"name": "Obesity"} |
| Target | 2350 | FOLR2 | {"name": "FOLR2", "id": "2350"} | CORRESPONDS_TO | - | Disease | - | Fatty Liver | {"name": "Fatty Liver"} |
| Target | 2353 | FOS | {"name": "FOS", "id": "2353"} | CORRESPONDS_TO | - | Disease | - | Obesity | {"name": "Obesity"} |
| Target | 3169 | FOXA1 | {"name": "FOXA1", "id": "3169"} | CORRESPONDS_TO | - | Disease | - | Fatty Liver | {"name": "Fatty Liver"} |
| Target | 3170 | FOXA2 | {"name": "FOXA2", "id": "3170"} | CORRESPONDS_TO | - | Disease | - | Type 2 Diabetes | {"name": "Type 2 Diabetes"} |
| Target | 2309 | FOXO3 | {"name": "FOXO3", "id": "2309"} | CORRESPONDS_TO | - | Disease | - | Obesity | {"name": "Obesity"} |
| Target | 2512 | FTL | {"name": "FTL", "id": "2512"} | CORRESPONDS_TO | - | Disease | - | Obesity | {"name": "Obesity"} |
| Target | 79068 | FTO | {"name": "FTO", "id": "79068"} | CORRESPONDS_TO | - | Disease | - | Type 2 Diabetes | {"name": "Type 2 Diabetes"} |
| Target | 79068 | FTO | {"name": "FTO", "id": "79068"} | CORRESPONDS_TO | - | Disease | - | Fatty Liver | {"name": "Fatty Liver"} |
| Target | 79068 | FTO | {"name": "FTO", "id": "79068"} | CORRESPONDS_TO | - | Disease | - | Obesity | {"name": "Obesity"} |
| Target | 57818 | G6PC2 | {"name": "G6PC2", "id": "57818"} | CORRESPONDS_TO | - | Disease | - | Type 2 Diabetes | {"name": "Type 2 Diabetes"} |
| Target | 2548 | GAA | {"name": "GAA", "id": "2548"} | CORRESPONDS_TO | - | Disease | - | Type 2 Diabetes | {"name": "Type 2 Diabetes"} |
| Target | 8522 | GAS7 | {"name": "GAS7", "id": "8522"} | CORRESPONDS_TO | - | Disease | - | Obesity | {"name": "Obesity"} |
| Target | 2641 | GCG | {"name": "GCG", "id": "2641"} | CORRESPONDS_TO | - | Disease | - | Type 2 Diabetes | {"name": "Type 2 Diabetes"} |
| Target | 2641 | GCG | {"name": "GCG", "id": "2641"} | CORRESPONDS_TO | - | Disease | - | Obesity | {"name": "Obesity"} |
| Target | 2642 | GCGR | {"name": "GCGR", "id": "2642"} | CORRESPONDS_TO | - | Disease | - | Type 2 Diabetes | {"name": "Type 2 Diabetes"} |
| Target | 2645 | GCK | {"name": "GCK", "id": "2645"} | CORRESPONDS_TO | - | Disease | - | Type 2 Diabetes | {"name": "Type 2 Diabetes"} |
| Target | 2646 | GCKR | {"name": "GCKR", "id": "2646"} | CORRESPONDS_TO | - | Disease | - | Type 2 Diabetes | {"name": "Type 2 Diabetes"} |
| Target | 2729 | GCLC | {"name": "GCLC", "id": "2729"} | CORRESPONDS_TO | - | Disease | - | Type 2 Diabetes | {"name": "Type 2 Diabetes"} |
| Target | 2730 | GCLM | {"name": "GCLM", "id": "2730"} | CORRESPONDS_TO | - | Disease | - | Type 2 Diabetes | {"name": "Type 2 Diabetes"} |
| Target | 2673 | GFPT1 | {"name": "GFPT1", "id": "2673"} | CORRESPONDS_TO | - | Disease | - | Obesity | {"name": "Obesity"} |
| Target | 2688 | GH1 | {"name": "GH1", "id": "2688"} | CORRESPONDS_TO | - | Disease | - | Obesity | {"name": "Obesity"} |
| Target | 51738 | GHRL | {"name": "GHRL", "id": "51738"} | CORRESPONDS_TO | - | Disease | - | Type 2 Diabetes | {"name": "Type 2 Diabetes"} |
| Target | 51738 | GHRL | {"name": "GHRL", "id": "51738"} | CORRESPONDS_TO | - | Disease | - | Obesity | {"name": "Obesity"} |

**Table S9: Complete node and relationship data used to construct the knowledge graph. This table summarizes interactions among FDA-approved drugs, formulation, herbs, diseases (type 2 diabetes, fatty liver, obesity), phytochemicals, and their target genes/proteins. Each entry includes source and target node details, edge type, and node annotations (Entrez Gene ID and gene symbol, where applicable).**

| Source Node Label | Source Node ID | Source Node Name | Source Node Properties | Edge Type | Edge Properties | Target Node Label | Target Node ID | Target Node Name | Target Node Properties |
|---|---|---|---|---|---|---|---|---|---|
| Target | 2695 | GIP | {"name": "GIP", "id": "2695"} | CORRESPONDS_TO | - | Disease | - | Type 2 Diabetes | {"name": "Type 2 Diabetes"} |
| Target | 2695 | GIP | {"name": "GIP", "id": "2695"} | CORRESPONDS_TO | - | Disease | - | Obesity | {"name": "Obesity"} |
| Target | 2731 | GLDC | {"name": "GLDC", "id": "2731"} | CORRESPONDS_TO | - | Disease | - | Obesity | {"name": "Obesity"} |
| Target | 169792 | GLIS3 | {"name": "GLIS3", "id": "169792"} | CORRESPONDS_TO | - | Disease | - | Type 2 Diabetes | {"name": "Type 2 Diabetes"} |
| Target | 2740 | GLP1R | {"name": "GLP1R", "id": "2740"} | CORRESPONDS_TO | - | Disease | - | Type 2 Diabetes | {"name": "Type 2 Diabetes"} |
| Target | 2740 | GLP1R | {"name": "GLP1R", "id": "2740"} | CORRESPONDS_TO | - | Disease | - | Obesity | {"name": "Obesity"} |
| Target | 2752 | GLUL | {"name": "GLUL", "id": "2752"} | CORRESPONDS_TO | - | Disease | - | Obesity | {"name": "Obesity"} |
| Target | 2778 | GNAS | {"name": "GNAS", "id": "2778"} | CORRESPONDS_TO | - | Disease | - | Obesity | {"name": "Obesity"} |
| Target | 2784 | GNB3 | {"name": "GNB3", "id": "2784"} | CORRESPONDS_TO | - | Disease | - | Type 2 Diabetes | {"name": "Type 2 Diabetes"} |
| Target | 2784 | GNB3 | {"name": "GNB3", "id": "2784"} | CORRESPONDS_TO | - | Disease | - | Obesity | {"name": "Obesity"} |
| Target | 27232 | GNMT | {"name": "GNMT", "id": "27232"} | CORRESPONDS_TO | - | Disease | - | Fatty Liver | {"name": "Fatty Liver"} |
| Target | 132789 | GNPDA2 | {"name": "GNPDA2", "id": "132789"} | CORRESPONDS_TO | - | Disease | - | Obesity | {"name": "Obesity"} |
| Target | 2813 | GP2 | {"name": "GP2", "id": "2813"} | CORRESPONDS_TO | - | Disease | - | Type 2 Diabetes | {"name": "Type 2 Diabetes"} |
| Target | 23171 | GPD1L | {"name": "GPD1L", "id": "23171"} | CORRESPONDS_TO | - | Disease | - | Fatty Liver | {"name": "Fatty Liver"} |
| Target | 2820 | GPD2 | {"name": "GPD2", "id": "2820"} | CORRESPONDS_TO | - | Disease | - | Type 2 Diabetes | {"name": "Type 2 Diabetes"} |
| Target | 2840 | GPR17 | {"name": "GPR17", "id": "2840"} | CORRESPONDS_TO | - | Disease | - | Obesity | {"name": "Obesity"} |
| Target | 2875 | GPT | {"name": "GPT", "id": "2875"} | CORRESPONDS_TO | - | Disease | - | Fatty Liver | {"name": "Fatty Liver"} |
| Target | 2876 | GPX1 | {"name": "GPX1", "id": "2876"} | CORRESPONDS_TO | - | Disease | - | Type 2 Diabetes | {"name": "Type 2 Diabetes"} |
| Target | 2876 | GPX1 | {"name": "GPX1", "id": "2876"} | CORRESPONDS_TO | - | Disease | - | Fatty Liver | {"name": "Fatty Liver"} |
| Target | 2876 | GPX1 | {"name": "GPX1", "id": "2876"} | CORRESPONDS_TO | - | Disease | - | Obesity | {"name": "Obesity"} |
| Target | 2878 | GPX3 | {"name": "GPX3", "id": "2878"} | CORRESPONDS_TO | - | Disease | - | Type 2 Diabetes | {"name": "Type 2 Diabetes"} |
| Target | 2878 | GPX3 | {"name": "GPX3", "id": "2878"} | CORRESPONDS_TO | - | Disease | - | Obesity | {"name": "Obesity"} |
| Target | 2879 | GPX4 | {"name": "GPX4", "id": "2879"} | CORRESPONDS_TO | - | Disease | - | Fatty Liver | {"name": "Fatty Liver"} |
| Target | 2888 | GRB14 | {"name": "GRB14", "id": "2888"} | CORRESPONDS_TO | - | Disease | - | Type 2 Diabetes | {"name": "Type 2 Diabetes"} |
| Target | 2938 | GSTA1 | {"name": "GSTA1", "id": "2938"} | CORRESPONDS_TO | - | Disease | - | Fatty Liver | {"name": "Fatty Liver"} |
| Target | 2944 | GSTM1 | {"name": "GSTM1", "id": "2944"} | CORRESPONDS_TO | - | Disease | - | Type 2 Diabetes | {"name": "Type 2 Diabetes"} |
| Target | 2944 | GSTM1 | {"name": "GSTM1", "id": "2944"} | CORRESPONDS_TO | - | Disease | - | Fatty Liver | {"name": "Fatty Liver"} |
| Target | 2950 | GSTP1 | {"name": "GSTP1", "id": "2950"} | CORRESPONDS_TO | - | Disease | - | Fatty Liver | {"name": "Fatty Liver"} |
| Target | 2952 | GSTT1 | {"name": "GSTT1", "id": "2952"} | CORRESPONDS_TO | - | Disease | - | Fatty Liver | {"name": "Fatty Liver"} |
| Target | 3033 | HADH | {"name": "HADH", "id": "3033"} | CORRESPONDS_TO | - | Disease | - | Obesity | {"name": "Obesity"} |
| Target | 3032 | HADHB | {"name": "HADHB", "id": "3032"} | CORRESPONDS_TO | - | Disease | - | Fatty Liver | {"name": "Fatty Liver"} |
| Target | 3038 | HAS3 | {"name": "HAS3", "id": "3038"} | CORRESPONDS_TO | - | Disease | - | Fatty Liver | {"name": "Fatty Liver"} |
| Target | 3039 | HBA1 | {"name": "HBA1", "id": "3039"} | CORRESPONDS_TO | - | Disease | - | Type 2 Diabetes | {"name": "Type 2 Diabetes"} |
| Target | 3060 | HCRT | {"name": "HCRT", "id": "3060"} | CORRESPONDS_TO | - | Disease | - | Obesity | {"name": "Obesity"} |
| Target | 283450 | HECTD4 | {"name": "HECTD4", "id": "283450"} | CORRESPONDS_TO | - | Disease | - | Type 2 Diabetes | {"name": "Type 2 Diabetes"} |
| Target | 3077 | HFE | {"name": "HFE", "id": "3077"} | CORRESPONDS_TO | - | Disease | - | Type 2 Diabetes | {"name": "Type 2 Diabetes"} |
| Target | 3077 | HFE | {"name": "HFE", "id": "3077"} | CORRESPONDS_TO | - | Disease | - | Fatty Liver | {"name": "Fatty Liver"} |
| Target | 3087 | HHEX | {"name": "HHEX", "id": "3087"} | CORRESPONDS_TO | - | Disease | - | Type 2 Diabetes | {"name": "Type 2 Diabetes"} |
| Target | 3087 | HHEX | {"name": "HHEX", "id": "3087"} | CORRESPONDS_TO | - | Disease | - | Fatty Liver | {"name": "Fatty Liver"} |
| Target | 3098 | HK1 | {"name": "HK1", "id": "3098"} | CORRESPONDS_TO | - | Disease | - | Type 2 Diabetes | {"name": "Type 2 Diabetes"} |
| Target | 3099 | HK2 | {"name": "HK2", "id": "3099"} | CORRESPONDS_TO | - | Disease | - | Obesity | {"name": "Obesity"} |
| Target | 10363 | HMG20A | {"name": "HMG20A", "id": "10363"} | CORRESPONDS_TO | - | Disease | - | Type 2 Diabetes | {"name": "Type 2 Diabetes"} |
| Target | 3159 | HMGA1 | {"name": "HMGA1", "id": "3159"} | CORRESPONDS_TO | - | Disease | - | Type 2 Diabetes | {"name": "Type 2 Diabetes"} |
| Target | 3148 | HMGB2 | {"name": "HMGB2", "id": "3148"} | CORRESPONDS_TO | - | Disease | - | Obesity | {"name": "Obesity"} |
| Target | 3156 | HMGCR | {"name": "HMGCR", "id": "3156"} | CORRESPONDS_TO | - | Disease | - | Type 2 Diabetes | {"name": "Type 2 Diabetes"} |
| Target | 3156 | HMGCR | {"name": "HMGCR", "id": "3156"} | CORRESPONDS_TO | - | Disease | - | Fatty Liver | {"name": "Fatty Liver"} |
| Target | 3162 | HMOX1 | {"name": "HMOX1", "id": "3162"} | CORRESPONDS_TO | - | Disease | - | Type 2 Diabetes | {"name": "Type 2 Diabetes"} |
| Target | 3162 | HMOX1 | {"name": "HMOX1", "id": "3162"} | CORRESPONDS_TO | - | Disease | - | Obesity | {"name": "Obesity"} |
| Target | 6927 | HNF1A | {"name": "HNF1A", "id": "6927"} | CORRESPONDS_TO | - | Disease | - | Type 2 Diabetes | {"name": "Type 2 Diabetes"} |
| Target | 6928 | HNF1B | {"name": "HNF1B", "id": "6928"} | CORRESPONDS_TO | - | Disease | - | Type 2 Diabetes | {"name": "Type 2 Diabetes"} |
| Target | 3172 | HNF4A | {"name": "HNF4A", "id": "3172"} | CORRESPONDS_TO | - | Disease | - | Type 2 Diabetes | {"name": "Type 2 Diabetes"} |
| Target | 3215 | HOXB5 | {"name": "HOXB5", "id": "3215"} | CORRESPONDS_TO | - | Disease | - | Obesity | {"name": "Obesity"} |
| Target | 3240 | HP | {"name": "HP", "id": "3240"} | CORRESPONDS_TO | - | Disease | - | Type 2 Diabetes | {"name": "Type 2 Diabetes"} |
| Target | 3263 | HPX | {"name": "HPX", "id": "3263"} | CORRESPONDS_TO | - | Disease | - | Type 2 Diabetes | {"name": "Type 2 Diabetes"} |
| Target | 3290 | HSD11B1 | {"name": "HSD11B1", "id": "3290"} | CORRESPONDS_TO | - | Disease | - | Obesity | {"name": "Obesity"} |
| Target | 3291 | HSD11B2 | {"name": "HSD11B2", "id": "3291"} | CORRESPONDS_TO | - | Disease | - | Obesity | {"name": "Obesity"} |
| Target | 3309 | HSPA5 | {"name": "HSPA5", "id": "3309"} | CORRESPONDS_TO | - | Disease | - | Obesity | {"name": "Obesity"} |
| Target | 3356 | HTR2A | {"name": "HTR2A", "id": "3356"} | CORRESPONDS_TO | - | Disease | - | Obesity | {"name": "Obesity"} |
| Target | 3358 | HTR2C | {"name": "HTR2C", "id": "3358"} | CORRESPONDS_TO | - | Disease | - | Obesity | {"name": "Obesity"} |
| Target | 3375 | IAPP | {"name": "IAPP", "id": "3375"} | CORRESPONDS_TO | - | Disease | - | Type 2 Diabetes | {"name": "Type 2 Diabetes"} |
| Target | 3383 | ICAM1 | {"name": "ICAM1", "id": "3383"} | CORRESPONDS_TO | - | Disease | - | Type 2 Diabetes | {"name": "Type 2 Diabetes"} |
| Target | 3383 | ICAM1 | {"name": "ICAM1", "id": "3383"} | CORRESPONDS_TO | - | Disease | - | Obesity | {"name": "Obesity"} |
| Target | 3397 | ID1 | {"name": "ID1", "id": "3397"} | CORRESPONDS_TO | - | Disease | - | Type 2 Diabetes | {"name": "Type 2 Diabetes"} |
| Target | 3416 | IDE | {"name": "IDE", "id": "3416"} | CORRESPONDS_TO | - | Disease | - | Type 2 Diabetes | {"name": "Type 2 Diabetes"} |
| Target | 3620 | IDO1 | {"name": "IDO1", "id": "3620"} | CORRESPONDS_TO | - | Disease | - | Obesity | {"name": "Obesity"} |
| Target | 10437 | IFI30 | {"name": "IFI30", "id": "10437"} | CORRESPONDS_TO | - | Disease | - | Fatty Liver | {"name": "Fatty Liver"} |
| Target | 3458 | IFNG | {"name": "IFNG", "id": "3458"} | CORRESPONDS_TO | - | Disease | - | Type 2 Diabetes | {"name": "Type 2 Diabetes"} |
| Target | 3479 | IGF1 | {"name": "IGF1", "id": "3479"} | CORRESPONDS_TO | - | Disease | - | Type 2 Diabetes | {"name": "Type 2 Diabetes"} |
| Target | 3479 | IGF1 | {"name": "IGF1", "id": "3479"} | CORRESPONDS_TO | - | Disease | - | Obesity | {"name": "Obesity"} |
| Target | 3481 | IGF2 | {"name": "IGF2", "id": "3481"} | CORRESPONDS_TO | - | Disease | - | Obesity | {"name": "Obesity"} |
| Target | 10644 | IGF2BP2 | {"name": "IGF2BP2", "id": "10644"} | CORRESPONDS_TO | - | Disease | - | Type 2 Diabetes | {"name": "Type 2 Diabetes"} |
| Target | 3485 | IGFBP2 | {"name": "IGFBP2", "id": "3485"} | CORRESPONDS_TO | - | Disease | - | Obesity | {"name": "Obesity"} |
| Target | 8517 | IKBKG | {"name": "IKBKG", "id": "8517"} | CORRESPONDS_TO | - | Disease | - | Fatty Liver | {"name": "Fatty Liver"} |
| Target | 3586 | IL10 | {"name": "IL10", "id": "3586"} | CORRESPONDS_TO | - | Disease | - | Type 2 Diabetes | {"name": "Type 2 Diabetes"} |
| Target | 3597 | IL13RA1 | {"name": "IL13RA1", "id": "3597"} | CORRESPONDS_TO | - | Disease | - | Type 2 Diabetes | {"name": "Type 2 Diabetes"} |
| Target | 3552 | IL1A | {"name": "IL1A", "id": "3552"} | CORRESPONDS_TO | - | Disease | - | Fatty Liver | {"name": "Fatty Liver"} |

**Table S9: Complete node and relationship data used to construct the knowledge graph. This table summarizes interactions among FDA-approved drugs, formulation, herbs, diseases (type 2 diabetes, fatty liver, obesity), phytochemicals, and their target genes/proteins. Each entry includes source and target node details, edge type, and node annotations (Entrez Gene ID and gene symbol, where applicable).**

| Source Node Label | Source Node ID | Source Node Name | Source Node Properties | Edge Type | Edge Properties | Target Node Label | Target Node ID | Target Node Name | Target Node Properties |
|---|---|---|---|---|---|---|---|---|---|
| Target | 3553 | IL1B | {"name": "IL1B", "id": "3553"} | CORRESPONDS_TO | - | Disease | - | Type 2 Diabetes | {"name": "Type 2 Diabetes"} |
| Target | 3553 | IL1B | {"name": "IL1B", "id": "3553"} | CORRESPONDS_TO | - | Disease | - | Fatty Liver | {"name": "Fatty Liver"} |
| Target | 3553 | IL1B | {"name": "IL1B", "id": "3553"} | CORRESPONDS_TO | - | Disease | - | Obesity | {"name": "Obesity"} |
| Target | 3557 | IL1RN | {"name": "IL1RN", "id": "3557"} | CORRESPONDS_TO | - | Disease | - | Fatty Liver | {"name": "Fatty Liver"} |
| Target | 3562 | IL3 | {"name": "IL3", "id": "3562"} | CORRESPONDS_TO | - | Disease | - | Fatty Liver | {"name": "Fatty Liver"} |
| Target | 3565 | IL4 | {"name": "IL4", "id": "3565"} | CORRESPONDS_TO | - | Disease | - | Fatty Liver | {"name": "Fatty Liver"} |
| Target | 3569 | IL6 | {"name": "IL6", "id": "3569"} | CORRESPONDS_TO | - | Disease | - | Type 2 Diabetes | {"name": "Type 2 Diabetes"} |
| Target | 3569 | IL6 | {"name": "IL6", "id": "3569"} | CORRESPONDS_TO | - | Disease | - | Fatty Liver | {"name": "Fatty Liver"} |
| Target | 3569 | IL6 | {"name": "IL6", "id": "3569"} | CORRESPONDS_TO | - | Disease | - | Obesity | {"name": "Obesity"} |
| Target | 3574 | IL7 | {"name": "IL7", "id": "3574"} | CORRESPONDS_TO | - | Disease | - | Obesity | {"name": "Obesity"} |
| Target | 56623 | INPP5E | {"name": "INPP5E", "id": "56623"} | CORRESPONDS_TO | - | Disease | - | Obesity | {"name": "Obesity"} |
| Target | 3636 | INPPL1 | {"name": "INPPL1", "id": "3636"} | CORRESPONDS_TO | - | Disease | - | Type 2 Diabetes | {"name": "Type 2 Diabetes"} |
| Target | 3636 | INPPL1 | {"name": "INPPL1", "id": "3636"} | CORRESPONDS_TO | - | Disease | - | Obesity | {"name": "Obesity"} |
| Target | 3630 | INS | {"name": "INS", "id": "3630"} | CORRESPONDS_TO | - | Disease | - | Type 2 Diabetes | {"name": "Type 2 Diabetes"} |
| Target | 3630 | INS | {"name": "INS", "id": "3630"} | CORRESPONDS_TO | - | Disease | - | Fatty Liver | {"name": "Fatty Liver"} |
| Target | 3630 | INS | {"name": "INS", "id": "3630"} | CORRESPONDS_TO | - | Disease | - | Obesity | {"name": "Obesity"} |
| Target | 723961 | INS-IGF2 | {"name": "INS-IGF2", "id": "723961"} | CORRESPONDS_TO | - | Disease | - | Type 2 Diabetes | {"name": "Type 2 Diabetes"} |
| Target | 3643 | INSR | {"name": "INSR", "id": "3643"} | CORRESPONDS_TO | - | Disease | - | Type 2 Diabetes | {"name": "Type 2 Diabetes"} |
| Target | 3643 | INSR | {"name": "INSR", "id": "3643"} | CORRESPONDS_TO | - | Disease | - | Obesity | {"name": "Obesity"} |
| Target | 3667 | IRS1 | {"name": "IRS1", "id": "3667"} | CORRESPONDS_TO | - | Disease | - | Type 2 Diabetes | {"name": "Type 2 Diabetes"} |
| Target | 3667 | IRS1 | {"name": "IRS1", "id": "3667"} | CORRESPONDS_TO | - | Disease | - | Obesity | {"name": "Obesity"} |
| Target | 8660 | IRS2 | {"name": "IRS2", "id": "8660"} | CORRESPONDS_TO | - | Disease | - | Type 2 Diabetes | {"name": "Type 2 Diabetes"} |
| Target | 3672 | ITGA1 | {"name": "ITGA1", "id": "3672"} | CORRESPONDS_TO | - | Disease | - | Type 2 Diabetes | {"name": "Type 2 Diabetes"} |
| Target | 3684 | ITGAM | {"name": "ITGAM", "id": "3684"} | CORRESPONDS_TO | - | Disease | - | Obesity | {"name": "Obesity"} |
| Target | 55600 | ITLN1 | {"name": "ITLN1", "id": "55600"} | CORRESPONDS_TO | - | Disease | - | Type 2 Diabetes | {"name": "Type 2 Diabetes"} |
| Target | 23338 | JADE2 | {"name": "JADE2", "id": "23338"} | CORRESPONDS_TO | - | Disease | - | Type 2 Diabetes | {"name": "Type 2 Diabetes"} |
| Target | 3717 | JAK2 | {"name": "JAK2", "id": "3717"} | CORRESPONDS_TO | - | Disease | - | Fatty Liver | {"name": "Fatty Liver"} |
| Target | 221895 | JAZF1 | {"name": "JAZF1", "id": "221895"} | CORRESPONDS_TO | - | Disease | - | Type 2 Diabetes | {"name": "Type 2 Diabetes"} |
| Target | 3767 | KCNJ11 | {"name": "KCNJ11", "id": "3767"} | CORRESPONDS_TO | - | Disease | - | Type 2 Diabetes | {"name": "Type 2 Diabetes"} |
| Target | 83795 | KCNK16 | {"name": "KCNK16", "id": "83795"} | CORRESPONDS_TO | - | Disease | - | Type 2 Diabetes | {"name": "Type 2 Diabetes"} |
| Target | 3778 | KCNMA1 | {"name": "KCNMA1", "id": "3778"} | CORRESPONDS_TO | - | Disease | - | Obesity | {"name": "Obesity"} |
| Target | 3784 | KCNQ1 | {"name": "KCNQ1", "id": "3784"} | CORRESPONDS_TO | - | Disease | - | Type 2 Diabetes | {"name": "Type 2 Diabetes"} |
| Target | 157855 | KCNU1 | {"name": "KCNU1", "id": "157855"} | CORRESPONDS_TO | - | Disease | - | Type 2 Diabetes | {"name": "Type 2 Diabetes"} |
| Target | 79047 | KCTD15 | {"name": "KCTD15", "id": "79047"} | CORRESPONDS_TO | - | Disease | - | Obesity | {"name": "Obesity"} |
| Target | 57498 | KIDINS220 | {"name": "KIDINS220", "id": "57498"} | CORRESPONDS_TO | - | Disease | - | Obesity | {"name": "Obesity"} |
| Target | 9365 | KL | {"name": "KL", "id": "9365"} | CORRESPONDS_TO | - | Disease | - | Type 2 Diabetes | {"name": "Type 2 Diabetes"} |
| Target | 152831 | KLB | {"name": "KLB", "id": "152831"} | CORRESPONDS_TO | - | Disease | - | Fatty Liver | {"name": "Fatty Liver"} |
| Target | 8462 | KLF11 | {"name": "KLF11", "id": "8462"} | CORRESPONDS_TO | - | Disease | - | Type 2 Diabetes | {"name": "Type 2 Diabetes"} |
| Target | 136259 | KLF14 | {"name": "KLF14", "id": "136259"} | CORRESPONDS_TO | - | Disease | - | Type 2 Diabetes | {"name": "Type 2 Diabetes"} |
| Target | 283455 | KSR2 | {"name": "KSR2", "id": "283455"} | CORRESPONDS_TO | - | Disease | - | Type 2 Diabetes | {"name": "Type 2 Diabetes"} |
| Target | 283455 | KSR2 | {"name": "KSR2", "id": "283455"} | CORRESPONDS_TO | - | Disease | - | Obesity | {"name": "Obesity"} |
| Target | 114294 | LACTB | {"name": "LACTB", "id": "114294"} | CORRESPONDS_TO | - | Disease | - | Obesity | {"name": "Obesity"} |
| Target | 284217 | LAMA1 | {"name": "LAMA1", "id": "284217"} | CORRESPONDS_TO | - | Disease | - | Fatty Liver | {"name": "Fatty Liver"} |
| Target | 3929 | LBP | {"name": "LBP", "id": "3929"} | CORRESPONDS_TO | - | Disease | - | Obesity | {"name": "Obesity"} |
| Target | 3949 | LDLR | {"name": "LDLR", "id": "3949"} | CORRESPONDS_TO | - | Disease | - | Fatty Liver | {"name": "Fatty Liver"} |
| Target | 3949 | LDLR | {"name": "LDLR", "id": "3949"} | CORRESPONDS_TO | - | Disease | - | Obesity | {"name": "Obesity"} |
| Target | 3952 | LEP | {"name": "LEP", "id": "3952"} | CORRESPONDS_TO | - | Disease | - | Type 2 Diabetes | {"name": "Type 2 Diabetes"} |
| Target | 3952 | LEP | {"name": "LEP", "id": "3952"} | CORRESPONDS_TO | - | Disease | - | Fatty Liver | {"name": "Fatty Liver"} |
| Target | 3952 | LEP | {"name": "LEP", "id": "3952"} | CORRESPONDS_TO | - | Disease | - | Obesity | {"name": "Obesity"} |
| Target | 3953 | LEPR | {"name": "LEPR", "id": "3953"} | CORRESPONDS_TO | - | Disease | - | Type 2 Diabetes | {"name": "Type 2 Diabetes"} |
| Target | 3953 | LEPR | {"name": "LEPR", "id": "3953"} | CORRESPONDS_TO | - | Disease | - | Fatty Liver | {"name": "Fatty Liver"} |
| Target | 3953 | LEPR | {"name": "LEPR", "id": "3953"} | CORRESPONDS_TO | - | Disease | - | Obesity | {"name": "Obesity"} |
| Target | 3976 | LIF | {"name": "LIF", "id": "3976"} | CORRESPONDS_TO | - | Disease | - | Fatty Liver | {"name": "Fatty Liver"} |
| Target | 3990 | LIPC | {"name": "LIPC", "id": "3990"} | CORRESPONDS_TO | - | Disease | - | Type 2 Diabetes | {"name": "Type 2 Diabetes"} |
| Target | 8513 | LIPF | {"name": "LIPF", "id": "8513"} | CORRESPONDS_TO | - | Disease | - | Obesity | {"name": "Obesity"} |
| Target | 4000 | LMNA | {"name": "LMNA", "id": "4000"} | CORRESPONDS_TO | - | Disease | - | Type 2 Diabetes | {"name": "Type 2 Diabetes"} |
| Target | 4023 | LPL | {"name": "LPL", "id": "4023"} | CORRESPONDS_TO | - | Disease | - | Type 2 Diabetes | {"name": "Type 2 Diabetes"} |
| Target | 4023 | LPL | {"name": "LPL", "id": "4023"} | CORRESPONDS_TO | - | Disease | - | Obesity | {"name": "Obesity"} |
| Target | 54585 | LZTFL1 | {"name": "LZTFL1", "id": "54585"} | CORRESPONDS_TO | - | Disease | - | Obesity | {"name": "Obesity"} |
| Target | 10296 | MAEA | {"name": "MAEA", "id": "10296"} | CORRESPONDS_TO | - | Disease | - | Type 2 Diabetes | {"name": "Type 2 Diabetes"} |
| Target | 54551 | MAGEL2 | {"name": "MAGEL2", "id": "54551"} | CORRESPONDS_TO | - | Disease | - | Obesity | {"name": "Obesity"} |
| Target | 9479 | MAPK8IP1 | {"name": "MAPK8IP1", "id": "9479"} | CORRESPONDS_TO | - | Disease | - | Type 2 Diabetes | {"name": "Type 2 Diabetes"} |
| Target | 4143 | MAT1A | {"name": "MAT1A", "id": "4143"} | CORRESPONDS_TO | - | Disease | - | Type 2 Diabetes | {"name": "Type 2 Diabetes"} |
| Target | 4143 | MAT1A | {"name": "MAT1A", "id": "4143"} | CORRESPONDS_TO | - | Disease | - | Fatty Liver | {"name": "Fatty Liver"} |
| Target | 4159 | MC3R | {"name": "MC3R", "id": "4159"} | CORRESPONDS_TO | - | Disease | - | Obesity | {"name": "Obesity"} |
| Target | 4160 | MC4R | {"name": "MC4R", "id": "4160"} | CORRESPONDS_TO | - | Disease | - | Type 2 Diabetes | {"name": "Type 2 Diabetes"} |
| Target | 4160 | MC4R | {"name": "MC4R", "id": "4160"} | CORRESPONDS_TO | - | Disease | - | Obesity | {"name": "Obesity"} |
| Target | 4199 | ME1 | {"name": "ME1", "id": "4199"} | CORRESPONDS_TO | - | Disease | - | Obesity | {"name": "Obesity"} |
| Target | 4233 | MET | {"name": "MET", "id": "4233"} | CORRESPONDS_TO | - | Disease | - | Fatty Liver | {"name": "Fatty Liver"} |
| Target | 8972 | MGAM | {"name": "MGAM", "id": "8972"} | CORRESPONDS_TO | - | Disease | - | Type 2 Diabetes | {"name": "Type 2 Diabetes"} |
| Target | 8195 | MKKS | {"name": "MKKS", "id": "8195"} | CORRESPONDS_TO | - | Disease | - | Obesity | {"name": "Obesity"} |
| Target | 4312 | MMP1 | {"name": "MMP1", "id": "4312"} | CORRESPONDS_TO | - | Disease | - | Fatty Liver | {"name": "Fatty Liver"} |
| Target | 4318 | MMP9 | {"name": "MMP9", "id": "4318"} | CORRESPONDS_TO | - | Disease | - | Obesity | {"name": "Obesity"} |
| Target | 4594 | MMUT | {"name": "MMUT", "id": "4594"} | CORRESPONDS_TO | - | Disease | - | Obesity | {"name": "Obesity"} |
| Target | 112609 | MRAP2 | {"name": "MRAP2", "id": "112609"} | CORRESPONDS_TO | - | Disease | - | Obesity | {"name": "Obesity"} |
| Target | 22808 | MRAS | {"name": "MRAS", "id": "22808"} | CORRESPONDS_TO | - | Disease | - | Type 2 Diabetes | {"name": "Type 2 Diabetes"} |

Table S9: Complete node and relationship data used to construct the knowledge graph. This table summarizes interactions among FDA-approved drugs, formulation, herbs, diseases (type 2 diabetes, fatty liver, obesity), phytochemicals, and their target genes/proteins. Each entry includes source and target node details, edge type, and node annotations (Entrez Gene ID and gene symbol, where applicable).

| Source Node Label | Source Node ID | Source Node Name | Source Node Properties | Edge Type | Edge Properties | Target Node Label | Target Node ID | Target Node Name | Target Node Properties |
|---|---|---|---|---|---|---|---|---|---|
| Target | 4360 | MRC1 | {"name": "MRC1", "id": "4360"} | CORRESPONDS_TO | - | Disease | - | Obesity | {"name": "Obesity"} |
| Target | 57591 | MRTFA | {"name": "MRTFA", "id": "57591"} | CORRESPONDS_TO | - | Disease | - | Fatty Liver | {"name": "Fatty Liver"} |
| Target | 4535 | MT-ND1\|ND1 | {"name": "MT-ND1\|ND1", "id": "4535"} | CORRESPONDS_TO | - | Disease | - | Type 2 Diabetes | {"name": "Type 2 Diabetes"} |
| Target | 4536 | MT-ND2 | {"name": "MT-ND2", "id": "4536"} | CORRESPONDS_TO | - | Disease | - | Type 2 Diabetes | {"name": "Type 2 Diabetes"} |
| Target | 4537 | MT-ND3 | {"name": "MT-ND3", "id": "4537"} | CORRESPONDS_TO | - | Disease | - | Type 2 Diabetes | {"name": "Type 2 Diabetes"} |
| Target | 4538 | MT-ND4 | {"name": "MT-ND4", "id": "4538"} | CORRESPONDS_TO | - | Disease | - | Type 2 Diabetes | {"name": "Type 2 Diabetes"} |
| Target | 4539 | MT-ND4L | {"name": "MT-ND4L", "id": "4539"} | CORRESPONDS_TO | - | Disease | - | Type 2 Diabetes | {"name": "Type 2 Diabetes"} |
| Target | 4540 | MT-ND5 | {"name": "MT-ND5", "id": "4540"} | CORRESPONDS_TO | - | Disease | - | Type 2 Diabetes | {"name": "Type 2 Diabetes"} |
| Target | 4541 | MT-ND6 | {"name": "MT-ND6", "id": "4541"} | CORRESPONDS_TO | - | Disease | - | Type 2 Diabetes | {"name": "Type 2 Diabetes"} |
| Target | 23788 | MTCH2 | {"name": "MTCH2", "id": "23788"} | CORRESPONDS_TO | - | Disease | - | Obesity | {"name": "Obesity"} |
| Target | 4524 | MTHFR | {"name": "MTHFR", "id": "4524"} | CORRESPONDS_TO | - | Disease | - | Type 2 Diabetes | {"name": "Type 2 Diabetes"} |
| Target | 4544 | MTNR1B | {"name": "MTNR1B", "id": "4544"} | CORRESPONDS_TO | - | Disease | - | Type 2 Diabetes | {"name": "Type 2 Diabetes"} |
| Target | 2475 | MTOR | {"name": "MTOR", "id": "2475"} | CORRESPONDS_TO | - | Disease | - | Type 2 Diabetes | {"name": "Type 2 Diabetes"} |
| Target | 4547 | MTTP | {"name": "MTTP", "id": "4547"} | CORRESPONDS_TO | - | Disease | - | Fatty Liver | {"name": "Fatty Liver"} |
| Target | 4547 | MTTP | {"name": "MTTP", "id": "4547"} | CORRESPONDS_TO | - | Disease | - | Obesity | {"name": "Obesity"} |
| Target | 4609 | MYC | {"name": "MYC", "id": "4609"} | CORRESPONDS_TO | - | Disease | - | Fatty Liver | {"name": "Fatty Liver"} |
| Target | 4615 | MYD88 | {"name": "MYD88", "id": "4615"} | CORRESPONDS_TO | - | Disease | - | Fatty Liver | {"name": "Fatty Liver"} |
| Target | 10135 | NAMPT | {"name": "NAMPT", "id": "10135"} | CORRESPONDS_TO | - | Disease | - | Obesity | {"name": "Obesity"} |
| Target | 4694 | NDUFA1 | {"name": "NDUFA1", "id": "4694"} | CORRESPONDS_TO | - | Disease | - | Type 2 Diabetes | {"name": "Type 2 Diabetes"} |
| Target | 4705 | NDUFA10 | {"name": "NDUFA10", "id": "4705"} | CORRESPONDS_TO | - | Disease | - | Type 2 Diabetes | {"name": "Type 2 Diabetes"} |
| Target | 126328 | NDUFA11 | {"name": "NDUFA11", "id": "126328"} | CORRESPONDS_TO | - | Disease | - | Type 2 Diabetes | {"name": "Type 2 Diabetes"} |
| Target | 55967 | NDUFA12 | {"name": "NDUFA12", "id": "55967"} | CORRESPONDS_TO | - | Disease | - | Type 2 Diabetes | {"name": "Type 2 Diabetes"} |
| Target | 51079 | NDUFA13 | {"name": "NDUFA13", "id": "51079"} | CORRESPONDS_TO | - | Disease | - | Type 2 Diabetes | {"name": "Type 2 Diabetes"} |
| Target | 4695 | NDUFA2 | {"name": "NDUFA2", "id": "4695"} | CORRESPONDS_TO | - | Disease | - | Type 2 Diabetes | {"name": "Type 2 Diabetes"} |
| Target | 4696 | NDUFA3 | {"name": "NDUFA3", "id": "4696"} | CORRESPONDS_TO | - | Disease | - | Type 2 Diabetes | {"name": "Type 2 Diabetes"} |
| Target | 4698 | NDUFA5 | {"name": "NDUFA5", "id": "4698"} | CORRESPONDS_TO | - | Disease | - | Type 2 Diabetes | {"name": "Type 2 Diabetes"} |
| Target | 4700 | NDUFA6 | {"name": "NDUFA6", "id": "4700"} | CORRESPONDS_TO | - | Disease | - | Type 2 Diabetes | {"name": "Type 2 Diabetes"} |
| Target | 4701 | NDUFA7 | {"name": "NDUFA7", "id": "4701"} | CORRESPONDS_TO | - | Disease | - | Type 2 Diabetes | {"name": "Type 2 Diabetes"} |
| Target | 4702 | NDUFA8 | {"name": "NDUFA8", "id": "4702"} | CORRESPONDS_TO | - | Disease | - | Type 2 Diabetes | {"name": "Type 2 Diabetes"} |
| Target | 4704 | NDUFA9 | {"name": "NDUFA9", "id": "4704"} | CORRESPONDS_TO | - | Disease | - | Type 2 Diabetes | {"name": "Type 2 Diabetes"} |
| Target | 4706 | NDUFAB1 | {"name": "NDUFAB1", "id": "4706"} | CORRESPONDS_TO | - | Disease | - | Type 2 Diabetes | {"name": "Type 2 Diabetes"} |
| Target | 51103 | NDUFAF1 | {"name": "NDUFAF1", "id": "51103"} | CORRESPONDS_TO | - | Disease | - | Type 2 Diabetes | {"name": "Type 2 Diabetes"} |
| Target | 91942 | NDUFAF2 | {"name": "NDUFAF2", "id": "91942"} | CORRESPONDS_TO | - | Disease | - | Type 2 Diabetes | {"name": "Type 2 Diabetes"} |
| Target | 25915 | NDUFAF3 | {"name": "NDUFAF3", "id": "25915"} | CORRESPONDS_TO | - | Disease | - | Type 2 Diabetes | {"name": "Type 2 Diabetes"} |
| Target | 29078 | NDUFAF4 | {"name": "NDUFAF4", "id": "29078"} | CORRESPONDS_TO | - | Disease | - | Type 2 Diabetes | {"name": "Type 2 Diabetes"} |
| Target | 4707 | NDUFB1 | {"name": "NDUFB1", "id": "4707"} | CORRESPONDS_TO | - | Disease | - | Type 2 Diabetes | {"name": "Type 2 Diabetes"} |
| Target | 4716 | NDUFB10 | {"name": "NDUFB10", "id": "4716"} | CORRESPONDS_TO | - | Disease | - | Type 2 Diabetes | {"name": "Type 2 Diabetes"} |
| Target | 54539 | NDUFB11 | {"name": "NDUFB11", "id": "54539"} | CORRESPONDS_TO | - | Disease | - | Type 2 Diabetes | {"name": "Type 2 Diabetes"} |
| Target | 4708 | NDUFB2 | {"name": "NDUFB2", "id": "4708"} | CORRESPONDS_TO | - | Disease | - | Type 2 Diabetes | {"name": "Type 2 Diabetes"} |
| Target | 4709 | NDUFB3 | {"name": "NDUFB3", "id": "4709"} | CORRESPONDS_TO | - | Disease | - | Type 2 Diabetes | {"name": "Type 2 Diabetes"} |
| Target | 4710 | NDUFB4 | {"name": "NDUFB4", "id": "4710"} | CORRESPONDS_TO | - | Disease | - | Type 2 Diabetes | {"name": "Type 2 Diabetes"} |
| Target | 4711 | NDUFB5 | {"name": "NDUFB5", "id": "4711"} | CORRESPONDS_TO | - | Disease | - | Type 2 Diabetes | {"name": "Type 2 Diabetes"} |
| Target | 4712 | NDUFB6 | {"name": "NDUFB6", "id": "4712"} | CORRESPONDS_TO | - | Disease | - | Type 2 Diabetes | {"name": "Type 2 Diabetes"} |
| Target | 4713 | NDUFB7 | {"name": "NDUFB7", "id": "4713"} | CORRESPONDS_TO | - | Disease | - | Type 2 Diabetes | {"name": "Type 2 Diabetes"} |
| Target | 4714 | NDUFB8 | {"name": "NDUFB8", "id": "4714"} | CORRESPONDS_TO | - | Disease | - | Type 2 Diabetes | {"name": "Type 2 Diabetes"} |
| Target | 4715 | NDUFB9 | {"name": "NDUFB9", "id": "4715"} | CORRESPONDS_TO | - | Disease | - | Type 2 Diabetes | {"name": "Type 2 Diabetes"} |
| Target | 4717 | NDUFC1 | {"name": "NDUFC1", "id": "4717"} | CORRESPONDS_TO | - | Disease | - | Type 2 Diabetes | {"name": "Type 2 Diabetes"} |
| Target | 4718 | NDUFC2 | {"name": "NDUFC2", "id": "4718"} | CORRESPONDS_TO | - | Disease | - | Type 2 Diabetes | {"name": "Type 2 Diabetes"} |
| Target | 4719 | NDUFS1 | {"name": "NDUFS1", "id": "4719"} | CORRESPONDS_TO | - | Disease | - | Type 2 Diabetes | {"name": "Type 2 Diabetes"} |
| Target | 4720 | NDUFS2 | {"name": "NDUFS2", "id": "4720"} | CORRESPONDS_TO | - | Disease | - | Type 2 Diabetes | {"name": "Type 2 Diabetes"} |
| Target | 4722 | NDUFS3 | {"name": "NDUFS3", "id": "4722"} | CORRESPONDS_TO | - | Disease | - | Type 2 Diabetes | {"name": "Type 2 Diabetes"} |
| Target | 4724 | NDUFS4 | {"name": "NDUFS4", "id": "4724"} | CORRESPONDS_TO | - | Disease | - | Type 2 Diabetes | {"name": "Type 2 Diabetes"} |
| Target | 4725 | NDUFS5 | {"name": "NDUFS5", "id": "4725"} | CORRESPONDS_TO | - | Disease | - | Type 2 Diabetes | {"name": "Type 2 Diabetes"} |
| Target | 4726 | NDUFS6 | {"name": "NDUFS6", "id": "4726"} | CORRESPONDS_TO | - | Disease | - | Type 2 Diabetes | {"name": "Type 2 Diabetes"} |
| Target | 374291 | NDUFS7 | {"name": "NDUFS7", "id": "374291"} | CORRESPONDS_TO | - | Disease | - | Type 2 Diabetes | {"name": "Type 2 Diabetes"} |
| Target | 4728 | NDUFS8 | {"name": "NDUFS8", "id": "4728"} | CORRESPONDS_TO | - | Disease | - | Type 2 Diabetes | {"name": "Type 2 Diabetes"} |
| Target | 4723 | NDUFV1 | {"name": "NDUFV1", "id": "4723"} | CORRESPONDS_TO | - | Disease | - | Type 2 Diabetes | {"name": "Type 2 Diabetes"} |
| Target | 4729 | NDUFV2 | {"name": "NDUFV2", "id": "4729"} | CORRESPONDS_TO | - | Disease | - | Type 2 Diabetes | {"name": "Type 2 Diabetes"} |
| Target | 4731 | NDUFV3 | {"name": "NDUFV3", "id": "4731"} | CORRESPONDS_TO | - | Disease | - | Type 2 Diabetes | {"name": "Type 2 Diabetes"} |
| Target | 257194 | NEGR1 | {"name": "NEGR1", "id": "257194"} | CORRESPONDS_TO | - | Disease | - | Obesity | {"name": "Obesity"} |
| Target | 79661 | NEIL1 | {"name": "NEIL1", "id": "79661"} | CORRESPONDS_TO | - | Disease | - | Fatty Liver | {"name": "Fatty Liver"} |
| Target | 79661 | NEIL1 | {"name": "NEIL1", "id": "79661"} | CORRESPONDS_TO | - | Disease | - | Obesity | {"name": "Obesity"} |
| Target | 4760 | NEUROD1 | {"name": "NEUROD1", "id": "4760"} | CORRESPONDS_TO | - | Disease | - | Type 2 Diabetes | {"name": "Type 2 Diabetes"} |
| Target | 50674 | NEUROG3 | {"name": "NEUROG3", "id": "50674"} | CORRESPONDS_TO | - | Disease | - | Type 2 Diabetes | {"name": "Type 2 Diabetes"} |
| Target | 4773 | NFATC2 | {"name": "NFATC2", "id": "4773"} | CORRESPONDS_TO | - | Disease | - | Type 2 Diabetes | {"name": "Type 2 Diabetes"} |
| Target | 4780 | NFE2L2 | {"name": "NFE2L2", "id": "4780"} | CORRESPONDS_TO | - | Disease | - | Fatty Liver | {"name": "Fatty Liver"} |
| Target | 4790 | NFKB1 | {"name": "NFKB1", "id": "4790"} | CORRESPONDS_TO | - | Disease | - | Type 2 Diabetes | {"name": "Type 2 Diabetes"} |
| Target | 4808 | NHLH2 | {"name": "NHLH2", "id": "4808"} | CORRESPONDS_TO | - | Disease | - | Obesity | {"name": "Obesity"} |
| Target | 4825 | NKX6-1 | {"name": "NKX6-1", "id": "4825"} | CORRESPONDS_TO | - | Disease | - | Type 2 Diabetes | {"name": "Type 2 Diabetes"} |
| Target | 114548 | NLRP3 | {"name": "NLRP3", "id": "114548"} | CORRESPONDS_TO | - | Disease | - | Fatty Liver | {"name": "Fatty Liver"} |
| Target | 4843 | NOS2 | {"name": "NOS2", "id": "4843"} | CORRESPONDS_TO | - | Disease | - | Type 2 Diabetes | {"name": "Type 2 Diabetes"} |
| Target | 4846 | NOS3 | {"name": "NOS3", "id": "4846"} | CORRESPONDS_TO | - | Disease | - | Type 2 Diabetes | {"name": "Type 2 Diabetes"} |
| Target | 4853 | NOTCH2 | {"name": "NOTCH2", "id": "4853"} | CORRESPONDS_TO | - | Disease | - | Type 2 Diabetes | {"name": "Type 2 Diabetes"} |
| Target | 4864 | NPC1 | {"name": "NPC1", "id": "4864"} | CORRESPONDS_TO | - | Disease | - | Obesity | {"name": "Obesity"} |
| Target | 4852 | NPY | {"name": "NPY", "id": "4852"} | CORRESPONDS_TO | - | Disease | - | Type 2 Diabetes | {"name": "Type 2 Diabetes"} |
| Target | 4852 | NPY | {"name": "NPY", "id": "4852"} | CORRESPONDS_TO | - | Disease | - | Obesity | {"name": "Obesity"} |

Table S9: Complete node and relationship data used to construct the knowledge graph. This table summarizes interactions among FDA-approved drugs, formulation, herbs, diseases (type 2 diabetes, fatty liver, obesity), phytochemicals, and their target genes/proteins. Each entry includes source and target node details, edge type, and node annotations (Entrez Gene ID and gene symbol, where applicable).

| Source Node Label | Source Node ID | Source Node Name | Source Node Properties | Edge Type | Edge Properties | Target Node Label | Target Node ID | Target Node Name | Target Node Properties |
|---|---|---|---|---|---|---|---|---|---|
| Target | 4886 | NPY1R | {"name": "NPY1R", "id": "4886"} | CORRESPONDS_TO | - | Disease | - | Obesity | {"name": "Obesity"} |
| Target | 4889 | NPY5R | {"name": "NPY5R", "id": "4889"} | CORRESPONDS_TO | - | Disease | - | Obesity | {"name": "Obesity"} |
| Target | 1728 | NQO1 | {"name": "NQO1", "id": "1728"} | CORRESPONDS_TO | - | Disease | - | Fatty Liver | {"name": "Fatty Liver"} |
| Target | 1728 | NQO1 | {"name": "NQO1", "id": "1728"} | CORRESPONDS_TO | - | Disease | - | Obesity | {"name": "Obesity"} |
| Target | 8431 | NR0B2 | {"name": "NR0B2", "id": "8431"} | CORRESPONDS_TO | - | Disease | - | Fatty Liver | {"name": "Fatty Liver"} |
| Target | 8431 | NR0B2 | {"name": "NR0B2", "id": "8431"} | CORRESPONDS_TO | - | Disease | - | Obesity | {"name": "Obesity"} |
| Target | 9572 | NR1D1 | {"name": "NR1D1", "id": "9572"} | CORRESPONDS_TO | - | Disease | - | Fatty Liver | {"name": "Fatty Liver"} |
| Target | 10062 | NR1H3 | {"name": "NR1H3", "id": "10062"} | CORRESPONDS_TO | - | Disease | - | Fatty Liver | {"name": "Fatty Liver"} |
| Target | 9971 | NR1H4 | {"name": "NR1H4", "id": "9971"} | CORRESPONDS_TO | - | Disease | - | Fatty Liver | {"name": "Fatty Liver"} |
| Target | 8856 | NR1I2 | {"name": "NR1I2", "id": "8856"} | CORRESPONDS_TO | - | Disease | - | Fatty Liver | {"name": "Fatty Liver"} |
| Target | 8856 | NR1I2 | {"name": "NR1I2", "id": "8856"} | CORRESPONDS_TO | - | Disease | - | Obesity | {"name": "Obesity"} |
| Target | 9970 | NR1I3 | {"name": "NR1I3", "id": "9970"} | CORRESPONDS_TO | - | Disease | - | Obesity | {"name": "Obesity"} |
| Target | 2908 | NR3C1 | {"name": "NR3C1", "id": "2908"} | CORRESPONDS_TO | - | Disease | - | Obesity | {"name": "Obesity"} |
| Target | 2494 | NR5A2 | {"name": "NR5A2", "id": "2494"} | CORRESPONDS_TO | - | Disease | - | Fatty Liver | {"name": "Fatty Liver"} |
| Target | 9315 | NREP | {"name": "NREP", "id": "9315"} | CORRESPONDS_TO | - | Disease | - | Fatty Liver | {"name": "Fatty Liver"} |
| Target | 4915 | NTRK2 | {"name": "NTRK2", "id": "4915"} | CORRESPONDS_TO | - | Disease | - | Obesity | {"name": "Obesity"} |
| Target | 10726 | NUDC | {"name": "NUDC", "id": "10726"} | CORRESPONDS_TO | - | Disease | - | Obesity | {"name": "Obesity"} |
| Target | 116150 | NUS1 | {"name": "NUS1", "id": "116150"} | CORRESPONDS_TO | - | Disease | - | Type 2 Diabetes | {"name": "Type 2 Diabetes"} |
| Target | 116150 | NUS1 | {"name": "NUS1", "id": "116150"} | CORRESPONDS_TO | - | Disease | - | Fatty Liver | {"name": "Fatty Liver"} |
| Target | 4968 | OGG1 | {"name": "OGG1", "id": "4968"} | CORRESPONDS_TO | - | Disease | - | Type 2 Diabetes | {"name": "Type 2 Diabetes"} |
| Target | 4968 | OGG1 | {"name": "OGG1", "id": "4968"} | CORRESPONDS_TO | - | Disease | - | Obesity | {"name": "Obesity"} |
| Target | 55690 | PACS1 | {"name": "PACS1", "id": "55690"} | CORRESPONDS_TO | - | Disease | - | Obesity | {"name": "Obesity"} |
| Target | 5066 | PAM | {"name": "PAM", "id": "5066"} | CORRESPONDS_TO | - | Disease | - | Type 2 Diabetes | {"name": "Type 2 Diabetes"} |
| Target | 142 | PARP1 | {"name": "PARP1", "id": "142"} | CORRESPONDS_TO | - | Disease | - | Obesity | {"name": "Obesity"} |
| Target | 5078 | PAX4 | {"name": "PAX4", "id": "5078"} | CORRESPONDS_TO | - | Disease | - | Type 2 Diabetes | {"name": "Type 2 Diabetes"} |
| Target | 5080 | PAX6 | {"name": "PAX6", "id": "5080"} | CORRESPONDS_TO | - | Disease | - | Type 2 Diabetes | {"name": "Type 2 Diabetes"} |
| Target | 56144 | PCDHA4 | {"name": "PCDHA4", "id": "56144"} | CORRESPONDS_TO | - | Disease | - | Obesity | {"name": "Obesity"} |
| Target | 5105 | PCK1 | {"name": "PCK1", "id": "5105"} | CORRESPONDS_TO | - | Disease | - | Obesity | {"name": "Obesity"} |
| Target | 5122 | PCSK1 | {"name": "PCSK1", "id": "5122"} | CORRESPONDS_TO | - | Disease | - | Obesity | {"name": "Obesity"} |
| Target | 5126 | PCSK2 | {"name": "PCSK2", "id": "5126"} | CORRESPONDS_TO | - | Disease | - | Type 2 Diabetes | {"name": "Type 2 Diabetes"} |
| Target | 5166 | PDK4 | {"name": "PDK4", "id": "5166"} | CORRESPONDS_TO | - | Disease | - | Fatty Liver | {"name": "Fatty Liver"} |
| Target | 3651 | PDX1 | {"name": "PDX1", "id": "3651"} | CORRESPONDS_TO | - | Disease | - | Type 2 Diabetes | {"name": "Type 2 Diabetes"} |
| Target | 10400 | PEMT | {"name": "PEMT", "id": "10400"} | CORRESPONDS_TO | - | Disease | - | Fatty Liver | {"name": "Fatty Liver"} |
| Target | 5184 | PEPD | {"name": "PEPD", "id": "5184"} | CORRESPONDS_TO | - | Disease | - | Type 2 Diabetes | {"name": "Type 2 Diabetes"} |
| Target | 8800 | PEX11A | {"name": "PEX11A", "id": "8800"} | CORRESPONDS_TO | - | Disease | - | Obesity | {"name": "Obesity"} |
| Target | 5209 | PFKFB3 | {"name": "PFKFB3", "id": "5209"} | CORRESPONDS_TO | - | Disease | - | Obesity | {"name": "Obesity"} |
| Target | 5290 | PIK3CA | {"name": "PIK3CA", "id": "5290"} | CORRESPONDS_TO | - | Disease | - | Obesity | {"name": "Obesity"} |
| Target | 5294 | PIK3CG | {"name": "PIK3CG", "id": "5294"} | CORRESPONDS_TO | - | Disease | - | Type 2 Diabetes | {"name": "Type 2 Diabetes"} |
| Target | 59338 | PLEKHA1 | {"name": "PLEKHA1", "id": "59338"} | CORRESPONDS_TO | - | Disease | - | Type 2 Diabetes | {"name": "Type 2 Diabetes"} |
| Target | 5346 | PLIN1 | {"name": "PLIN1", "id": "5346"} | CORRESPONDS_TO | - | Disease | - | Type 2 Diabetes | {"name": "Type 2 Diabetes"} |
| Target | 5346 | PLIN1 | {"name": "PLIN1", "id": "5346"} | CORRESPONDS_TO | - | Disease | - | Obesity | {"name": "Obesity"} |
| Target | 123 | PLIN2 | {"name": "PLIN2", "id": "123"} | CORRESPONDS_TO | - | Disease | - | Fatty Liver | {"name": "Fatty Liver"} |
| Target | 5367 | PMCH | {"name": "PMCH", "id": "5367"} | CORRESPONDS_TO | - | Disease | - | Obesity | {"name": "Obesity"} |
| Target | 5406 | PNLIP | {"name": "PNLIP", "id": "5406"} | CORRESPONDS_TO | - | Disease | - | Obesity | {"name": "Obesity"} |
| Target | 57104 | PNPLA2 | {"name": "PNPLA2", "id": "57104"} | CORRESPONDS_TO | - | Disease | - | Obesity | {"name": "Obesity"} |
| Target | 80339 | PNPLA3 | {"name": "PNPLA3", "id": "80339"} | CORRESPONDS_TO | - | Disease | - | Type 2 Diabetes | {"name": "Type 2 Diabetes"} |
| Target | 80339 | PNPLA3 | {"name": "PNPLA3", "id": "80339"} | CORRESPONDS_TO | - | Disease | - | Fatty Liver | {"name": "Fatty Liver"} |
| Target | 5443 | POMC | {"name": "POMC", "id": "5443"} | CORRESPONDS_TO | - | Disease | - | Fatty Liver | {"name": "Fatty Liver"} |
| Target | 5443 | POMC | {"name": "POMC", "id": "5443"} | CORRESPONDS_TO | - | Disease | - | Obesity | {"name": "Obesity"} |
| Target | 5444 | PON1 | {"name": "PON1", "id": "5444"} | CORRESPONDS_TO | - | Disease | - | Type 2 Diabetes | {"name": "Type 2 Diabetes"} |
| Target | 5444 | PON1 | {"name": "PON1", "id": "5444"} | CORRESPONDS_TO | - | Disease | - | Fatty Liver | {"name": "Fatty Liver"} |
| Target | 5465 | PPARA | {"name": "PPARA", "id": "5465"} | CORRESPONDS_TO | - | Disease | - | Type 2 Diabetes | {"name": "Type 2 Diabetes"} |
| Target | 5465 | PPARA | {"name": "PPARA", "id": "5465"} | CORRESPONDS_TO | - | Disease | - | Fatty Liver | {"name": "Fatty Liver"} |
| Target | 5465 | PPARA | {"name": "PPARA", "id": "5465"} | CORRESPONDS_TO | - | Disease | - | Obesity | {"name": "Obesity"} |
| Target | 5467 | PPARD | {"name": "PPARD", "id": "5467"} | CORRESPONDS_TO | - | Disease | - | Type 2 Diabetes | {"name": "Type 2 Diabetes"} |
| Target | 5467 | PPARD | {"name": "PPARD", "id": "5467"} | CORRESPONDS_TO | - | Disease | - | Fatty Liver | {"name": "Fatty Liver"} |
| Target | 5467 | PPARD | {"name": "PPARD", "id": "5467"} | CORRESPONDS_TO | - | Disease | - | Obesity | {"name": "Obesity"} |
| Target | 5468 | PPARG | {"name": "PPARG", "id": "5468"} | CORRESPONDS_TO | - | Disease | - | Type 2 Diabetes | {"name": "Type 2 Diabetes"} |
| Target | 5468 | PPARG | {"name": "PPARG", "id": "5468"} | CORRESPONDS_TO | - | Disease | - | Fatty Liver | {"name": "Fatty Liver"} |
| Target | 5468 | PPARG | {"name": "PPARG", "id": "5468"} | CORRESPONDS_TO | - | Disease | - | Obesity | {"name": "Obesity"} |
| Target | 10891 | PPARGC1A | {"name": "PPARGC1A", "id": "10891"} | CORRESPONDS_TO | - | Disease | - | Type 2 Diabetes | {"name": "Type 2 Diabetes"} |
| Target | 10891 | PPARGC1A | {"name": "PPARGC1A", "id": "10891"} | CORRESPONDS_TO | - | Disease | - | Fatty Liver | {"name": "Fatty Liver"} |
| Target | 10891 | PPARGC1A | {"name": "PPARGC1A", "id": "10891"} | CORRESPONDS_TO | - | Disease | - | Obesity | {"name": "Obesity"} |
| Target | 133522 | PPARGC1B | {"name": "PPARGC1B", "id": "133522"} | CORRESPONDS_TO | - | Disease | - | Obesity | {"name": "Obesity"} |
| Target | 151742 | PPM1L | {"name": "PPM1L", "id": "151742"} | CORRESPONDS_TO | - | Disease | - | Obesity | {"name": "Obesity"} |
| Target | 5506 | PPP1R3A | {"name": "PPP1R3A", "id": "5506"} | CORRESPONDS_TO | - | Disease | - | Type 2 Diabetes | {"name": "Type 2 Diabetes"} |
| Target | 10935 | PRDX3 | {"name": "PRDX3", "id": "10935"} | CORRESPONDS_TO | - | Disease | - | Fatty Liver | {"name": "Fatty Liver"} |
| Target | 5551 | PRF1 | {"name": "PRF1", "id": "5551"} | CORRESPONDS_TO | - | Disease | - | Fatty Liver | {"name": "Fatty Liver"} |
| Target | 5563 | PRKAA2 | {"name": "PRKAA2", "id": "5563"} | CORRESPONDS_TO | - | Disease | - | Type 2 Diabetes | {"name": "Type 2 Diabetes"} |
| Target | 5566 | PRKACA | {"name": "PRKACA", "id": "5566"} | CORRESPONDS_TO | - | Disease | - | Fatty Liver | {"name": "Fatty Liver"} |
| Target | 5573 | PRKAR1A | {"name": "PRKAR1A", "id": "5573"} | CORRESPONDS_TO | - | Disease | - | Obesity | {"name": "Obesity"} |
| Target | 5577 | PRKAR2B | {"name": "PRKAR2B", "id": "5577"} | CORRESPONDS_TO | - | Disease | - | Obesity | {"name": "Obesity"} |
| Target | 5578 | PRKCA | {"name": "PRKCA", "id": "5578"} | CORRESPONDS_TO | - | Disease | - | Fatty Liver | {"name": "Fatty Liver"} |
| Target | 5579 | PRKCB | {"name": "PRKCB", "id": "5579"} | CORRESPONDS_TO | - | Disease | - | Type 2 Diabetes | {"name": "Type 2 Diabetes"} |
| Target | 5580 | PRKCD | {"name": "PRKCD", "id": "5580"} | CORRESPONDS_TO | - | Disease | - | Fatty Liver | {"name": "Fatty Liver"} |
| Target | 5581 | PRKCE | {"name": "PRKCE", "id": "5581"} | CORRESPONDS_TO | - | Disease | - | Fatty Liver | {"name": "Fatty Liver"} |

Table S9: Complete node and relationship data used to construct the knowledge graph. This table summarizes interactions among FDA-approved drugs, formulation, herbs, diseases (type 2 diabetes, fatty liver, obesity), phytochemicals, and their target genes/proteins. Each entry includes source and target node details, edge type, and node annotations (Entrez Gene ID and gene symbol, where applicable).

| Source Node Label | Source Node ID | Source Node Name | Source Node Properties | Edge Type | Edge Properties | Target Node Label | Target Node ID | Target Node Name | Target Node Properties |
|---|---|---|---|---|---|---|---|---|---|
| Target | 5583 | PRKCH | {"name": "PRKCH", "id": "5583"} | CORRESPONDS_TO | - | Disease | - | Obesity | {"name": "Obesity"} |
| Target | 54496 | PRMT7 | {"name": "PRMT7", "id": "54496"} | CORRESPONDS_TO | - | Disease | - | Obesity | {"name": "Obesity"} |
| Target | 5629 | PROX1 | {"name": "PROX1", "id": "5629"} | CORRESPONDS_TO | - | Disease | - | Type 2 Diabetes | {"name": "Type 2 Diabetes"} |
| Target | 5686 | PSMA5 | {"name": "PSMA5", "id": "5686"} | CORRESPONDS_TO | - | Disease | - | Fatty Liver | {"name": "Fatty Liver"} |
| Target | 9861 | PSMD6 | {"name": "PSMD6", "id": "9861"} | CORRESPONDS_TO | - | Disease | - | Type 2 Diabetes | {"name": "Type 2 Diabetes"} |
| Target | 5728 | PTEN | {"name": "PTEN", "id": "5728"} | CORRESPONDS_TO | - | Disease | - | Fatty Liver | {"name": "Fatty Liver"} |
| Target | 5743 | PTGS2 | {"name": "PTGS2", "id": "5743"} | CORRESPONDS_TO | - | Disease | - | Obesity | {"name": "Obesity"} |
| Target | 5757 | PTMA | {"name": "PTMA", "id": "5757"} | CORRESPONDS_TO | - | Disease | - | Fatty Liver | {"name": "Fatty Liver"} |
| Target | 5770 | PTPN1 | {"name": "PTPN1", "id": "5770"} | CORRESPONDS_TO | - | Disease | - | Type 2 Diabetes | {"name": "Type 2 Diabetes"} |
| Target | 5770 | PTPN1 | {"name": "PTPN1", "id": "5770"} | CORRESPONDS_TO | - | Disease | - | Obesity | {"name": "Obesity"} |
| Target | 5697 | PYY | {"name": "PYY", "id": "5697"} | CORRESPONDS_TO | - | Disease | - | Obesity | {"name": "Obesity"} |
| Target | 23011 | RAB21 | {"name": "RAB21", "id": "23011"} | CORRESPONDS_TO | - | Disease | - | Obesity | {"name": "Obesity"} |
| Target | 5897 | RAG2 | {"name": "RAG2", "id": "5897"} | CORRESPONDS_TO | - | Disease | - | Fatty Liver | {"name": "Fatty Liver"} |
| Target | 10743 | RAI1 | {"name": "RAI1", "id": "10743"} | CORRESPONDS_TO | - | Disease | - | Obesity | {"name": "Obesity"} |
| Target | 10411 | RAPGEF3 | {"name": "RAPGEF3", "id": "10411"} | CORRESPONDS_TO | - | Disease | - | Obesity | {"name": "Obesity"} |
| Target | 5919 | RARRES2 | {"name": "RARRES2", "id": "5919"} | CORRESPONDS_TO | - | Disease | - | Obesity | {"name": "Obesity"} |
| Target | 5950 | RBP4 | {"name": "RBP4", "id": "5950"} | CORRESPONDS_TO | - | Disease | - | Type 2 Diabetes | {"name": "Type 2 Diabetes"} |
| Target | 3516 | RBPJ | {"name": "RBPJ", "id": "3516"} | CORRESPONDS_TO | - | Disease | - | Type 2 Diabetes | {"name": "Type 2 Diabetes"} |
| Target | 11317 | RBPJL | {"name": "RBPJL", "id": "11317"} | CORRESPONDS_TO | - | Disease | - | Type 2 Diabetes | {"name": "Type 2 Diabetes"} |
| Target | 5962 | RDX | {"name": "RDX", "id": "5962"} | CORRESPONDS_TO | - | Disease | - | Fatty Liver | {"name": "Fatty Liver"} |
| Target | 56729 | RETN | {"name": "RETN", "id": "56729"} | CORRESPONDS_TO | - | Disease | - | Type 2 Diabetes | {"name": "Type 2 Diabetes"} |
| Target | 56729 | RETN | {"name": "RETN", "id": "56729"} | CORRESPONDS_TO | - | Disease | - | Obesity | {"name": "Obesity"} |
| Target | 6049 | RNF6 | {"name": "RNF6", "id": "6049"} | CORRESPONDS_TO | - | Disease | - | Type 2 Diabetes | {"name": "Type 2 Diabetes"} |
| Target | 6277 | S100A6 | {"name": "S100A6", "id": "6277"} | CORRESPONDS_TO | - | Disease | - | Type 2 Diabetes | {"name": "Type 2 Diabetes"} |
| Target | 6303 | SAT1 | {"name": "SAT1", "id": "6303"} | CORRESPONDS_TO | - | Disease | - | Obesity | {"name": "Obesity"} |
| Target | 949 | SCARB1 | {"name": "SCARB1", "id": "949"} | CORRESPONDS_TO | - | Disease | - | Fatty Liver | {"name": "Fatty Liver"} |
| Target | 6319 | SCD | {"name": "SCD", "id": "6319"} | CORRESPONDS_TO | - | Disease | - | Fatty Liver | {"name": "Fatty Liver"} |
| Target | 6319 | SCD | {"name": "SCD", "id": "6319"} | CORRESPONDS_TO | - | Disease | - | Obesity | {"name": "Obesity"} |
| Target | 6344 | SCTR | {"name": "SCTR", "id": "6344"} | CORRESPONDS_TO | - | Disease | - | Type 2 Diabetes | {"name": "Type 2 Diabetes"} |
| Target | 9672 | SDC3 | {"name": "SDC3", "id": "9672"} | CORRESPONDS_TO | - | Disease | - | Obesity | {"name": "Obesity"} |
| Target | 10806 | SDCCAG8 | {"name": "SDCCAG8", "id": "10806"} | CORRESPONDS_TO | - | Disease | - | Obesity | {"name": "Obesity"} |
| Target | 5265 | SERPINA1 | {"name": "SERPINA1", "id": "5265"} | CORRESPONDS_TO | - | Disease | - | Fatty Liver | {"name": "Fatty Liver"} |
| Target | 866 | SERPINA6 | {"name": "SERPINA6", "id": "866"} | CORRESPONDS_TO | - | Disease | - | Fatty Liver | {"name": "Fatty Liver"} |
| Target | 5055 | SERPINB2 | {"name": "SERPINB2", "id": "5055"} | CORRESPONDS_TO | - | Disease | - | Fatty Liver | {"name": "Fatty Liver"} |
| Target | 5054 | SERPINE1 | {"name": "SERPINE1", "id": "5054"} | CORRESPONDS_TO | - | Disease | - | Type 2 Diabetes | {"name": "Type 2 Diabetes"} |
| Target | 5054 | SERPINE1 | {"name": "SERPINE1", "id": "5054"} | CORRESPONDS_TO | - | Disease | - | Fatty Liver | {"name": "Fatty Liver"} |
| Target | 5054 | SERPINE1 | {"name": "SERPINE1", "id": "5054"} | CORRESPONDS_TO | - | Disease | - | Obesity | {"name": "Obesity"} |
| Target | 6424 | SFRP4 | {"name": "SFRP4", "id": "6424"} | CORRESPONDS_TO | - | Disease | - | Type 2 Diabetes | {"name": "Type 2 Diabetes"} |
| Target | 25970 | SH2B1 | {"name": "SH2B1", "id": "25970"} | CORRESPONDS_TO | - | Disease | - | Obesity | {"name": "Obesity"} |
| Target | 6462 | SHBG | {"name": "SHBG", "id": "6462"} | CORRESPONDS_TO | - | Disease | - | Type 2 Diabetes | {"name": "Type 2 Diabetes"} |
| Target | 6462 | SHBG | {"name": "SHBG", "id": "6462"} | CORRESPONDS_TO | - | Disease | - | Obesity | {"name": "Obesity"} |
| Target | 6492 | SIM1 | {"name": "SIM1", "id": "6492"} | CORRESPONDS_TO | - | Disease | - | Obesity | {"name": "Obesity"} |
| Target | 23411 | SIRT1 | {"name": "SIRT1", "id": "23411"} | CORRESPONDS_TO | - | Disease | - | Type 2 Diabetes | {"name": "Type 2 Diabetes"} |
| Target | 23411 | SIRT1 | {"name": "SIRT1", "id": "23411"} | CORRESPONDS_TO | - | Disease | - | Fatty Liver | {"name": "Fatty Liver"} |
| Target | 23411 | SIRT1 | {"name": "SIRT1", "id": "23411"} | CORRESPONDS_TO | - | Disease | - | Obesity | {"name": "Obesity"} |
| Target | 23410 | SIRT3 | {"name": "SIRT3", "id": "23410"} | CORRESPONDS_TO | - | Disease | - | Obesity | {"name": "Obesity"} |
| Target | 51548 | SIRT6 | {"name": "SIRT6", "id": "51548"} | CORRESPONDS_TO | - | Disease | - | Fatty Liver | {"name": "Fatty Liver"} |
| Target | 51547 | SIRT7 | {"name": "SIRT7", "id": "51547"} | CORRESPONDS_TO | - | Disease | - | Fatty Liver | {"name": "Fatty Liver"} |
| Target | 284111 | SLC13A5 | {"name": "SLC13A5", "id": "284111"} | CORRESPONDS_TO | - | Disease | - | Fatty Liver | {"name": "Fatty Liver"} |
| Target | 9194 | SLC16A7 | {"name": "SLC16A7", "id": "9194"} | CORRESPONDS_TO | - | Disease | - | Obesity | {"name": "Obesity"} |
| Target | 6571 | SLC18A2 | {"name": "SLC18A2", "id": "6571"} | CORRESPONDS_TO | - | Disease | - | Obesity | {"name": "Obesity"} |
| Target | 6506 | SLC1A2 | {"name": "SLC1A2", "id": "6506"} | CORRESPONDS_TO | - | Disease | - | Type 2 Diabetes | {"name": "Type 2 Diabetes"} |
| Target | 6580 | SLC22A1 | {"name": "SLC22A1", "id": "6580"} | CORRESPONDS_TO | - | Disease | - | Obesity | {"name": "Obesity"} |
| Target | 6582 | SLC22A2 | {"name": "SLC22A2", "id": "6582"} | CORRESPONDS_TO | - | Disease | - | Obesity | {"name": "Obesity"} |
| Target | 6581 | SLC22A3 | {"name": "SLC22A3", "id": "6581"} | CORRESPONDS_TO | - | Disease | - | Type 2 Diabetes | {"name": "Type 2 Diabetes"} |
| Target | 6581 | SLC22A3 | {"name": "SLC22A3", "id": "6581"} | CORRESPONDS_TO | - | Disease | - | Obesity | {"name": "Obesity"} |
| Target | 9376 | SLC22A8 | {"name": "SLC22A8", "id": "9376"} | CORRESPONDS_TO | - | Disease | - | Fatty Liver | {"name": "Fatty Liver"} |
| Target | 6513 | SLC2A1 | {"name": "SLC2A1", "id": "6513"} | CORRESPONDS_TO | - | Disease | - | Type 2 Diabetes | {"name": "Type 2 Diabetes"} |
| Target | 6514 | SLC2A2 | {"name": "SLC2A2", "id": "6514"} | CORRESPONDS_TO | - | Disease | - | Type 2 Diabetes | {"name": "Type 2 Diabetes"} |
| Target | 6514 | SLC2A2 | {"name": "SLC2A2", "id": "6514"} | CORRESPONDS_TO | - | Disease | - | Fatty Liver | {"name": "Fatty Liver"} |
| Target | 6517 | SLC2A4 | {"name": "SLC2A4", "id": "6517"} | CORRESPONDS_TO | - | Disease | - | Type 2 Diabetes | {"name": "Type 2 Diabetes"} |
| Target | 6517 | SLC2A4 | {"name": "SLC2A4", "id": "6517"} | CORRESPONDS_TO | - | Disease | - | Obesity | {"name": "Obesity"} |
| Target | 169026 | SLC30A8 | {"name": "SLC30A8", "id": "169026"} | CORRESPONDS_TO | - | Disease | - | Type 2 Diabetes | {"name": "Type 2 Diabetes"} |
| Target | 55244 | SLC47A1 | {"name": "SLC47A1", "id": "55244"} | CORRESPONDS_TO | - | Disease | - | Type 2 Diabetes | {"name": "Type 2 Diabetes"} |
| Target | 6523 | SLC5A1 | {"name": "SLC5A1", "id": "6523"} | CORRESPONDS_TO | - | Disease | - | Type 2 Diabetes | {"name": "Type 2 Diabetes"} |
| Target | 6524 | SLC5A2 | {"name": "SLC5A2", "id": "6524"} | CORRESPONDS_TO | - | Disease | - | Type 2 Diabetes | {"name": "Type 2 Diabetes"} |
| Target | 11254 | SLC6A14 | {"name": "SLC6A14", "id": "11254"} | CORRESPONDS_TO | - | Disease | - | Obesity | {"name": "Obesity"} |
| Target | 6530 | SLC6A2 | {"name": "SLC6A2", "id": "6530"} | CORRESPONDS_TO | - | Disease | - | Obesity | {"name": "Obesity"} |
| Target | 6531 | SLC6A3 | {"name": "SLC6A3", "id": "6531"} | CORRESPONDS_TO | - | Disease | - | Obesity | {"name": "Obesity"} |
| Target | 4090 | SMAD5 | {"name": "SMAD5", "id": "4090"} | CORRESPONDS_TO | - | Disease | - | Type 2 Diabetes | {"name": "Type 2 Diabetes"} |
| Target | 132332 | SMIM43 | {"name": "SMIM43", "id": "132332"} | CORRESPONDS_TO | - | Disease | - | Type 2 Diabetes | {"name": "Type 2 Diabetes"} |
| Target | 6616 | SNAP25 | {"name": "SNAP25", "id": "6616"} | CORRESPONDS_TO | - | Disease | - | Type 2 Diabetes | {"name": "Type 2 Diabetes"} |
| Target | 6647 | SOD1 | {"name": "SOD1", "id": "6647"} | CORRESPONDS_TO | - | Disease | - | Type 2 Diabetes | {"name": "Type 2 Diabetes"} |
| Target | 6647 | SOD1 | {"name": "SOD1", "id": "6647"} | CORRESPONDS_TO | - | Disease | - | Fatty Liver | {"name": "Fatty Liver"} |
| Target | 6647 | SOD1 | {"name": "SOD1", "id": "6647"} | CORRESPONDS_TO | - | Disease | - | Obesity | {"name": "Obesity"} |
| Target | 6648 | SOD2 | {"name": "SOD2", "id": "6648"} | CORRESPONDS_TO | - | Disease | - | Type 2 Diabetes | {"name": "Type 2 Diabetes"} |

**Table S9: Complete node and relationship data used to construct the knowledge graph. This table summarizes interactions among FDA-approved drugs, formulation, herbs, diseases (type 2 diabetes, fatty liver, obesity), phytochemicals, and their target genes/proteins. Each entry includes source and target node details, edge type, and node annotations (Entrez Gene ID and gene symbol, where applicable).**

| Source Node Label | Source Node ID | Source Node Name | Source Node Properties | Edge Type | Edge Properties | Target Node Label | Target Node ID | Target Node Name | Target Node Properties |
|---|---|---|---|---|---|---|---|---|---|
| Target | 6648 | SOD2 | {"name": "SOD2", "id": "6648"} | CORRESPONDS_TO | - | Disease | - | Fatty Liver | {"name": "Fatty Liver"} |
| Target | 6648 | SOD2 | {"name": "SOD2", "id": "6648"} | CORRESPONDS_TO | - | Disease | - | Obesity | {"name": "Obesity"} |
| Target | 6720 | SREBF1 | {"name": "SREBF1", "id": "6720"} | CORRESPONDS_TO | - | Disease | - | Type 2 Diabetes | {"name": "Type 2 Diabetes"} |
| Target | 6720 | SREBF1 | {"name": "SREBF1", "id": "6720"} | CORRESPONDS_TO | - | Disease | - | Fatty Liver | {"name": "Fatty Liver"} |
| Target | 6720 | SREBF1 | {"name": "SREBF1", "id": "6720"} | CORRESPONDS_TO | - | Disease | - | Obesity | {"name": "Obesity"} |
| Target | 6480 | ST6GAL1 | {"name": "ST6GAL1", "id": "6480"} | CORRESPONDS_TO | - | Disease | - | Type 2 Diabetes | {"name": "Type 2 Diabetes"} |
| Target | 6774 | STAT3 | {"name": "STAT3", "id": "6774"} | CORRESPONDS_TO | - | Disease | - | Type 2 Diabetes | {"name": "Type 2 Diabetes"} |
| Target | 6774 | STAT3 | {"name": "STAT3", "id": "6774"} | CORRESPONDS_TO | - | Disease | - | Obesity | {"name": "Obesity"} |
| Target | 8614 | STC2 | {"name": "STC2", "id": "8614"} | CORRESPONDS_TO | - | Disease | - | Fatty Liver | {"name": "Fatty Liver"} |
| Target | 412 | STS | {"name": "STS", "id": "412"} | CORRESPONDS_TO | - | Disease | - | Fatty Liver | {"name": "Fatty Liver"} |
| Target | 412 | STS | {"name": "STS", "id": "412"} | CORRESPONDS_TO | - | Disease | - | Obesity | {"name": "Obesity"} |
| Target | 8803 | SUCLA2 | {"name": "SUCLA2", "id": "8803"} | CORRESPONDS_TO | - | Disease | - | Obesity | {"name": "Obesity"} |
| Target | 8801 | SUCLG2 | {"name": "SUCLG2", "id": "8801"} | CORRESPONDS_TO | - | Disease | - | Obesity | {"name": "Obesity"} |
| Target | 134864 | TAAR1 | {"name": "TAAR1", "id": "134864"} | CORRESPONDS_TO | - | Disease | - | Obesity | {"name": "Obesity"} |
| Target | 23216 | TBC1D1 | {"name": "TBC1D1", "id": "23216"} | CORRESPONDS_TO | - | Disease | - | Obesity | {"name": "Obesity"} |
| Target | 9882 | TBC1D4 | {"name": "TBC1D4", "id": "9882"} | CORRESPONDS_TO | - | Disease | - | Type 2 Diabetes | {"name": "Type 2 Diabetes"} |
| Target | 6934 | TCF7L2 | {"name": "TCF7L2", "id": "6934"} | CORRESPONDS_TO | - | Disease | - | Type 2 Diabetes | {"name": "Type 2 Diabetes"} |
| Target | 7018 | TF | {"name": "TF", "id": "7018"} | CORRESPONDS_TO | - | Disease | - | Fatty Liver | {"name": "Fatty Liver"} |
| Target | 7018 | TF | {"name": "TF", "id": "7018"} | CORRESPONDS_TO | - | Disease | - | Obesity | {"name": "Obesity"} |
| Target | 7037 | TFRC | {"name": "TFRC", "id": "7037"} | CORRESPONDS_TO | - | Disease | - | Obesity | {"name": "Obesity"} |
| Target | 7040 | TGFB1 | {"name": "TGFB1", "id": "7040"} | CORRESPONDS_TO | - | Disease | - | Type 2 Diabetes | {"name": "Type 2 Diabetes"} |
| Target | 7040 | TGFB1 | {"name": "TGFB1", "id": "7040"} | CORRESPONDS_TO | - | Disease | - | Fatty Liver | {"name": "Fatty Liver"} |
| Target | 63892 | THADA | {"name": "THADA", "id": "63892"} | CORRESPONDS_TO | - | Disease | - | Type 2 Diabetes | {"name": "Type 2 Diabetes"} |
| Target | 7076 | TIMP1 | {"name": "TIMP1", "id": "7076"} | CORRESPONDS_TO | - | Disease | - | Type 2 Diabetes | {"name": "Type 2 Diabetes"} |
| Target | 7099 | TLR4 | {"name": "TLR4", "id": "7099"} | CORRESPONDS_TO | - | Disease | - | Type 2 Diabetes | {"name": "Type 2 Diabetes"} |
| Target | 7099 | TLR4 | {"name": "TLR4", "id": "7099"} | CORRESPONDS_TO | - | Disease | - | Fatty Liver | {"name": "Fatty Liver"} |
| Target | 53345 | TM6SF2 | {"name": "TM6SF2", "id": "53345"} | CORRESPONDS_TO | - | Disease | - | Fatty Liver | {"name": "Fatty Liver"} |
| Target | 129787 | TMEM18 | {"name": "TMEM18", "id": "129787"} | CORRESPONDS_TO | - | Disease | - | Type 2 Diabetes | {"name": "Type 2 Diabetes"} |
| Target | 129787 | TMEM18 | {"name": "TMEM18", "id": "129787"} | CORRESPONDS_TO | - | Disease | - | Obesity | {"name": "Obesity"} |
| Target | 91147 | TMEM67 | {"name": "TMEM67", "id": "91147"} | CORRESPONDS_TO | - | Disease | - | Obesity | {"name": "Obesity"} |
| Target | 7124 | TNF | {"name": "TNF", "id": "7124"} | CORRESPONDS_TO | - | Disease | - | Type 2 Diabetes | {"name": "Type 2 Diabetes"} |
| Target | 7124 | TNF | {"name": "TNF", "id": "7124"} | CORRESPONDS_TO | - | Disease | - | Fatty Liver | {"name": "Fatty Liver"} |
| Target | 7124 | TNF | {"name": "TNF", "id": "7124"} | CORRESPONDS_TO | - | Disease | - | Obesity | {"name": "Obesity"} |
| Target | 7132 | TNFRSF1A | {"name": "TNFRSF1A", "id": "7132"} | CORRESPONDS_TO | - | Disease | - | Type 2 Diabetes | {"name": "Type 2 Diabetes"} |
| Target | 7133 | TNFRSF1B | {"name": "TNFRSF1B", "id": "7133"} | CORRESPONDS_TO | - | Disease | - | Type 2 Diabetes | {"name": "Type 2 Diabetes"} |
| Target | 7133 | TNFRSF1B | {"name": "TNFRSF1B", "id": "7133"} | CORRESPONDS_TO | - | Disease | - | Fatty Liver | {"name": "Fatty Liver"} |
| Target | 7133 | TNFRSF1B | {"name": "TNFRSF1B", "id": "7133"} | CORRESPONDS_TO | - | Disease | - | Obesity | {"name": "Obesity"} |
| Target | 10221 | TRIB1 | {"name": "TRIB1", "id": "10221"} | CORRESPONDS_TO | - | Disease | - | Fatty Liver | {"name": "Fatty Liver"} |
| Target | 57761 | TRIB3 | {"name": "TRIB3", "id": "57761"} | CORRESPONDS_TO | - | Disease | - | Fatty Liver | {"name": "Fatty Liver"} |
| Target | 7442 | TRPV1 | {"name": "TRPV1", "id": "7442"} | CORRESPONDS_TO | - | Disease | - | Obesity | {"name": "Obesity"} |
| Target | 123016 | TTC8 | {"name": "TTC8", "id": "123016"} | CORRESPONDS_TO | - | Disease | - | Obesity | {"name": "Obesity"} |
| Target | 7275 | TUB | {"name": "TUB", "id": "7275"} | CORRESPONDS_TO | - | Disease | - | Obesity | {"name": "Obesity"} |
| Target | 7325 | UBE2E2 | {"name": "UBE2E2", "id": "7325"} | CORRESPONDS_TO | - | Disease | - | Type 2 Diabetes | {"name": "Type 2 Diabetes"} |
| Target | 7350 | UCP1 | {"name": "UCP1", "id": "7350"} | CORRESPONDS_TO | - | Disease | - | Type 2 Diabetes | {"name": "Type 2 Diabetes"} |
| Target | 7350 | UCP1 | {"name": "UCP1", "id": "7350"} | CORRESPONDS_TO | - | Disease | - | Obesity | {"name": "Obesity"} |
| Target | 7351 | UCP2 | {"name": "UCP2", "id": "7351"} | CORRESPONDS_TO | - | Disease | - | Type 2 Diabetes | {"name": "Type 2 Diabetes"} |
| Target | 7351 | UCP2 | {"name": "UCP2", "id": "7351"} | CORRESPONDS_TO | - | Disease | - | Fatty Liver | {"name": "Fatty Liver"} |
| Target | 7351 | UCP2 | {"name": "UCP2", "id": "7351"} | CORRESPONDS_TO | - | Disease | - | Obesity | {"name": "Obesity"} |
| Target | 7352 | UCP3 | {"name": "UCP3", "id": "7352"} | CORRESPONDS_TO | - | Disease | - | Type 2 Diabetes | {"name": "Type 2 Diabetes"} |
| Target | 7352 | UCP3 | {"name": "UCP3", "id": "7352"} | CORRESPONDS_TO | - | Disease | - | Obesity | {"name": "Obesity"} |
| Target | 7385 | UQCRC2 | {"name": "UQCRC2", "id": "7385"} | CORRESPONDS_TO | - | Disease | - | Obesity | {"name": "Obesity"} |
| Target | 84196 | USP48 | {"name": "USP48", "id": "84196"} | CORRESPONDS_TO | - | Disease | - | Type 2 Diabetes | {"name": "Type 2 Diabetes"} |
| Target | 7425 | VGF | {"name": "VGF", "id": "7425"} | CORRESPONDS_TO | - | Disease | - | Obesity | {"name": "Obesity"} |
| Target | 7436 | VLDLR | {"name": "VLDLR", "id": "7436"} | CORRESPONDS_TO | - | Disease | - | Fatty Liver | {"name": "Fatty Liver"} |
| Target | 7436 | VLDLR | {"name": "VLDLR", "id": "7436"} | CORRESPONDS_TO | - | Disease | - | Obesity | {"name": "Obesity"} |
| Target | 157680 | VPS13B | {"name": "VPS13B", "id": "157680"} | CORRESPONDS_TO | - | Disease | - | Obesity | {"name": "Obesity"} |
| Target | 9559 | VPS26A | {"name": "VPS26A", "id": "9559"} | CORRESPONDS_TO | - | Disease | - | Type 2 Diabetes | {"name": "Type 2 Diabetes"} |
| Target | 7466 | WFS1 | {"name": "WFS1", "id": "7466"} | CORRESPONDS_TO | - | Disease | - | Type 2 Diabetes | {"name": "Type 2 Diabetes"} |
| Target | 89780 | WNT3A | {"name": "WNT3A", "id": "89780"} | CORRESPONDS_TO | - | Disease | - | Obesity | {"name": "Obesity"} |
| Target | 7486 | WRN | {"name": "WRN", "id": "7486"} | CORRESPONDS_TO | - | Disease | - | Type 2 Diabetes | {"name": "Type 2 Diabetes"} |
| Target | 7494 | XBP1 | {"name": "XBP1", "id": "7494"} | CORRESPONDS_TO | - | Disease | - | Fatty Liver | {"name": "Fatty Liver"} |
| Target | 51043 | ZBTB7B | {"name": "ZBTB7B", "id": "51043"} | CORRESPONDS_TO | - | Disease | - | Obesity | {"name": "Obesity"} |
| Target | 84872 | ZC3H10 | {"name": "ZC3H10", "id": "84872"} | CORRESPONDS_TO | - | Disease | - | Obesity | {"name": "Obesity"} |
| Target | 51530 | ZC3HC1 | {"name": "ZC3HC1", "id": "51530"} | CORRESPONDS_TO | - | Disease | - | Type 2 Diabetes | {"name": "Type 2 Diabetes"} |
| Target | 60685 | ZFAND3 | {"name": "ZFAND3", "id": "60685"} | CORRESPONDS_TO | - | Disease | - | Type 2 Diabetes | {"name": "Type 2 Diabetes"} |
| Target | 463 | ZFHX3 | {"name": "ZFHX3", "id": "463"} | CORRESPONDS_TO | - | Disease | - | Obesity | {"name": "Obesity"} |
| Target | 23217 | ZFR2 | {"name": "ZFR2", "id": "23217"} | CORRESPONDS_TO | - | Disease | - | Obesity | {"name": "Obesity"} |
| Target | 169841 | ZNF169 | {"name": "ZNF169", "id": "169841"} | CORRESPONDS_TO | - | Disease | - | Obesity | {"name": "Obesity"} |

**Table S10: Cypher queries used for knowledge graph-based analyses. This table lists the Cypher queries implemented within the Neo4j environment to retrieve and analyze relationships among formulations, herbs, phytochemicals, molecular targets, FDA-approved drugs, and the three disease conditions considered in this study: type 2 diabetes, fatty liver disease, and obesity. The queries were designed to identify shared targets between formulations and FDA-approved drugs, overlaps between formulation-associated and disease-associated target sets, shared targets across disease pairs and all three disease conditions, and disease-associated targets modulated by both formulations.**

| Tasks | Cypher Query |
|---|---|
| Target Modulated by Formulations and FDA approved Drugs | "MATCH (f:Formulation)-[:CONTAINS_HERB]->(h:Herb)-[:CONTAINS_CHEMICAL]->(c:Chemical)-[:INTERACTS_WITH]->(t:Target)<-[:AFFECTS]-(d:Drug)<br>OPTIONAL MATCH (d)-[:TREATS]->(dis:Disease)<br>RETURN f.name AS Formulation,<br> h.name AS Herb,<br> c.name AS Phytochemical,<br> t.symbol AS Target,<br> d.name AS Drug,<br> dis.name AS Disease" |
| Intersection Between Formulation and DiseaseTarget Sets | MATCH (f:Formulation)-[:CONTAINS_HERB]->(:Herb)-[:CONTAINS_CHEMICAL]->(:Chemical)-[:INTERACTS_WITH]->(t:Target)-[:CORRESPONDS_TO]->(dis:Disease)<br>WHERE dis.name IN ['Obesity', 'Type 2 Diabetes', 'Fatty Liver']<br>WITH f.name AS Formulation, dis.name AS Disease, collect(DISTINCT toString(t.entrez_id)) AS Overlapping_Genes<br>RETURN Formulation,<br> Disease,<br> size(Overlapping_Genes) AS Overlap_Count,<br> Overlapping_Genes |
| Target involved in atleast 2 disease and modulated by both formulations | MATCH (f:Formulation)-[:CONTAINS_HERB]->(:Herb)-[:CONTAINS_CHEMICAL]->(:Chemical)-[:INTERACTS_WITH]->(t:Target)<br>WITH t, collect(DISTINCT coalesce(f.name, f.formulation_id, f.id)) AS Formulations<br>WHERE size(Formulations) > 1<br>MATCH (t)-[r]-(dis:Disease)<br>WHERE type(r) =~ '(?i)corresponds_to'<br>WITH t, Formulations, collect(DISTINCT coalesce(dis.name, dis.disease_id, dis.id)) AS Diseases<br>WHERE size(Diseases) >= 2<br>RETURN coalesce(t.symbol, t.target_id, t.id, t.name) AS Target,<br> Formulations,<br> size(Diseases) AS Disease_Count,<br> Diseases AS Diseases_Involved<br>ORDER BY Disease_Count DESC, Target ASC |
| Overlap of assocaited target between disease pairs | MATCH (d1:Disease)-[]-(t:Target)-[]-(d2:Disease)<br>WHERE d1.name <> d2.name<br>WITH d1, d2, count(DISTINCT t) AS shared_targets<br>RETURN d1.name AS Disease_Name,<br> d2.name AS Overlaps_With,<br> shared_targets AS Overlap_Count<br>ORDER BY Disease_Name ASC, Overlap_Count DESC |
| Overlap between targets of all three diseases | MATCH (d1:Disease {name: 'Type 2 Diabetes'})-[]-(t:Target)-[]-(d2:Disease {name: 'Obesity'}),<br> (t)-[]-(d3:Disease {name: 'Fatty Liver'})<br>RETURN t.symbol AS Core_Gene_Symbol,<br> t.entrez_id AS Entrez_ID<br>ORDER BY t.symbol ASC |
| Overlap of phytochemical between the formulation having at least one target | WITH "Nishamalaki" AS form1Name, "Arogyavardhini" AS form2Name<br>MATCH (f1:Formulation {name: form1Name})-[:CONTAINS_HERB]-(:Herb)-[:CONTAINS_CHEMICAL]-(p:Chemical)<br>MATCH (f2:Formulation {name: form2Name})-[:CONTAINS_HERB]-(:Herb)-[:CONTAINS_CHEMICAL]-(p)<br>MATCH (p)-[]-(t:Target)<br>RETURN<br> count(DISTINCT p.name) AS TotalSharedTargetAssociatedPhytos,<br> collect(DISTINCT p.name) AS SharedPhytochemicals |
| Overlap of target between both formulation | WITH "Nishamalaki" AS form1Name, "Arogyavardhini" AS form2Name<br>MATCH (f1:Formulation {name: form1Name})-[:CONTAINS_HERB]-(:Herb)-[:CONTAINS_CHEMICAL]-(:Chemical)-[]-(t:Target)<br>MATCH (f2:Formulation {name: form2Name})-[:CONTAINS_HERB]-(:Herb)-[:CONTAINS_CHEMICAL]-(:Chemical)-[]-(t)<br>RETURN<br> count(DISTINCT t) AS TotalSharedTargets,<br> collect(DISTINCT coalesce(t.symbol, t.name)) AS SharedTargetNames |

**Table S11: Statistical overlap between disease-associated gene sets. For each pair of disease conditions, the table reports the sizes of the corresponding disease-associated gene sets, the number of overlapping genes, and the associated statistical significance. p-values were calculated using a right-tailed hypergeometric test to assess whether the observed overlaps were greater than expected by chance. Multiple testing correction was performed using the Benjamini-Hochberg false discovery rate (FDR) method, and the corresponding adjusted p-values are reported.**

| Disease 1 | Disease 2 | Gene set size (Disease 1) | Gene set size (Disease 2) | Overlap size | p-value | FDR-adjusted p-value | Significant (FDR<0.05) |
|---|---|---|---|---|---|---|---|
| Type 2 diabetes | Fatty liver | 279 | 149 | 41 | 1.941975636e-41 | 1.941975636e-41 | True |
| Type 2 diabetes | Obesity | 279 | 253 | 62 | 2.113351443e-59 | 6.34005433e-59 | True |
| Fatty liver | Obesity | 149 | 253 | 46 | 3.577383508e-51 | 5.366075263e-51 | True |

**Table S12: Statistical overlap between formulation-associated target sets and disease-associated gene sets. For each formulation-disease pair, the table reports the sizes of the formulation-associated target set and the corresponding disease-associated gene set, the number of overlapping genes, and the associated statistical significance. p-values were calculated using a right-tailed hypergeometric test to evaluate whether the observed overlaps were greater than expected by chance. Multiple testing correction was performed using the Benjamini-Hochberg false discovery rate (FDR) method, and the corresponding adjusted p-values are reported.**

| Fromulation | Disease | Formulation target set size | Disease-associated gene set size | Overlap size | p-value | FDR-adjusted p-value | Significant (FDR<0.05) |
|---|---|---|---|---|---|---|---|
| Arogyavardhini | Type 2 diabetes | 656 | 279 | 40 | 6.675570149e-15 | 8.010684179e-15 | True |
| Nishamalaki | Type 2 diabetes | 617 | 279 | 38 | 2.476004808e-14 | 2.476004808e-14 | True |
| Arogyavardhini | Fatty liver | 656 | 149 | 36 | 3.638899453e-21 | 7.277798907e-21 | True |
| Nishamalaki | Fatty liver | 617 | 149 | 33 | 4.595620109e-19 | 6.893430163e-19 | True |
| Arogyavardhini | Obesity | 656 | 253 | 49 | 8.804248052e-24 | 2.641274416e-23 | True |
| Nishamalaki | Obesity | 617 | 253 | 50 | 7.167902528e-26 | 4.300741517e-25 | True |

**Table S13: Knowledge graph-based identification of disease-associated targets shared between the formulations and at least two diseases. The table summarizes targets that are modulated by both formulations and are associated with at least two of the three disease conditions considered in this study: type 2 diabetes, fatty liver disease, and obesity.**

| Gene Symbol | Formulations | Diseases |
|---|---|---|
| PPARA | Arogyavardhini, Nishamalaki | Type 2 diabetes, Fatty liver, Obesity |
| PPARG | Arogyavardhini, Nishamalaki | Type 2 diabetes, Fatty liver, Obesity |
| AKT1 | Arogyavardhini, Nishamalaki | Type 2 diabetes, Fatty liver, Obesity |
| SIRT1 | Arogyavardhini, Nishamalaki | Type 2 diabetes, Fatty liver, Obesity |
| TNF | Arogyavardhini, Nishamalaki | Type 2 diabetes, Fatty liver, Obesity |
| PPARD | Arogyavardhini, Nishamalaki | Type 2 diabetes, Fatty liver, Obesity |
| SERPINE1 | Arogyavardhini, Nishamalaki | Type 2 diabetes, Fatty liver, Obesity |
| ACE | Arogyavardhini, Nishamalaki | Type 2 diabetes, Fatty liver, Obesity |
| CYP1A2 | Arogyavardhini, Nishamalaki | Type 2 diabetes, Fatty liver |
| HMGCR | Arogyavardhini, Nishamalaki | Type 2 diabetes, Fatty liver |
| PON1 | Arogyavardhini, Nishamalaki | Type 2 diabetes, Fatty liver |
| SLC2A2 | Arogyavardhini, Nishamalaki | Type 2 diabetes, Fatty liver |
| CYP1B1 | Arogyavardhini, Nishamalaki | Fatty liver, Obesity |
| CYP2E1 | Arogyavardhini, Nishamalaki | Fatty liver, Obesity |
| ESR1 | Arogyavardhini, Nishamalaki | Fatty liver, Obesity |
| F2 | Arogyavardhini, Nishamalaki | Fatty liver, Obesity |
| ADRB3 | Arogyavardhini, Nishamalaki | Type 2 diabetes, Obesity |
| INSR | Arogyavardhini, Nishamalaki | Type 2 diabetes, Obesity |
| MC4R | Arogyavardhini, Nishamalaki | Type 2 diabetes. Obesity |
| PTPN1 | Arogyavardhini, Nishamalaki | Type 2 diabetes, Obesity |
| STAT3 | Arogyavardhini, Nishamalaki | Type 2 diabetes, Obesity |

**Table S14. Formulation – herb – phytochemical – target – FDA-approved drug – disease associations linking formulations to therapeutically relevant drugs. This table presents formulation, herbs, their constituent phytochemicals, associated molecular targets, and corresponding FDA-approved drugs, highlighting target-level overlaps between formulations and FDA-approved drugs**

| Formulation | Herb | Phytochemical | Gene Symbol | Mimicked Drug | Disease |
|---|---|---|---|---|---|
| Arogyavardhini | Commiphora wightii | 5-[3-(1,3-Benzodioxol-5-yl)-1,3,3a,4,6,6a-hexahydrofuro[3,4-c]furan-6-yl]-1,3-benzodioxole | GLP1R | Pramlintide | Type 2 diabetes |
| Arogyavardhini | Commiphora wightii | 5-[3-(1,3-Benzodioxol-5-yl)-1,3,3a,4,6,6a-hexahydrofuro[3,4-c]furan-6-yl]-1,3-benzodioxole | GLP1R | Semaglutide | Type 2 diabetes |
| Arogyavardhini | Commiphora wightii | 5-[3-(1,3-Benzodioxol-5-yl)-1,3,3a,4,6,6a-hexahydrofuro[3,4-c]furan-6-yl]-1,3-benzodioxole | GLP1R | Tirzepatide | Type 2 diabetes |
| Arogyavardhini | Terminalia chebula | Docosanoic acid | PPARG | Pioglitazone | Type 2 diabetes |
| Arogyavardhini | Azadirachta indica | Isorhamnetin | PPARG | Pioglitazone | Type 2 diabetes |
| Arogyavardhini | Terminalia chebula | Linoleic acid | PPARG | Pioglitazone | Type 2 diabetes |
| Arogyavardhini | Phyllanthus emblica | Riboflavin | PPARG | Pioglitazone | Type 2 diabetes |
| Arogyavardhini | Commiphora wightii | Pluviatilol | PPARG | Pioglitazone | Type 2 diabetes |
| Arogyavardhini | Phyllanthus emblica | Quercetin | PPARG | Pioglitazone | Type 2 diabetes |
| Arogyavardhini | Azadirachta indica | Quercetin | PPARG | Pioglitazone | Type 2 diabetes |
| Arogyavardhini | Terminalia chebula | Arachidic acid | PPARG | Pioglitazone | Type 2 diabetes |
| Arogyavardhini | Terminalia chebula | Stearic acid | PPARG | Pioglitazone | Type 2 diabetes |
| Arogyavardhini | Phyllanthus emblica | Catechol | PPARG | Pioglitazone | Type 2 diabetes |
| Arogyavardhini | Terminalia chebula | Oleic acid | PPARG | Pioglitazone | Type 2 diabetes |
| Arogyavardhini | Commiphora wightii | E-Guggulsterone | PPARG | Pioglitazone | Type 2 diabetes |
| Arogyavardhini | Terminalia chebula | Palmitic acid | PPARG | Pioglitazone | Type 2 diabetes |
| Arogyavardhini | Terminalia chebula | Caffeic acid | DPP4 | Alogliptin | Type 2 diabetes |
| Arogyavardhini | Terminalia bellirica | Gallic acid | DPP4 | Alogliptin | Type 2 diabetes |
| Arogyavardhini | Terminalia chebula | Gallic acid | DPP4 | Alogliptin | Type 2 diabetes |
| Arogyavardhini | Phyllanthus emblica | Gallic acid | DPP4 | Alogliptin | Type 2 diabetes |
| Arogyavardhini | Terminalia chebula | 4-Hydroxycinnamic acid | DPP4 | Alogliptin | Type 2 diabetes |
| Arogyavardhini | Terminalia chebula | beta-Sitosterol | DPP4 | Alogliptin | Type 2 diabetes |
| Arogyavardhini | Azadirachta indica | beta-Sitosterol | DPP4 | Alogliptin | Type 2 diabetes |
| Arogyavardhini | Commiphora wightii | beta-Sitosterol | DPP4 | Alogliptin | Type 2 diabetes |
| Arogyavardhini | Terminalia bellirica | beta-Sitosterol | DPP4 | Alogliptin | Type 2 diabetes |
| Arogyavardhini | Phyllanthus emblica | Quercetin | DPP4 | Alogliptin | Type 2 diabetes |
| Arogyavardhini | Azadirachta indica | Quercetin | DPP4 | Alogliptin | Type 2 diabetes |
| Arogyavardhini | Terminalia chebula | Caffeic acid | DPP4 | Linagliptin | Type 2 diabetes |
| Arogyavardhini | Terminalia bellirica | Gallic acid | DPP4 | Linagliptin | Type 2 diabetes |
| Arogyavardhini | Terminalia chebula | Gallic acid | DPP4 | Linagliptin | Type 2 diabetes |
| Arogyavardhini | Phyllanthus emblica | Gallic acid | DPP4 | Linagliptin | Type 2 diabetes |
| Arogyavardhini | Terminalia chebula | 4-Hydroxycinnamic acid | DPP4 | Linagliptin | Type 2 diabetes |
| Arogyavardhini | Terminalia chebula | beta-Sitosterol | DPP4 | Linagliptin | Type 2 diabetes |
| Arogyavardhini | Azadirachta indica | beta-Sitosterol | DPP4 | Linagliptin | Type 2 diabetes |
| Arogyavardhini | Commiphora wightii | beta-Sitosterol | DPP4 | Linagliptin | Type 2 diabetes |
| Arogyavardhini | Terminalia bellirica | beta-Sitosterol | DPP4 | Linagliptin | Type 2 diabetes |
| Arogyavardhini | Phyllanthus emblica | Quercetin | DPP4 | Linagliptin | Type 2 diabetes |
| Arogyavardhini | Azadirachta indica | Quercetin | DPP4 | Linagliptin | Type 2 diabetes |
| Arogyavardhini | Terminalia chebula | Caffeic acid | DPP4 | Saxagliptin | Type 2 diabetes |
| Arogyavardhini | Terminalia bellirica | Gallic acid | DPP4 | Saxagliptin | Type 2 diabetes |
| Arogyavardhini | Terminalia chebula | Gallic acid | DPP4 | Saxagliptin | Type 2 diabetes |
| Arogyavardhini | Phyllanthus emblica | Gallic acid | DPP4 | Saxagliptin | Type 2 diabetes |
| Arogyavardhini | Terminalia chebula | 4-Hydroxycinnamic acid | DPP4 | Saxagliptin | Type 2 diabetes |
| Arogyavardhini | Terminalia chebula | beta-Sitosterol | DPP4 | Saxagliptin | Type 2 diabetes |
| Arogyavardhini | Azadirachta indica | beta-Sitosterol | DPP4 | Saxagliptin | Type 2 diabetes |
| Arogyavardhini | Commiphora wightii | beta-Sitosterol | DPP4 | Saxagliptin | Type 2 diabetes |
| Arogyavardhini | Terminalia bellirica | beta-Sitosterol | DPP4 | Saxagliptin | Type 2 diabetes |
| Arogyavardhini | Phyllanthus emblica | Quercetin | DPP4 | Saxagliptin | Type 2 diabetes |
| Arogyavardhini | Azadirachta indica | Quercetin | DPP4 | Saxagliptin | Type 2 diabetes |
| Arogyavardhini | Terminalia chebula | Caffeic acid | DPP4 | Sitagliptin | Type 2 diabetes |
| Arogyavardhini | Terminalia bellirica | Gallic acid | DPP4 | Sitagliptin | Type 2 diabetes |
| Arogyavardhini | Terminalia chebula | Gallic acid | DPP4 | Sitagliptin | Type 2 diabetes |
| Arogyavardhini | Phyllanthus emblica | Gallic acid | DPP4 | Sitagliptin | Type 2 diabetes |
| Arogyavardhini | Terminalia chebula | 4-Hydroxycinnamic acid | DPP4 | Sitagliptin | Type 2 diabetes |
| Arogyavardhini | Terminalia chebula | beta-Sitosterol | DPP4 | Sitagliptin | Type 2 diabetes |
| Arogyavardhini | Azadirachta indica | beta-Sitosterol | DPP4 | Sitagliptin | Type 2 diabetes |
| Arogyavardhini | Commiphora wightii | beta-Sitosterol | DPP4 | Sitagliptin | Type 2 diabetes |
| Arogyavardhini | Terminalia bellirica | beta-Sitosterol | DPP4 | Sitagliptin | Type 2 diabetes |
| Arogyavardhini | Phyllanthus emblica | Quercetin | DPP4 | Sitagliptin | Type 2 diabetes |
| Arogyavardhini | Azadirachta indica | Quercetin | DPP4 | Sitagliptin | Type 2 diabetes |
| Arogyavardhini | Phyllanthus emblica | Ellagic acid | INSR | Metformin | Type 2 diabetes |
| Arogyavardhini | Terminalia bellirica | Ellagic acid | INSR | Metformin | Type 2 diabetes |
| Arogyavardhini | Terminalia chebula | Ellagic acid | INSR | Metformin | Type 2 diabetes |
| Arogyavardhini | Terminalia chebula | Vanillic acid | INSR | Metformin | Type 2 diabetes |
| Arogyavardhini | Picrorhiza kurroa | Vanillic acid | INSR | Metformin | Type 2 diabetes |
| Arogyavardhini | Phyllanthus emblica | Quercetin | MGAM | Miglitol | Type 2 diabetes |
| Arogyavardhini | Azadirachta indica | Quercetin | MGAM | Miglitol | Type 2 diabetes |
| Arogyavardhini | Azadirachta indica | Scopoletin | GAA | Miglitol | Type 2 diabetes |
| Arogyavardhini | Phyllanthus emblica | Quercetin | GAA | Miglitol | Type 2 diabetes |
| Arogyavardhini | Azadirachta indica | Quercetin | GAA | Miglitol | Type 2 diabetes |

**Table S14. Formulation – herb – phytochemical – target – FDA-approved drug – disease associations linking formulations to therapeutically relevant drugs. This table presents formulation, herbs, their constituent phytochemicals, associated molecular targets, and corresponding FDA-approved drugs, highlighting target-level overlaps between formulations and FDA-approved drugs**

| Formulation | Herb | Phytochemical | Gene Symbol | Mimicked Drug | Disease |
|---|---|---|---|---|---|
| Arogyavardhini | Phyllanthus emblica | Methyl gallate | GAA | Miglitol | Type 2 diabetes |
| Arogyavardhini | Terminalia chebula | Caffeic acid | GAA | Miglitol | Type 2 diabetes |
| Arogyavardhini | Phyllanthus emblica | Ellagic acid | GAA | Miglitol | Type 2 diabetes |
| Arogyavardhini | Terminalia bellirica | Ellagic acid | GAA | Miglitol | Type 2 diabetes |
| Arogyavardhini | Terminalia chebula | Ellagic acid | GAA | Miglitol | Type 2 diabetes |
| Arogyavardhini | Terminalia bellirica | Mannitol | HMGCR | Pravastatin | Fatty liver |
| Arogyavardhini | Terminalia chebula | Ferulic acid | HMGCR | Pravastatin | Fatty liver |
| Arogyavardhini | Commiphora wightii | Ferulic acid | HMGCR | Pravastatin | Fatty liver |
| Arogyavardhini | Terminalia chebula | Ascorbic acid | HMGCR | Pravastatin | Fatty liver |
| Arogyavardhini | Phyllanthus emblica | Ascorbic acid | HMGCR | Pravastatin | Fatty liver |
| Arogyavardhini | Phyllanthus emblica | Pyrogallol | HMGCR | Pravastatin | Fatty liver |
| Arogyavardhini | Phyllanthus emblica | Catechol | HMGCR | Pravastatin | Fatty liver |
| Arogyavardhini | Terminalia bellirica | Mannitol | HMGCR | Lovastatin | Fatty liver |
| Arogyavardhini | Terminalia chebula | Ferulic acid | HMGCR | Lovastatin | Fatty liver |
| Arogyavardhini | Commiphora wightii | Ferulic acid | HMGCR | Lovastatin | Fatty liver |
| Arogyavardhini | Terminalia chebula | Ascorbic acid | HMGCR | Lovastatin | Fatty liver |
| Arogyavardhini | Phyllanthus emblica | Ascorbic acid | HMGCR | Lovastatin | Fatty liver |
| Arogyavardhini | Phyllanthus emblica | Pyrogallol | HMGCR | Lovastatin | Fatty liver |
| Arogyavardhini | Phyllanthus emblica | Catechol | HMGCR | Lovastatin | Fatty liver |
| Arogyavardhini | Terminalia bellirica | Mannitol | HMGCR | Simvastatin | Fatty liver |
| Arogyavardhini | Terminalia chebula | Ferulic acid | HMGCR | Simvastatin | Fatty liver |
| Arogyavardhini | Commiphora wightii | Ferulic acid | HMGCR | Simvastatin | Fatty liver |
| Arogyavardhini | Terminalia chebula | Ascorbic acid | HMGCR | Simvastatin | Fatty liver |
| Arogyavardhini | Phyllanthus emblica | Ascorbic acid | HMGCR | Simvastatin | Fatty liver |
| Arogyavardhini | Phyllanthus emblica | Pyrogallol | HMGCR | Simvastatin | Fatty liver |
| Arogyavardhini | Phyllanthus emblica | Catechol | HMGCR | Simvastatin | Fatty liver |
| Arogyavardhini | Terminalia bellirica | Mannitol | HMGCR | Atorvastatin | Fatty liver |
| Arogyavardhini | Terminalia chebula | Ferulic acid | HMGCR | Atorvastatin | Fatty liver |
| Arogyavardhini | Commiphora wightii | Ferulic acid | HMGCR | Atorvastatin | Fatty liver |
| Arogyavardhini | Terminalia chebula | Ascorbic acid | HMGCR | Atorvastatin | Fatty liver |
| Arogyavardhini | Phyllanthus emblica | Ascorbic acid | HMGCR | Atorvastatin | Fatty liver |
| Arogyavardhini | Phyllanthus emblica | Pyrogallol | HMGCR | Atorvastatin | Fatty liver |
| Arogyavardhini | Phyllanthus emblica | Catechol | HMGCR | Atorvastatin | Fatty liver |
| Arogyavardhini | Terminalia bellirica | Mannitol | HMGCR | Pitavastatin | Fatty liver |
| Arogyavardhini | Terminalia chebula | Ferulic acid | HMGCR | Pitavastatin | Fatty liver |
| Arogyavardhini | Commiphora wightii | Ferulic acid | HMGCR | Pitavastatin | Fatty liver |
| Arogyavardhini | Terminalia chebula | Ascorbic acid | HMGCR | Pitavastatin | Fatty liver |
| Arogyavardhini | Phyllanthus emblica | Ascorbic acid | HMGCR | Pitavastatin | Fatty liver |
| Arogyavardhini | Phyllanthus emblica | Pyrogallol | HMGCR | Pitavastatin | Fatty liver |
| Arogyavardhini | Phyllanthus emblica | Catechol | HMGCR | Pitavastatin | Fatty liver |
| Arogyavardhini | Azadirachta indica | beta-Caryophyllene | PPARA | Fenofibrate | Fatty liver |
| Arogyavardhini | Commiphora wightii | beta-Caryophyllene | PPARA | Fenofibrate | Fatty liver |
| Arogyavardhini | Terminalia chebula | Oleic acid | PPARA | Fenofibrate | Fatty liver |
| Arogyavardhini | Terminalia chebula | Linoleic acid | PPARA | Fenofibrate | Fatty liver |
| Arogyavardhini | Terminalia chebula | Palmitic acid | PPARA | Fenofibrate | Fatty liver |
| Arogyavardhini | Commiphora wightii | E-Guggulsterone | PPARA | Fenofibrate | Fatty liver |
| Arogyavardhini | Commiphora wightii | Pluviatilol | PPARA | Fenofibrate | Fatty liver |
| Arogyavardhini | Terminalia chebula | Docosanoic acid | PPARA | Fenofibrate | Fatty liver |
| Arogyavardhini | Terminalia chebula | Arachidic acid | PPARA | Fenofibrate | Fatty liver |
| Arogyavardhini | Terminalia chebula | Stearic acid | PPARA | Fenofibrate | Fatty liver |
| Arogyavardhini | Azadirachta indica | beta-Caryophyllene | PPARA | Fenofibric acid | Fatty liver |
| Arogyavardhini | Commiphora wightii | beta-Caryophyllene | PPARA | Fenofibric acid | Fatty liver |
| Arogyavardhini | Terminalia chebula | Oleic acid | PPARA | Fenofibric acid | Fatty liver |
| Arogyavardhini | Terminalia chebula | Linoleic acid | PPARA | Fenofibric acid | Fatty liver |
| Arogyavardhini | Terminalia chebula | Palmitic acid | PPARA | Fenofibric acid | Fatty liver |
| Arogyavardhini | Commiphora wightii | E-Guggulsterone | PPARA | Fenofibric acid | Fatty liver |
| Arogyavardhini | Commiphora wightii | Pluviatilol | PPARA | Fenofibric acid | Fatty liver |
| Arogyavardhini | Terminalia chebula | Docosanoic acid | PPARA | Fenofibric acid | Fatty liver |
| Arogyavardhini | Terminalia chebula | Arachidic acid | PPARA | Fenofibric acid | Fatty liver |
| Arogyavardhini | Terminalia chebula | Stearic acid | PPARA | Fenofibric acid | Fatty liver |
| Arogyavardhini | Terminalia bellirica | Mannitol | MC4R | Setmelanotide | Obesity |
| Arogyavardhini | Phyllanthus emblica | Pyrogallol | MC4R | Setmelanotide | Obesity |
| Arogyavardhini | Phyllanthus emblica | Catechol | MC4R | Setmelanotide | Obesity |
| Arogyavardhini | Terminalia chebula | Ascorbic acid | MC4R | Setmelanotide | Obesity |
| Arogyavardhini | Phyllanthus emblica | Ascorbic acid | MC4R | Setmelanotide | Obesity |
| Arogyavardhini | Phyllanthus emblica | Quercetin | MC4R | Setmelanotide | Obesity |
| Arogyavardhini | Azadirachta indica | Quercetin | MC4R | Setmelanotide | Obesity |
| Arogyavardhini | Terminalia chebula | Ferulic acid | MC4R | Setmelanotide | Obesity |
| Arogyavardhini | Commiphora wightii | Ferulic acid | MC4R | Setmelanotide | Obesity |
| Arogyavardhini | Phyllanthus emblica | Quercetin | PNLIP | Orlistat | Obesity |

**Table S14. Formulation – herb – phytochemical – target – FDA-approved drug – disease associations linking formulations to therapeutically relevant drugs. This table presents formulation, herbs, their constituent phytochemicals, associated molecular targets, and corresponding FDA-approved drugs, highlighting target-level overlaps between formulations and FDA-approved drugs**

| Formulation | Herb | Phytochemical | Gene Symbol | Mimicked Drug | Disease |
|---|---|---|---|---|---|
| Arogyavardhini | Azadirachta indica | Quercetin | PNLIP | Orlistat | Obesity |
| Arogyavardhini | Terminalia bellirica | Mannitol | SLC6A3 | Amphetamine | Obesity |
| Arogyavardhini | Phyllanthus emblica | Catechol | SLC6A3 | Amphetamine | Obesity |
| Arogyavardhini | Terminalia chebula | Stearic acid | SLC6A3 | Amphetamine | Obesity |
| Arogyavardhini | Terminalia chebula | Oleic acid | SLC6A3 | Amphetamine | Obesity |
| Arogyavardhini | Terminalia chebula | Ascorbic acid | SLC6A3 | Amphetamine | Obesity |
| Arogyavardhini | Phyllanthus emblica | Ascorbic acid | SLC6A3 | Amphetamine | Obesity |
| Arogyavardhini | Phyllanthus emblica | Quercetin | SLC6A3 | Amphetamine | Obesity |
| Arogyavardhini | Azadirachta indica | Quercetin | SLC6A3 | Amphetamine | Obesity |
| Arogyavardhini | Phyllanthus emblica | Pyrogallol | SLC6A3 | Amphetamine | Obesity |
| Arogyavardhini | Azadirachta indica | beta-Carotene | SLC6A3 | Amphetamine | Obesity |
| Arogyavardhini | Terminalia chebula | Ferulic acid | SLC6A3 | Amphetamine | Obesity |
| Arogyavardhini | Commiphora wightii | Ferulic acid | SLC6A3 | Amphetamine | Obesity |
| Arogyavardhini | Phyllanthus emblica | Pyrogallol | SLC6A2 | Amphetamine | Obesity |
| Arogyavardhini | Phyllanthus emblica | Quercetin | SLC6A2 | Amphetamine | Obesity |
| Arogyavardhini | Azadirachta indica | Quercetin | SLC6A2 | Amphetamine | Obesity |
| Arogyavardhini | Terminalia chebula | Oleic acid | SLC6A2 | Amphetamine | Obesity |
| Arogyavardhini | Terminalia chebula | Stearic acid | SLC6A2 | Amphetamine | Obesity |
| Arogyavardhini | Terminalia chebula | Ascorbic acid | SLC6A2 | Amphetamine | Obesity |
| Arogyavardhini | Phyllanthus emblica | Ascorbic acid | SLC6A2 | Amphetamine | Obesity |
| Arogyavardhini | Terminalia bellirica | Mannitol | SLC6A2 | Amphetamine | Obesity |
| Arogyavardhini | Phyllanthus emblica | Catechol | SLC6A2 | Amphetamine | Obesity |
| Arogyavardhini | Azadirachta indica | beta-Carotene | SLC6A2 | Amphetamine | Obesity |
| Arogyavardhini | Terminalia chebula | Ferulic acid | SLC6A2 | Amphetamine | Obesity |
| Arogyavardhini | Commiphora wightii | Ferulic acid | SLC6A2 | Amphetamine | Obesity |
| Arogyavardhini | Terminalia bellirica | Mannitol | SLC6A3 | Diethylpropion | Obesity |
| Arogyavardhini | Phyllanthus emblica | Catechol | SLC6A3 | Diethylpropion | Obesity |
| Arogyavardhini | Terminalia chebula | Stearic acid | SLC6A3 | Diethylpropion | Obesity |
| Arogyavardhini | Terminalia chebula | Oleic acid | SLC6A3 | Diethylpropion | Obesity |
| Arogyavardhini | Terminalia chebula | Ascorbic acid | SLC6A3 | Diethylpropion | Obesity |
| Arogyavardhini | Phyllanthus emblica | Ascorbic acid | SLC6A3 | Diethylpropion | Obesity |
| Arogyavardhini | Phyllanthus emblica | Quercetin | SLC6A3 | Diethylpropion | Obesity |
| Arogyavardhini | Azadirachta indica | Quercetin | SLC6A3 | Diethylpropion | Obesity |
| Arogyavardhini | Phyllanthus emblica | Pyrogallol | SLC6A3 | Diethylpropion | Obesity |
| Arogyavardhini | Azadirachta indica | beta-Carotene | SLC6A3 | Diethylpropion | Obesity |
| Arogyavardhini | Terminalia chebula | Ferulic acid | SLC6A3 | Diethylpropion | Obesity |
| Arogyavardhini | Commiphora wightii | Ferulic acid | SLC6A3 | Diethylpropion | Obesity |
| Arogyavardhini | Phyllanthus emblica | Pyrogallol | SLC6A2 | Diethylpropion | Obesity |
| Arogyavardhini | Phyllanthus emblica | Quercetin | SLC6A2 | Diethylpropion | Obesity |
| Arogyavardhini | Azadirachta indica | Quercetin | SLC6A2 | Diethylpropion | Obesity |
| Arogyavardhini | Terminalia chebula | Oleic acid | SLC6A2 | Diethylpropion | Obesity |
| Arogyavardhini | Terminalia chebula | Stearic acid | SLC6A2 | Diethylpropion | Obesity |
| Arogyavardhini | Terminalia chebula | Ascorbic acid | SLC6A2 | Diethylpropion | Obesity |
| Arogyavardhini | Phyllanthus emblica | Ascorbic acid | SLC6A2 | Diethylpropion | Obesity |
| Arogyavardhini | Terminalia bellirica | Mannitol | SLC6A2 | Diethylpropion | Obesity |
| Arogyavardhini | Phyllanthus emblica | Catechol | SLC6A2 | Diethylpropion | Obesity |
| Arogyavardhini | Azadirachta indica | beta-Carotene | SLC6A2 | Diethylpropion | Obesity |
| Arogyavardhini | Terminalia chebula | Ferulic acid | SLC6A2 | Diethylpropion | Obesity |
| Arogyavardhini | Commiphora wightii | Ferulic acid | SLC6A2 | Diethylpropion | Obesity |
| Arogyavardhini | Terminalia bellirica | Mannitol | SLC6A3 | Metamfetamine | Obesity |
| Arogyavardhini | Phyllanthus emblica | Catechol | SLC6A3 | Metamfetamine | Obesity |
| Arogyavardhini | Terminalia chebula | Stearic acid | SLC6A3 | Metamfetamine | Obesity |
| Arogyavardhini | Terminalia chebula | Oleic acid | SLC6A3 | Metamfetamine | Obesity |
| Arogyavardhini | Terminalia chebula | Ascorbic acid | SLC6A3 | Metamfetamine | Obesity |
| Arogyavardhini | Phyllanthus emblica | Ascorbic acid | SLC6A3 | Metamfetamine | Obesity |
| Arogyavardhini | Phyllanthus emblica | Quercetin | SLC6A3 | Metamfetamine | Obesity |
| Arogyavardhini | Azadirachta indica | Quercetin | SLC6A3 | Metamfetamine | Obesity |
| Arogyavardhini | Phyllanthus emblica | Pyrogallol | SLC6A3 | Metamfetamine | Obesity |
| Arogyavardhini | Azadirachta indica | beta-Carotene | SLC6A3 | Metamfetamine | Obesity |
| Arogyavardhini | Terminalia chebula | Ferulic acid | SLC6A3 | Metamfetamine | Obesity |
| Arogyavardhini | Commiphora wightii | Ferulic acid | SLC6A3 | Metamfetamine | Obesity |
| Arogyavardhini | Terminalia bellirica | Mannitol | SLC6A3 | Phendimetrazine | Obesity |
| Arogyavardhini | Phyllanthus emblica | Catechol | SLC6A3 | Phendimetrazine | Obesity |
| Arogyavardhini | Terminalia chebula | Stearic acid | SLC6A3 | Phendimetrazine | Obesity |
| Arogyavardhini | Terminalia chebula | Oleic acid | SLC6A3 | Phendimetrazine | Obesity |
| Arogyavardhini | Terminalia chebula | Ascorbic acid | SLC6A3 | Phendimetrazine | Obesity |
| Arogyavardhini | Phyllanthus emblica | Ascorbic acid | SLC6A3 | Phendimetrazine | Obesity |
| Arogyavardhini | Phyllanthus emblica | Quercetin | SLC6A3 | Phendimetrazine | Obesity |
| Arogyavardhini | Azadirachta indica | Quercetin | SLC6A3 | Phendimetrazine | Obesity |
| Arogyavardhini | Phyllanthus emblica | Pyrogallol | SLC6A3 | Phendimetrazine | Obesity |

**Table S14. Formulation – herb – phytochemical – target – FDA-approved drug – disease associations linking formulations to therapeutically relevant drugs. This table presents formulation, herbs, their constituent phytochemicals, associated molecular targets, and corresponding FDA-approved drugs, highlighting target-level overlaps between formulations and FDA-approved drugs**

| Formulation | Herb | Phytochemical | Gene Symbol | Mimicked Drug | Disease |
|---|---|---|---|---|---|
| Arogyavardhini | Azadirachta indica | beta-Carotene | SLC6A3 | Phendimetrazine | Obesity |
| Arogyavardhini | Terminalia chebula | Ferulic acid | SLC6A3 | Phendimetrazine | Obesity |
| Arogyavardhini | Commiphora wightii | Ferulic acid | SLC6A3 | Phendimetrazine | Obesity |
| Arogyavardhini | Phyllanthus emblica | Pyrogallol | SLC6A2 | Phendimetrazine | Obesity |
| Arogyavardhini | Phyllanthus emblica | Quercetin | SLC6A2 | Phendimetrazine | Obesity |
| Arogyavardhini | Azadirachta indica | Quercetin | SLC6A2 | Phendimetrazine | Obesity |
| Arogyavardhini | Terminalia chebula | Oleic acid | SLC6A2 | Phendimetrazine | Obesity |
| Arogyavardhini | Terminalia chebula | Stearic acid | SLC6A2 | Phendimetrazine | Obesity |
| Arogyavardhini | Terminalia chebula | Ascorbic acid | SLC6A2 | Phendimetrazine | Obesity |
| Arogyavardhini | Phyllanthus emblica | Ascorbic acid | SLC6A2 | Phendimetrazine | Obesity |
| Arogyavardhini | Terminalia bellirica | Mannitol | SLC6A2 | Phendimetrazine | Obesity |
| Arogyavardhini | Phyllanthus emblica | Catechol | SLC6A2 | Phendimetrazine | Obesity |
| Arogyavardhini | Azadirachta indica | beta-Carotene | SLC6A2 | Phendimetrazine | Obesity |
| Arogyavardhini | Terminalia chebula | Ferulic acid | SLC6A2 | Phendimetrazine | Obesity |
| Arogyavardhini | Commiphora wightii | Ferulic acid | SLC6A2 | Phendimetrazine | Obesity |
| Arogyavardhini | Phyllanthus emblica | Pyrogallol | SLC6A2 | Benzphetamine | Obesity |
| Arogyavardhini | Phyllanthus emblica | Quercetin | SLC6A2 | Benzphetamine | Obesity |
| Arogyavardhini | Azadirachta indica | Quercetin | SLC6A2 | Benzphetamine | Obesity |
| Arogyavardhini | Terminalia chebula | Oleic acid | SLC6A2 | Benzphetamine | Obesity |
| Arogyavardhini | Terminalia chebula | Stearic acid | SLC6A2 | Benzphetamine | Obesity |
| Arogyavardhini | Terminalia chebula | Ascorbic acid | SLC6A2 | Benzphetamine | Obesity |
| Arogyavardhini | Phyllanthus emblica | Ascorbic acid | SLC6A2 | Benzphetamine | Obesity |
| Arogyavardhini | Terminalia bellirica | Mannitol | SLC6A2 | Benzphetamine | Obesity |
| Arogyavardhini | Phyllanthus emblica | Catechol | SLC6A2 | Benzphetamine | Obesity |
| Arogyavardhini | Azadirachta indica | beta-Carotene | SLC6A2 | Benzphetamine | Obesity |
| Arogyavardhini | Terminalia chebula | Ferulic acid | SLC6A2 | Benzphetamine | Obesity |
| Arogyavardhini | Commiphora wightii | Ferulic acid | SLC6A2 | Benzphetamine | Obesity |
| Arogyavardhini | Phyllanthus emblica | Pyrogallol | SLC6A2 | Phentermine | Obesity |
| Arogyavardhini | Phyllanthus emblica | Quercetin | SLC6A2 | Phentermine | Obesity |
| Arogyavardhini | Azadirachta indica | Quercetin | SLC6A2 | Phentermine | Obesity |
| Arogyavardhini | Terminalia chebula | Oleic acid | SLC6A2 | Phentermine | Obesity |
| Arogyavardhini | Terminalia chebula | Stearic acid | SLC6A2 | Phentermine | Obesity |
| Arogyavardhini | Terminalia chebula | Ascorbic acid | SLC6A2 | Phentermine | Obesity |
| Arogyavardhini | Phyllanthus emblica | Ascorbic acid | SLC6A2 | Phentermine | Obesity |
| Arogyavardhini | Terminalia bellirica | Mannitol | SLC6A2 | Phentermine | Obesity |
| Arogyavardhini | Phyllanthus emblica | Catechol | SLC6A2 | Phentermine | Obesity |
| Arogyavardhini | Azadirachta indica | beta-Carotene | SLC6A2 | Phentermine | Obesity |
| Arogyavardhini | Terminalia chebula | Ferulic acid | SLC6A2 | Phentermine | Obesity |
| Arogyavardhini | Commiphora wightii | Ferulic acid | SLC6A2 | Phentermine | Obesity |
| Nishamalaki | Curcuma longa | Piperine | PPARG | Pioglitazone | Type 2 diabetes |
| Nishamalaki | Phyllanthus emblica | Riboflavin | PPARG | Pioglitazone | Type 2 diabetes |
| Nishamalaki | Curcuma longa | Decanoic acid | PPARG | Pioglitazone | Type 2 diabetes |
| Nishamalaki | Phyllanthus emblica | Quercetin | PPARG | Pioglitazone | Type 2 diabetes |
| Nishamalaki | Phyllanthus emblica | Catechol | PPARG | Pioglitazone | Type 2 diabetes |
| Nishamalaki | Curcuma longa | Methyleugenol | PPARG | Pioglitazone | Type 2 diabetes |
| Nishamalaki | Curcuma longa | Curcumin | PPARG | Pioglitazone | Type 2 diabetes |
| Nishamalaki | Phyllanthus emblica | Gallic acid | DPP4 | Alogliptin | Type 2 diabetes |
| Nishamalaki | Curcuma longa | beta-Sitosterol | DPP4 | Alogliptin | Type 2 diabetes |
| Nishamalaki | Phyllanthus emblica | Quercetin | DPP4 | Alogliptin | Type 2 diabetes |
| Nishamalaki | Phyllanthus emblica | Gallic acid | DPP4 | Linagliptin | Type 2 diabetes |
| Nishamalaki | Curcuma longa | beta-Sitosterol | DPP4 | Linagliptin | Type 2 diabetes |
| Nishamalaki | Phyllanthus emblica | Quercetin | DPP4 | Linagliptin | Type 2 diabetes |
| Nishamalaki | Phyllanthus emblica | Gallic acid | DPP4 | Saxagliptin | Type 2 diabetes |
| Nishamalaki | Curcuma longa | beta-Sitosterol | DPP4 | Saxagliptin | Type 2 diabetes |
| Nishamalaki | Phyllanthus emblica | Quercetin | DPP4 | Saxagliptin | Type 2 diabetes |
| Nishamalaki | Phyllanthus emblica | Gallic acid | DPP4 | Sitagliptin | Type 2 diabetes |
| Nishamalaki | Curcuma longa | beta-Sitosterol | DPP4 | Sitagliptin | Type 2 diabetes |
| Nishamalaki | Phyllanthus emblica | Quercetin | DPP4 | Sitagliptin | Type 2 diabetes |
| Nishamalaki | Phyllanthus emblica | Ellagic acid | INSR | Metformin | Type 2 diabetes |
| Nishamalaki | Phyllanthus emblica | Quercetin | MGAM | Miglitol | Type 2 diabetes |
| Nishamalaki | Curcuma longa | Curcumin | GAA | Miglitol | Type 2 diabetes |
| Nishamalaki | Phyllanthus emblica | Quercetin | GAA | Miglitol | Type 2 diabetes |
| Nishamalaki | Phyllanthus emblica | Methyl gallate | GAA | Miglitol | Type 2 diabetes |
| Nishamalaki | Phyllanthus emblica | Ellagic acid | GAA | Miglitol | Type 2 diabetes |
| Nishamalaki | Curcuma longa | Geraniol | HMGCR | Pravastatin | Fatty liver |
| Nishamalaki | Curcuma longa | Safrole | HMGCR | Pravastatin | Fatty liver |
| Nishamalaki | Phyllanthus emblica | Ascorbic acid | HMGCR | Pravastatin | Fatty liver |
| Nishamalaki | Curcuma longa | Eucalyptol | HMGCR | Pravastatin | Fatty liver |
| Nishamalaki | Curcuma longa | Eugenol | HMGCR | Pravastatin | Fatty liver |
| Nishamalaki | Phyllanthus emblica | Pyrogallol | HMGCR | Pravastatin | Fatty liver |

**Table S14. Formulation – herb – phytochemical – target – FDA-approved drug – disease associations linking formulations to therapeutically relevant drugs. This table presents formulation, herbs, their constituent phytochemicals, associated molecular targets, and corresponding FDA-approved drugs, highlighting target-level overlaps between formulations and FDA-approved drugs**

| Formulation | Herb | Phytochemical | Gene Symbol | Mimicked Drug | Disease |
|---|---|---|---|---|---|
| Nishamalaki | Phyllanthus emblica | Catechol | HMGCR | Pravastatin | Fatty liver |
| Nishamalaki | Curcuma longa | Geraniol | HMGCR | Lovastatin | Fatty liver |
| Nishamalaki | Curcuma longa | Safrole | HMGCR | Lovastatin | Fatty liver |
| Nishamalaki | Phyllanthus emblica | Ascorbic acid | HMGCR | Lovastatin | Fatty liver |
| Nishamalaki | Curcuma longa | Eucalyptol | HMGCR | Lovastatin | Fatty liver |
| Nishamalaki | Curcuma longa | Eugenol | HMGCR | Lovastatin | Fatty liver |
| Nishamalaki | Phyllanthus emblica | Pyrogallol | HMGCR | Lovastatin | Fatty liver |
| Nishamalaki | Phyllanthus emblica | Catechol | HMGCR | Lovastatin | Fatty liver |
| Nishamalaki | Curcuma longa | Geraniol | HMGCR | Simvastatin | Fatty liver |
| Nishamalaki | Curcuma longa | Safrole | HMGCR | Simvastatin | Fatty liver |
| Nishamalaki | Phyllanthus emblica | Ascorbic acid | HMGCR | Simvastatin | Fatty liver |
| Nishamalaki | Curcuma longa | Eucalyptol | HMGCR | Simvastatin | Fatty liver |
| Nishamalaki | Curcuma longa | Eugenol | HMGCR | Simvastatin | Fatty liver |
| Nishamalaki | Phyllanthus emblica | Pyrogallol | HMGCR | Simvastatin | Fatty liver |
| Nishamalaki | Phyllanthus emblica | Catechol | HMGCR | Simvastatin | Fatty liver |
| Nishamalaki | Curcuma longa | Geraniol | HMGCR | Atorvastatin | Fatty liver |
| Nishamalaki | Curcuma longa | Safrole | HMGCR | Atorvastatin | Fatty liver |
| Nishamalaki | Phyllanthus emblica | Ascorbic acid | HMGCR | Atorvastatin | Fatty liver |
| Nishamalaki | Curcuma longa | Eucalyptol | HMGCR | Atorvastatin | Fatty liver |
| Nishamalaki | Curcuma longa | Eugenol | HMGCR | Atorvastatin | Fatty liver |
| Nishamalaki | Phyllanthus emblica | Pyrogallol | HMGCR | Atorvastatin | Fatty liver |
| Nishamalaki | Phyllanthus emblica | Catechol | HMGCR | Atorvastatin | Fatty liver |
| Nishamalaki | Curcuma longa | Geraniol | HMGCR | Pitavastatin | Fatty liver |
| Nishamalaki | Curcuma longa | Safrole | HMGCR | Pitavastatin | Fatty liver |
| Nishamalaki | Phyllanthus emblica | Ascorbic acid | HMGCR | Pitavastatin | Fatty liver |
| Nishamalaki | Curcuma longa | Eucalyptol | HMGCR | Pitavastatin | Fatty liver |
| Nishamalaki | Curcuma longa | Eugenol | HMGCR | Pitavastatin | Fatty liver |
| Nishamalaki | Phyllanthus emblica | Pyrogallol | HMGCR | Pitavastatin | Fatty liver |
| Nishamalaki | Phyllanthus emblica | Catechol | HMGCR | Pitavastatin | Fatty liver |
| Nishamalaki | Curcuma longa | beta-Caryophyllene | PPARA | Fenofibrate | Fatty liver |
| Nishamalaki | Curcuma longa | Decanoic acid | PPARA | Fenofibrate | Fatty liver |
| Nishamalaki | Curcuma longa | Methyleugenol | PPARA | Fenofibrate | Fatty liver |
| Nishamalaki | Curcuma longa | beta-Caryophyllene | PPARA | Fenofibric acid | Fatty liver |
| Nishamalaki | Curcuma longa | Decanoic acid | PPARA | Fenofibric acid | Fatty liver |
| Nishamalaki | Curcuma longa | Methyleugenol | PPARA | Fenofibric acid | Fatty liver |
| Nishamalaki | Curcuma longa | Safrole | MC4R | Setmelanotide | Obesity |
| Nishamalaki | Curcuma longa | Geraniol | MC4R | Setmelanotide | Obesity |
| Nishamalaki | Phyllanthus emblica | Pyrogallol | MC4R | Setmelanotide | Obesity |
| Nishamalaki | Phyllanthus emblica | Catechol | MC4R | Setmelanotide | Obesity |
| Nishamalaki | Phyllanthus emblica | Ascorbic acid | MC4R | Setmelanotide | Obesity |
| Nishamalaki | Curcuma longa | Eugenol | MC4R | Setmelanotide | Obesity |
| Nishamalaki | Phyllanthus emblica | Quercetin | MC4R | Setmelanotide | Obesity |
| Nishamalaki | Curcuma longa | Eucalyptol | MC4R | Setmelanotide | Obesity |
| Nishamalaki | Phyllanthus emblica | Quercetin | PNLIP | Orlistat | Obesity |
| Nishamalaki | Curcuma longa | Thymol | SLC6A3 | Amphetamine | Obesity |
| Nishamalaki | Phyllanthus emblica | Catechol | SLC6A3 | Amphetamine | Obesity |
| Nishamalaki | Curcuma longa | Safrole | SLC6A3 | Amphetamine | Obesity |
| Nishamalaki | Curcuma longa | Eucalyptol | SLC6A3 | Amphetamine | Obesity |
| Nishamalaki | Curcuma longa | Geraniol | SLC6A3 | Amphetamine | Obesity |
| Nishamalaki | Curcuma longa | Eugenol | SLC6A3 | Amphetamine | Obesity |
| Nishamalaki | Phyllanthus emblica | Ascorbic acid | SLC6A3 | Amphetamine | Obesity |
| Nishamalaki | Phyllanthus emblica | Quercetin | SLC6A3 | Amphetamine | Obesity |
| Nishamalaki | Phyllanthus emblica | Pyrogallol | SLC6A3 | Amphetamine | Obesity |
| Nishamalaki | Curcuma longa | Eucalyptol | SLC6A2 | Amphetamine | Obesity |
| Nishamalaki | Phyllanthus emblica | Pyrogallol | SLC6A2 | Amphetamine | Obesity |
| Nishamalaki | Phyllanthus emblica | Quercetin | SLC6A2 | Amphetamine | Obesity |
| Nishamalaki | Phyllanthus emblica | Ascorbic acid | SLC6A2 | Amphetamine | Obesity |
| Nishamalaki | Curcuma longa | Geraniol | SLC6A2 | Amphetamine | Obesity |
| Nishamalaki | Phyllanthus emblica | Catechol | SLC6A2 | Amphetamine | Obesity |
| Nishamalaki | Curcuma longa | Safrole | SLC6A2 | Amphetamine | Obesity |
| Nishamalaki | Curcuma longa | Eugenol | SLC6A2 | Amphetamine | Obesity |
| Nishamalaki | Curcuma longa | Thymol | SLC6A2 | Amphetamine | Obesity |
| Nishamalaki | Curcuma longa | Thymol | SLC6A3 | Diethylpropion | Obesity |
| Nishamalaki | Phyllanthus emblica | Catechol | SLC6A3 | Diethylpropion | Obesity |
| Nishamalaki | Curcuma longa | Safrole | SLC6A3 | Diethylpropion | Obesity |
| Nishamalaki | Curcuma longa | Eucalyptol | SLC6A3 | Diethylpropion | Obesity |
| Nishamalaki | Curcuma longa | Geraniol | SLC6A3 | Diethylpropion | Obesity |
| Nishamalaki | Curcuma longa | Eugenol | SLC6A3 | Diethylpropion | Obesity |
| Nishamalaki | Phyllanthus emblica | Ascorbic acid | SLC6A3 | Diethylpropion | Obesity |
| Nishamalaki | Phyllanthus emblica | Quercetin | SLC6A3 | Diethylpropion | Obesity |

**Table S14. Formulation – herb – phytochemical – target – FDA-approved drug – disease associations linking formulations to therapeutically relevant drugs. This table presents formulation, herbs, their constituent phytochemicals, associated molecular targets, and corresponding FDA-approved drugs, highlighting target-level overlaps between formulations and FDA-approved drugs**

| Formulation | Herb | Phytochemical | Gene Symbol | Mimicked Drug | Disease |
|---|---|---|---|---|---|
| Nishamalaki | Phyllanthus emblica | Pyrogallol | SLC6A3 | Diethylpropion | Obesity |
| Nishamalaki | Curcuma longa | Eucalyptol | SLC6A2 | Diethylpropion | Obesity |
| Nishamalaki | Phyllanthus emblica | Pyrogallol | SLC6A2 | Diethylpropion | Obesity |
| Nishamalaki | Phyllanthus emblica | Quercetin | SLC6A2 | Diethylpropion | Obesity |
| Nishamalaki | Phyllanthus emblica | Ascorbic acid | SLC6A2 | Diethylpropion | Obesity |
| Nishamalaki | Curcuma longa | Geraniol | SLC6A2 | Diethylpropion | Obesity |
| Nishamalaki | Phyllanthus emblica | Catechol | SLC6A2 | Diethylpropion | Obesity |
| Nishamalaki | Curcuma longa | Safrole | SLC6A2 | Diethylpropion | Obesity |
| Nishamalaki | Curcuma longa | Eugenol | SLC6A2 | Diethylpropion | Obesity |
| Nishamalaki | Curcuma longa | Thymol | SLC6A2 | Diethylpropion | Obesity |
| Nishamalaki | Curcuma longa | Thymol | SLC6A3 | Metamfetamine | Obesity |
| Nishamalaki | Phyllanthus emblica | Catechol | SLC6A3 | Metamfetamine | Obesity |
| Nishamalaki | Curcuma longa | Safrole | SLC6A3 | Metamfetamine | Obesity |
| Nishamalaki | Curcuma longa | Eucalyptol | SLC6A3 | Metamfetamine | Obesity |
| Nishamalaki | Curcuma longa | Geraniol | SLC6A3 | Metamfetamine | Obesity |
| Nishamalaki | Curcuma longa | Eugenol | SLC6A3 | Metamfetamine | Obesity |
| Nishamalaki | Phyllanthus emblica | Ascorbic acid | SLC6A3 | Metamfetamine | Obesity |
| Nishamalaki | Phyllanthus emblica | Quercetin | SLC6A3 | Metamfetamine | Obesity |
| Nishamalaki | Phyllanthus emblica | Pyrogallol | SLC6A3 | Metamfetamine | Obesity |
| Nishamalaki | Curcuma longa | Thymol | SLC6A3 | Phendimetrazine | Obesity |
| Nishamalaki | Phyllanthus emblica | Catechol | SLC6A3 | Phendimetrazine | Obesity |
| Nishamalaki | Curcuma longa | Safrole | SLC6A3 | Phendimetrazine | Obesity |
| Nishamalaki | Curcuma longa | Eucalyptol | SLC6A3 | Phendimetrazine | Obesity |
| Nishamalaki | Curcuma longa | Geraniol | SLC6A3 | Phendimetrazine | Obesity |
| Nishamalaki | Curcuma longa | Eugenol | SLC6A3 | Phendimetrazine | Obesity |
| Nishamalaki | Phyllanthus emblica | Ascorbic acid | SLC6A3 | Phendimetrazine | Obesity |
| Nishamalaki | Phyllanthus emblica | Quercetin | SLC6A3 | Phendimetrazine | Obesity |
| Nishamalaki | Phyllanthus emblica | Pyrogallol | SLC6A3 | Phendimetrazine | Obesity |
| Nishamalaki | Curcuma longa | Eucalyptol | SLC6A2 | Phendimetrazine | Obesity |
| Nishamalaki | Phyllanthus emblica | Pyrogallol | SLC6A2 | Phendimetrazine | Obesity |
| Nishamalaki | Phyllanthus emblica | Quercetin | SLC6A2 | Phendimetrazine | Obesity |
| Nishamalaki | Phyllanthus emblica | Ascorbic acid | SLC6A2 | Phendimetrazine | Obesity |
| Nishamalaki | Curcuma longa | Geraniol | SLC6A2 | Phendimetrazine | Obesity |
| Nishamalaki | Phyllanthus emblica | Catechol | SLC6A2 | Phendimetrazine | Obesity |
| Nishamalaki | Curcuma longa | Safrole | SLC6A2 | Phendimetrazine | Obesity |
| Nishamalaki | Curcuma longa | Eugenol | SLC6A2 | Phendimetrazine | Obesity |
| Nishamalaki | Curcuma longa | Thymol | SLC6A2 | Phendimetrazine | Obesity |
| Nishamalaki | Curcuma longa | Eucalyptol | SLC6A2 | Benzphetamine | Obesity |
| Nishamalaki | Phyllanthus emblica | Pyrogallol | SLC6A2 | Benzphetamine | Obesity |
| Nishamalaki | Phyllanthus emblica | Quercetin | SLC6A2 | Benzphetamine | Obesity |
| Nishamalaki | Phyllanthus emblica | Ascorbic acid | SLC6A2 | Benzphetamine | Obesity |
| Nishamalaki | Curcuma longa | Geraniol | SLC6A2 | Benzphetamine | Obesity |
| Nishamalaki | Phyllanthus emblica | Catechol | SLC6A2 | Benzphetamine | Obesity |
| Nishamalaki | Curcuma longa | Safrole | SLC6A2 | Benzphetamine | Obesity |
| Nishamalaki | Curcuma longa | Eugenol | SLC6A2 | Benzphetamine | Obesity |
| Nishamalaki | Curcuma longa | Thymol | SLC6A2 | Benzphetamine | Obesity |
| Nishamalaki | Curcuma longa | Eucalyptol | SLC6A2 | Phentermine | Obesity |
| Nishamalaki | Phyllanthus emblica | Pyrogallol | SLC6A2 | Phentermine | Obesity |
| Nishamalaki | Phyllanthus emblica | Quercetin | SLC6A2 | Phentermine | Obesity |
| Nishamalaki | Phyllanthus emblica | Ascorbic acid | SLC6A2 | Phentermine | Obesity |
| Nishamalaki | Curcuma longa | Geraniol | SLC6A2 | Phentermine | Obesity |
| Nishamalaki | Phyllanthus emblica | Catechol | SLC6A2 | Phentermine | Obesity |
| Nishamalaki | Curcuma longa | Safrole | SLC6A2 | Phentermine | Obesity |
| Nishamalaki | Curcuma longa | Eugenol | SLC6A2 | Phentermine | Obesity |
| Nishamalaki | Curcuma longa | Thymol | SLC6A2 | Phentermine | Obesity |

**Table S15: Identification top hub proteins across formulation-disease combinations using network pharmacology analysis. Each sub-table corresponds to a specific formulation–disease pair and reports the top ten proteins identified by five different network centrality measures.**

| **Arogyavardhini-Type 2 diabetes** | | | | | | | | | | | |
|---|---|---|---|---|---|---|---|---|---|---|---|
| MCC | | MNC | | Degree | | Closeness centrality | | Betweenness centralit y | | Consensus | |
| Gene symbol | Entrez Gene ID | Gene symbol | Entrez Gene ID | Gene symbol | Entrez Gene ID | Gene symbol | Entrez Gene ID | Gene symbol | Entrez Gene ID | Gene symbol | Entrez Gene ID |
| PPARA | 5465 | PPARA | 5465 | PPARA | 5465 | PPARA | 5465 | PPARA | 5465 | AKT1 | 207 |
| PPARG | 5468 | PPARG | 5468 | PPARG | 5468 | PPARG | 5468 | PPARG | 5468 | ALB | 213 |
| ALB | 213 | ALB | 213 | ALB | 213 | ALB | 213 | ALB | 213 | NFKB1 | 4790 |
| EGFR | 1956 | EGFR | 1956 | EGFR | 1956 | EGFR | 1956 | MGAM | 8972 | PPARA | 5465 |
| STAT3 | 6774 | STAT3 | 6774 | STAT3 | 6774 | STAT3 | 6774 | ACE | 1636 | PPARG | 5468 |
| MTOR | 2475 | MTOR | 2475 | MTOR | 2475 | ACE | 1636 | AKT1 | 207 | SIRT1 | 23411 |
| AKT1 | 207 | AKT1 | 207 | AKT1 | 207 | AKT1 | 207 | GLP1R | 2740 | TNF | 7124 |
| SIRT1 | 23411 | SIRT1 | 23411 | SIRT1 | 23411 | SIRT1 | 23411 | SIRT1 | 23411 | | |
| NFKB1 | 4790 | NFKB1 | 4790 | NFKB1 | 4790 | NFKB1 | 4790 | NFKB1 | 4790 | | |
| TNF | 7124 | TNF | 7124 | TNF | 7124 | TNF | 7124 | TNF | 7124 | | |
| Arogyavardhini-Fatty liver | | | | | | | | | | | |
| MCC | | MNC | | Degree | | Closeness centrality | | Betweenness centralit y | | Consensus | |
| Gene symbol | Entrez Gene ID | Gene symbol | Entrez Gene ID | Gene symbol | Entrez Gene ID | Gene symbol | Entrez Gene ID | Gene symbol | Entrez Gene ID | Gene symbol | Entrez Gene ID |
| HMGCR | 3156 | PPARG | 5468 | PPARG | 5468 | PPARG | 5468 | PPARG | 5468 | AKT1 | 207 |
| PPARG | 5468 | ESR1 | 2099 | ESR1 | 2099 | CYP2E1 | 1571 | CYP2E1 | 1571 | ESR1 | 2099 |
| SERPINE1 | 5054 | SIRT1 | 23411 | SIRT1 | 23411 | ESR1 | 2099 | ESR1 | 2099 | GPT | 2875 |
| ESR1 | 2099 | NR1H4 | 9971 | NR1H4 | 9971 | SIRT1 | 23411 | SIRT1 | 23411 | PPARA | 5465 |
| SIRT1 | 23411 | NFE2L2 | 4780 | NFE2L2 | 4780 | NR1H4 | 9971 | NFE2L2 | 4780 | PPARG | 5468 |
| ACE | 1636 | AKT1 | 207 | AKT1 | 207 | NFE2L2 | 4780 | AKT1 | 207 | SIRT1 | 23411 |
| AKT1 | 207 | PPARA | 5465 | PPARA | 5465 | AKT1 | 207 | PPARA | 5465 | TNF | 7124 |
| PPARA | 5465 | GPT | 2875 | GPT | 2875 | PPARA | 5465 | GPT | 2875 | | |
| GPT | 2875 | CYP1A2 | 1544 | CYP1A2 | 1544 | GPT | 2875 | CYP1A2 | 1544 | | |
| TNF | 7124 | TNF | 7124 | TNF | 7124 | TNF | 7124 | TNF | 7124 | | |
| Arogyavardhini-Obesity | | | | | | | | | | | |
| MCC | | MNC | | Degree | | Closeness centrality | | Betweenness centralit y | | Consensus | |
| Gene symbol | Entrez Gene ID | Gene symbol | Entrez Gene ID | Gene symbol | Entrez Gene ID | Gene symbol | Entrez Gene ID | Gene symbol | Entrez Gene ID | Gene symbol | Entrez Gene ID |
| PPARA | 5465 | PPARA | 5465 | PPARA | 5465 | PPARA | 5465 | TNF | 7124 | AKT1 | 207 |
| STAT3 | 6774 | ACE | 1636 | ACE | 1636 | ACE | 1636 | PTGS2 | 5743 | PPARA | 5465 |
| PTGS2 | 5743 | STAT3 | 6774 | STAT3 | 6774 | STAT3 | 6774 | PPARG | 5468 | PPARG | 5468 |
| SIRT1 | 23411 | PTGS2 | 5743 | NR3C1 | 2908 | NR3C1 | 2908 | PPARA | 5465 | PTGS2 | 5743 |
| TNF | 7124 | SIRT1 | 23411 | PTGS2 | 5743 | PTGS2 | 5743 | NR3C1 | 2908 | TNF | 7124 |
| AKT1 | 207 | TNF | 7124 | SIRT1 | 23411 | SIRT1 | 23411 | NR1I2 | 8856 | | |
| HSPA5 | 3309 | AKT1 | 207 | TNF | 7124 | TNF | 7124 | GLP1R | 2740 | | |
| MMP9 | 4318 | MMP9 | 4318 | AKT1 | 207 | AKT1 | 207 | F2 | 2147 | | |
| PPARG | 5468 | PPARG | 5468 | PPARG | 5468 | PPARG | 5468 | AKT1 | 207 | | |
| ESR1 | 2099 | ESR1 | 2099 | ESR1 | 2099 | ESR1 | 2099 | ACE | 1636 | | |

| **Nishamalaki-Type 2 diabetes** | | | | | | | | | | | |
|---|---|---|---|---|---|---|---|---|---|---|---|
| MCC | | MNC | | Degree | | Closeness centrality | | Betweenness centralit y | | Consensus | |
| Gene symbol | Entrez Gene ID | Gene symbol | Entrez Gene ID | Gene symbol | Entrez Gene ID | Gene symbol | Entrez Gene ID | Gene symbol | Entrez Gene ID | Gene symbol | Entrez Gene ID |
| MTOR | 2475 | SIRT1 | 23411 | SIRT1 | 23411 | SIRT1 | 23411 | SIRT1 | 23411 | ACE | 1636 |
| SIRT1 | 23411 | AKT1 | 207 | AKT1 | 207 | AKT1 | 207 | ATP2A2 | 488 | AKT1 | 207 |
| AKT1 | 207 | EGFR | 1956 | EGFR | 1956 | EGFR | 1956 | AKT1 | 207 | ALB | 213 |
| NFKB1 | 4790 | NFKB1 | 4790 | NFKB1 | 4790 | NFKB1 | 4790 | EGFR | 1956 | PPARA | 5465 |
| TNF | 7124 | TNF | 7124 | TNF | 7124 | TNF | 7124 | MGAM | 8972 | PPARG | 5468 |
| ACE | 1636 | ACE | 1636 | ACE | 1636 | ACE | 1636 | TNF | 7124 | SIRT1 | 23411 |
| PPARA | 5465 | PPARA | 5465 | PPARA | 5465 | PPARA | 5465 | ACE | 1636 | TNF | 7124 |
| ALB | 213 | ALB | 213 | ALB | 213 | ALB | 213 | PPARA | 5465 | | |
| PPARG | 5468 | PPARG | 5468 | PPARG | 5468 | PPARG | 5468 | ALB | 213 | | |
| STAT3 | 6774 | STAT3 | 6774 | STAT3 | 6774 | STAT3 | 6774 | PPARG | 5468 | | |
| Nishamalaki-Fatty liver | | | | | | | | | | | |
| MCC | | MNC | | Degree | | Closeness centrality | | Betweenness centralit y | | Consensus | |
| Gene symbol | Entrez Gene ID | Gene symbol | Entrez Gene ID | Gene symbol | Entrez Gene ID | Gene symbol | Entrez Gene ID | Gene symbol | Entrez Gene ID | Gene symbol | Entrez Gene ID |
| TNF | 7124 | TNF | 7124 | TNF | 7124 | TNF | 7124 | TNF | 7124 | AKT1 | 207 |
| SERPINE1 | 5054 | PPARG | 5468 | PPARG | 5468 | PPARG | 5468 | PPARG | 5468 | ESR1 | 2099 |
| PPARG | 5468 | CYP1A2 | 1544 | CYP1A2 | 1544 | ESR1 | 2099 | CYP1A2 | 1544 | NR1H4 | 9971 |
| ESR1 | 2099 | ESR1 | 2099 | ESR1 | 2099 | SIRT1 | 23411 | ESR1 | 2099 | PPARA | 5465 |
| ACE | 1636 | SIRT1 | 23411 | SIRT1 | 23411 | AKT1 | 207 | SIRT1 | 23411 | PPARG | 5468 |
| SIRT1 | 23411 | AKT1 | 207 | AKT1 | 207 | CYP2E1 | 1571 | AKT1 | 207 | SIRT1 | 23411 |
| AKT1 | 207 | NR1H4 | 9971 | NR1H4 | 9971 | NR1H4 | 9971 | CYP2E1 | 1571 | TNF | 7124 |
| NR1H4 | 9971 | NFE2L2 | 4780 | NFE2L2 | 4780 | NFE2L2 | 4780 | NR1H4 | 9971 | | |
| PPARA | 5465 | PPARA | 5465 | PPARA | 5465 | PPARA | 5465 | NFE2L2 | 4780 | | |
| HMGCR | 3156 | HMGCR | 3156 | HMGCR | 3156 | HMGCR | 3156 | PPARA | 5465 | | |
| Nishamalaki-Obesity | | | | | | | | | | | |
| MCC | | MNC | | Degree | | Closeness centrality | | Betweenness centralit y | | Consensus | |
| Gene symbol | Entrez Gene ID | Gene symbol | Entrez Gene ID | Gene symbol | Entrez Gene ID | Gene symbol | Entrez Gene ID | Gene symbol | Entrez Gene ID | Gene symbol | Entrez Gene ID |
| STAT3 | 6774 | STAT3 | 6774 | STAT3 | 6774 | STAT3 | 6774 | MC4R | 4160 | AKT1 | 207 |
| PTGS2 | 5743 | ACE | 1636 | ACE | 1636 | ACE | 1636 | ACE | 1636 | ESR1 | 2099 |
| TNF | 7124 | PTGS2 | 5743 | PTGS2 | 5743 | PTGS2 | 5743 | PTGS2 | 5743 | PPARA | 5465 |
| AKT1 | 207 | TNF | 7124 | TNF | 7124 | TNF | 7124 | EP300 | 2033 | PPARG | 5468 |
| HSPA5 | 3309 | AKT1 | 207 | AKT1 | 207 | AKT1 | 207 | TNF | 7124 | PTGS2 | 5743 |
| MMP9 | 4318 | PPARG | 5468 | PPARG | 5468 | PPARG | 5468 | AKT1 | 207 | TNF | 7124 |
| PPARG | 5468 | SIRT1 | 23411 | SIRT1 | 23411 | SIRT1 | 23411 | PPARG | 5468 | | |
| SIRT1 | 23411 | ESR1 | 2099 | ESR1 | 2099 | ESR1 | 2099 | ESR1 | 2099 | | |
| ESR1 | 2099 | PPARA | 5465 | PPARA | 5465 | PPARA | 5465 | PPARA | 5465 | | |
| PPARA | 5465 | NR3C1 | 2908 | NR3C1 | 2908 | NR3C1 | 2908 | NR3C1 | 2908 | | |

**Table S16: Recurrence of consensus hub proteins across formulation-disease networks identified through network pharmacology analyses. The table lists proteins that appeared as consensus hubs in one or more formulation-disease networks, where consensus hubs were defined as proteins consistently ranked among the top ten nodes by all five network centrality measures applied in this study. For each protein, the table reports the corresponding formulations and disease conditions in which the protein was identified as a hub.**

| Gene Symbol | Entrez Gene ID | Formulations | Diseases |
|---|---|---|---|
| PPARA | 5465 | Arogyavardhini, Nishamalaki | Type 2 diabetes, Fatty liver, Obesity |
| PPARG | 5468 | Arogyavardhini, Nishamalaki | Type 2 diabetes, Fatty liver, Obesity |
| AKT1 | 207 | Arogyavardhini, Nishamalaki | Type 2 diabetes, Fatty liver, Obesity |
| TNF | 7124 | Arogyavardhini, Nishamalaki | Type 2 diabetes, Fatty liver, Obesity |
| SIRT1 | 23411 | Arogyavardhini, Nishamalaki | Type 2 diabetes, Fatty liver |
| ESR1 | 2099 | Arogyavardhini, Nishamalaki | Fatty liver, Obesity (Nishamalaki) |
| ALB | 213 | Arogyavardhini, Nishamalaki | Type 2 diabetes |
| NFKB1 | 4790 | Arogyavardhini | Type 2 diabetes |
| ACE | 1636 | Nishamalaki | Type 2 diabetes |
| GPT | 2875 | Arogyavardhini | Fatty liver |
| NR1H4 | 9971 | Nishamalaki | Fatty liver |
| PTGS2 | 5743 | Arogyavardhini, Nishamalaki | Obesity |

| Group | n | <5% | 5–33% | 34–66% | >66% | <5% | 5–33% | 34–66% | >66% |
|---|---|---|---|---|---|---|---|---|---|
| Control | 4 | 3 (75%) | 1 (25%) | 0 (0%) | 0 (0%) | 4 (100%) | 0 (0%) | 0 (0%) | 0 (0%) |
| HFD | 4 | 0 (0%) | 1 (33%) | 2 (67%) | 0 (0%) | 0 (0%) | 3 (100%) | 0 (0%) | 0 (0%) |
| HFD+AVR | 4 | 4 (100%) | 0 (0%) | 0 (0%) | 0 (0%) | 4 (100%) | 0 (0%) | 0 (0%) | 0 (0%) |
| HFD+NA | 4 | 2 (50%) | 2 (50%) | 0 (0%) | 0 (0%) | 3 (75%) | 1 (25%) | 0 (0%) | 0 (0%) |